\documentclass[12pt]{article}
\pdfoutput = 1
\usepackage{graphicx}
\usepackage{adjustbox}
\usepackage{fancybox}
\usepackage{colortbl}
\usepackage{dcolumn}
\usepackage{multirow}
\usepackage{multicol}
\usepackage{amssymb}
\usepackage{amsthm}
\usepackage{amsfonts,amsmath}
\usepackage{appendix}
\usepackage{tikz} 
\usetikzlibrary{fit,positioning}
\usepackage[section]{placeins}
\usepackage{cite}
\usepackage{lineno}
\usepackage{caption}
\usepackage{subcaption}
\usepackage{float}
\usepackage{graphicx}
\usepackage{hyperref}
\usepackage{colortbl}
\usepackage{authblk}
\usepackage{array}
\usepackage{adjustbox}
\usepackage{cite}
\usepackage{multirow}
\usepackage{nameref,hyperref}
\usepackage{multicol}

\title{Wavelet Analysis of Dengue Incidence and its Correlation with Weather and Vegetation Variables in Costa Rica}

\date{March 2020}

\author[1,3,*, +]{Yury E. Garc\'ia}
\author[2, +]{Luis A. Barboza}
\author[2, +]{Fabio Sanchez}
\author[1, +]{Paola V\'asquez}
\author[2, +]{Juan G. Calvo}

\affil[1]{Universidad de Costa Rica, Centro de Investigaci\'on en Matem\'atica Pura y Aplicada, San Jos\'e, Costa Rica}

\affil[1]{Universidad de Costa Rica, Centro de Investigaci\'on en Matem\'atica Pura y Aplicada - Escuela de Matem\'atica, San Jos\'e, Costa Rica}

\affil[3]{Department of Public Health Sciences, 
University of California Davis, CA, USA}

\affil[*]{ygarciapuerta@ucdavis.edu}

\affil[+]{these authors contributed equally to this work}

\date{}
\begin{document}

\maketitle
\begin{abstract}
Dengue  represents a serious public health problem in tropical and subtropical regions worldwide. The number of dengue cases and its geographical expansion has increased in recent decades, driven mostly after by social and environmental factors. In Costa Rica, it has been endemic since it was first introduced in 1993. In this article, wavelet analyzes (wavelet power spectrum and wavelet coherence) were performed to detect and quantify dengue periodicity and describe patterns of synchrony between dengue incidence and climatic and environmental factors: Normalized Difference Water Index, Enhanced Vegetation Index, Normalized Difference Vegetation Index, Tropical North Atlantic indices, Land Surface Temperature, and El Niño Southern Oscillation indices in 32 different cantons, using dengue surveillance from 2000 to 2019.  Results showed that the dengue dominant cycles  are in periods of 1, 2, and 3 years. The wavelet coherence analysis showed that the vegetation indices are correlated with dengue incidence in places located in the central and Northern Pacific of the country in the period of 1 year. Climatic variables such as El Ni\~no 3, 3.4, 4, showed a strong correlation with dengue incidence in the period of 3 years and the Tropical North Atlantic is correlated with dengue incidence in the period of 1 year. Land Surface Temperature showed a strong correlation with dengue time series in the 32 cantons. 
\end{abstract}

\section{Introduction}

Dengue is a mosquito-borne viral infection caused by four antigenically distinct dengue virus serotypes (DENV-1, DENV-2, DENV-3, DENV-4) and is transmitted by the bite of an infected female mosquito \textit{Aedes aegypti} as the main vector, and which is more common in rural areas, with \textit{Aedes albopictus} as a secondary vector. Dengue is a flu-like illness that affects individuals of all ages that could cause significant health, economic, and social burden in a population \cite{shepard2011economic}. The clinical profile of patients can range from asymptomatic infection to severe cases.

In recent years, the complex interaction of biological, socioeconomic, environmental, and climatic factors has facilitated the rapid emergence of this viral infection throughout the world, becoming endemic and a relevant public health problem in more than 100 countries \cite{zeng2021global}. In the last decades, the number of dengue cases reported to the World Health Organization (WHO) has increased from 505,430 cases in 2000 to more than 4.2 million in 2019  \cite{DengueanWHO,zeng2021global}.

Seasonal case patterns and vector abundance suggest that dengue transmission is sensitive to climatic and environmental factors \cite{morin2013climate, kolivras2010changes,christophrs1960aedes}. Climatic conditions can alter spatial and temporal dynamics of vector ecology, potentially increasing vector ranges, lengthening the duration of vector activity, and increasing the mosquito’s infectious period \cite{morin2013climate}. Precipitation provides habitats for the aquatic stages of the mosquito life cycle and strongly influences vector distribution \cite{kolivras2010changes, christophrs1960aedes, tun477effects}. On the other hand, water temperature plays a significant role in mosquito reproduction since this directly affects it survival at all stages of their life cycle \cite{ebi2016dengue}. Temperature increases are also linked with a faster rate of viral replication within the vector and with a shorter extrinsic incubation period \cite{morin2013climate}. Furthermore, higher humidity is associated with an increase in \textit{Ae. aegypti} feeding activity, which enhances the spread of the diseases.

The complexity of dengue transmission has driven many studies to assess its correlation with meteorological and ecological variables \cite{prabodanie2020coherence,ehelepola2015study,beltran2014spatiotemporal, cuong2016quantifying,thai2010dengue,johansson2009multiyear,cazelles2005nonstationary,simoes2013modeling,jury2008climate}. Most of these works evaluated the effects and correlation between dengue incidence and climate variables \cite{nitatpattana2007potential,nakhapakorn2020assessment}. There are other climate factors, like seasonal vegetation dynamics which may also influence the biology of the vector populations at relatively local scales affecting the dengue dynamics \cite{chaves2021modeling,mudele2021modeling,troyo2009urban}. Barrera et al. \cite{barrera2006ecological} suggested that dense vegetation can promote \textit{Ae. aegypti} pupal productivity by contributing organic material to the habitat, and influence water temperature and evaporation by creating shades. All these results show a correlation between dengue incidence and various climatic and environmental variables. 

This study focuses on Costa Rica, where, after intense prevention and control campaigns, the mosquito \textit{Ae. aegypti} was eradicated in 1961. In  1970, the lack of continuity in active surveillance caused the mosquito to be found again on the Pacific coast. By September 1993, the first dengue case was reported in the country, and since then, dengue infections have been documented annually.  The patterns and periodicity of transmission are different across the country. Despite the coastal regions being the most affected areas, trends observed over the years show variations in transmission peaks in all affected areas, posing a challenge for public health authorities to allocate and optimize available resources. In 2019, Costa Rica reported three times more cases than the previous year \cite{PAHO1}. The number of reported cases in 2019 is the highest in the history of dengue in Latin America \cite{PAHO1}.


This work aimed to study the incidence patterns of dengue in 32 different municipalities of Costa Rica and its correlation with climatic and vegetation variables using wavelet analysis. This method decomposes a time series in both the time and frequency domains, revealing how different periods change over time into non-stationary signals. Furthermore, wavelet analysis allows conclusions to be drawn about the synchronicity of the series for some periods. First, the wavelet power spectrum is computed to identify the periodicity present in the time series of dengue incidence. Then, using coherence analysis, we quantify associations between monthly dengue cases and local meteorological covariates, and examine the decomposed frequency components to identify possible time delays (phase differences) and significant associations. 

Epidemic data is typically noisy, complex, and non-stationary. Changes in the periodicity over time are due to external factors or inherent characteristics of the disease. It is important to understand the mechanistic processes that lead to the spread of dengue and the external factors with which they can relate. Wavelet analysis allow a retrospective study to characterize outbreaks over time, which provides important guidelines for future modeling approaches in which explicit mechanisms can be incorporated.

This article is organized as follows: Section \ref{Methods}; describes the data and methods, Section \ref{Results}; present the results, and Section \ref{discussion} gives our final thoughts and discussion.

\section{Methods}
\label{Methods}

\subsection{Data}

For this study, we used data for 32 different locations, from January 2000 to December 2019. This data corresponds to weekly dengue incidence records provided by the Ministry of Health of Costa Rica. The data is aggregated monthly and is square-root transformed and standardize \cite{cazelles2005nonstationary}.

Costa Rica is a mountainous country located between the Caribbean Sea and Pacific Ocean. Thus, we use indices related with both ocean anomalies. El Niño Southern Oscillation (ENSO) is a natural phenomenon in the ocean-atmospheric system over the tropical Pacific; it is responsible for some anomalies such as changes in the sea surface temperatures (SST) \cite{hanley2003quantitative}. For this research, El Niño 1+2, El Niño 3, El Niño 4, and El Niño 3.4 indices \cite{CDG} were used. Data were accessed at [\url{https://www.cpc.ncep.noaa.gov/data/indices/ersst5.nino.mth.81-10.ascii}]. Information related to anomalies in the Atlantic Ocean is measured with the TNA index, abbreviation of Tropical North Atlantic, provides information on sea surface temperatures in the North-Eastern Tropical Atlantic Ocean \cite{TNA,wu2002tropical}. Data are available at [\url{https://www.esrl.noaa.gov/psd/data/correlation/tna.data}].

Normalized Difference Water Index (NDWI) represents the changes in the liquid water content of the leaves, giving a picture of the water stress of the vegetationr \cite{parselia2019satellite,gao1996ndwi,NDWI}. The time series of the Enhanced Vegetation Index (EVI) and the Normalized Difference Vegetation Index (NDVI) provide spatial and temporal information about vegetation \cite{GisGeography}. The NDVI has been widely used for remote sensing of vegetation and has been applied to mosquito studies and vector-borne diseases \cite{troyo2009urban,estallo2012effectiveness,lacaux2007classification}. EVI responds to canopy structural variations, including leaf area index (LAI), canopy type, plant physiognomy, and canopy architecture, and NDVI is more sensitive to chlorophyll. The two vegetation indices complement each other. \cite{matsushita2007sensitivity}. Land Surface Temperature (LST) is defined as a measure of how hot the surface of the Earth would feel to the touch \cite{yu2018land}. These last indices where download using the R package MODIS Tools, which extracts the information from MODIS Land Products Subsets [\url{https://modis.ornl.gov/data/modis_webservice.html}]. The  MOD13Q1 product was used to EVI, NDVI, and NDVI data, and the MOD11A2 product to the night and day LST data.

\subsection{Wavelets analysis}


Wavelets have been used to study time series with different purposes: to evaluate the main characteristics of non-stationary time series \cite{talagala2015wavelet, cazelles2005nonstationary,chaves2006climate, nagao2008decreases,johansson2009multiyear,cuong2013spatiotemporal,simoes2013modeling}, to analyze spatial patterns \cite{cazelles2005nonstationary,thai2010dengue}, to study the relationship between population and environmental time series; finding the phase and/or synchrony patterns \cite{chaves2006climate,johansson2009multiyear,cazelles2005nonstationary}, and to study multiple time series \cite{d2012wavelets,aghabozorgi2015time}. Here, we used wavelet analysis to study patterns in dengue incidence in Costa Rica and identify its correlation with specific climate and vegetation variables in 32 different locations.

\subsubsection{Wavelet power spectra}

The wavelet analysis is based on a wavelet transform 
defined as
\begin{equation}
W_x(s,\tau)=\dfrac{1}{\sqrt{s}}\int_{-\infty}^{\infty}x(t)\Psi^*\left(\dfrac{t-\tau}{s}\right)dt  
=\int_{-\infty}^{\infty}x(t)\Psi^{*}_{s,\tau}(t)dt
\end{equation}    
where $*$ denotes the complex conjugate form and $\Psi_{s,\tau}(t)$ represent a family of functions derived from a single function called the ``mother wavelet''. The signal is decomposed in these functions which can be expressed in terms of two parameters, one for the time position $\tau$, and the other for the scale of the wavelets $s$.  

\begin{equation}
\Psi_{s,\tau}(t) = \dfrac{1}{\sqrt{s}}\Psi\left(\dfrac{t-\tau}{s}\right)    
\end{equation}

For this study we use the Morlet mother
wavelet \cite{cazelles2007time, cazelles2008wavelet}:

\begin{equation}
\Psi(t) = \pi^{-1/4}e^{i\omega t}e^{\frac{-t^2}{2}}
\end{equation}
The local amplitude of any periodic component of the time series and how it evolves with time, can be retrieved from the modulus of its wavelet transform \cite{rosch2016waveletcomp}.

\begin{equation}
Ampl(s,\tau) = \dfrac{1}{s^{1/2}}|W(s,\tau)|    
\end{equation}

The square of the amplitude has an interpretation as time-frequency (or time-period) called the wavelet power spectrum \cite{carmona1998practical}.

\begin{equation}
Power(s,\tau) = \dfrac{1}{s}|W(s,\tau)|^2    
\end{equation}

\subsubsection{Wavelet coherence and phase difference}
To quantify the time series synchronization between dengue and the different climate and vegetation variables, we compute the wavelet coherence given by

\begin{equation}\
R_{x,y}(s,\tau) =\dfrac{|<W_{x,y}(s,\tau)>|^2}{|<W_x(s,\tau)>|^2|<W_y(s,\tau)>|^2}    
\end{equation}

The angle brackets indicate smoothing in both time and frequency, $W_x(s, \tau) $ and $W_y(s,\tau)$ are the wavelet transform of the series $x(t)$ and $y(t)$, respectively. $W_{x,y}(s,\tau)=W_x(s,\tau)W_y^*(s,\tau)$ is the cross-wavelet transform \cite{cazelles2005nonstationary}. The value of $R_{x,y}(s,\tau)$ range between 0 and 1, where 1 represent a perfect linear relationship between the time series $x(t)$ and $y(t)$. Additional to this analysis, it is possible to compute the phase difference associated to the two signals, which gives information about series synchronization (i.e., in phase or out of phase). The Morlet wavelet is a complex wavelet, so the phase difference can be computed in terms of the real ($\mathcal{R}$) and the imaginary ($\mathcal{I}$) part, as shown in (\ref{eq:phase})

\begin{equation}
\Phi_{x,y}(s,\tau) = \dfrac{\mathcal{I}(<W_{x,y}(s,\tau)>)}{\mathcal{R}(<W_{x,y}(s,\tau)>)}
\label{eq:phase}
\end{equation}

The instantaneous time lag between the time series $x(t)$ and $y(t)$ is also computed \cite{cazelles2005nonstationary}.

Furthermore, Torrence and Compo \cite{torrence1998practical} offer a description of wavelet analysis and Cazellez et al. offer a perspective of the use of these techniques in ecological \cite{cazelles2008wavelet, cazelles2007time} or epidemic scenarios \cite{cazelles2014wavelet}.

All of the computations were done by using R version 2.4 (\url{ www.R-project.org}) \cite{team2013r} and the wavelet time series analyses were done with the algorithms implemented in the R package \textit{WaveletComp version 1.1} \cite{rosch2016waveletcomp}. All significance levels were based on 1000 bootstrap series and, to test the null hypothesis of \textit{non-periodicity}, significance is evaluated with simulation algorithms in which surrogate time series are provided with a similar spectrum (AR time series). \cite{rosch2016waveletcomp,roesch2014package}.

\section{Results}
\label{Results}

\subsection{Wavelet analysis – dengue periodicity}

\begin{figure}
\captionsetup[subfigure]{labelformat=empty}
\includegraphics[scale=0.9]{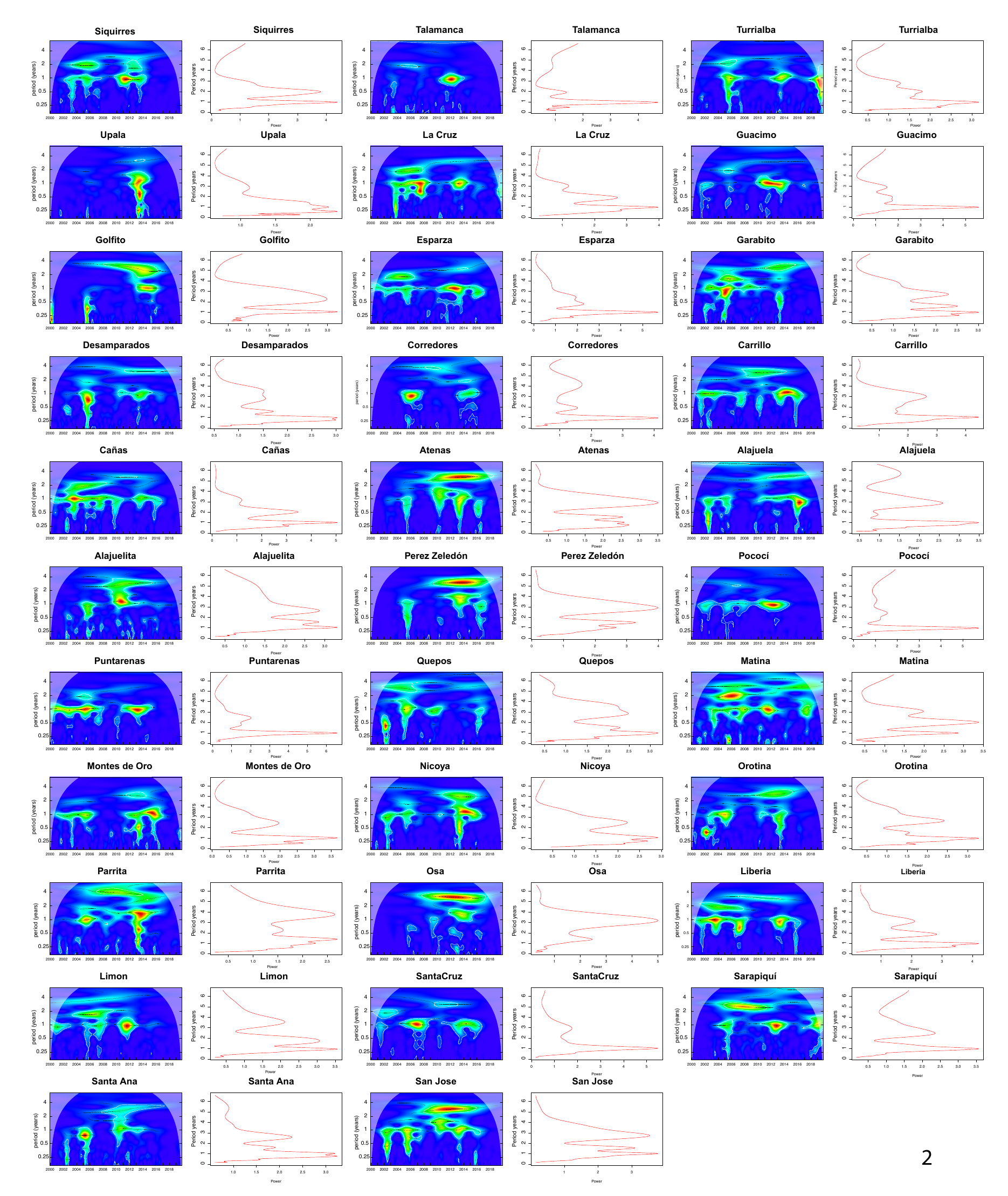}
\caption{Wavelet power spectra (color map) of dengue incidence from 2000 to 2019, in 32 municipalities of Costa Rica (periodicity on y-axis, time on x-axis). Data to compute the wavelet power spectrum of dengue cases is square rooted and standardize; colors code for increasing power intensity, from blue to red; $95\%$ confidence levels are encircled by white lines, and shaded areas indicate the presence of significant edge effects. On the right side of each wavelet power spectra is the classical Fourier spectrum of the time series.}
\label{fig:WaveletCanton}
\end{figure}

In general, wavelet time series analysis revealed that the dominant periods for the incidence of dengue, in the 32 municipalities, are in the bands of 1, 2, and 3-yr, in accord with the periodicity range of dengue in Vietnam \cite{thai2010dengue}, Per\'u \cite{chowell2011influence} and Thailand \cite{cummings2004travelling, cazelles2005nonstationary}. Periods of 4 and 6-yr are observed in Siquirres, Talamanca, Turrialba, Upala, Golfito, Alajuela, Puntarenas, Matina, Osa, Santa Ana, Lim\'on, and Atenas, but this must be interpreted cautiously because of the short length of the time series. These periods are not stationary, they vary over time and space. Before 2008, the cycles of 1 and 2-yr leading in most locations. But, after 2008, the prevalent dengue periodicity change to cycles of 1 and 3-yr (see Figure \ref{fig:WaveletCanton}). The years 2008-2010 mark a decrease period in the variability of dengue incidence. 

\begin{figure}
\captionsetup[subfigure]{labelformat=empty}
\subfloat[]{\includegraphics[scale=0.2]{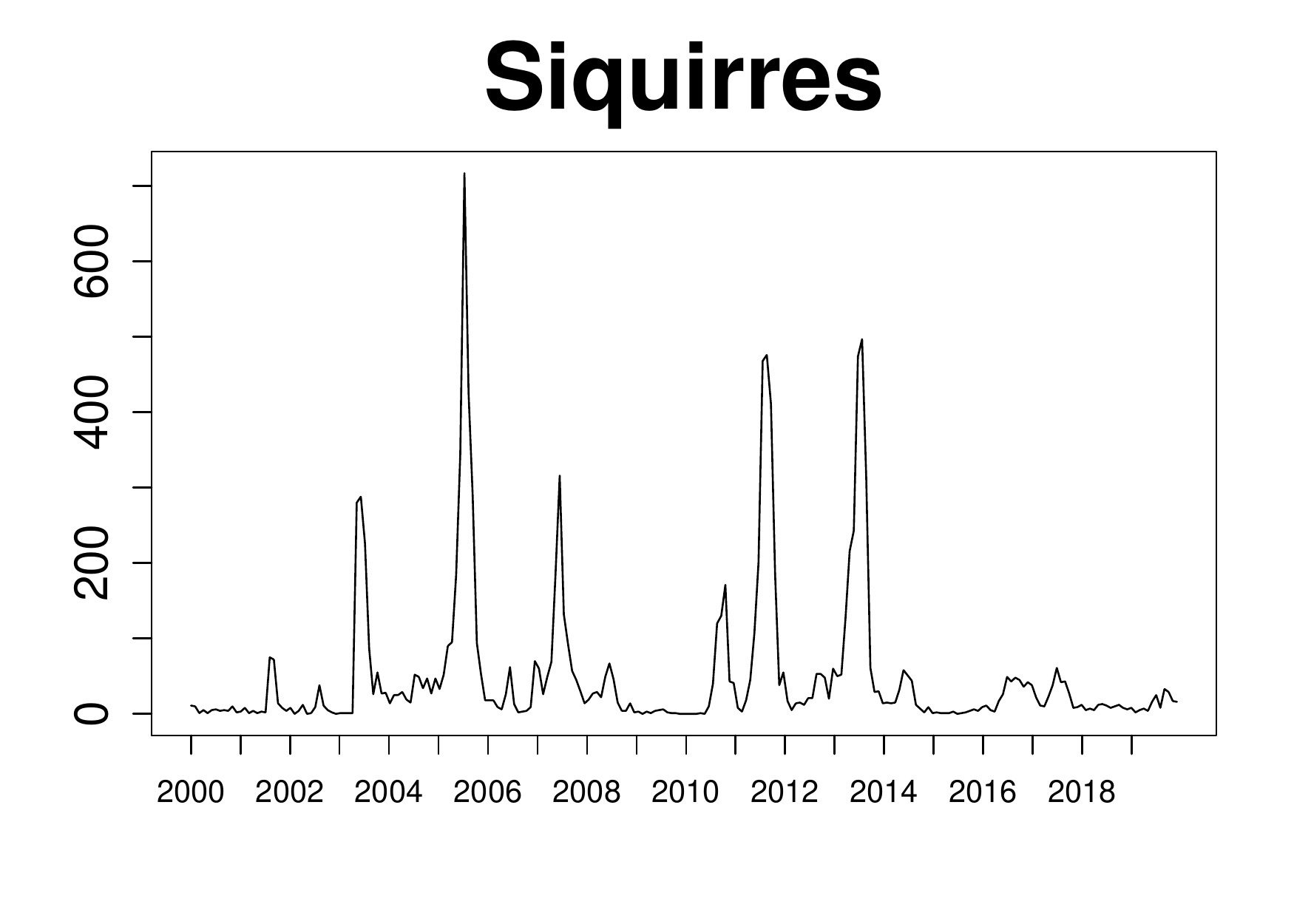}}\vspace{-0.15cm}%
\subfloat[]{\includegraphics[scale=0.2]{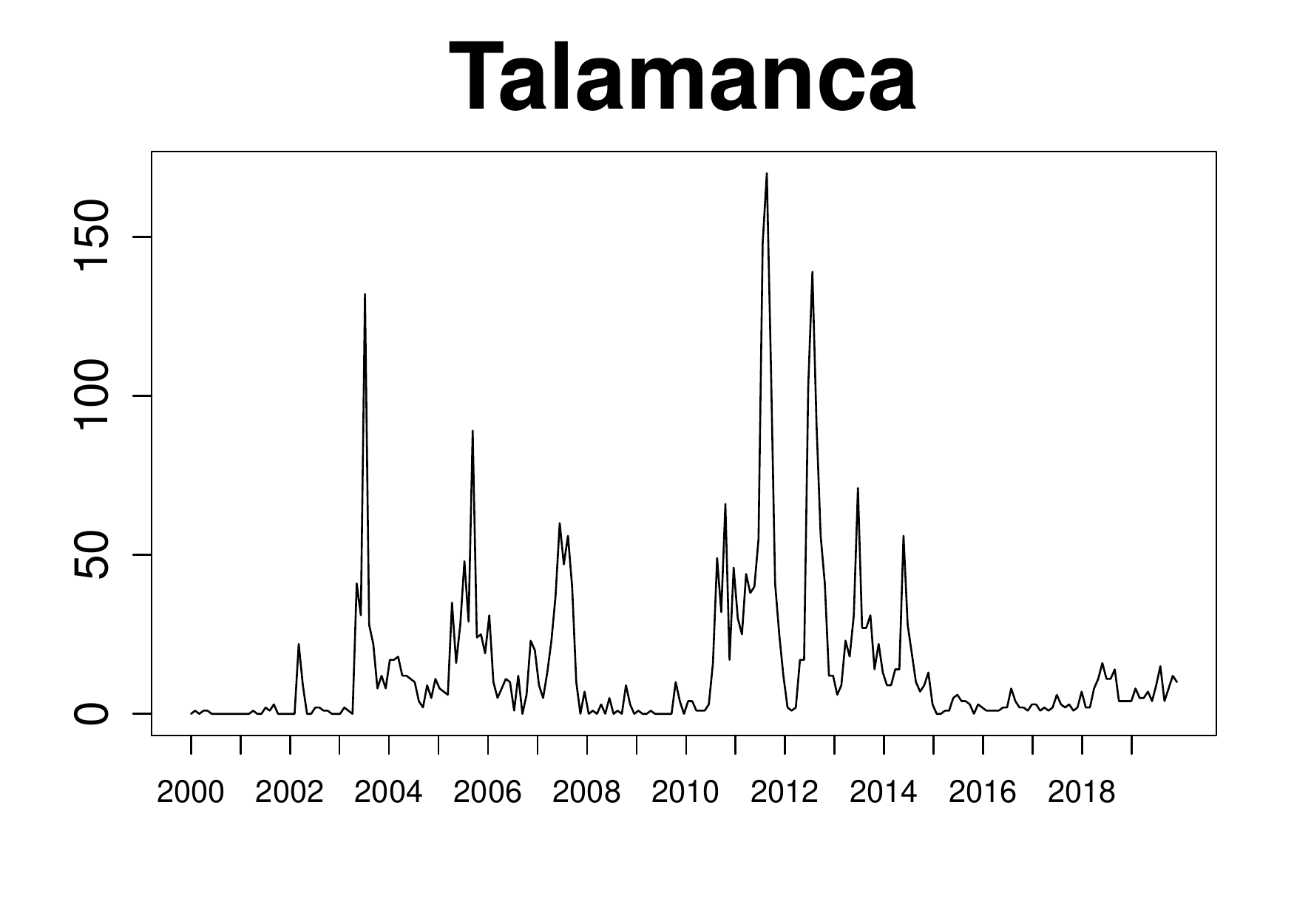}}\vspace{-0.15cm}%
\subfloat[]{\includegraphics[scale=0.2]{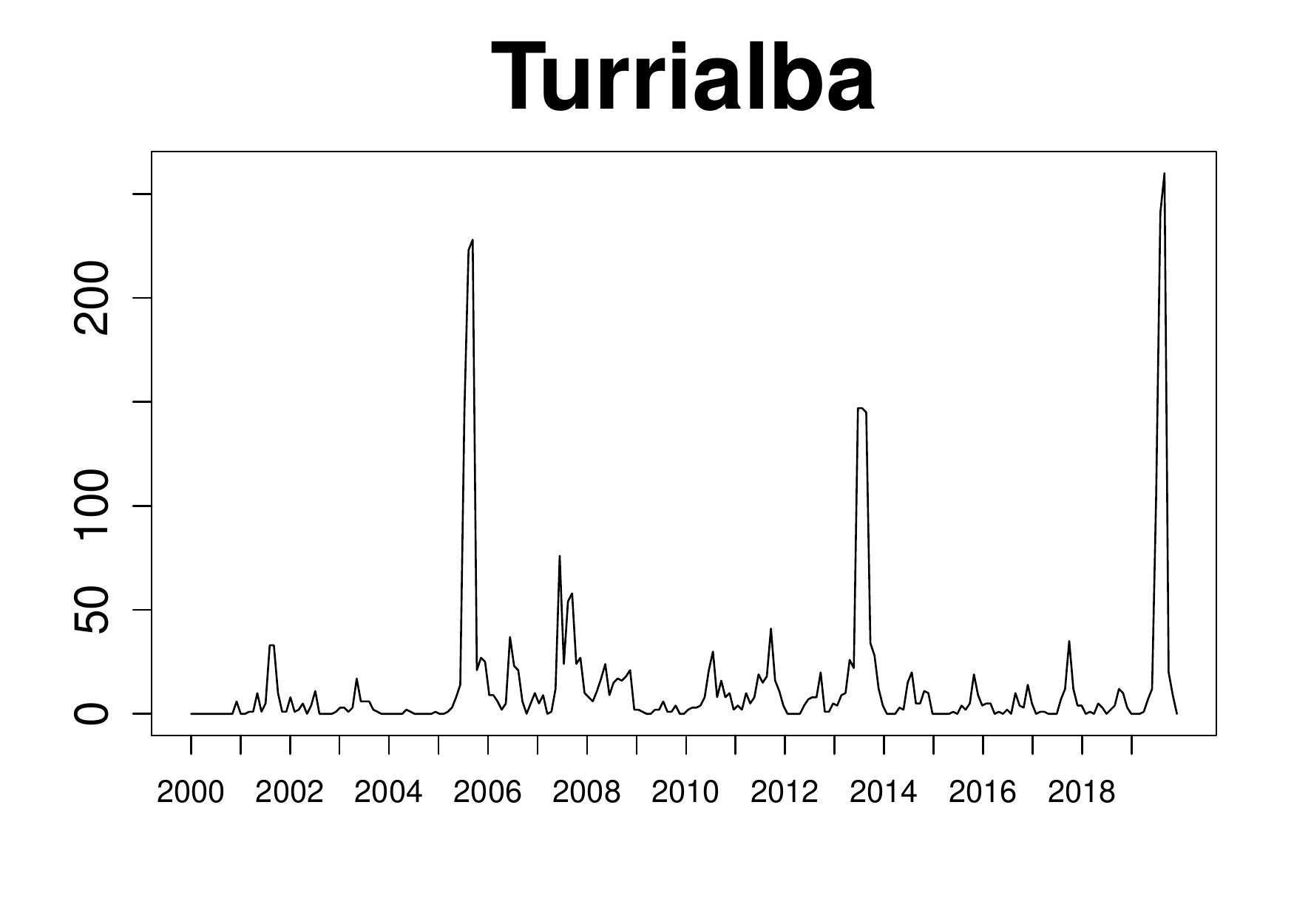}}\vspace{-0.15cm}%
\subfloat[]{\includegraphics[scale=0.2]{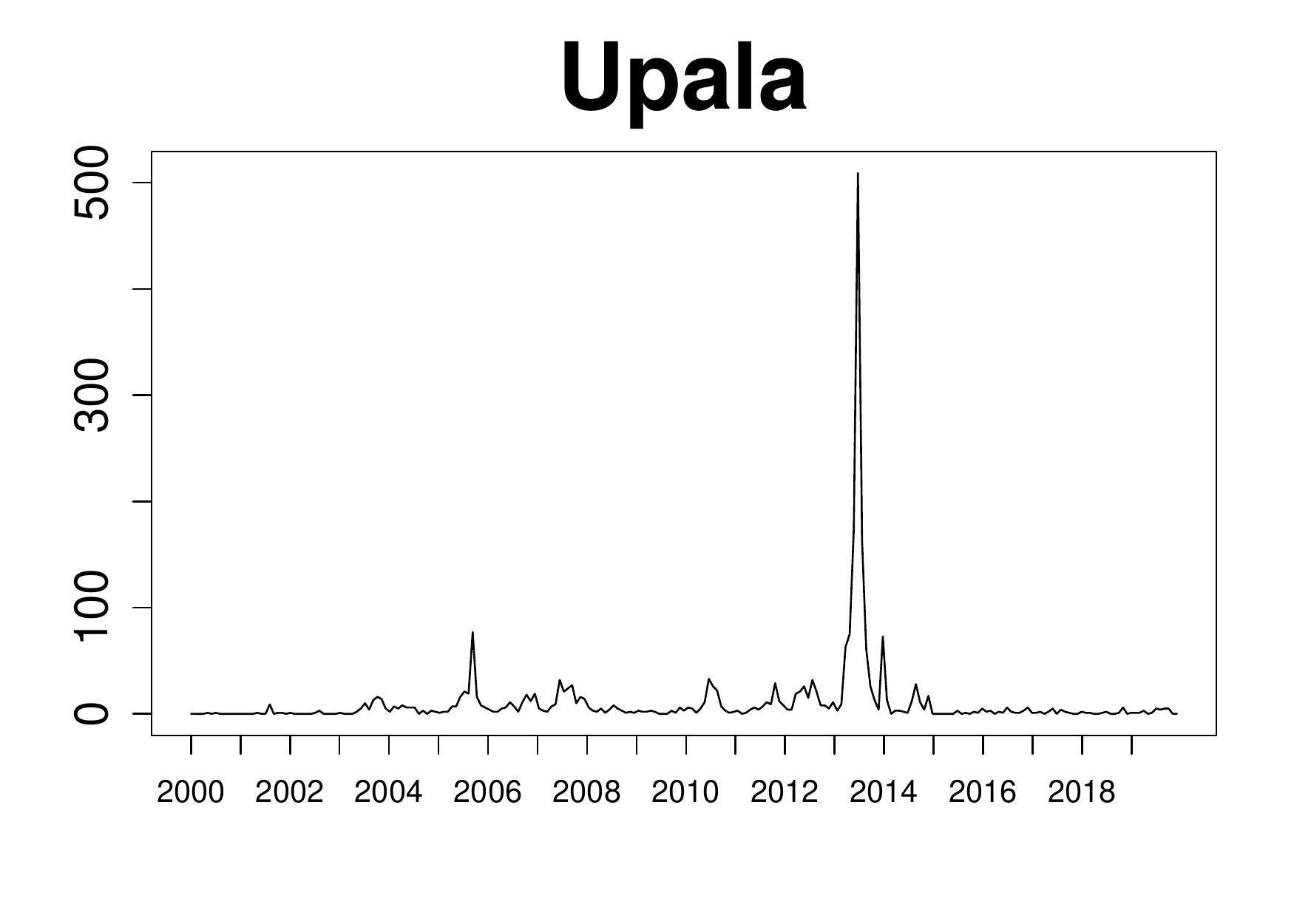}}\vspace{-0.15cm}%
\subfloat[]{\includegraphics[scale=0.2]{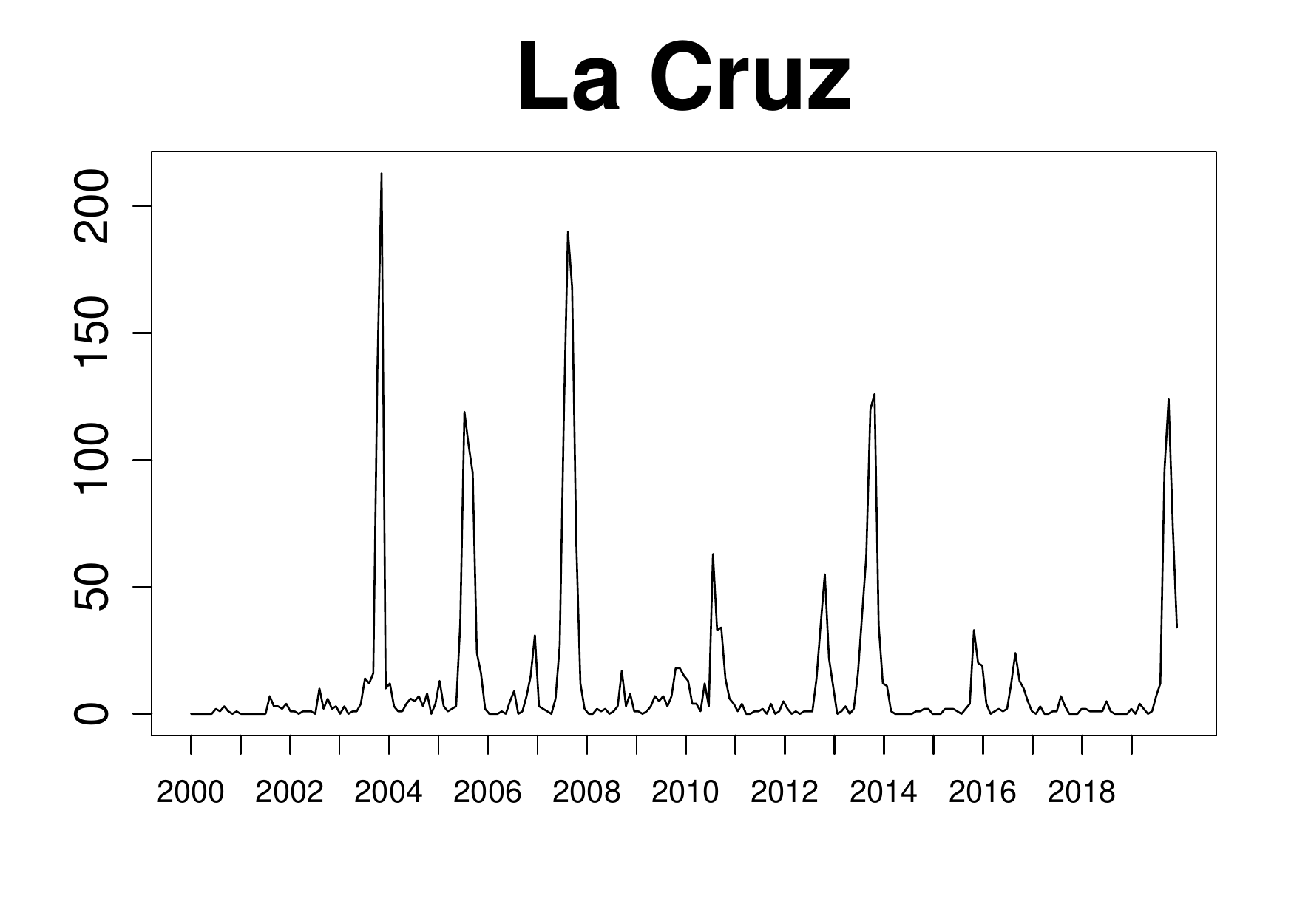}}\vspace{-0.15cm}\\
\subfloat[]{\includegraphics[scale=0.2]{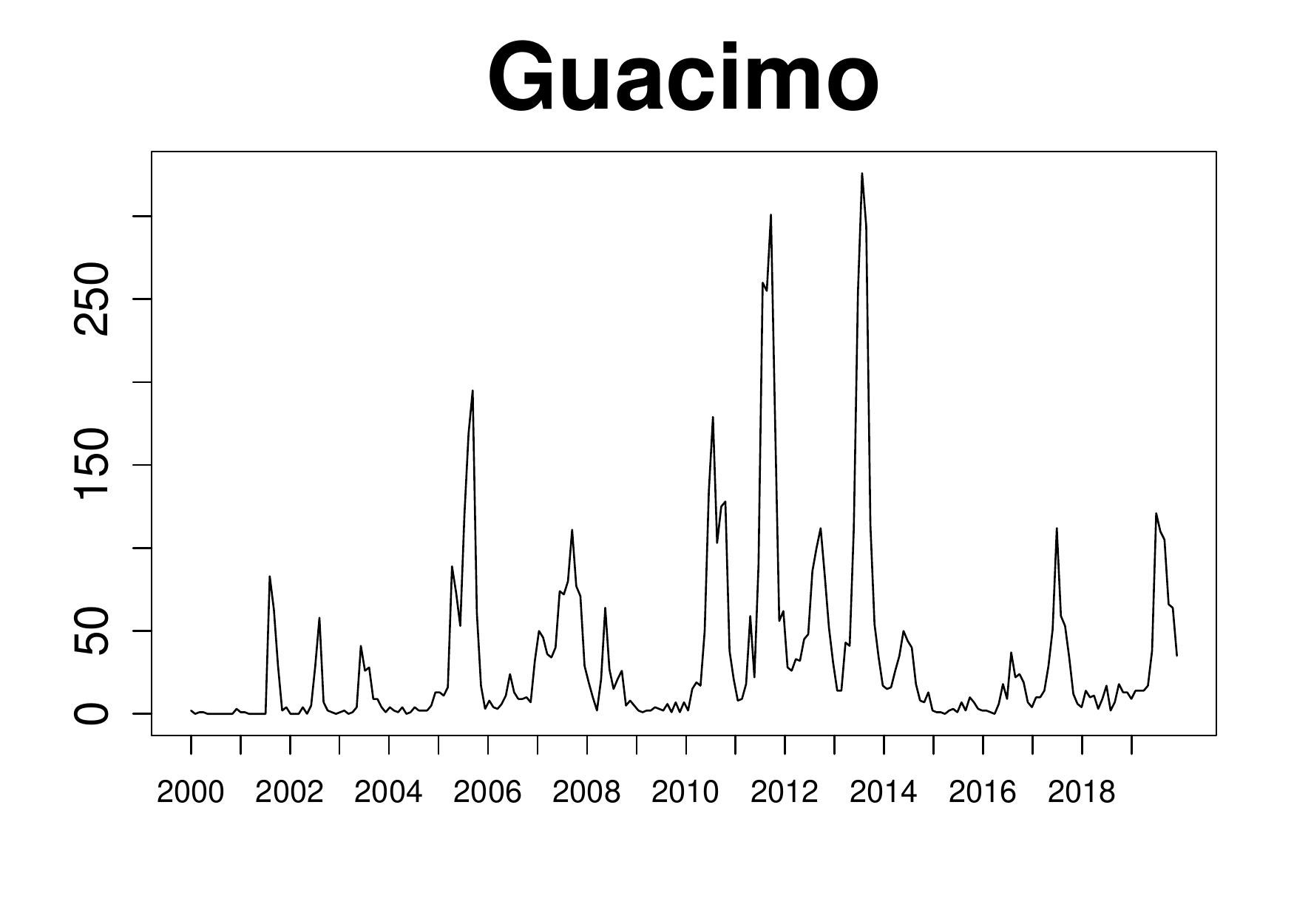}}\vspace{-0.15cm}%
\subfloat[]{\includegraphics[scale=0.2]{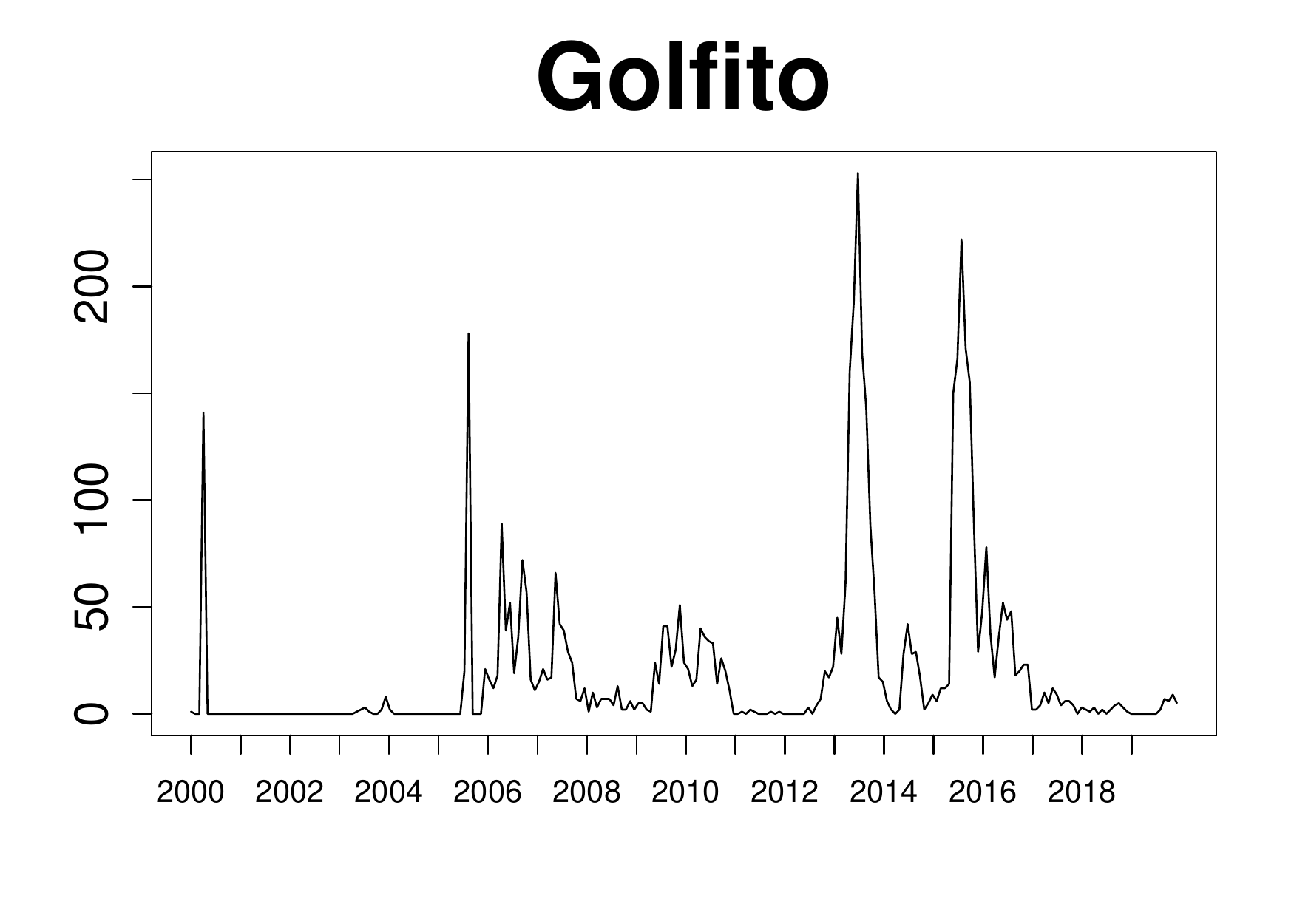}}\vspace{-0.15cm}%
\subfloat[]{\includegraphics[scale=0.2]{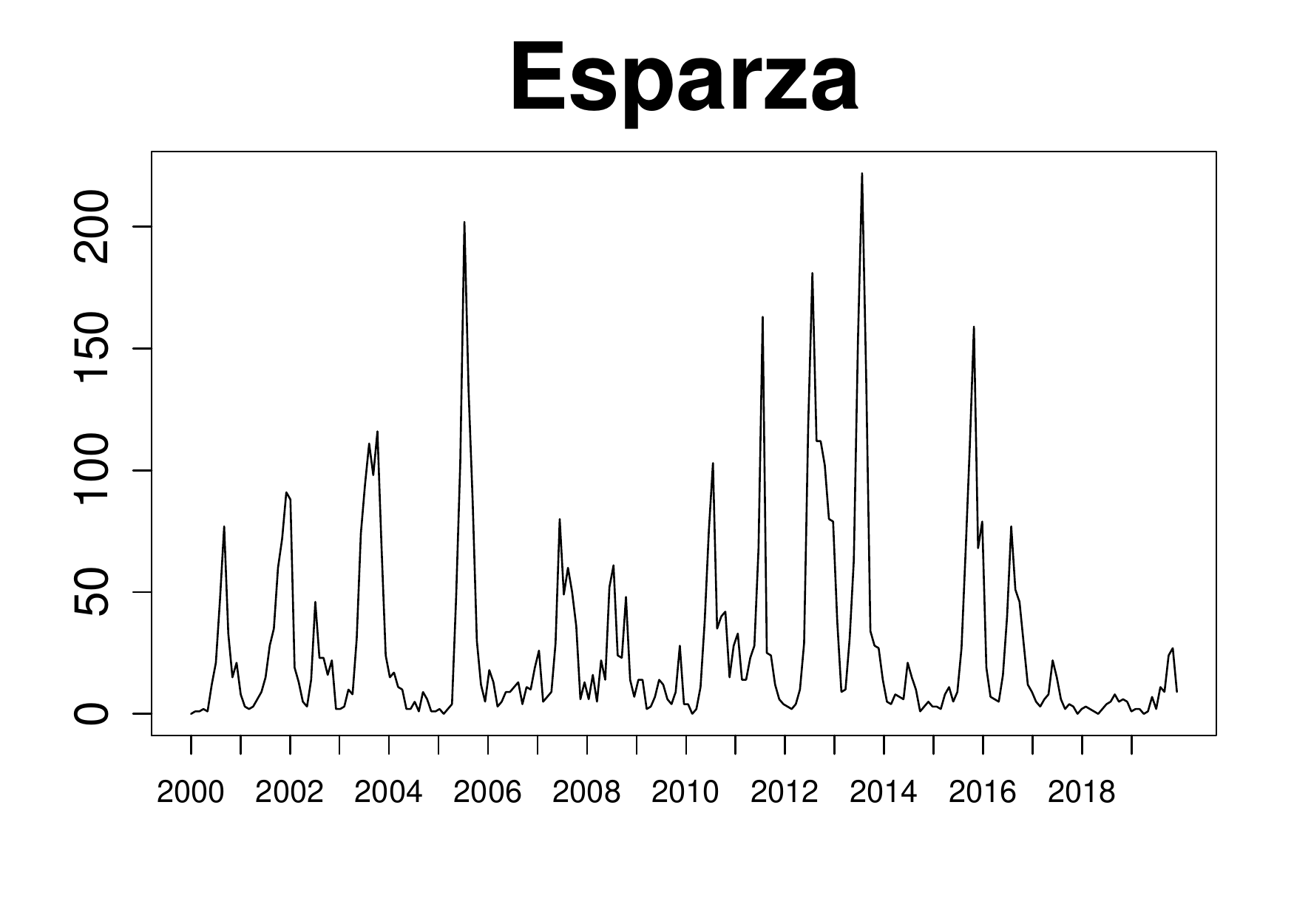}}\vspace{-0.15cm}%
\subfloat[]{\includegraphics[scale=0.2]{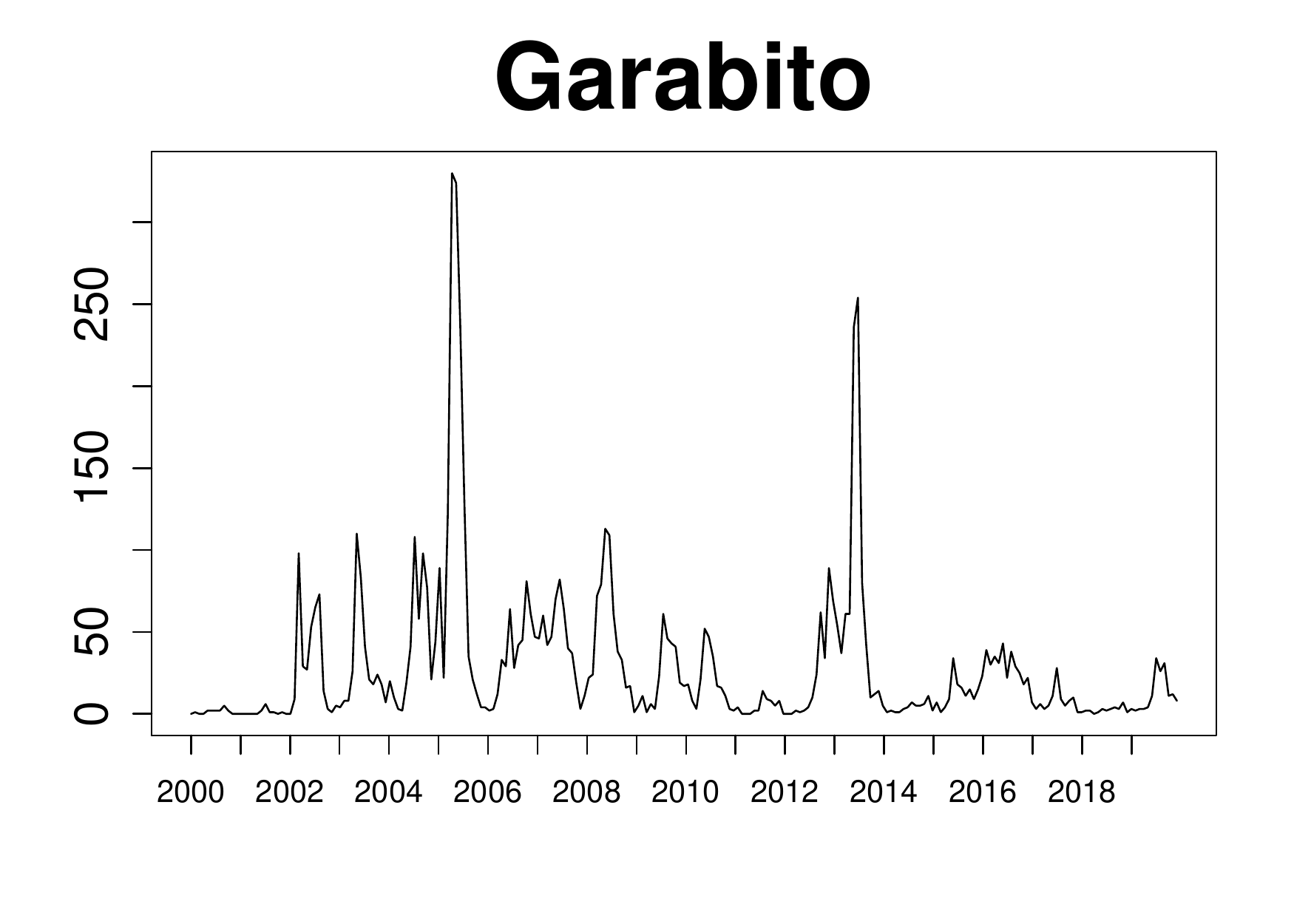}}\vspace{-0.15cm}%
\subfloat[]{\includegraphics[scale=0.2]{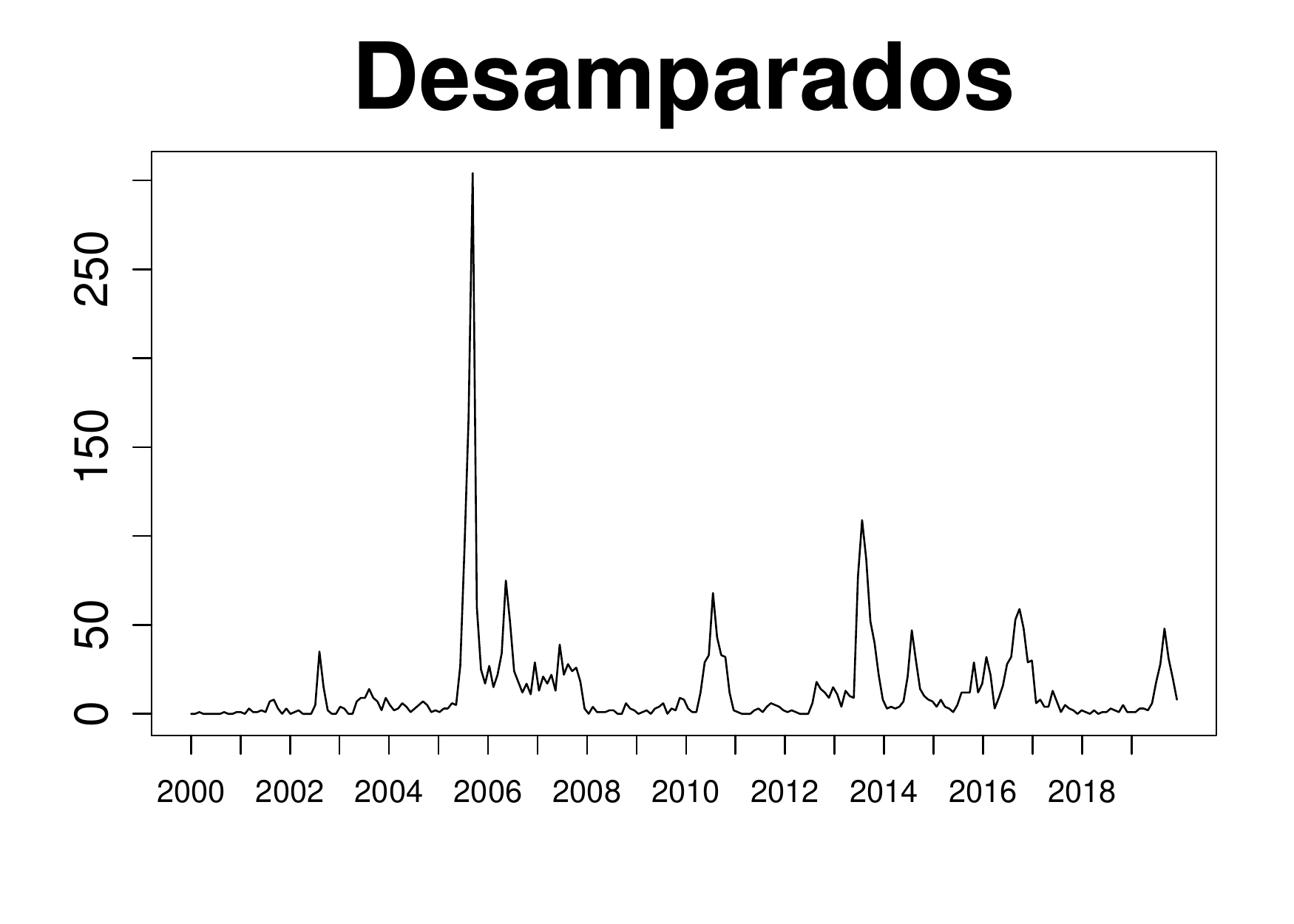}}\vspace{-0.15cm}\\
\subfloat[]{\includegraphics[scale=0.2]{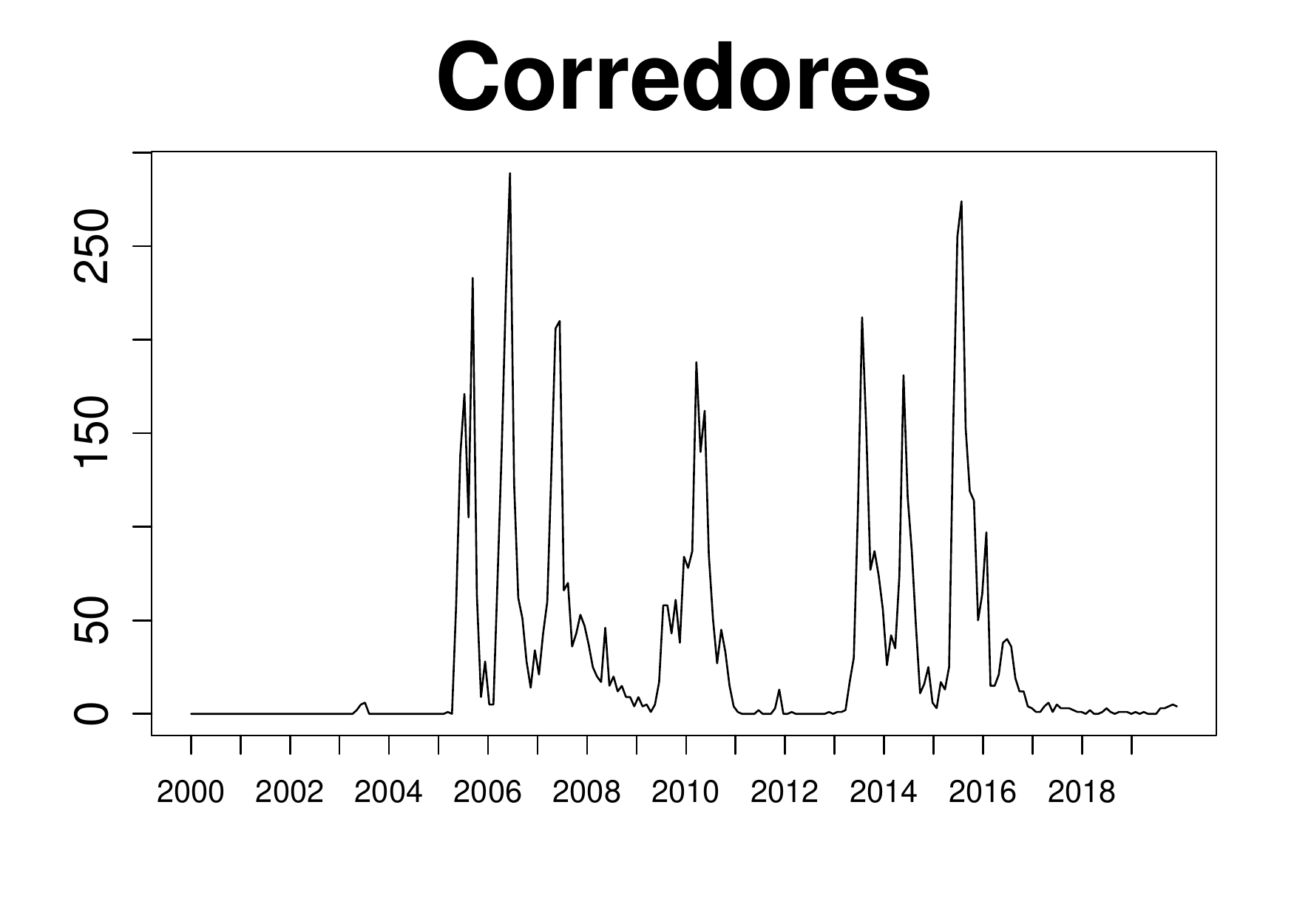}}\vspace{-0.15cm}%
\subfloat[]{\includegraphics[scale=0.2]{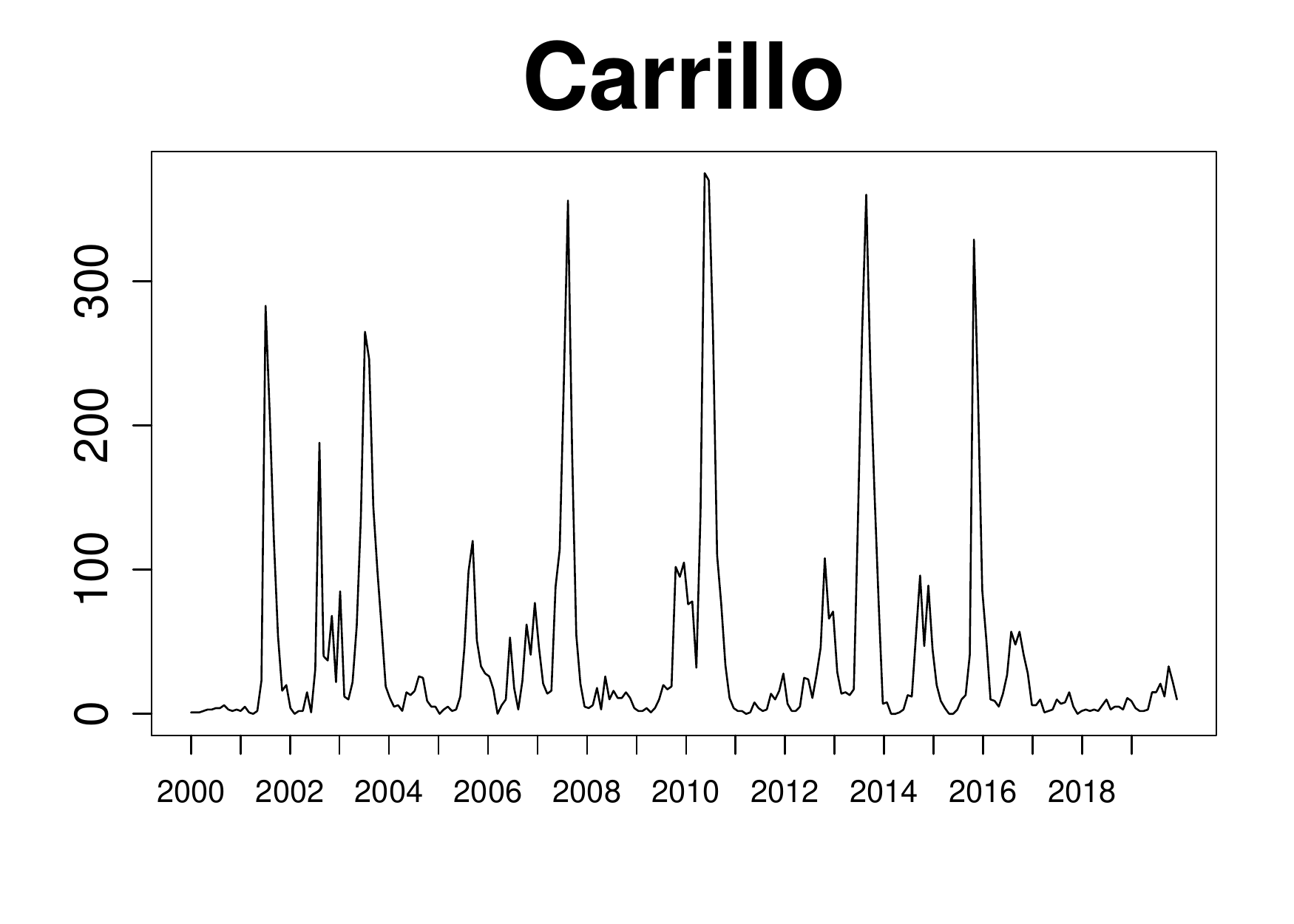}}\vspace{-0.15cm}%
\subfloat[]{\includegraphics[scale=0.2]{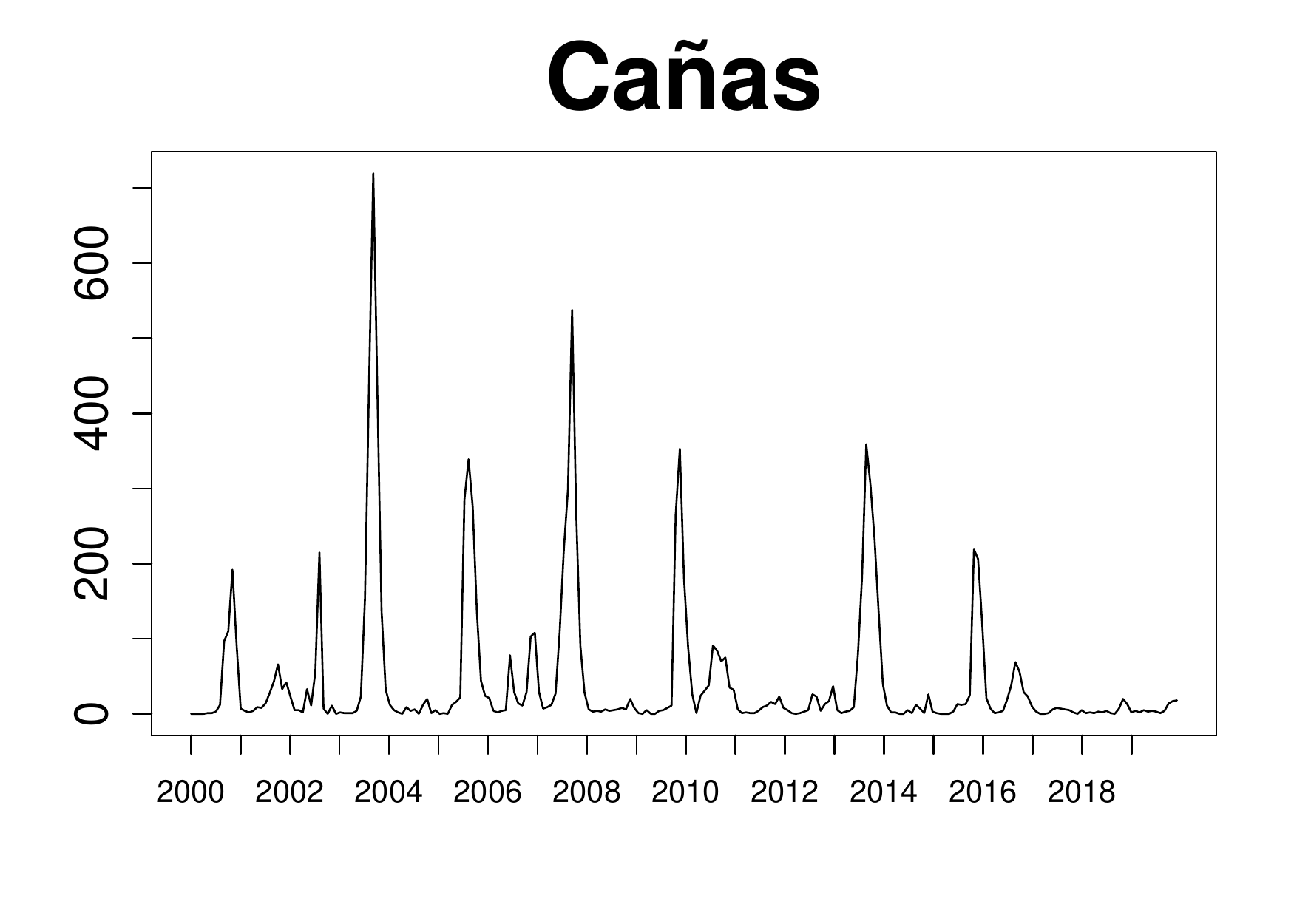}}\vspace{-0.15cm}%
\subfloat[]{\includegraphics[scale=0.2]{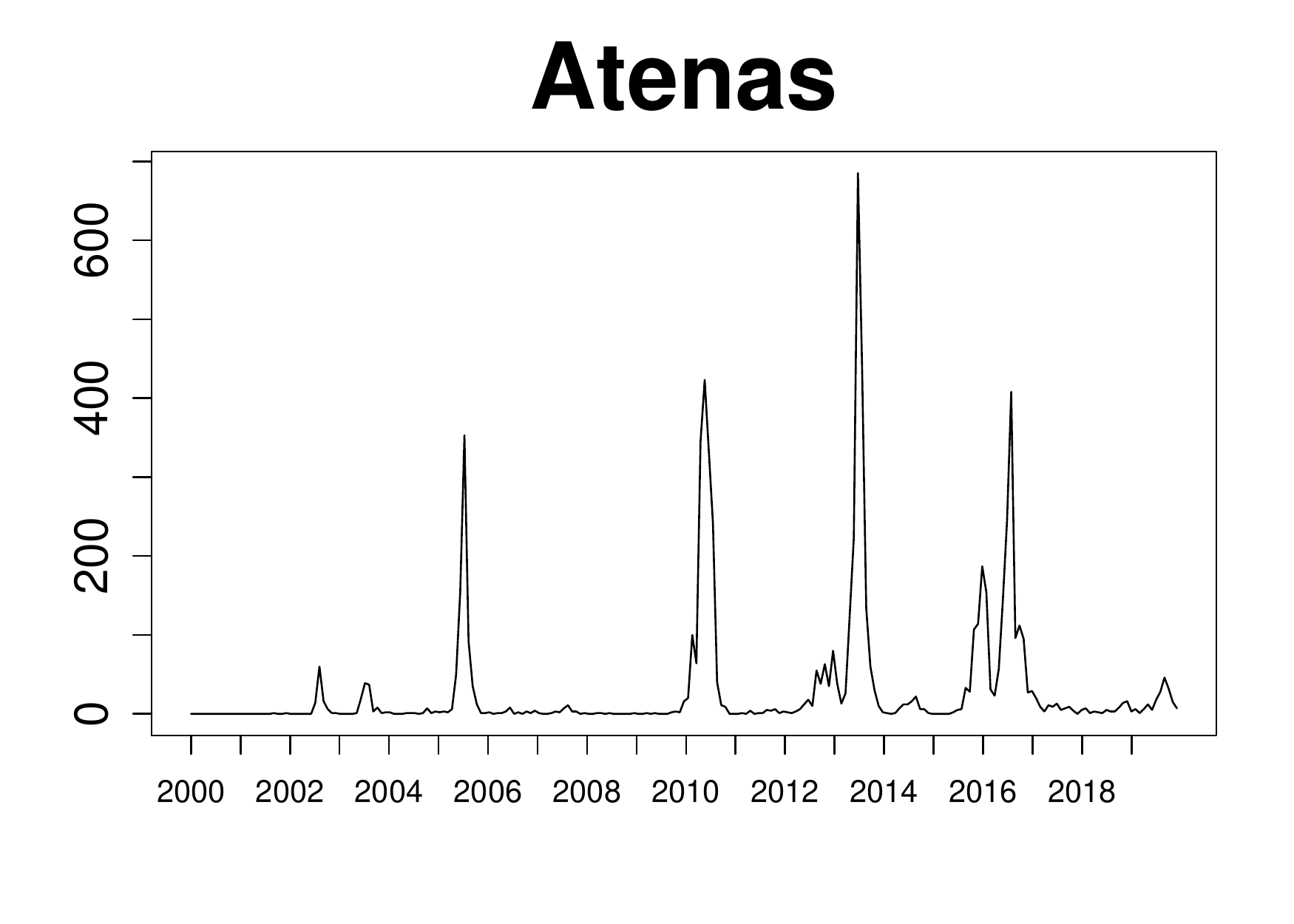}}\vspace{-0.15cm}%
\subfloat[]{\includegraphics[scale=0.2]{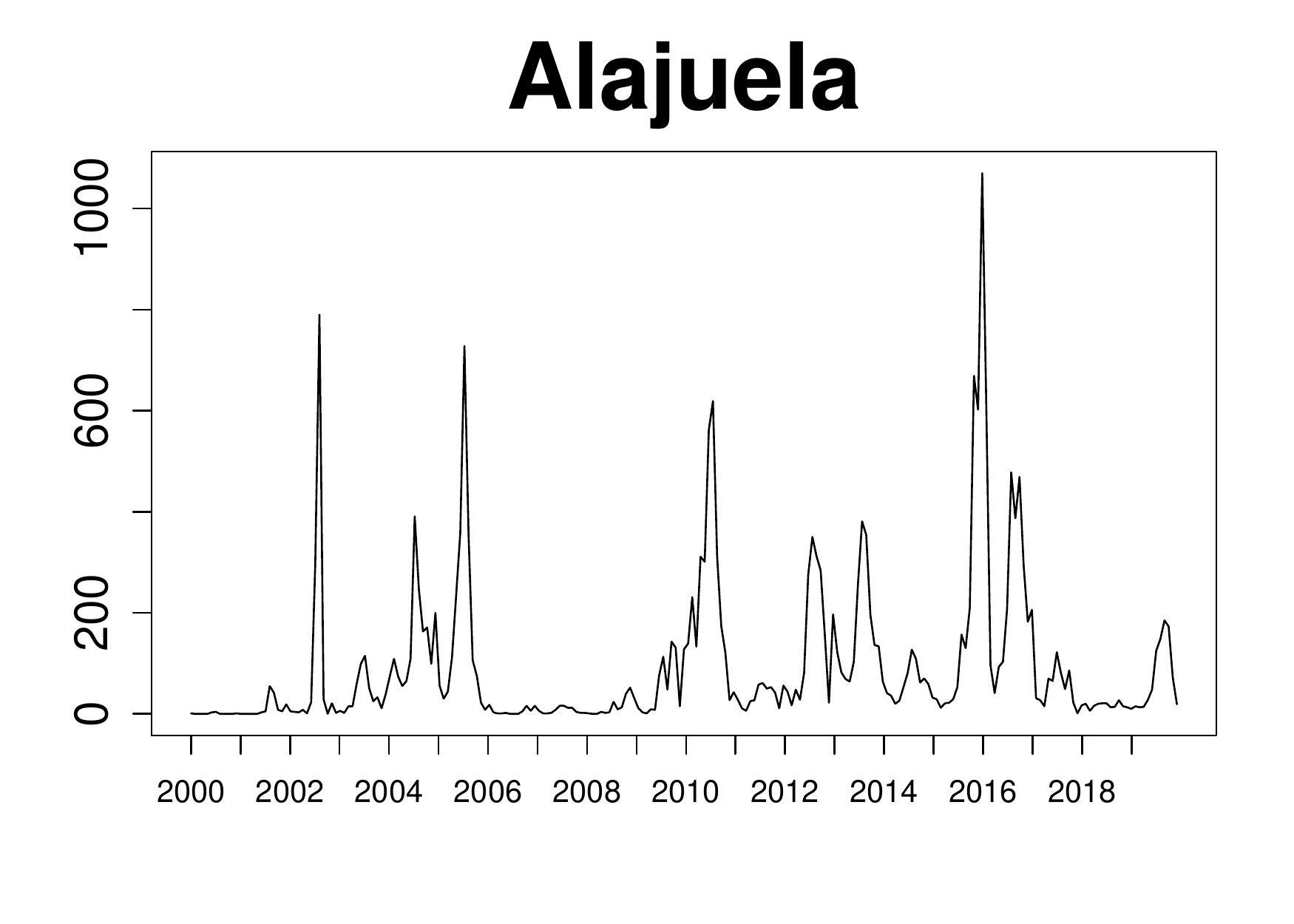}}\vspace{-0.15cm}
\subfloat[]{\includegraphics[scale=0.2]{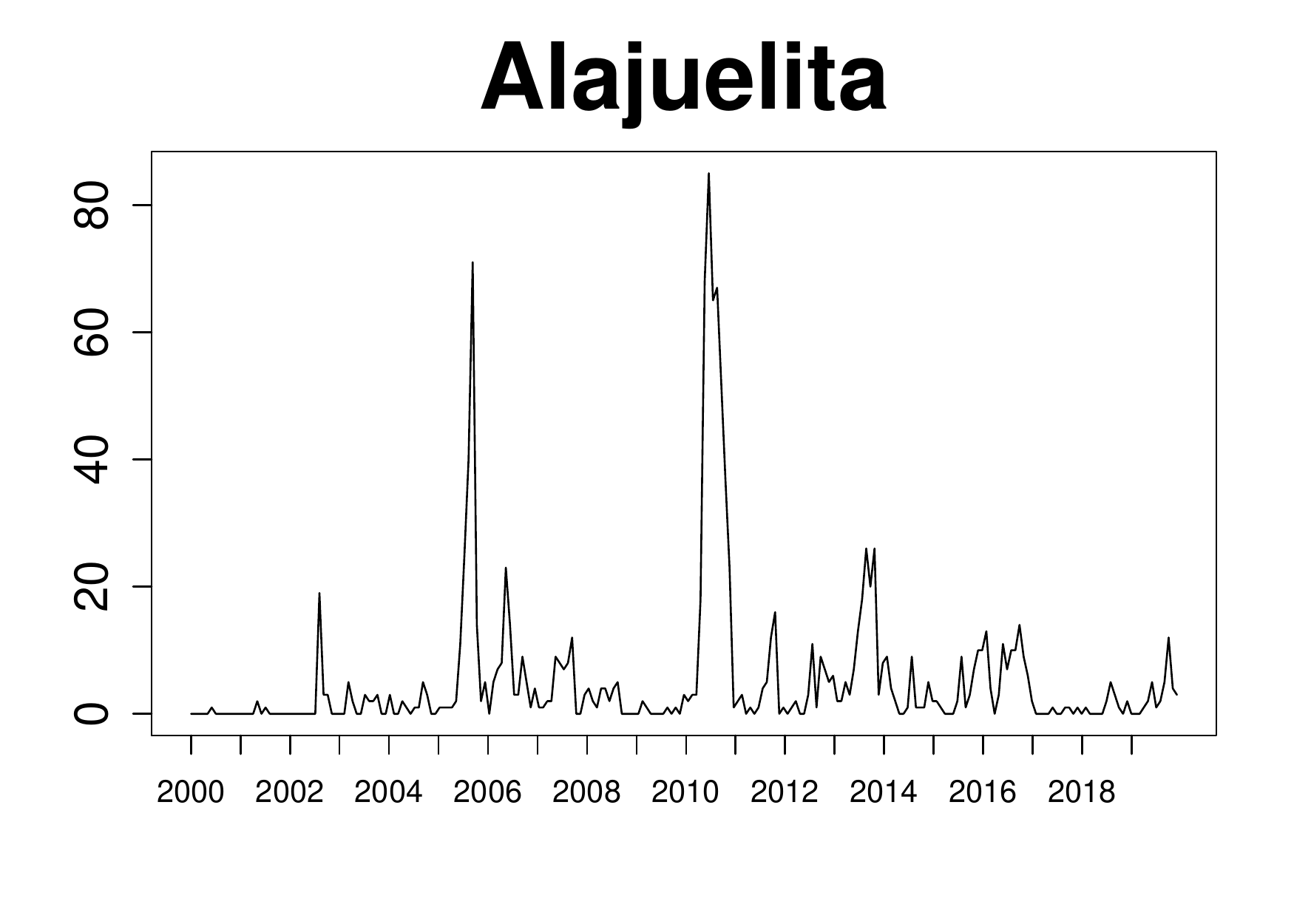}}\vspace{-0.15cm}%
\subfloat[]{\includegraphics[scale=0.2]{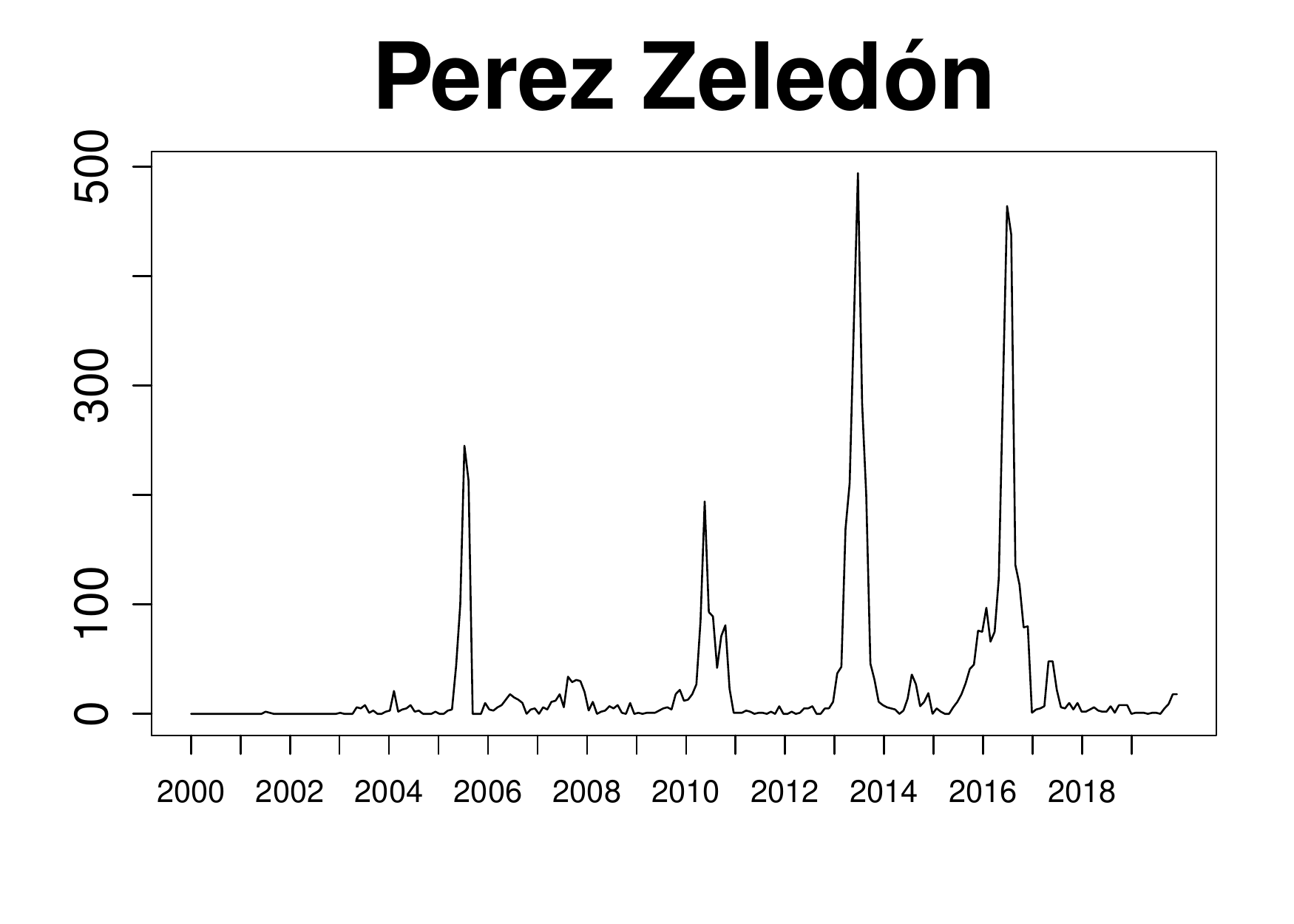}}\vspace{-0.15cm}%
\subfloat[]{\includegraphics[scale=0.2]{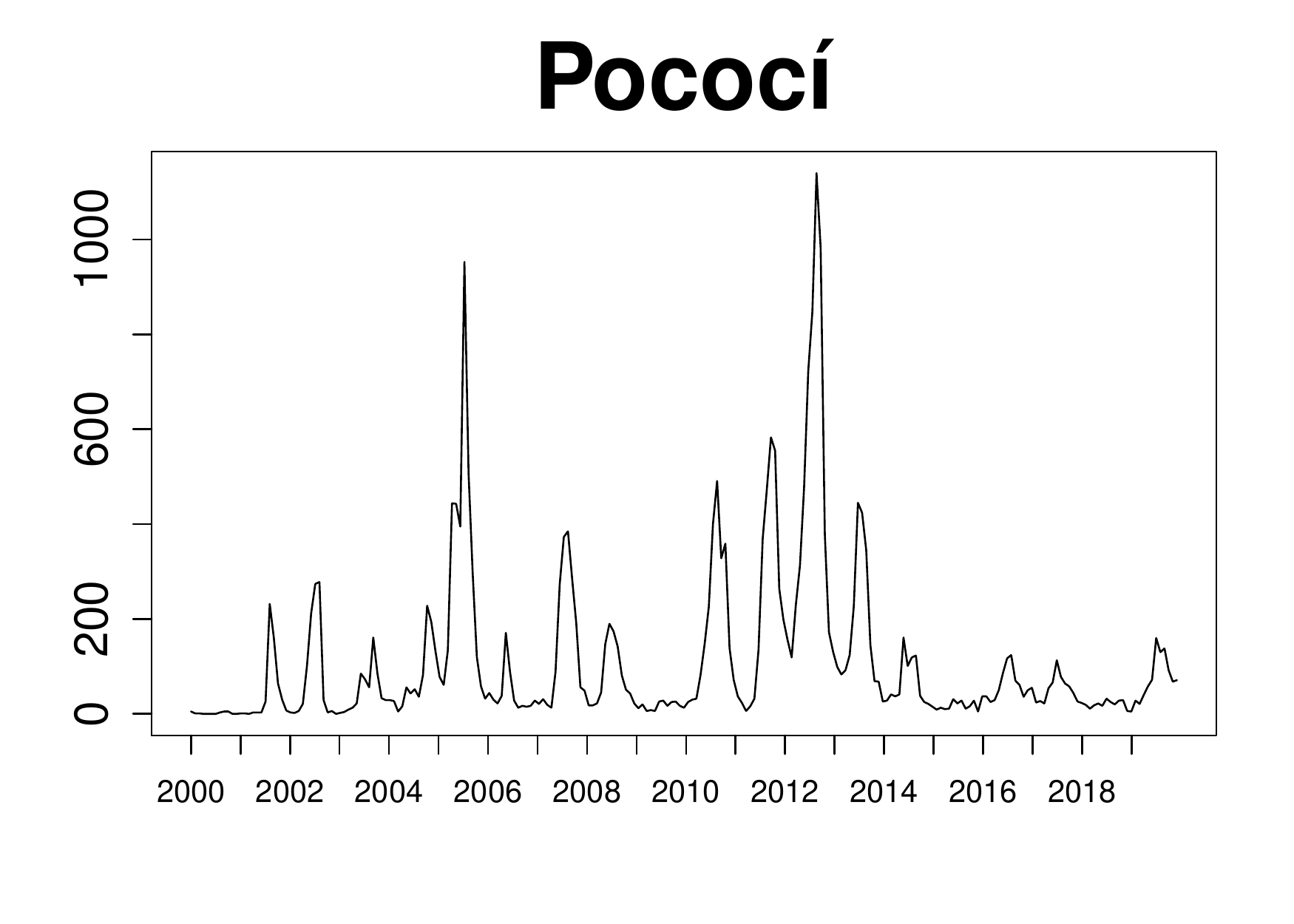}}\vspace{-0.15cm}%
\subfloat[]{\includegraphics[scale=0.2]{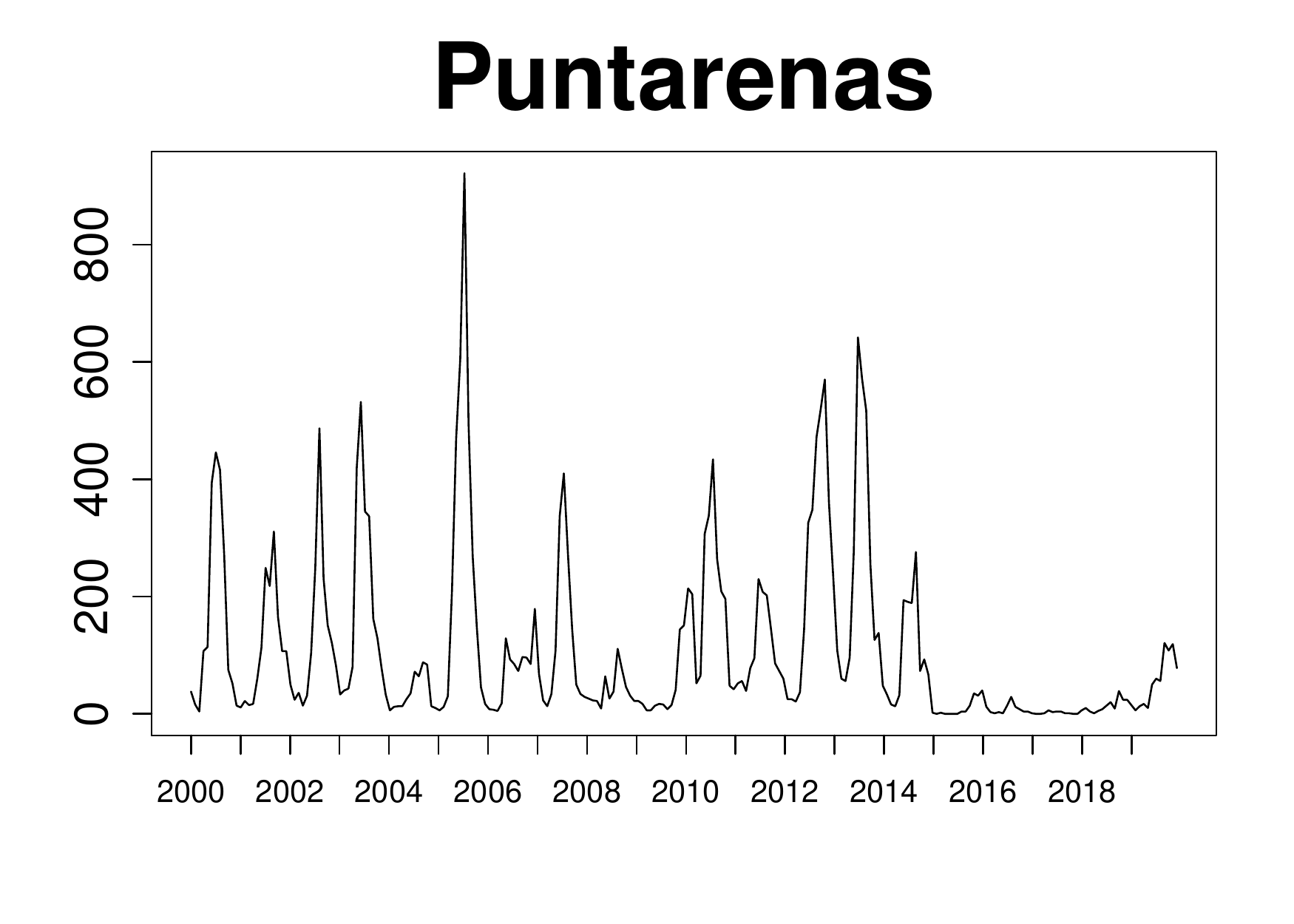}}\vspace{-0.15cm}%
\subfloat[]{\includegraphics[scale=0.2]{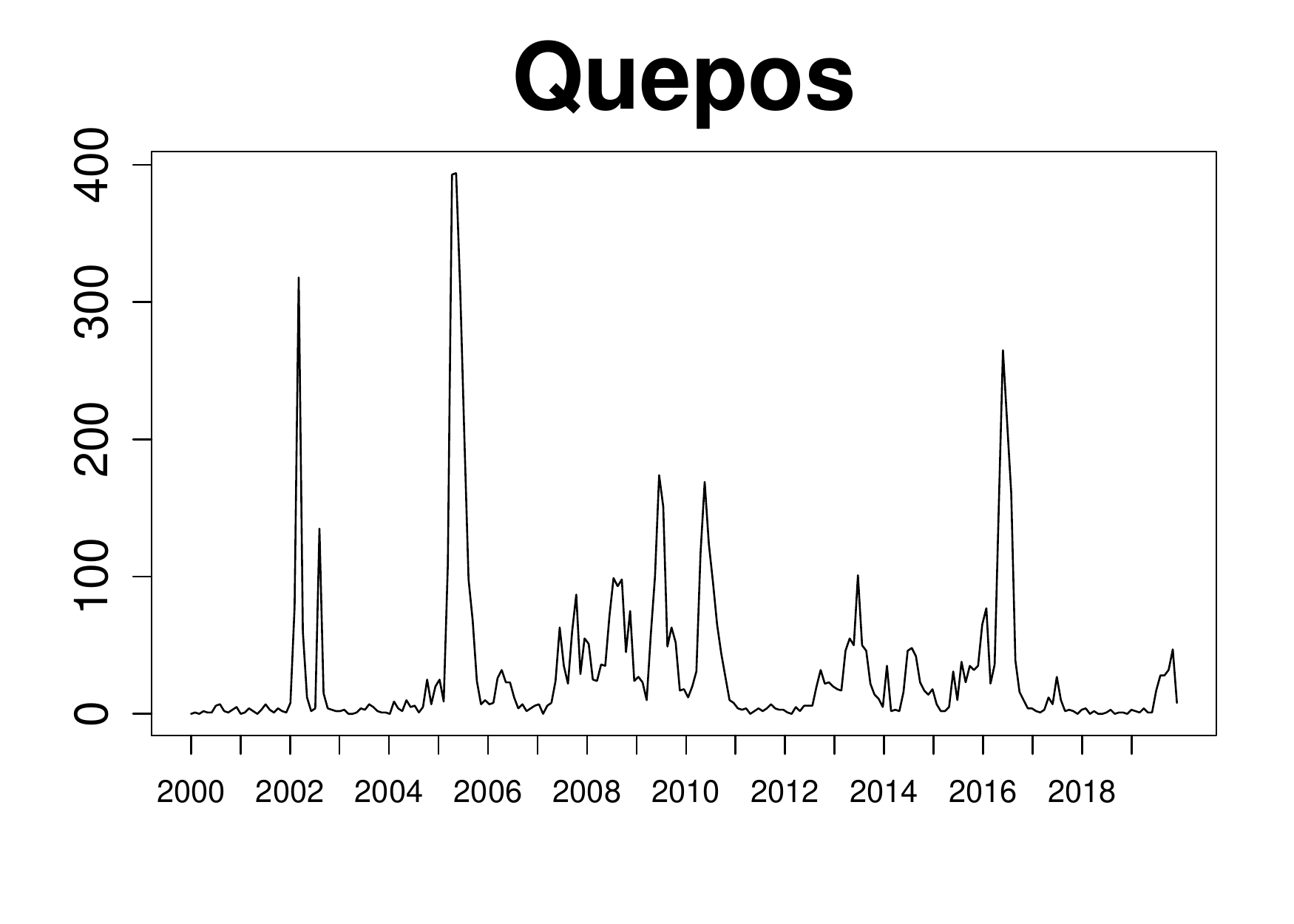}}\vspace{-0.15cm}\\
\subfloat[]{\includegraphics[scale=0.2]{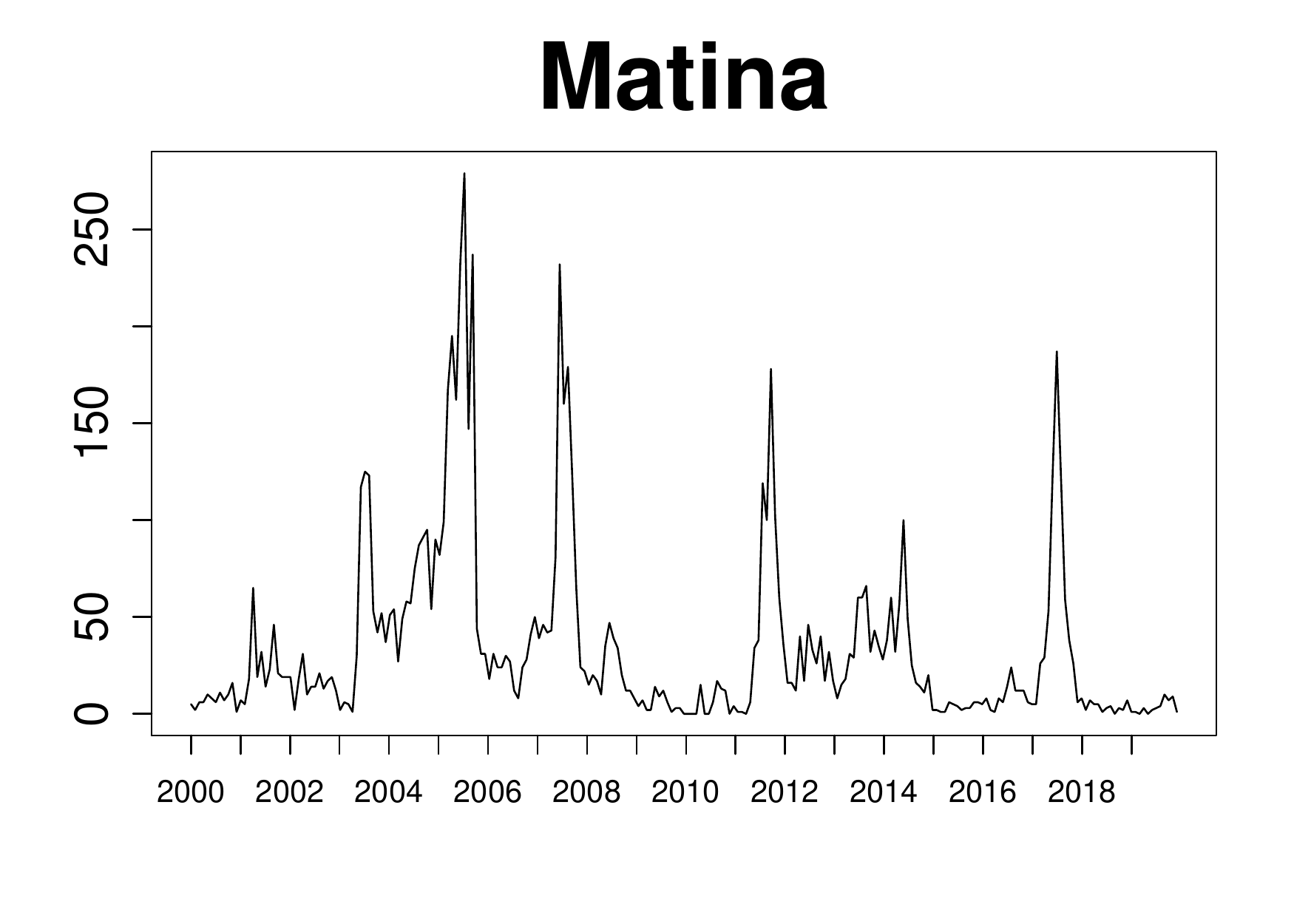}}\vspace{-0.15cm}%
\subfloat[]{\includegraphics[scale=0.2]{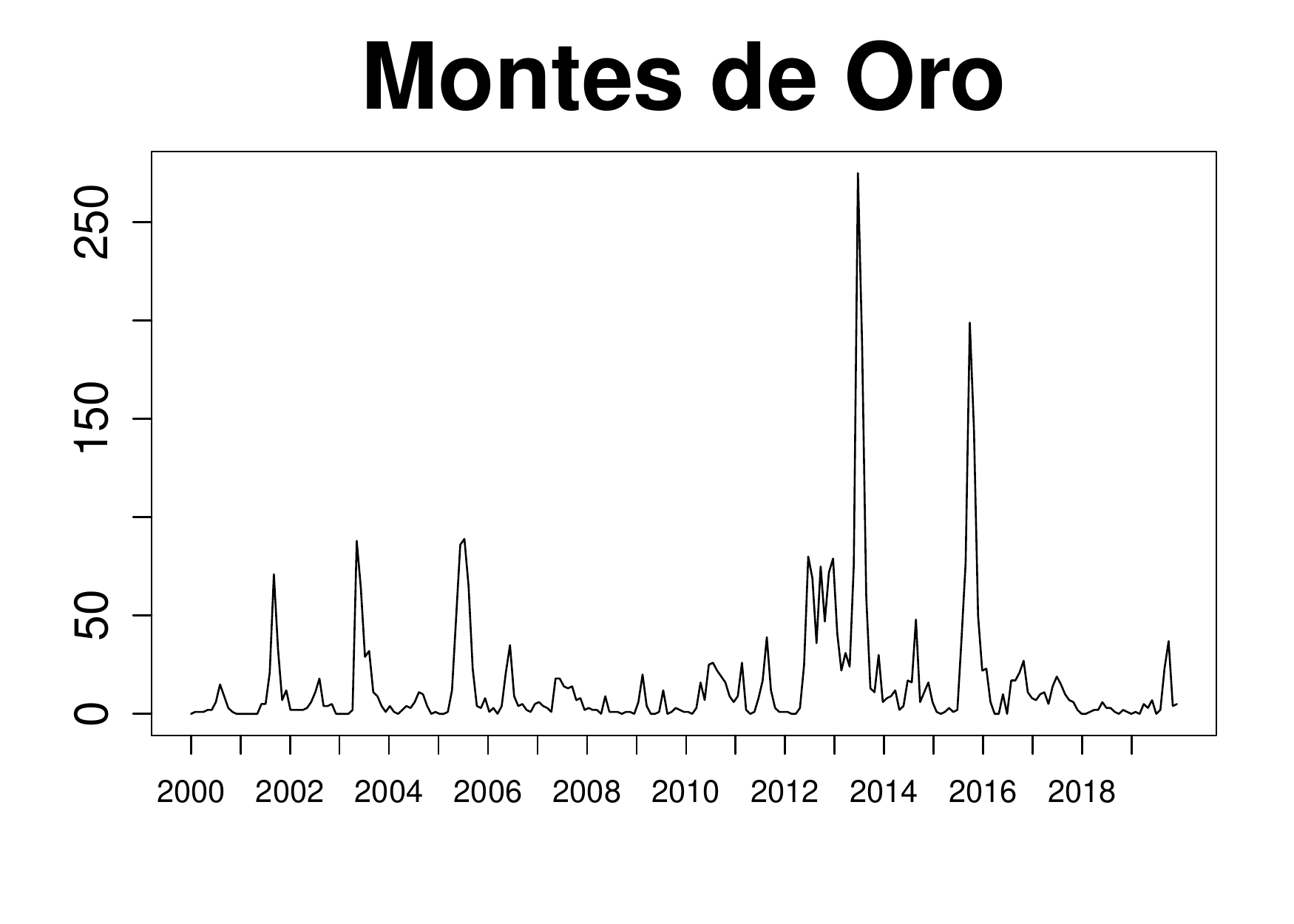}}\vspace{-0.15cm}%
\subfloat[]{\includegraphics[scale=0.2]{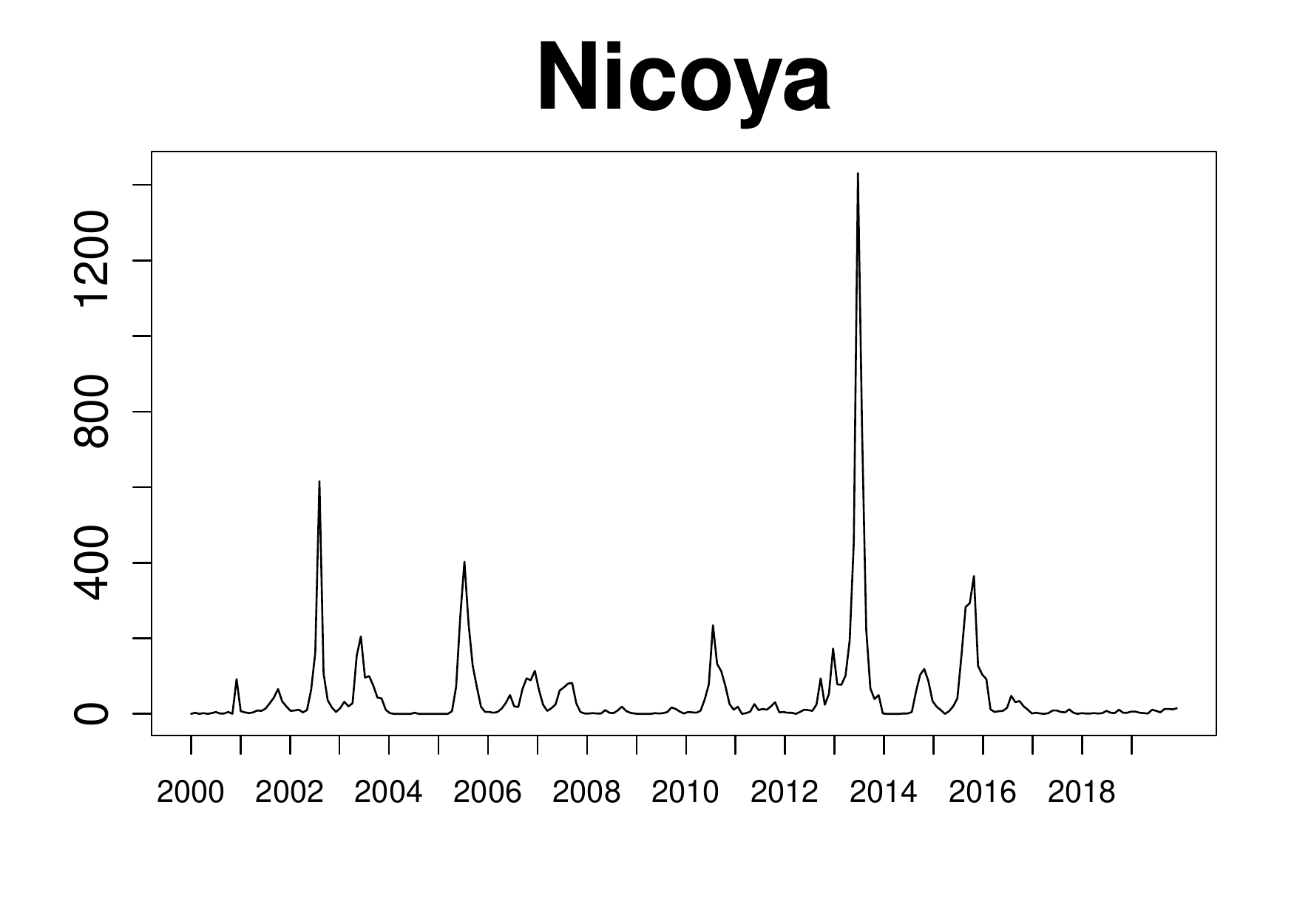}}\vspace{-0.15cm}%
\subfloat[]{\includegraphics[scale=0.2]{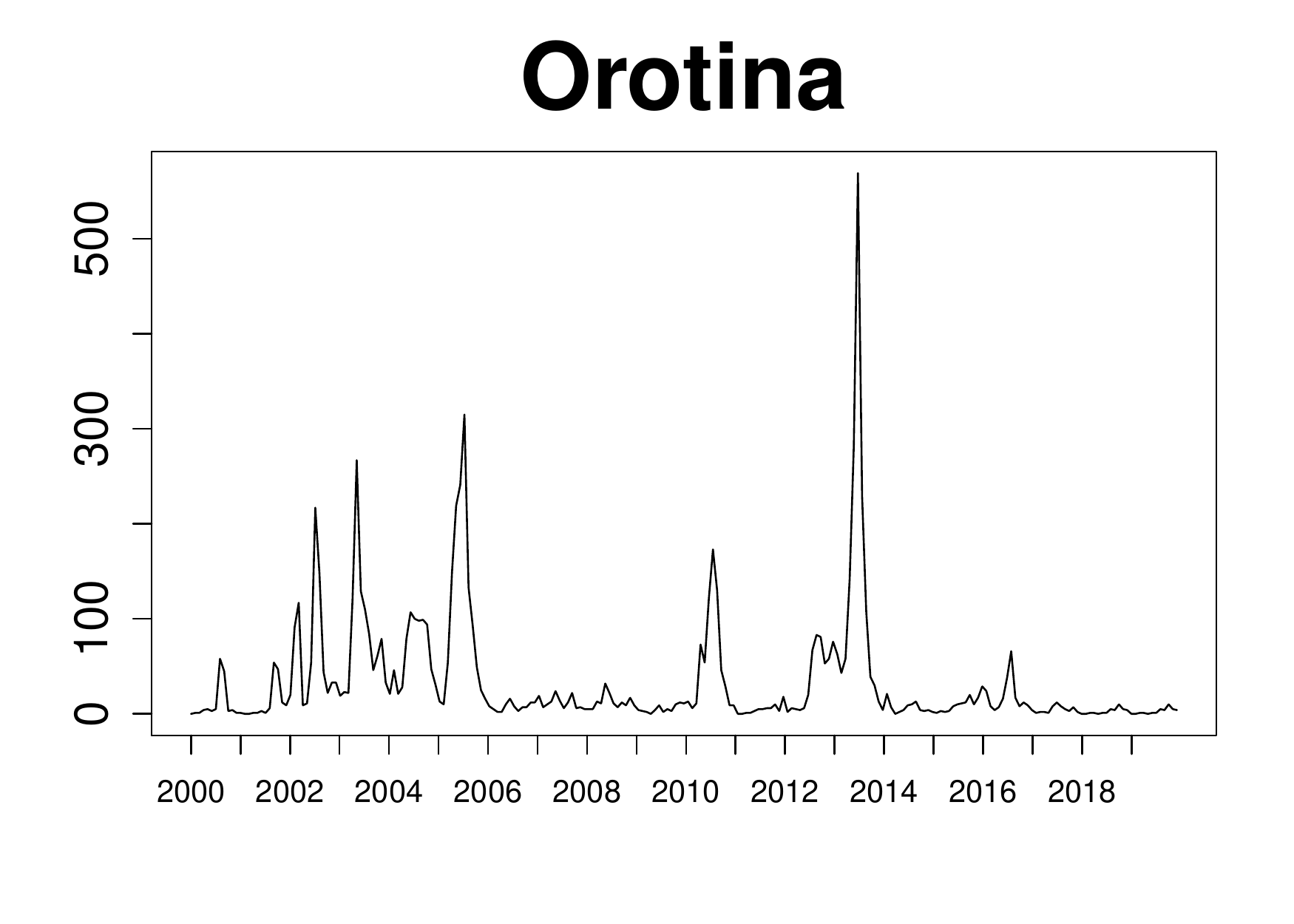}}\vspace{-0.15cm}%
\subfloat[]{\includegraphics[scale=0.2]{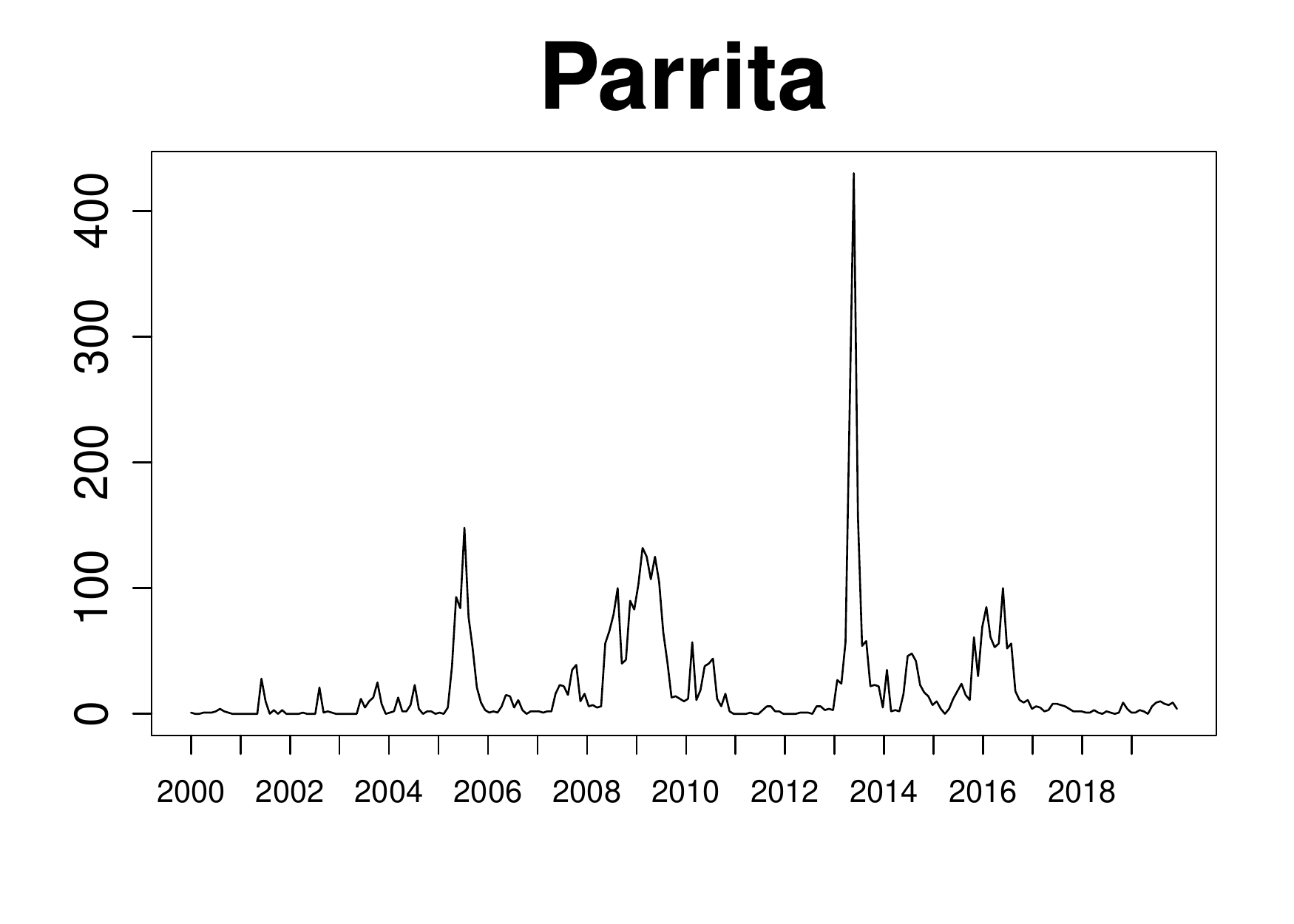}}\vspace{-0.15cm}\\
\subfloat[]{\includegraphics[scale=0.2]{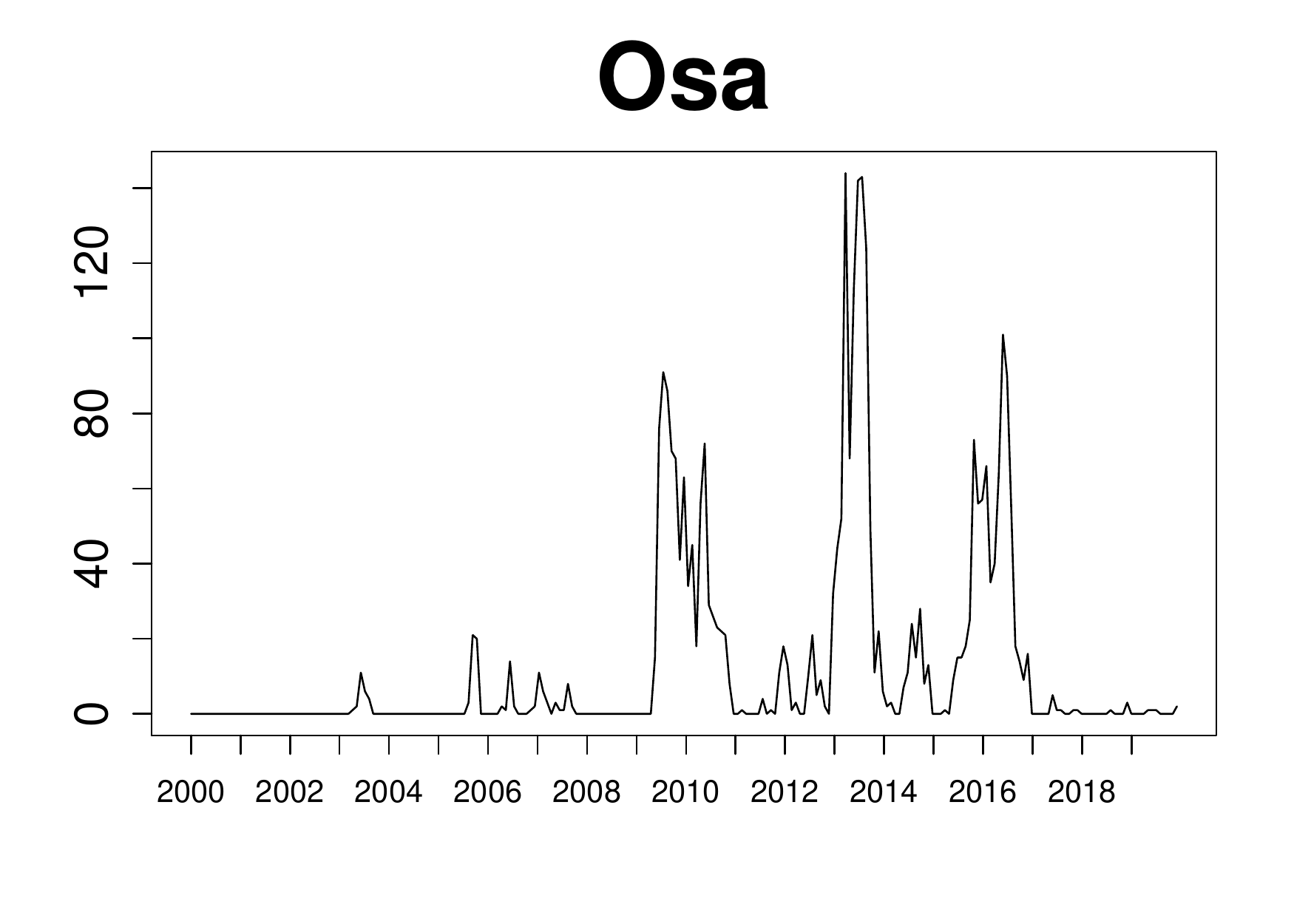}}\vspace{-0.15cm}%
\subfloat[]{\includegraphics[scale=0.2]{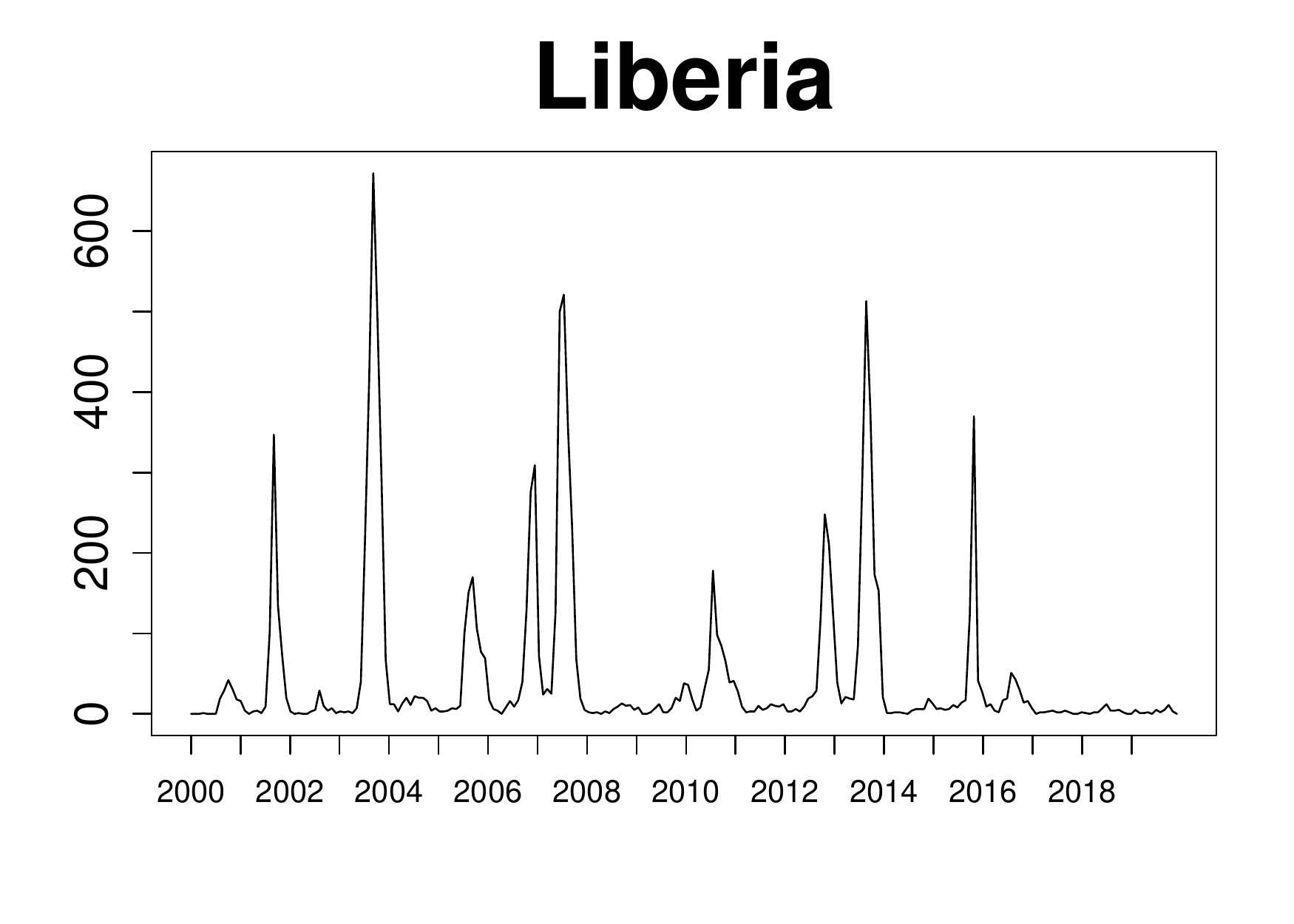}}\vspace{-0.15cm}%
\subfloat[]{\includegraphics[scale=0.2]{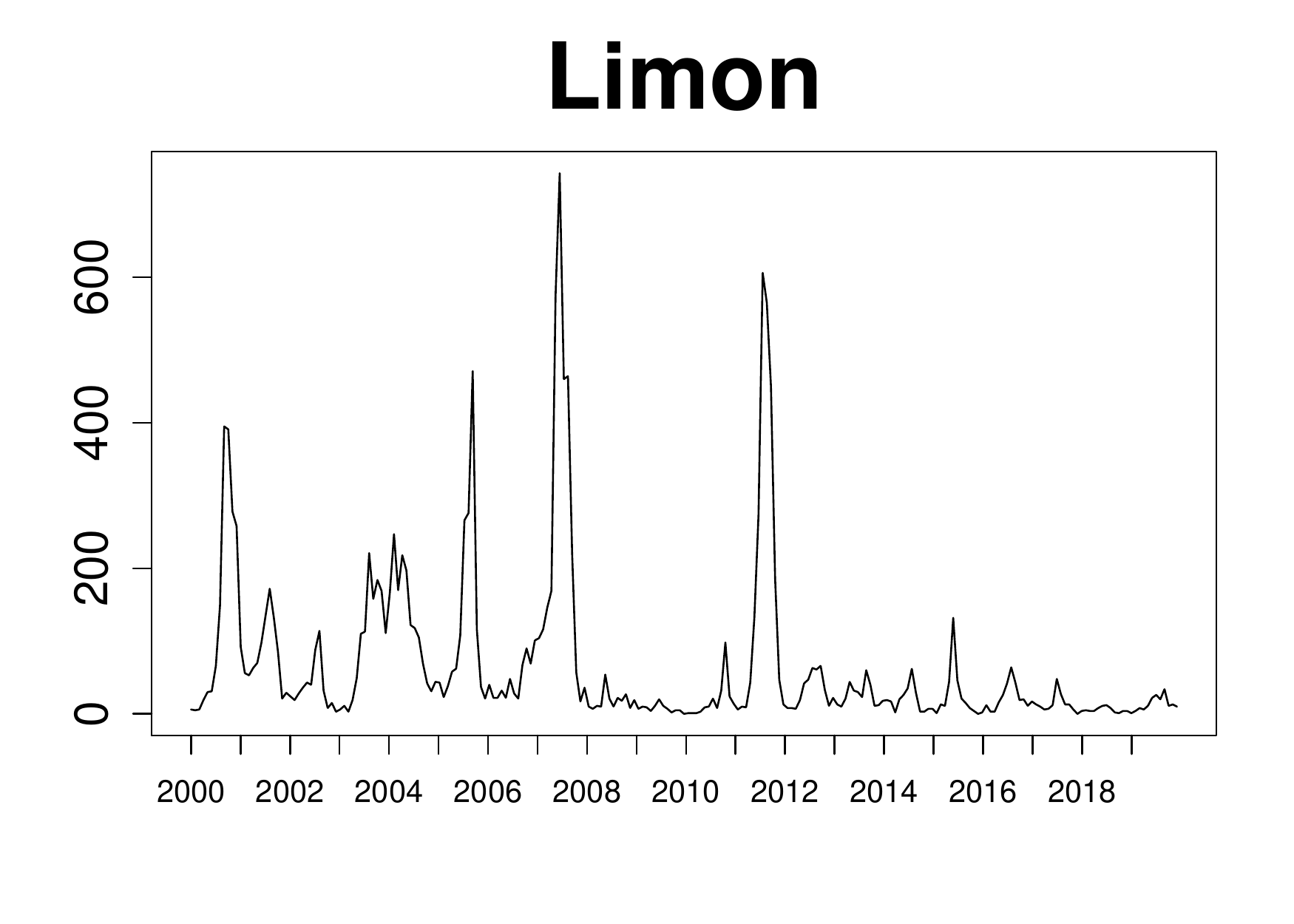}}\vspace{-0.15cm}%
\subfloat[]{\includegraphics[scale=0.2]{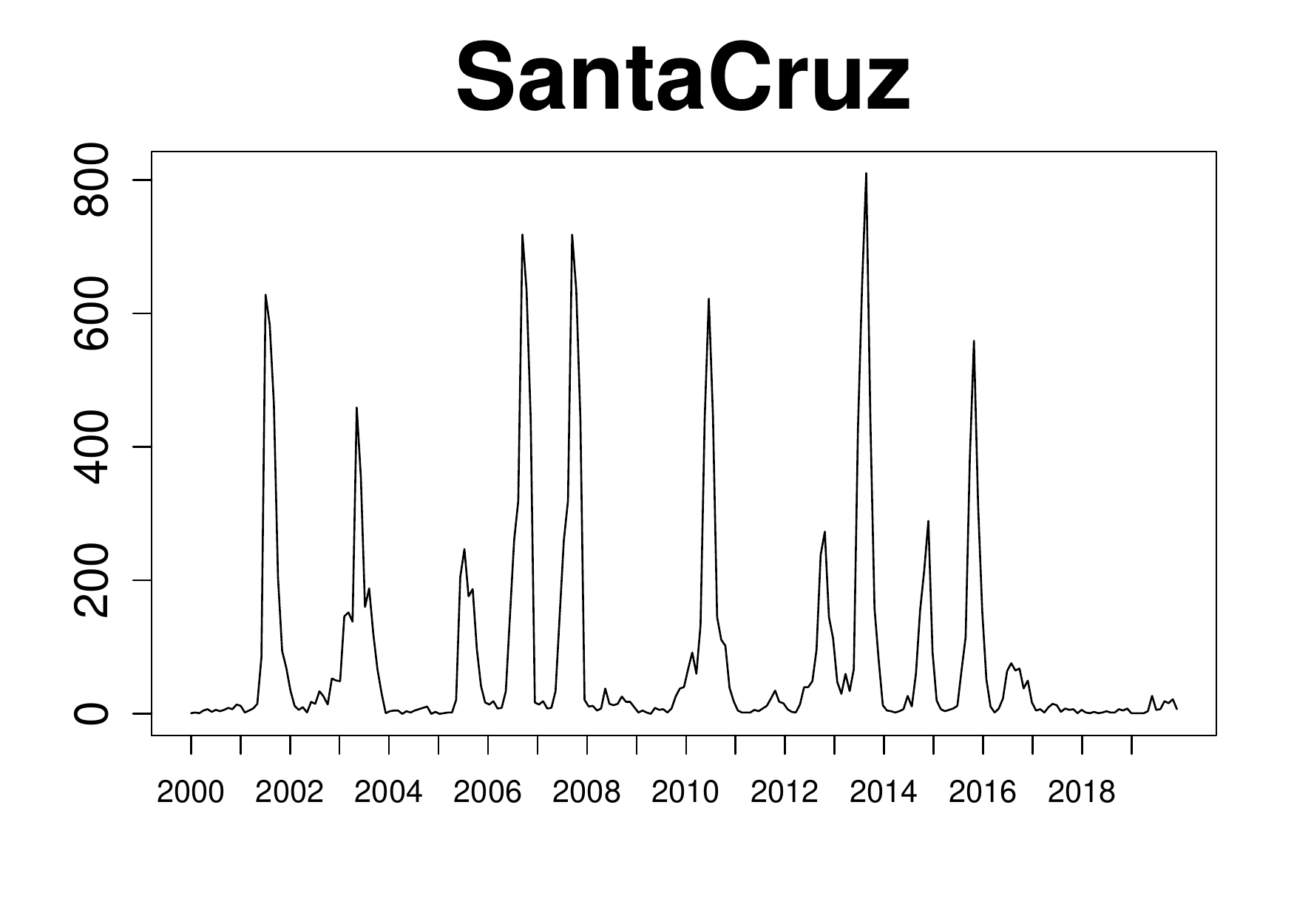}}\vspace{-0.15cm}%
\subfloat[]{\includegraphics[scale=0.2]{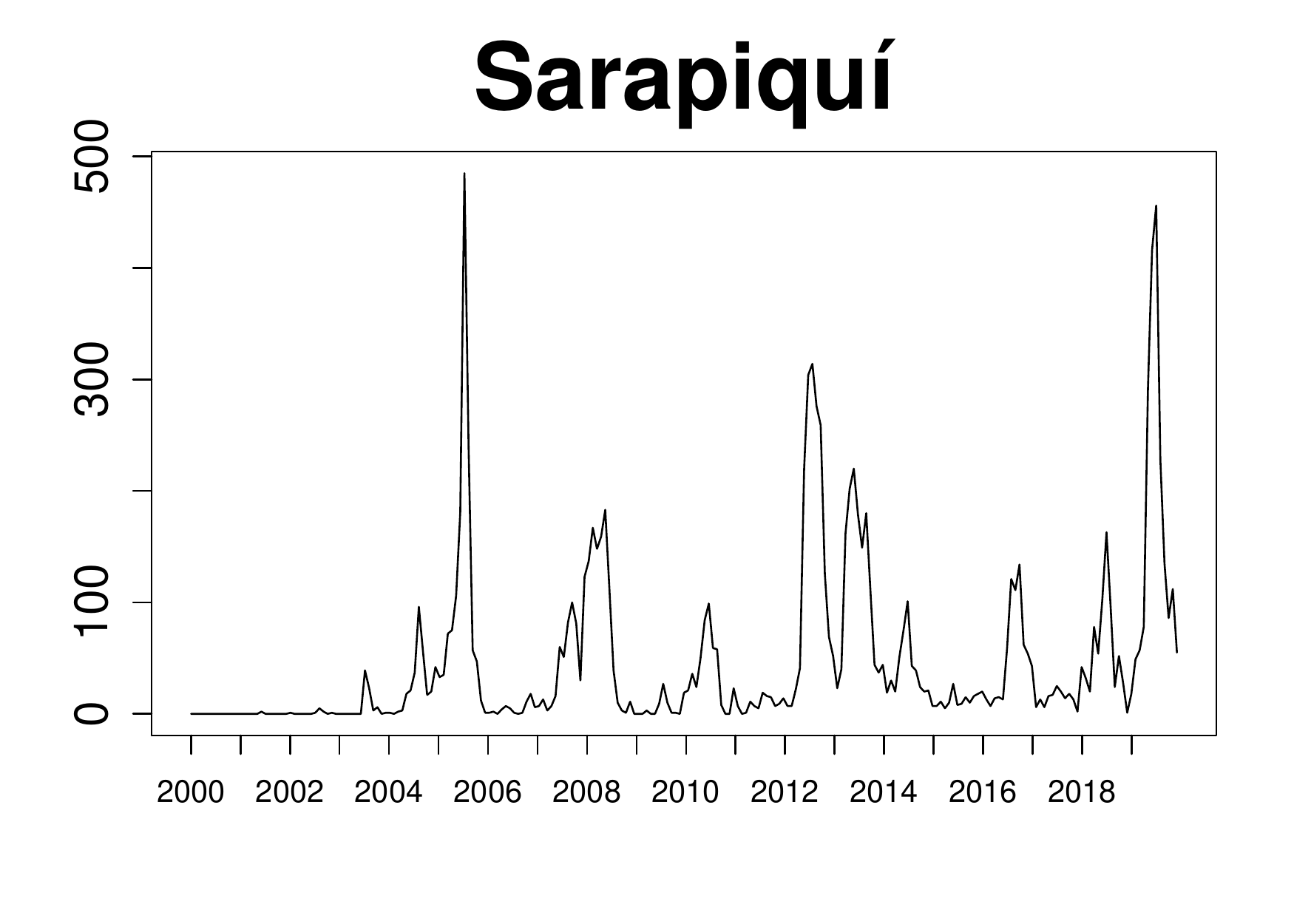}}\vspace{-0.15cm}\\
\subfloat[]{\includegraphics[scale=0.2]{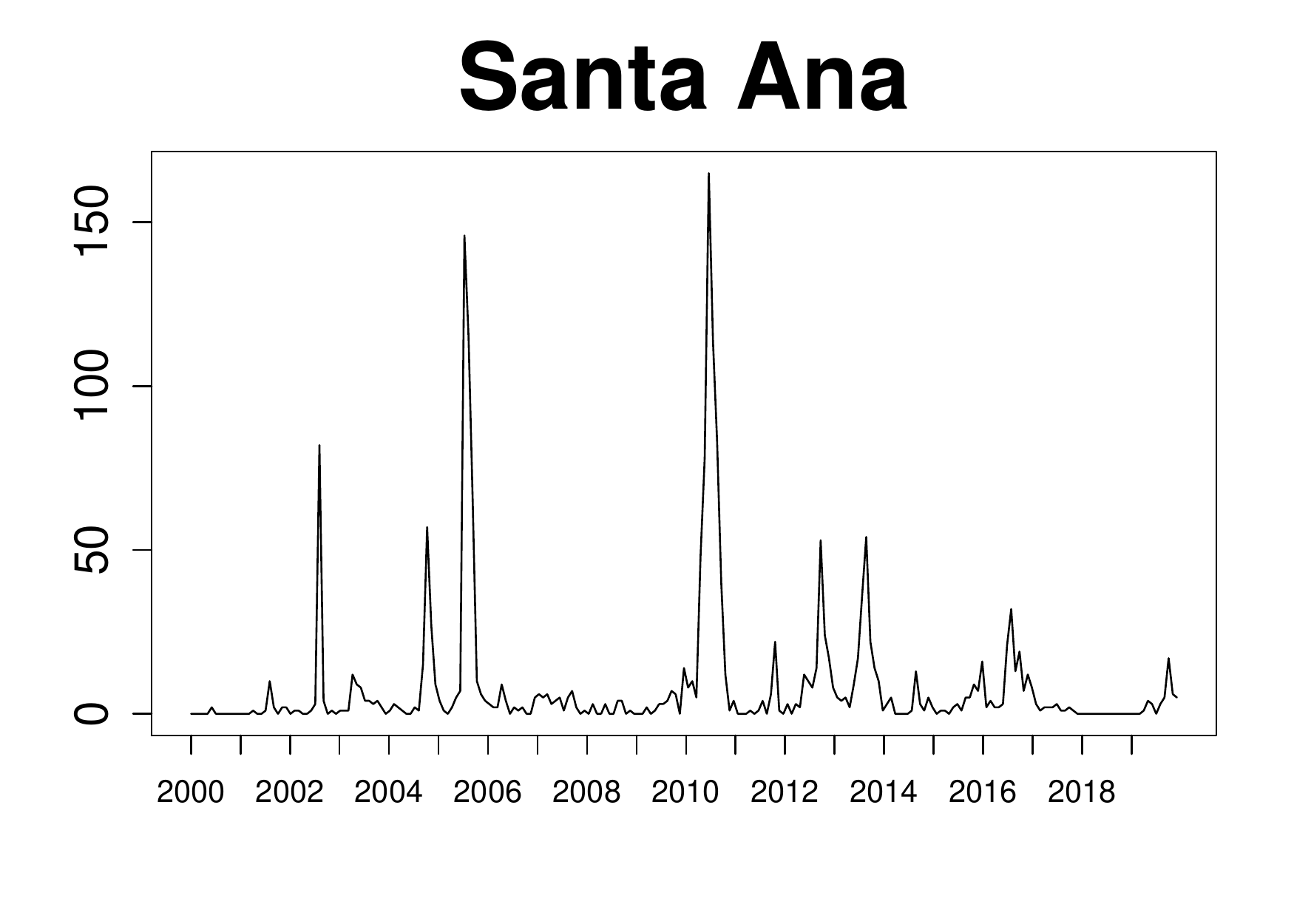}}\vspace{-0.15cm}%
\subfloat[]{\includegraphics[scale=0.2]{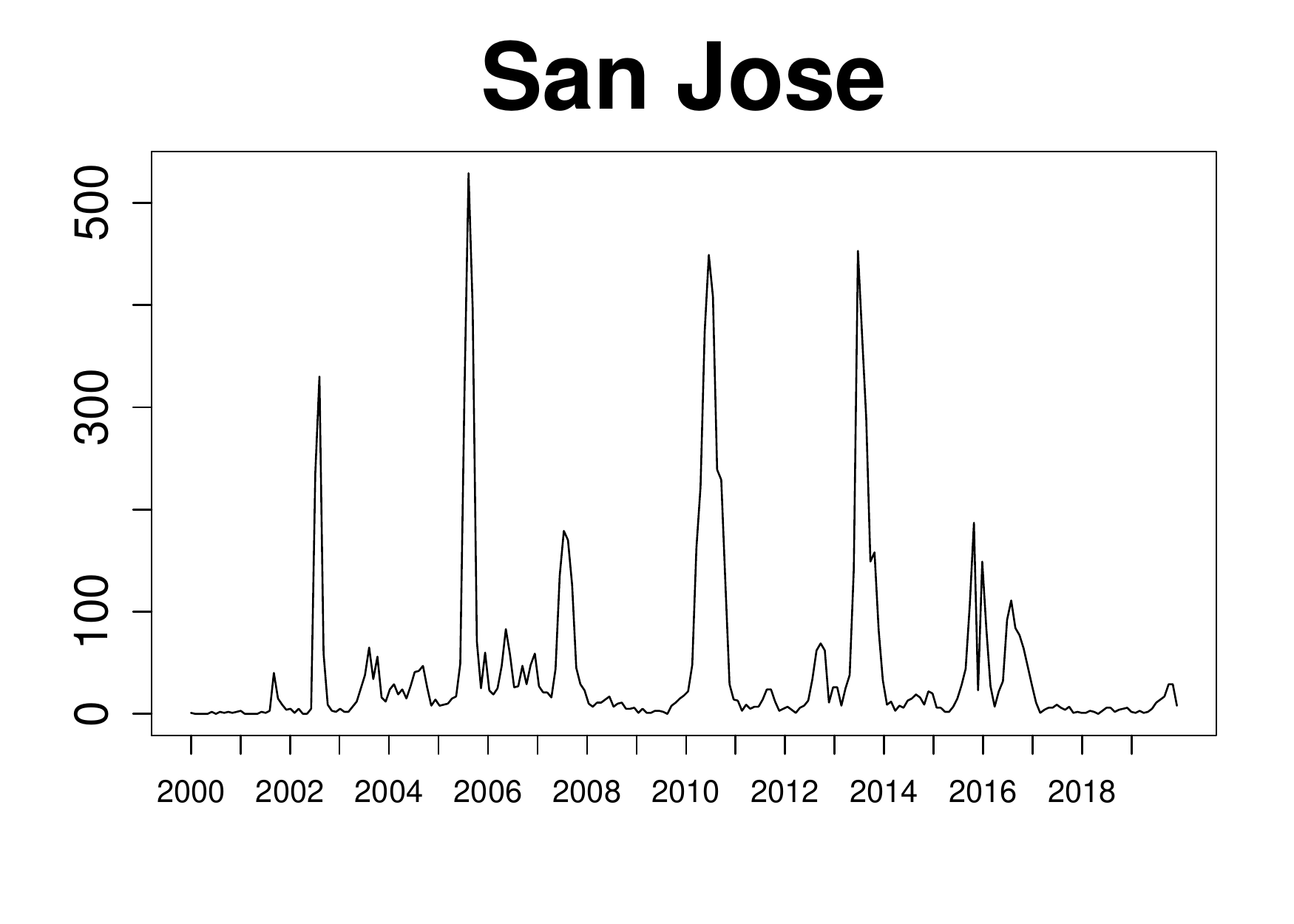}}\vspace{-0.15cm}
\caption{Time series of monthly data for dengue incidence from 2000 to 2019 in 32 municipalities of Costa Rica.}
\label{Fig:DengueCases}
\end{figure}


\subsection{Coherence and phase analyses – synchrony between dengue incidence, climate and vegetation variables}

For each time series, the wavelet power spectrum is first calculated to determine the dominant periods individually and then the coherence between each covariate and the incidence of dengue is calculated.

\subsubsection*{Enhanced Vegetation Index (EVI)}

The index of vegetation shows a continues band for the period of 1-yr in the municipalities of La Cruz, Esparza, Santa Cruz, San Jos\'e, Santa Ana, Perez Zeled\'on, Puntarenas, Carrillo, Ca\~nas, Atenas, Alajuelita, Alajuela, Nicoya, Orotina, Desamparados, and Liberia. On the other hand, Siquirries, Turrialba, Pococ\'i, Matina, Montes de Oro, Osa, Corredores, Guacimo, Lim\'on have short time interval whit a significant periodicity on the scale of 1-yr. Golfito, Talamanca, Parrita, Upala, Quepos, Garabito, and Sarapiqu\'i do not reach significance in any period when simulations are compared to the null hypothesis.

Wavelet coherence analysis shows that EVI is significantly associated with dengue incidence in the 1-yr periodic band, for all municipalities except Quepos, Parrita, Upala, Golfito, and Garabito. For these places where there is no marked annual seasonality for the vegetation indices (Turrialba, Matina, Osa, Corredores, Guacimo, Lim\'on, Talamanca, and Sarapiqu\'i) the correlation with dengue incidence present small areas of high significance (see Figure S1).  

Phase differences (see supplementary material) suggest that the annual wave of dengue and EVI are in phase in all localities with an approximate lag of 3 months. Something to highlight is that in Osa, a continuous band for the vegetation index is recorded in the period of 1-yr after 2011, and one year later, a band period with the same scale for the incidence of dengue is shown with a strong correlation with EVI from 2012 to 2017.
 
\subsubsection*{Normalized Difference Vegetation Index (NDVI)}

For NDVI, the wavelet analysis shows a marked and significant annual periodicity over time in the municipalities of La Cruz, Esparza, Santa Cruz, Santa Ana, Carrillo, Ca\~nas, Alajuela, Nicoya, Libera, Puntarenas, Atenas, and Orotina. For the rest of the localities, there is no correlation or the significant area is present in short periods throughout the time.

Analyzing the direct association between the incidence of dengue and the NDVI, a significant local coherence is observed in the 1-yr band for all municipalities listed above, where the annual periodicity of the NDVI is relevant. The correlation between dengue cases and DNVI for Siquirres, Desamparados, Peréz Zeledón, Pococí, Montes de Oro, Corredores, Sarapiquí, Guacimo, Talamanca, and Upala happen only in small interval of time. For Alajuenlita, Matina, Turrialba, Osa, Quepos, Parrita, Limón, Golfito, and Garabito there is not correlation (see Figure S2).

The phase difference (see supplementary material) reveals that the time series are in-phase. With total sync or an approximate delay of 3 months in Alajuela, Atenas, Ca\~nas, Carrillo, Desamparados, Esparza, La Cruz, Liberia, Nicoya, Puntarenas, Santa Ana, and Santa Cruz, where an increase in the vegetation index is followed by a rise in dengue cases.

\subsubsection*{Normalized Difference Water Index (NDWI)}

Similar to the results for EVI and NDVI, the wavelet coherence analysis for NDWI shows a high and significant coherence for the 1-yr periodic mode with the places located in the center and the North Pacific of the country (La Cruz, Esparza, Desamparados, Carrillo, Ca\~nas, Atenas, Alajuela, Alajuelita, Pococ\'i, Puntarenas, Montes de Oro, Nicoya, Orotina, Liberia, Santa Cruz, Santa Ana, and San José). For those places, the wavelet analysis of the NDWI time series shows a continuous high significant area in the period of one year throughout time. A weaker correlation is observed with the municipalities of Siquirres, Turrialba, Parrita, Corredores, Golfito, Sarapiquí, Matina, Limón, and Upala. There is not correlation in the municipalities of Talamanca, Guacimo, Garabito, and Quepos (see Figure S3). 

The two-time series are in phase with a delay time of zero (full synchronization) at three months. NDWI is the leading time series in cities located in the North Pacific and Central Regions. While the leading series is not clearly defined in most of the municipalities of the Atlantic and the South Pacific.

\begin{figure}[H]
\centering
\subfloat[EVI]{\label{fig1:EVI}\includegraphics[scale=0.35]{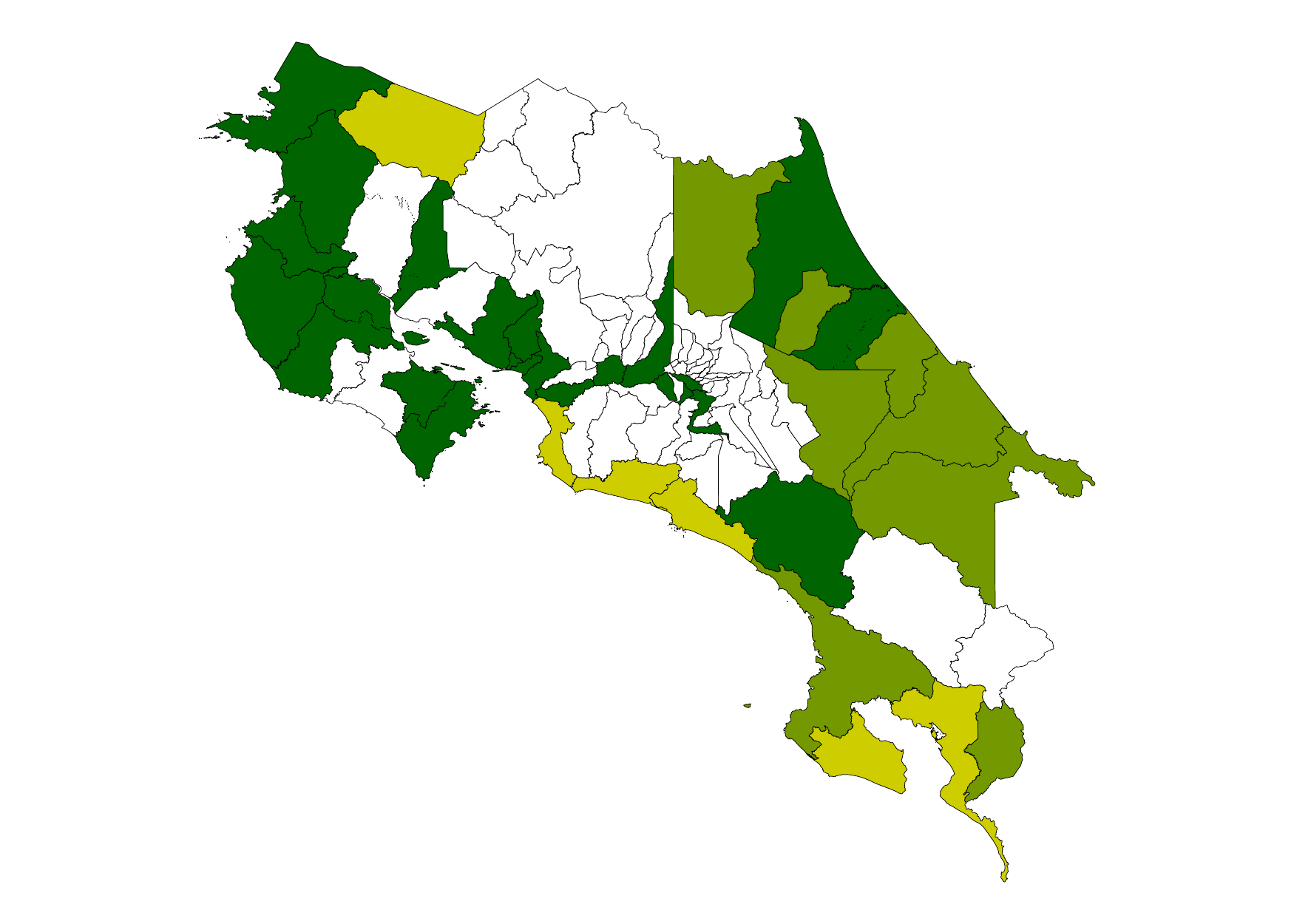}}\hspace*{-1.3cm}
\subfloat[NDVI]{\label{fig1:NDVI}\includegraphics[scale=0.35]{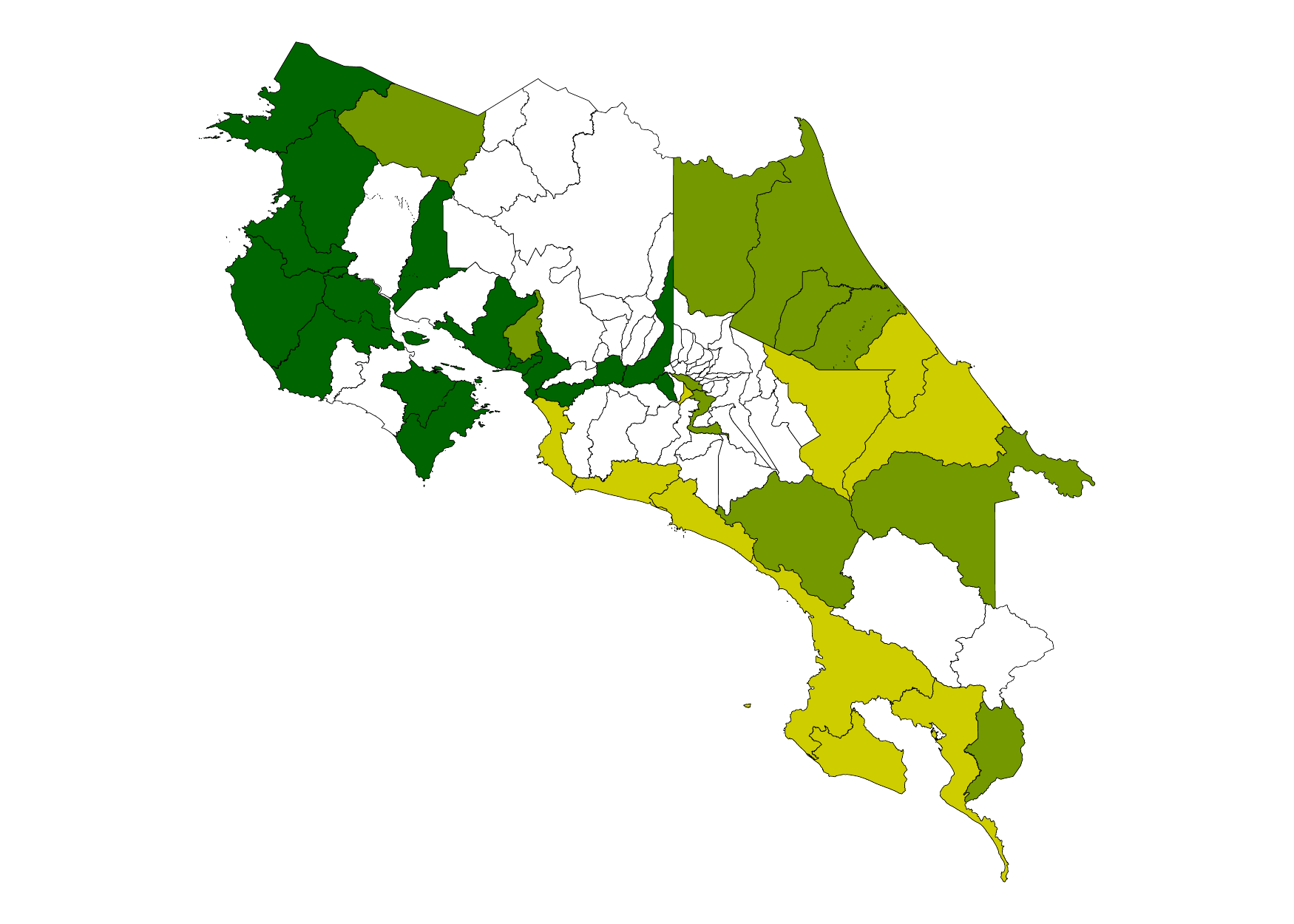}}\hspace*{-1.3cm}
\subfloat[NDWI]{\label{fig1:NDWI}\includegraphics[scale=0.35]{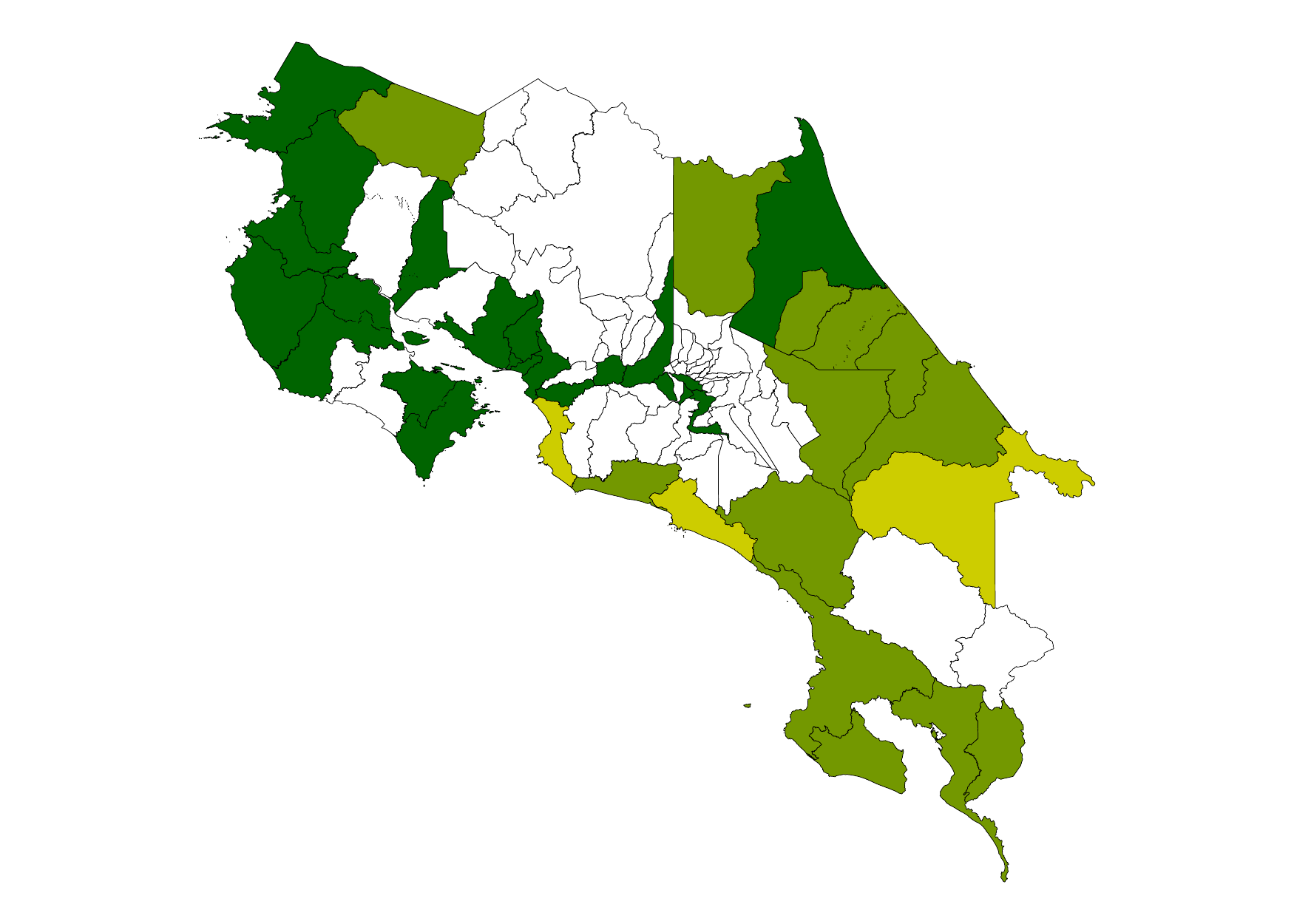}}
\caption{Map of the wavelet coherence between the incidence of dengue and the corresponding index (EVI, NDVI, and NDWI). Dark green corresponds to those places where the area of high significance is more continues over time. Light green corresponds to those places where the correlation is only in short periods and yellow to those where there are no significant areas.} 
\label{Map:VegetationIndices}
\end{figure}

\subsubsection*{Land Surface Temperature (LST)}

Wavelet power spectrum analysis shows a strong annual component for LSN and LSD in all municipalities. For LSN, a continuous band in the 1-yr is present in all cantons, while for LSD only a continuous band is observed for La Cruz, Esparza, Desamparados, Carrillo, Liberia, Osa, and Pococ\'i and a non-stationary behavior, in the same scale, for the rest of localities.

Wavelet coherence evinces a significant association of dengue incidence with seasonal temperature in the dominant wave of 1-yr. The lapse of time over which these patterns are correlated and are statistically significant varied across the time and geographic locations (see Figures S4-S5). 

The phase difference for the 1-yr band indicates that land surface temperature and dengue incidence signal are out of phase. The positive peak of dengue lags behind the negative peak of the land surface temperature with a difference of 3 (in-phase) to 6 (completely mirror images of each other) months. This pattern is regular in all municipalities over time, except for Lim\'on, Corredores, Ca\~nas, Atenas, Perez Zeled\'on, Quepos, Parrita, Garabito, and Upala, where the incidence of dengue leads over LSD, LSN or both (supplementary material).

\subsubsection*{Tropical North Atlantic (TNA)}

\begin{figure}[H]
\centering
\subfloat{\includegraphics[scale=0.4]{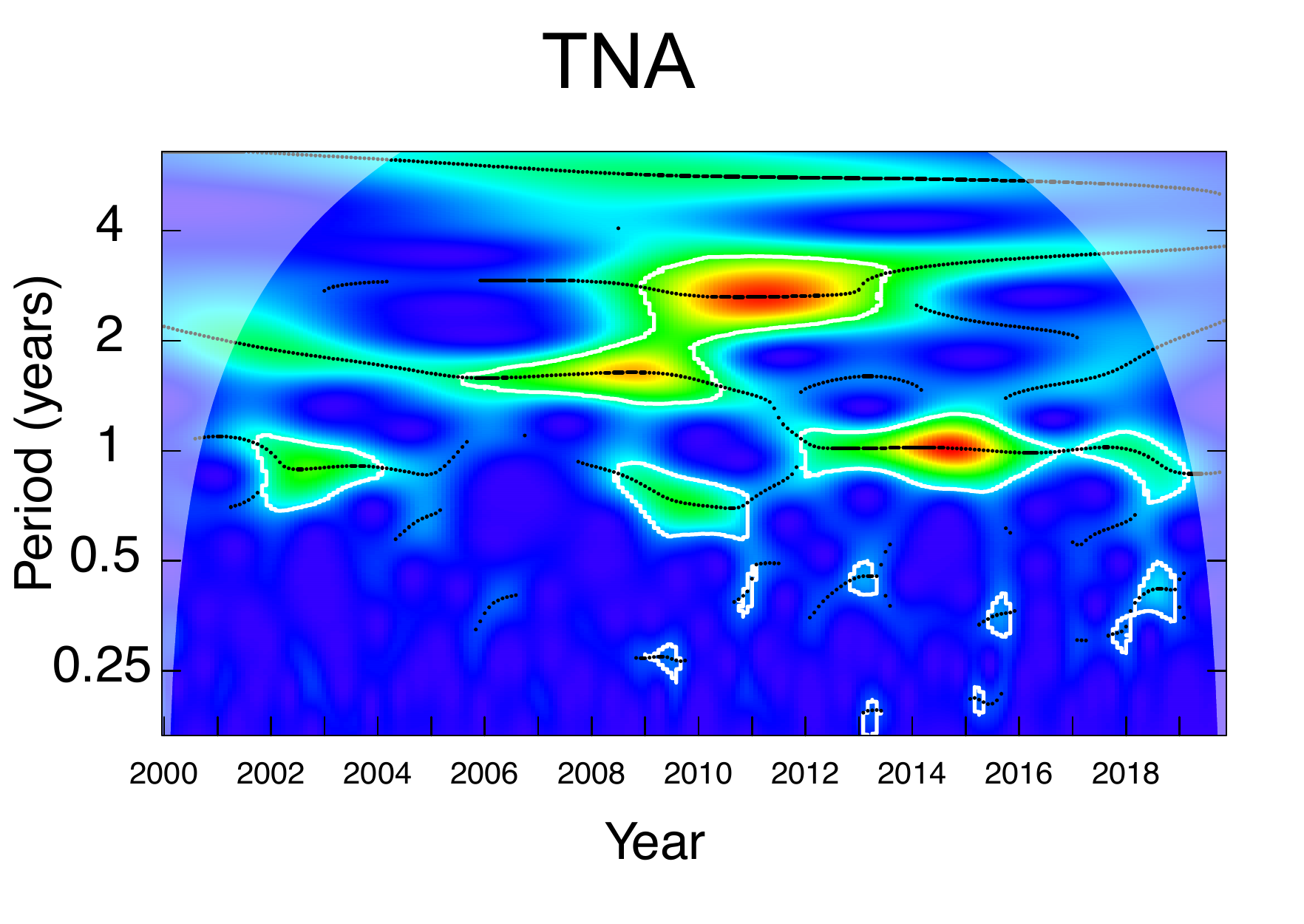}}
\subfloat{\includegraphics[scale=0.4]{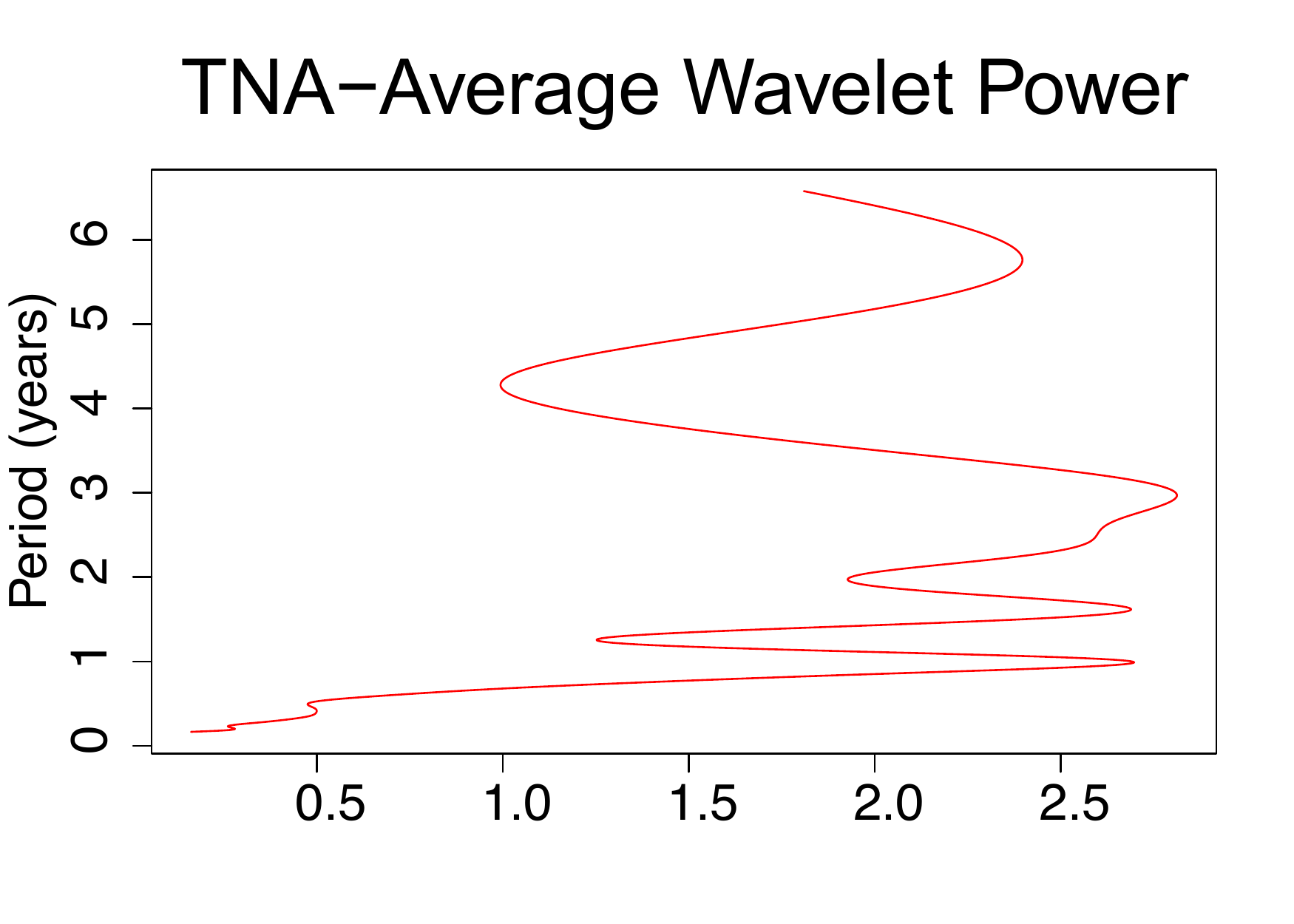}}
\caption{ (Left) Wavelet power spectrum of Tropical North Atlantic index from 2000 to 2019 (periodicity on y axis, time on x axis). Data to compute The wavelet power spectrum of TNA is standardize; colors code for increasing power intensity, from blue to red; $95\%$ confidence level are encircled by white lines, and shaded areas indicate the presence of significant edge effects. (Right) Average wavelet power spectrum.}
\label{Fig:Wavelet_TNA}
\end{figure}

The wavelet power spectrum for the TNA data shows a dominant period in the 1-yr, 18-month (1.5-yr), 3-yr, and 6-yr range (see Figure \ref{Fig:Wavelet_TNA}). The period of 6-yr should be taken carefully due to the series length. 

The wavelet coherence analysis shows a significant correlation between the TNA index and the incidence of dengue in all municipalities (except for Matina and Quepos), especially in the band corresponding to the periods of 1 and 3-ys (see Figure S6). After 2011, there is a significant correlation between the time series in the one-year band for all localities except Alajuelita and Garabito. 

For Desamparados, Santa Cruz, San Jos\'e, Santa Ana, P\'erez Zeled\'on, Carrillo, Atenas, Alajuelita, Alajuela, Nicoya, Orotina, Osa, Montes de Oro, Golfito, Upala, Garabito, and Sarapiqu\'i, the phase difference reveals that the dengue wave is in perfect synchronization with the TNA wave in the 3-yr band from 2009 to 2014, approximately. For all locations, TNA and dengue incidence signals are in phase, but the lagging and leading wave changes with geographic location and over time (see supplementary material).

\subsubsection*{Sea surface temperatures (ENSO SST)}

\begin{figure}[H]
\captionsetup[subfigure]{labelformat=empty}
\centering
\subfloat[]{\includegraphics[scale=0.22]{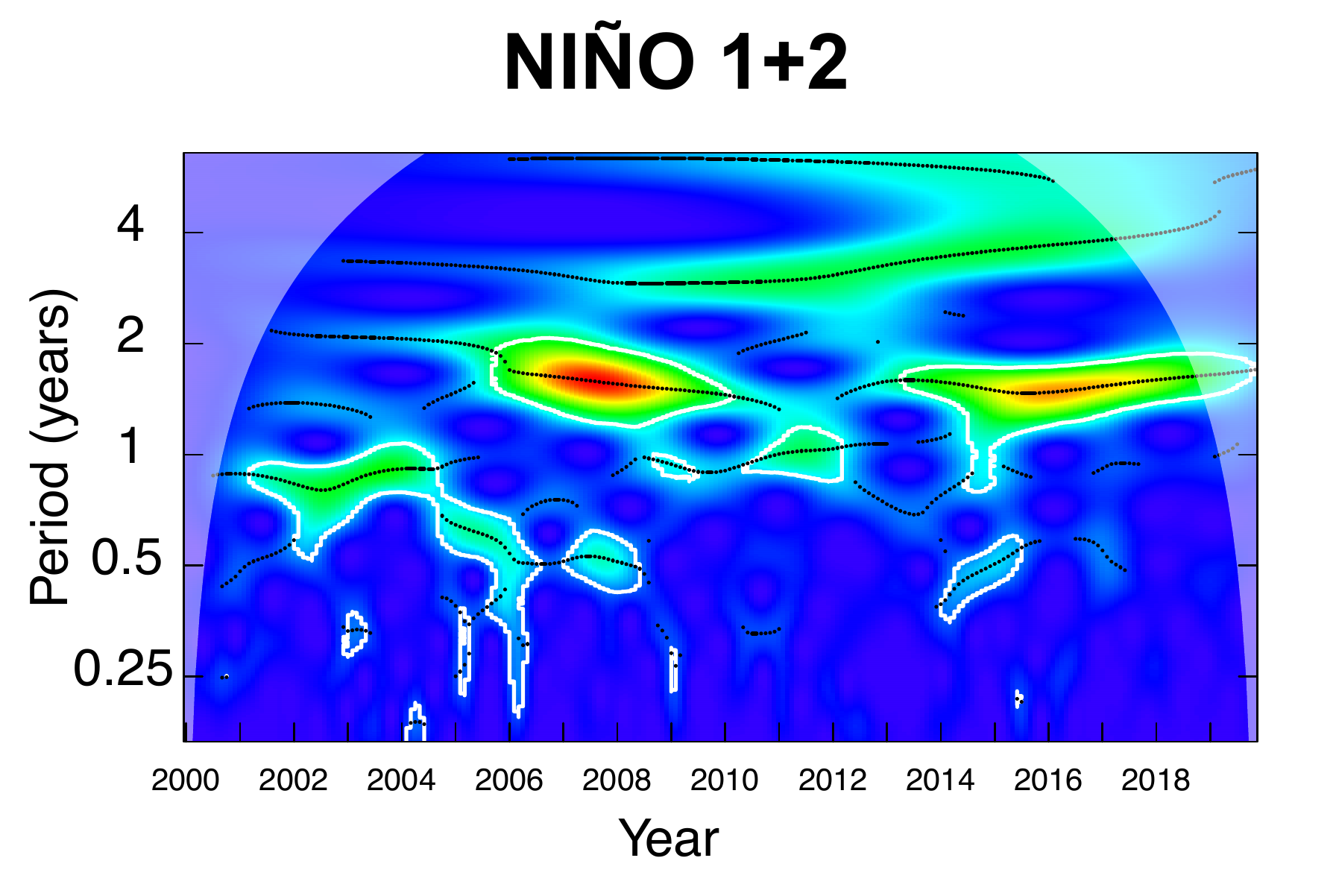}}\vspace{-0.15cm}%
\subfloat[]{\includegraphics[scale=0.22]{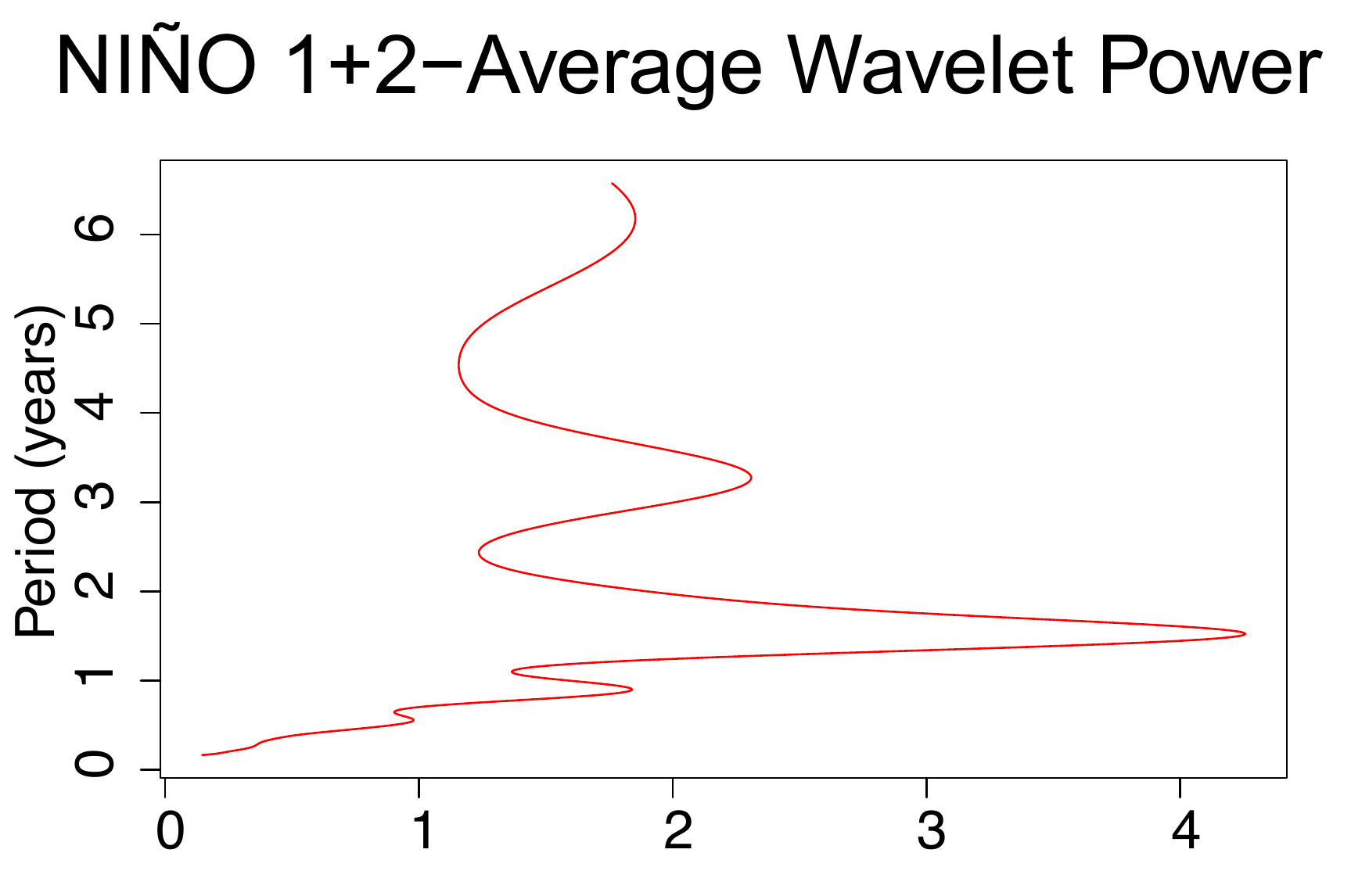}}\vspace{-0.15cm}%
\subfloat[]{\includegraphics[scale=0.22]{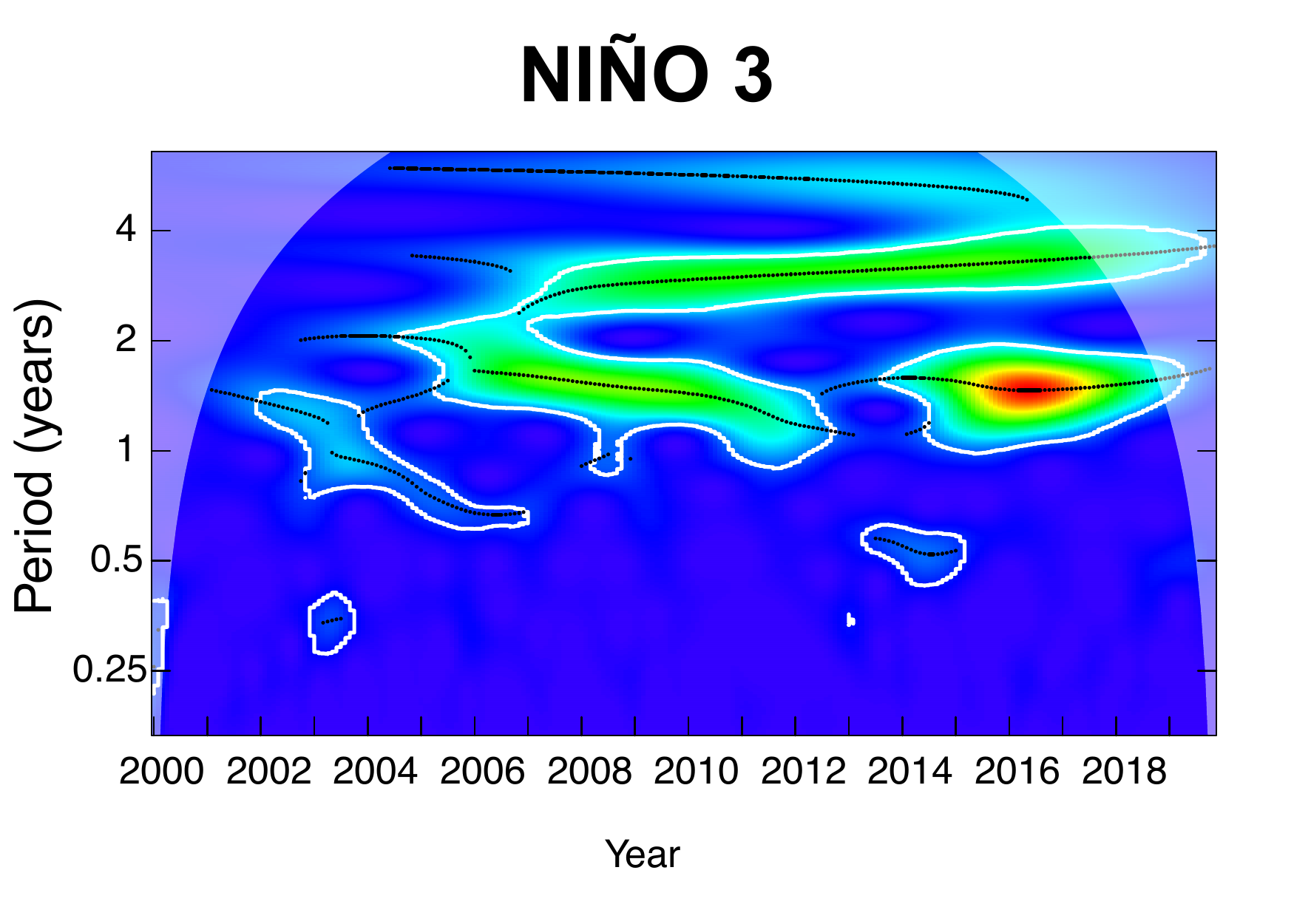}}\vspace{-0.15cm}%
\subfloat[]{\includegraphics[scale=0.22]{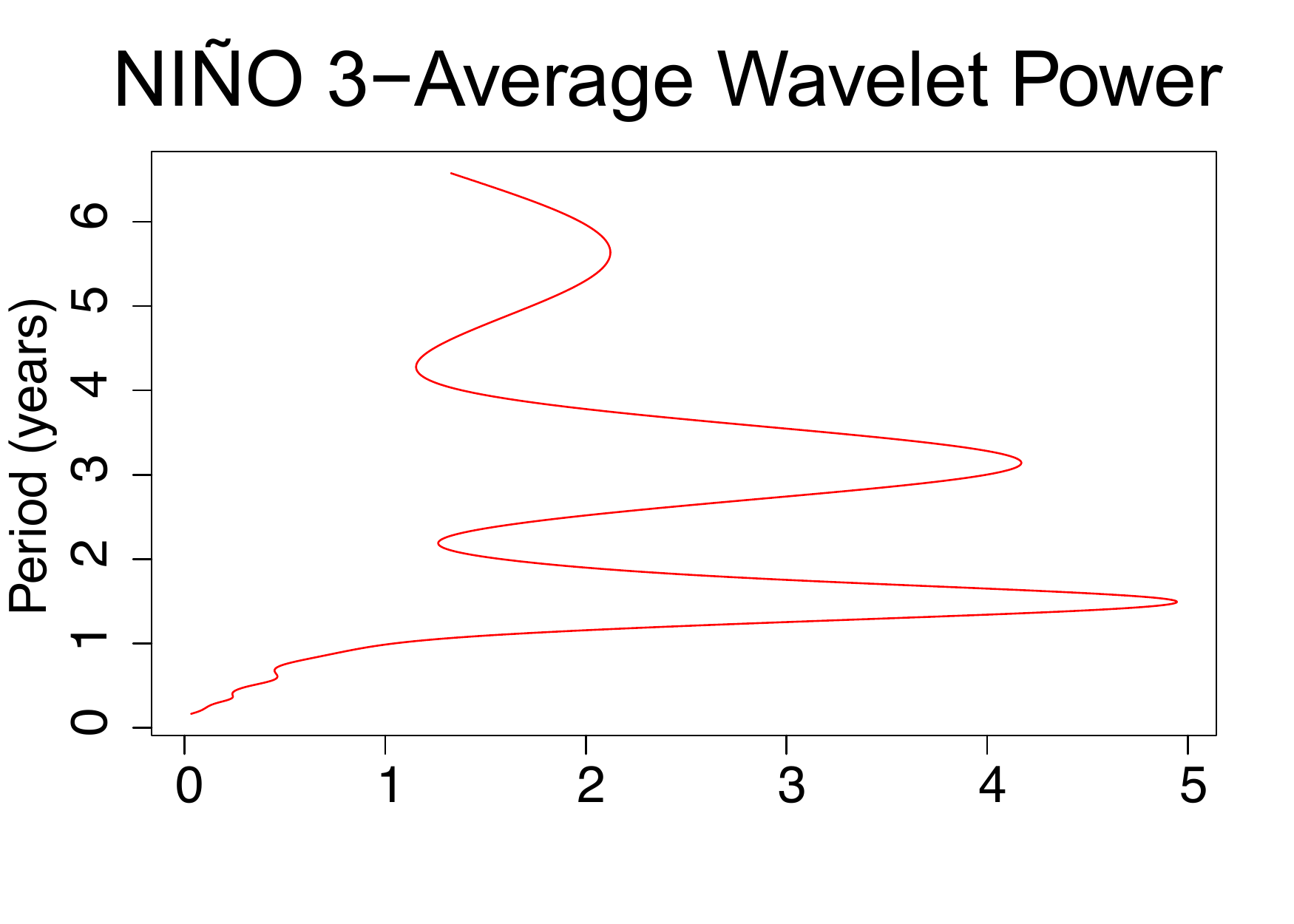}}\vspace{-0.15cm}\\
\subfloat[]{\includegraphics[scale=0.22]{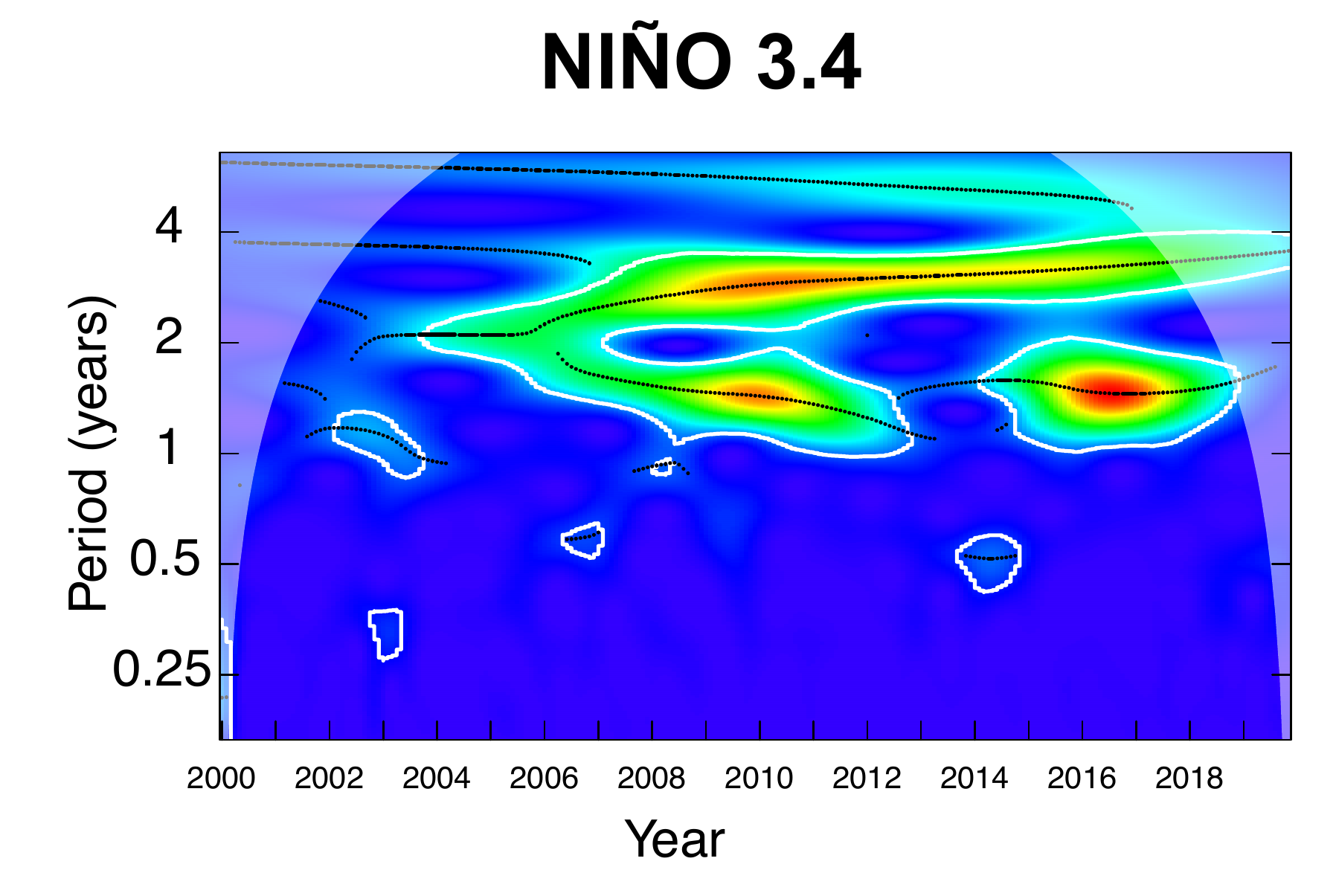}}\vspace{-0.15cm}%
\subfloat[]{\includegraphics[scale=0.22]{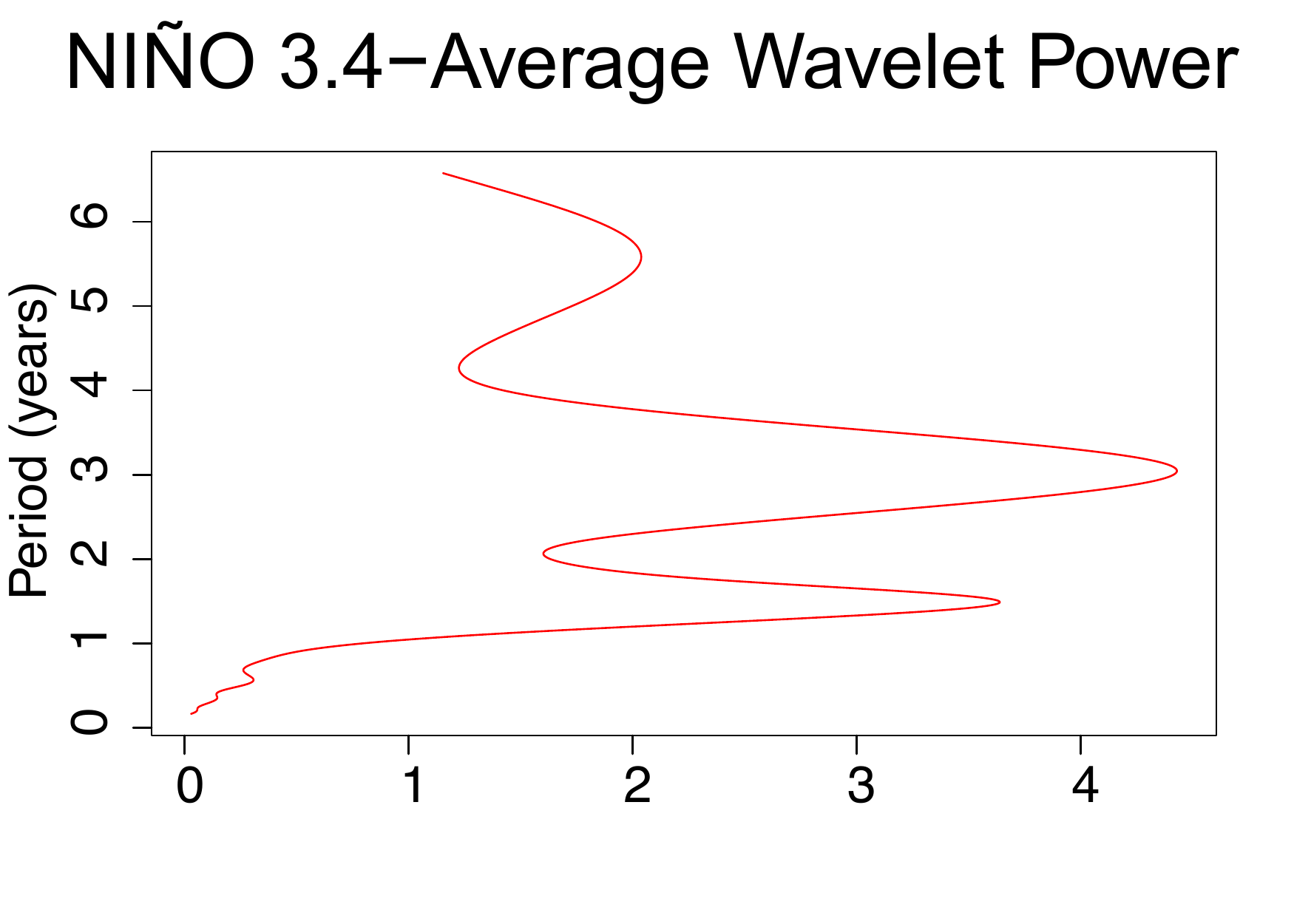}}\vspace{-0.15cm}%
\subfloat[]{\includegraphics[scale=0.22]{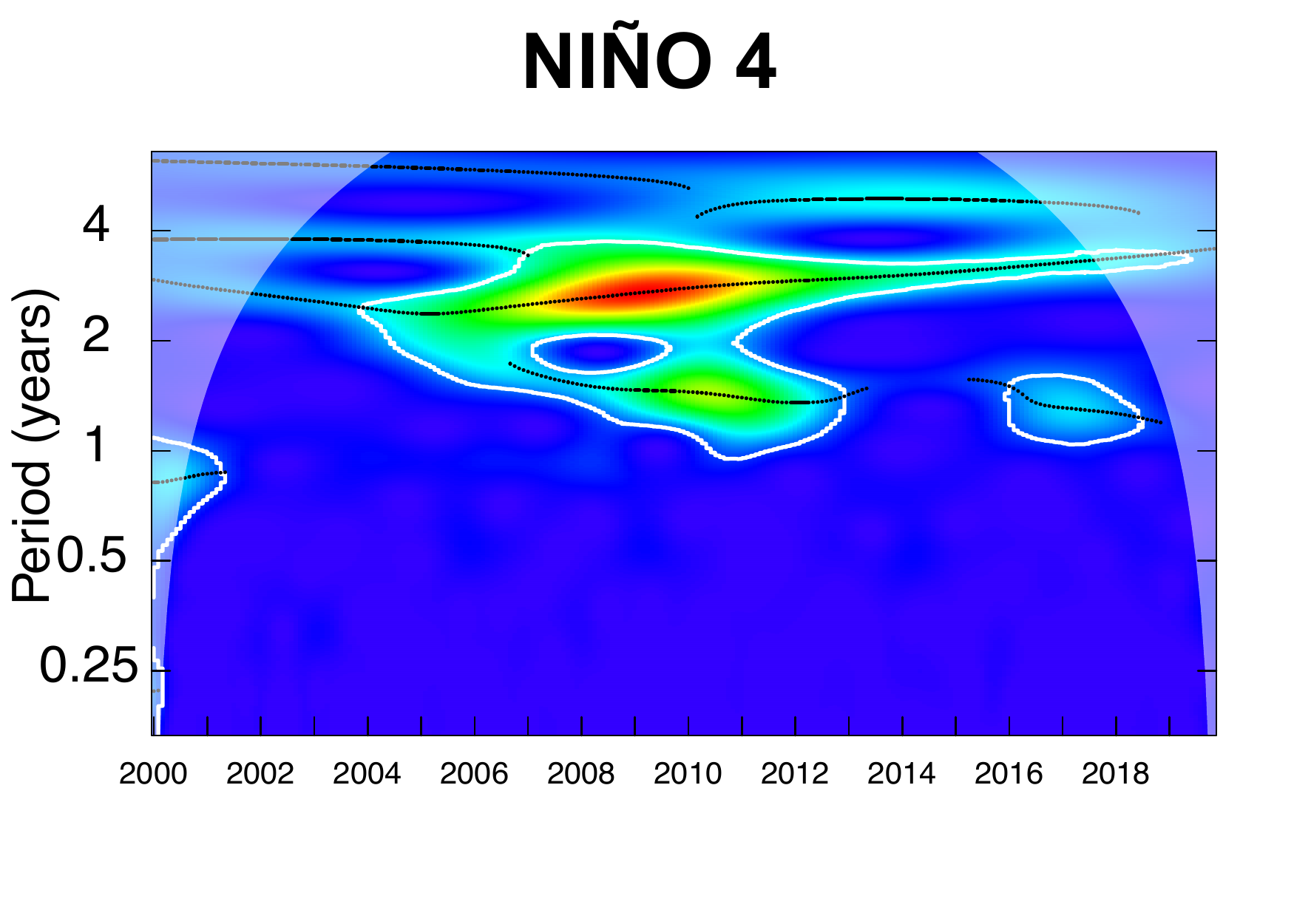}}
\vspace{-0.15cm}%
\subfloat[]{\includegraphics[scale=0.22]{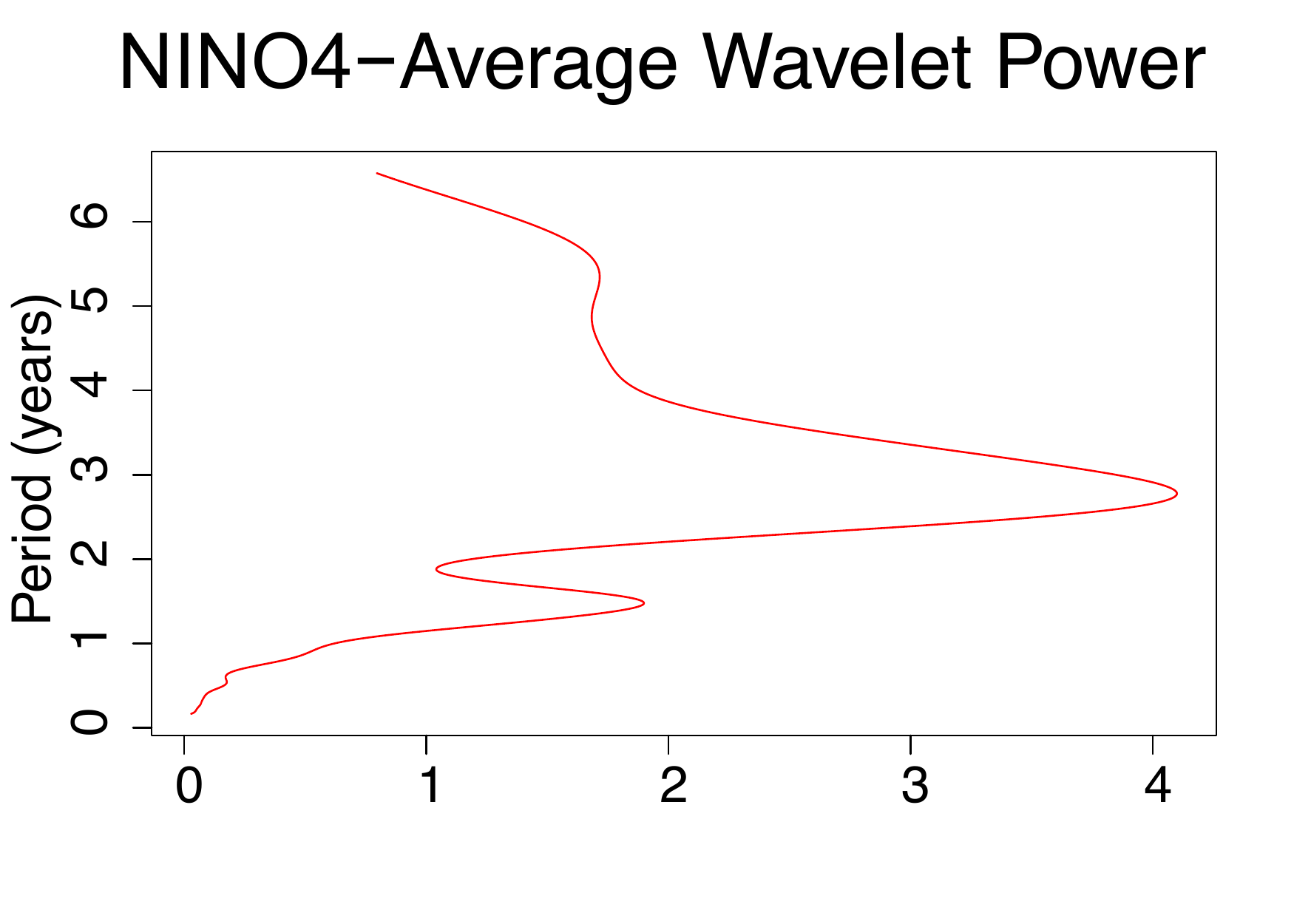}}
\vspace{-0.15cm}
\caption{(Color Maps) Wavelet power spectrum of Sea surface temperatures (ENSO SST) indeces from 2000 to 2019 (periodicity on y axis, time on x axis). Data to compute the wavelet power spectrum of ENSO SST indices is standardize; colors code for increasing power intensity, from blue to red; $95\%$ confidence level are encircled by white lines, and shaded areas indicate the presence of significant edge effects. (Red line) Average wavelet power spectrum for El Ni\~no 1+2, El Ni\~no, El Ni\~no 3.4, and El Ni\~no 4, respectively.}             
\label{fig:Wavelet_ENSO}
\end{figure}

The wavelet power spectrum was performed for the four different Ni\~no indices: Ni\~no 1 + 2, Ni\~no 3, Ni\~ no 3.4, Ni\~ no 4 (see Figure \ref{fig:Wavelet_ENSO}). The results show that the dominant non-stationary peaks, for the four indices, are in the 1-yr, 1.5-yr, and 3-yr bands (see Figure \ref{fig:Wavelet_ENSO}).

\begin{figure}[H]
\centering
\subfloat[El Ni\~no 3]{\label{fig1:NINO3}\includegraphics[scale=0.32]{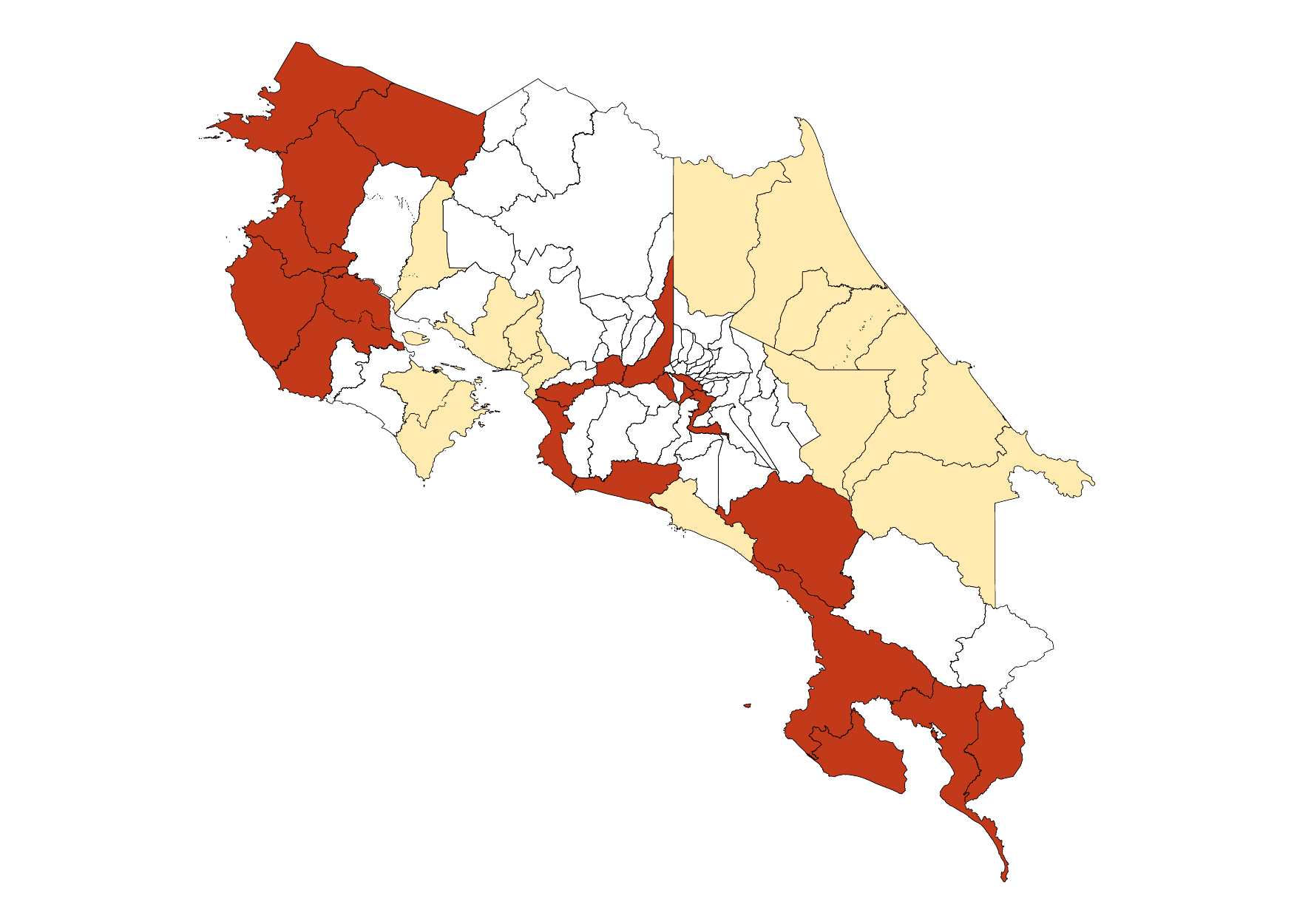}}
\subfloat[El Ni\~no 3.4]{\label{fig1:NINO34}\includegraphics[scale=0.32]{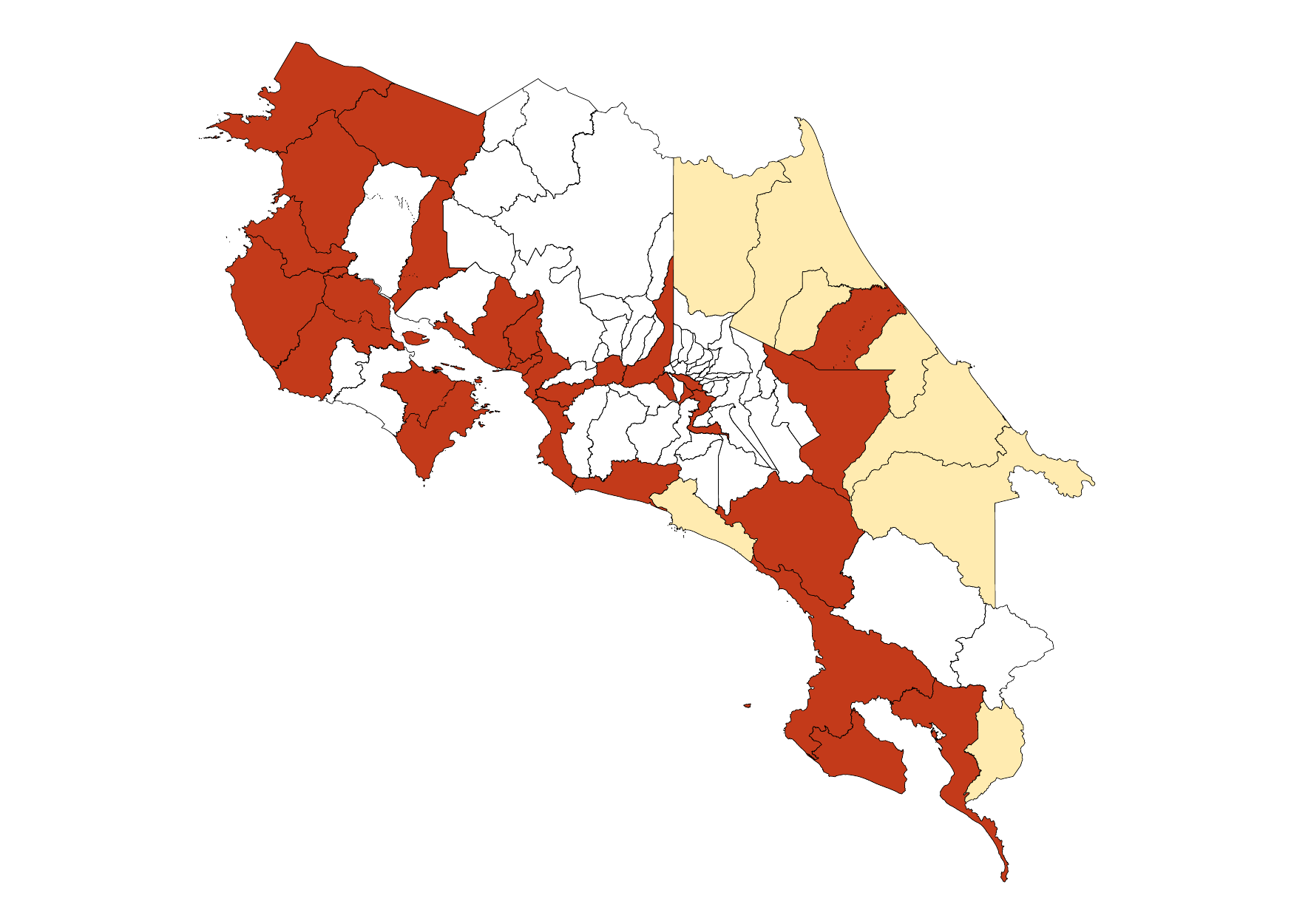}}
\subfloat[El Ni\~no 4]{\label{fig1:NINO4}\includegraphics[scale=0.32]{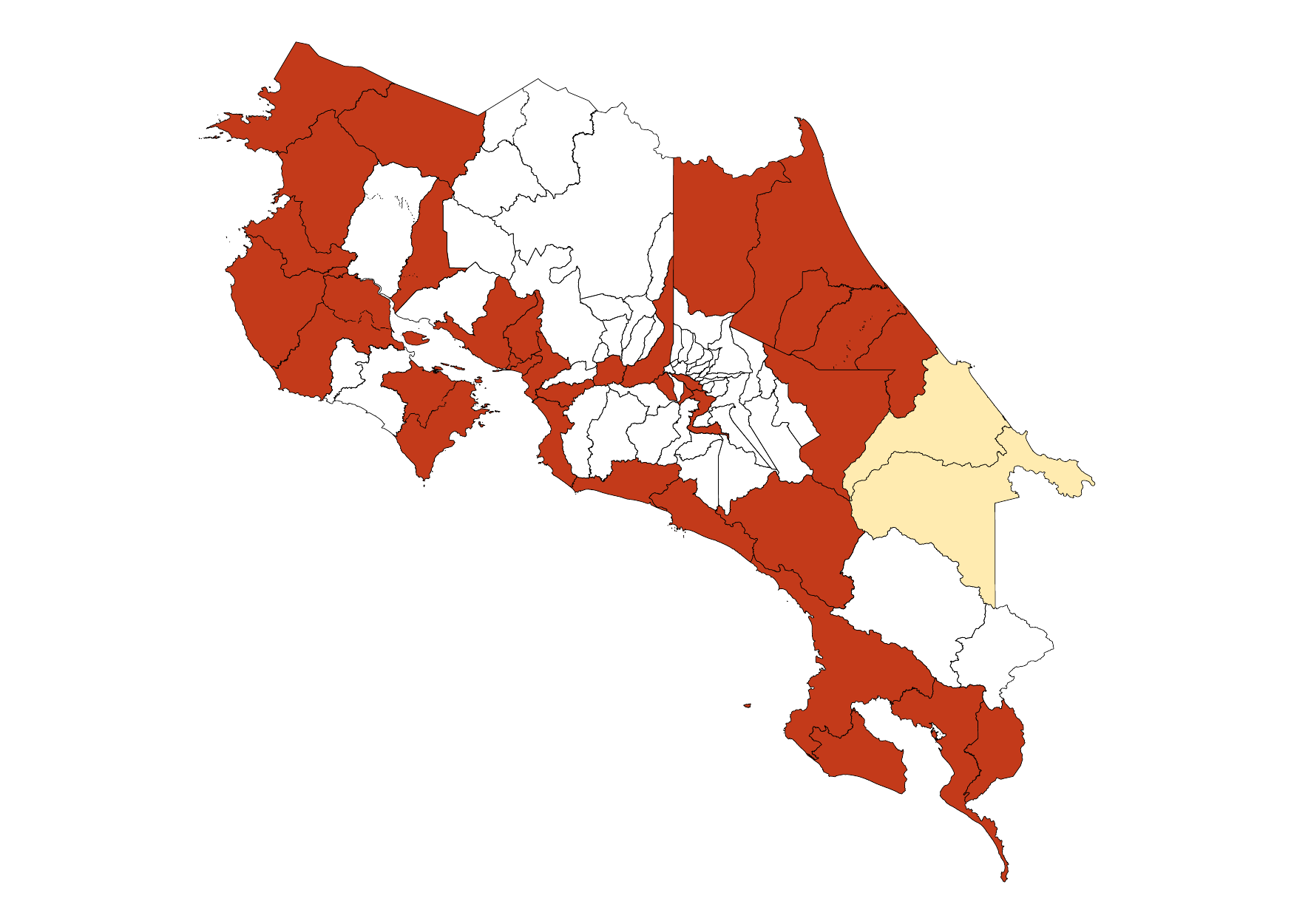}}
\caption{Map of the wavelet coherence between the incidence of dengue and the corresponding index (El Ni\~no 3, El Ni\~no 3.4, El Ni\~no 4). Red color corresponds to those localities that have an area of high significance in the band of 3-yr. The cream color corresponds to those places where there is a significant area for periods different to 3-yr.} 
\label{fig:Map2}
\end{figure}

The coherence between the incidence of dengue and the Ni\~no 1 + 2 index is not clear. There is possible to identify small areas of significance between this and the dengue incidence in some places in the Pacific (Liberia, Puntarenas, Esparza, Garabito, Montes de Oro, Nicoya, Orotina and Carrillo) and the Atlantic (Matina, Siquirris, Pococ\'i, and Lim\'on) in the band of 1 year, from 2001 to 2007 approximately. On the other hand, Ni\~no 3, 3.4, and 4 are correlated with dengue incidence in the period of 1, 1.5, 2, and 3-yr. The correlation is non-stationary and vary over time and geographically. However, there is a strong correlation between those indices and dengue incidence in most of the localities in the period of 3-yr (see Figure \ref{fig:Map2}). 


\begin{table}
\centering
\begin{adjustbox}{width=1\textwidth}
\begin{tabular}{llllllllll}
\rowcolor{gray}
\hline
\centering \textbf{Canton} & \centering\textbf{EVI} & \centering \textbf{NDVI} & \centering \textbf{NDWI}& \textbf{LSD}& \textbf{LSN}&\textbf{TNA} & \textbf{NI\~NO 3}&\centering \textbf{NI\~NO 34}& \textbf{NI\~NO 4}\\
\hline
La Cruz& x&	x&	x& x& x&P1 & P3& P3& P3\\
\rowcolor{gray!8}
Esparza   & x&	x& x& x& x& P1 & -- & P3 & P3\\
Puntarenas&	x&	x& x& x& x&	P1 & -- & P3 & P3\\
\rowcolor{gray!8}
Carrillo  &	x&	x& x& x& x&	P1 & P3 & P3 & P3\\
Cañas	  & x&	x& x& x& x&	P1 & -- & P3 & P3\\
\rowcolor{gray!8}
Atenas	  & x&	x& x& x& x&	P1 & P3 & P3 & P3\\
Alajuela  & x&	x& x& x& x&	P1 & P3 & P3 & P3\\
\rowcolor{gray!8}
Nicoya	  & x&	x& x& x& x&	P1 & P3 & P3 & P3\\
Orotina   & x&	x& x& x& x&	P1 & P3 & P3 & P3\\
\rowcolor{gray!8}
Liberia	  & x&	x& x& x& x&	P1 & P3 & P3 & P3\\
Santa Cruz& x&	x& x& x& x&	P1 & P3 & P3 & P3\\
\rowcolor{gray!8}
Santa Ana &	x&	x& x& x& x&	P1 & P3 & P3 & P3\\
San José  & x& 2001/2007-2013/2017&  x& x& x& P1 & P3 & P3 & P3\\
\rowcolor{gray!8}
Alajuelita &x &  & x& x& x& -- & P3 & P3 & P3\\
Matina	   &2010/2013 &  &  & x& x&    & -- & -- & P3\\
\rowcolor{gray!8}
Turrialba &2010/2014&  & 2004/2008-2011/2015& x& x&	P1 & -- & P3 &	P3\\
Siquirres & x& 2011/2013 &  2003/2008-2010/2014& x& x&	P1& -- &	P3&	P3\\
\rowcolor{gray!8}
Desamparados &x	&2012/2017 &	x&	x&	x&	P1 & P3 & P3& P3\\
Perez Zeledón&x	&2004/2006-2013/2018& x & x & x&	P1 & P3& P3& P3\\
\rowcolor{gray!8}
Pococí&	x& 2001/2003-2011/2014&	x&	x&	x&P1& -- & --& P3\\
Montes de oro &x& 2000/2003-2013/2017& x& x& x&	P1 & --& P3&P3\\
\rowcolor{gray!8}
Corredores &2005/2007 - 2013/2017&	2013/2015&	2005/2008-2015/2017& x & x &	P1&	P3&	--&	P3\\
Limón &	2010/2013  && 2007/2013&	x &	x &P1 & -- & --	& --\\
\rowcolor{gray!8}
Talamanca &	2011/2014&	2011/2014&  & x& x& P1&	P1 &-- & --\\
Sarapiquí&	2011/2014& 2012/2015& 2011/2015& x&	x& P1& --&--&P3\\
\rowcolor{gray!8}
Guacimo	&2010/2015&	2002-2011/2014&	2010/2014&	x&	x&	P1&	--&	--&	P3\\
Osa	&2012/2017 & & 2013/2016 & x&x&P1 &	P3	& P3 &	P3\\
\rowcolor{gray!8}
Quepos  & &  & &x &x & & -- & -- &	P3\\
Parrita	& &  &2013/2014	& x & x & P1 &	P3 & P3 &P3\\
\rowcolor{gray!8}
Upala  & &2013 & 2012/2015 & x& x&P1 &P3&P3&P3\\
Golfito	& & & 2013/2017	   & x& x &P1& P3& P3&P3\\
\rowcolor{gray!8}
Garabito& &  & & x & x	&--	& P3& P3&	P3\\
\hline
\end{tabular}
\end{adjustbox}
\caption{(x) represent those places where there is a significant correlation between the incidence of dengue and the corresponding index. When the correlation time interval is short, the dates for the area of high significance are indicated. P1 represents places where there is a notable correlation in the 1-yr band for the incidence of dengue and TNA after 2011. P3 corresponds to those places where there is a strong correlation between the ENSO variables and the incidence of dengue in a band of 3-yr.}
\label{tab:SummaryResults1}
\end{table}

\begin{table}
\centering
\caption{Synchronization and lag results summary.}
\begin{adjustbox}{width=1\textwidth}
\begin{tabular}{|p{4cm}|p{15cm}|p{3cm}|p{3cm}|}
\rowcolor{gray}
\hline
\centering \textbf{Index} & \centering\textbf{Period and Phase} & \multicolumn{2}{c}{\textbf{Lag}}\\
\hline
\begin{minipage}[t]{\linewidth}%
\vspace{0.2cm}
EVI \\
NDVI\\
NDWI 
\vspace{0.2cm}
\end{minipage}
&
\begin{minipage}[t]{\linewidth}%
\vspace{0.2cm}
\centering
Vegetation indices showed a strong correlation with dengue incidence in the 1-yr period. The time series move in-phase. 
\end{minipage}
&
\multicolumn{2}{c|}{
\begin{minipage}[t]{6cm}
\vspace{0.2cm}
The lag is approximately 3 months, however, there are periods when both signals are in sync.
\end{minipage}}\\
\hline
\vspace{0.2cm}
LST (LSN-LSD)
&
\begin{minipage}[t]{\linewidth}%
\vspace{0.2cm}
LSN and LSD are correlated with dengue indices in the period of 1-yr. Time series move out of phase.
\vspace{0.2cm}
\end{minipage}
&
\multicolumn{2}{c|}{
3-6 Months}\\
\hline
\vspace{0.2cm}
TNA
&
\begin{minipage}[t]{\linewidth}%
\vspace{0.2cm}
The TNA series has dominant periods of 1, 1.5, and 3 years. It is in phase with the time series of dengue cases and there is a strong correlation with those in the periods of 1-yr and 3-yrs.
\vspace{0.2cm}
\end{minipage}
&
\vspace{0.2cm}
\centering
3 Months in the band of 1-yr
&
\vspace{0.2cm}
\centering\arraybackslash
6 Months in the band of 3-yr\\
\cline{1-4} 
\rowcolor{gray!8}
\vspace{0.2cm}
ENSO SST
&
\begin{minipage}[t]{\linewidth}%
\vspace{0.2cm}
\centering
Non-stationary period in the band of 1-yr and 3 yr.
\vspace{0.2cm}
\end{minipage}
&
\centering
Lag in the period of 1-yr.
&
\centering\arraybackslash
Lag in the period of 3-yr. \\
\cline{1-4} 
Ni\~no 3   & & 3 Months & 15 Months\\
\cline{3-4}
Ni\~no 4   & & 3-6 Months & 15 Months\\
\hline
\end{tabular}
\end{adjustbox}
\label{tab:SummaryResults2}
\end{table}

\section{Discussion}
\label{discussion}

Wavelet analysis was performed over monthly time series of dengue incidence and six different satellite climate and vegetation variables in Costa Rica (i) to detect and quantify dengue periodicity in 32 different cantons and (ii) to associate the relationship between dengue incidence, climate and vegetation variables in those places.

Wavelet power spectra showed that the dominant periods of dengue in the 32 locations in Costa Rica are in the 1, 2, and 3-yr bands. These periods are not stationary; they vary over time and space. After 2008, a high area of significance in the period of 3-yr appears in most places which, according to the wavelet coherence analysis, dengue incidence correlates with the Ni\~no 3, 3.4, 4, and TNA index in the same period.

The wavelet coherence analysis shows a strong correlation between the vegetation indices (EVI, NDVI, and NDWI) and the incidence of dengue in cantons located mainly in the center or North Pacific of the country. For cantons in the Atlantic or the South Pacific, the areas of significance are small or do not exist. The time series are in phase with a lag of 3 months, but there are some periods where the two series are in sync. The LSN and LSD indices are strongly correlated with the incidence of dengue in all municipalities. The time series are out-phase with a lag ranging from 3 to 6 months. The TNA is correlated with dengue cases in the periods of 1-yr and 3-ys. After 2011, there is a marked correlation in the 1-year band for all places except Matina, Quepos, Alajuelita, and Garabito. Ni\~no 3, 3.4, and 4 show a significant correlation in different band periods, but the correlation in the 3-yr band is dominant in most localities. On the other hand, the correlation between the Niño 1 + 2 index and the incidence of dengue is less clear.

Weather and climate influence disease ecology on many levels creating complex dynamics. Although it is known that dengue dynamics can be affected by various climatic and vegetation variables, not all of them must be present for successful spread. Human factors, such as behavior, immunity, and socioeconomic factors, could also contribute to dengue spread and complexity. This analysis showed that even in the same country, the variables correlated with dengue cases could change, highlighting the importance of analyzing a localized way.

As far as we know, this is the first study for Costa Rica that incorporates a wide variety of climatic variables and considers the cantons of interest to the health authorities. This type of analysis provides a better understanding of the local dynamics of diseases, which allows for improved prediction models and health interventions at specific sites. It can be enriched with the incorporation of new socioeconomic variables and with a cluster approach, which allows characterizing the regions and identifying the conditions and characteristics shared by the different cantons.

\bibliography{Bibliography}{}

\begin{thebibliography}{10}

\bibitem{aghabozorgi2015time}
Saeed Aghabozorgi, Ali~Seyed Shirkhorshidi, and Teh~Ying Wah.
\newblock Time-series clustering--a decade review.
\newblock {\em Information Systems}, 53:16--38, 2015.

\bibitem{barrera2006ecological}
Roberto Barrera, Manuel Amador, and Gary~G Clark.
\newblock Ecological factors influencing aedes aegypti (diptera: Culicidae)
  productivity in artificial containers in \uppercase{S}alinas,
  \uppercase{P}uerto \uppercase{R}ico.
\newblock {\em Journal of medical entomology}, 43(3):484--492, 2006.

\bibitem{carmona1998practical}
Ren{\'e} Carmona, Wen-Liang Hwang, and Bruno Torresani.
\newblock {\em Practical Time-Frequency Analysis: Gabor and wavelet transforms,
  with an implementation in S}.
\newblock Academic Press, 1998.

\bibitem{cazelles2014wavelet}
Bernard Cazelles, K{\'e}vin Cazelles, and Mario Chavez.
\newblock Wavelet analysis in ecology and epidemiology: impact of statistical
  tests.
\newblock {\em Journal of the Royal Society Interface}, 11(91):20130585, 2014.

\bibitem{cazelles2008wavelet}
Bernard Cazelles, Mario Chavez, Dominique Berteaux, Fr{\'e}d{\'e}ric
  M{\'e}nard, Jon~Olav Vik, St{\'e}phanie Jenouvrier, and Nils~C Stenseth.
\newblock Wavelet analysis of ecological time series.
\newblock {\em Oecologia}, 156(2):287--304, 2008.

\bibitem{cazelles2007time}
Bernard Cazelles, Mario Chavez, Guillaume Constantin~de Magny, Jean-Francois
  Gu{\'e}gan, and Simon Hales.
\newblock Time-dependent spectral analysis of epidemiological time-series with
  wavelets.
\newblock {\em Journal of the Royal Society Interface}, 4(15):625--636, 2007.

\bibitem{cazelles2005nonstationary}
Bernard Cazelles, Mario Chavez, Anthony~J McMichael, and Simon Hales.
\newblock Nonstationary influence of el nino on the synchronous dengue
  epidemics in thailand.
\newblock {\em PLoS medicine}, 2(4), 2005.

\bibitem{chaves2021modeling}
Luis~Fernando Chaves, Jos{\'e} Angel~Valer{\'\i}n Cordero, Gabriela Delgado,
  Carlos Aguilar-Avenda{\~n}o, Ezequ{\'\i}as Maynes, Jos{\'e}
  Manuel~Guti{\'e}rrez Alvarado, Melissa~Ram{\'\i}rez Rojas, Luis~Mario Romero,
  and Rodrigo~Mar{\'\i}n Rodr{\'\i}guez.
\newblock Modeling the association between aedes aegypti ovitrap egg counts,
  multi-scale remotely sensed environmental data and arboviral cases at
  puntarenas, costa rica (2017--2018).
\newblock {\em Current Research in Parasitology and Vector-Borne Diseases},
  1:100014, 2021.

\bibitem{chaves2006climate}
Luis~Fernando Chaves and Mercedes Pascual.
\newblock Climate cycles and forecasts of cutaneous leishmaniasis, a
  nonstationary vector-borne disease.
\newblock {\em PLoS medicine}, 3(8), 2006.

\bibitem{chowell2011influence}
Gerardo Chowell, Bernard Cazelles, Helene Broutin, and Cesar~V Munayco.
\newblock The influence of geographic and climate factors on the timing of
  dengue epidemics in per{\'u}, 1994-2008.
\newblock {\em BMC infectious diseases}, 11(1):164, 2011.

\bibitem{christophrs1960aedes}
Sir~SR Christophrs et~al.
\newblock A{\"e}des aegypt{\`\i} (l.) the yellow fever mosquito; its life
  history, bionomics and structure.
\newblock {\em A{\"e}des aegypt{\`\i} (L.) the yellow fever mosquito; its life
  history, bionomics and structure.}, 1960.

\bibitem{cummings2004travelling}
Derek~AT Cummings, Rafael~A Irizarry, Norden~E Huang, Timothy~P Endy, Ananda
  Nisalak, Kumnuan Ungchusak, and Donald~S Burke.
\newblock Travelling waves in the occurrence of dengue haemorrhagic fever in
  thailand.
\newblock {\em Nature}, 427(6972):344--347, 2004.

\bibitem{cuong2016quantifying}
Hoang~Quoc Cuong, Nguyen~Tran Hien, Tran~Nhu Duong, Tran~Vu Phong, Nguyen~Nhat
  Cam, Jeremy Farrar, Vu~Sinh Nam, Thai~TD Khoa, and Peter Horby.
\newblock Quantifying the emergence of dengue in hanoi, vietnam: 1998--2000.
\newblock {\em PLoS Negl Trop Dis}, 9(5), 2016.

\bibitem{cuong2013spatiotemporal}
Hoang~Quoc Cuong, Nguyen~Thanh Vu, Bernard Cazelles, Maciej~F Boni, Khoa~TD
  Thai, Maia~A Rabaa, Luong~Chan Quang, Cameron~P Simmons, Tran~Ngoc Huu, and
  Katherine~L Anders.
\newblock Spatiotemporal dynamics of dengue epidemics, southern vietnam.
\newblock {\em Emerging infectious diseases}, 19(6):945, 2013.

\bibitem{d2012wavelets}
Pierpaolo D'Urso and Elizabeth~Ann Maharaj.
\newblock Wavelets-based clustering of multivariate time series.
\newblock {\em Fuzzy Sets and Systems}, 193:33--61, 2012.

\bibitem{NDWI}
{E}arth~{O}bserving {S}ystem.
\newblock Ndvi.
\newblock \url{https://eos.com/ndwi/}.

\bibitem{ebi2016dengue}
Kristie~L Ebi and Joshua Nealon.
\newblock Dengue in a changing climate.
\newblock {\em Environmental research}, 151:115--123, 2016.

\bibitem{ehelepola2015study}
NDB Ehelepola, Kusalika Ariyaratne, WMNP Buddhadasa, Sunil Ratnayake, and
  Malani Wickramasinghe.
\newblock A study of the correlation between dengue and weather in kandy city,
  sri lanka (2003-2012) and lessons learned.
\newblock {\em Infectious diseases of poverty}, 4(1):1--15, 2015.

\bibitem{estallo2012effectiveness}
Elizabet~Lilia Estallo, Francisco~Felipe Luduena-Almeida, Andr{\'e}s~Mario
  Visintin, Carlos~Marcelo Scavuzzo, Mario~Alberto Lamfri, Mar{\'\i}a~Virginia
  Introini, Mario Zaidenberg, and Walter~Ricardo Almir{\'o}n.
\newblock Effectiveness of normalized difference water index in modelling aedes
  aegypti house index.
\newblock {\em International journal of remote sensing}, 33(13):4254--4265,
  2012.

\bibitem{CDG}
National~Center for Atmospheric~Research.
\newblock {Climate {D}ata {G}uide}.
\newblock
  \url{https://climatedataguide.ucar.edu/climate-data/nino-sst-indices-nino-12-3-34-4-oni-and-tni}.

\bibitem{gao1996ndwi}
Bo-Cai Gao.
\newblock Ndwi—a normalized difference water index for remote sensing of
  vegetation liquid water from space.
\newblock {\em Remote sensing of environment}, 58(3):257--266, 1996.

\bibitem{GisGeography}
GISGeography.
\newblock {W}hat is {NDVI} ({N}ormalized {D}ifference {V}egetation index?
\newblock
  \url{https://gisgeography.com/ndvi-normalized-difference-vegetation-index/},
  febrary 2018.

\bibitem{hanley2003quantitative}
Deborah~E Hanley, Mark~A Bourassa, James~J O'Brien, Shawn~R Smith, and
  Elizabeth~R Spade.
\newblock A quantitative evaluation of enso indices.
\newblock {\em Journal of Climate}, 16(8):1249--1258, 2003.

\bibitem{johansson2009multiyear}
Michael~A Johansson, Derek~AT Cummings, and Gregory~E Glass.
\newblock Multiyear climate variability and dengue—el nino southern
  oscillation, weather, and dengue incidence in puerto rico, mexico, and
  thailand: a longitudinal data analysis.
\newblock {\em PLoS medicine}, 6(11), 2009.

\bibitem{jury2008climate}
Mark~R Jury.
\newblock Climate influence on dengue epidemics in puerto rico.
\newblock {\em International journal of environmental health research},
  18(5):323--334, 2008.

\bibitem{kolivras2010changes}
Korine~N Kolivras.
\newblock Changes in dengue risk potential in hawaii, usa, due to climate
  variability and change.
\newblock {\em Climate Research}, 42(1):1--11, 2010.

\bibitem{lacaux2007classification}
JP~Lacaux, YM~Tourre, Cecile Vignolles, JA~Ndione, and M~Lafaye.
\newblock Classification of ponds from high-spatial resolution remote sensing:
  Application to rift valley fever epidemics in senegal.
\newblock {\em Remote Sensing of Environment}, 106(1):66--74, 2007.

\bibitem{matsushita2007sensitivity}
Bunkei Matsushita, Wei Yang, Jin Chen, Yuyichi Onda, and Guoyu Qiu.
\newblock Sensitivity of the enhanced vegetation index (evi) and normalized
  difference vegetation index (ndvi) to topographic effects: a case study in
  high-density cypress forest.
\newblock {\em Sensors}, 7(11):2636--2651, 2007.

\bibitem{morin2013climate}
Cory~W Morin, Andrew~C Comrie, and Kacey Ernst.
\newblock Climate and dengue transmission: evidence and implications.
\newblock {\em Environmental health perspectives}, 121(11-12):1264--1272, 2013.

\bibitem{mudele2021modeling}
Oladimeji Mudele, Alejandro~C Frery, Lucas~FR Zanandrez, Alvaro~E Eiras, and
  Paolo Gamba.
\newblock Modeling dengue vector population with earth observation data and a
  generalized linear model.
\newblock {\em Acta Tropica}, 215:105809, 2021.

\bibitem{nagao2008decreases}
Yoshiro Nagao and Katia Koelle.
\newblock Decreases in dengue transmission may act to increase the incidence of
  dengue hemorrhagic fever.
\newblock {\em Proceedings of the National Academy of Sciences},
  105(6):2238--2243, 2008.

\bibitem{nakhapakorn2020assessment}
Kanchana Nakhapakorn, Warisara Sancharoen, Auemphorn Mutchimwong, Supet
  Jirakajohnkool, Rattapon Onchang, Chawarat Rotejanaprasert, Kraichat
  Tantrakarnapa, and Richard Paul.
\newblock Assessment of urban land surface temperature and vertical city
  associated with dengue incidences.
\newblock {\em Remote Sensing}, 12(22):3802, 2020.

\bibitem{nitatpattana2007potential}
Narong Nitatpattana, Pratap Singhasivanon, Honda Kiyoshi, Haja Andrianasolo,
  Sutee Yoksan, J~Gonzalez, and Philippe Barbazan.
\newblock Potential association of dengue hemorrhagic fever incidence and
  remote senses land surface temperature, thailand, 1998.
\newblock {\em Southeast Asian journal of tropical medicine and public health},
  38(3):427, 2007.

\bibitem{TNA}
The~State of~the Ocean~Climate.
\newblock Tropical northern atlantic index (tna).
\newblock \url{https://stateoftheocean.osmc.noaa.gov/sur/atl/tna.php}.

\bibitem{PAHO1}
Pan-American~Health Organization.
\newblock Epidemiological update: Dengue - 7 february 2020 - paho/who | pan
  american health organization.
\newblock
  \url{https://www.paho.org/en/documents/epidemiological-update-dengue-7-february-2020}.
\newblock (Accessed on 03/27/2021).

\bibitem{parselia2019satellite}
Elisavet Parselia, Charalampos Kontoes, Alexia Tsouni, Christos
  Hadjichristodoulou, Ioannis Kioutsioukis, Gkikas Magiorkinis, and Nikolaos~I
  Stilianakis.
\newblock Satellite earth observation data in epidemiological modeling of
  malaria, dengue and west nile virus: A scoping review.
\newblock {\em Remote Sensing}, 11(16):1862, 2019.

\bibitem{prabodanie2020coherence}
RA~Ranga Prabodanie, Sergei Schreider, Bernard Cazelles, and Lewi Stone.
\newblock Coherence of dengue incidence and climate in the wet and dry zones of
  sri lanka.
\newblock {\em Science of The Total Environment}, 724:138269, 2020.

\bibitem{roesch2014package}
Angi Roesch, Harald Schmidbauer, and Maintainer~Angi Roesch.
\newblock Package ‘waveletcomp’.
\newblock {\em The Comprehensive R Archive Network 2014}, 2014.

\bibitem{rosch2016waveletcomp}
Angi R{\"o}sch and Harald Schmidbauer.
\newblock Waveletcomp 1.1: A guided tour through the r package.
\newblock {\em URL: http://www. hsstat.
  com/projects/WaveletComp/WaveletComp\_guided\_tour. pdf}, 2016.

\bibitem{shepard2011economic}
Donald~S Shepard, Laurent Coudeville, Yara~A Halasa, Betzana Zambrano, and
  Gustavo~H Dayan.
\newblock Economic impact of dengue illness in the americas.
\newblock {\em The American journal of tropical medicine and hygiene},
  84(2):200--207, 2011.

\bibitem{simoes2013modeling}
Taynana~C Simoes, Claudia~T Code{\c{c}}o, Aline~A Nobre, and Alvaro~E Eiras.
\newblock Modeling the non-stationary climate dependent temporal dynamics of
  aedes aegypti.
\newblock {\em PLoS One}, 8(8), 2013.

\bibitem{beltran2014spatiotemporal}
Anna~M Stewart-Ibarra, {\'A}ngel~G Mu{\~n}oz, Sadie~J Ryan,
  Efra{\'\i}n~Beltr{\'a}n Ayala, Mercy~J Borbor-Cordova, Julia~L Finkelstein,
  Ra{\'u}l Mej{\'\i}a, Tania Ordo{\~n}ez, G~Cristina Recalde-Coronel, and
  Keytia Rivero.
\newblock Spatiotemporal clustering, climate periodicity, and social-ecological
  risk factors for dengue during an outbreak in machala, ecuador, in 2010.
\newblock {\em BMC infectious diseases}, 14(1):1--16, 2014.

\bibitem{talagala2015wavelet}
Thiyanga Talagala and Ravindra Lokupitiya.
\newblock Wavelet analysis of dengue transmission pattern in sri lanka.
\newblock {\em Int J Mosquito Res}, 2(4):13--18, 2015.

\bibitem{team2013r}
R~Core Team et~al.
\newblock R: A language and environment for statistical computing.
\newblock {\em URL:http://www.R-project.org/}, 2013.

\bibitem{thai2010dengue}
Khoa~TD Thai, Bernard Cazelles, Nam Van~Nguyen, Long~Thi Vo, Maciej~F Boni,
  Jeremy Farrar, Cameron~P Simmons, H~Rogier van Doorn, and Peter~J de~Vries.
\newblock Dengue dynamics in binh thuan province, southern vietnam:
  periodicity, synchronicity and climate variability.
\newblock {\em PLoS neglected tropical diseases}, 4(7), 2010.

\bibitem{torrence1998practical}
Christopher Torrence and Gilbert~P Compo.
\newblock A practical guide to wavelet analysis.
\newblock {\em Bulletin of the American Meteorological society}, 79(1):61--78,
  1998.

\bibitem{troyo2009urban}
Adriana Troyo, Douglas~O Fuller, Olger Calder{\'o}n-Arguedas, Mayra~E Solano,
  and John~C Beier.
\newblock Urban structure and dengue incidence in puntarenas, costa rica.
\newblock {\em Singapore journal of tropical geography}, 30(2):265--282, 2009.

\bibitem{tun477effects}
W~Tun-Lin, TR~Burkot, and BH~Kay.
\newblock Effects of temperature and larval diet on development 476 rates and
  survival of the dengue vector aedes aegypti in north queensland, australia.
\newblock {\em Med}, 477:31--37, 2000.

\bibitem{DengueanWHO}
WHO.
\newblock Dengue and severe dengue.
\newblock
  \url{https://www.who.int/news-room/fact-sheets/detail/dengue-and-severe-dengue}.
\newblock (Accessed on 03/22/2021).

\bibitem{wu2002tropical}
Lixin Wu and Zhengyu Liu.
\newblock Is tropical atlantic variability driven by the north atlantic
  oscillation?
\newblock {\em Geophysical research letters}, 29(13):31--1, 2002.

\bibitem{yu2018land}
Y~Yu, Y~Liu, and P~Yu.
\newblock Land surface temperature product development for jpss and goes-r
  missions. "in comprenhesive remote sensing".
\newblock {\em Earth’s Energy Budget}, 5, 2018.

\bibitem{zeng2021global}
Zhilin Zeng, Juan Zhan, Liyuan Chen, Huilong Chen, and Sheng Cheng.
\newblock Global, regional, and national dengue burden from 1990 to 2017: A
  systematic analysis based on the global burden of disease study 2017.
\newblock {\em EClinicalMedicine}, 32:100712, 2021.

\end{thebibliography}
\bibliographystyle{plain}


\newpage
\vspace{1cm}
\begin{center}
\Large Supplementary Material: Wavelet Analysis of Dengue Incidence and its Correlation with Weather and Vegetation Variables in Costa Rica
\end{center}

\vspace{0.5cm}
\begin{center}
Yury E. Garc\'ia, Luis A. Barboza, Fabio Sanchez,\\
Paola V\'asquez, Juan G. Calvo.
\end{center}

\vspace{0.5cm}
\section*{Wavelet coherence and average cross-wavelet power between dengue incidence and EVI}
\begin{figure}[H]
\caption*{\textbf{Figure S1:} Wavelet coherence (color map) between dengue incidence from 2000 to 2019, and EVI in 32 municipalities of Costa Rica (periodicity on y-axis, time on x-axis). Colors code for increasing power intensity, from blue to red; $95\%$ confidence levels are encircled by white lines, and shaded areas indicate the presence of significant edge effects. On the right side of each wavelet coherence is the average cross-wavelet power (Red line). The arrows indicate whether the two series are in-phase or out-phase.}
\captionsetup[subfigure]{labelformat=empty}
\subfloat[]{\includegraphics[scale=0.23]{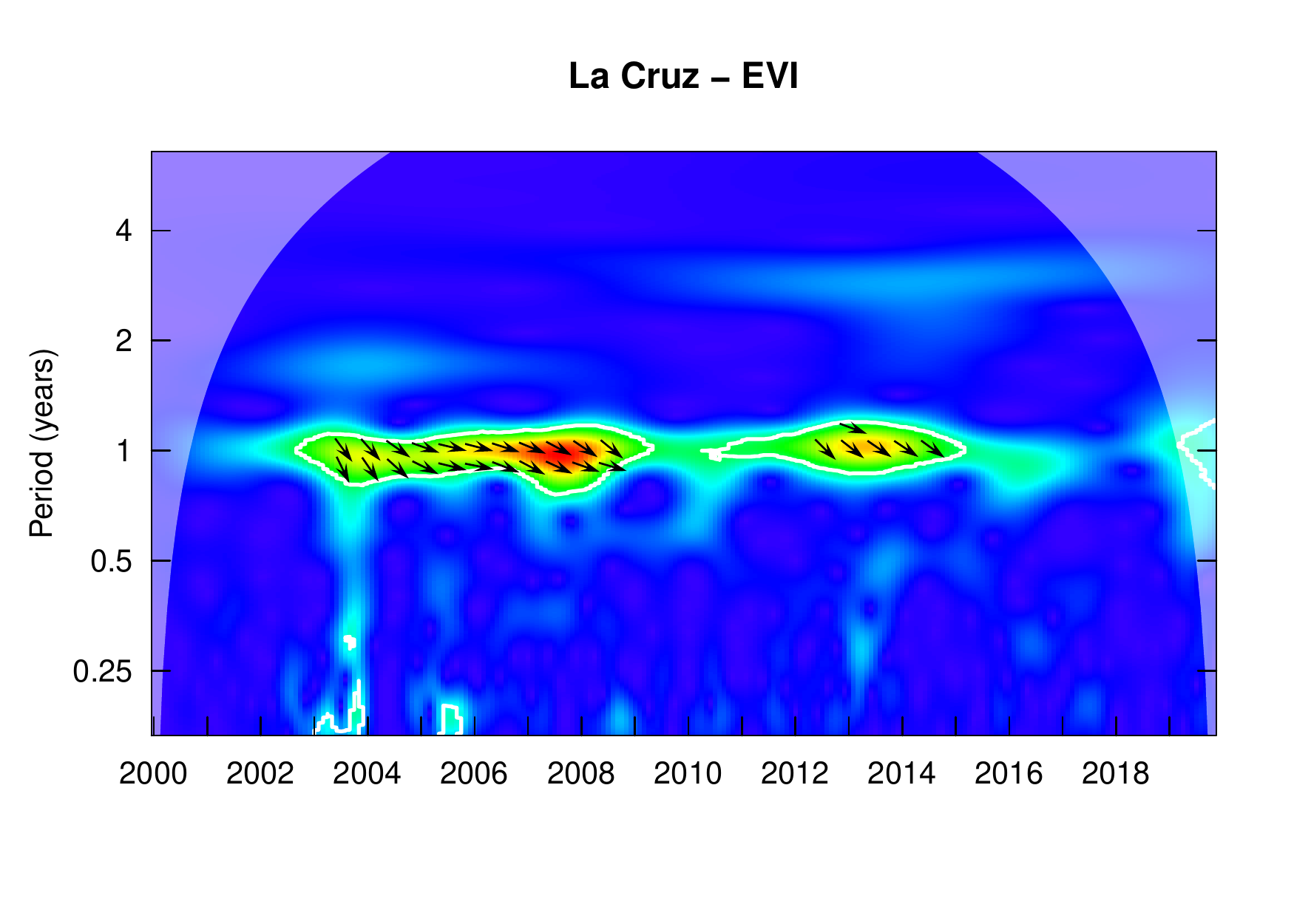}}\vspace{-0.15cm}%
\subfloat[]{\includegraphics[scale=0.23]{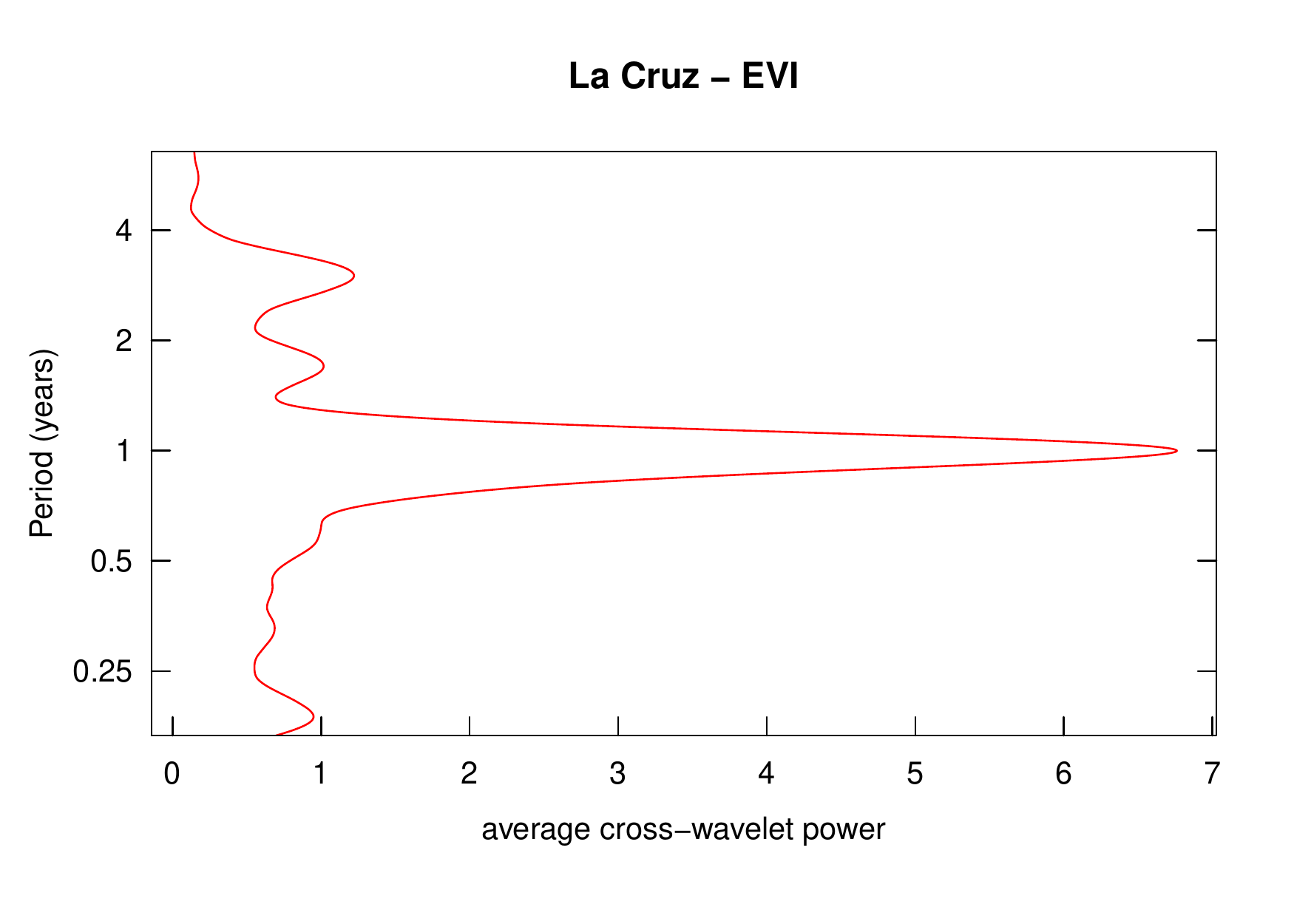}}\vspace{-0.15cm}%
\subfloat[]{\includegraphics[scale=0.23]{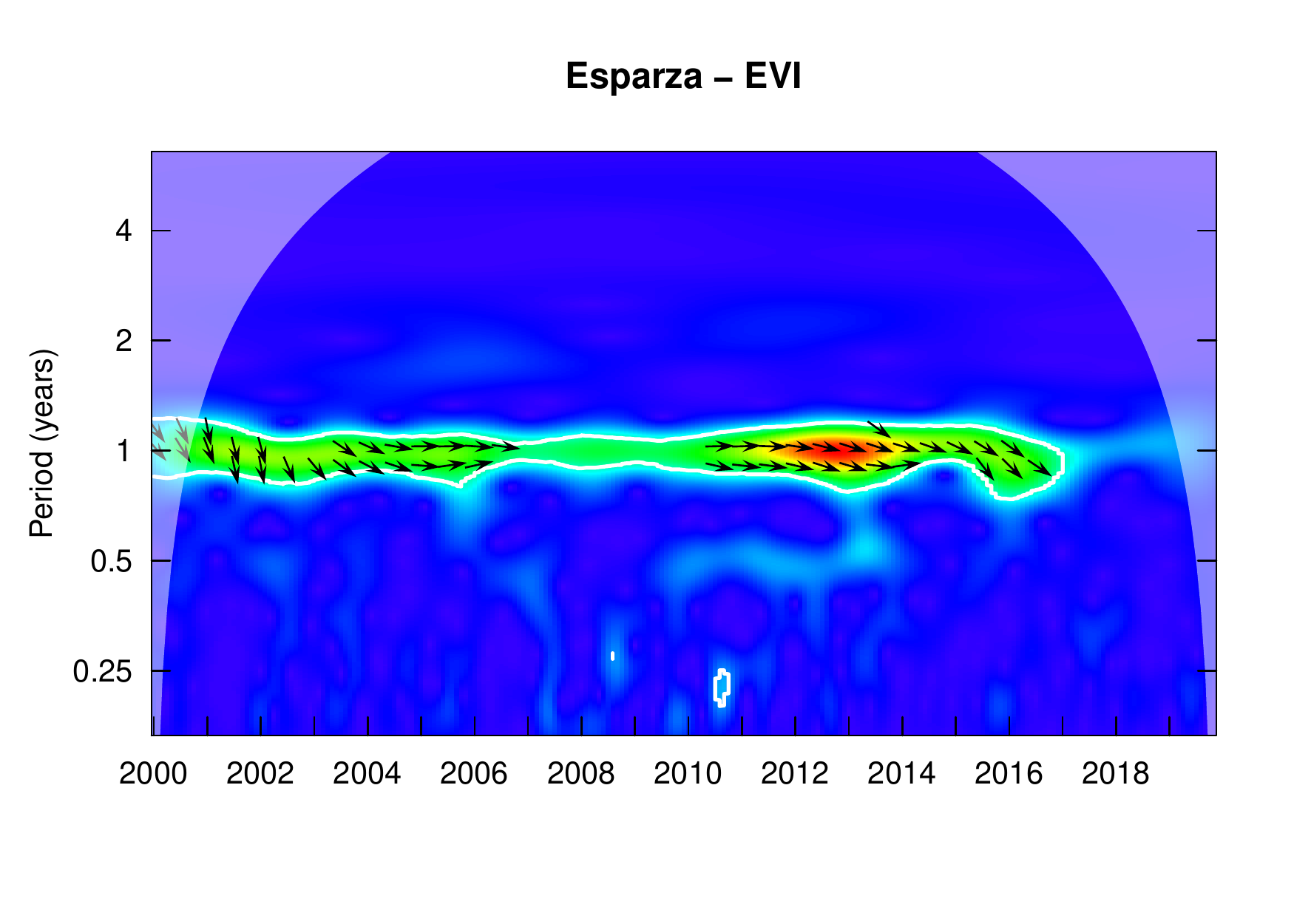}}\vspace{-0.15cm}%
\subfloat[]{\includegraphics[scale=0.23]{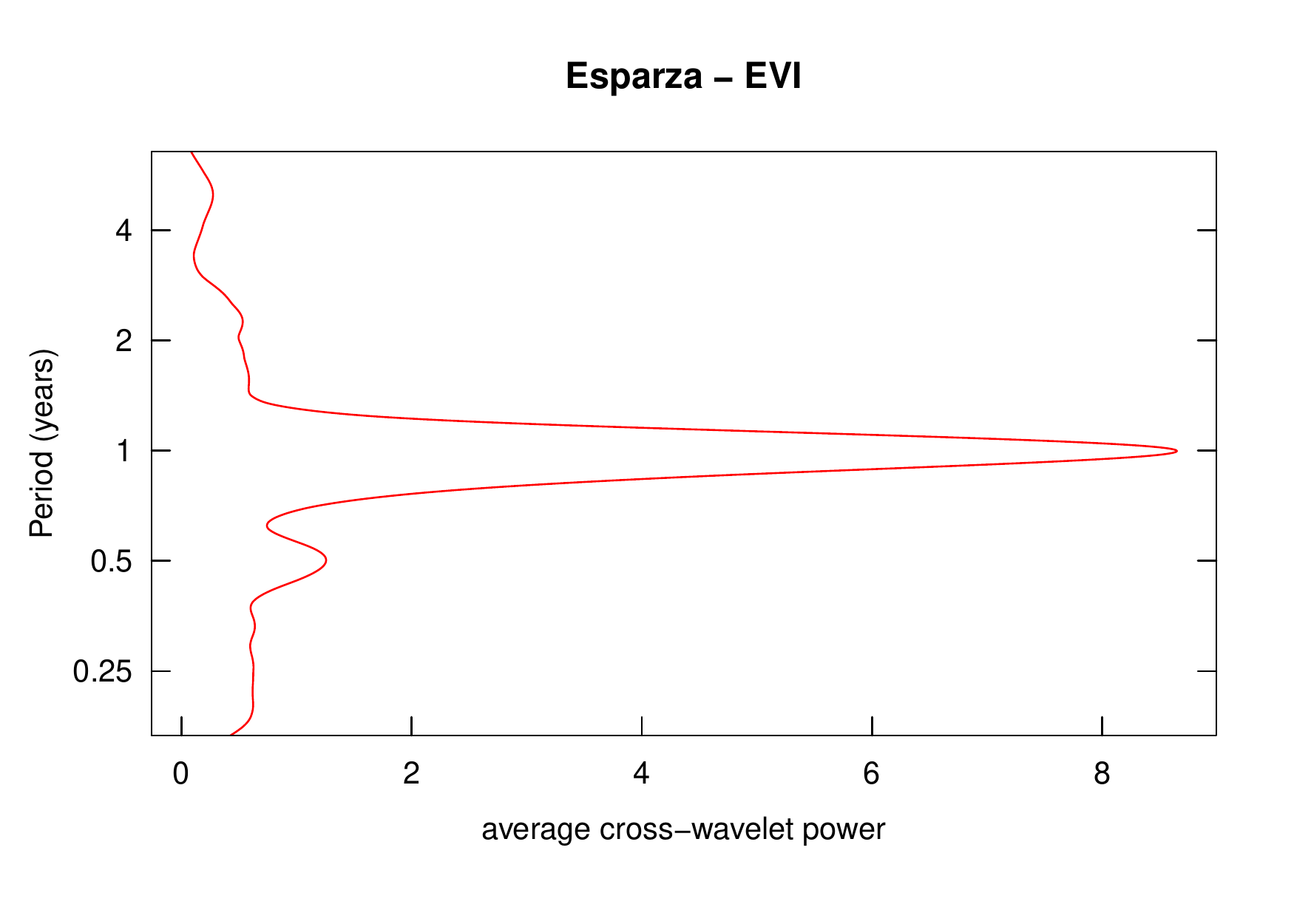}}\vspace{-0.15cm}\\
\subfloat[]{\includegraphics[scale=0.23]{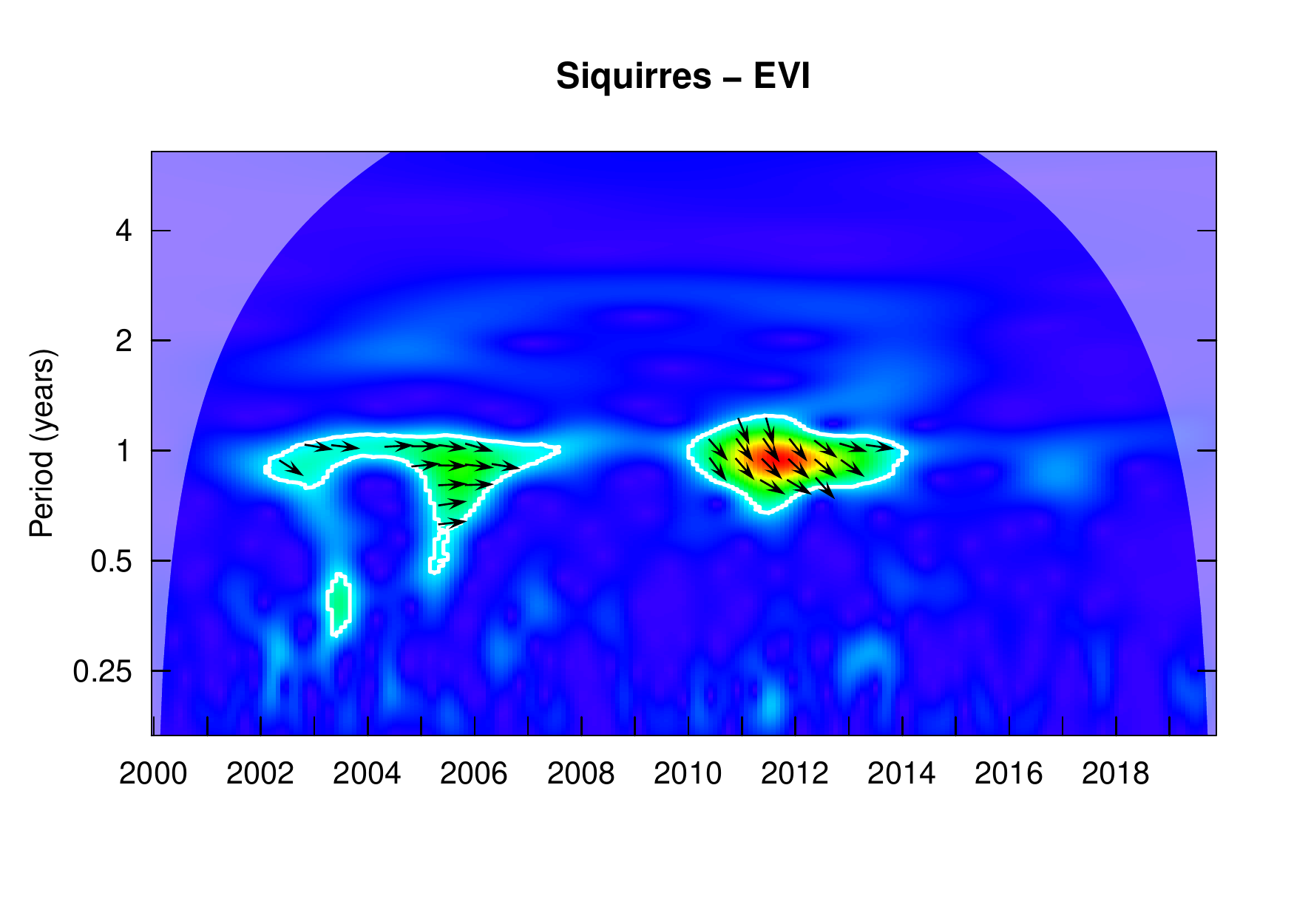}}\vspace{-0.15cm}%
\subfloat[]{\includegraphics[scale=0.23]{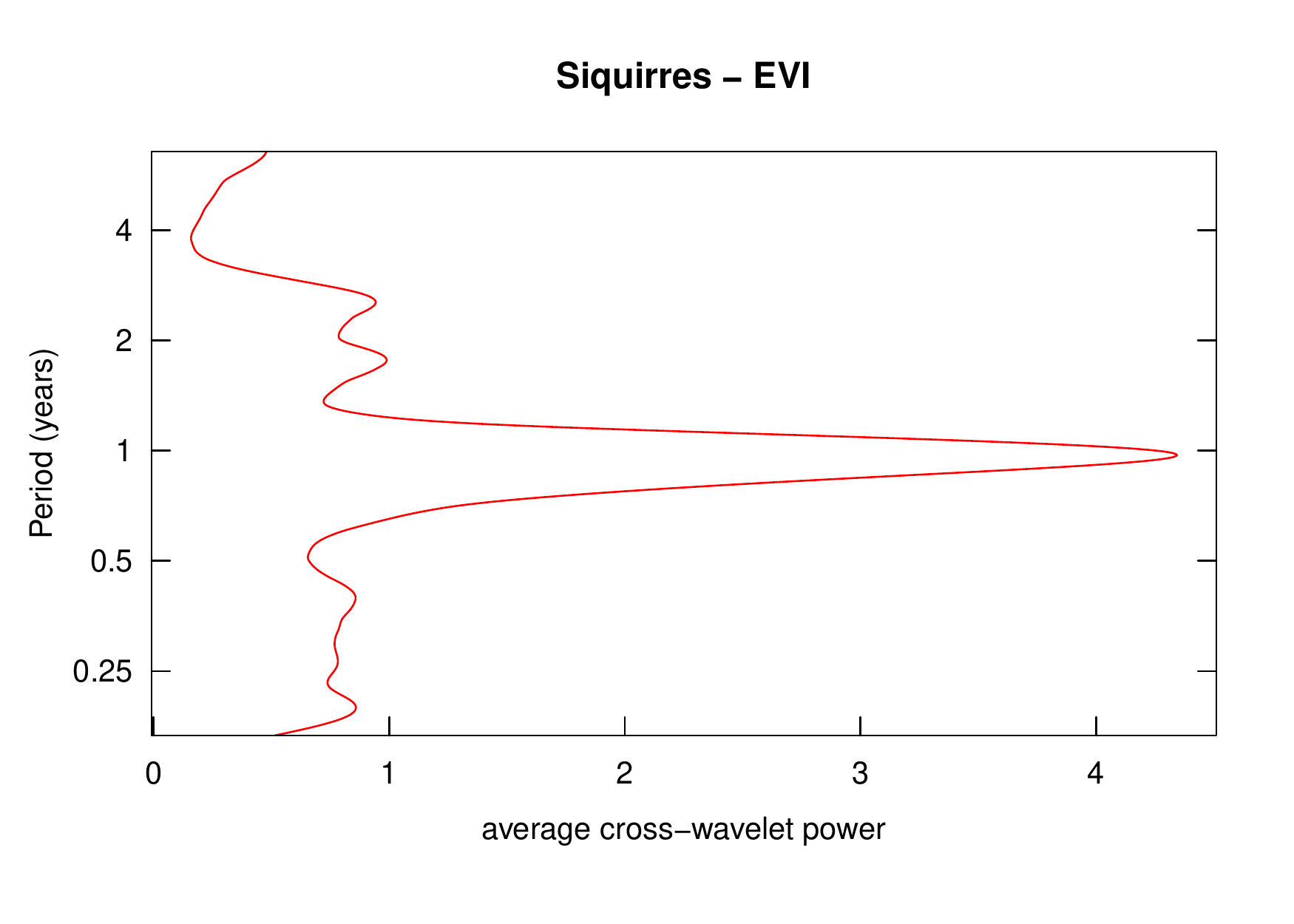}}\vspace{-0.15cm}%
\subfloat[]{\includegraphics[scale=0.23]{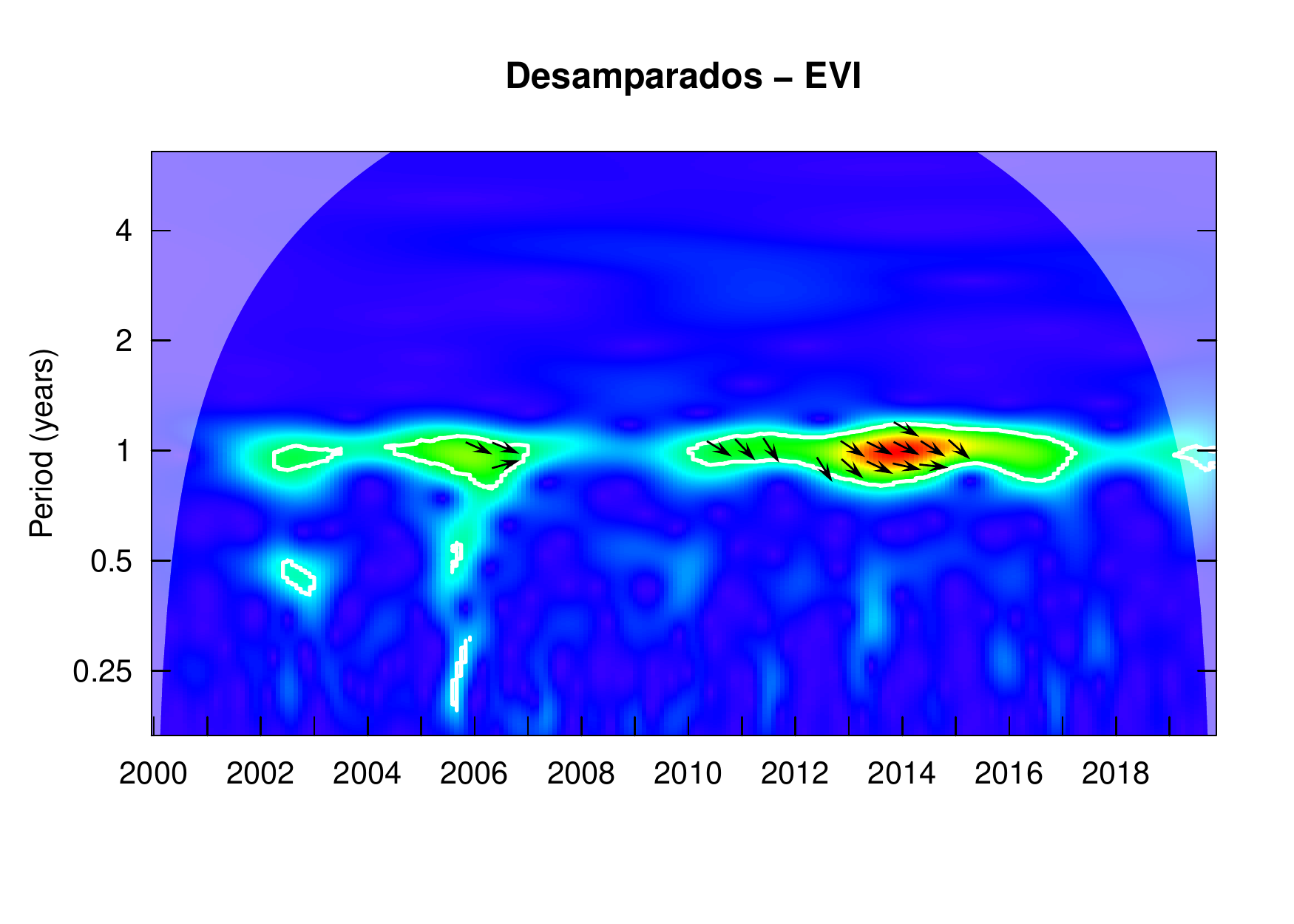}}\vspace{-0.15cm}%
\subfloat[]{\includegraphics[scale=0.23]{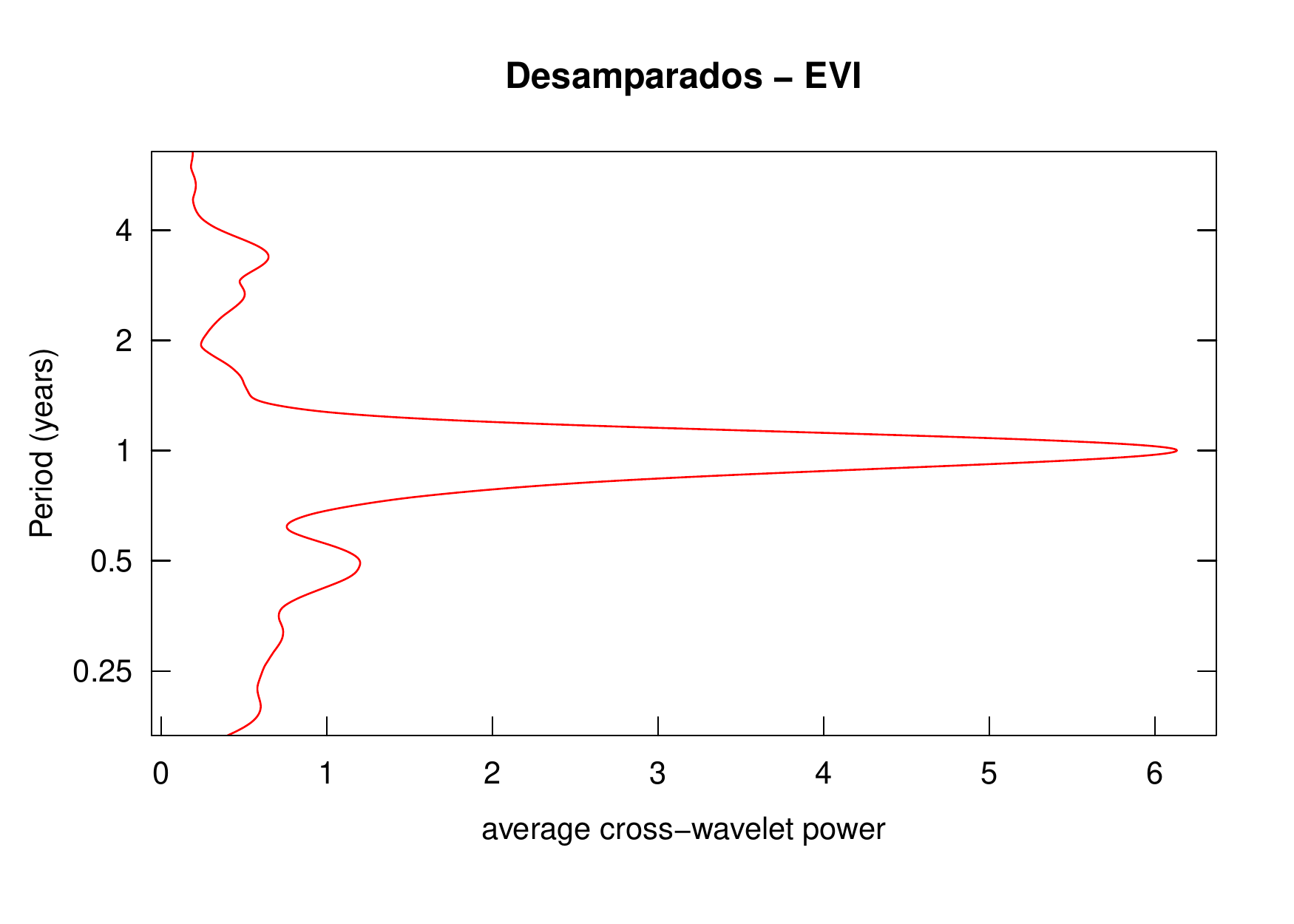}}\vspace{-0.15cm}\\
\subfloat[]{\includegraphics[scale=0.23]{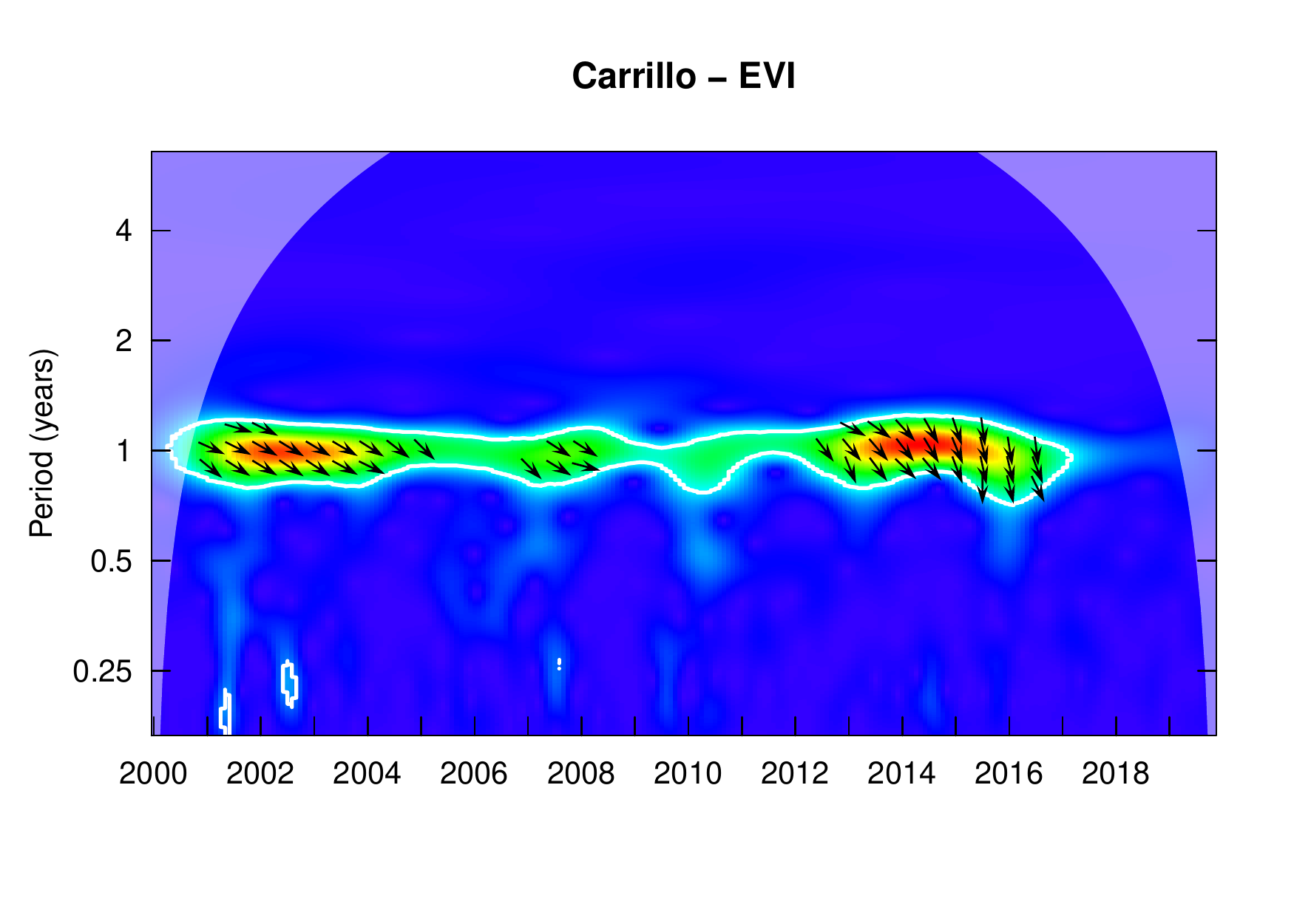}}\vspace{-0.15cm}%
\subfloat[]{\includegraphics[scale=0.23]{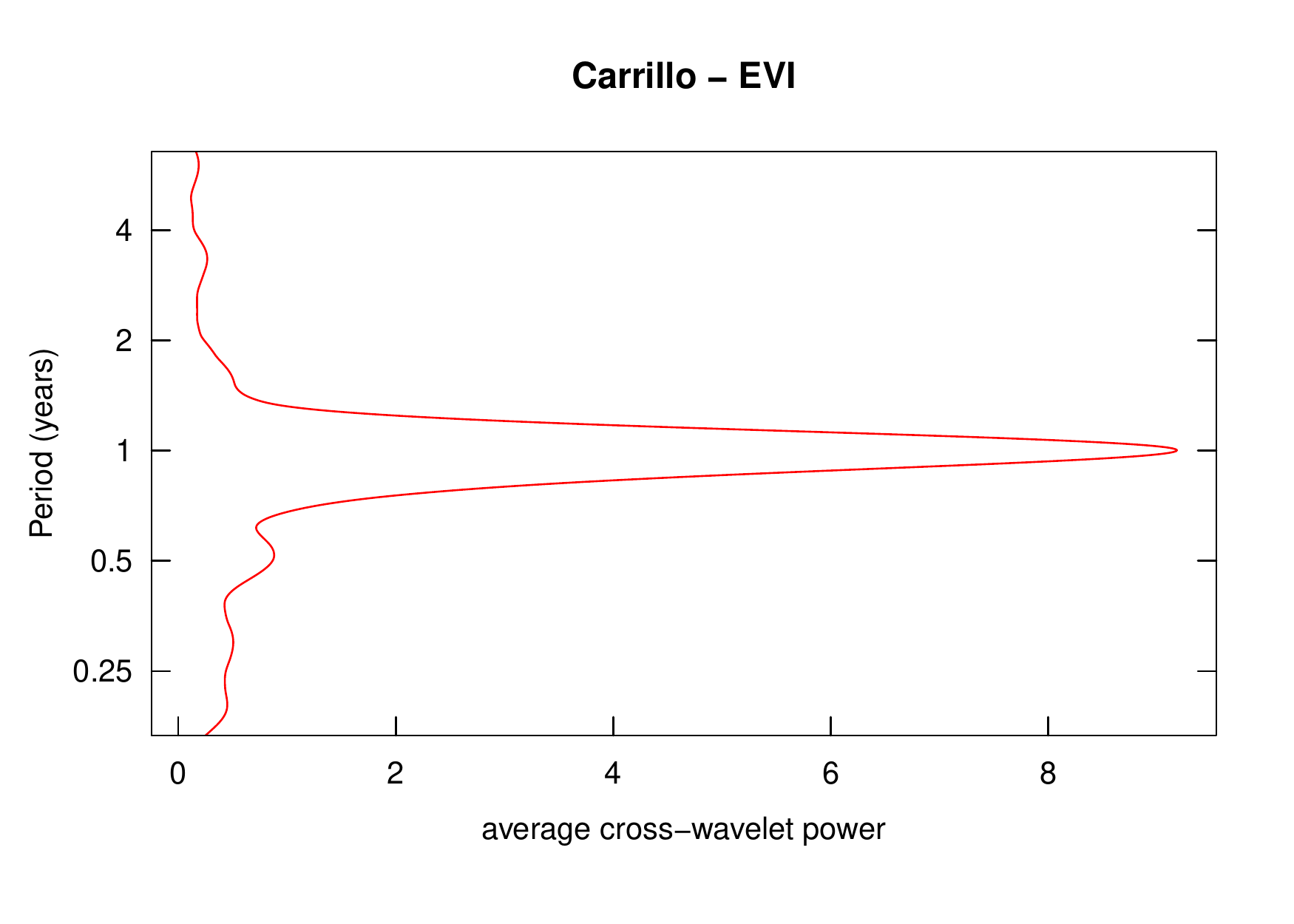}}\vspace{-0.15cm}%
\subfloat[]{\includegraphics[scale=0.23]{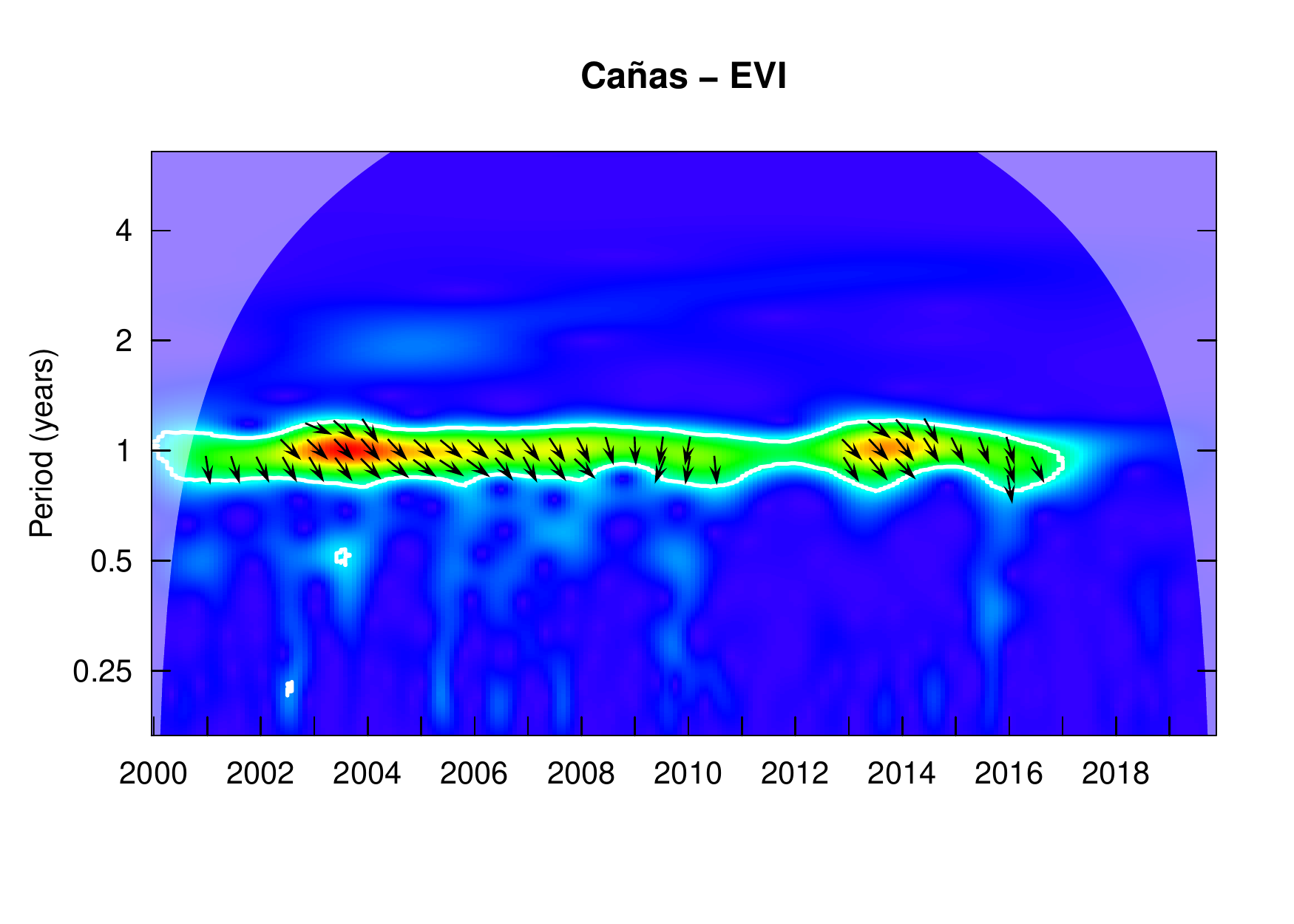}}\vspace{-0.15cm}%
\subfloat[]{\includegraphics[scale=0.23]{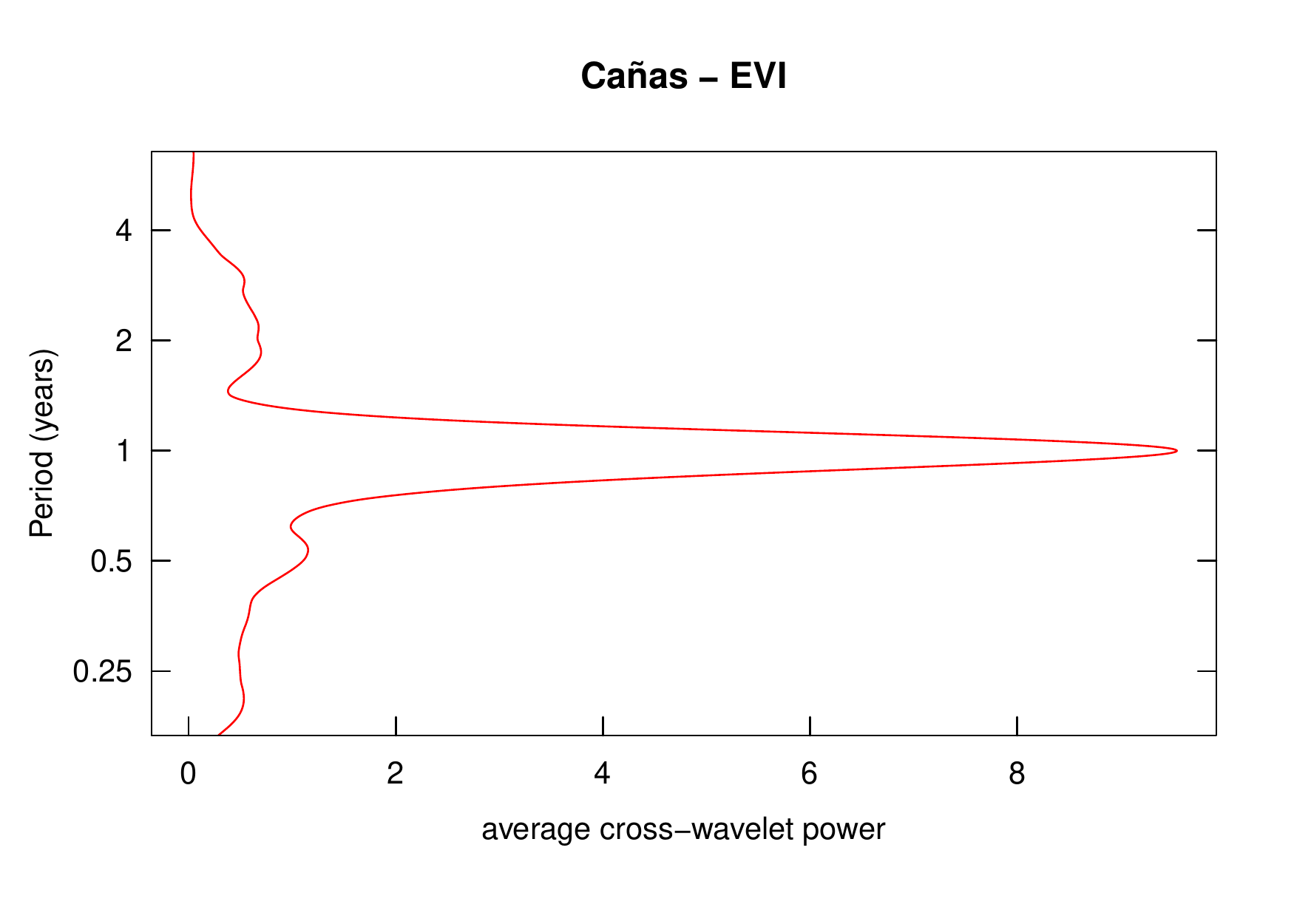}}\vspace{-0.15cm}
\subfloat[]{\includegraphics[scale=0.23]{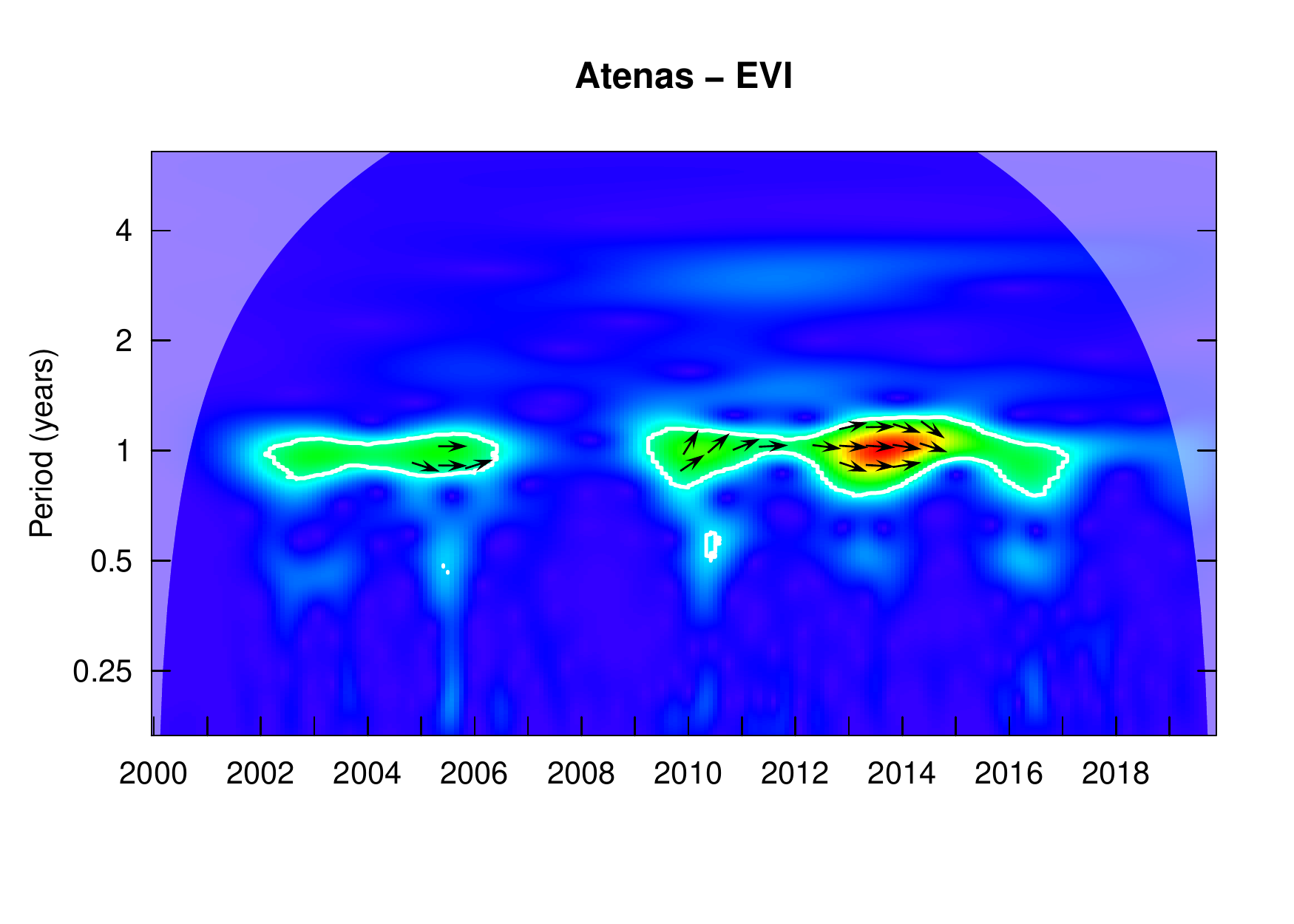}}\vspace{-0.15cm}%
\subfloat[]{\includegraphics[scale=0.23]{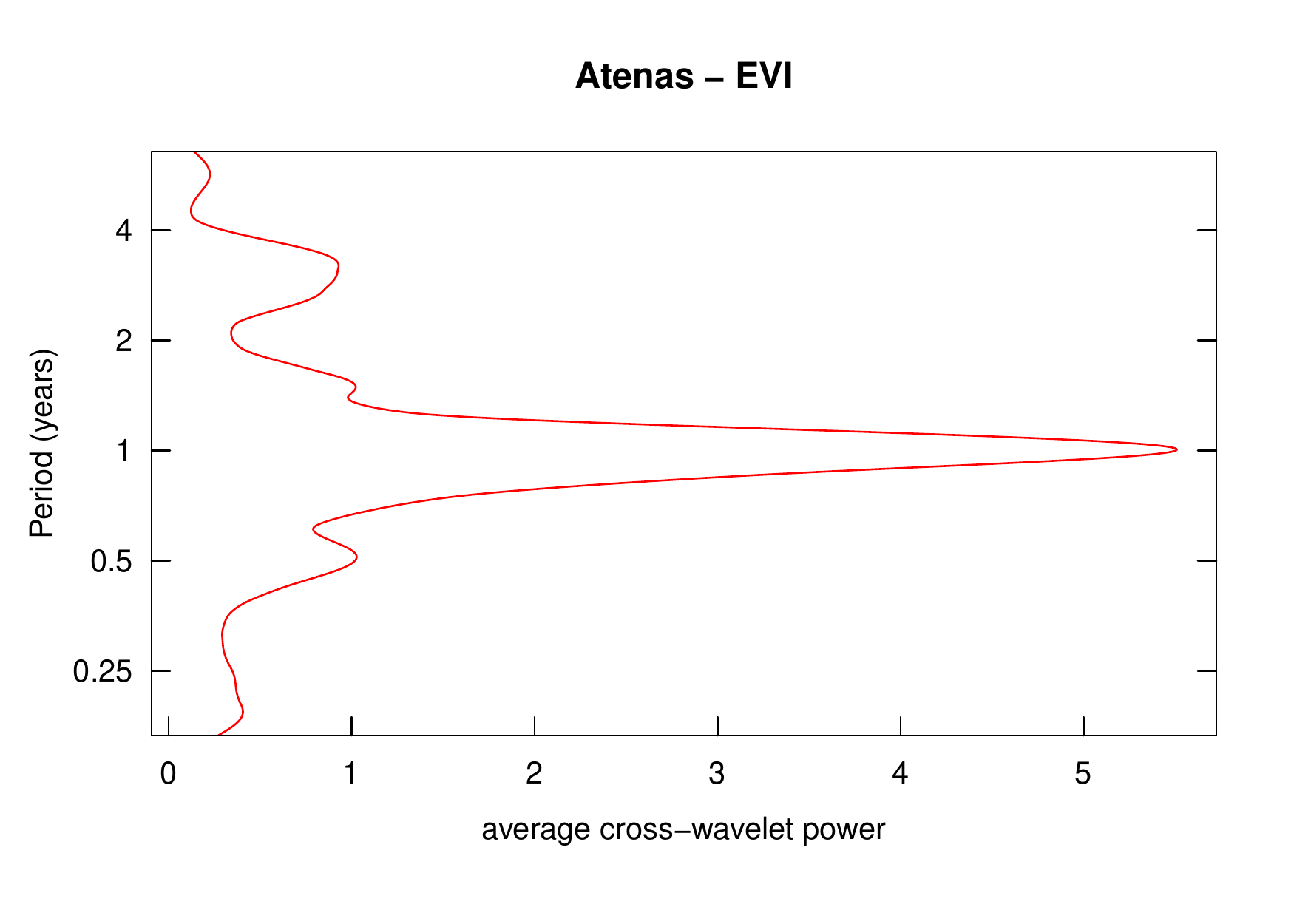}}\vspace{-0.15cm}%
\subfloat[]{\includegraphics[scale=0.23]{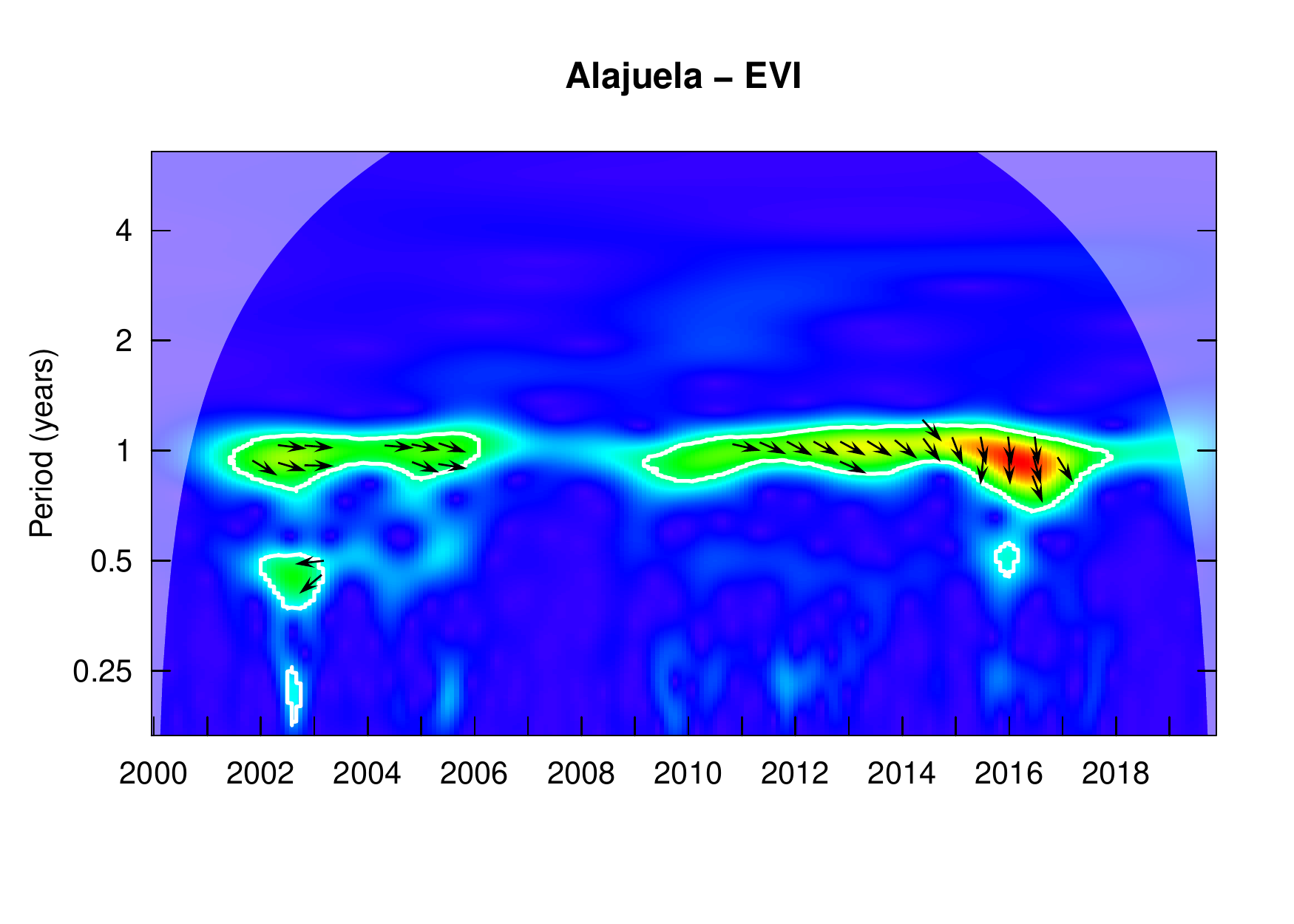}}\vspace{-0.15cm}%
\subfloat[]{\includegraphics[scale=0.23]{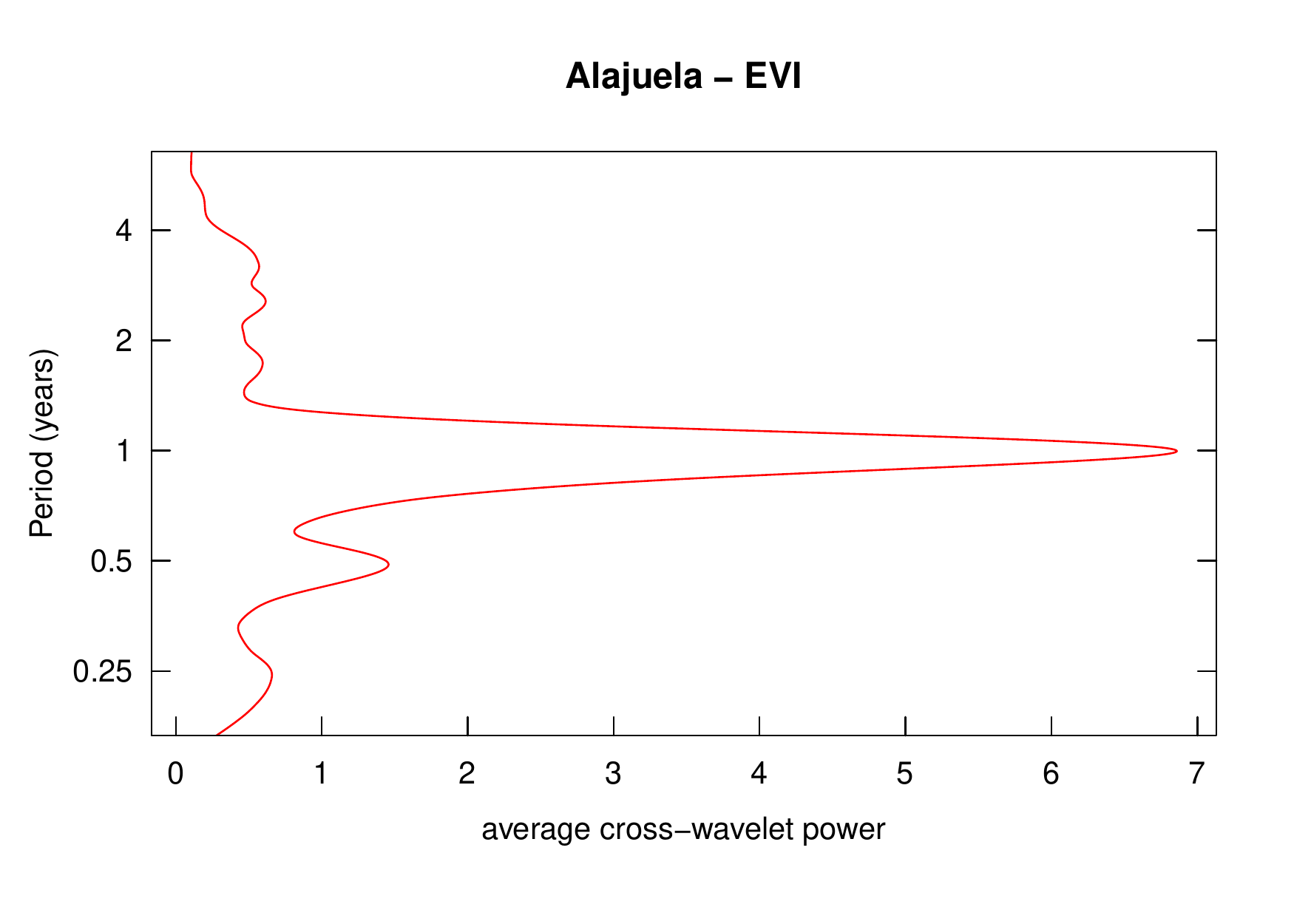}}\vspace{-0.15cm}\\
\end{figure}

\begin{figure}[H]
\captionsetup[subfigure]{labelformat=empty}
\subfloat[]{\includegraphics[scale=0.23]{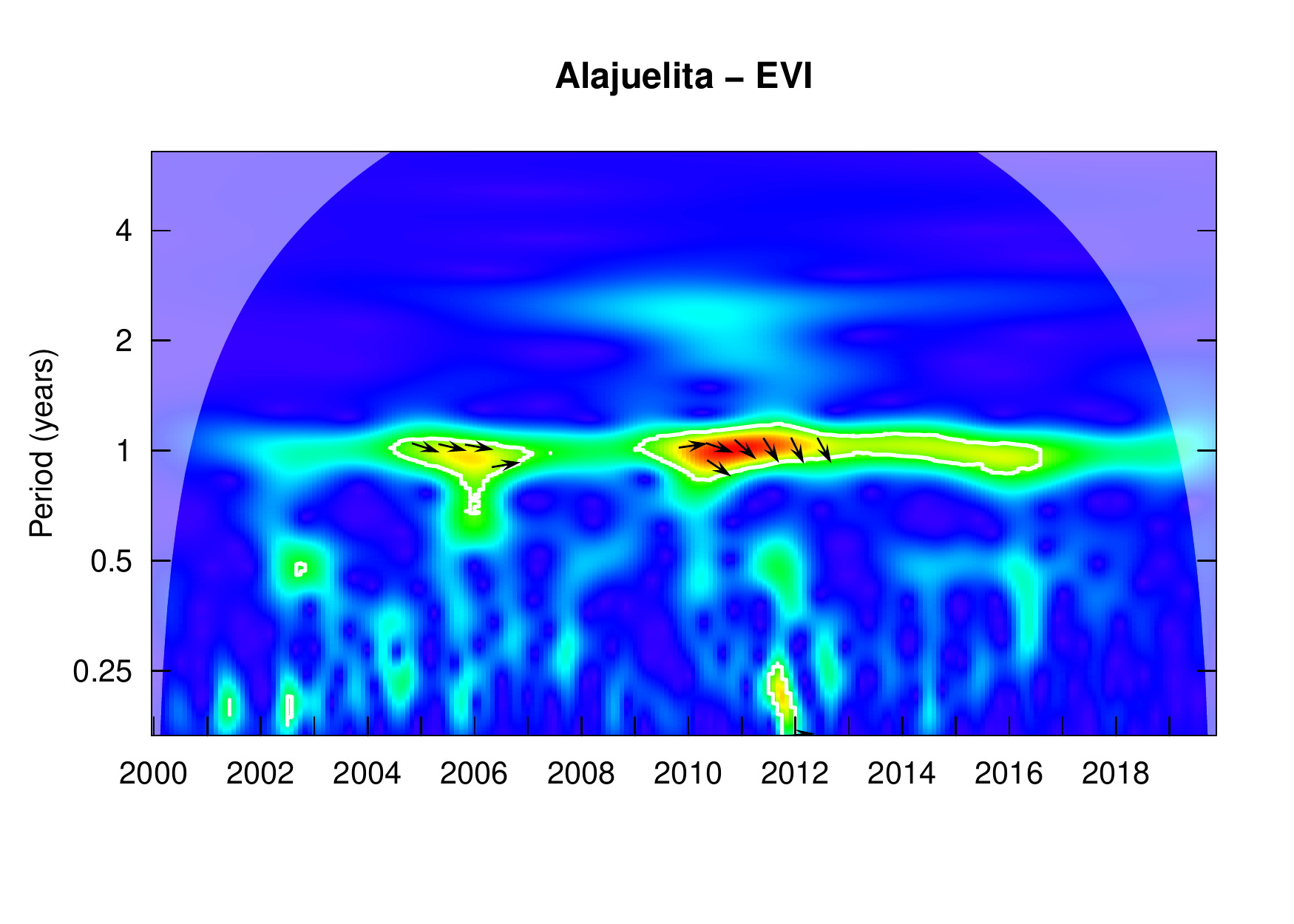}}\vspace{-0.15cm}%
\subfloat[]{\includegraphics[scale=0.23]{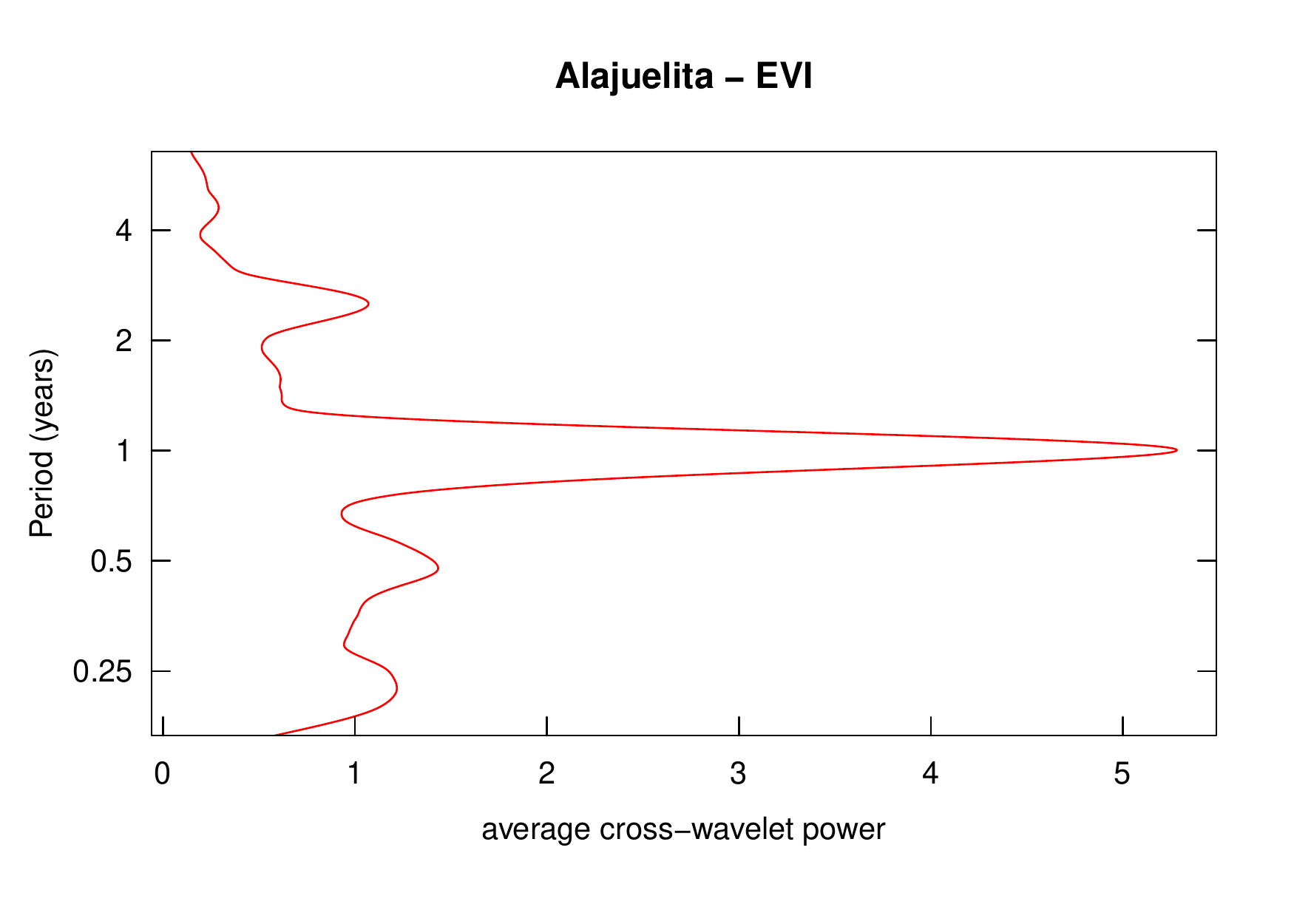}}\vspace{-0.15cm}%
\subfloat[]{\includegraphics[scale=0.23]{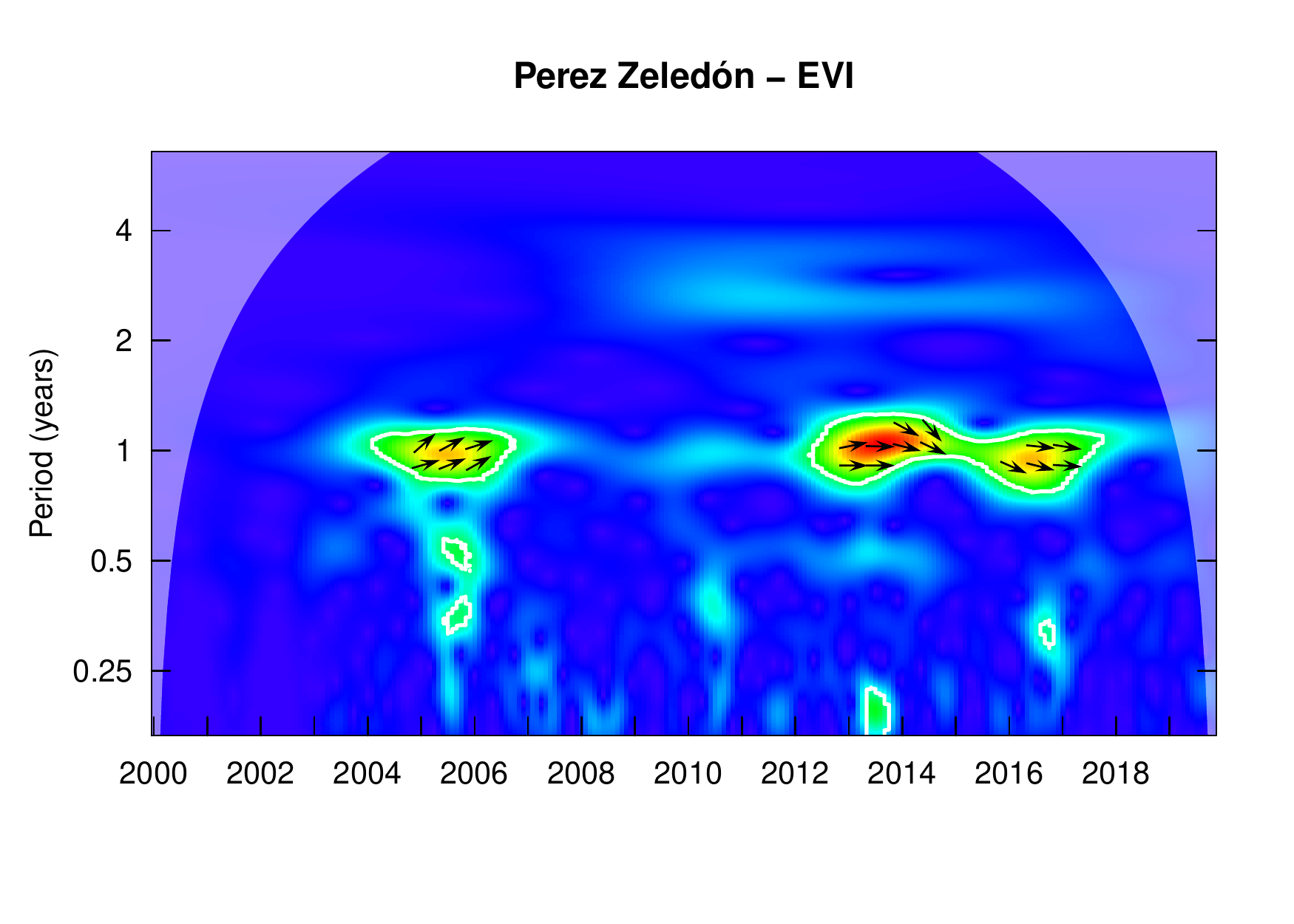}}\vspace{-0.15cm}%
\subfloat[]{\includegraphics[scale=0.23]{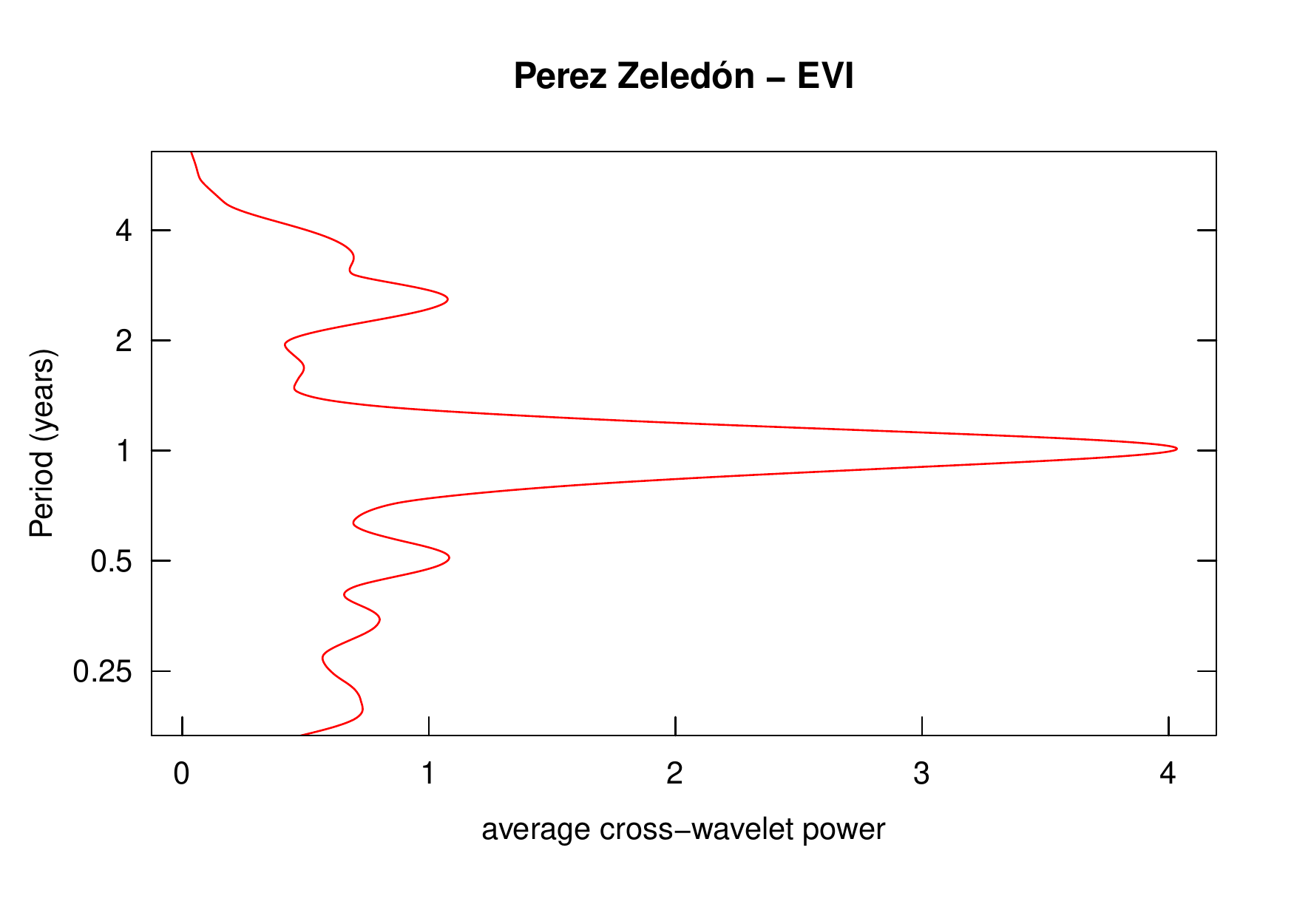}}\vspace{-0.15cm}\\
\subfloat[]{\includegraphics[scale=0.23]{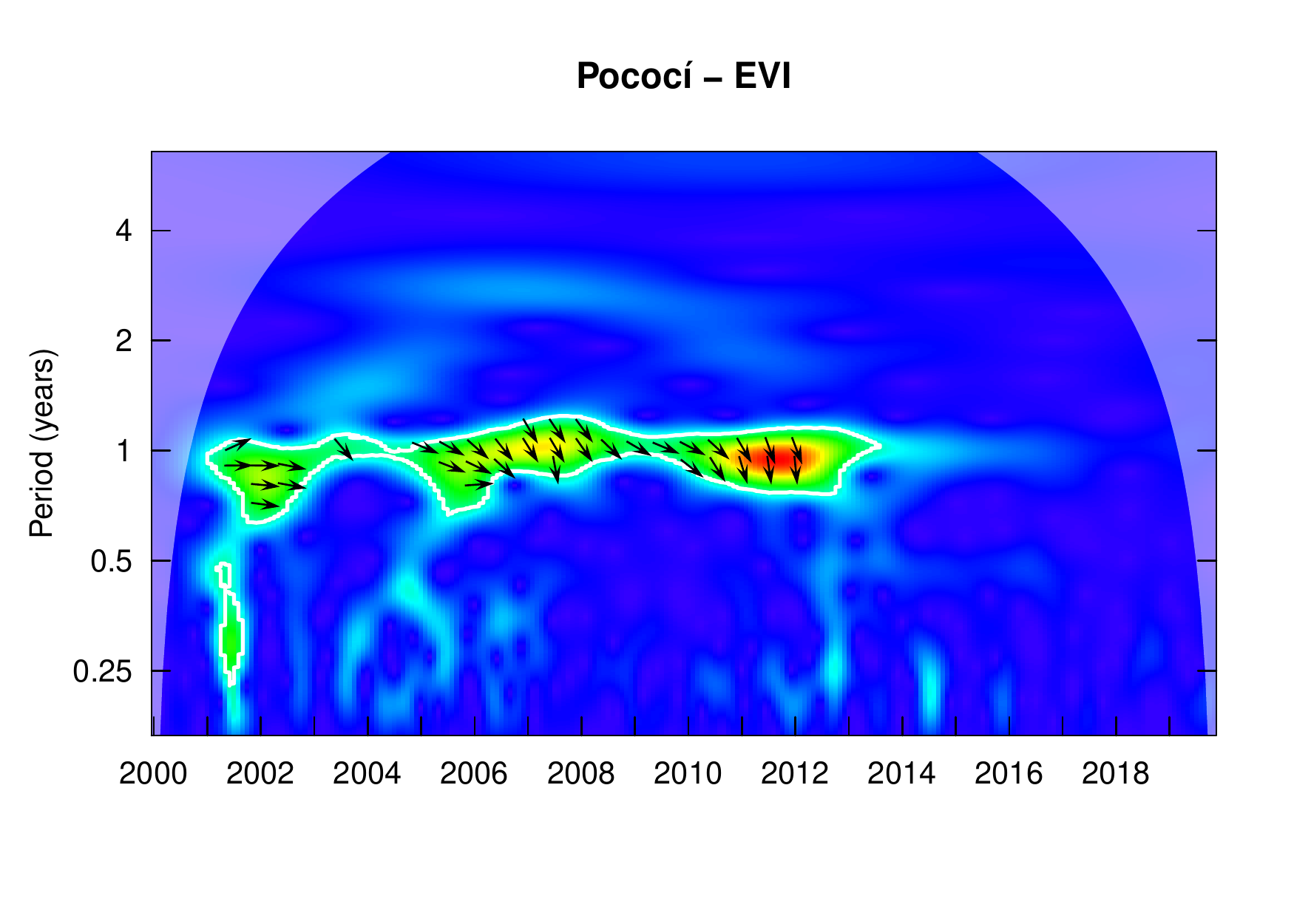}}\vspace{-0.15cm}%
\subfloat[]{\includegraphics[scale=0.23]{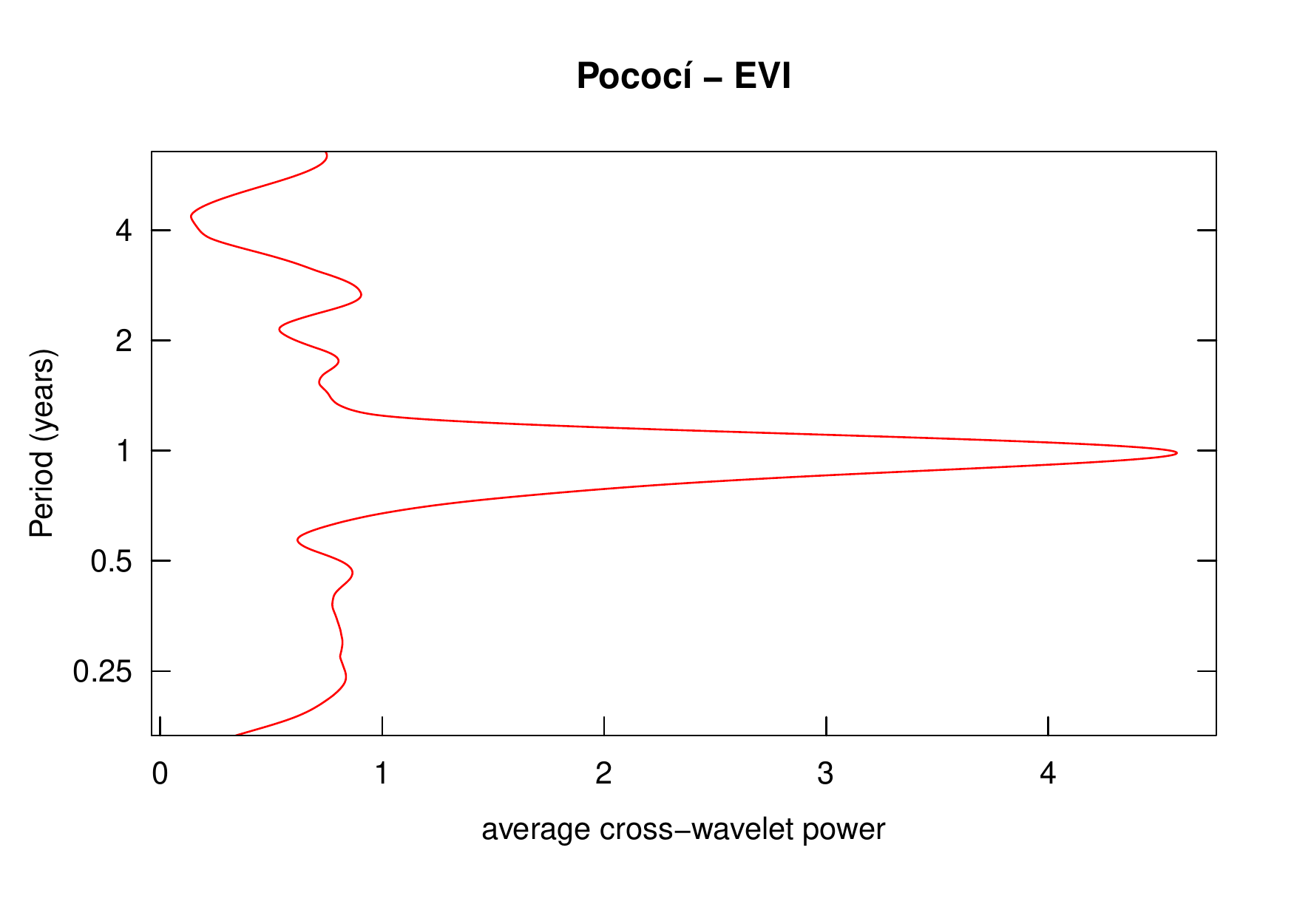}}\vspace{-0.15cm}%
\subfloat[]{\includegraphics[scale=0.23]{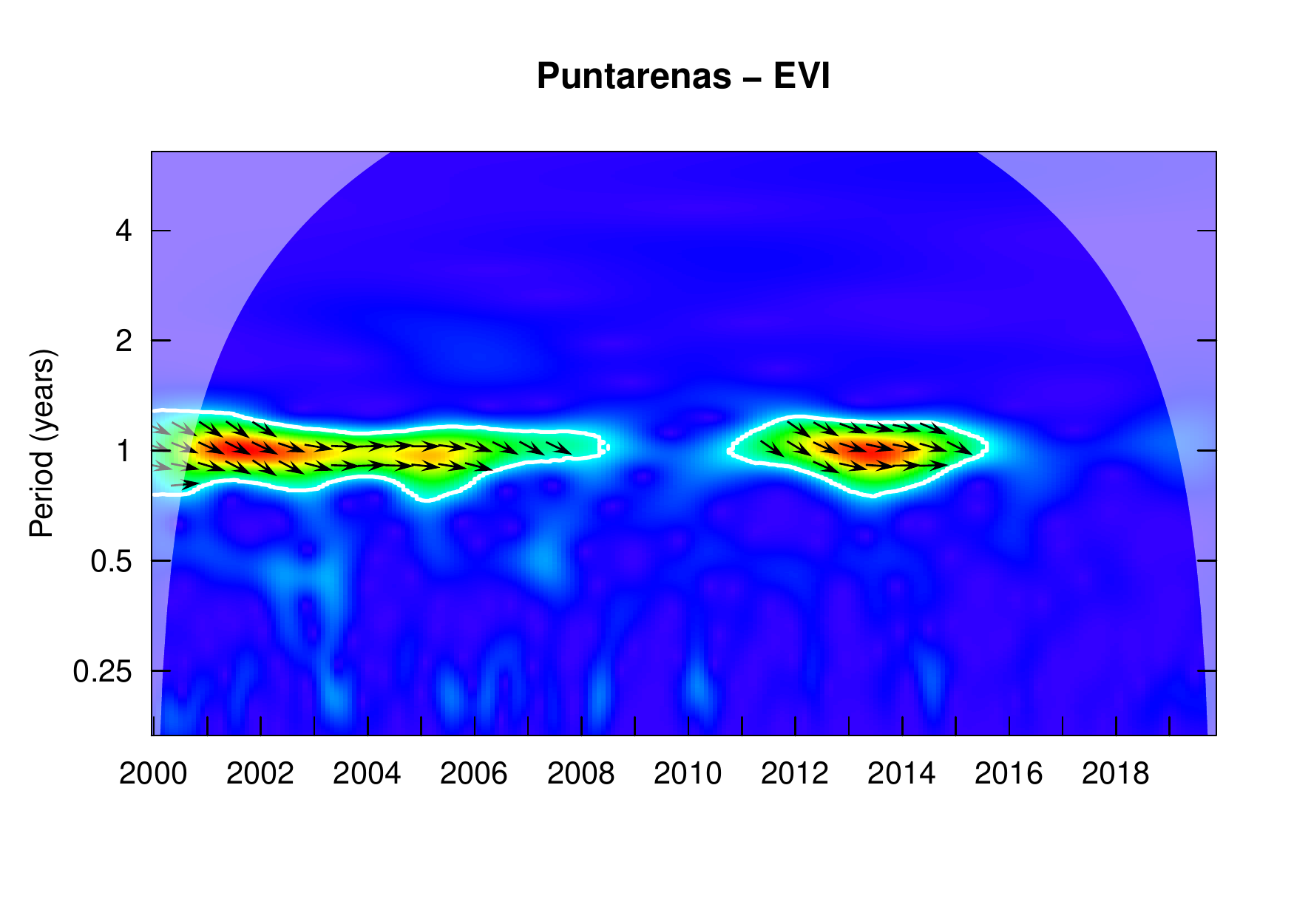}}\vspace{-0.15cm}%
\subfloat[]{\includegraphics[scale=0.23]{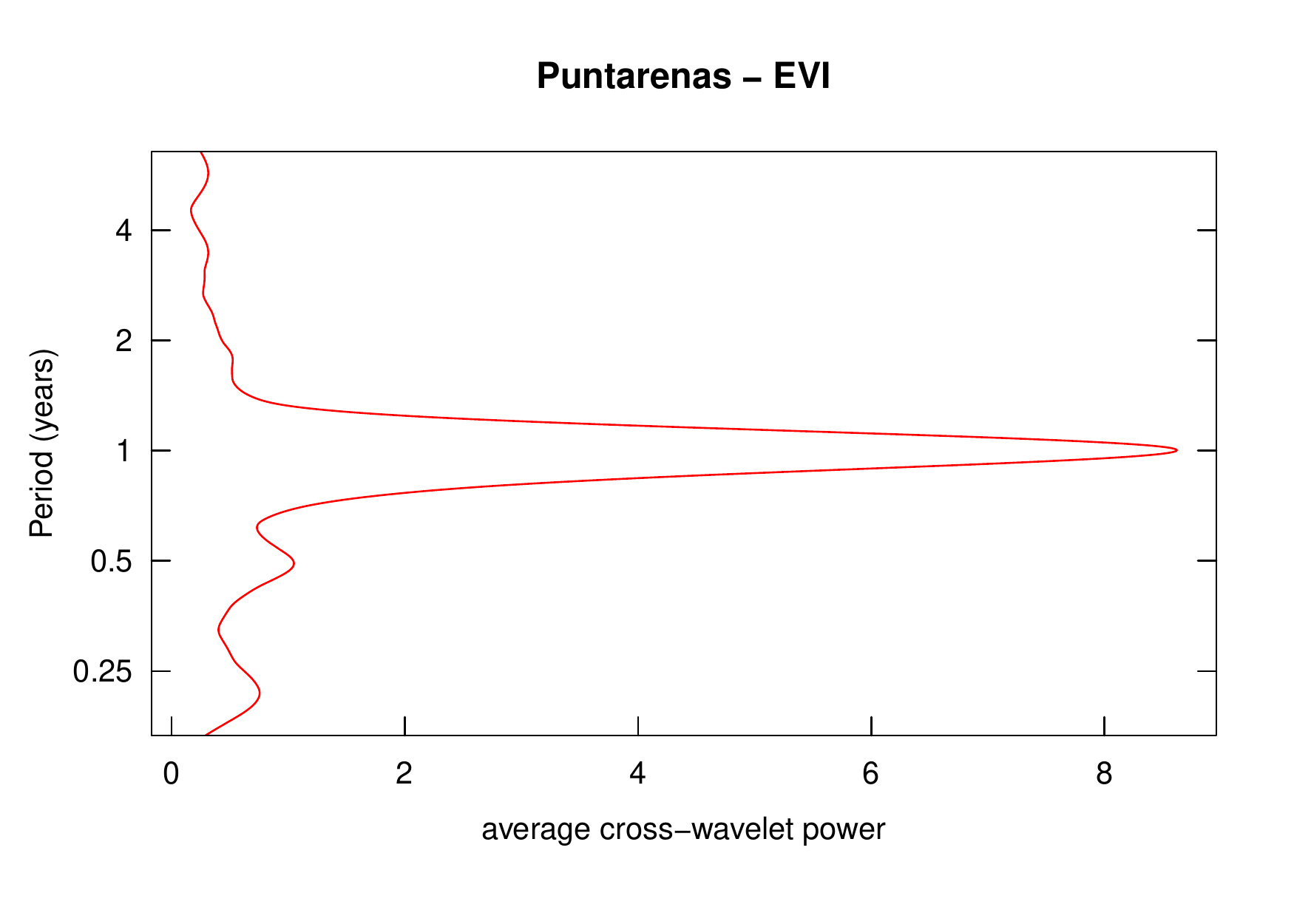}}\vspace{-0.15cm}\\
\subfloat[]{\includegraphics[scale=0.23]{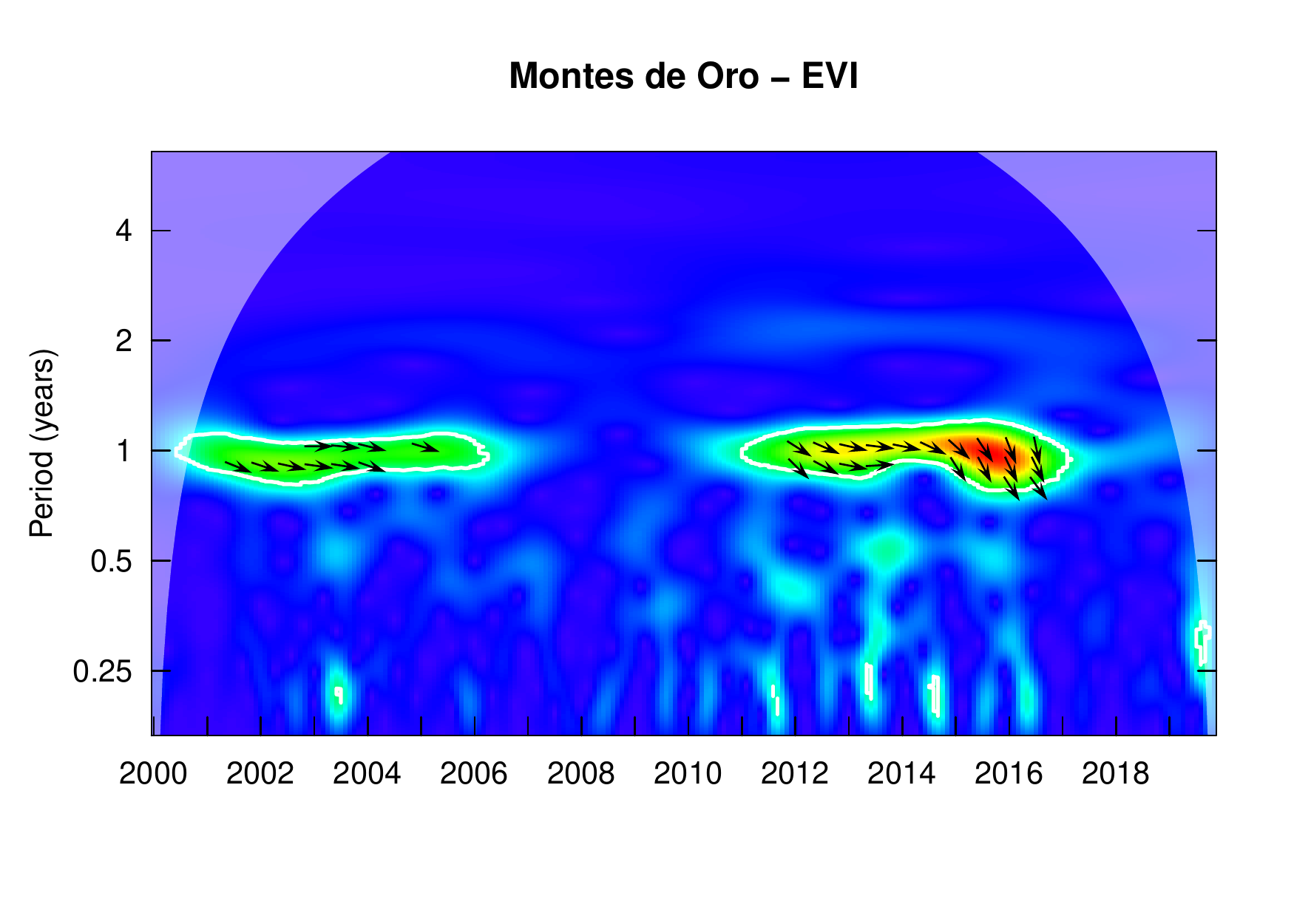}}\vspace{-0.15cm}%
\subfloat[]{\includegraphics[scale=0.23]{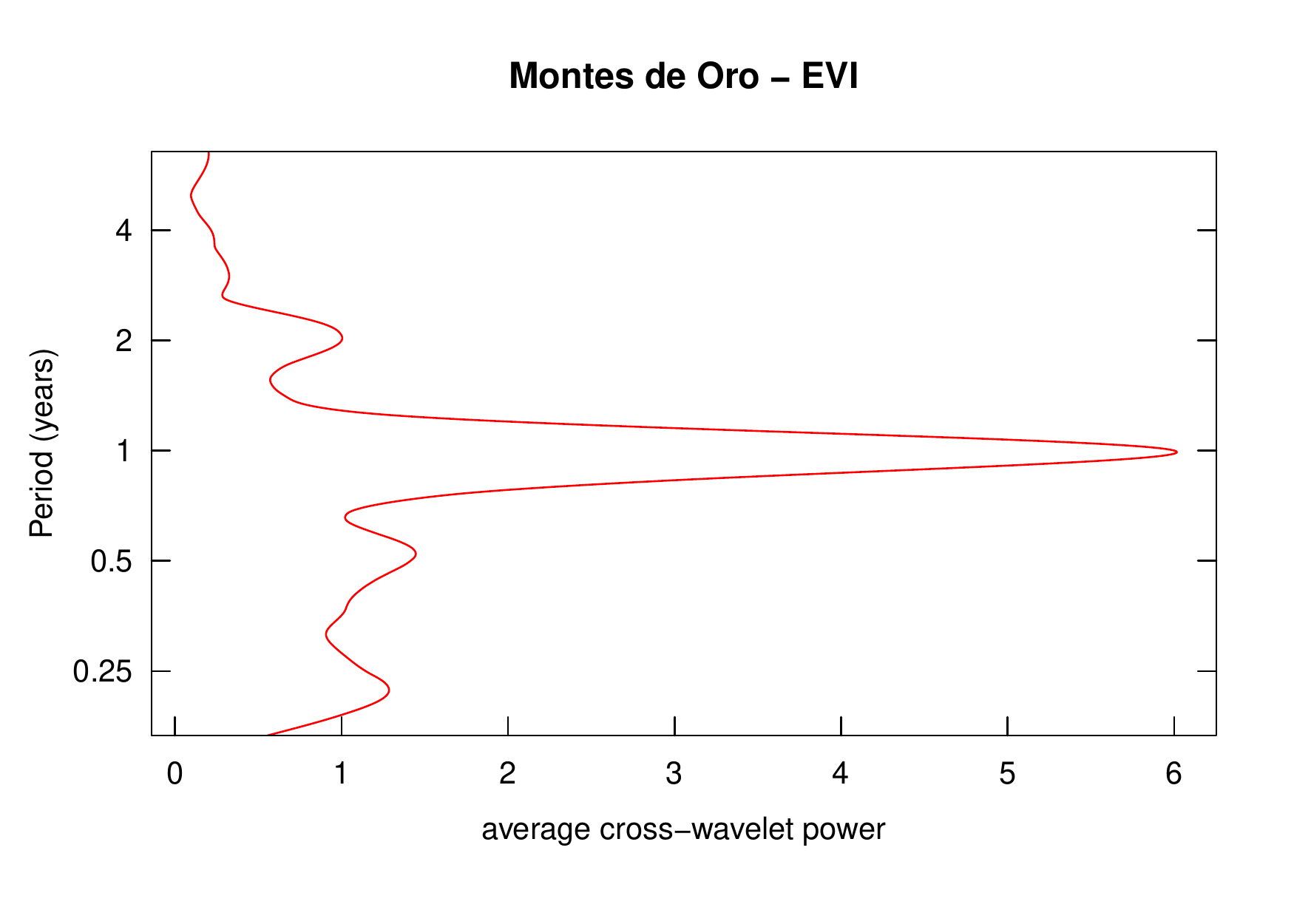}}\vspace{-0.15cm}%
\subfloat[]{\includegraphics[scale=0.23]{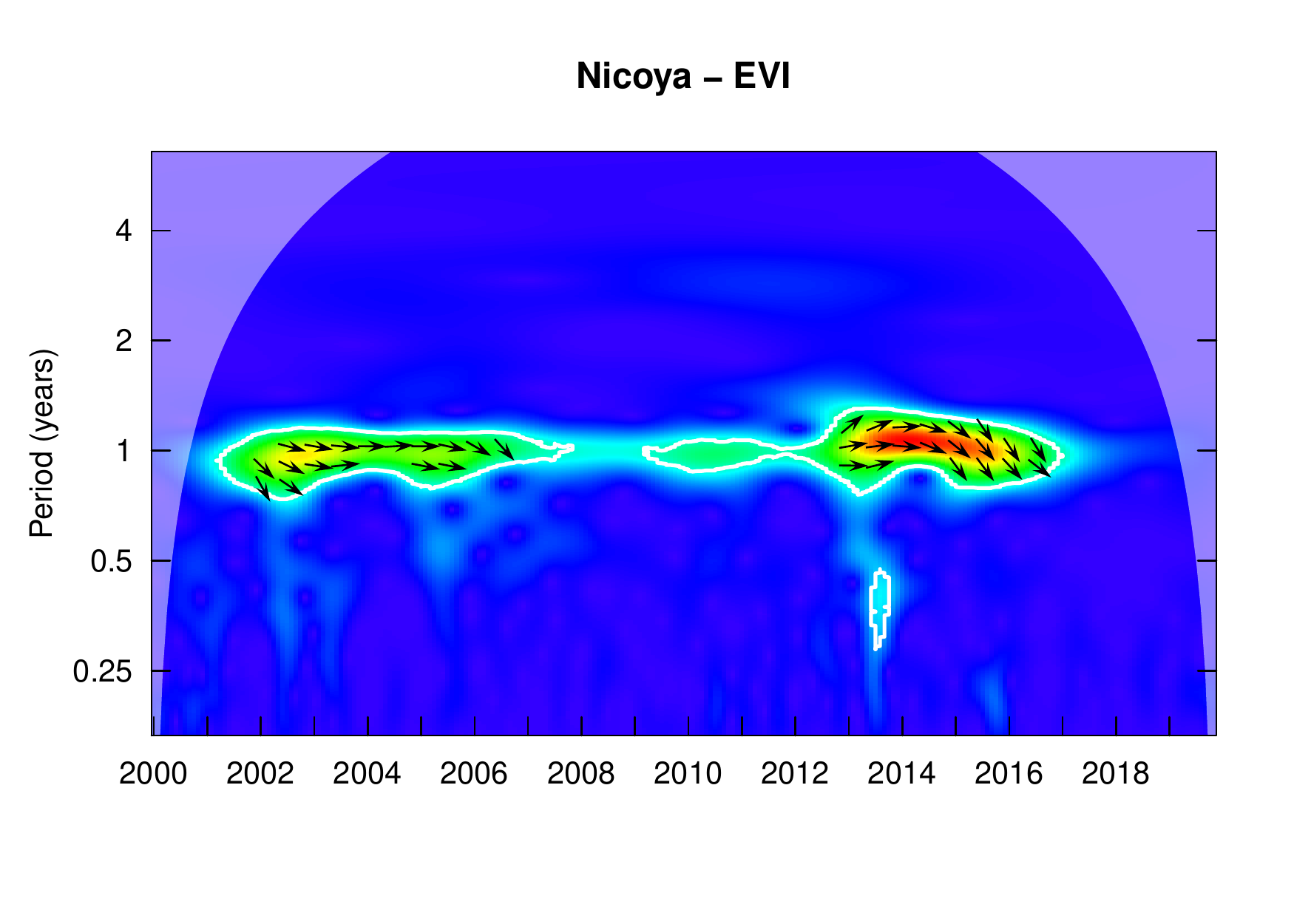}}\vspace{-0.15cm}%
\subfloat[]{\includegraphics[scale=0.23]{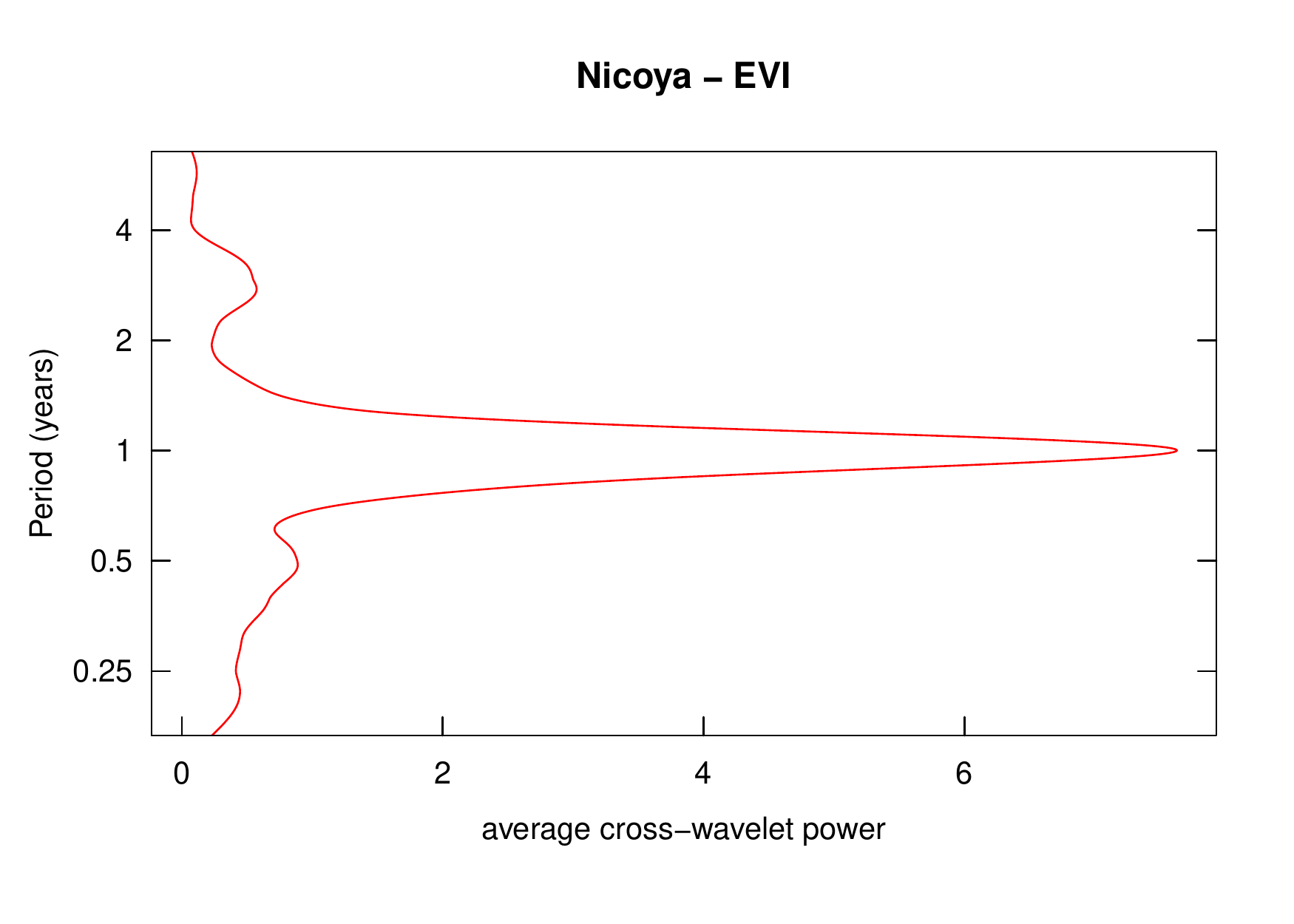}}\vspace{-0.15cm}\\
\subfloat[]{\includegraphics[scale=0.23]{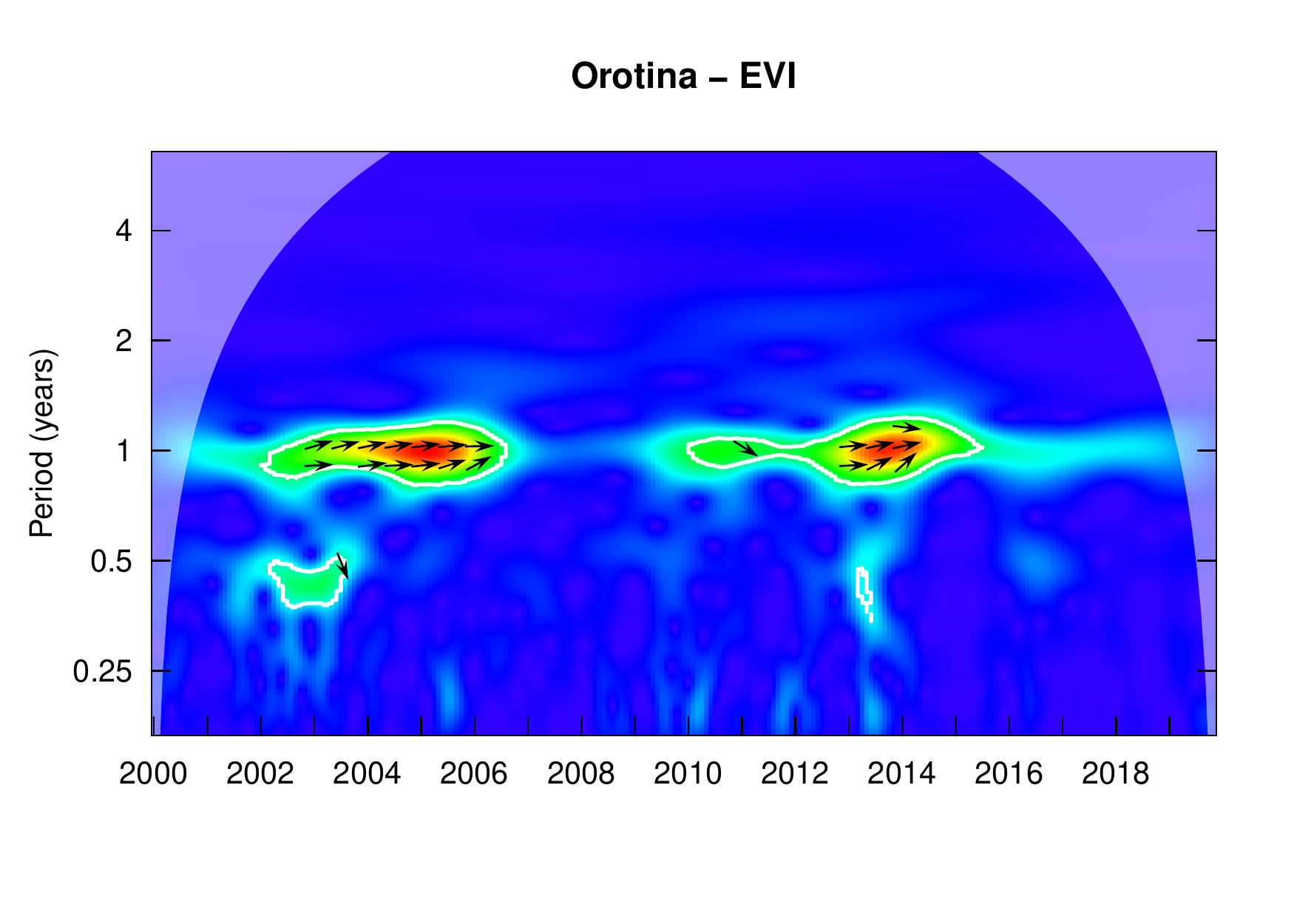}}\vspace{-0.15cm}%
\subfloat[]{\includegraphics[scale=0.23]{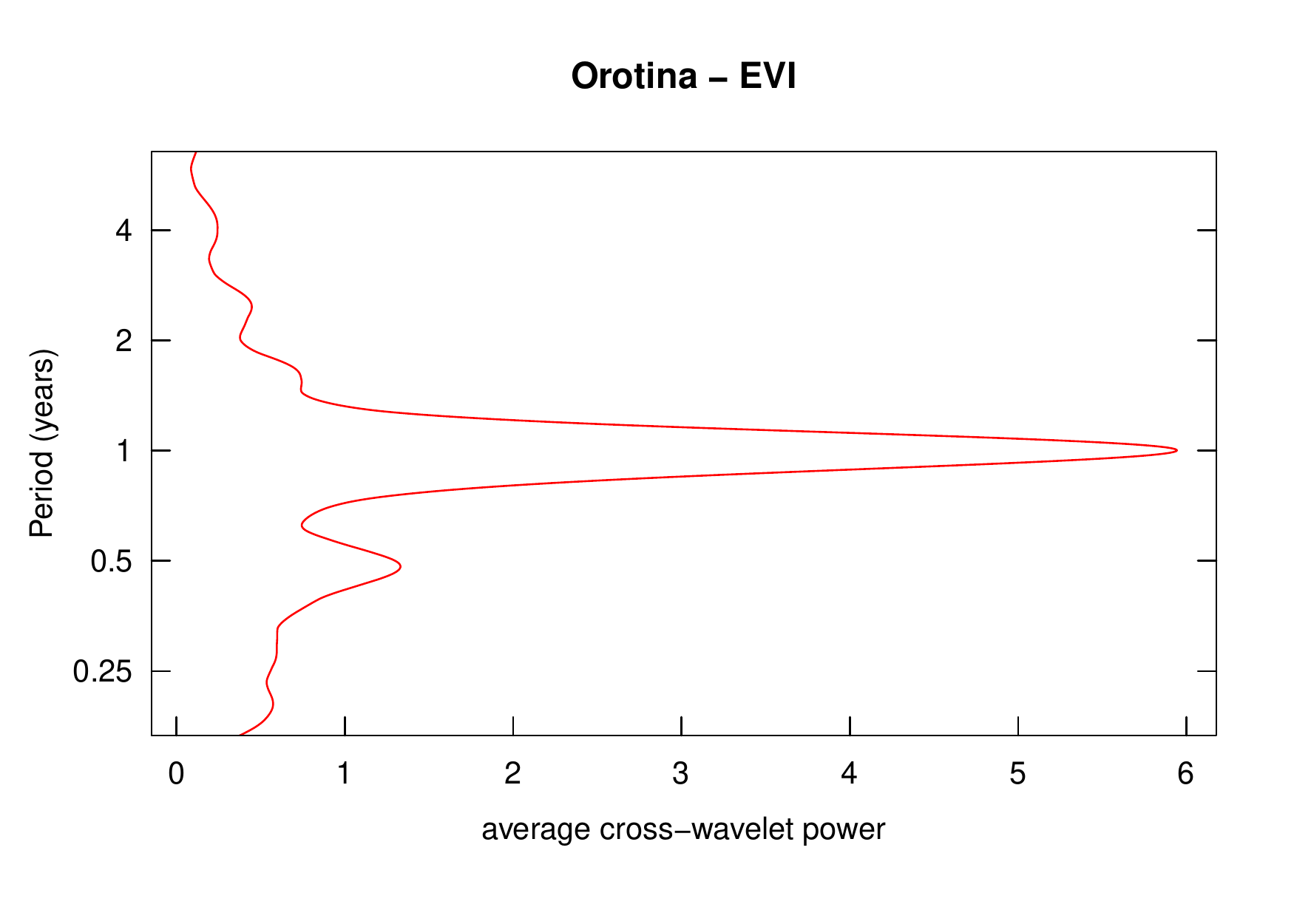}}\vspace{-0.15cm}
\subfloat[]{\includegraphics[scale=0.23]{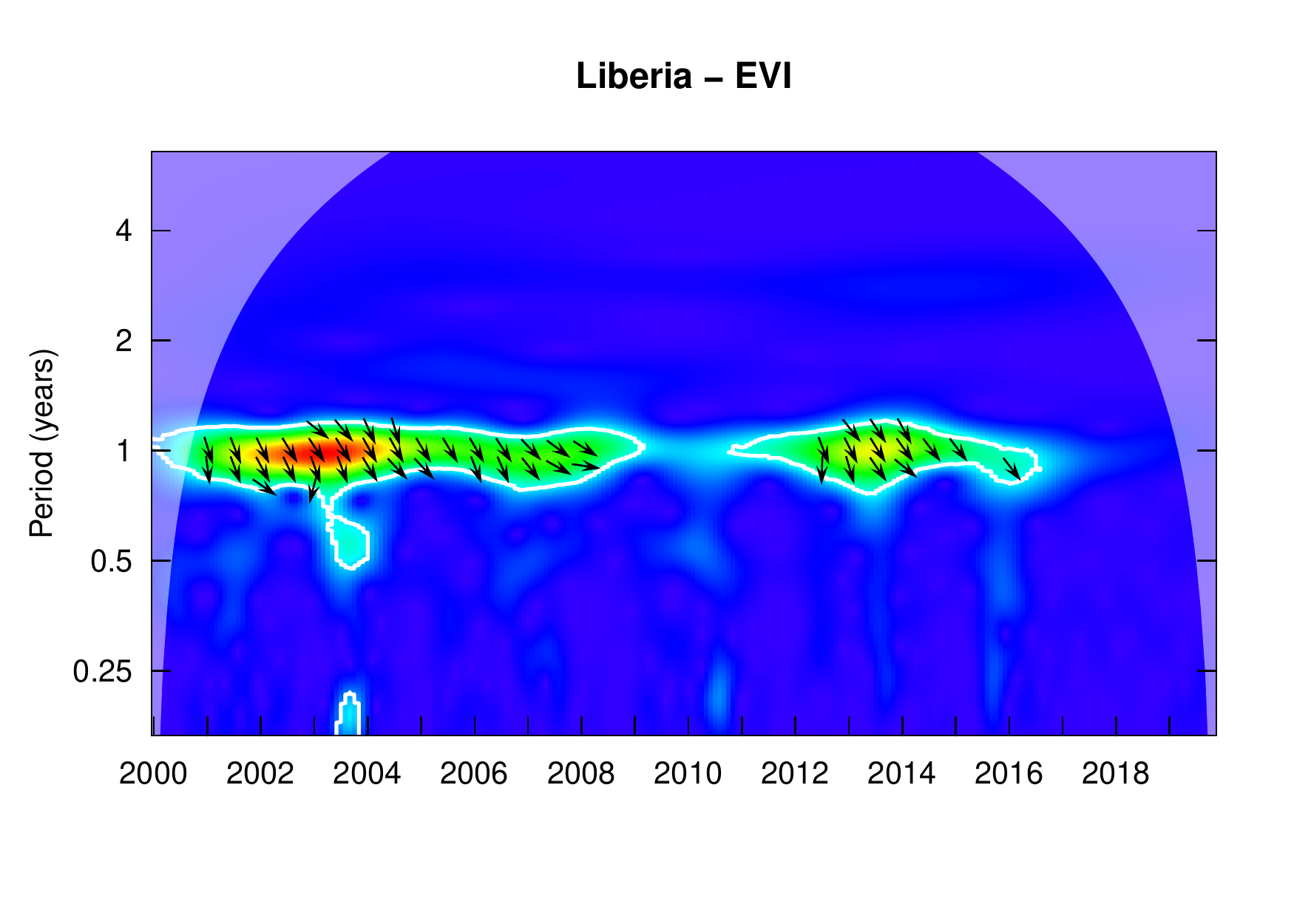}}\vspace{-0.15cm}%
\subfloat[]{\includegraphics[scale=0.23]{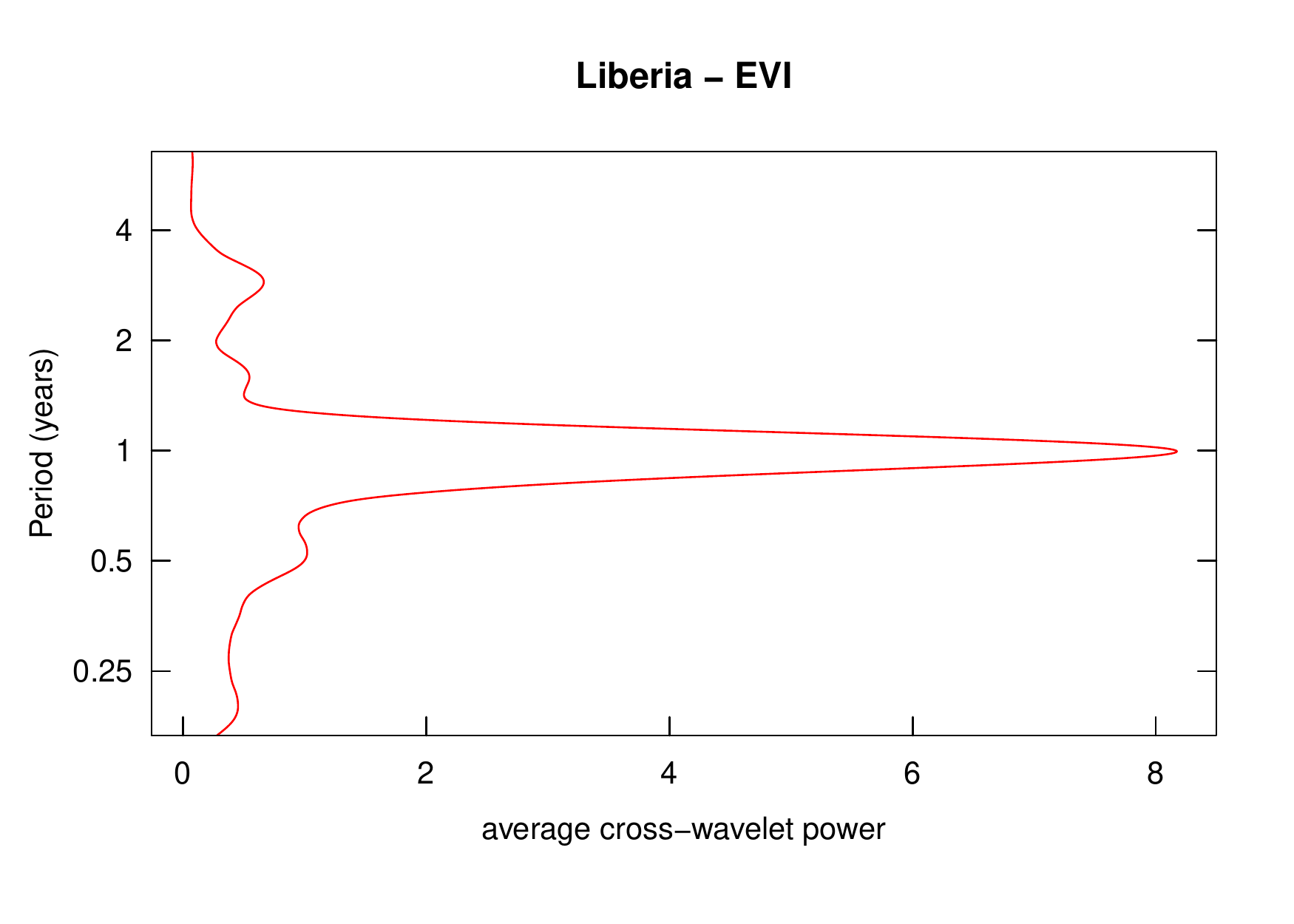}}\vspace{-0.15cm}\\
\subfloat[]{\includegraphics[scale=0.23]{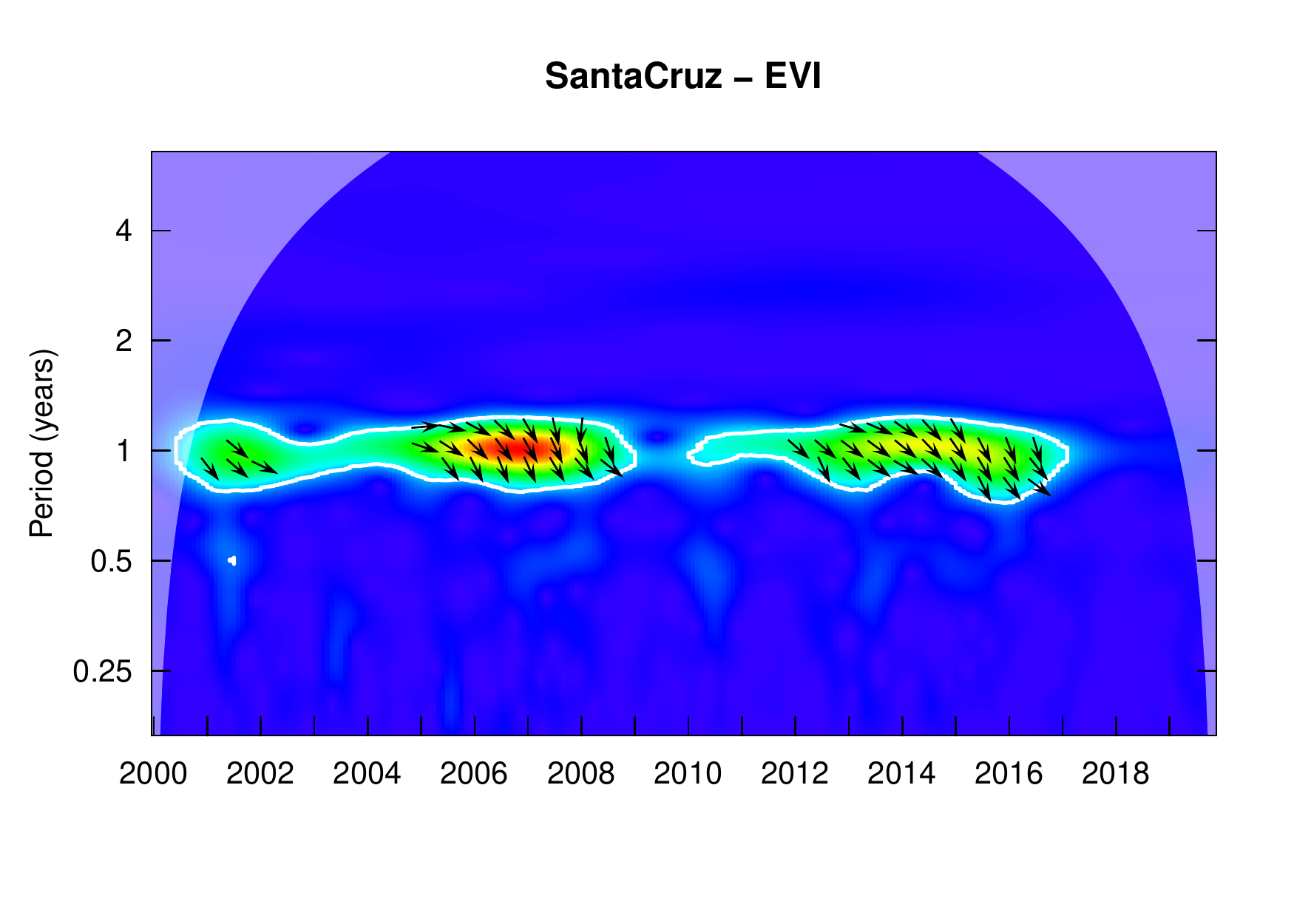}}\vspace{-0.15cm}%
\subfloat[]{\includegraphics[scale=0.23]{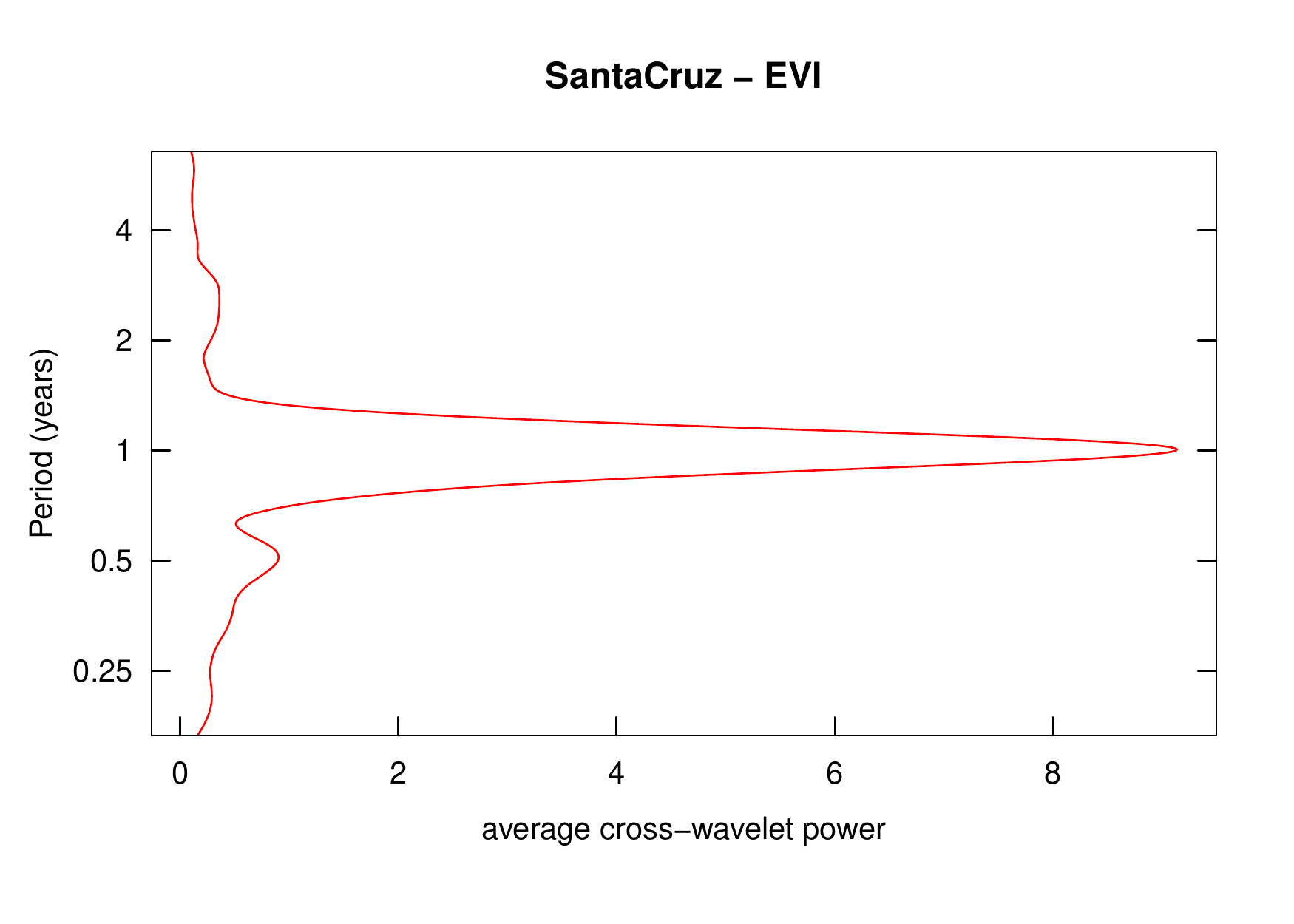}}\vspace{-0.15cm}%
\subfloat[]{\includegraphics[scale=0.23]{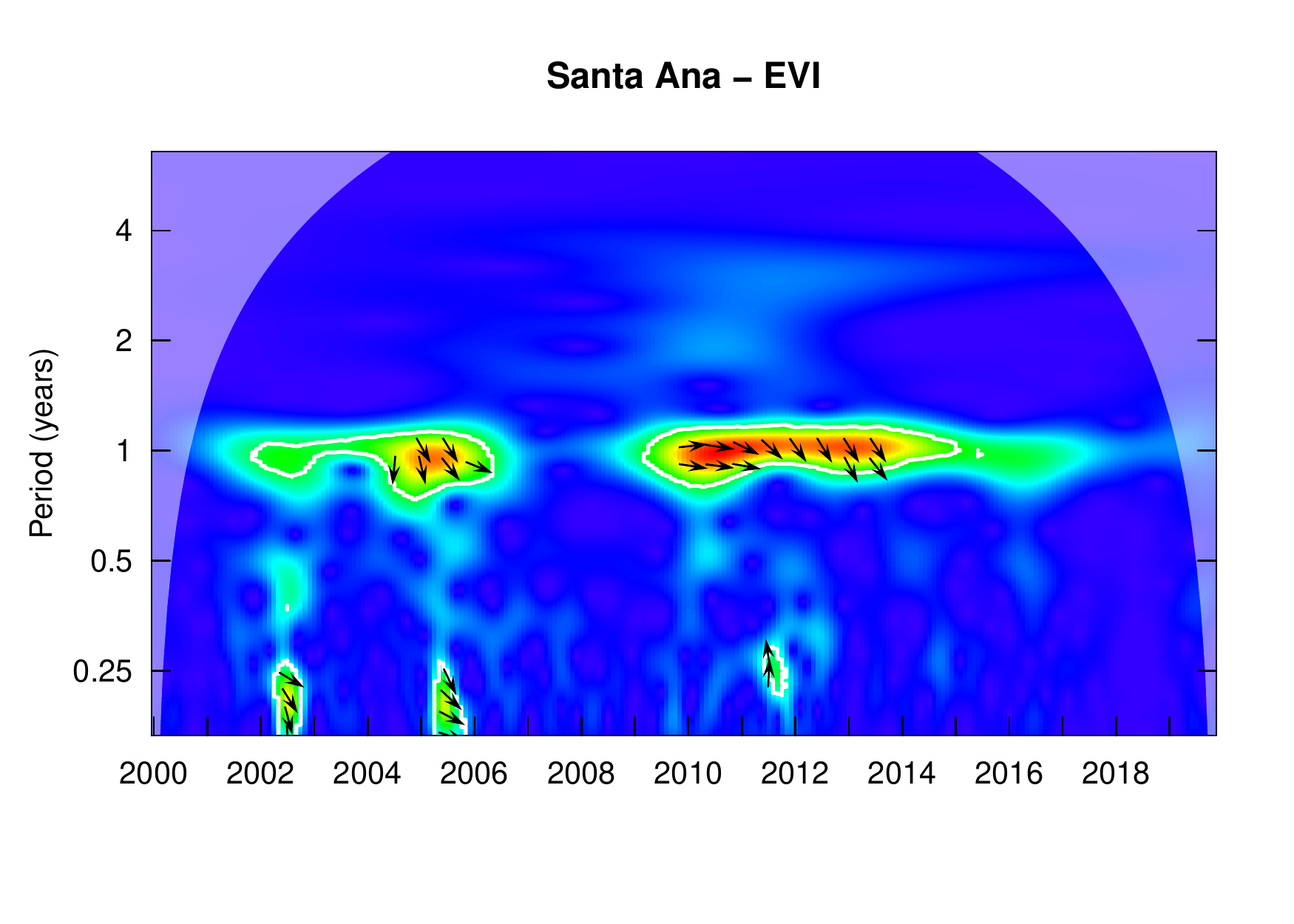}}\vspace{-0.15cm}%
\subfloat[]{\includegraphics[scale=0.23]{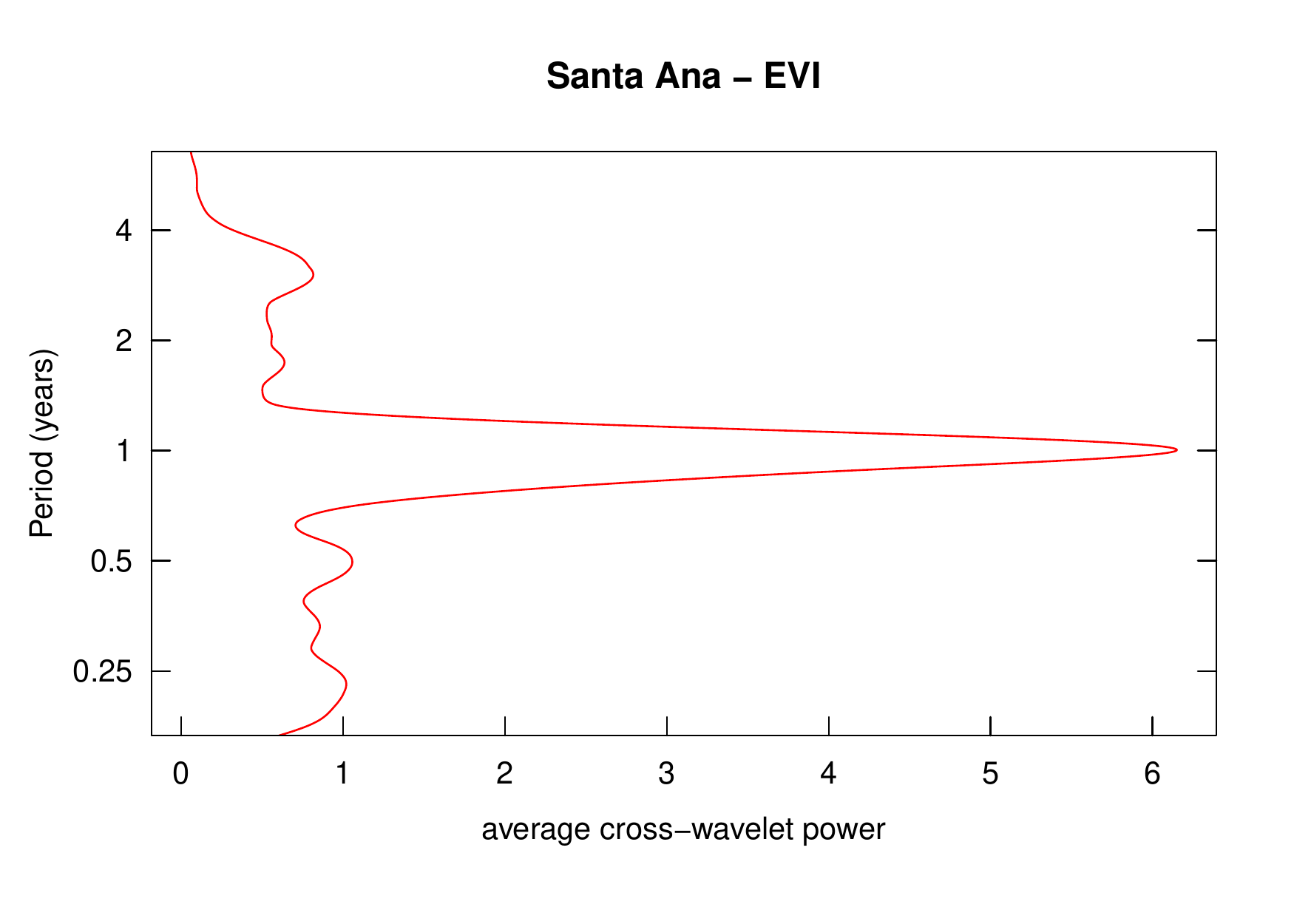}}\vspace{-0.15cm}\\
\subfloat[]{\includegraphics[scale=0.23]{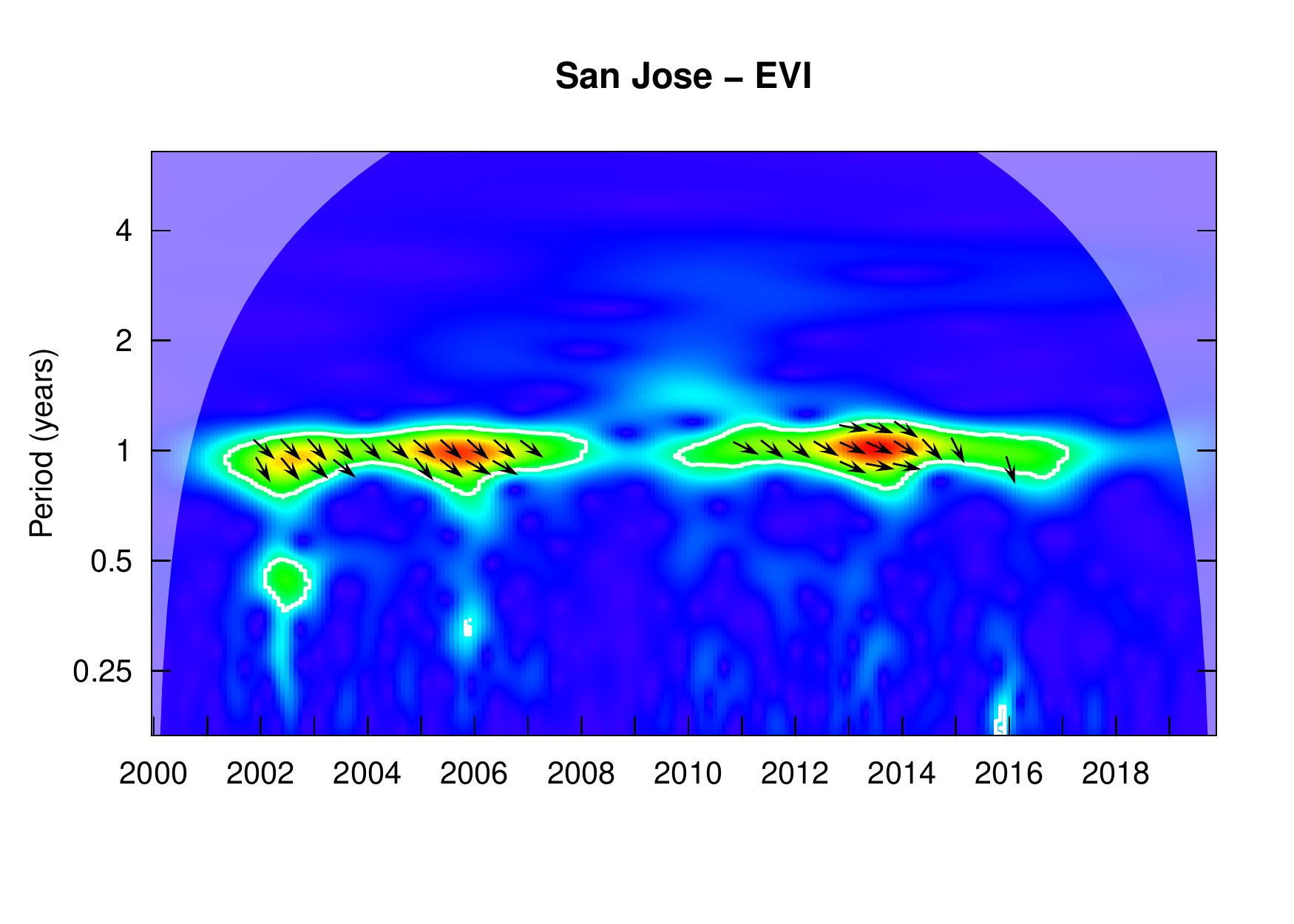}}\vspace{-0.15cm}%
\subfloat[]{\includegraphics[scale=0.23]{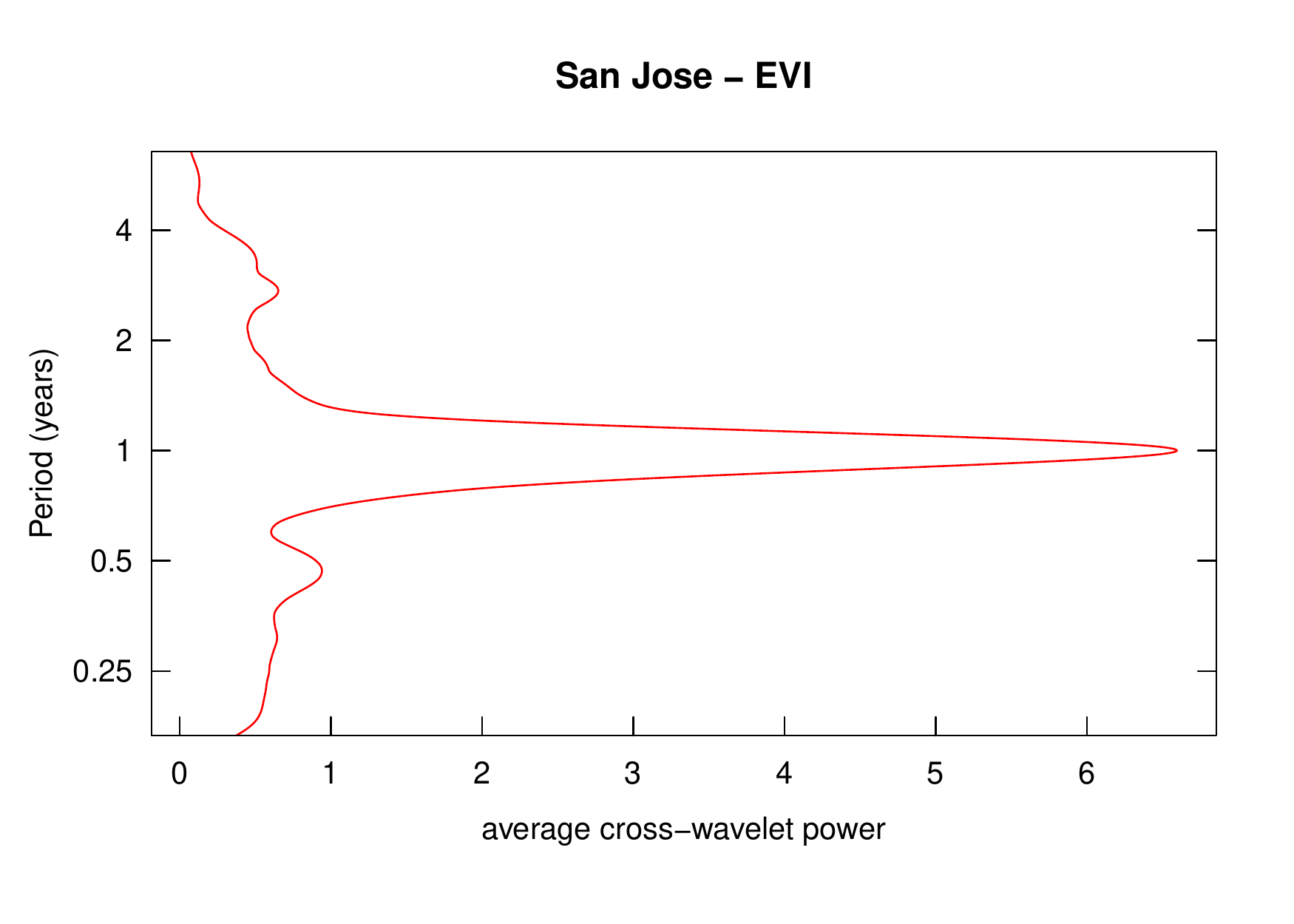}}\vspace{-0.15cm}%
\subfloat[]{\includegraphics[scale=0.23]{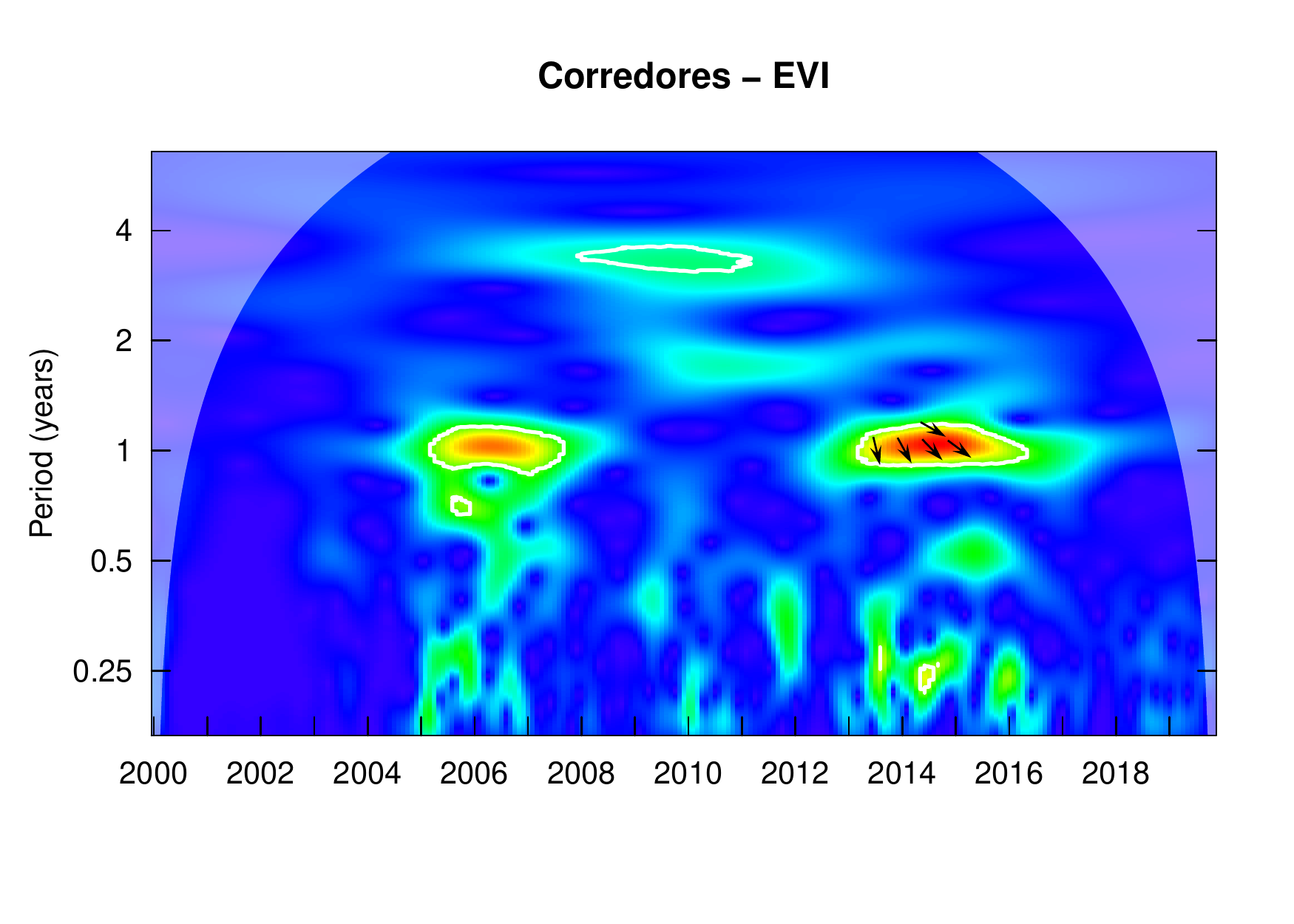}}\vspace{-0.15cm}%
\subfloat[]{\includegraphics[scale=0.23]{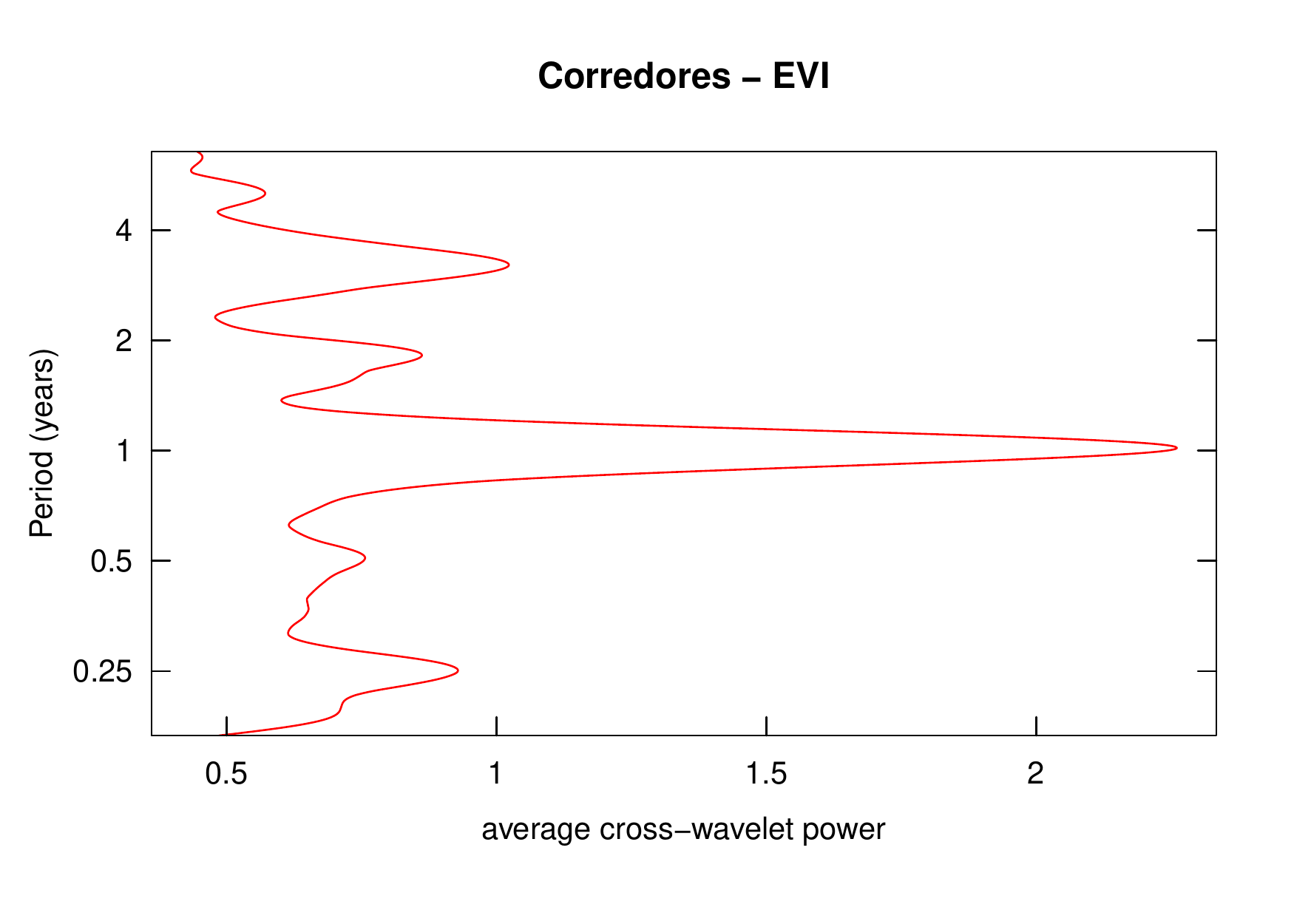}}\vspace{-0.15cm}\\
\subfloat[]{\includegraphics[scale=0.23]{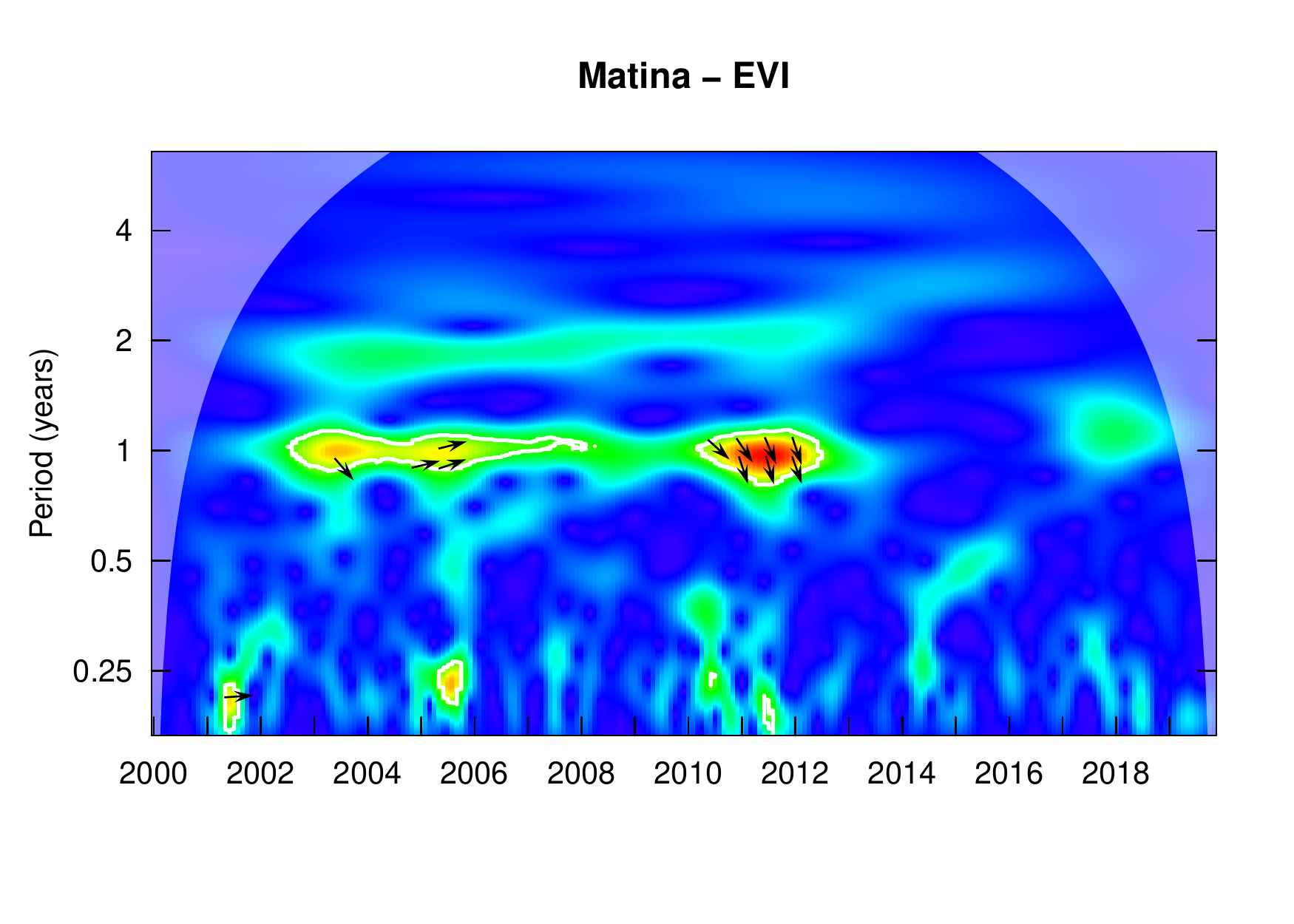}}\vspace{-0.15cm}%
\subfloat[]{\includegraphics[scale=0.23]{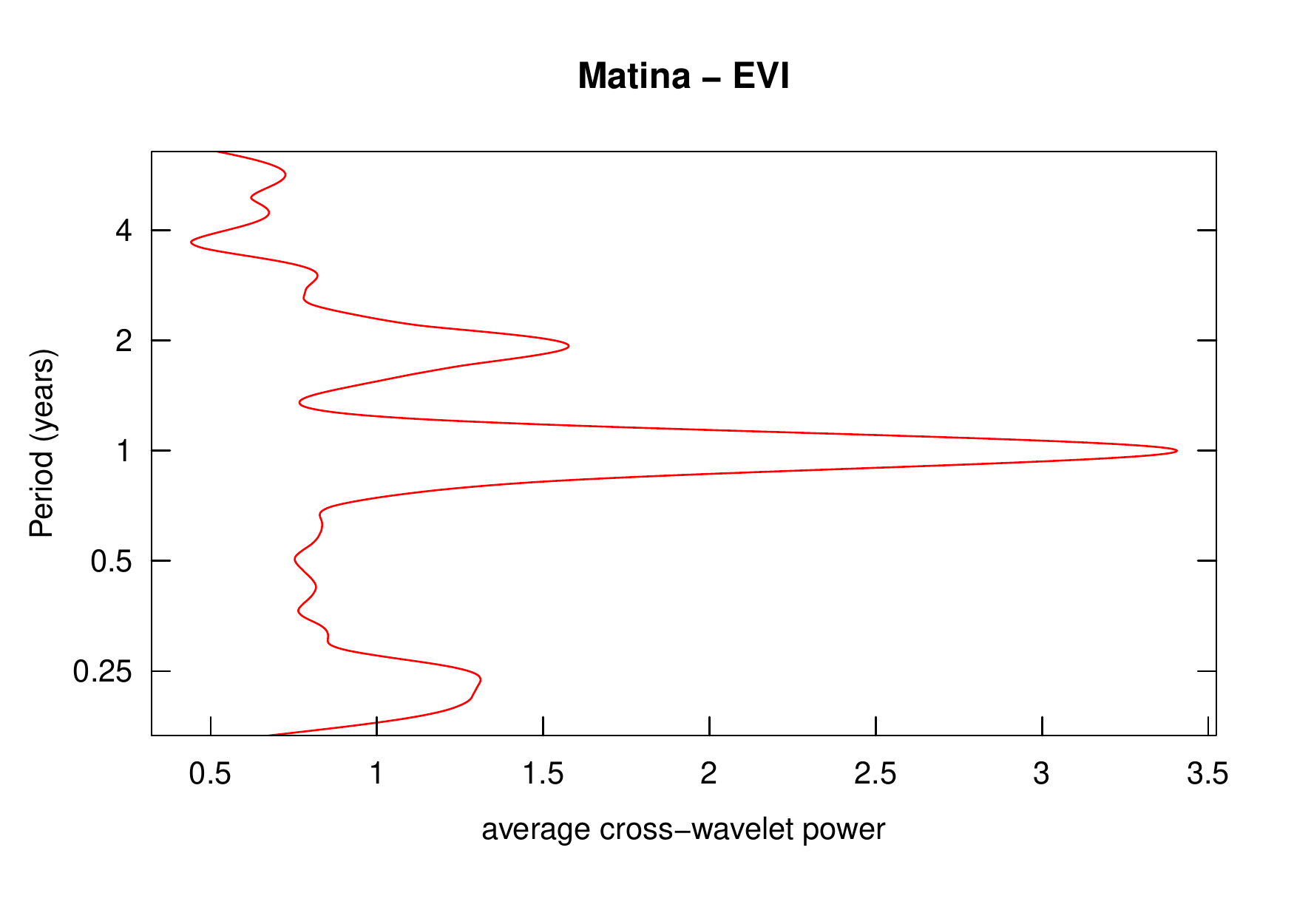}}\vspace{-0.15cm}%
\subfloat[]{\includegraphics[scale=0.23]{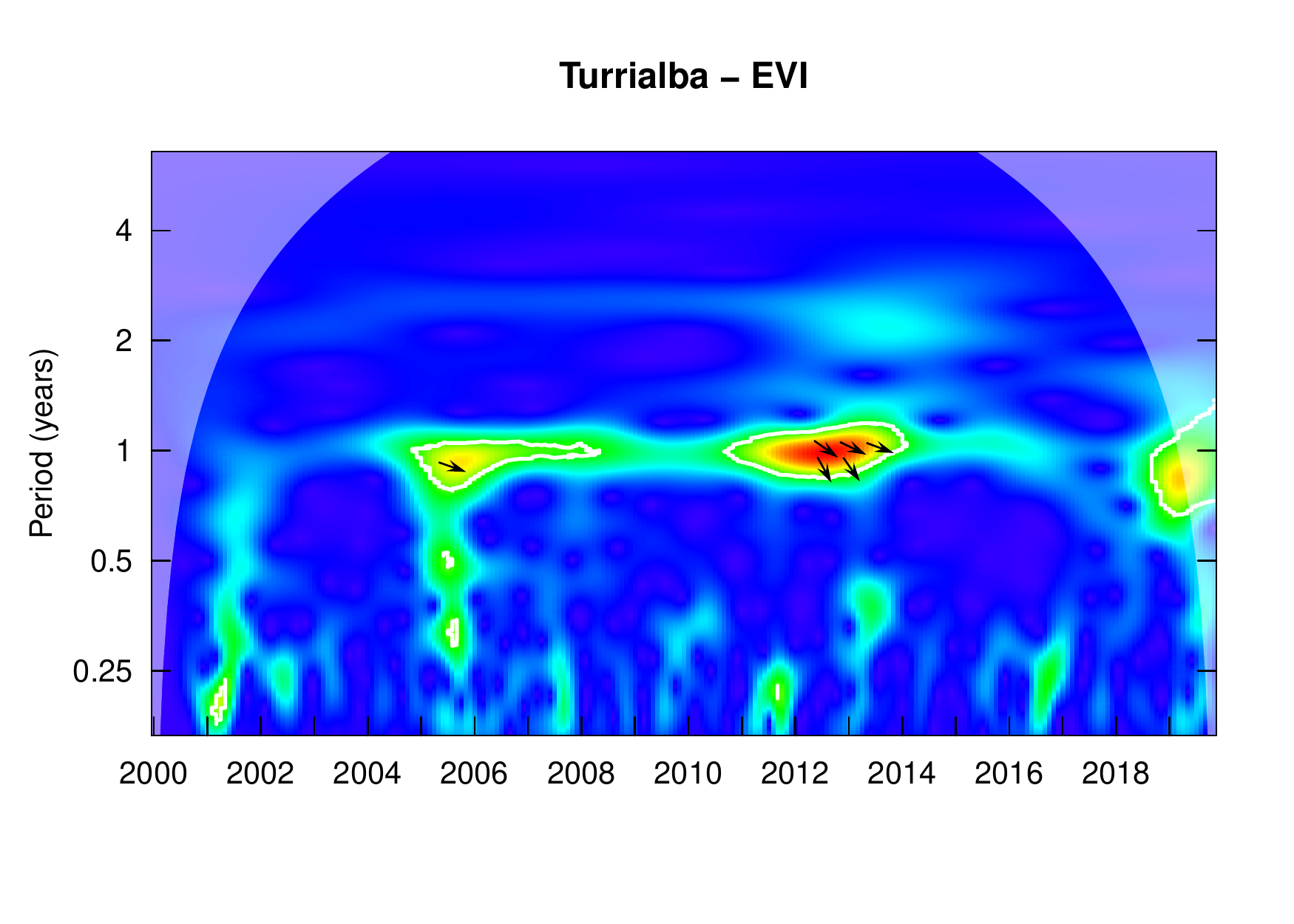}}\vspace{-0.15cm}%
\subfloat[]{\includegraphics[scale=0.23]{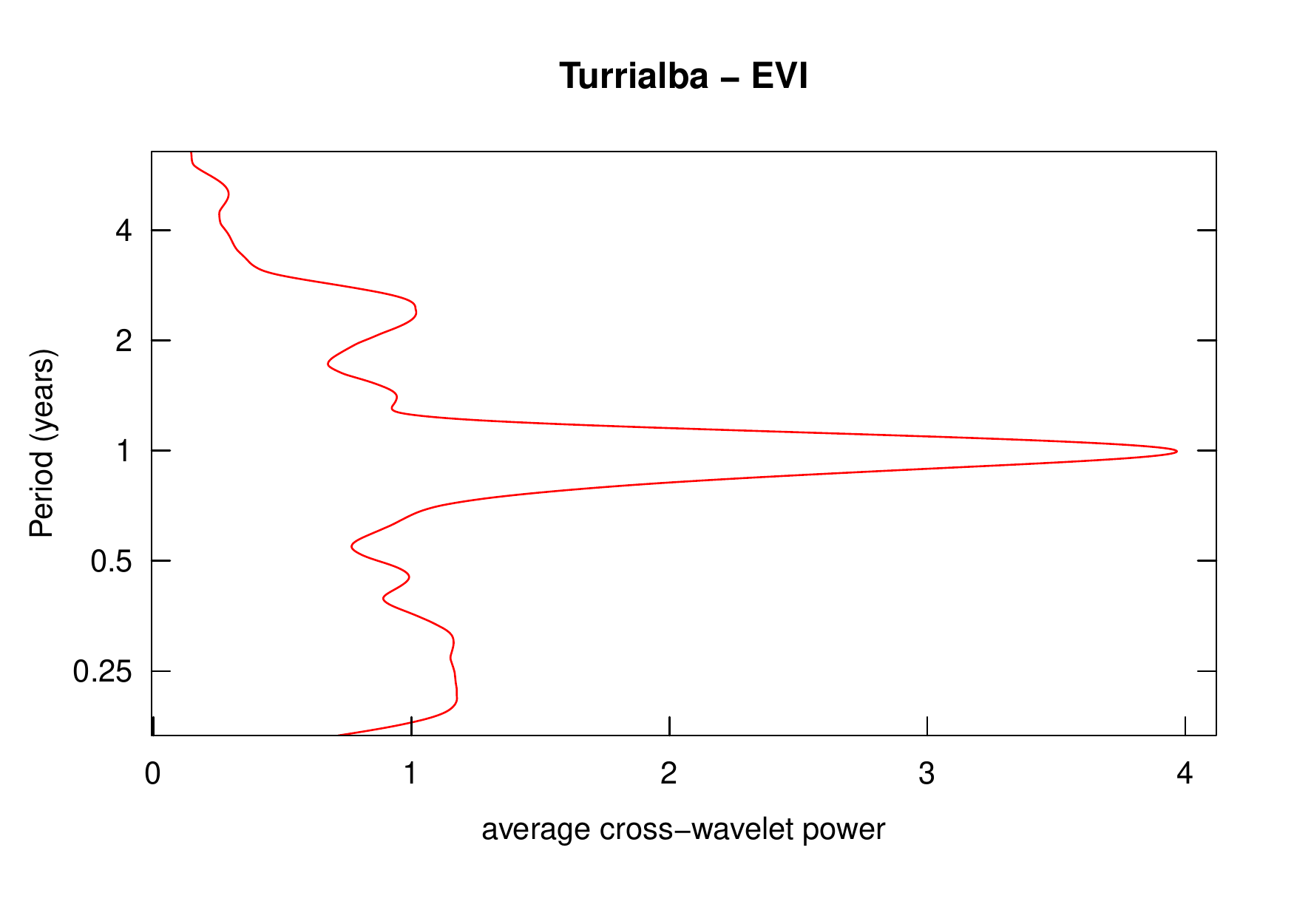}}\vspace{-0.15cm}
\subfloat[]{\includegraphics[scale=0.23]{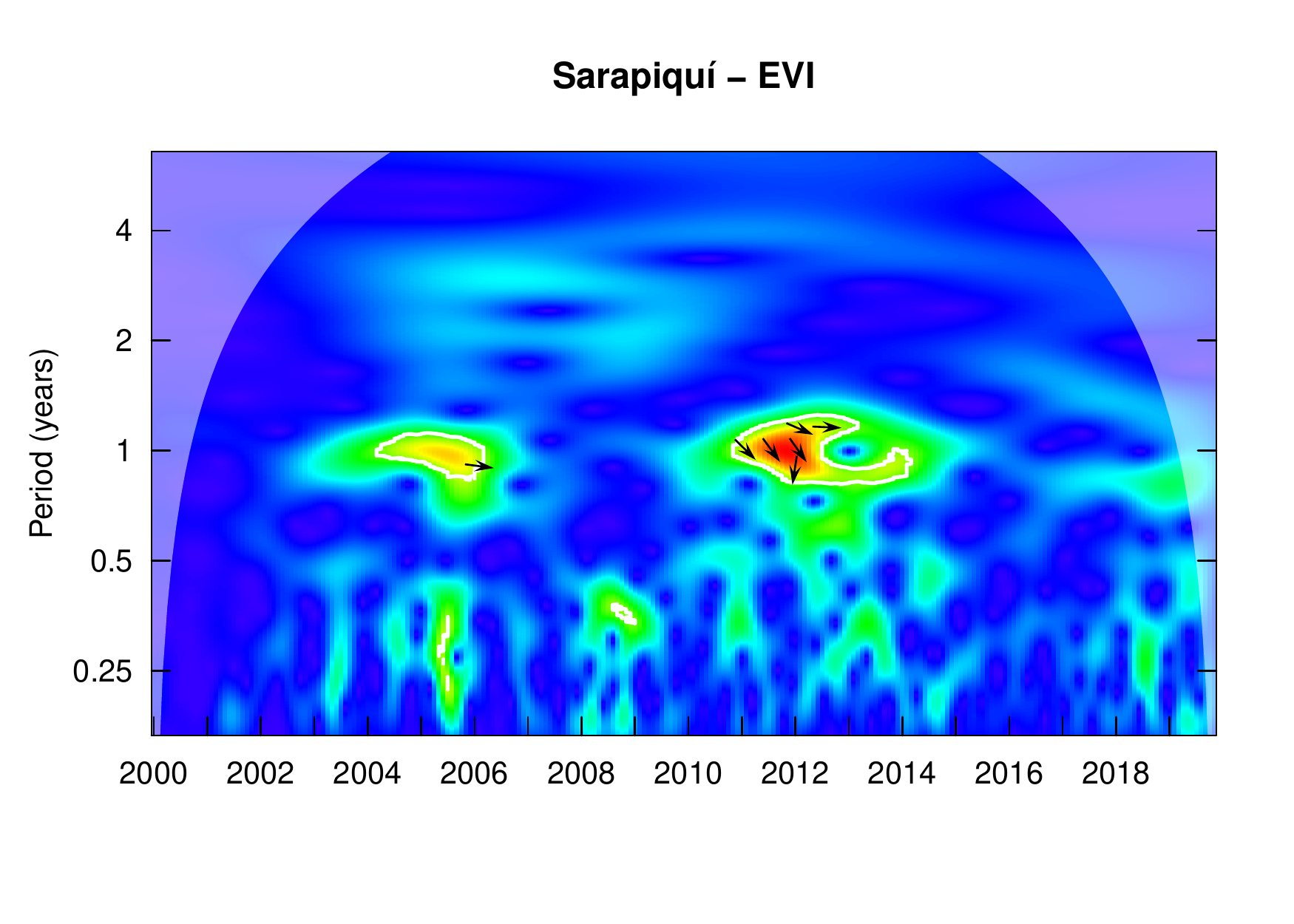}}\vspace{-0.15cm}%
\subfloat[]{\includegraphics[scale=0.23]{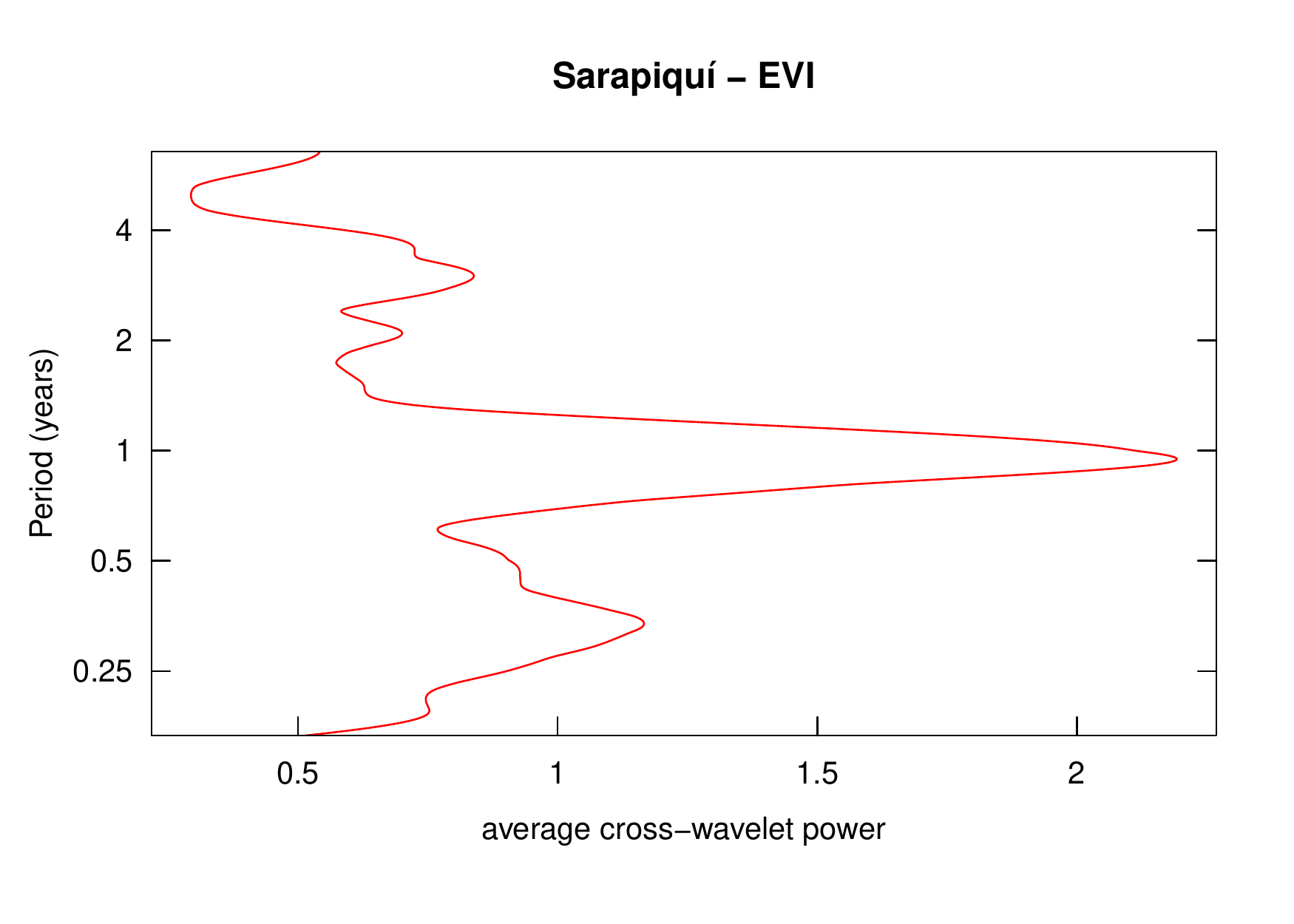}}\vspace{-0.15cm}%
\subfloat[]{\includegraphics[scale=0.23]{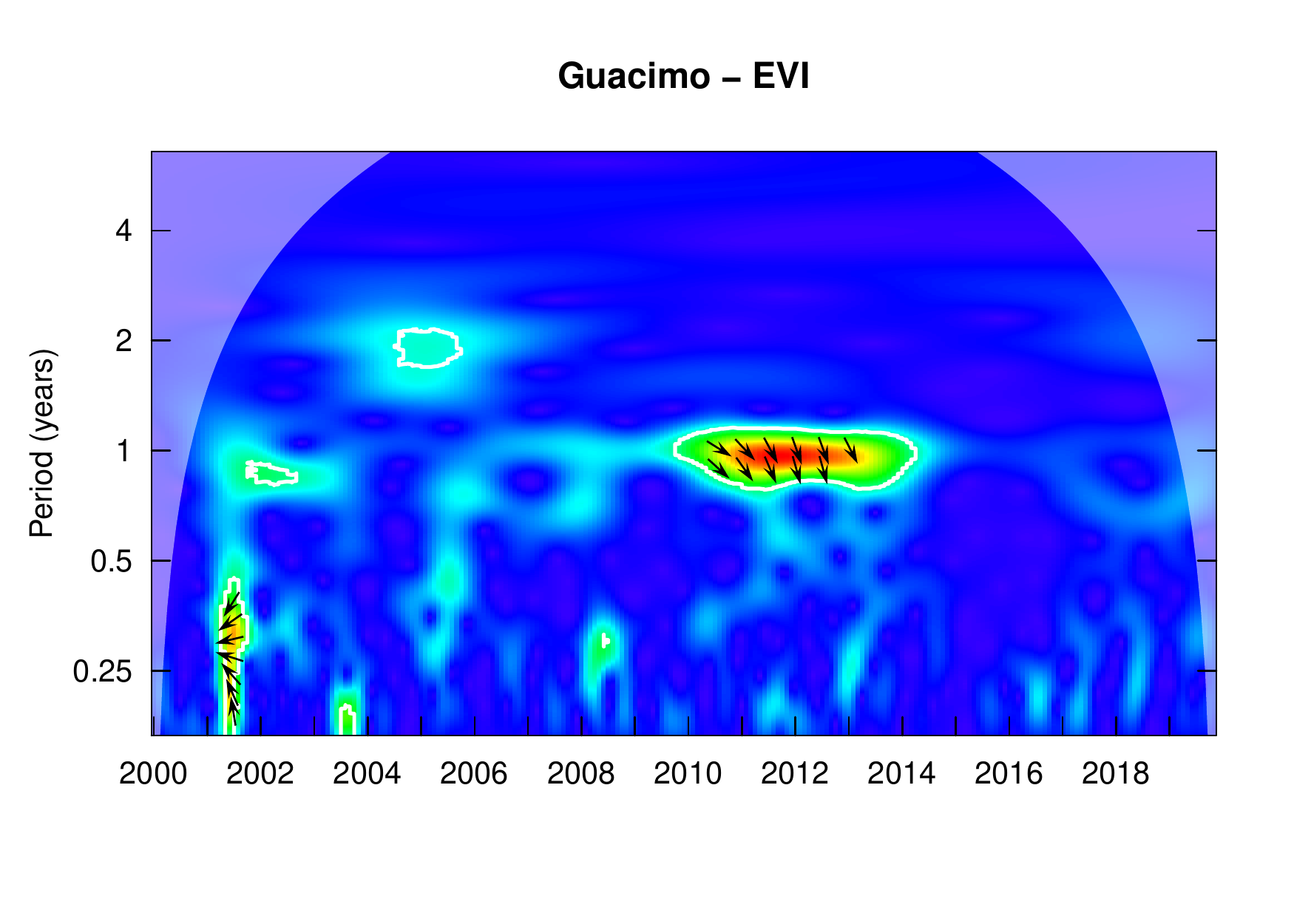}}\vspace{-0.15cm}%
\subfloat[]{\includegraphics[scale=0.23]{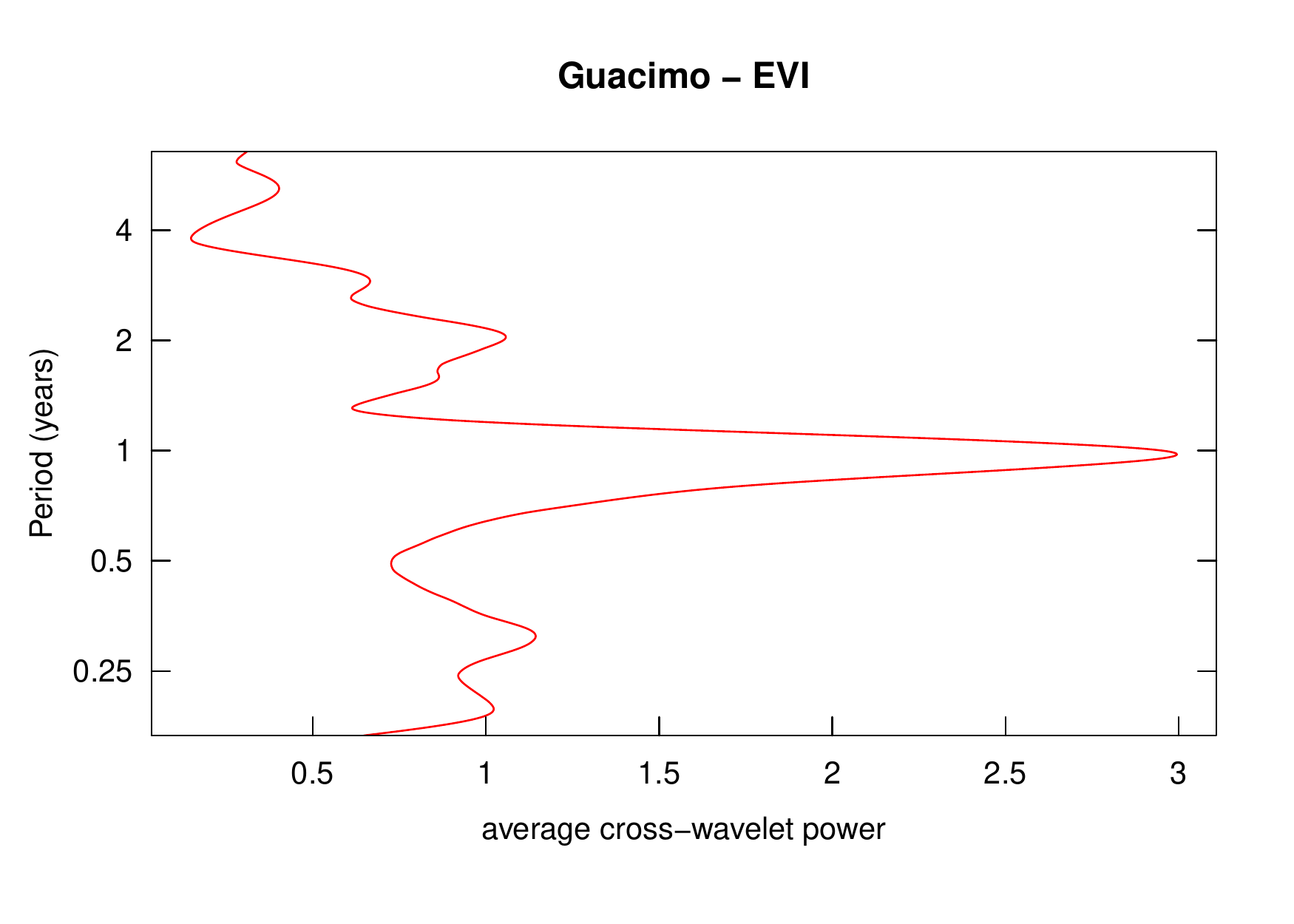}}\vspace{-0.15cm}\\
\end{figure}

\begin{figure}[H]
\captionsetup[subfigure]{labelformat=empty}
\subfloat[]{\includegraphics[scale=0.23]{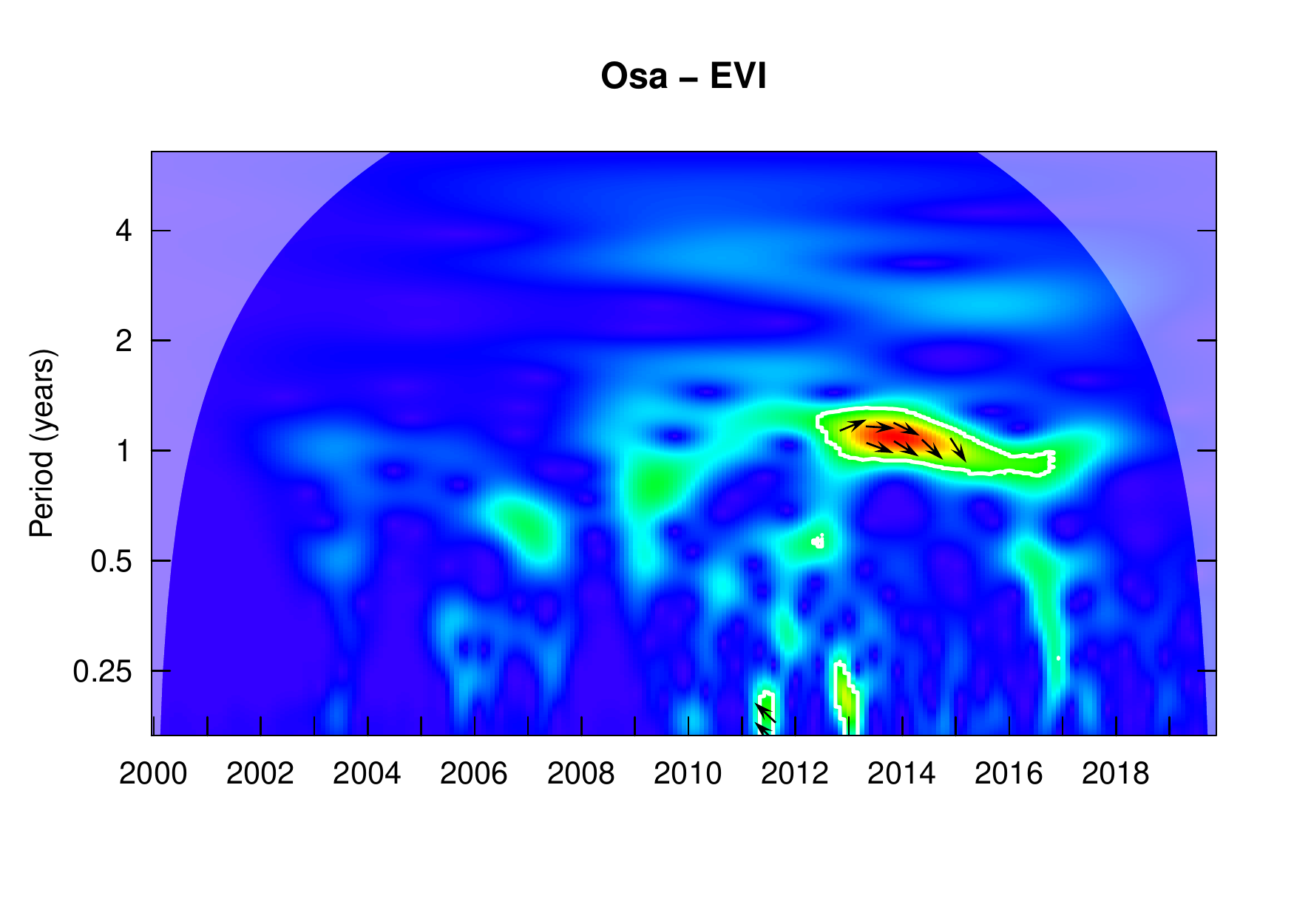}}\vspace{-0.15cm}%
\subfloat[]{\includegraphics[scale=0.23]{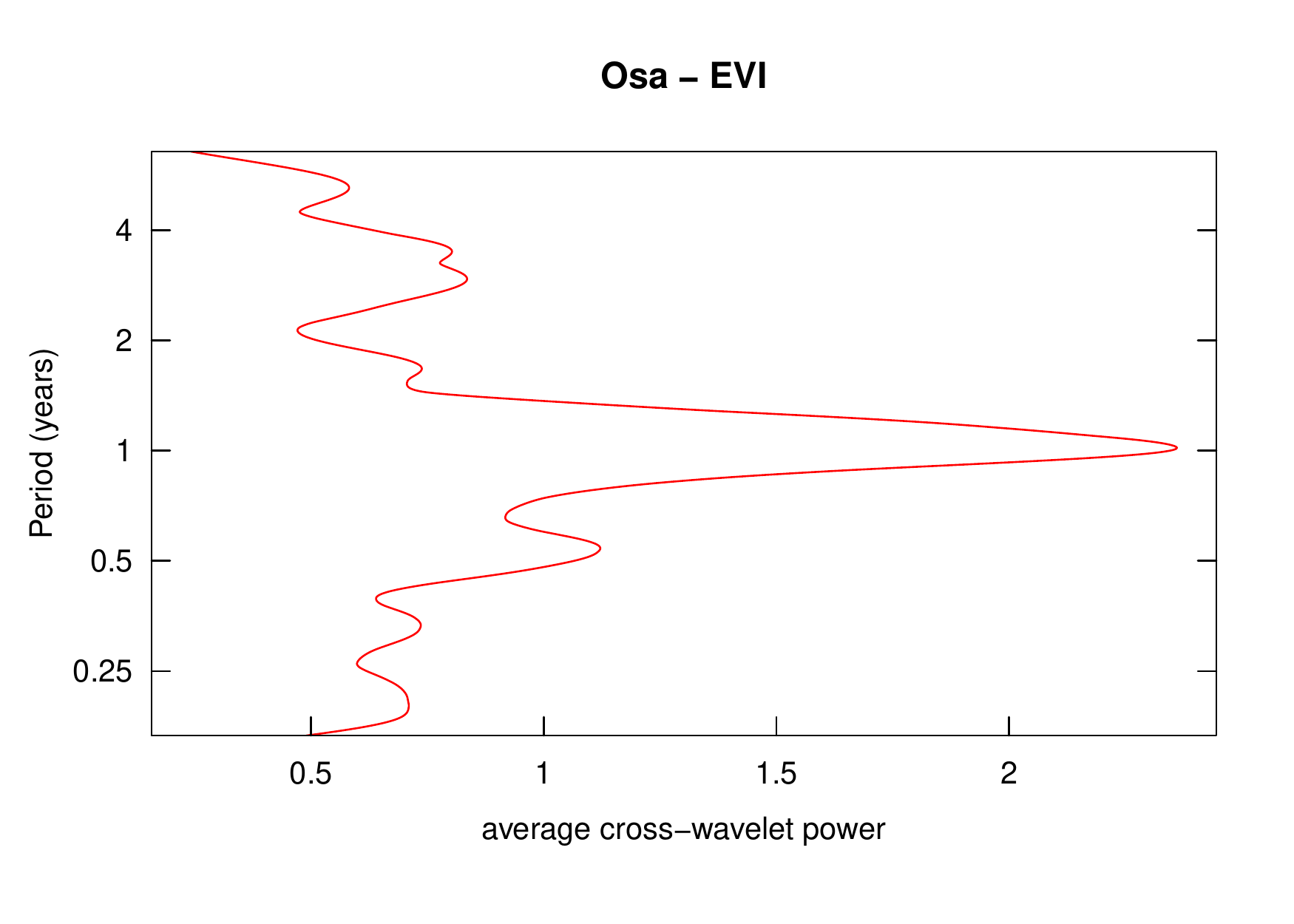}}\vspace{-0.15cm}%
\subfloat[]{\includegraphics[scale=0.23]{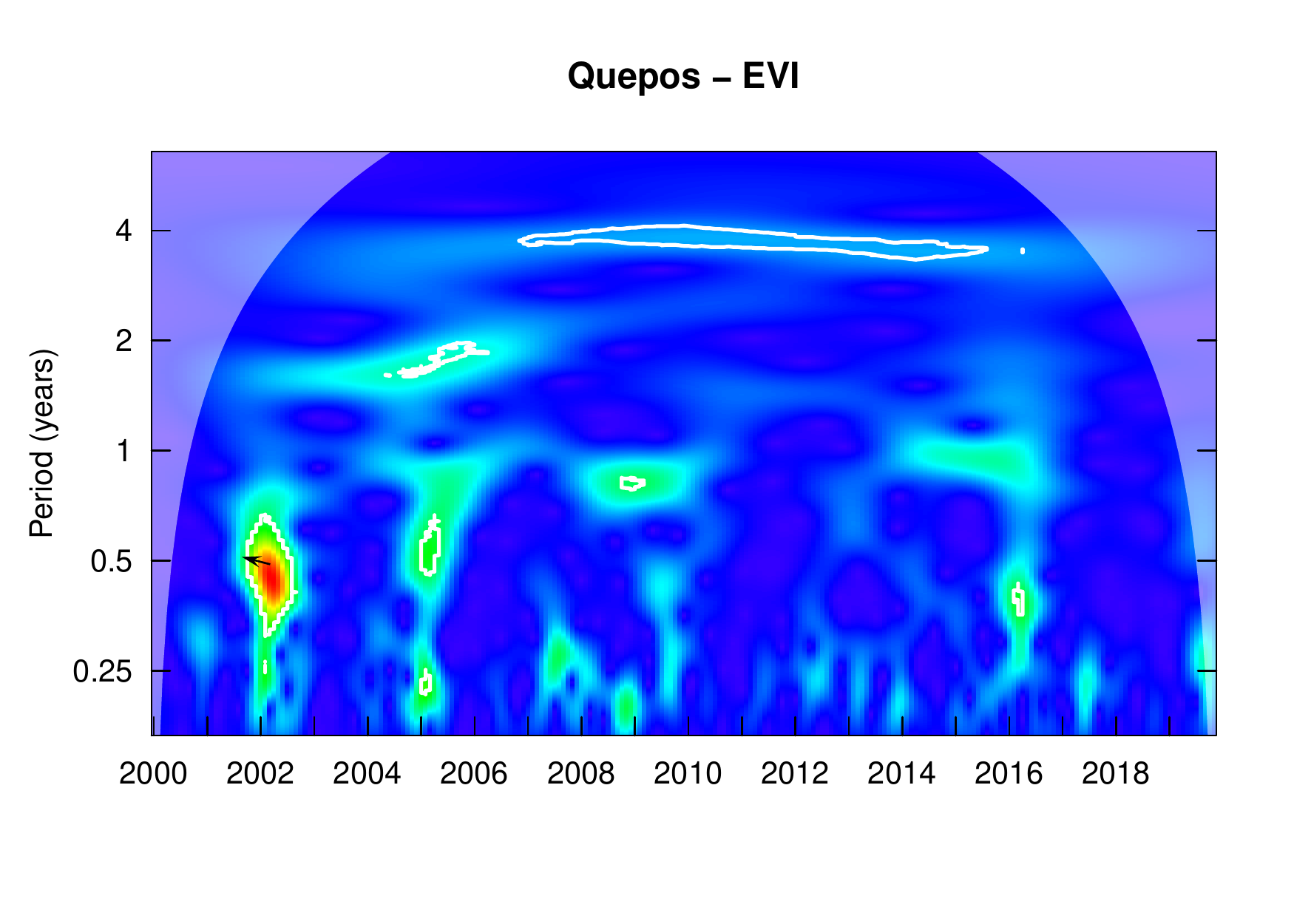}}\vspace{-0.15cm}%
\subfloat[]{\includegraphics[scale=0.23]{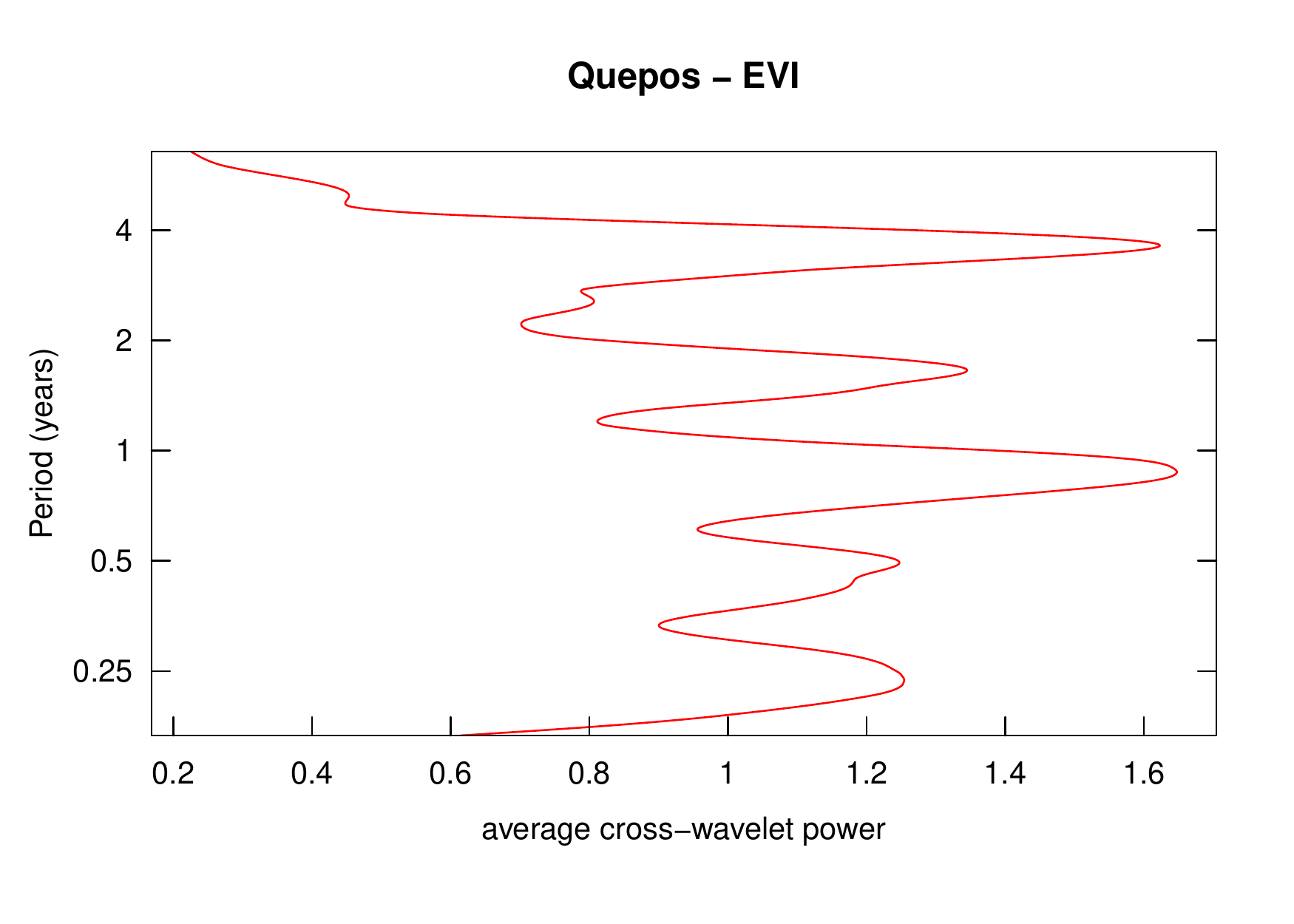}}\vspace{-0.15cm}\\
\subfloat[]{\includegraphics[scale=0.23]{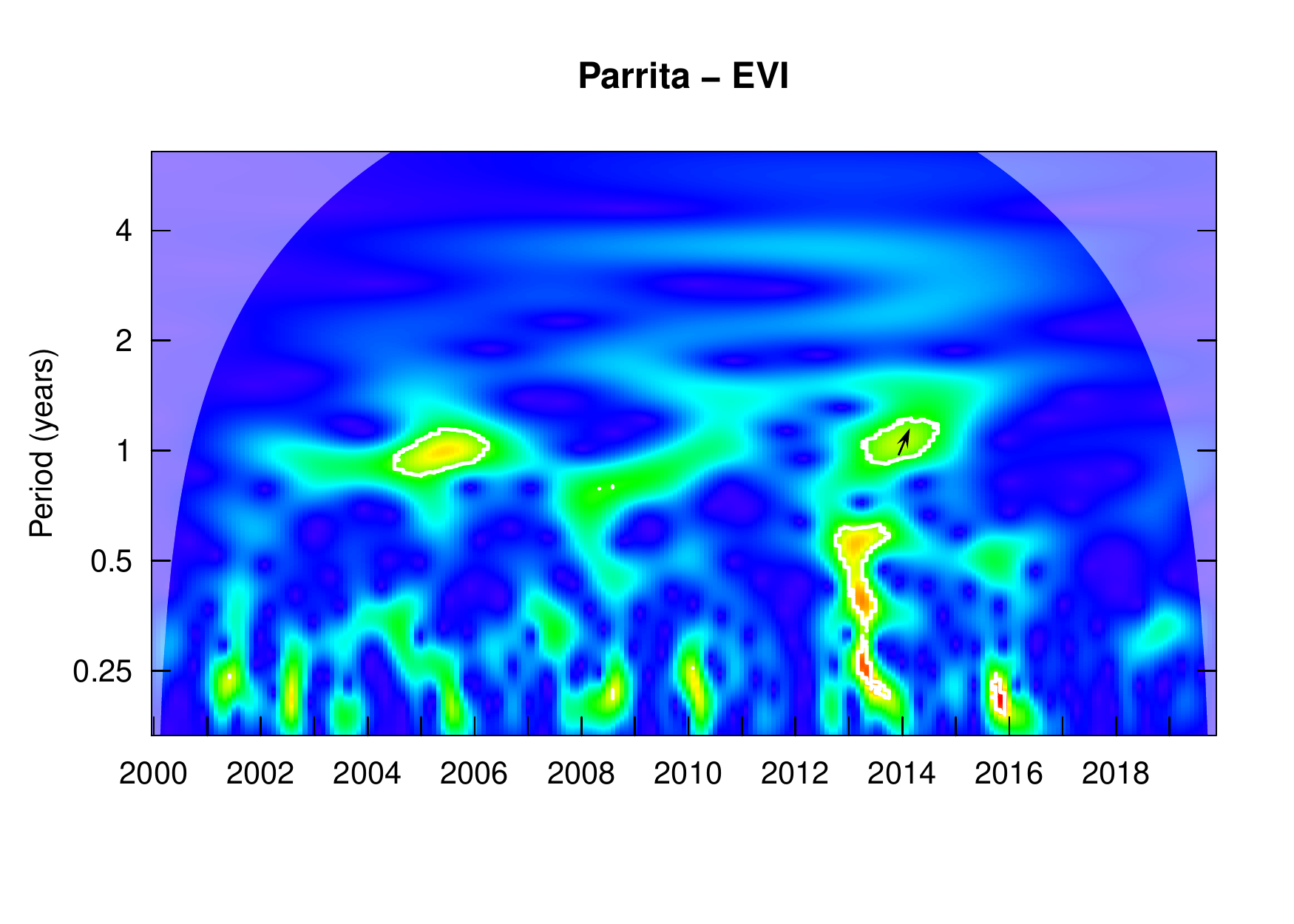}}\vspace{-0.15cm}%
\subfloat[]{\includegraphics[scale=0.23]{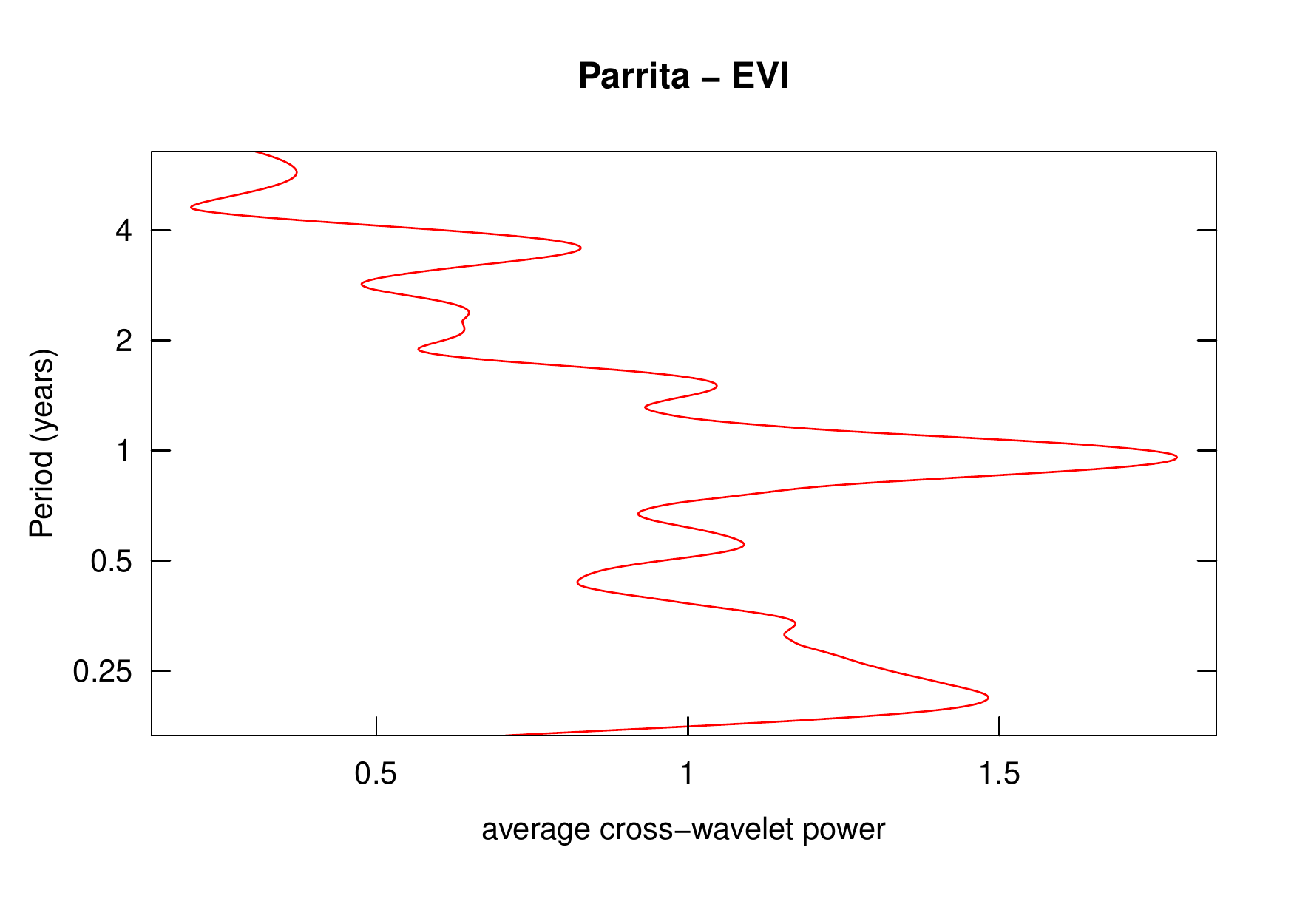}}\vspace{-0.15cm}%
\subfloat[]{\includegraphics[scale=0.23]{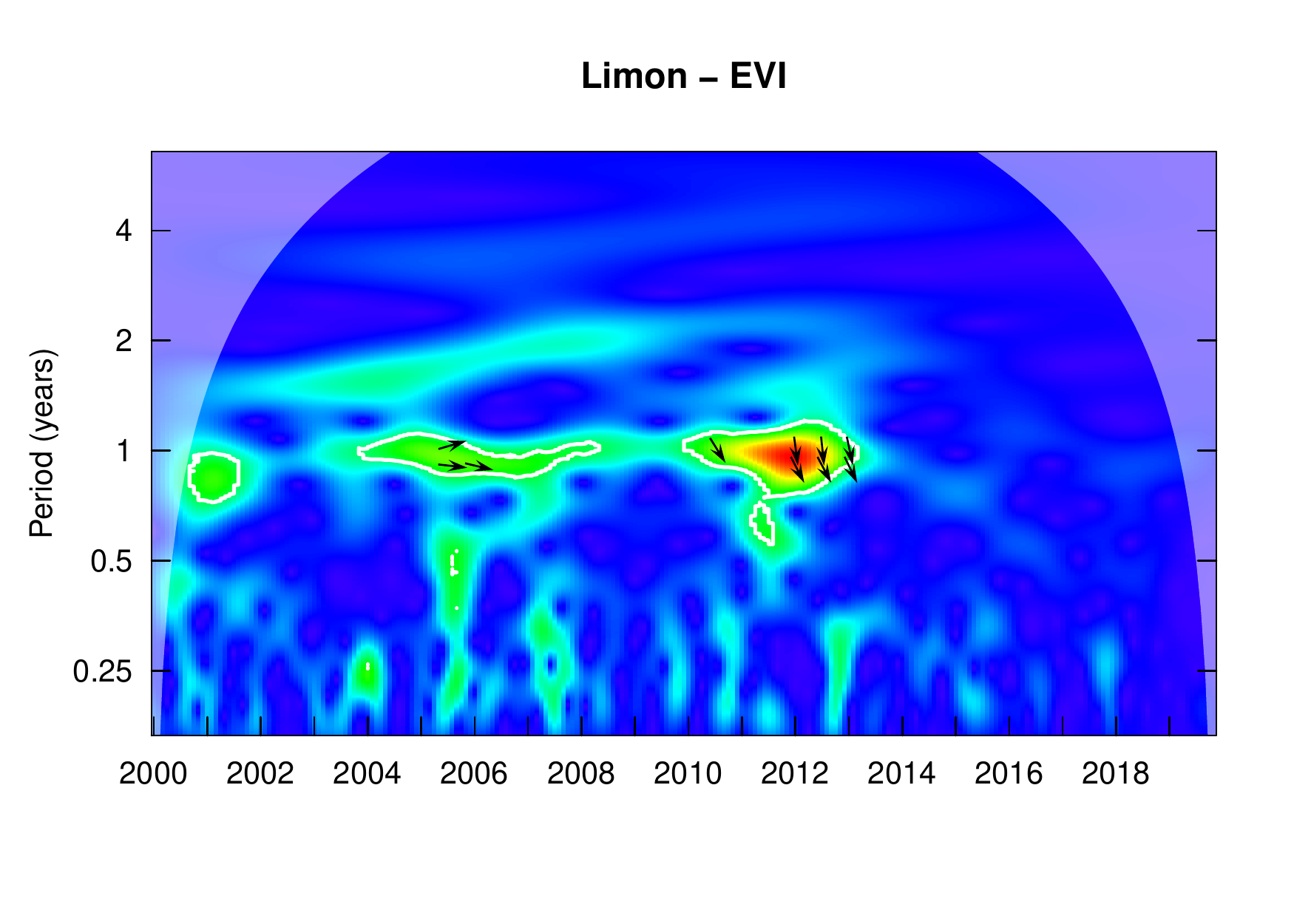}}\vspace{-0.15cm}%
\subfloat[]{\includegraphics[scale=0.23]{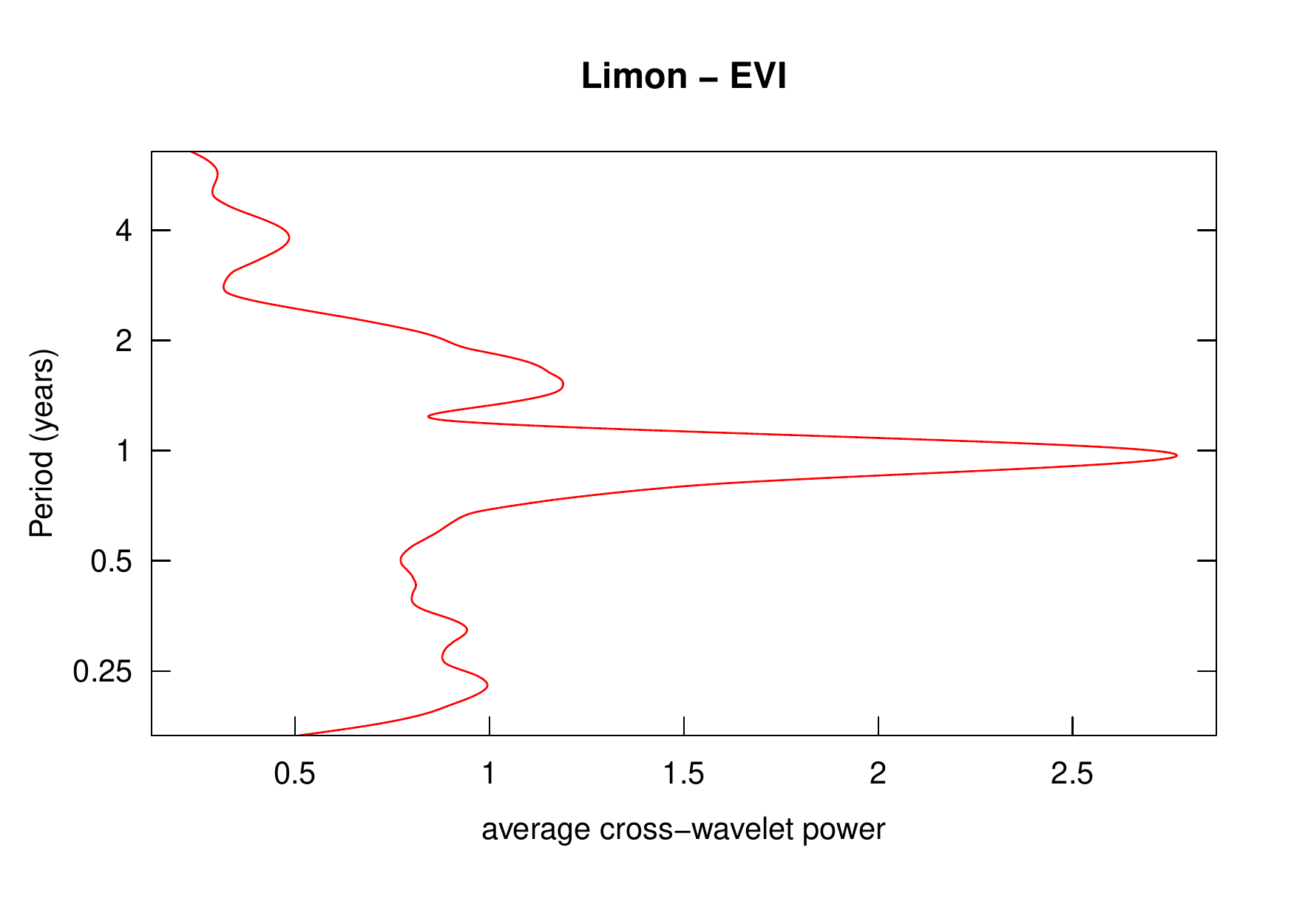}}\vspace{-0.15cm}\\
\subfloat[]{\includegraphics[scale=0.23]{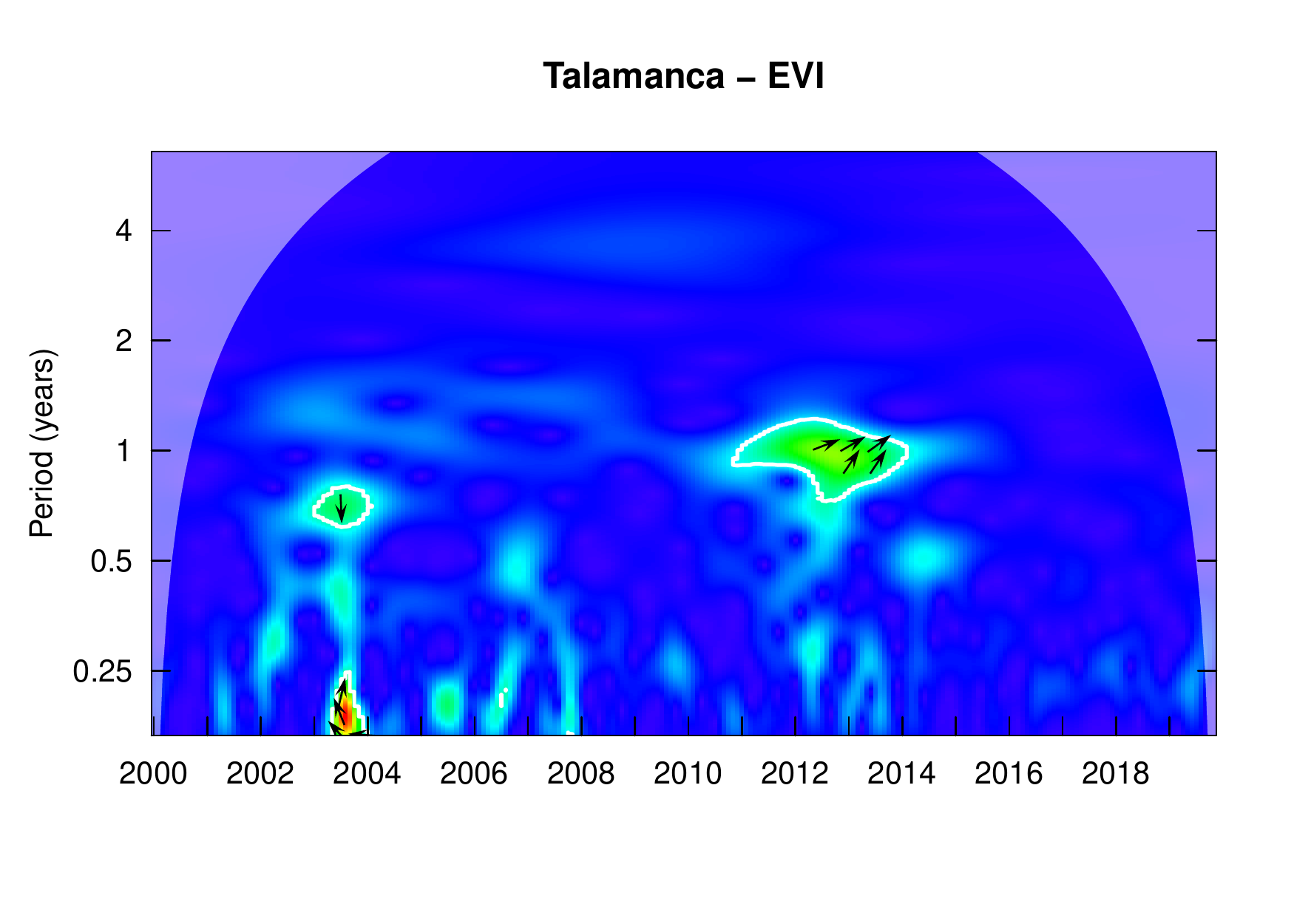}}\vspace{-0.15cm}%
\subfloat[]{\includegraphics[scale=0.23]{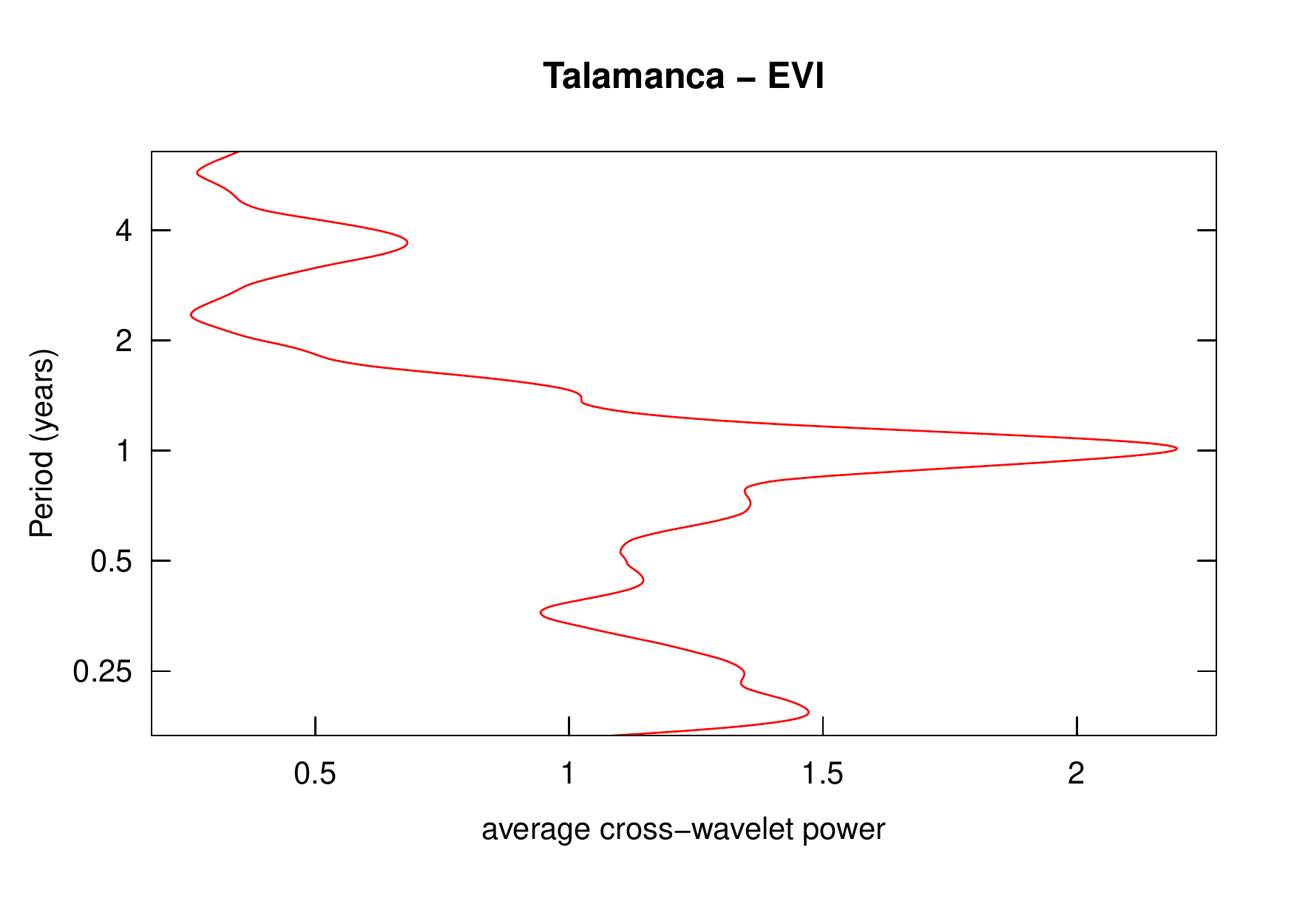}}\vspace{-0.15cm}%
\subfloat[]{\includegraphics[scale=0.23]{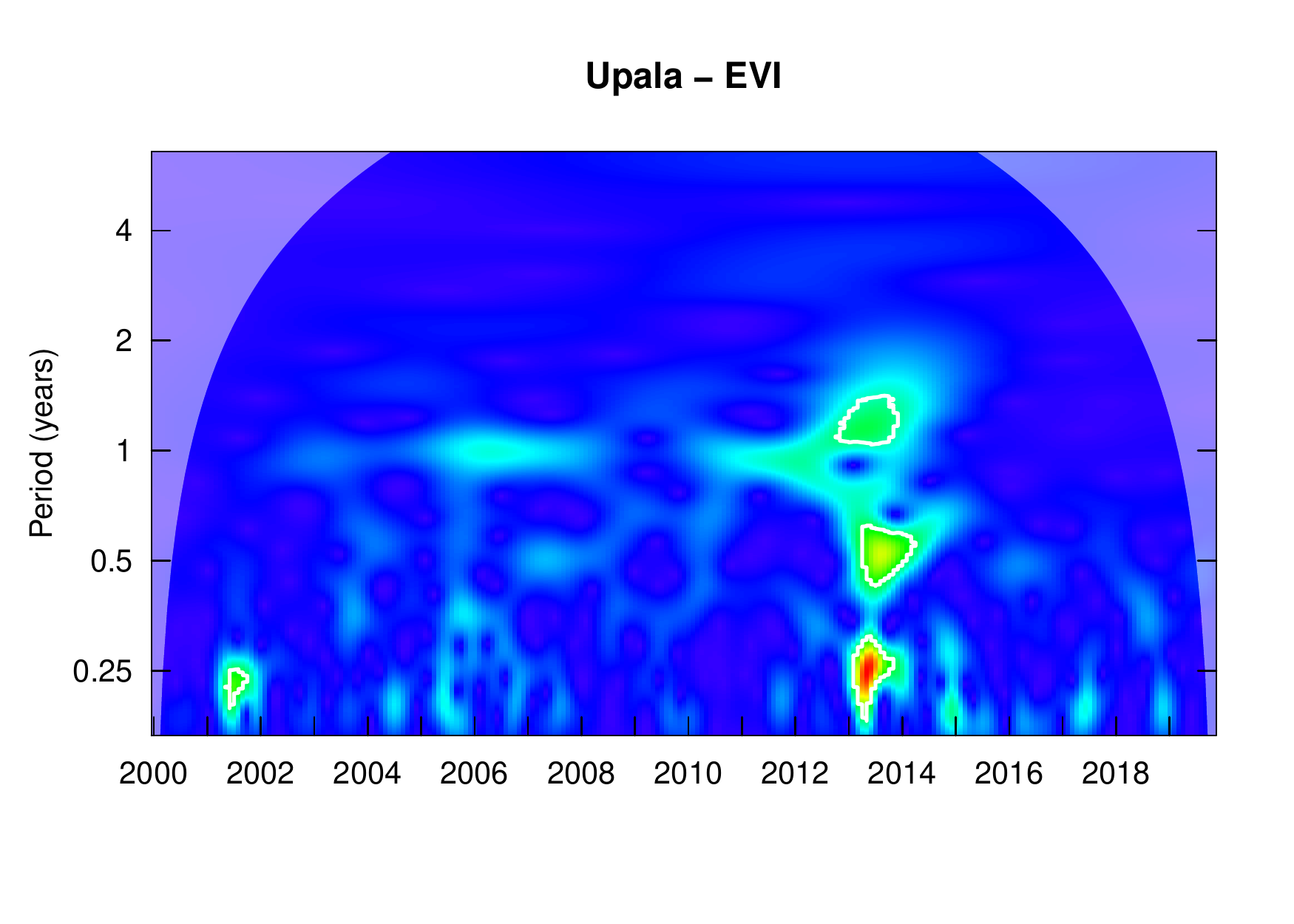}}\vspace{-0.15cm}%
\subfloat[]{\includegraphics[scale=0.23]{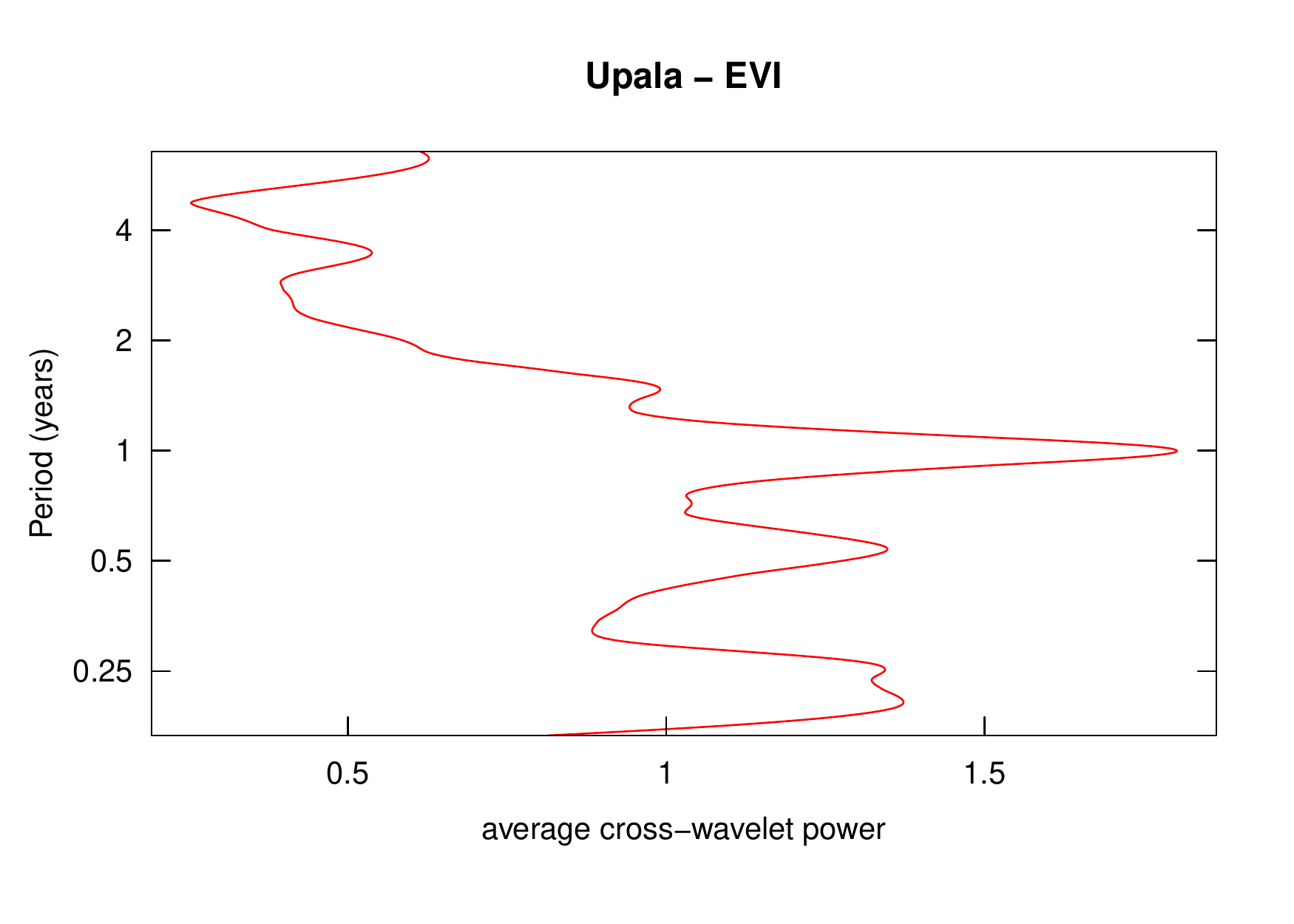}}\vspace{-0.15cm}\\
\subfloat[]{\includegraphics[scale=0.23]{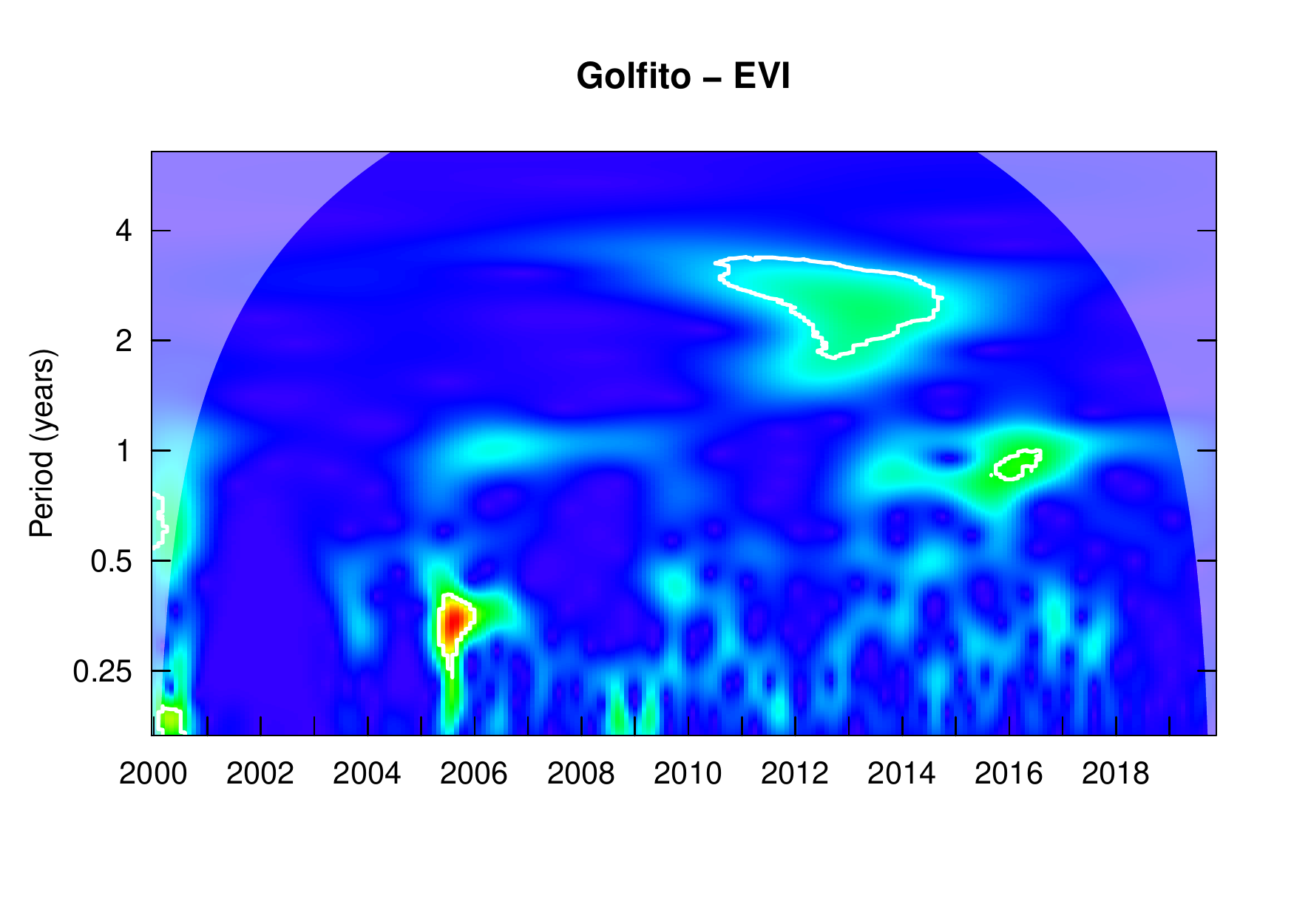}}\vspace{-0.15cm}%
\subfloat[]{\includegraphics[scale=0.23]{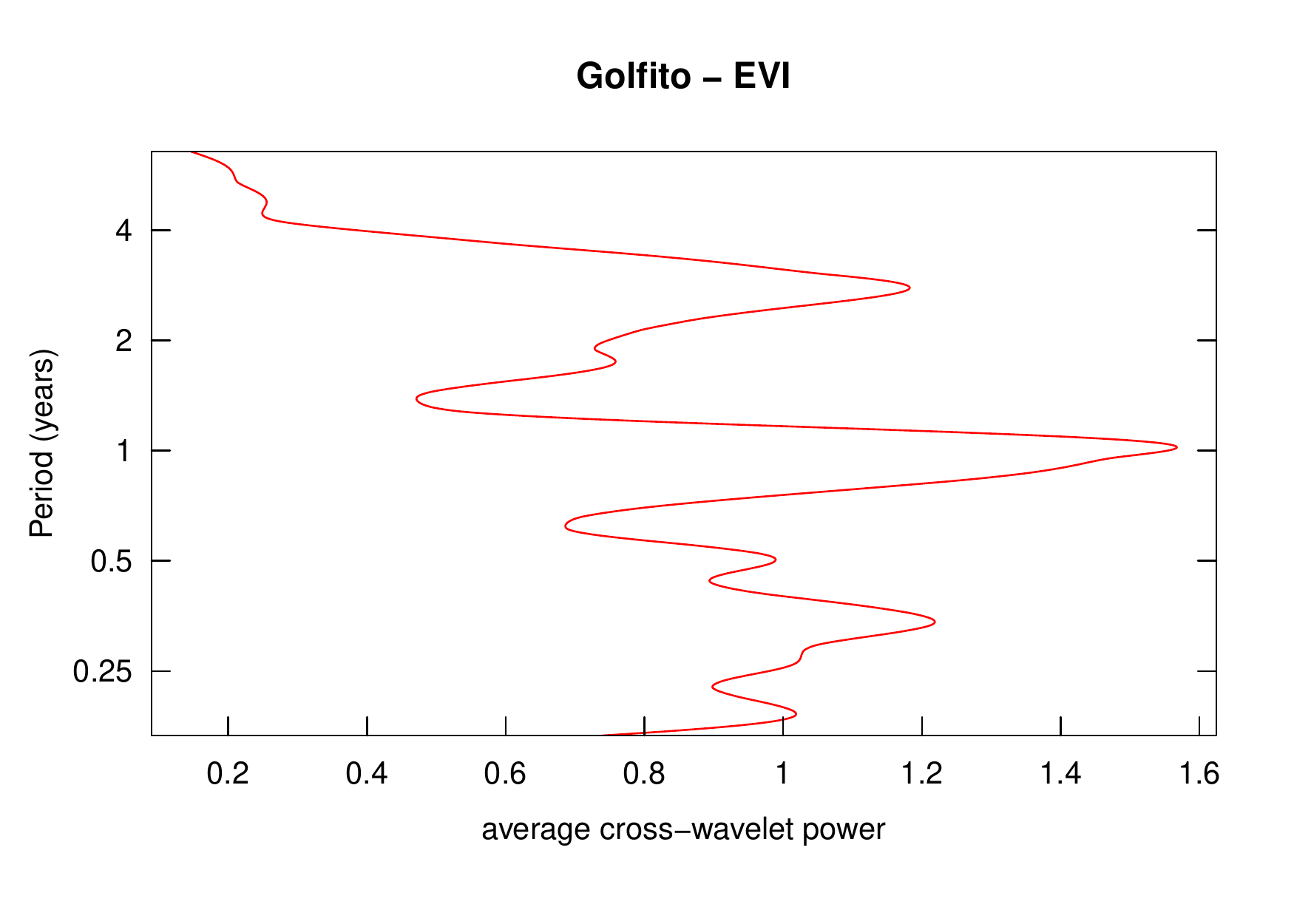}}\vspace{-0.15cm}%
\subfloat[]{\includegraphics[scale=0.23]{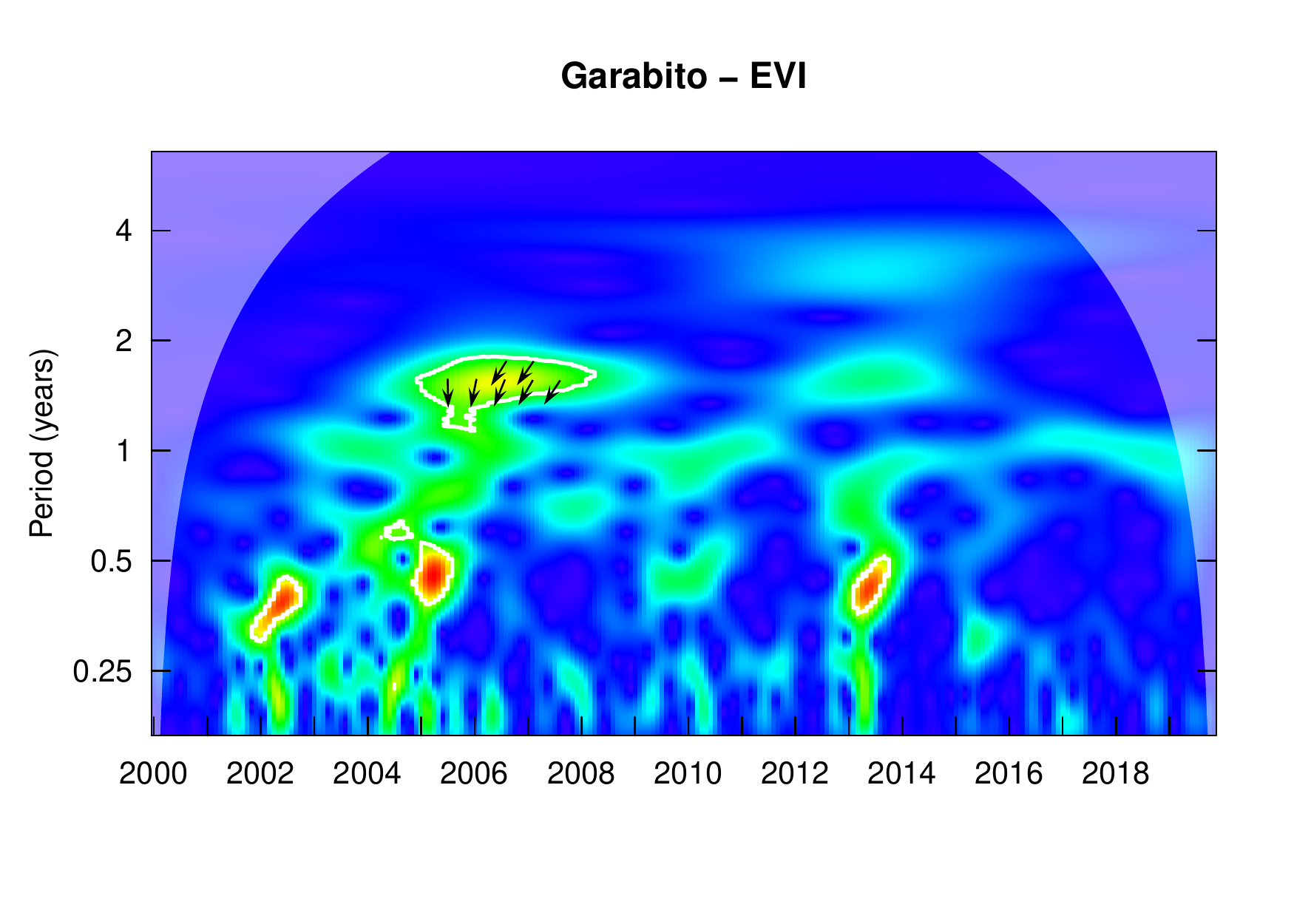}}\vspace{-0.15cm}%
\subfloat[]{\includegraphics[scale=0.23]{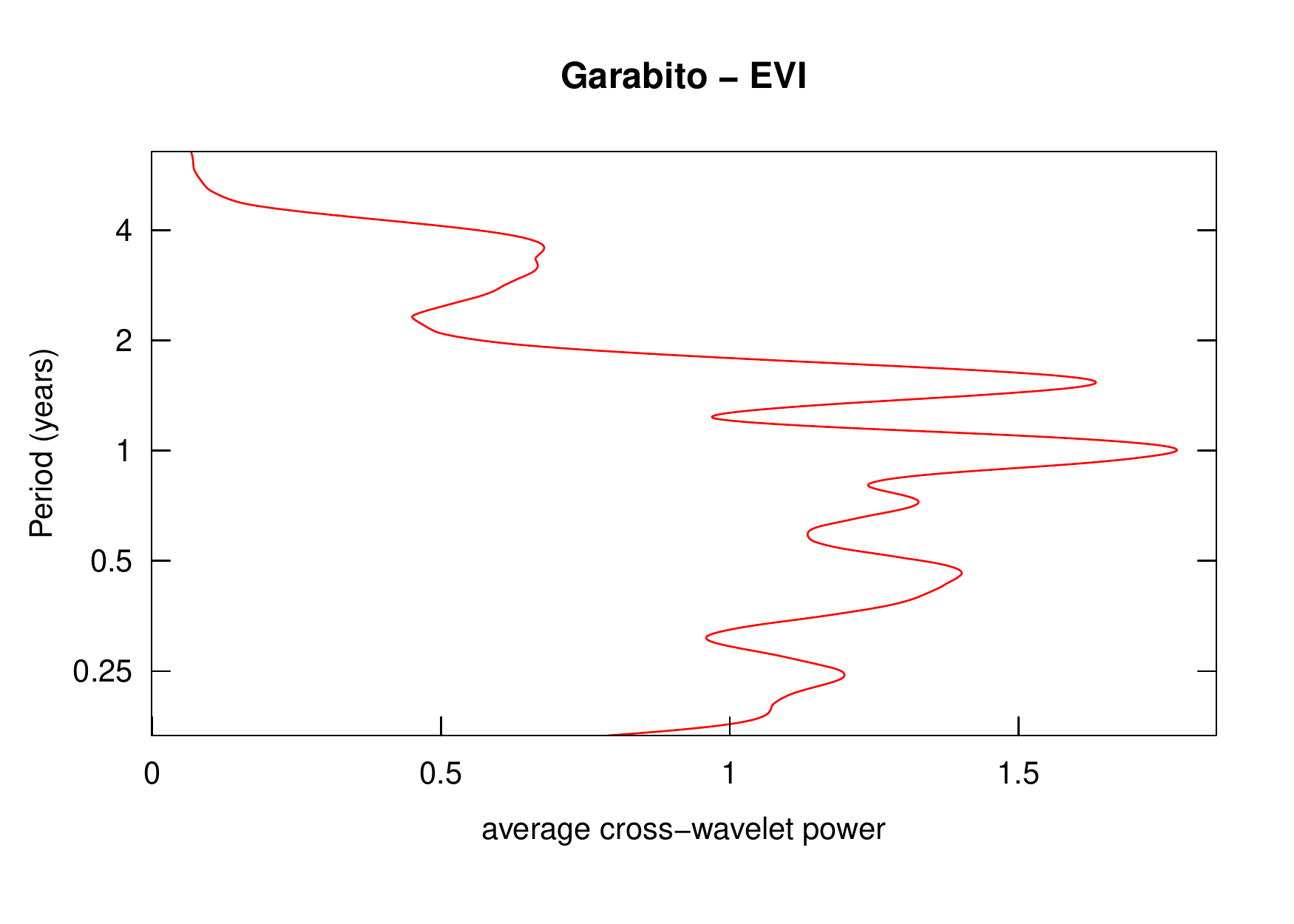}}\vspace{-0.15cm}
\end{figure}

\section*{Wavelet coherence and average cross-wavelet power between dengue incidence and NDVI}

\begin{figure}[H]
\captionsetup[subfigure]{labelformat=empty}
\caption*{\textbf{Figure S2:} Wavelet coherence (color map) between dengue incidence from 2000 to 2019, and NDVI in 32 municipalities of Costa Rica (periodicity on y-axis, time on x-axis). Colors code for increasing power intensity, from blue to red; $95\%$ confidence levels are encircled by white lines, and shaded areas indicate the presence of significant edge effects. Average cross-wavelet power (Red line). The arrows indicate if time series are in-phase or out-phase.}
\subfloat[]{\includegraphics[scale=0.23]{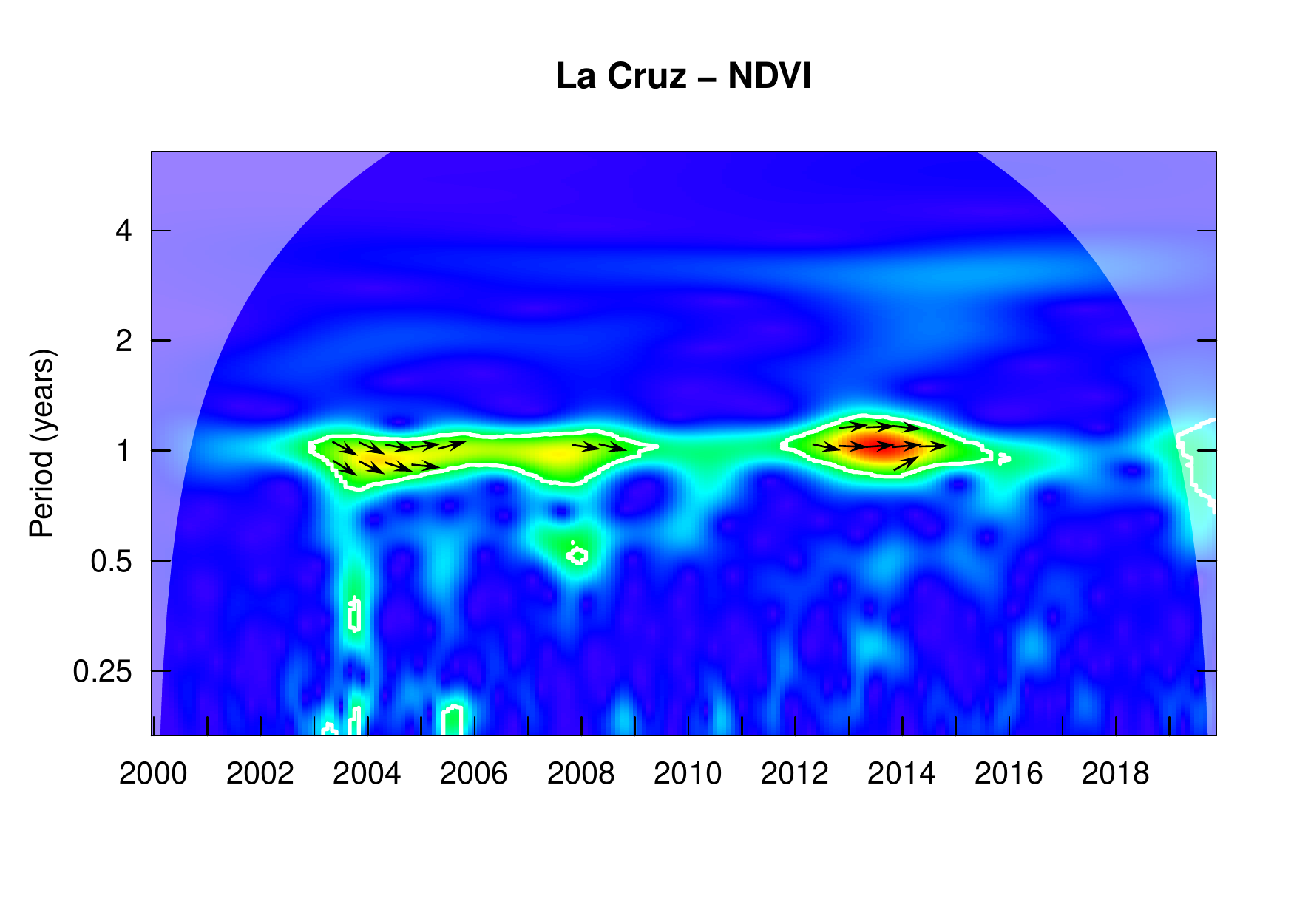}}\vspace{-0.15cm}%
\subfloat[]{\includegraphics[scale=0.23]{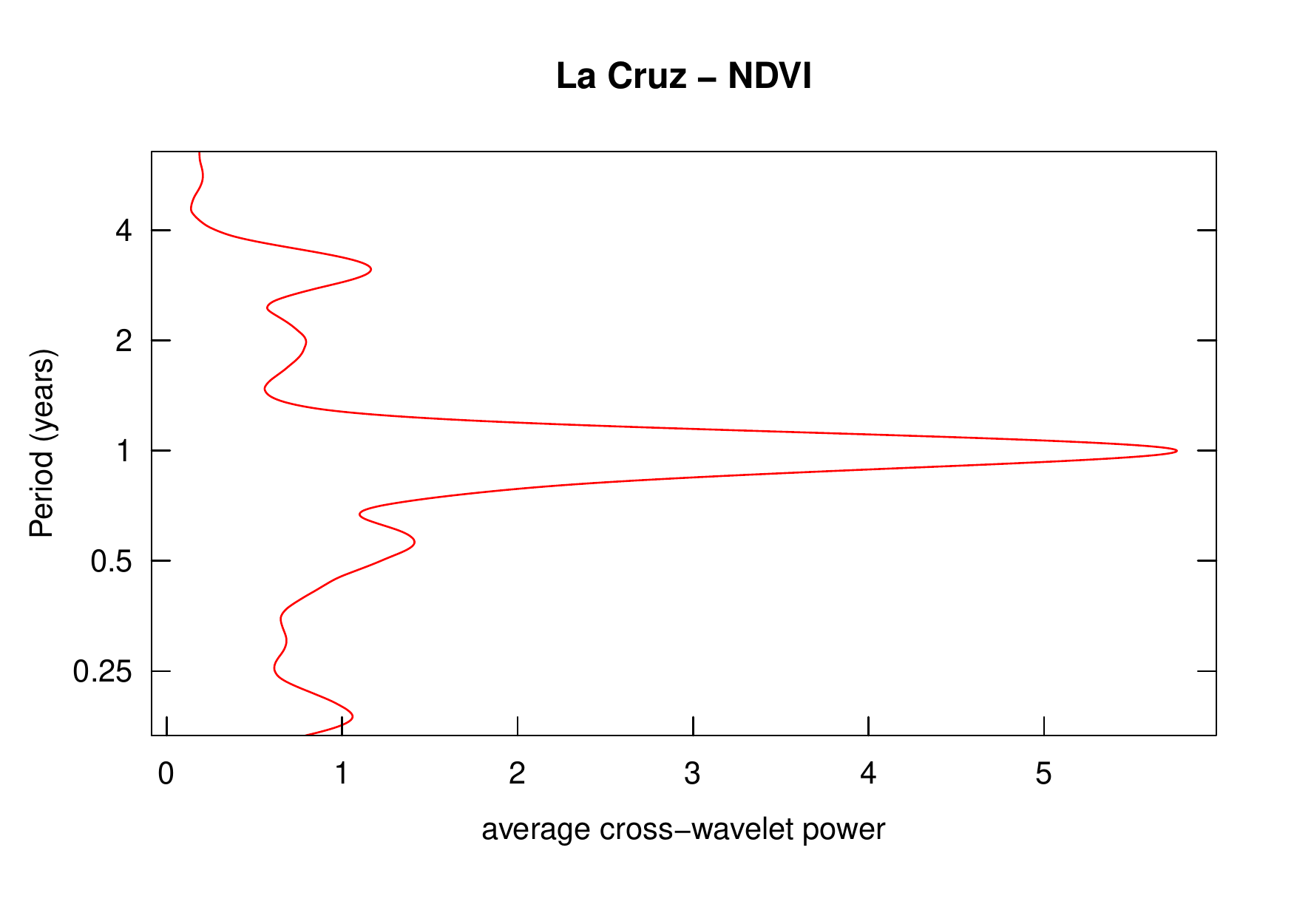}}\vspace{-0.15cm}
\subfloat[]{\includegraphics[scale=0.23]{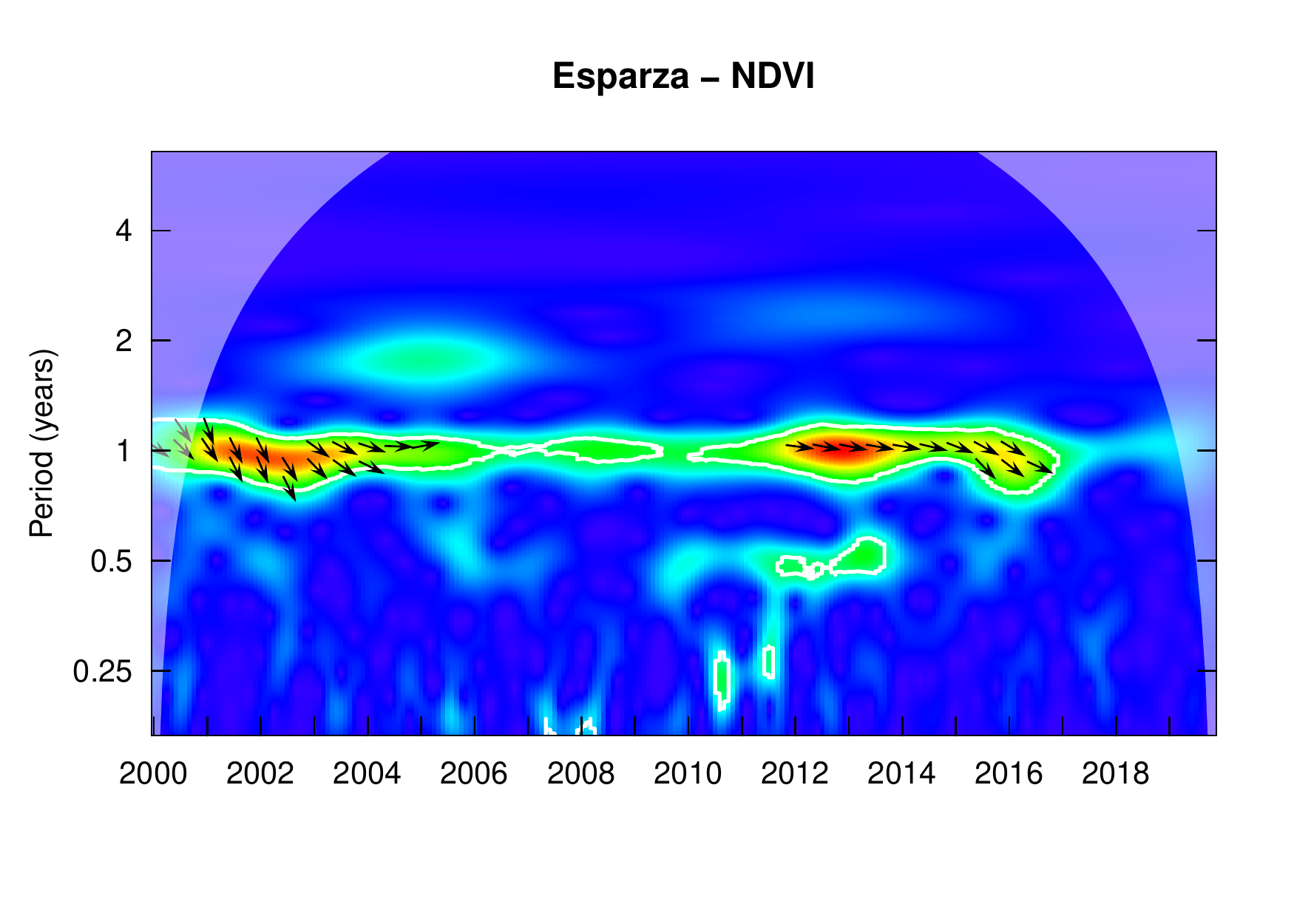}}\vspace{-0.15cm}%
\subfloat[]{\includegraphics[scale=0.23]{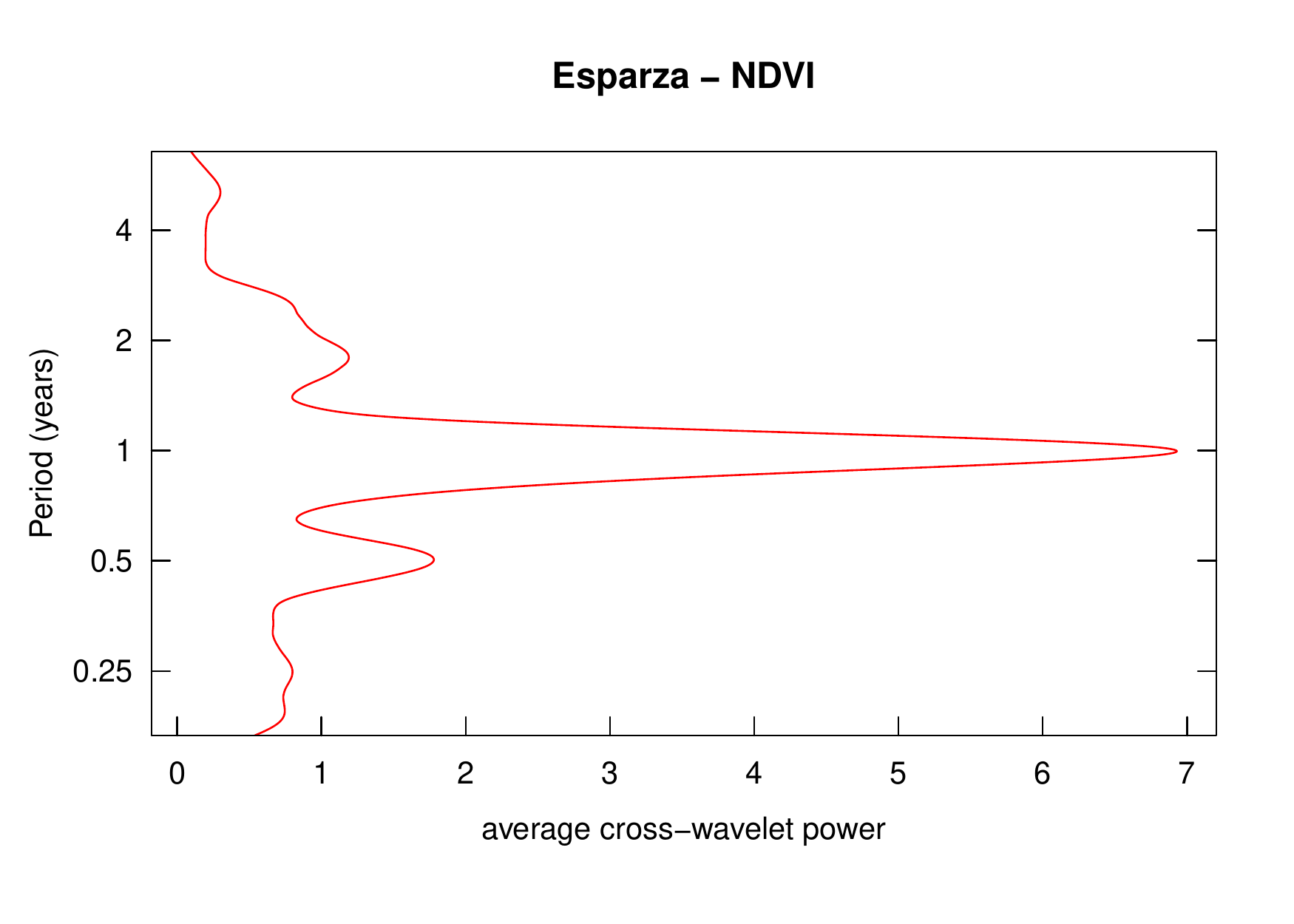}}\vspace{-0.15cm}
\end{figure}

\begin{figure}[H]
\captionsetup[subfigure]{labelformat=empty}
\subfloat[]{\includegraphics[scale=0.23]{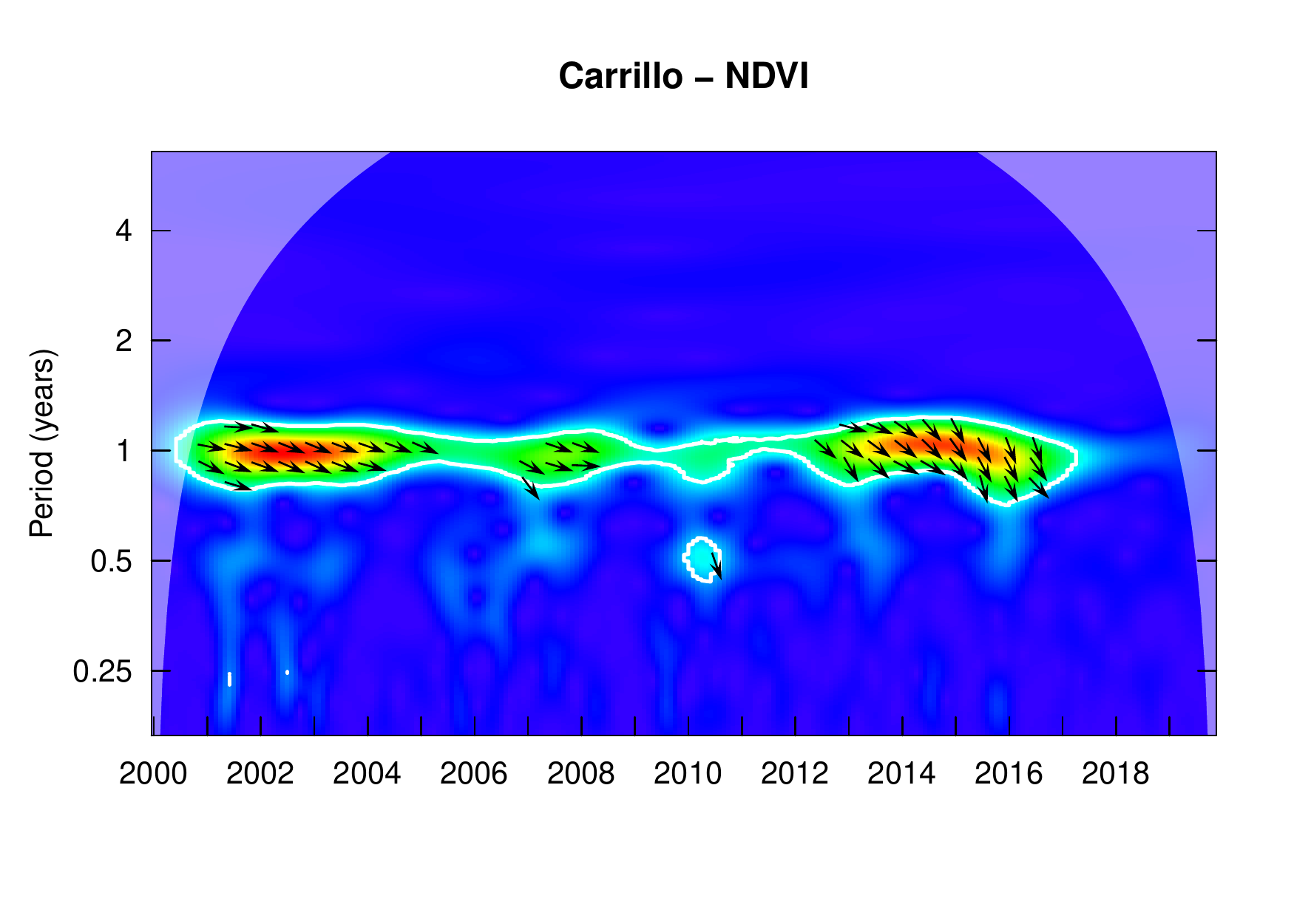}}\vspace{-0.15cm}%
\subfloat[]{\includegraphics[scale=0.23]{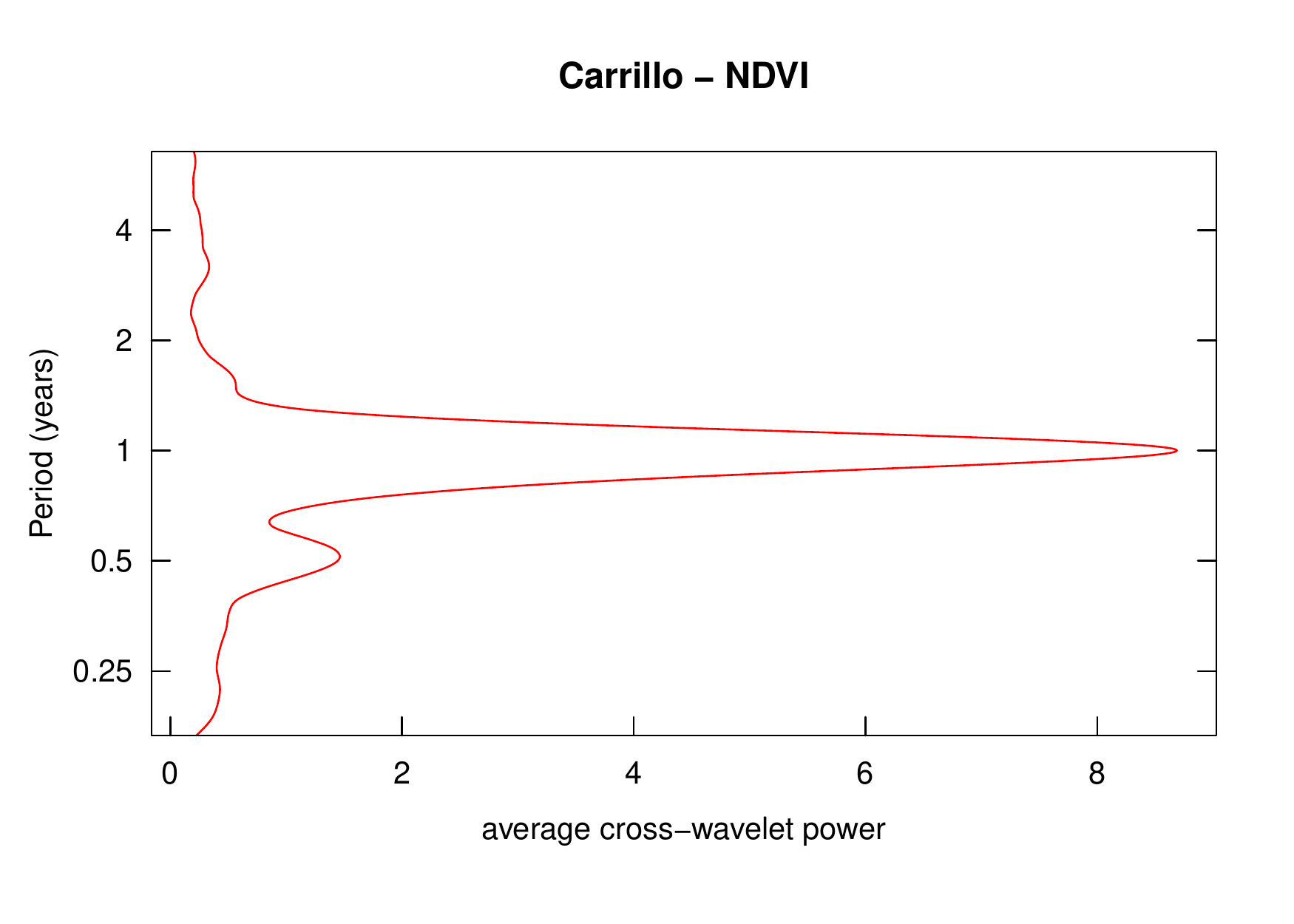}}\vspace{-0.15cm}%
\subfloat[]{\includegraphics[scale=0.23]{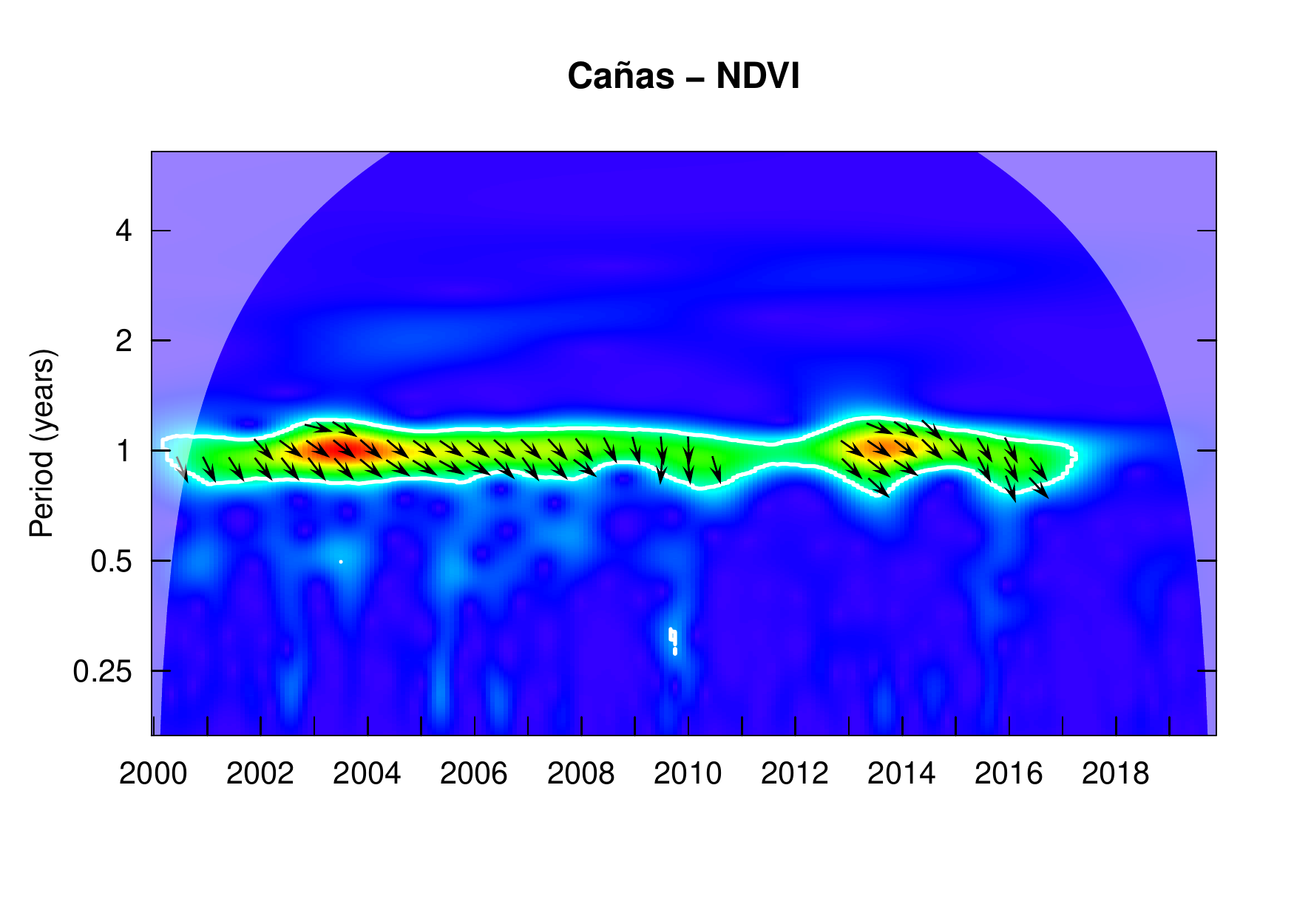}}\vspace{-0.15cm}%
\subfloat[]{\includegraphics[scale=0.23]{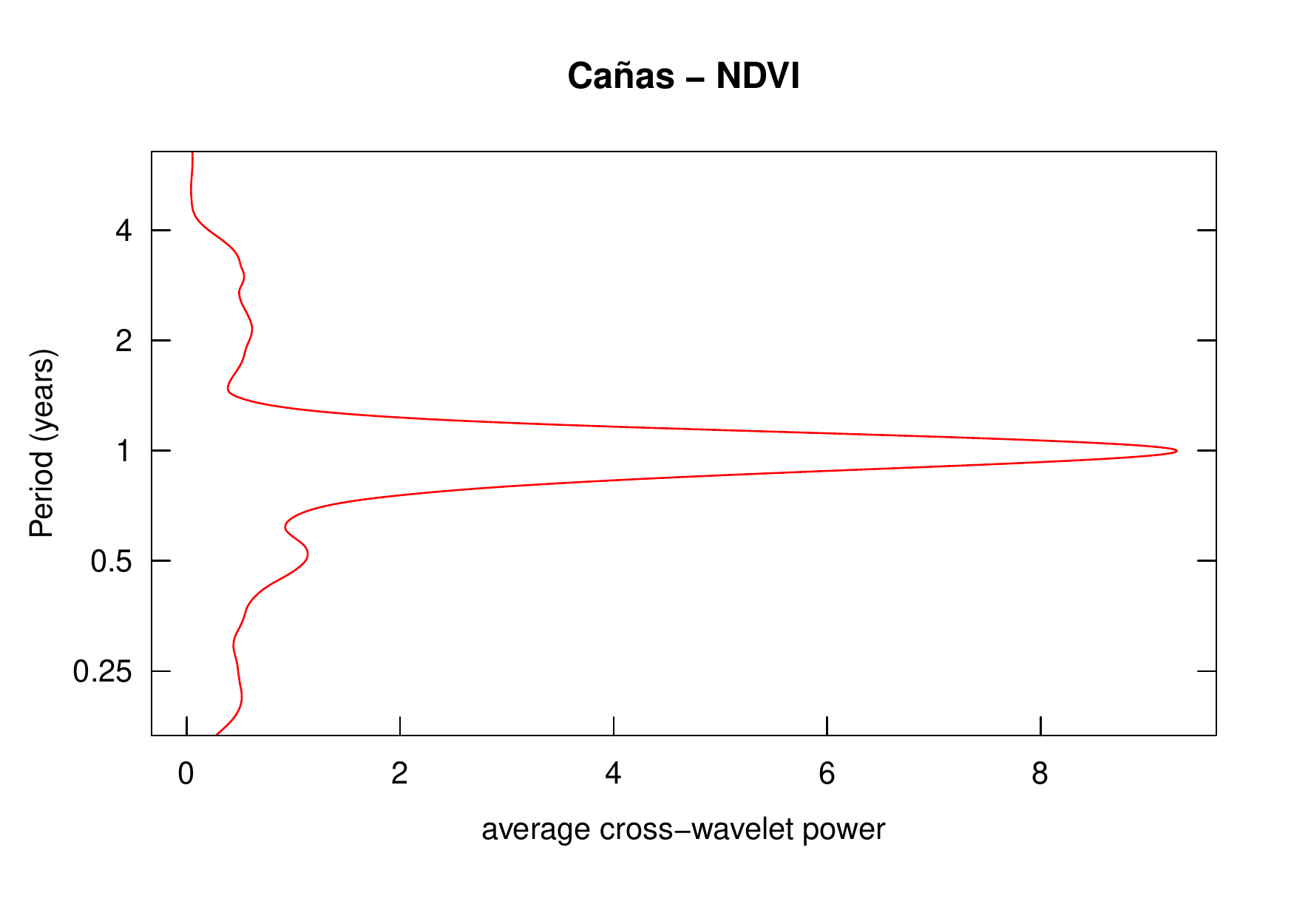}}\vspace{-0.15cm}\\
\subfloat[]{\includegraphics[scale=0.23]{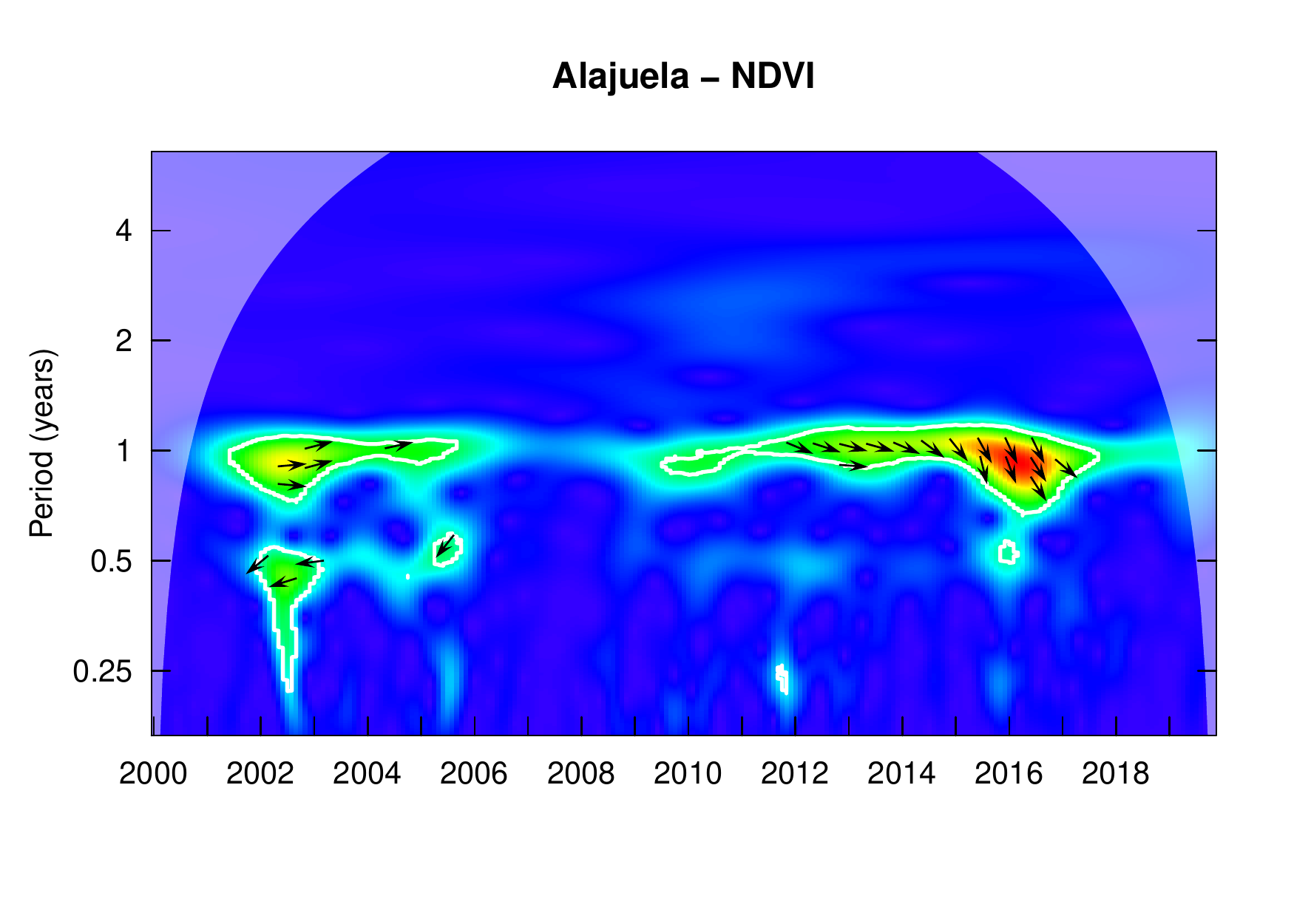}}\vspace{-0.15cm}%
\subfloat[]{\includegraphics[scale=0.23]{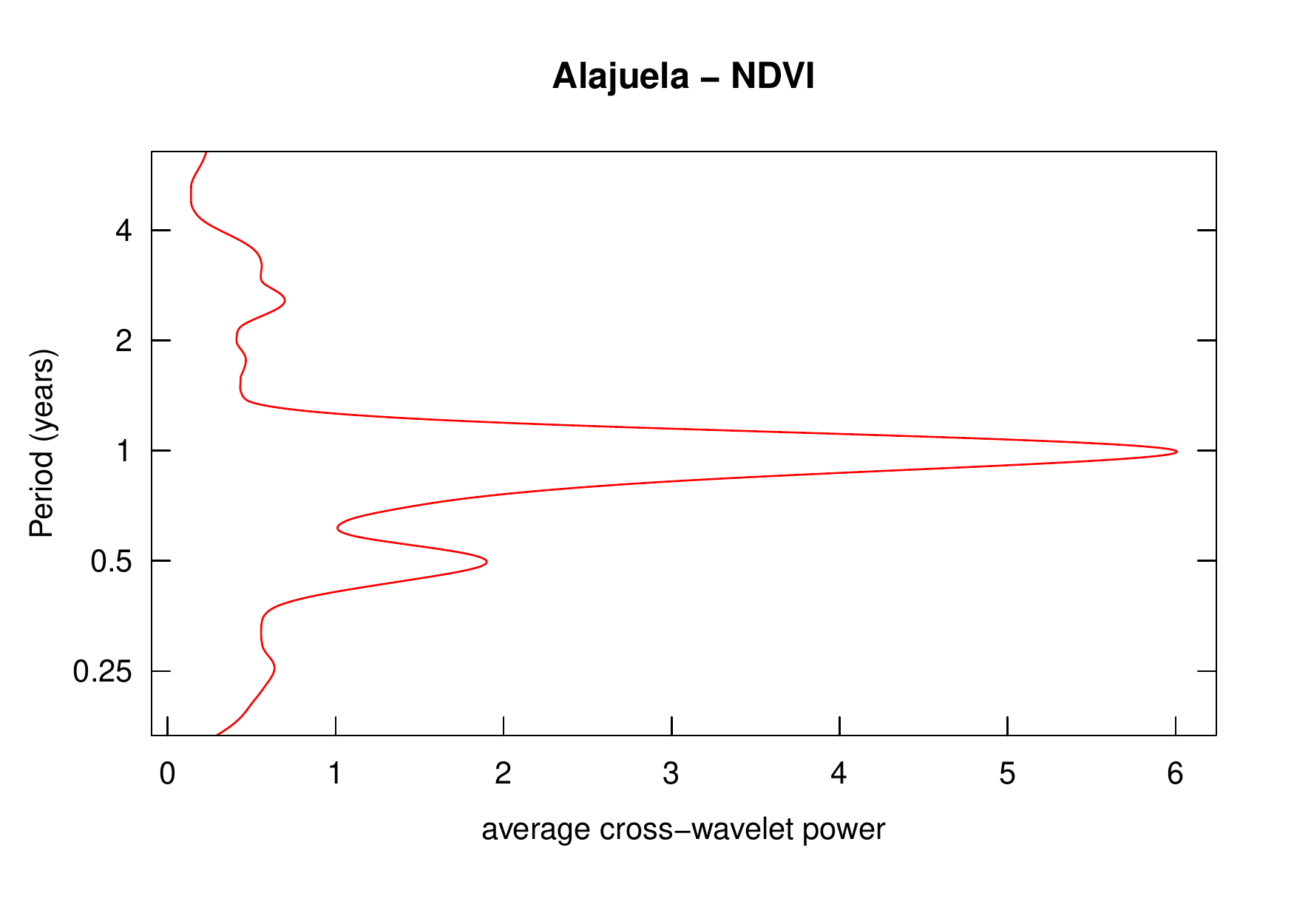}}\vspace{-0.15cm}%
\subfloat[]{\includegraphics[scale=0.23]{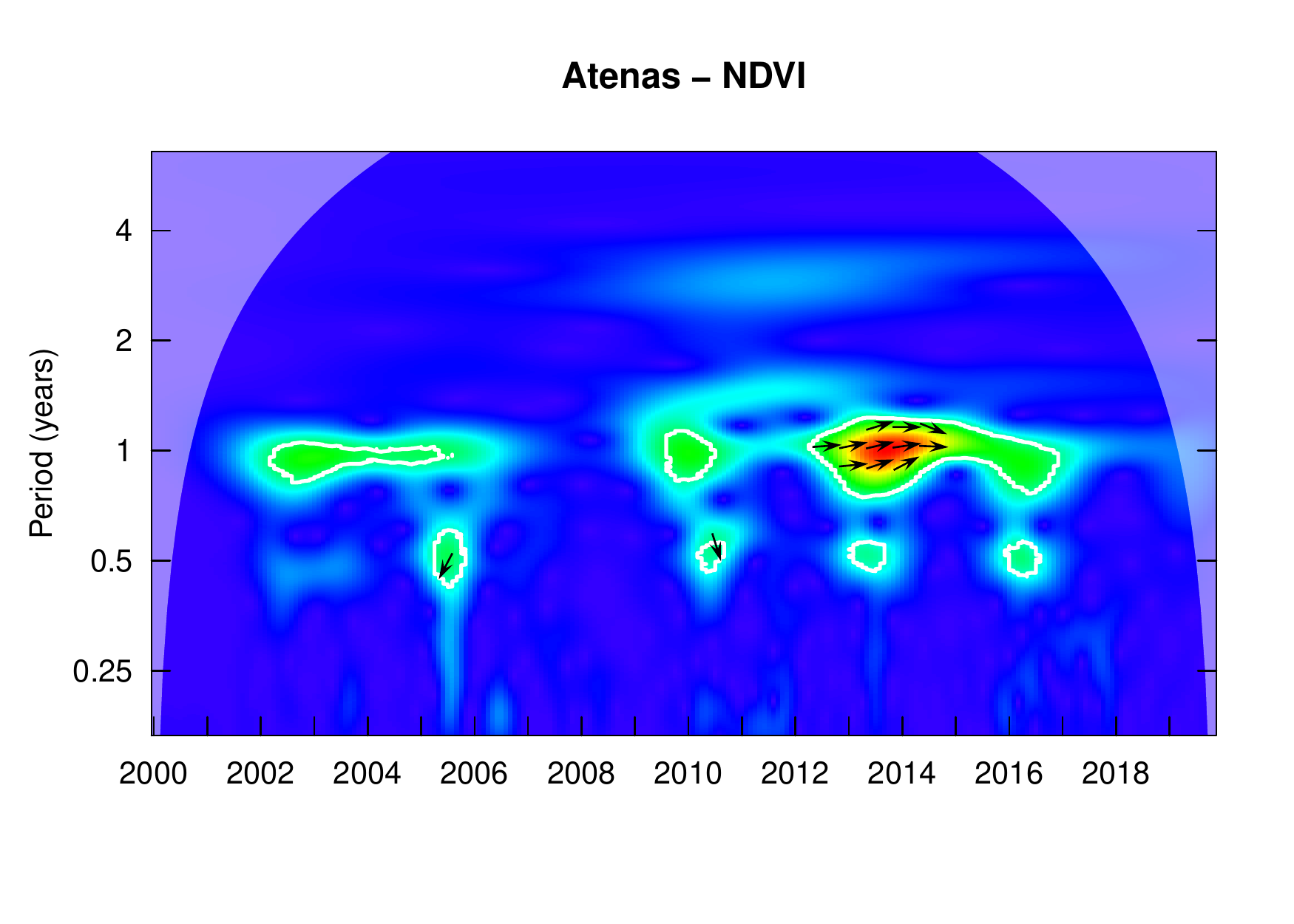}}\vspace{-0.15cm}%
\subfloat[]{\includegraphics[scale=0.23]{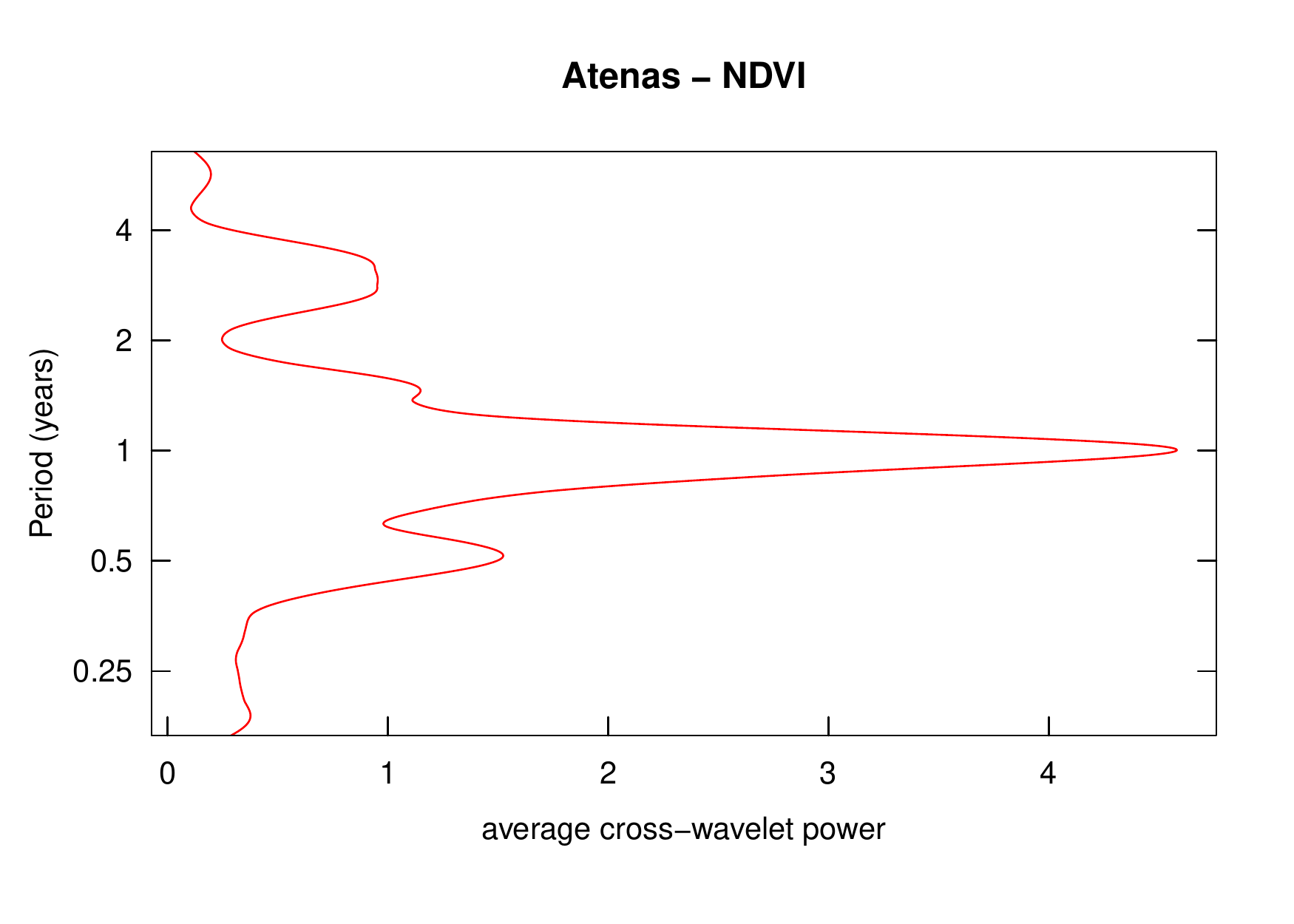}}\vspace{-0.15cm}\\
\subfloat[]{\includegraphics[scale=0.23]{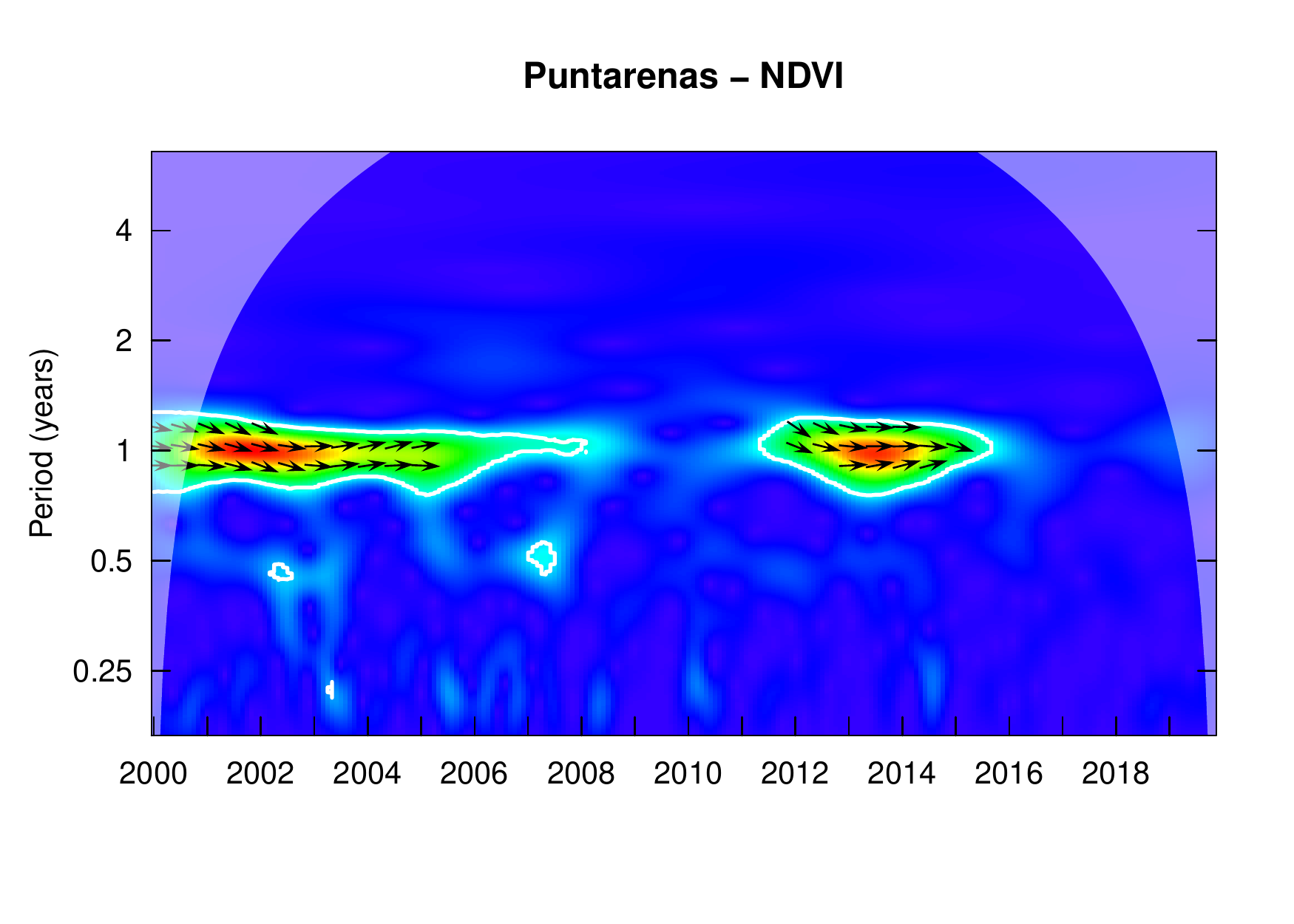}}\vspace{-0.15cm}%
\subfloat[]{\includegraphics[scale=0.23]{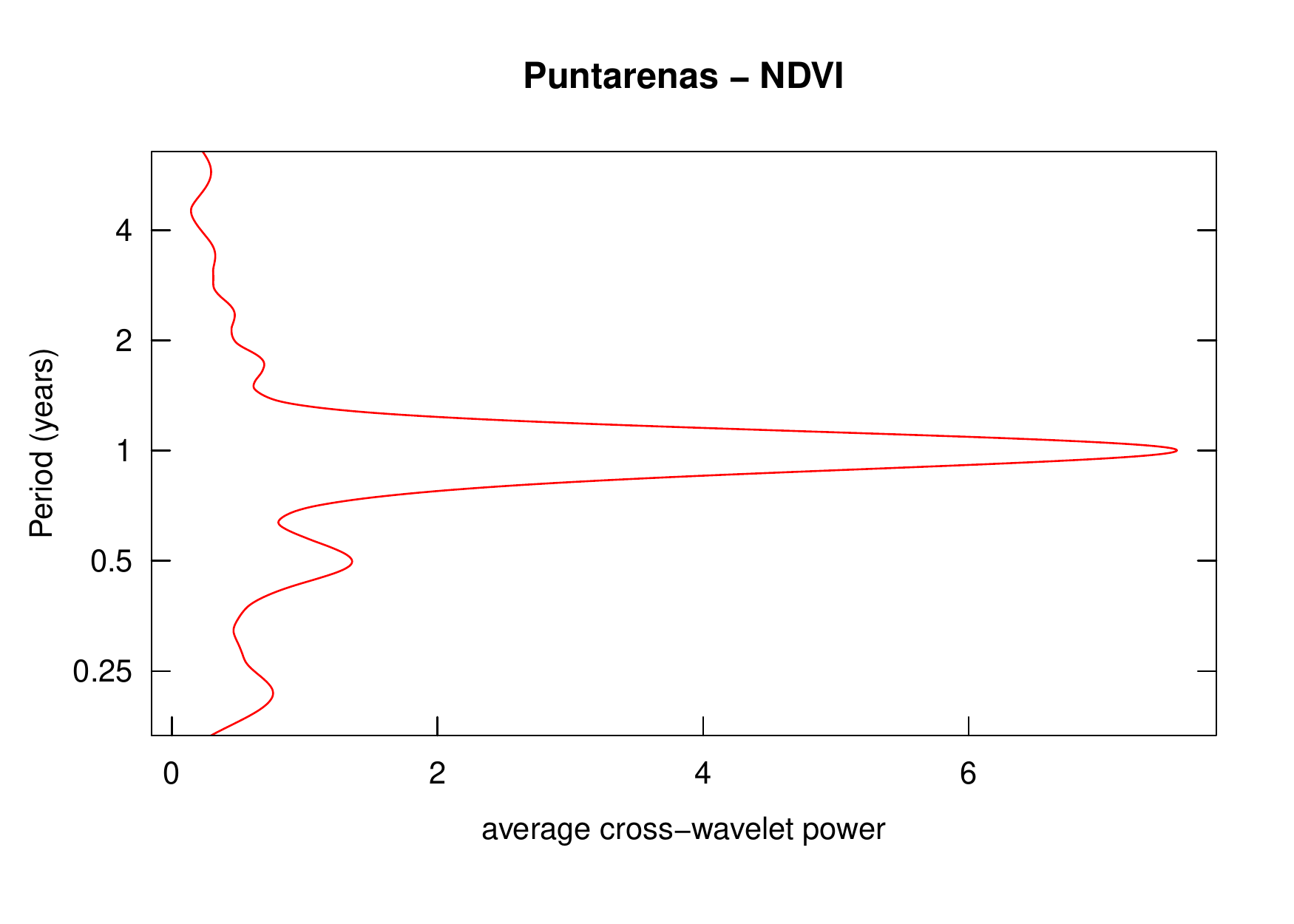}}\vspace{-0.15cm}%
\subfloat[]{\includegraphics[scale=0.23]{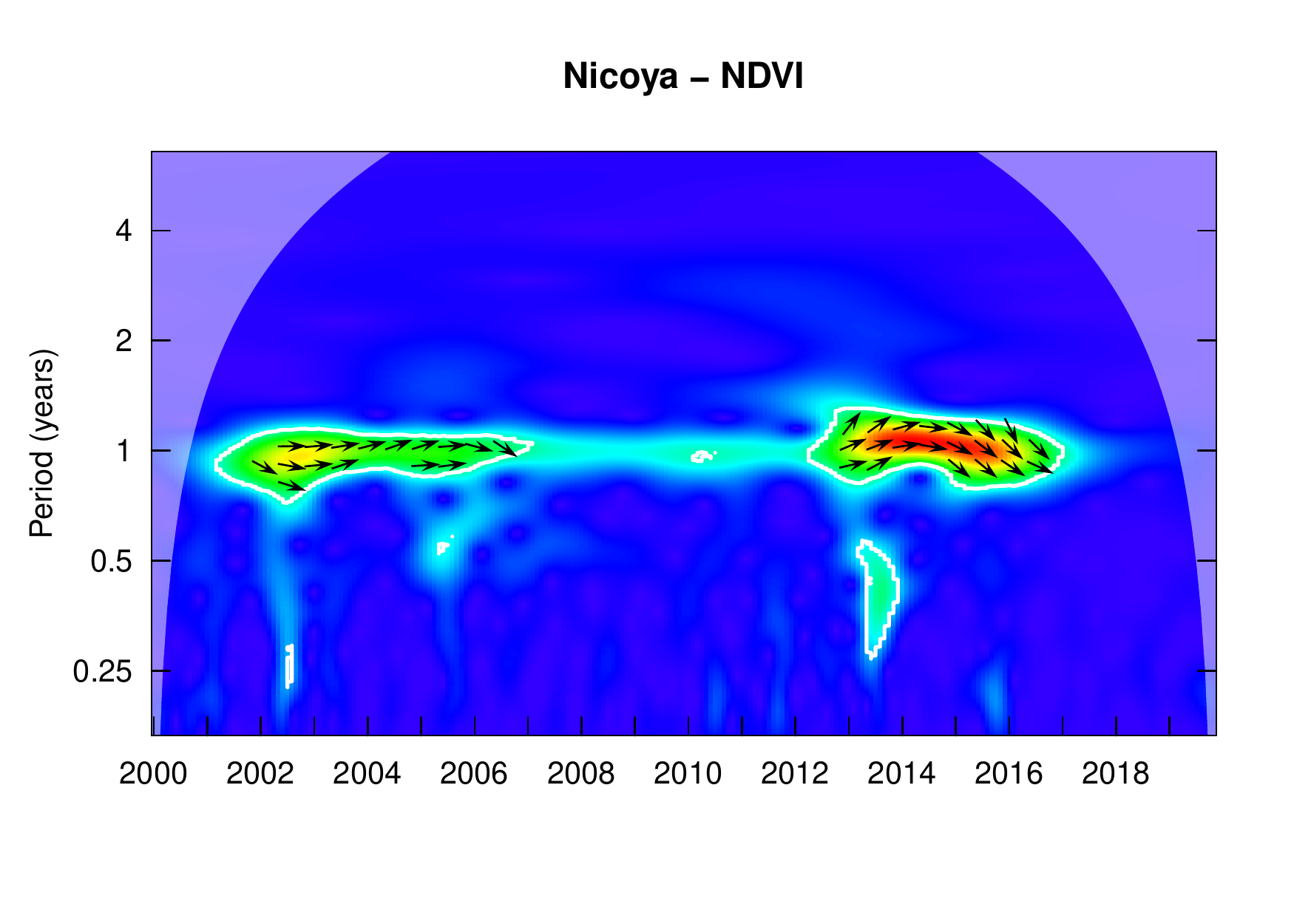}}\vspace{-0.15cm}%
\subfloat[]{\includegraphics[scale=0.23]{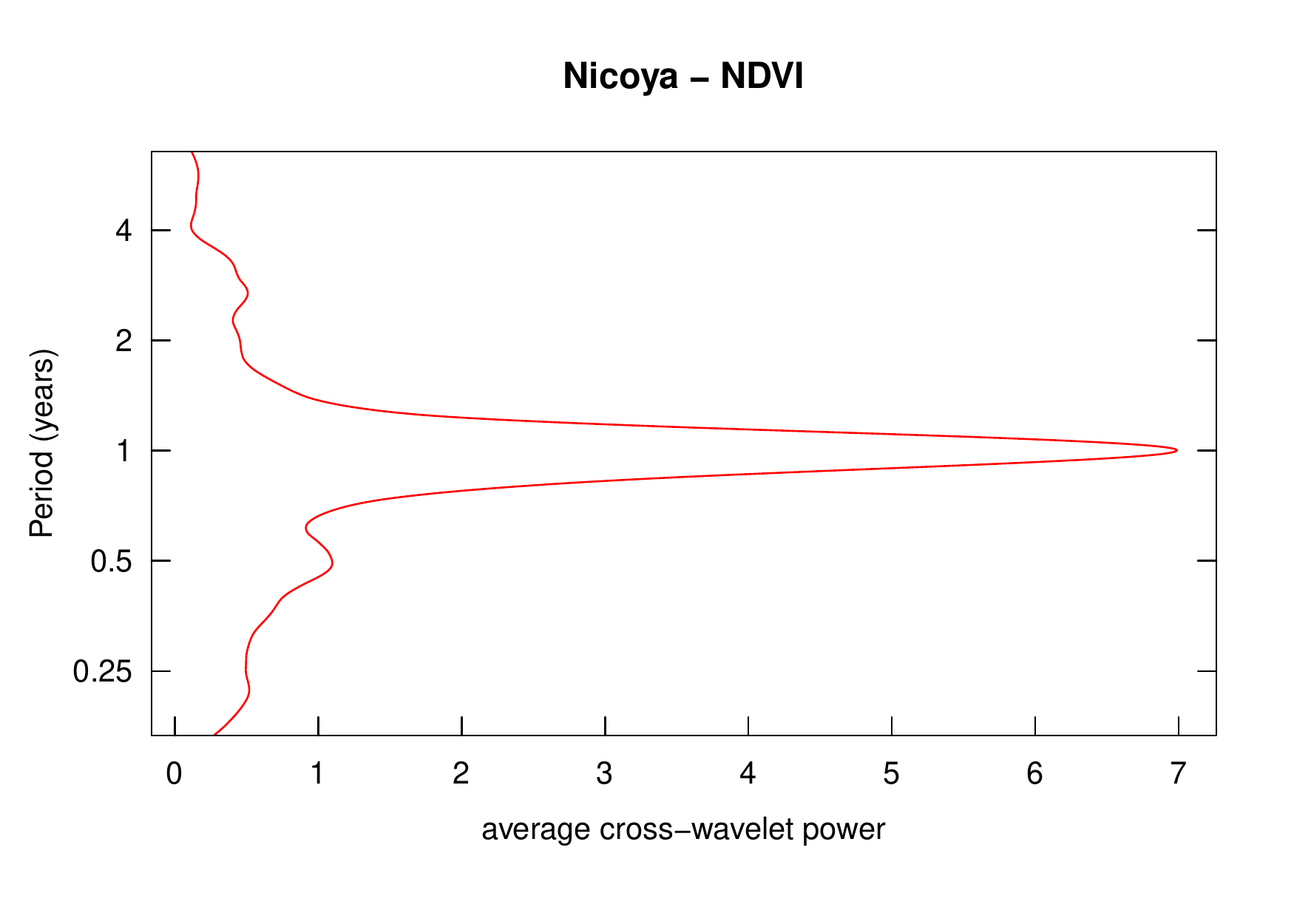}}\vspace{-0.15cm}\\
\subfloat[]{\includegraphics[scale=0.23]{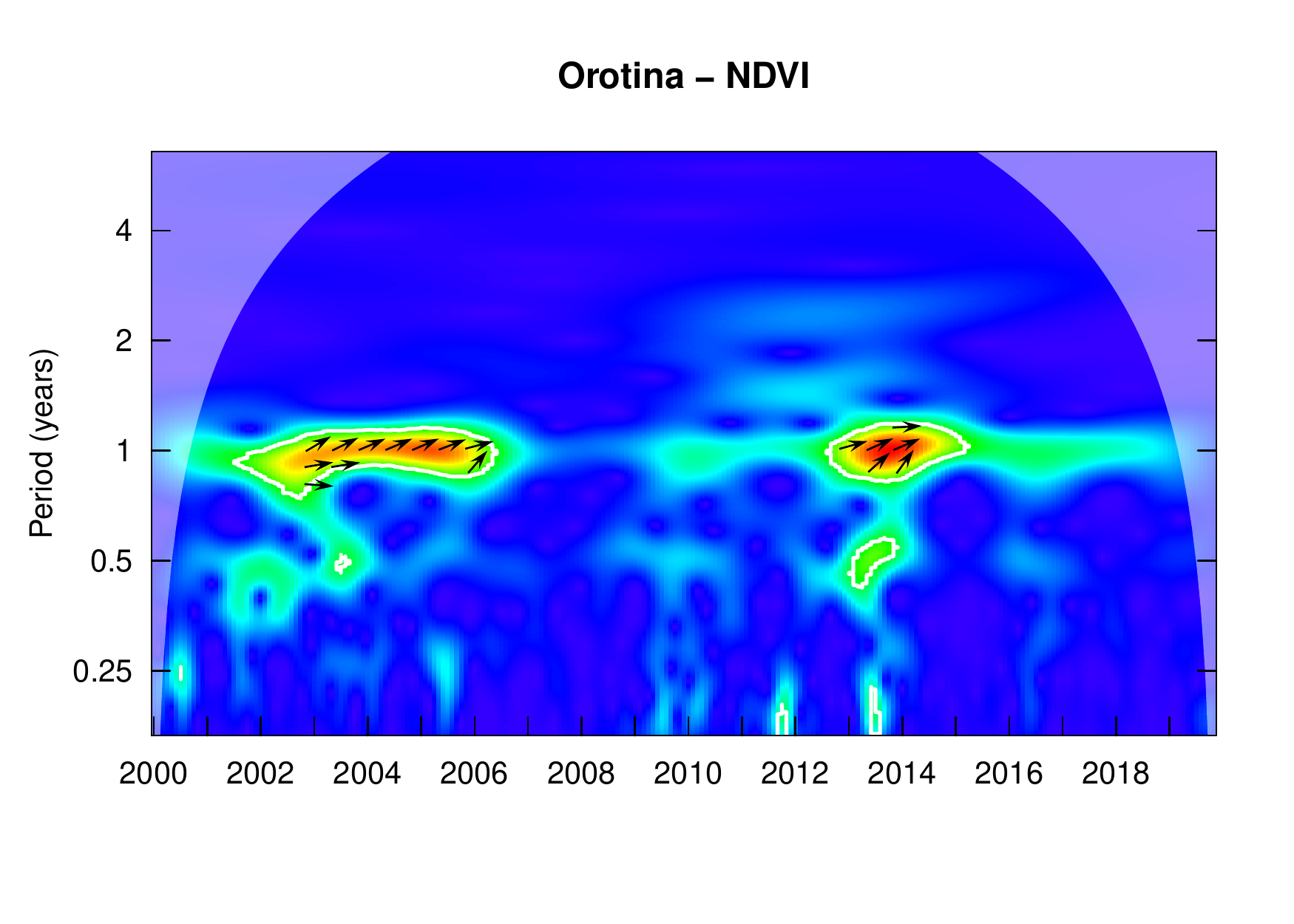}}\vspace{-0.15cm}%
\subfloat[]{\includegraphics[scale=0.23]{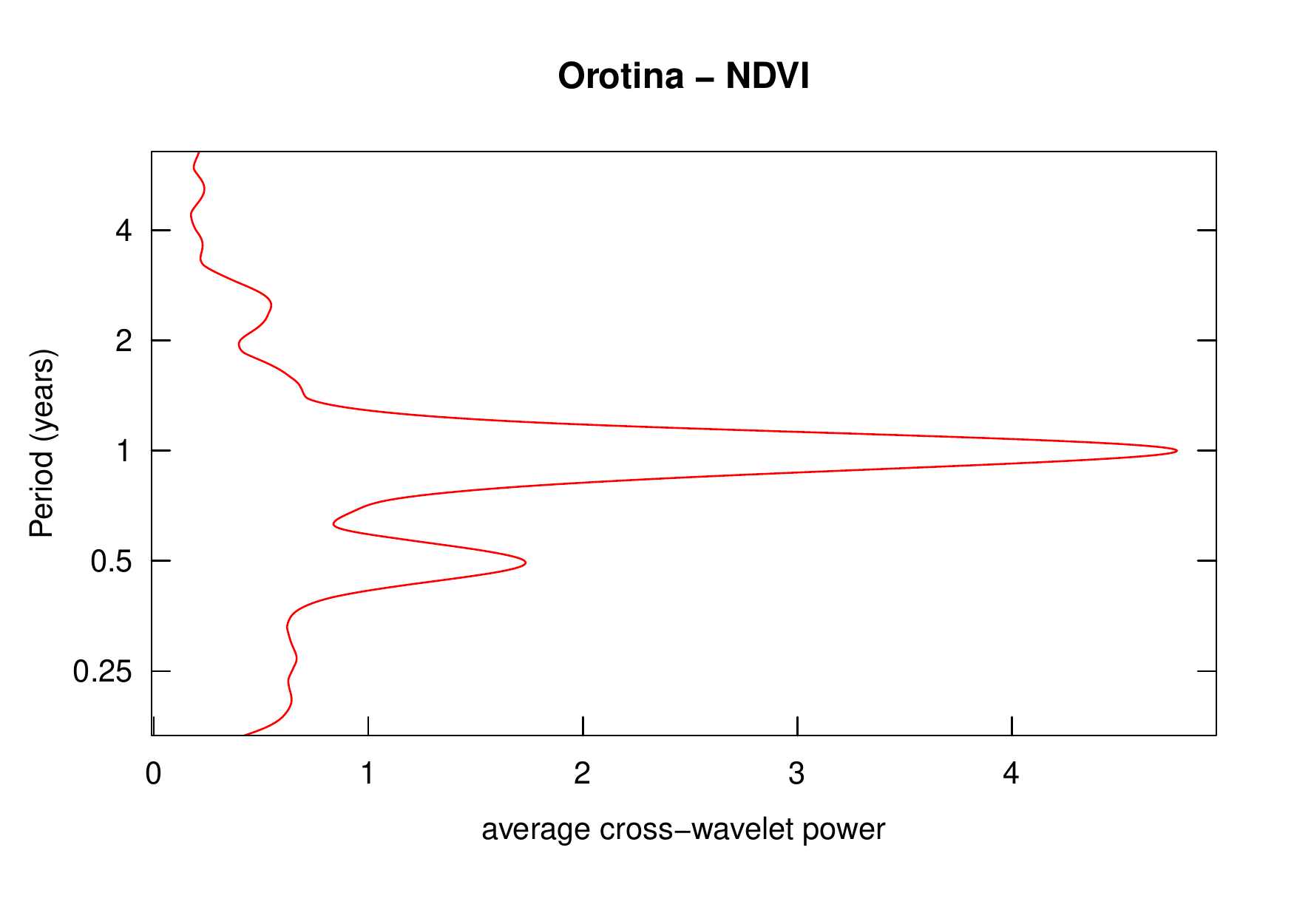}}\vspace{-0.15cm}%
\subfloat[]{\includegraphics[scale=0.23]{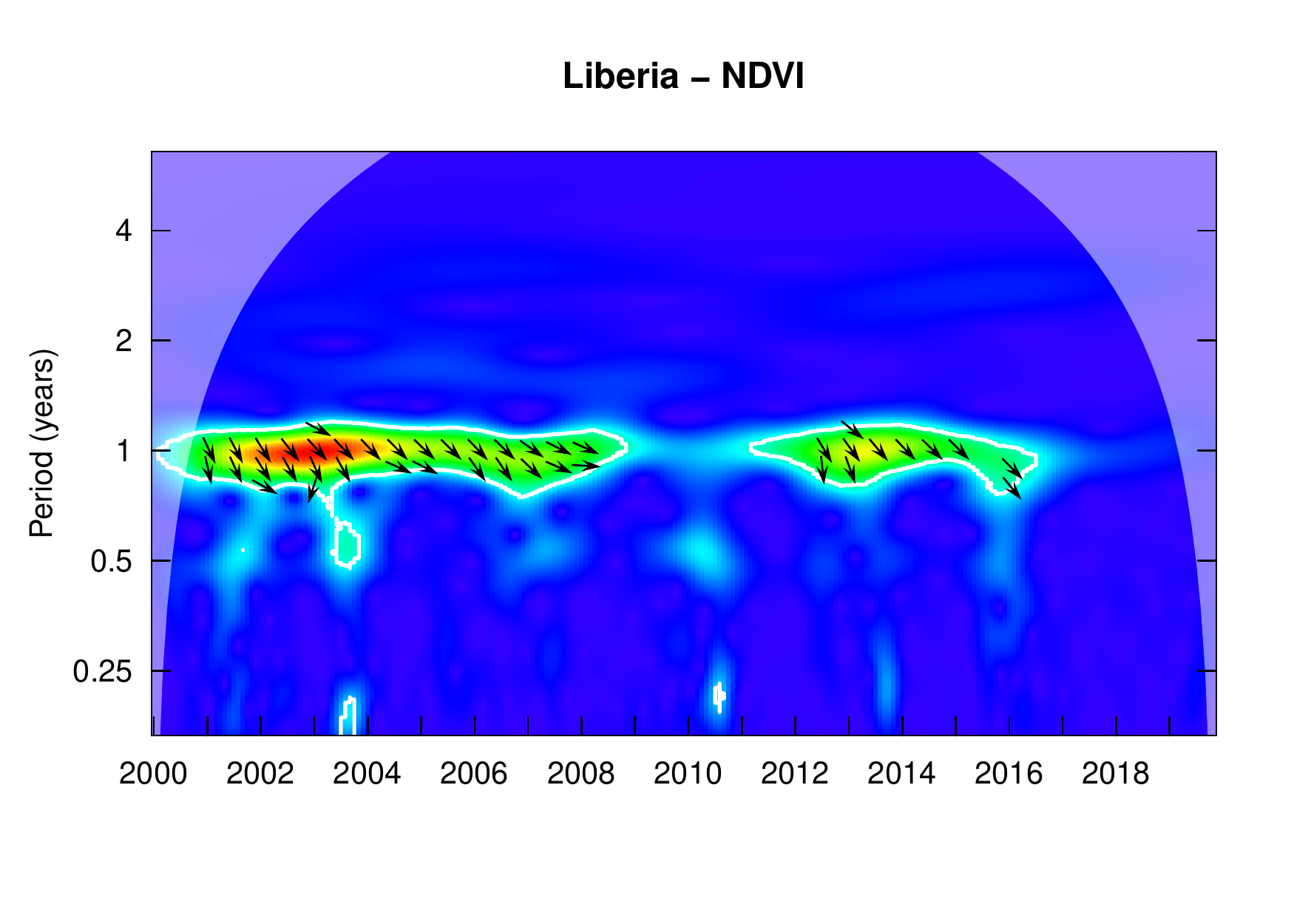}}\vspace{-0.15cm}%
\subfloat[]{\includegraphics[scale=0.23]{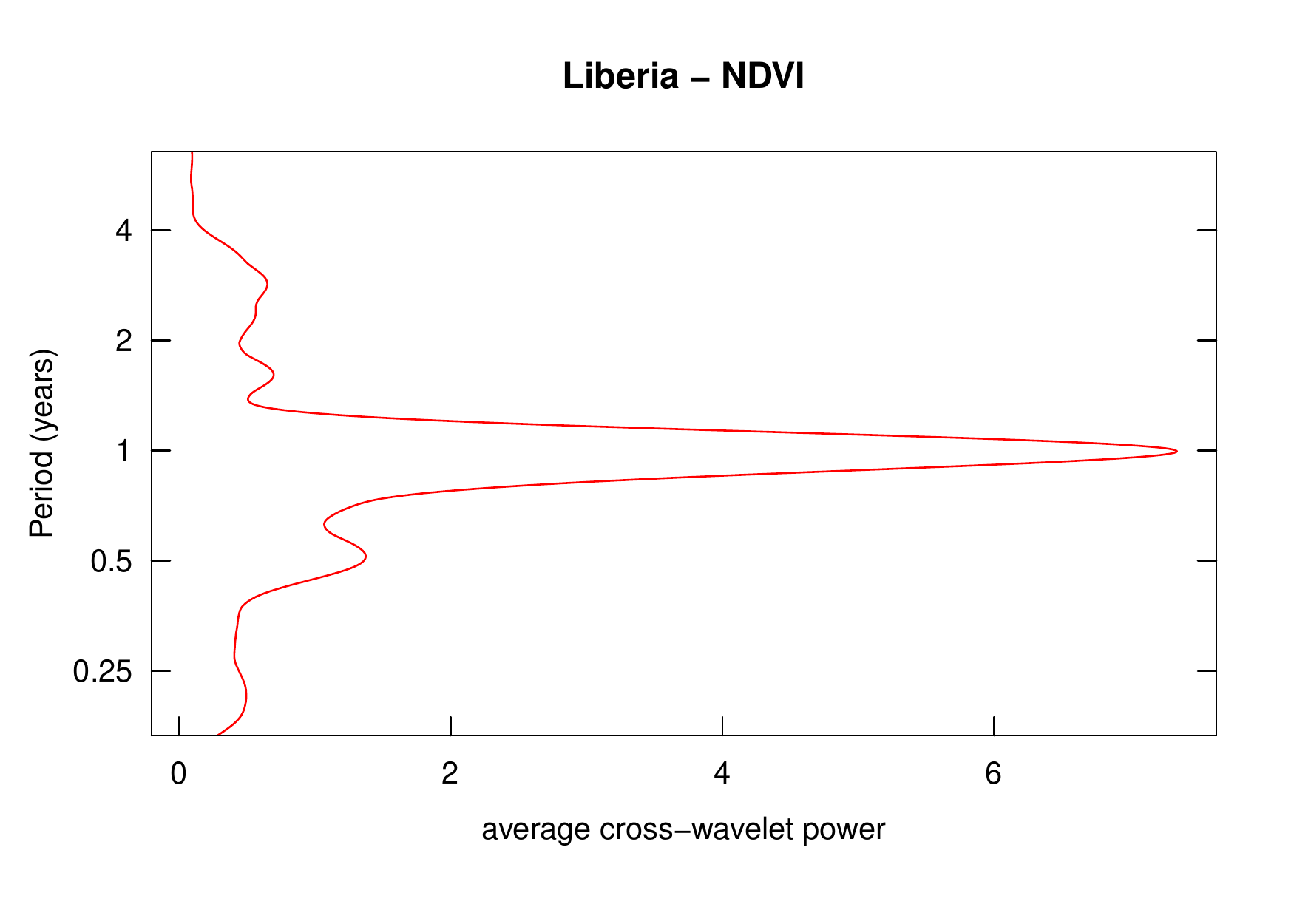}}\vspace{-0.15cm}\\
\subfloat[]{\includegraphics[scale=0.23]{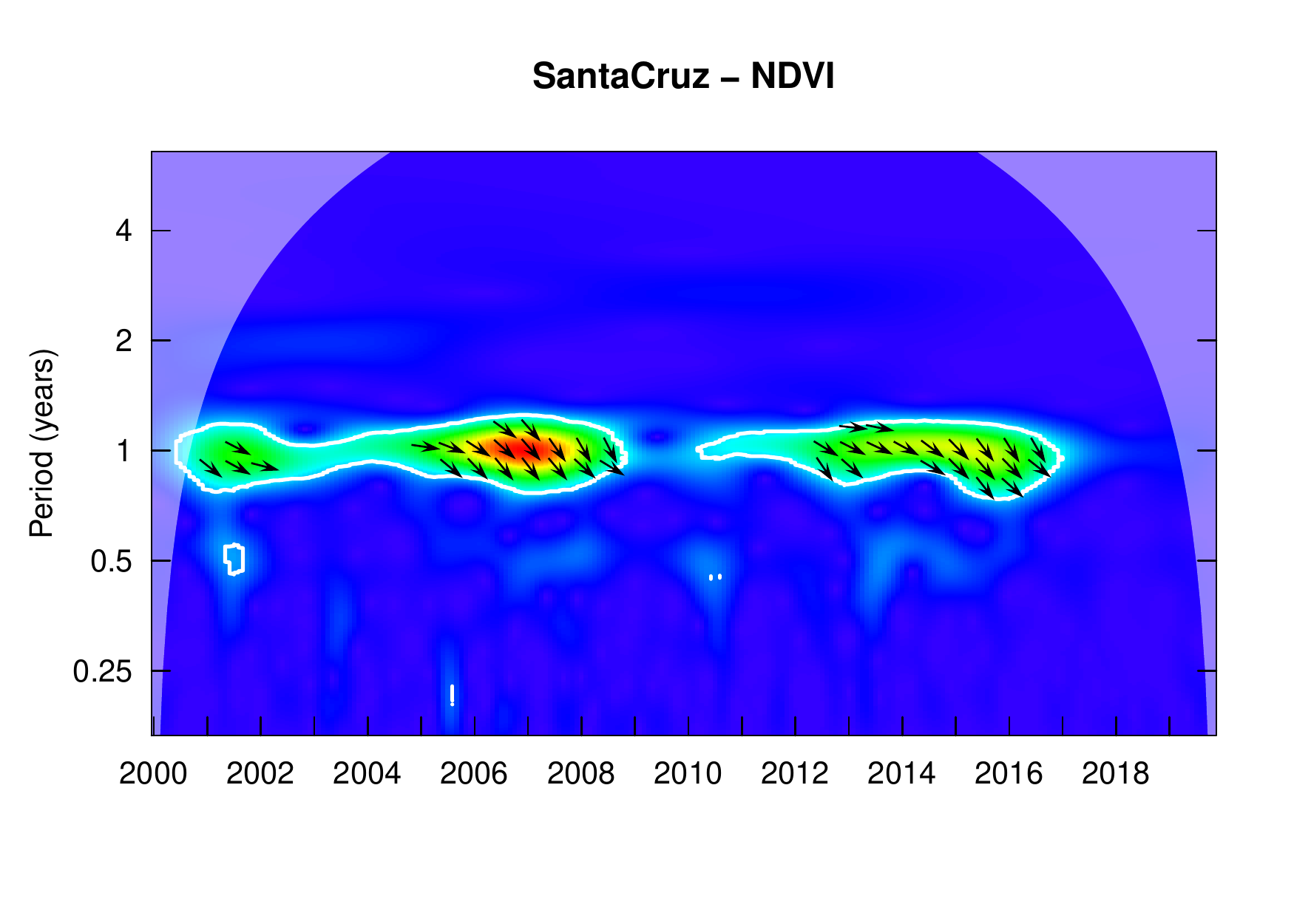}}\vspace{-0.15cm}%
\subfloat[]{\includegraphics[scale=0.23]{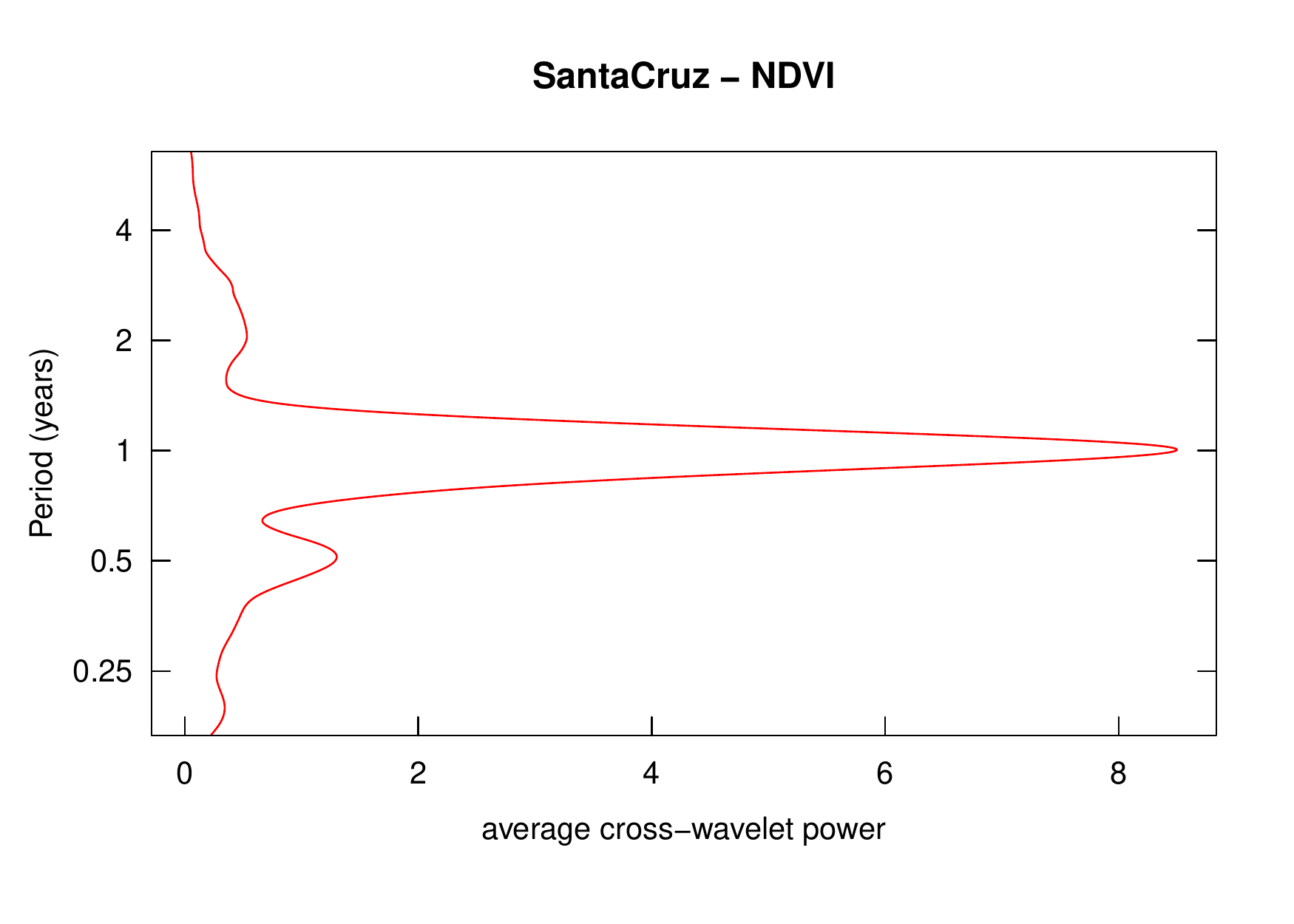}}\vspace{-0.15cm}%
\subfloat[]{\includegraphics[scale=0.23]{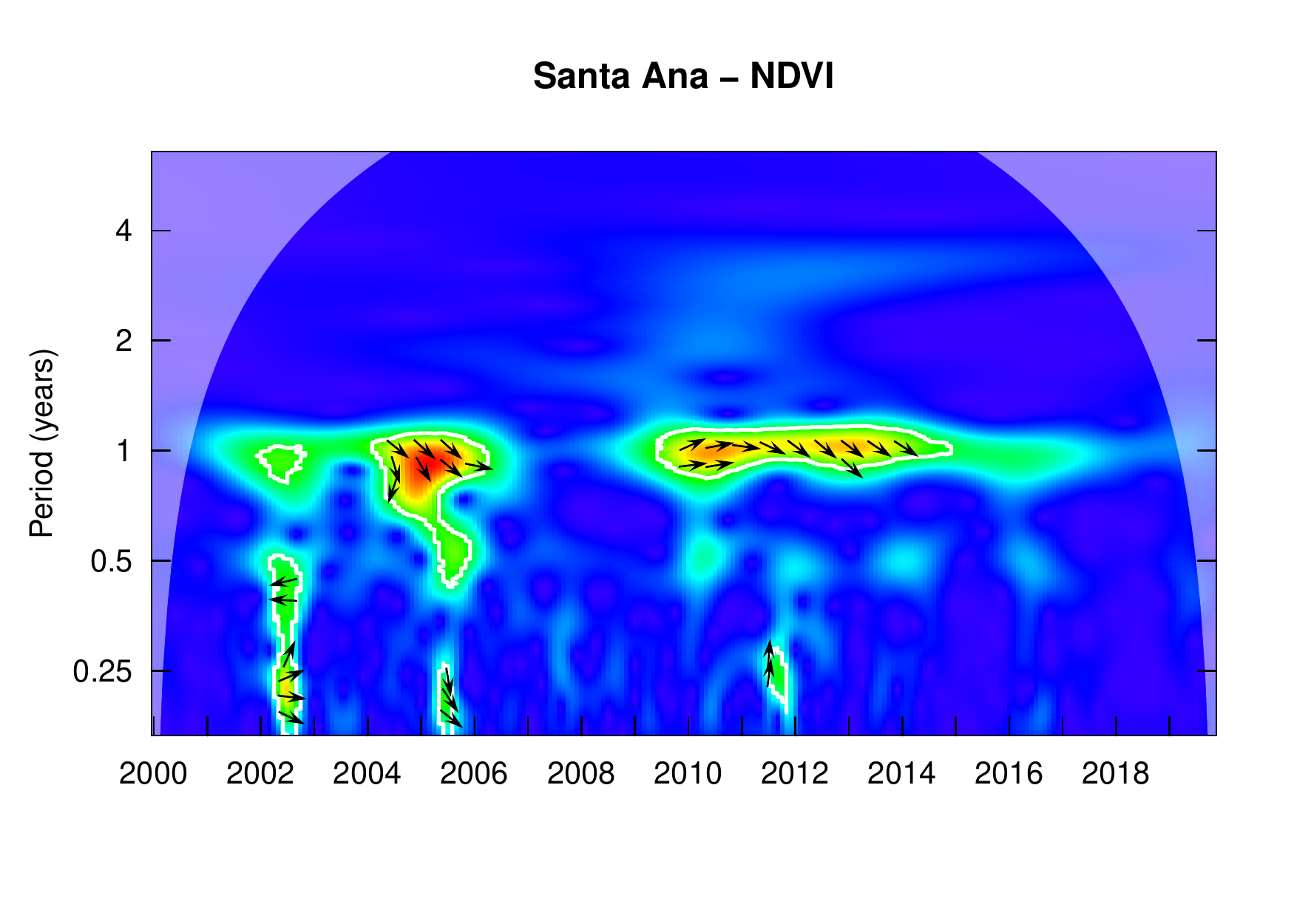}}\vspace{-0.15cm}%
\subfloat[]{\includegraphics[scale=0.23]{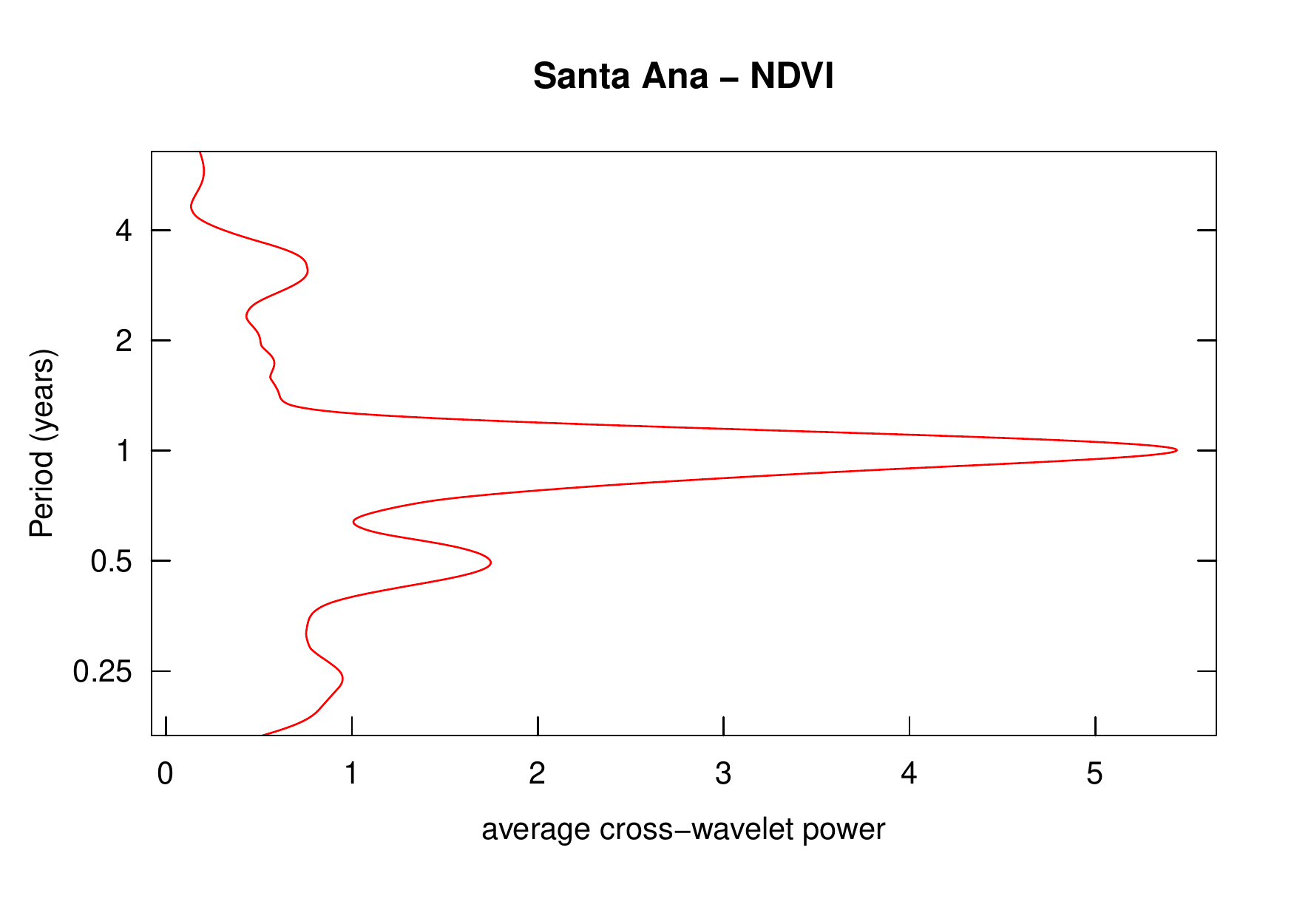}}\vspace{-0.15cm}\\
\subfloat[]{\includegraphics[scale=0.23]{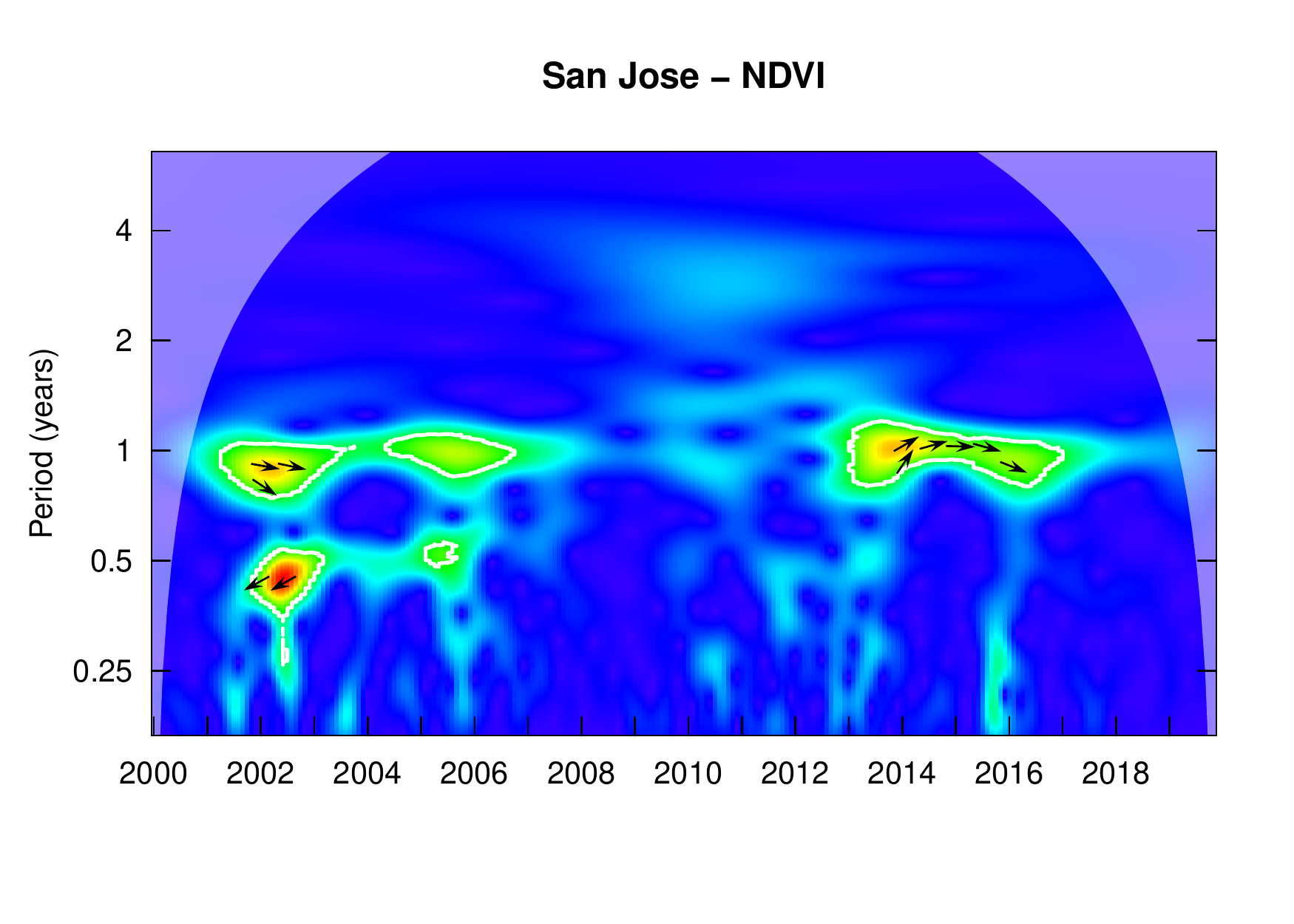}}\vspace{-0.15cm}%
\subfloat[]{\includegraphics[scale=0.23]{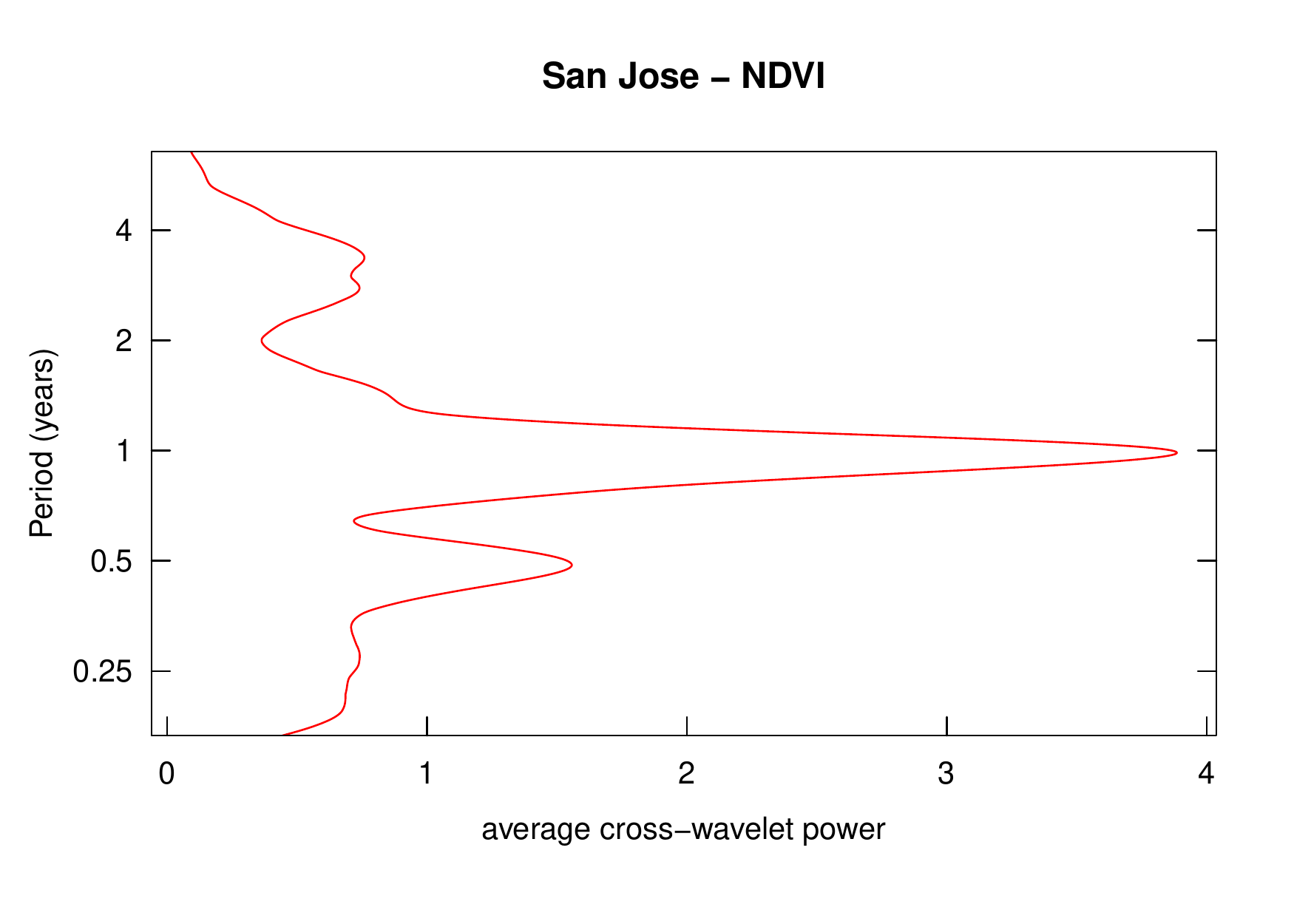}}\vspace{-0.15cm}%
\subfloat[]{\includegraphics[scale=0.23]{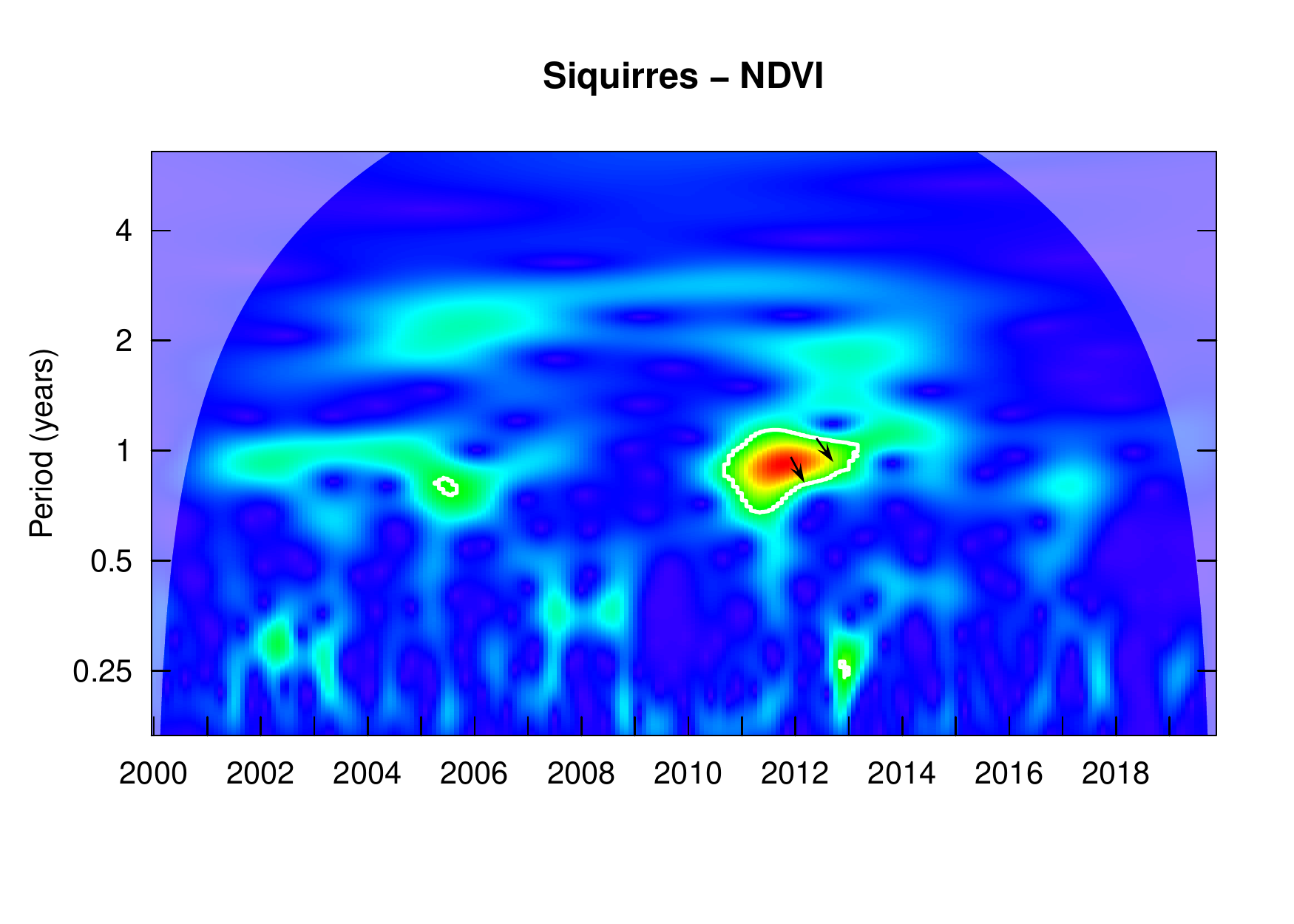}}\vspace{-0.15cm}%
\subfloat[]{\includegraphics[scale=0.23]{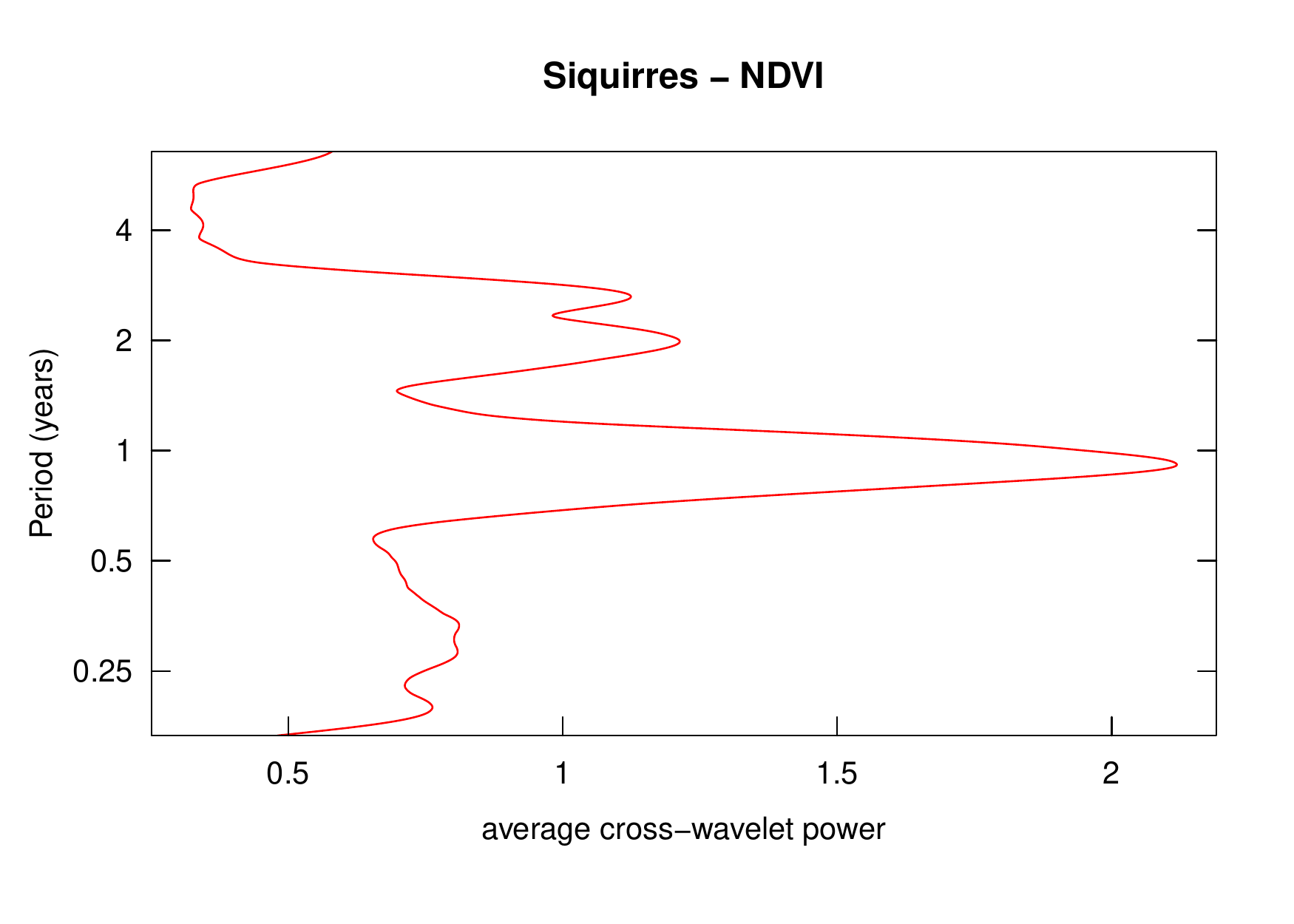}}\vspace{-0.15cm}\\
\subfloat[]{\includegraphics[scale=0.23]{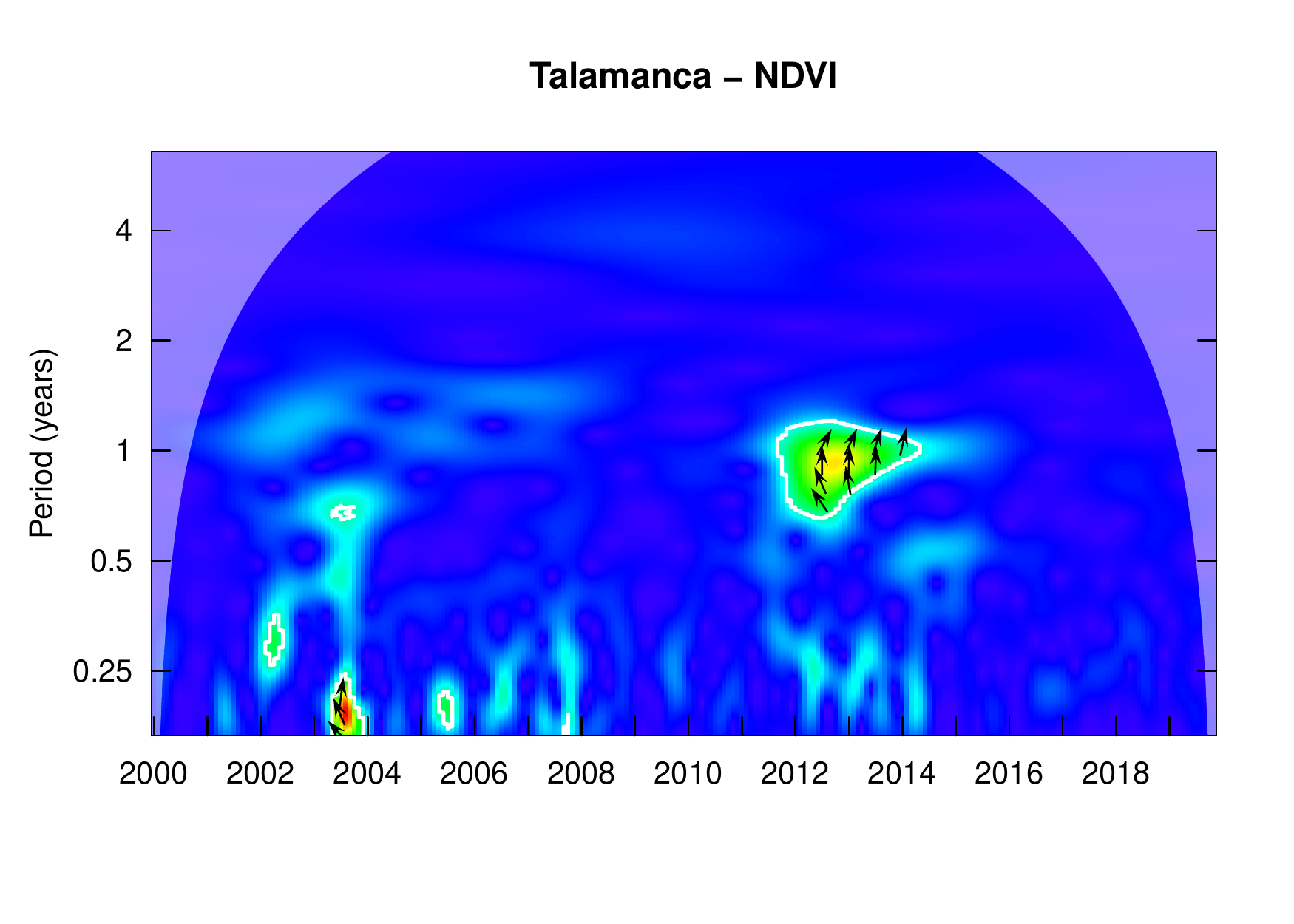}}\vspace{-0.15cm}%
\subfloat[]{\includegraphics[scale=0.23]{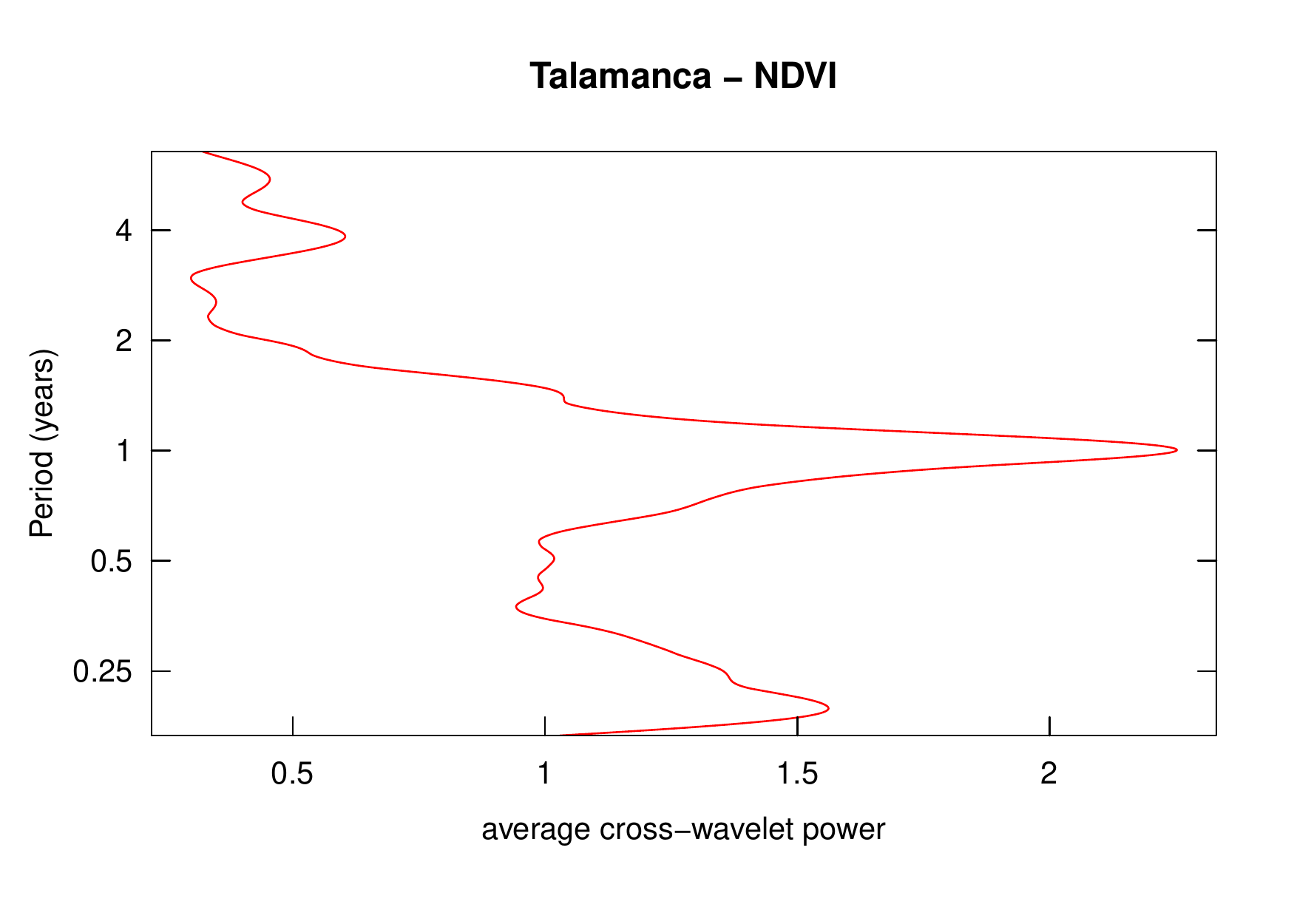}}\vspace{-0.15cm}%
\subfloat[]{\includegraphics[scale=0.23]{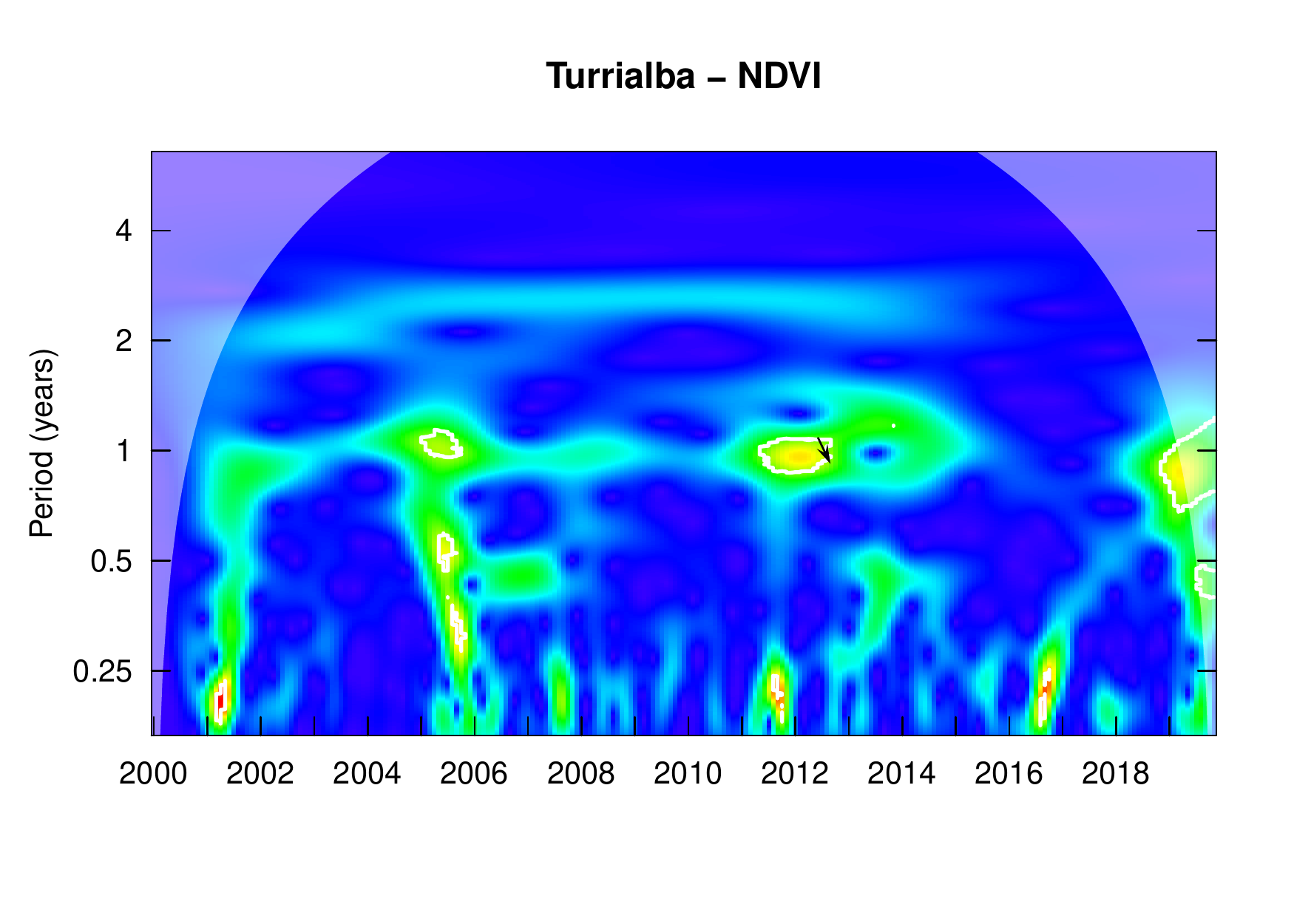}}\vspace{-0.15cm}%
\subfloat[]{\includegraphics[scale=0.23]{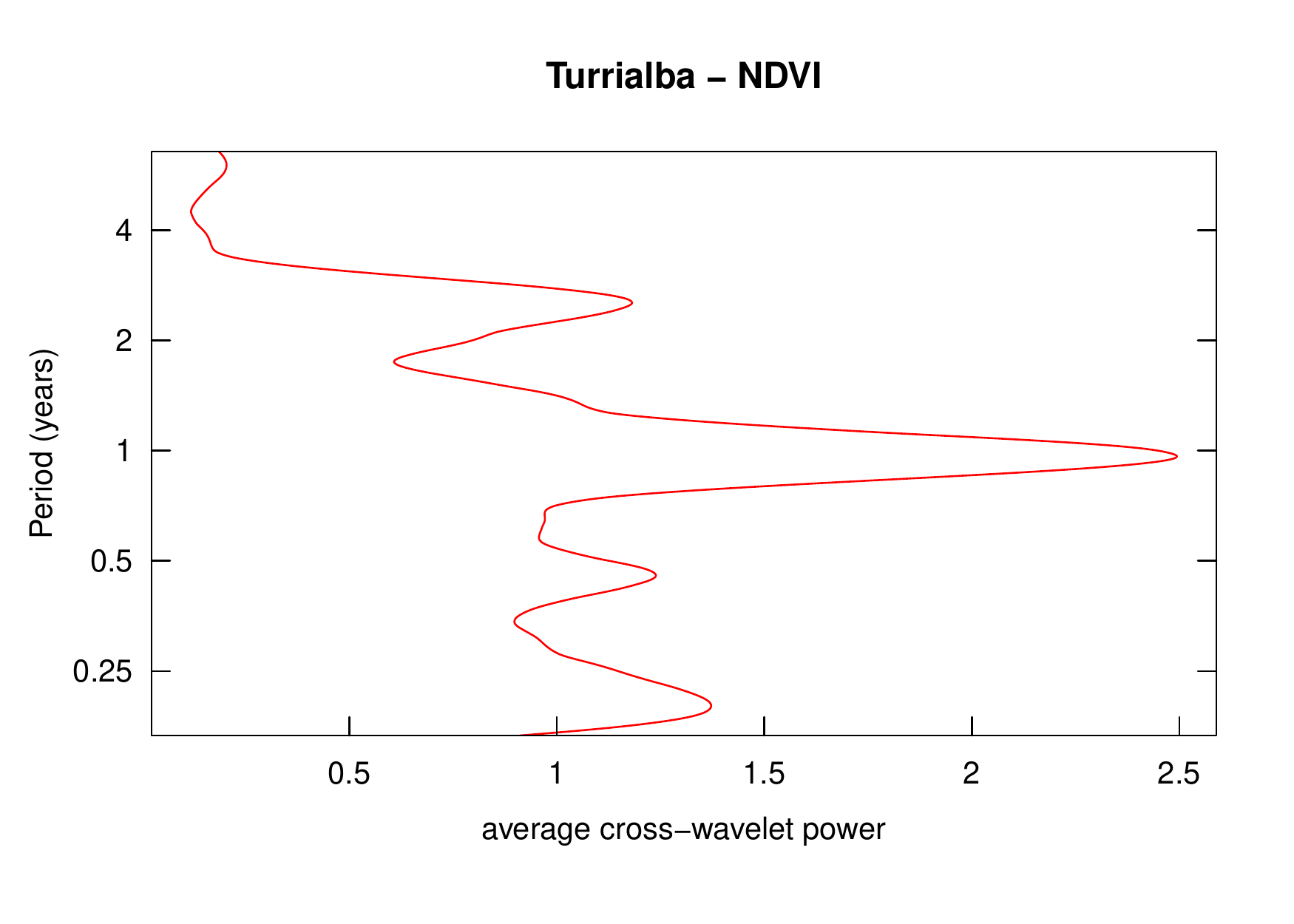}}\vspace{-0.15cm}\\
\subfloat[]{\includegraphics[scale=0.23]{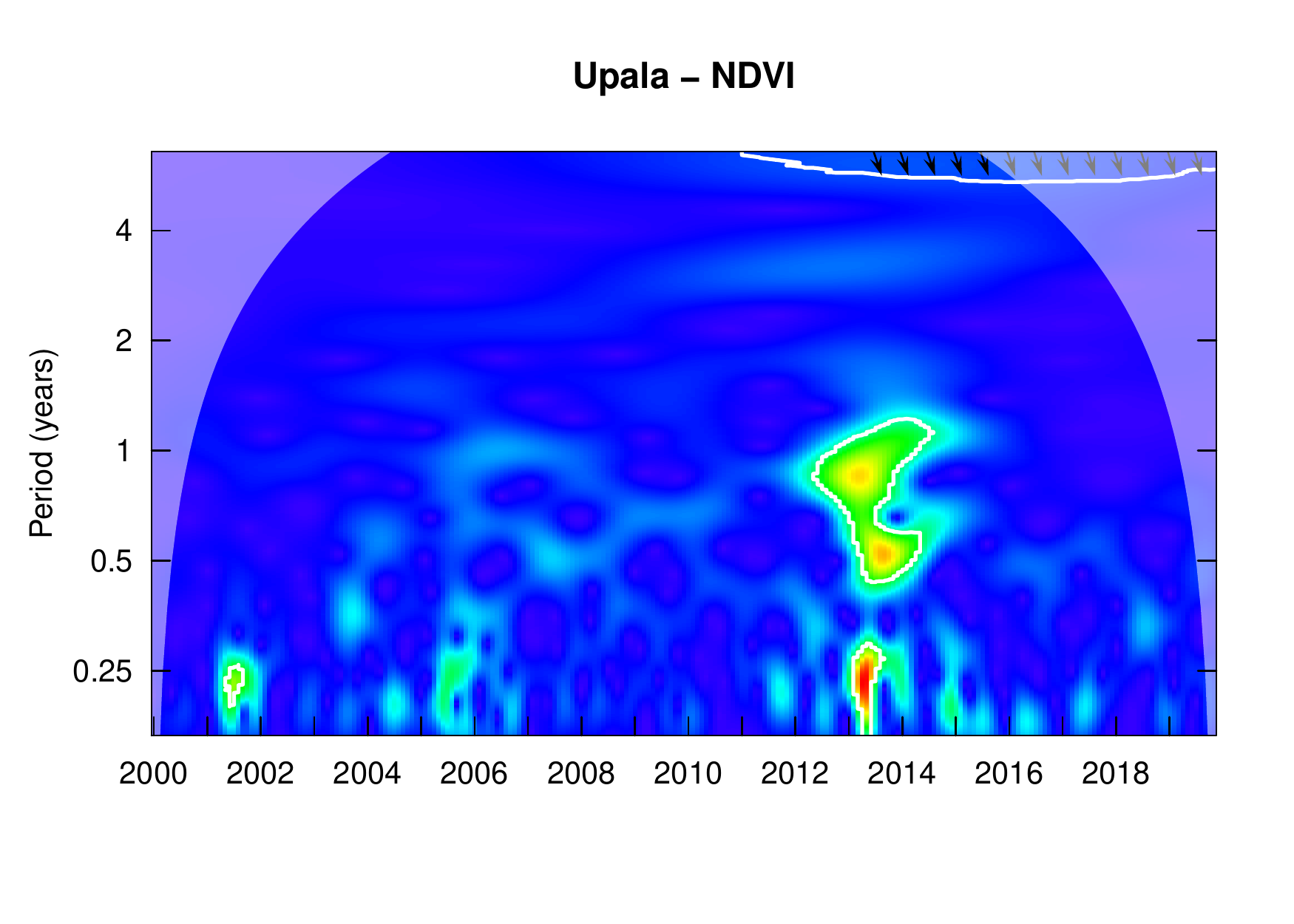}}\vspace{-0.15cm}%
\subfloat[]{\includegraphics[scale=0.23]{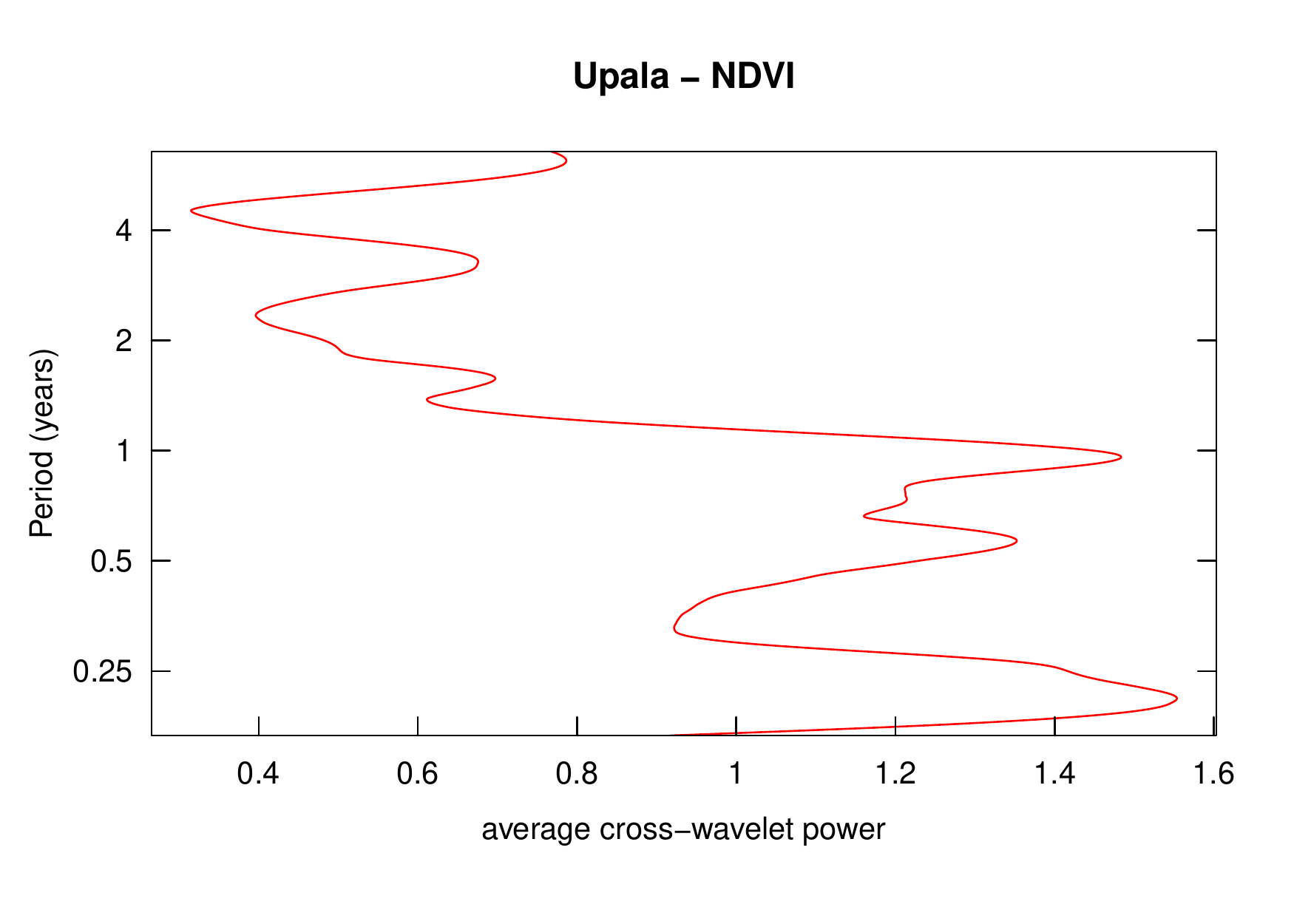}}\vspace{-0.15cm}
\subfloat[]{\includegraphics[scale=0.23]{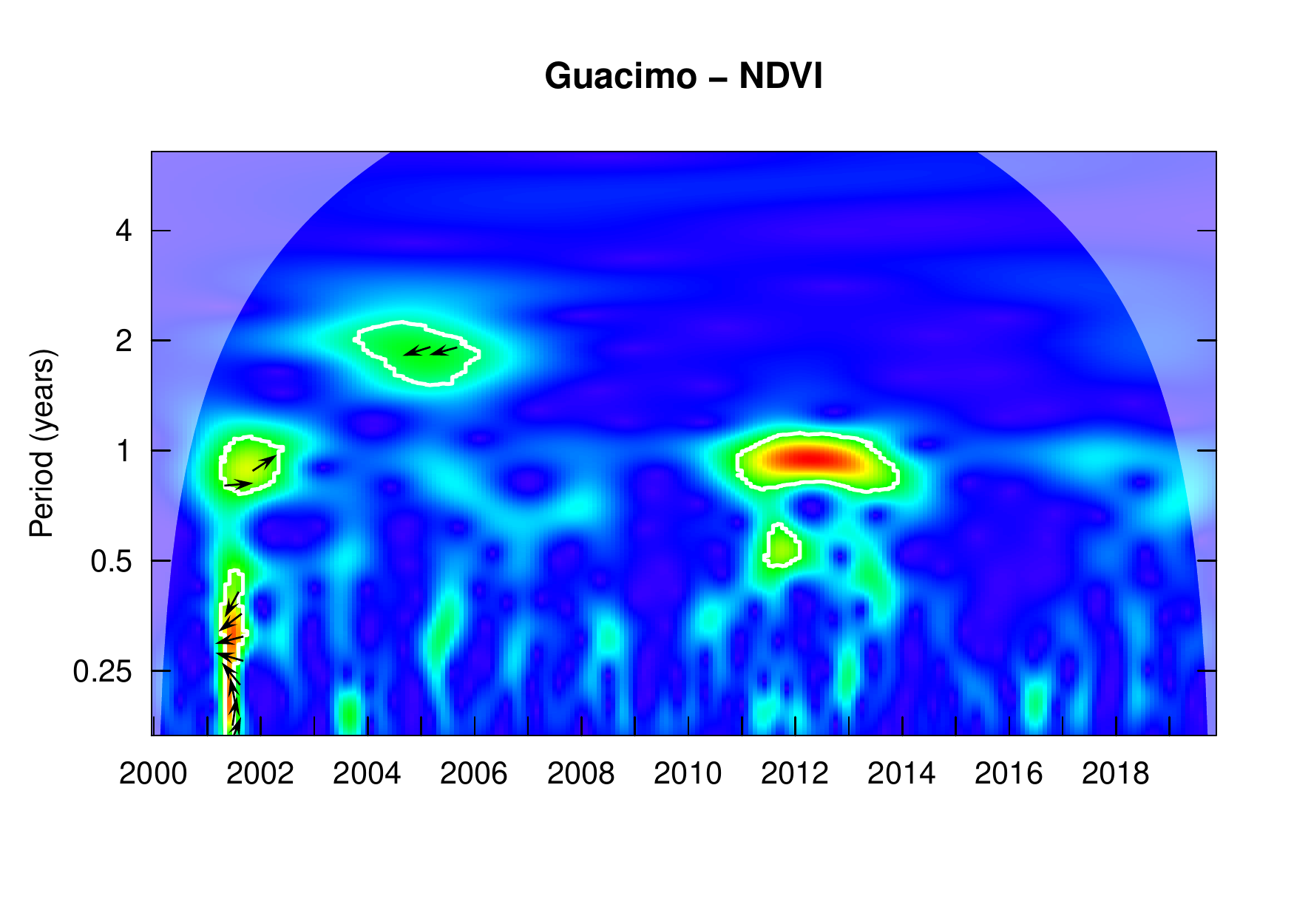}}\vspace{-0.15cm}%
\subfloat[]{\includegraphics[scale=0.23]{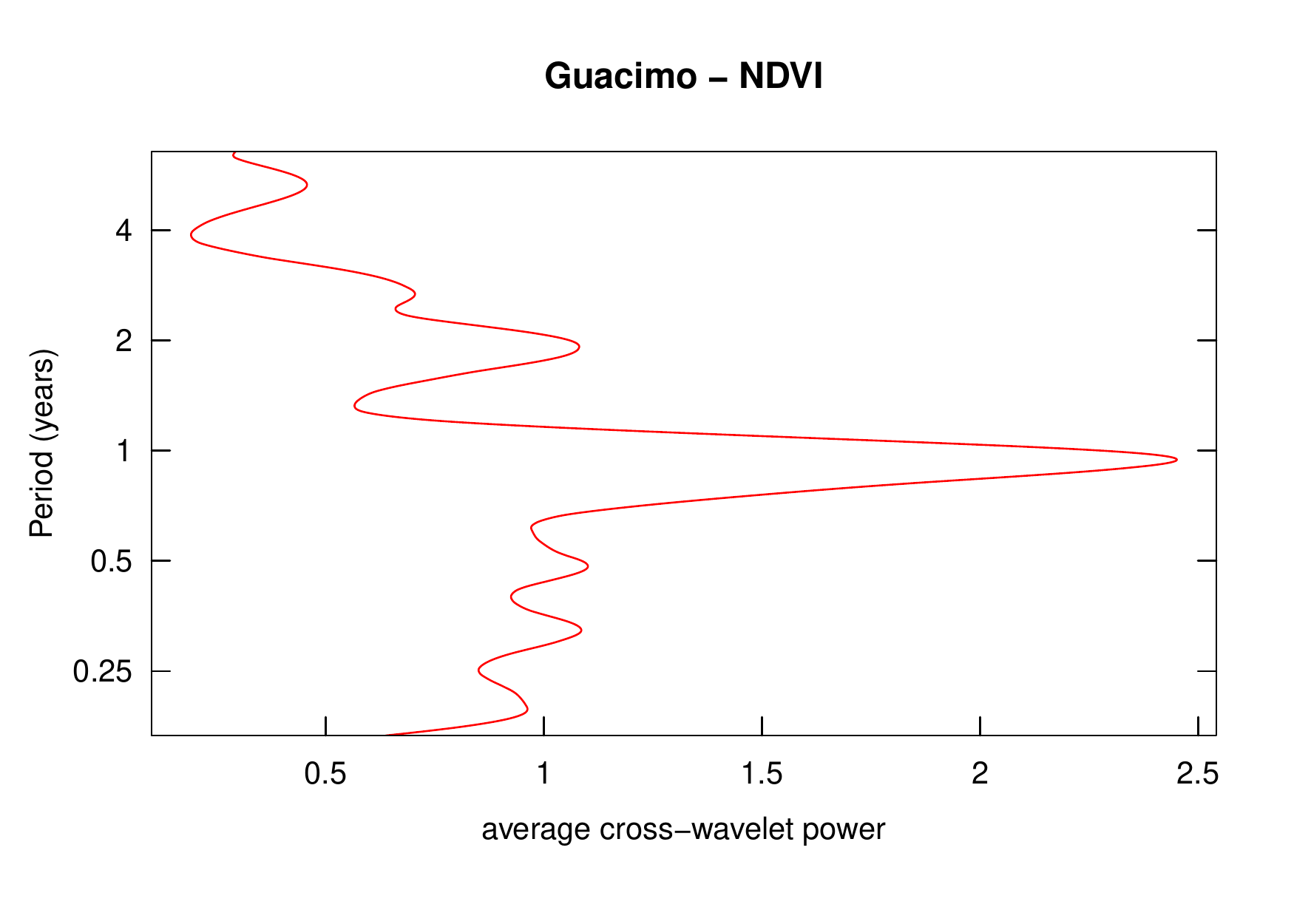}}\vspace{-0.15cm}
\caption*{}
\end{figure}

\begin{figure}[H]
\captionsetup[subfigure]{labelformat=empty}
\subfloat[]{\includegraphics[scale=0.23]{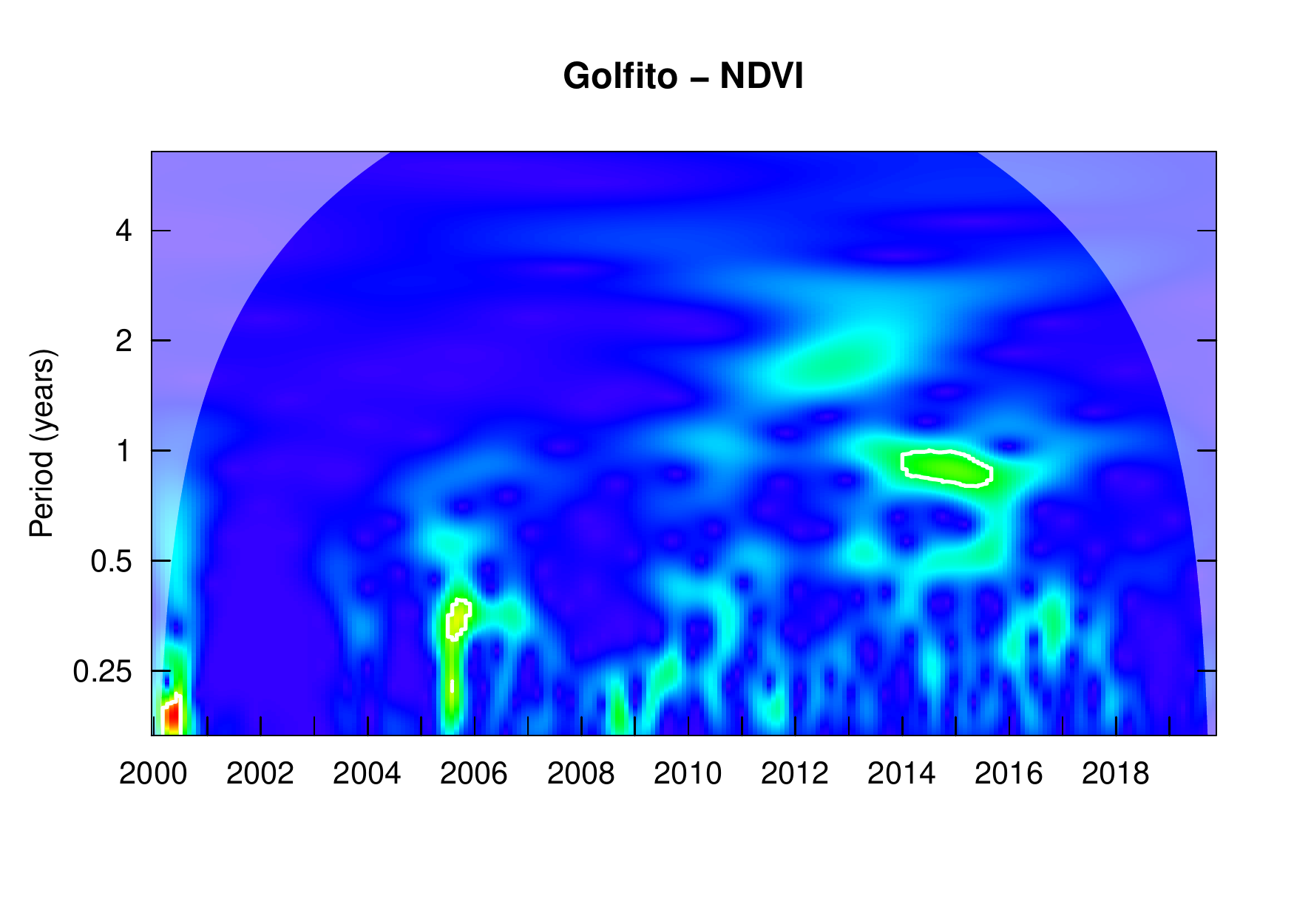}}\vspace{-0.15cm}%
\subfloat[]{\includegraphics[scale=0.23]{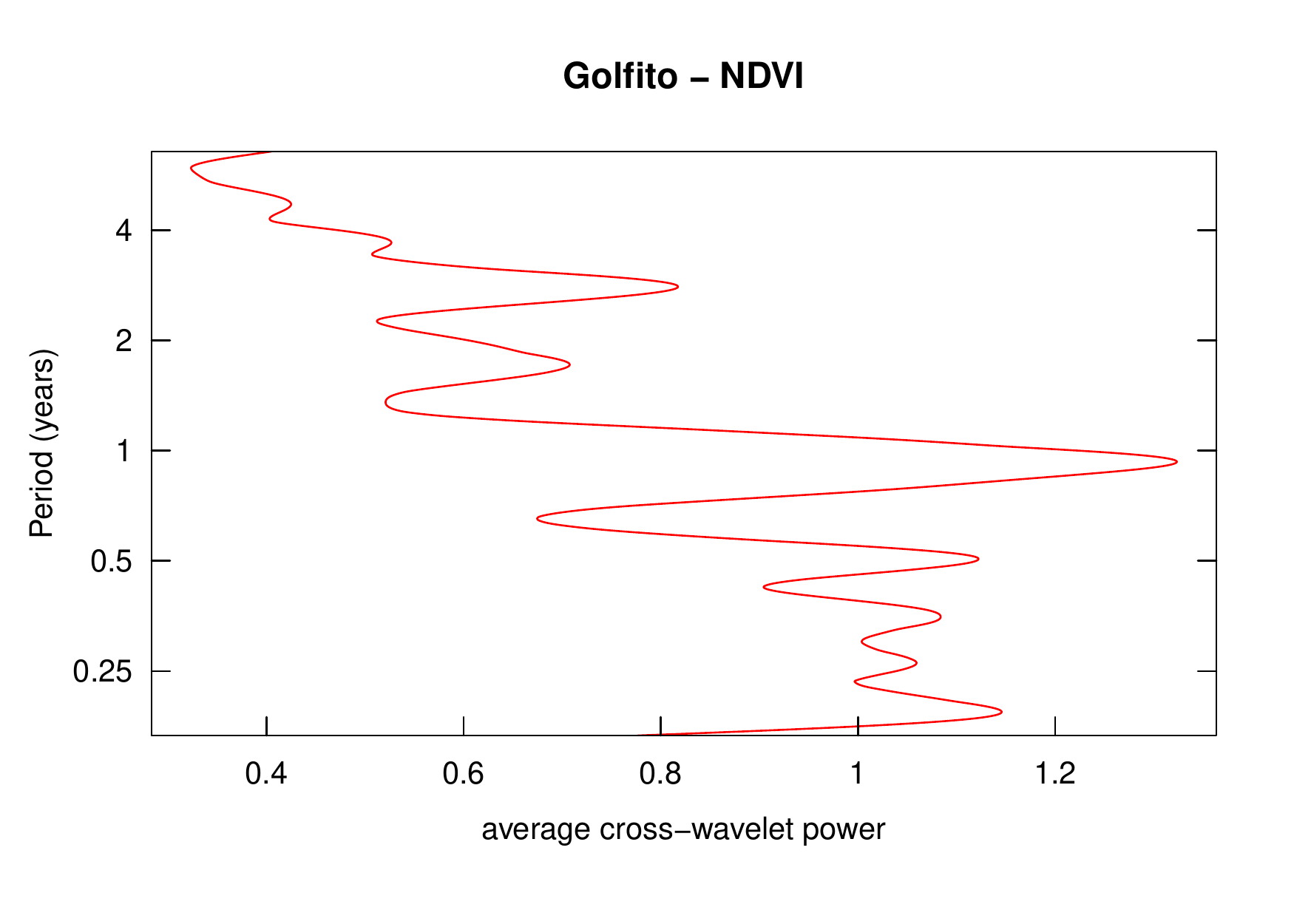}}\vspace{-0.15cm}%
\subfloat[]{\includegraphics[scale=0.23]{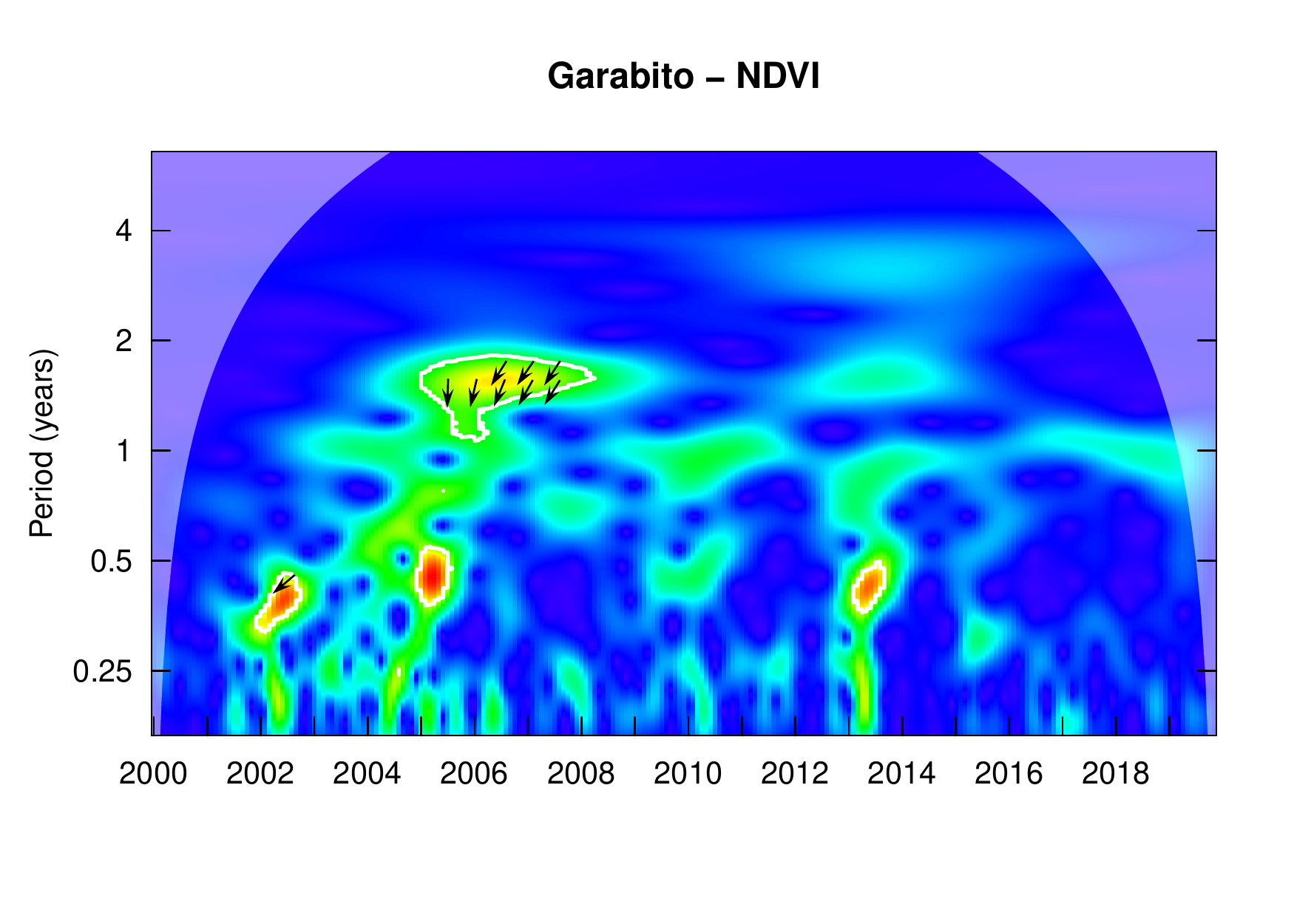}}\vspace{-0.15cm}%
\subfloat[]{\includegraphics[scale=0.23]{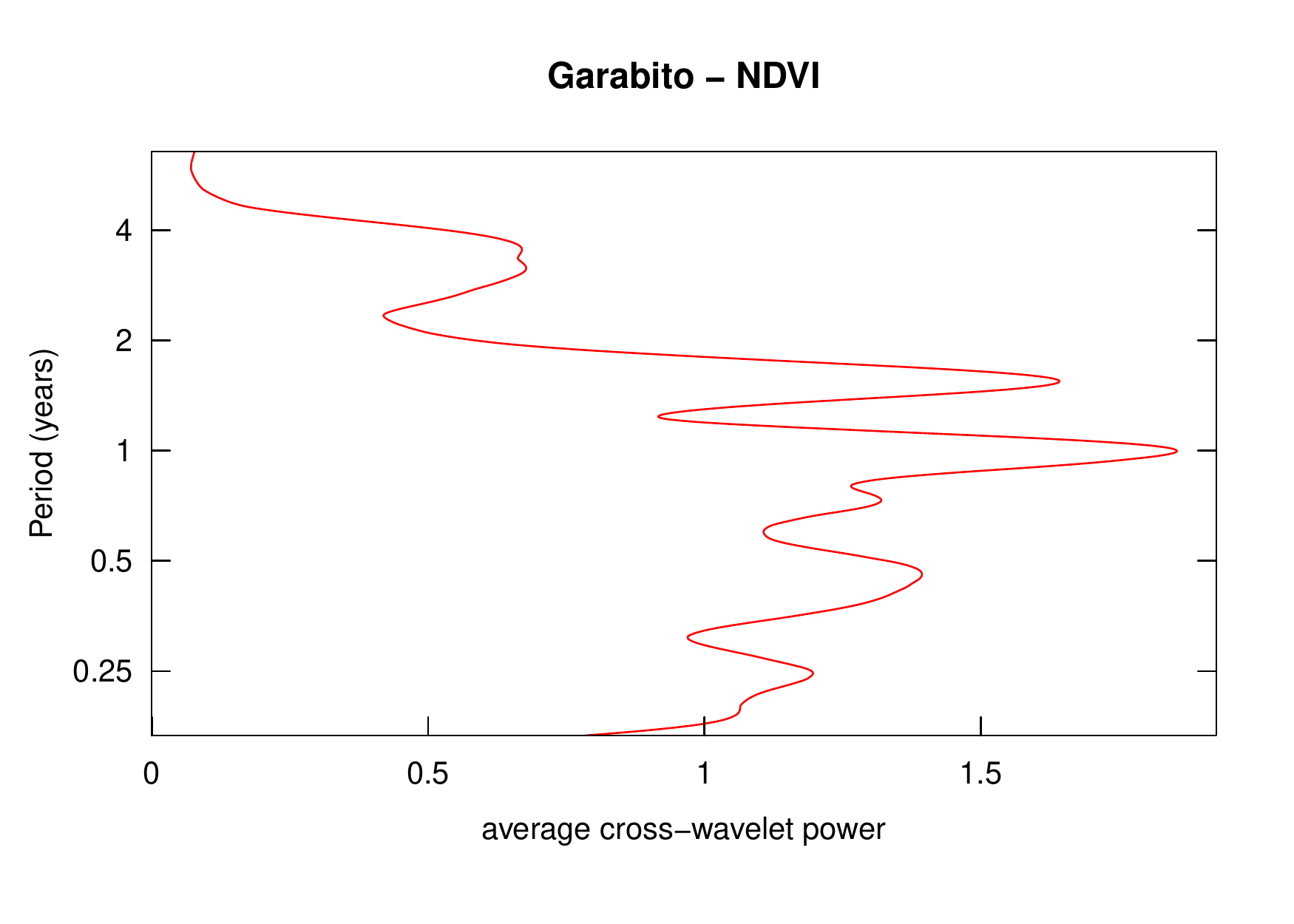}}\vspace{-0.15cm}\\
\subfloat[]{\includegraphics[scale=0.23]{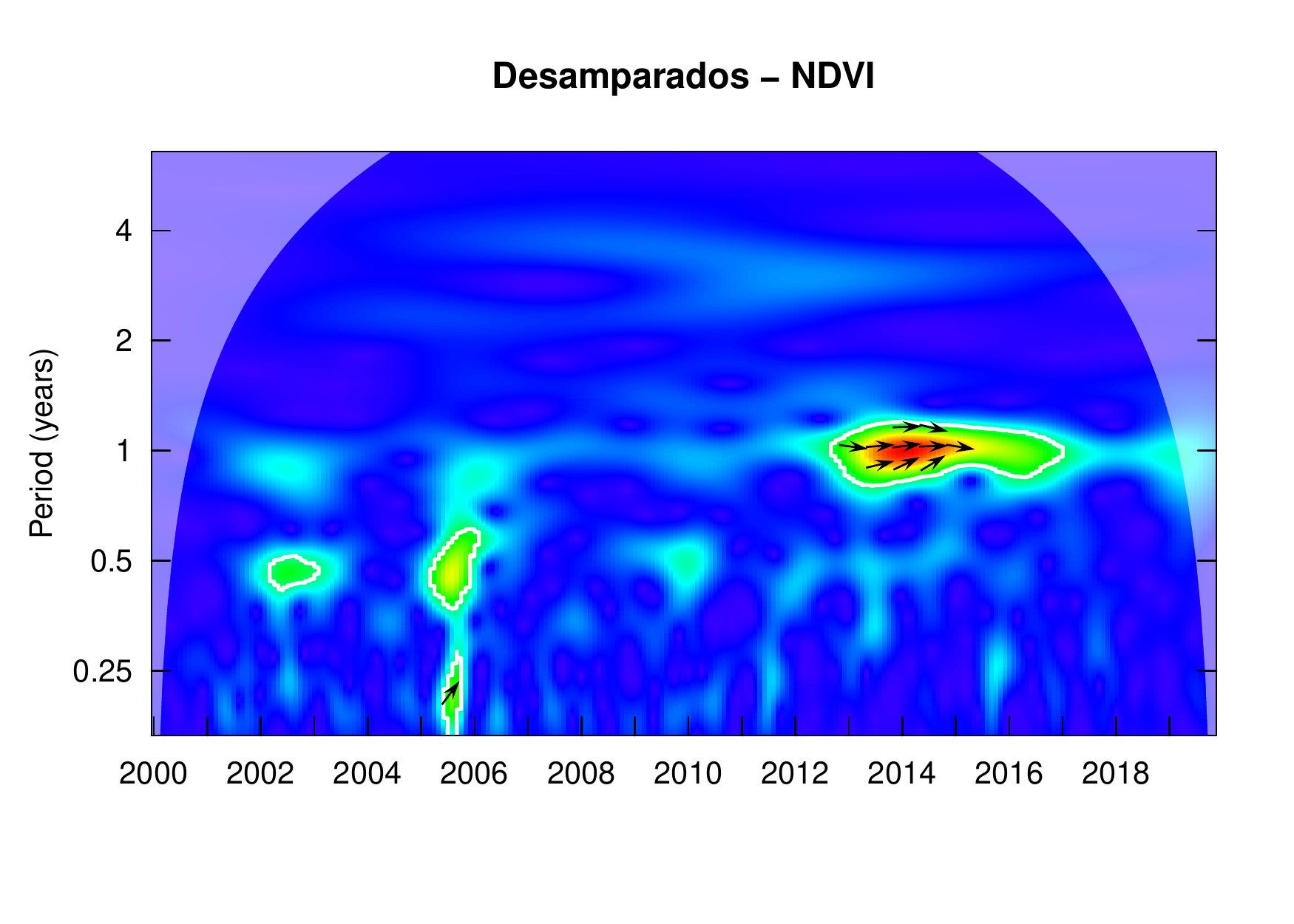}}\vspace{-0.15cm}%
\subfloat[]{\includegraphics[scale=0.23]{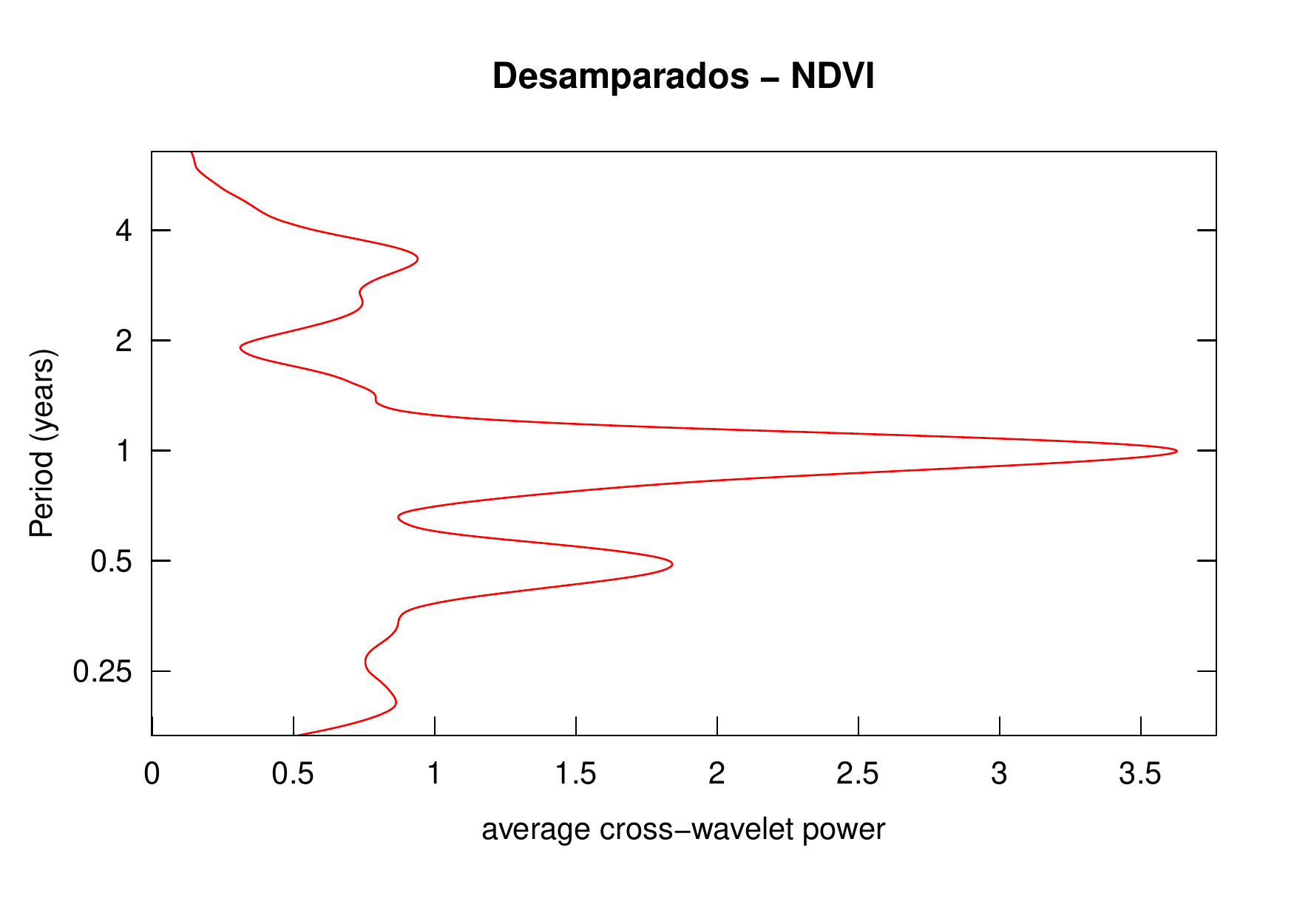}}\vspace{-0.15cm}%
\subfloat[]{\includegraphics[scale=0.23]{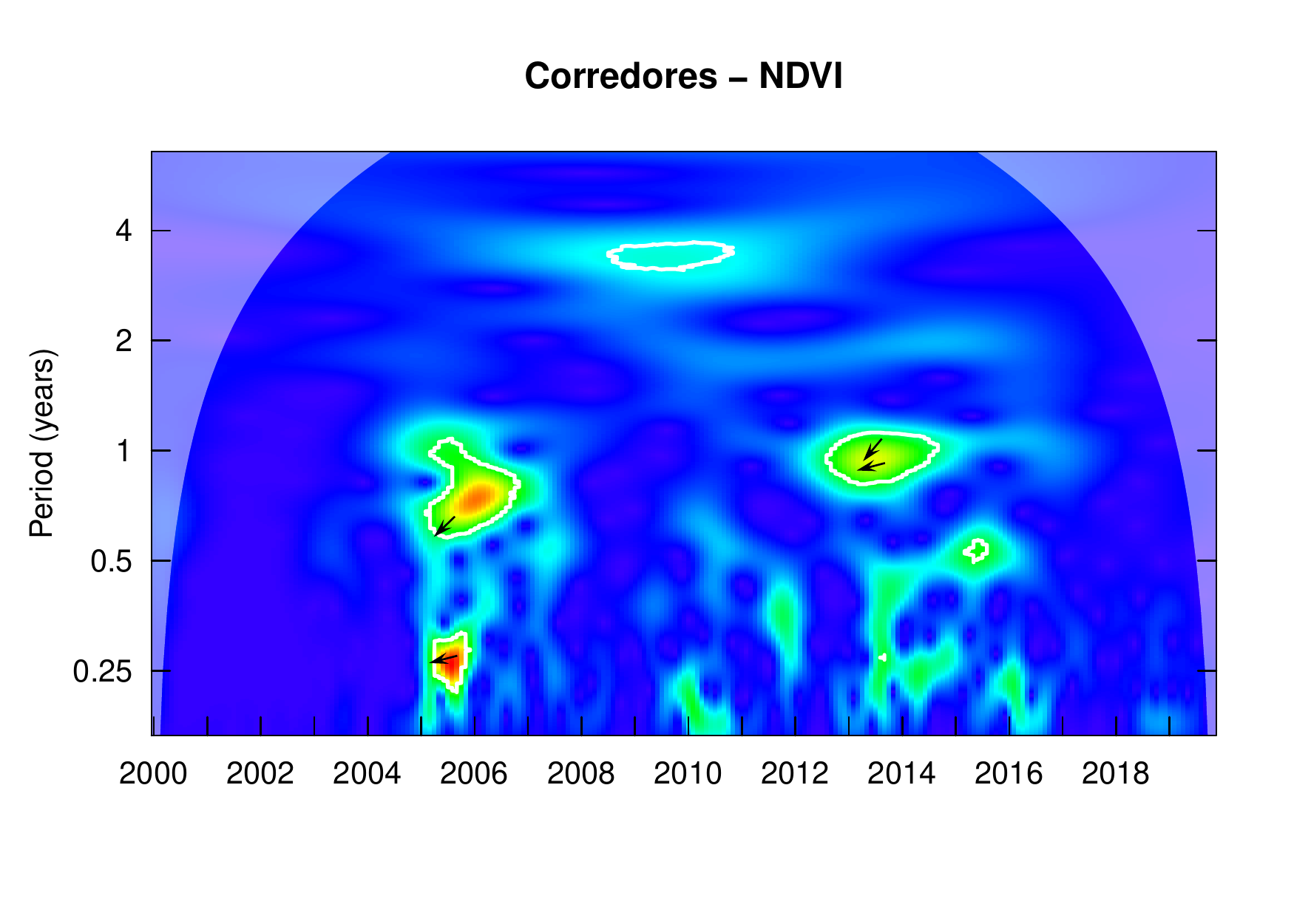}}\vspace{-0.15cm}%
\subfloat[]{\includegraphics[scale=0.23]{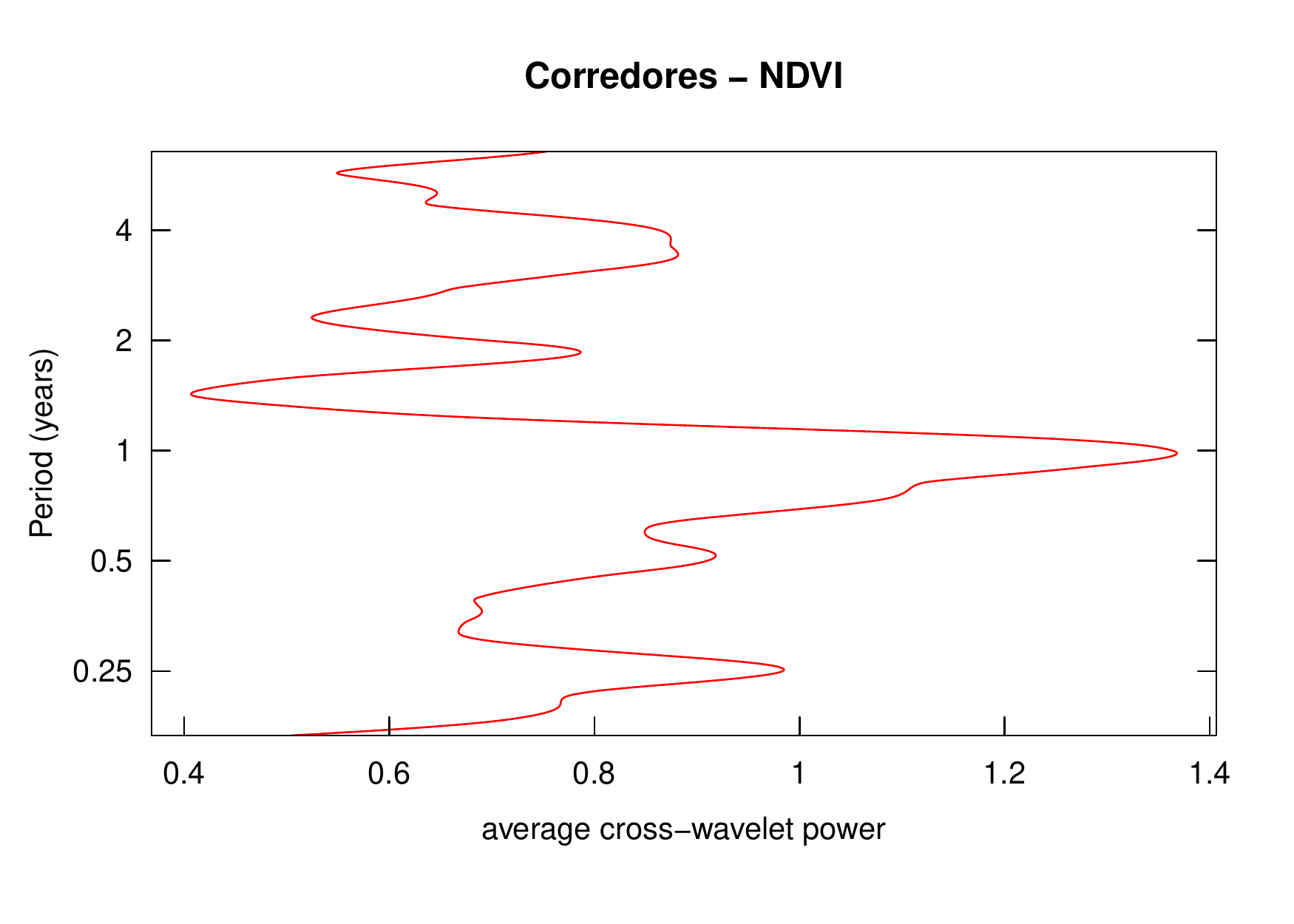}}\vspace{-0.15cm}\\
\subfloat[]{\includegraphics[scale=0.23]{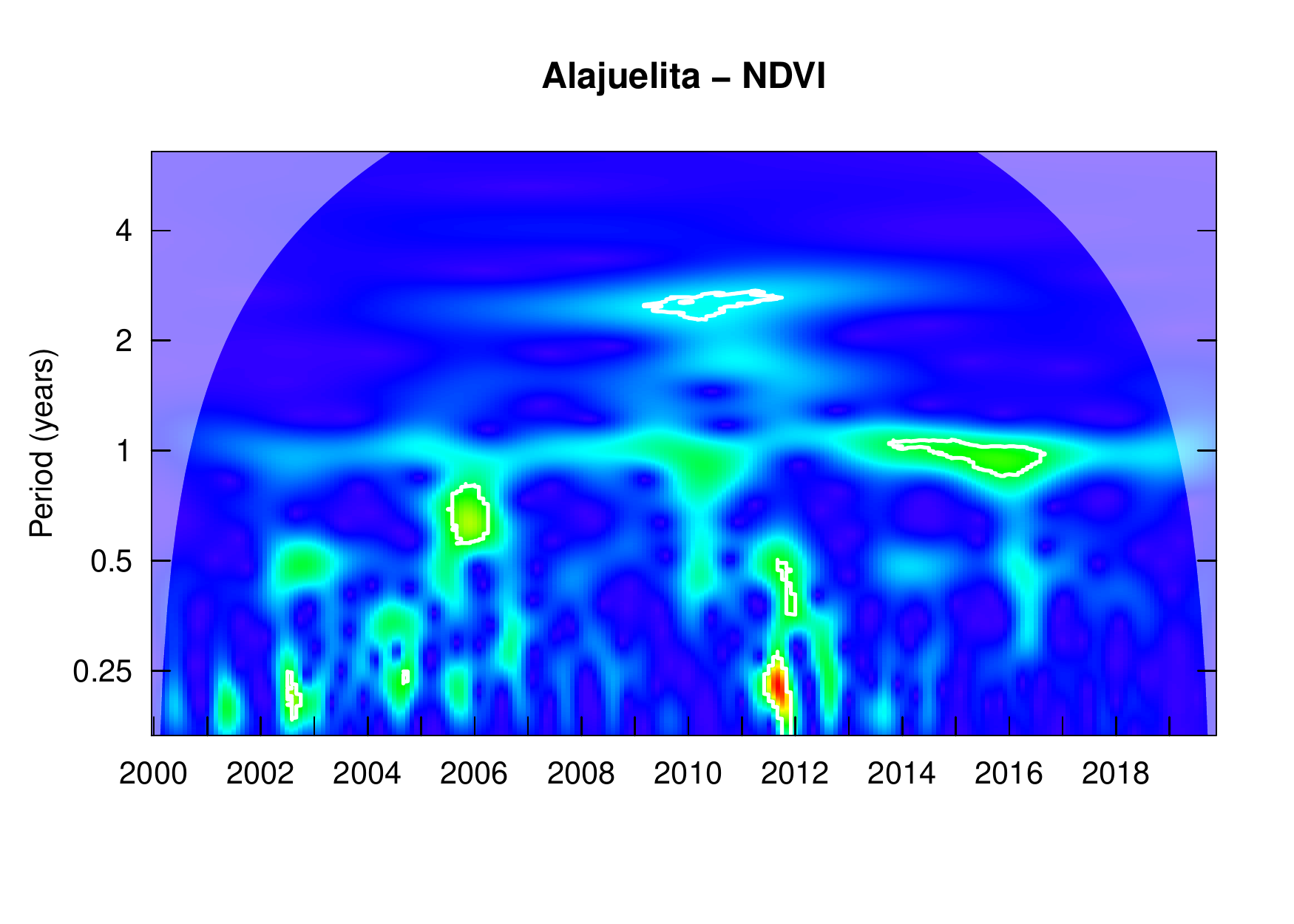}}\vspace{-0.15cm}%
\subfloat[]{\includegraphics[scale=0.23]{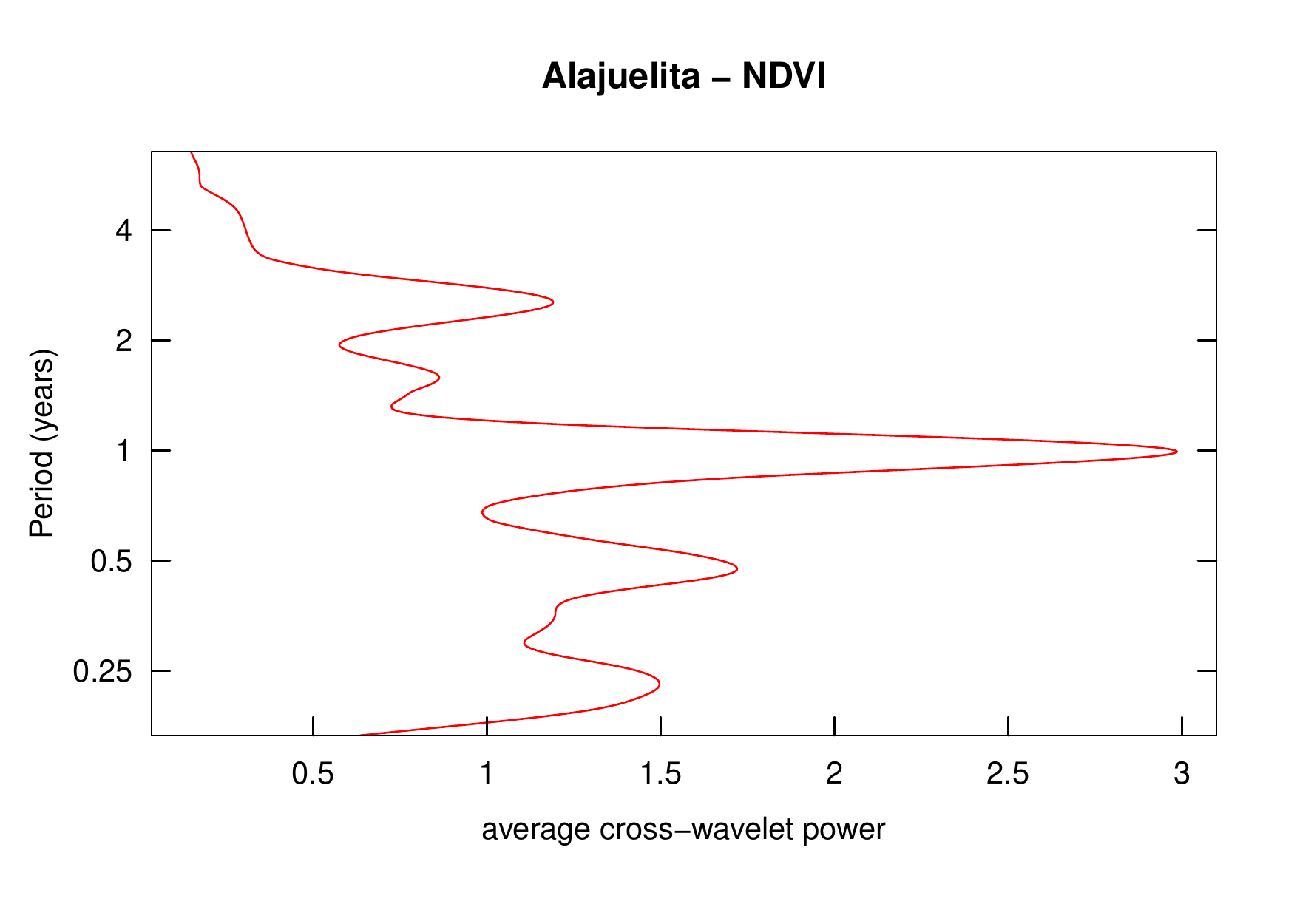}}\vspace{-0.15cm}%
\subfloat[]{\includegraphics[scale=0.23]{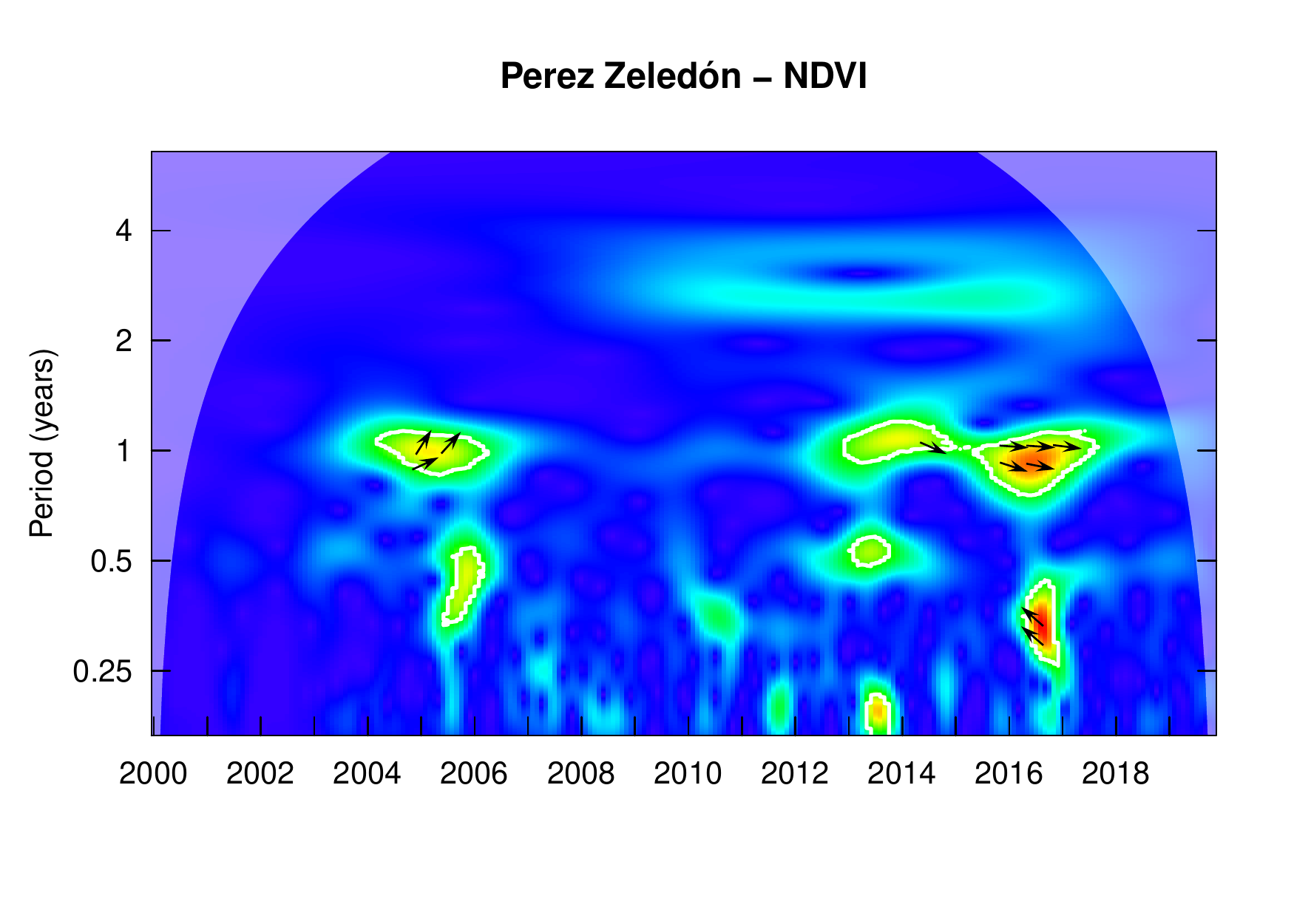}}\vspace{-0.15cm}%
\subfloat[]{\includegraphics[scale=0.23]{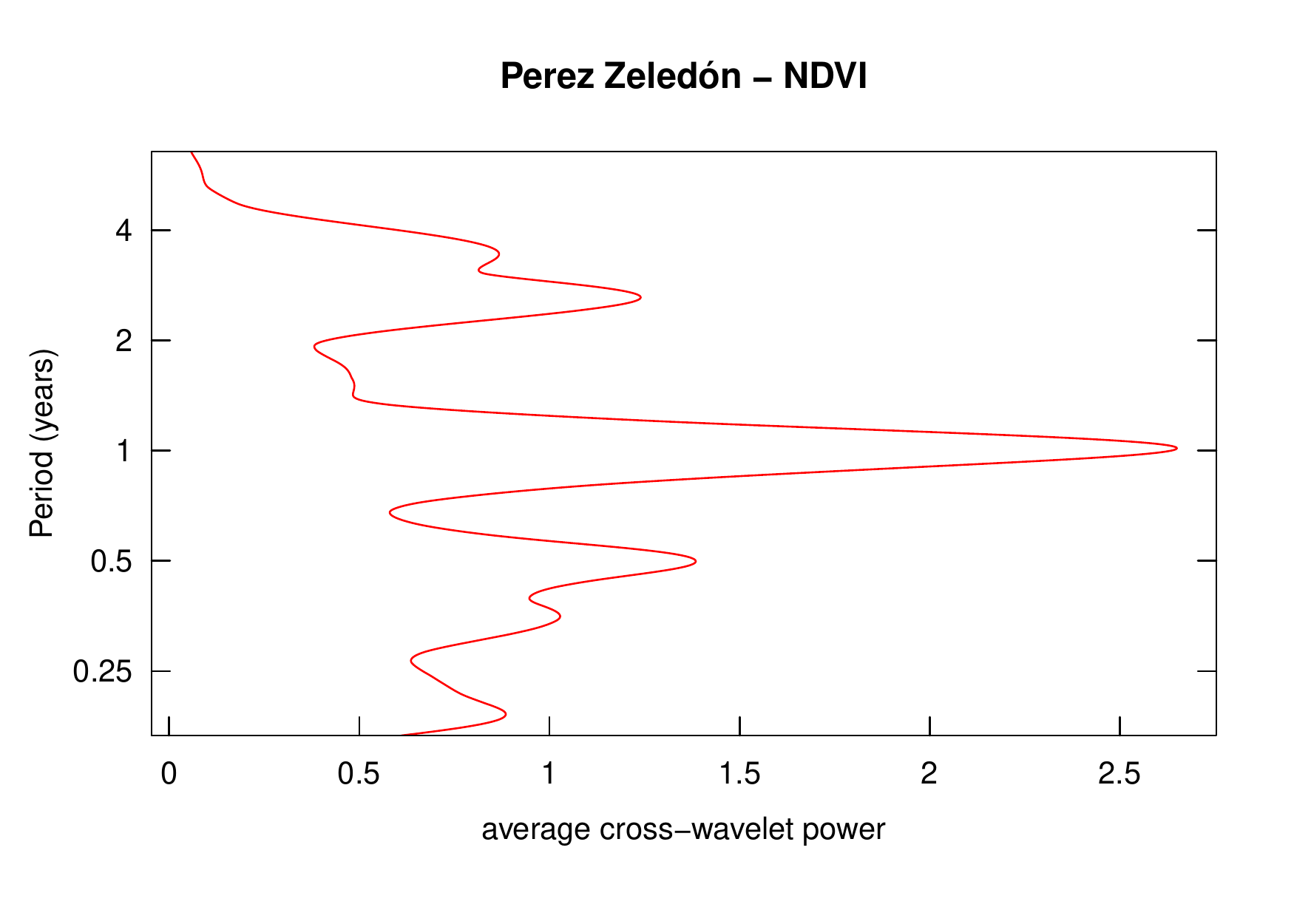}}\vspace{-0.15cm}\\
\subfloat[]{\includegraphics[scale=0.23]{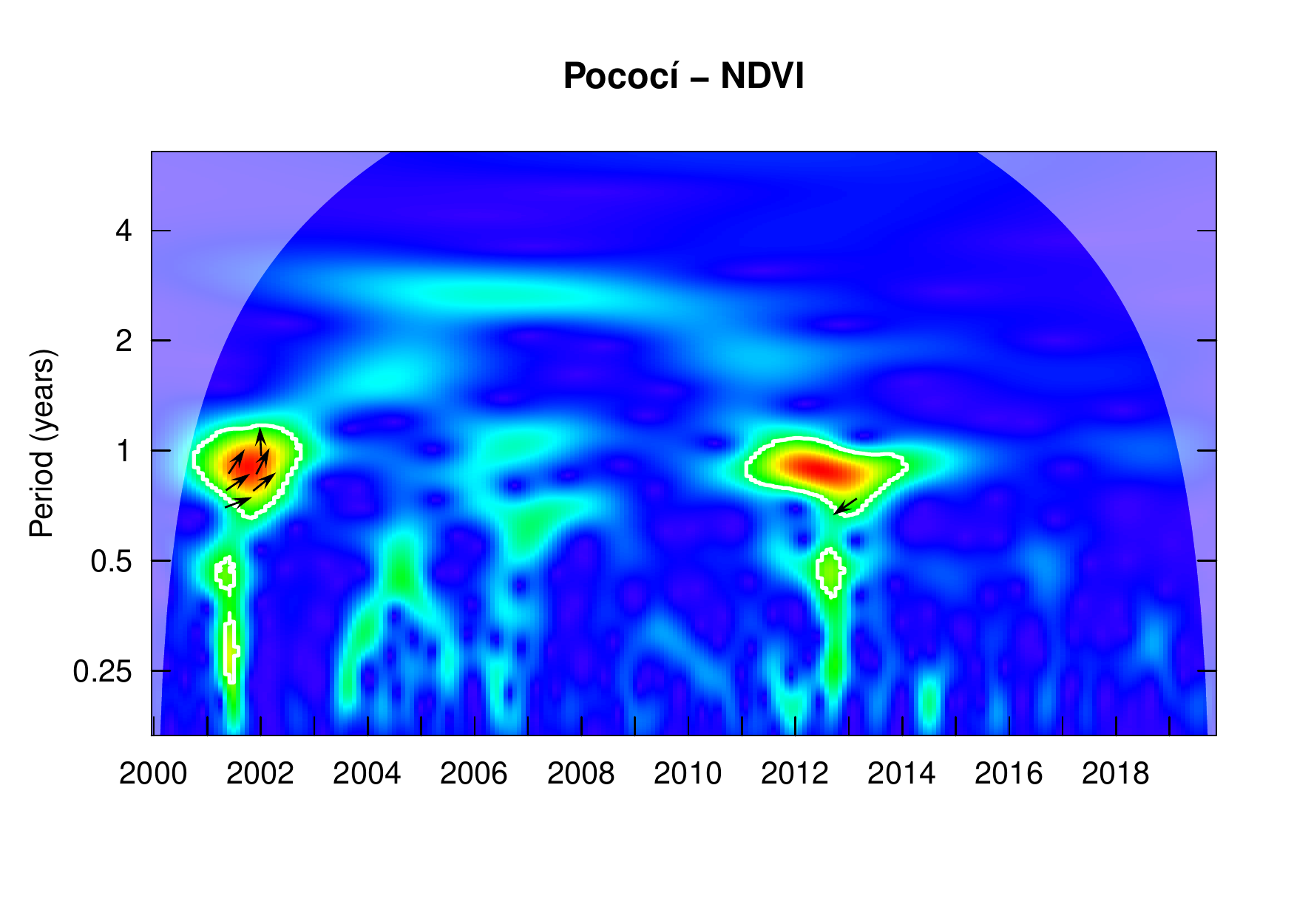}}\vspace{-0.15cm}%
\subfloat[]{\includegraphics[scale=0.23]{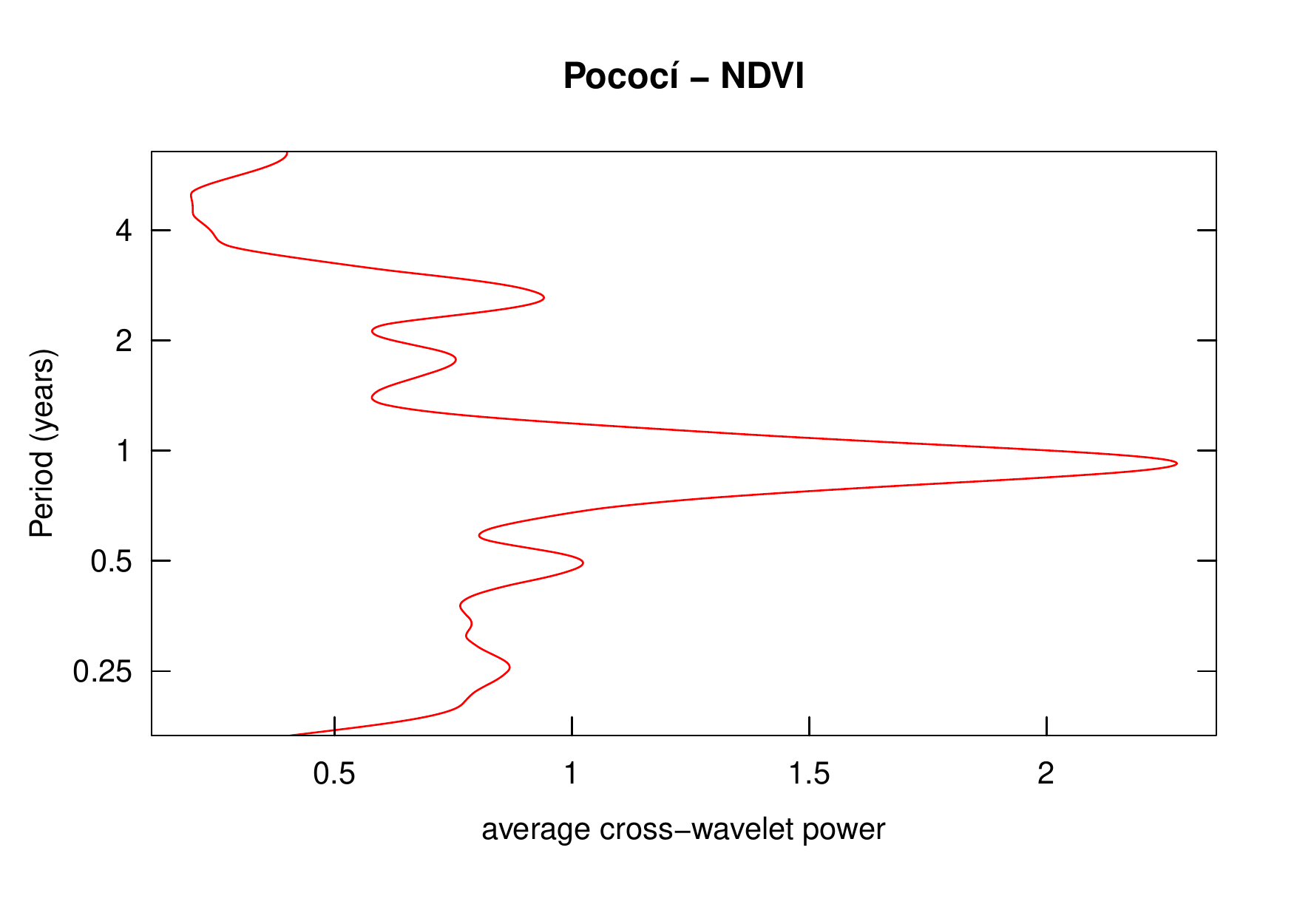}}\vspace{-0.15cm}%
\subfloat[]{\includegraphics[scale=0.23]{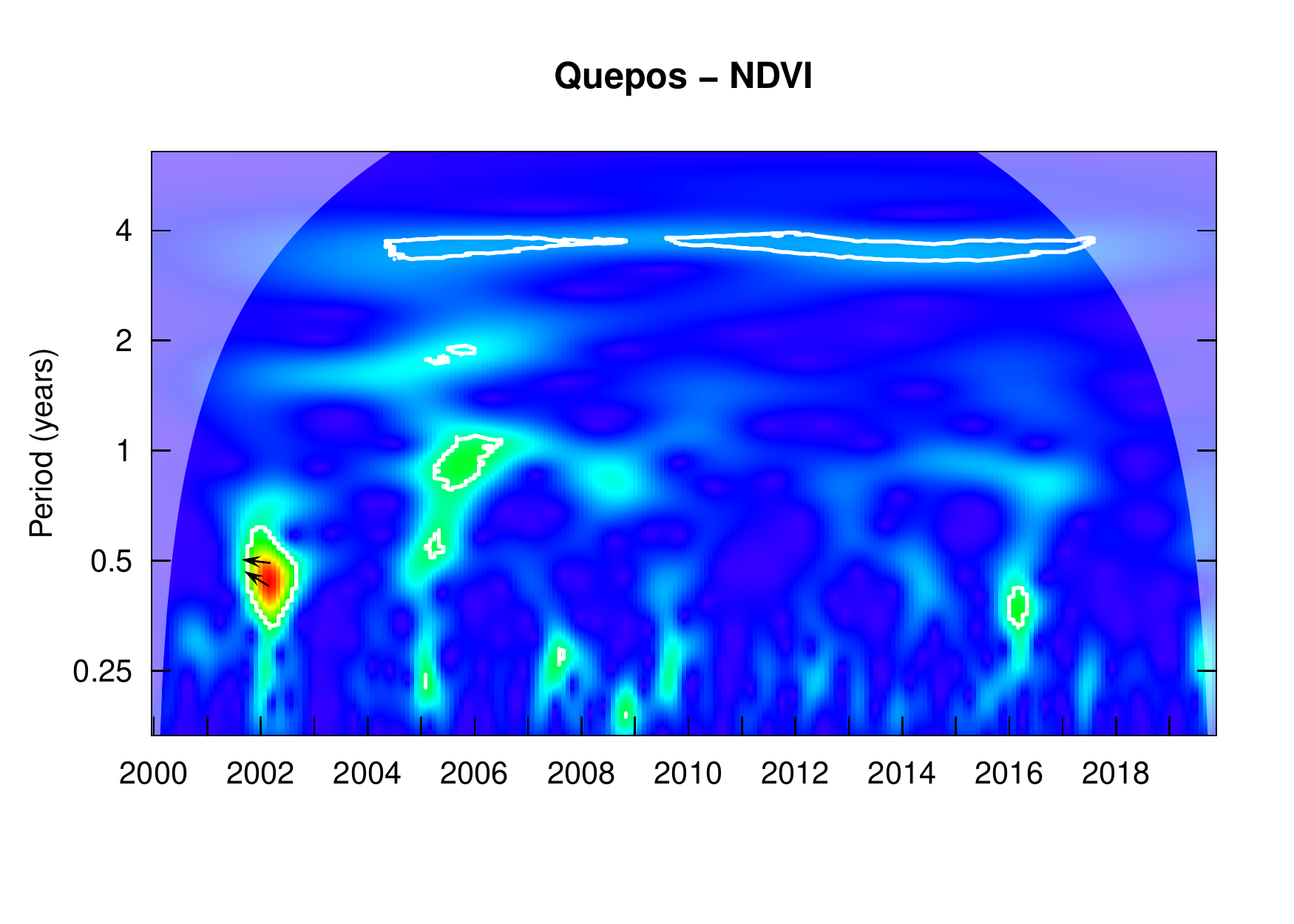}}\vspace{-0.15cm}%
\subfloat[]{\includegraphics[scale=0.23]{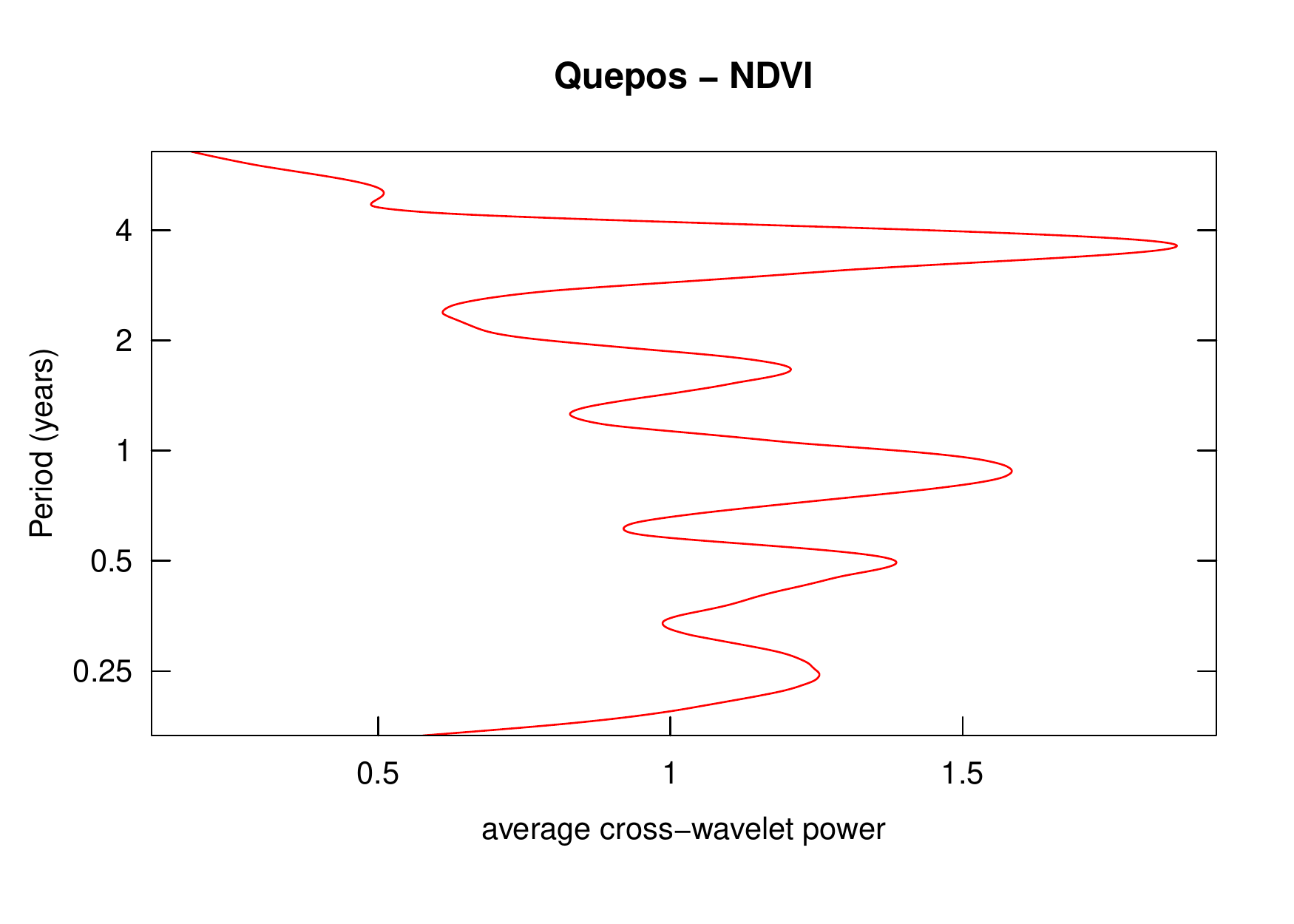}}\vspace{-0.15cm}\\
\subfloat[]{\includegraphics[scale=0.23]{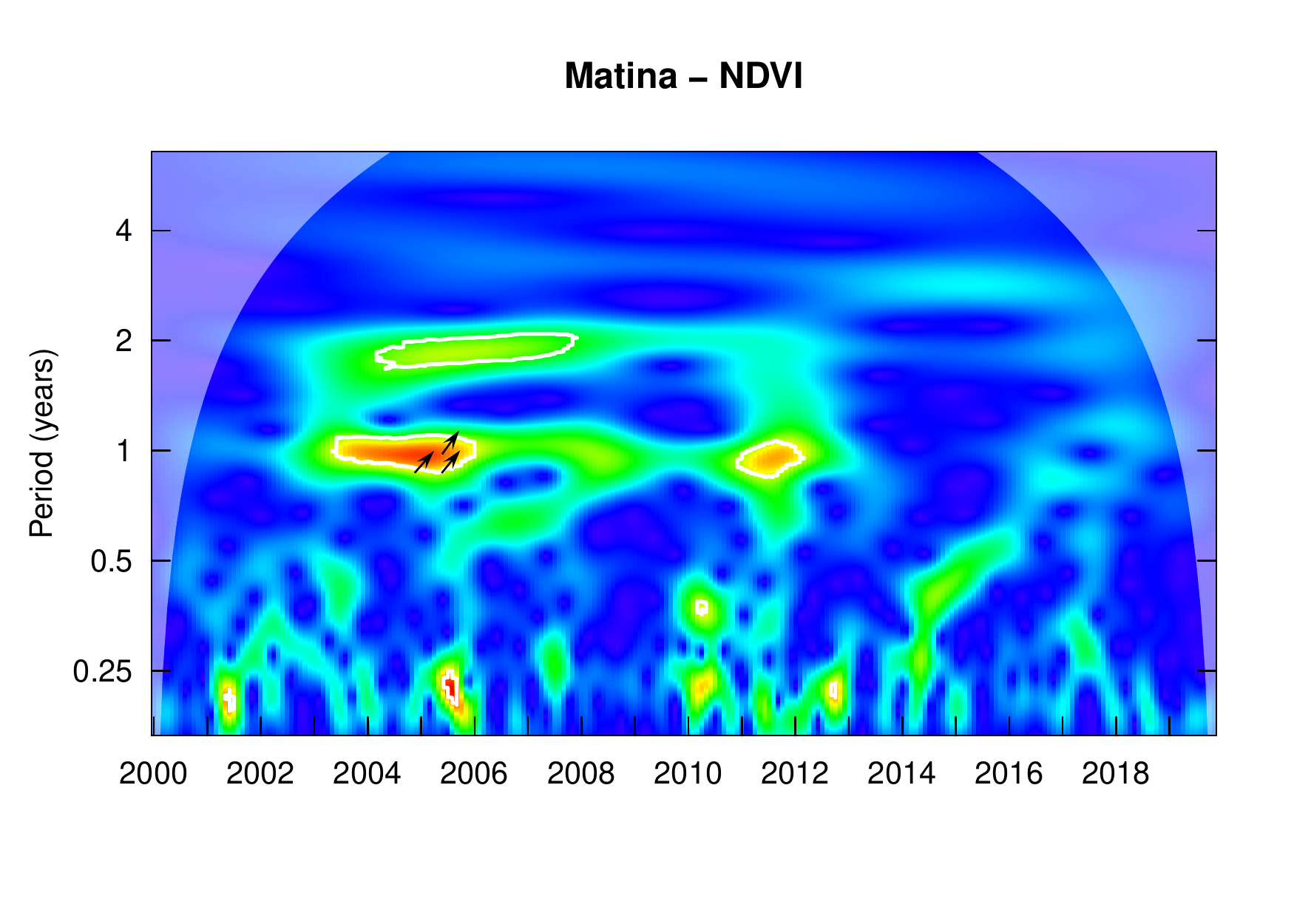}}\vspace{-0.15cm}%
\subfloat[]{\includegraphics[scale=0.23]{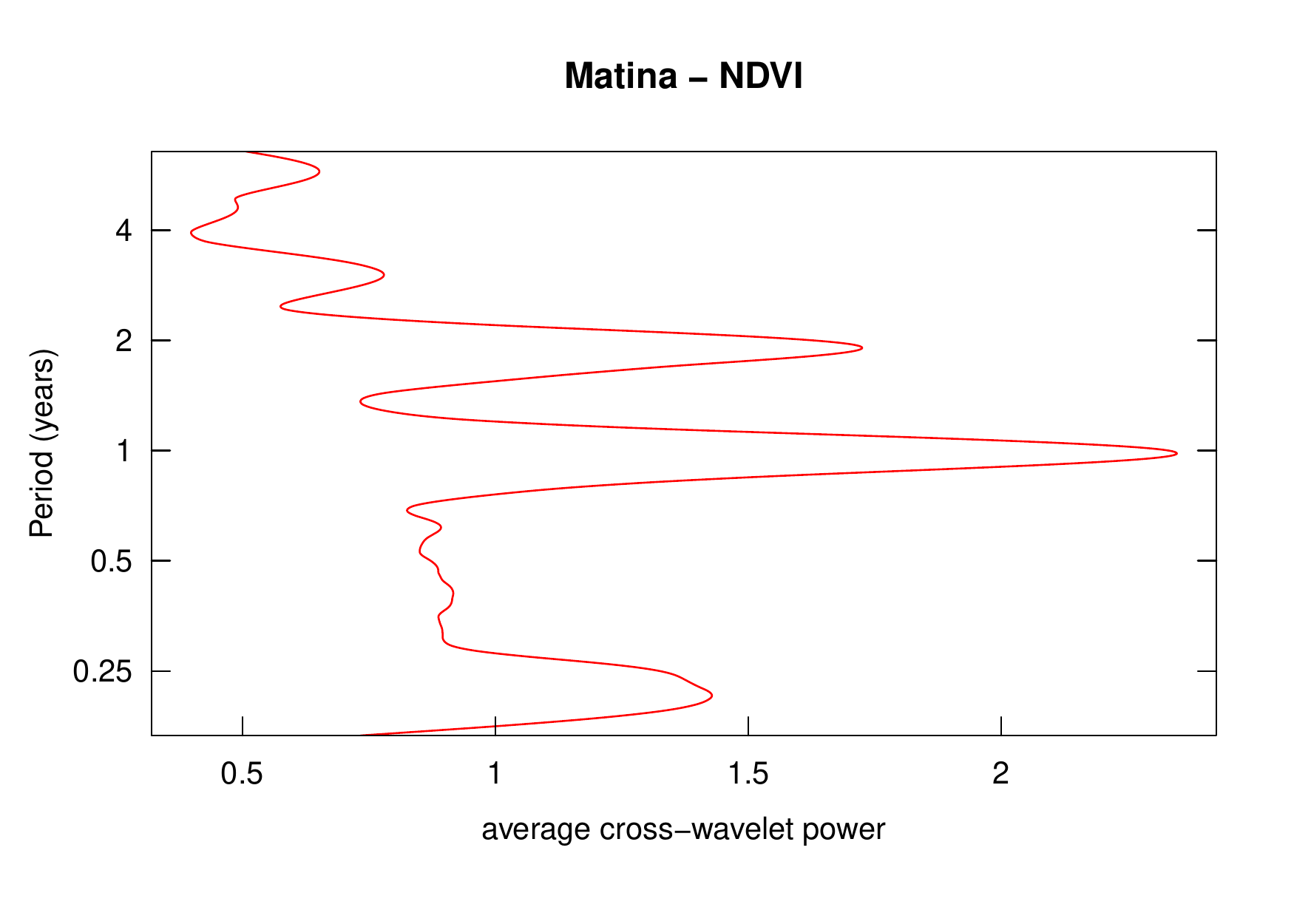}}\vspace{-0.15cm}%
\subfloat[]{\includegraphics[scale=0.23]{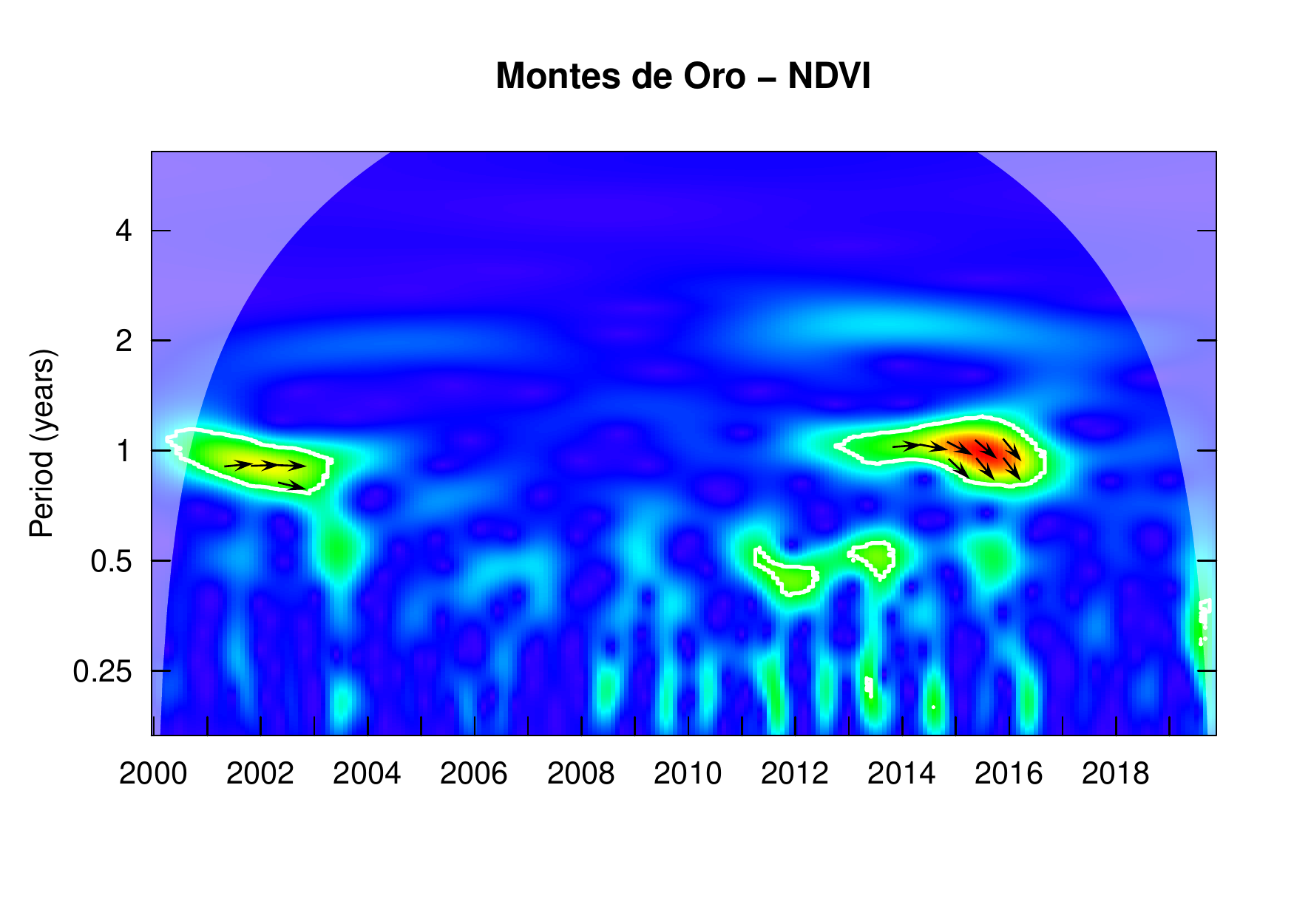}}\vspace{-0.15cm}%
\subfloat[]{\includegraphics[scale=0.23]{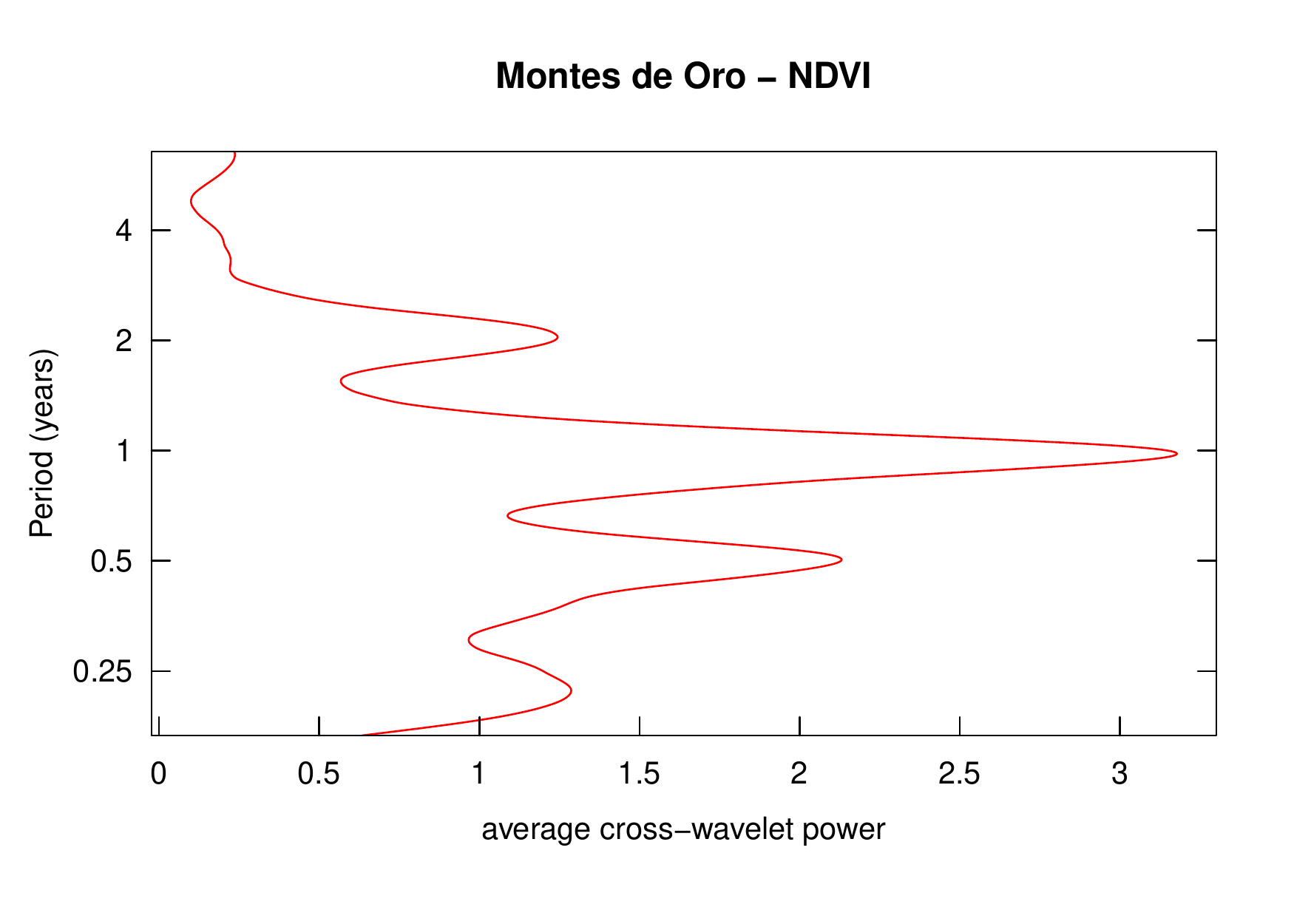}}\vspace{-0.15cm}\\
\subfloat[]{\includegraphics[scale=0.23]{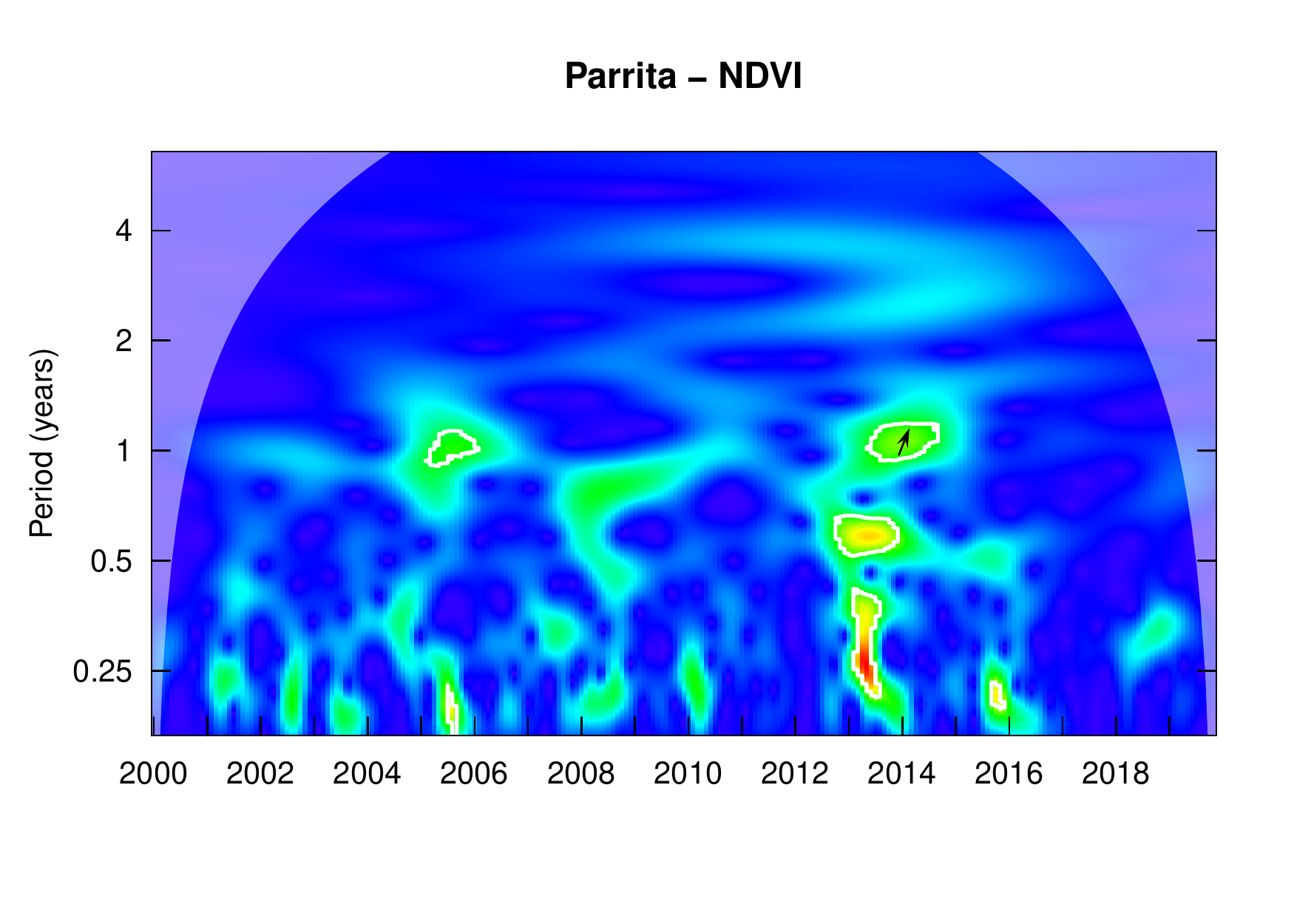}}\vspace{-0.15cm}%
\subfloat[]{\includegraphics[scale=0.23]{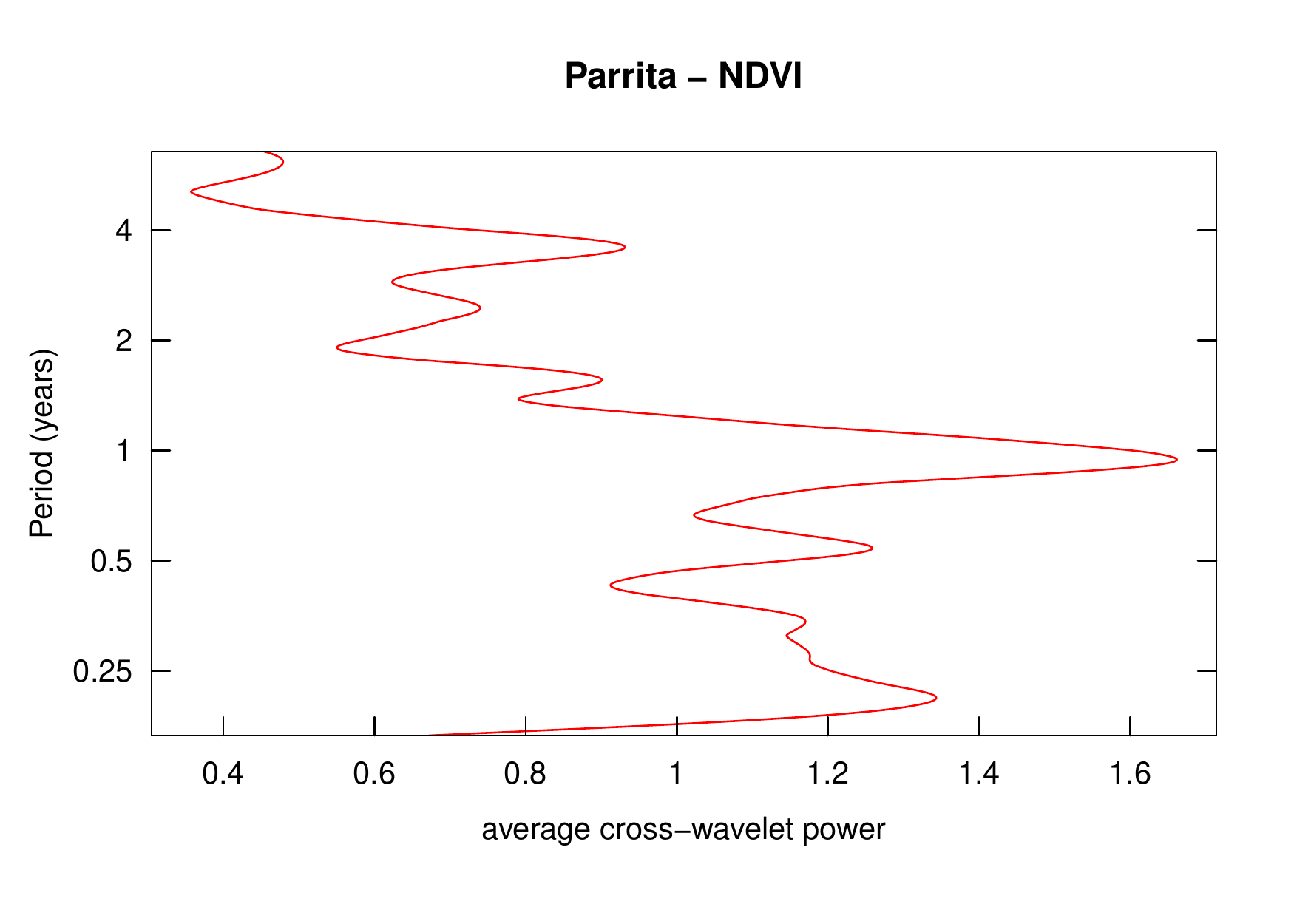}}\vspace{-0.15cm}%
\subfloat[]{\includegraphics[scale=0.23]{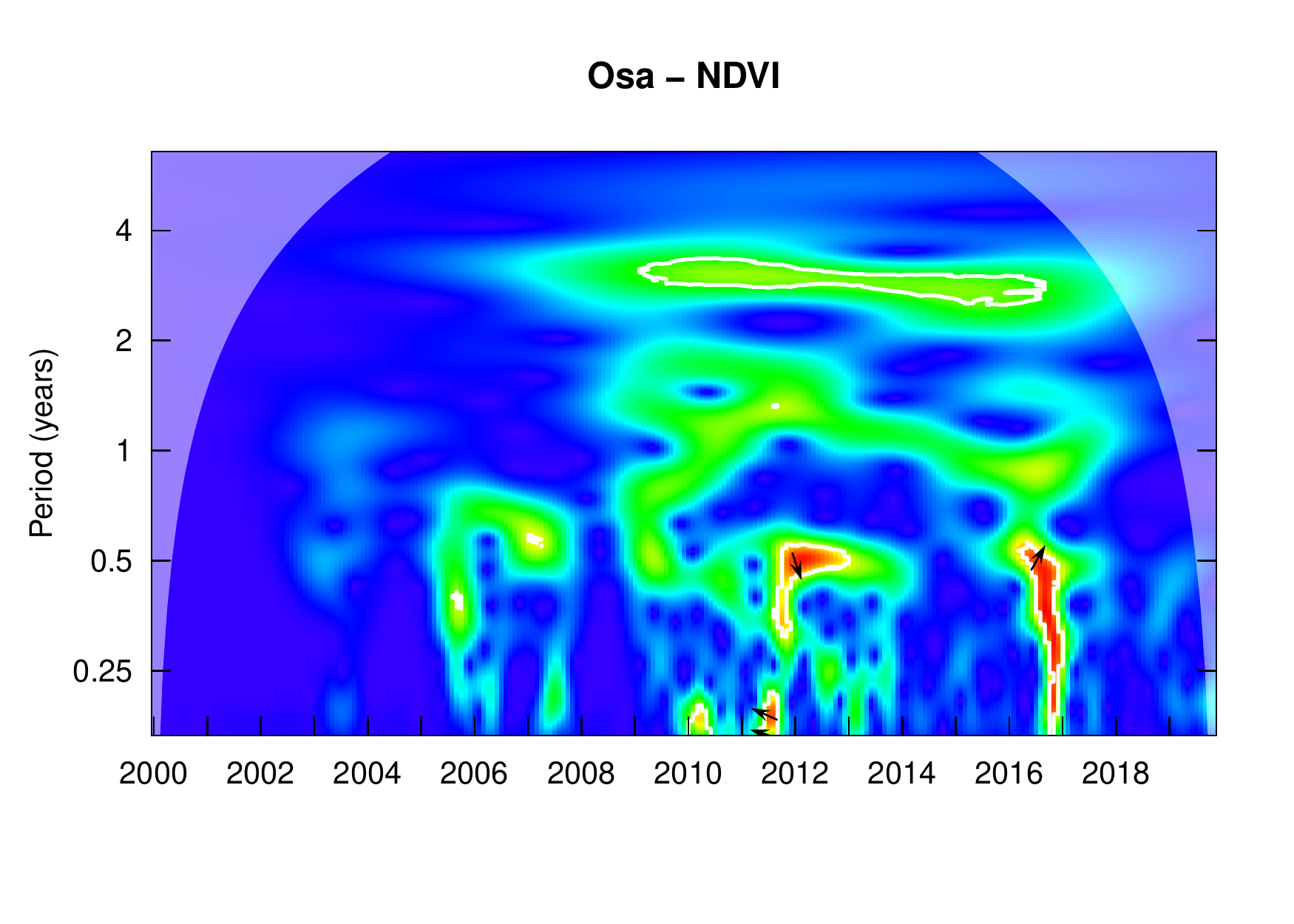}}\vspace{-0.15cm}%
\subfloat[]{\includegraphics[scale=0.23]{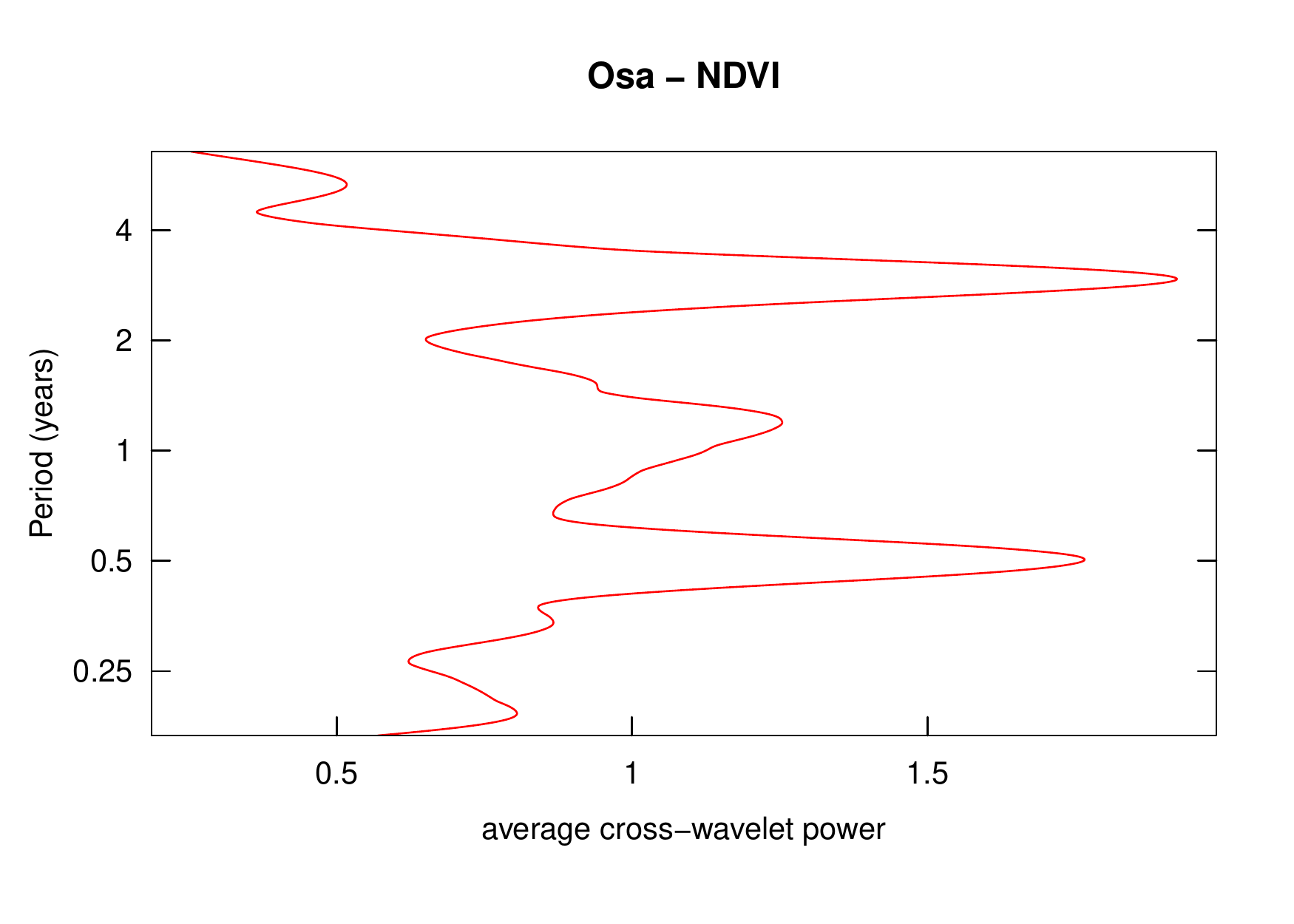}}\vspace{-0.15cm}\\
\subfloat[]{\includegraphics[scale=0.23]{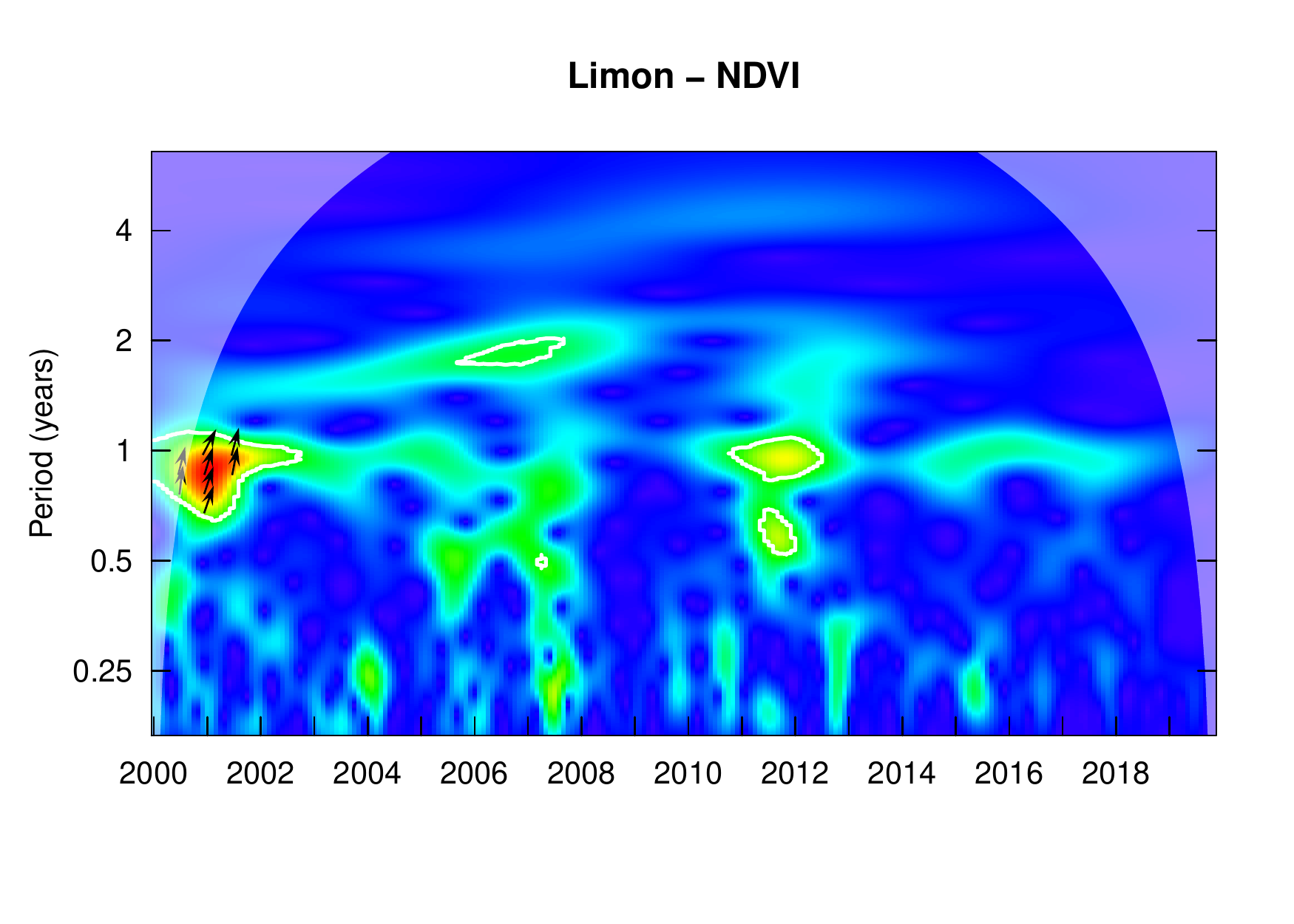}}\vspace{-0.15cm}%
\subfloat[]{\includegraphics[scale=0.23]{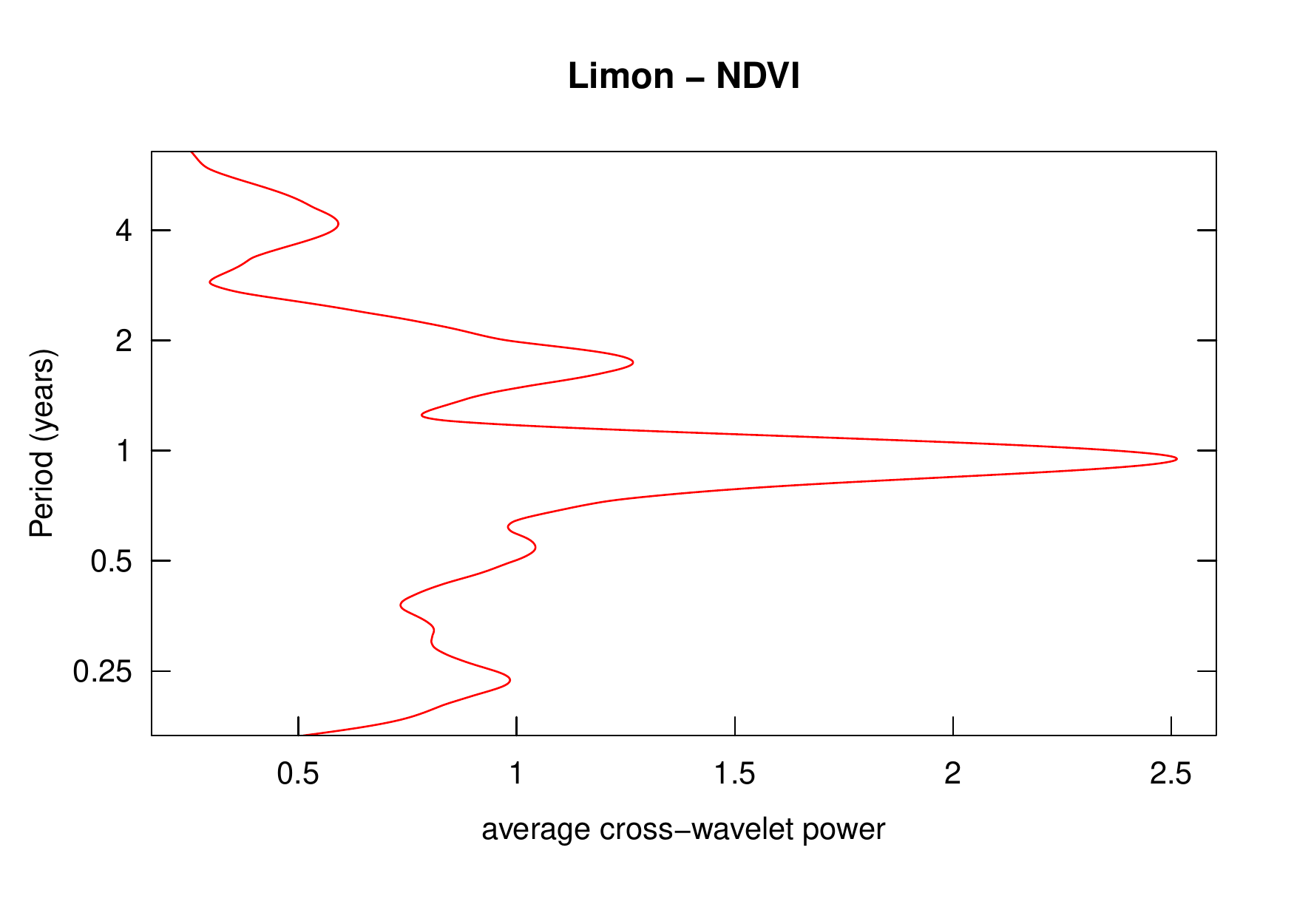}}\vspace{-0.15cm}
\subfloat[]{\includegraphics[scale=0.23]{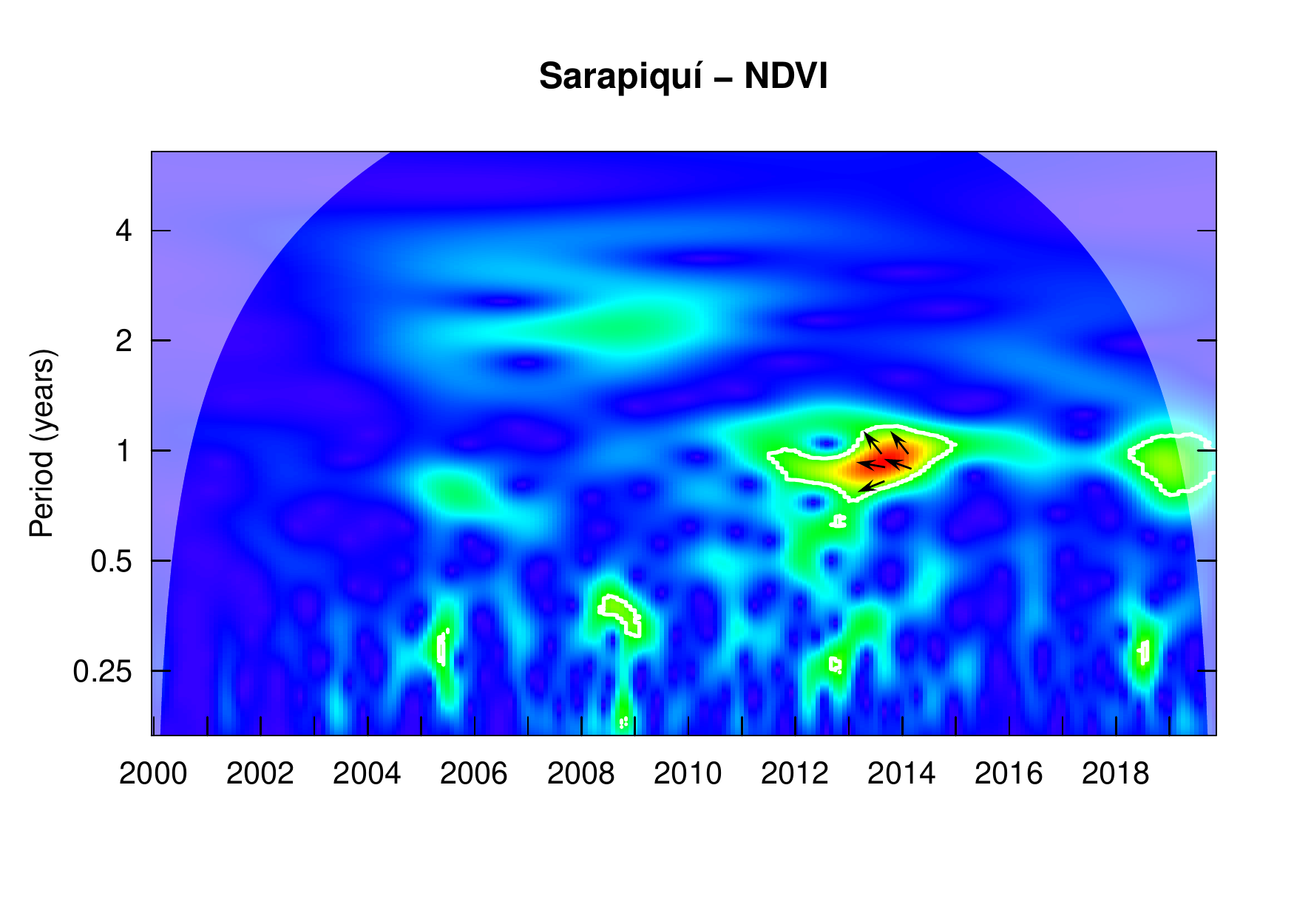}}\vspace{-0.15cm}%
\subfloat[]{\includegraphics[scale=0.23]{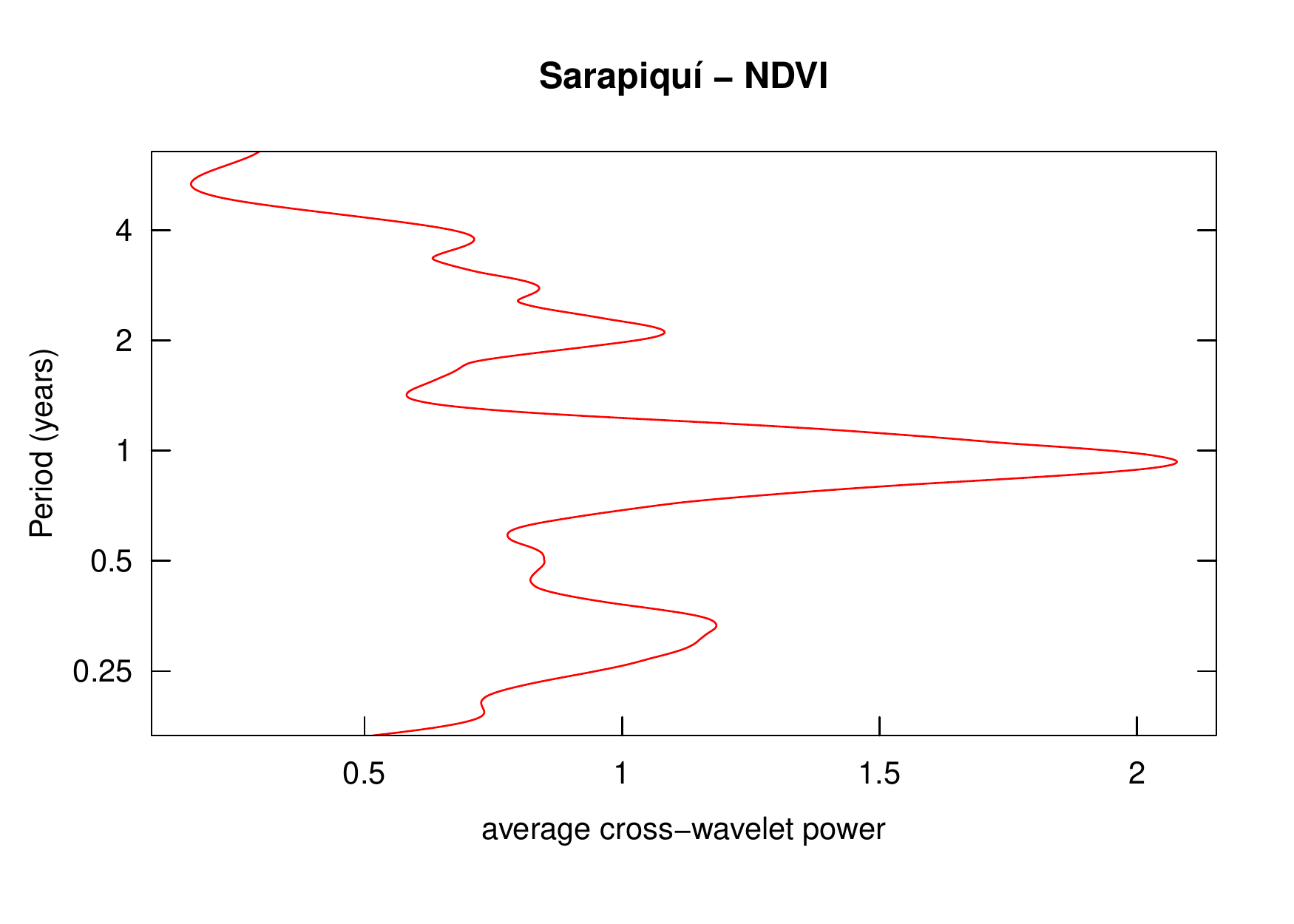}}\vspace{-0.15cm}
\caption*{}
\end{figure}

\section*{Wavelet coherence and average cross-wavelet power between dengue incidence and NDWI}

\begin{figure}[H]
\captionsetup[subfigure]{labelformat=empty}
\caption*{\textbf{Figure S3:} Wavelet coherence (color map) between dengue incidence from 2000 to 2019, and NDWI in 32 municipalities of Costa Rica (periodicity on y-axis, time on x-axis). Colors code for increasing power intensity, from blue to red; $95\%$ confidence levels are encircled by white lines, and shaded areas indicate the presence of significant edge effects. On the right side of each wavelet coherence is the average cross-wavelet power (Red line). The arrows indicate whether the two series are in-phase or out-phase.}
\subfloat[]{\includegraphics[scale=0.23]{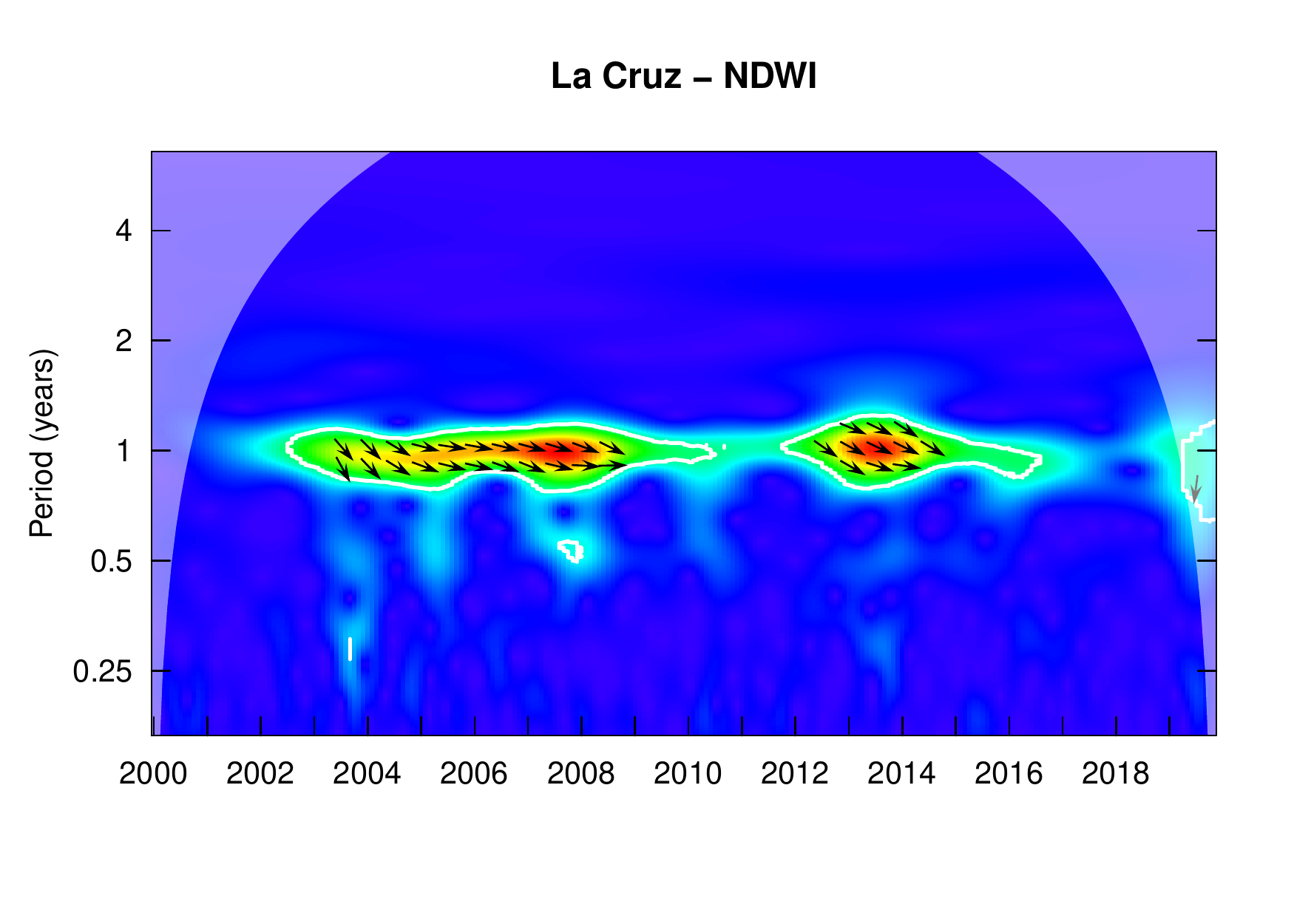}}\vspace{-0.15cm}%
\subfloat[]{\includegraphics[scale=0.23]{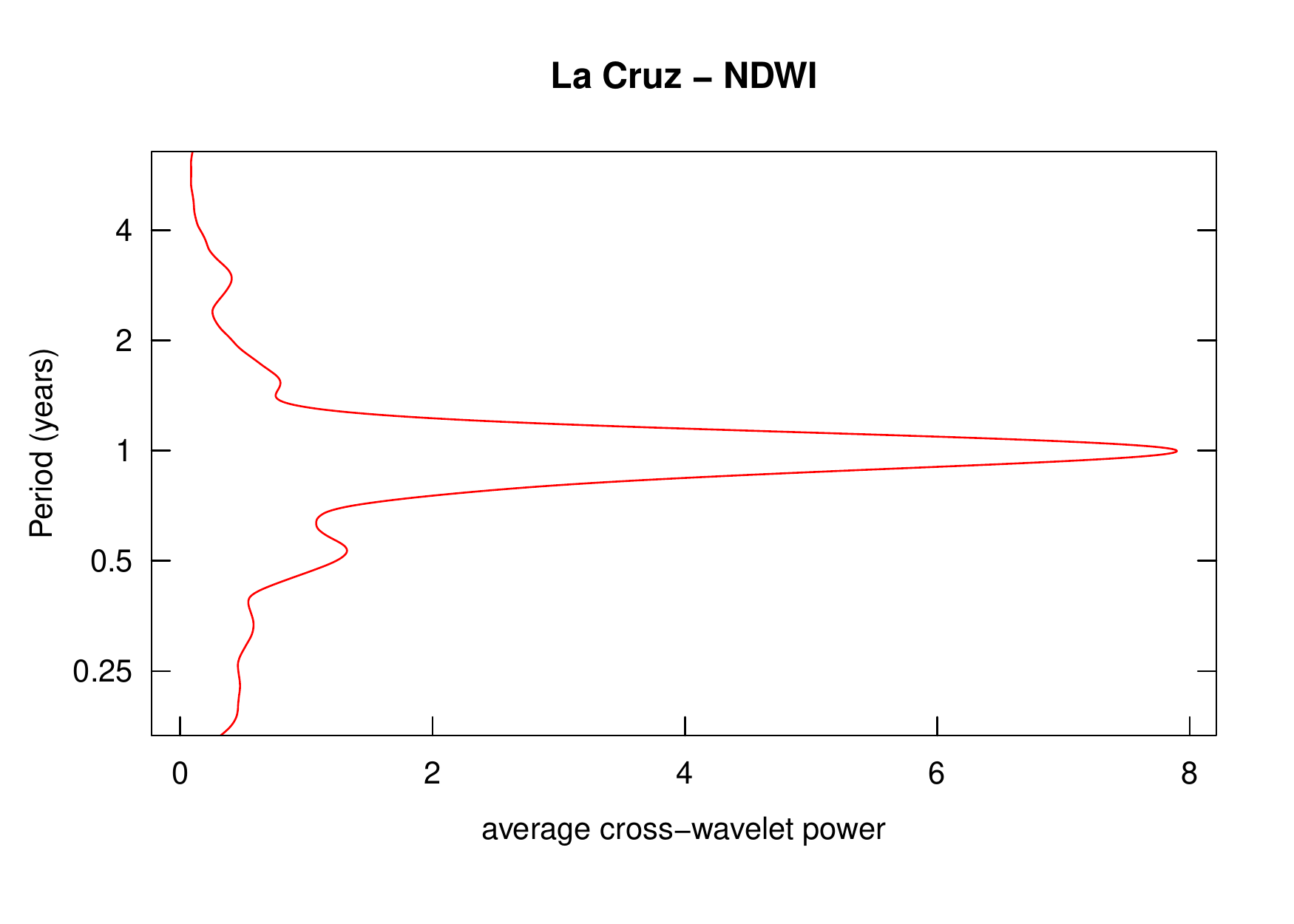}}\vspace{-0.15cm}%
\subfloat[]{\includegraphics[scale=0.23]{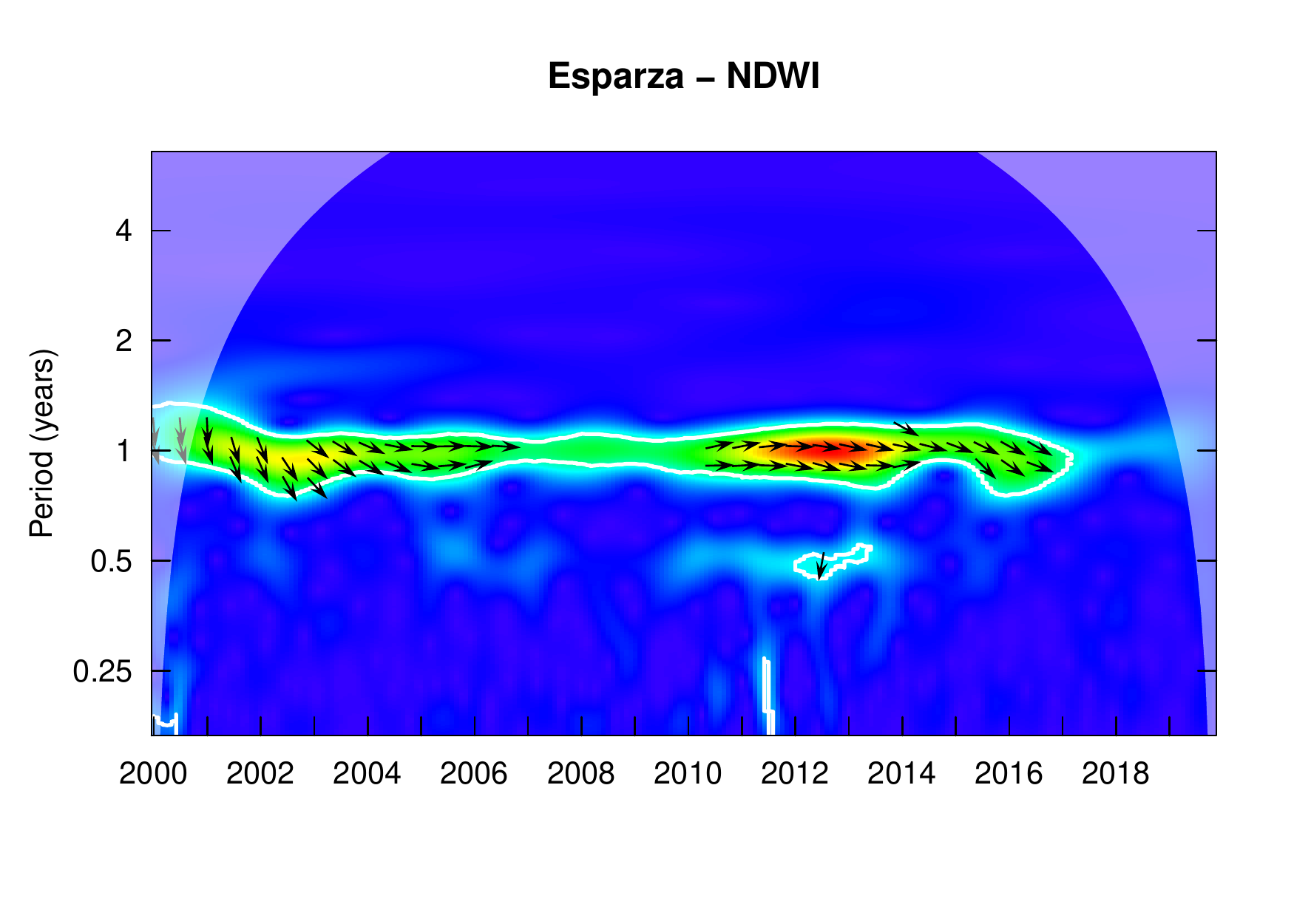}}\vspace{-0.15cm}%
\subfloat[]{\includegraphics[scale=0.23]{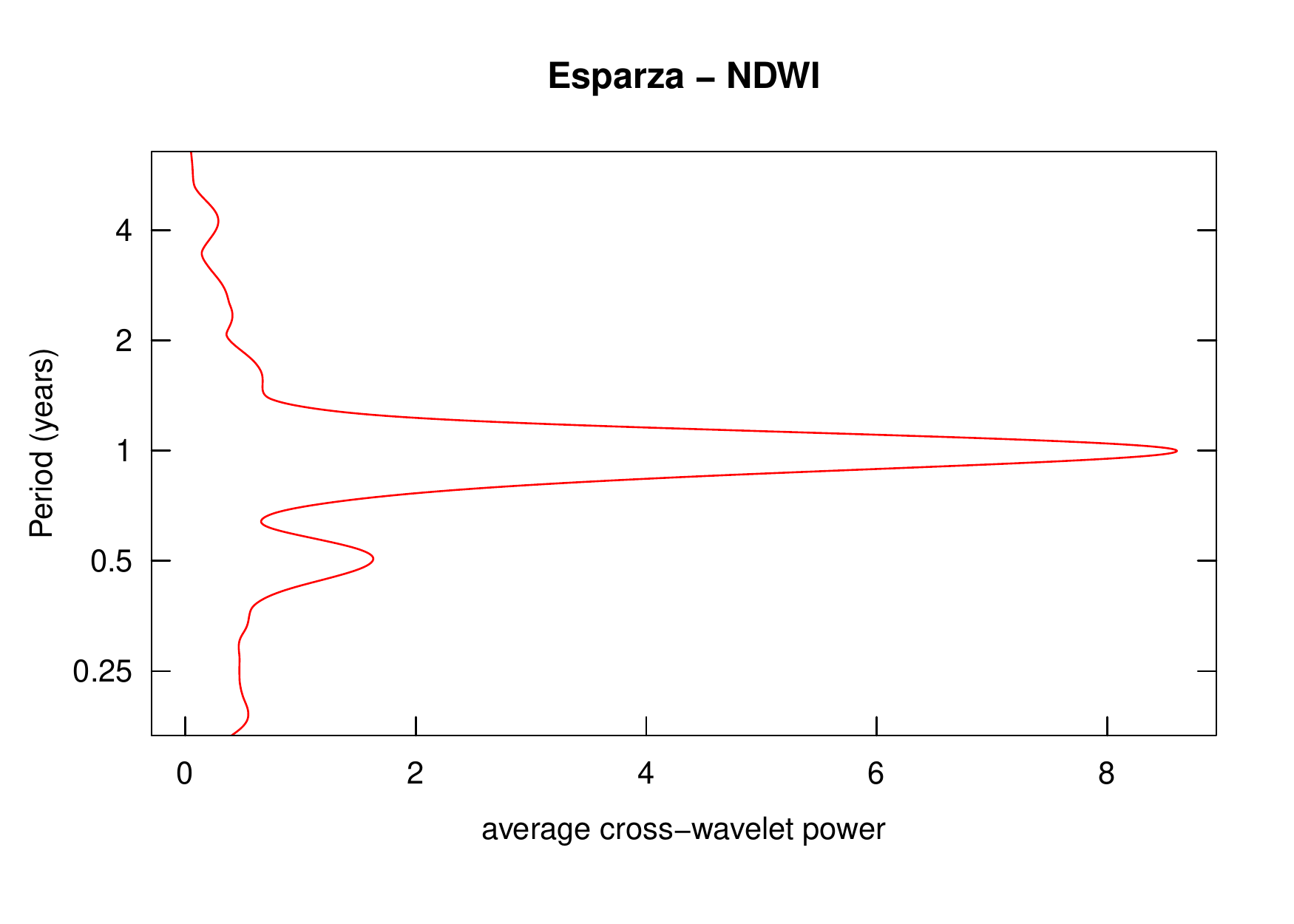}}\vspace{-0.15cm}\\
\subfloat[]{\includegraphics[scale=0.23]{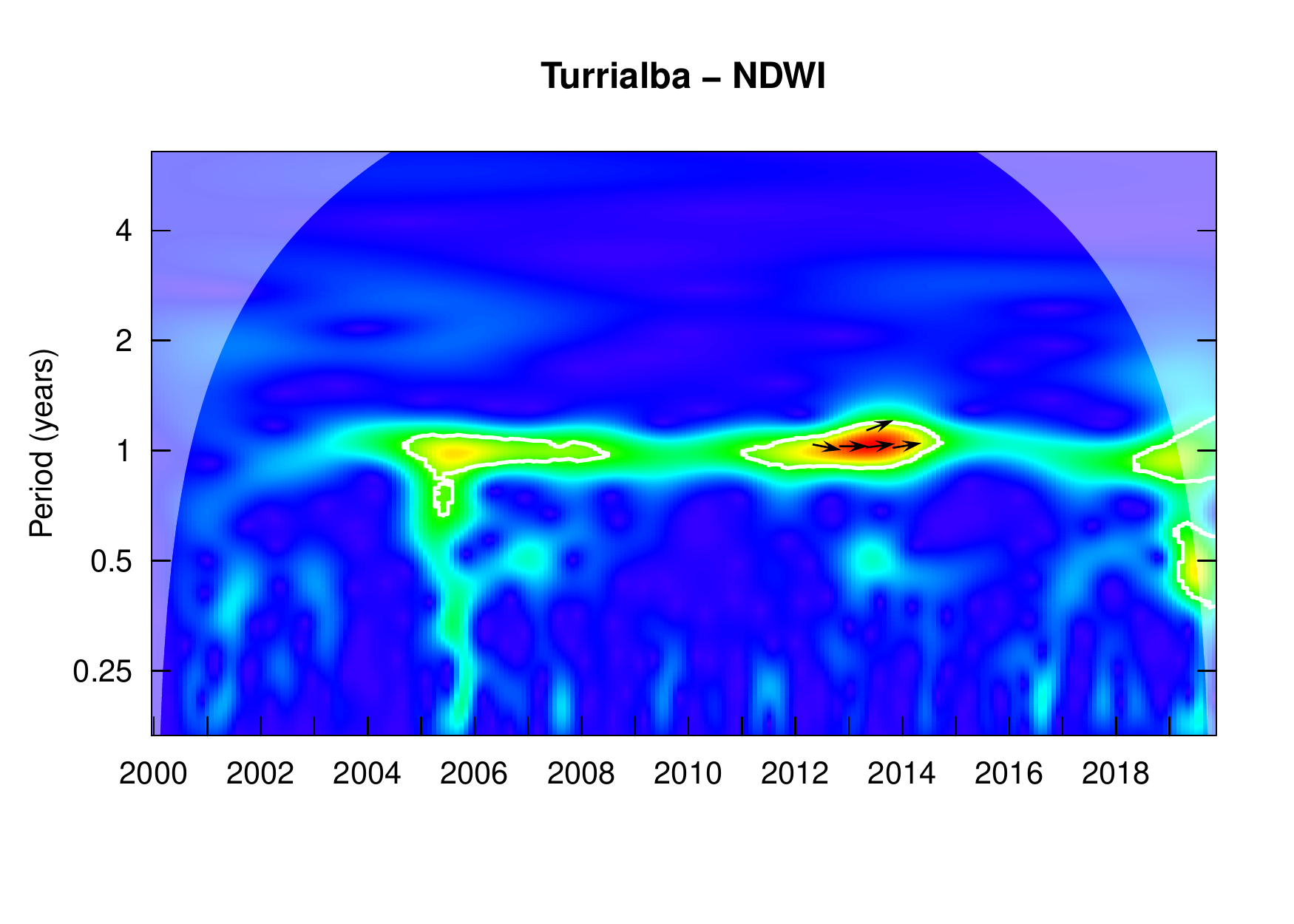}}\vspace{-0.15cm}%
\subfloat[]{\includegraphics[scale=0.23]{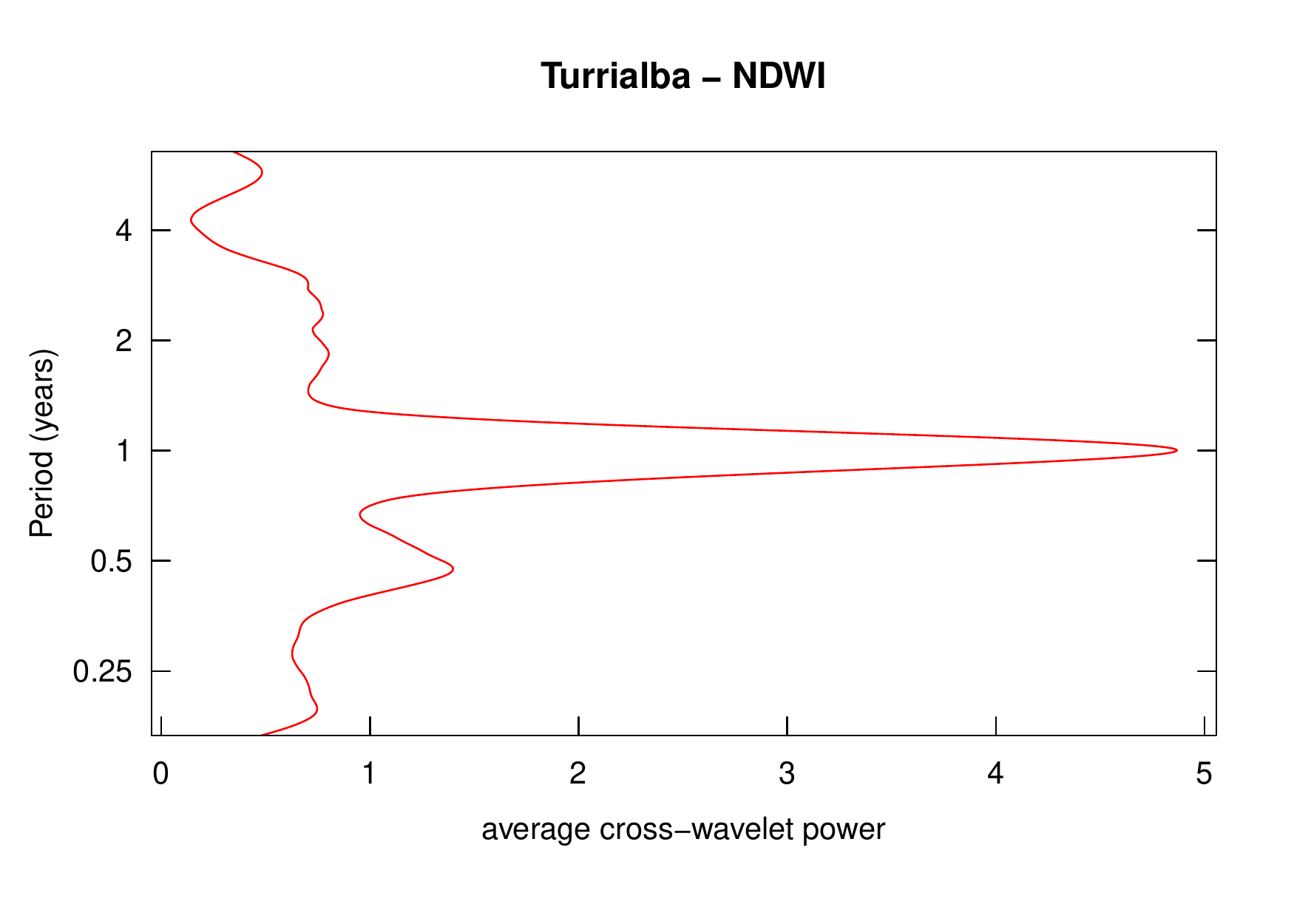}}\vspace{-0.15cm}%
\subfloat[]{\includegraphics[scale=0.23]{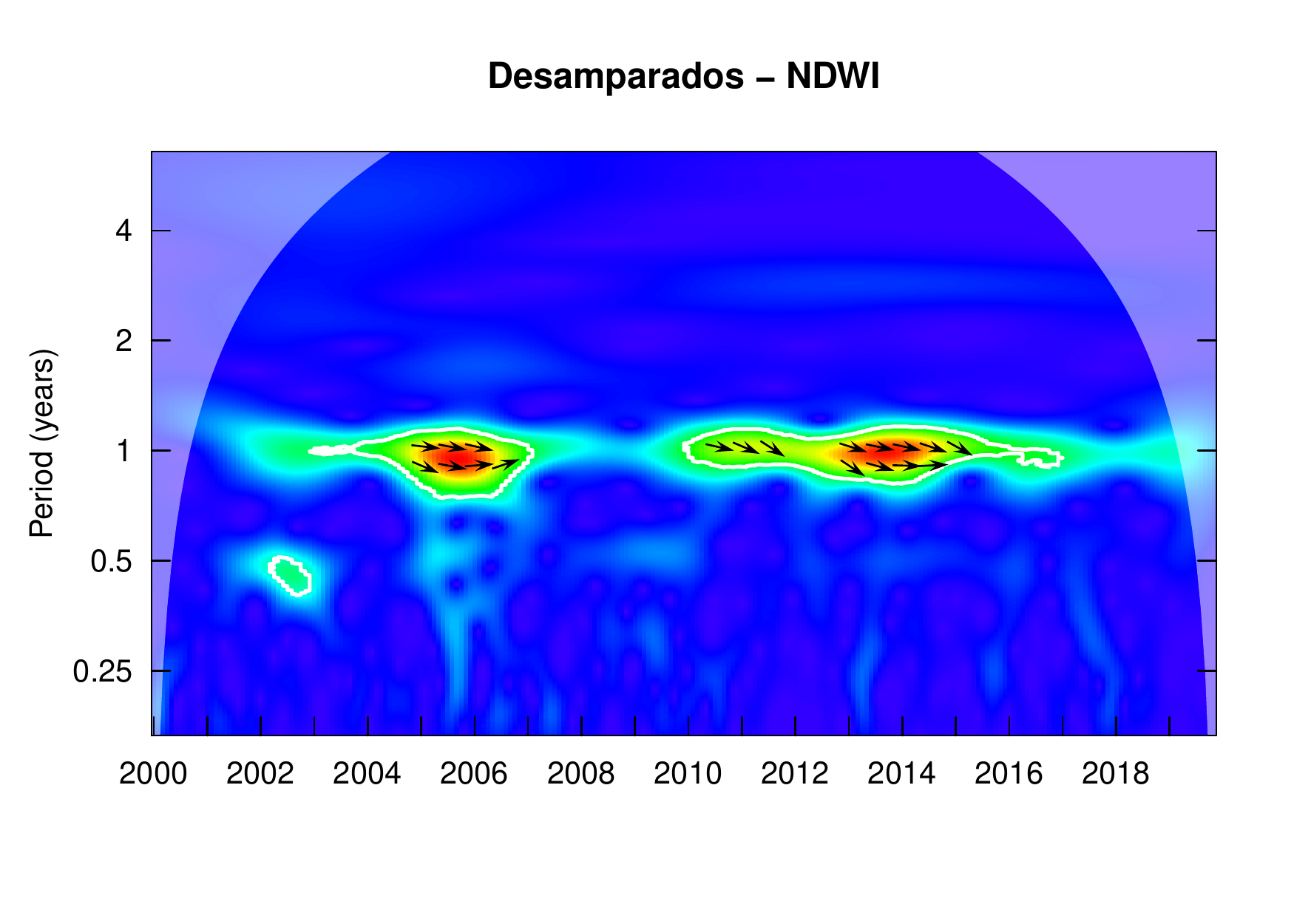}}\vspace{-0.15cm}%
\subfloat[]{\includegraphics[scale=0.23]{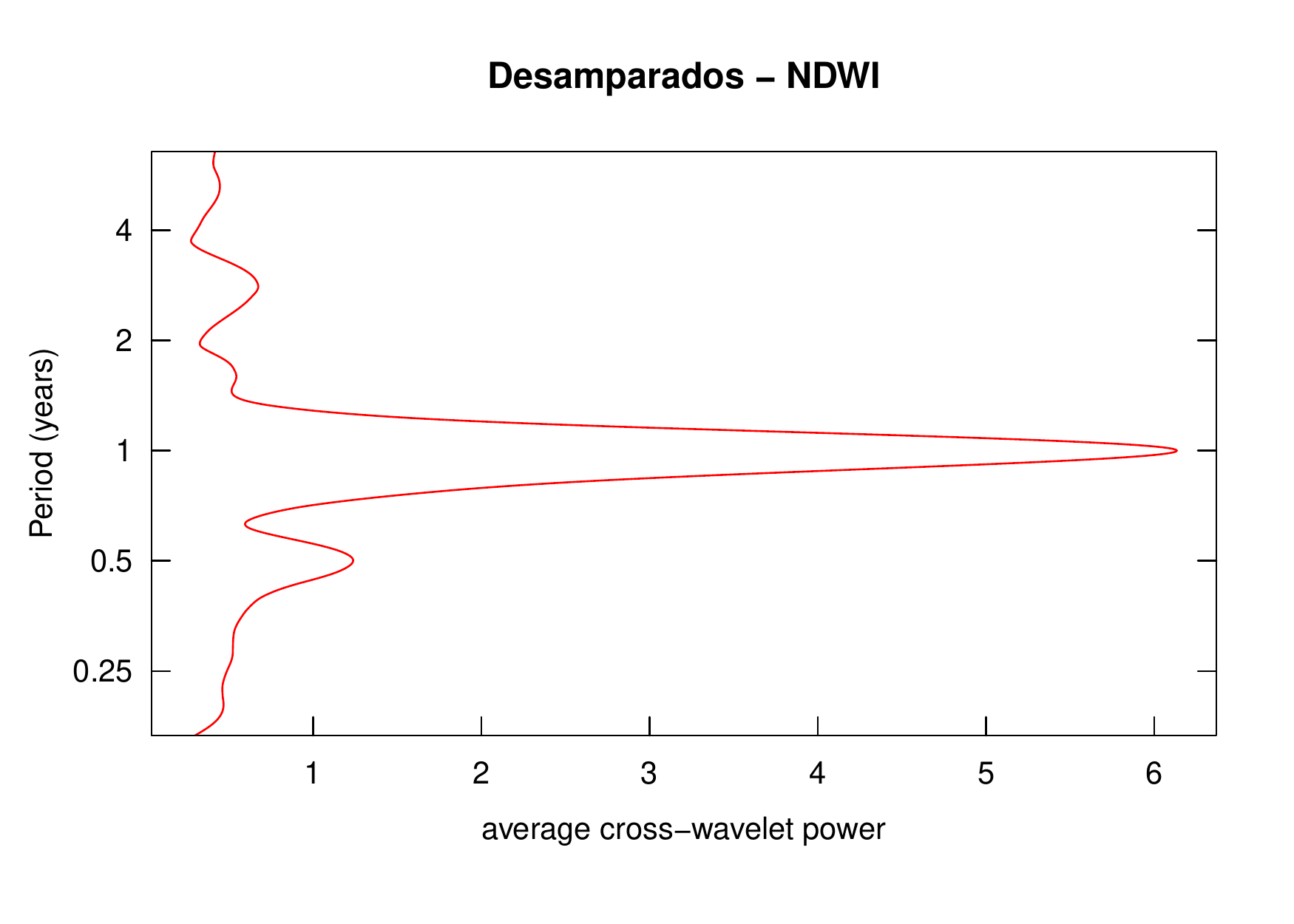}}\vspace{-0.15cm}\\
\subfloat[]{\includegraphics[scale=0.23]{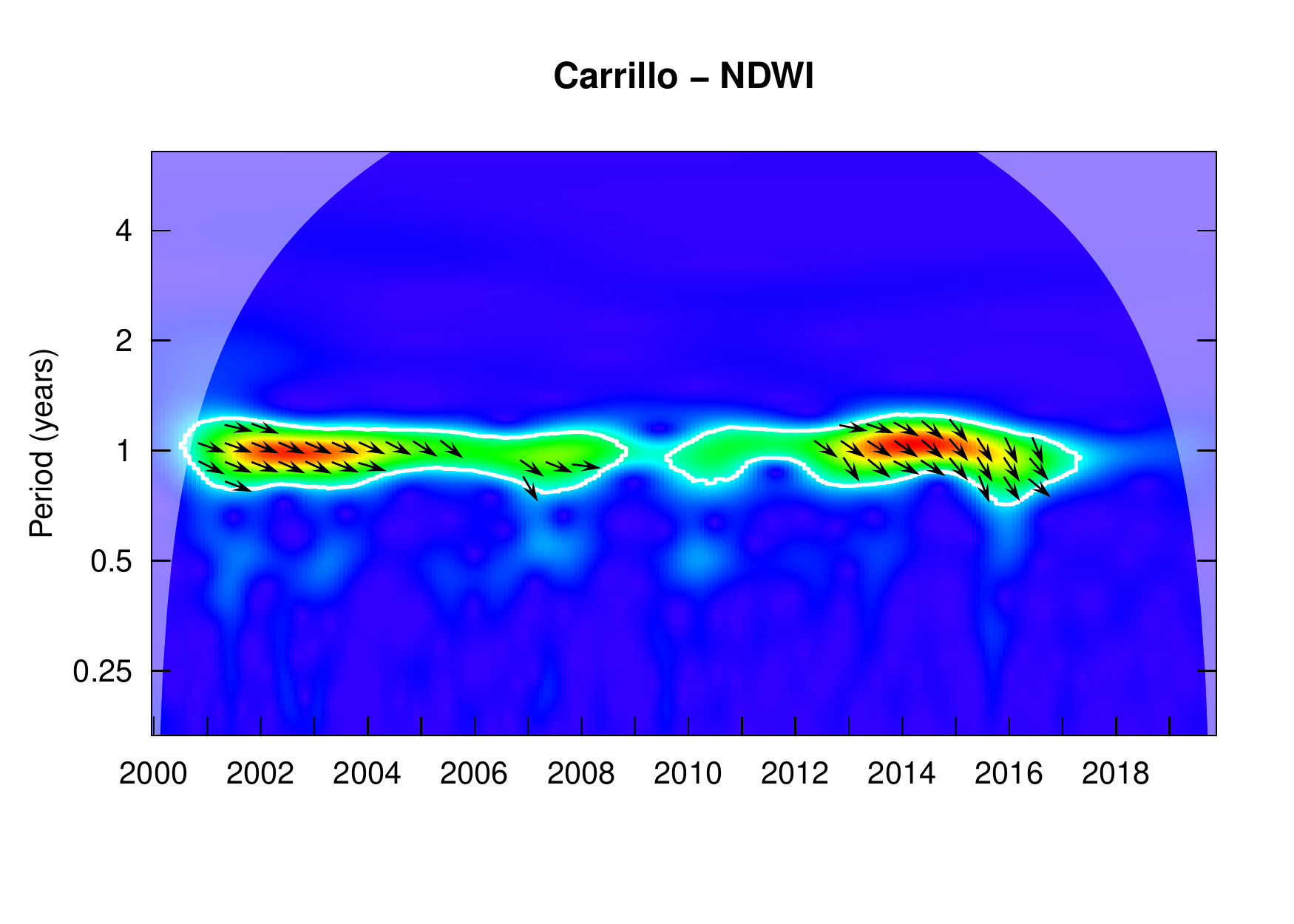}}\vspace{-0.15cm}%
\subfloat[]{\includegraphics[scale=0.23]{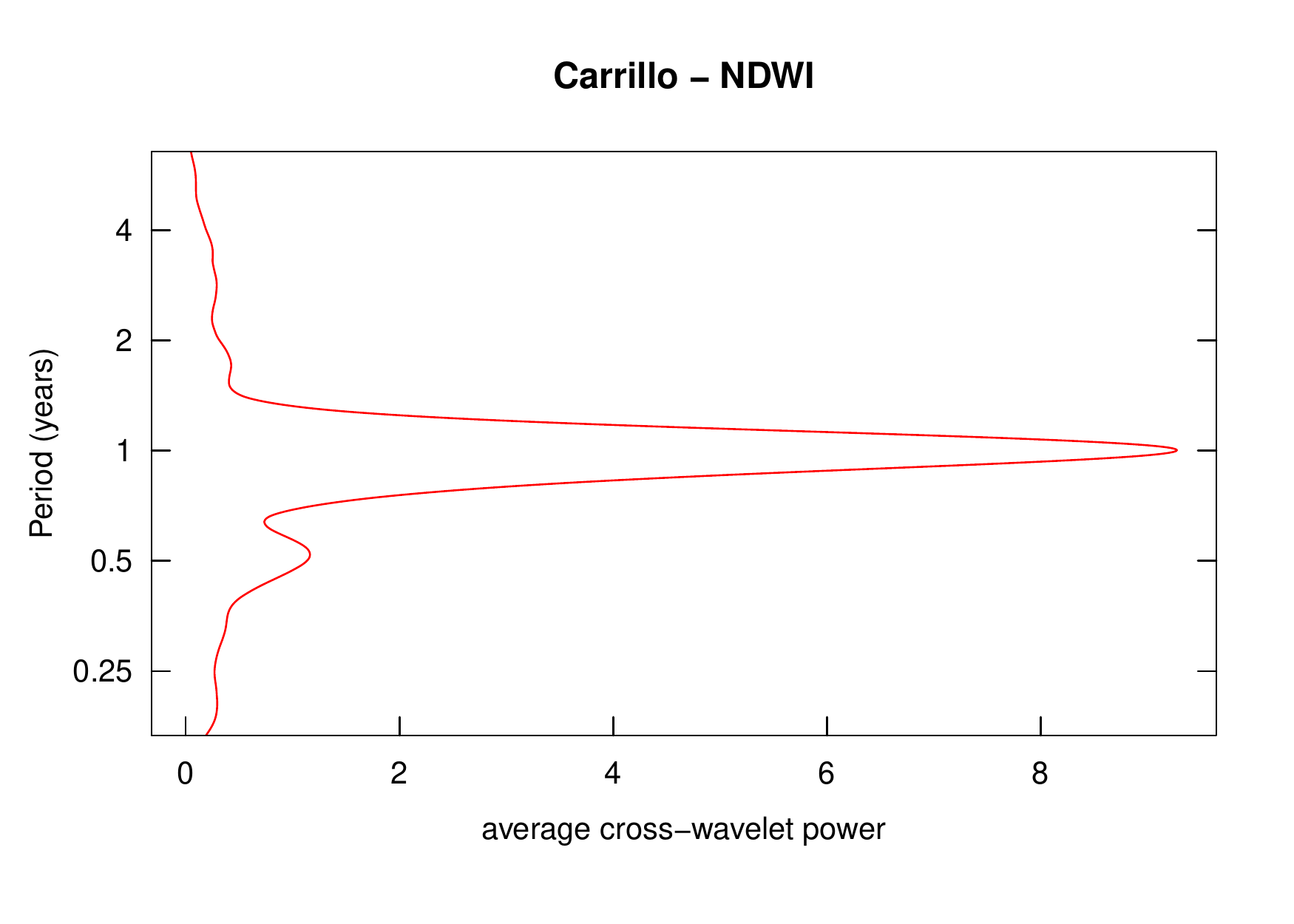}}\vspace{-0.15cm}%
\subfloat[]{\includegraphics[scale=0.23]{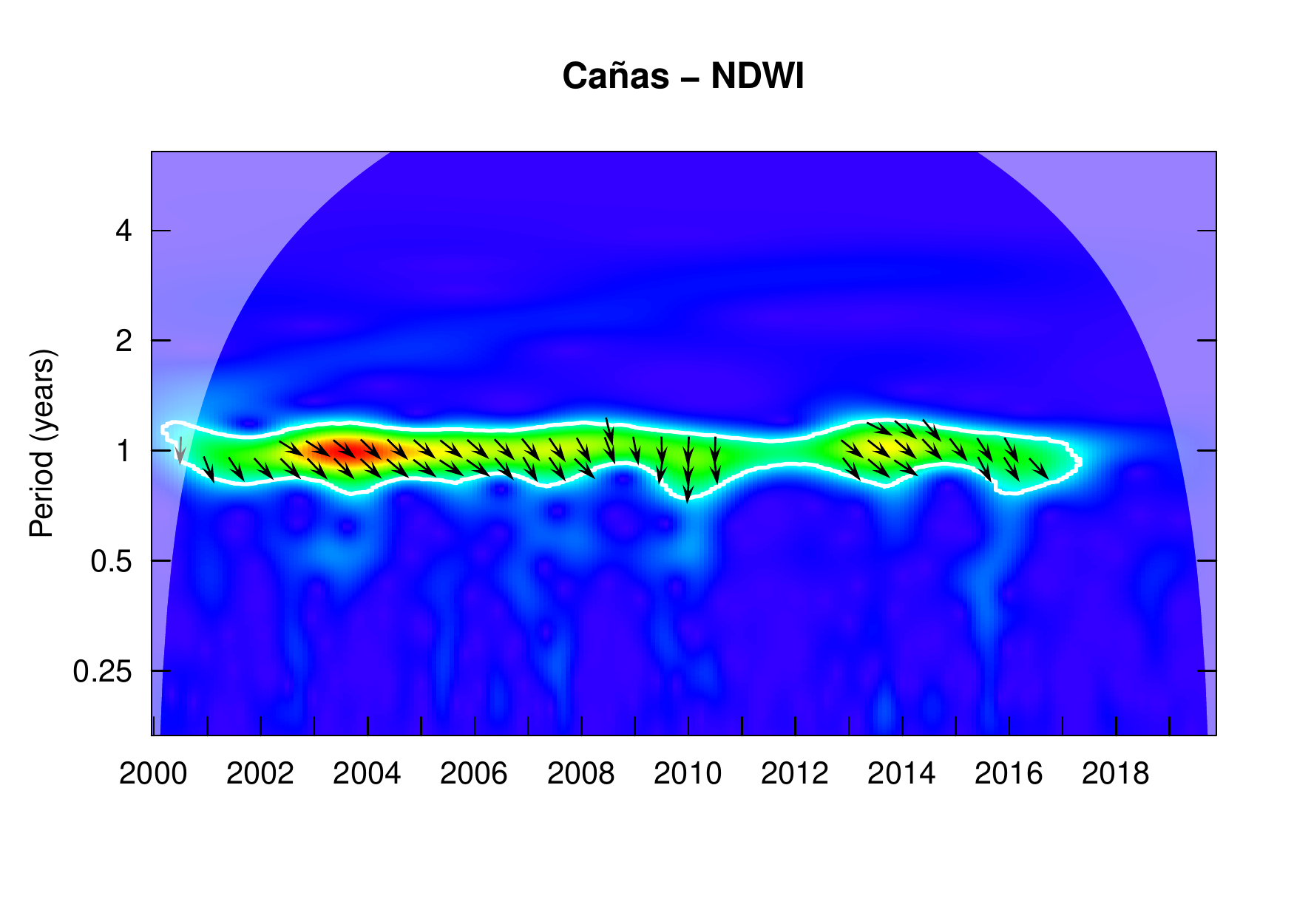}}\vspace{-0.15cm}%
\subfloat[]{\includegraphics[scale=0.23]{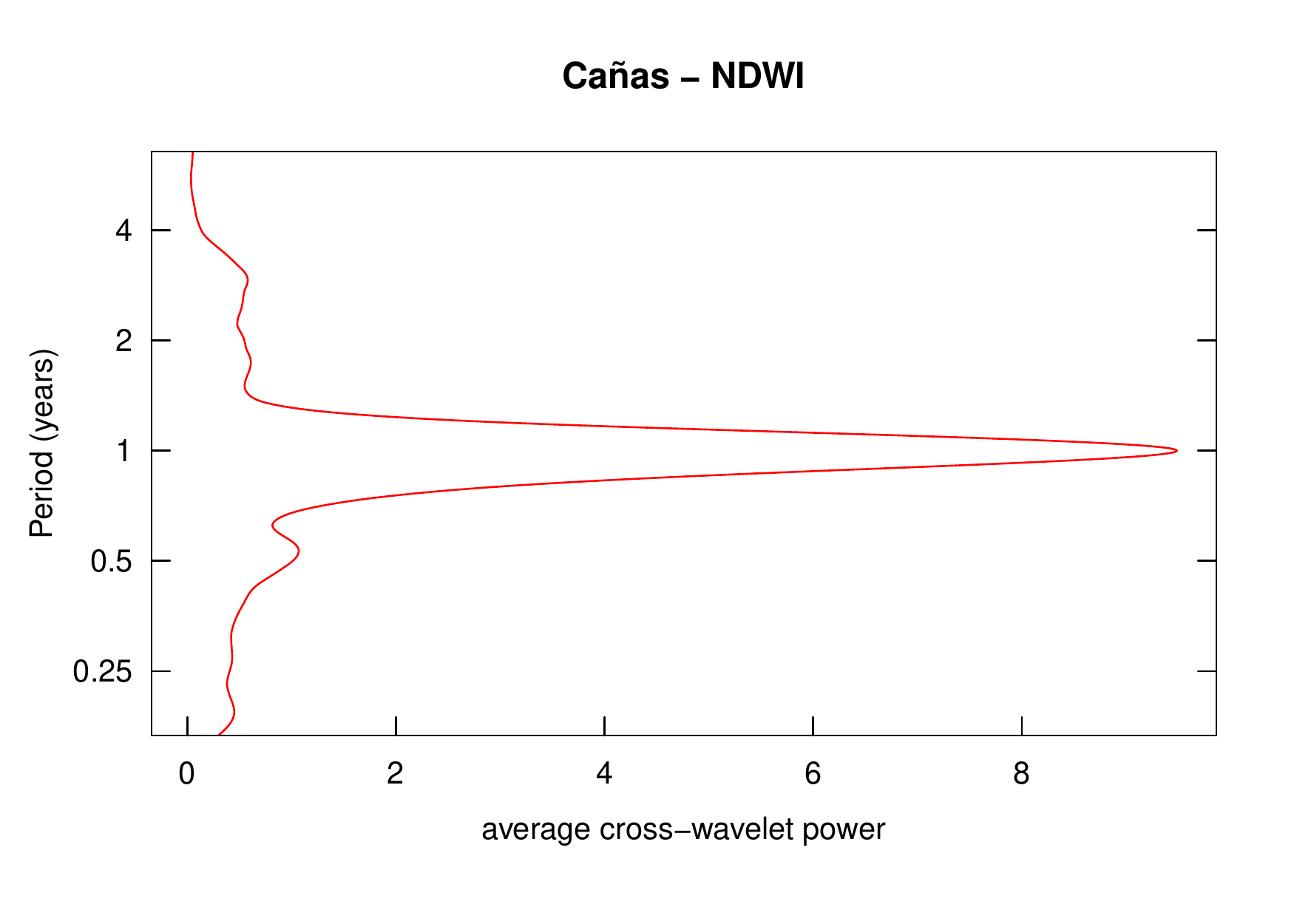}}\vspace{-0.15cm}\\
\subfloat[]{\includegraphics[scale=0.23]{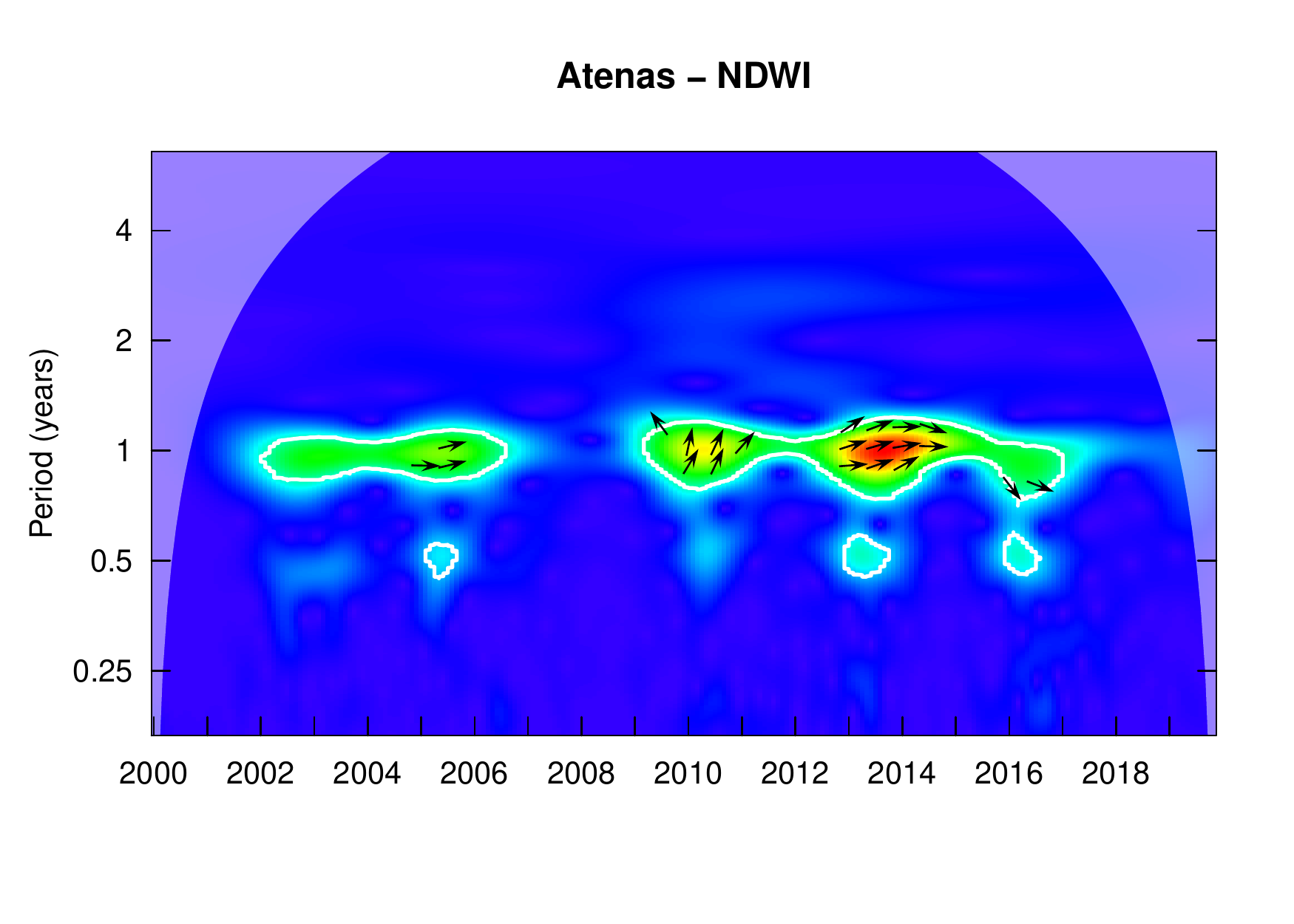}}\vspace{-0.15cm}%
\subfloat[]{\includegraphics[scale=0.23]{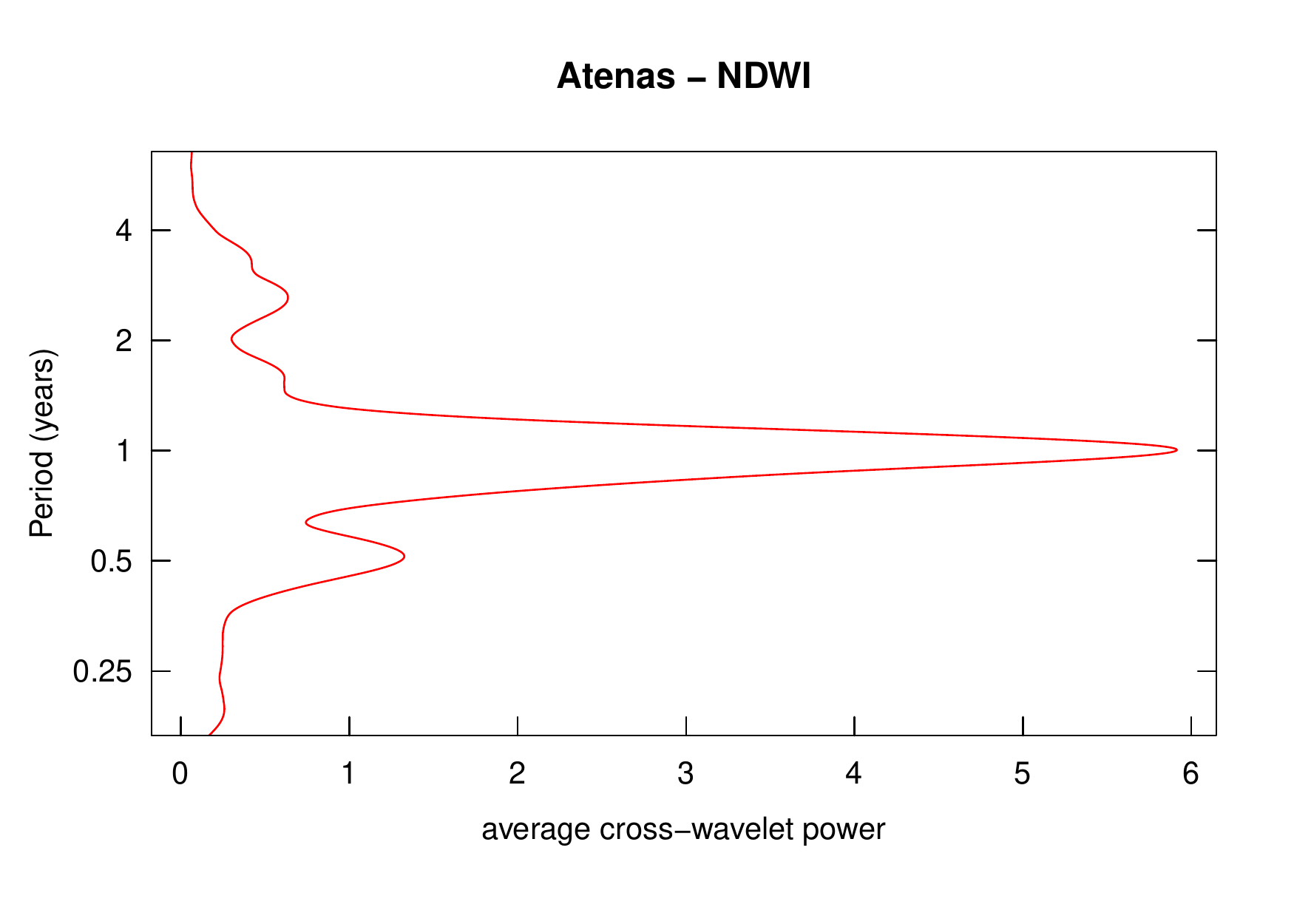}}\vspace{-0.15cm}%
\subfloat[]{\includegraphics[scale=0.23]{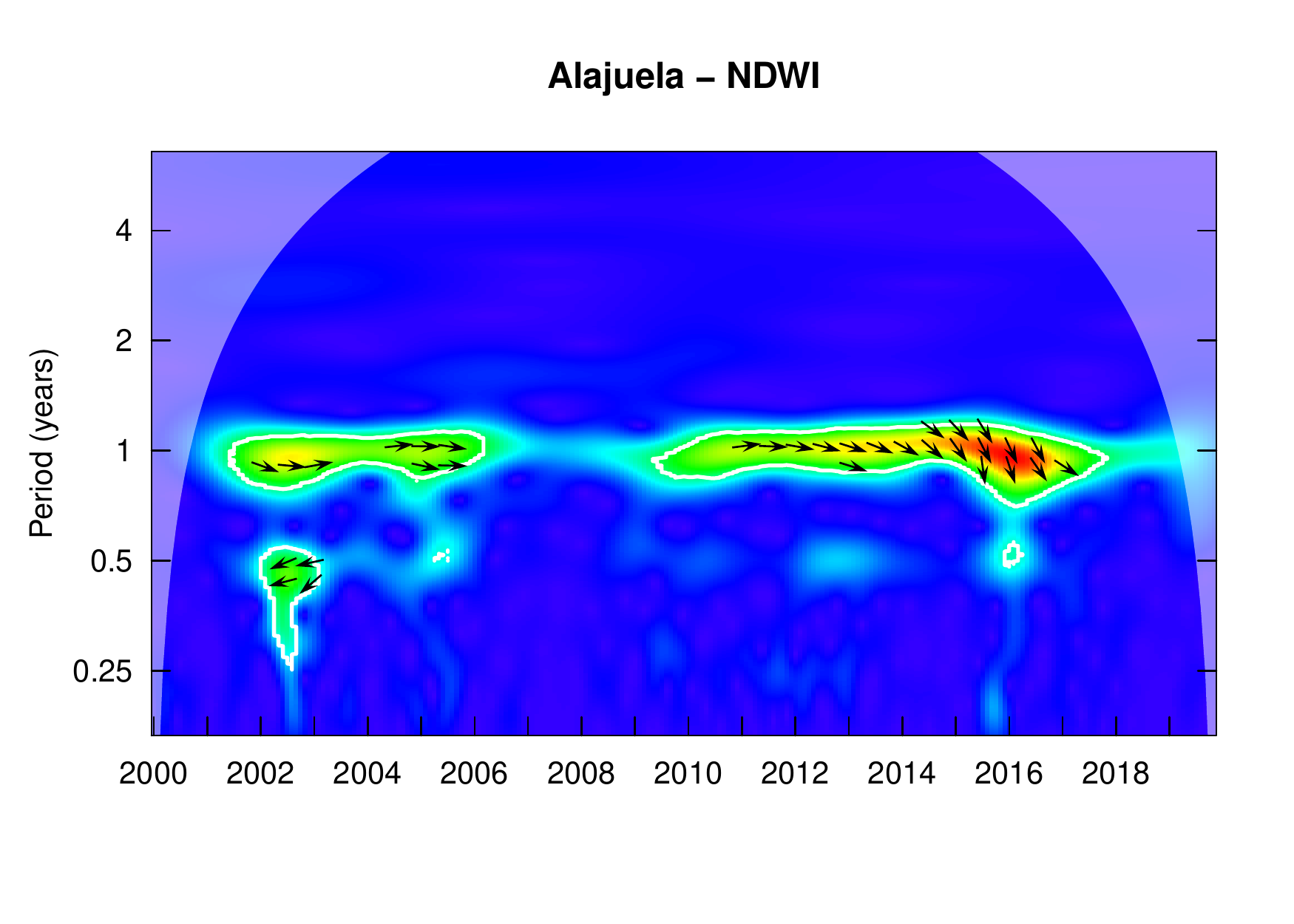}}\vspace{-0.15cm}%
\subfloat[]{\includegraphics[scale=0.23]{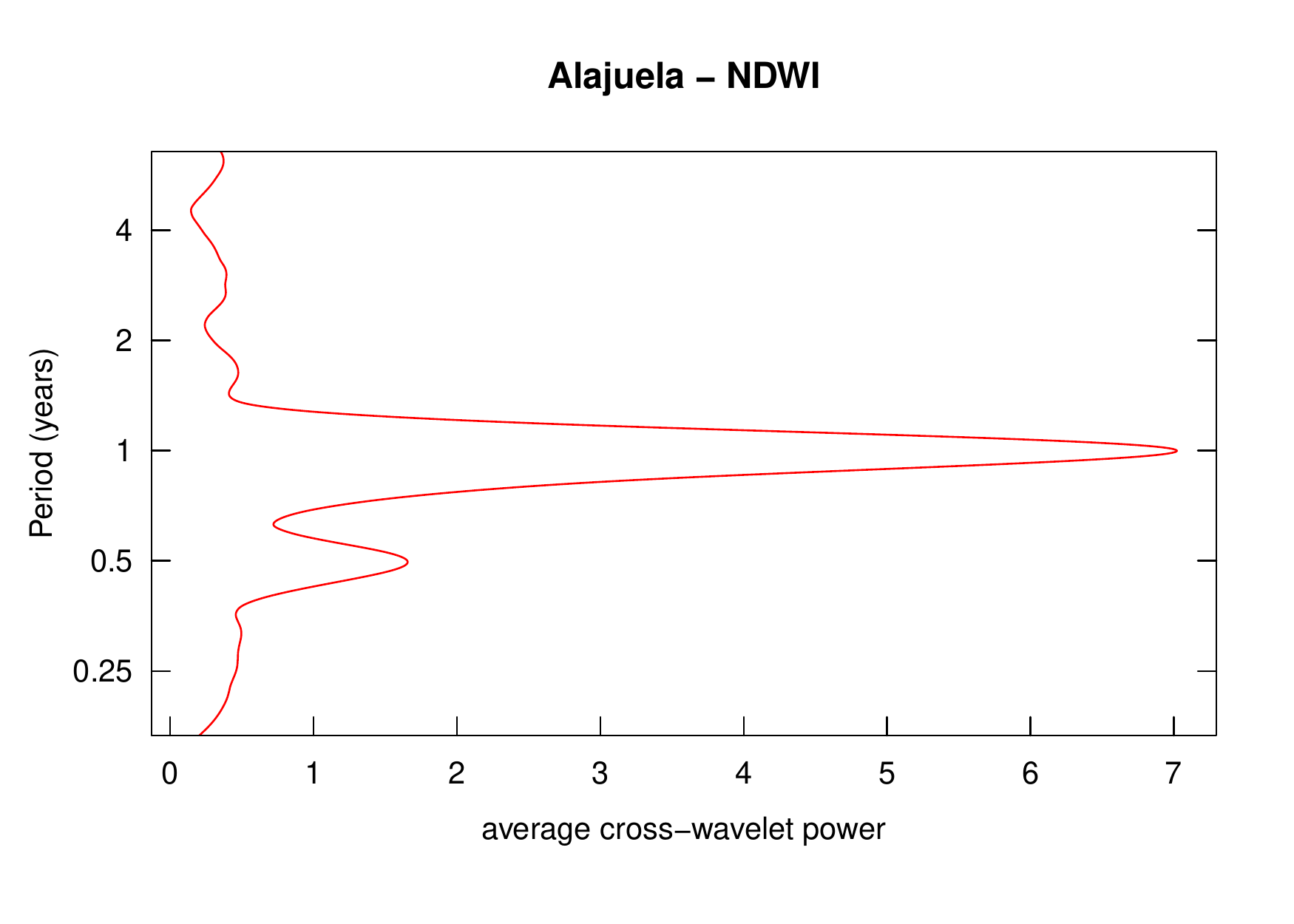}}\vspace{-0.15cm}\\
\subfloat[]{\includegraphics[scale=0.23]{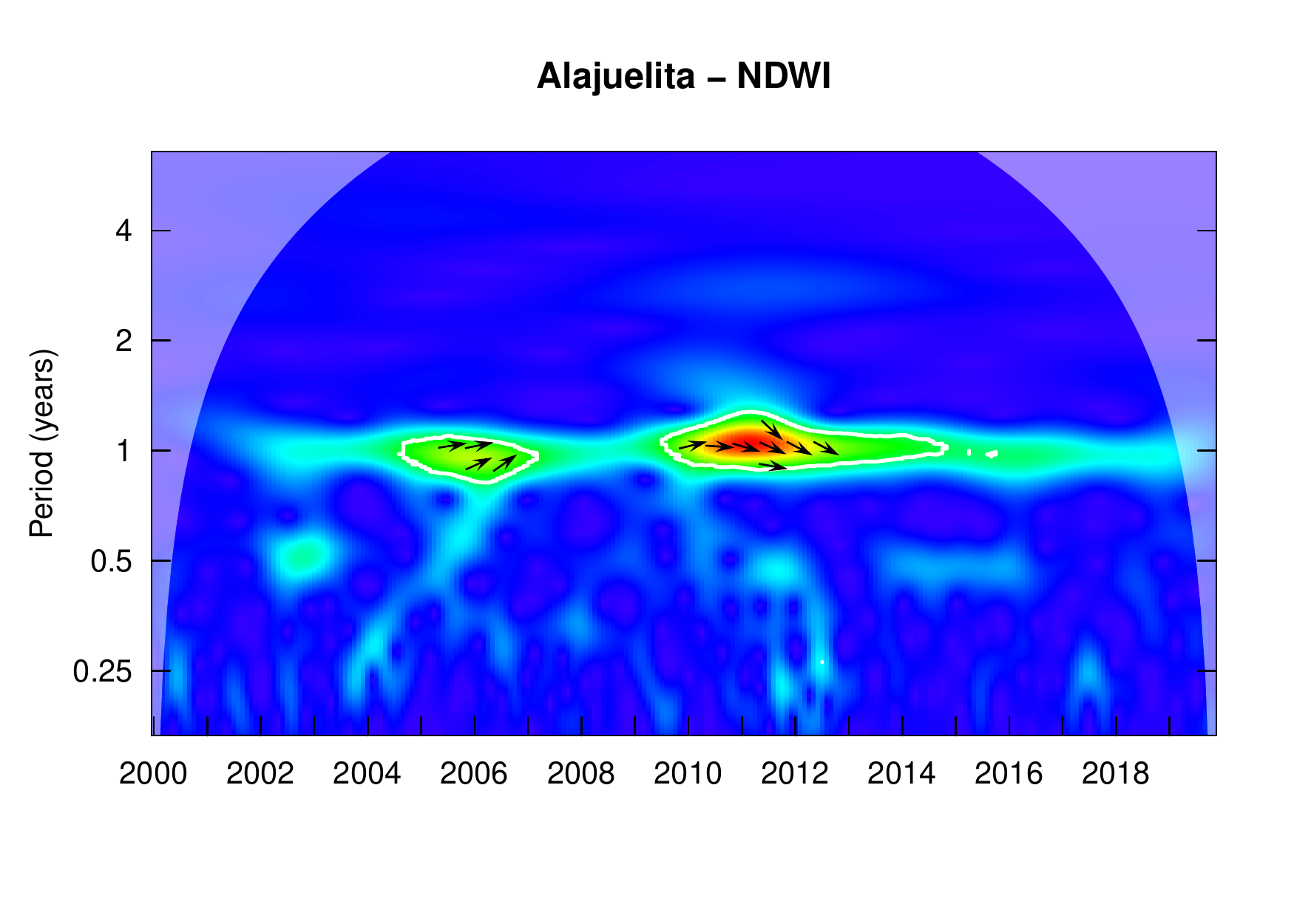}}\vspace{-0.15cm}%
\subfloat[]{\includegraphics[scale=0.23]{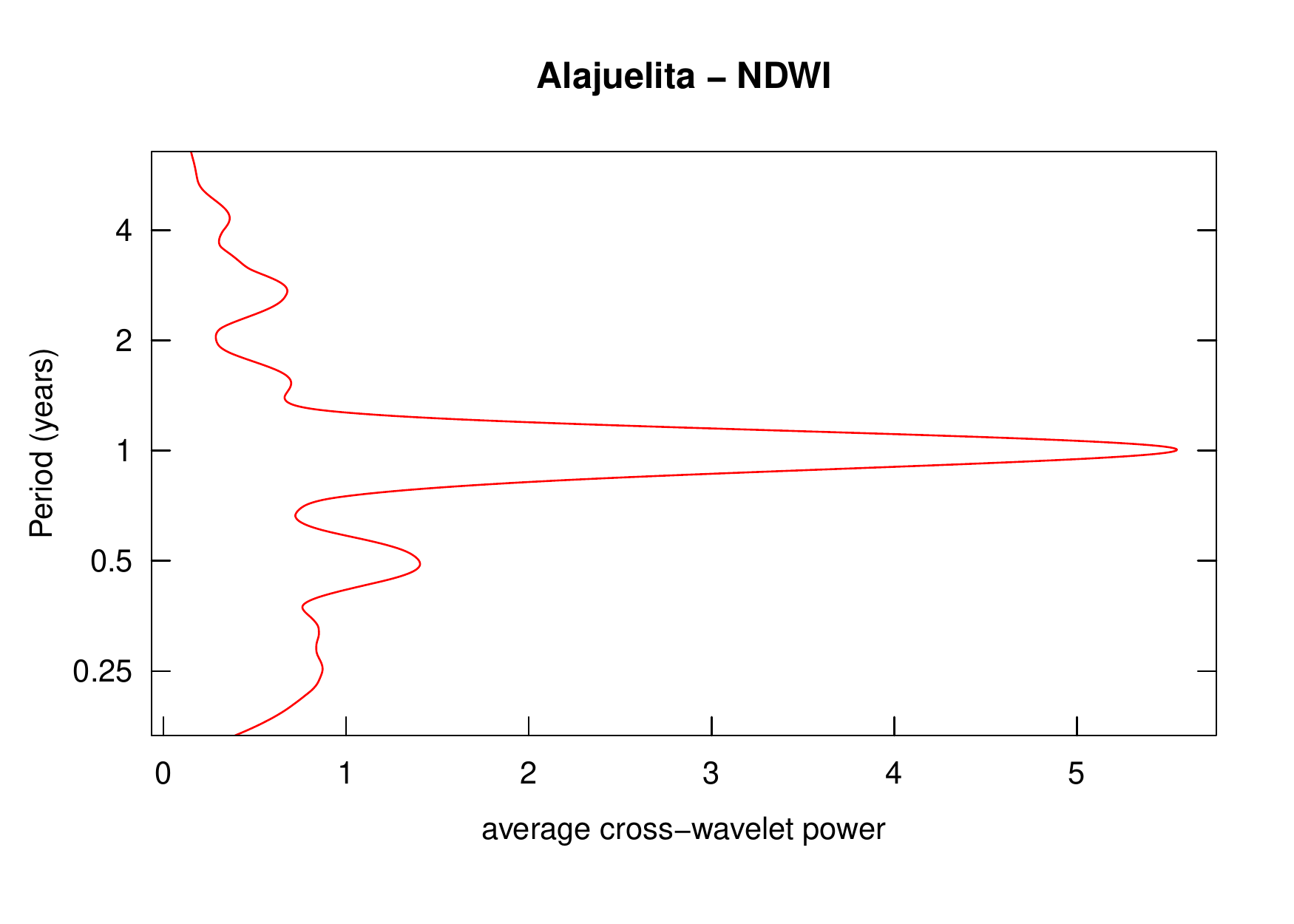}}\vspace{-0.15cm}%
\subfloat[]{\includegraphics[scale=0.23]{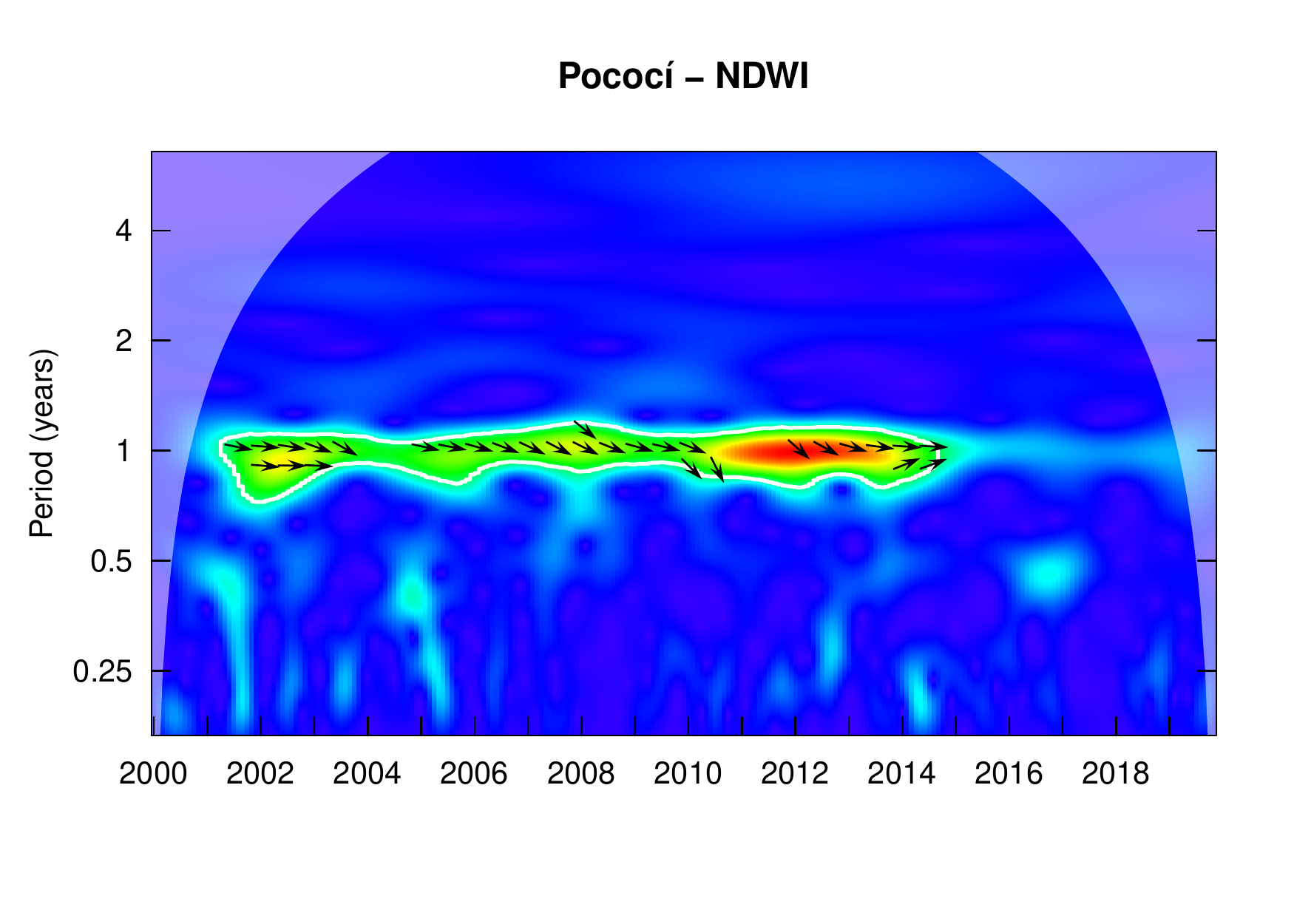}}\vspace{-0.15cm}%
\subfloat[]{\includegraphics[scale=0.23]{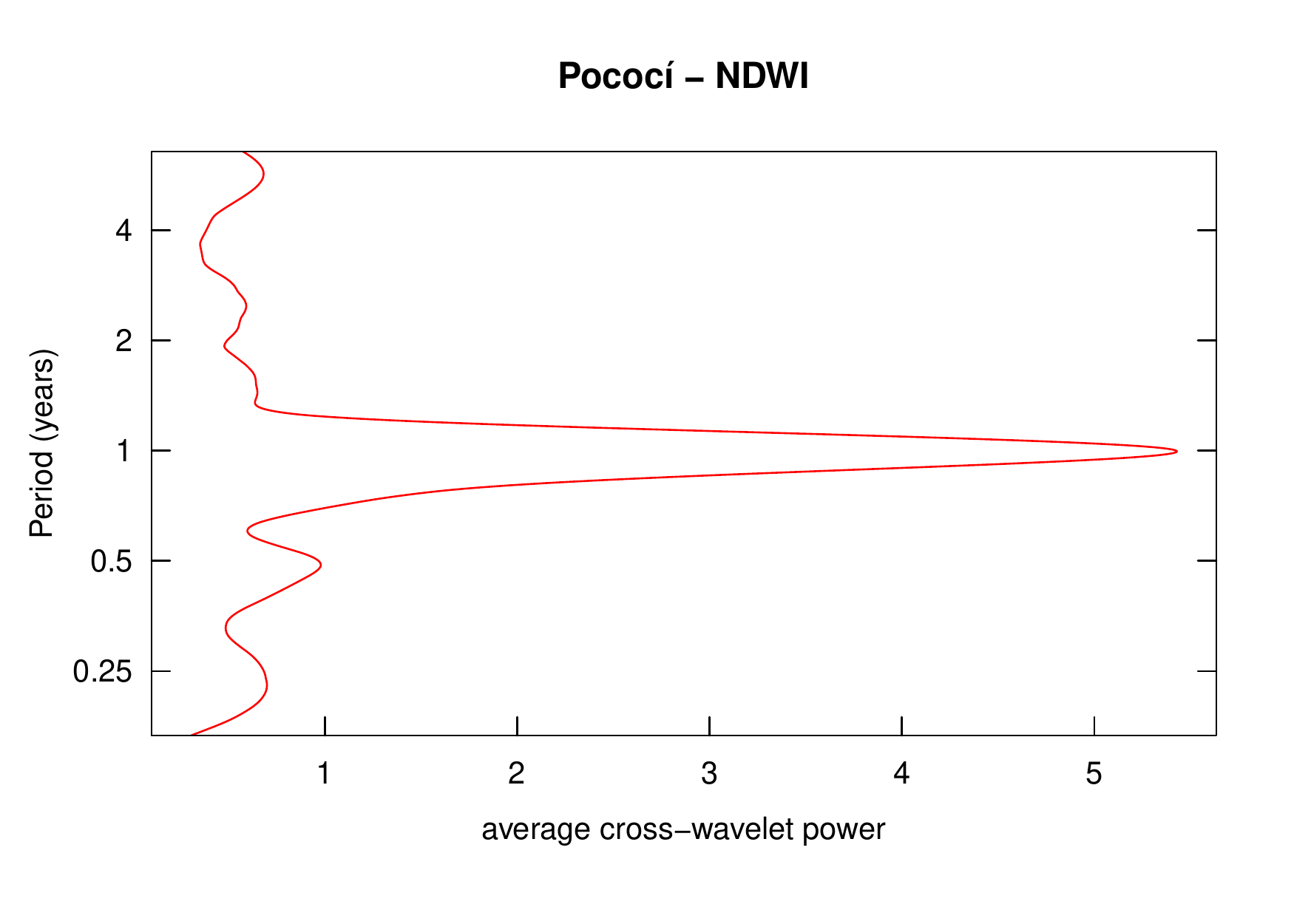}}\vspace{-0.15cm}\\
\subfloat[]{\includegraphics[scale=0.23]{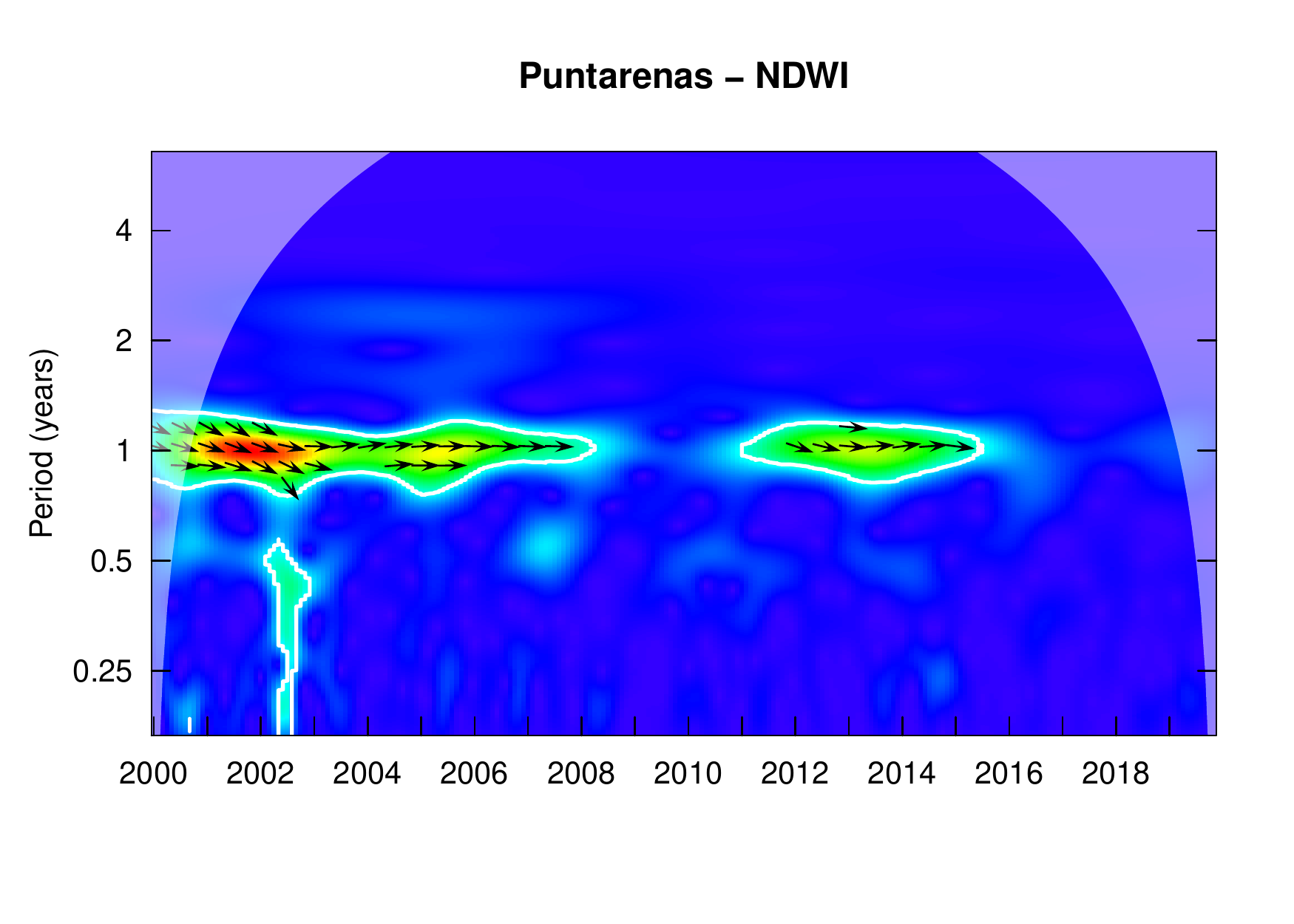}}\vspace{-0.15cm}%
\subfloat[]{\includegraphics[scale=0.23]{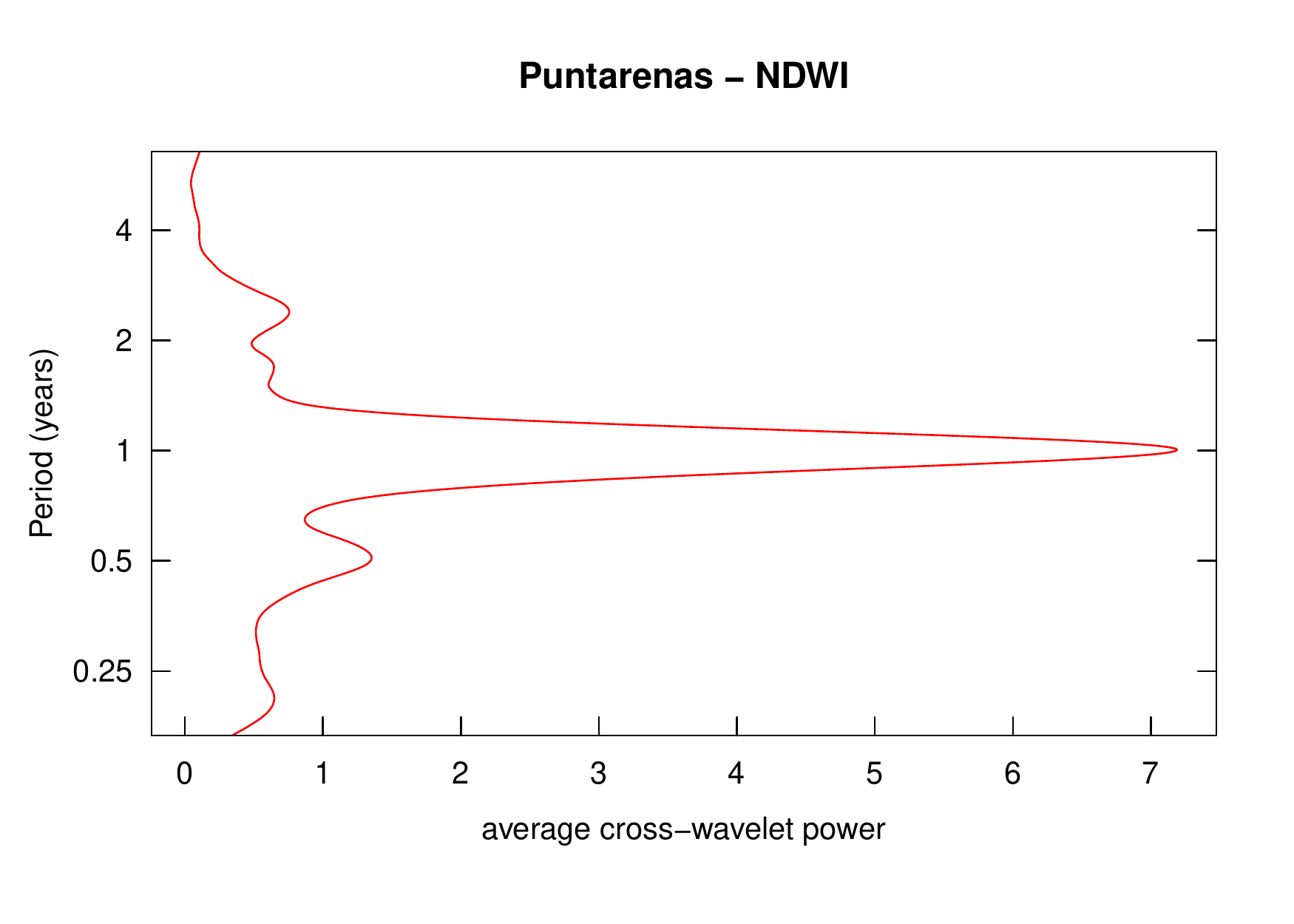}}\vspace{-0.15cm}%
\subfloat[]{\includegraphics[scale=0.23]{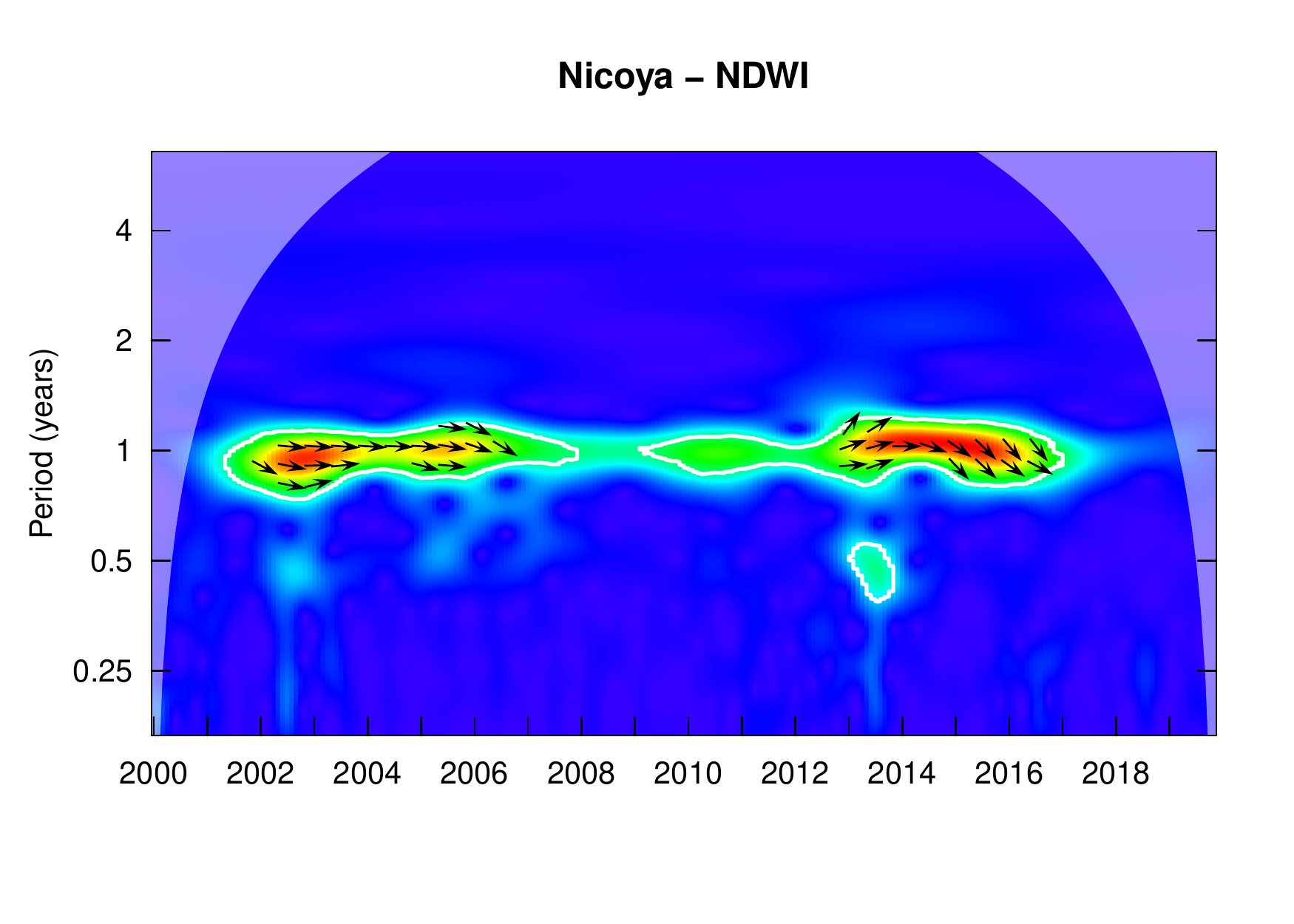}}\vspace{-0.15cm}%
\subfloat[]{\includegraphics[scale=0.23]{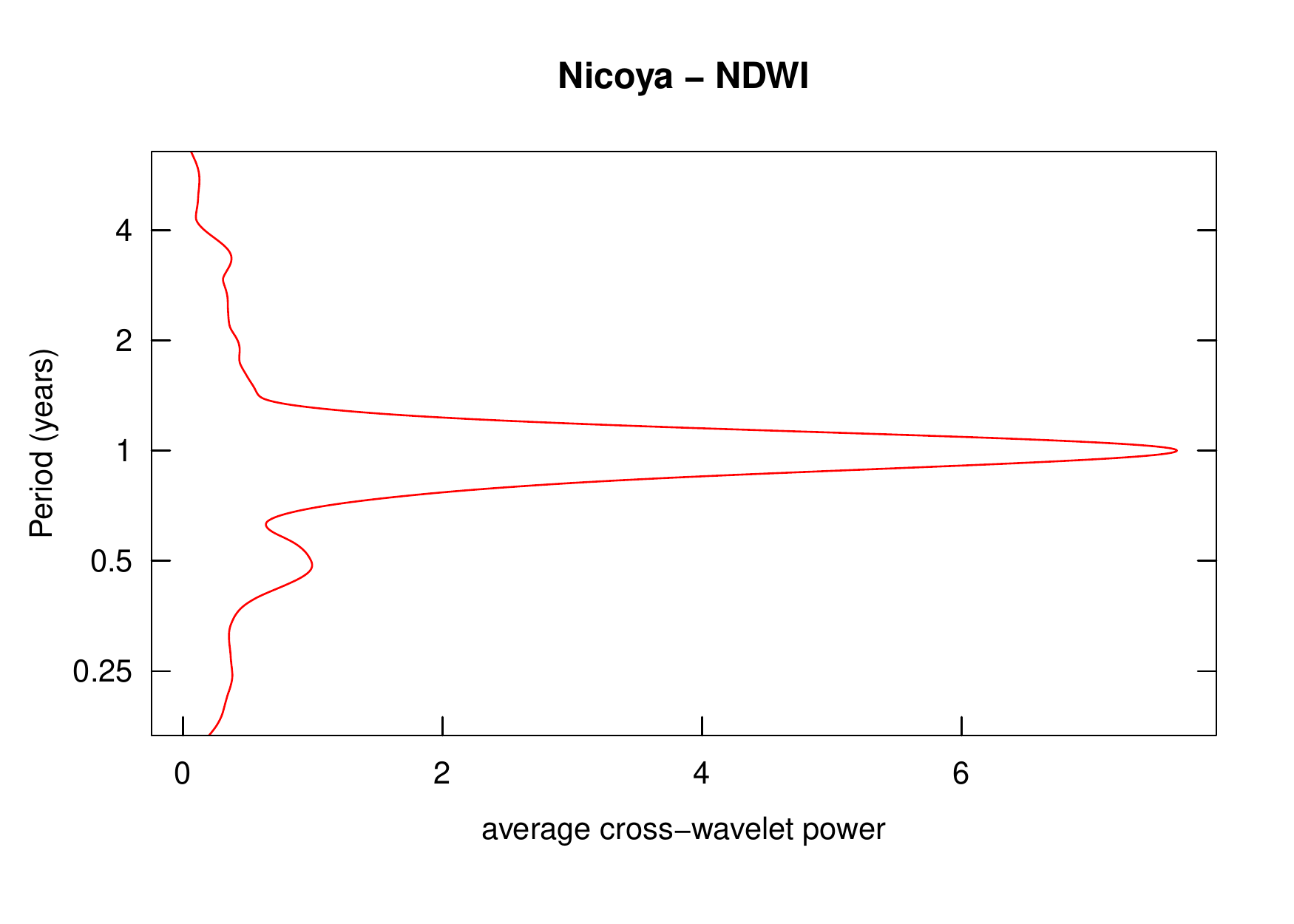}}\vspace{-0.15cm}\\
\caption*{}
\end{figure}

\begin{figure}[H]
\captionsetup[subfigure]{labelformat=empty}
\subfloat[]{\includegraphics[scale=0.23]{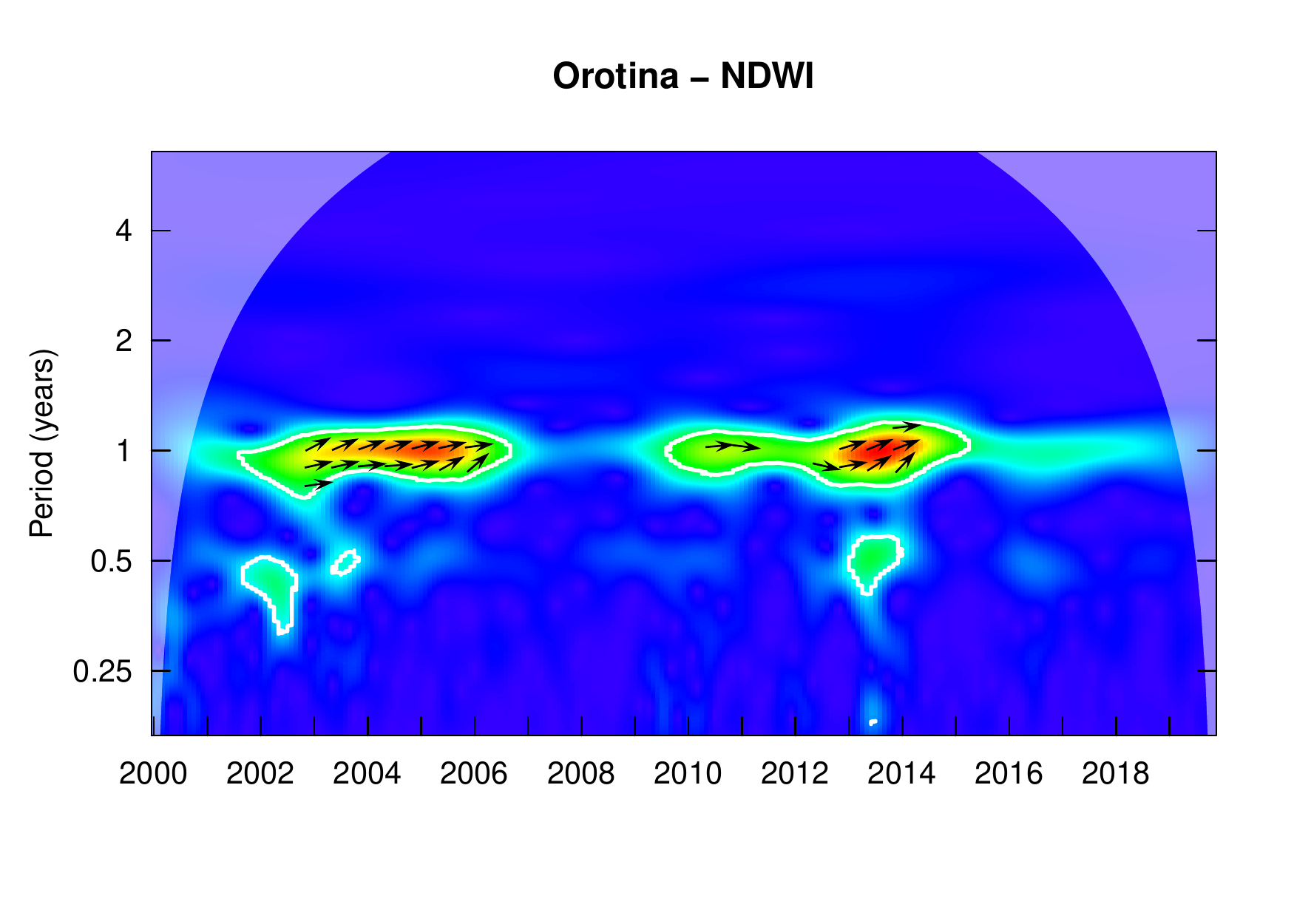}}\vspace{-0.15cm}%
\subfloat[]{\includegraphics[scale=0.23]{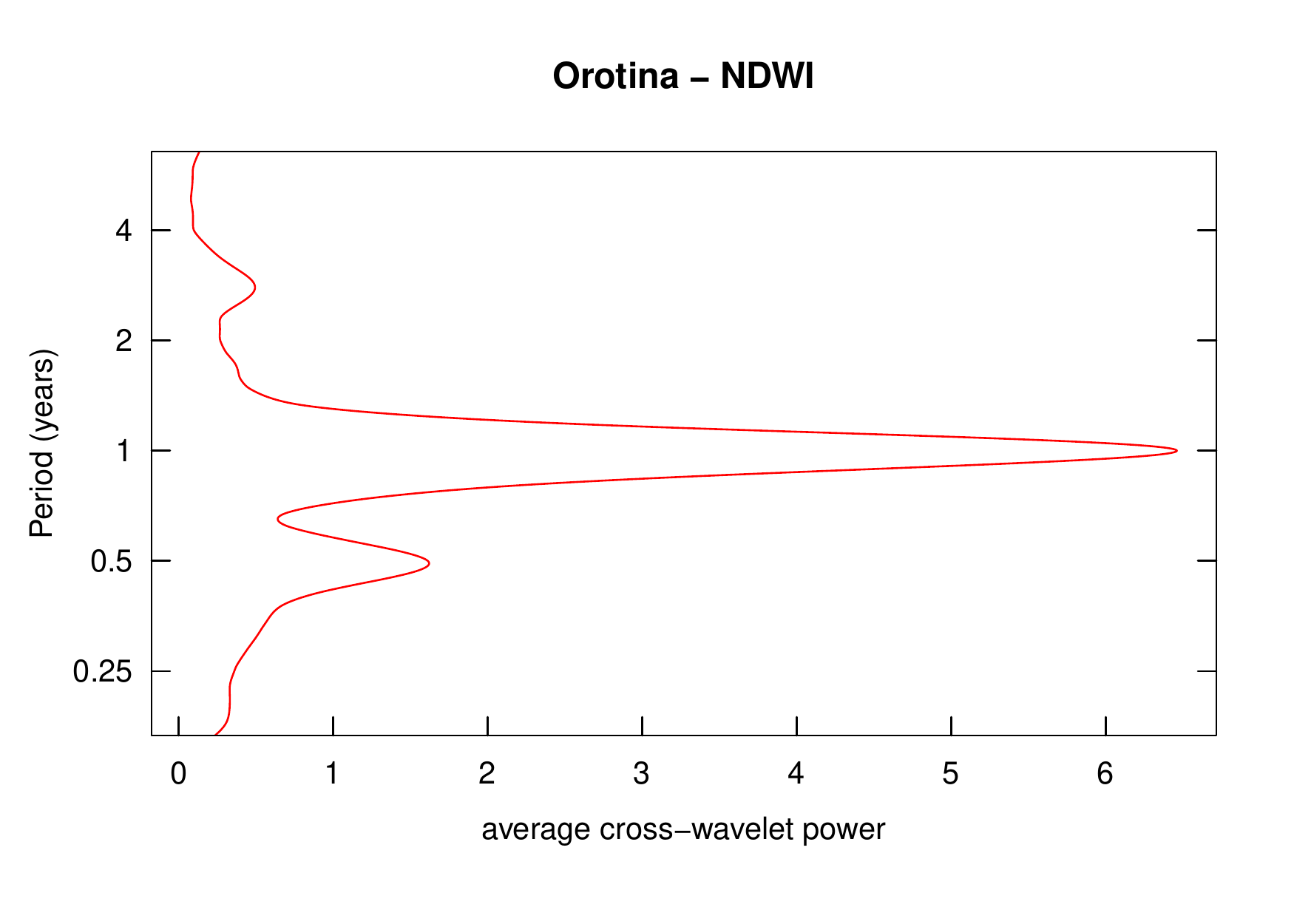}}\vspace{-0.15cm}%
\subfloat[]{\includegraphics[scale=0.23]{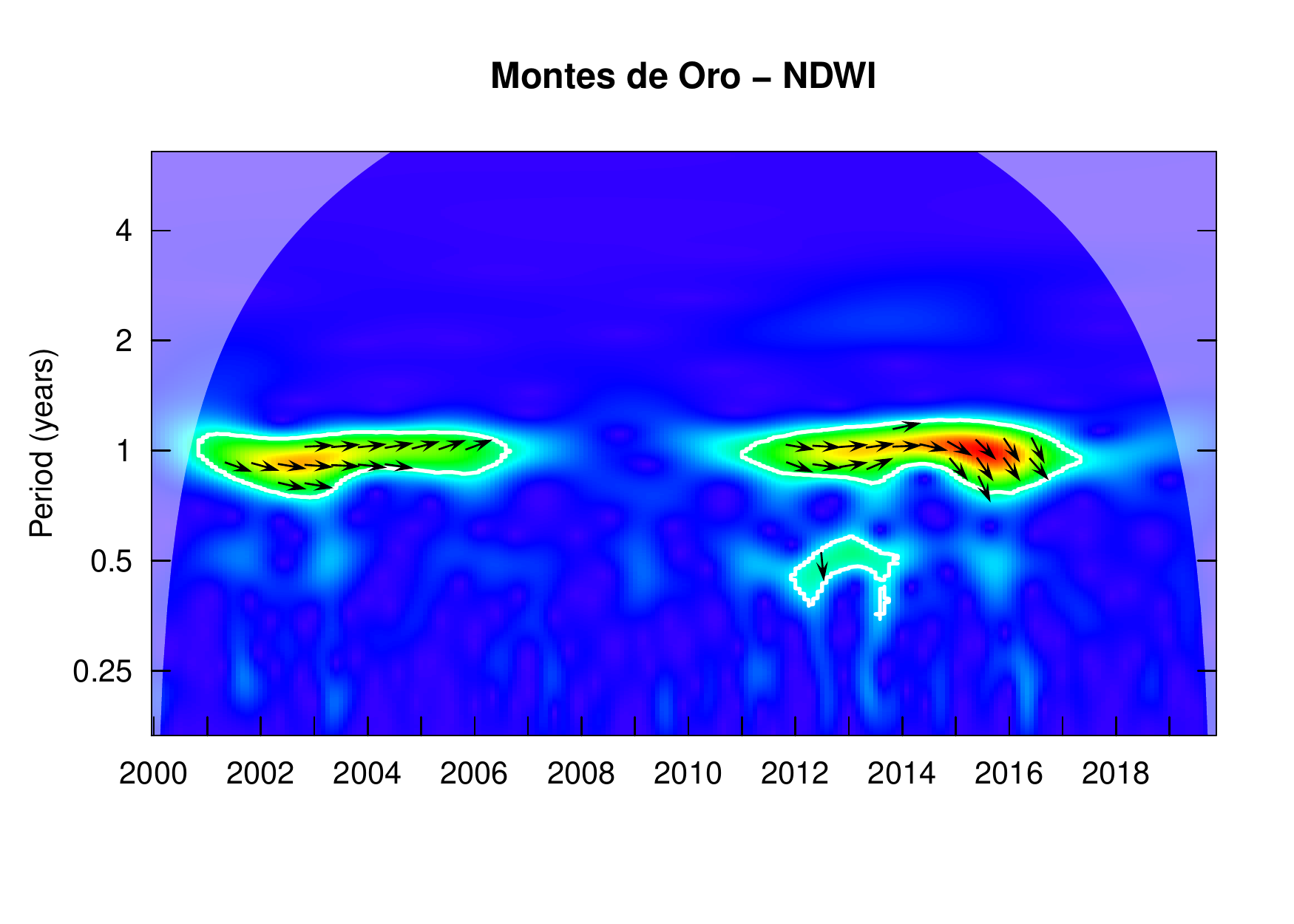}}\vspace{-0.15cm}%
\subfloat[]{\includegraphics[scale=0.23]{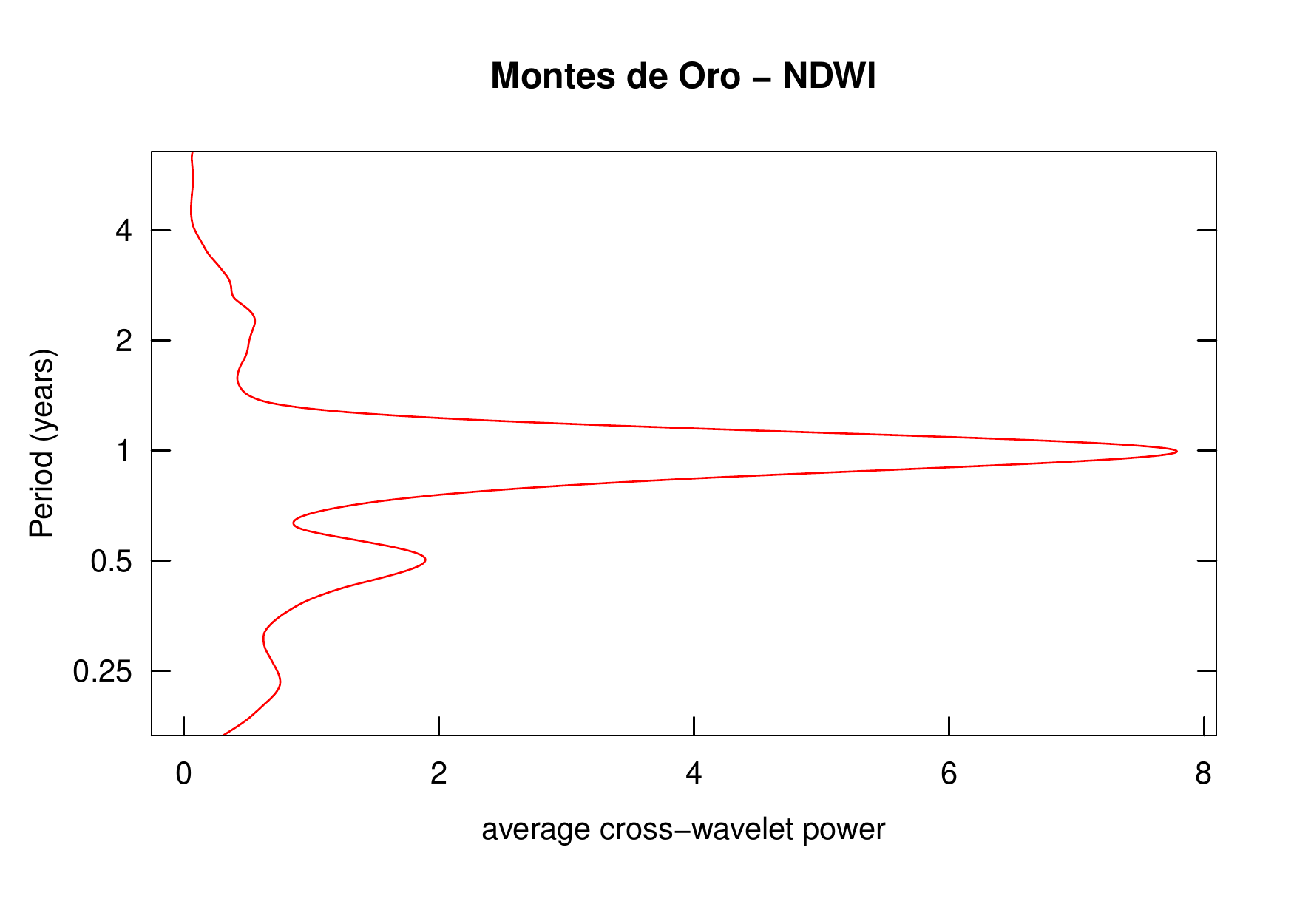}}\vspace{-0.15cm}\\
\subfloat[]{\includegraphics[scale=0.23]{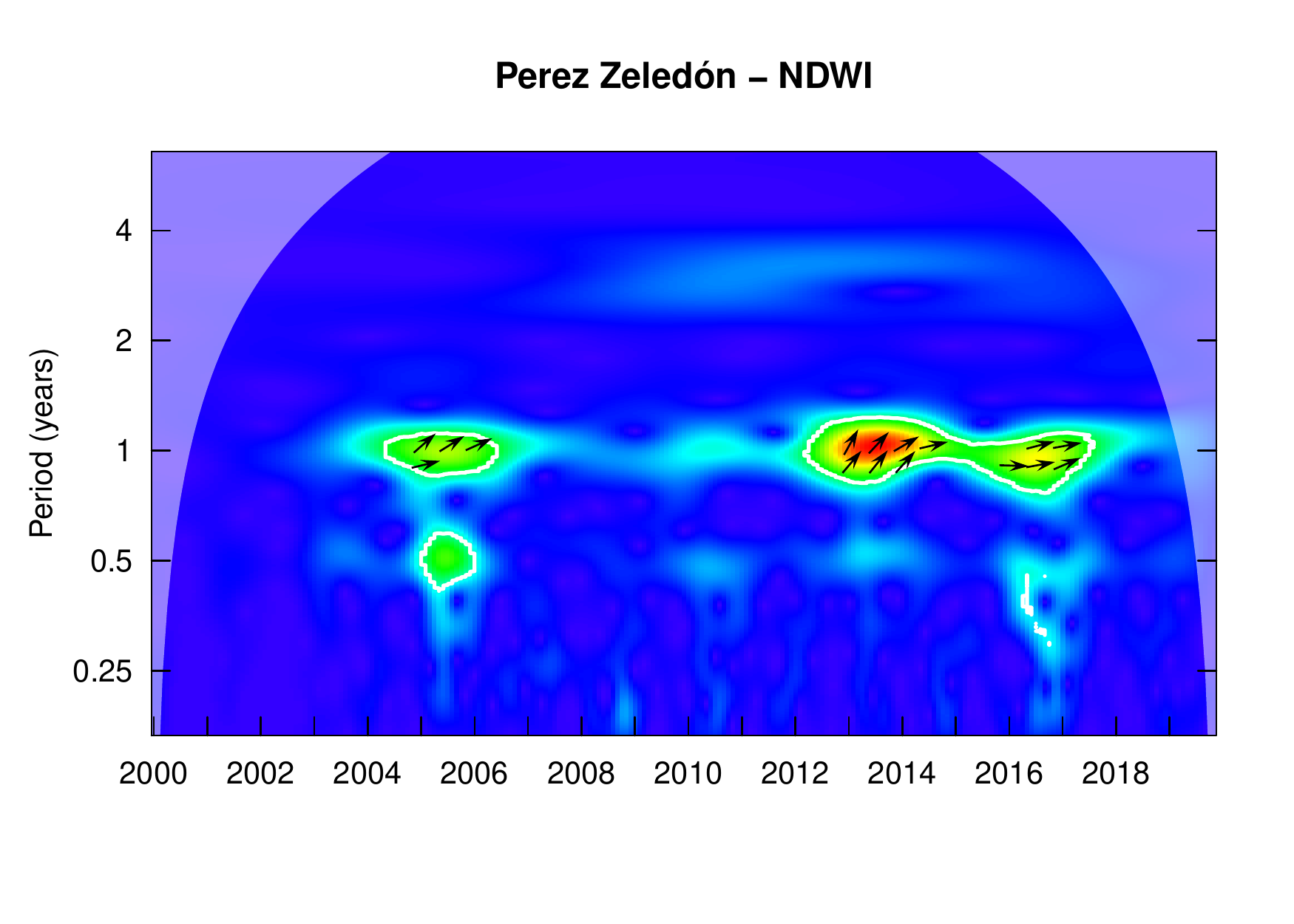}}\vspace{-0.15cm}%
\subfloat[]{\includegraphics[scale=0.23]{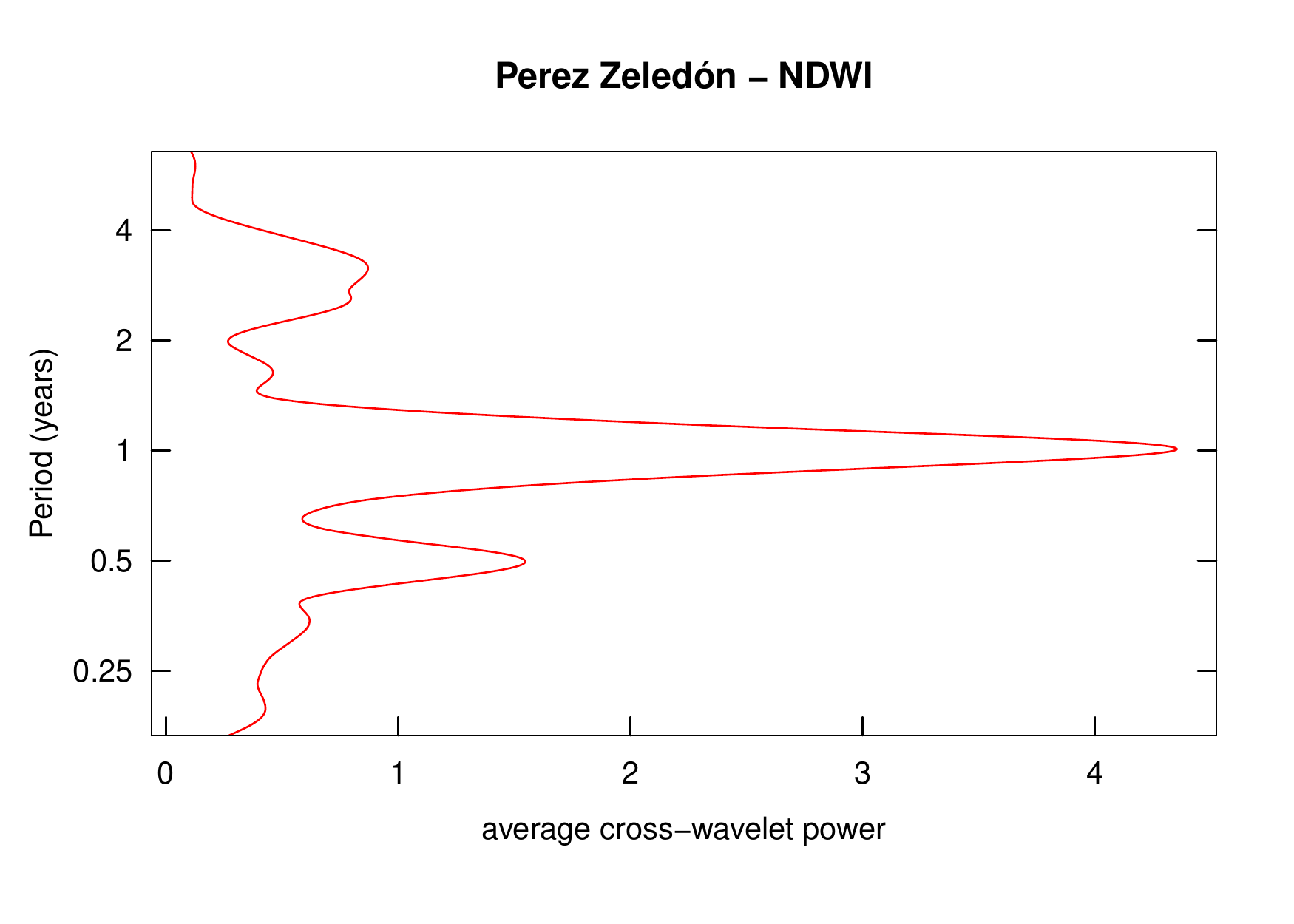}}\vspace{-0.15cm}%
\subfloat[]{\includegraphics[scale=0.23]{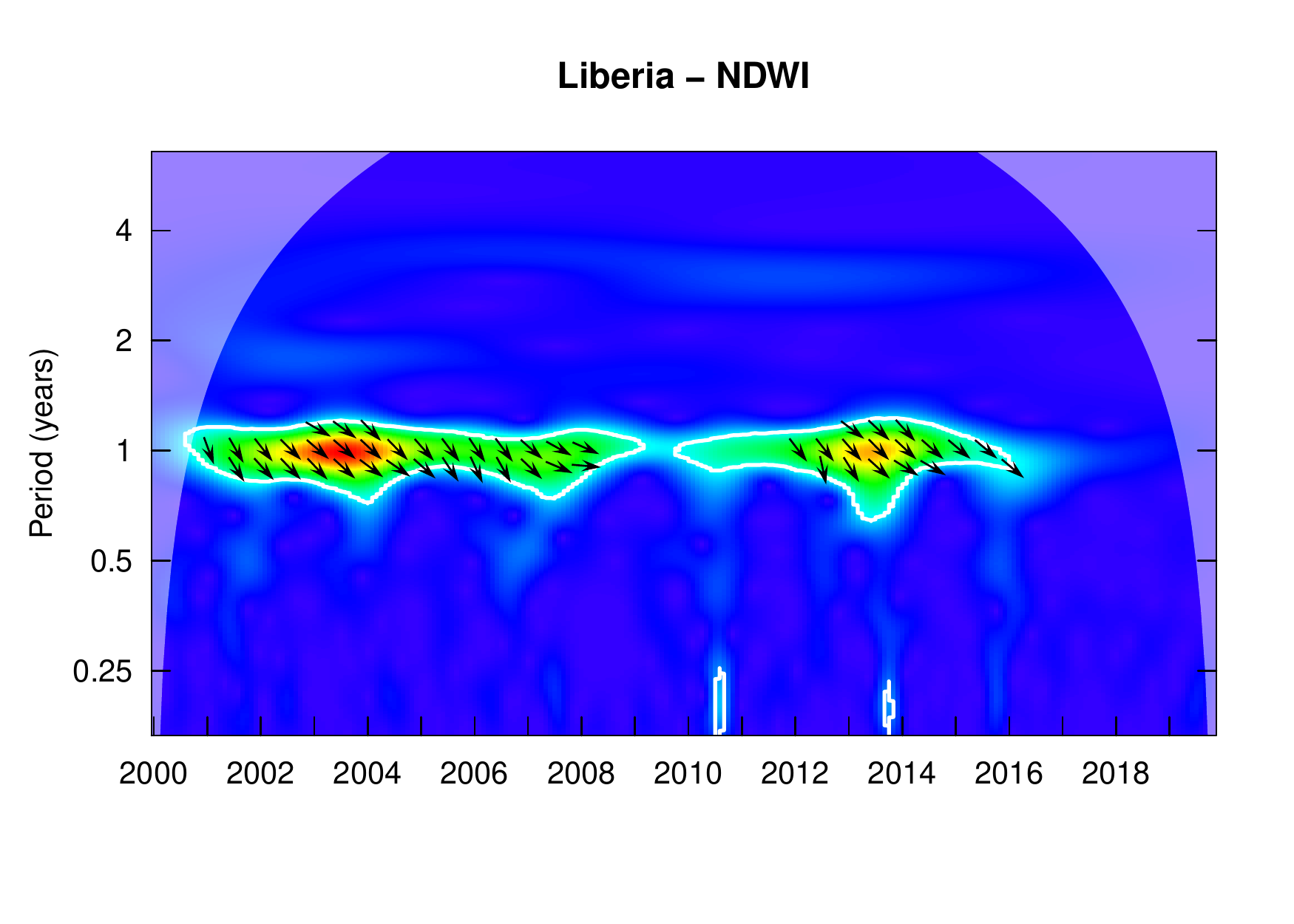}}\vspace{-0.15cm}%
\subfloat[]{\includegraphics[scale=0.23]{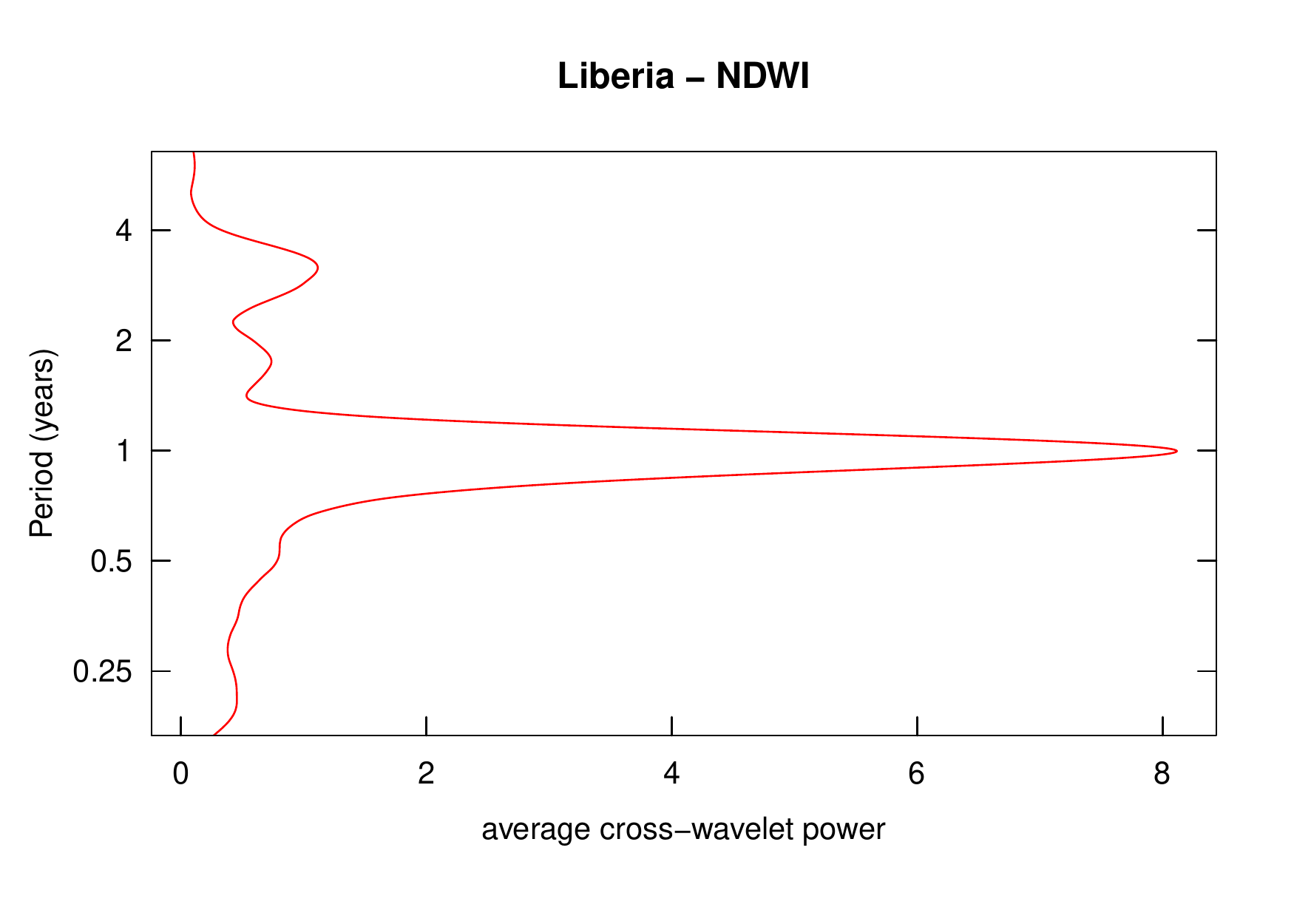}}\vspace{-0.15cm}\\
\subfloat[]{\includegraphics[scale=0.23]{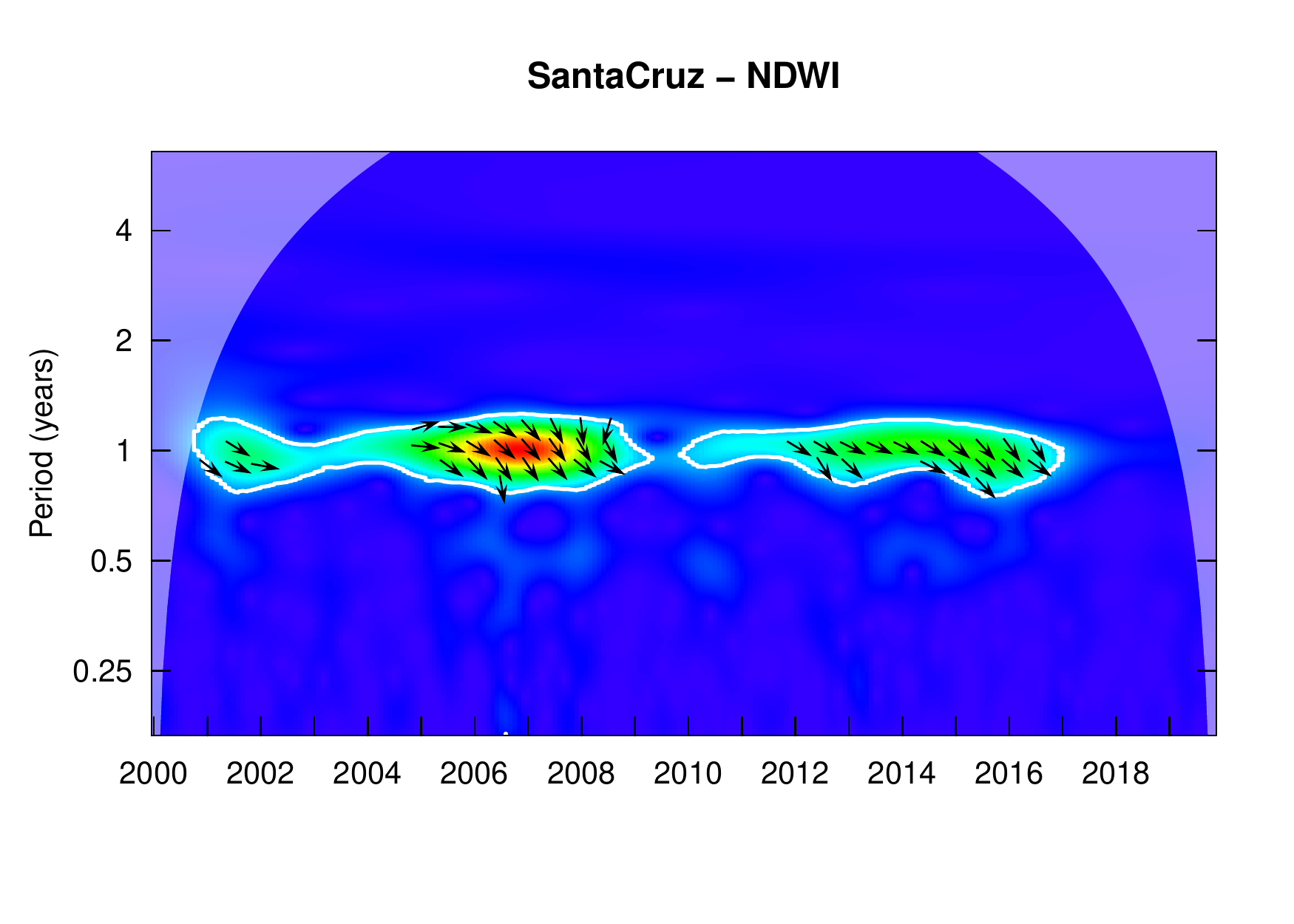}}\vspace{-0.15cm}%
\subfloat[]{\includegraphics[scale=0.23]{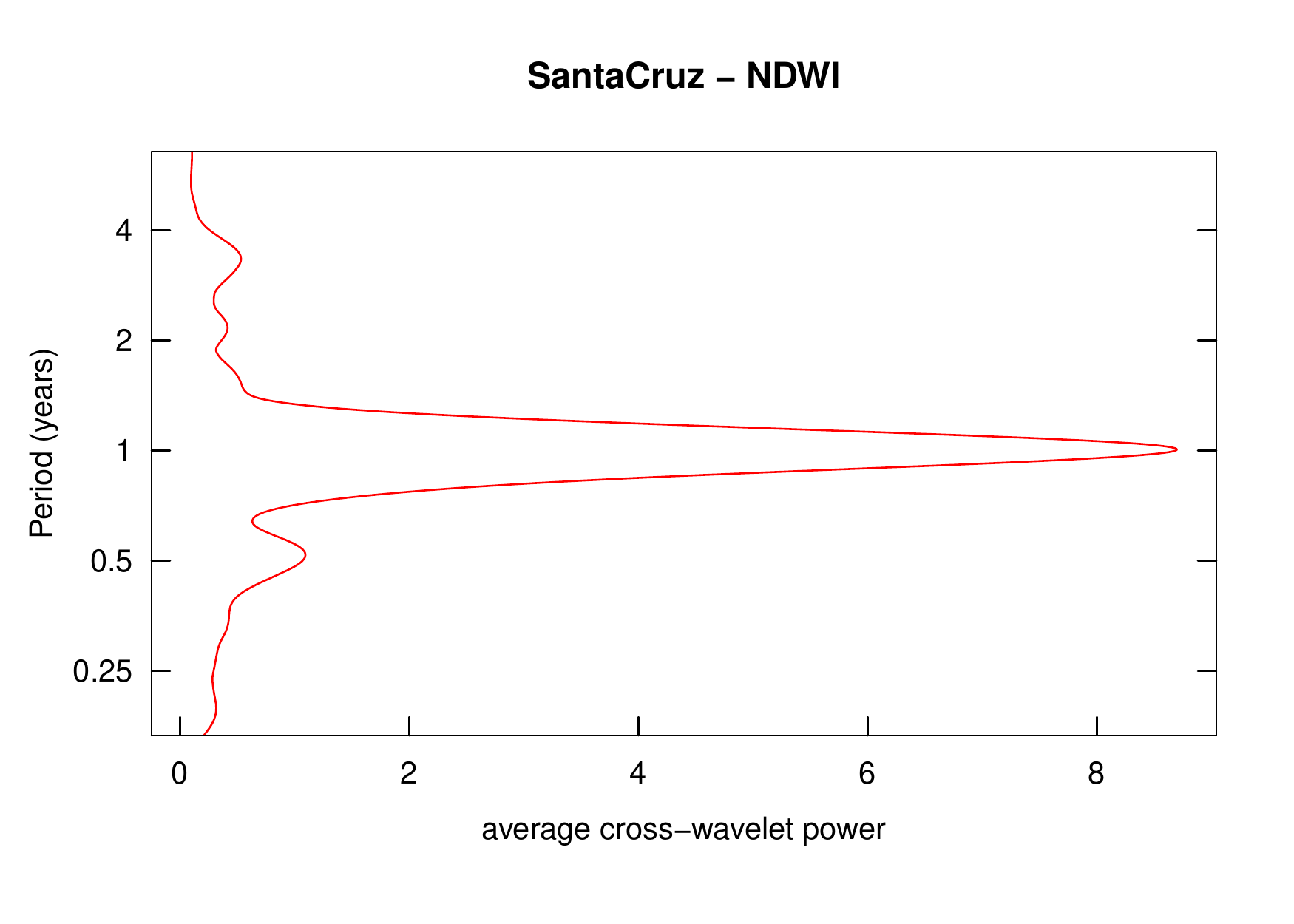}}\vspace{-0.15cm}%
\subfloat[]{\includegraphics[scale=0.23]{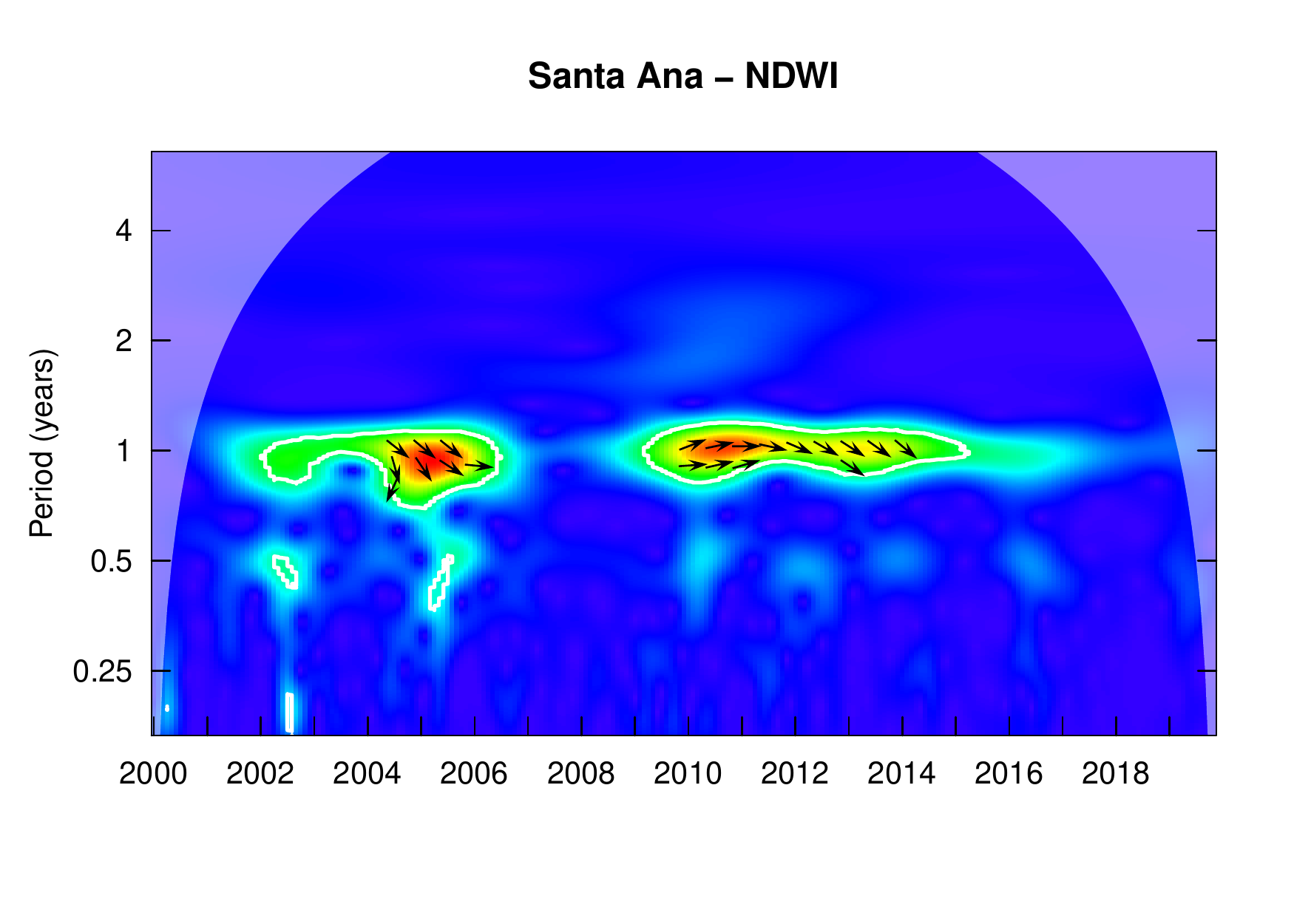}}\vspace{-0.15cm}%
\subfloat[]{\includegraphics[scale=0.23]{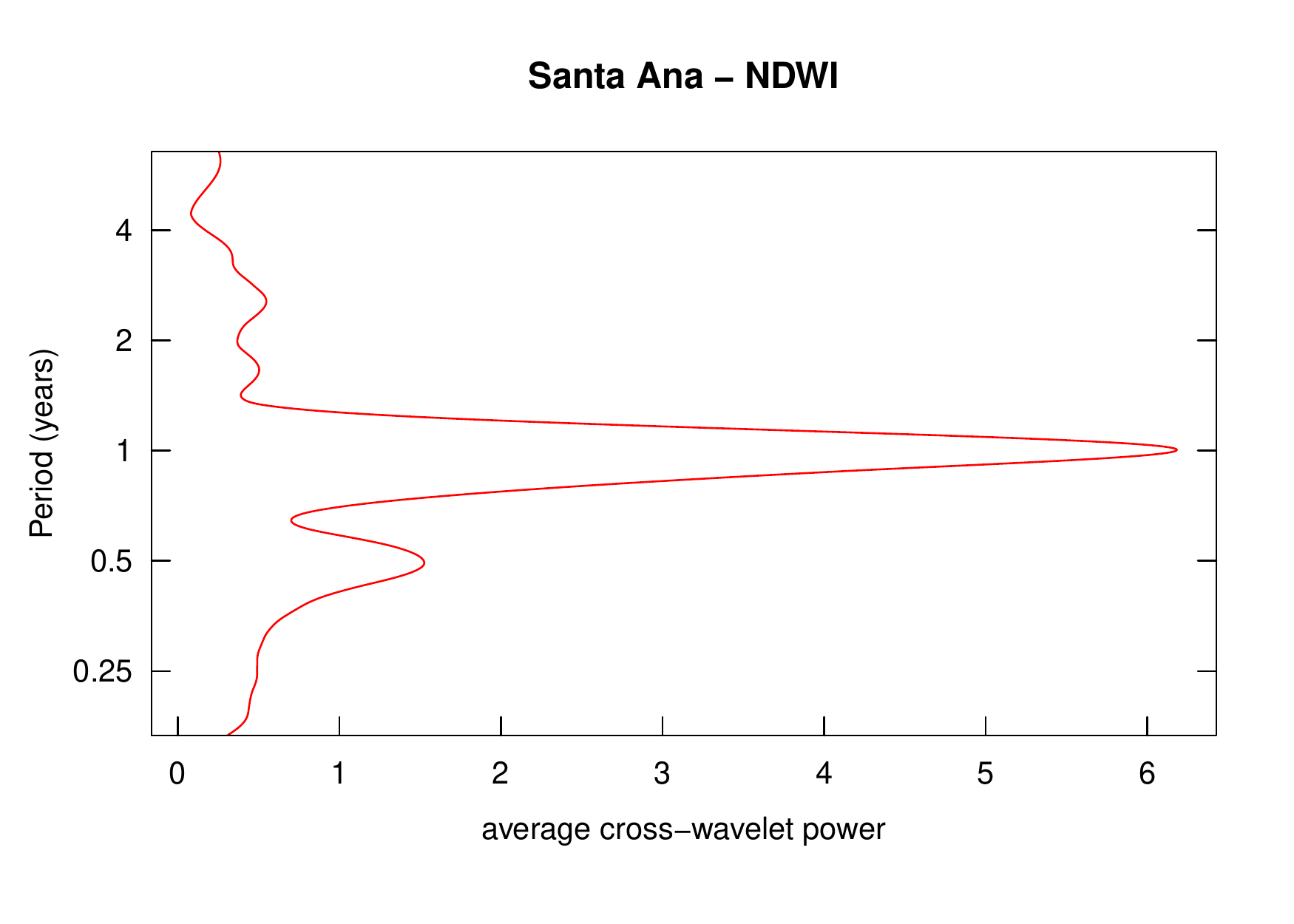}}\vspace{-0.15cm}\\
\subfloat[]{\includegraphics[scale=0.23]{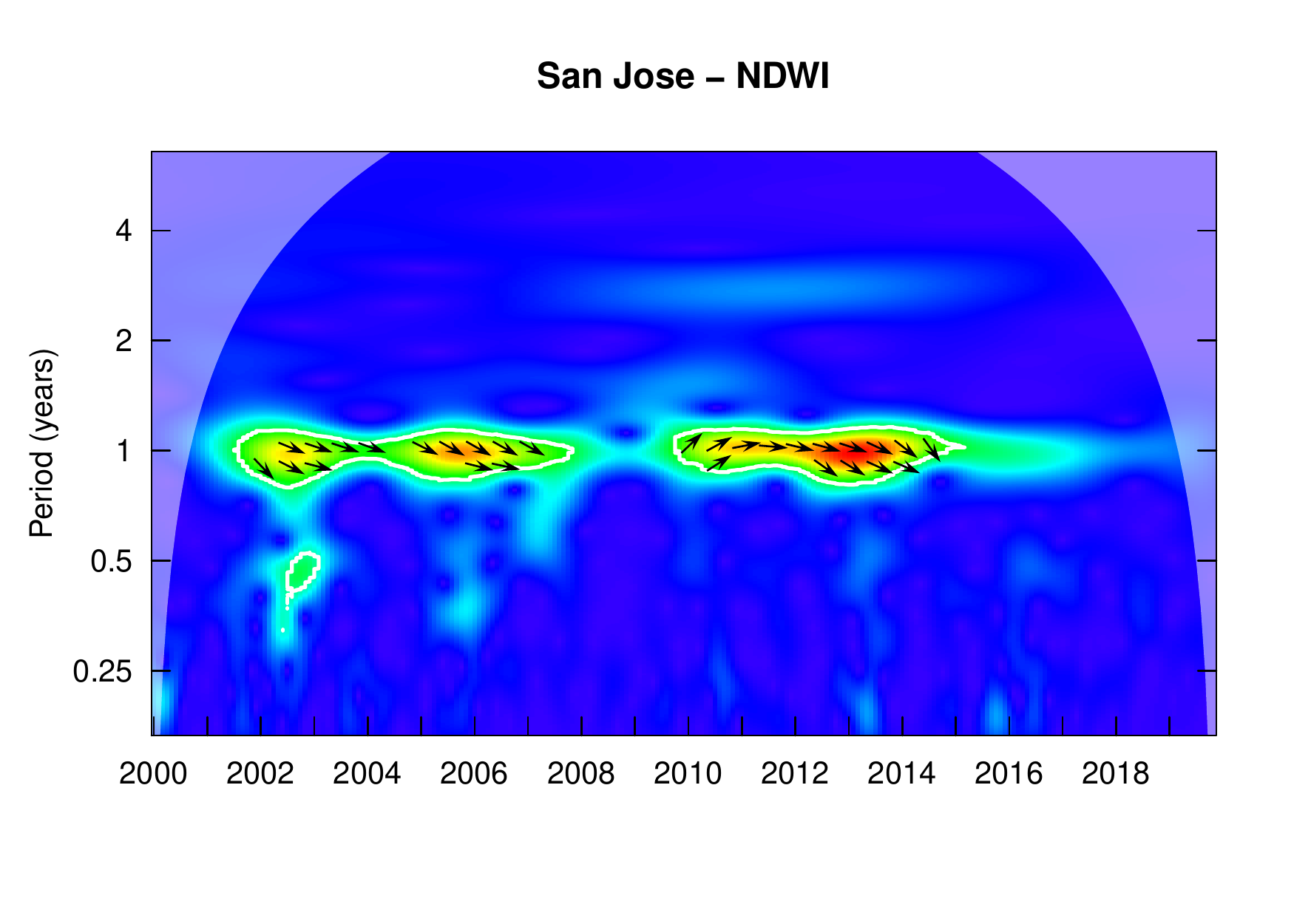}}\vspace{-0.15cm}%
\subfloat[]{\includegraphics[scale=0.23]{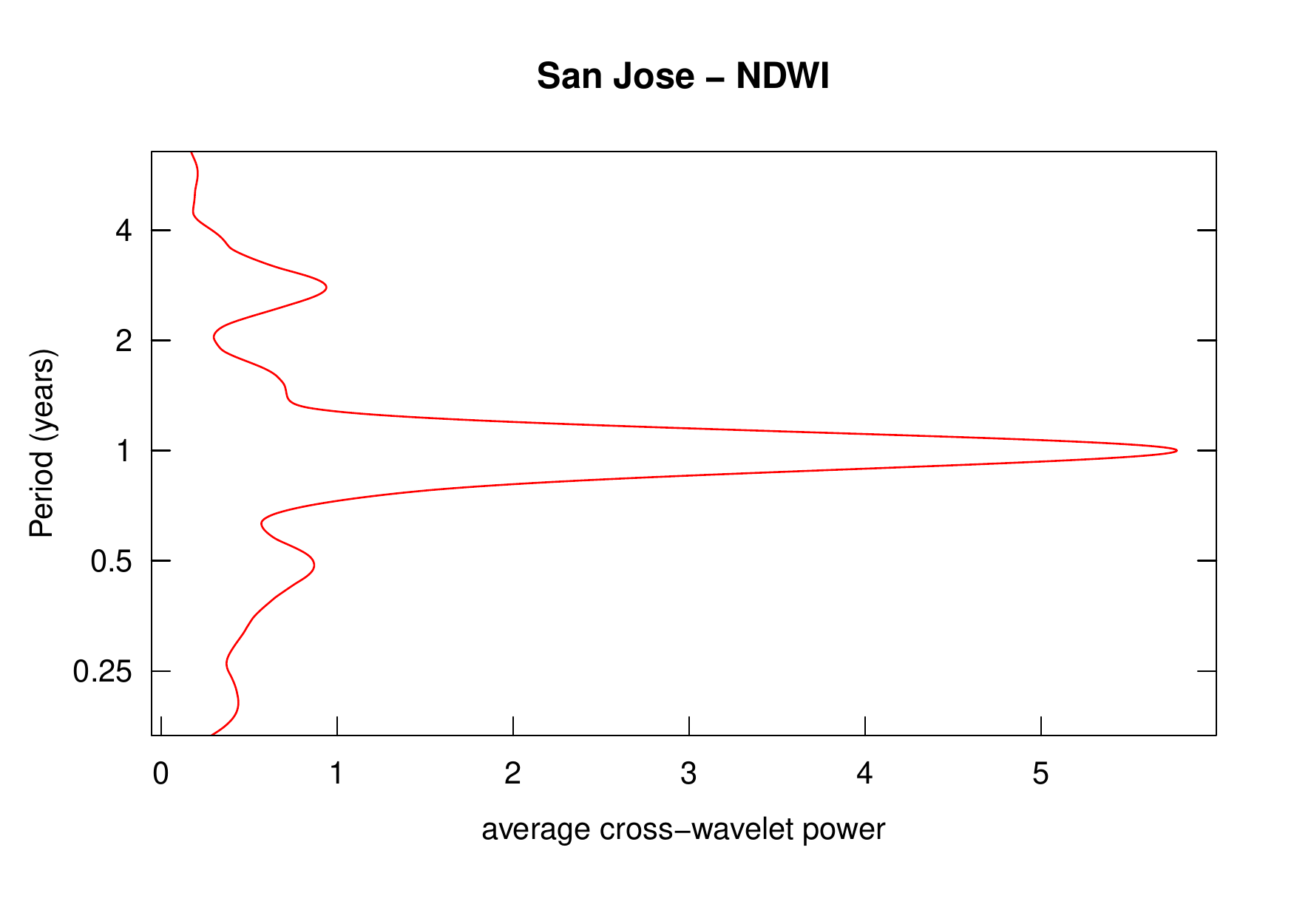}}\vspace{-0.15cm}%
\subfloat[]{\includegraphics[scale=0.23]{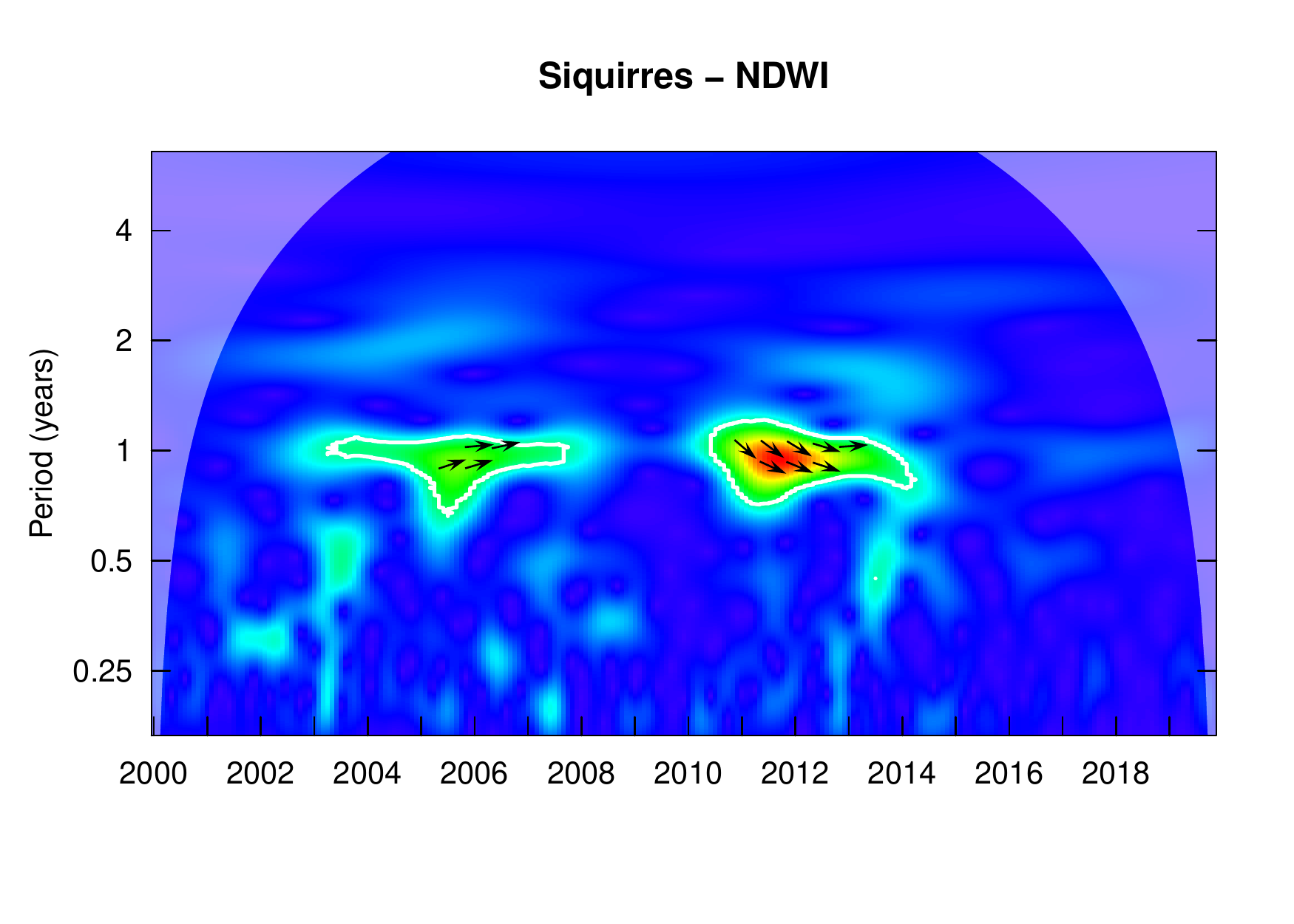}}\vspace{-0.15cm}%
\subfloat[]{\includegraphics[scale=0.23]{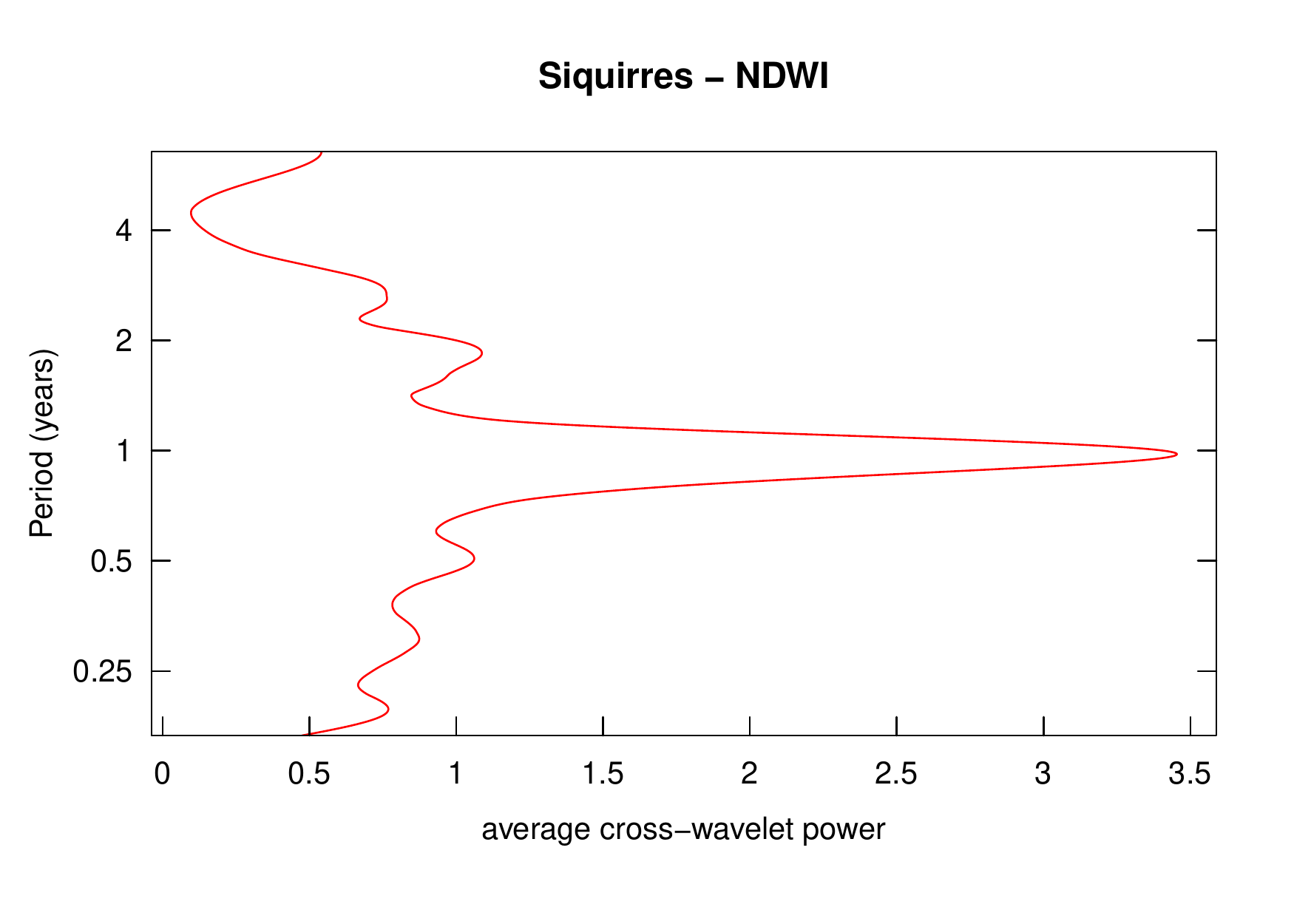}}\vspace{-0.15cm}\\
\subfloat[]{\includegraphics[scale=0.23]{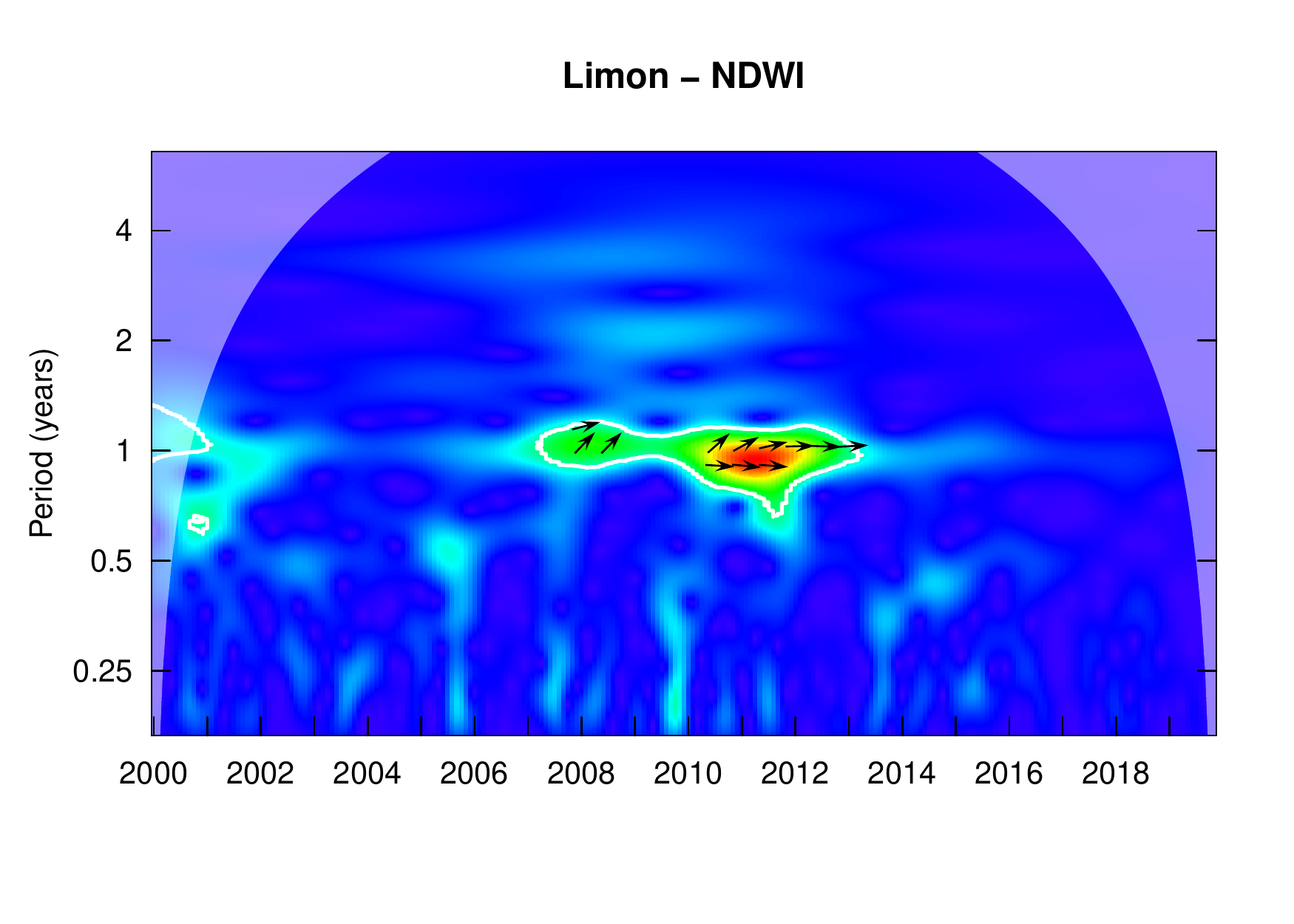}}\vspace{-0.15cm}%
\subfloat[]{\includegraphics[scale=0.23]{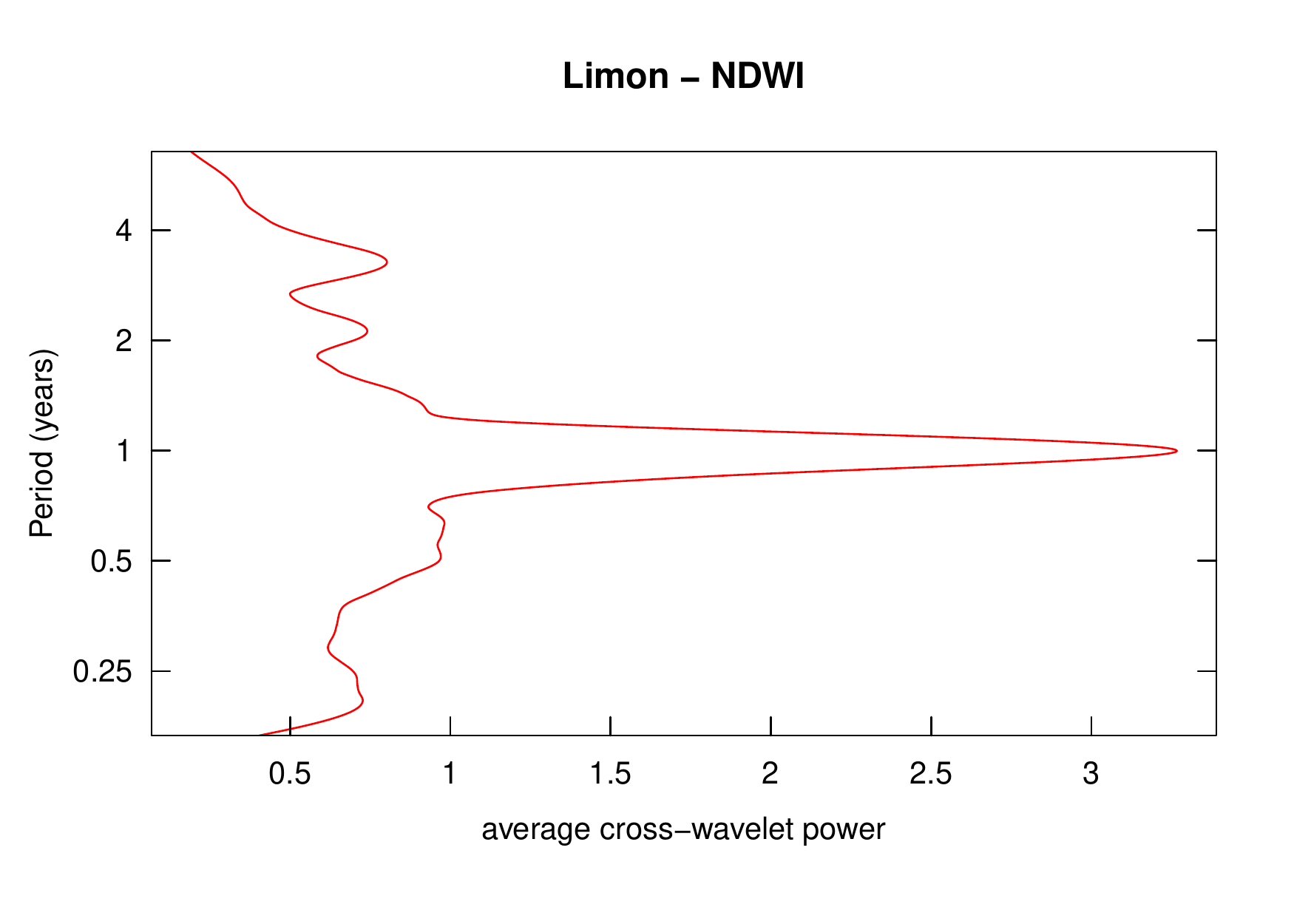}}\vspace{-0.15cm}%
\subfloat[]{\includegraphics[scale=0.23]{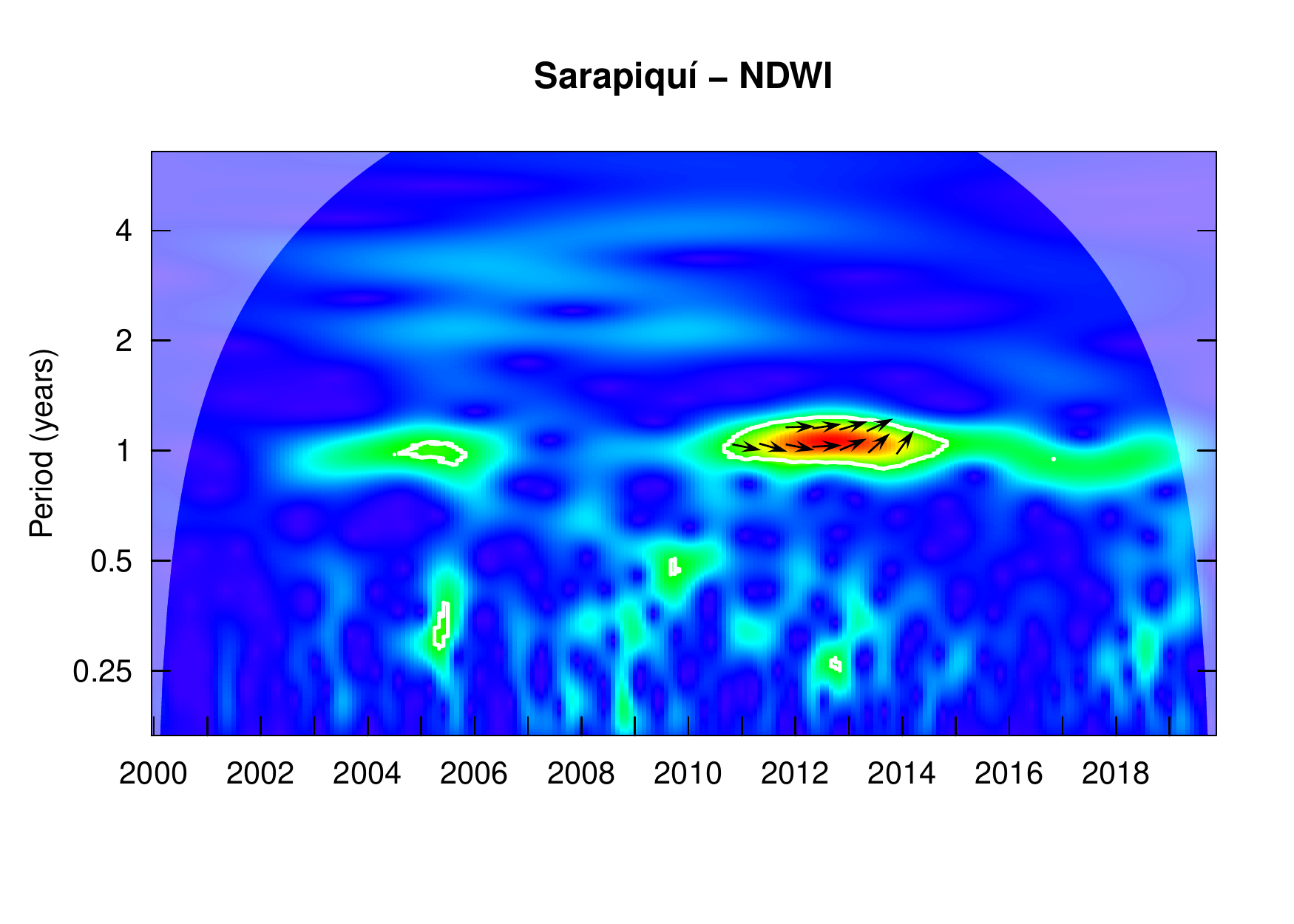}}\vspace{-0.15cm}%
\subfloat[]{\includegraphics[scale=0.23]{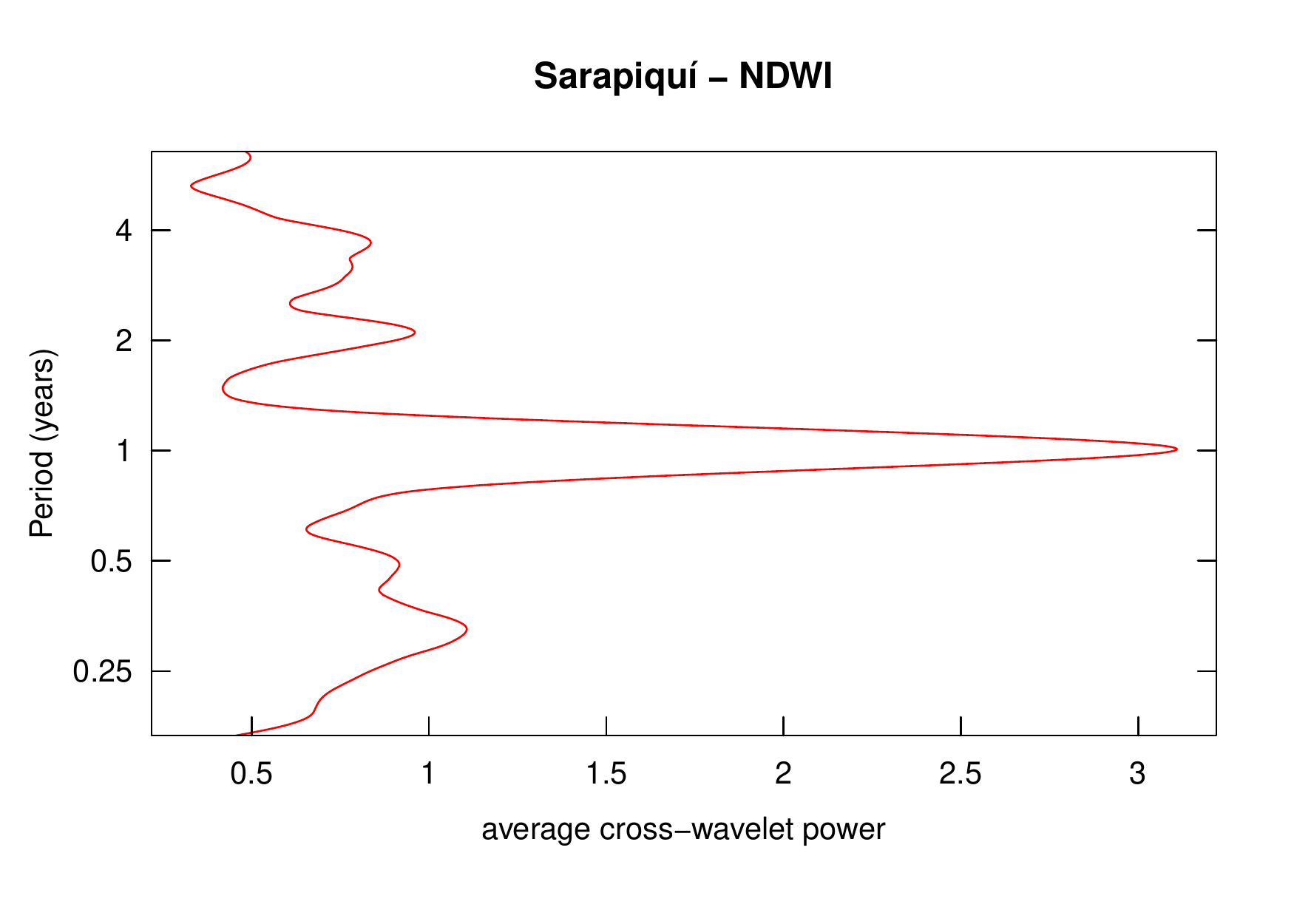}}\vspace{-0.15cm}\\
\subfloat[]{\includegraphics[scale=0.23]{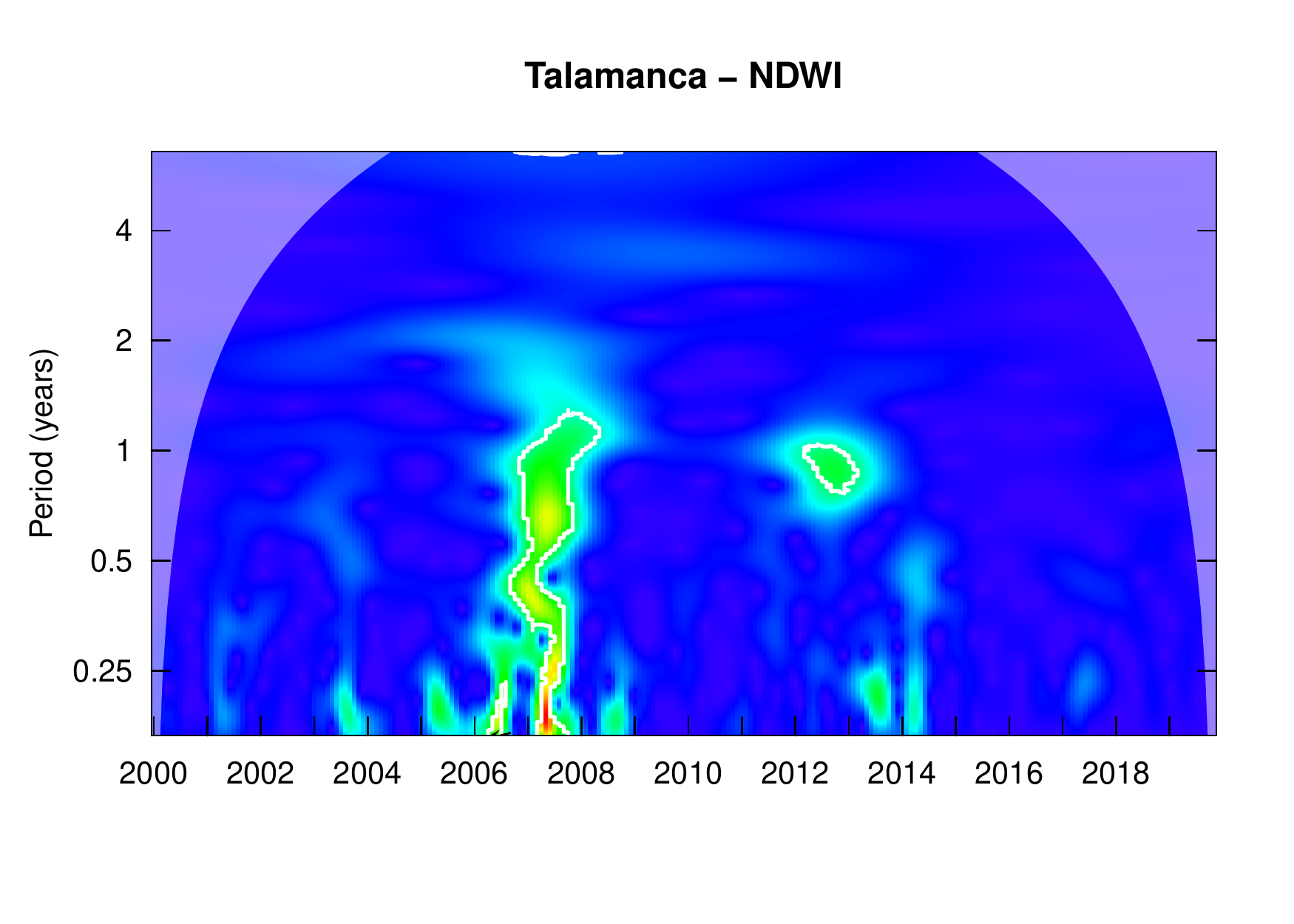}}\vspace{-0.15cm}%
\subfloat[]{\includegraphics[scale=0.23]{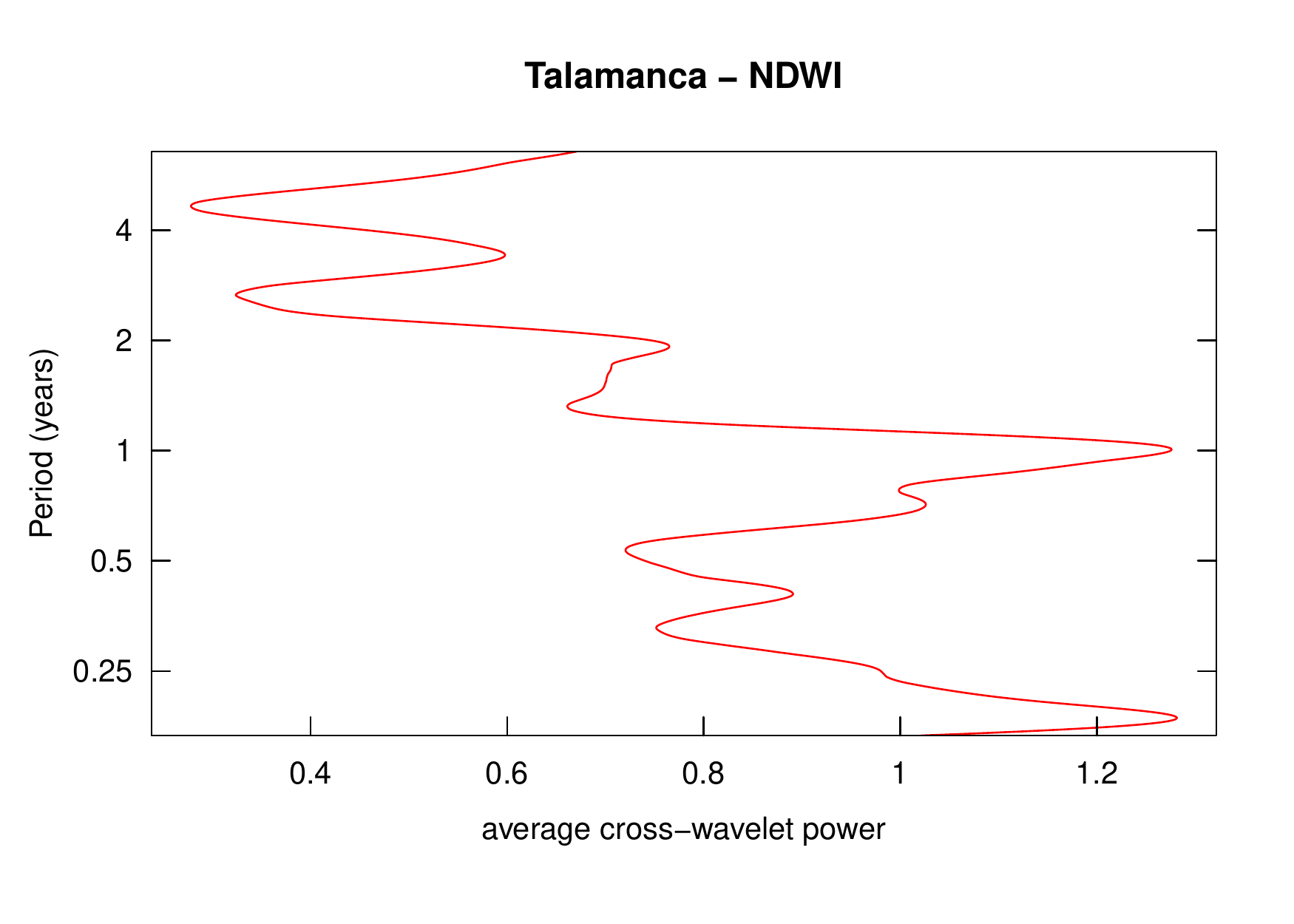}}\vspace{-0.15cm}%
\subfloat[]{\includegraphics[scale=0.23]{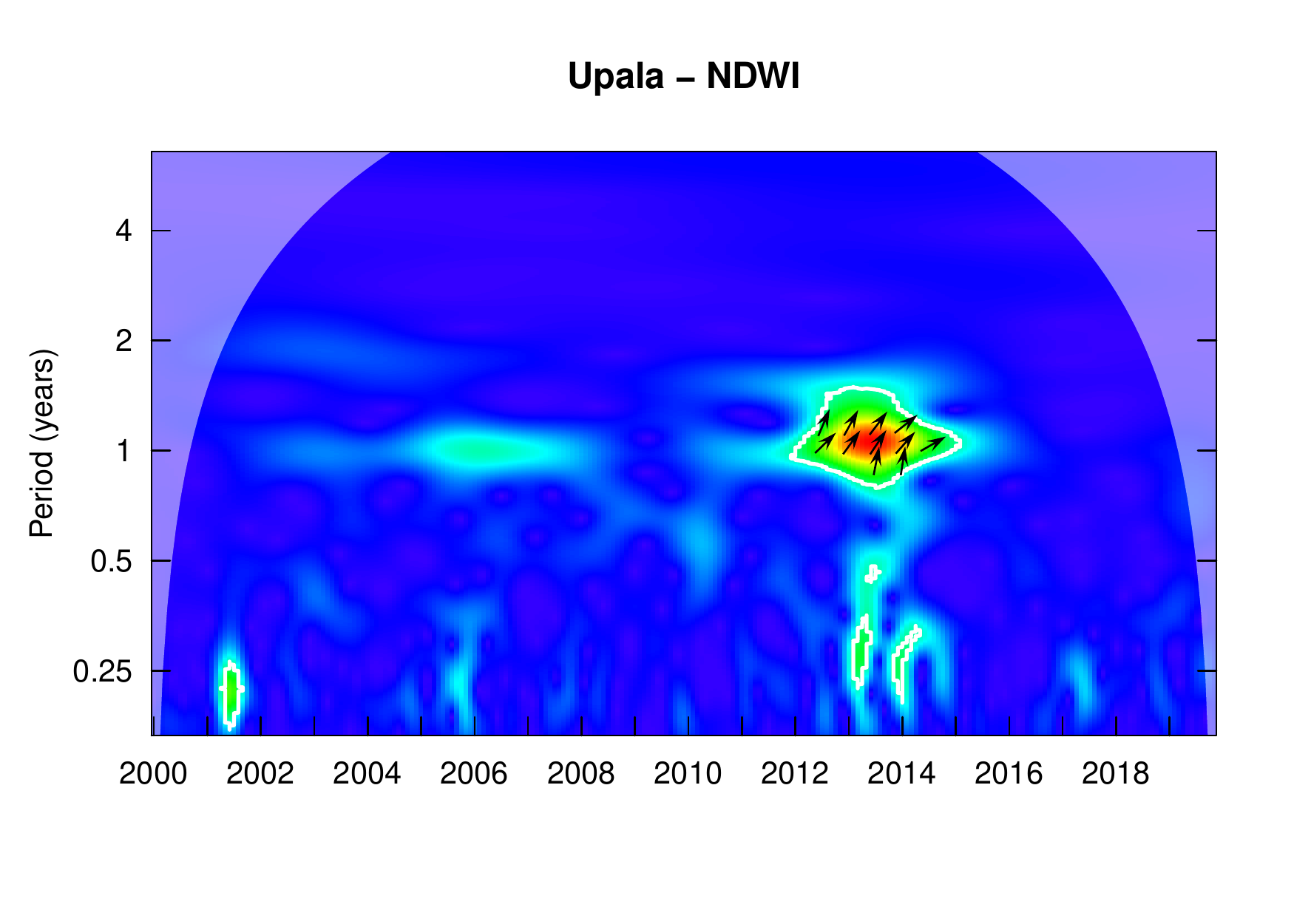}}\vspace{-0.15cm}%
\subfloat[]{\includegraphics[scale=0.23]{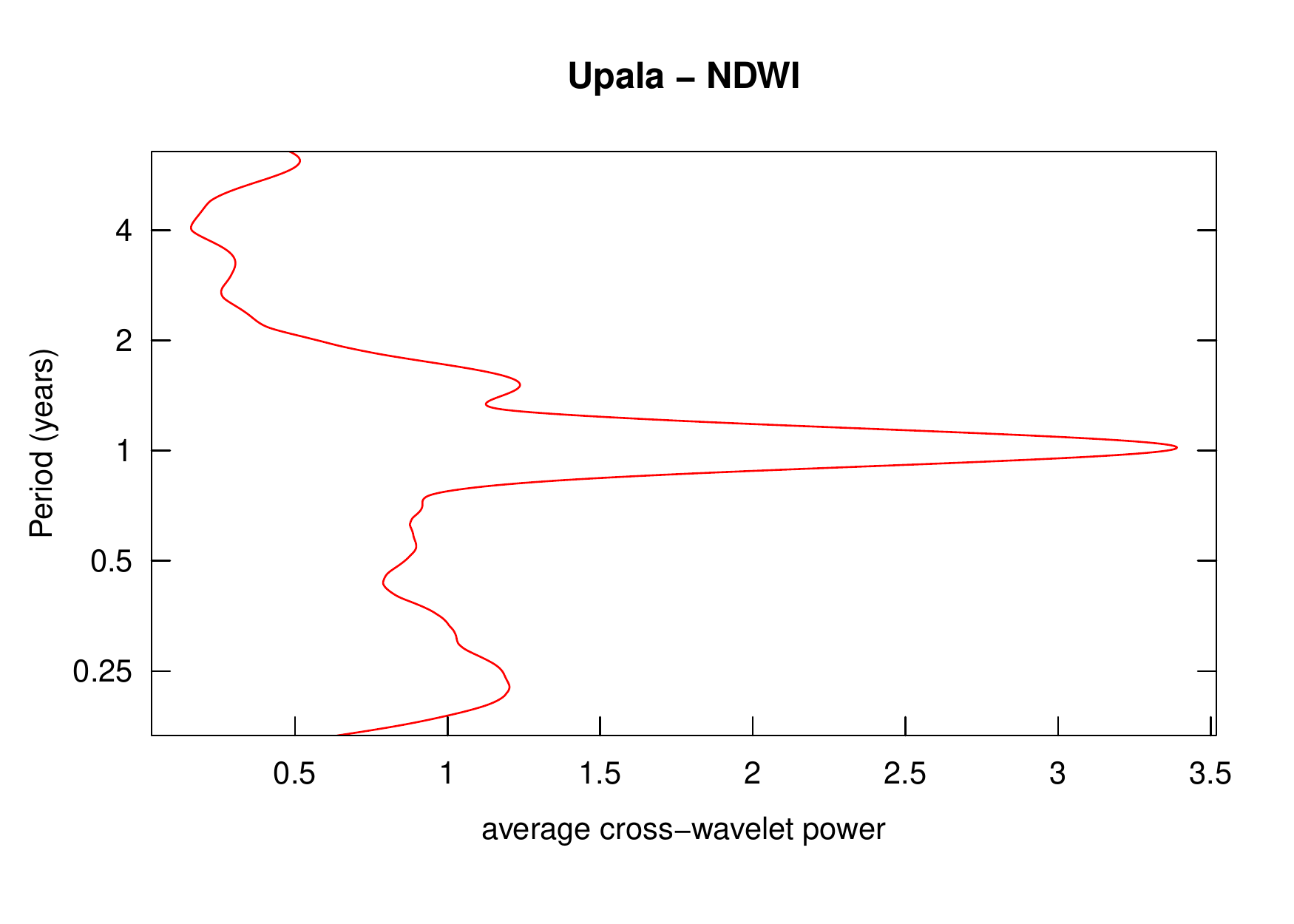}}\vspace{-0.15cm}\\
\subfloat[]{\includegraphics[scale=0.23]{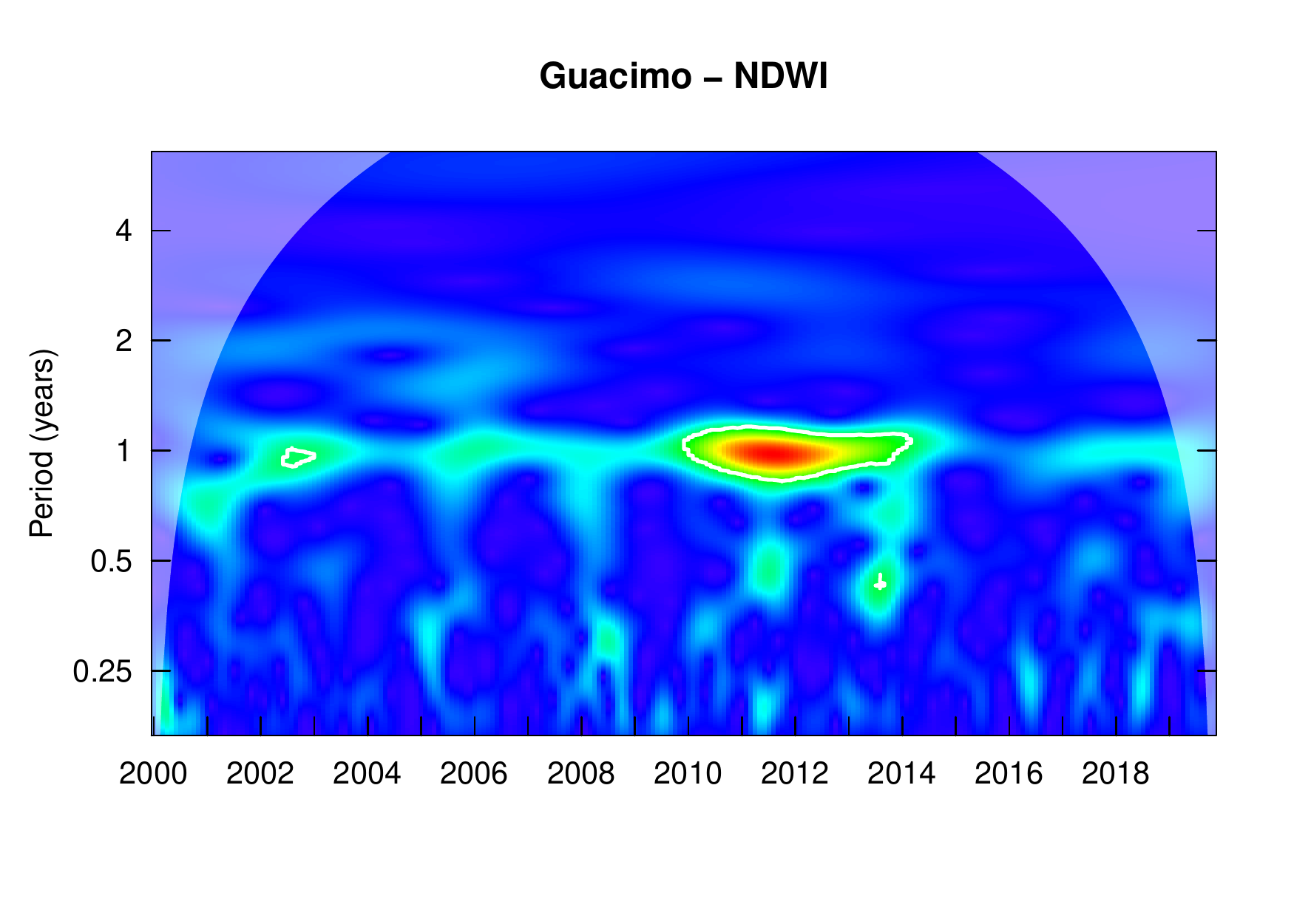}}\vspace{-0.15cm}%
\subfloat[]{\includegraphics[scale=0.23]{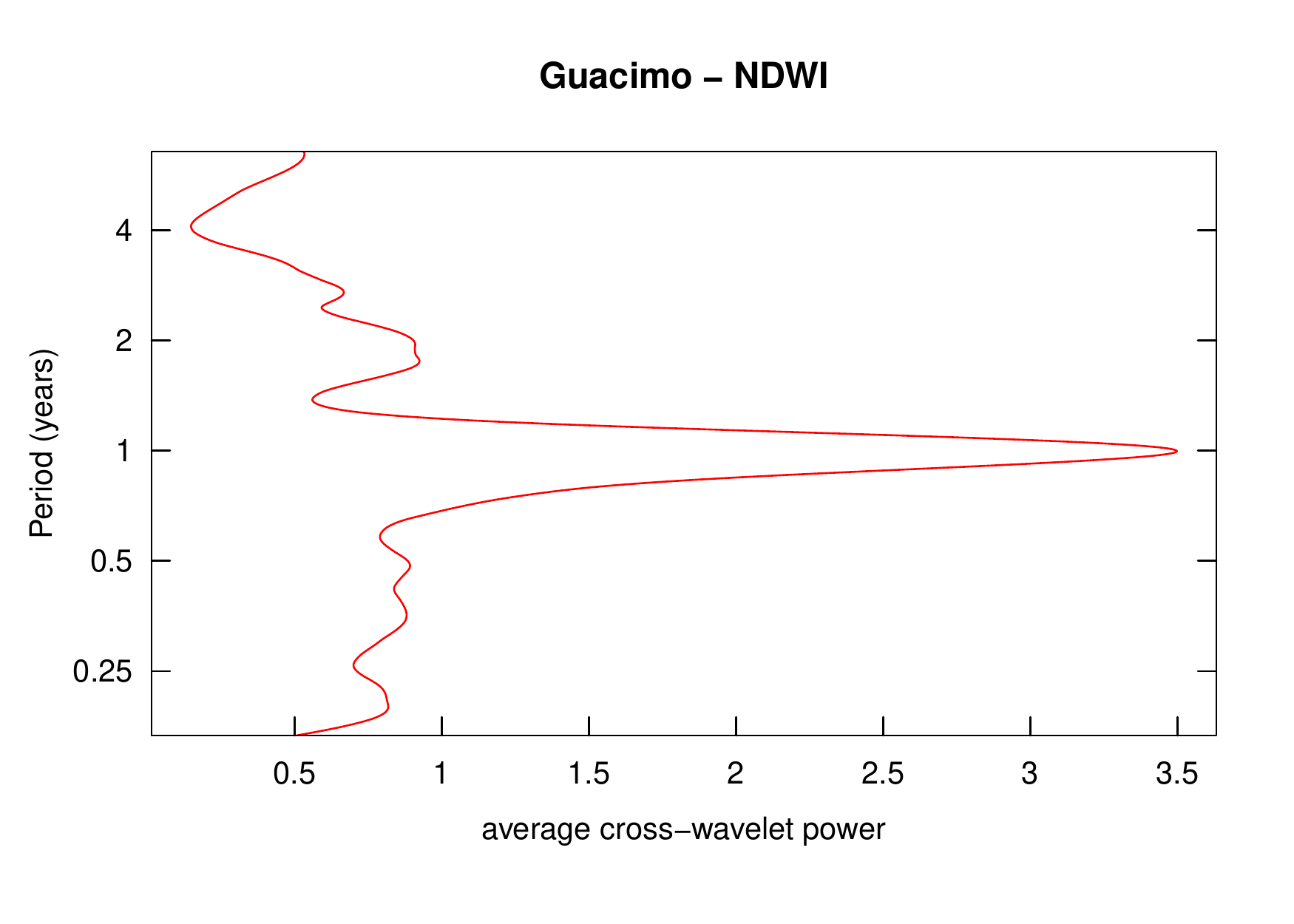}}\vspace{-0.15cm}%
\subfloat[]{\includegraphics[scale=0.23]{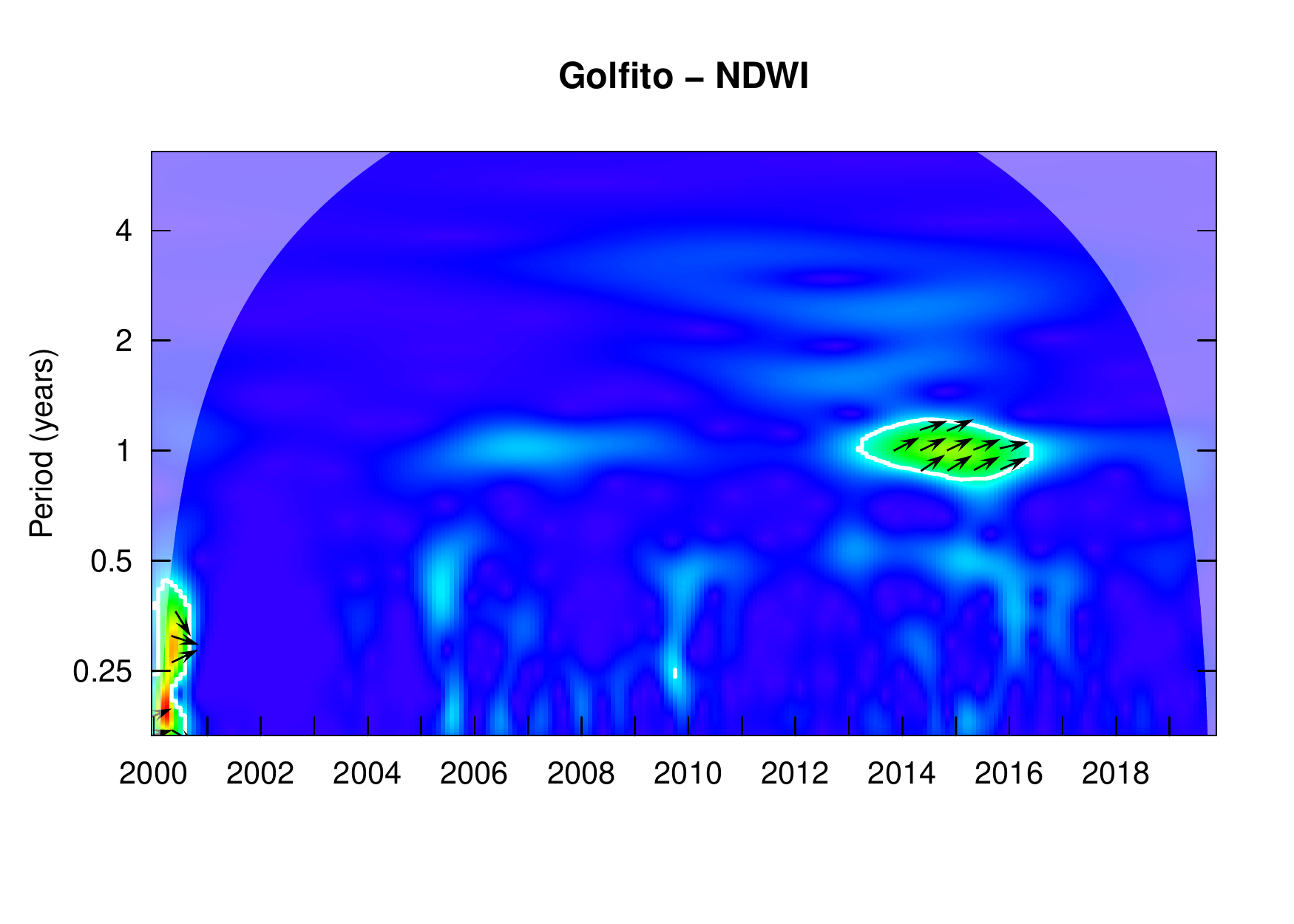}}\vspace{-0.15cm}%
\subfloat[]{\includegraphics[scale=0.23]{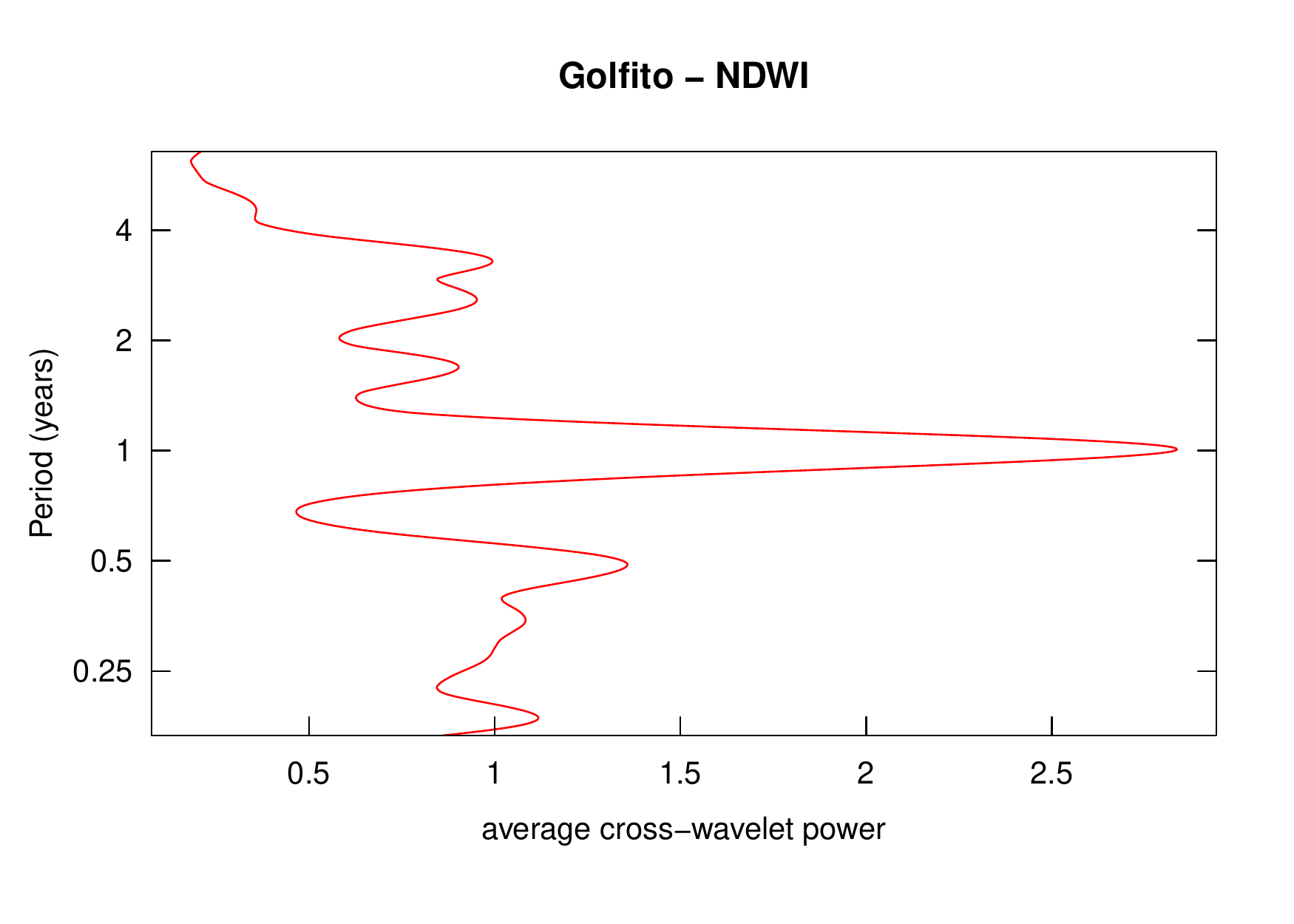}}\vspace{-0.15cm}\\
\subfloat[]{\includegraphics[scale=0.23]{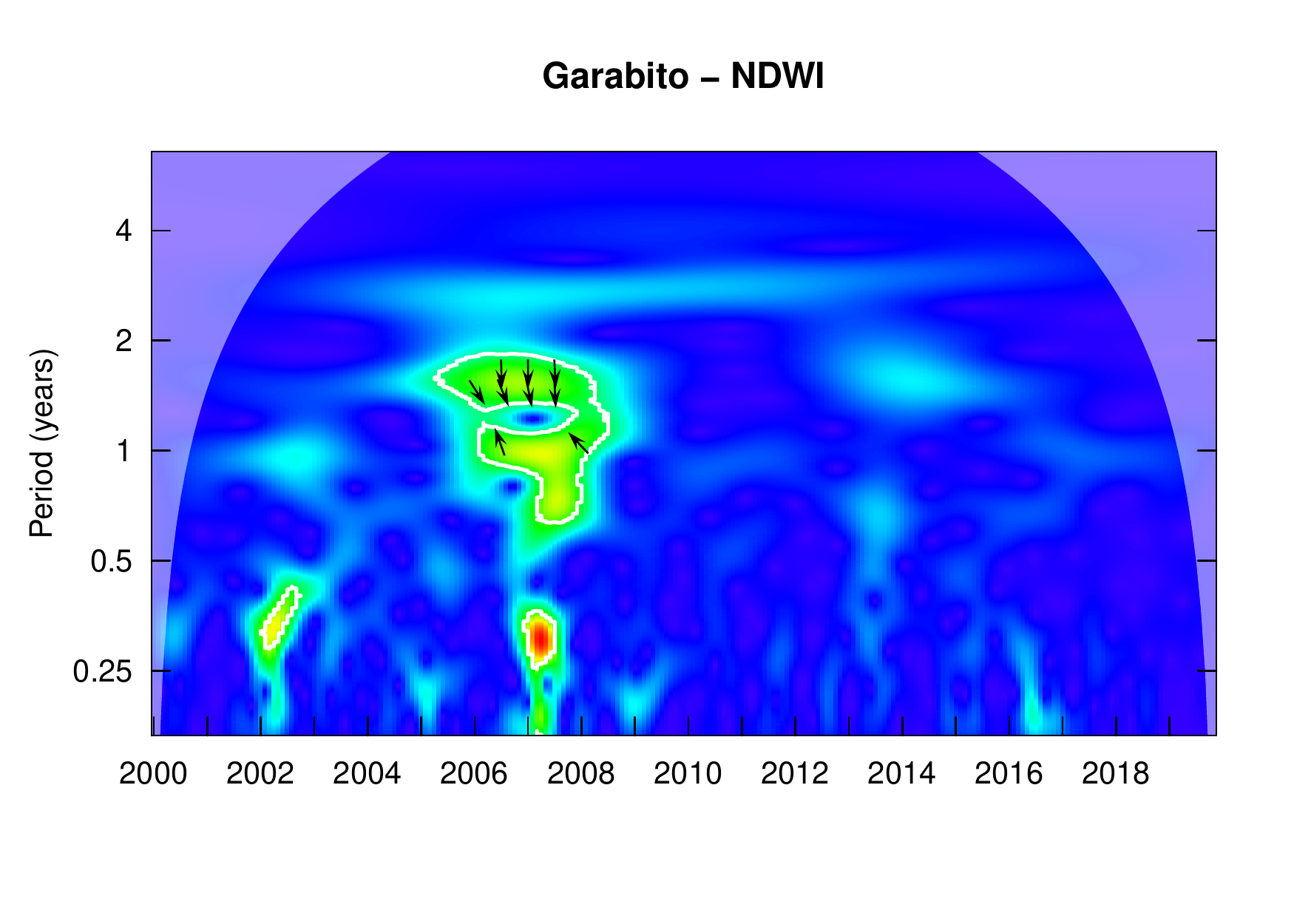}}\vspace{-0.15cm}%
\subfloat[]{\includegraphics[scale=0.23]{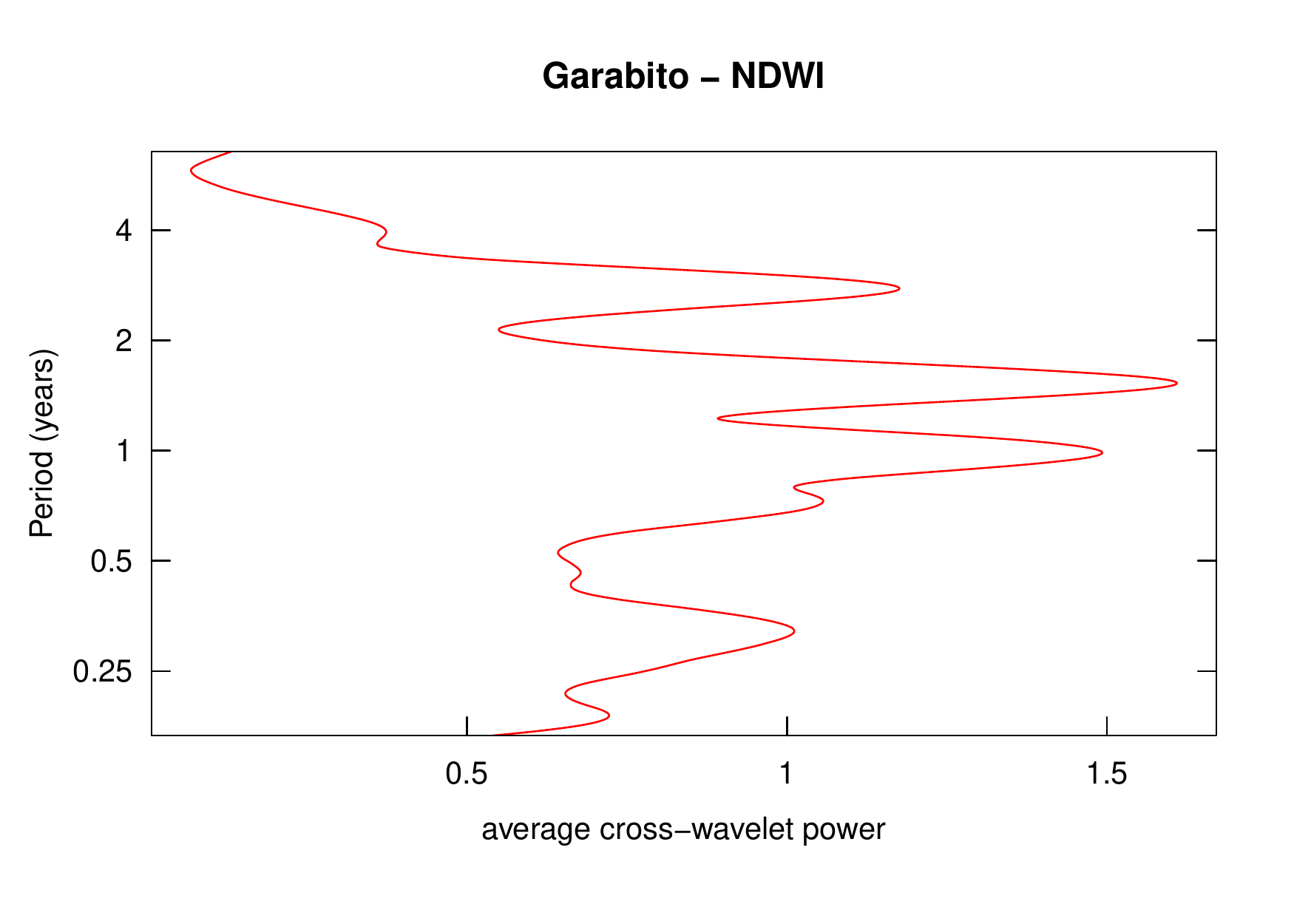}}\vspace{-0.15cm}%
\subfloat[]{\includegraphics[scale=0.23]{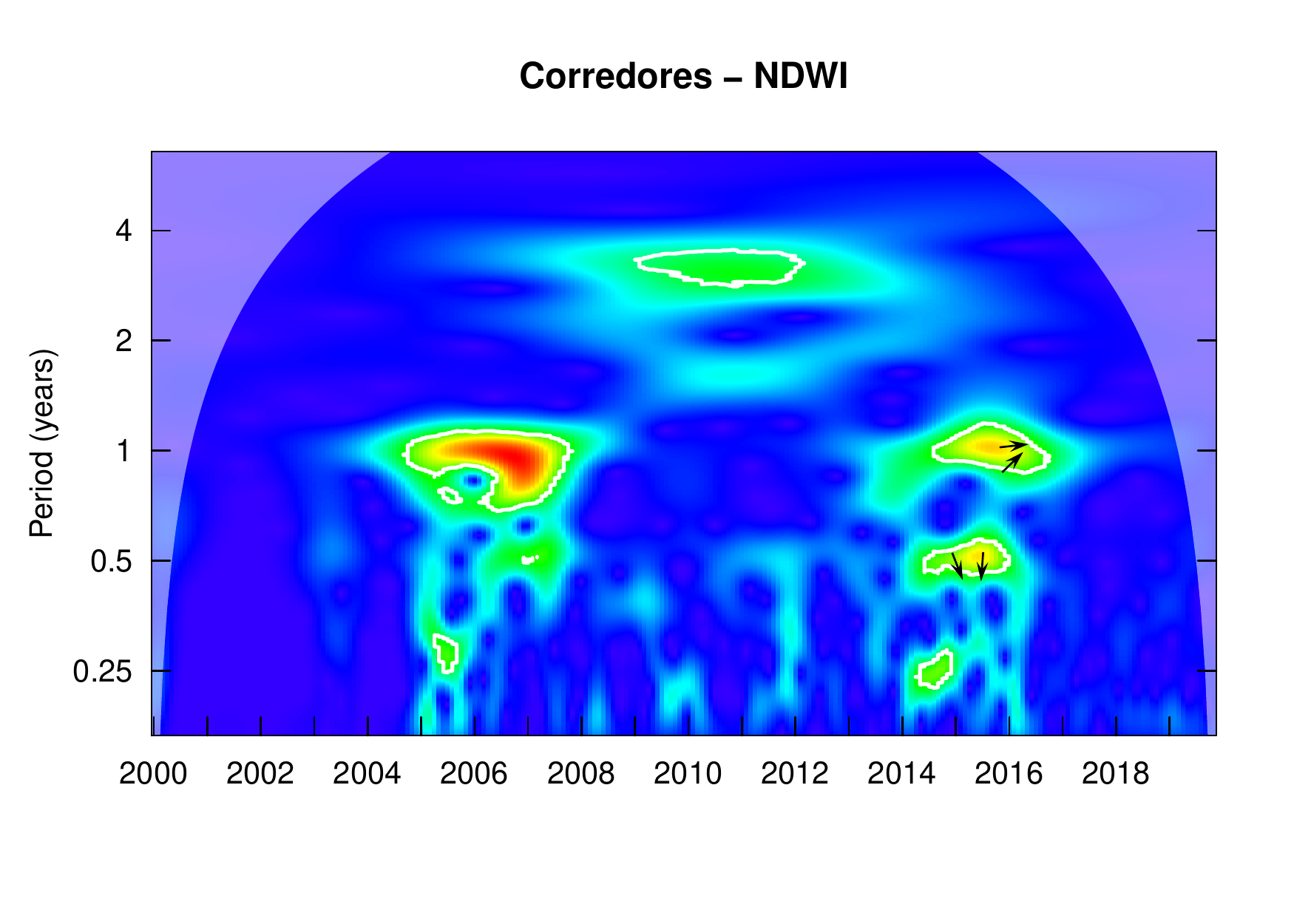}}\vspace{-0.15cm}%
\subfloat[]{\includegraphics[scale=0.23]{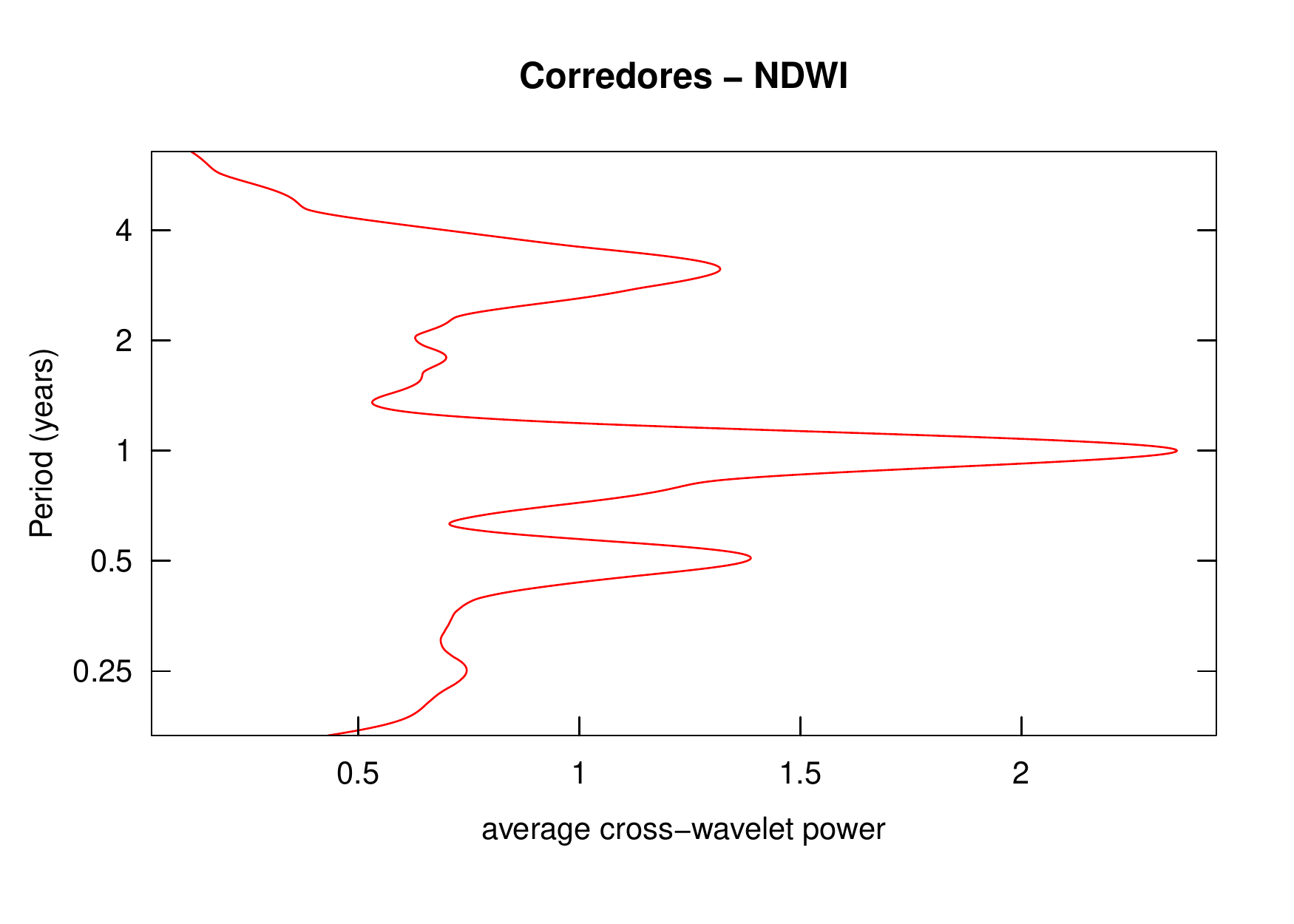}}\vspace{-0.15cm}\\
\caption*{}
\end{figure}

\begin{figure}[H]
\captionsetup[subfigure]{labelformat=empty}
\subfloat[]{\includegraphics[scale=0.23]{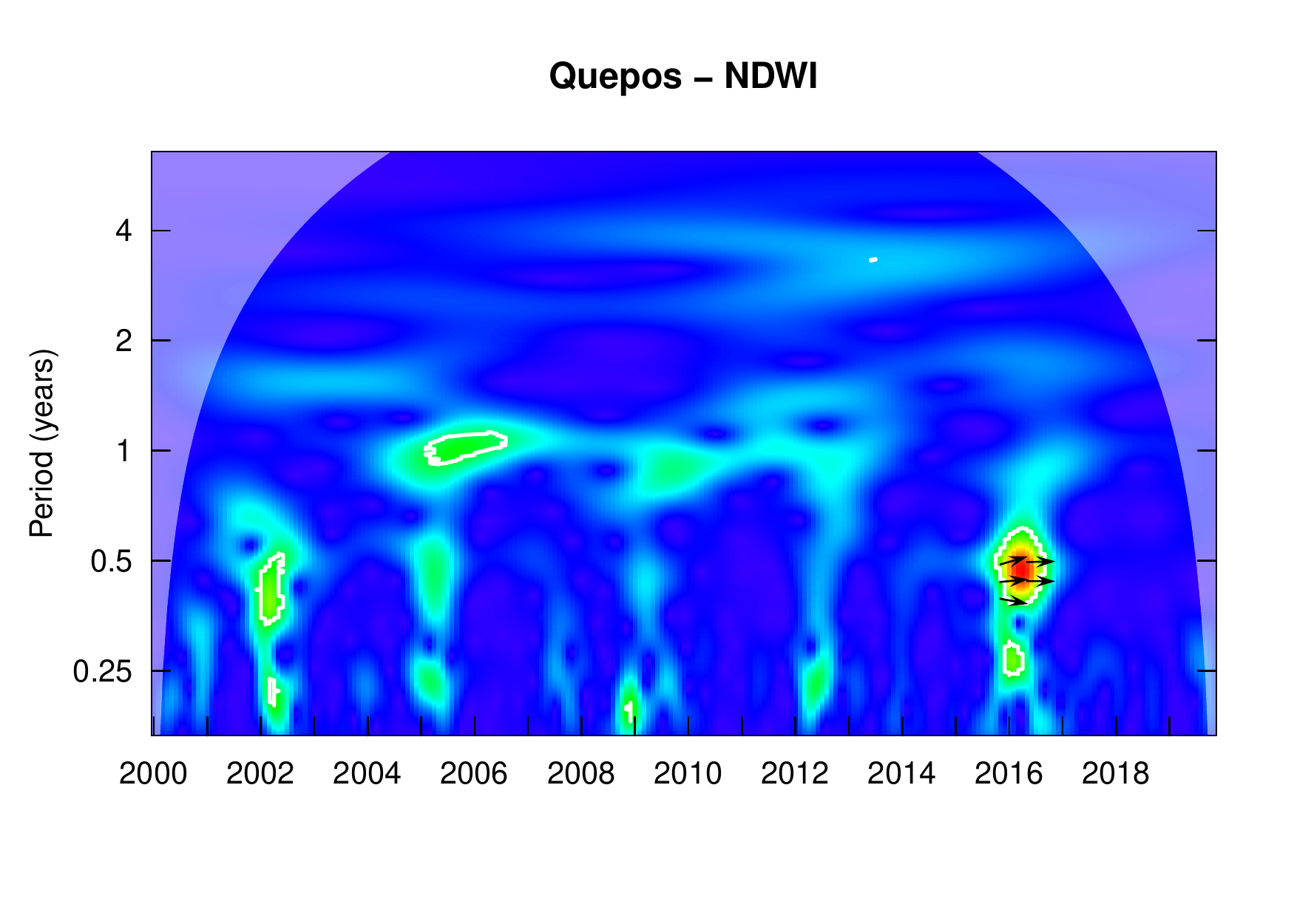}}\vspace{-0.15cm}%
\subfloat[]{\includegraphics[scale=0.23]{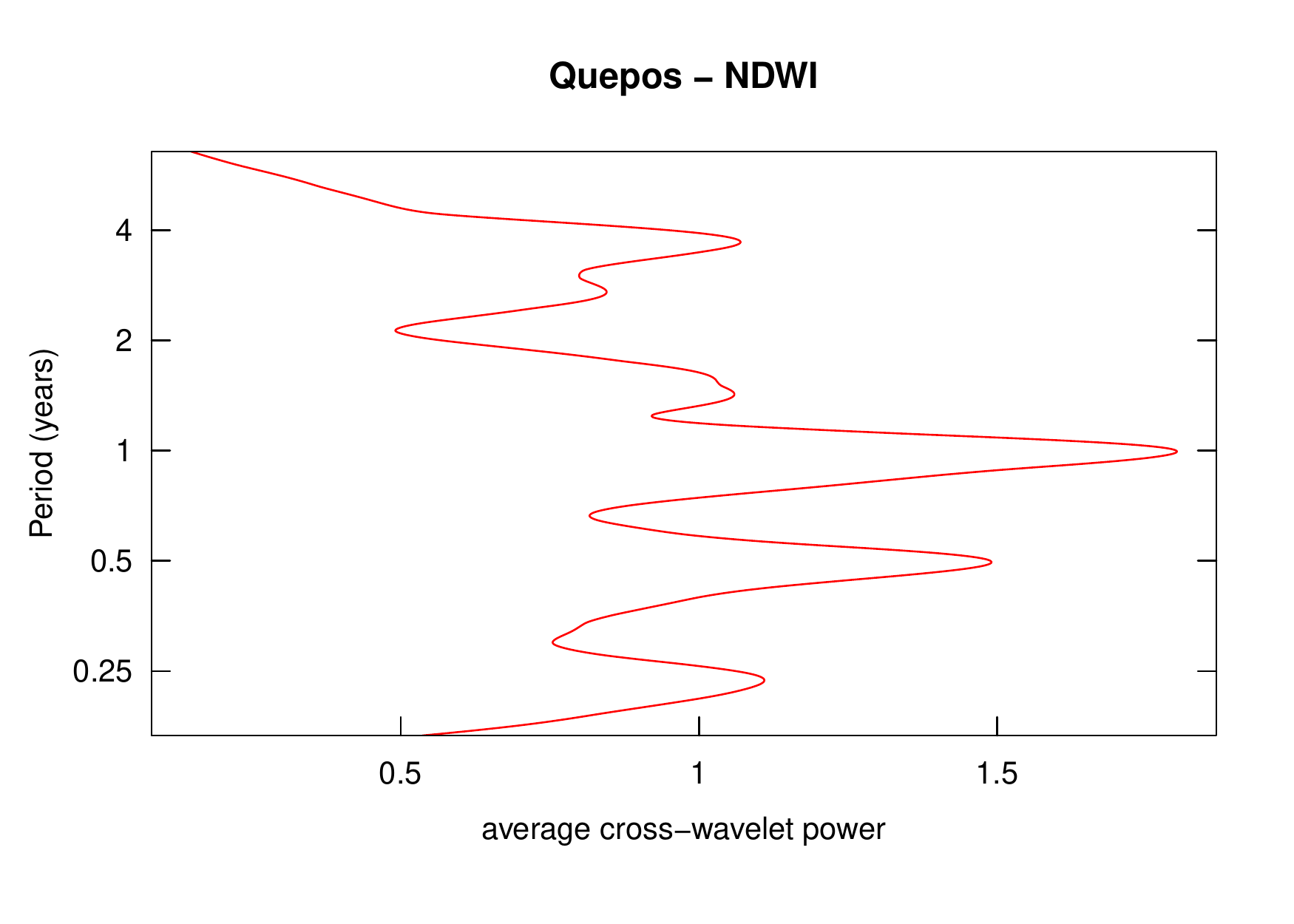}}\vspace{-0.15cm}%
\subfloat[]{\includegraphics[scale=0.23]{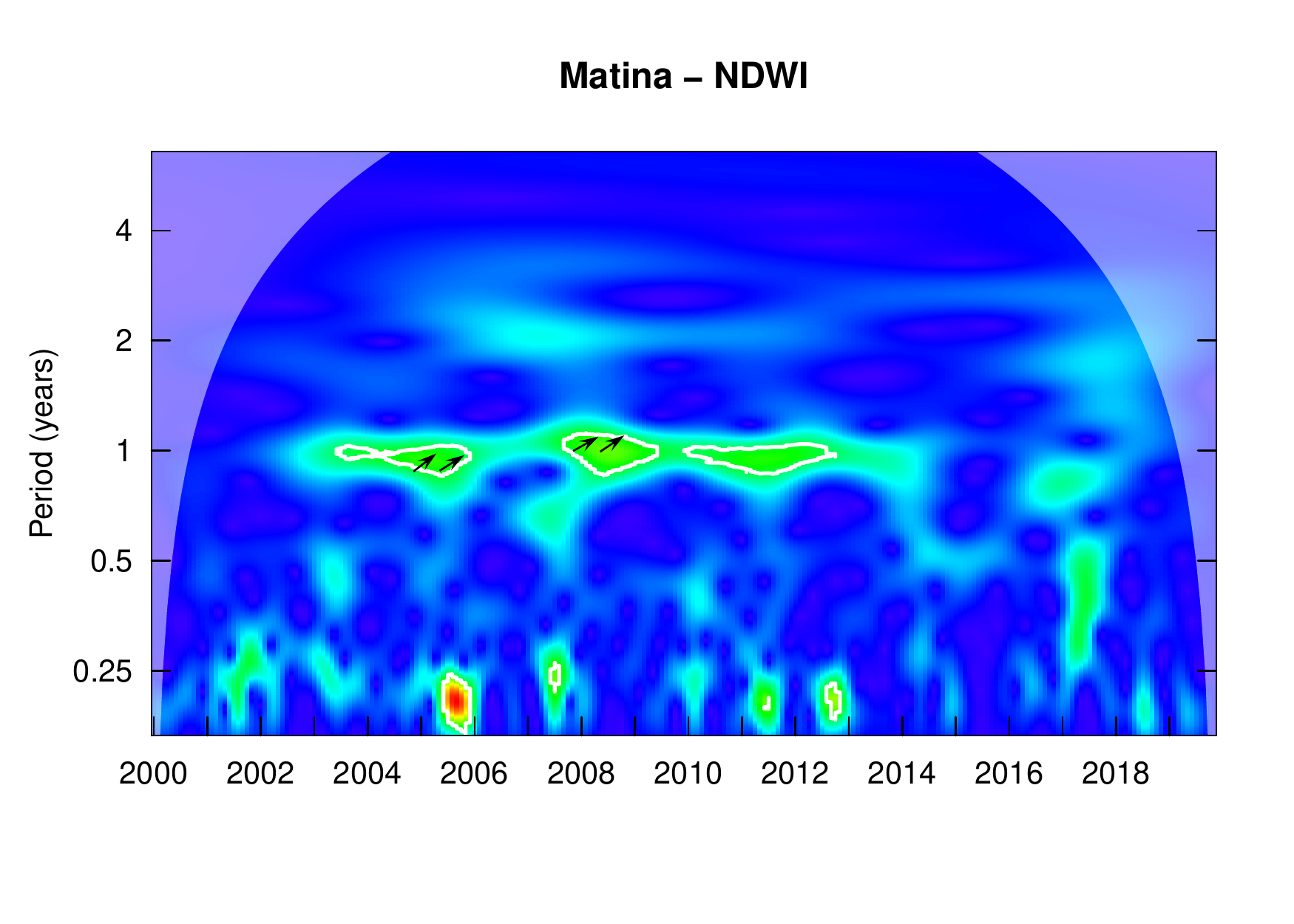}}\vspace{-0.15cm}%
\subfloat[]{\includegraphics[scale=0.23]{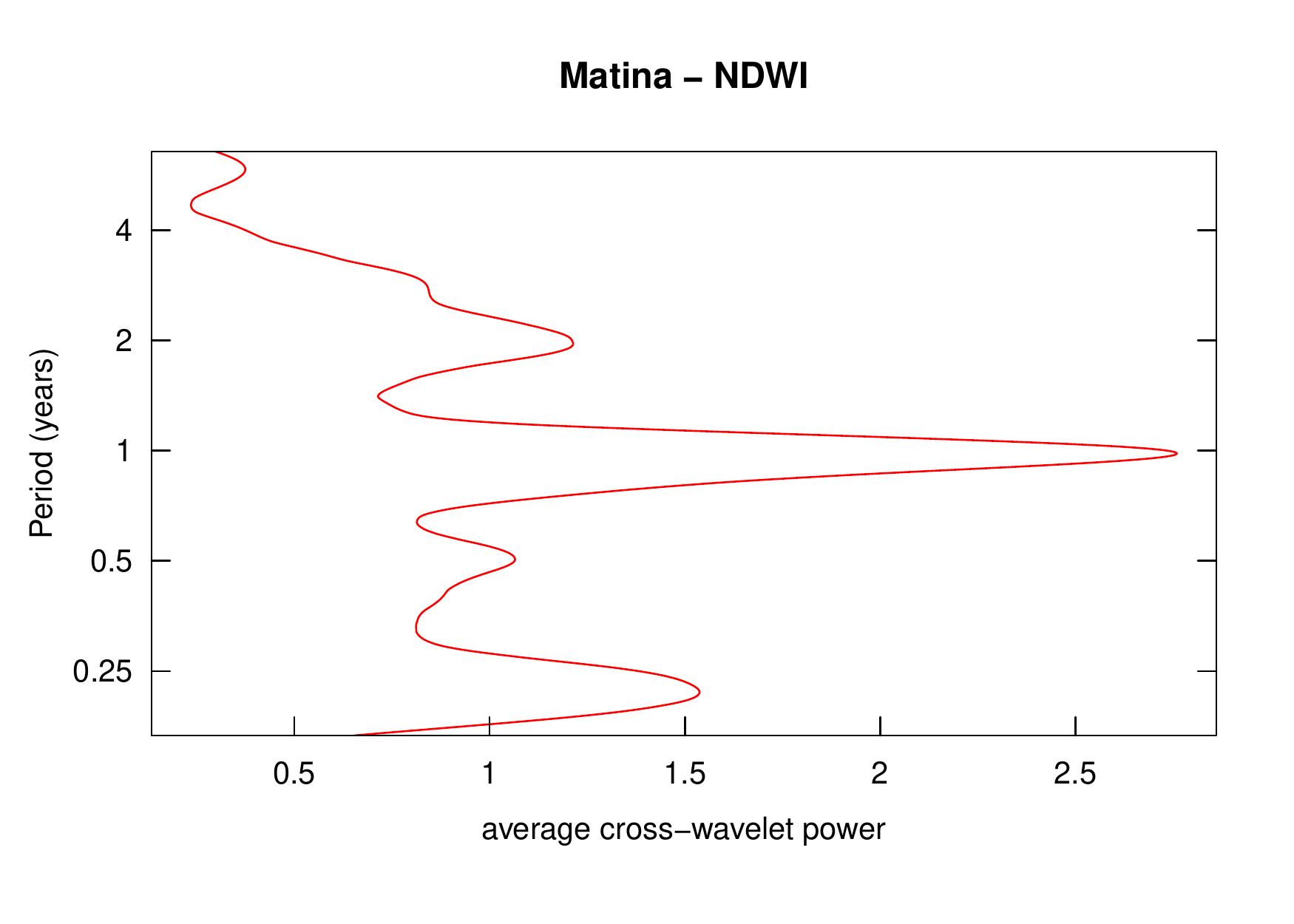}}\vspace{-0.15cm}\\
\subfloat[]{\includegraphics[scale=0.23]{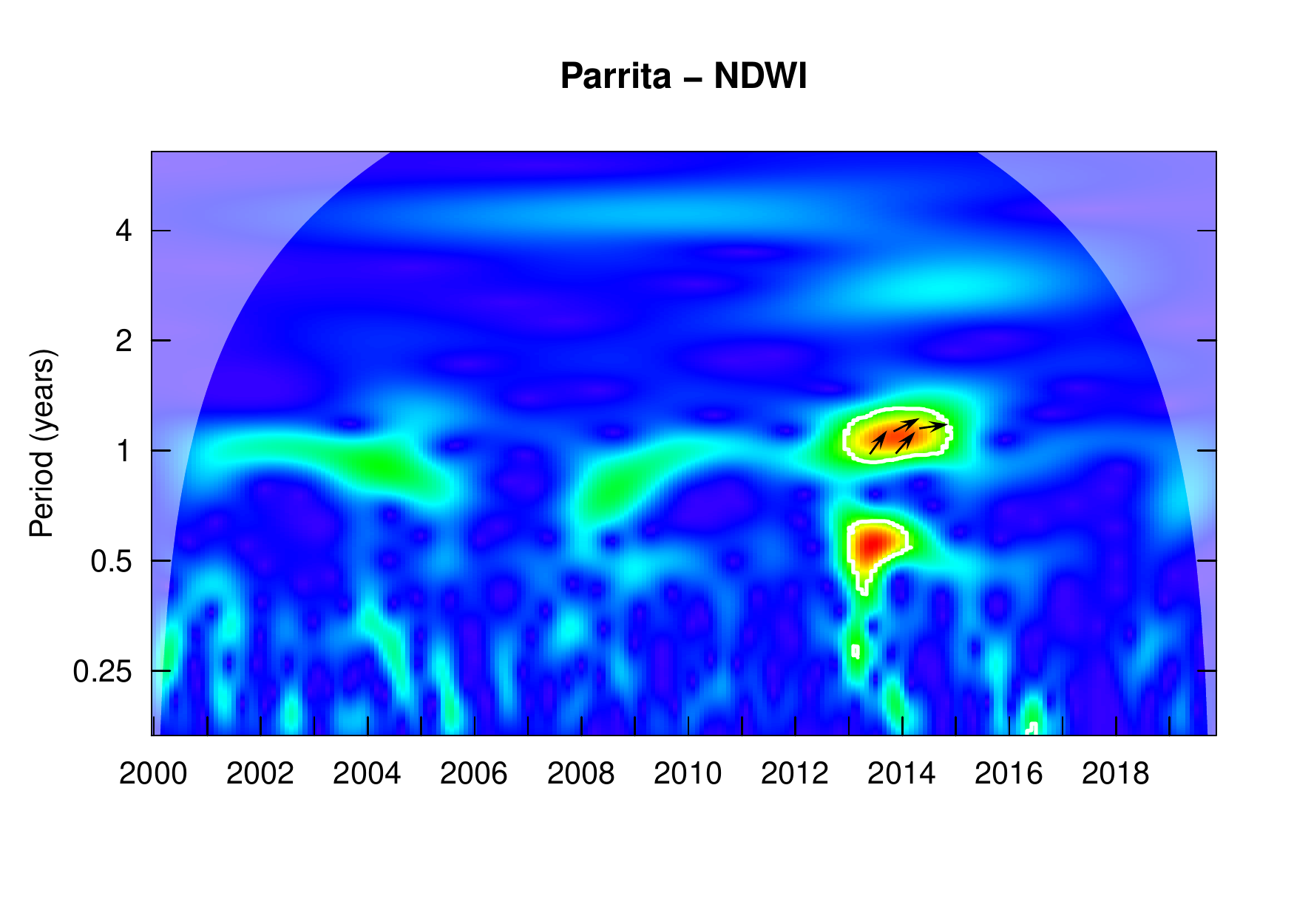}}\vspace{-0.15cm}%
\subfloat[]{\includegraphics[scale=0.23]{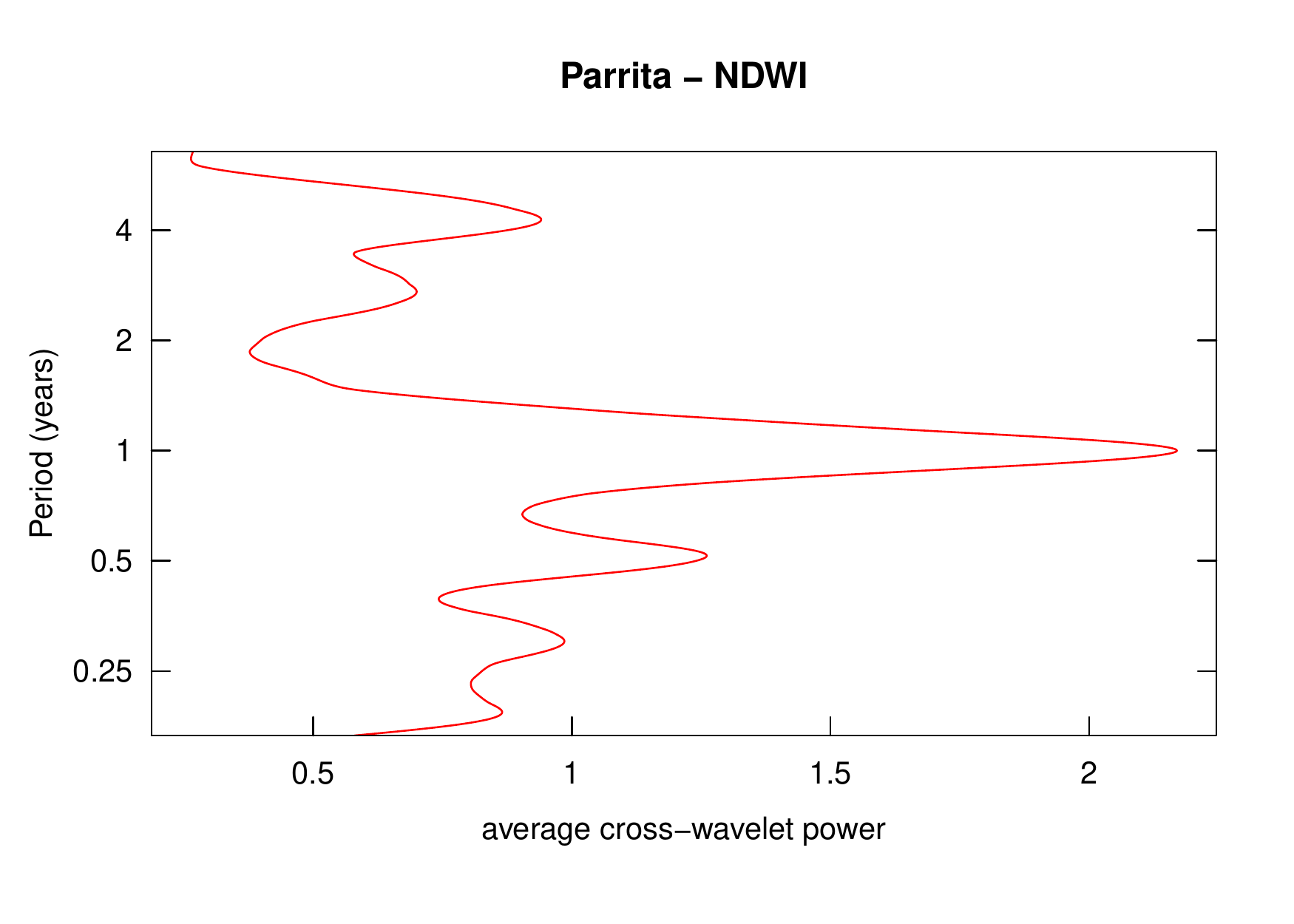}}\vspace{-0.15cm}%
\subfloat[]{\includegraphics[scale=0.23]{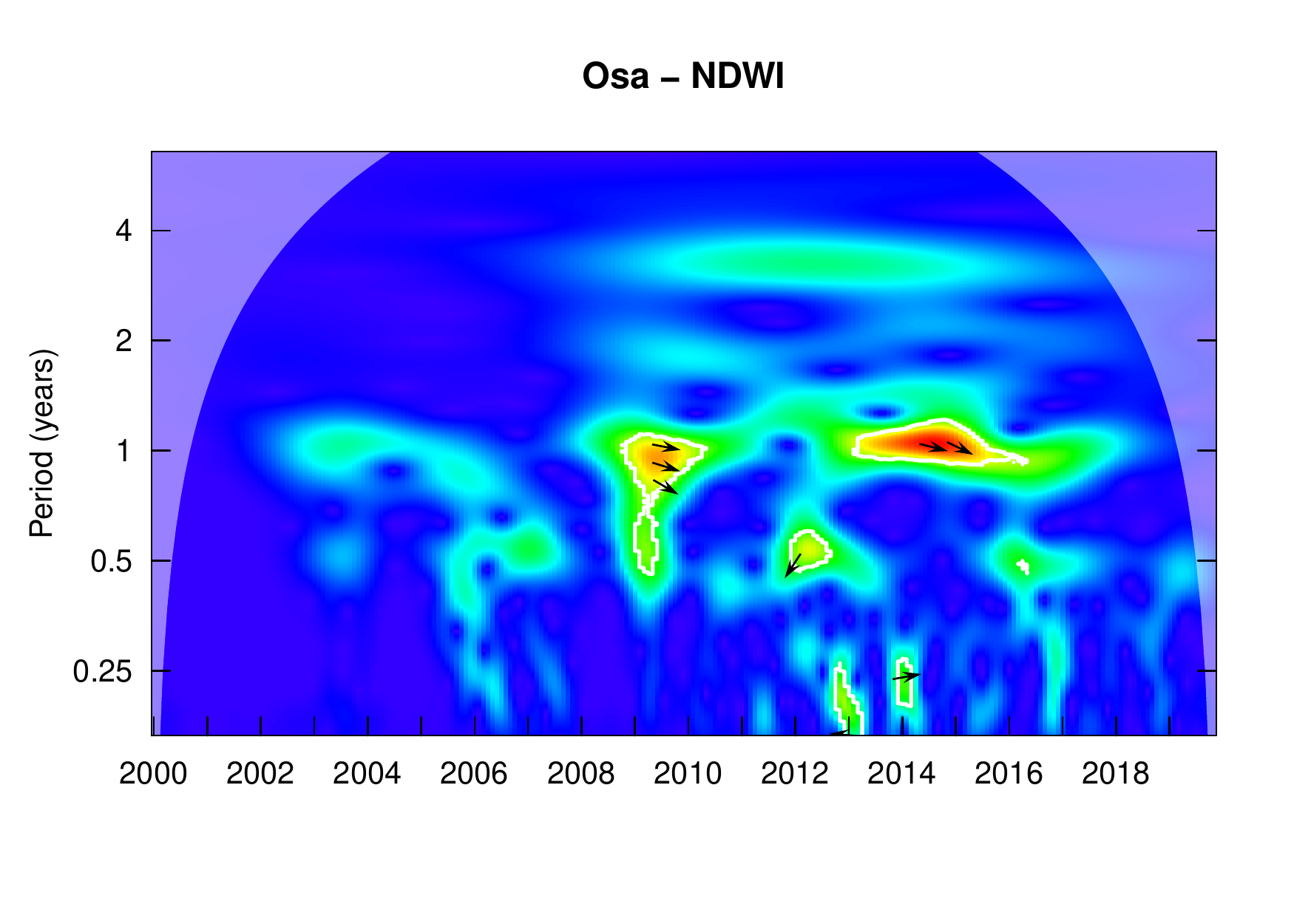}}\vspace{-0.15cm}%
\subfloat[]{\includegraphics[scale=0.23]{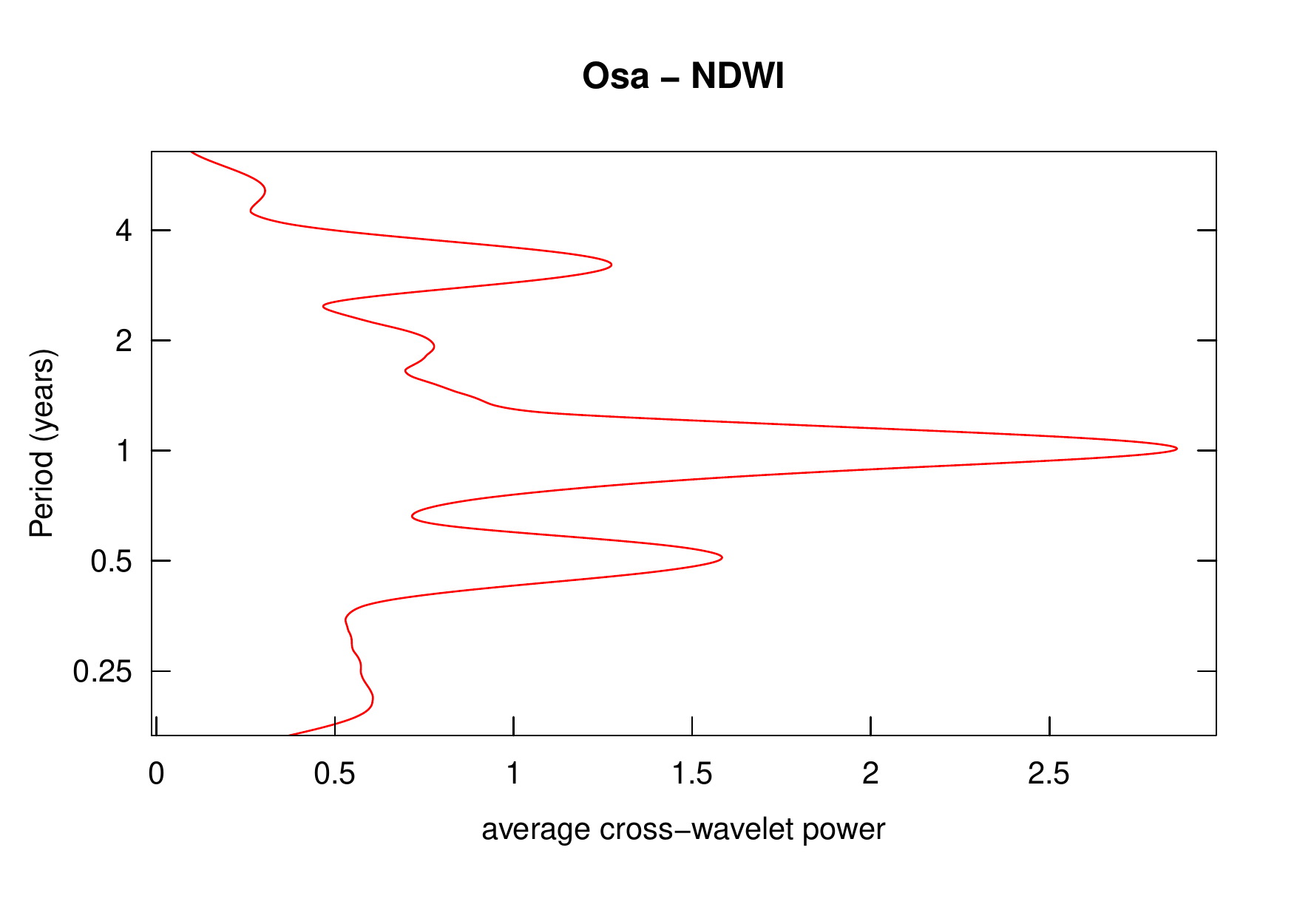}}\vspace{-0.15cm}
\caption*{}
\end{figure}

\section*{Wavelet coherence and average cross-wavelet power between dengue incidence and LSD}

\begin{figure}[H]
\captionsetup[subfigure]{labelformat=empty}
\caption*{\textbf{Figure S4:} Wavelet coherence (color map) between dengue incidence from 2000 to 2019, and LSD in 32 municipalities of Costa Rica (periodicity on y-axis, time on x-axis). Colors code for increasing power intensity, from blue to red; $95\%$ confidence levels are encircled by white lines, and shaded areas indicate the presence of significant edge effects. On the right side of each wavelet coherence is the average cross-wavelet power (Red line). The arrows indicate whether the two series are in-phase or out-phase.}
\subfloat[]{\includegraphics[scale=0.23]{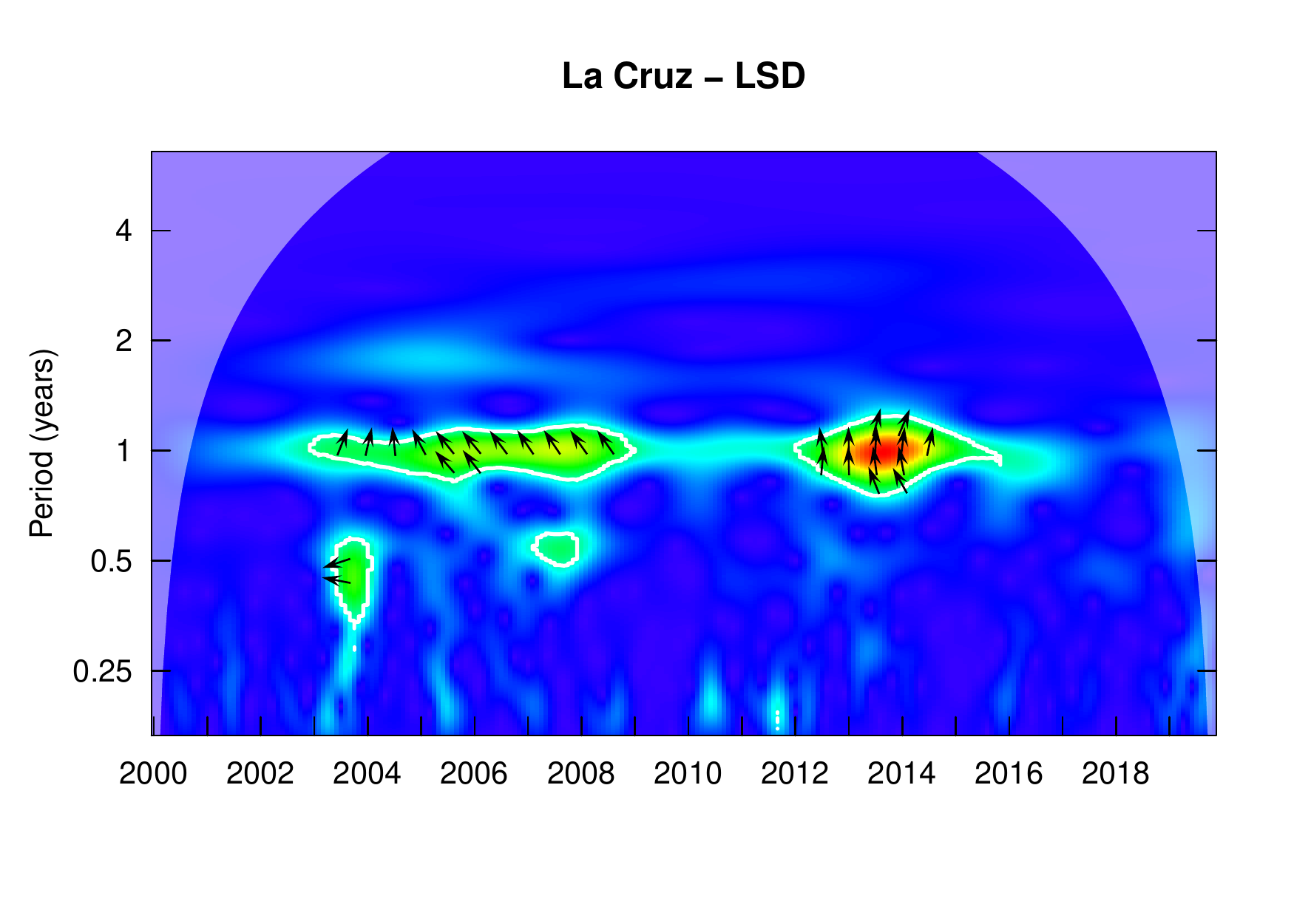}}\vspace{-0.15cm}%
\subfloat[]{\includegraphics[scale=0.23]{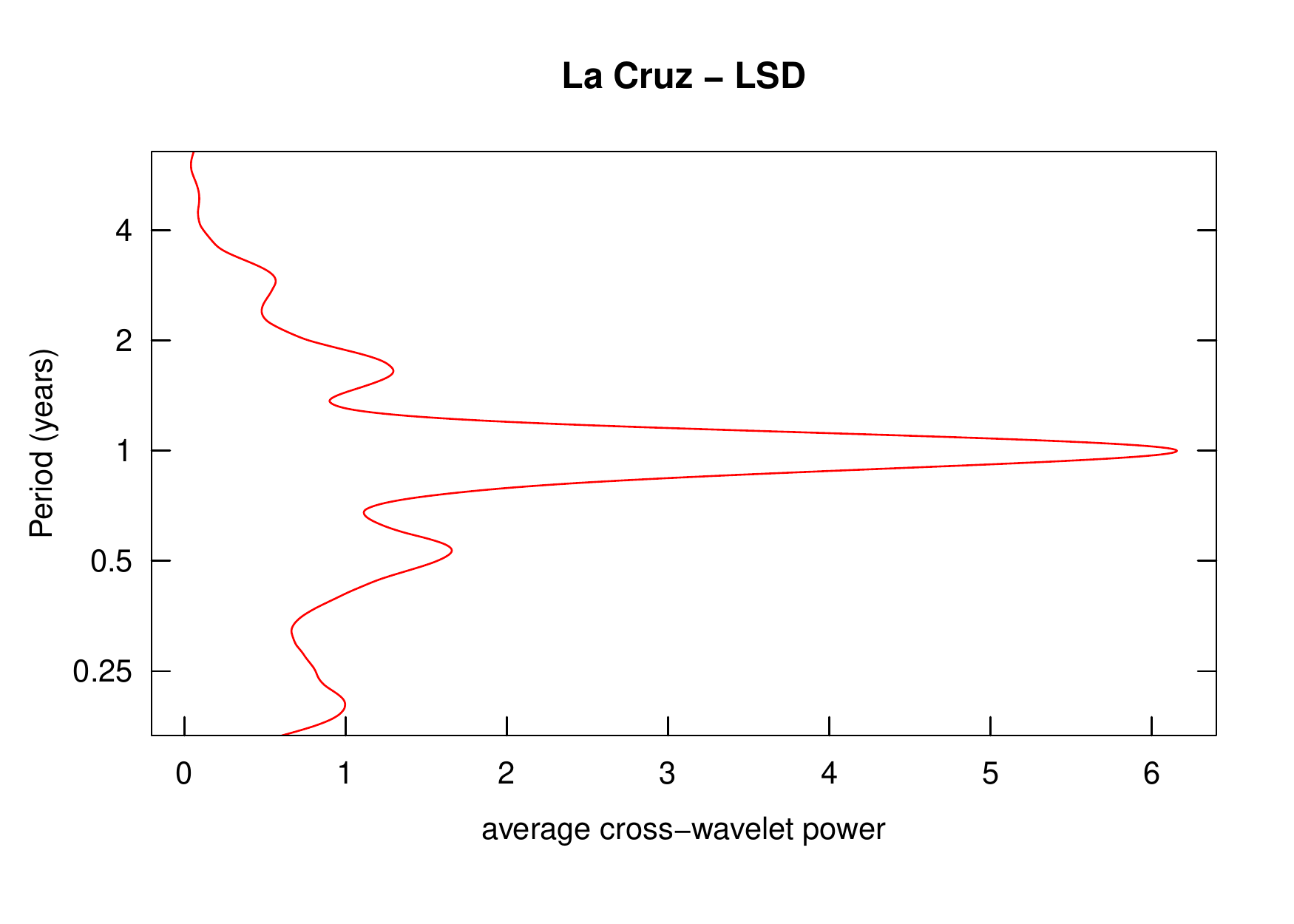}}\vspace{-0.15cm}%
\subfloat[]{\includegraphics[scale=0.23]{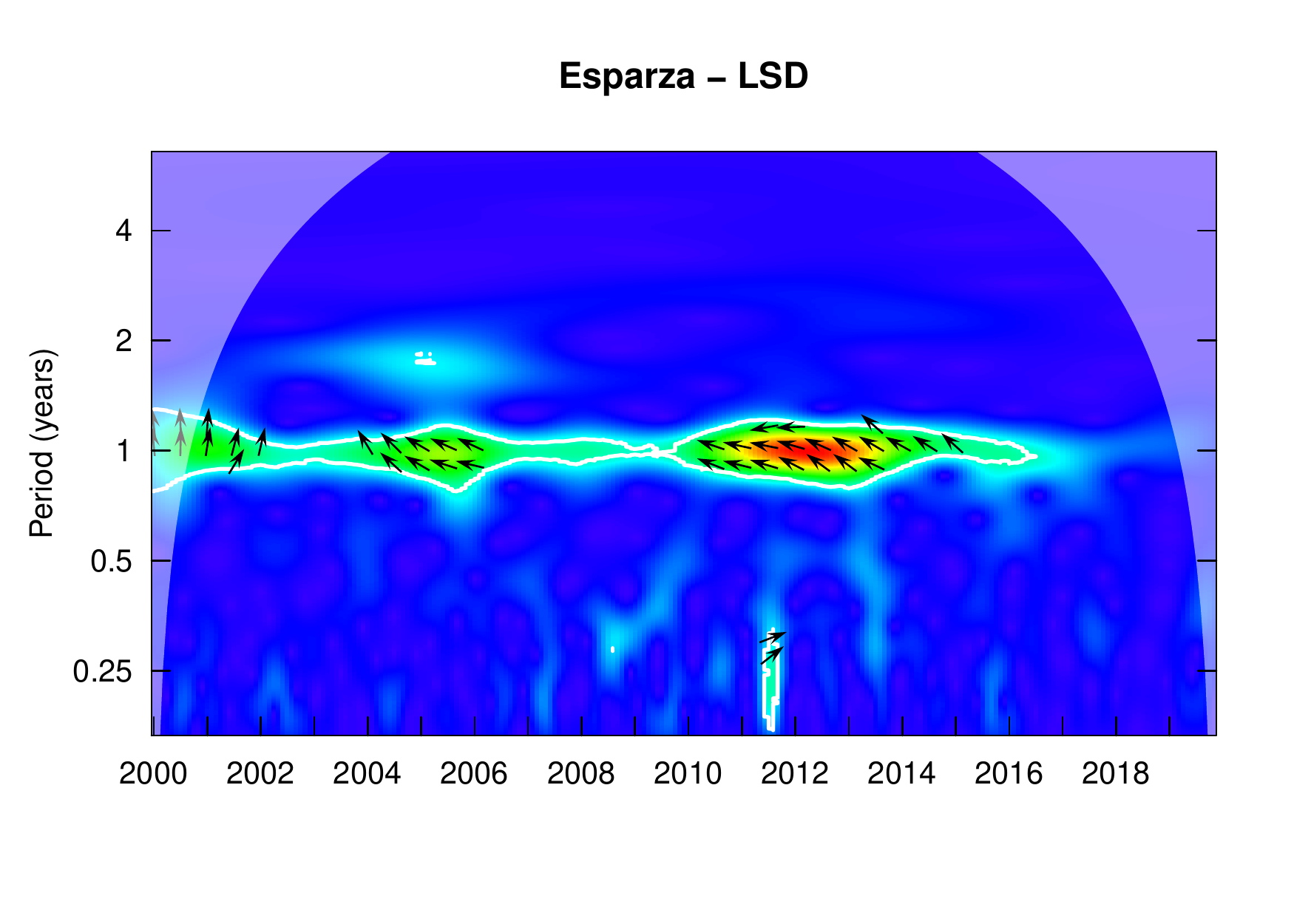}}\vspace{-0.15cm}%
\subfloat[]{\includegraphics[scale=0.23]{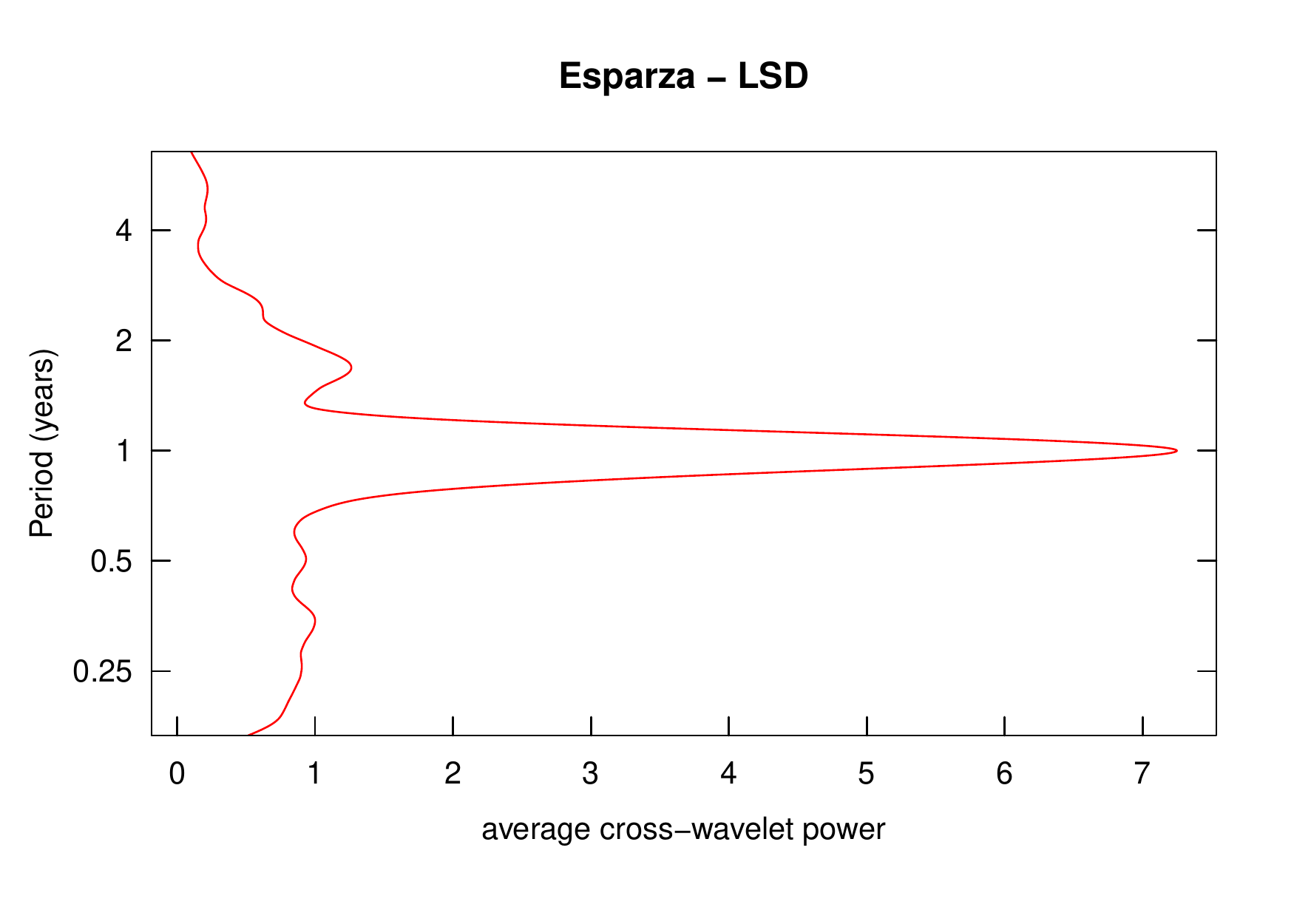}}\vspace{-0.15cm}\\
\subfloat[]{\includegraphics[scale=0.23]{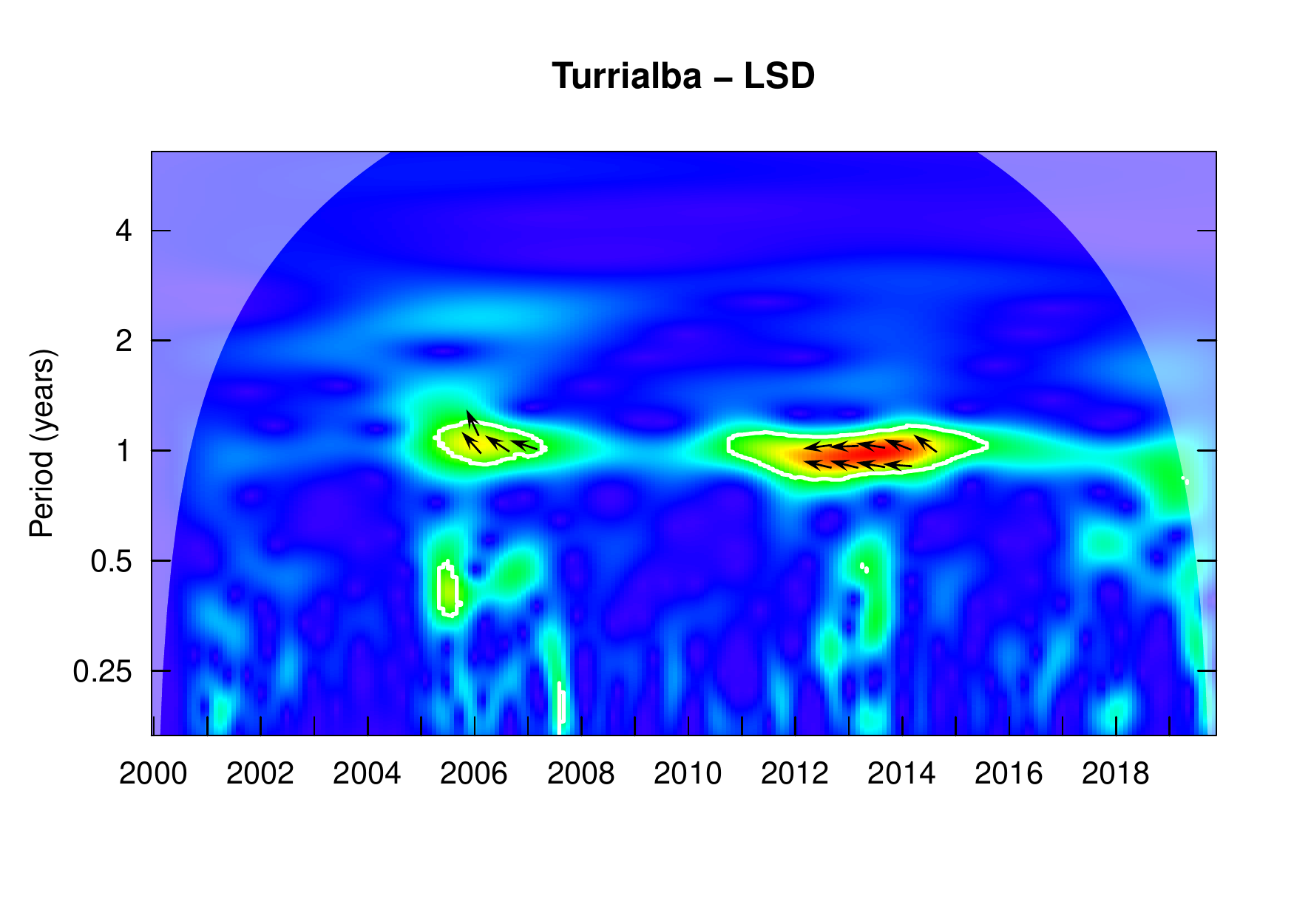}}\vspace{-0.15cm}%
\subfloat[]{\includegraphics[scale=0.23]{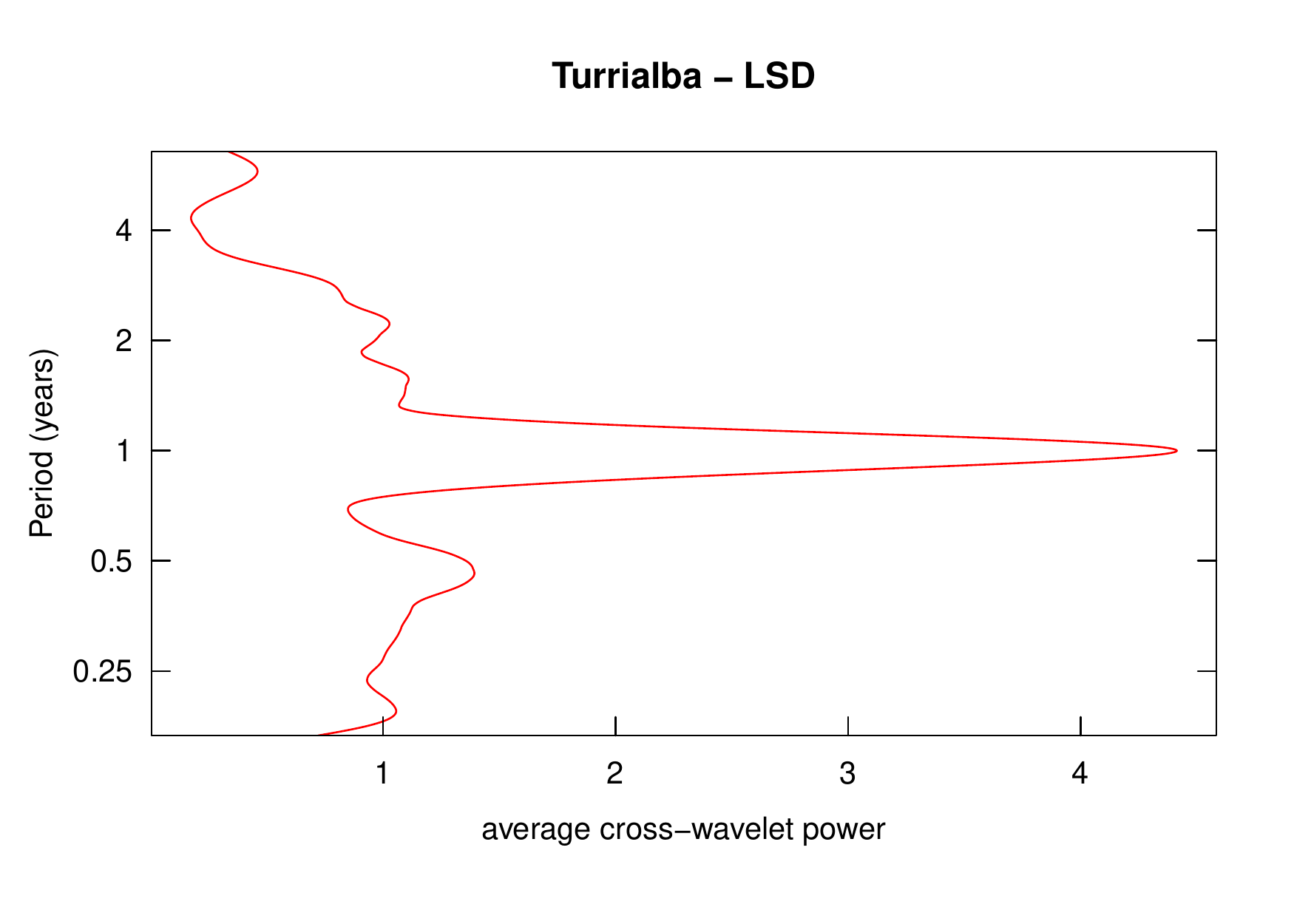}}\vspace{-0.15cm}%
\subfloat[]{\includegraphics[scale=0.23]{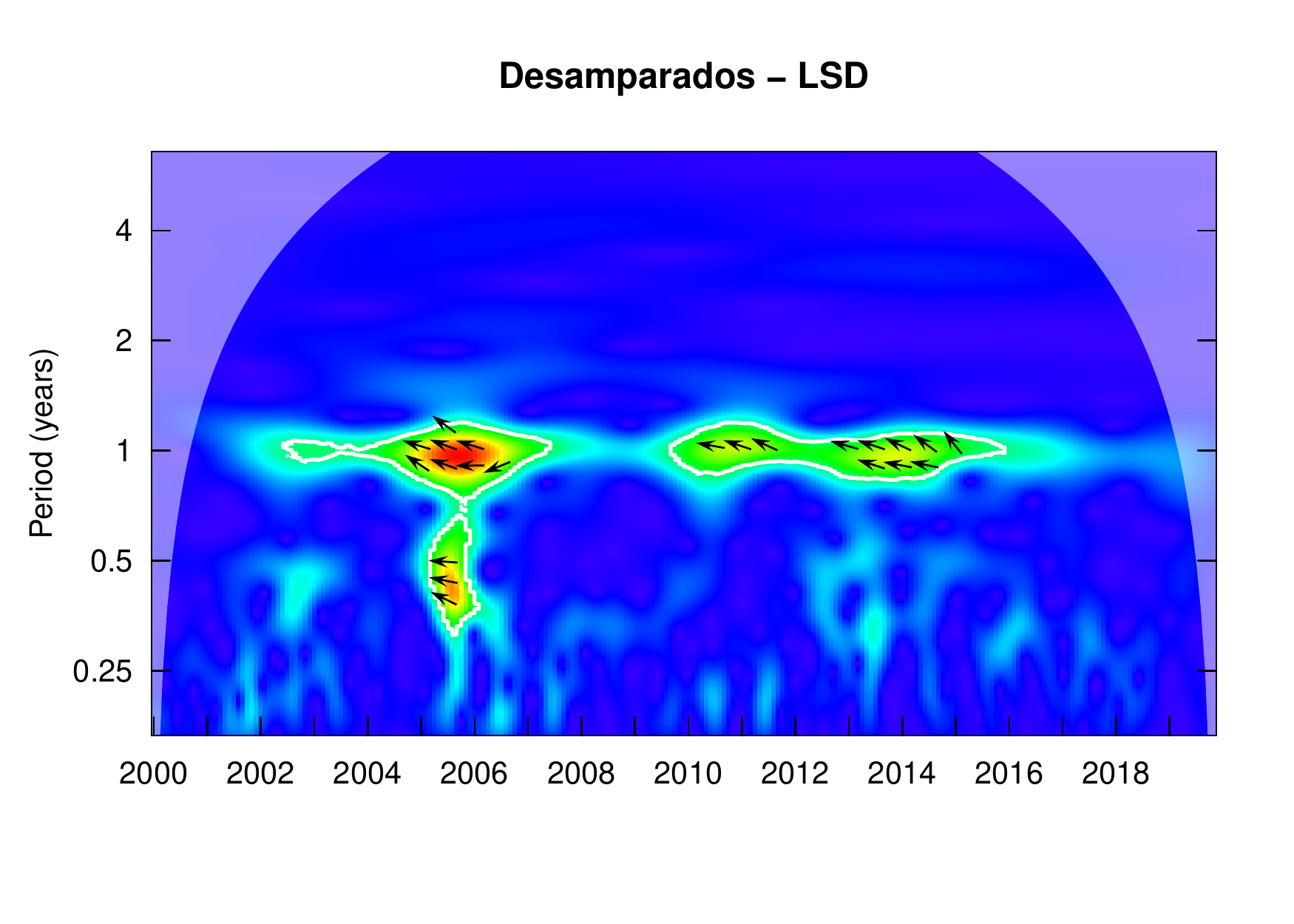}}\vspace{-0.15cm}%
\subfloat[]{\includegraphics[scale=0.23]{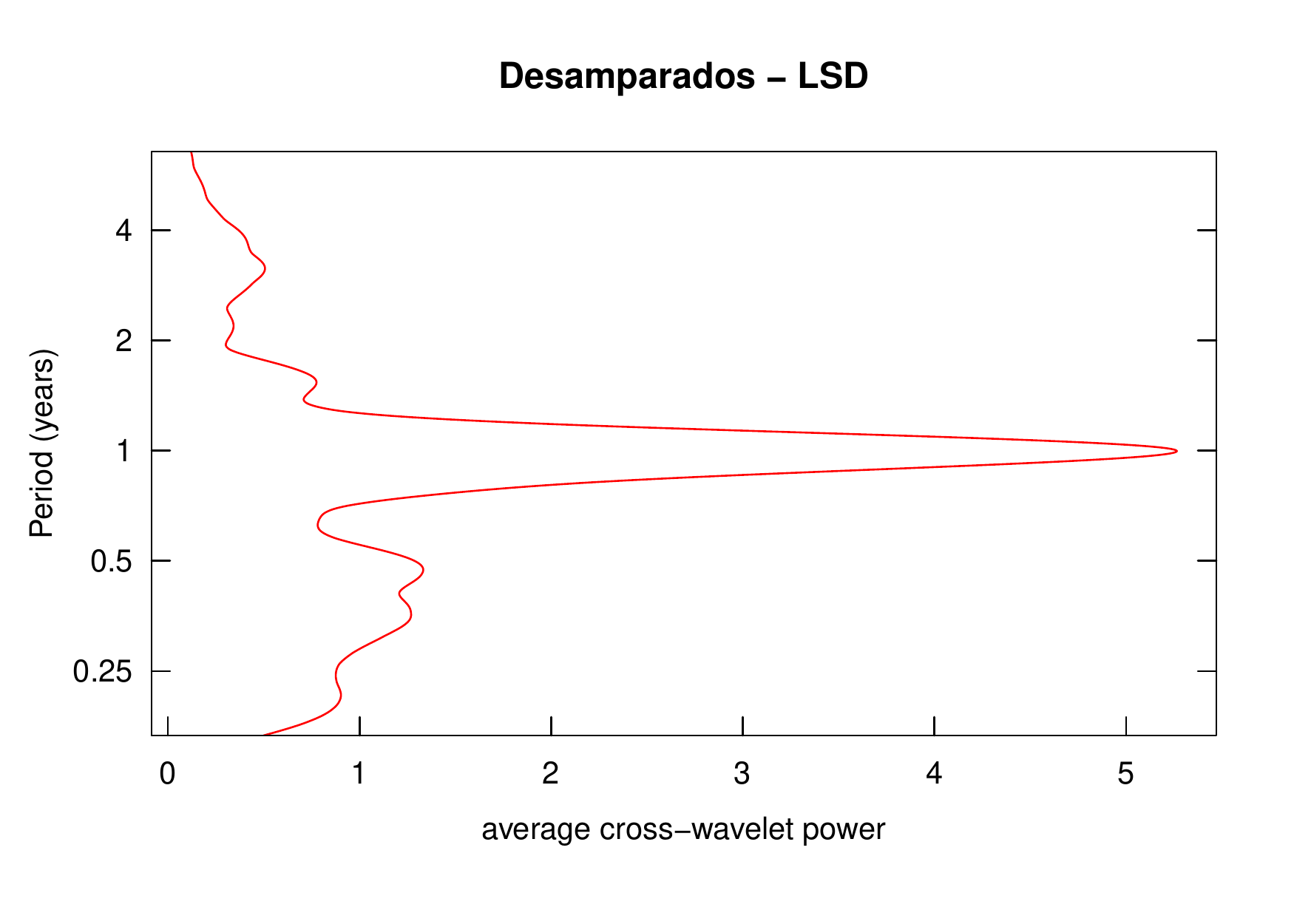}}\vspace{-0.15cm}\\
\subfloat[]{\includegraphics[scale=0.23]{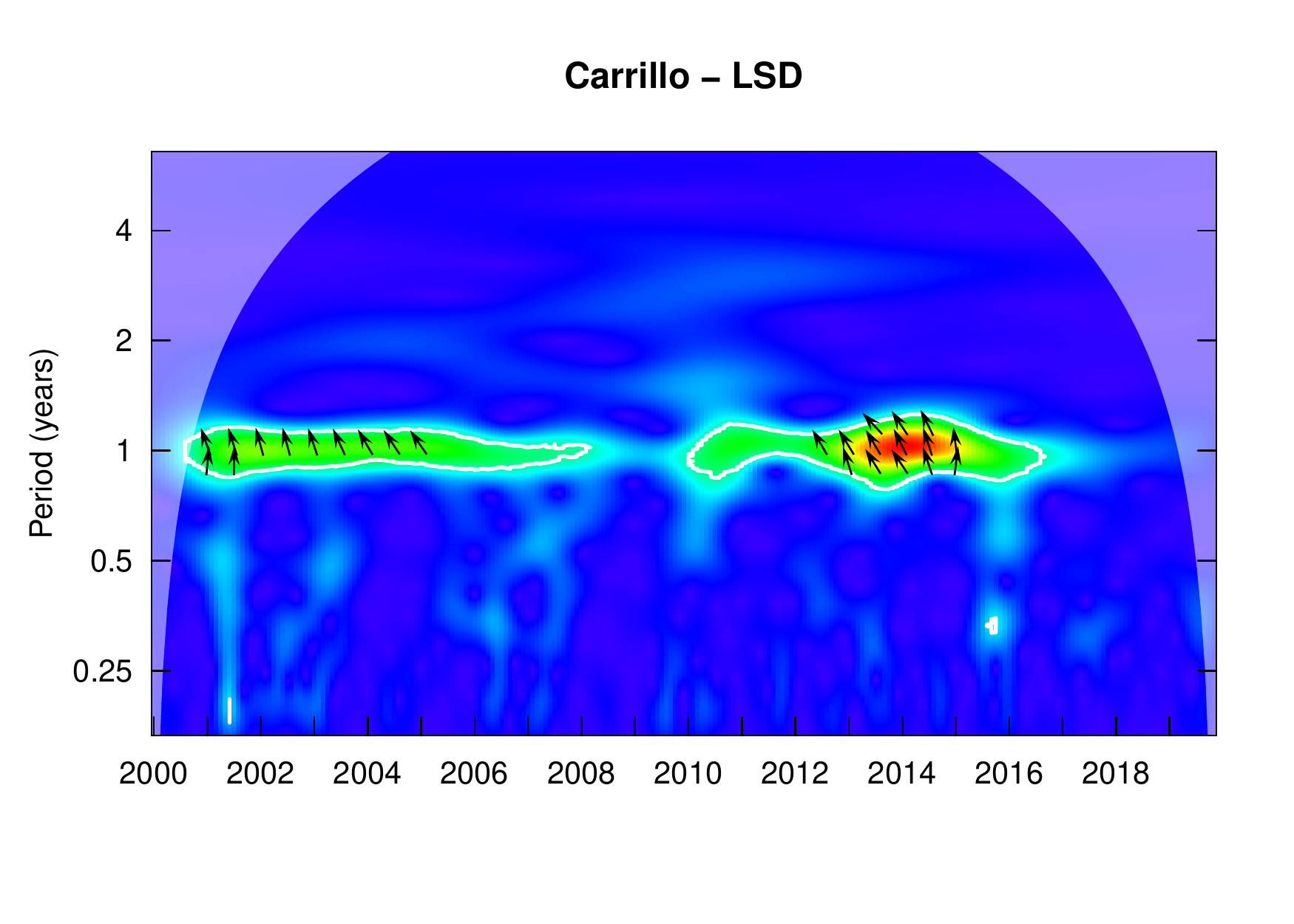}}\vspace{-0.15cm}%
\subfloat[]{\includegraphics[scale=0.23]{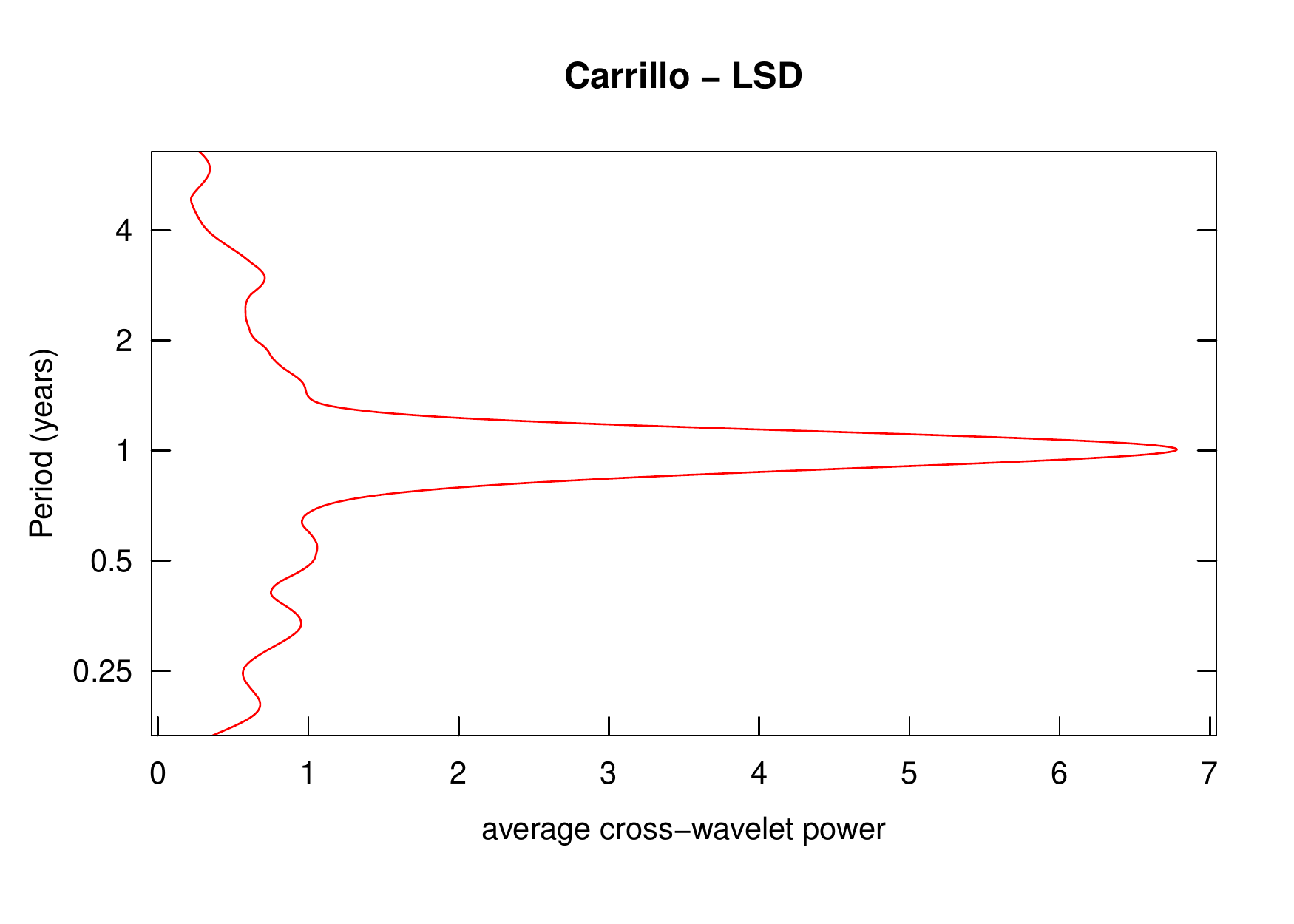}}\vspace{-0.15cm}%
\subfloat[]{\includegraphics[scale=0.23]{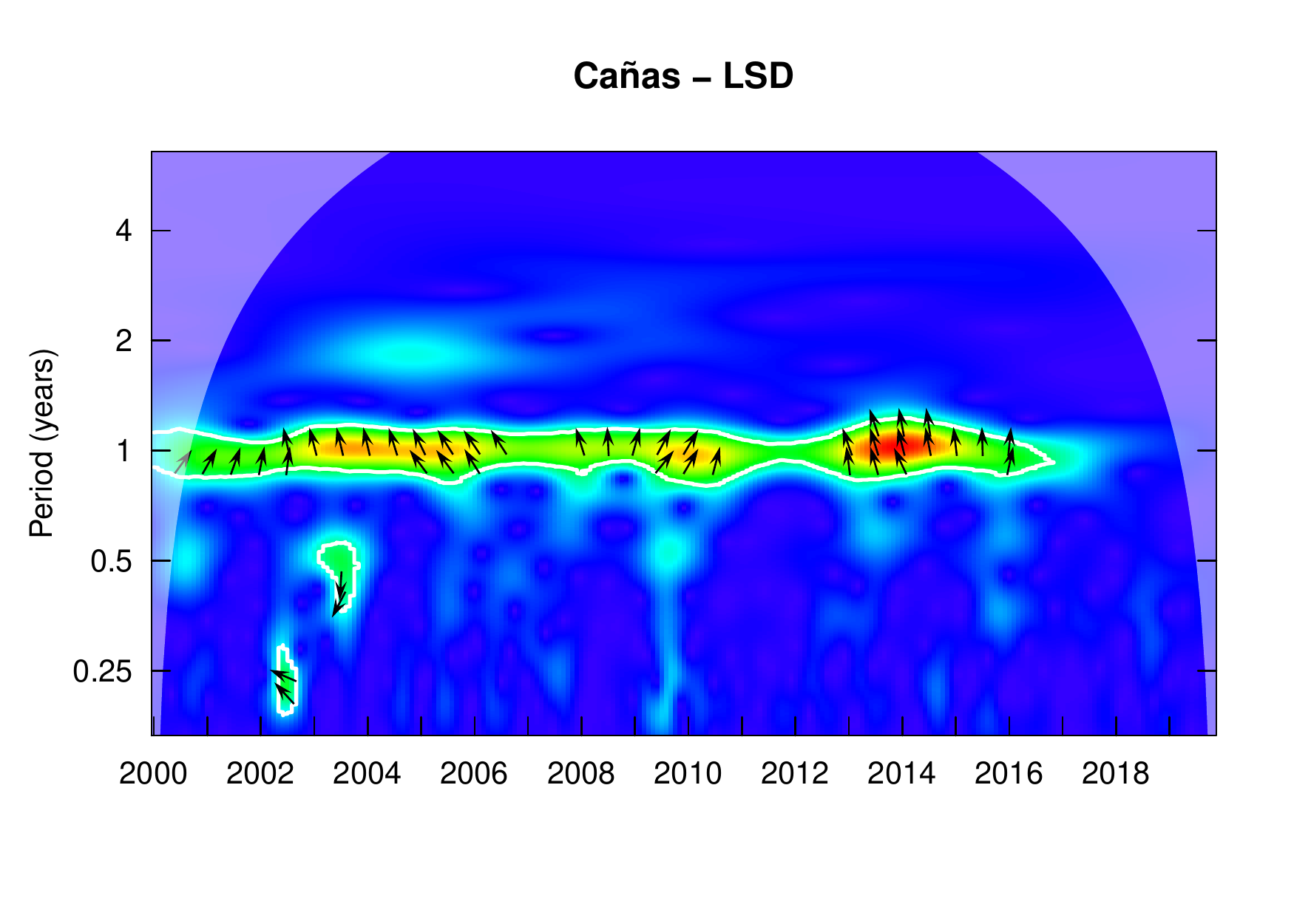}}\vspace{-0.15cm}%
\subfloat[]{\includegraphics[scale=0.23]{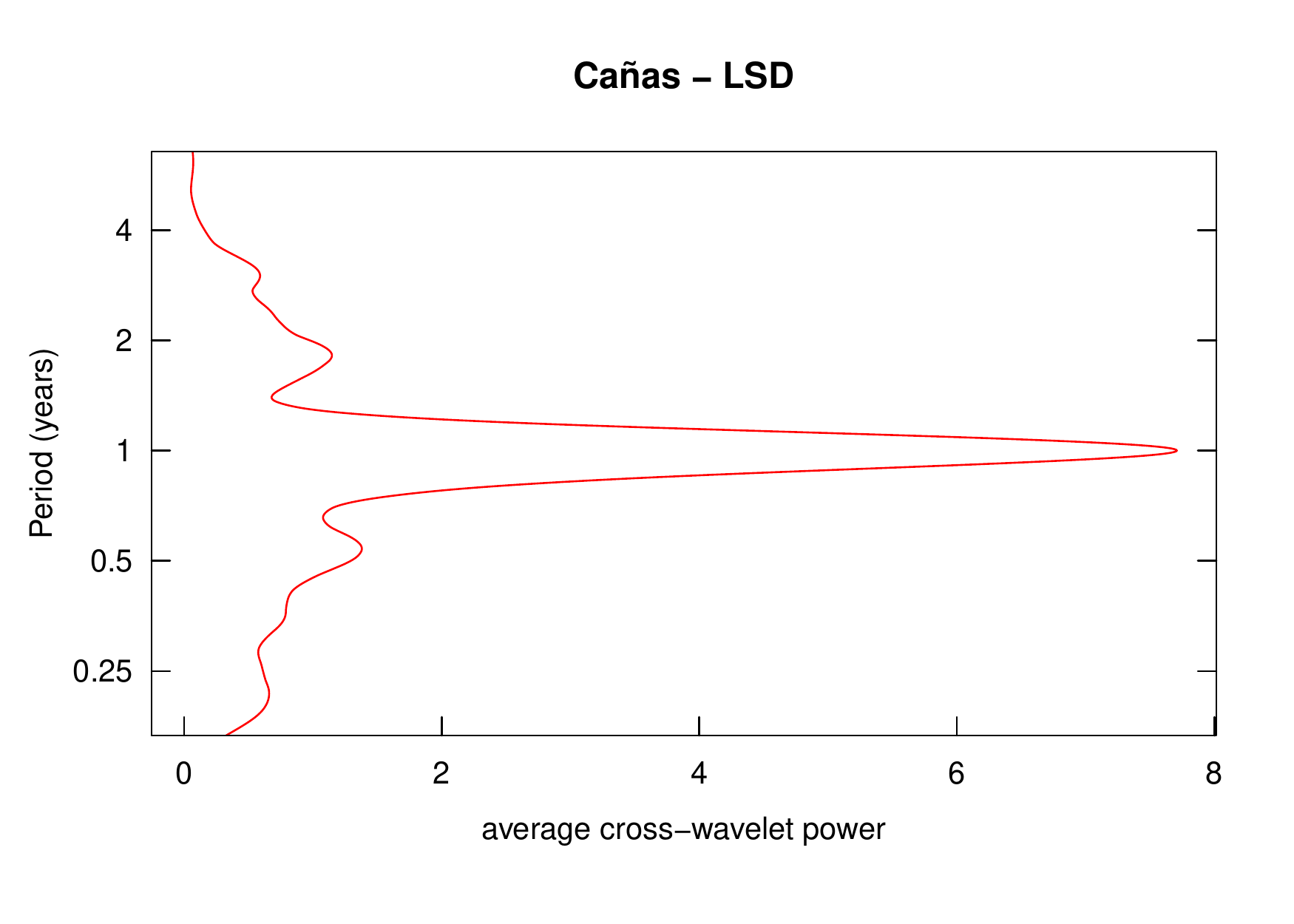}}\vspace{-0.15cm}\\
\caption*{}
\end{figure}

\begin{figure}[H]
\captionsetup[subfigure]{labelformat=empty}
\subfloat[]{\includegraphics[scale=0.23]{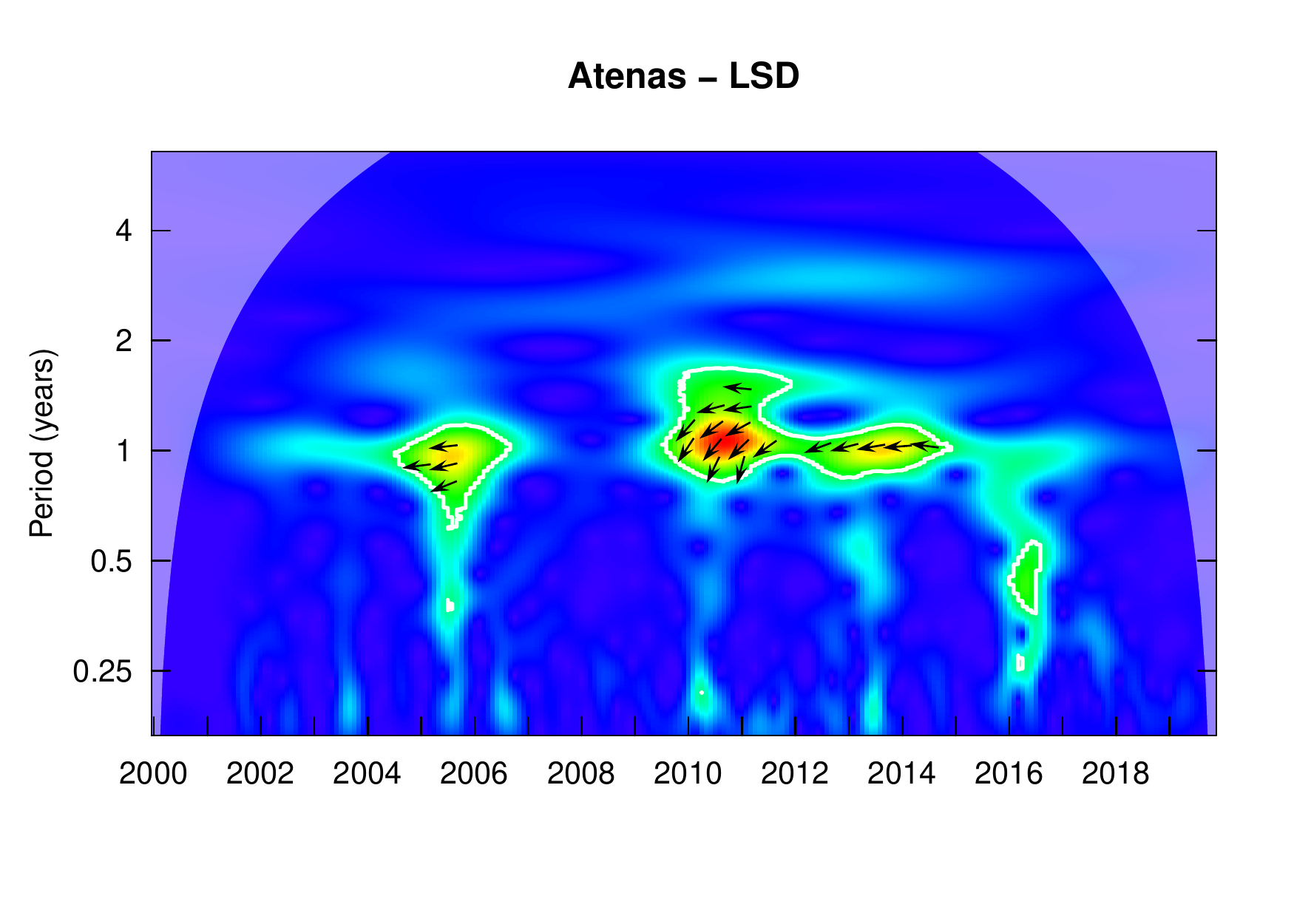}}\vspace{-0.15cm}%
\subfloat[]{\includegraphics[scale=0.23]{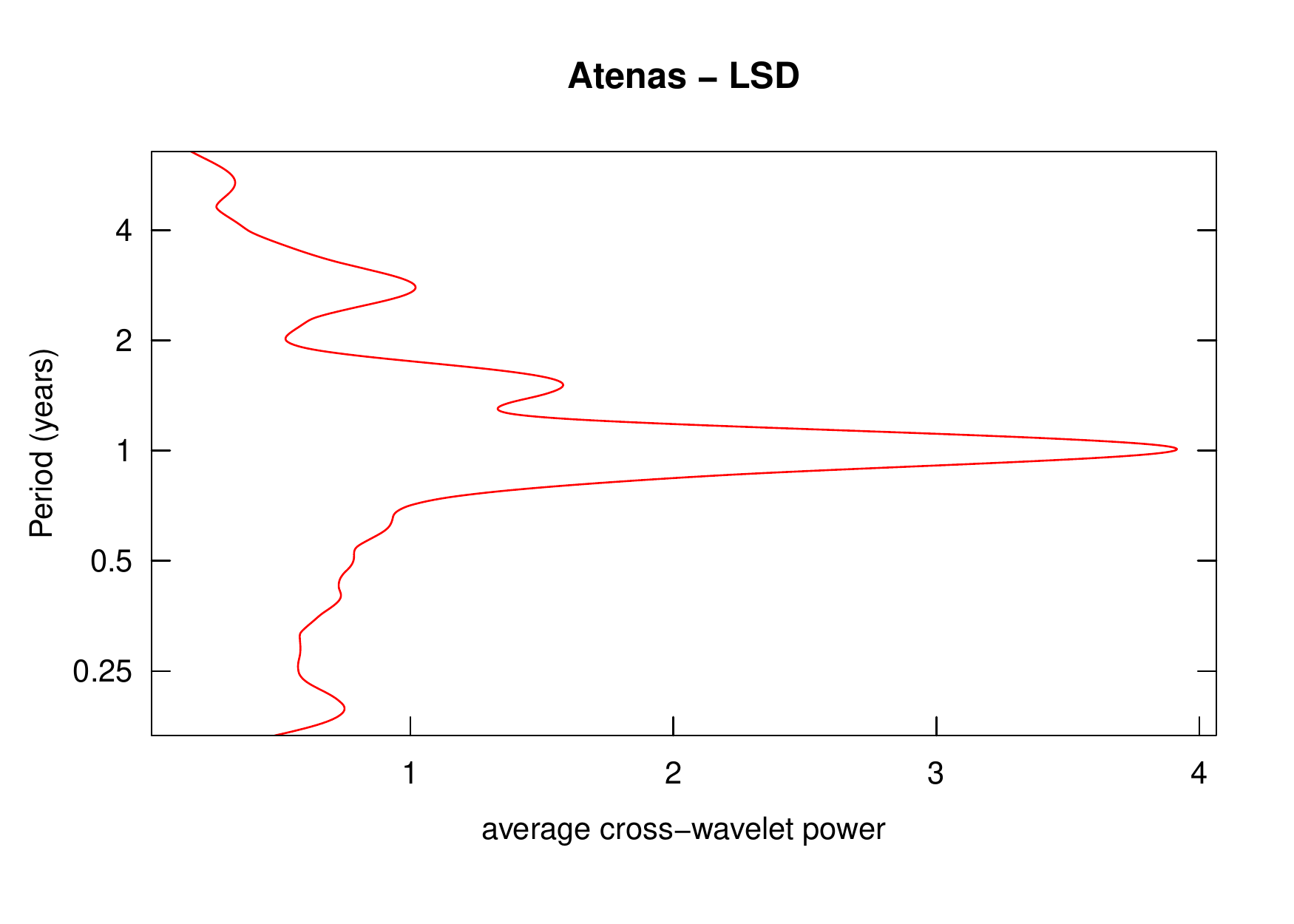}}\vspace{-0.15cm}%
\subfloat[]{\includegraphics[scale=0.23]{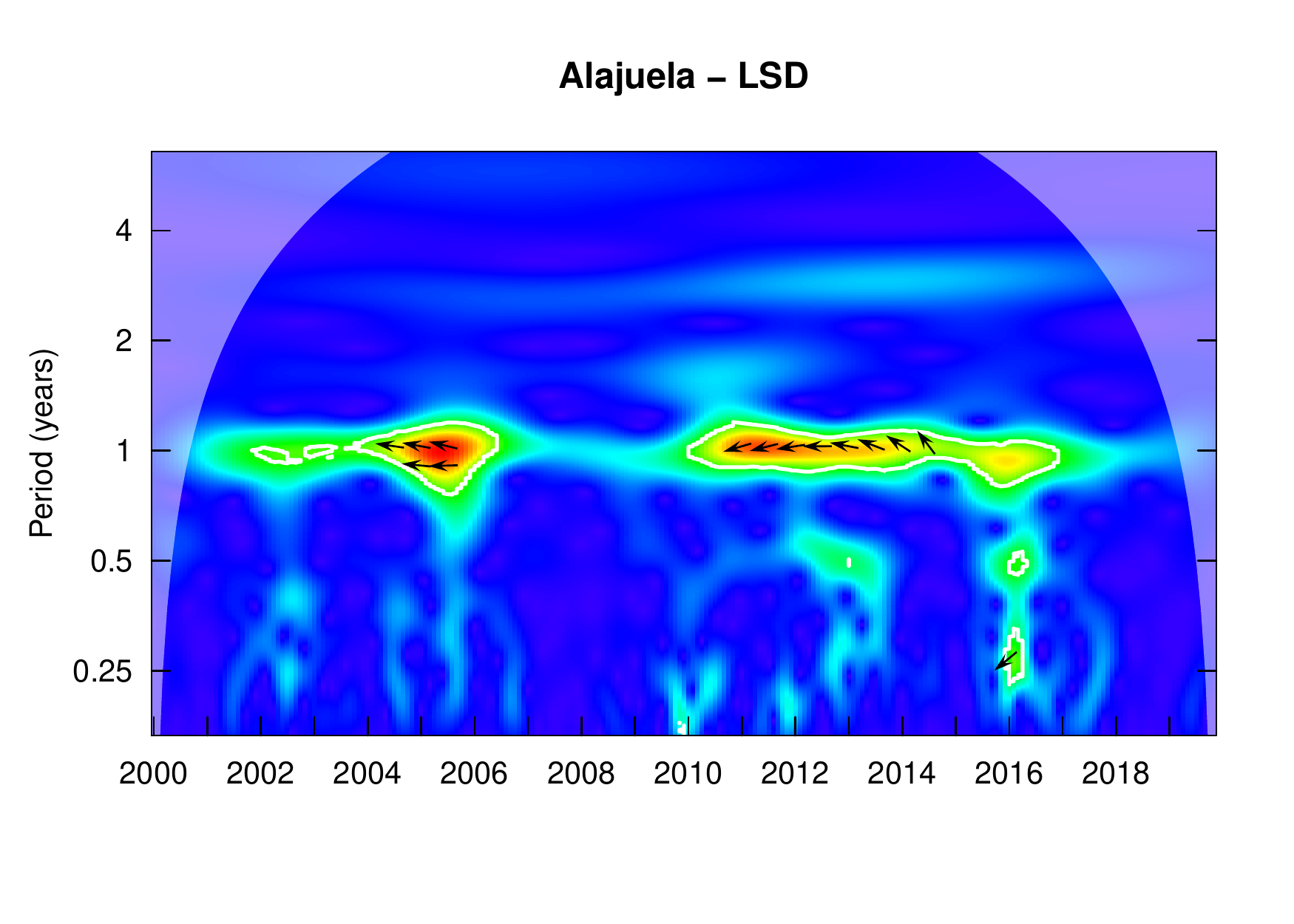}}\vspace{-0.15cm}%
\subfloat[]{\includegraphics[scale=0.23]{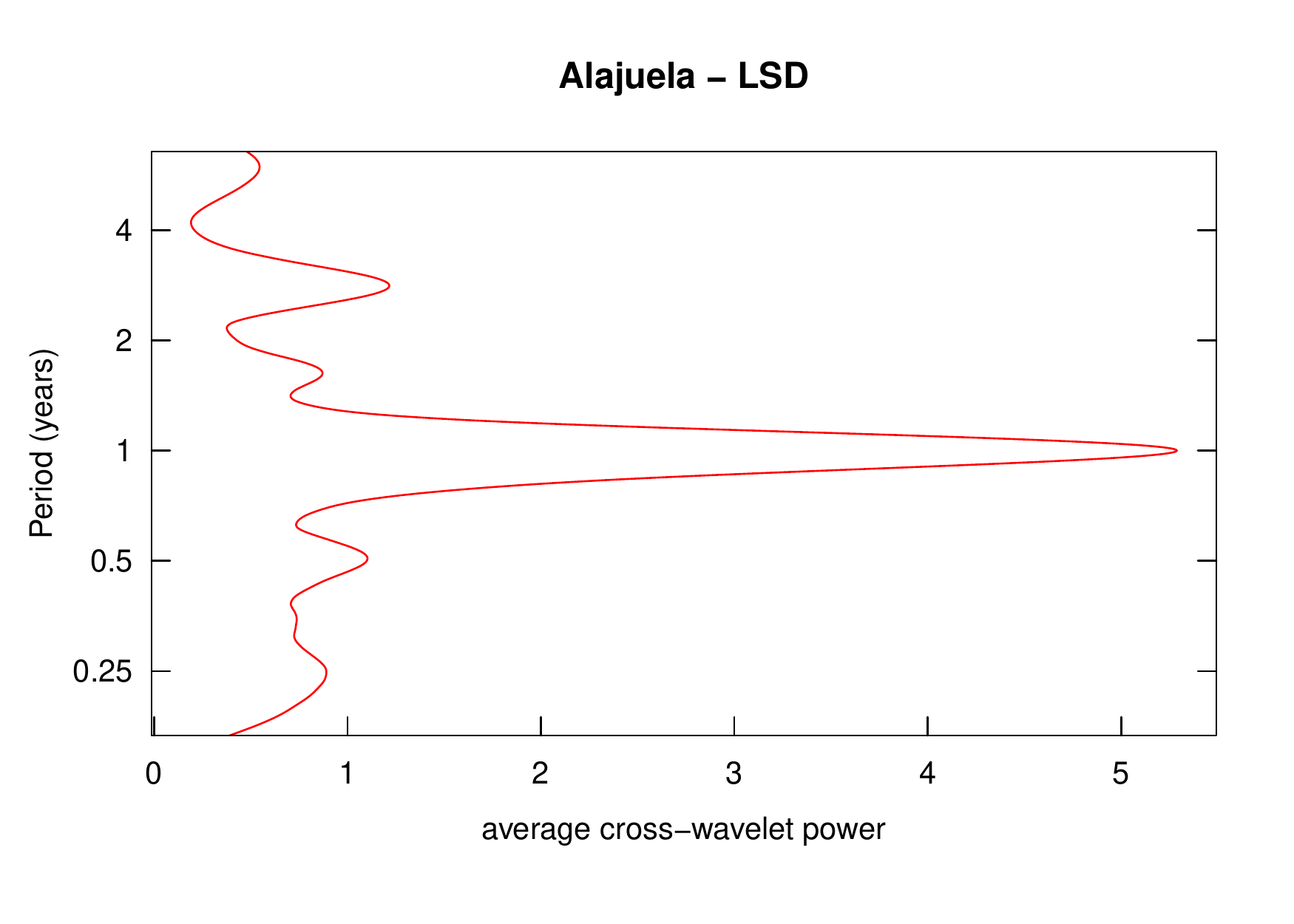}}\vspace{-0.15cm}\\
\subfloat[]{\includegraphics[scale=0.23]{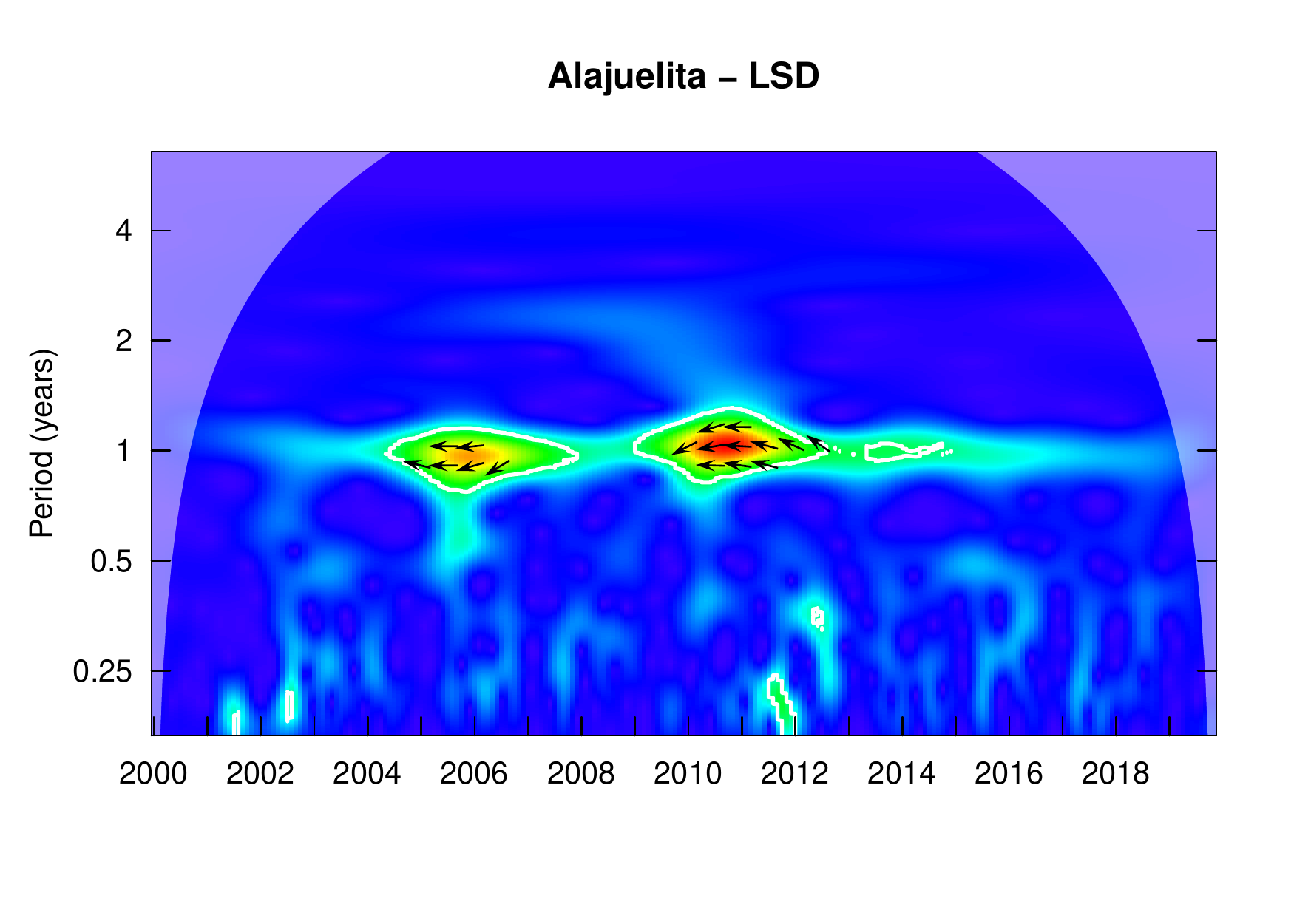}}\vspace{-0.15cm}%
\subfloat[]{\includegraphics[scale=0.23]{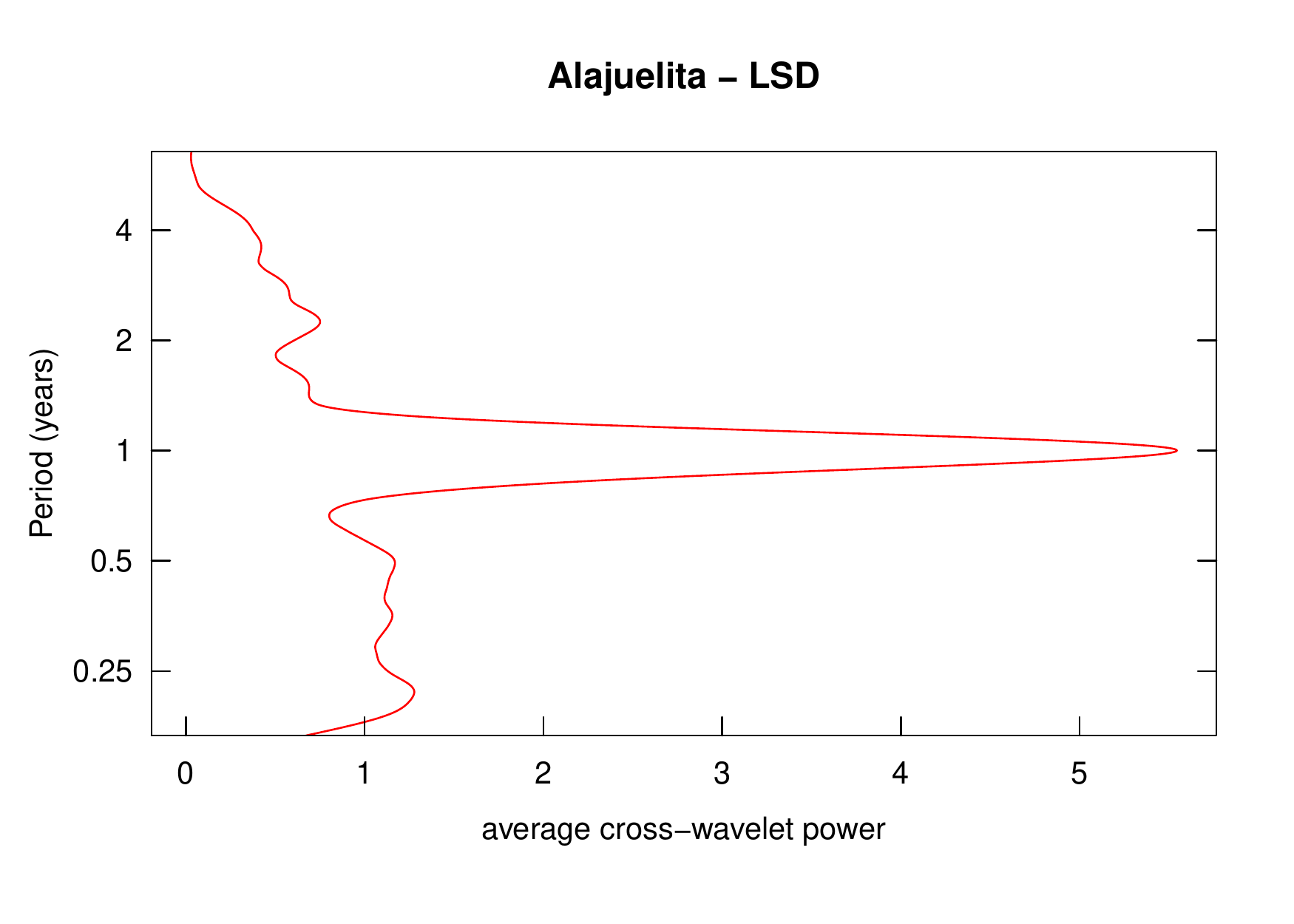}}\vspace{-0.15cm}%
\subfloat[]{\includegraphics[scale=0.23]{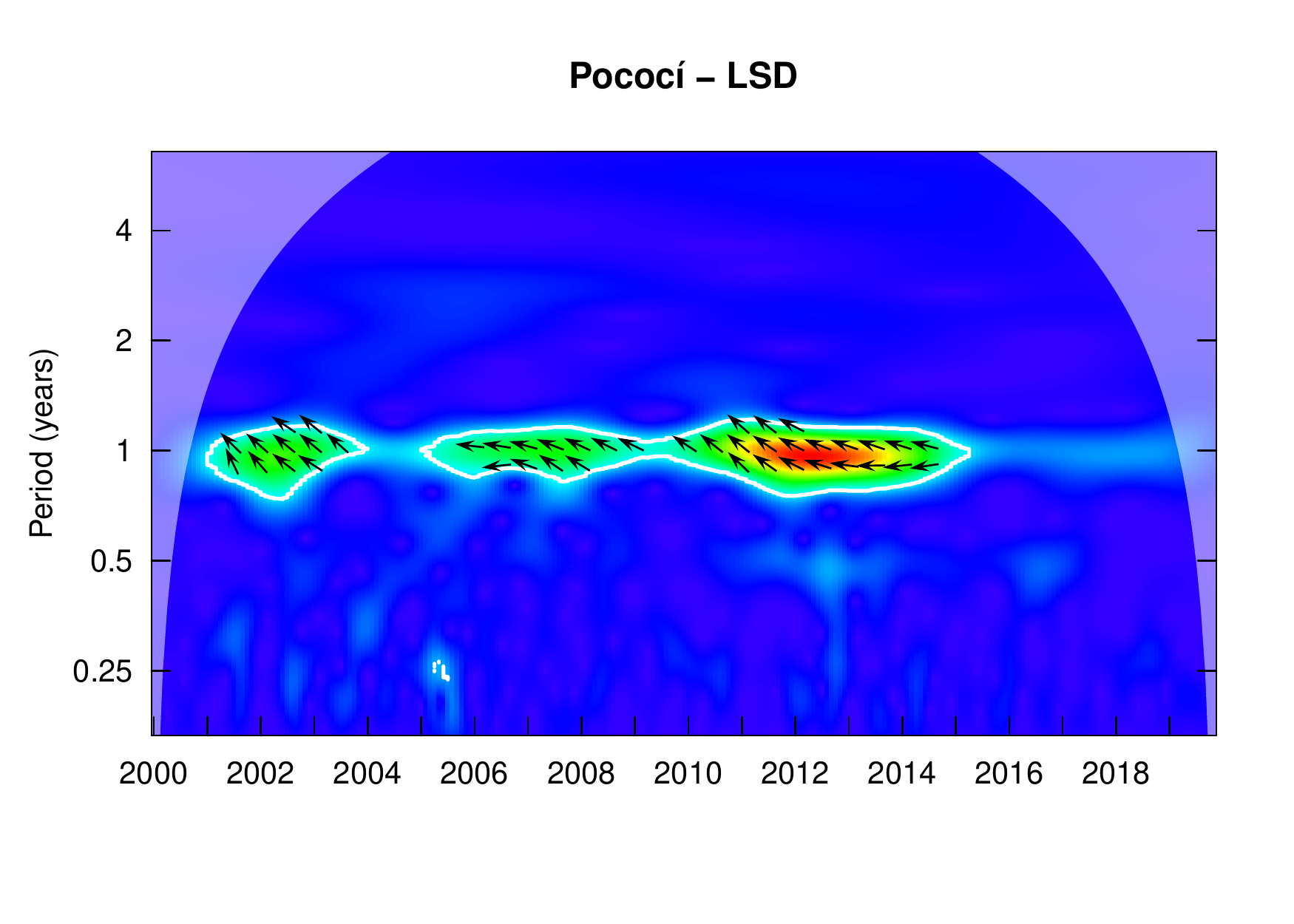}}\vspace{-0.15cm}%
\subfloat[]{\includegraphics[scale=0.23]{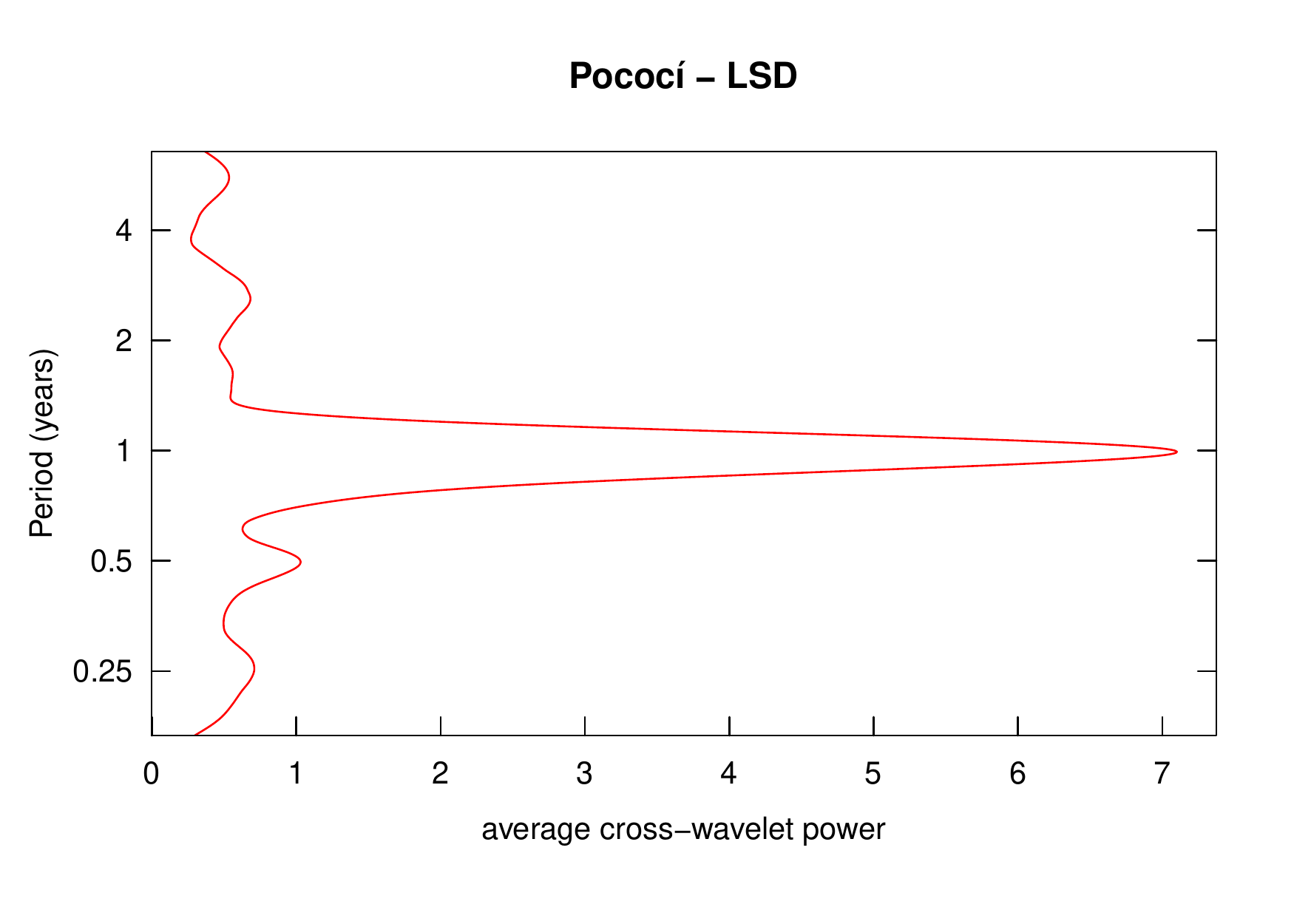}}\vspace{-0.15cm}\\
\subfloat[]{\includegraphics[scale=0.23]{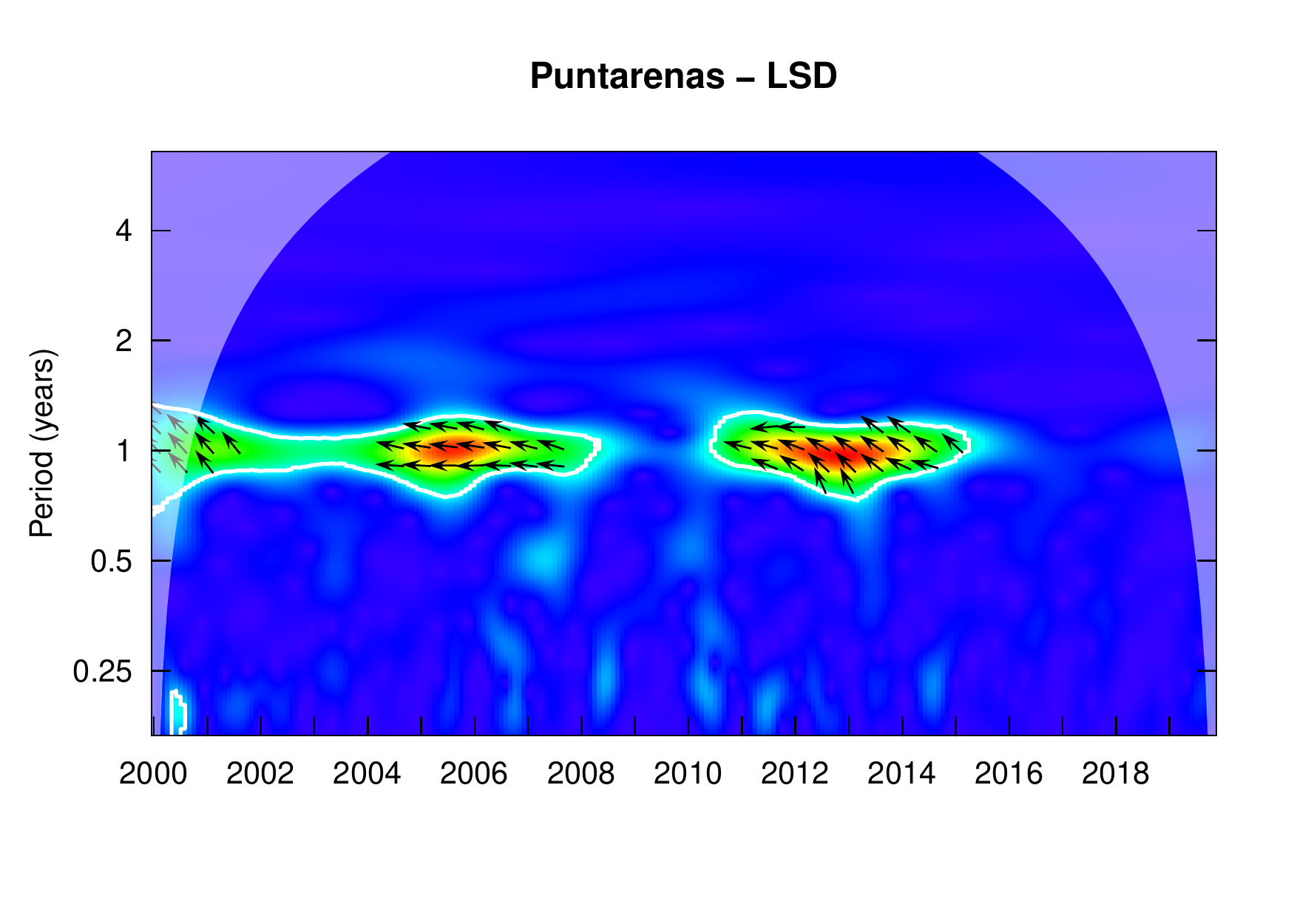}}\vspace{-0.15cm}%
\subfloat[]{\includegraphics[scale=0.23]{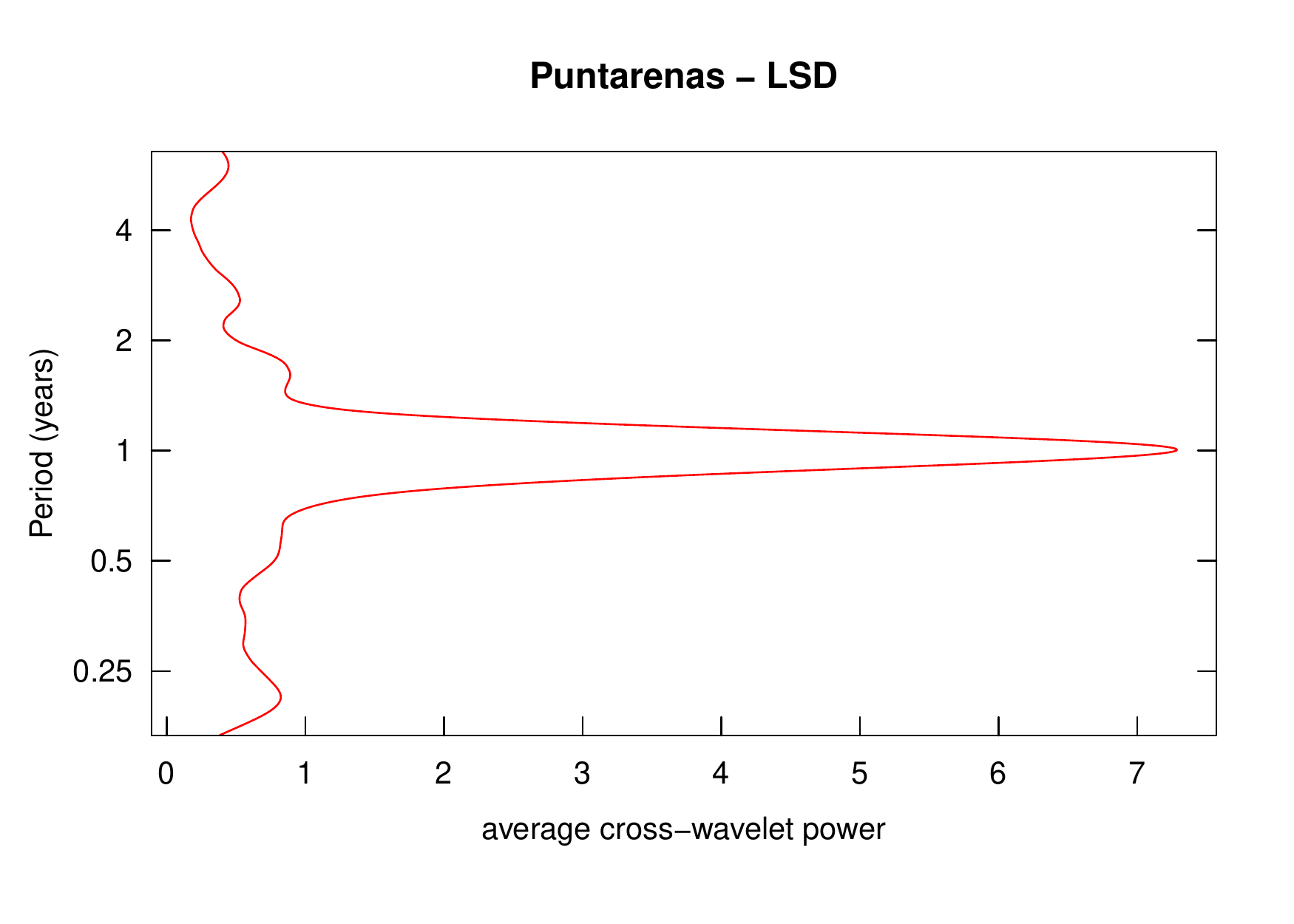}}\vspace{-0.15cm}%
\subfloat[]{\includegraphics[scale=0.23]{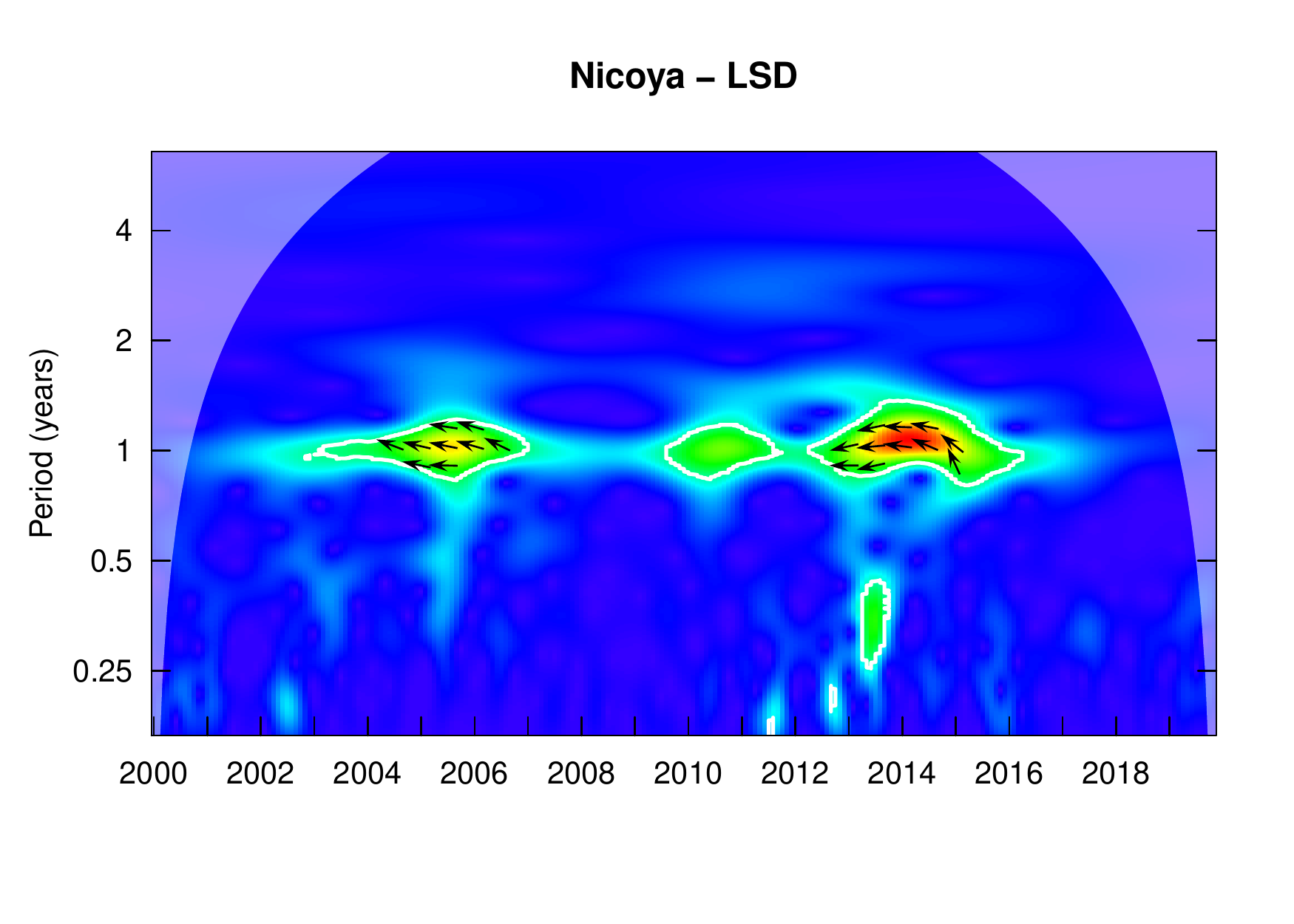}}\vspace{-0.15cm}%
\subfloat[]{\includegraphics[scale=0.23]{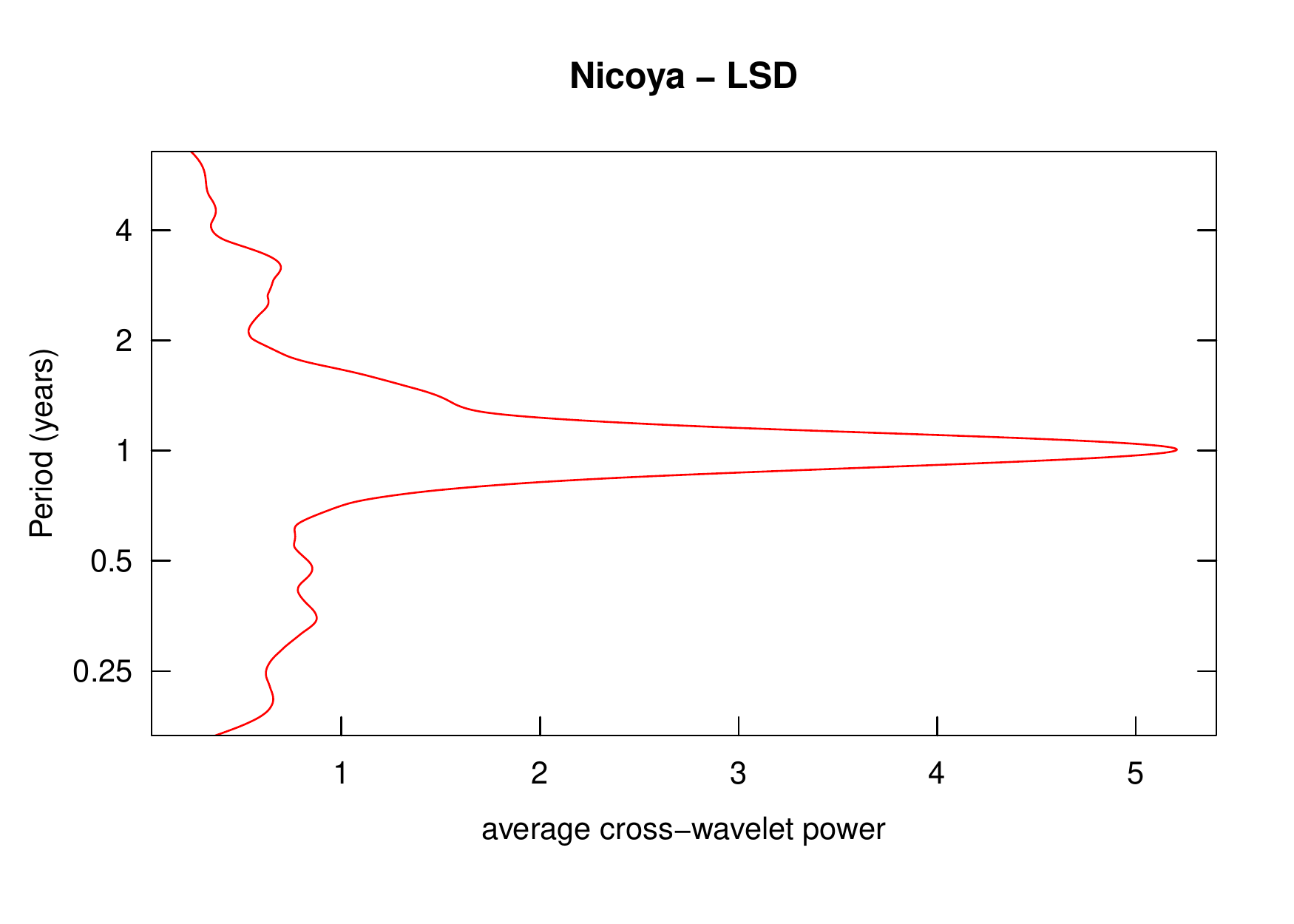}}\vspace{-0.15cm}\\
\subfloat[]{\includegraphics[scale=0.23]{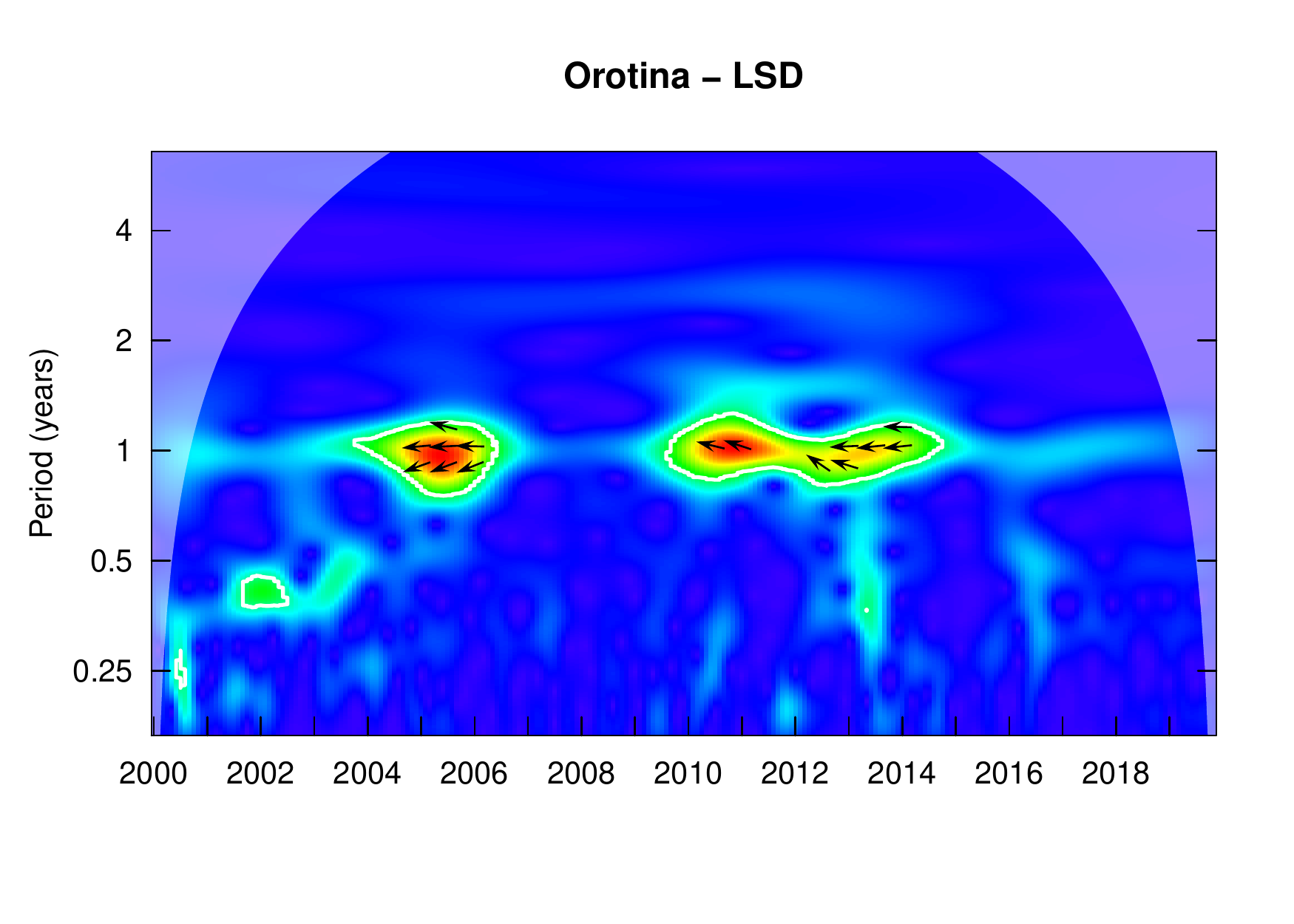}}\vspace{-0.15cm}%
\subfloat[]{\includegraphics[scale=0.23]{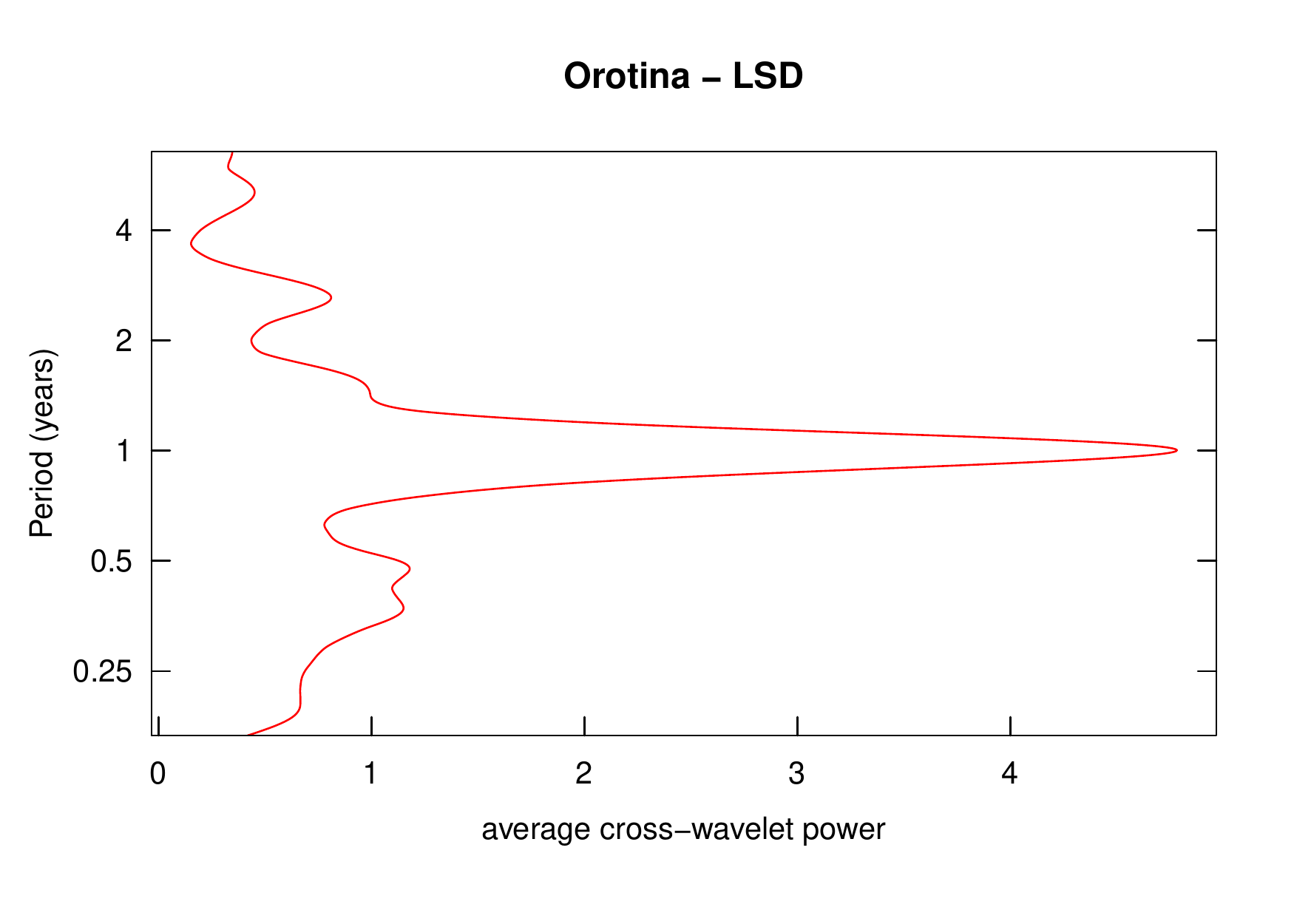}}\vspace{-0.15cm}%
\subfloat[]{\includegraphics[scale=0.23]{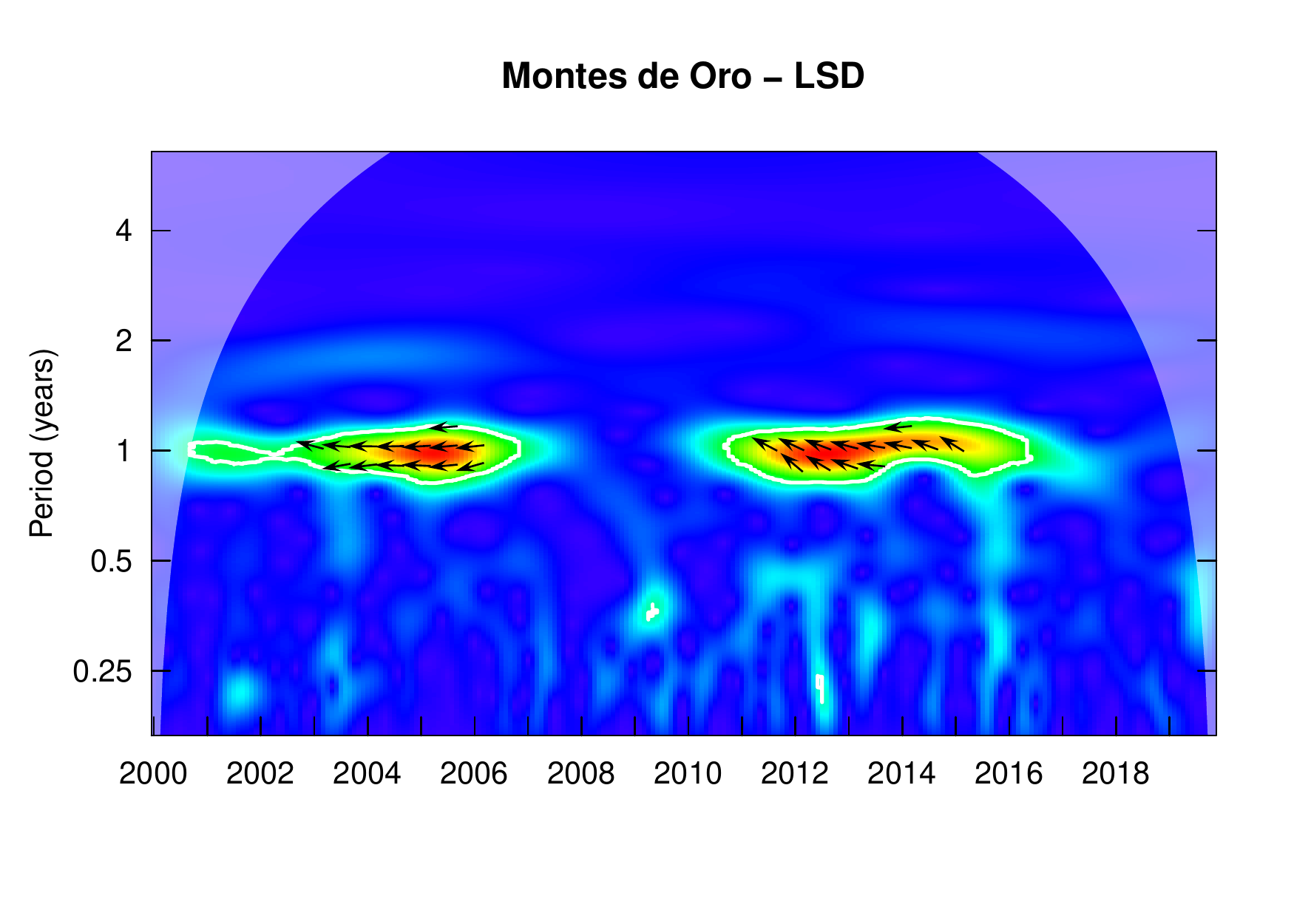}}\vspace{-0.15cm}%
\subfloat[]{\includegraphics[scale=0.23]{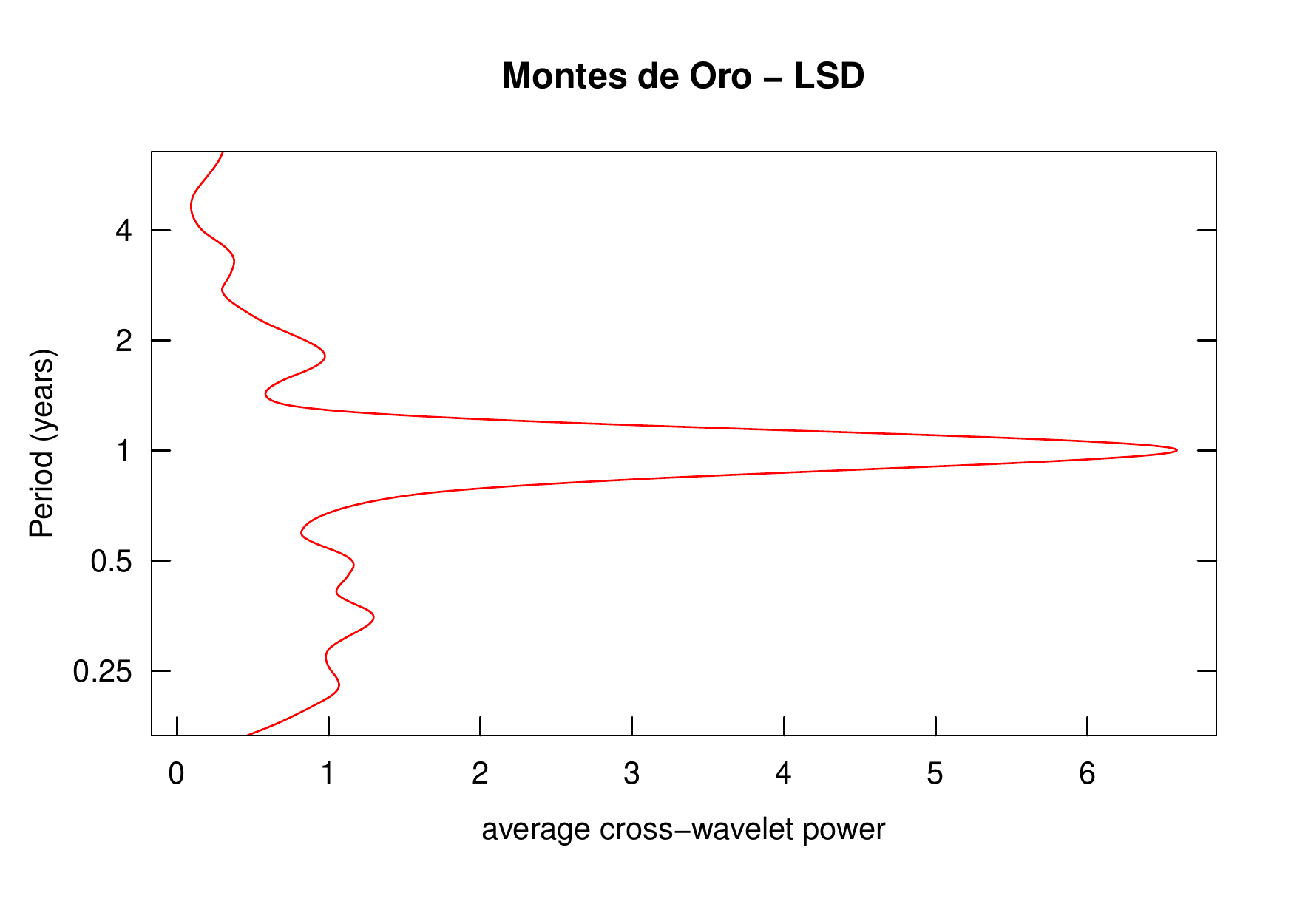}}\vspace{-0.15cm}\\
\subfloat[]{\includegraphics[scale=0.23]{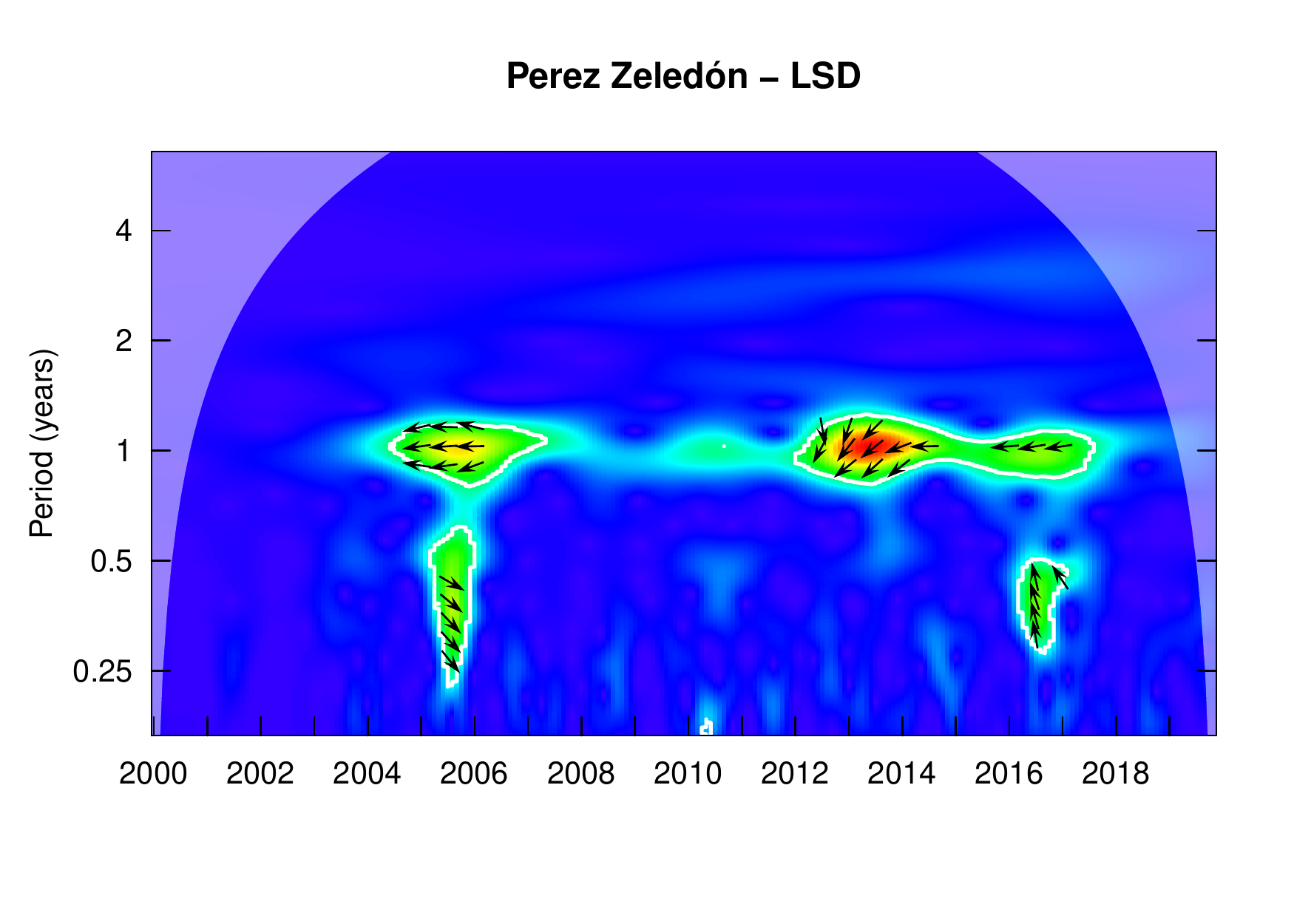}}\vspace{-0.15cm}%
\subfloat[]{\includegraphics[scale=0.23]{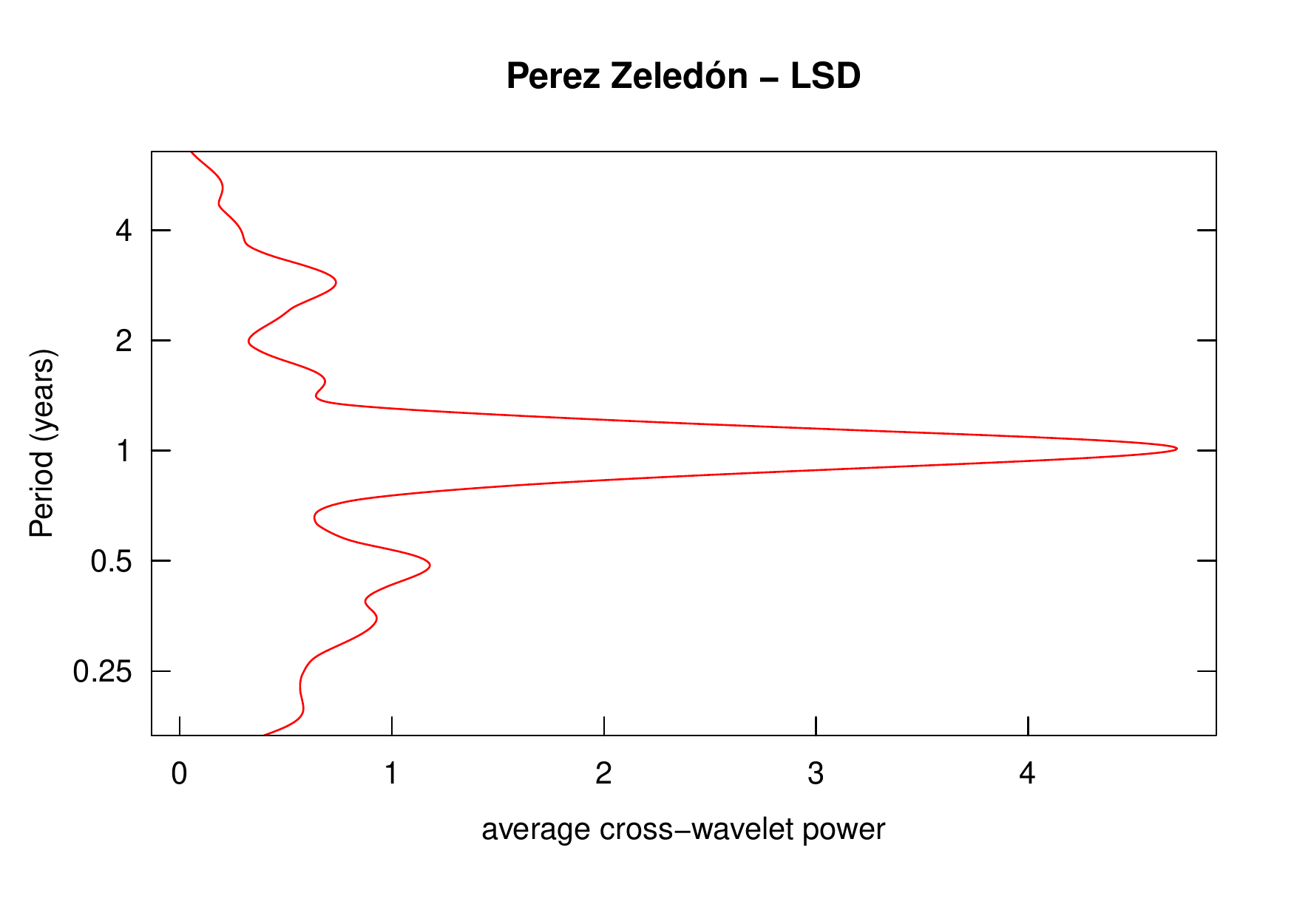}}\vspace{-0.15cm}%
\subfloat[]{\includegraphics[scale=0.23]{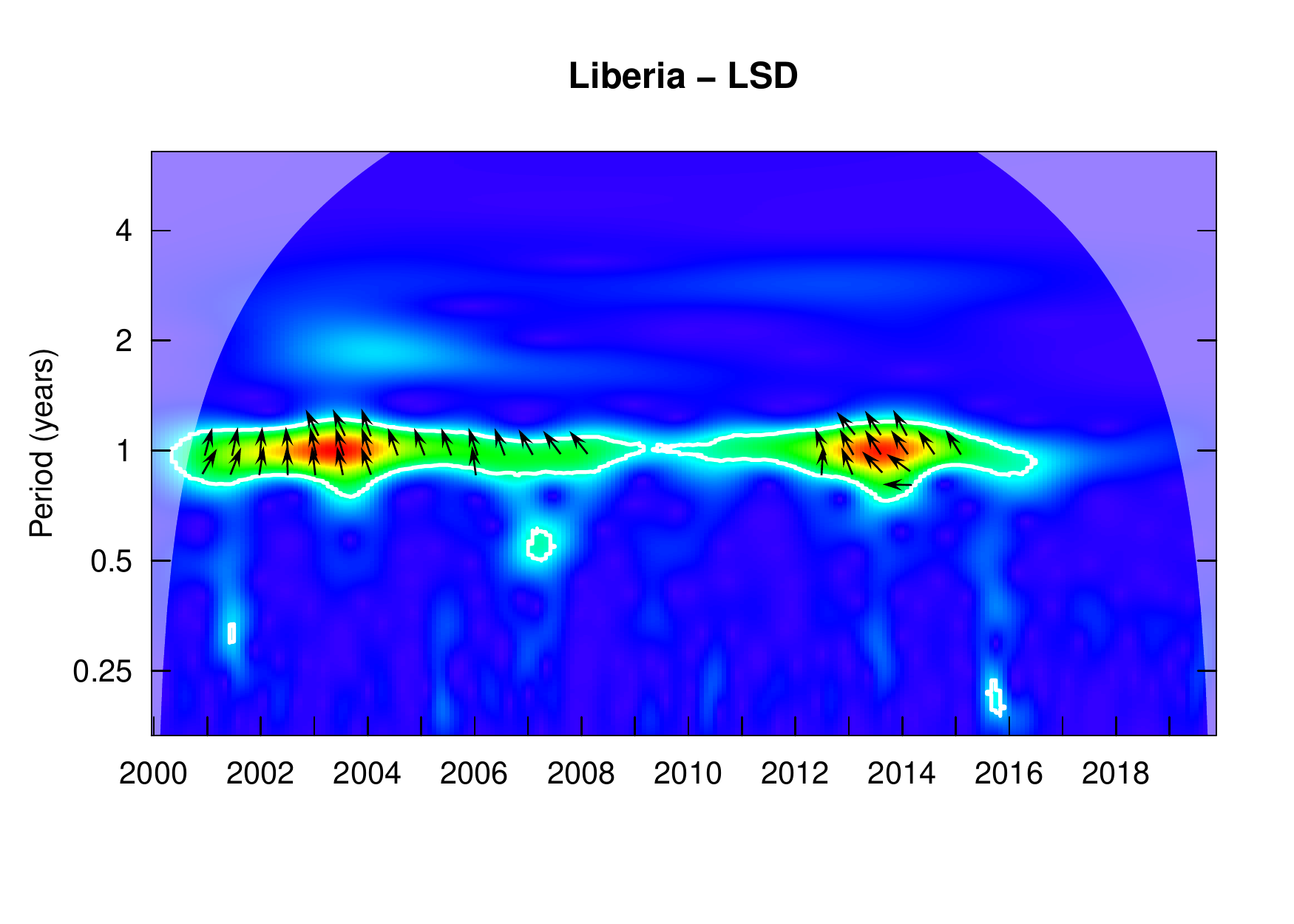}}\vspace{-0.15cm}%
\subfloat[]{\includegraphics[scale=0.23]{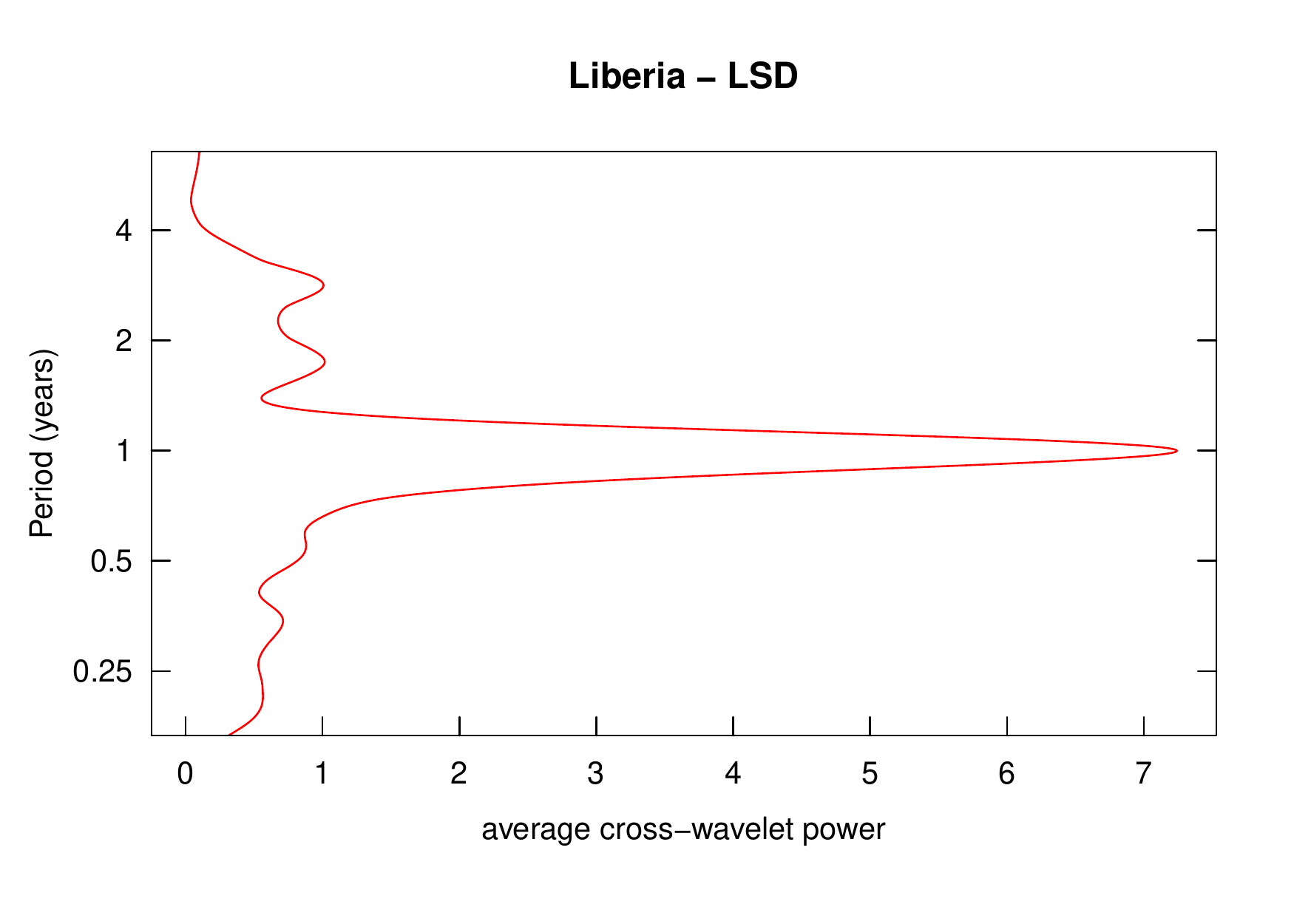}}\vspace{-0.15cm}\\
\subfloat[]{\includegraphics[scale=0.23]{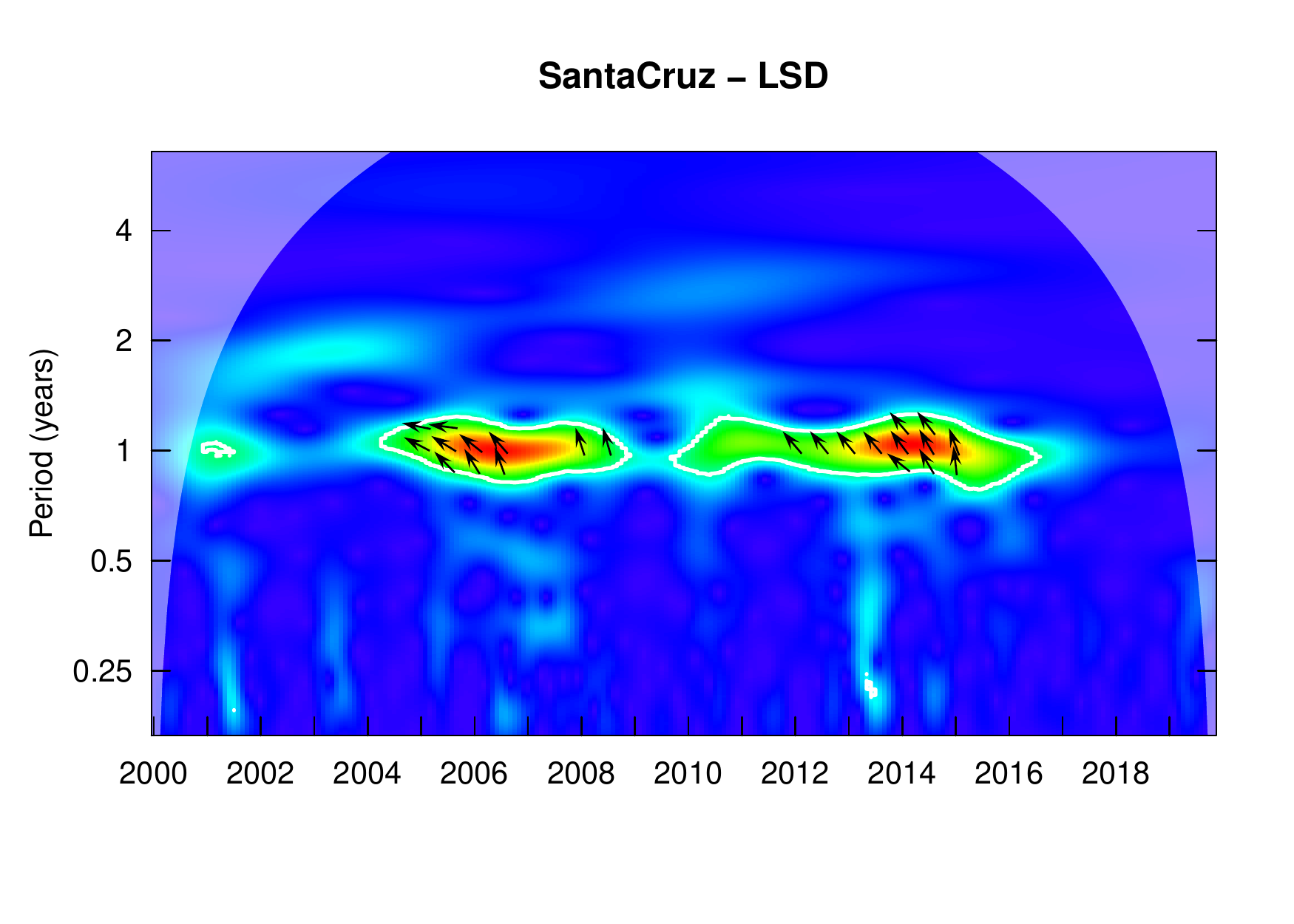}}\vspace{-0.15cm}%
\subfloat[]{\includegraphics[scale=0.23]{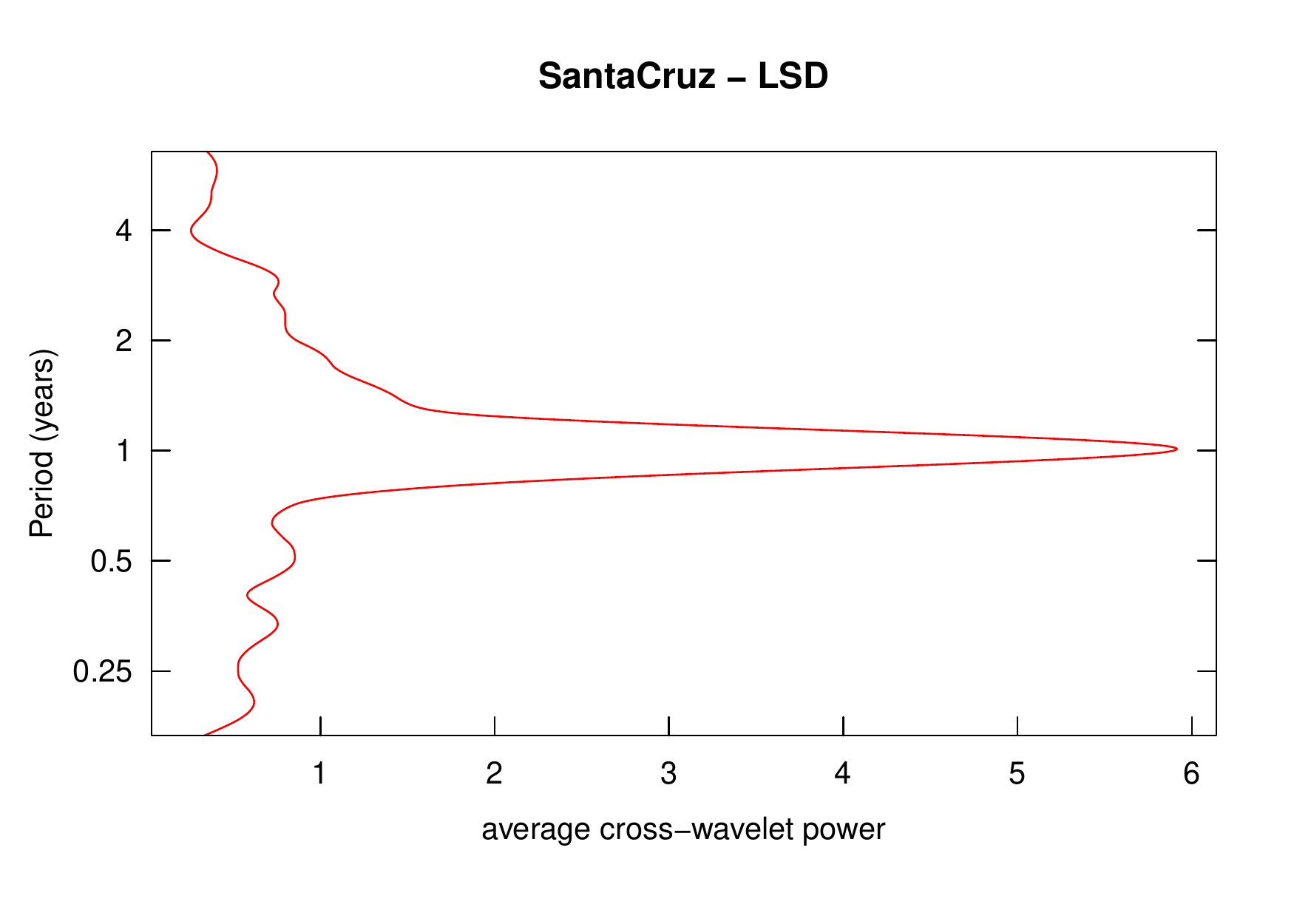}}\vspace{-0.15cm}%
\subfloat[]{\includegraphics[scale=0.23]{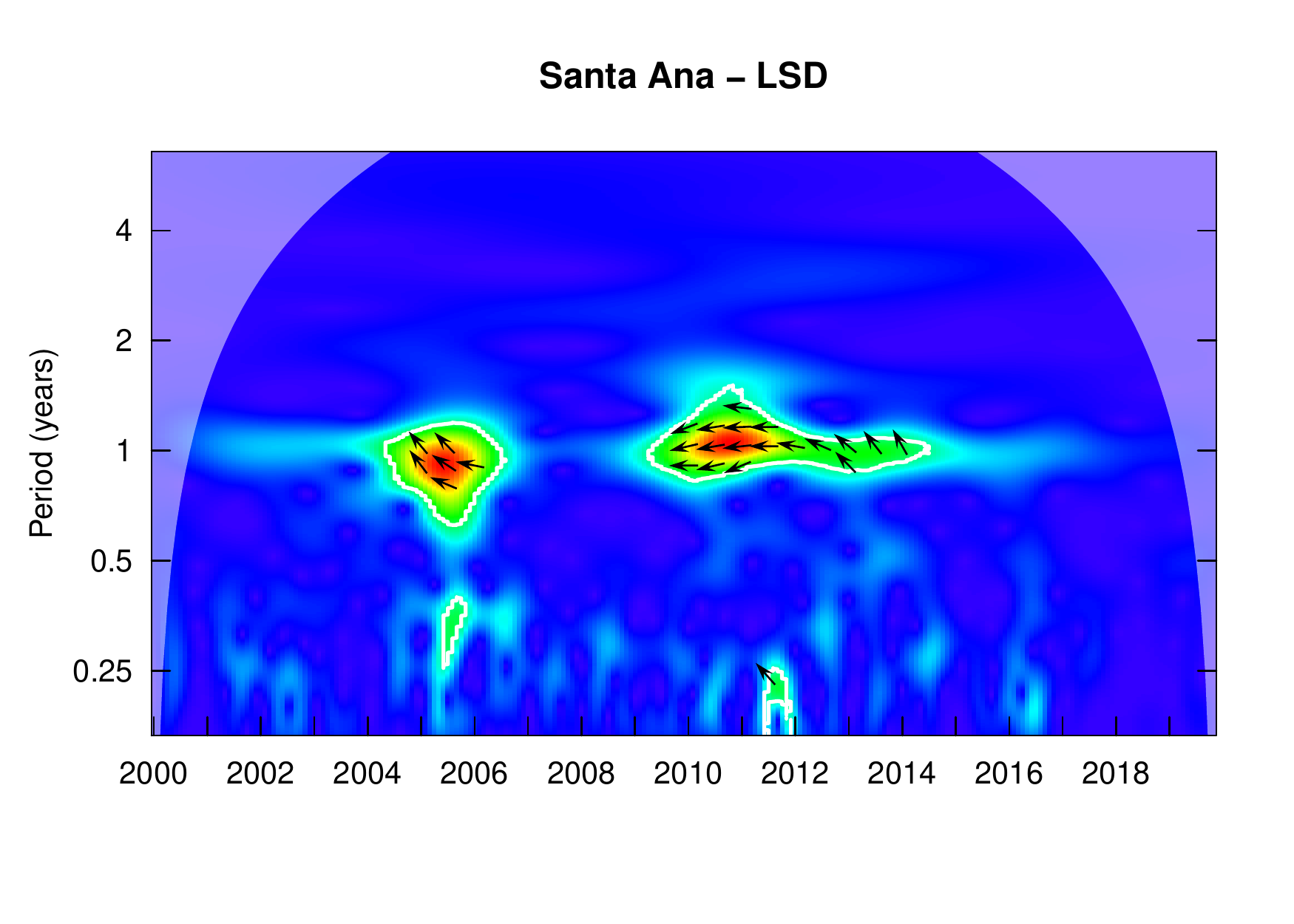}}\vspace{-0.15cm}%
\subfloat[]{\includegraphics[scale=0.23]{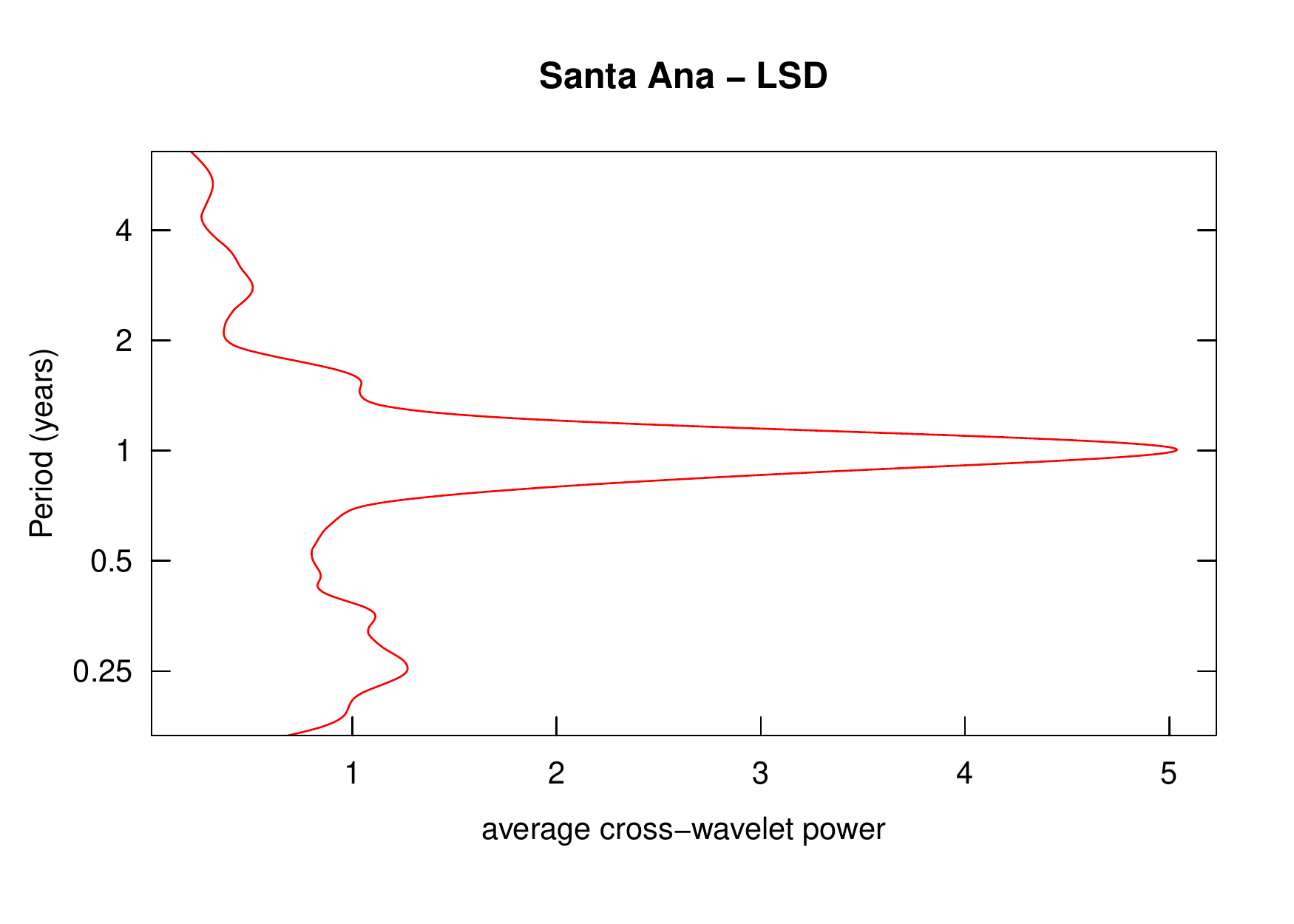}}\vspace{-0.15cm}\\
\subfloat[]{\includegraphics[scale=0.23]{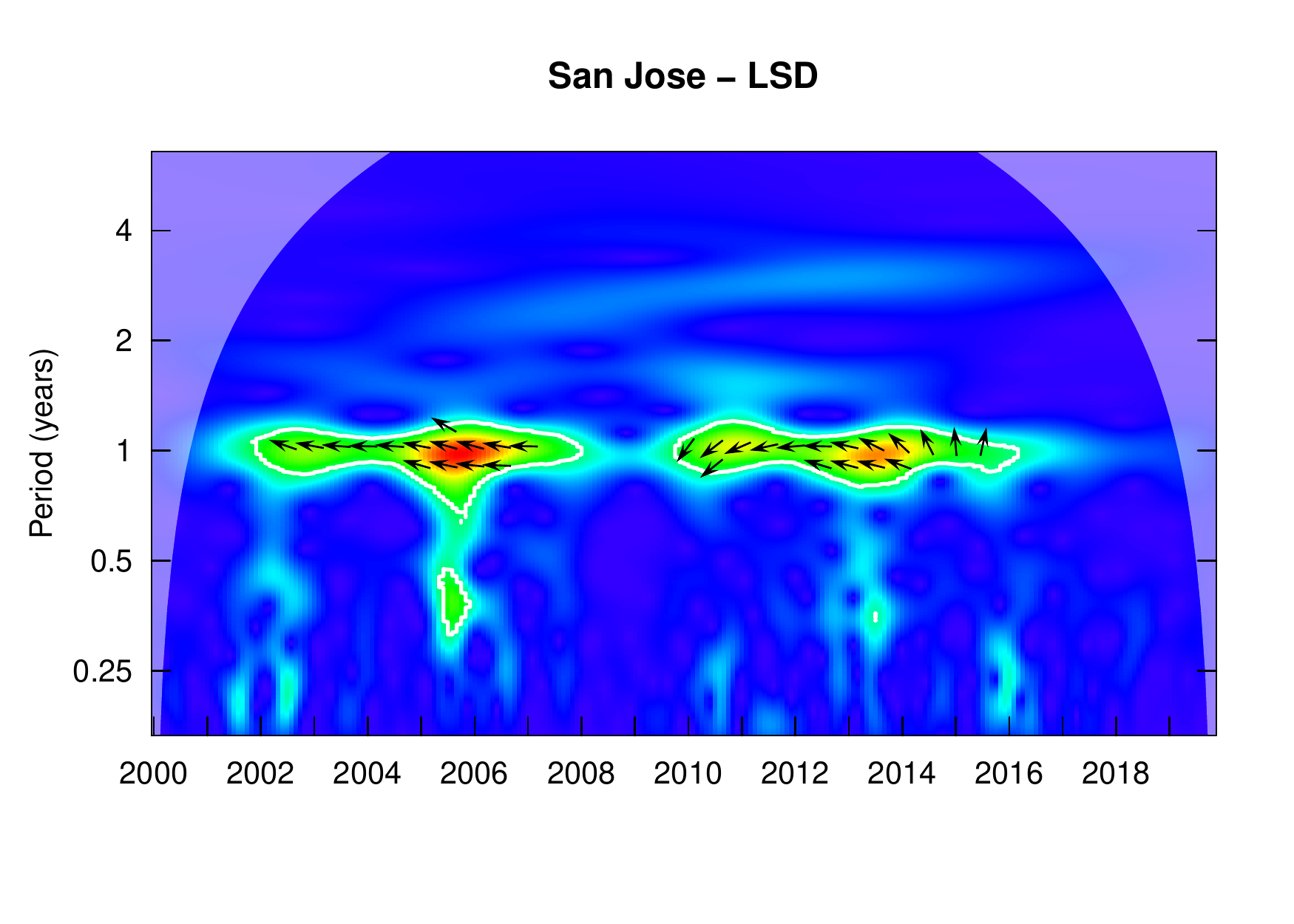}}\vspace{-0.15cm}%
\subfloat[]{\includegraphics[scale=0.23]{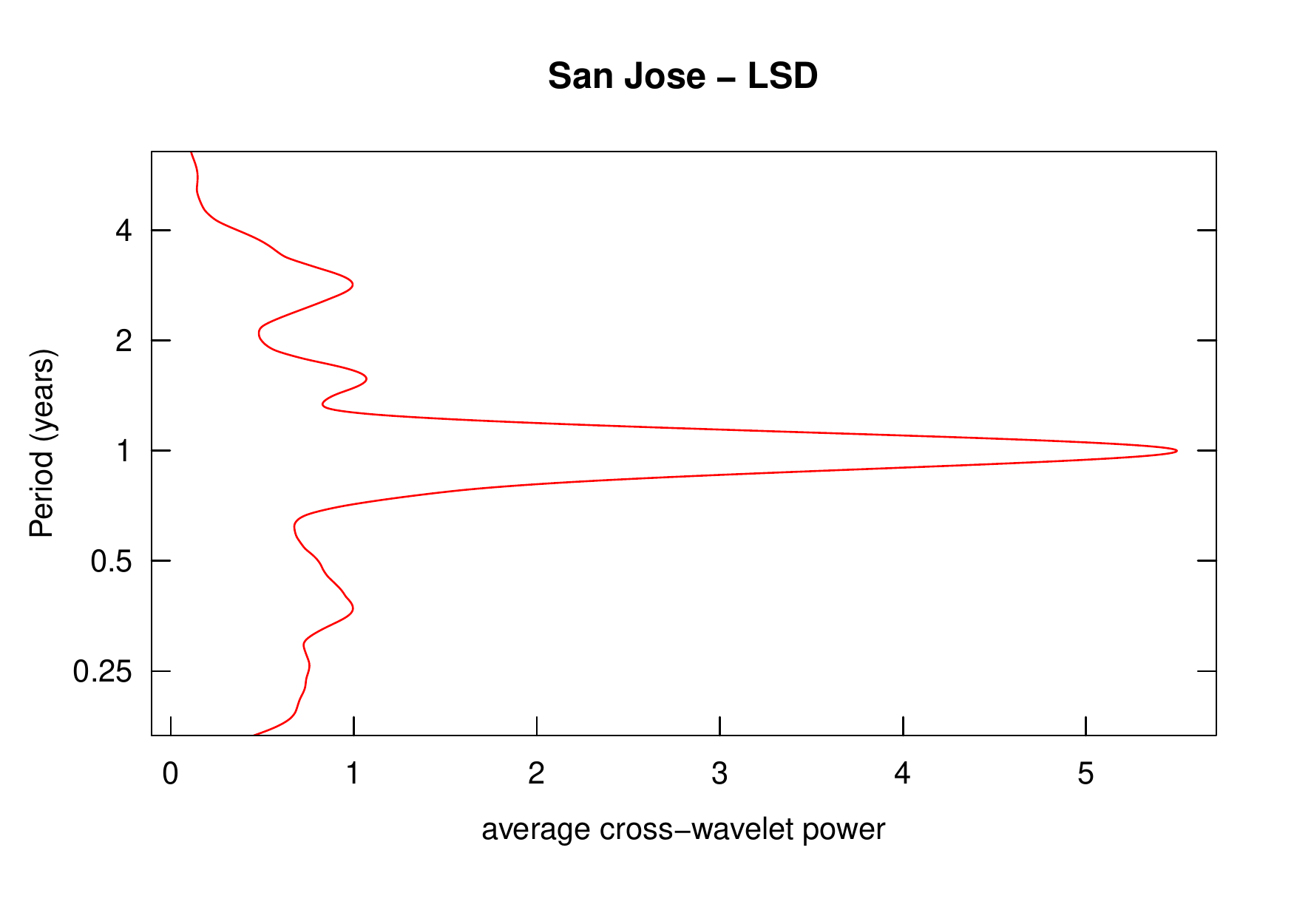}}\vspace{-0.15cm}%
\subfloat[]{\includegraphics[scale=0.23]{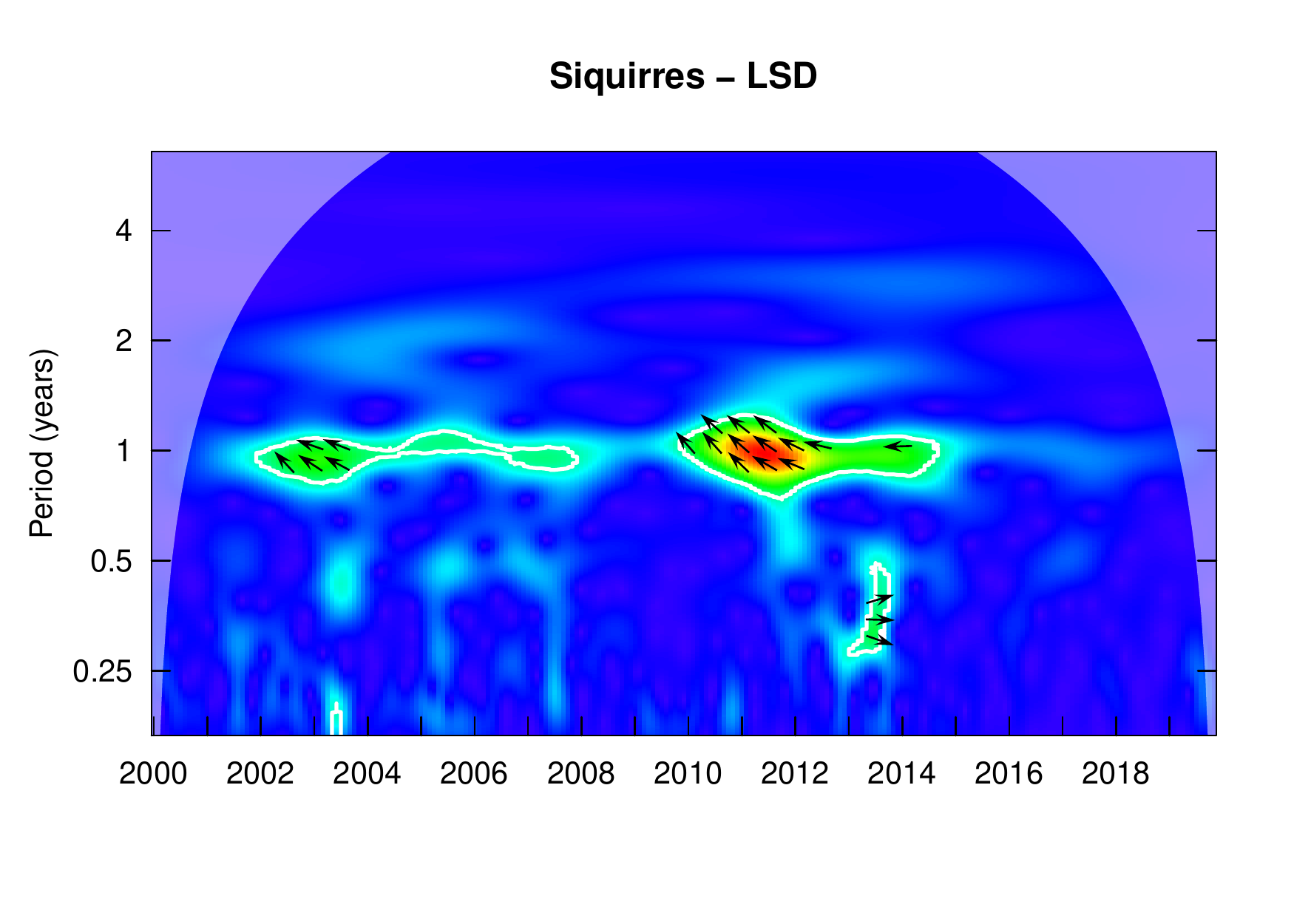}}\vspace{-0.15cm}%
\subfloat[]{\includegraphics[scale=0.23]{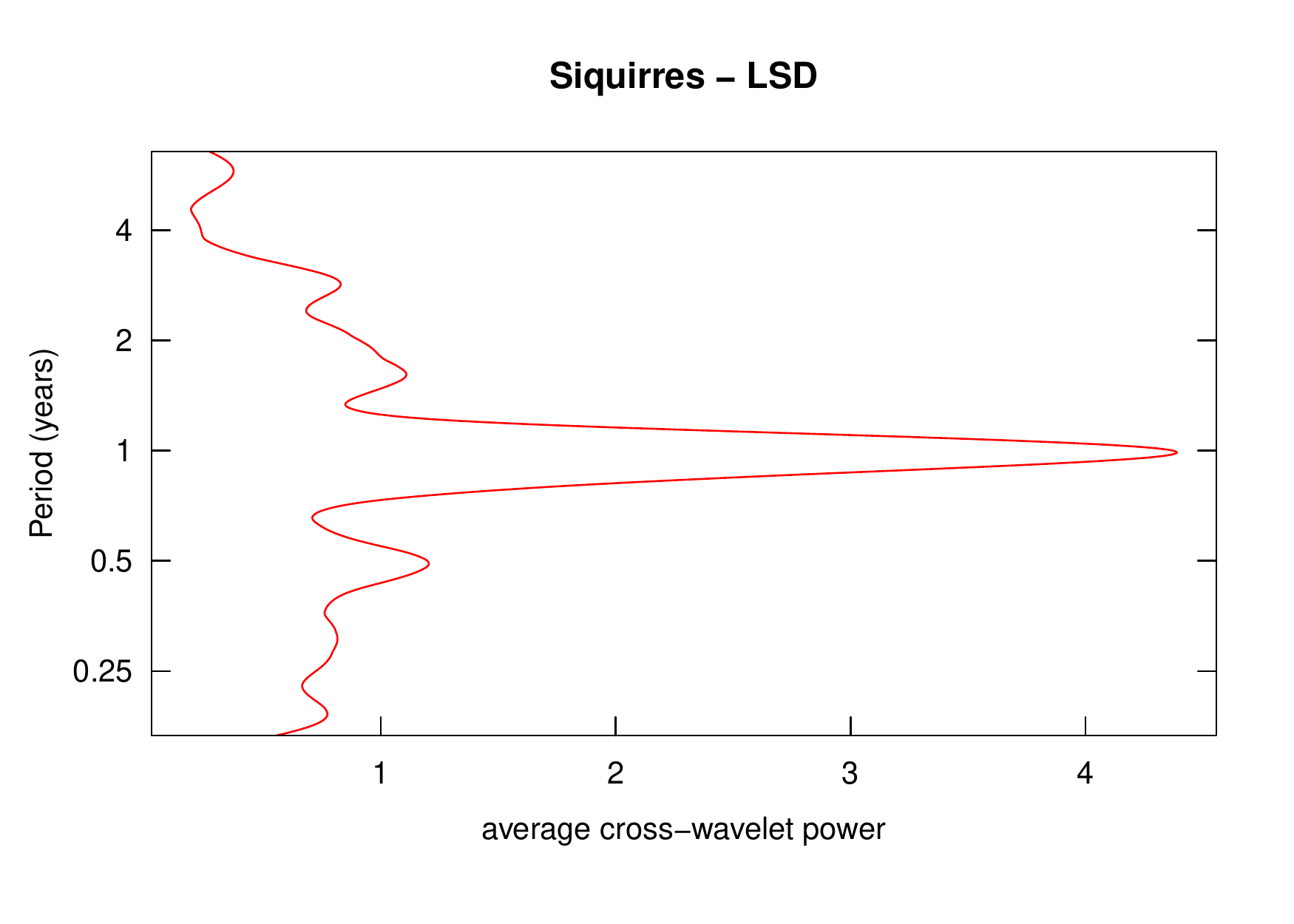}}\vspace{-0.15cm}\\
\subfloat[]{\includegraphics[scale=0.23]{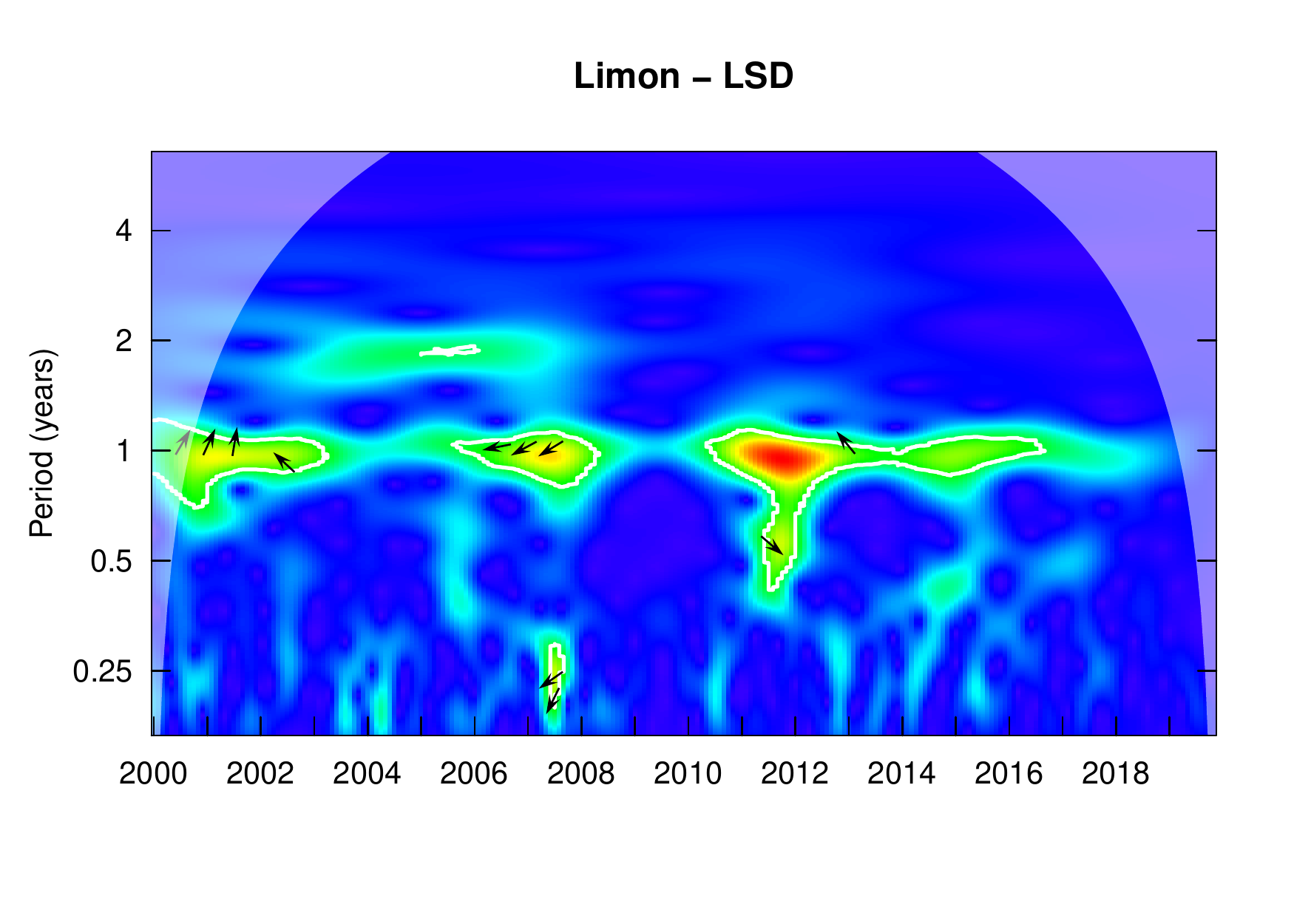}}\vspace{-0.15cm}%
\subfloat[]{\includegraphics[scale=0.23]{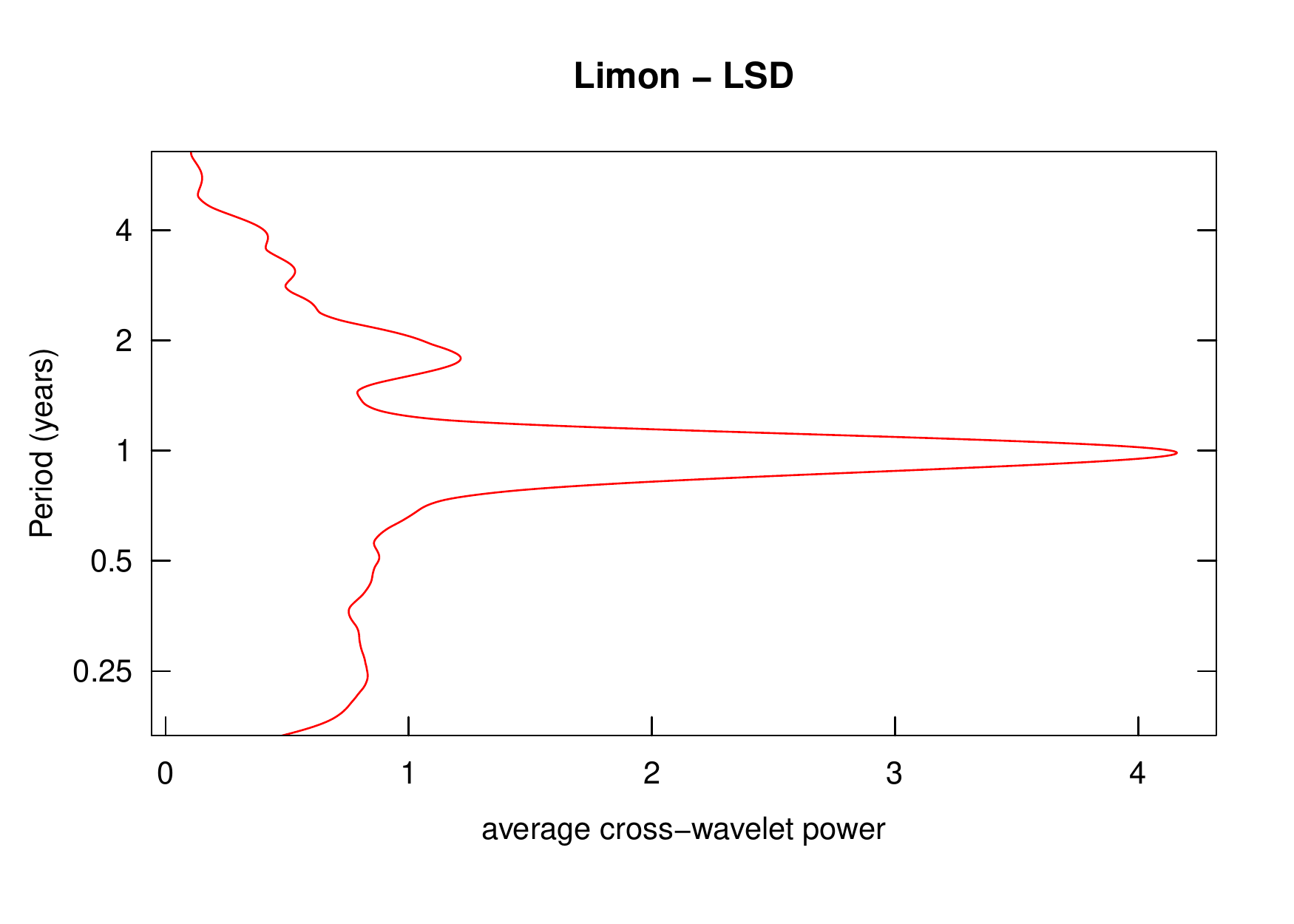}}\vspace{-0.15cm}%
\subfloat[]{\includegraphics[scale=0.23]{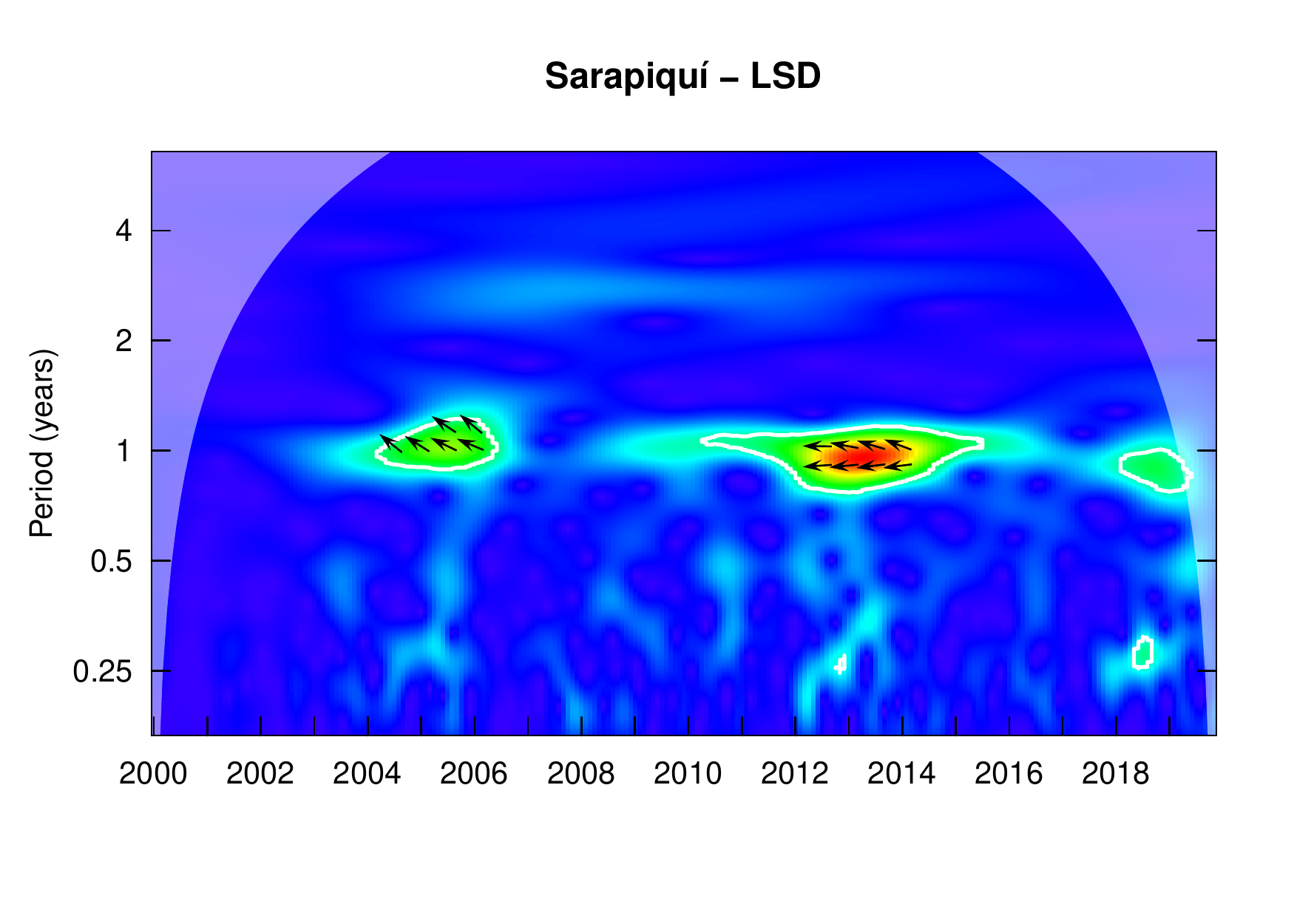}}\vspace{-0.15cm}%
\subfloat[]{\includegraphics[scale=0.23]{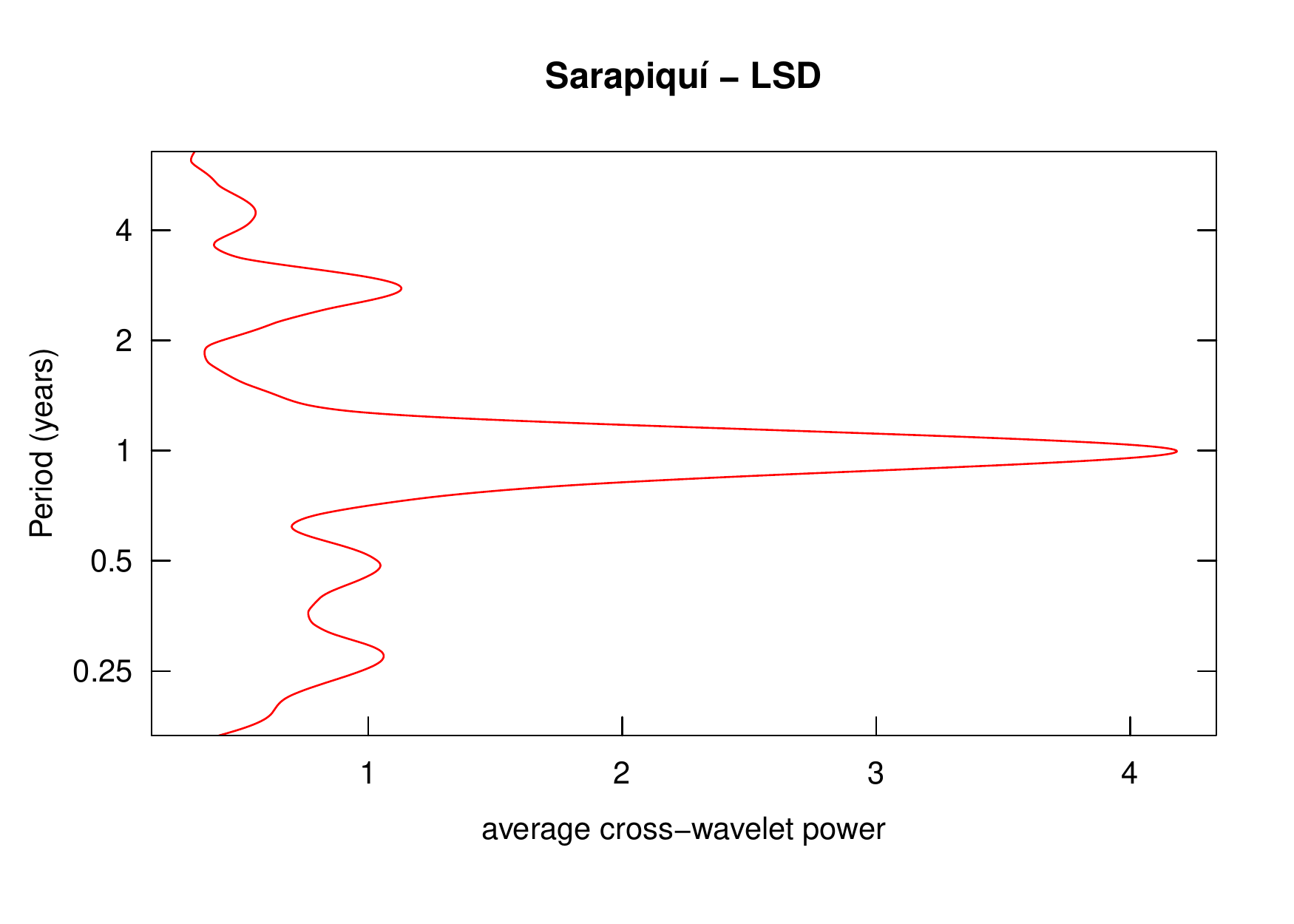}}\vspace{-0.15cm}\\
\caption*{}
\end{figure}

\begin{figure}[H]
\captionsetup[subfigure]{labelformat=empty}
\subfloat[]{\includegraphics[scale=0.23]{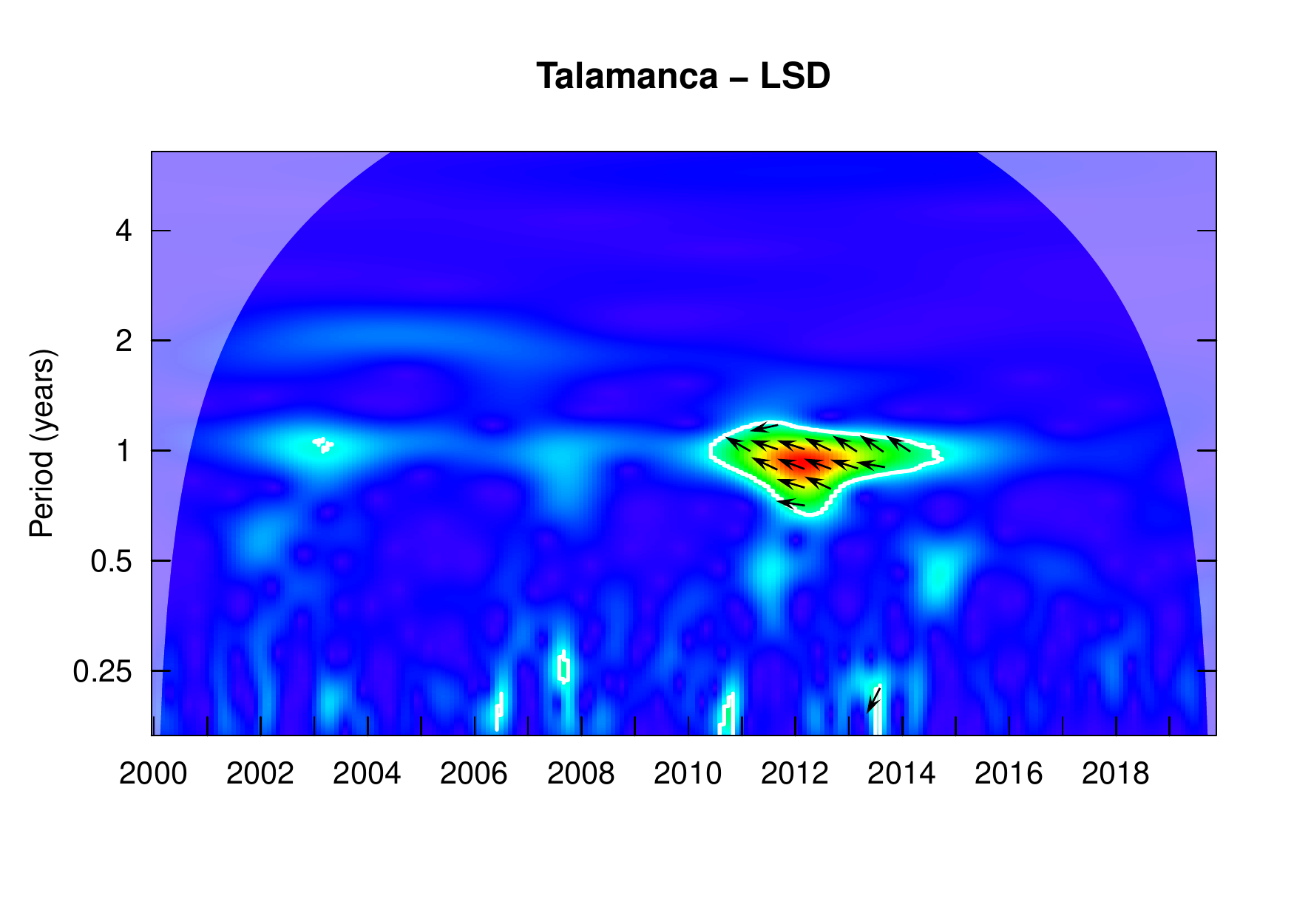}}\vspace{-0.15cm}%
\subfloat[]{\includegraphics[scale=0.23]{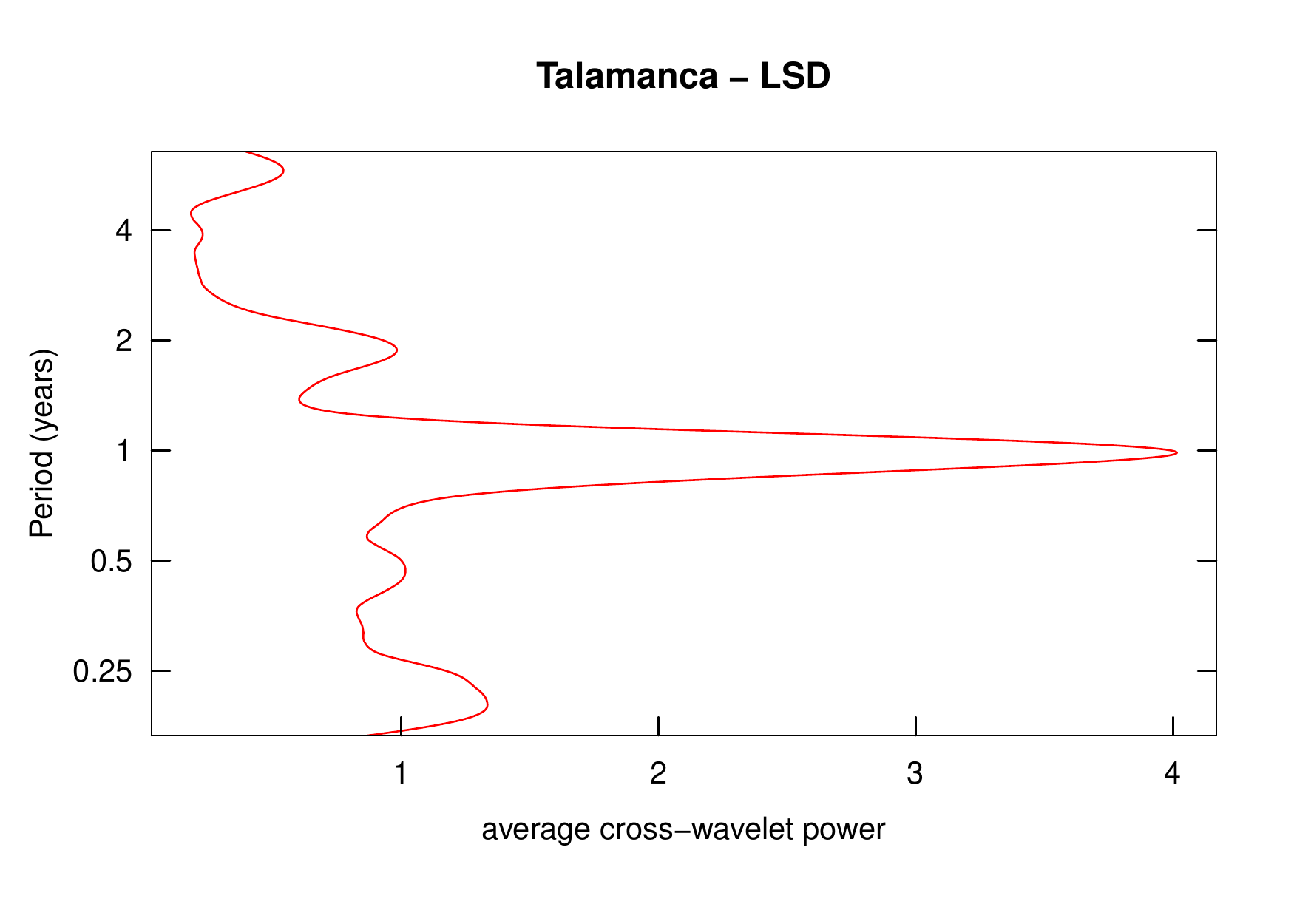}}\vspace{-0.15cm}%
\subfloat[]{\includegraphics[scale=0.23]{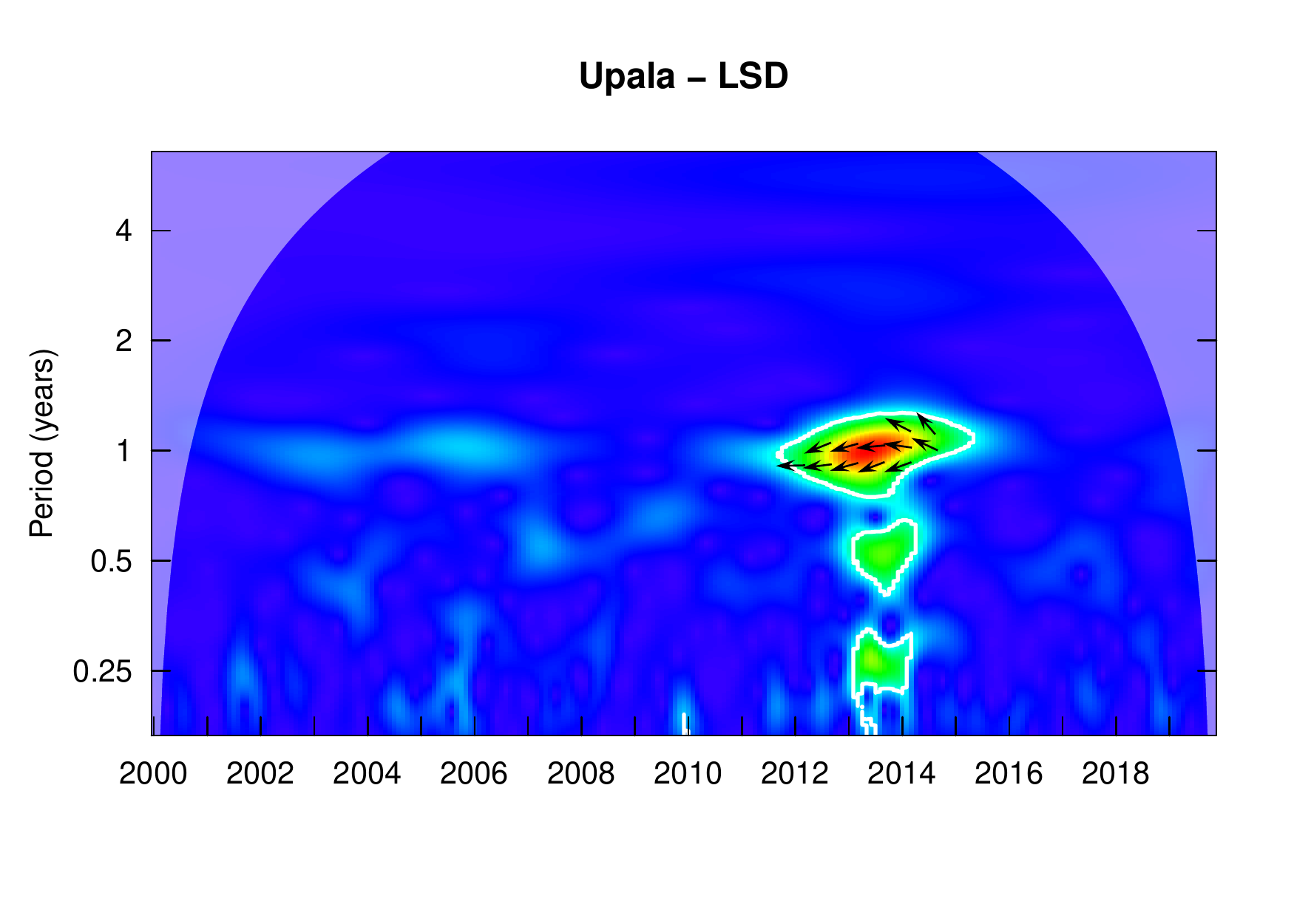}}\vspace{-0.15cm}%
\subfloat[]{\includegraphics[scale=0.23]{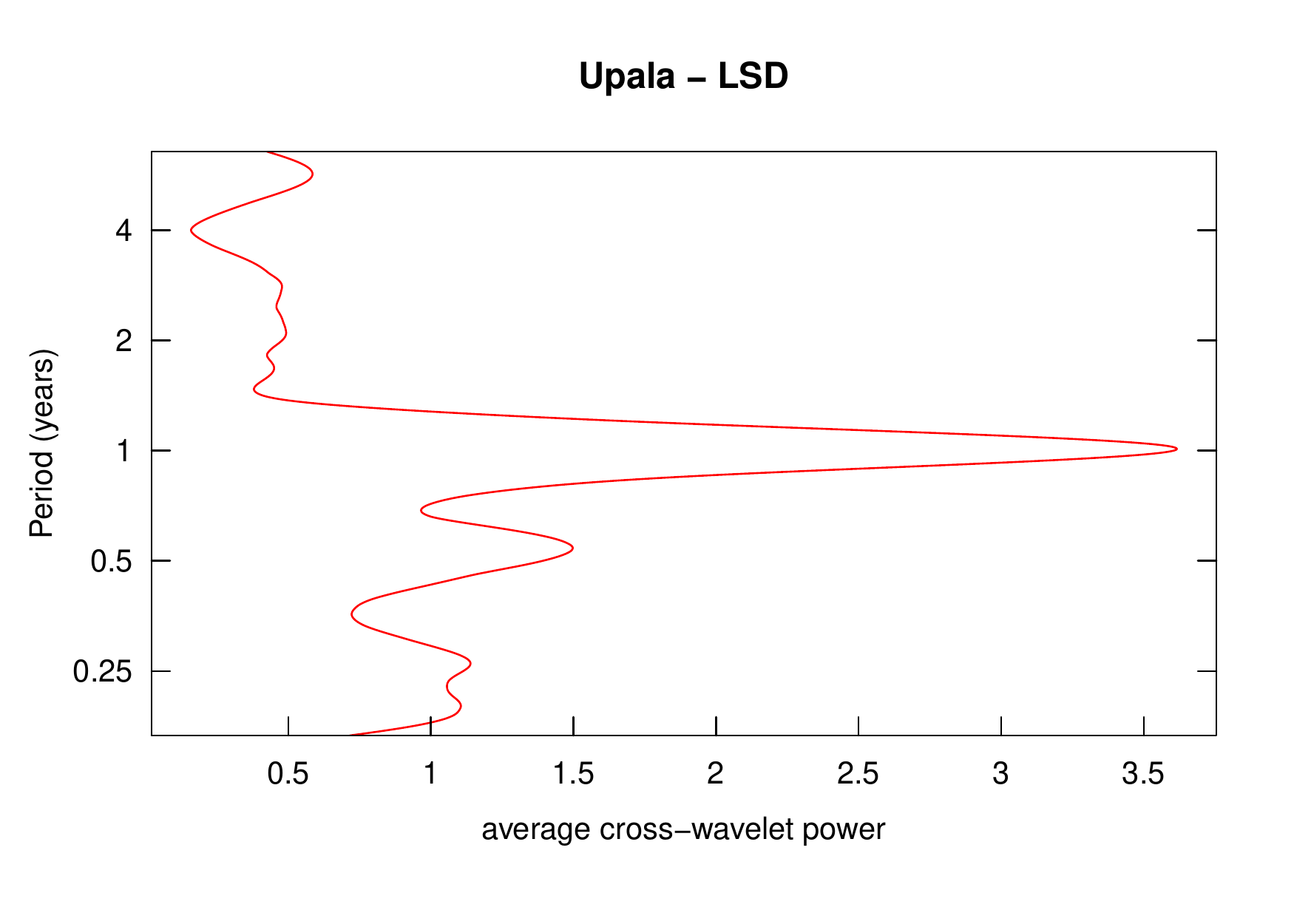}}\vspace{-0.15cm}\\
\subfloat[]{\includegraphics[scale=0.23]{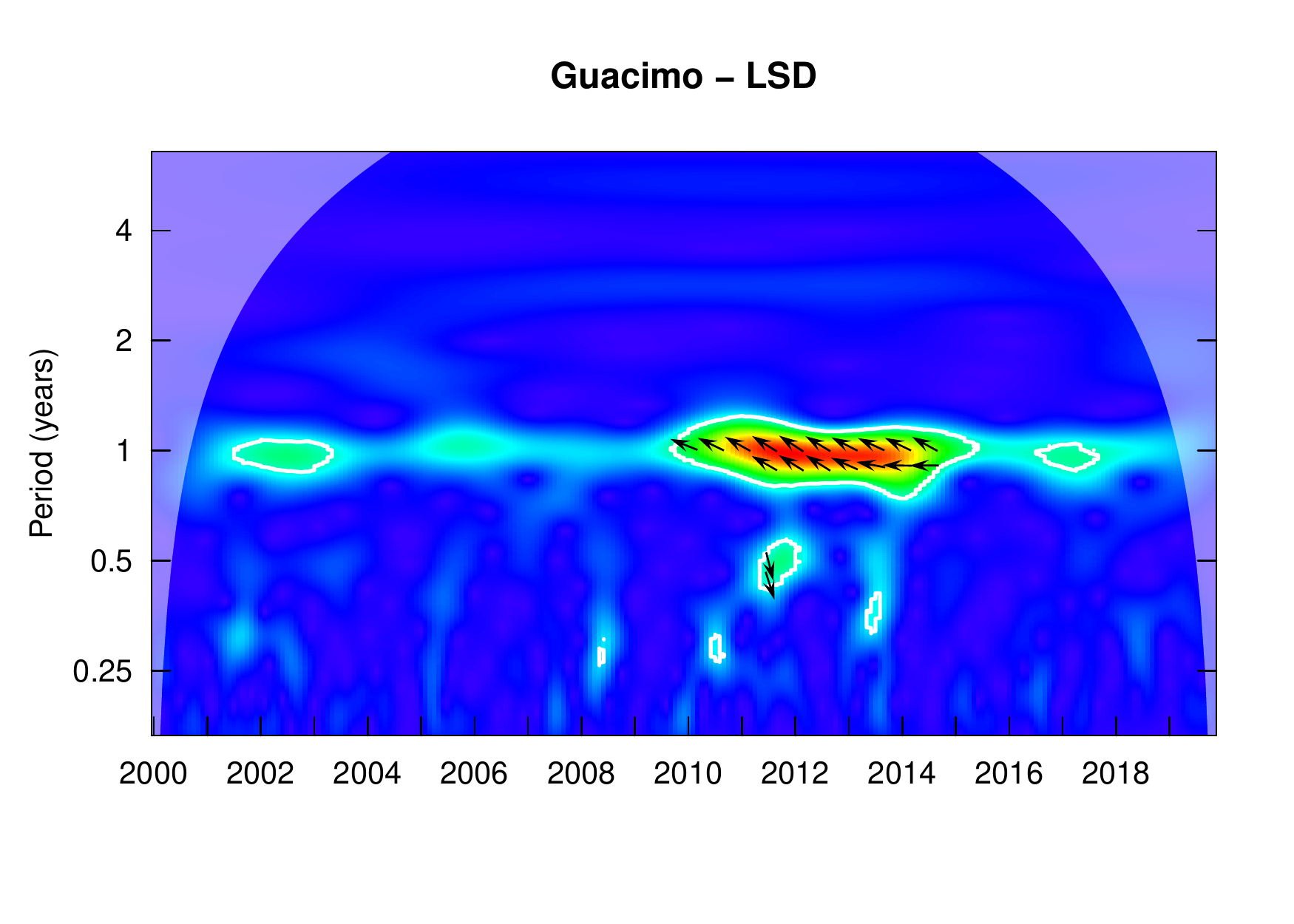}}\vspace{-0.15cm}%
\subfloat[]{\includegraphics[scale=0.23]{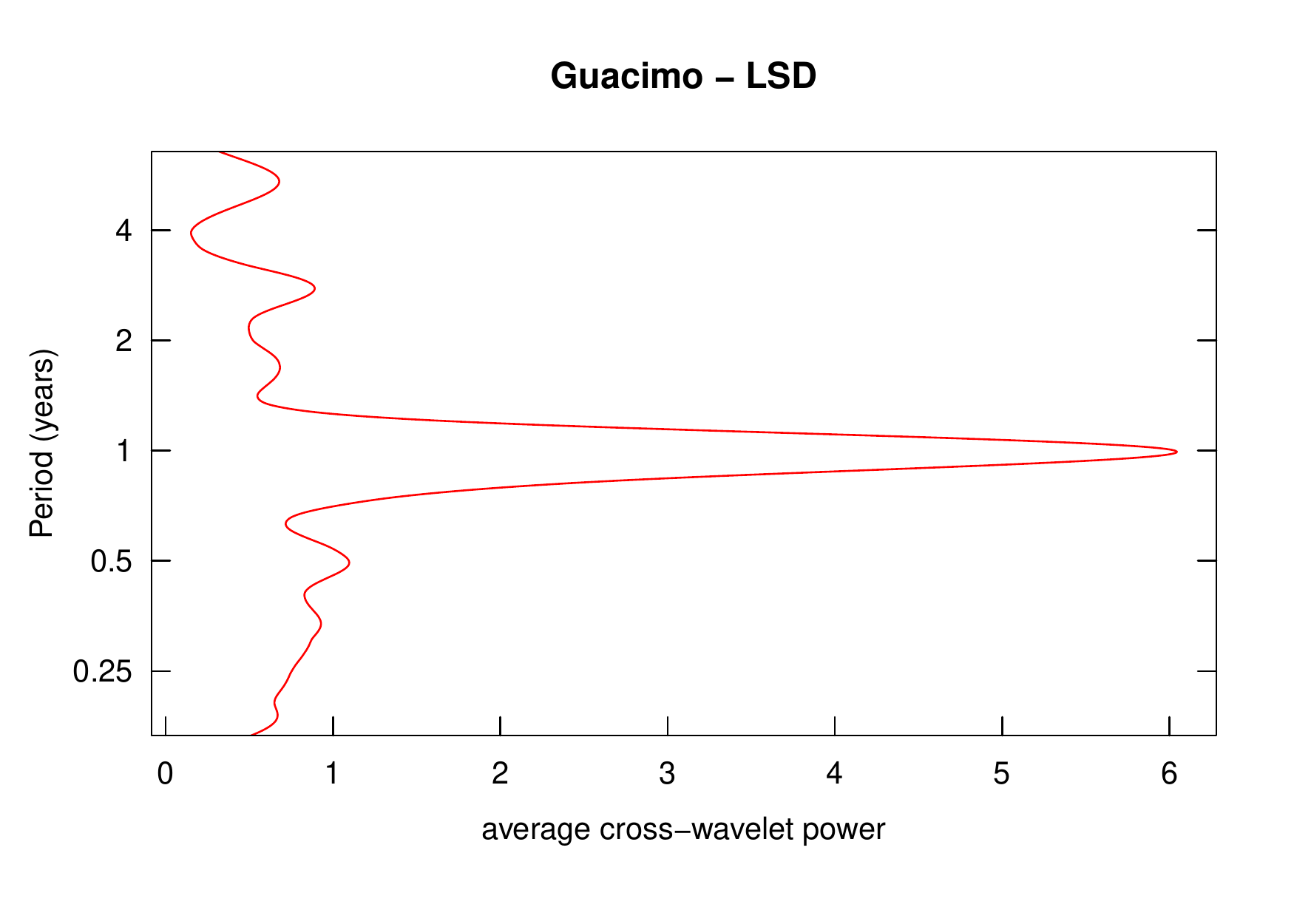}}\vspace{-0.15cm}%
\subfloat[]{\includegraphics[scale=0.23]{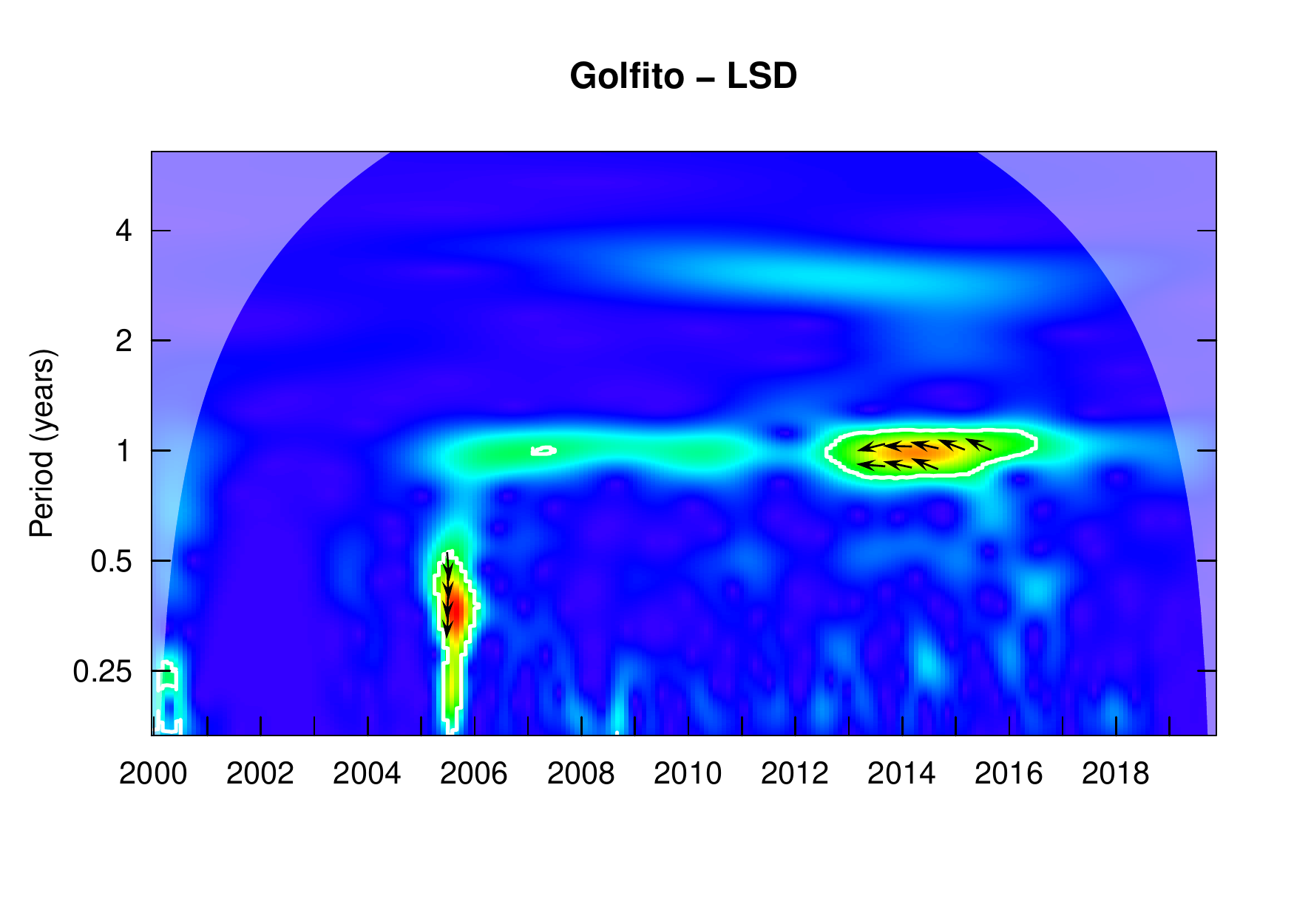}}\vspace{-0.15cm}%
\subfloat[]{\includegraphics[scale=0.23]{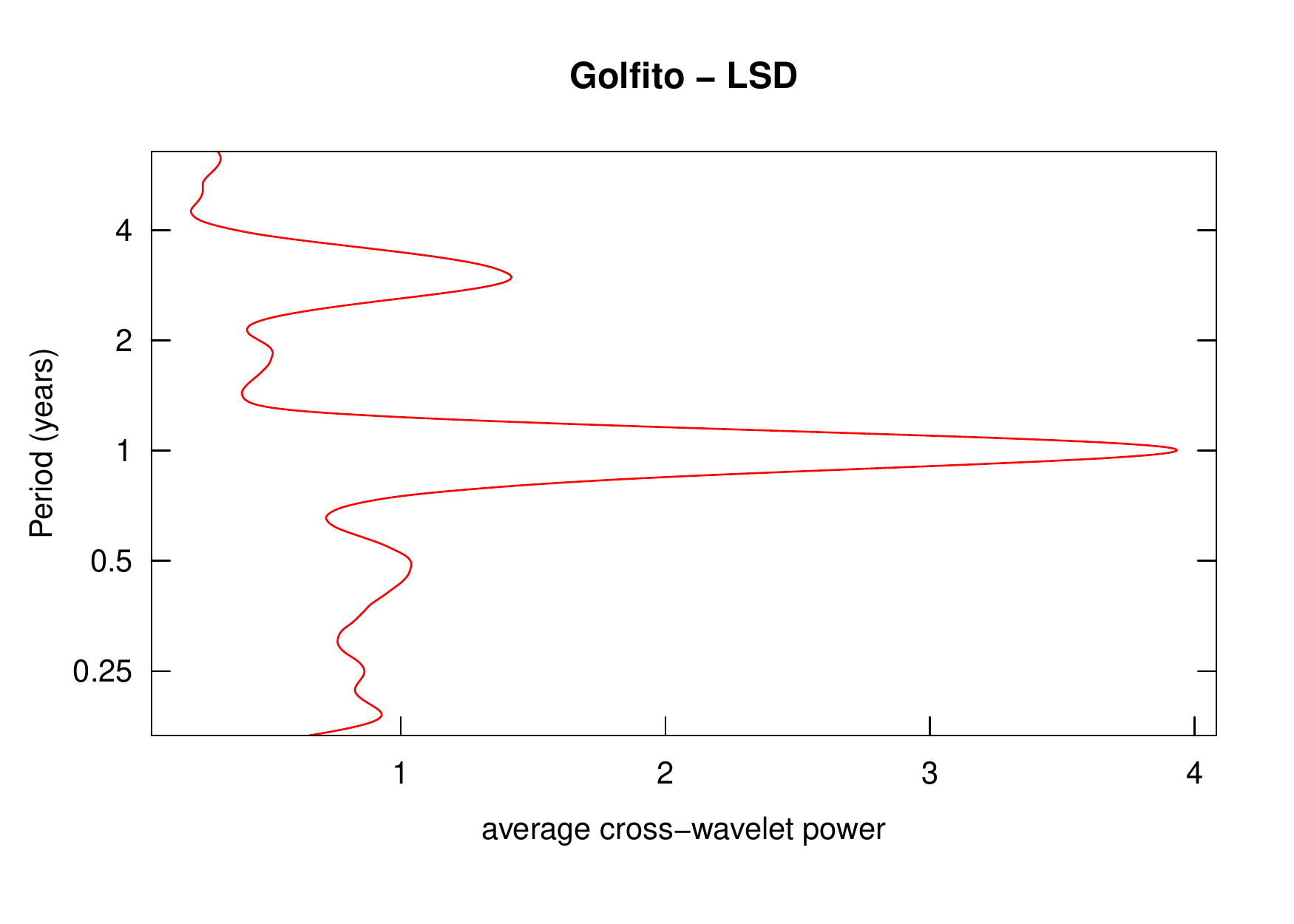}}\vspace{-0.15cm}\\
\subfloat[]{\includegraphics[scale=0.23]{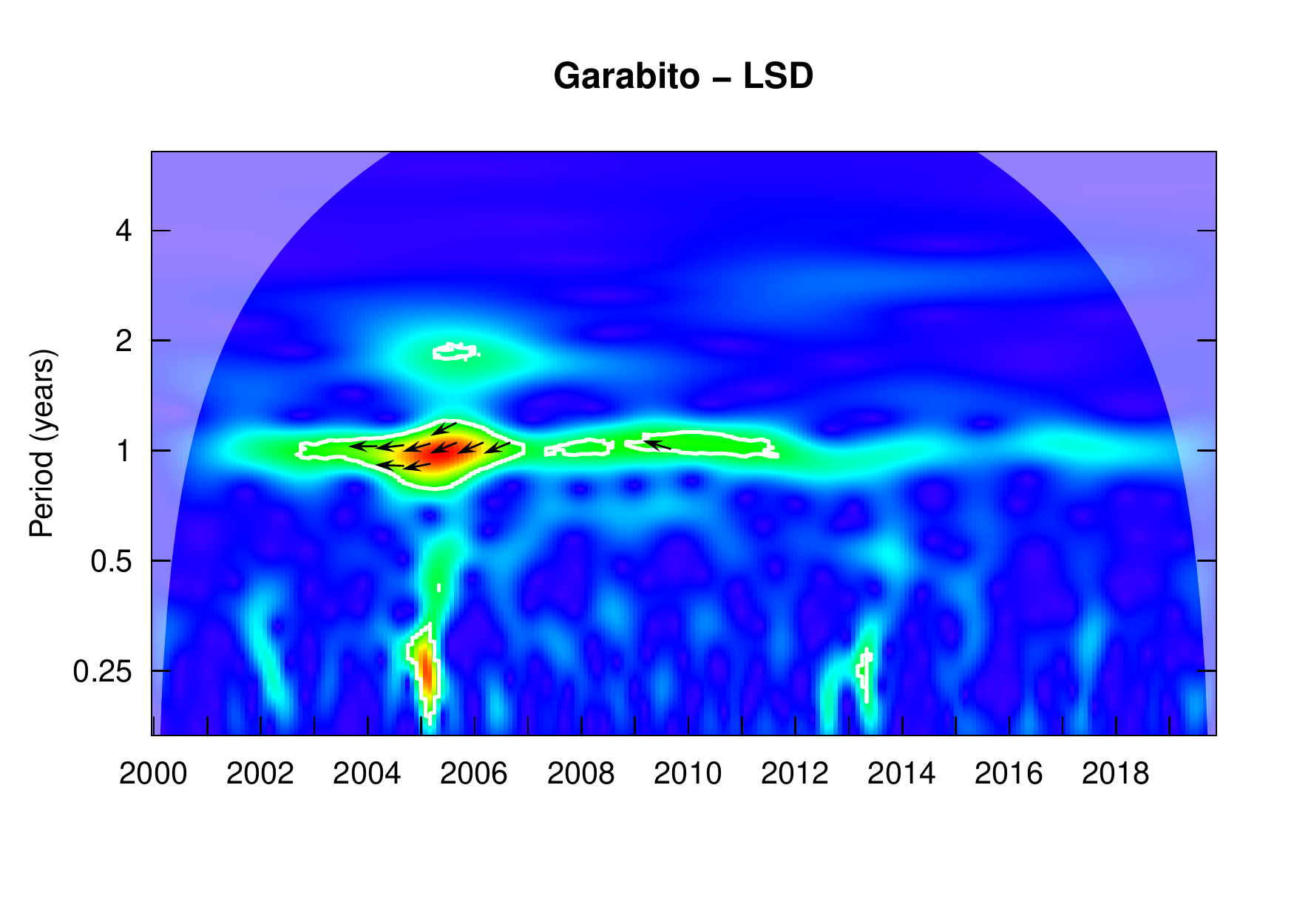}}\vspace{-0.15cm}%
\subfloat[]{\includegraphics[scale=0.23]{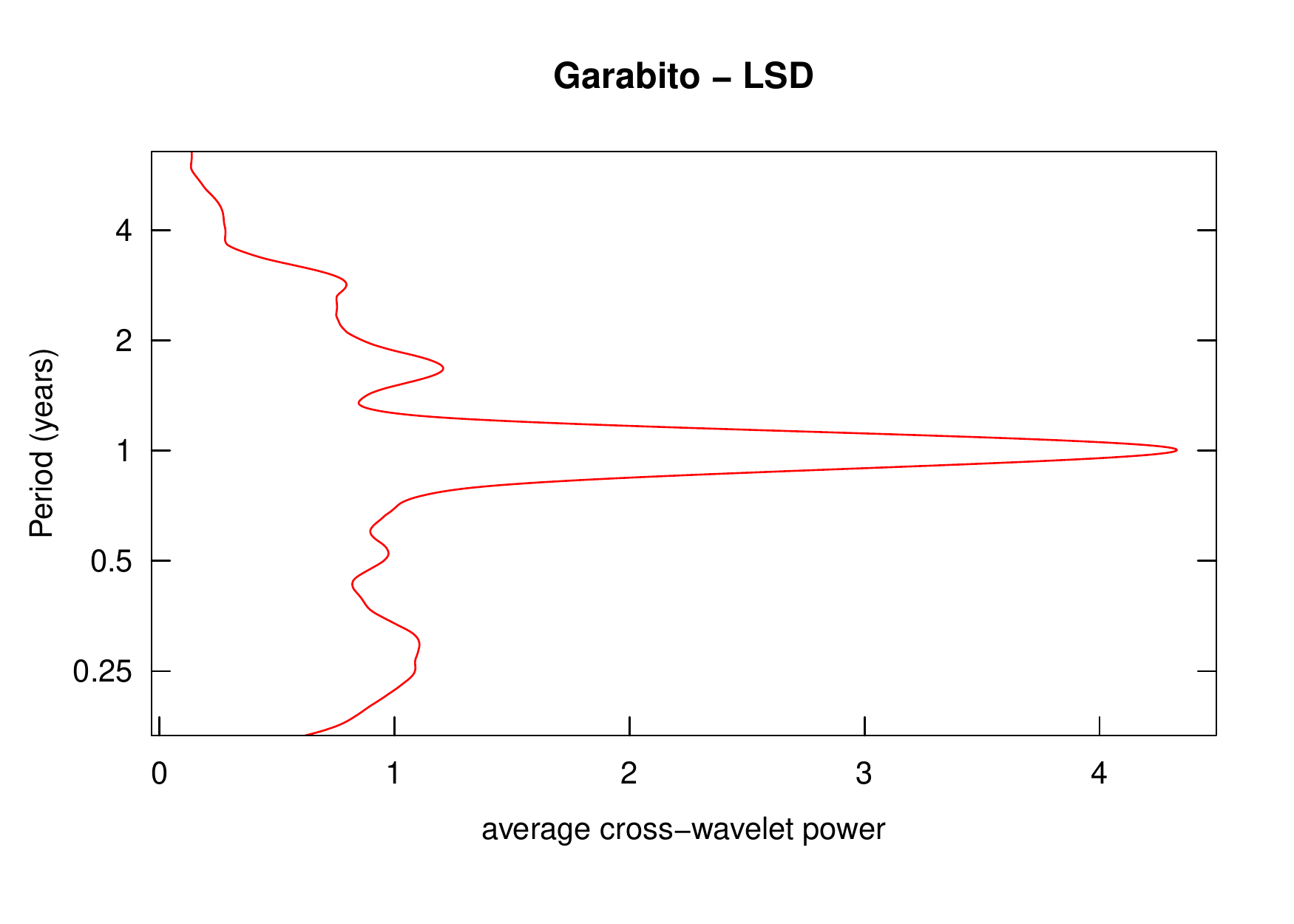}}\vspace{-0.15cm}%
\subfloat[]{\includegraphics[scale=0.23]{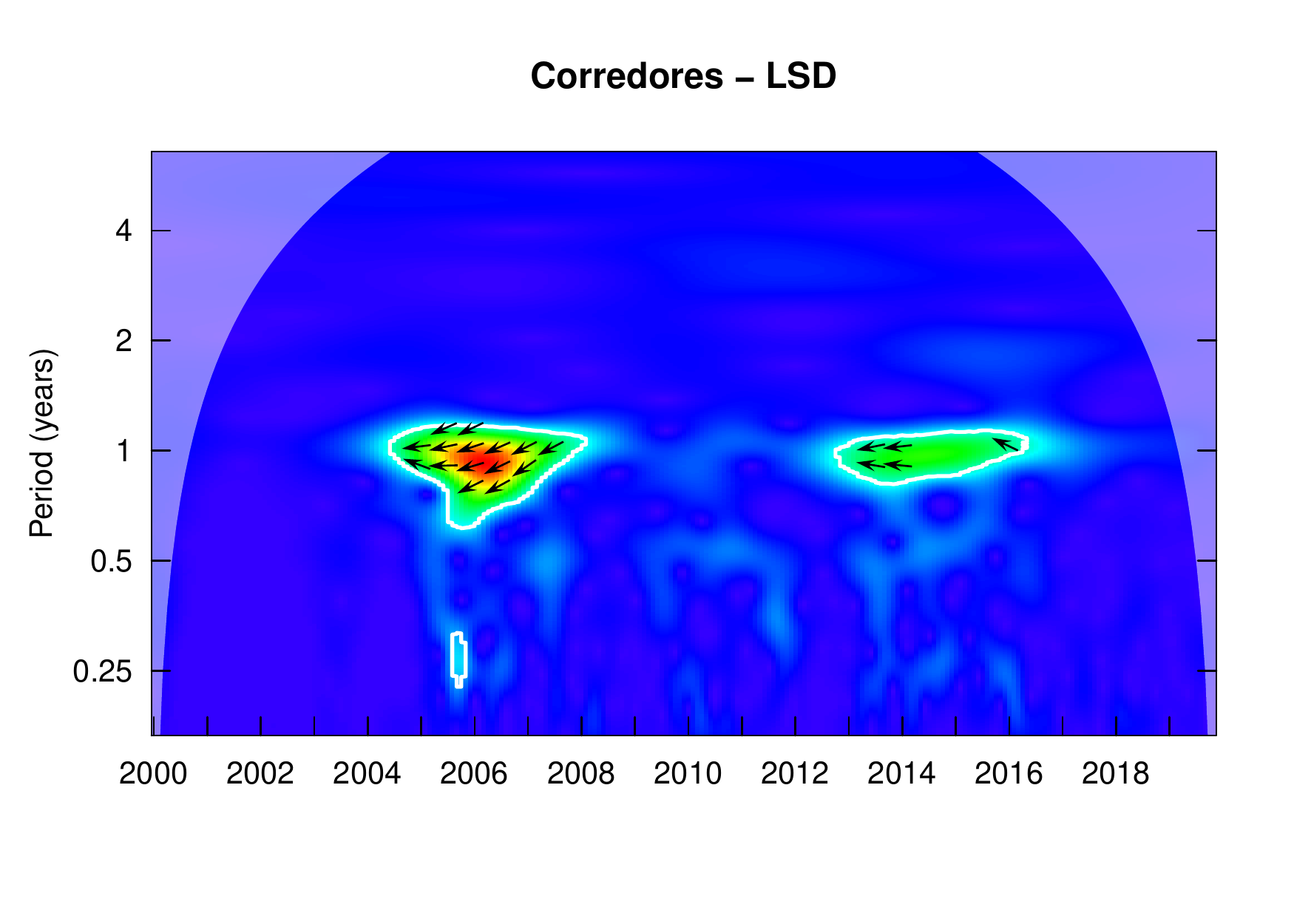}}\vspace{-0.15cm}%
\subfloat[]{\includegraphics[scale=0.23]{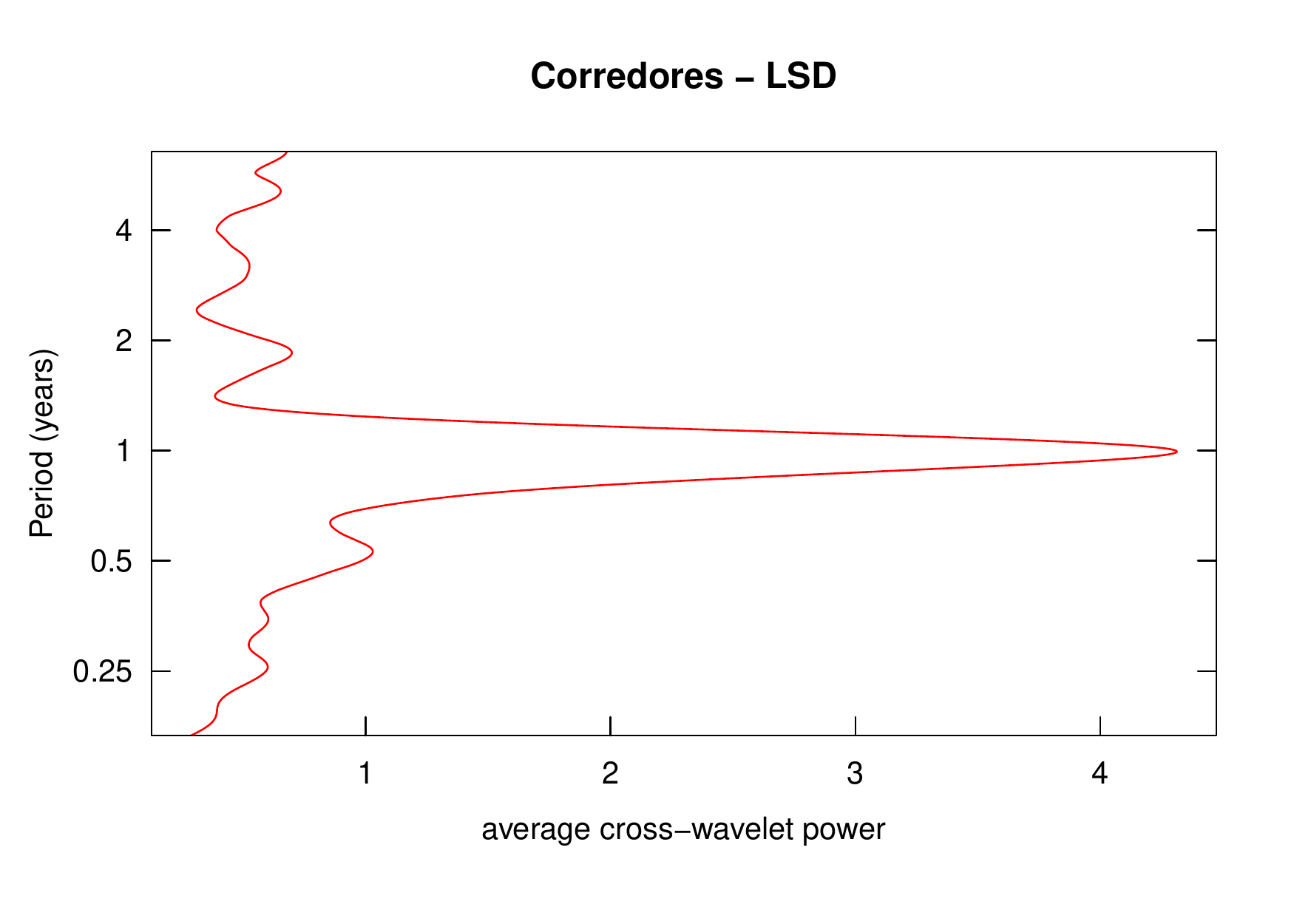}}\vspace{-0.15cm}\\
\subfloat[]{\includegraphics[scale=0.23]{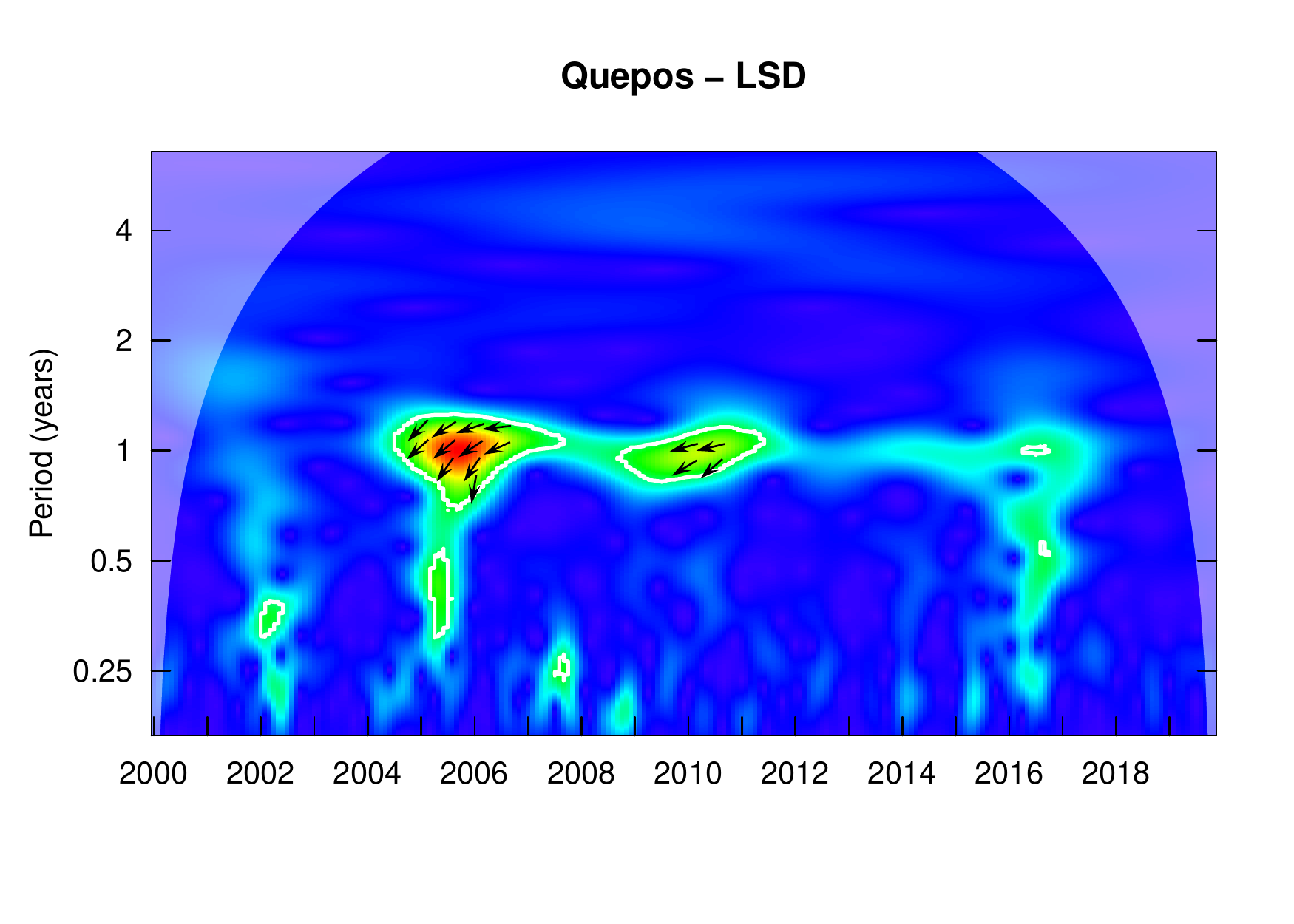}}\vspace{-0.15cm}%
\subfloat[]{\includegraphics[scale=0.23]{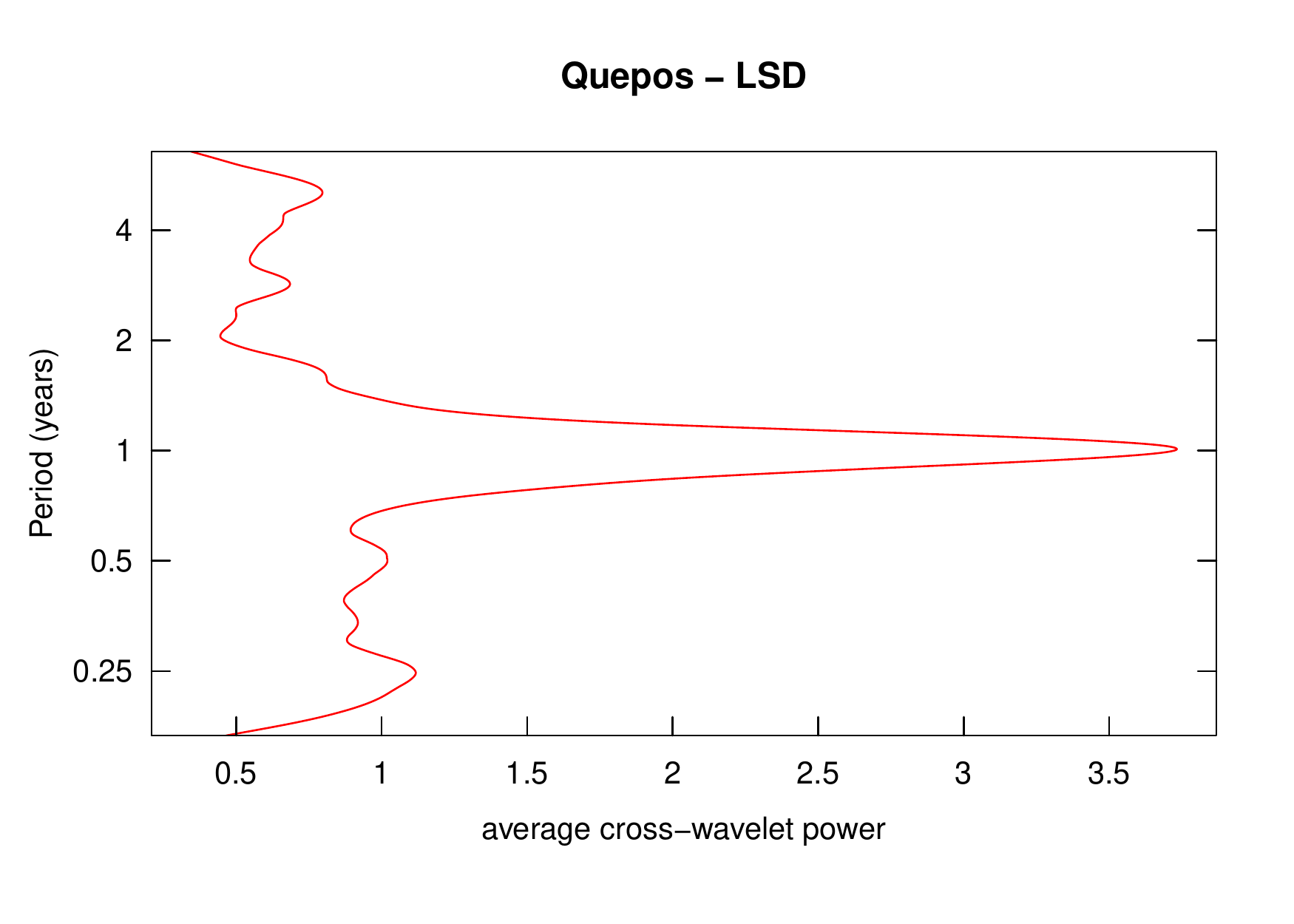}}\vspace{-0.15cm}%
\subfloat[]{\includegraphics[scale=0.23]{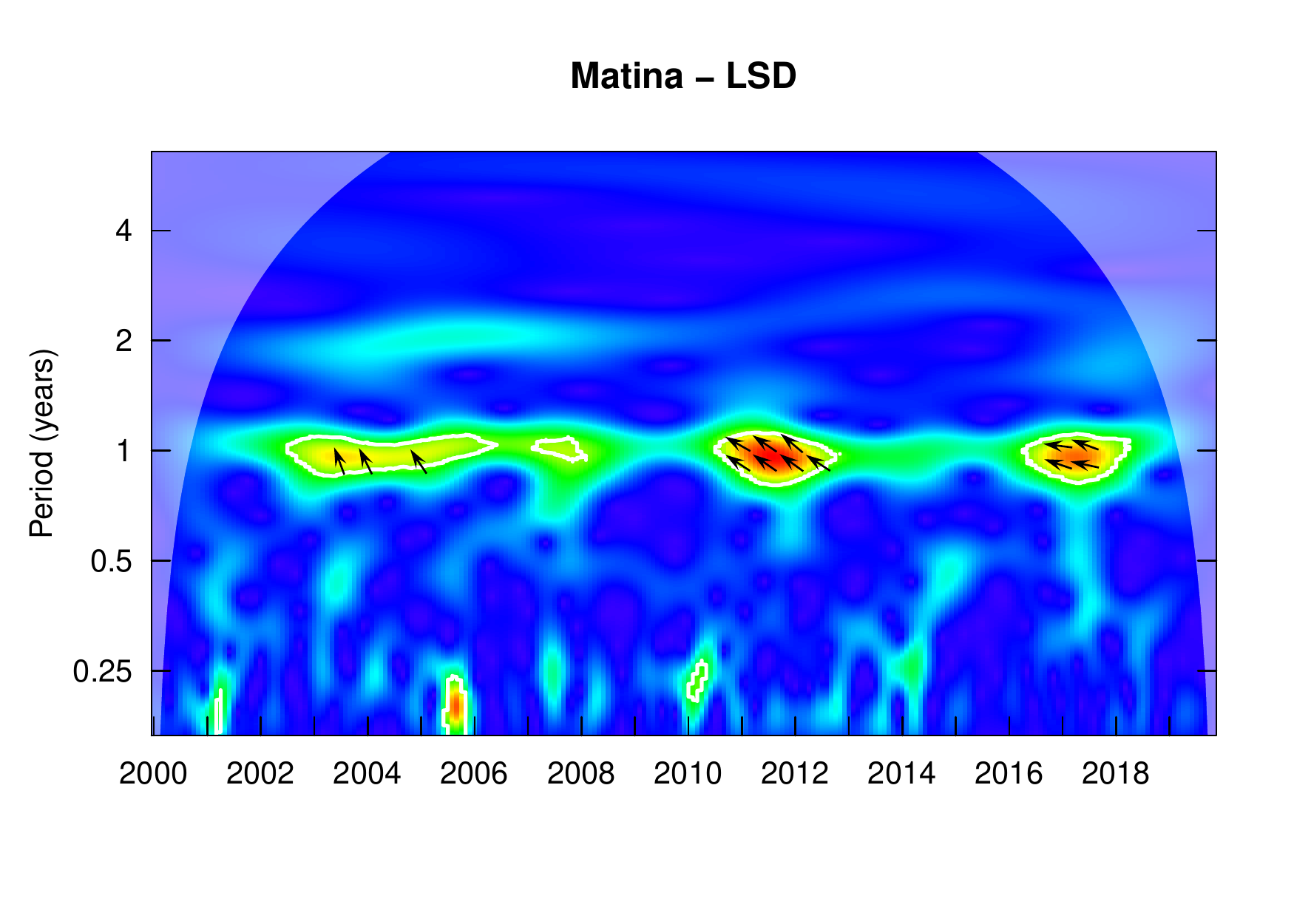}}\vspace{-0.15cm}%
\subfloat[]{\includegraphics[scale=0.23]{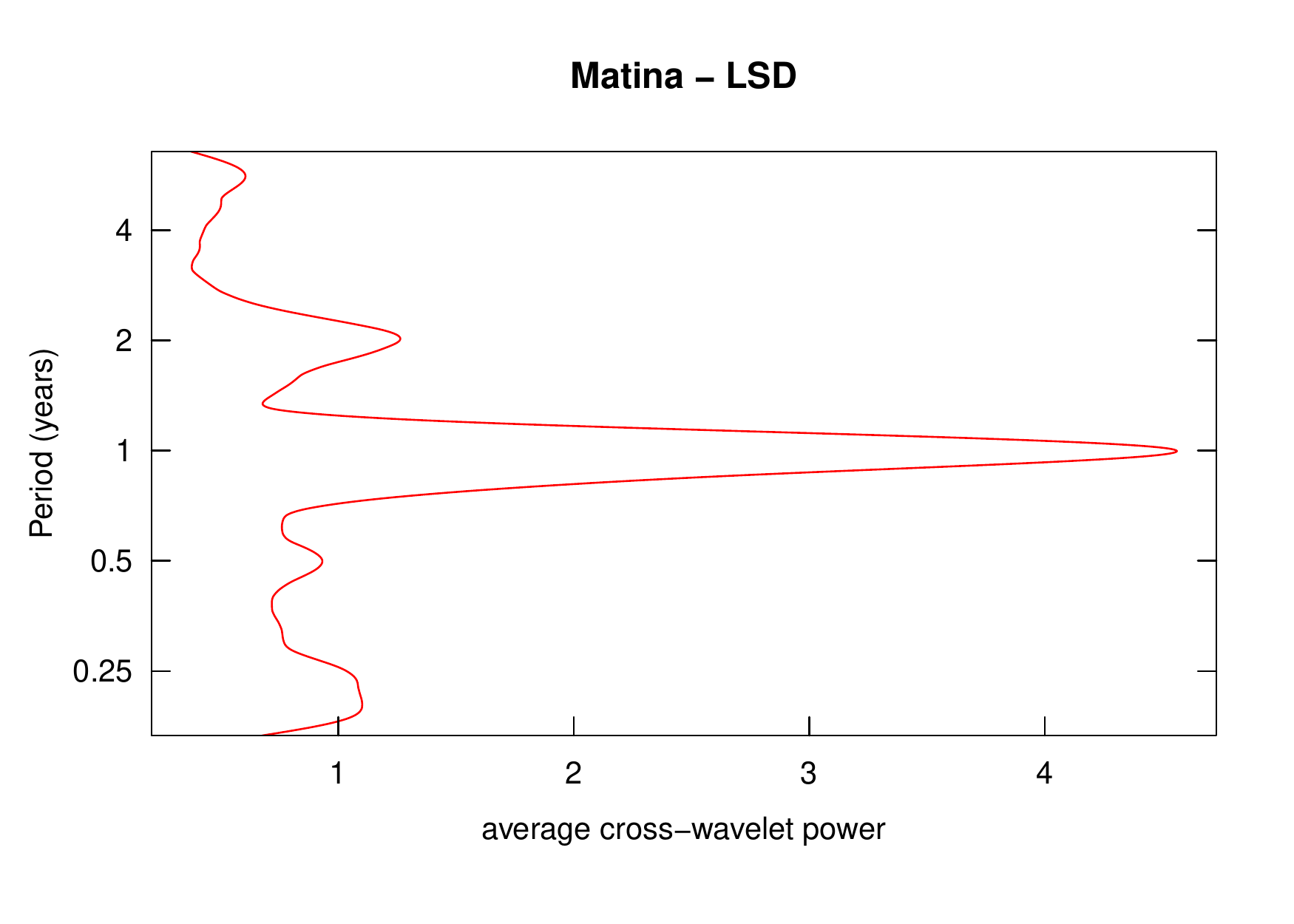}}\vspace{-0.15cm}\\
\subfloat[]{\includegraphics[scale=0.23]{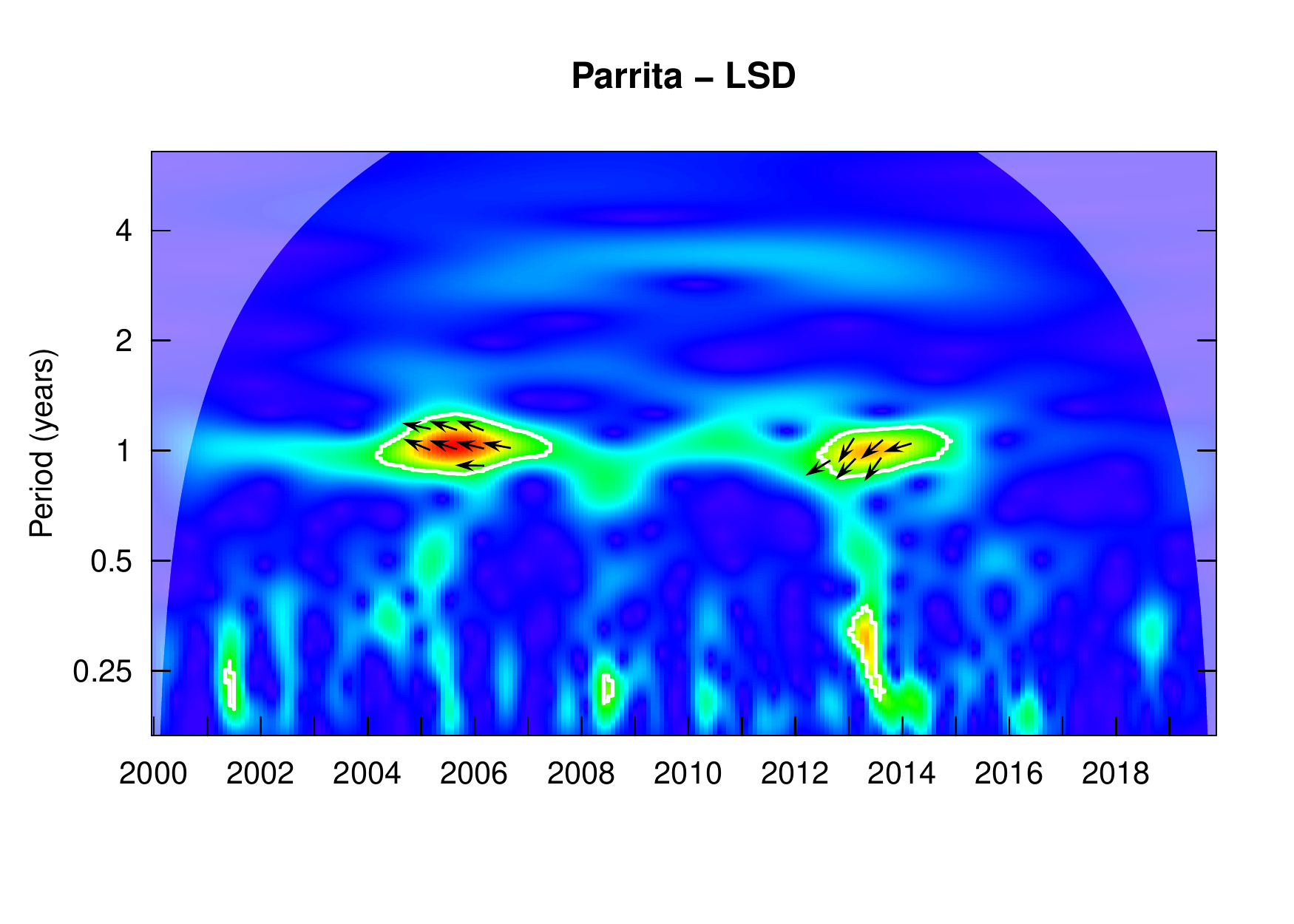}}\vspace{-0.15cm}%
\subfloat[]{\includegraphics[scale=0.23]{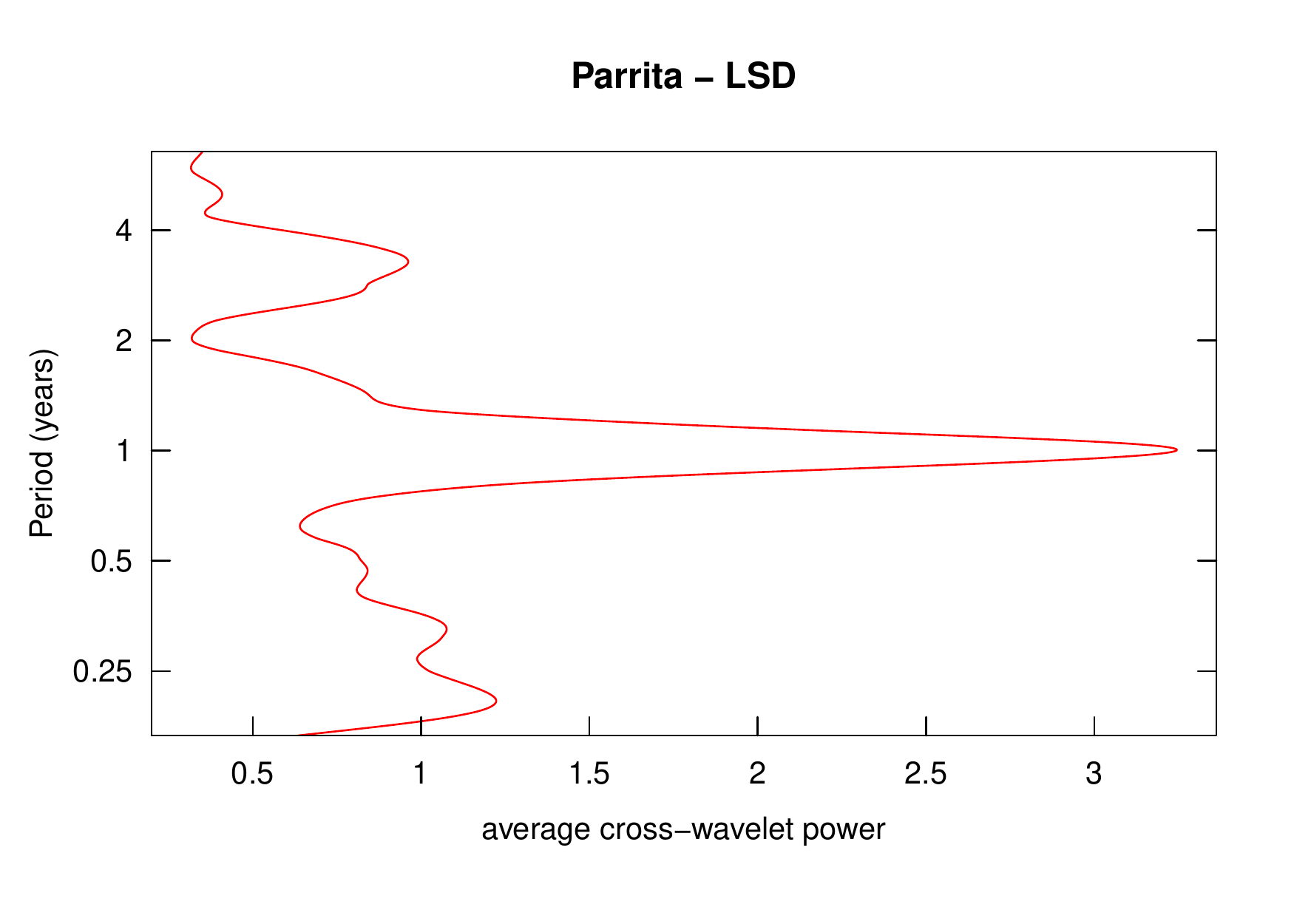}}\vspace{-0.15cm}%
\subfloat[]{\includegraphics[scale=0.23]{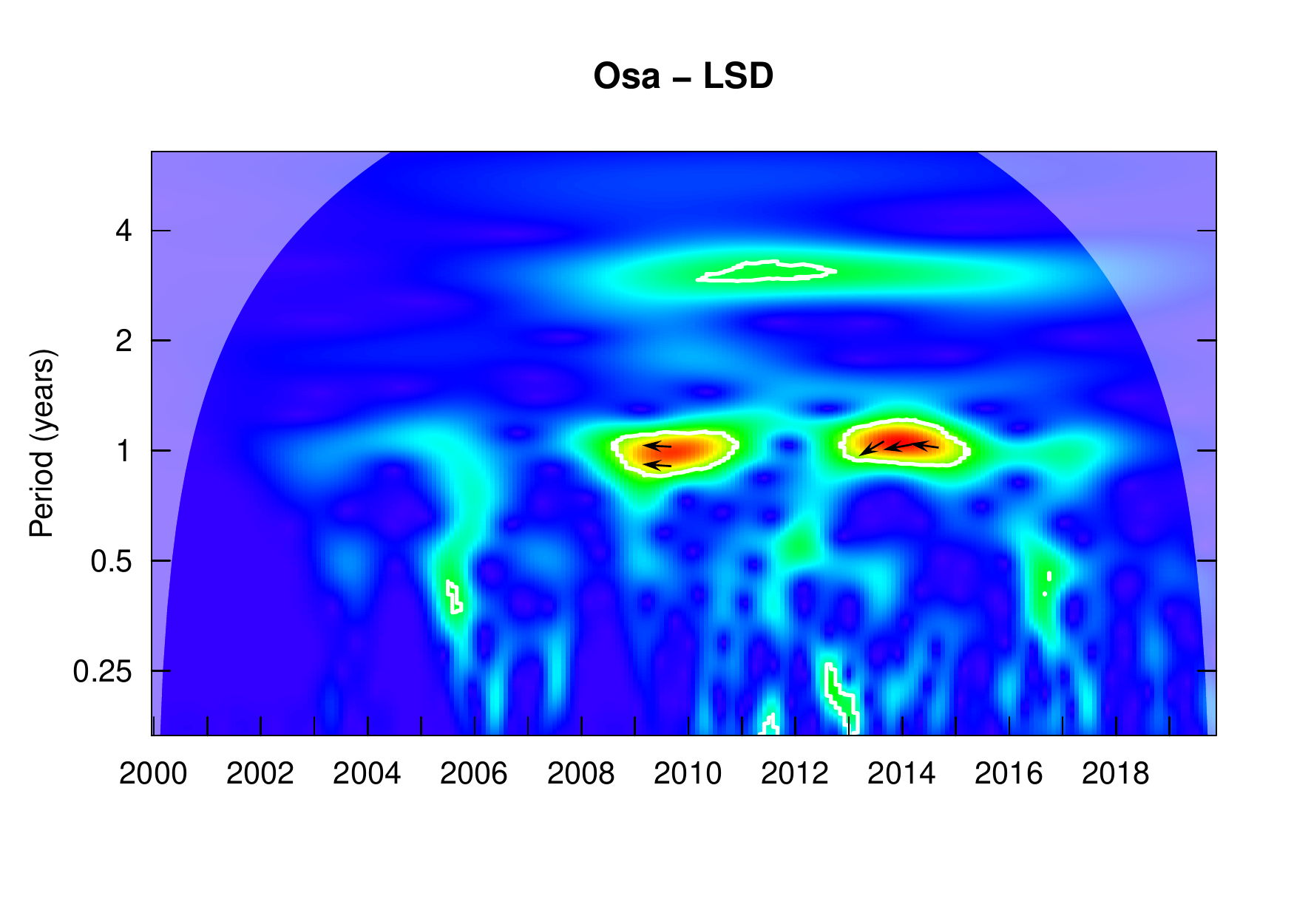}}\vspace{-0.15cm}%
\subfloat[]{\includegraphics[scale=0.23]{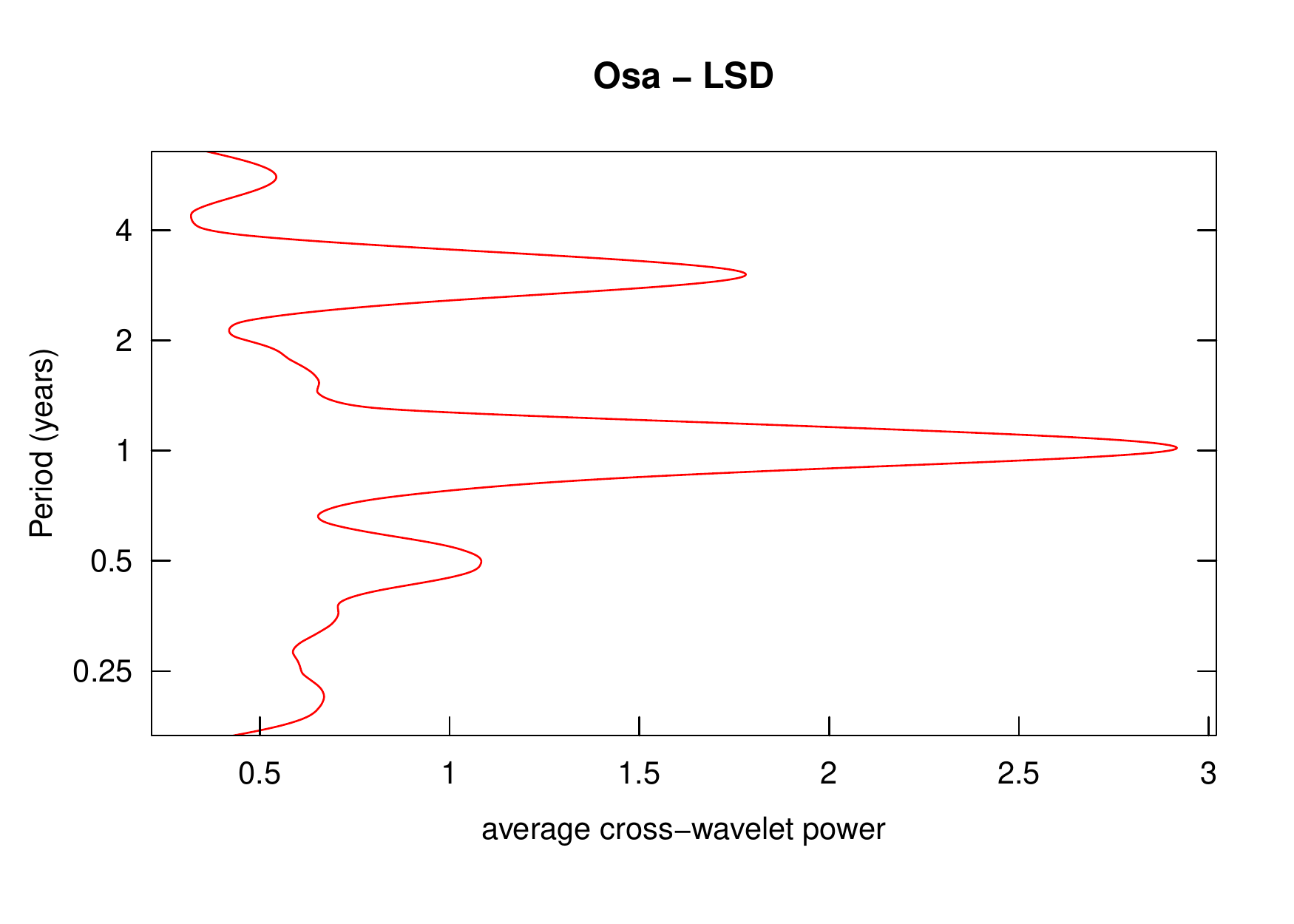}}\vspace{-0.15cm}
\caption*{}
\end{figure}

\section*{Wavelet coherence and average cross-wavelet power between dengue incidence and LSN}

\begin{figure}[H]
\captionsetup[subfigure]{labelformat=empty}
\caption*{\textbf{Figure S5:} Wavelet coherence (color map) between dengue incidence from 2000 to 2019, and LSN in 32 municipalities of Costa Rica (periodicity on y-axis, time on x-axis). Colors code for increasing power intensity, from blue to red; $95\%$ confidence levels are encircled by white lines, and shaded areas indicate the presence of significant edge effects. On the right side of each wavelet coherence is the average cross-wavelet power (Red line). The arrows indicate whether the two series are in-phase or out-phase.}
\subfloat[]{\includegraphics[scale=0.23]{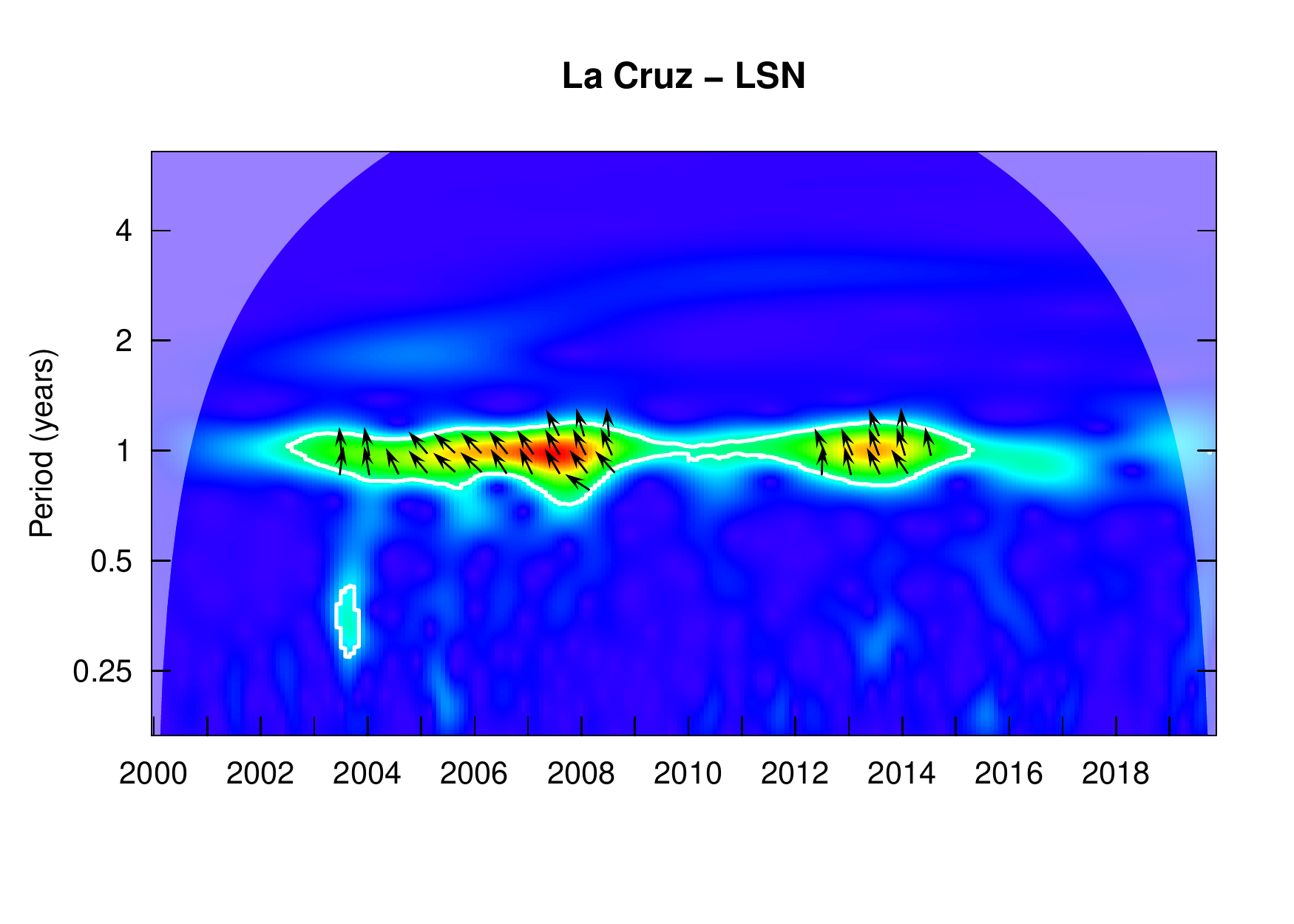}}\vspace{-0.15cm}%
\subfloat[]{\includegraphics[scale=0.23]{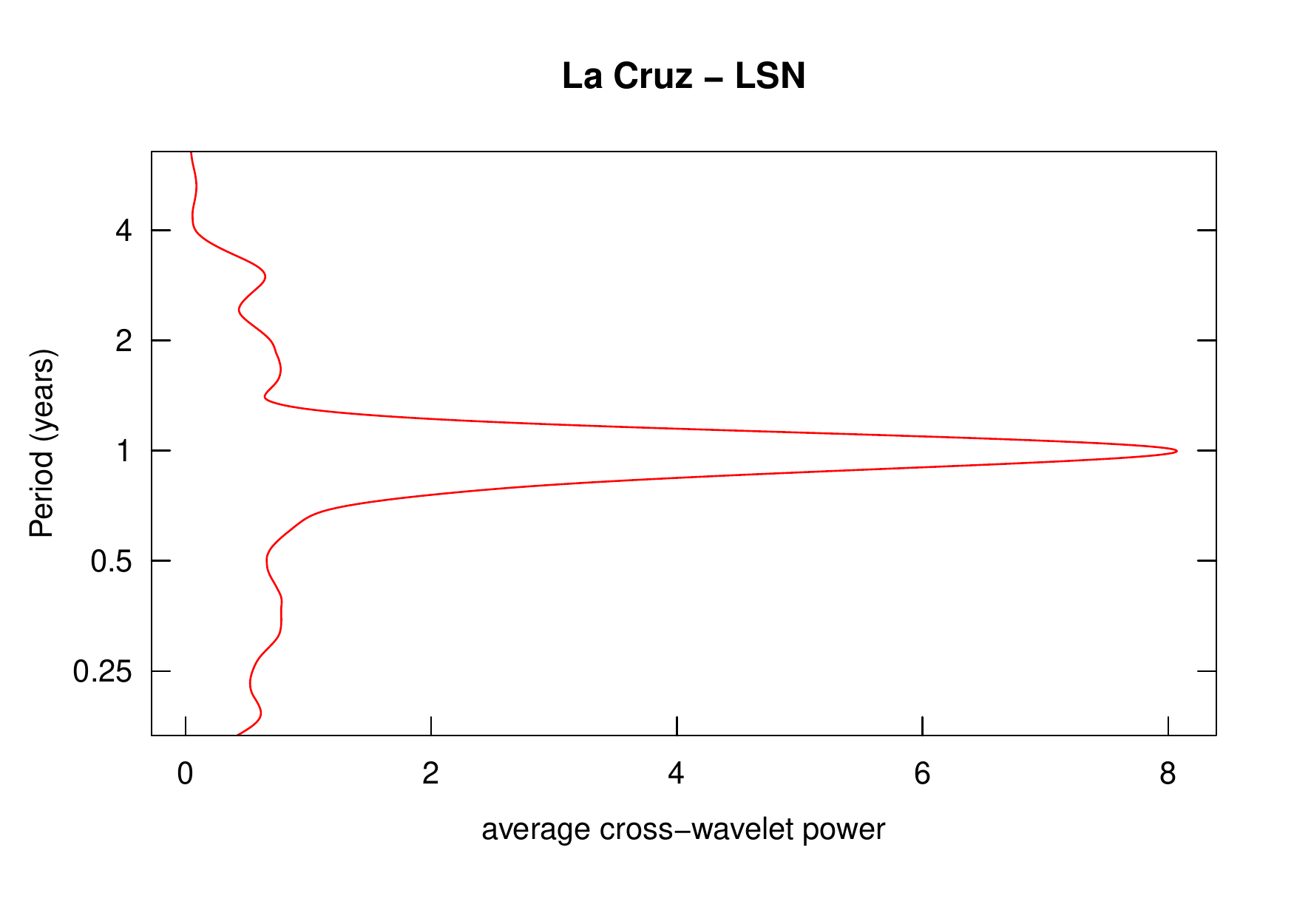}}\vspace{-0.15cm}%
\subfloat[]{\includegraphics[scale=0.23]{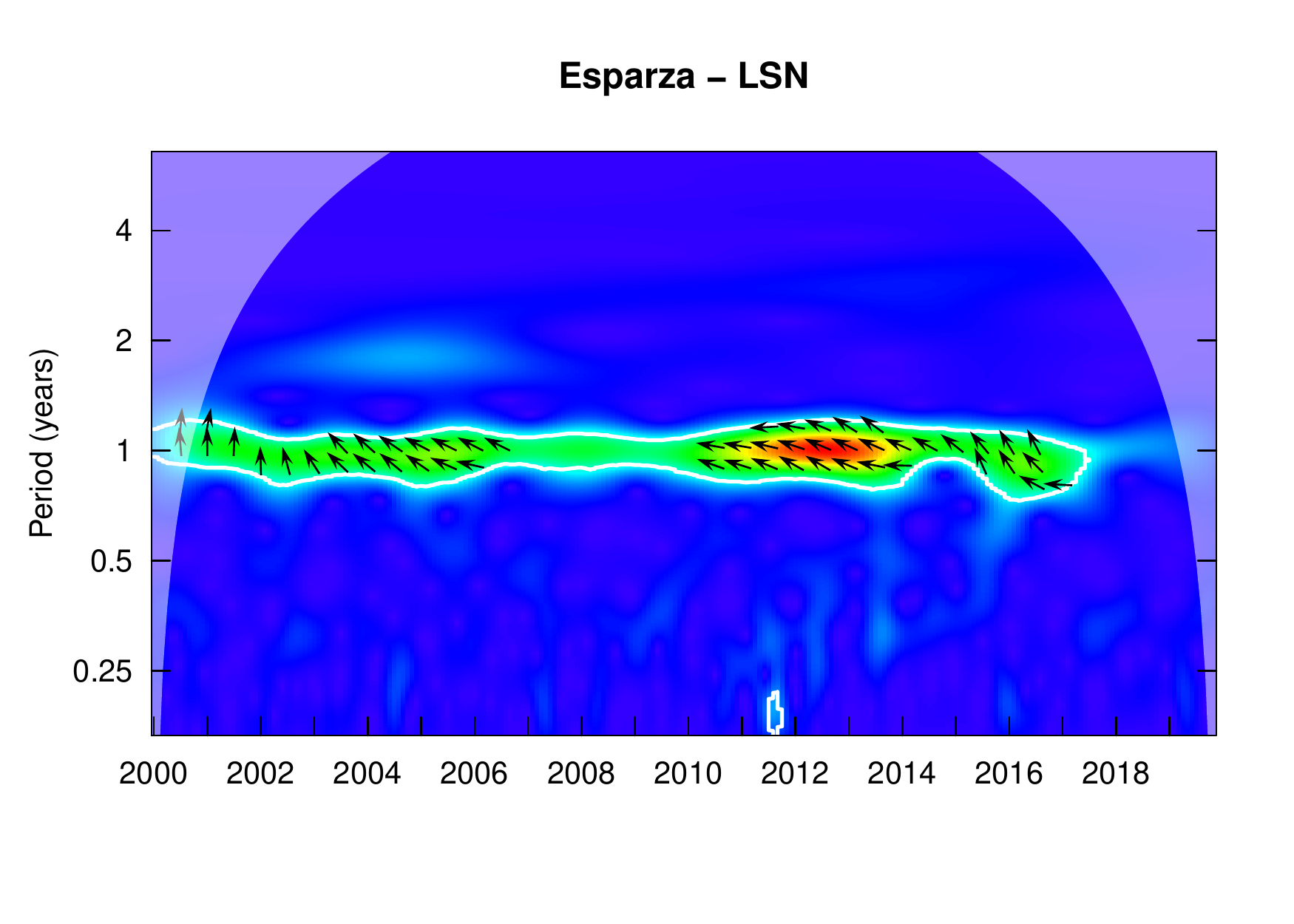}}\vspace{-0.15cm}%
\subfloat[]{\includegraphics[scale=0.23]{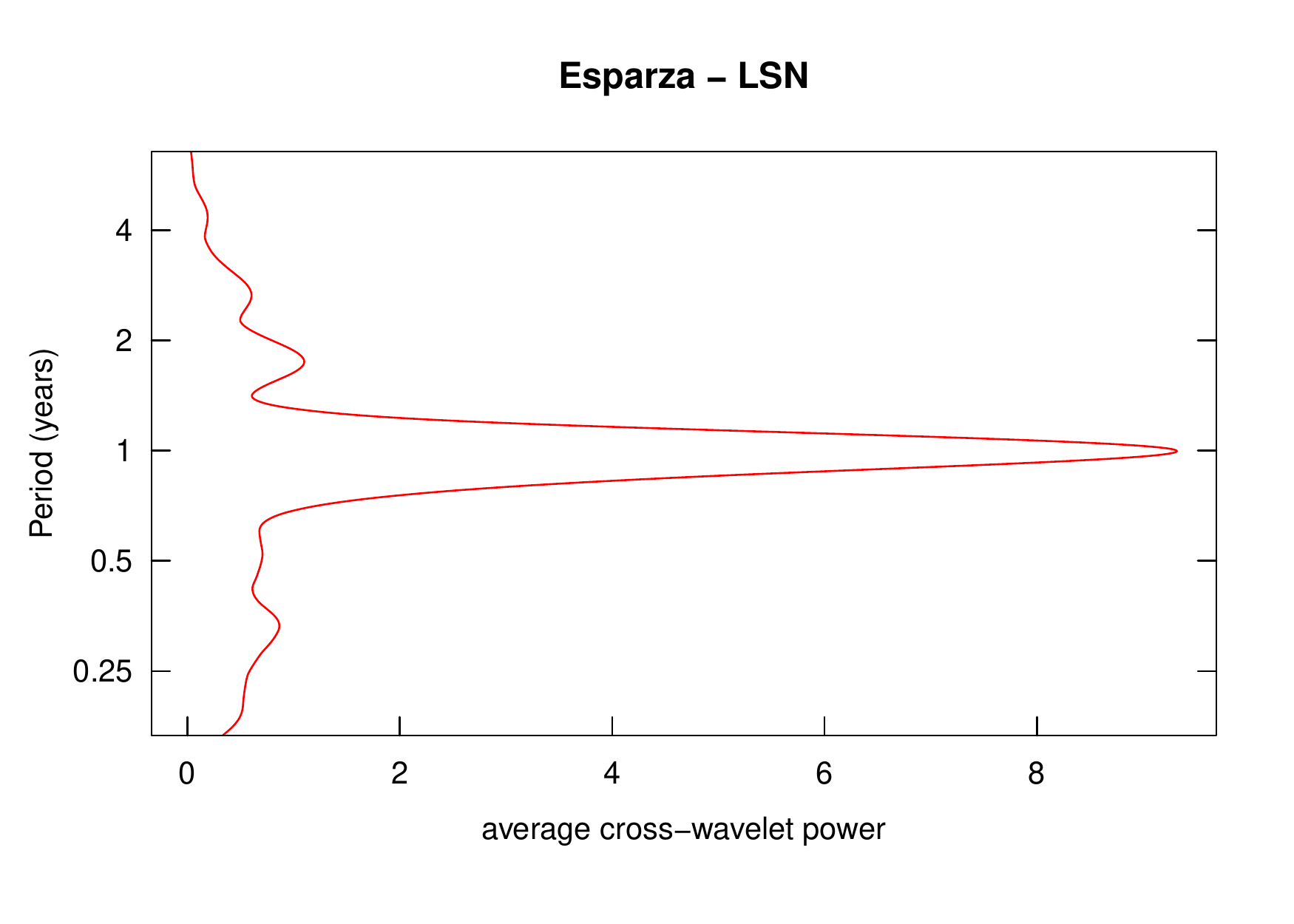}}\vspace{-0.15cm}\\
\subfloat[]{\includegraphics[scale=0.23]{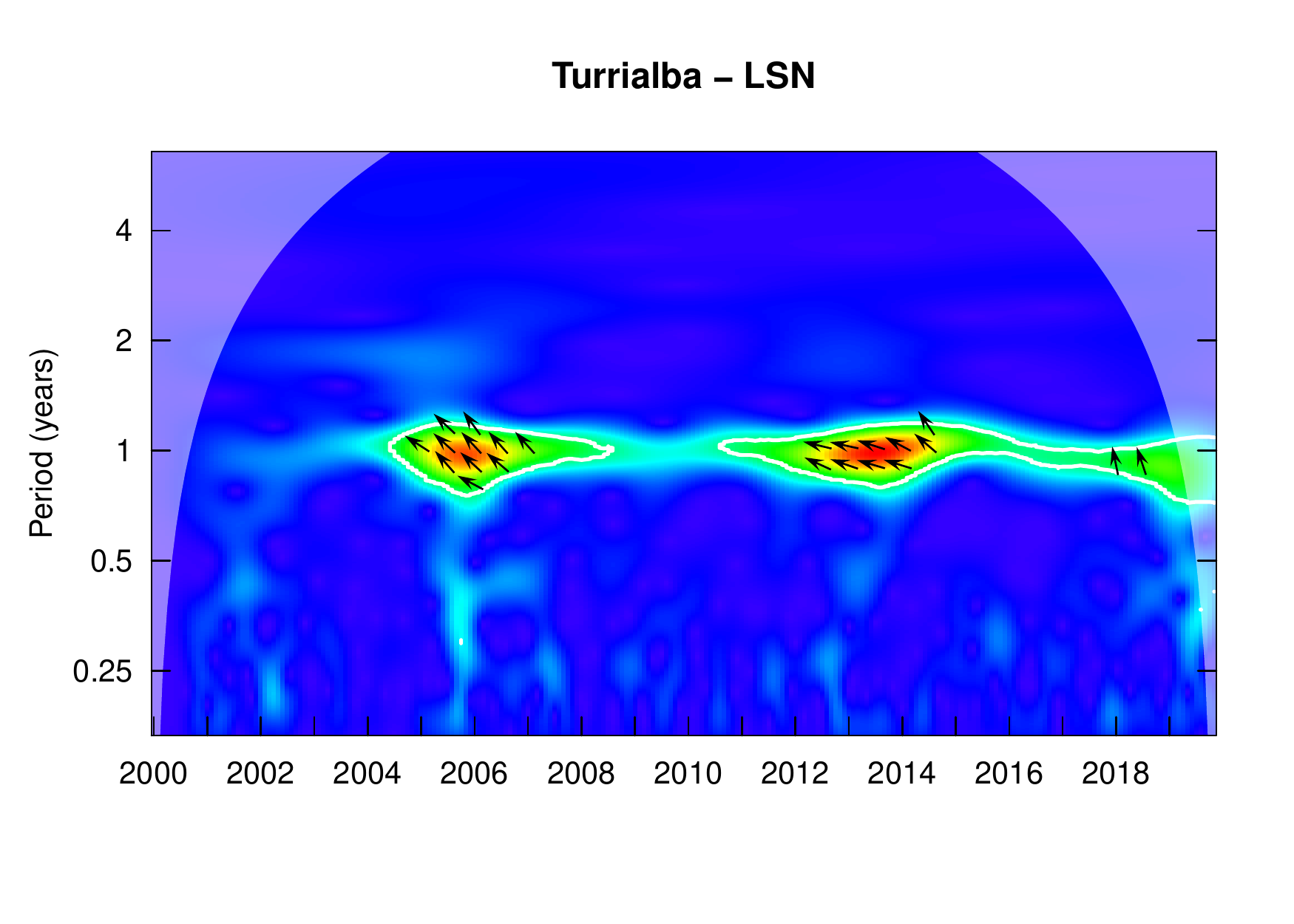}}\vspace{-0.15cm}%
\subfloat[]{\includegraphics[scale=0.23]{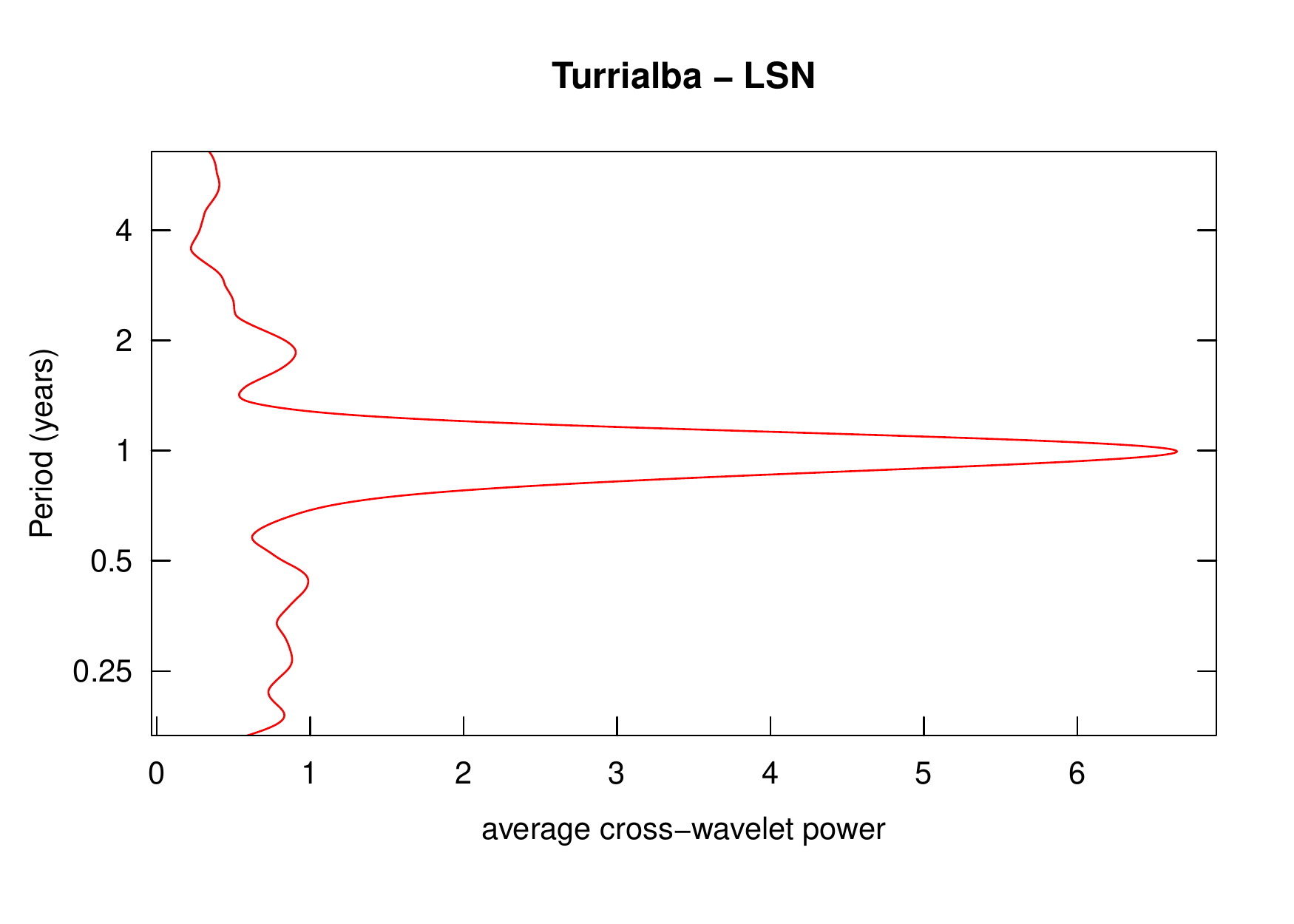}}\vspace{-0.15cm}%
\subfloat[]{\includegraphics[scale=0.23]{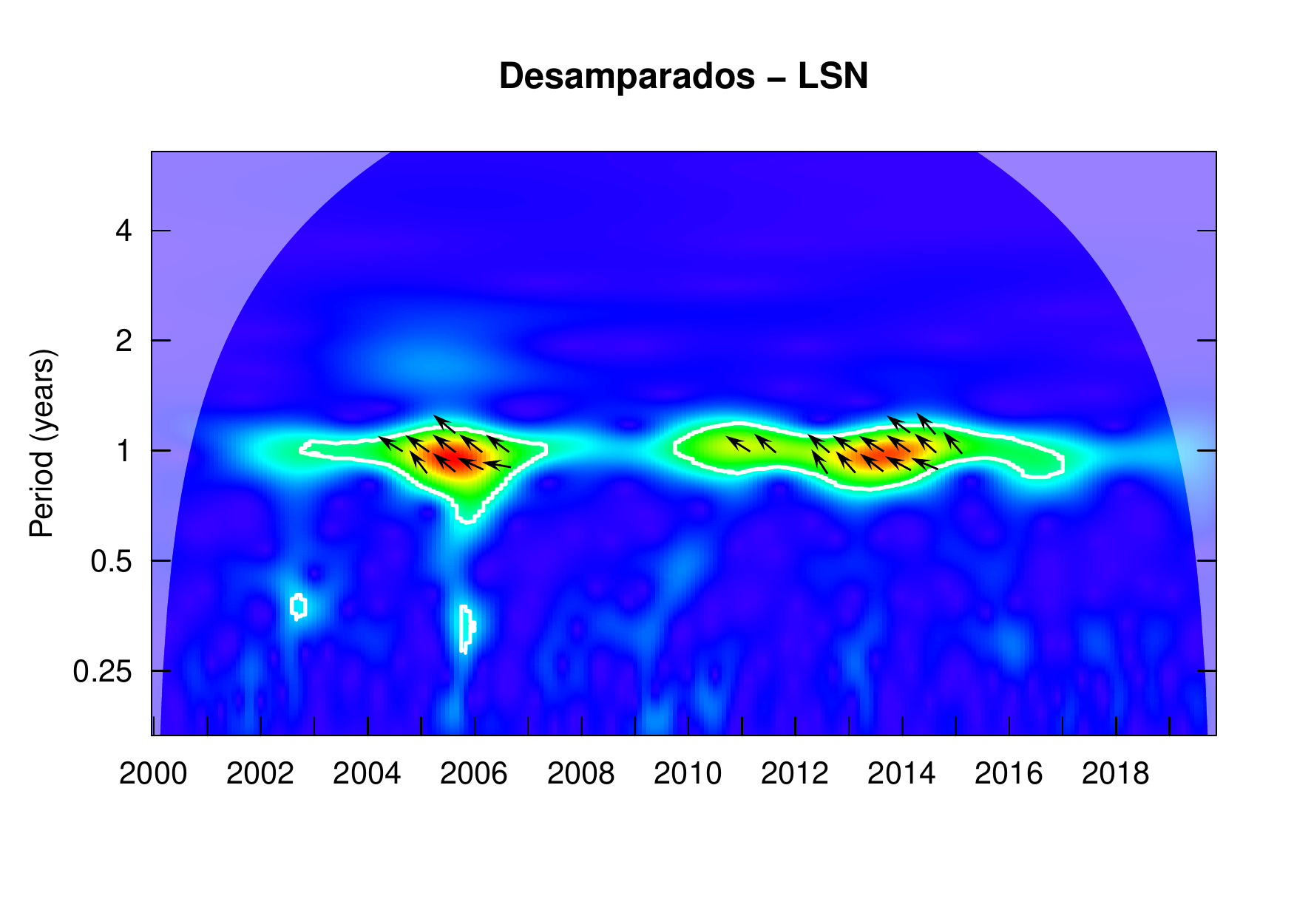}}\vspace{-0.15cm}%
\subfloat[]{\includegraphics[scale=0.23]{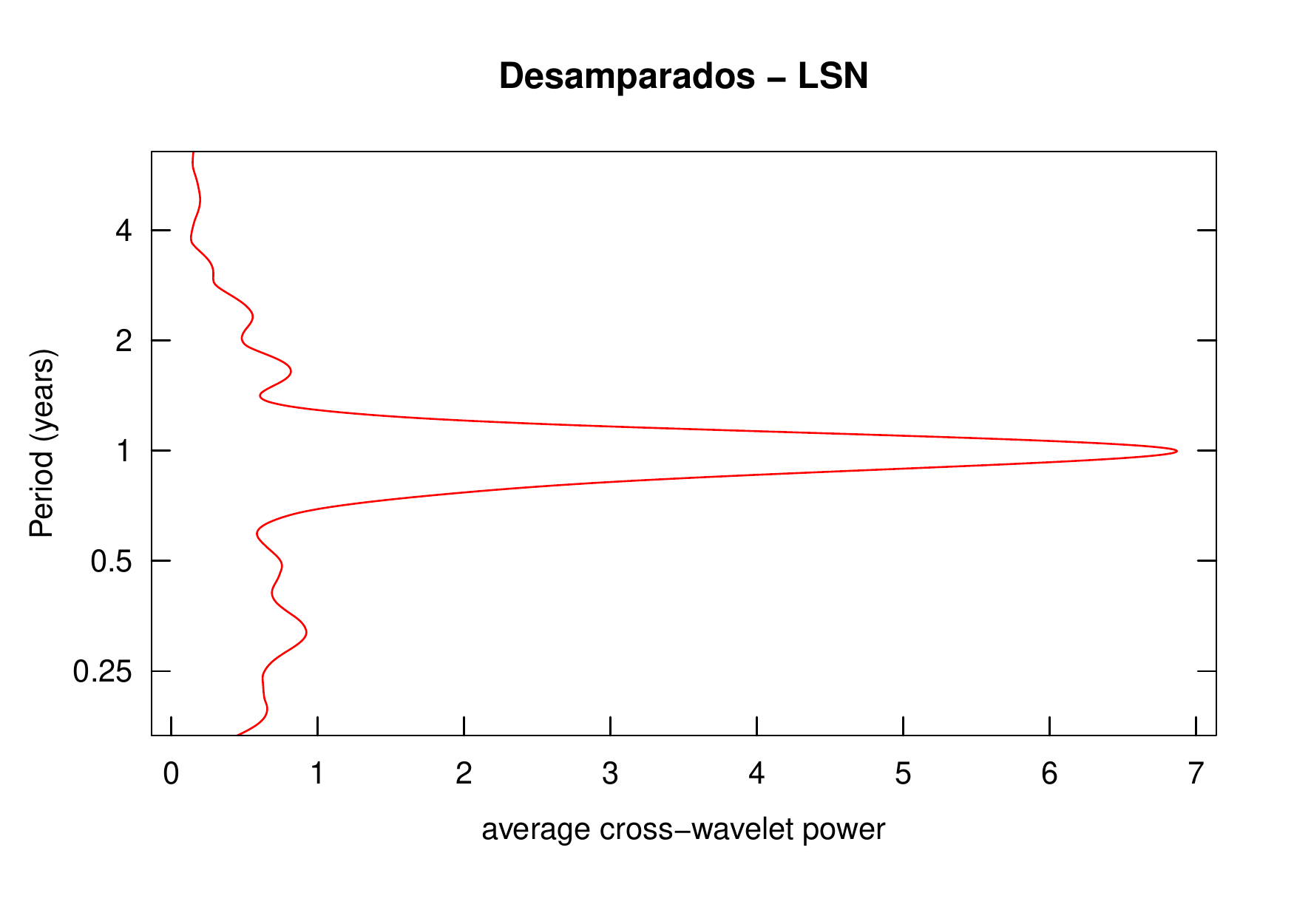}}\vspace{-0.15cm}
\subfloat[]{\includegraphics[scale=0.23]{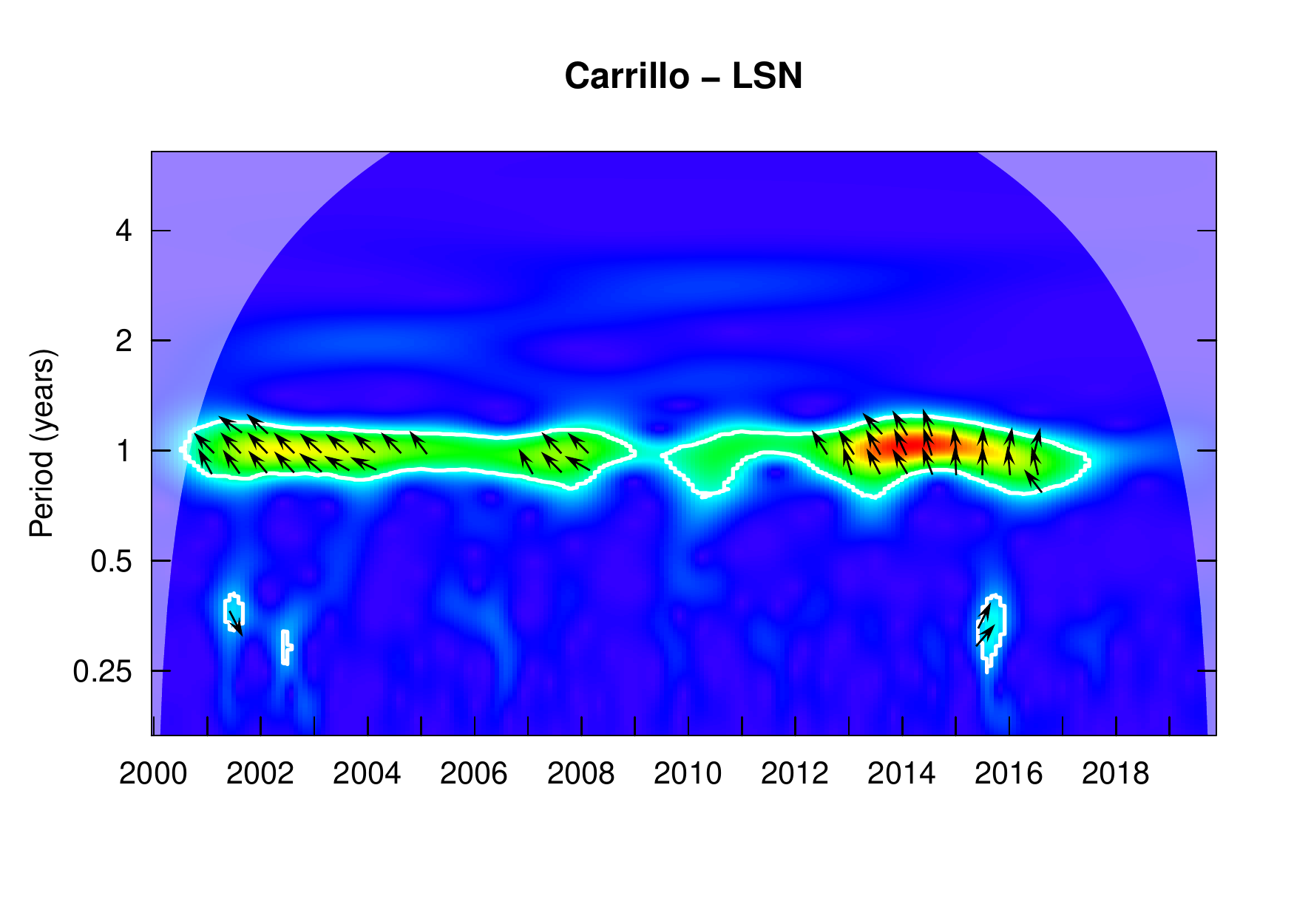}}\vspace{-0.15cm}%
\subfloat[]{\includegraphics[scale=0.23]{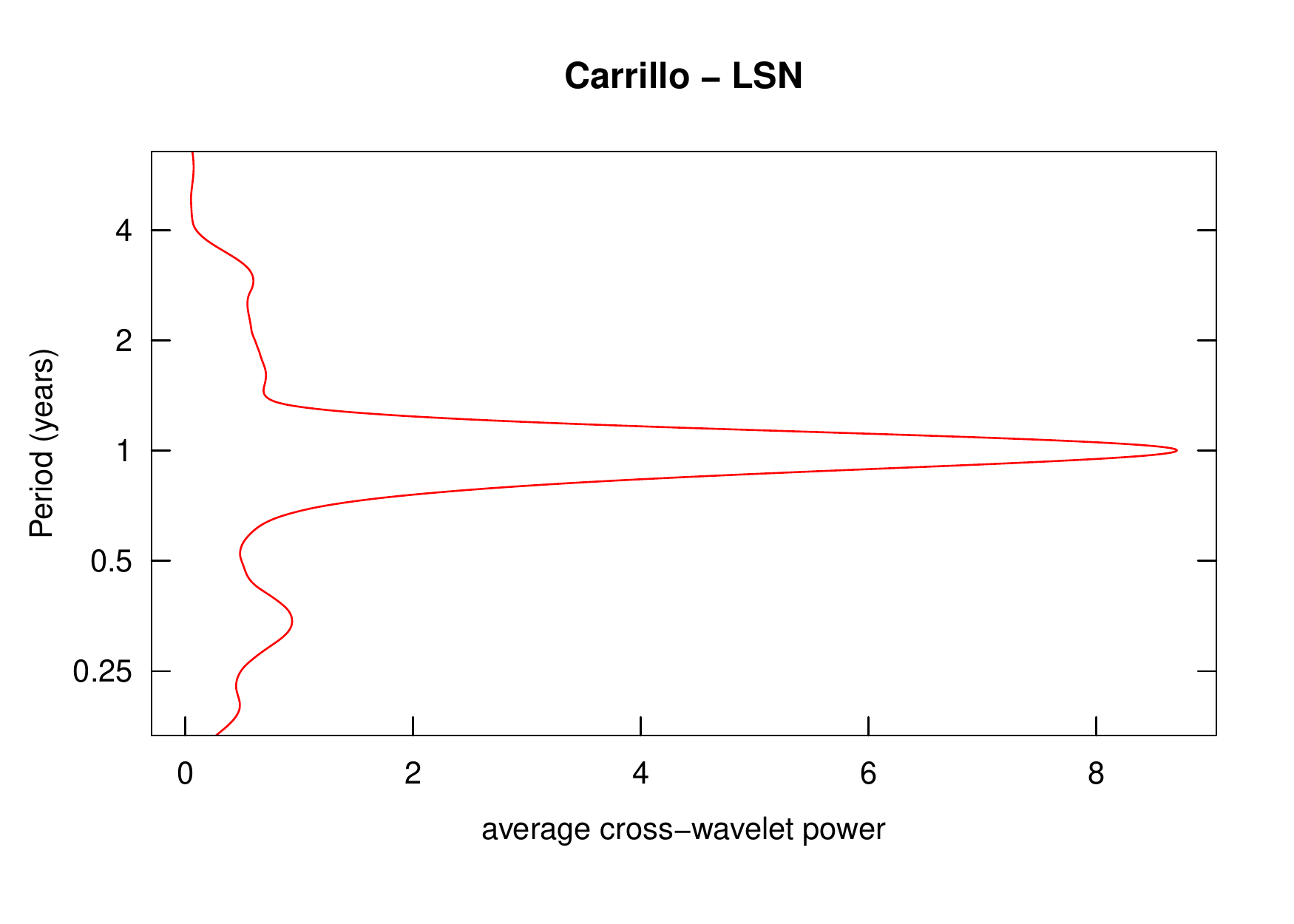}}\vspace{-0.15cm}%
\subfloat[]{\includegraphics[scale=0.23]{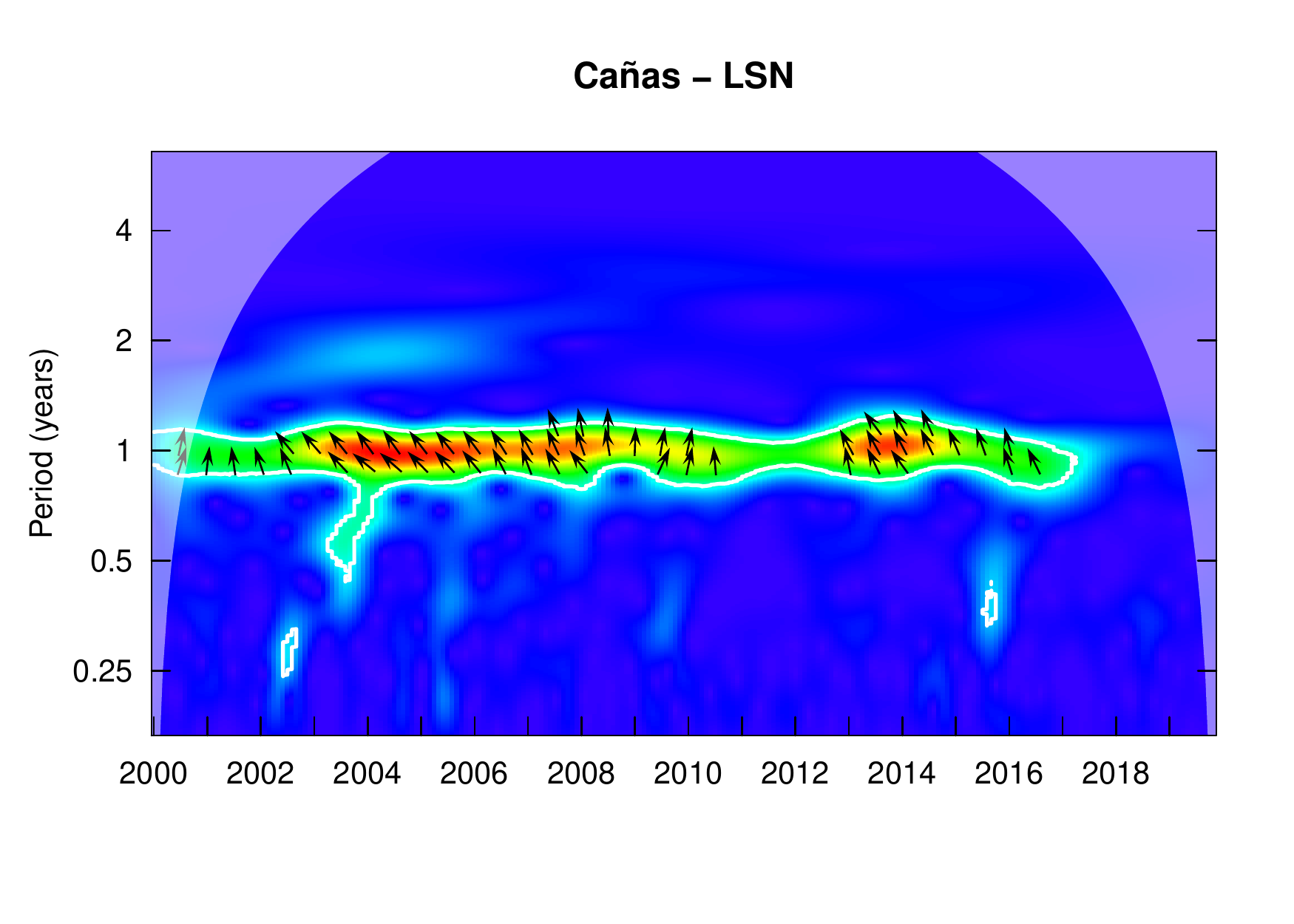}}\vspace{-0.15cm}%
\subfloat[]{\includegraphics[scale=0.23]{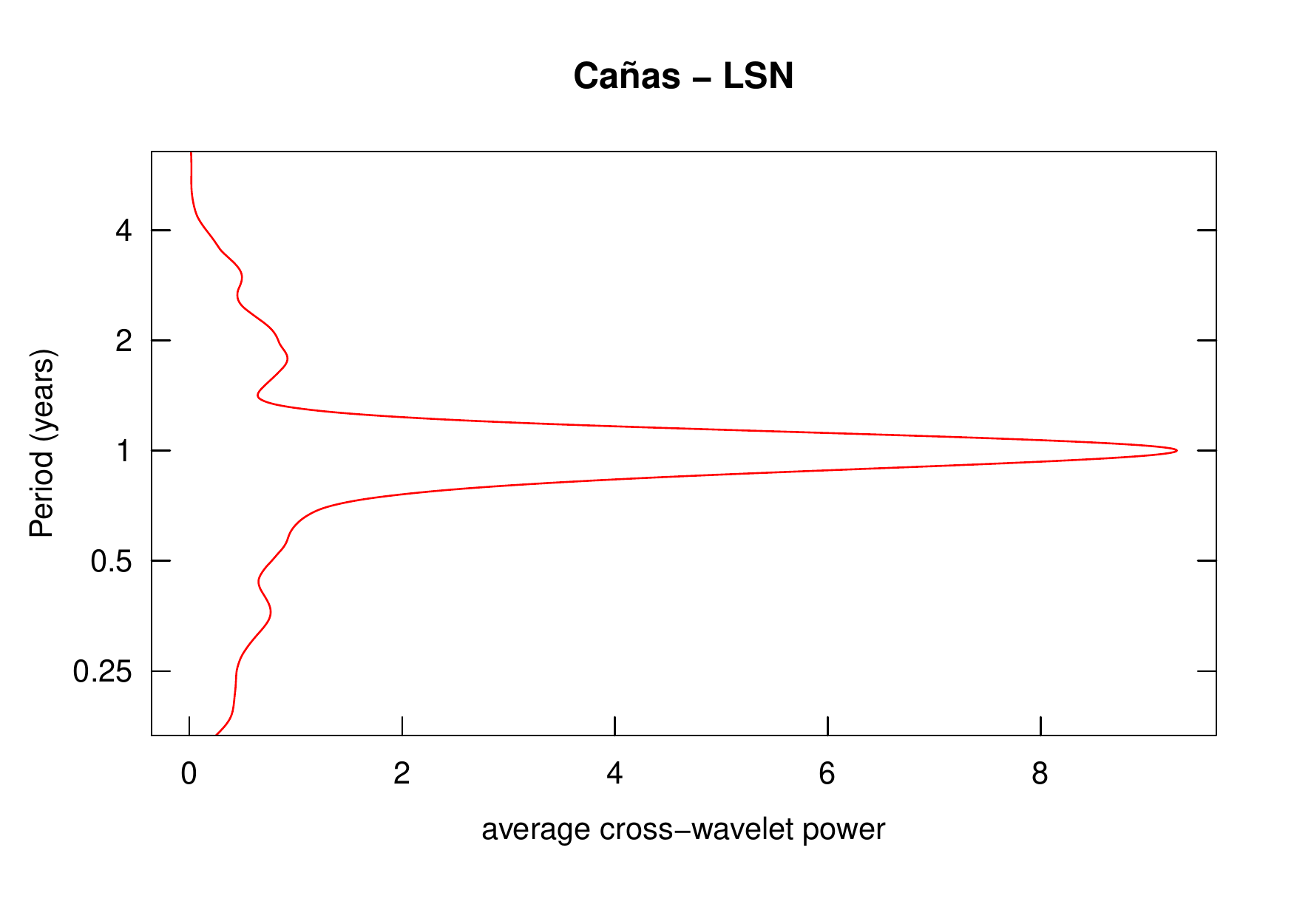}}\vspace{-0.15cm}\\
\subfloat[]{\includegraphics[scale=0.23]{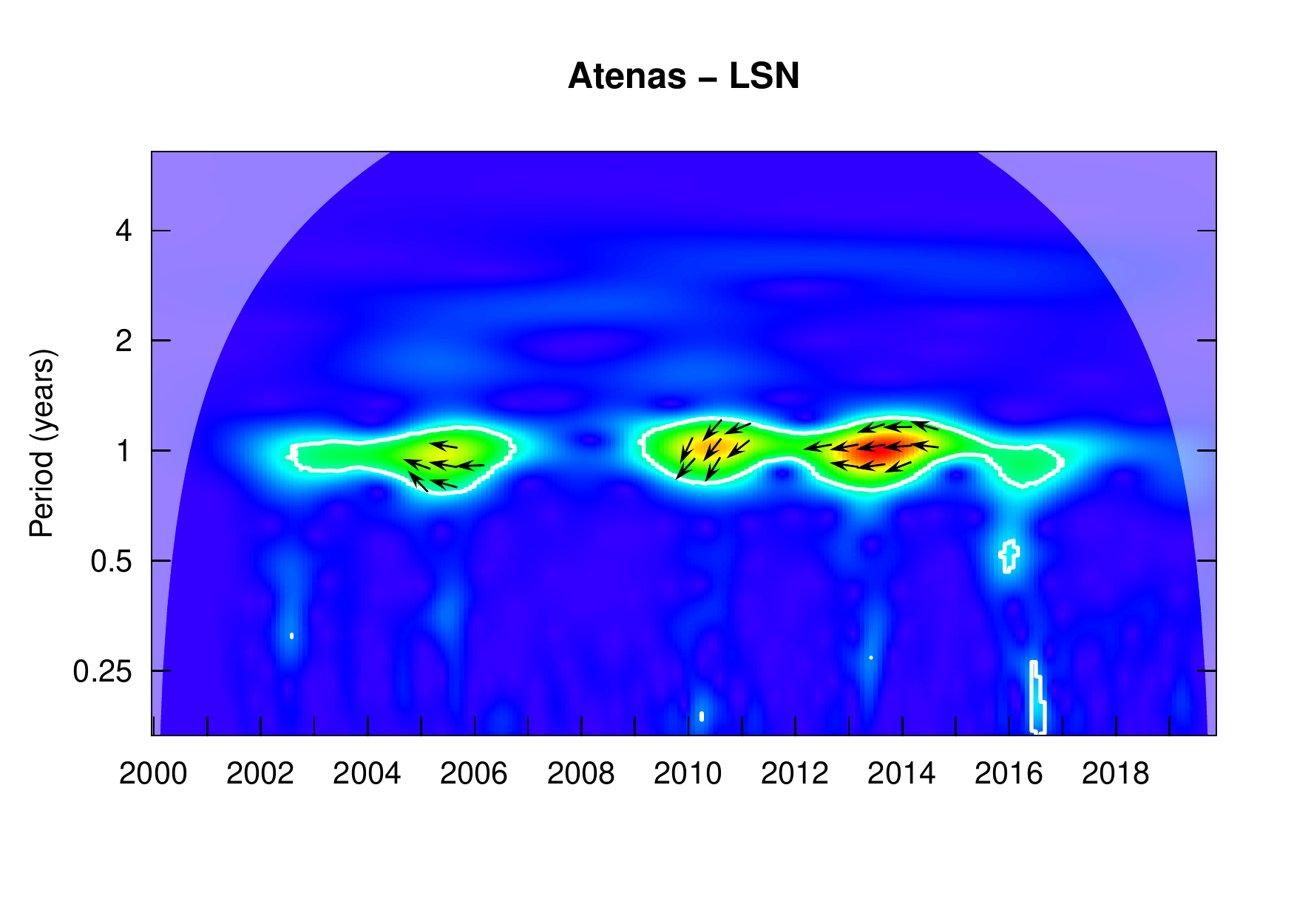}}\vspace{-0.15cm}%
\subfloat[]{\includegraphics[scale=0.23]{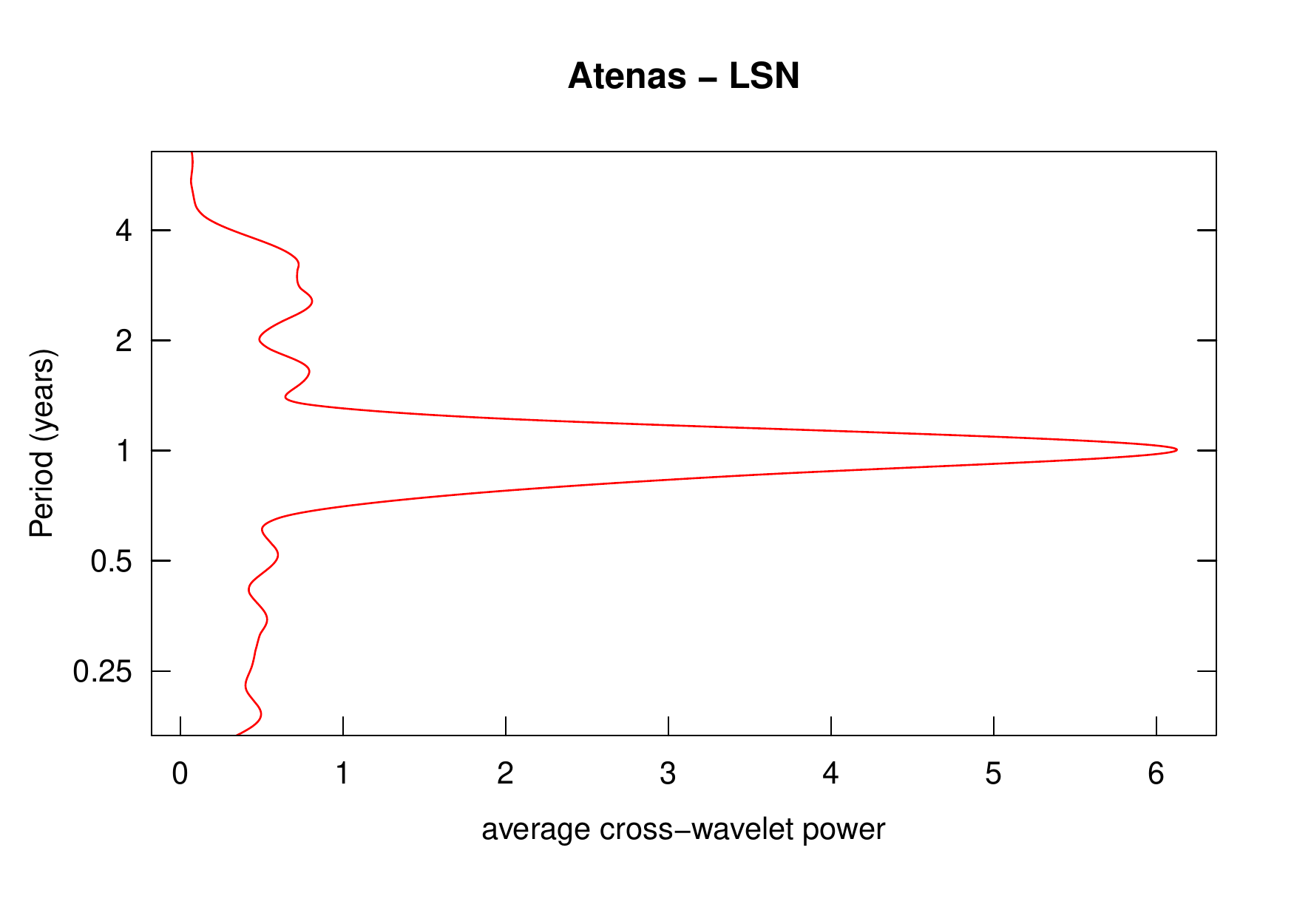}}\vspace{-0.15cm}%
\subfloat[]{\includegraphics[scale=0.23]{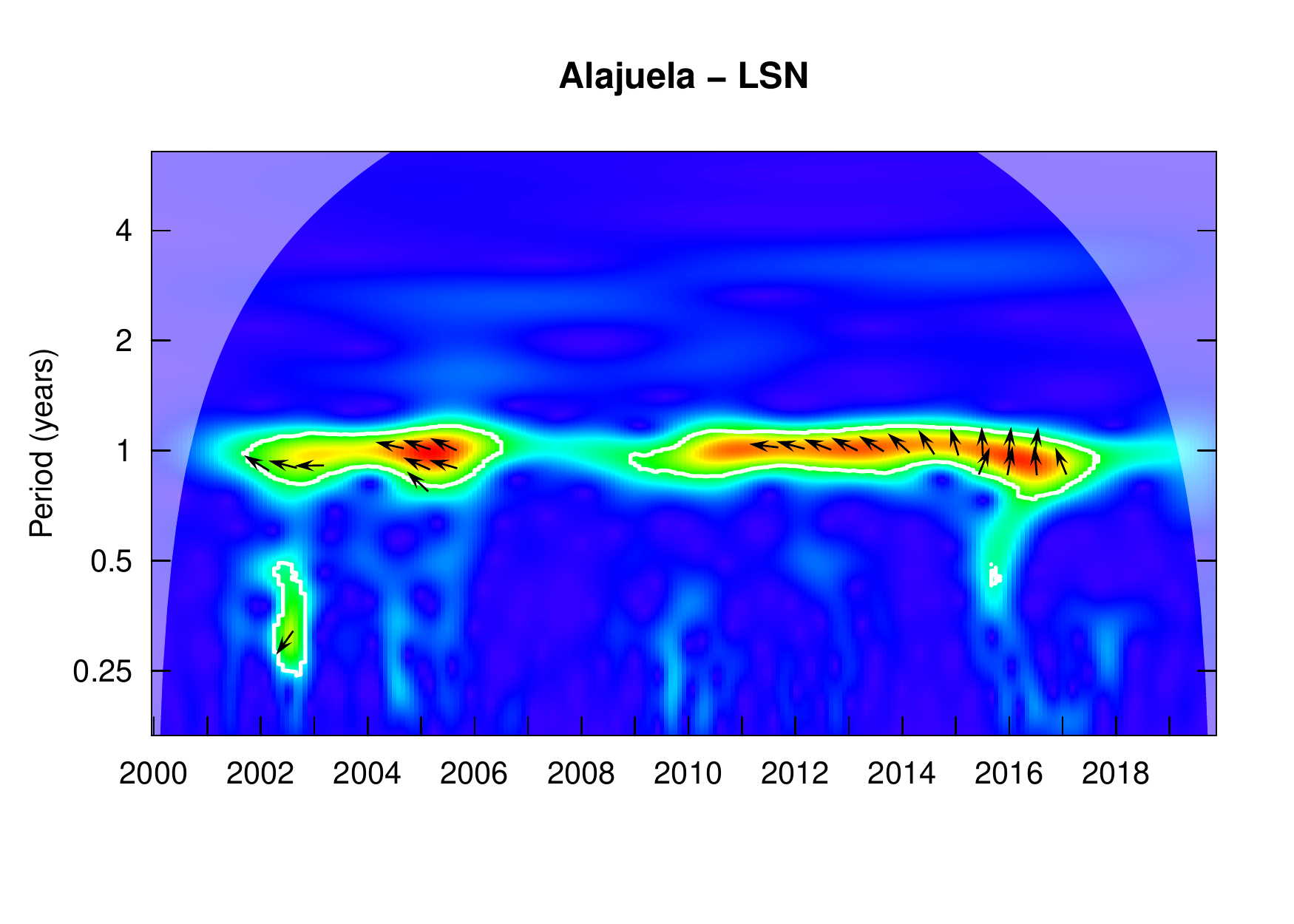}}\vspace{-0.15cm}%
\subfloat[]{\includegraphics[scale=0.23]{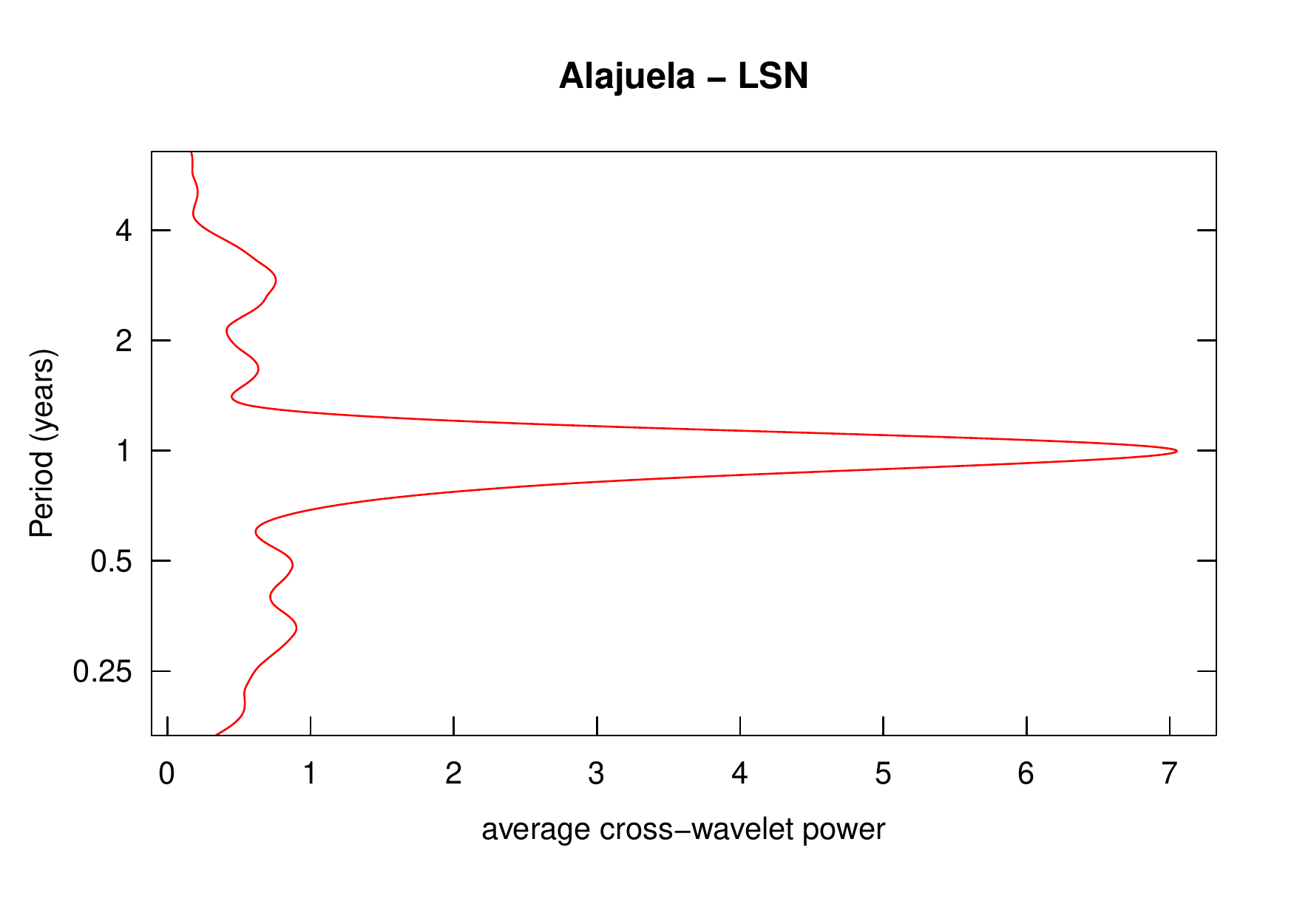}}\vspace{-0.15cm}\\
\subfloat[]{\includegraphics[scale=0.23]{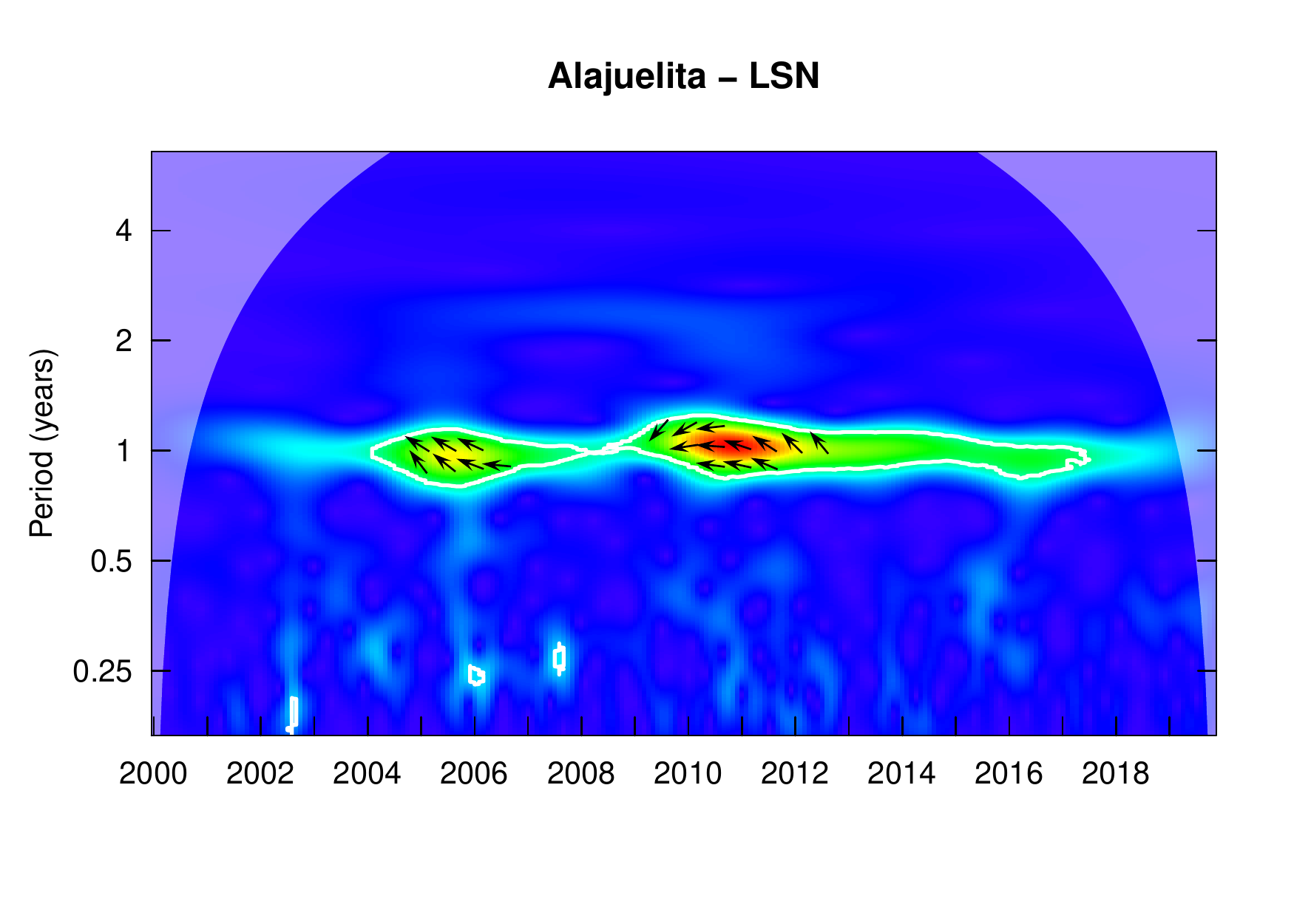}}\vspace{-0.15cm}%
\subfloat[]{\includegraphics[scale=0.23]{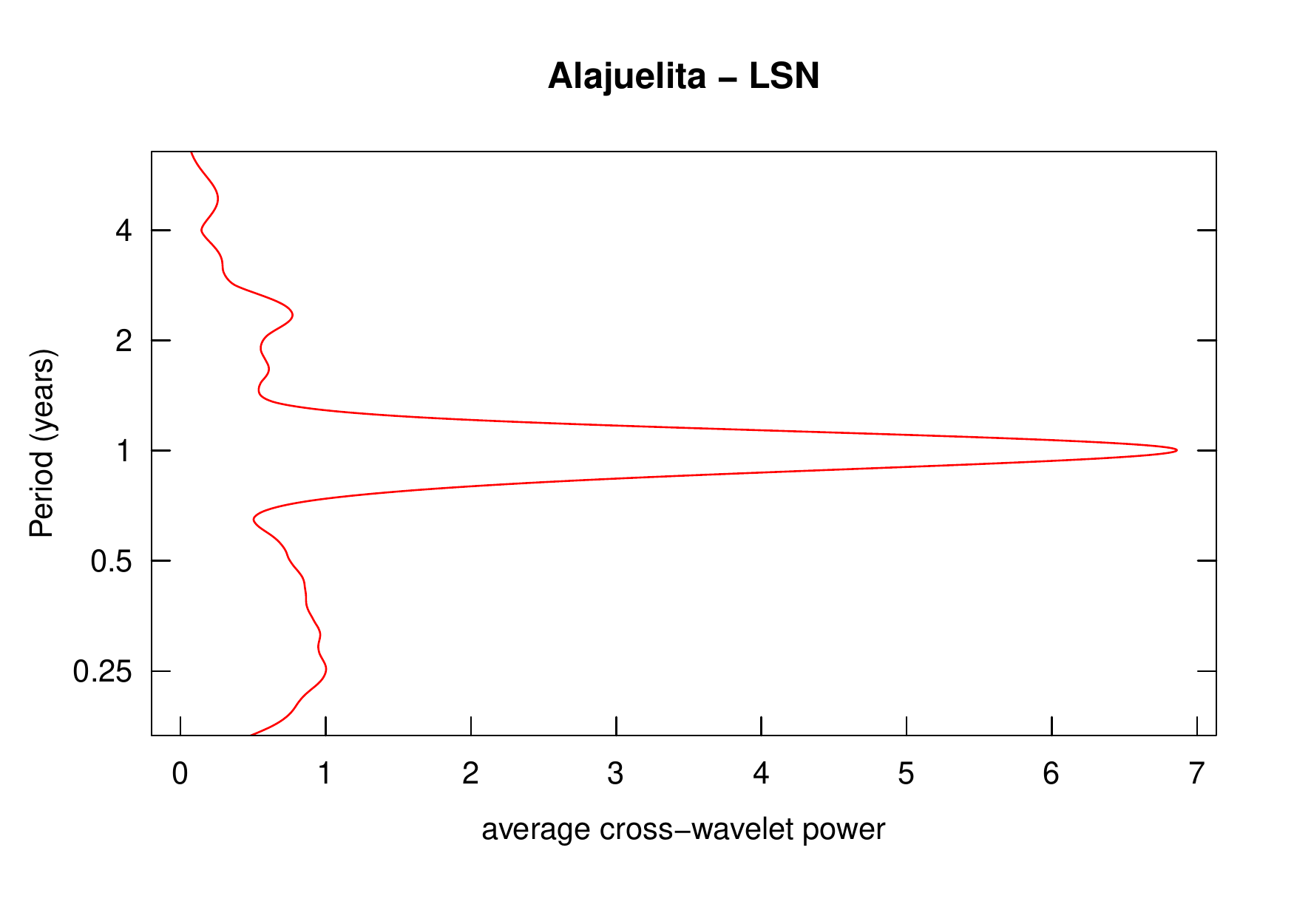}}\vspace{-0.15cm}%
\subfloat[]{\includegraphics[scale=0.23]{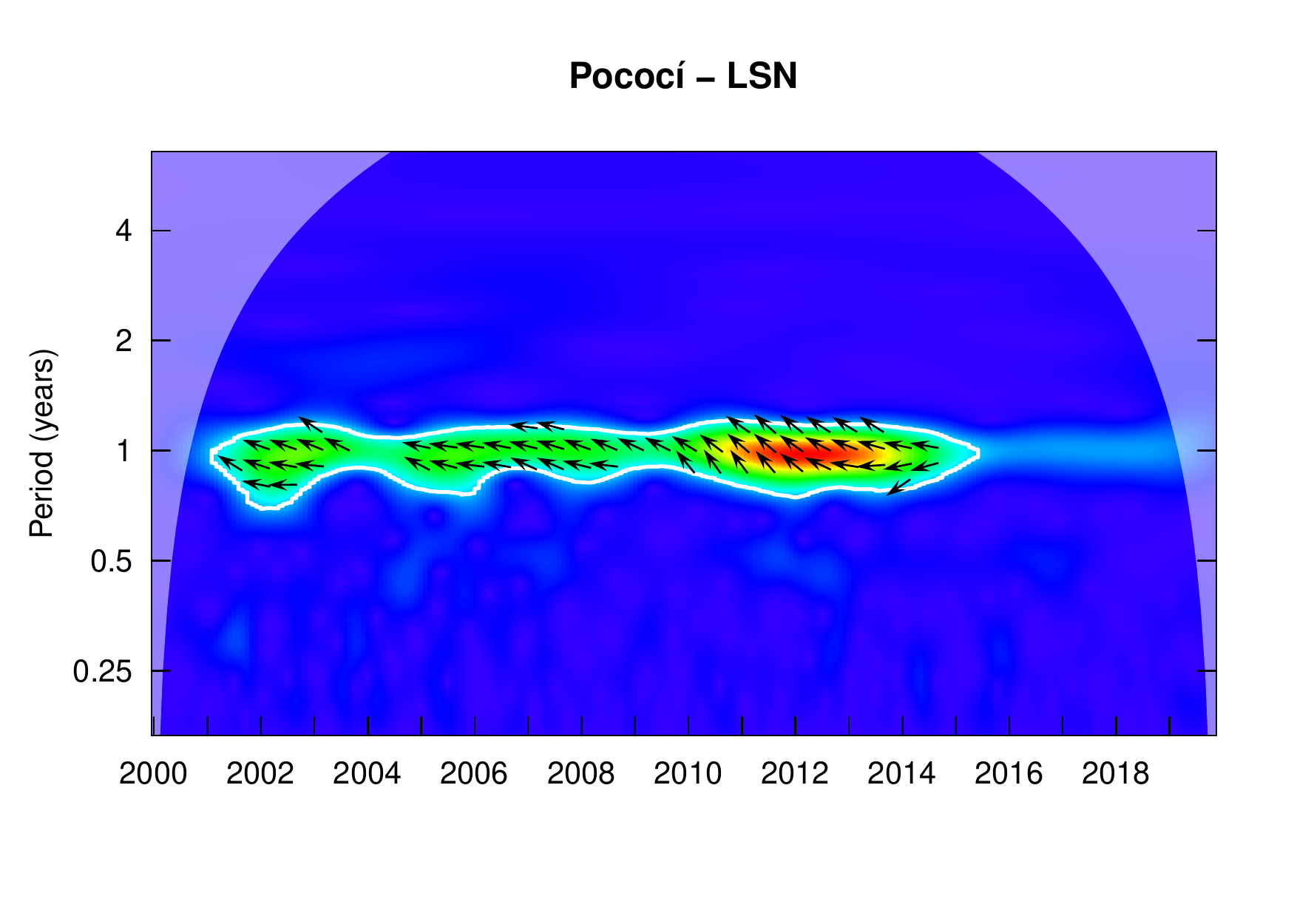}}\vspace{-0.15cm}%
\subfloat[]{\includegraphics[scale=0.23]{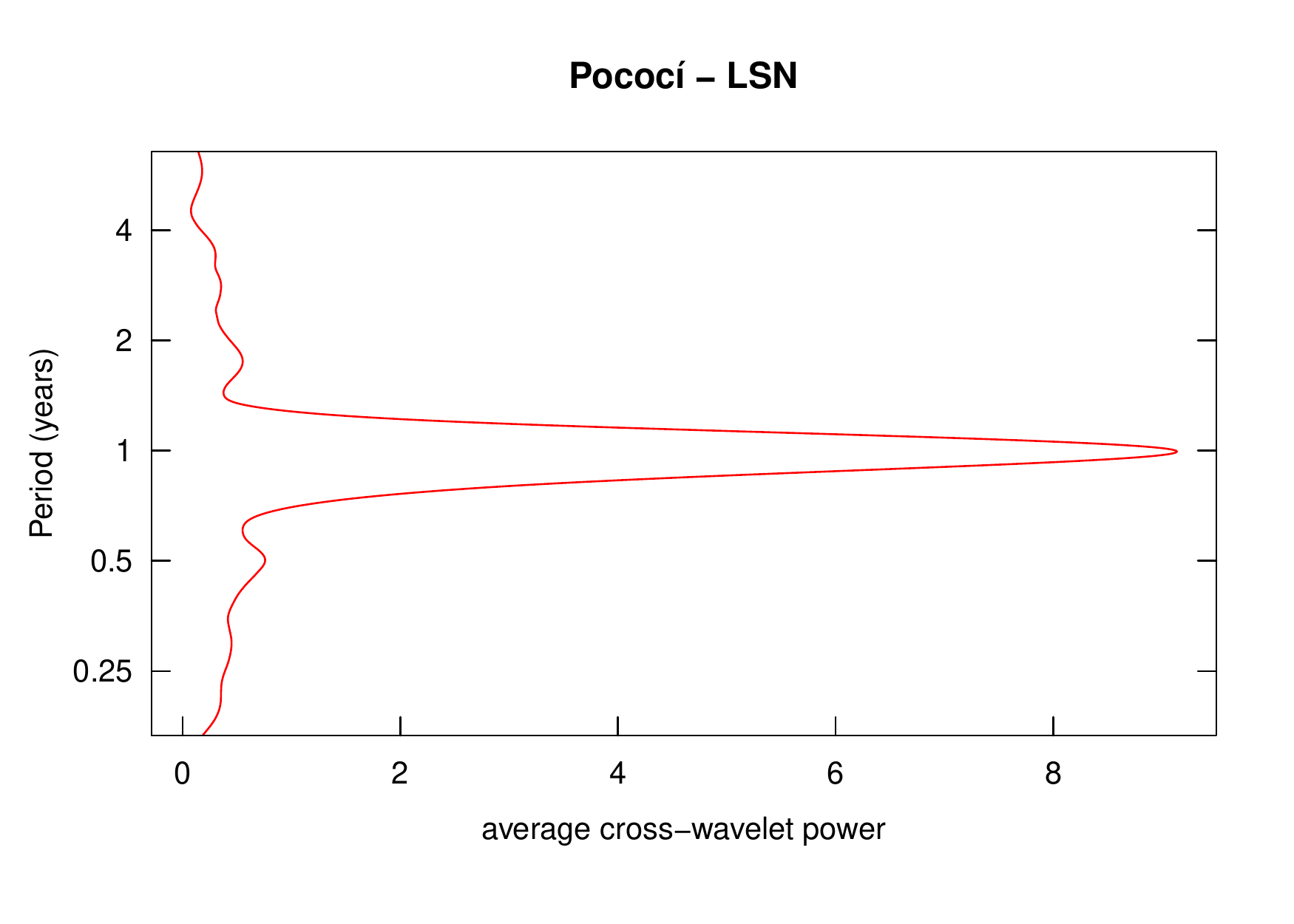}}\vspace{-0.15cm}\\
\subfloat[]{\includegraphics[scale=0.23]{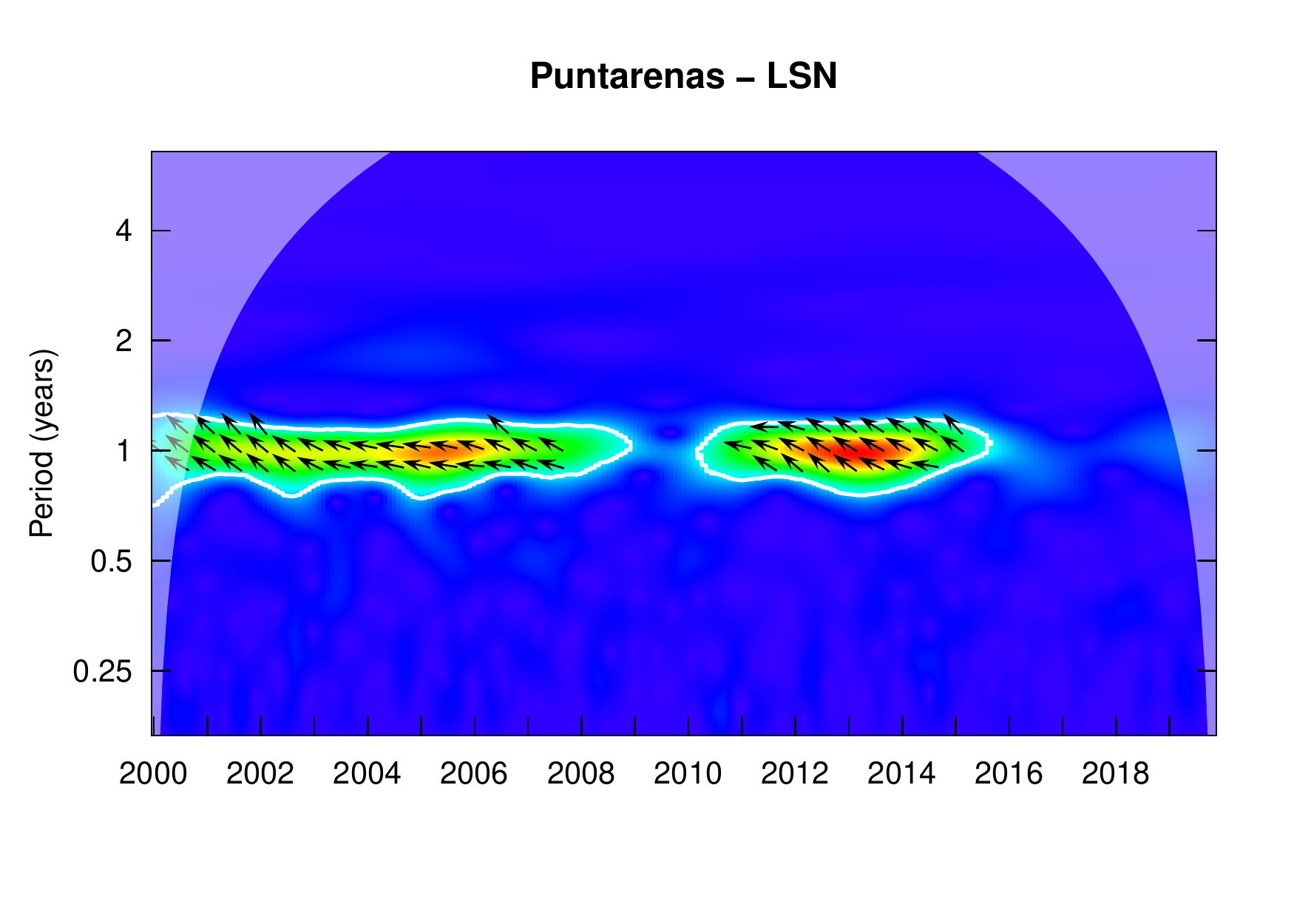}}\vspace{-0.15cm}%
\subfloat[]{\includegraphics[scale=0.23]{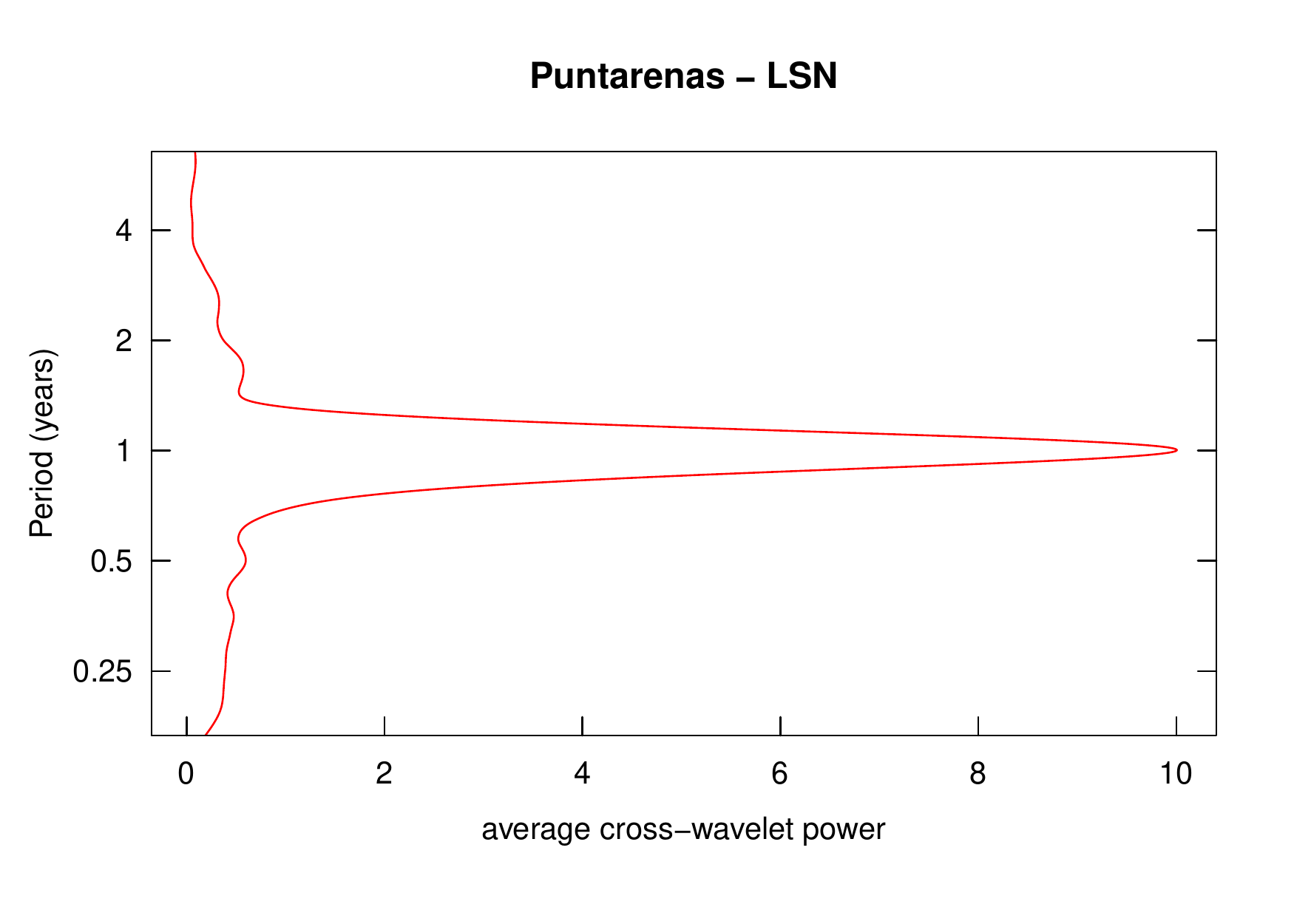}}\vspace{-0.15cm}%
\subfloat[]{\includegraphics[scale=0.23]{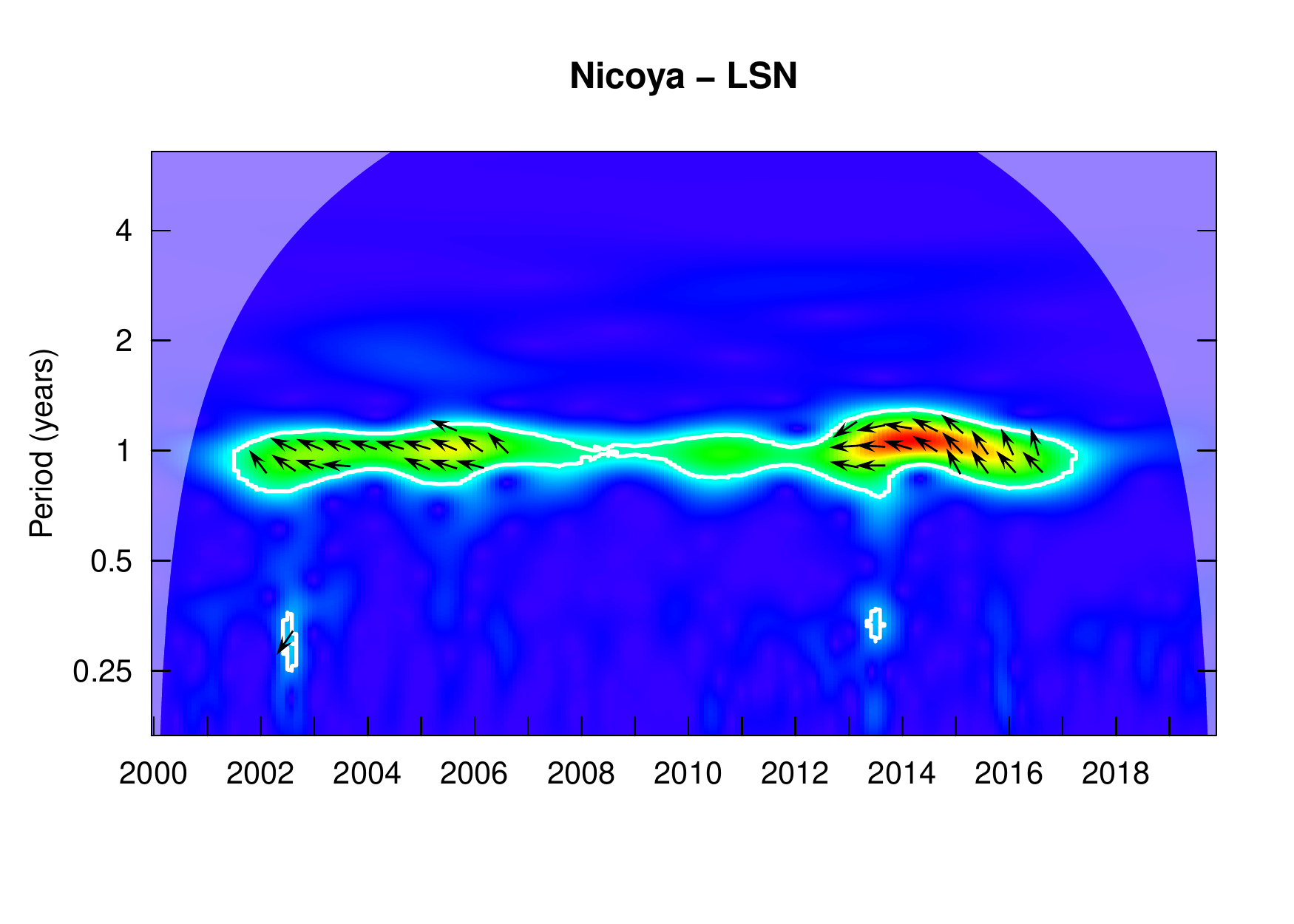}}\vspace{-0.15cm}%
\subfloat[]{\includegraphics[scale=0.23]{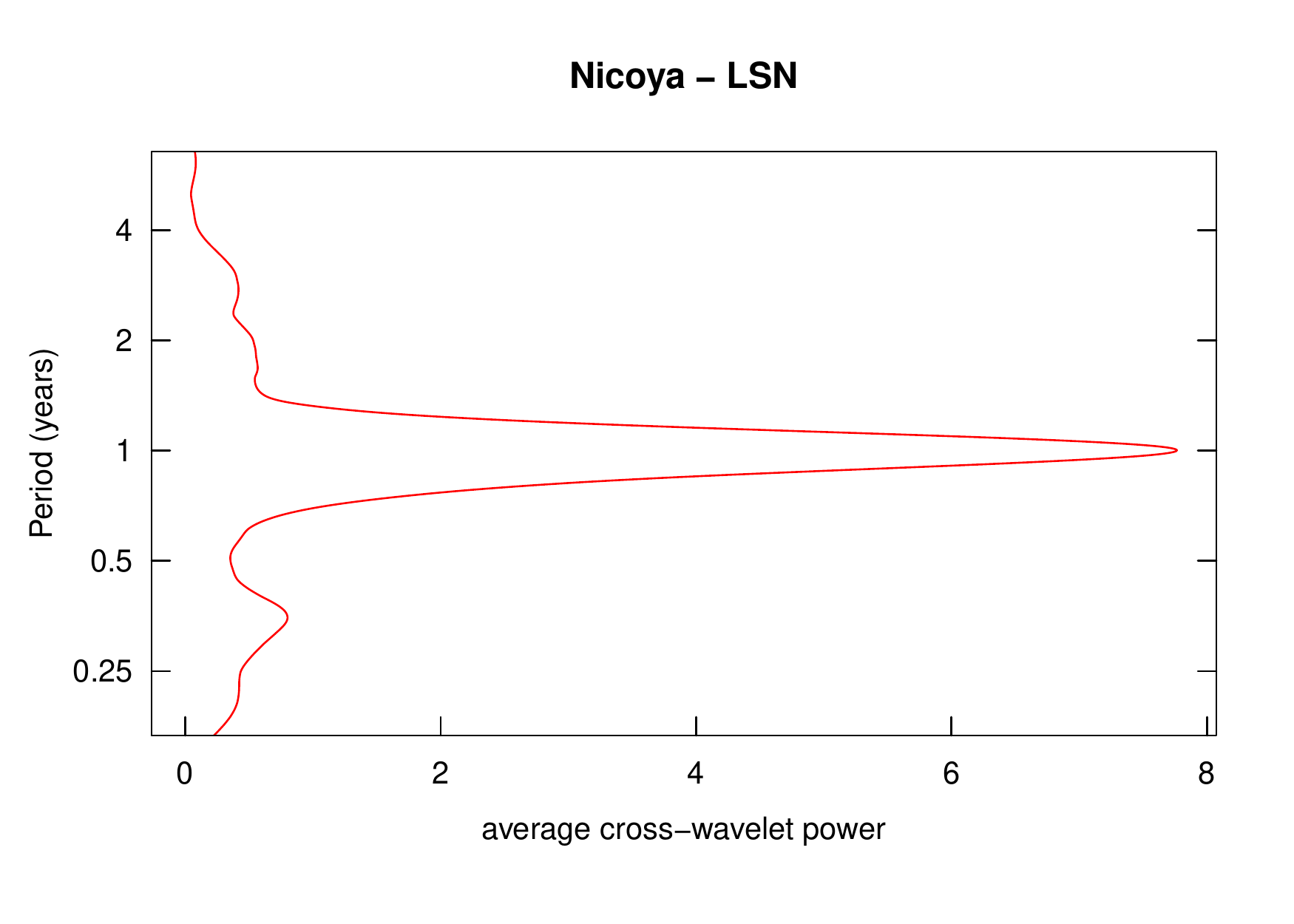}}\vspace{-0.15cm}\\
\caption*{}
\end{figure}

\begin{figure}[H]
\captionsetup[subfigure]{labelformat=empty}
\subfloat[]{\includegraphics[scale=0.23]{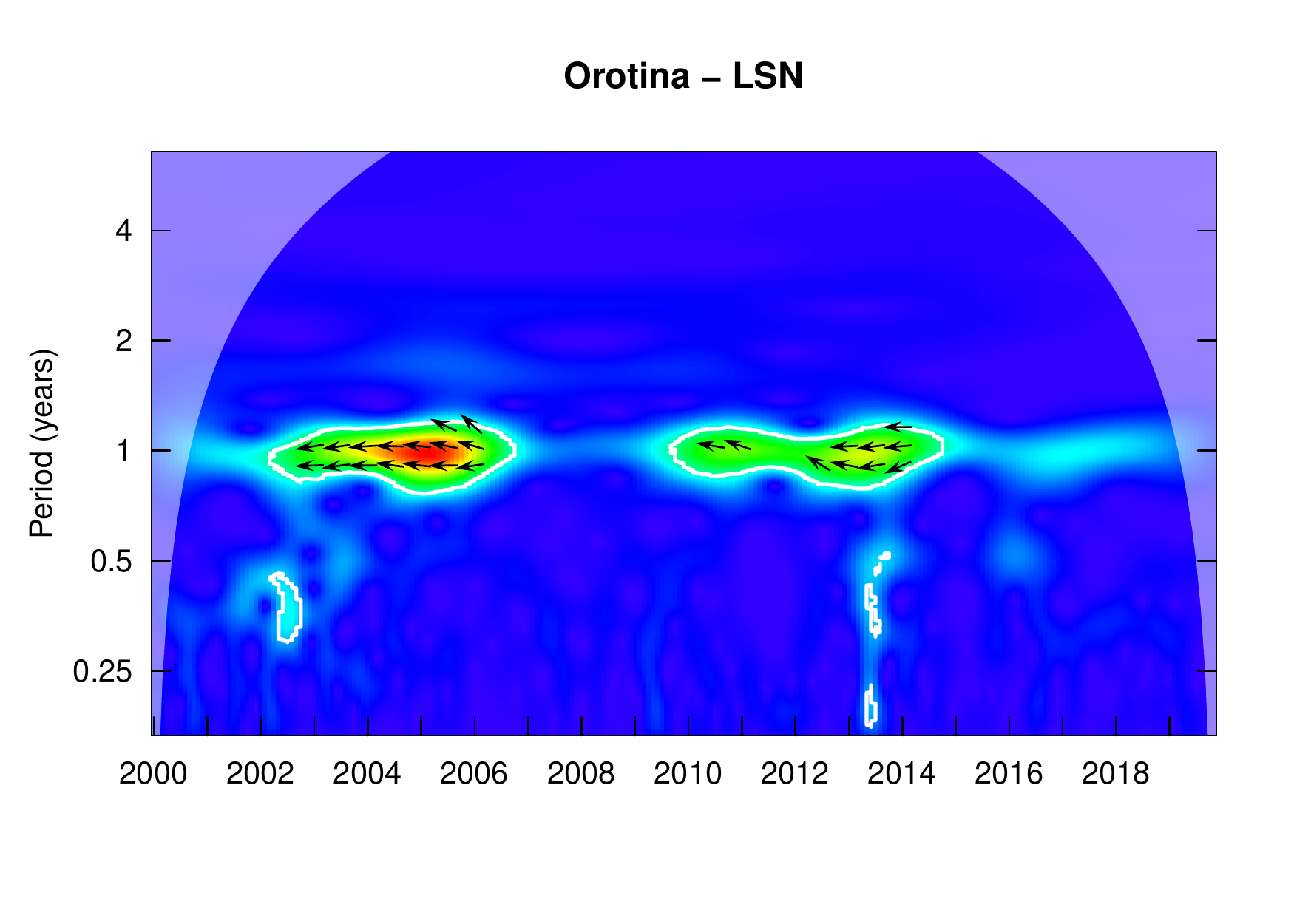}}\vspace{-0.15cm}%
\subfloat[]{\includegraphics[scale=0.23]{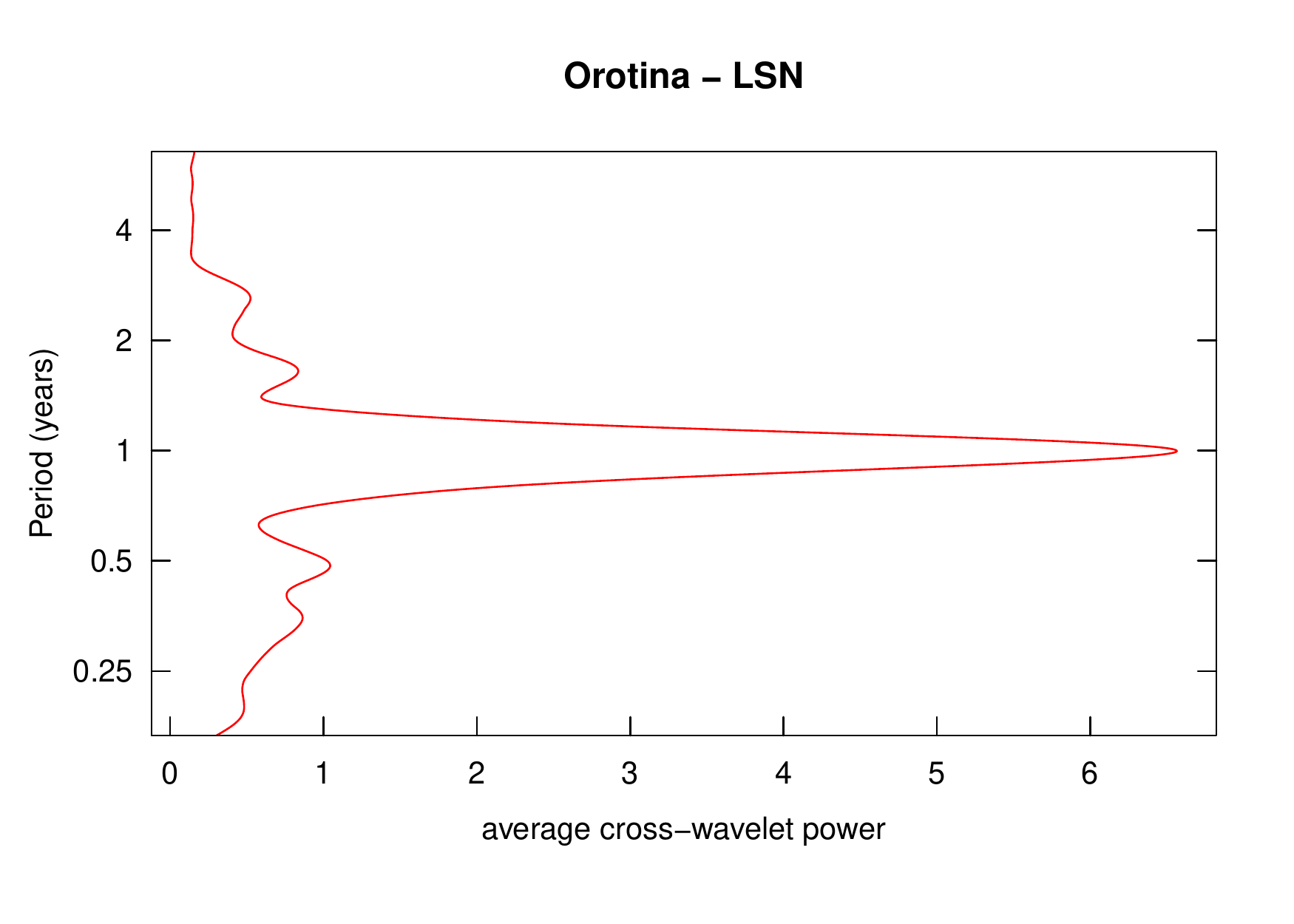}}\vspace{-0.15cm}%
\subfloat[]{\includegraphics[scale=0.23]{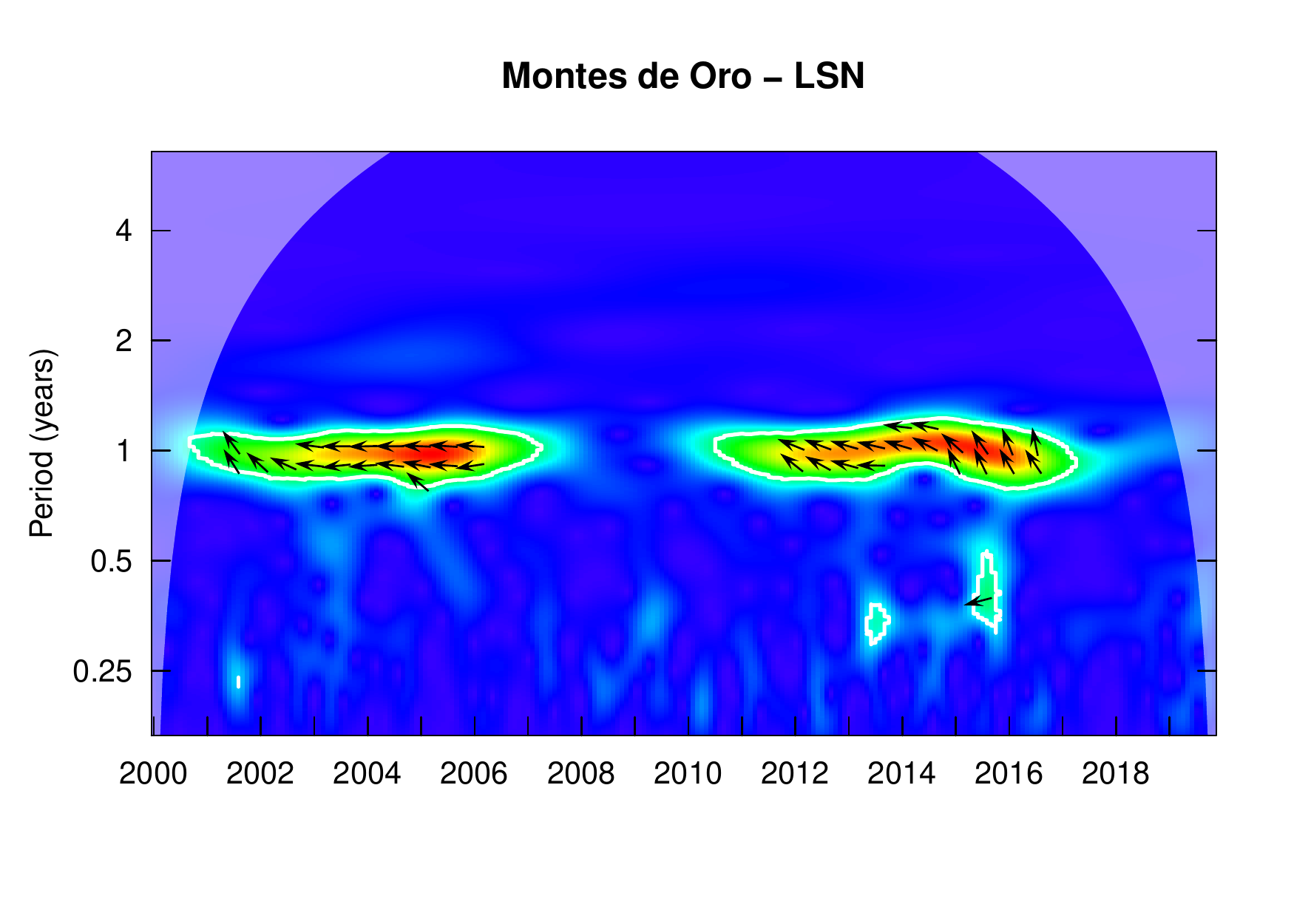}}\vspace{-0.15cm}%
\subfloat[]{\includegraphics[scale=0.23]{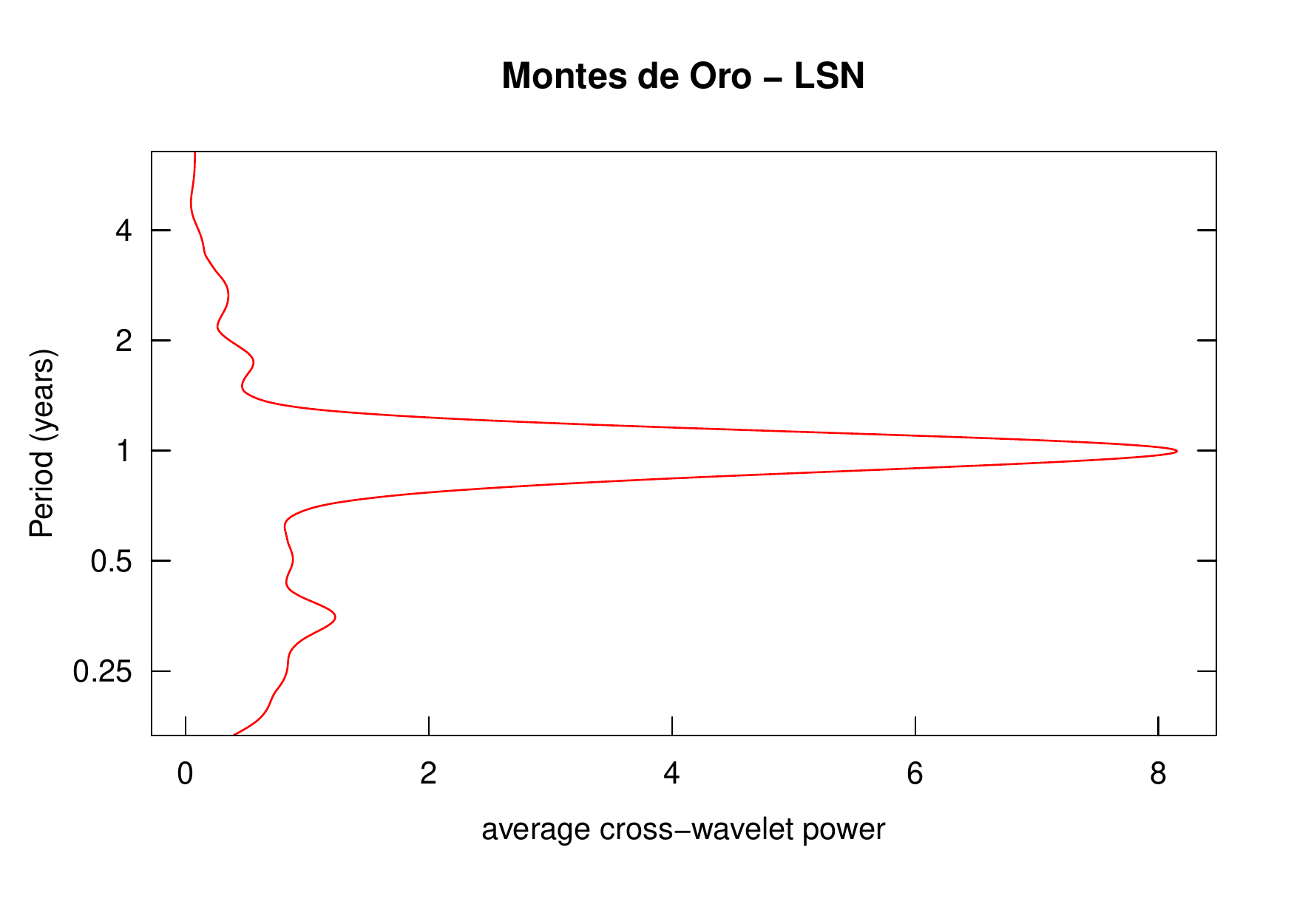}}\vspace{-0.15cm}\\
\subfloat[]{\includegraphics[scale=0.23]{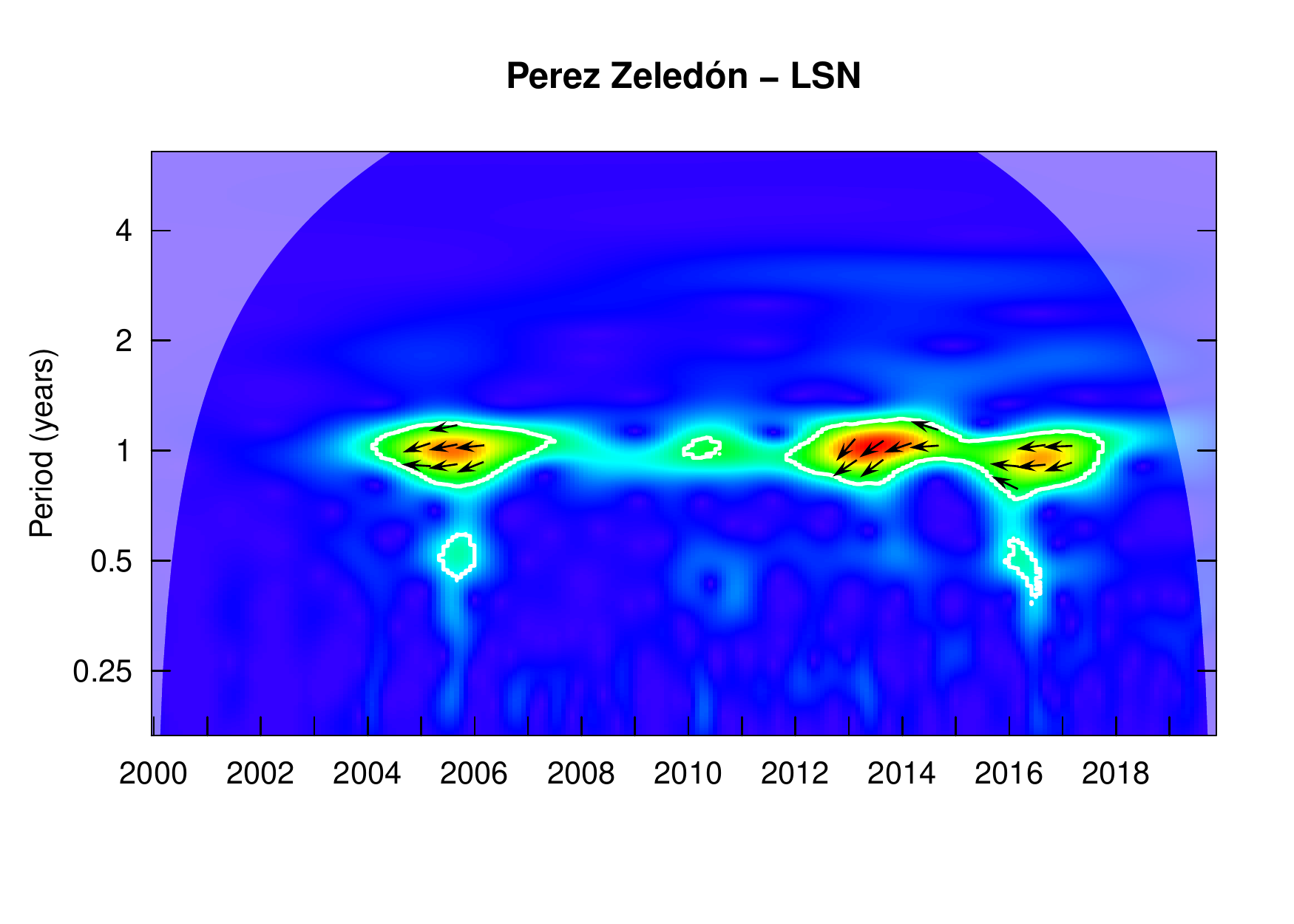}}\vspace{-0.15cm}%
\subfloat[]{\includegraphics[scale=0.23]{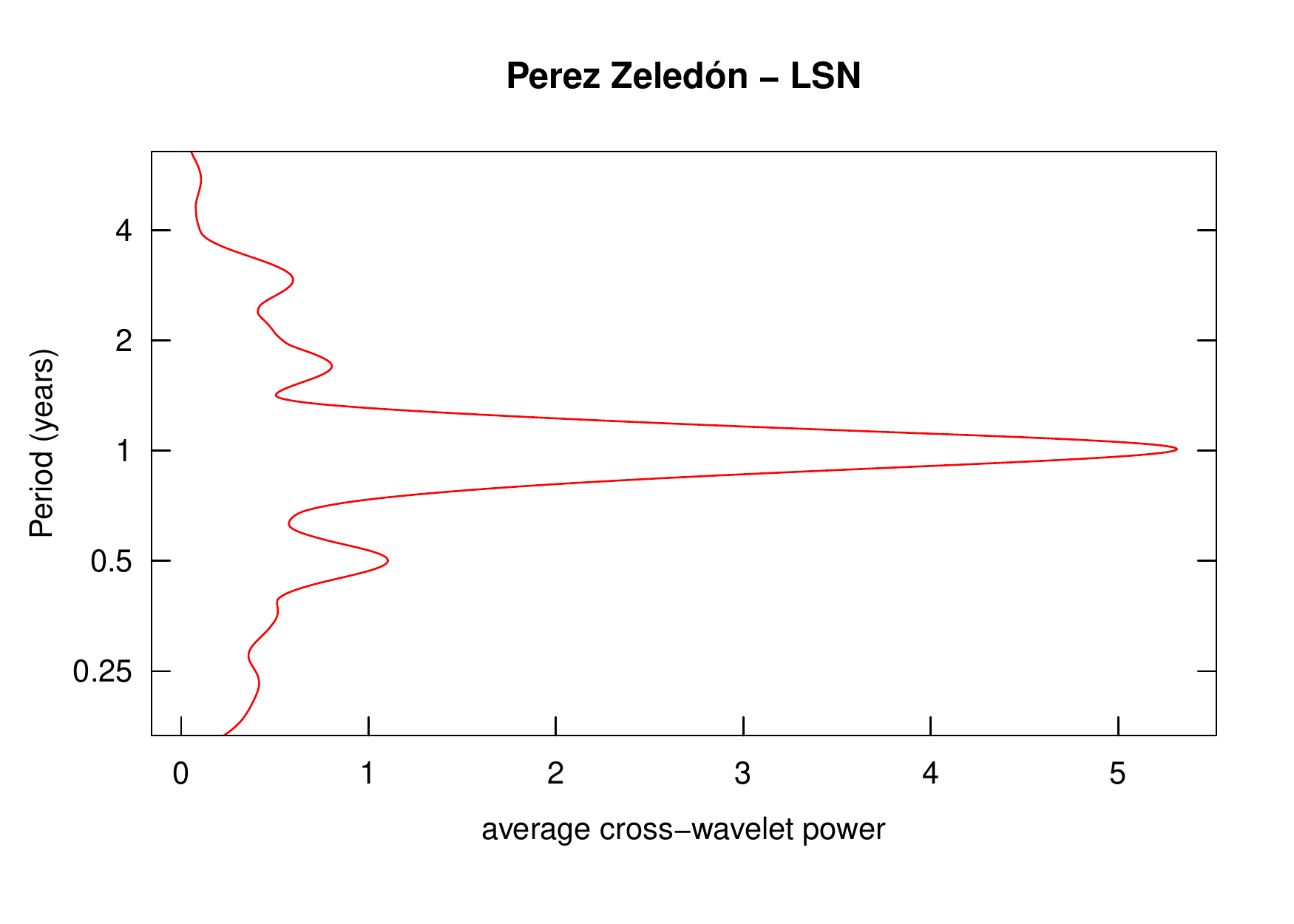}}\vspace{-0.15cm}%
\subfloat[]{\includegraphics[scale=0.23]{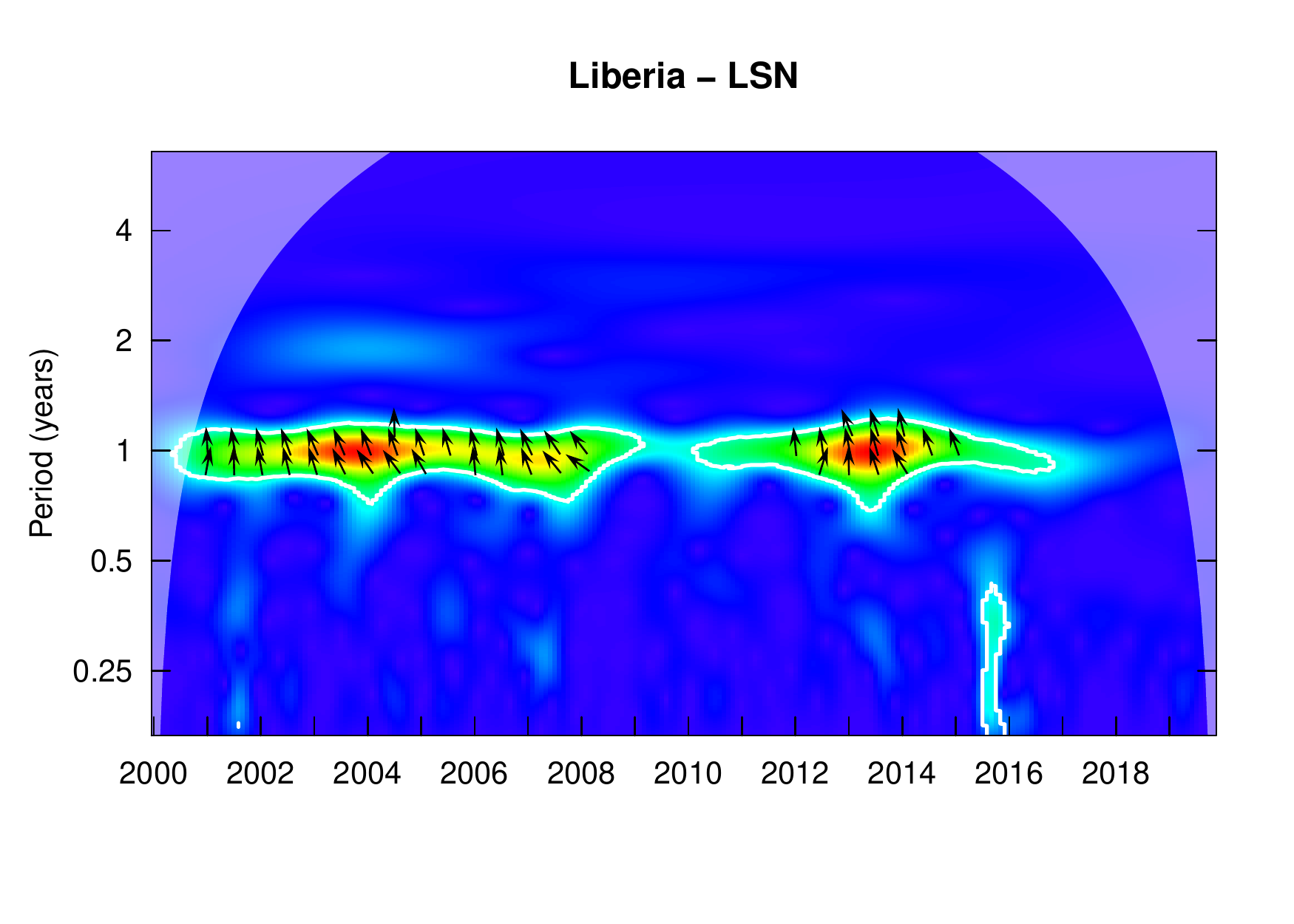}}\vspace{-0.15cm}%
\subfloat[]{\includegraphics[scale=0.23]{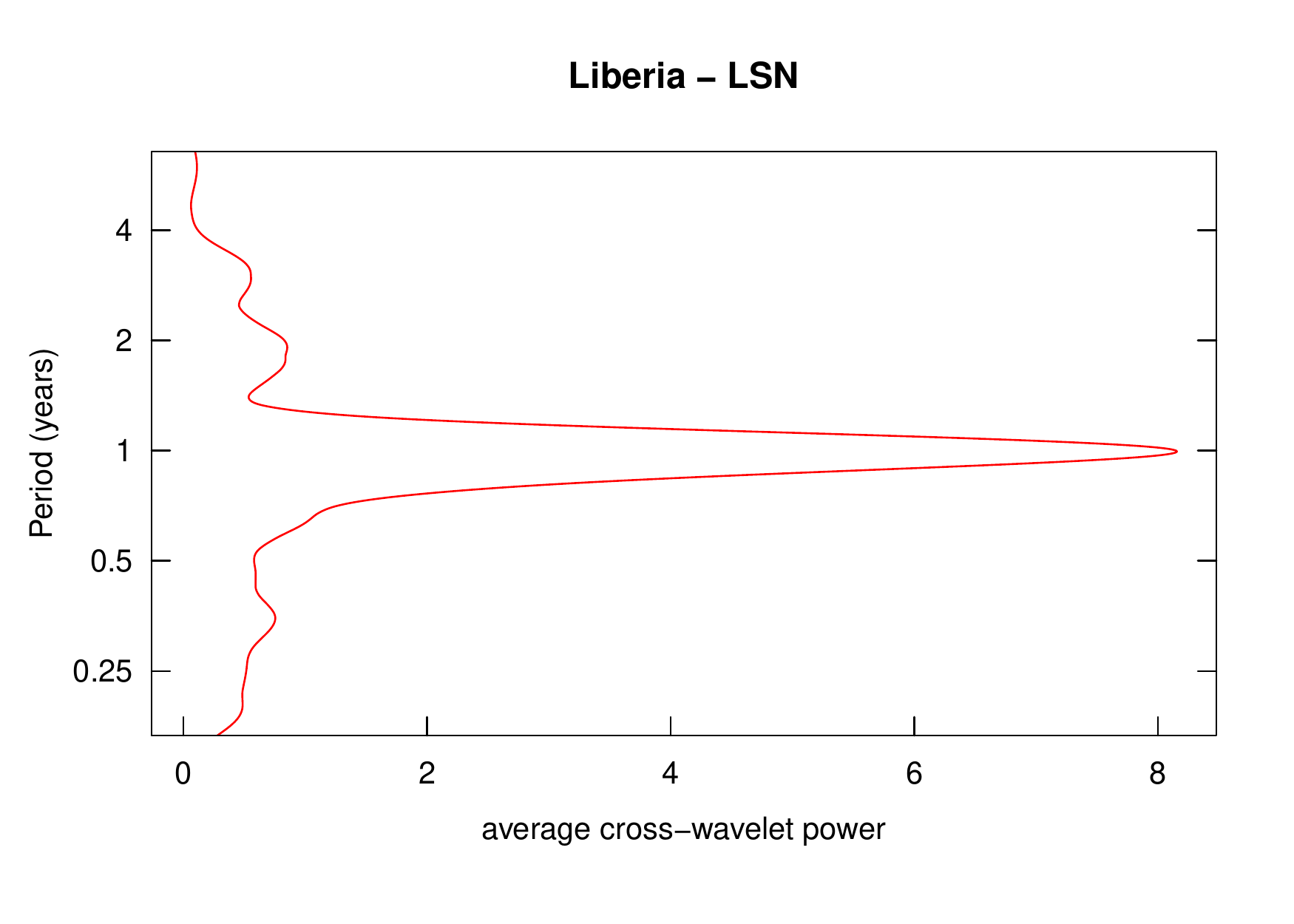}}\vspace{-0.15cm}\\
\subfloat[]{\includegraphics[scale=0.23]{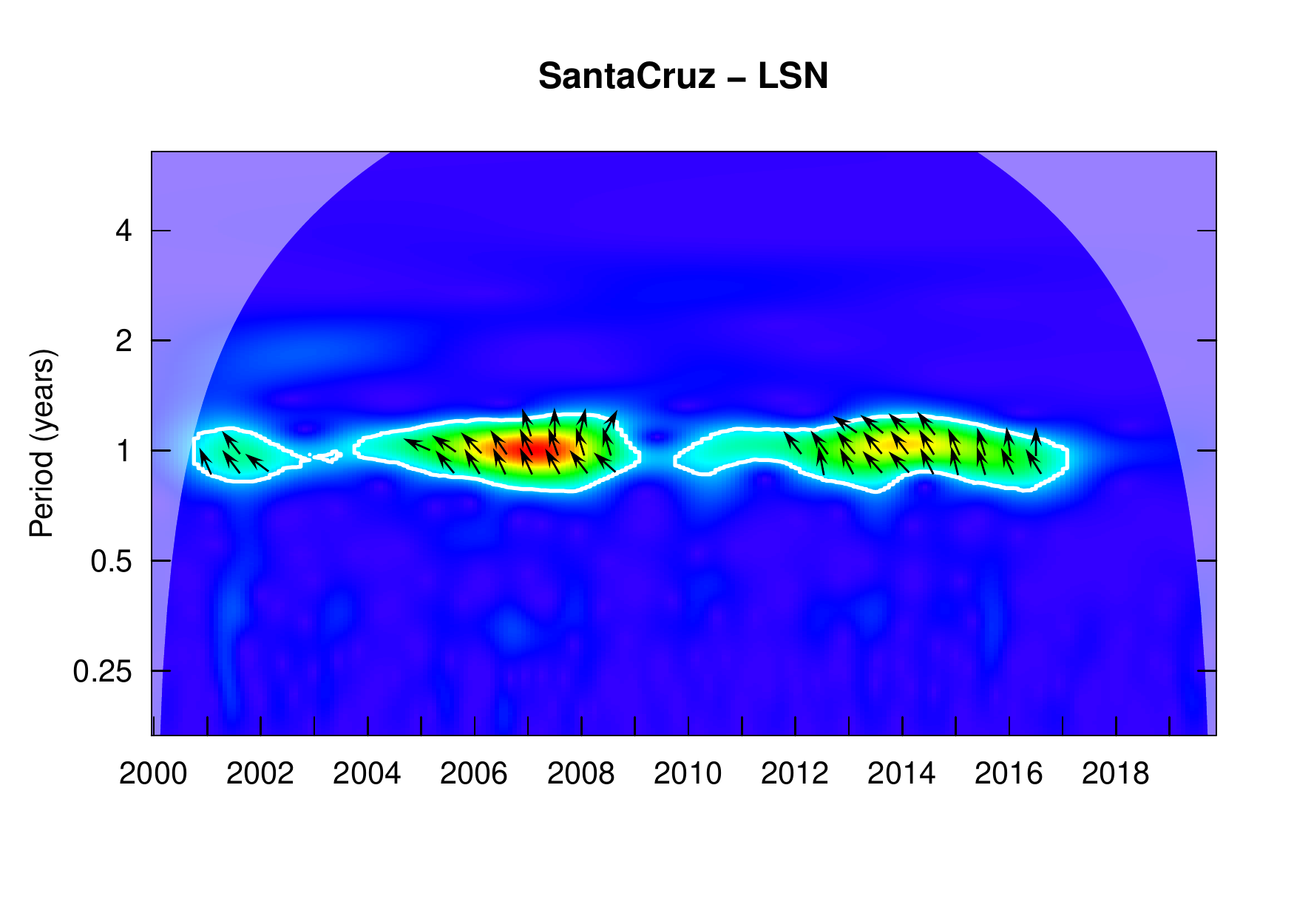}}\vspace{-0.15cm}%
\subfloat[]{\includegraphics[scale=0.23]{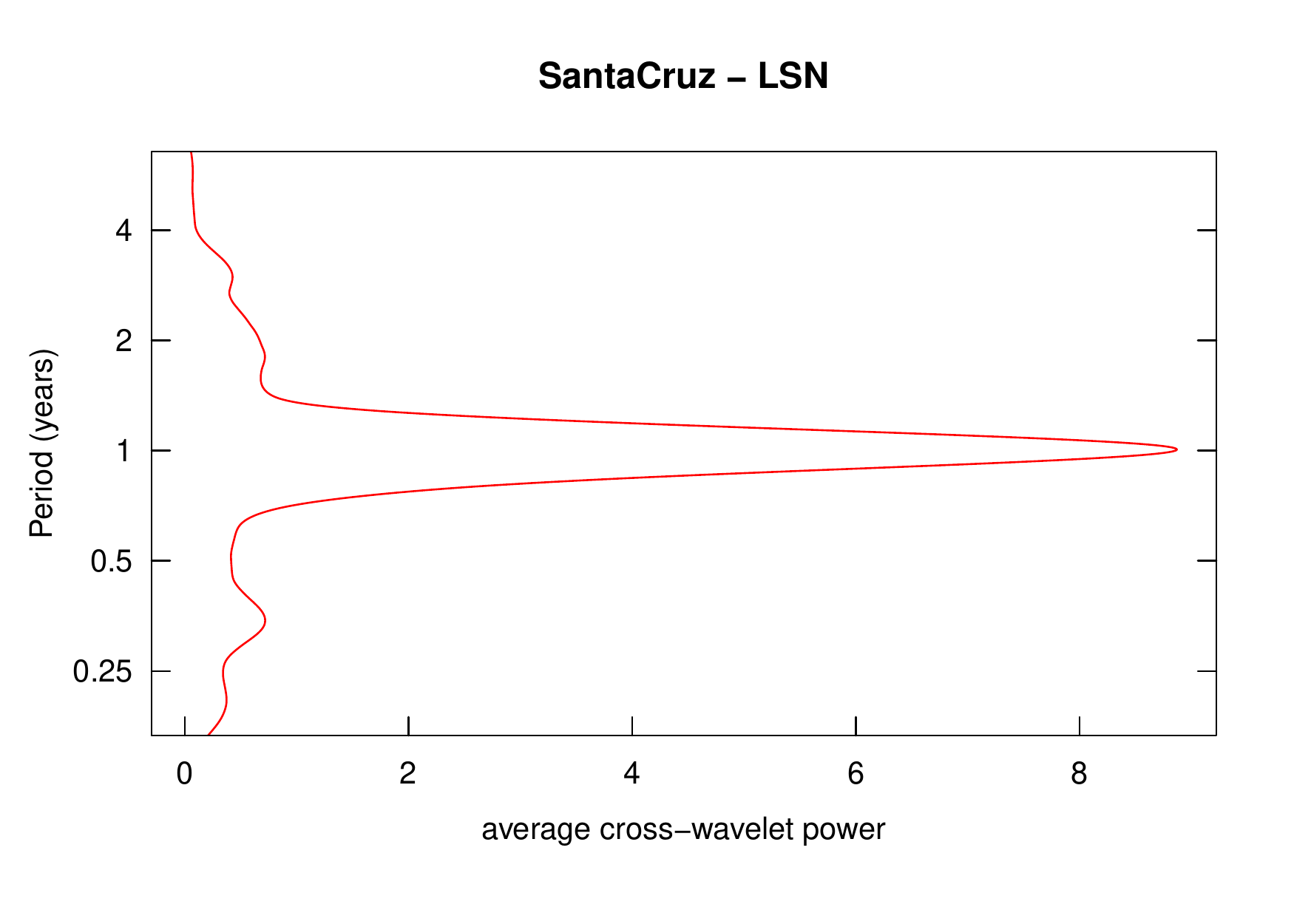}}\vspace{-0.15cm}%
\subfloat[]{\includegraphics[scale=0.23]{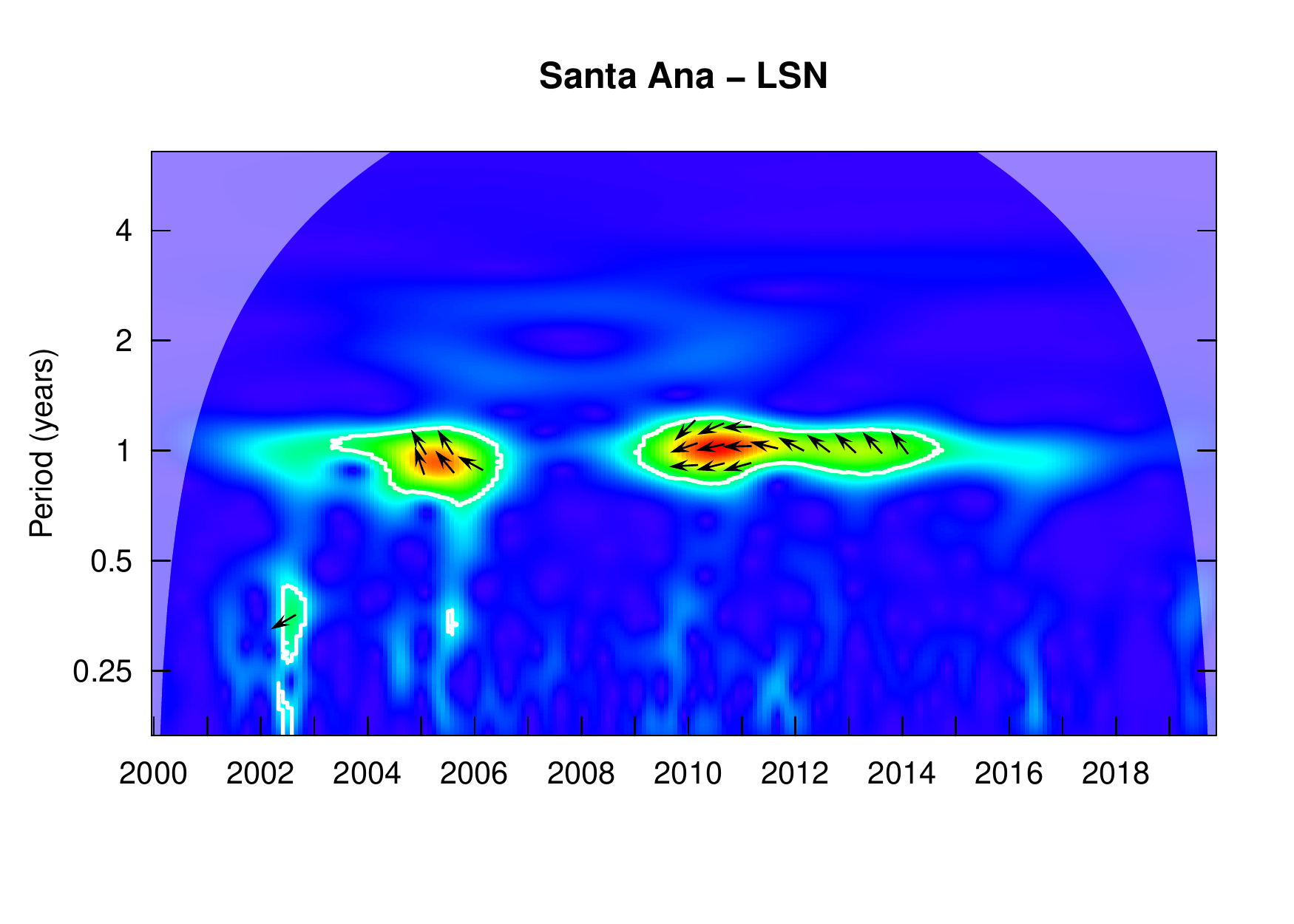}}\vspace{-0.15cm}%
\subfloat[]{\includegraphics[scale=0.23]{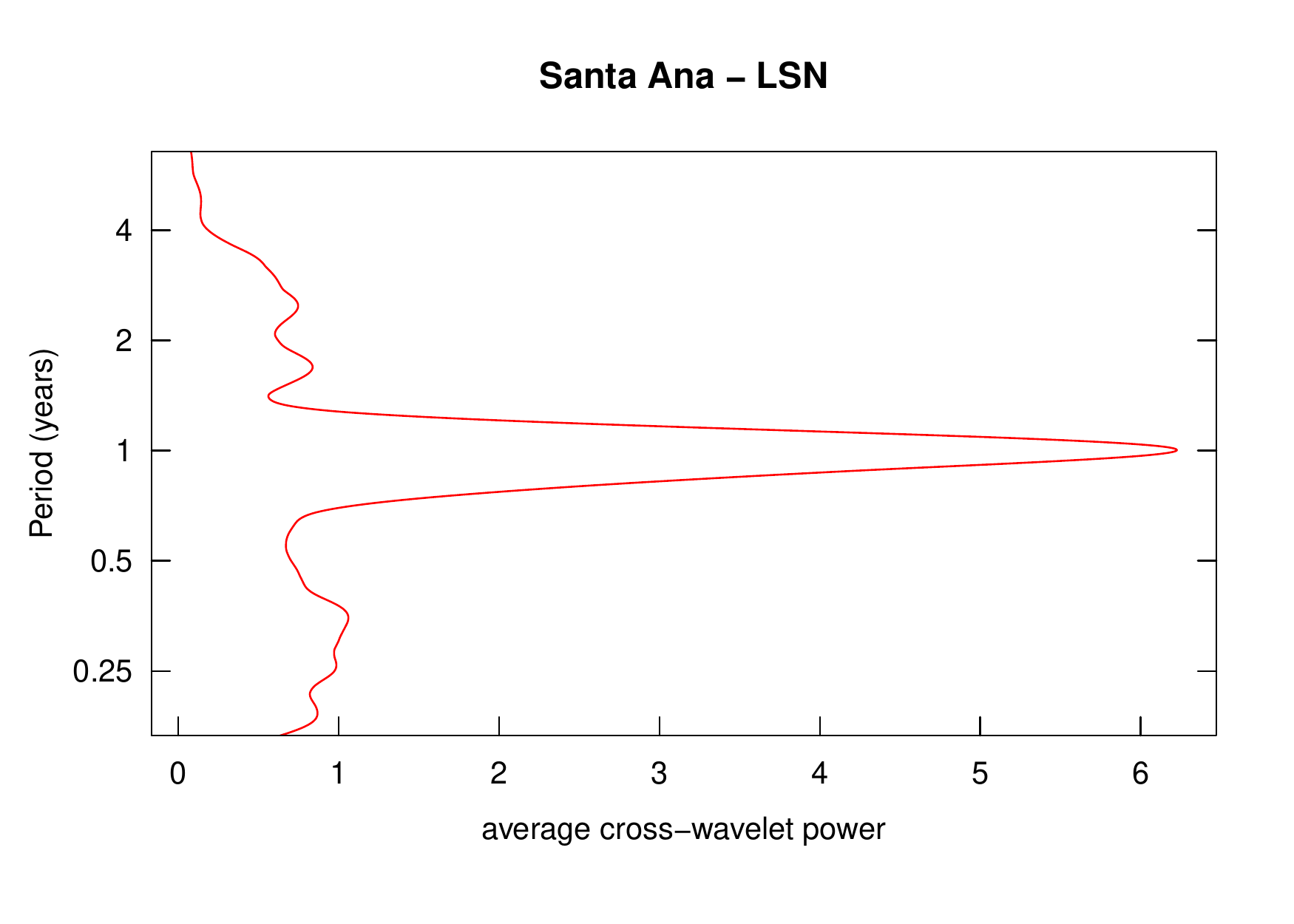}}\vspace{-0.15cm}\\
\subfloat[]{\includegraphics[scale=0.23]{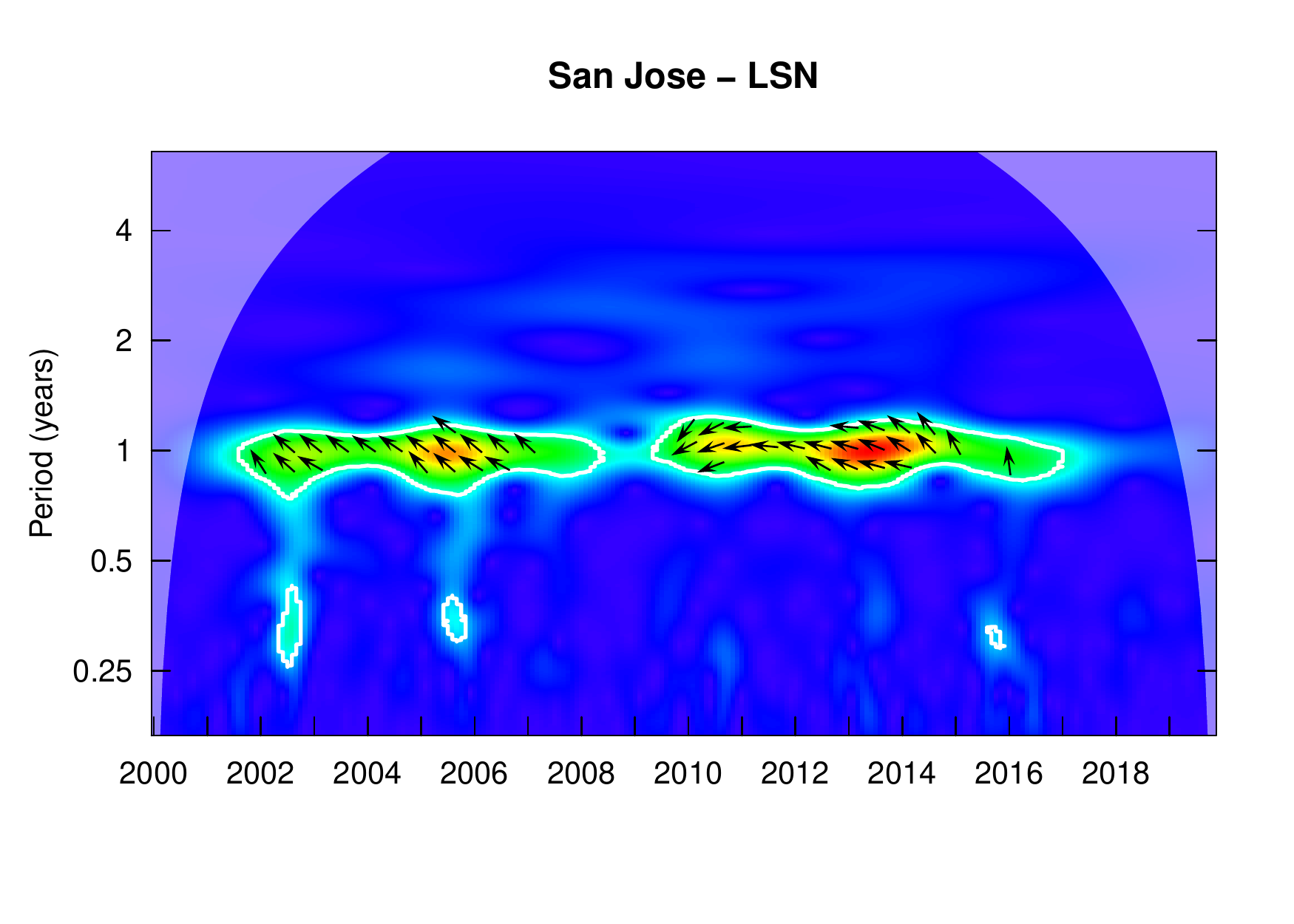}}\vspace{-0.15cm}%
\subfloat[]{\includegraphics[scale=0.23]{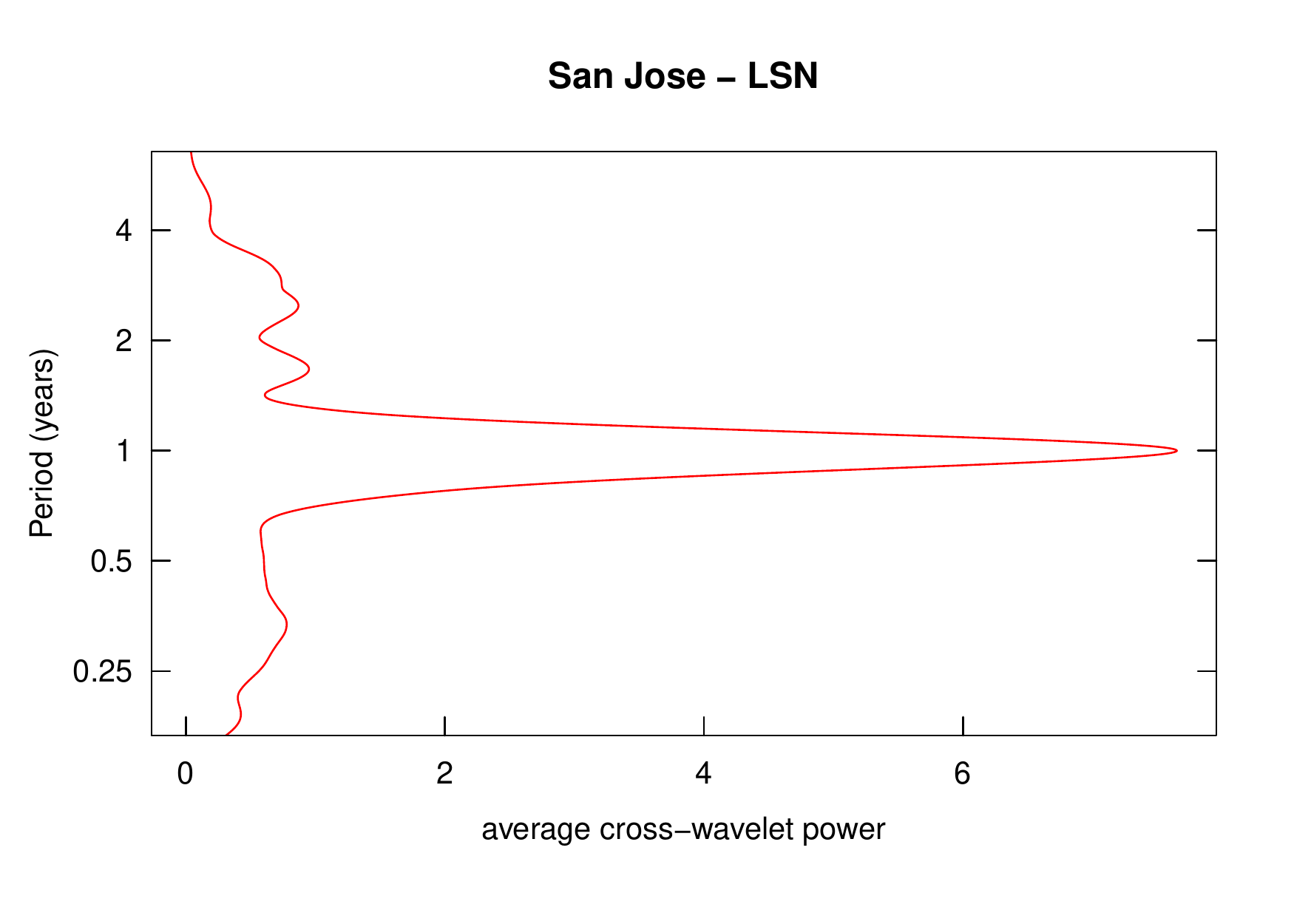}}\vspace{-0.15cm}%
\subfloat[]{\includegraphics[scale=0.23]{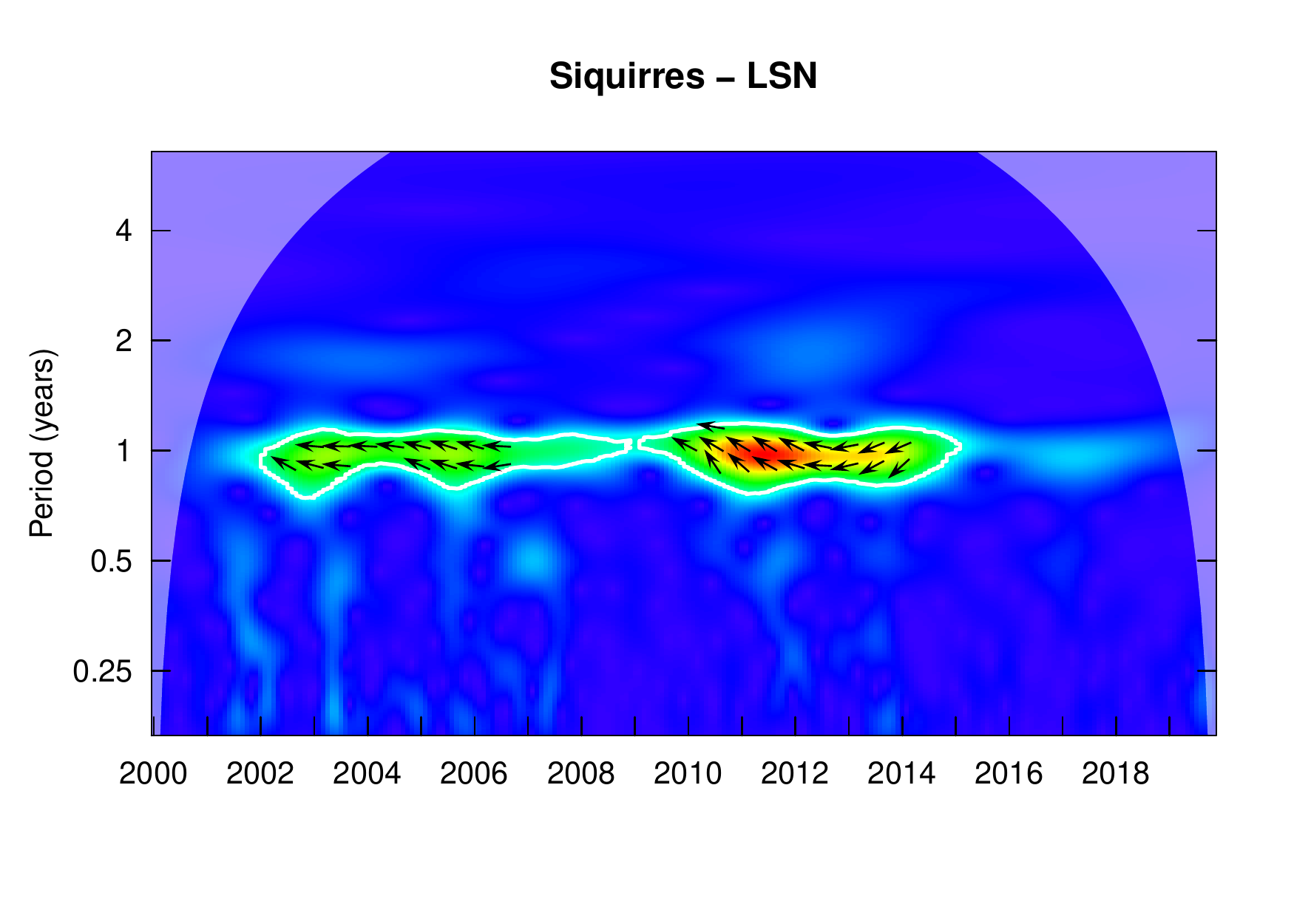}}\vspace{-0.15cm}%
\subfloat[]{\includegraphics[scale=0.23]{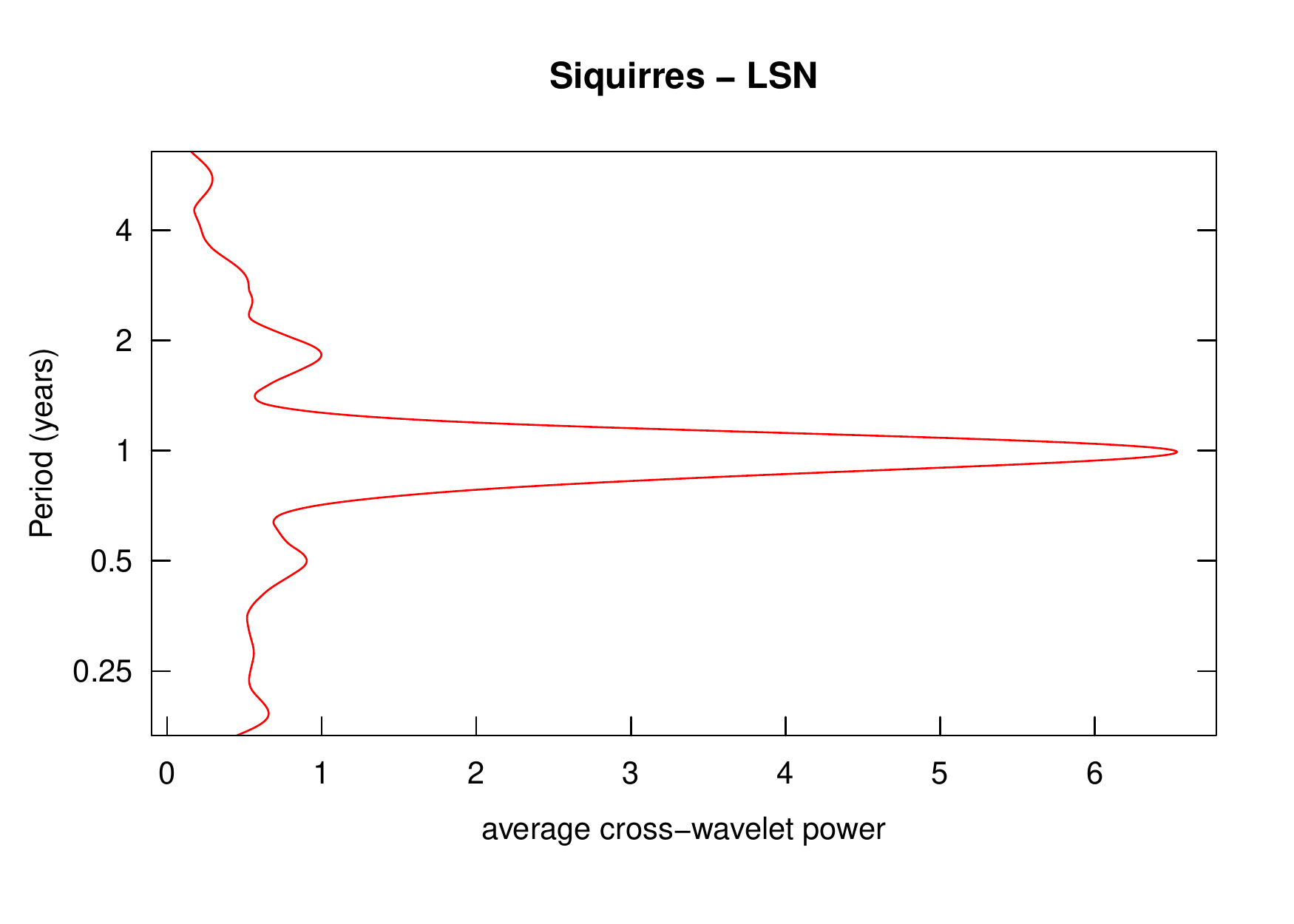}}\vspace{-0.15cm}\\
\subfloat[]{\includegraphics[scale=0.23]{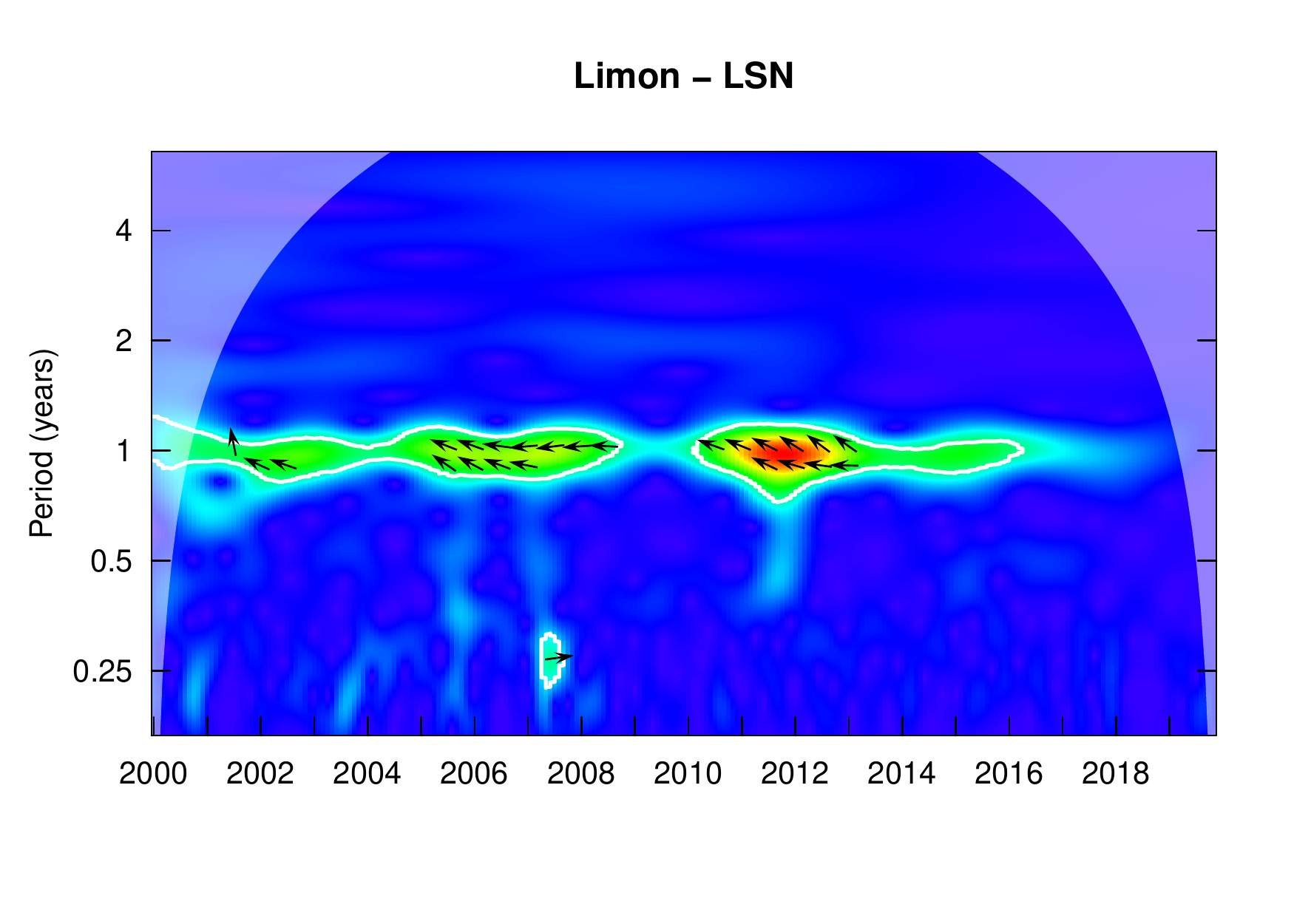}}\vspace{-0.15cm}%
\subfloat[]{\includegraphics[scale=0.23]{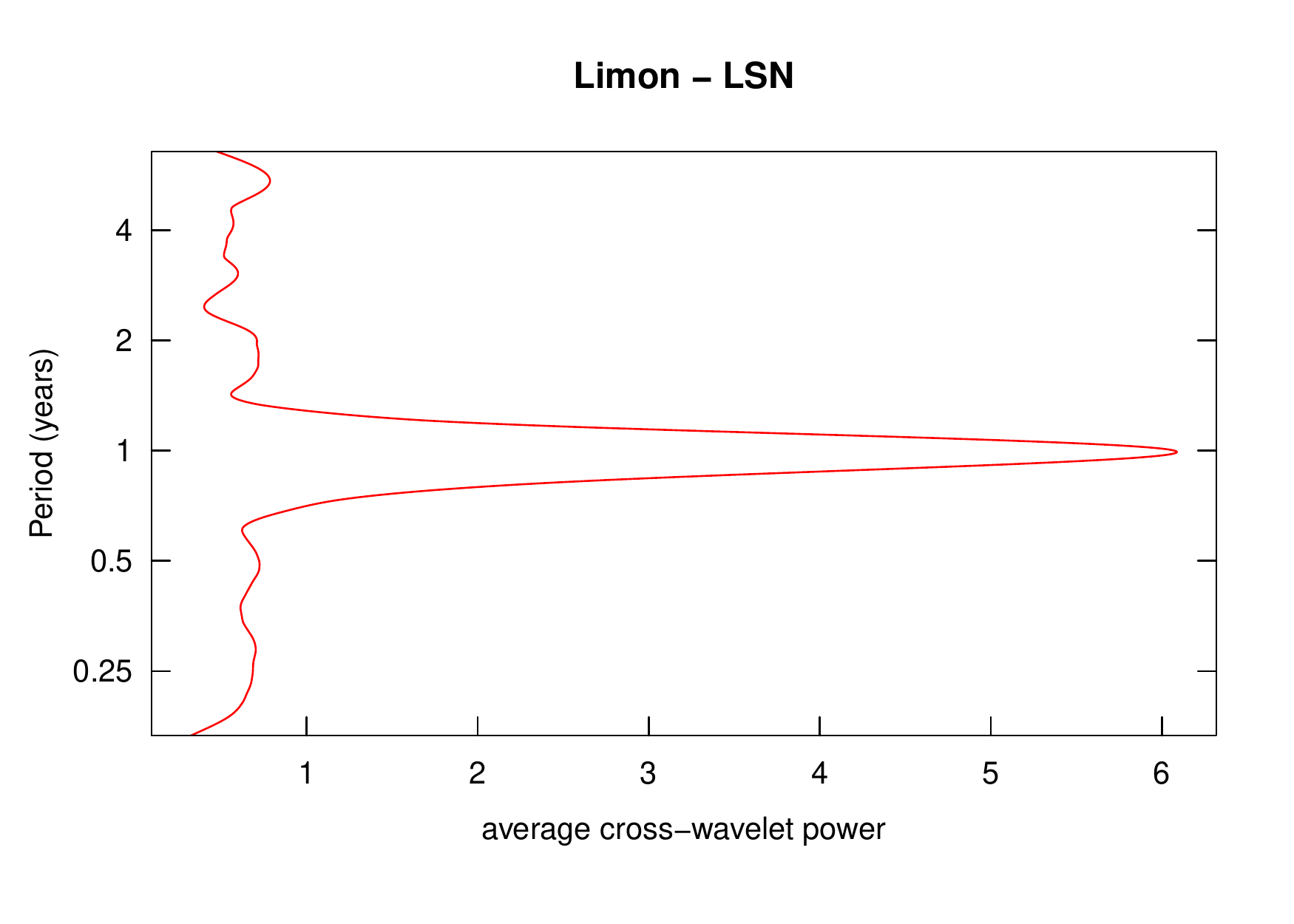}}\vspace{-0.15cm}%
\subfloat[]{\includegraphics[scale=0.23]{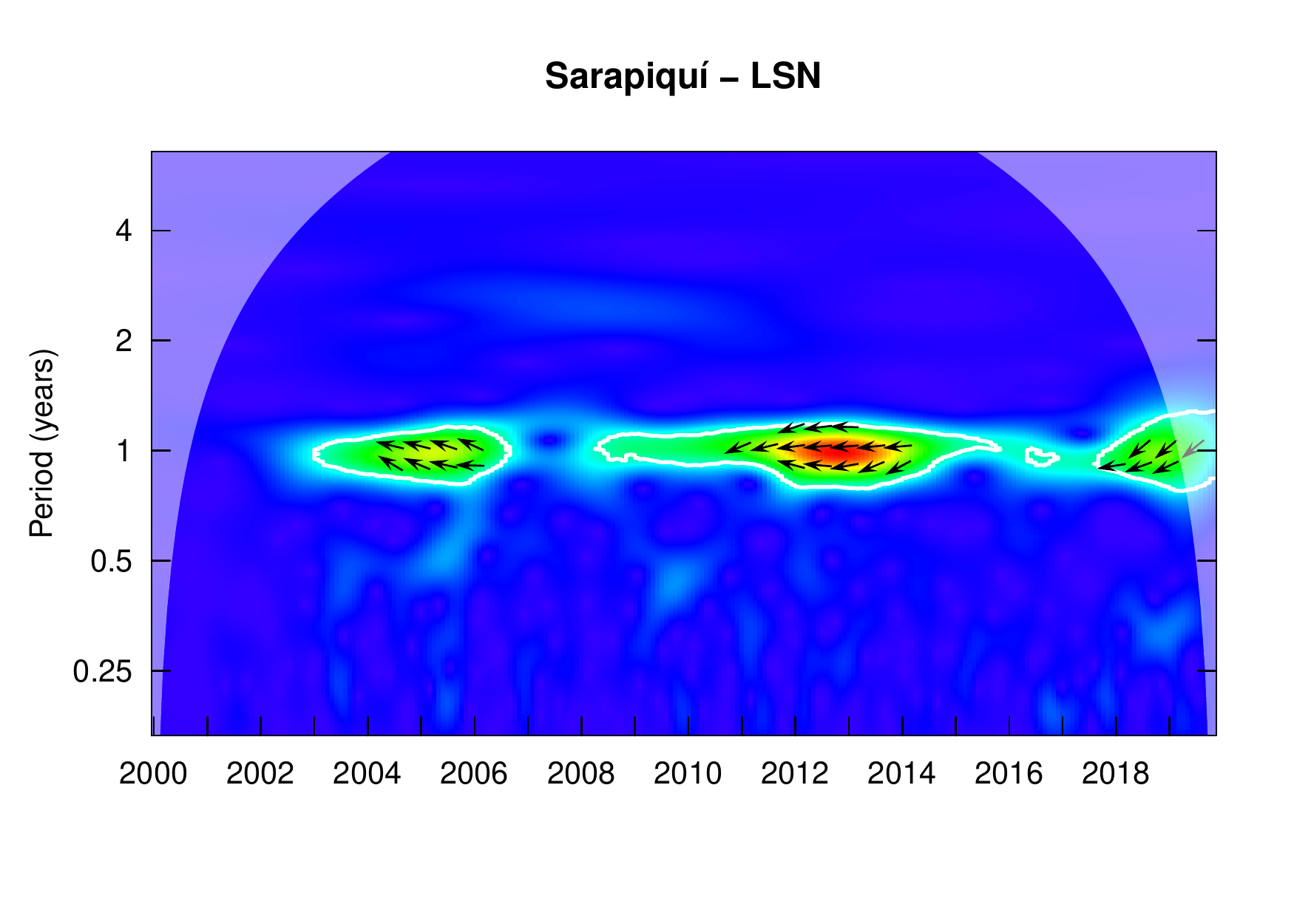}}\vspace{-0.15cm}%
\subfloat[]{\includegraphics[scale=0.23]{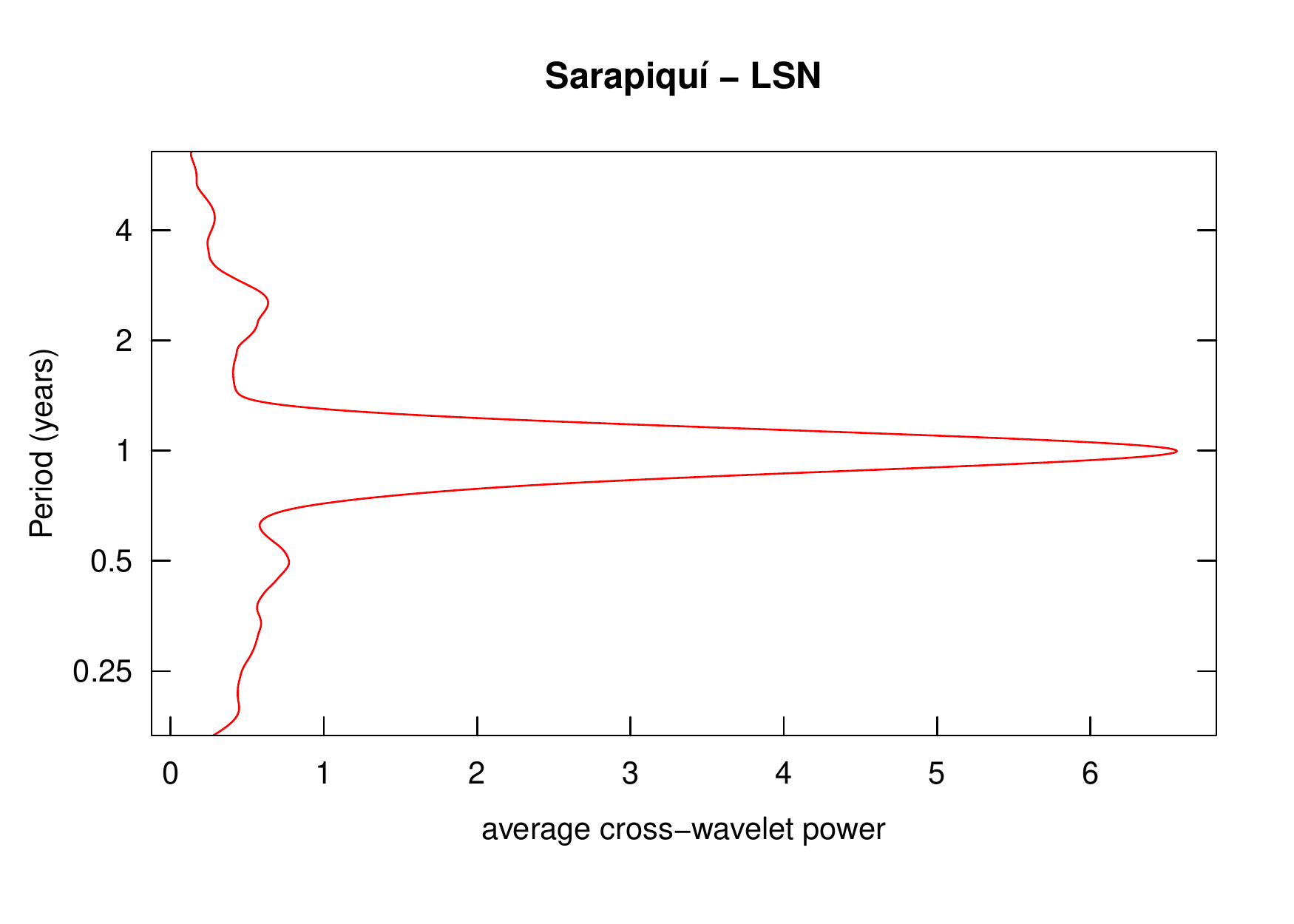}}\vspace{-0.15cm}\\
\subfloat[]{\includegraphics[scale=0.23]{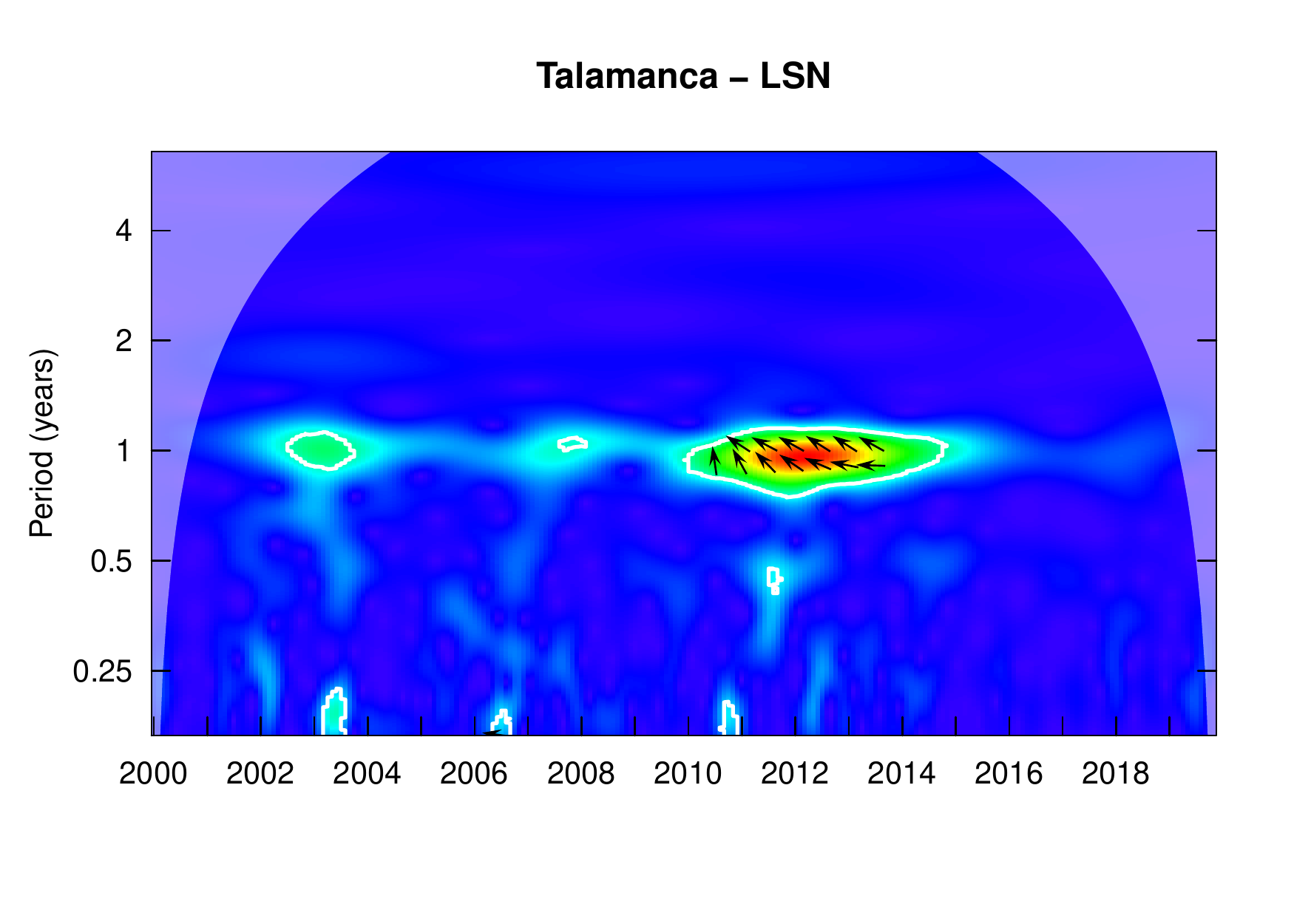}}\vspace{-0.15cm}%
\subfloat[]{\includegraphics[scale=0.23]{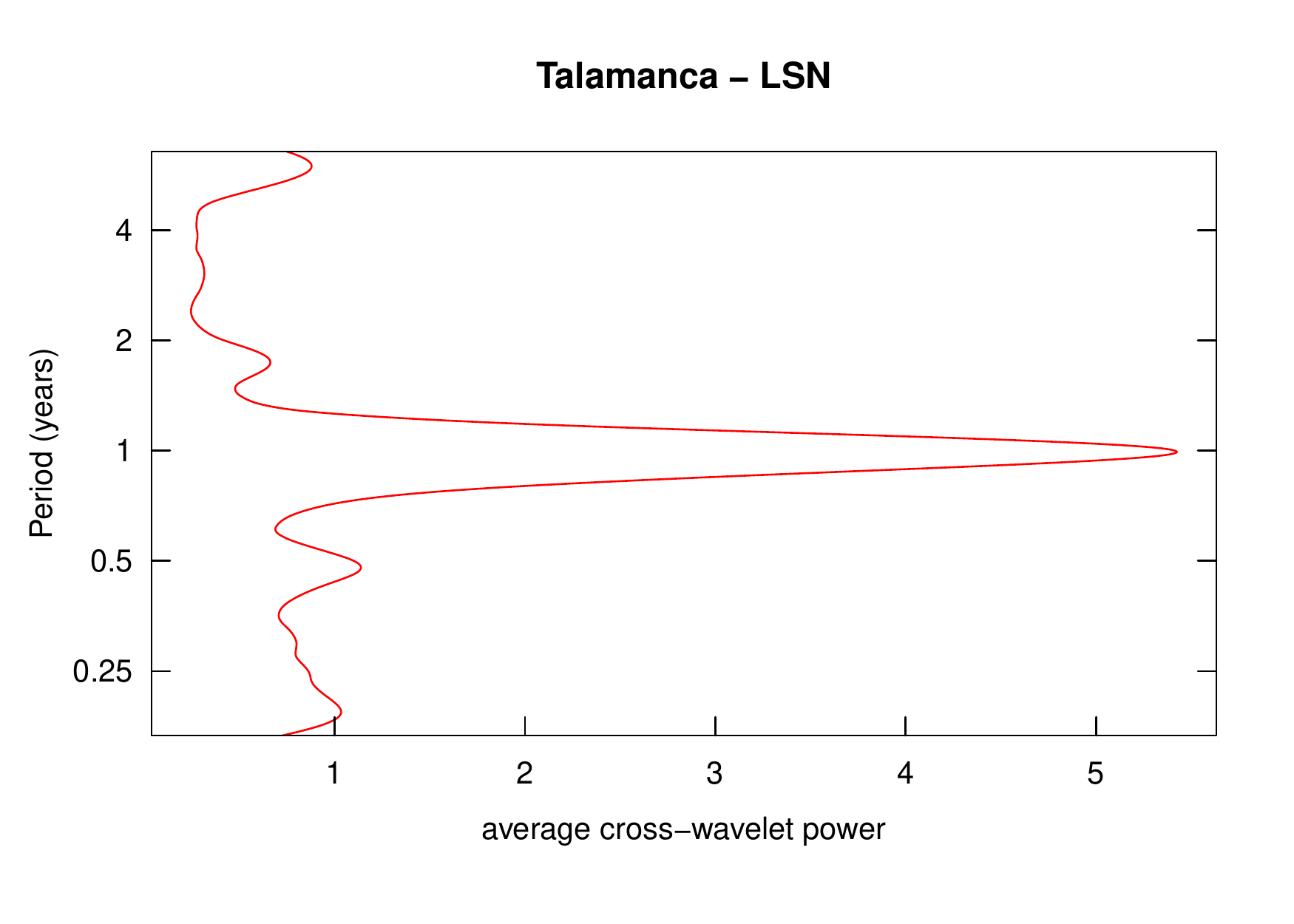}}\vspace{-0.15cm}
\subfloat[]{\includegraphics[scale=0.23]{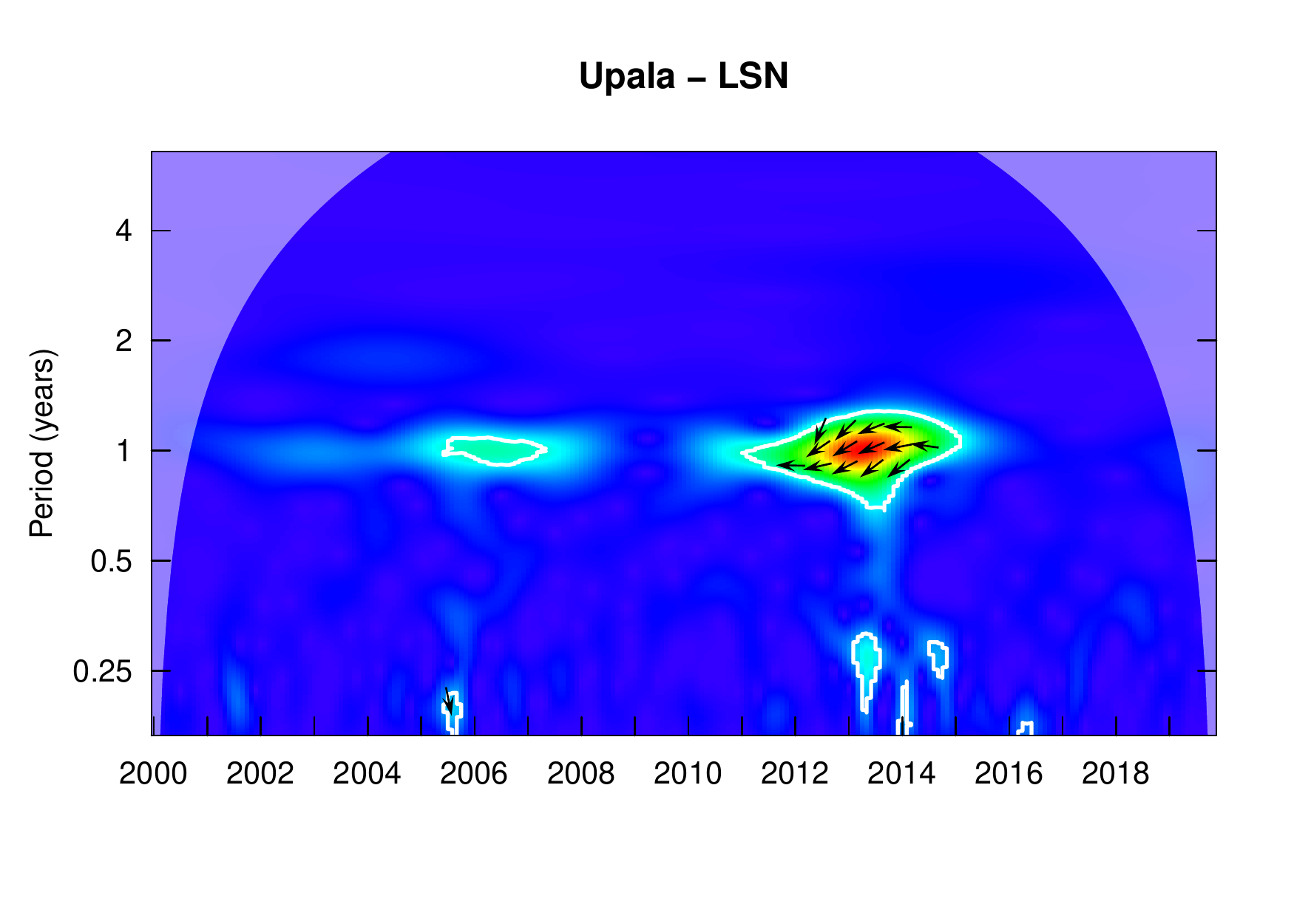}}\vspace{-0.15cm}%
\subfloat[]{\includegraphics[scale=0.23]{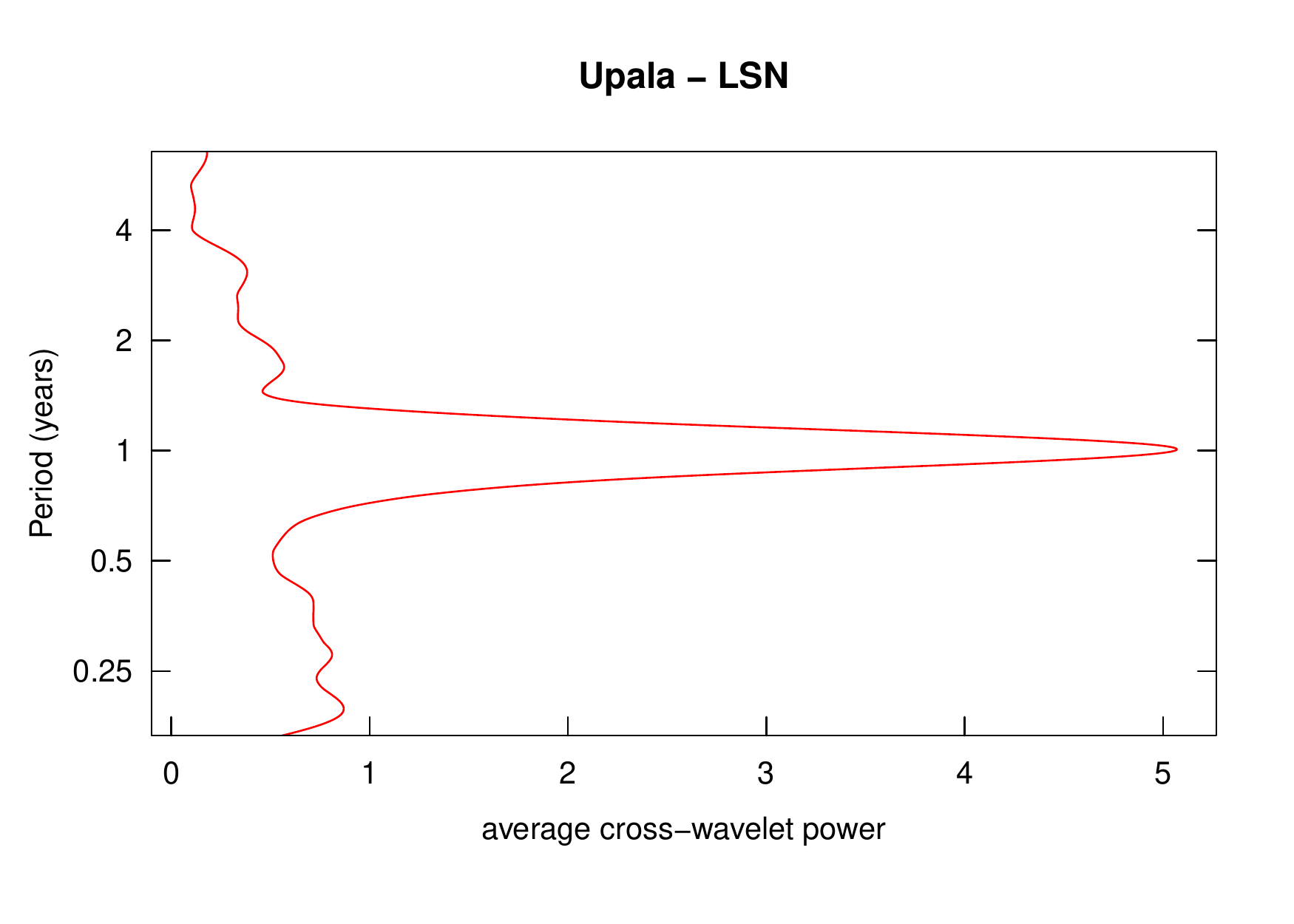}}\vspace{-0.15cm}\\
\subfloat[]{\includegraphics[scale=0.23]{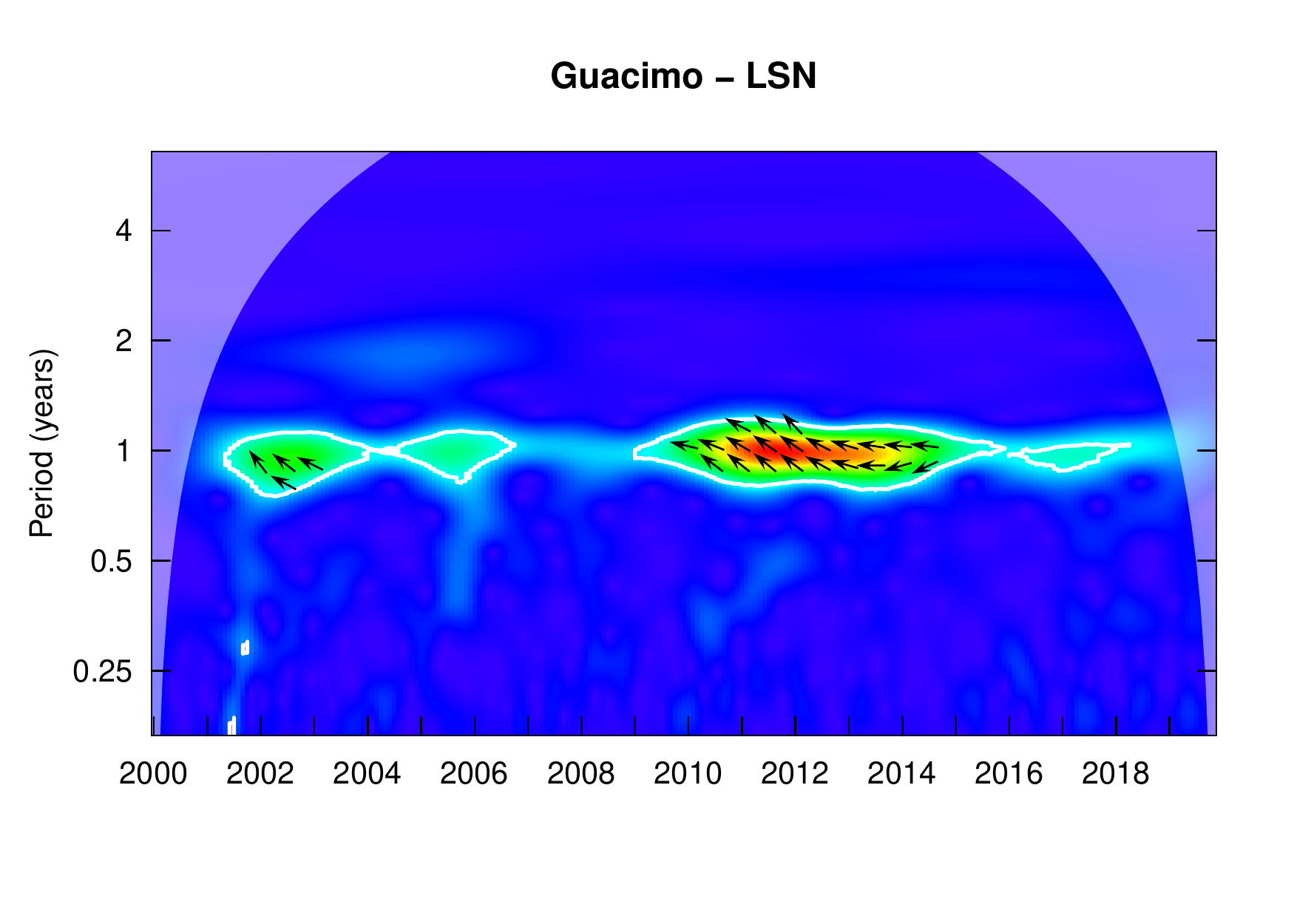}}\vspace{-0.15cm}%
\subfloat[]{\includegraphics[scale=0.23]{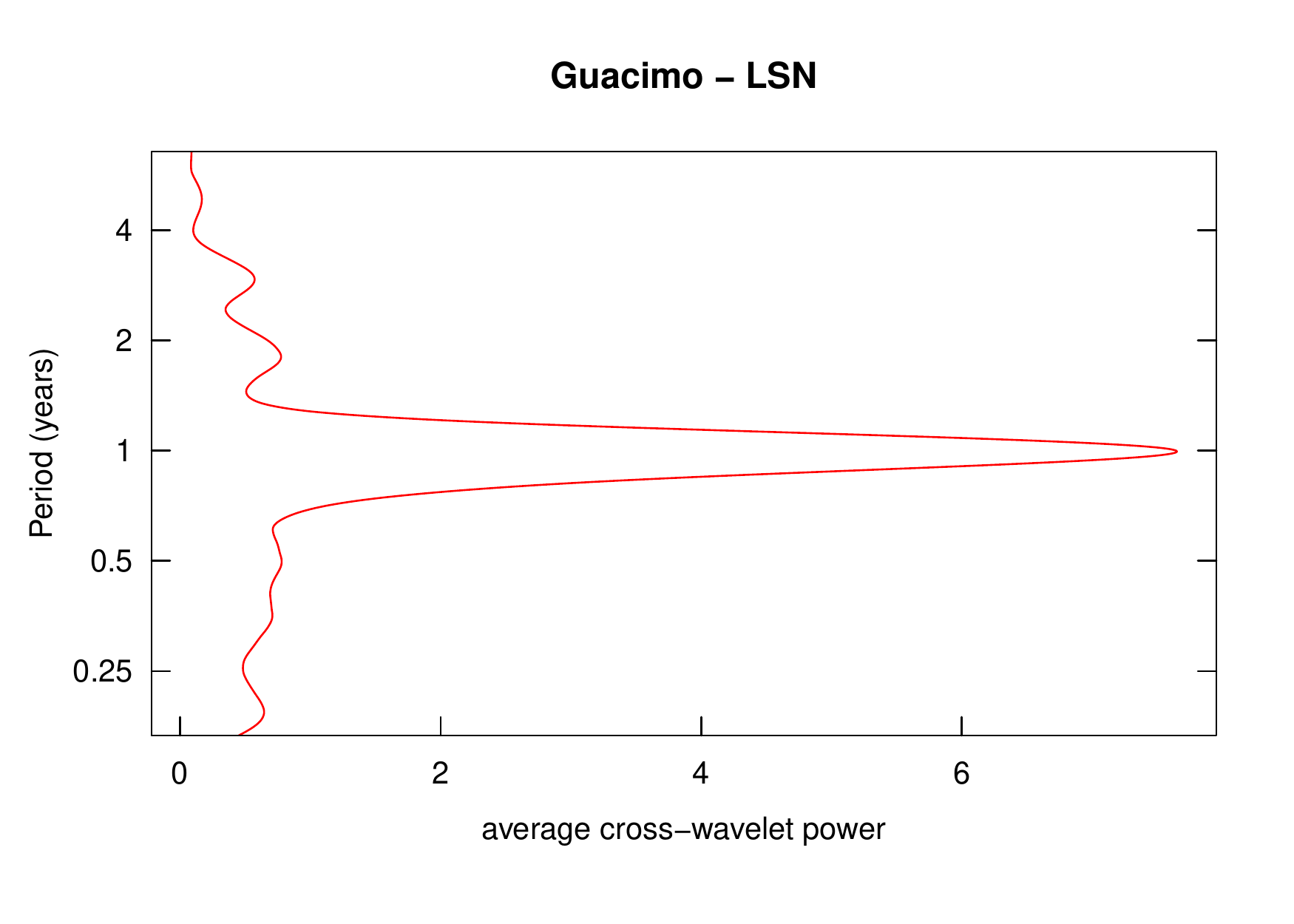}}\vspace{-0.15cm}%
\subfloat[]{\includegraphics[scale=0.23]{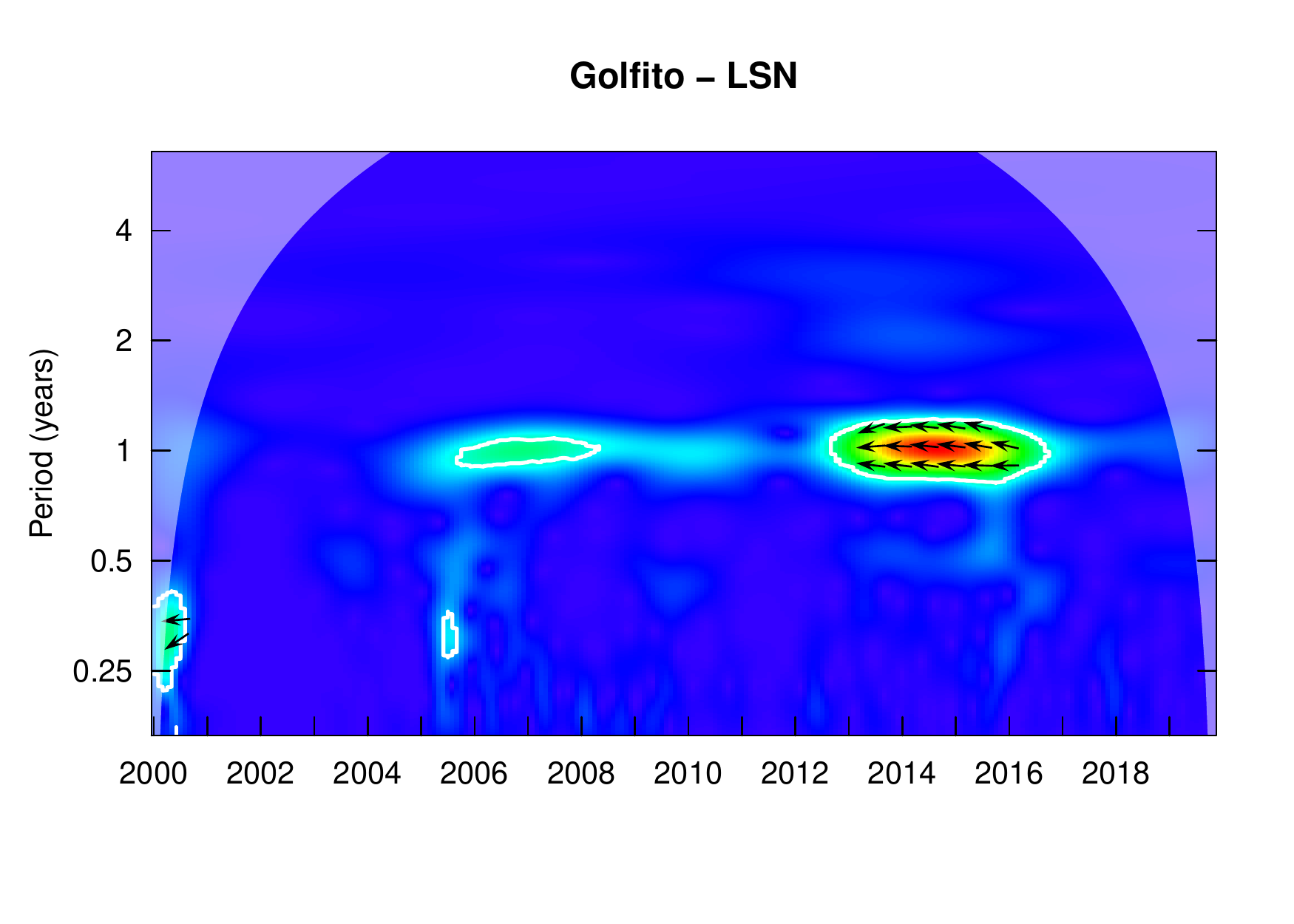}}\vspace{-0.15cm}%
\subfloat[]{\includegraphics[scale=0.23]{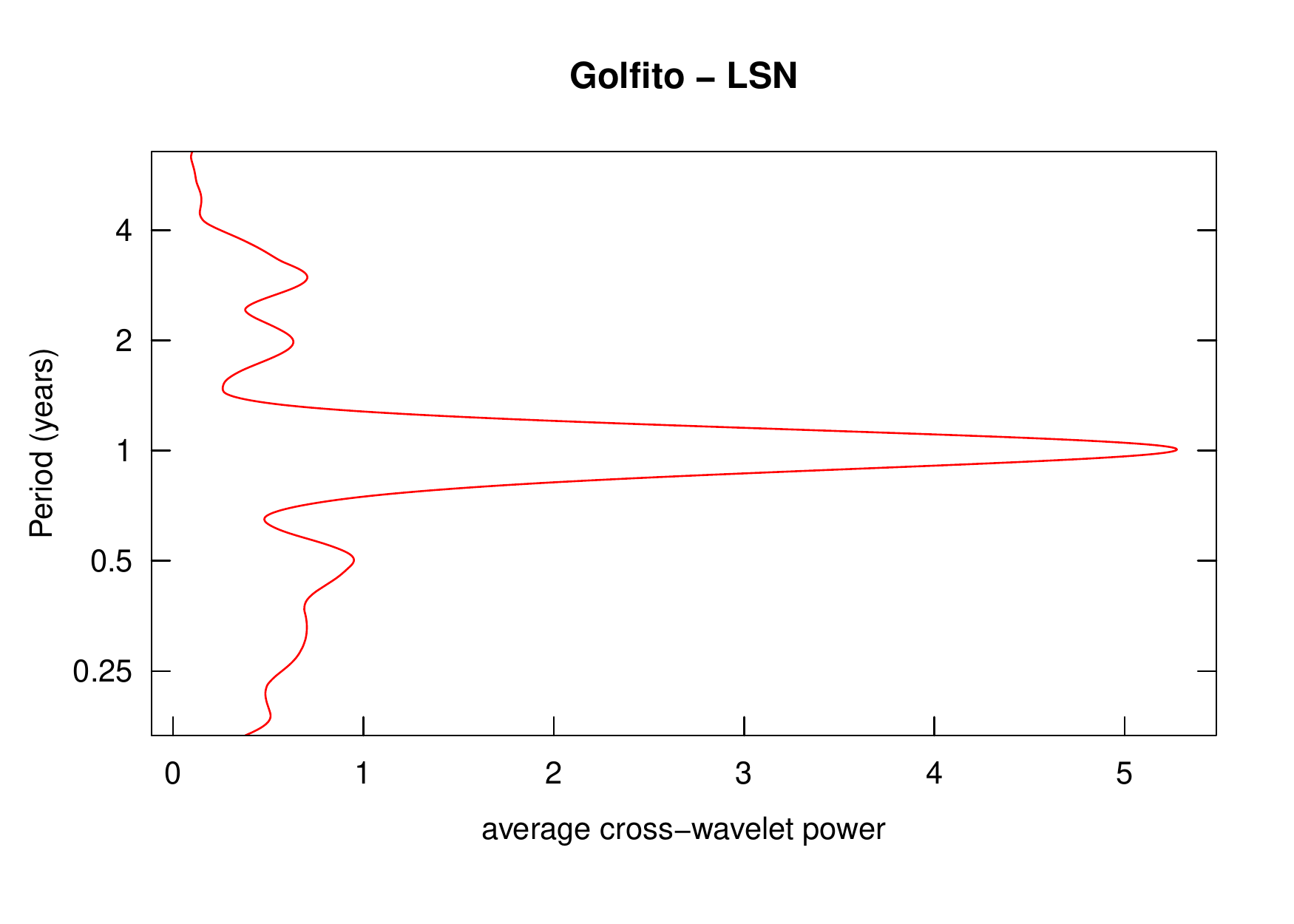}}\vspace{-0.15cm}\\
\subfloat[]{\includegraphics[scale=0.23]{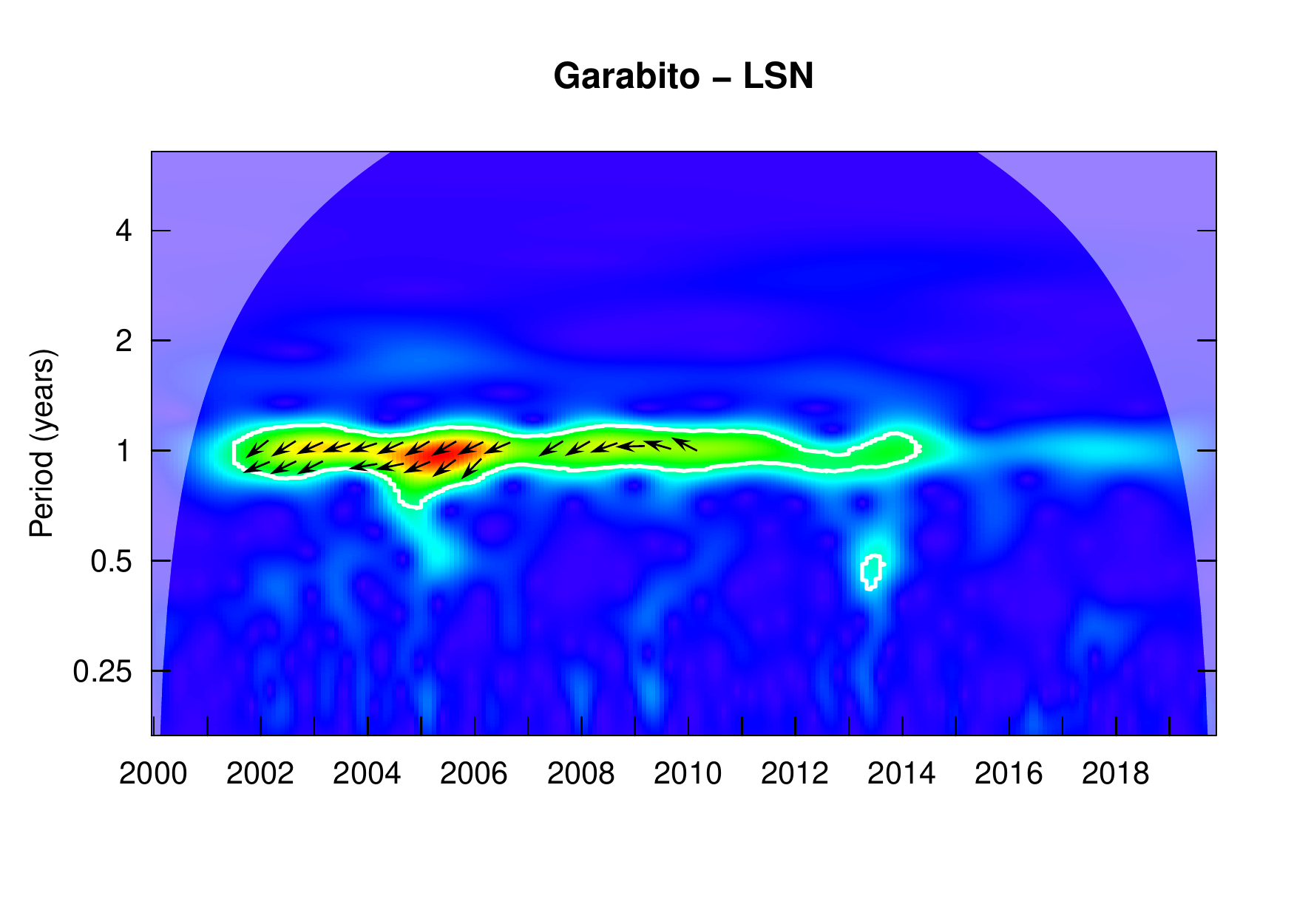}}\vspace{-0.15cm}%
\subfloat[]{\includegraphics[scale=0.23]{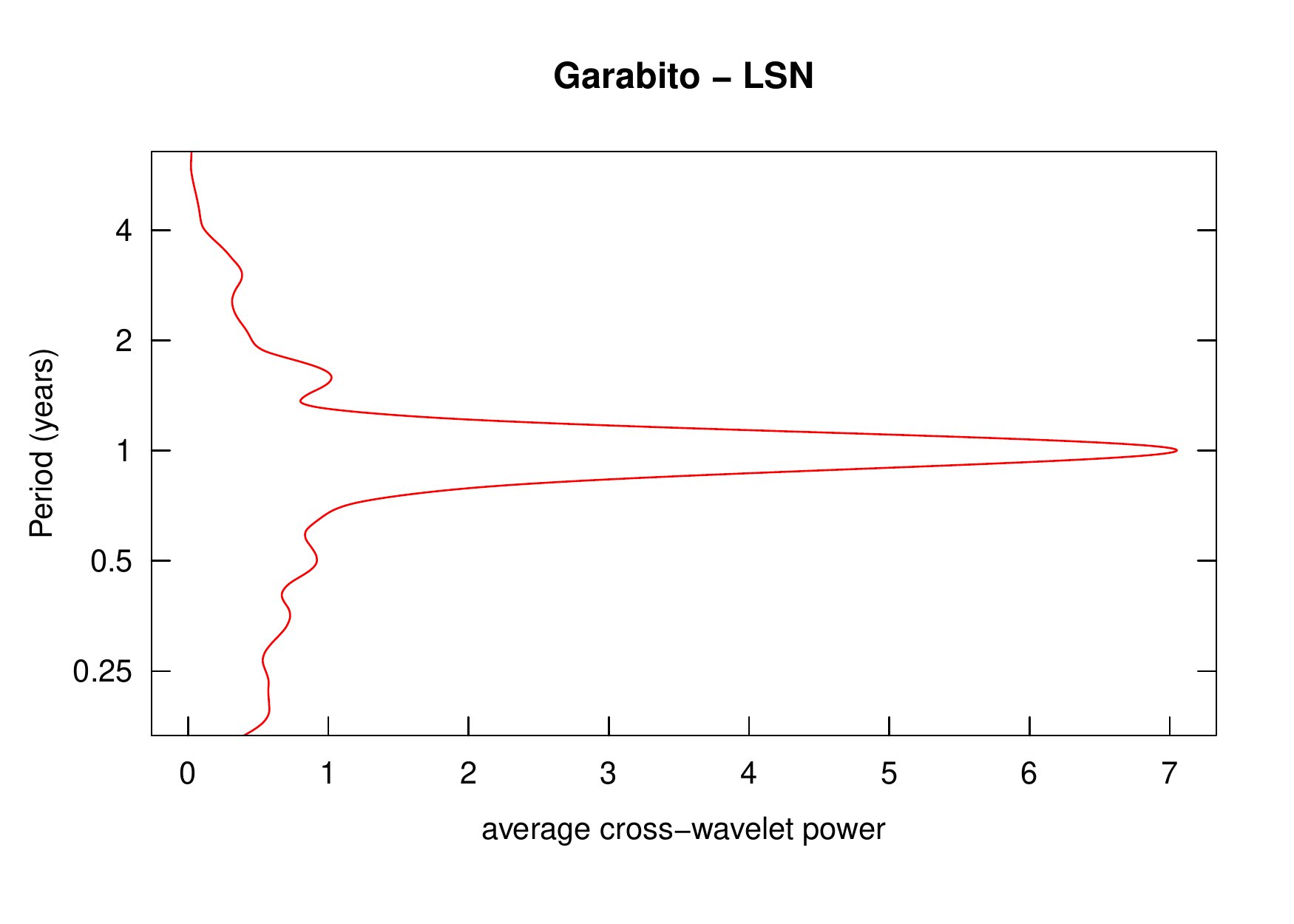}}\vspace{-0.15cm}%
\subfloat[]{\includegraphics[scale=0.23]{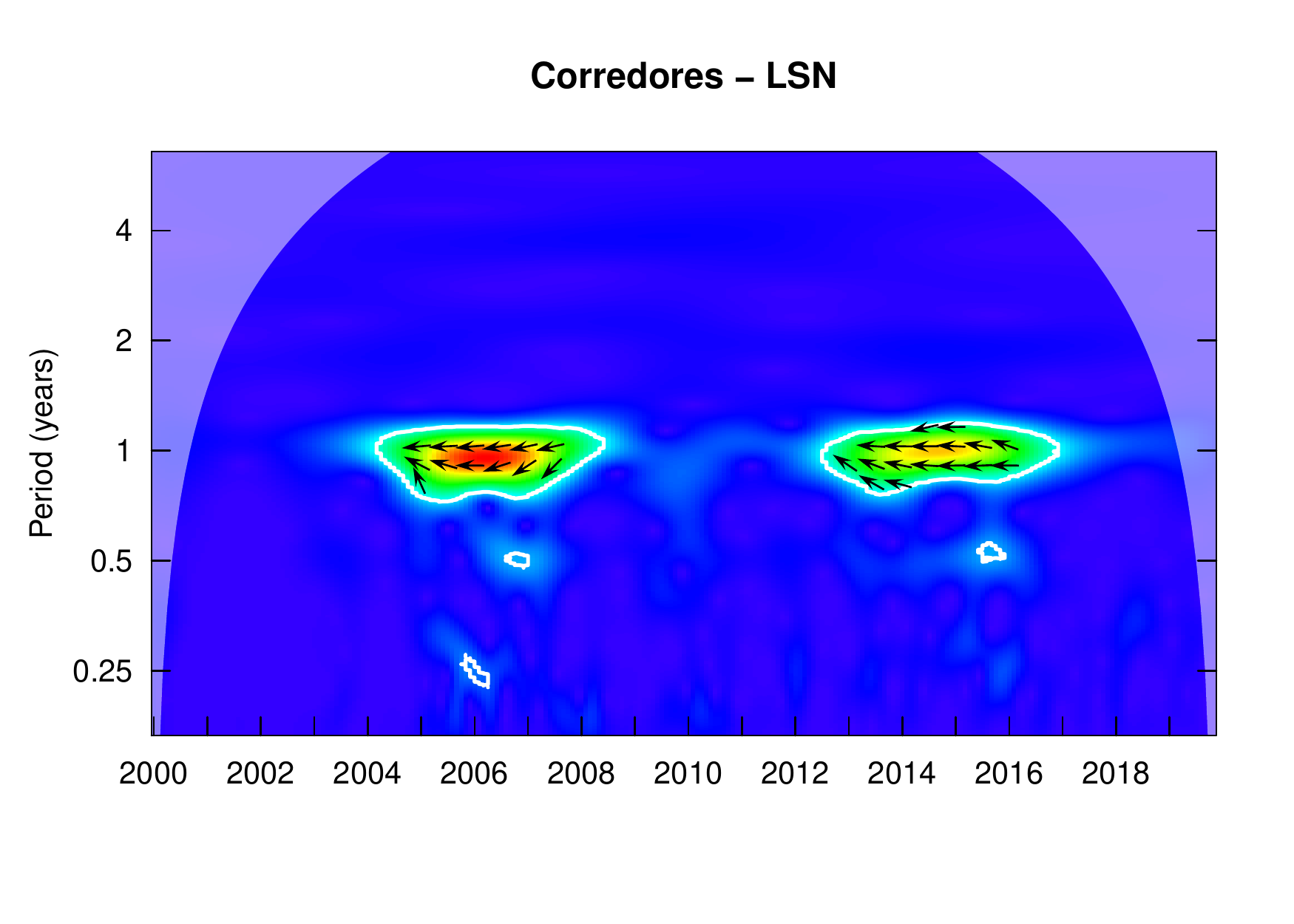}}\vspace{-0.15cm}%
\subfloat[]{\includegraphics[scale=0.23]{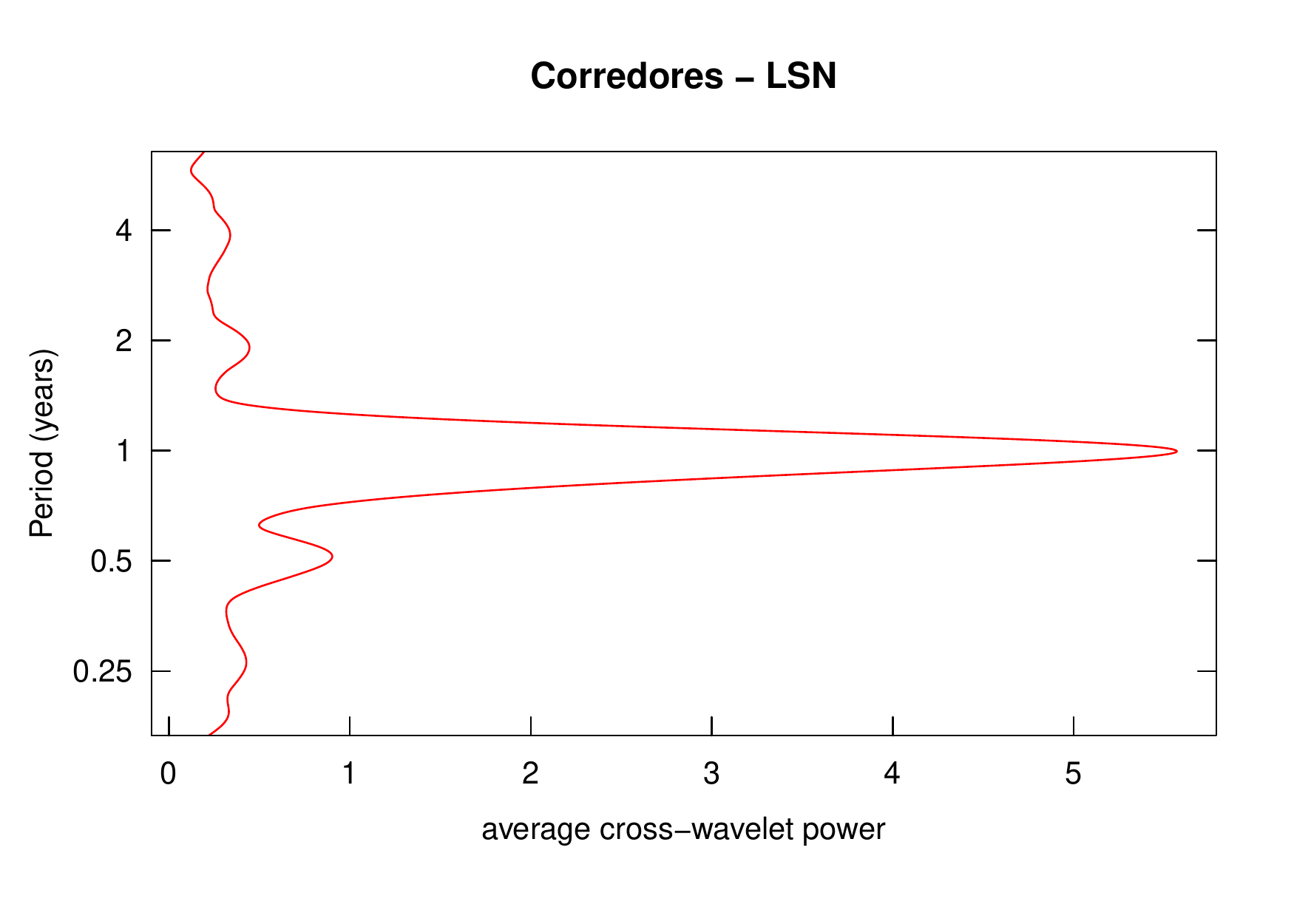}}\vspace{-0.15cm}\\
\caption*{}
\end{figure}

\begin{figure}[H]
\captionsetup[subfigure]{labelformat=empty}
\subfloat[]{\includegraphics[scale=0.23]{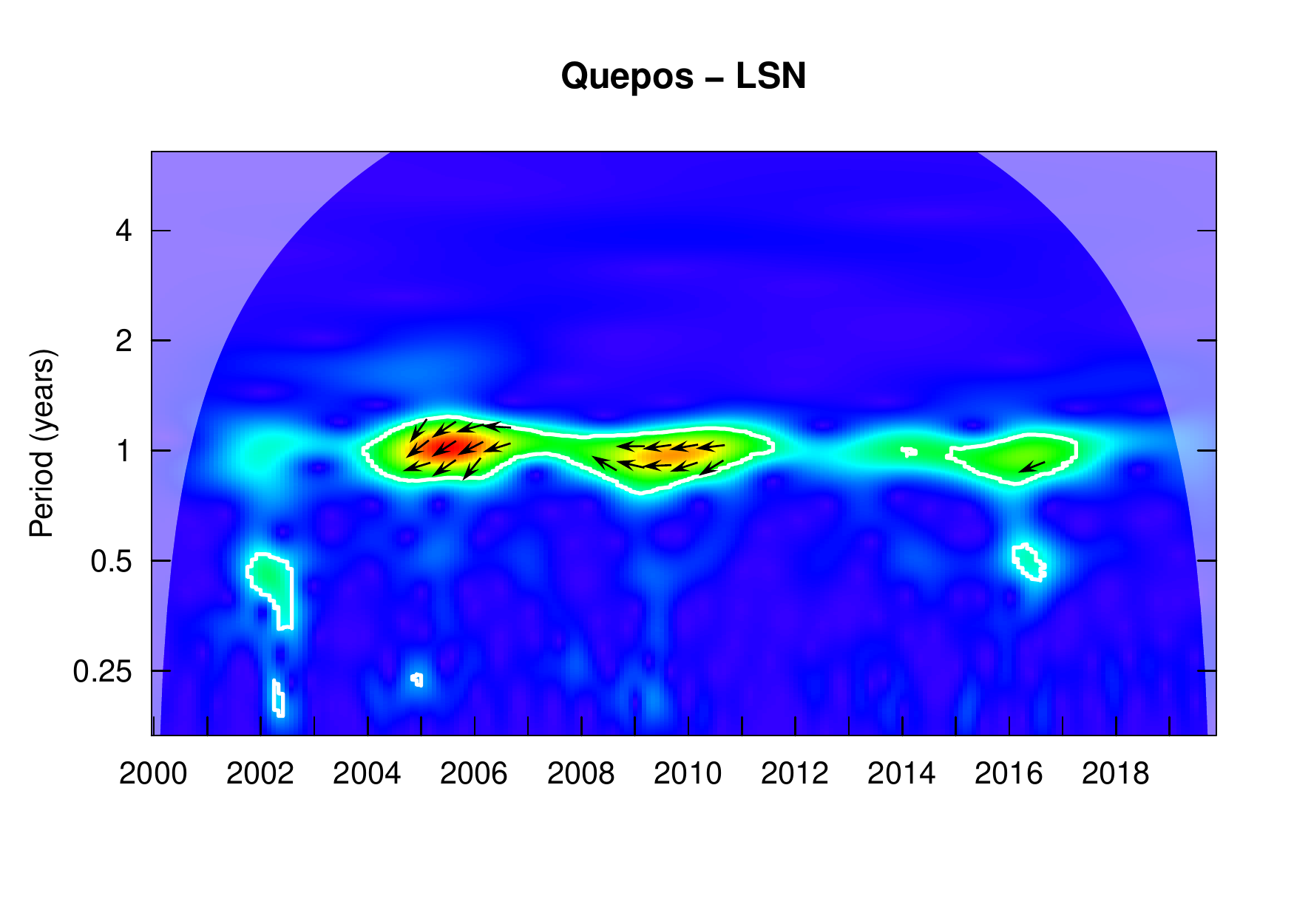}}\vspace{-0.15cm}%
\subfloat[]{\includegraphics[scale=0.23]{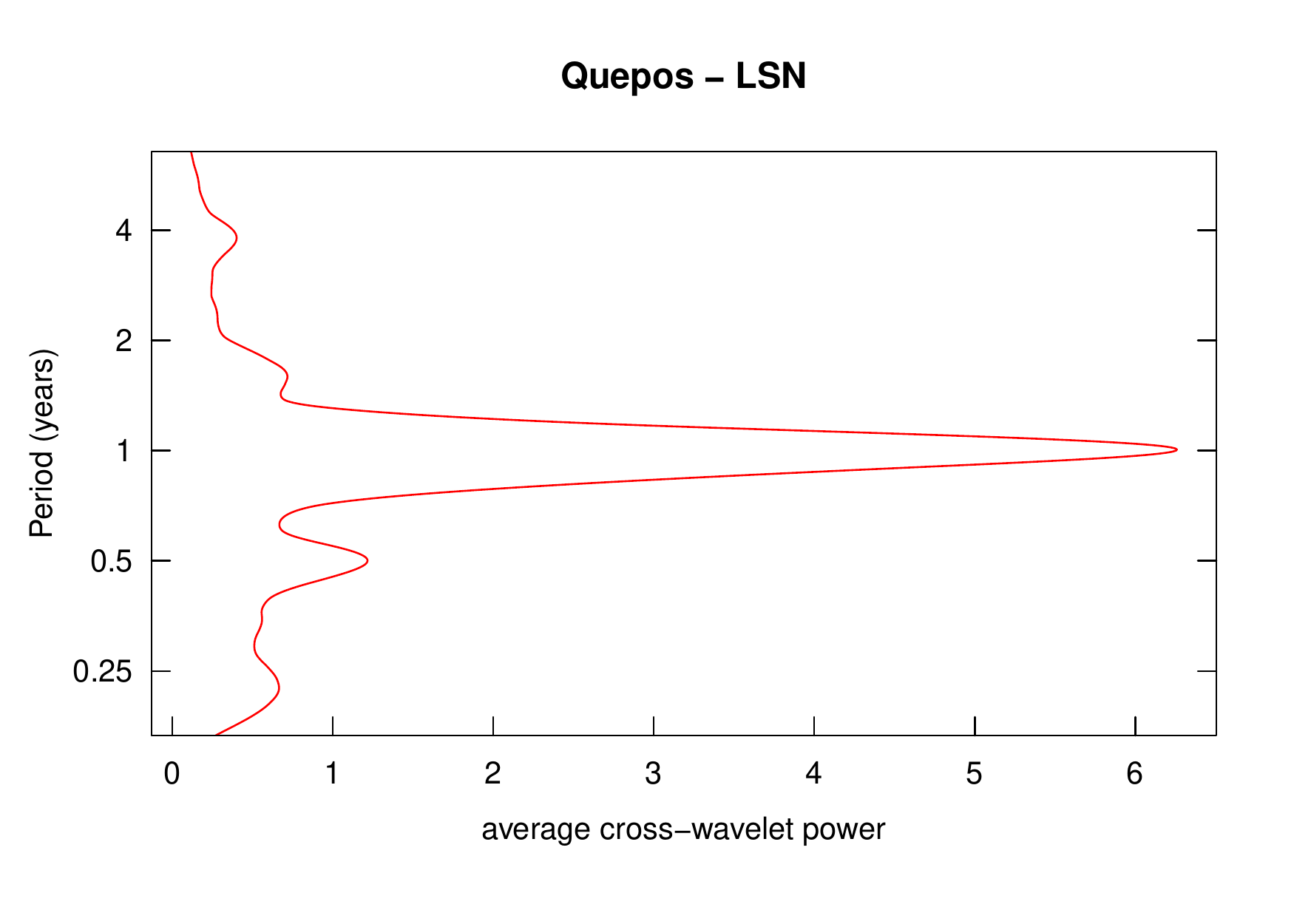}}\vspace{-0.15cm}%
\subfloat[]{\includegraphics[scale=0.23]{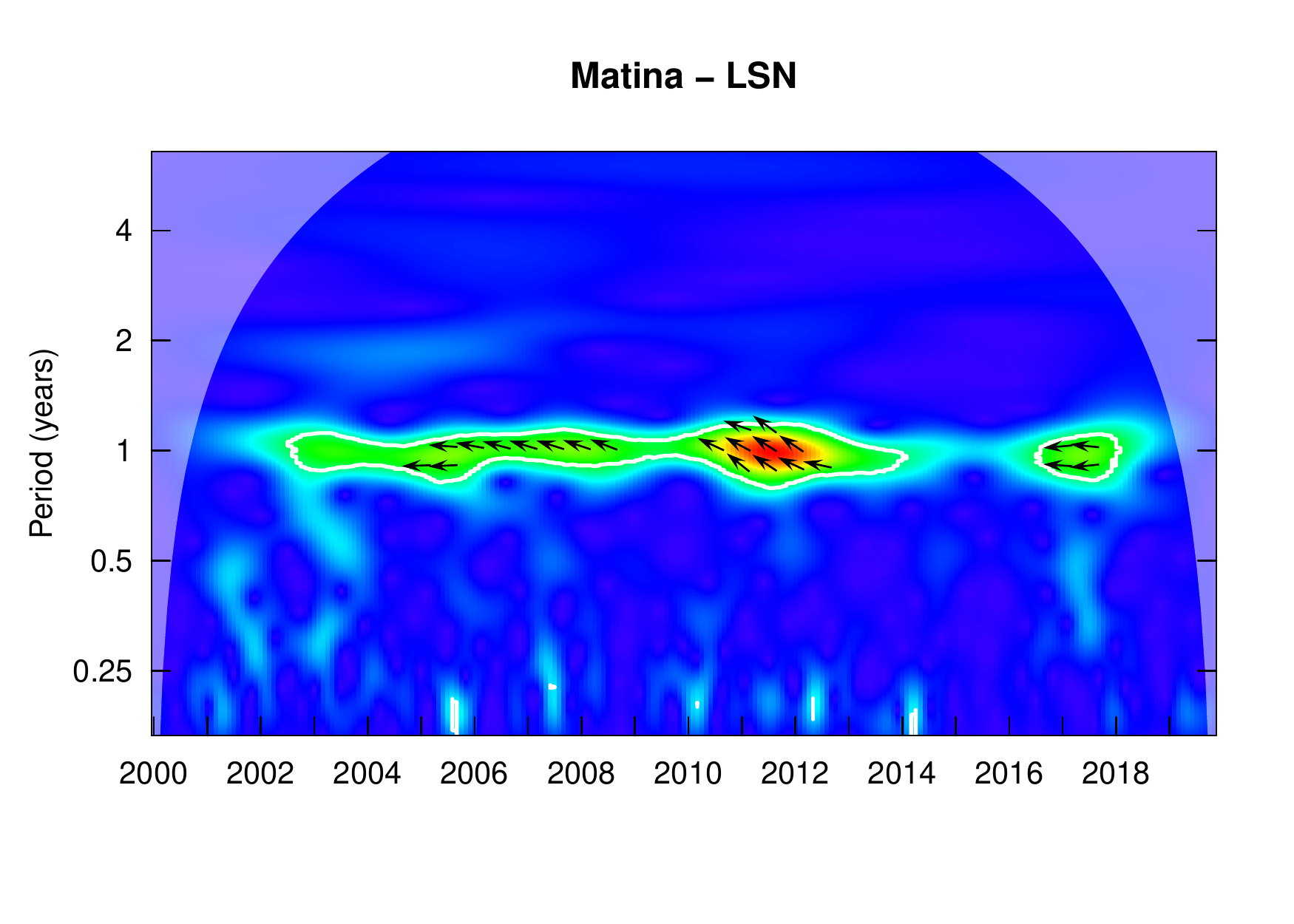}}\vspace{-0.15cm}%
\subfloat[]{\includegraphics[scale=0.23]{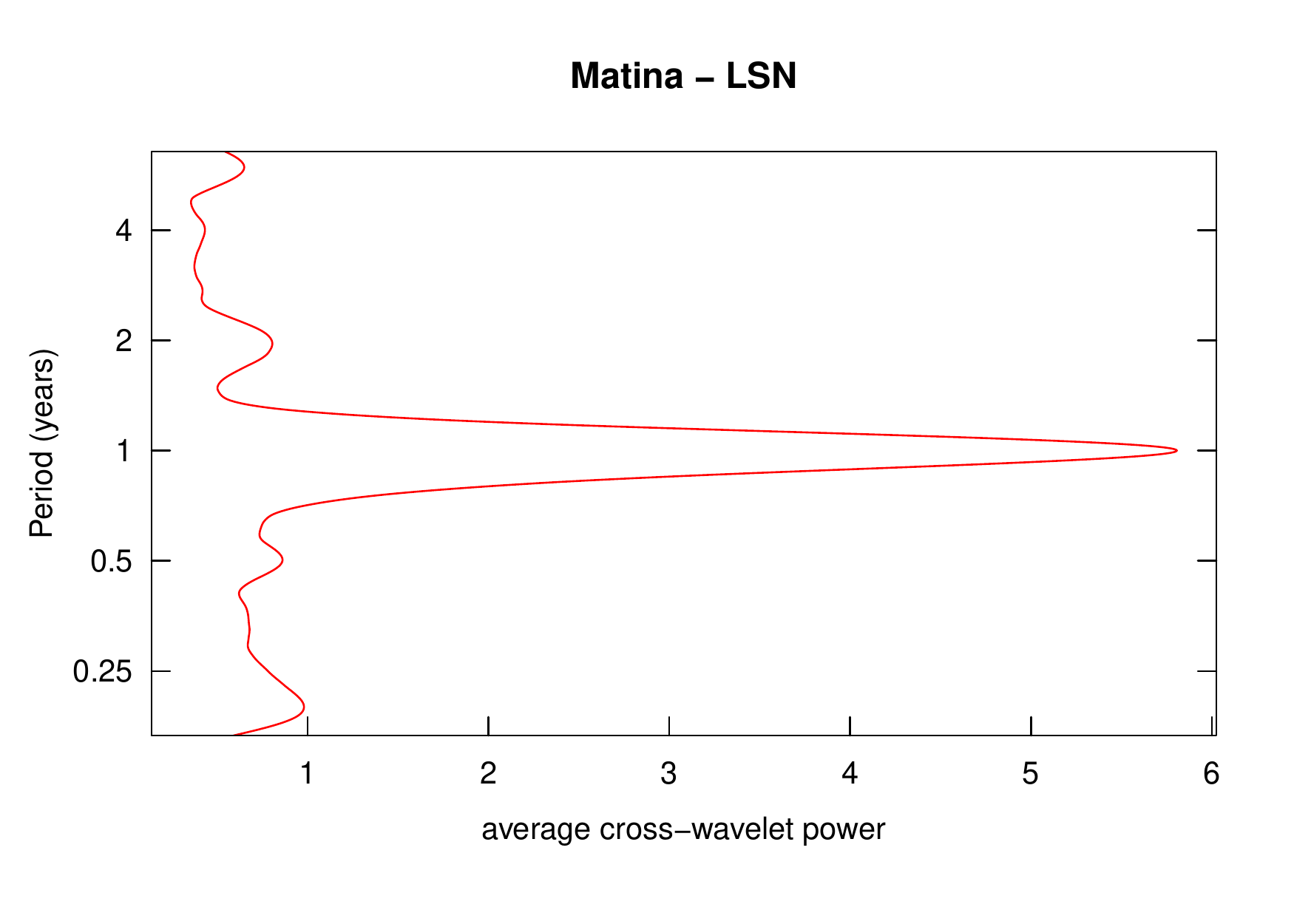}}\vspace{-0.15cm}\\
\subfloat[]{\includegraphics[scale=0.23]{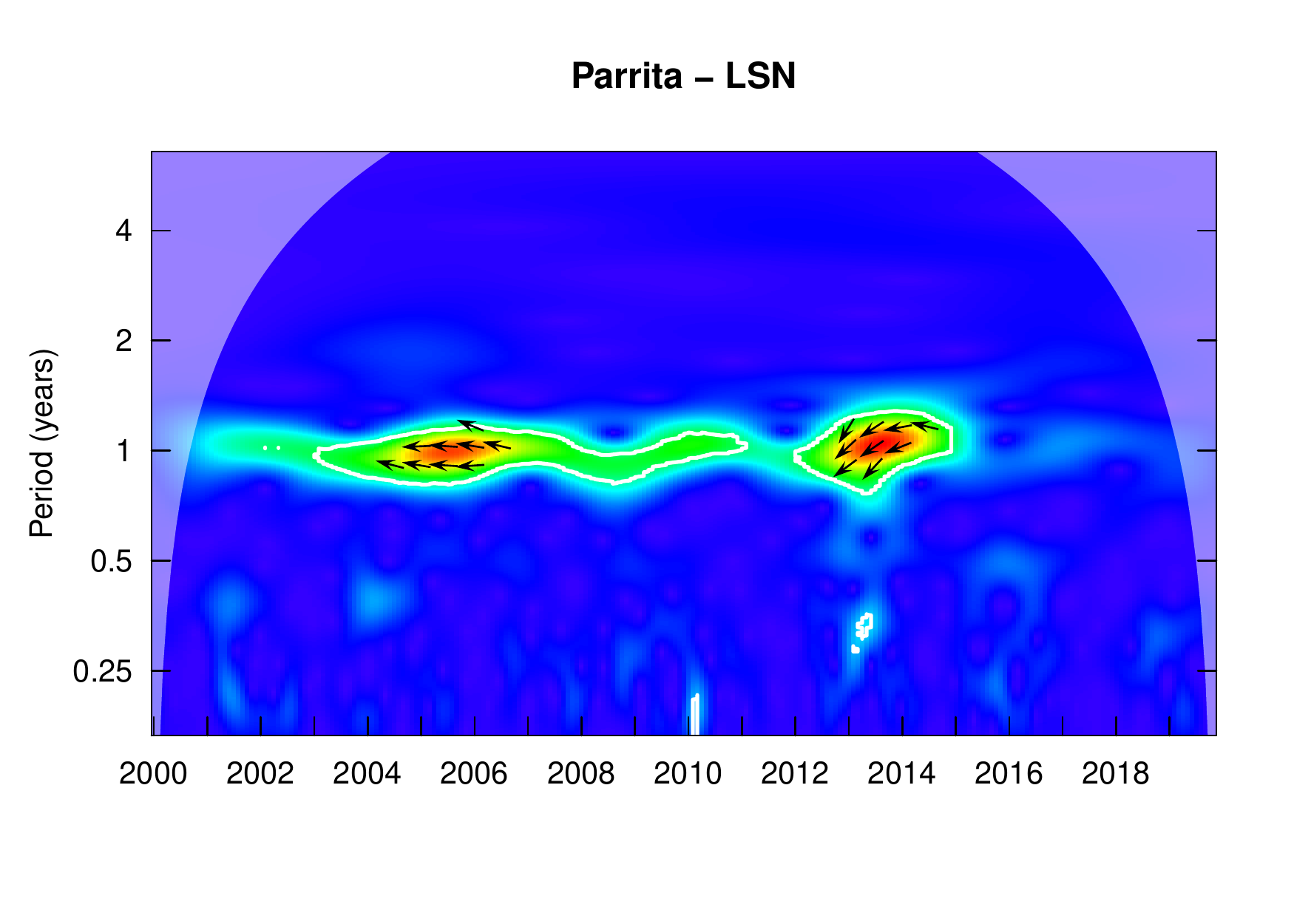}}\vspace{-0.15cm}%
\subfloat[]{\includegraphics[scale=0.23]{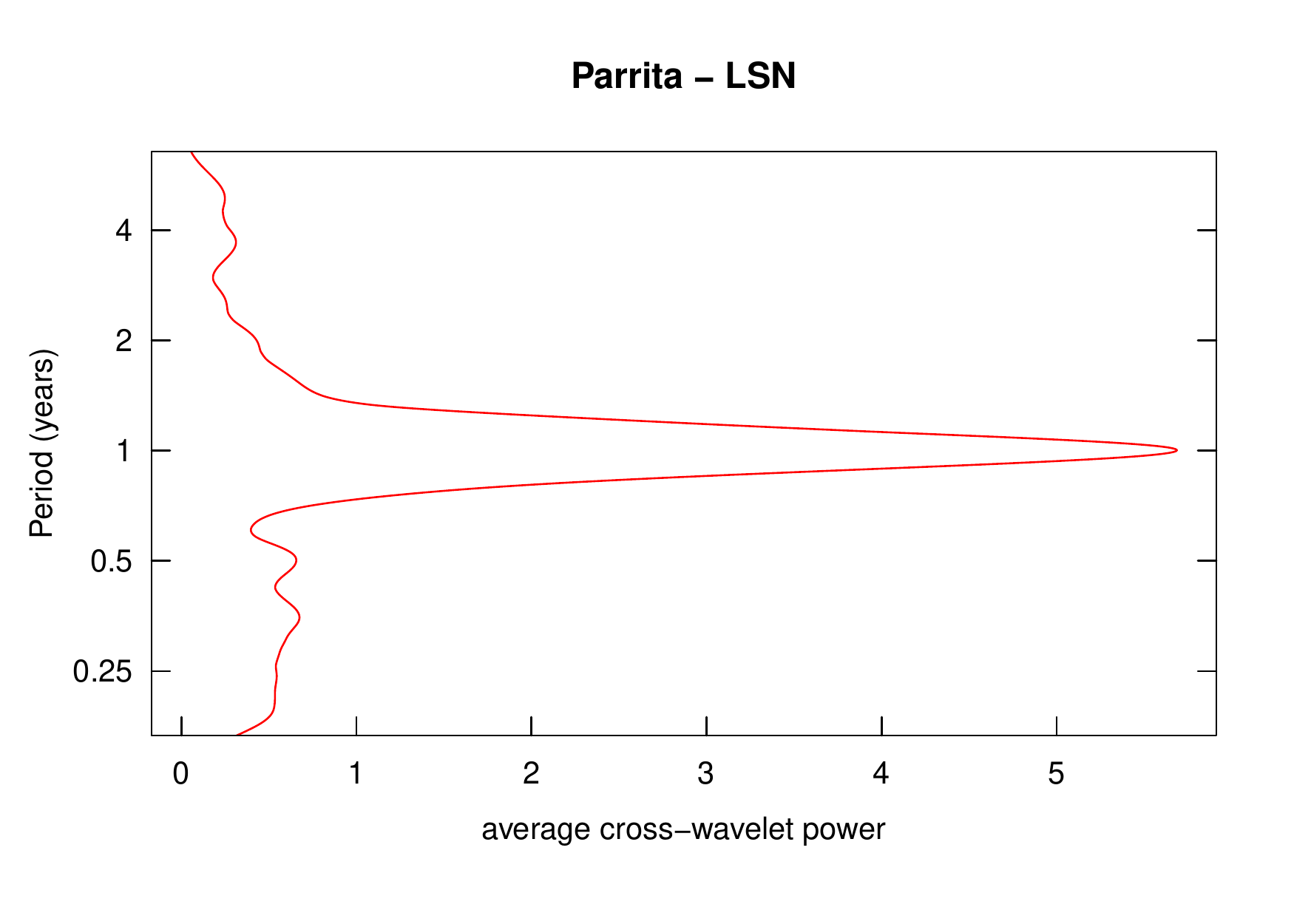}}\vspace{-0.15cm}%
\subfloat[]{\includegraphics[scale=0.23]{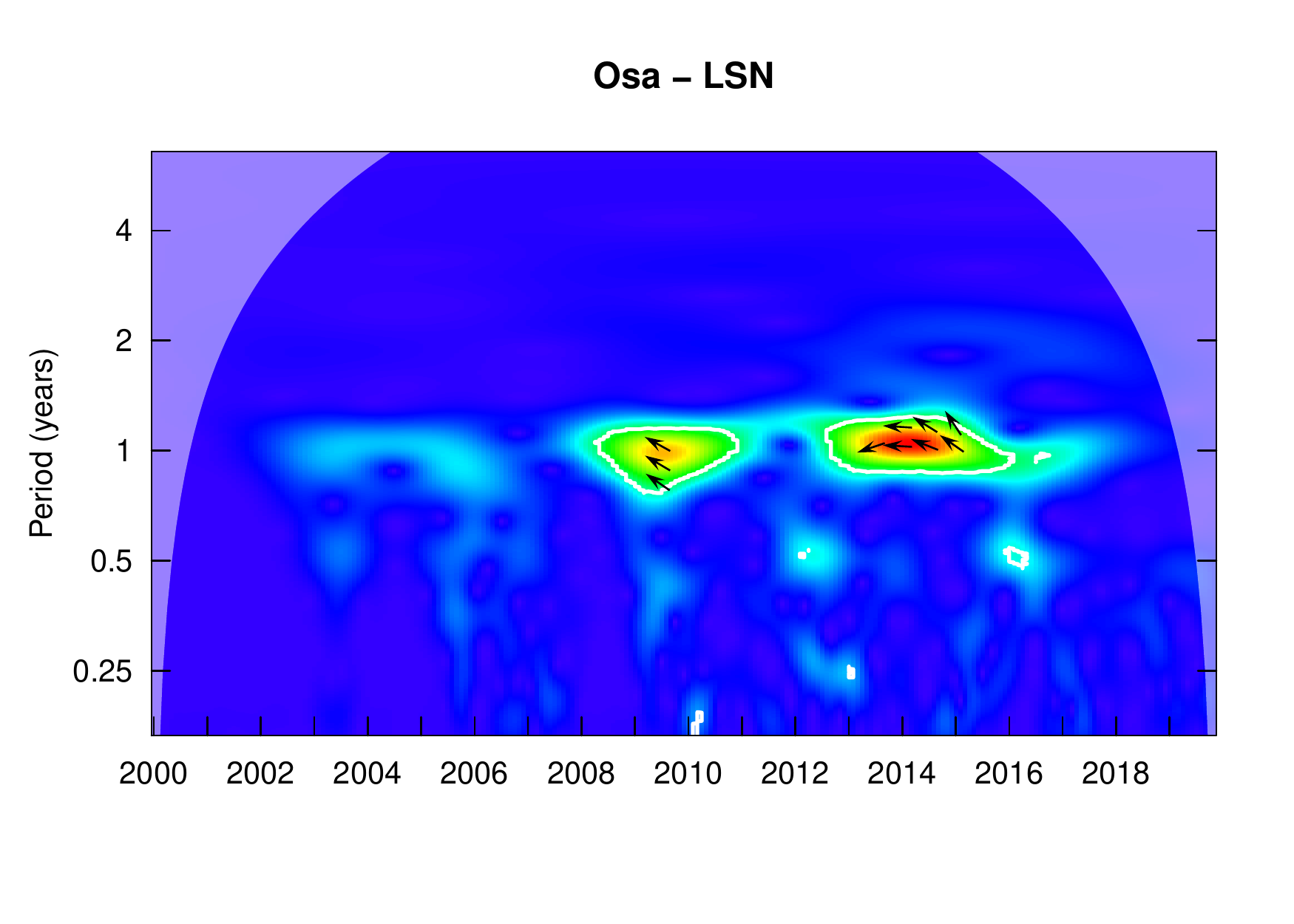}}\vspace{-0.15cm}%
\subfloat[]{\includegraphics[scale=0.23]{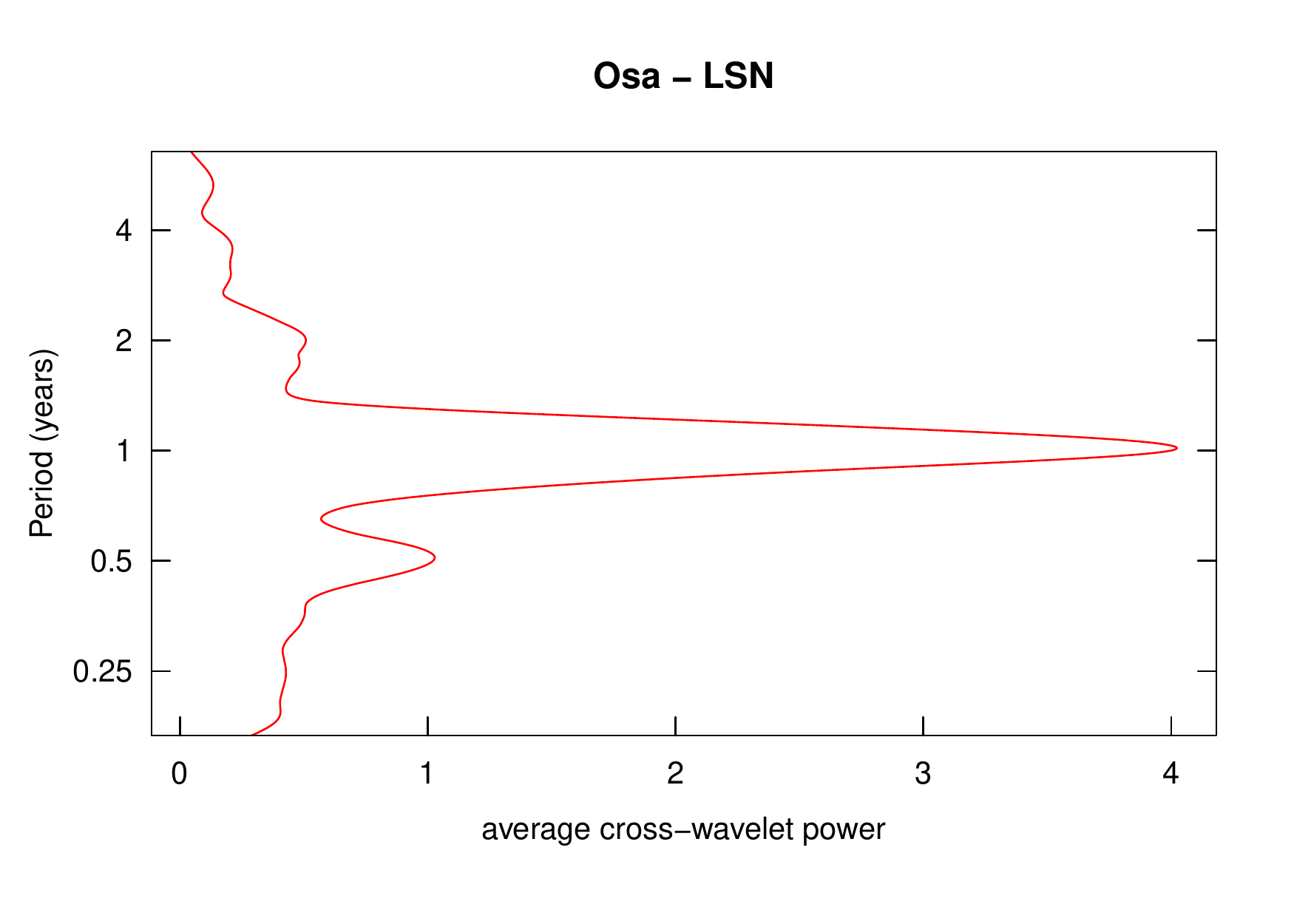}}\vspace{-0.15cm}
\caption*{}
\end{figure}

\section*{Wavelet coherence and average cross-wavelet power between dengue incidence and TNA}

\begin{figure}[H]
\captionsetup[subfigure]{labelformat=empty}
\caption*{\textbf{Figure S6:} Wavelet coherence (color map) between dengue incidence from 2000 to 2019, and TNA in 32 municipalities of Costa Rica (periodicity on y-axis, time on x-axis). Colors code for increasing power intensity, from blue to red; $95\%$ confidence levels are encircled by white lines, and shaded areas indicate the presence of significant edge effects. On the right side of each wavelet coherence is the average cross-wavelet power (Red line). The arrows indicate whether the two series are in-phase or out-phase.}
\subfloat[]{\includegraphics[scale=0.23]{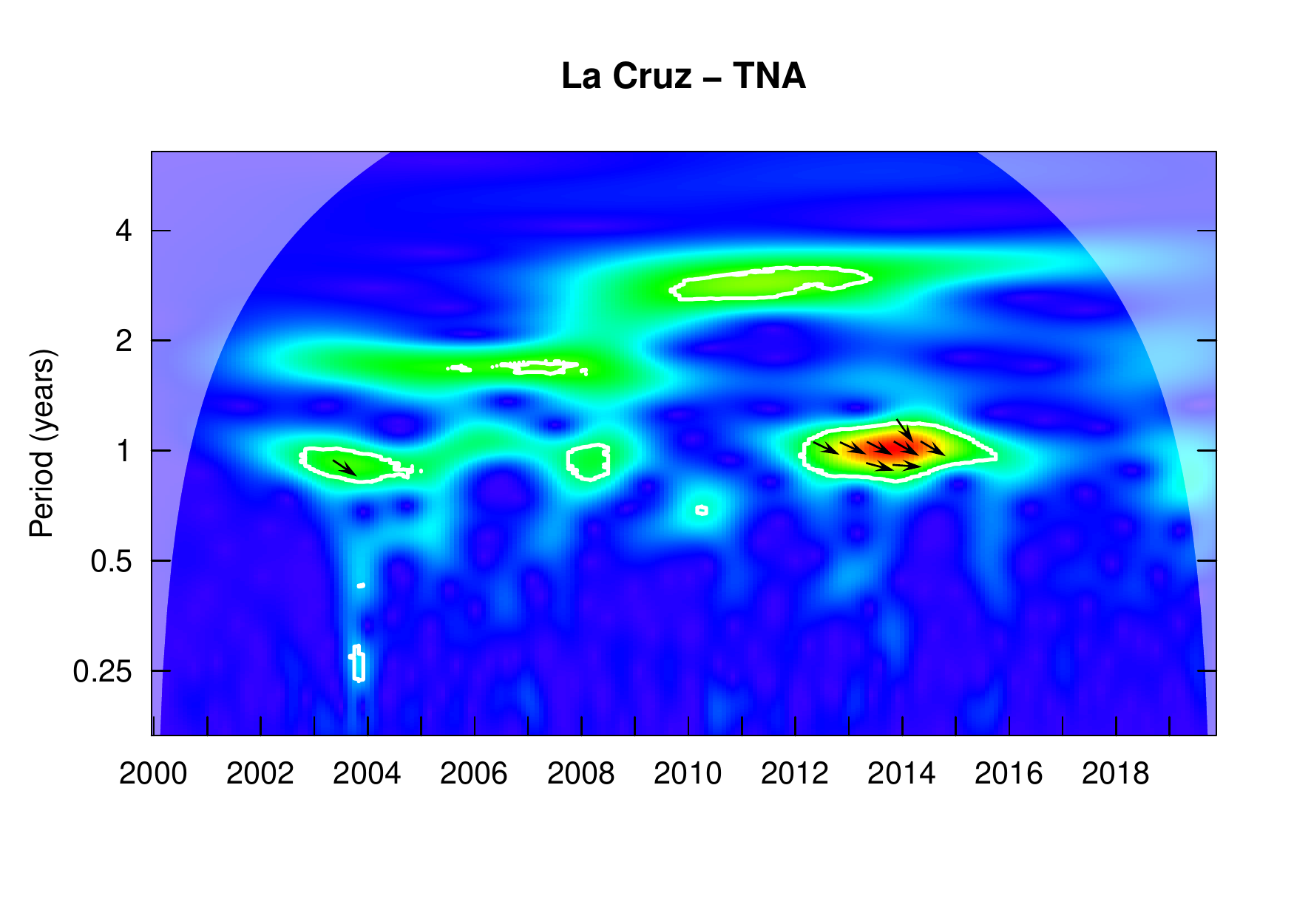}}\vspace{-0.15cm}%
\subfloat[]{\includegraphics[scale=0.23]{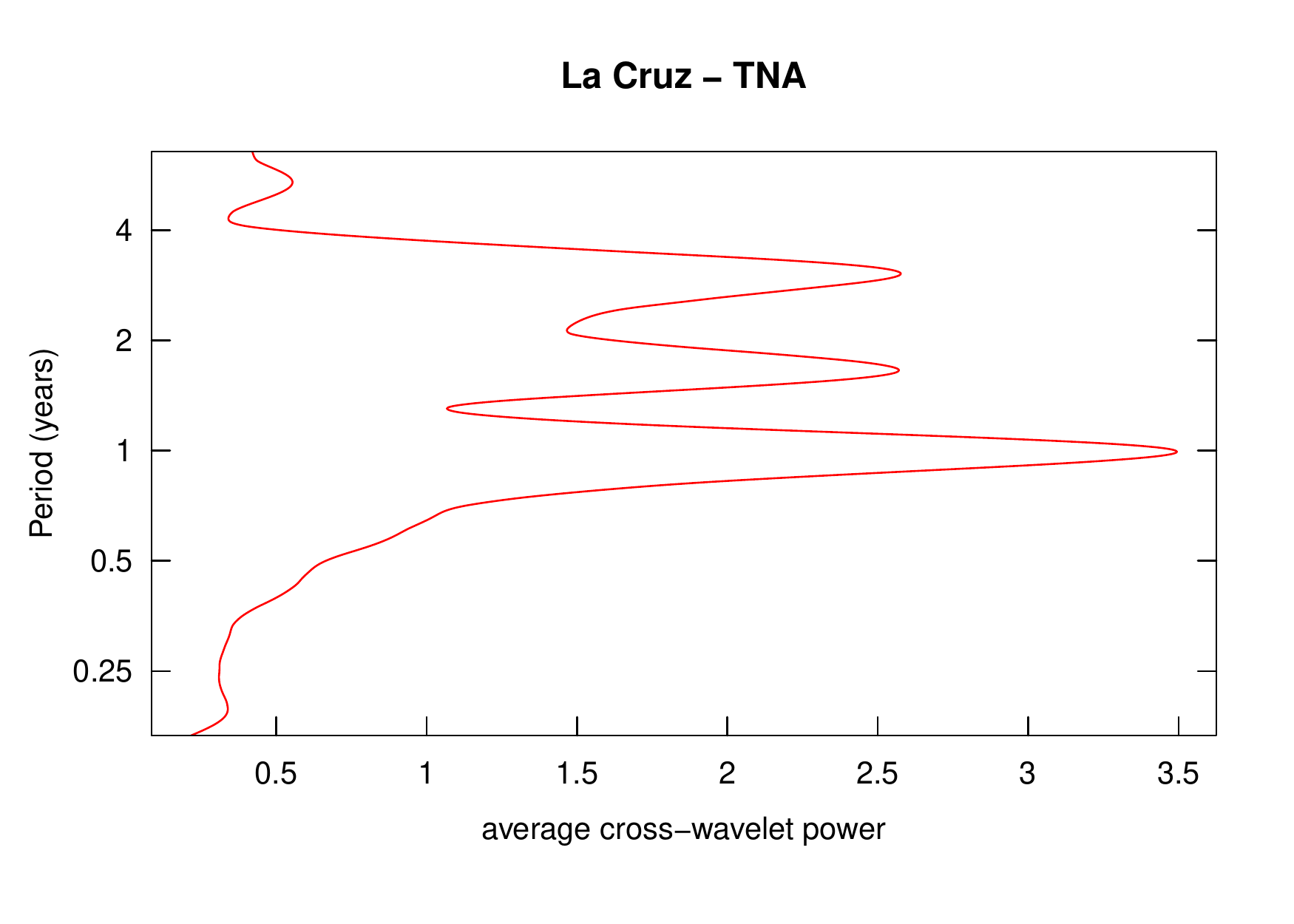}}\vspace{-0.15cm}%
\subfloat[]{\includegraphics[scale=0.23]{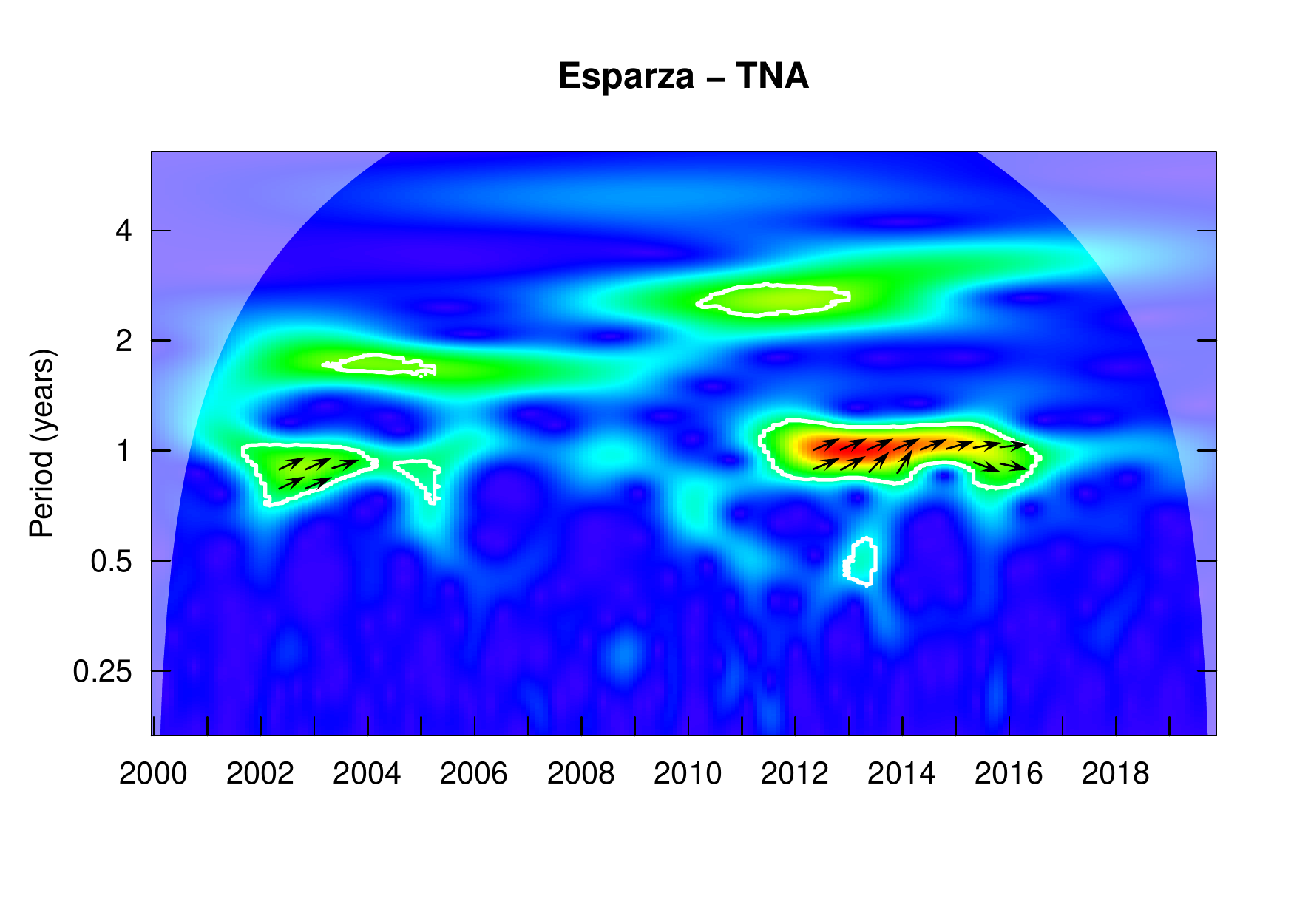}}\vspace{-0.15cm}%
\subfloat[]{\includegraphics[scale=0.23]{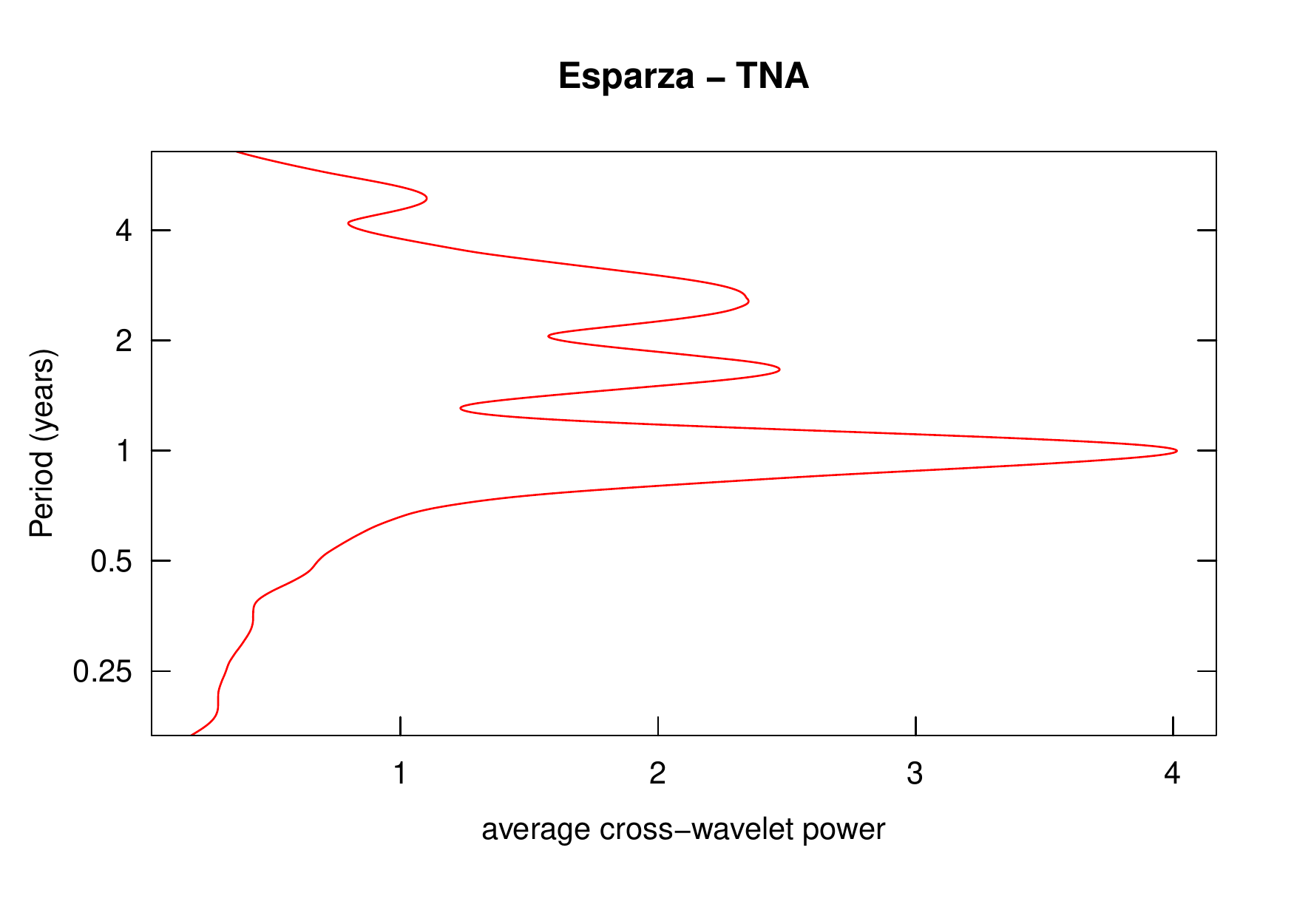}}\vspace{-0.15cm}\\
\subfloat[]{\includegraphics[scale=0.23]{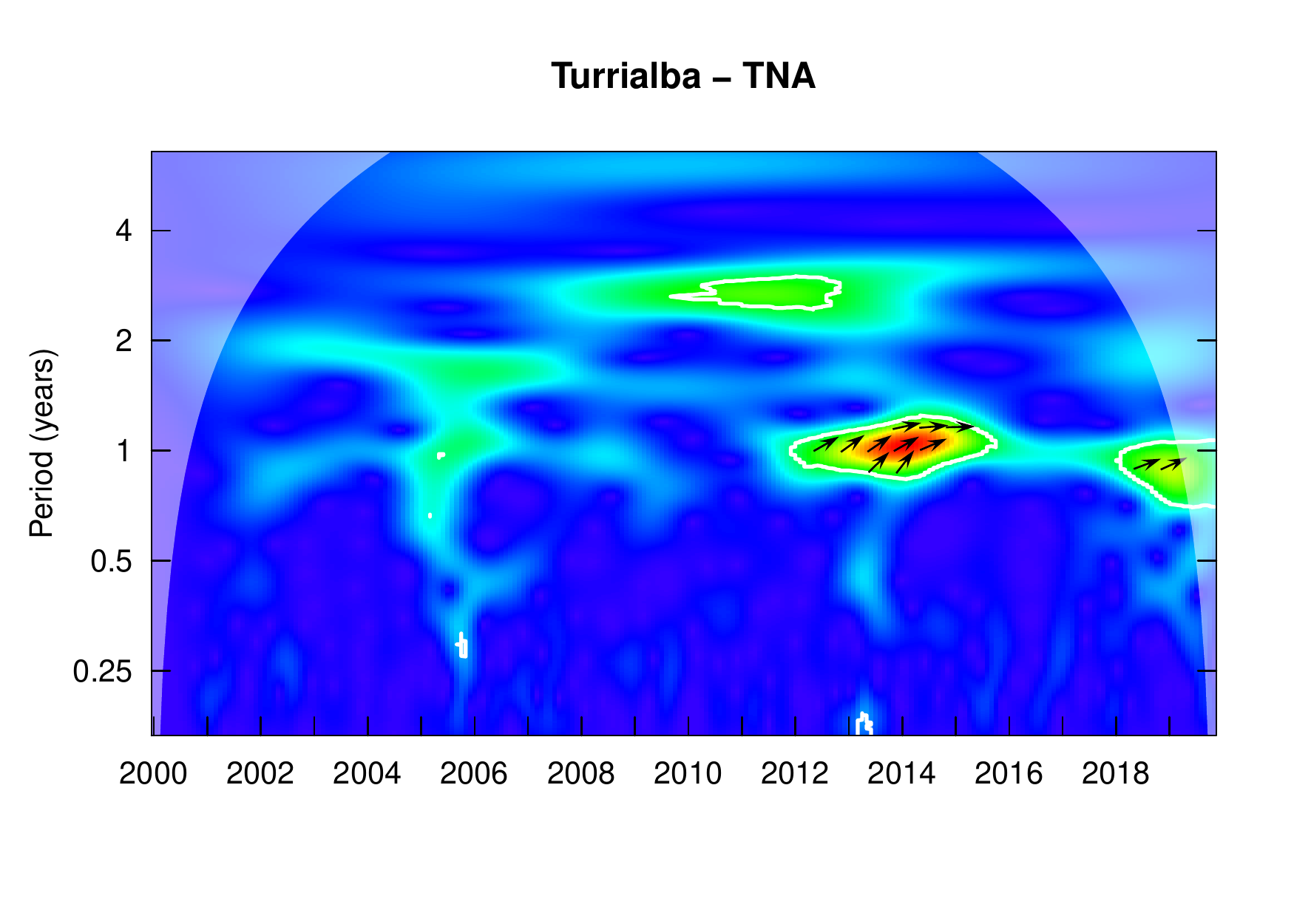}}\vspace{-0.15cm}%
\subfloat[]{\includegraphics[scale=0.23]{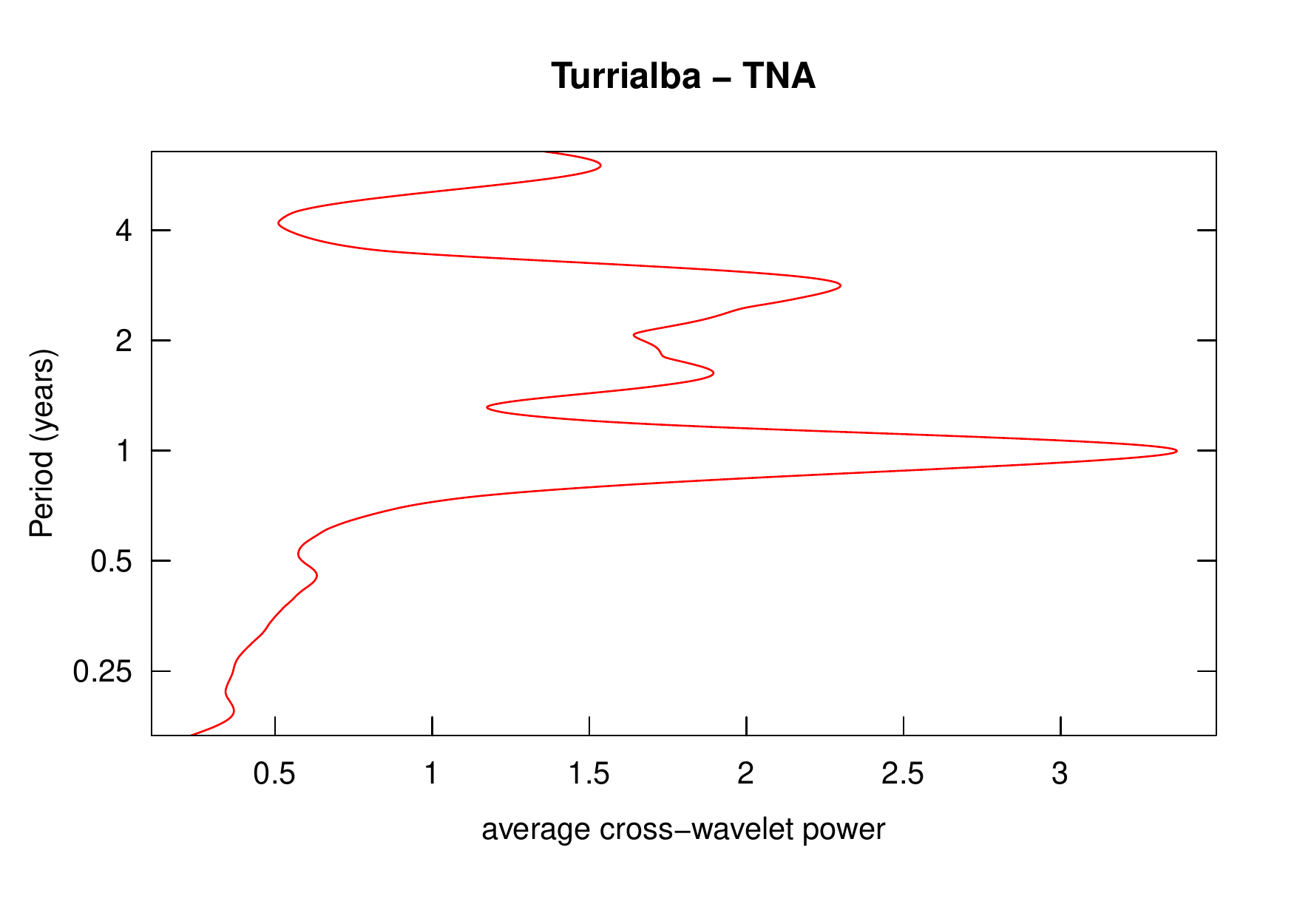}}\vspace{-0.15cm}%
\subfloat[]{\includegraphics[scale=0.23]{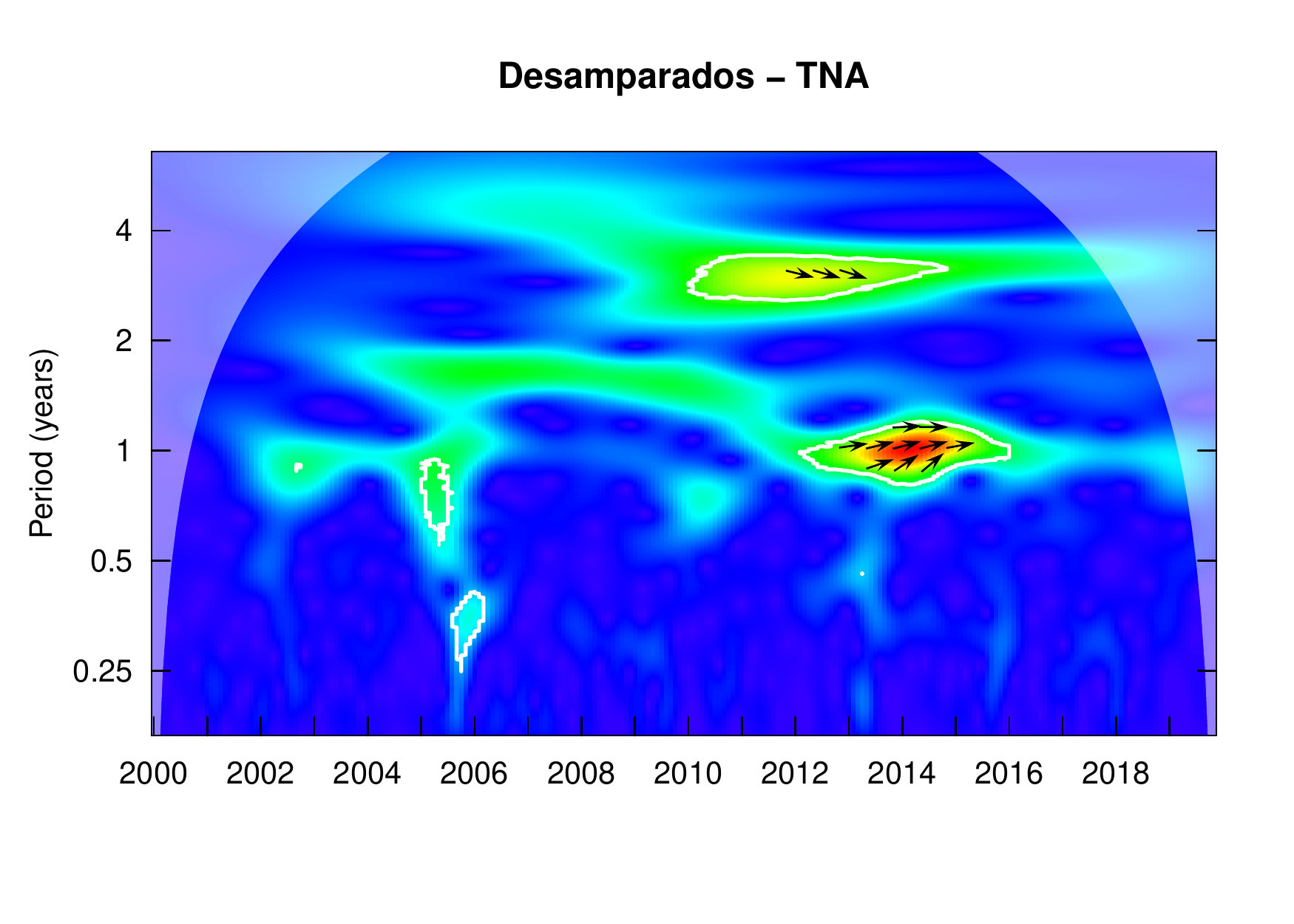}}\vspace{-0.15cm}%
\subfloat[]{\includegraphics[scale=0.23]{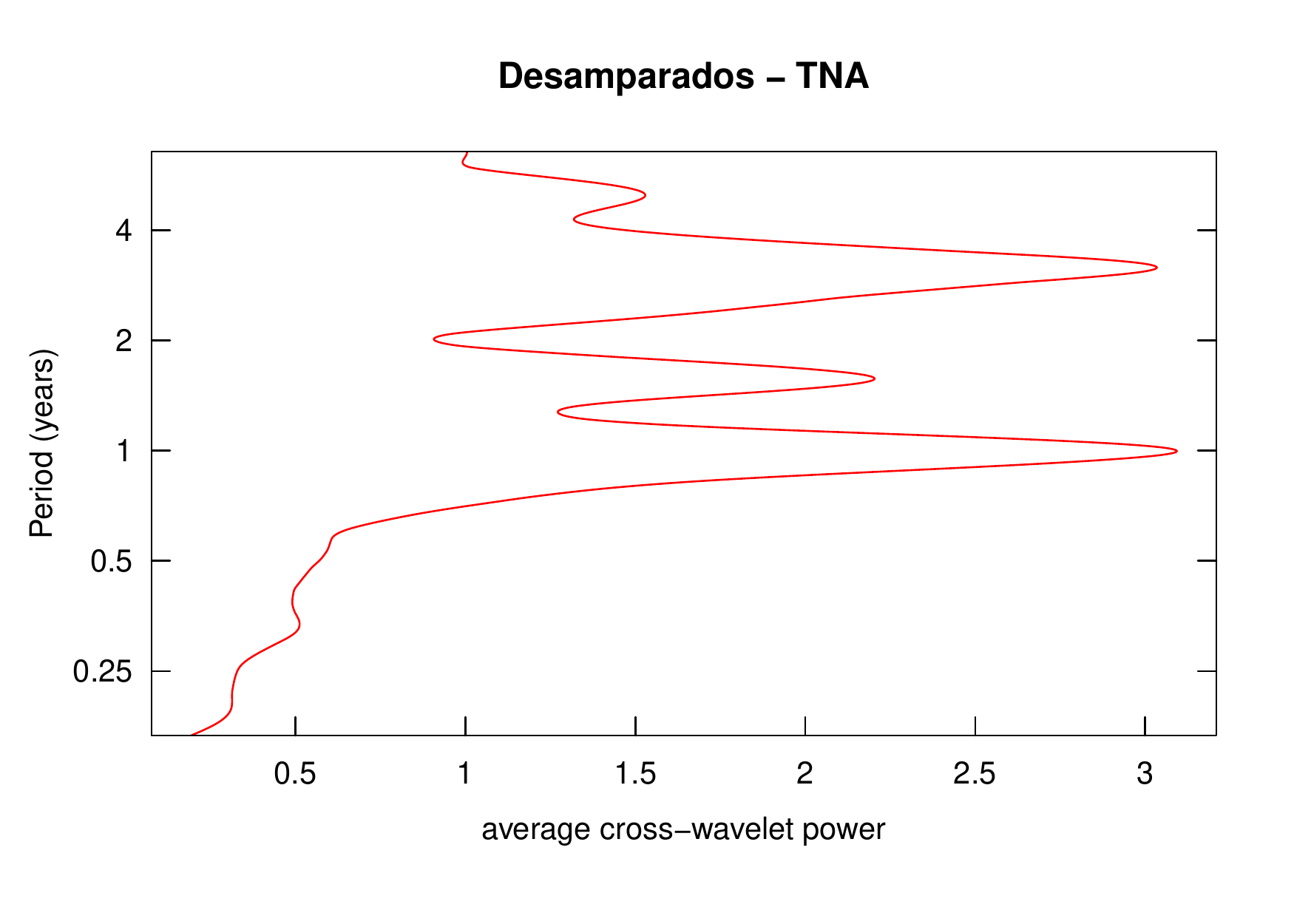}}\vspace{-0.15cm}\\
\subfloat[]{\includegraphics[scale=0.23]{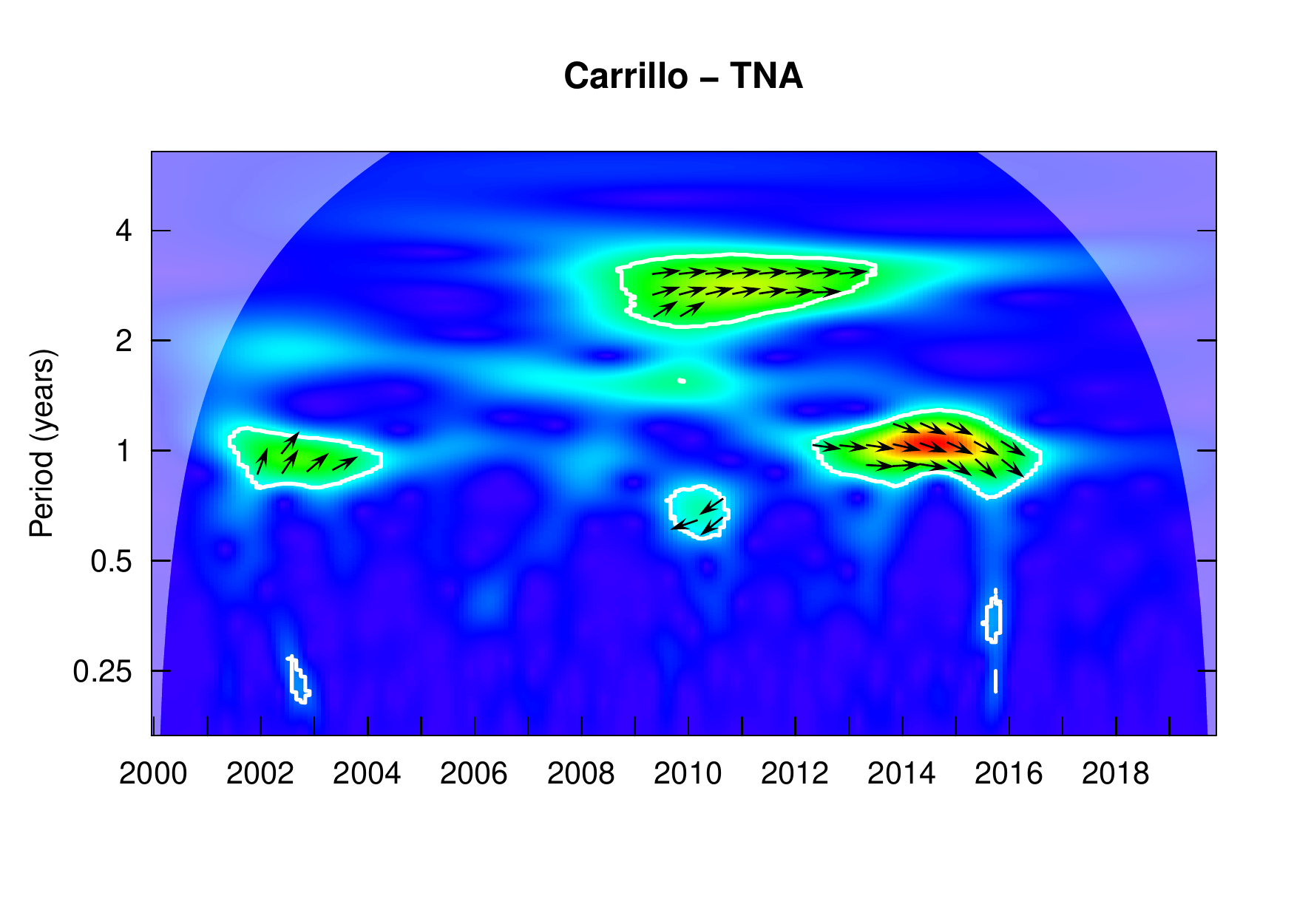}}\vspace{-0.15cm}%
\subfloat[]{\includegraphics[scale=0.23]{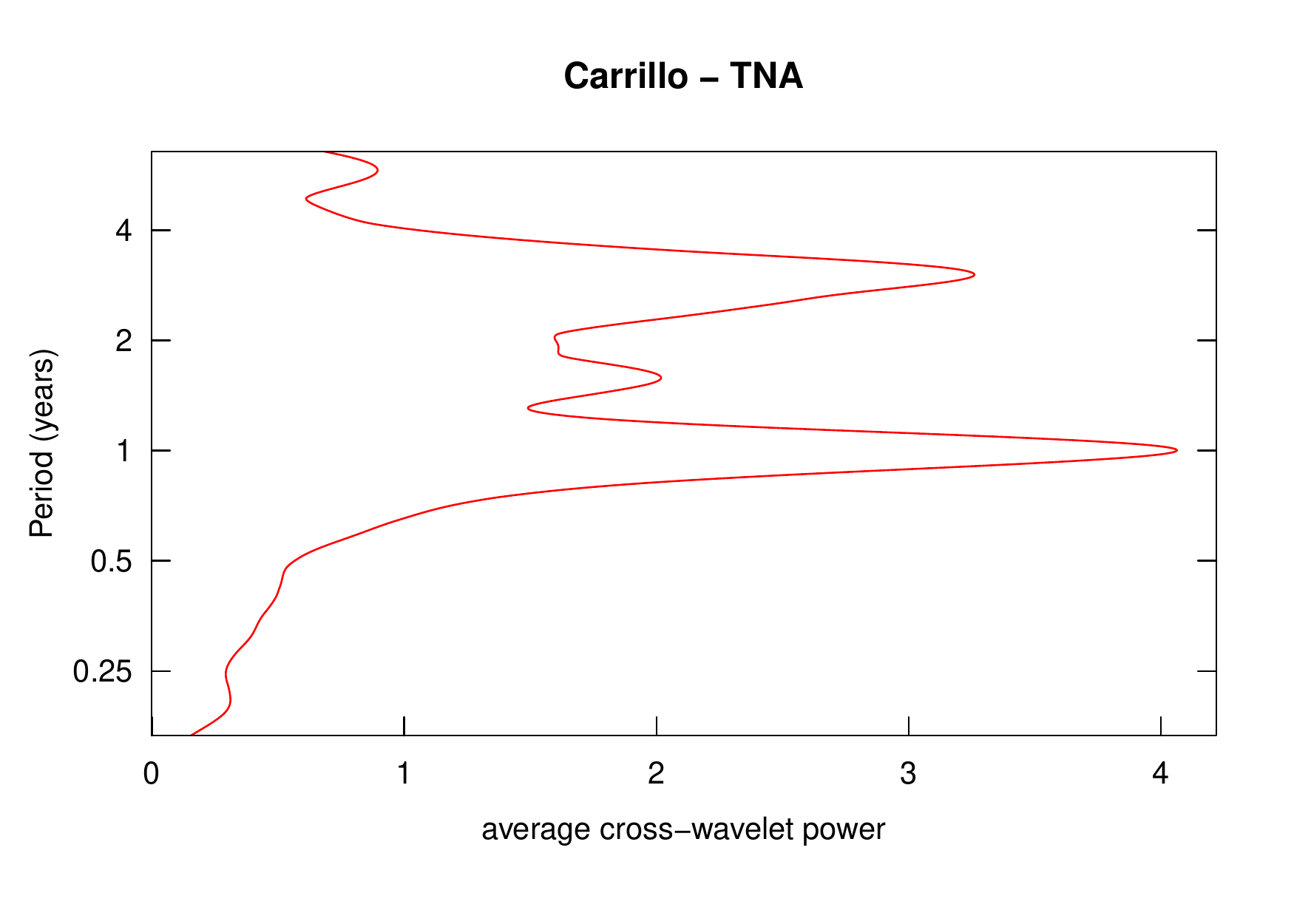}}\vspace{-0.15cm}%
\subfloat[]{\includegraphics[scale=0.23]{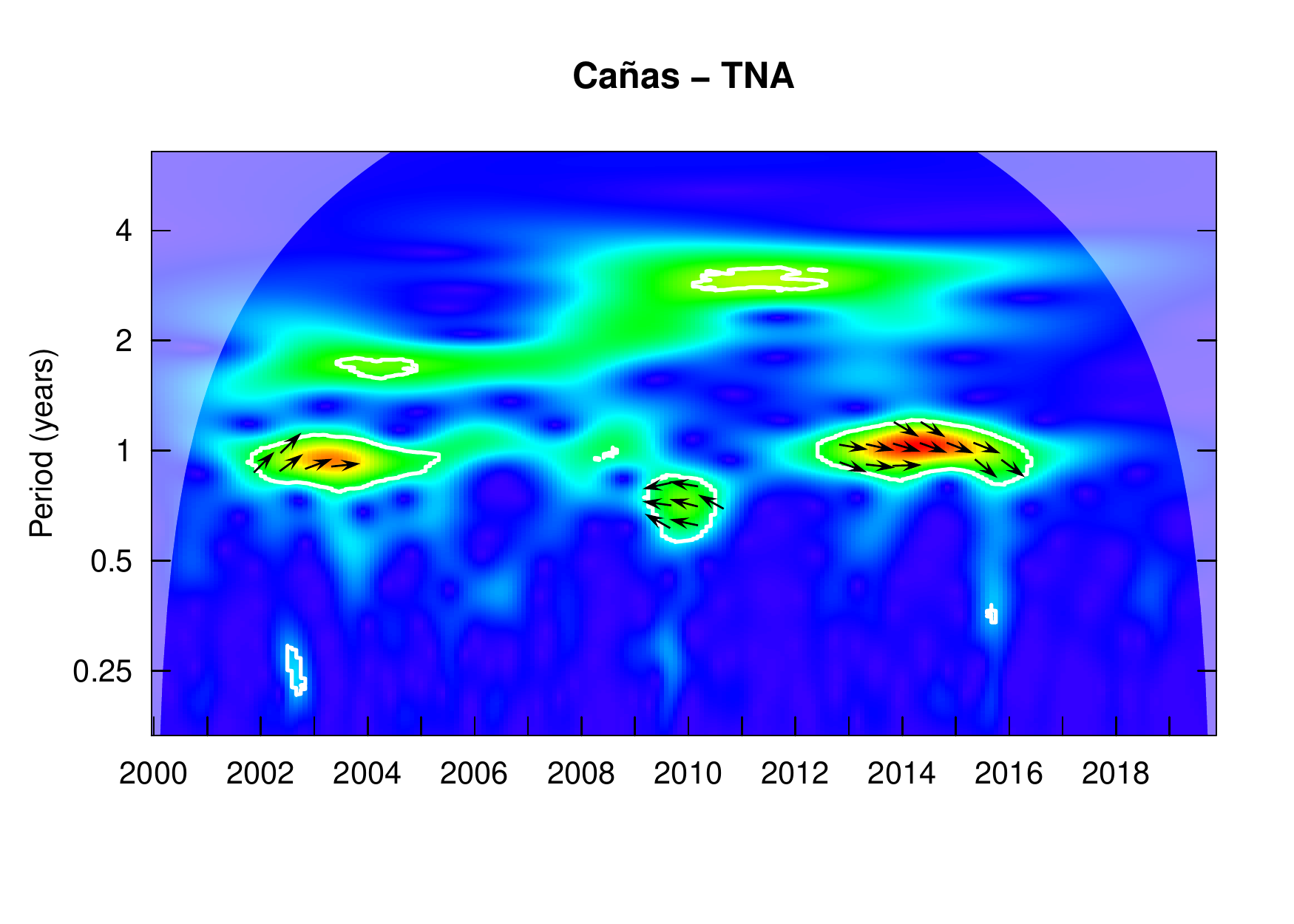}}\vspace{-0.15cm}%
\subfloat[]{\includegraphics[scale=0.23]{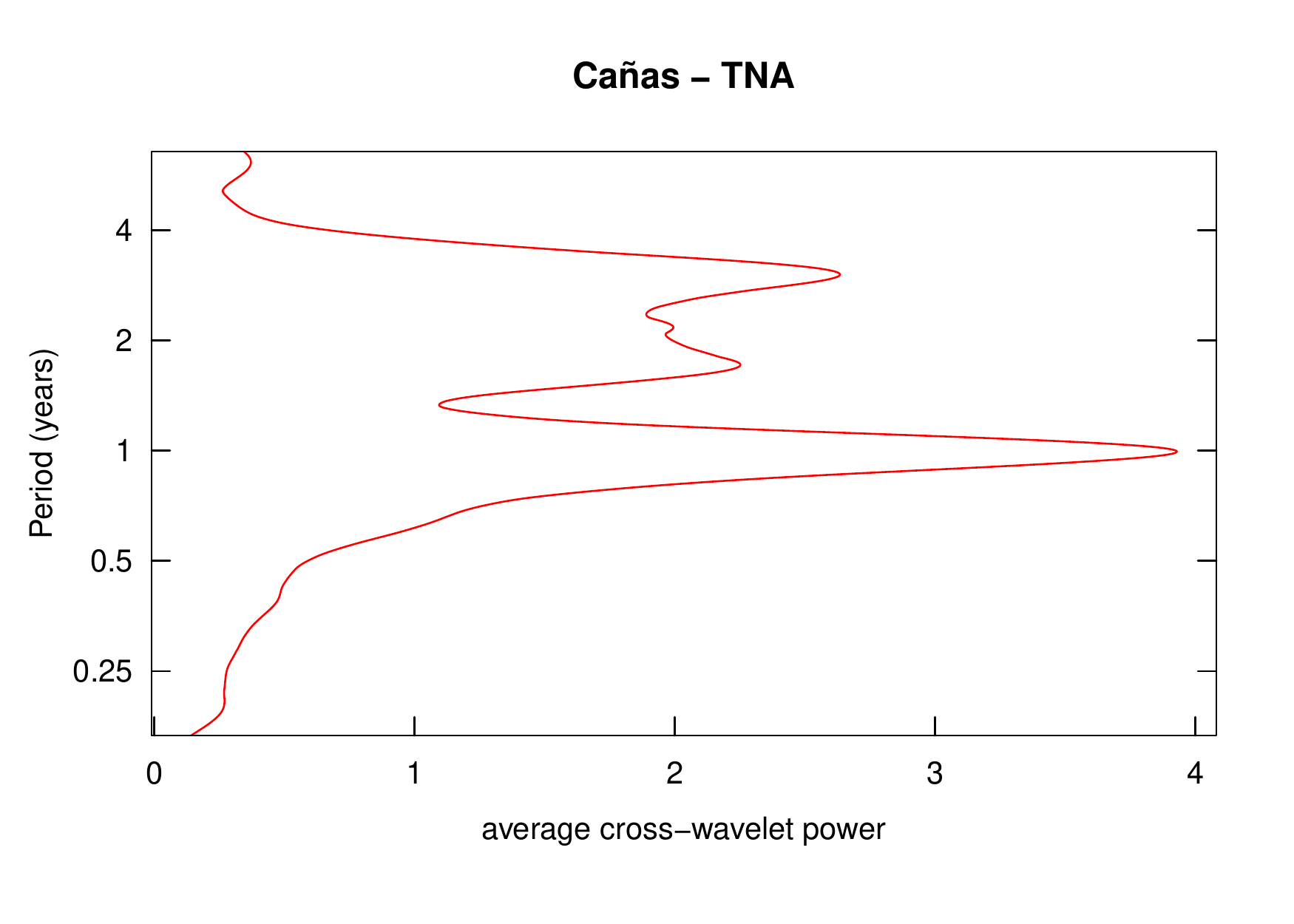}}\vspace{-0.15cm}\\
\caption*{}
\end{figure}

\begin{figure}[H]
\captionsetup[subfigure]{labelformat=empty}
\subfloat[]{\includegraphics[scale=0.23]{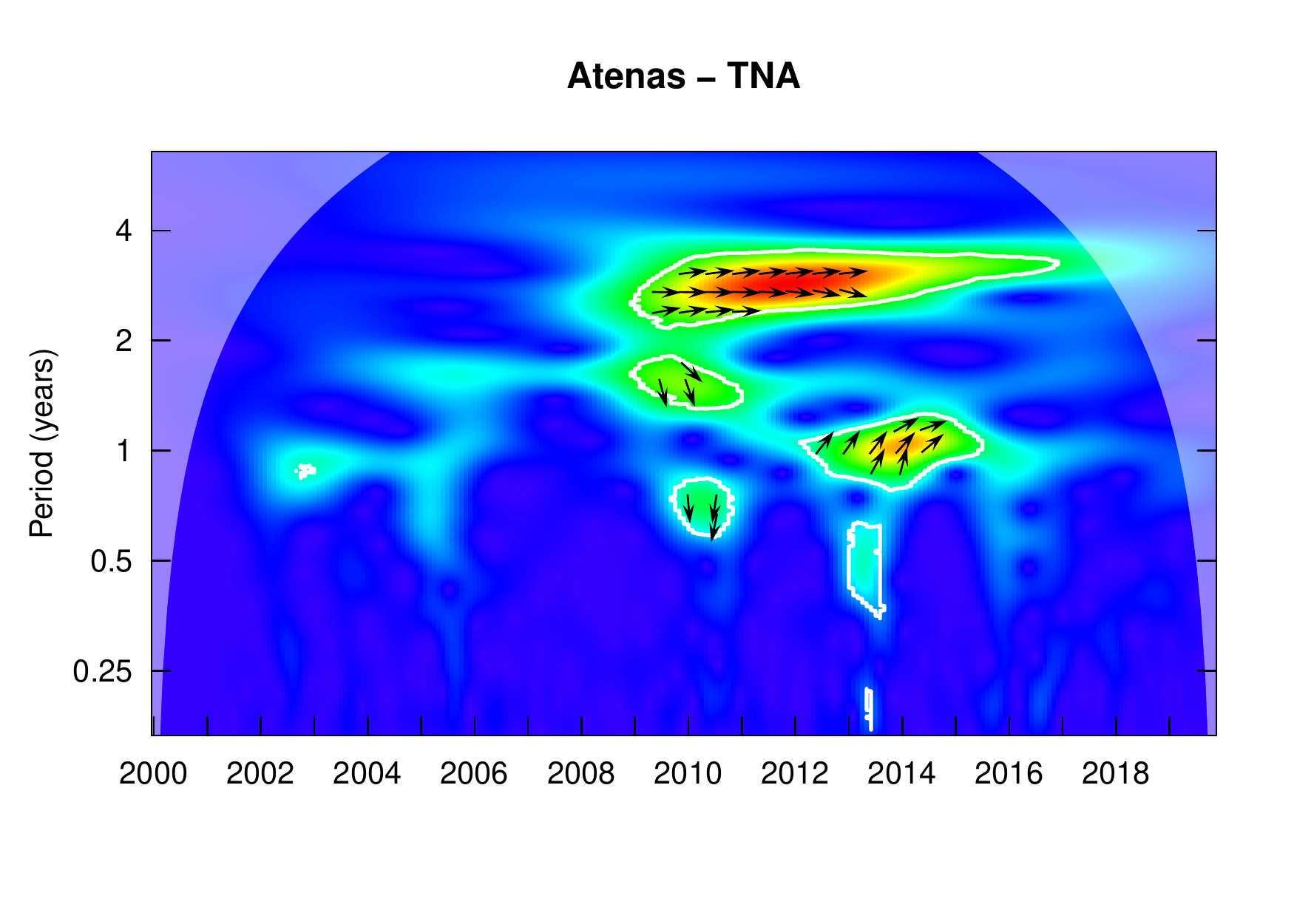}}\vspace{-0.15cm}%
\subfloat[]{\includegraphics[scale=0.23]{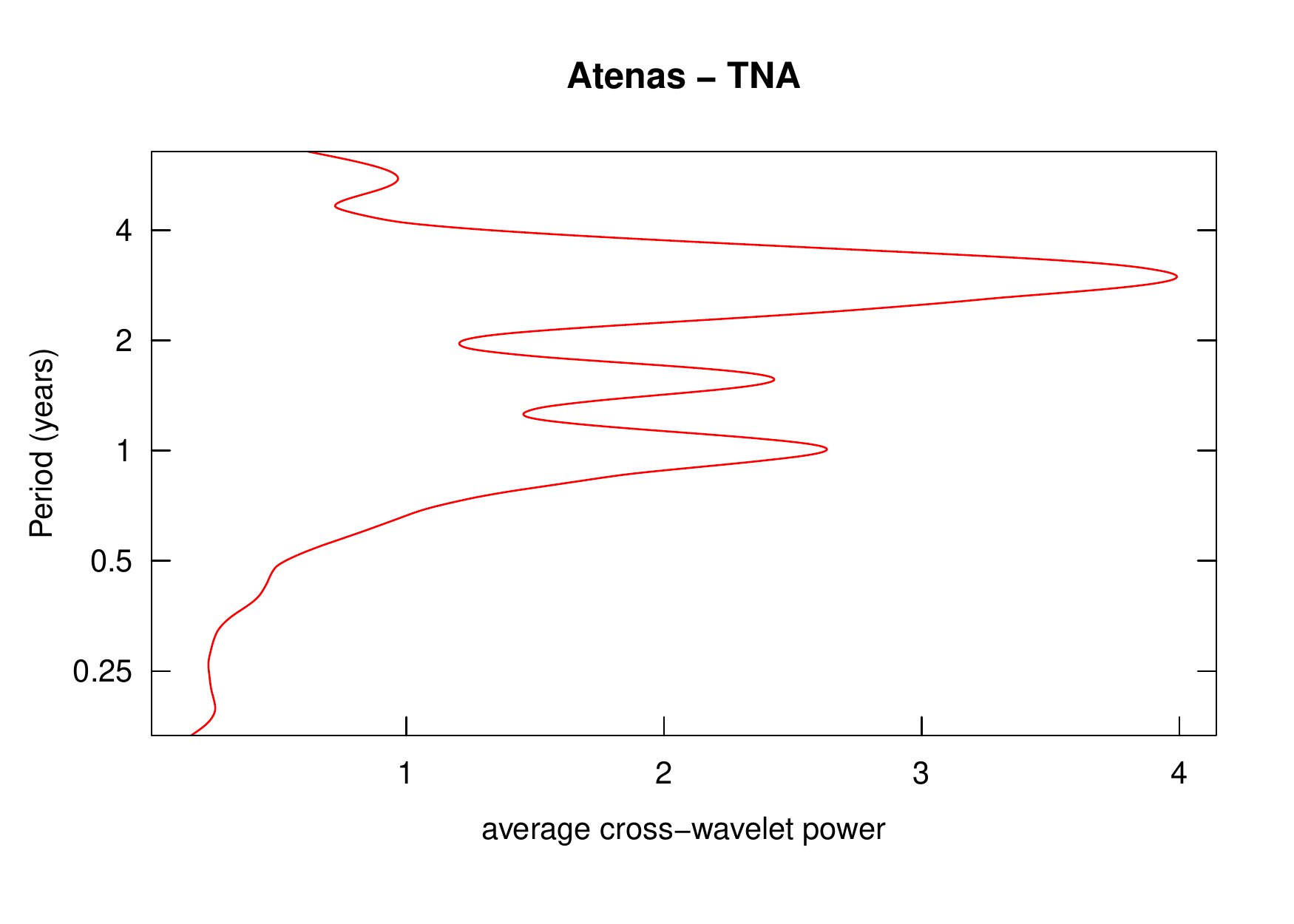}}\vspace{-0.15cm}%
\subfloat[]{\includegraphics[scale=0.23]{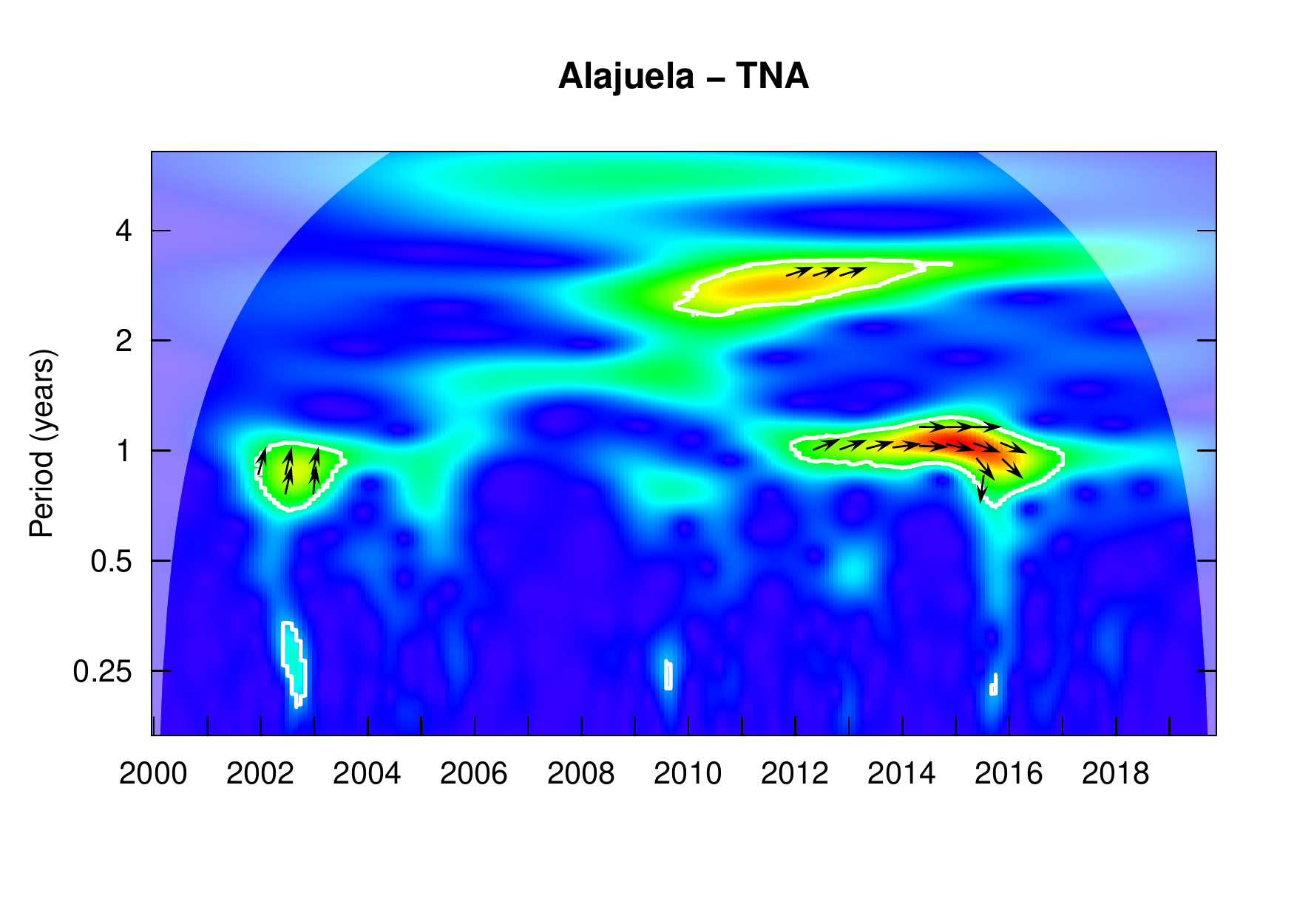}}\vspace{-0.15cm}%
\subfloat[]{\includegraphics[scale=0.23]{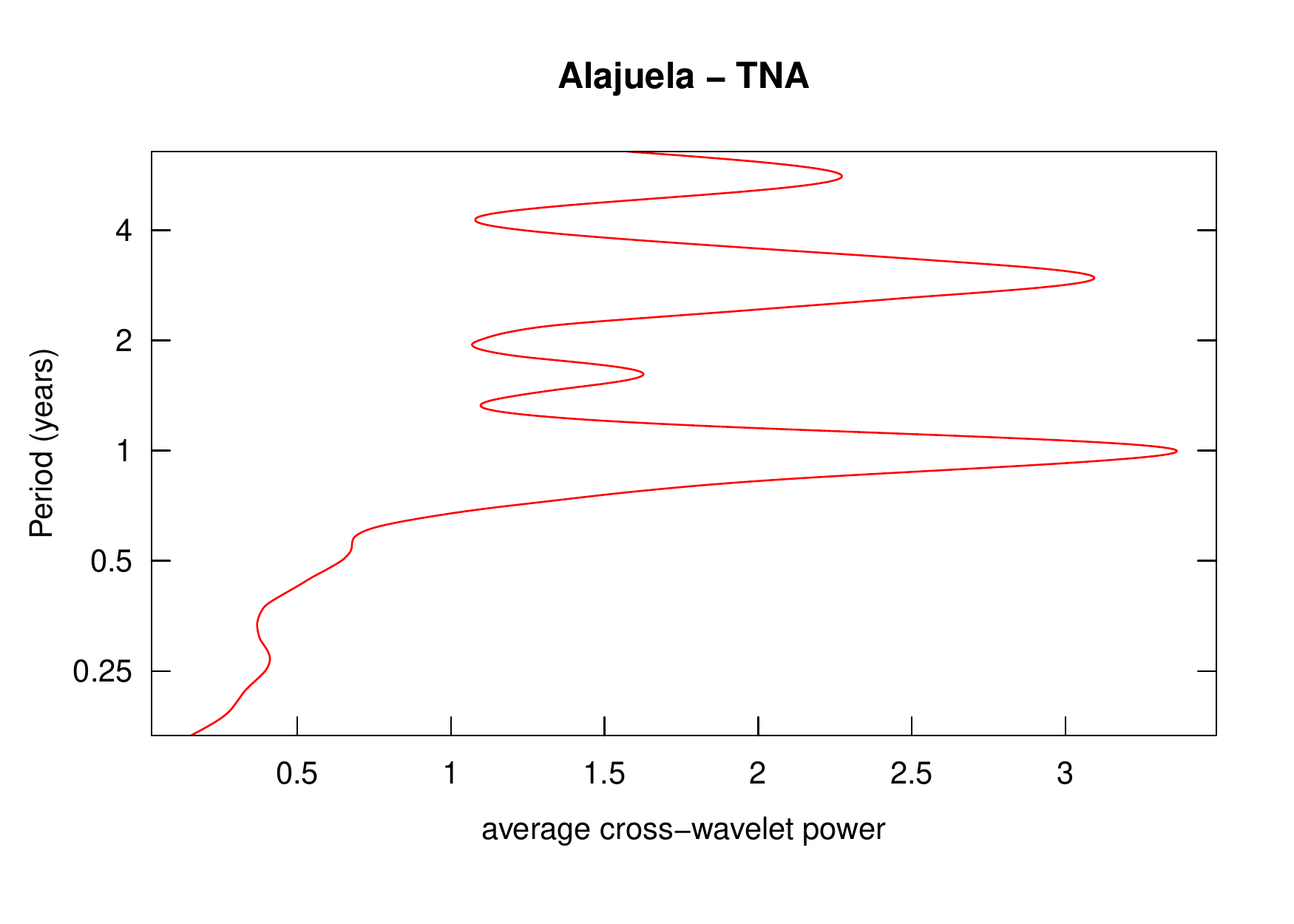}}\vspace{-0.15cm}\\
\subfloat[]{\includegraphics[scale=0.23]{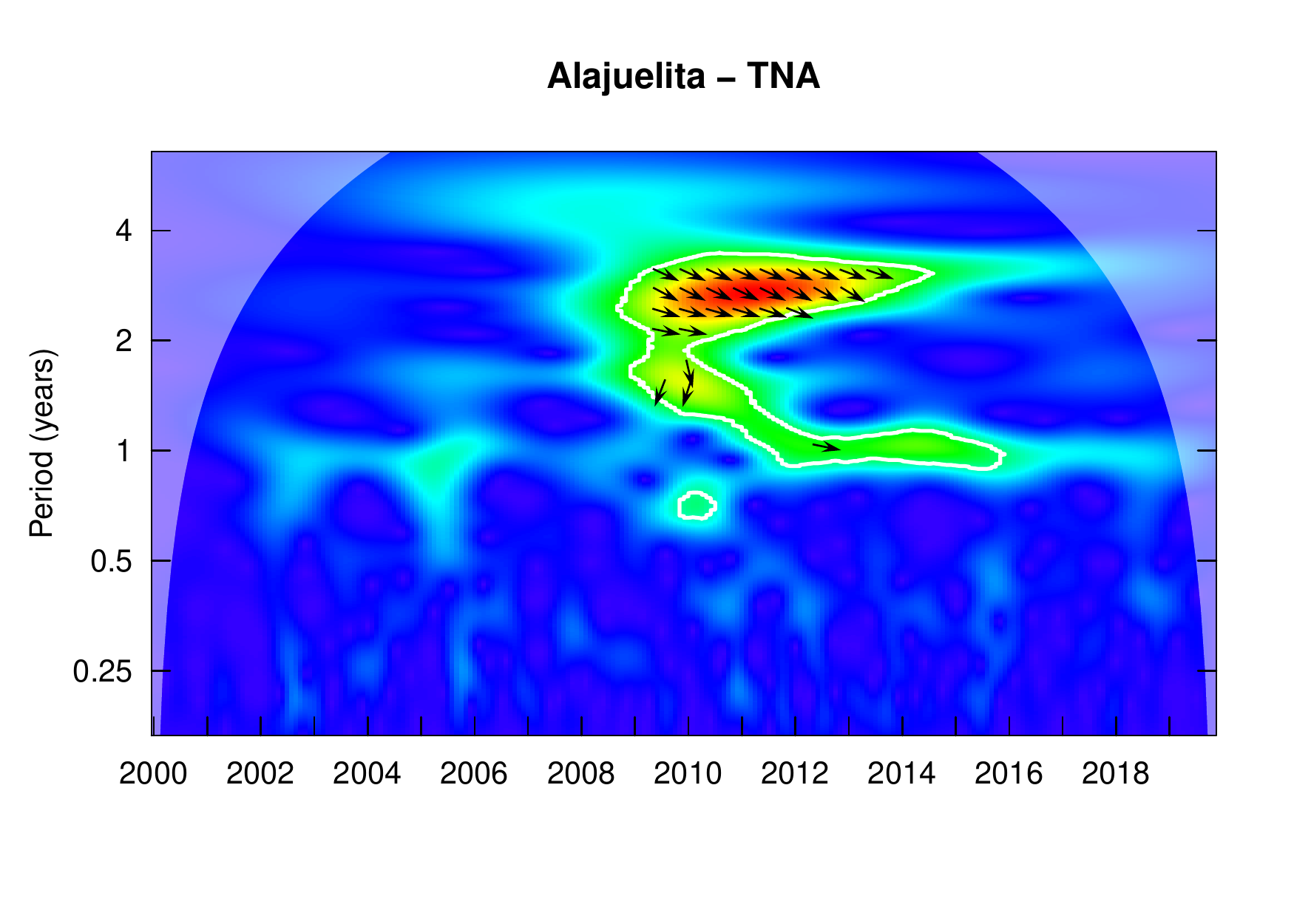}}\vspace{-0.15cm}%
\subfloat[]{\includegraphics[scale=0.23]{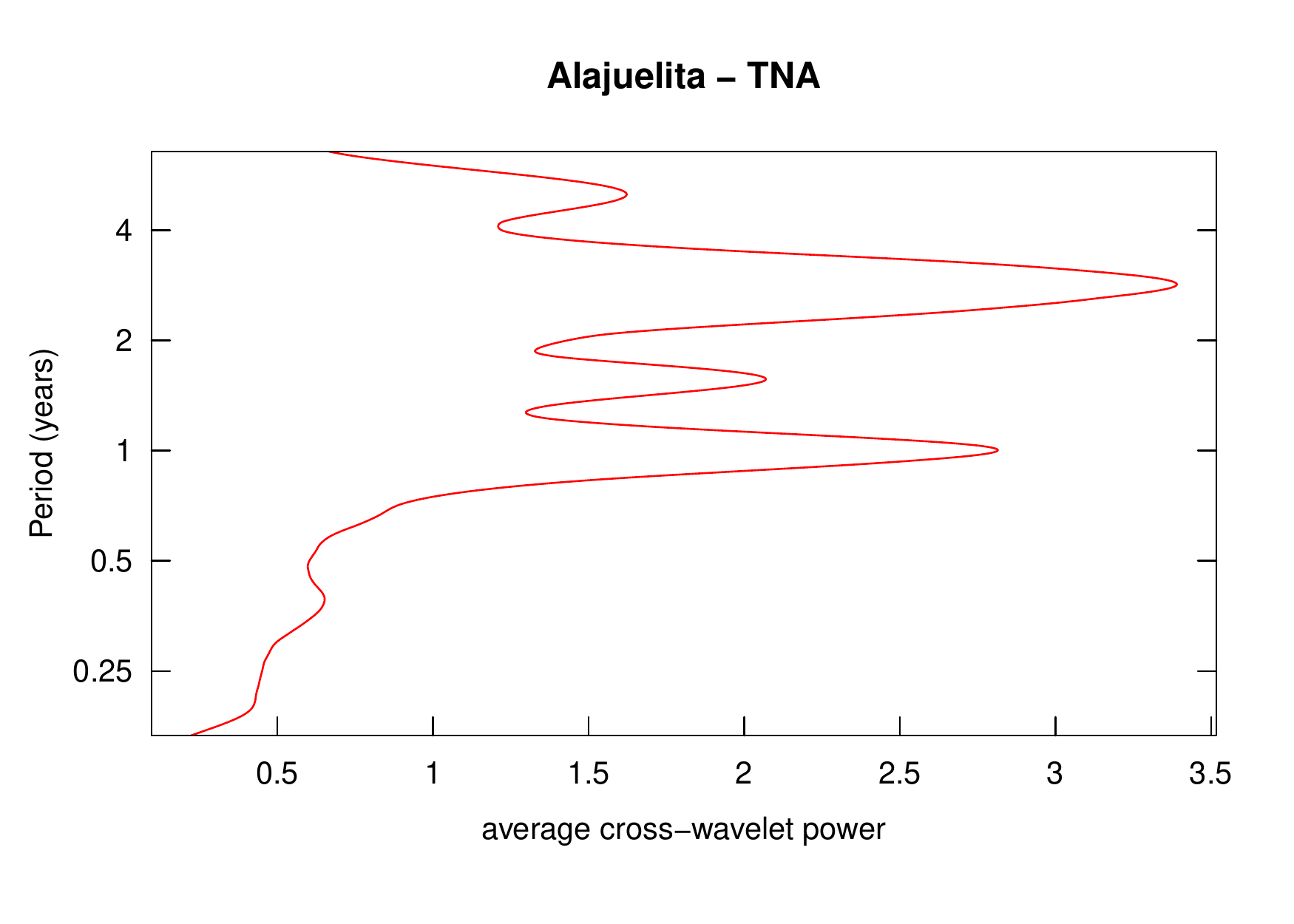}}\vspace{-0.15cm}%
\subfloat[]{\includegraphics[scale=0.23]{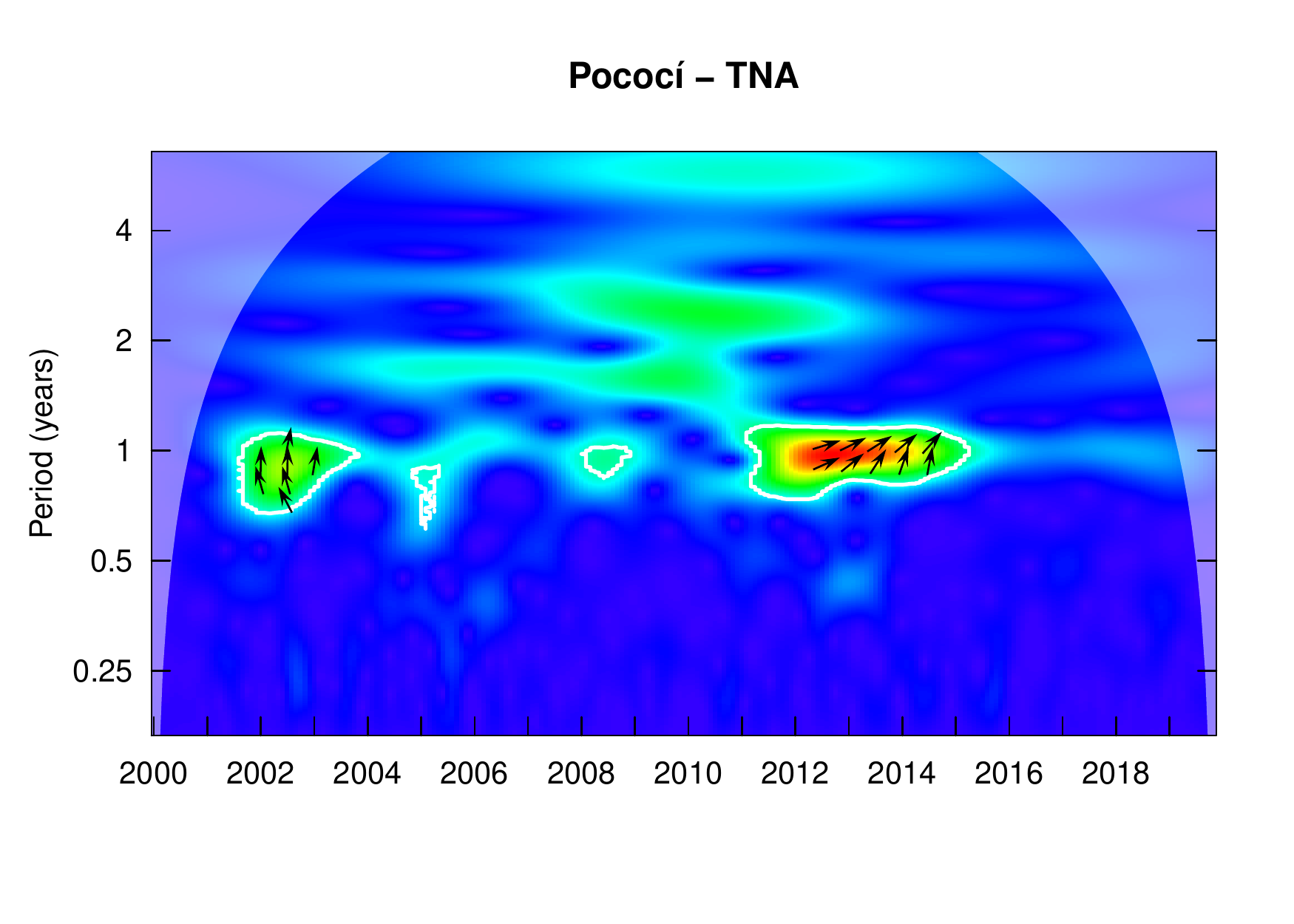}}\vspace{-0.15cm}%
\subfloat[]{\includegraphics[scale=0.23]{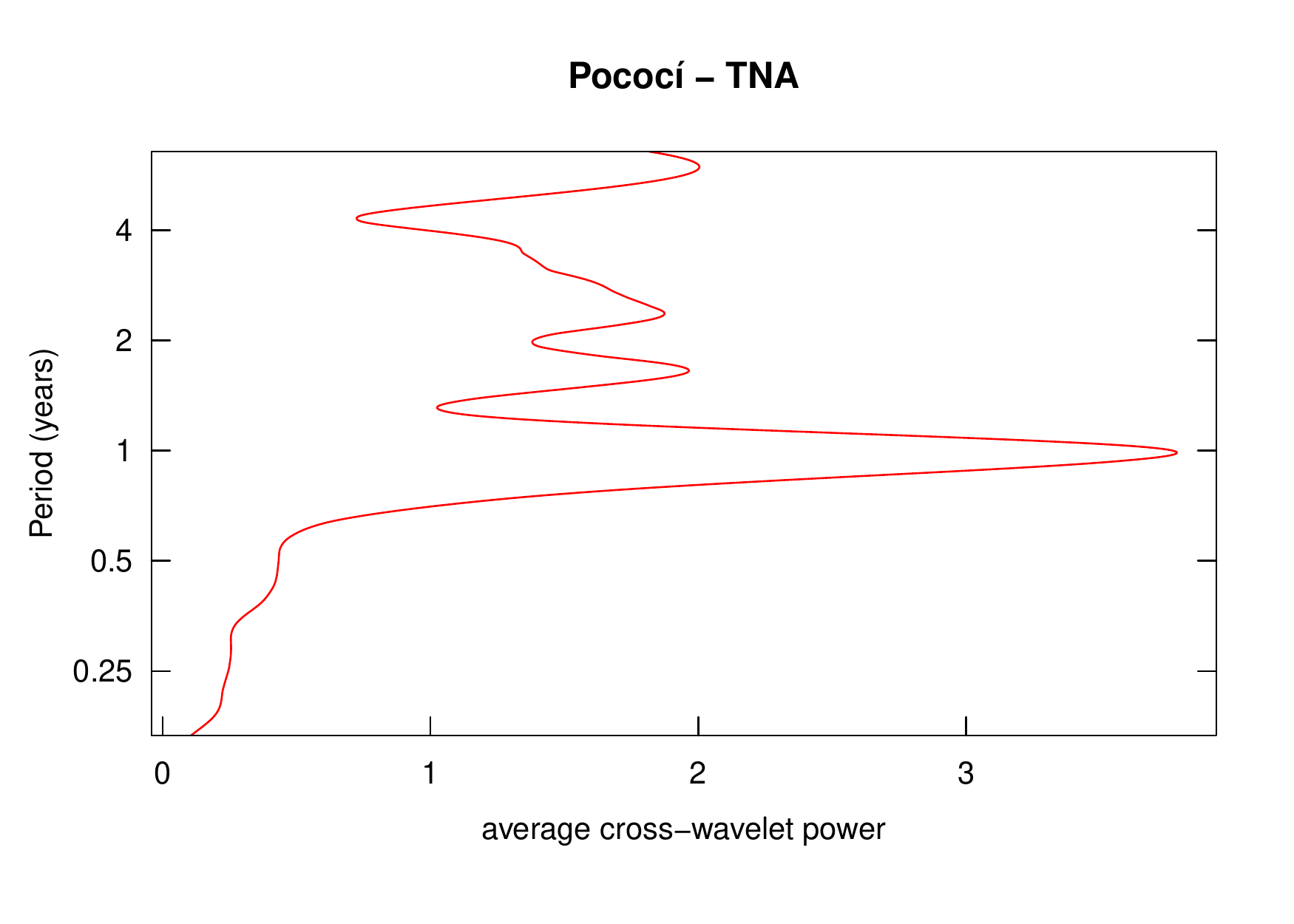}}\vspace{-0.15cm}\\
\subfloat[]{\includegraphics[scale=0.23]{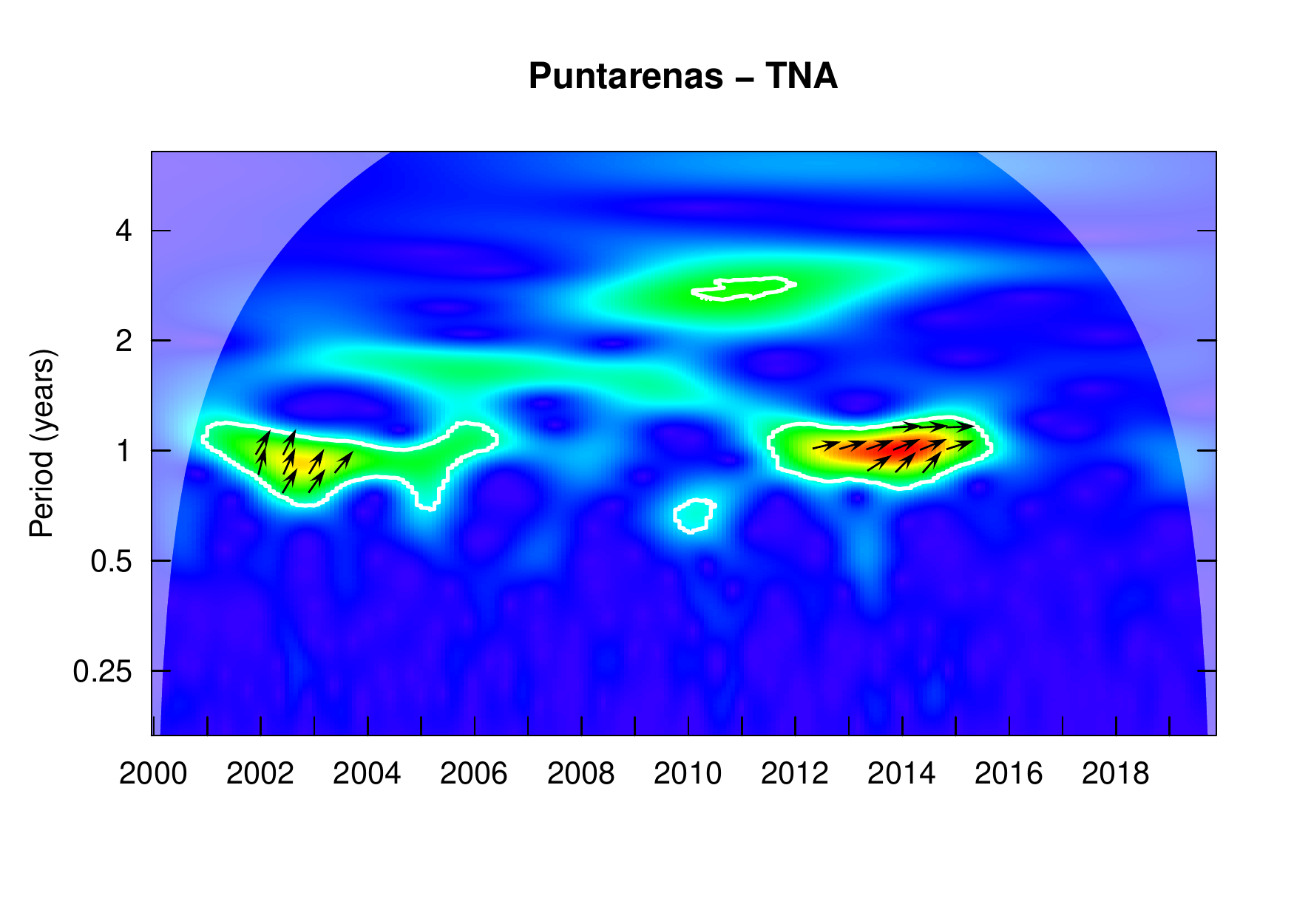}}\vspace{-0.15cm}%
\subfloat[]{\includegraphics[scale=0.23]{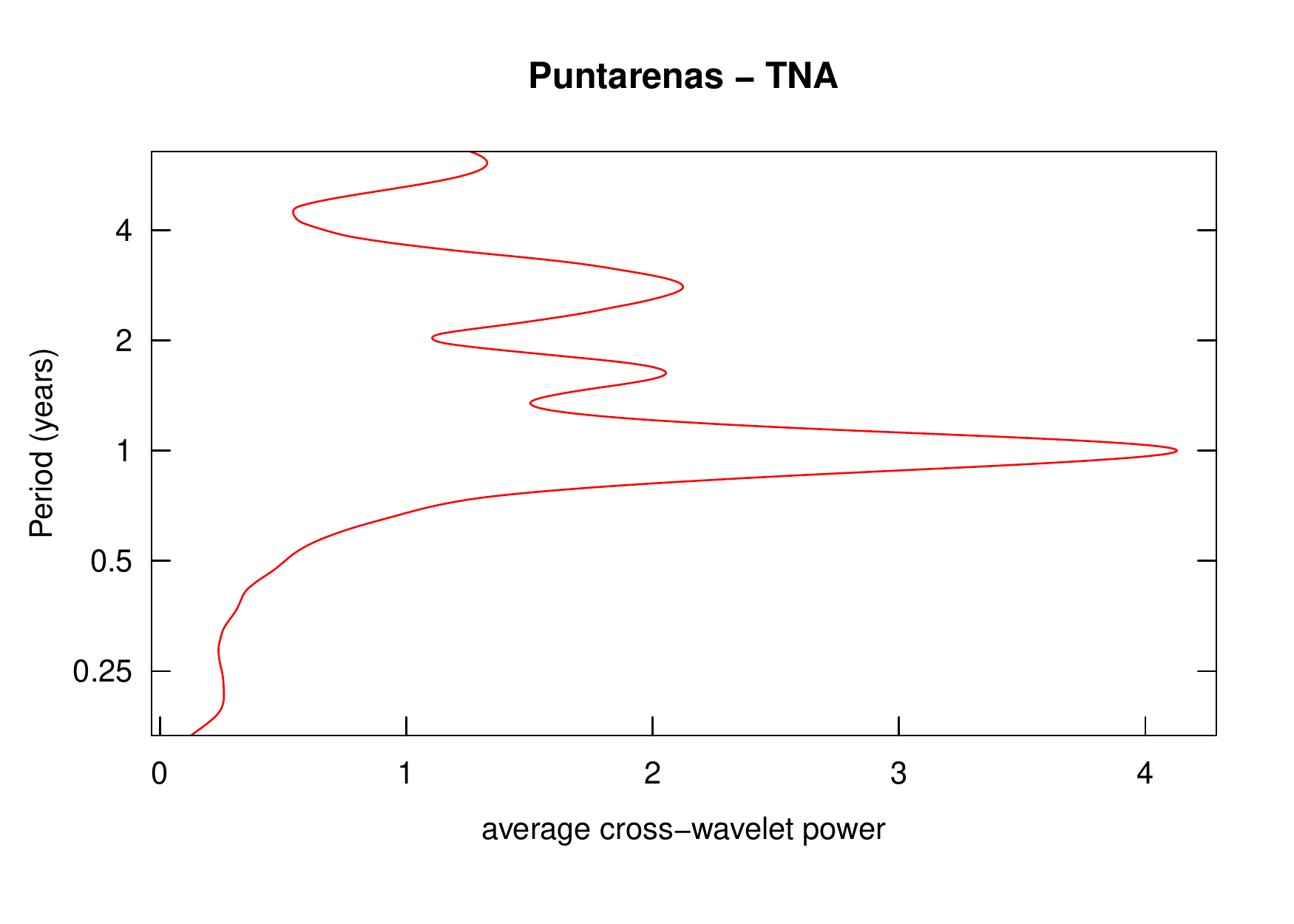}}\vspace{-0.15cm}%
\subfloat[]{\includegraphics[scale=0.23]{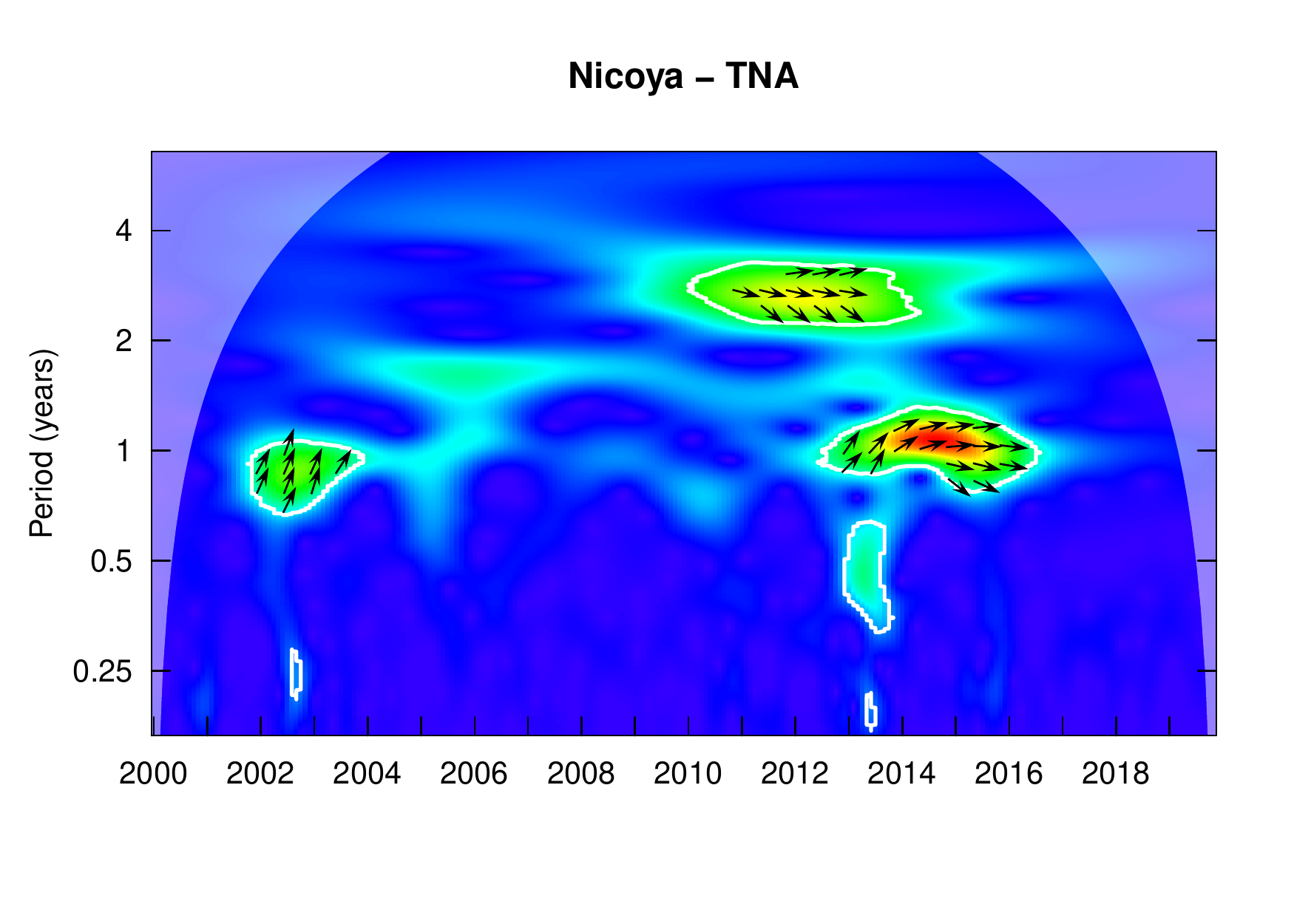}}\vspace{-0.15cm}%
\subfloat[]{\includegraphics[scale=0.23]{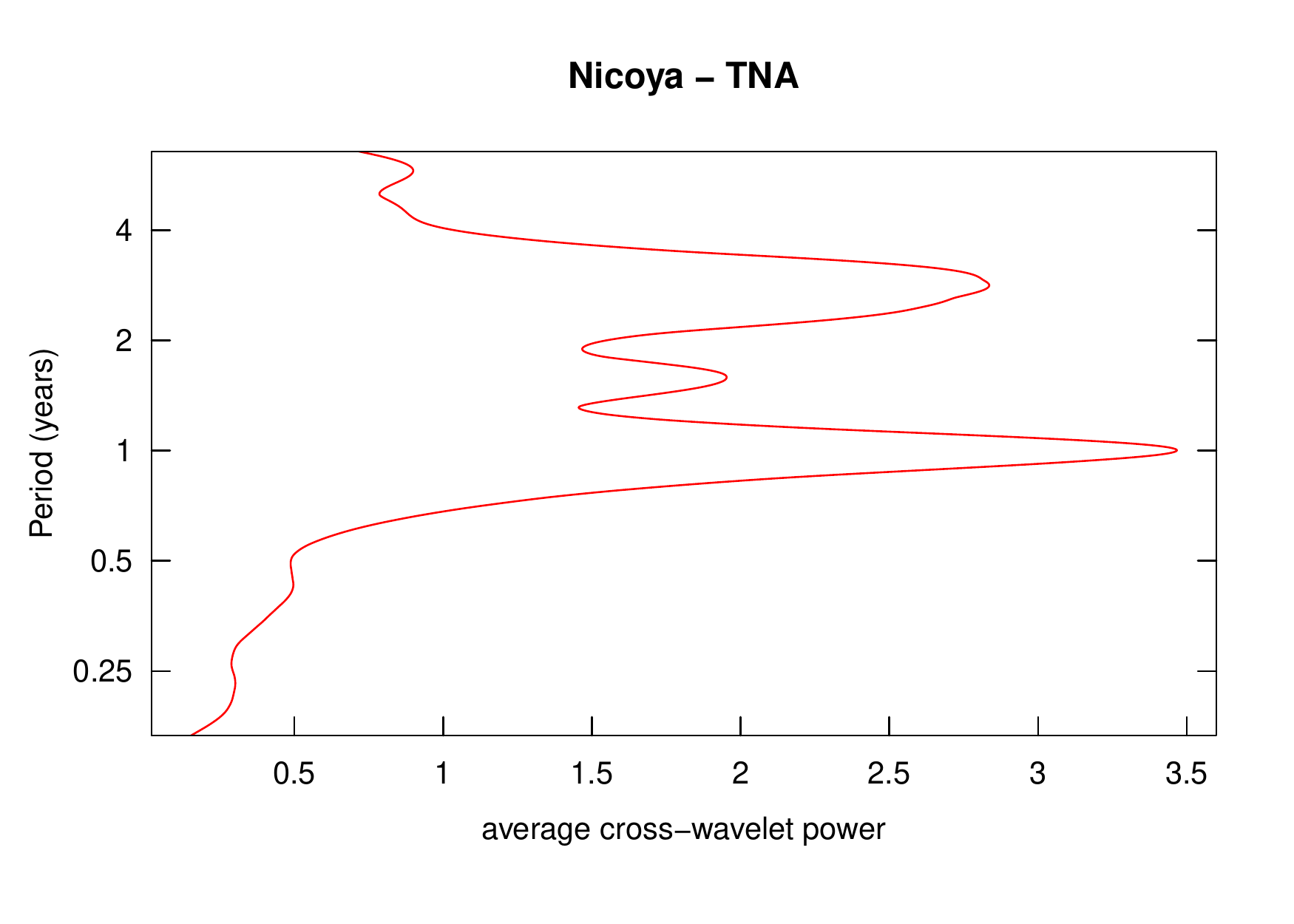}}\vspace{-0.15cm}\\
\subfloat[]{\includegraphics[scale=0.23]{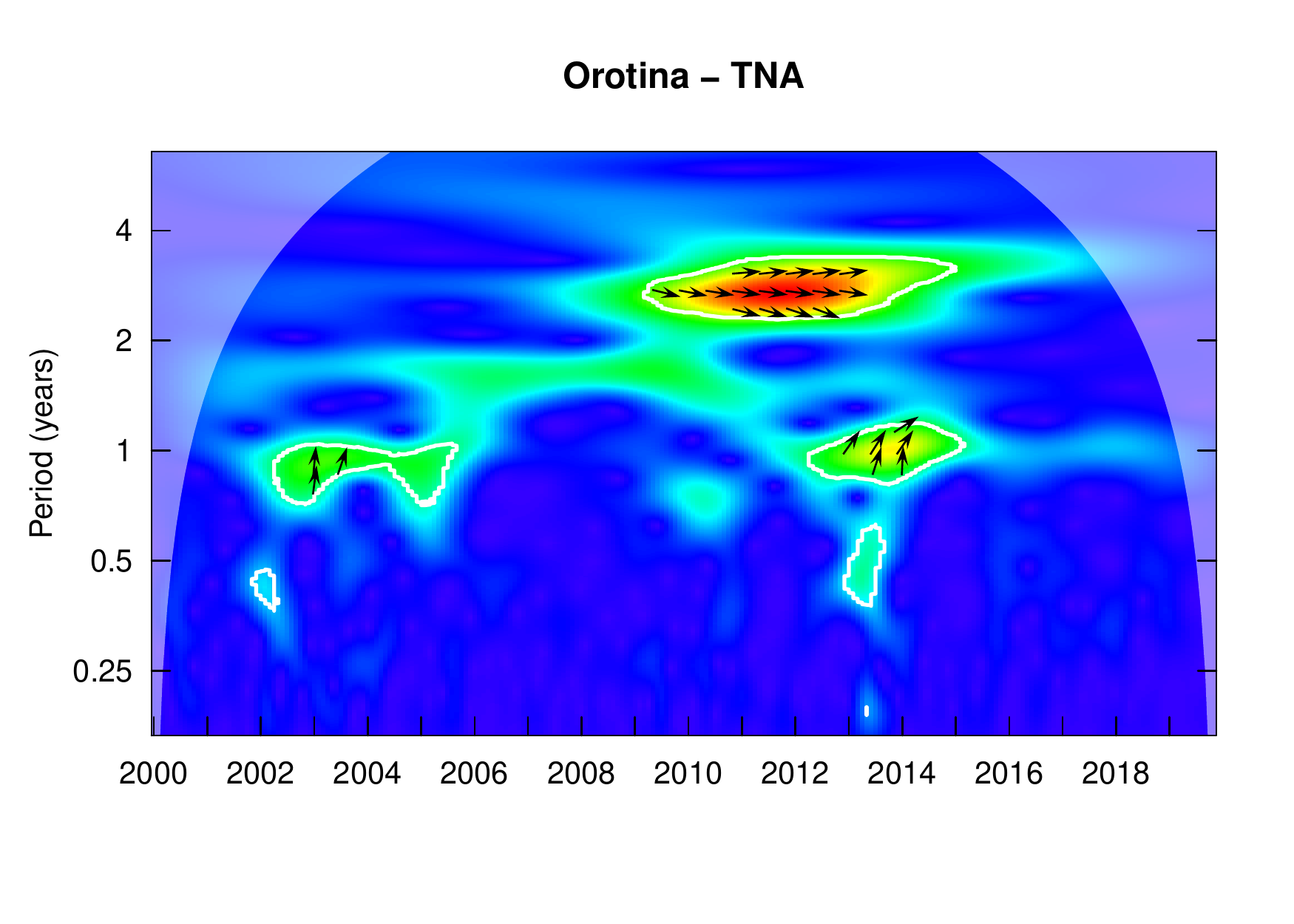}}\vspace{-0.15cm}%
\subfloat[]{\includegraphics[scale=0.23]{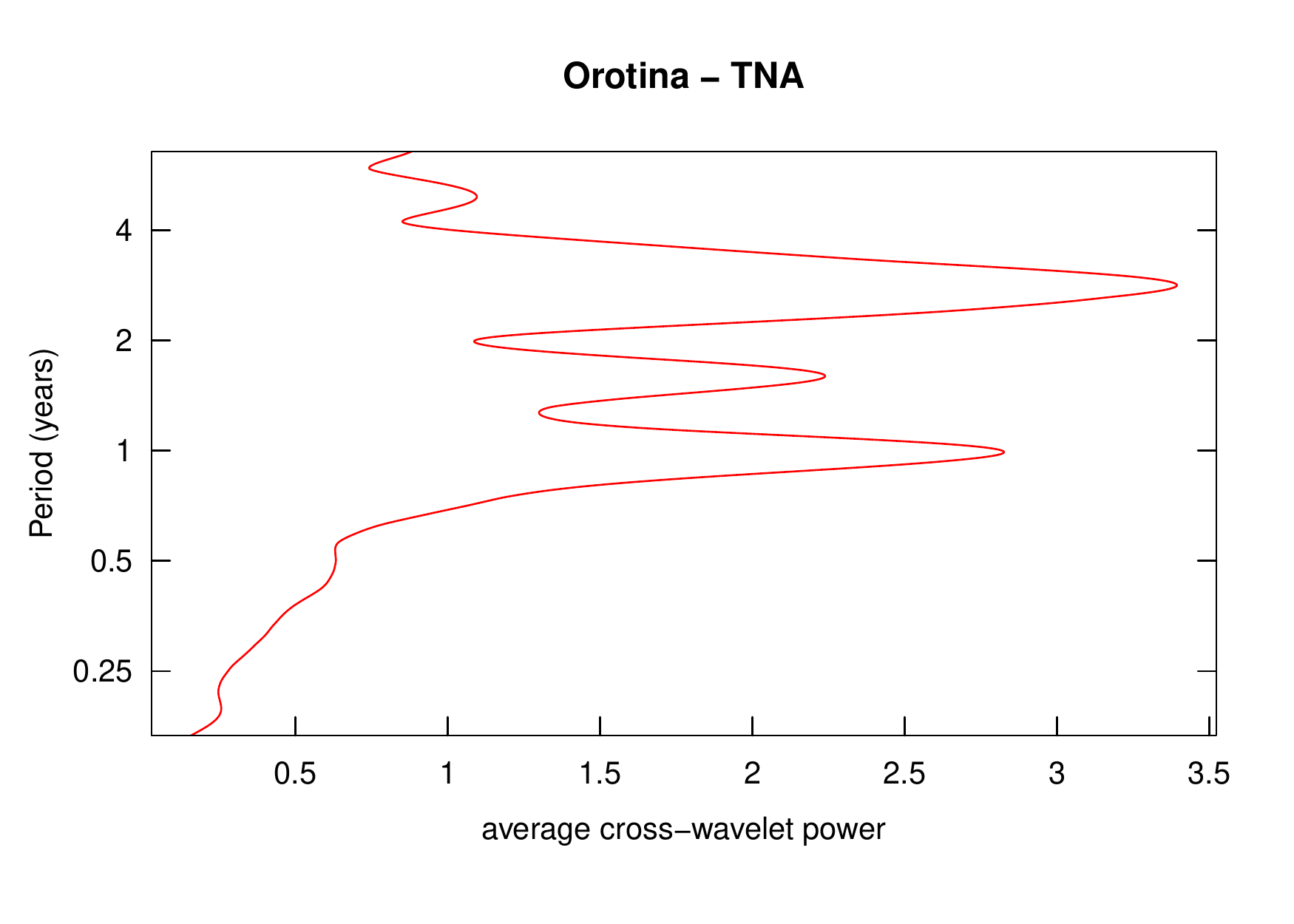}}\vspace{-0.15cm}%
\subfloat[]{\includegraphics[scale=0.23]{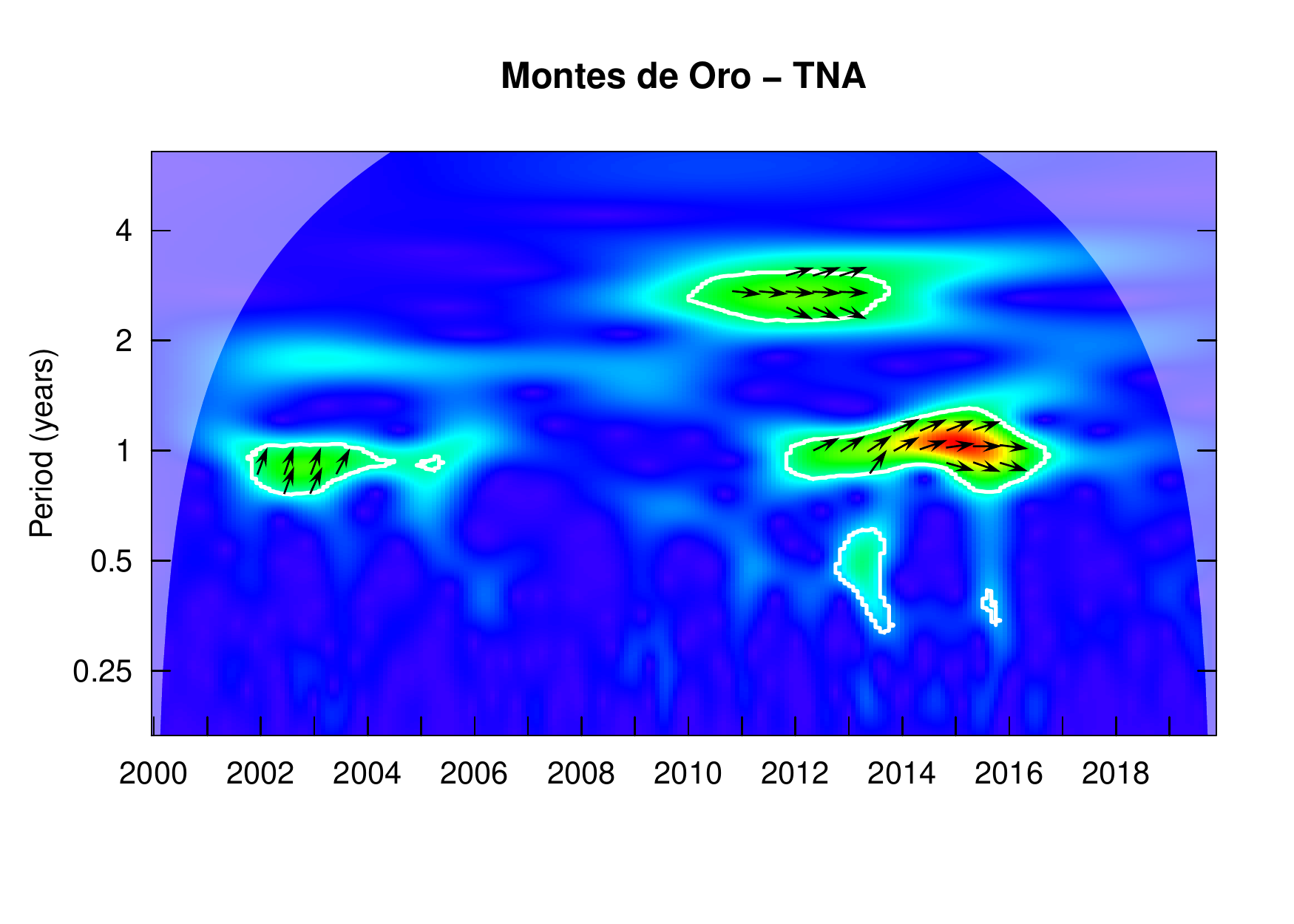}}\vspace{-0.15cm}%
\subfloat[]{\includegraphics[scale=0.23]{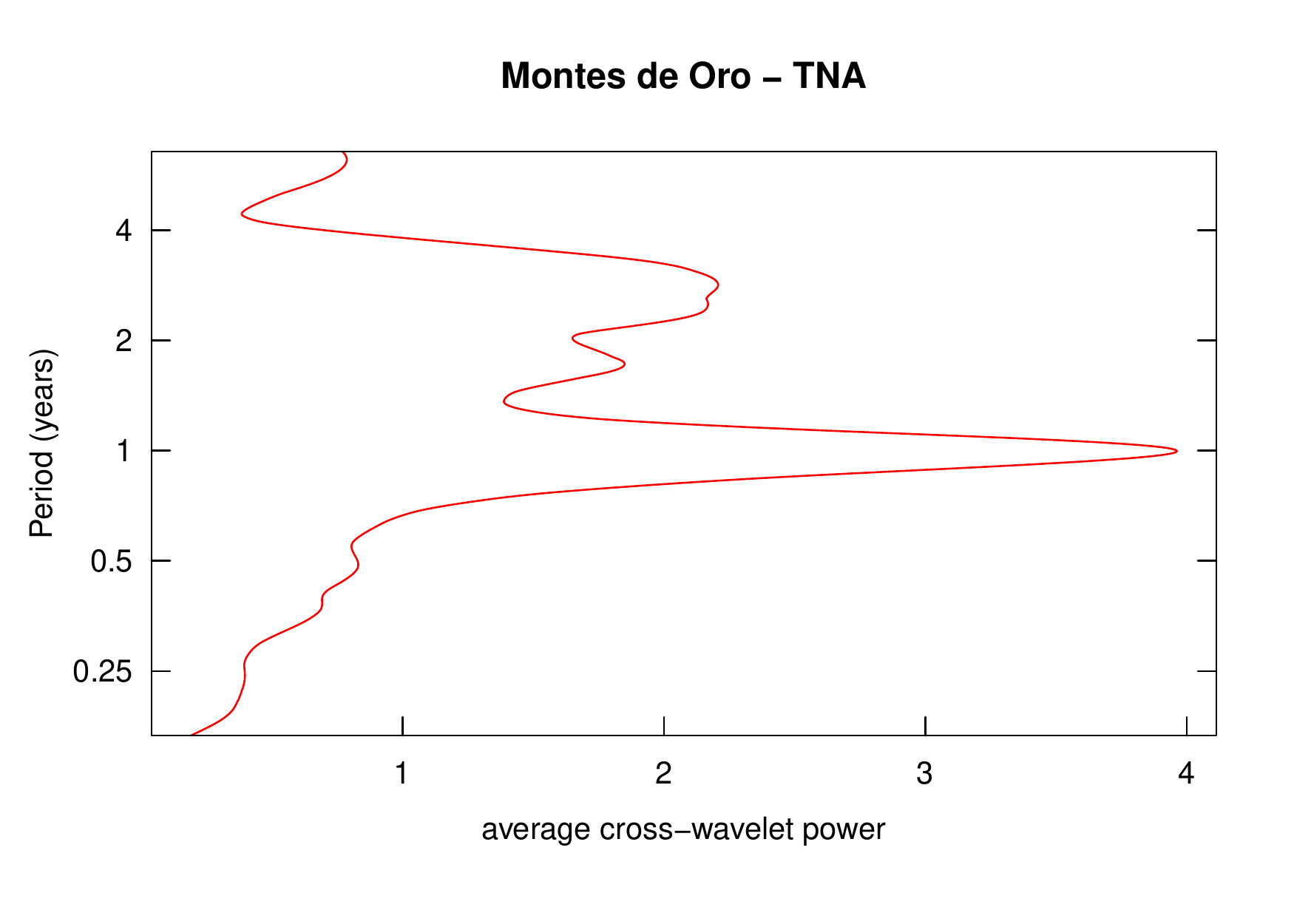}}\vspace{-0.15cm}\\
\subfloat[]{\includegraphics[scale=0.23]{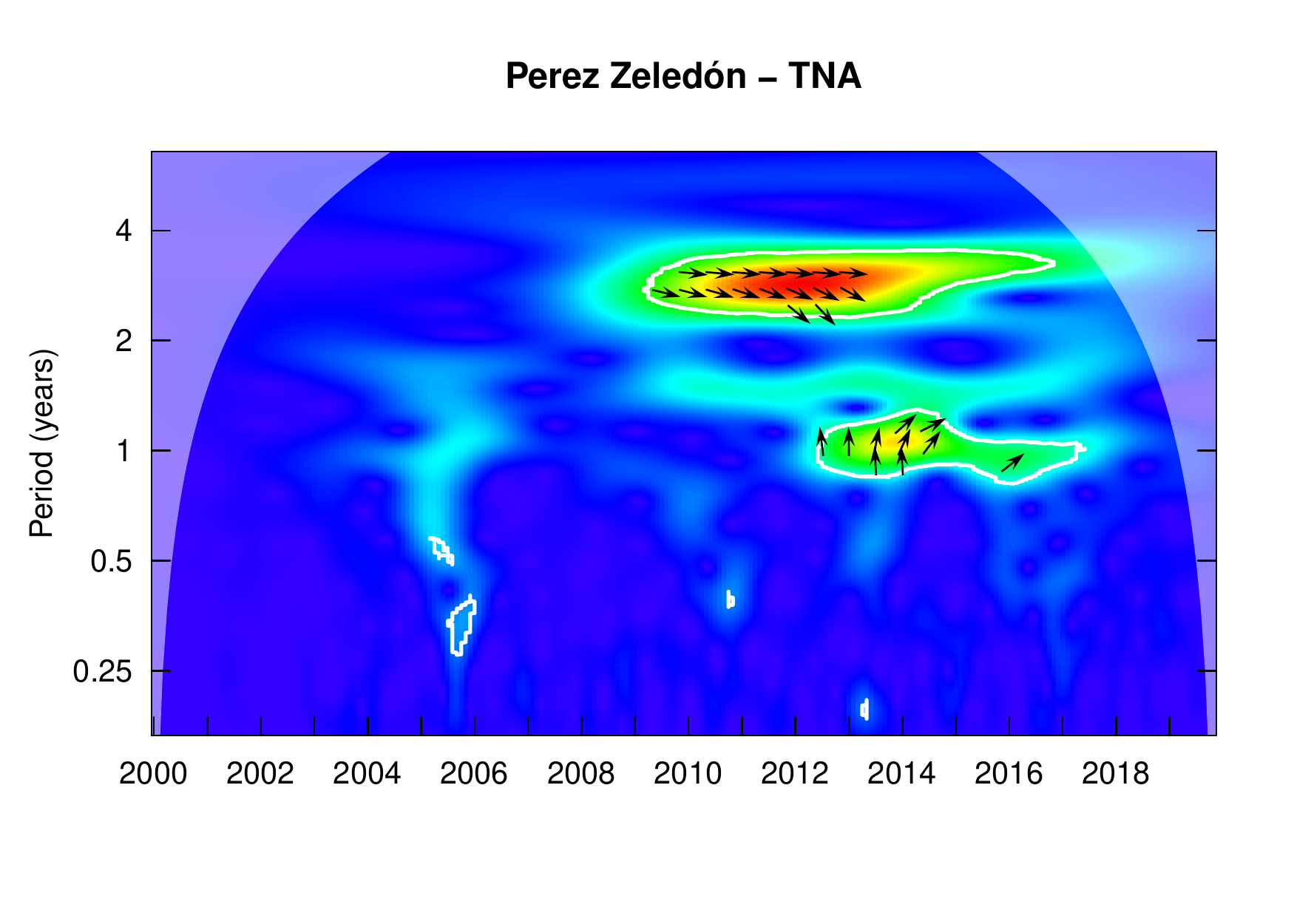}}\vspace{-0.15cm}%
\subfloat[]{\includegraphics[scale=0.23]{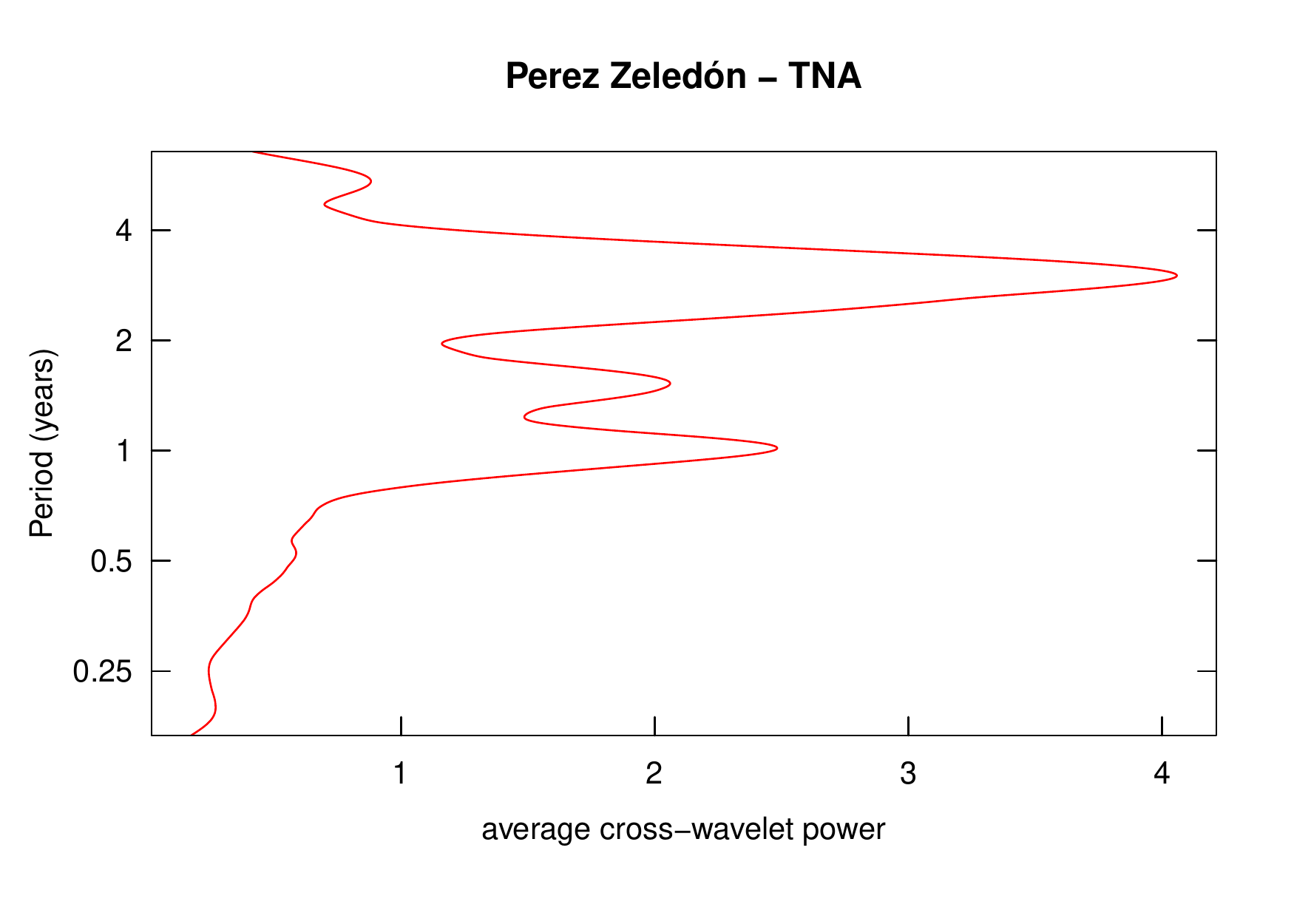}}\vspace{-0.15cm}%
\subfloat[]{\includegraphics[scale=0.23]{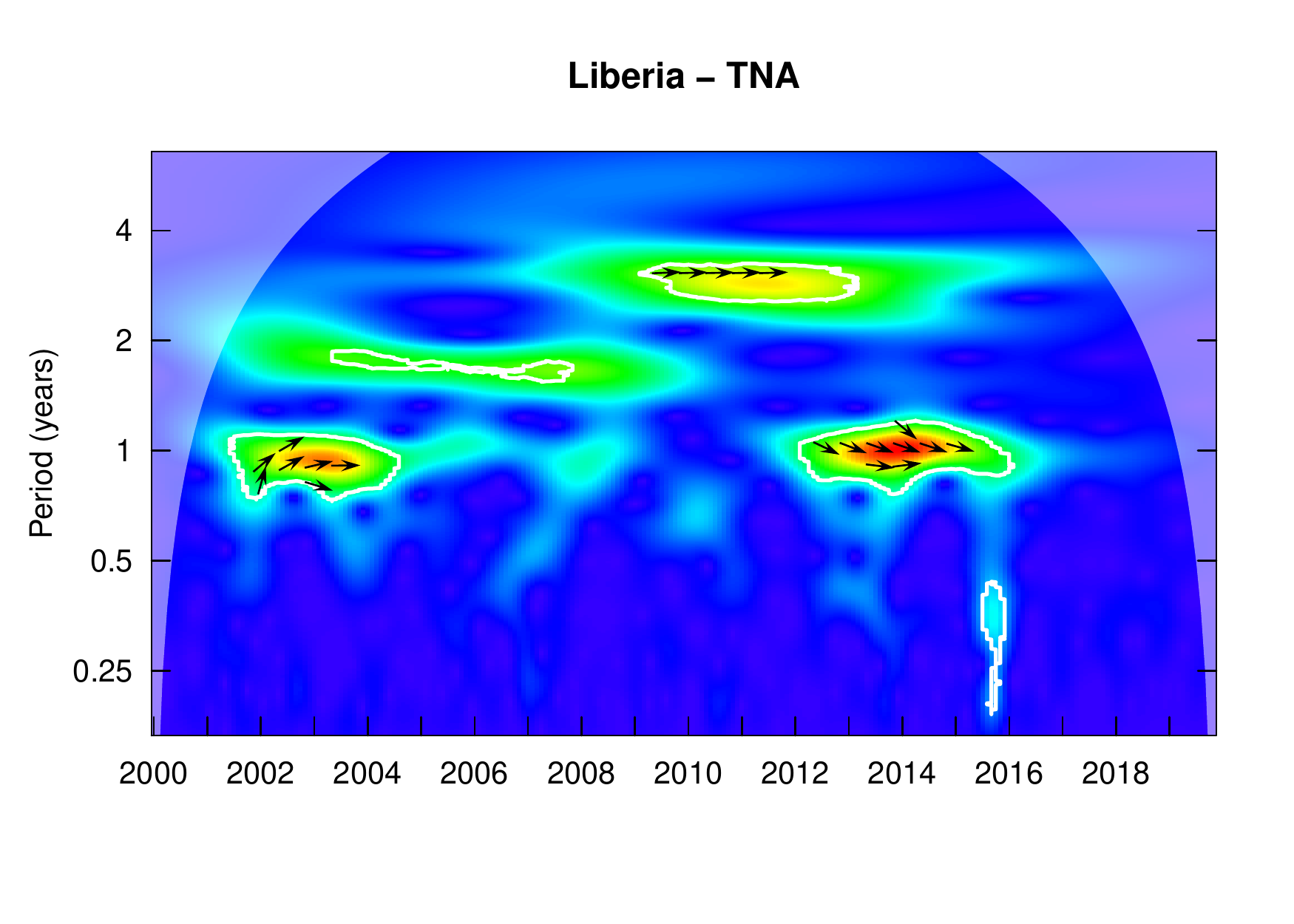}}\vspace{-0.15cm}%
\subfloat[]{\includegraphics[scale=0.23]{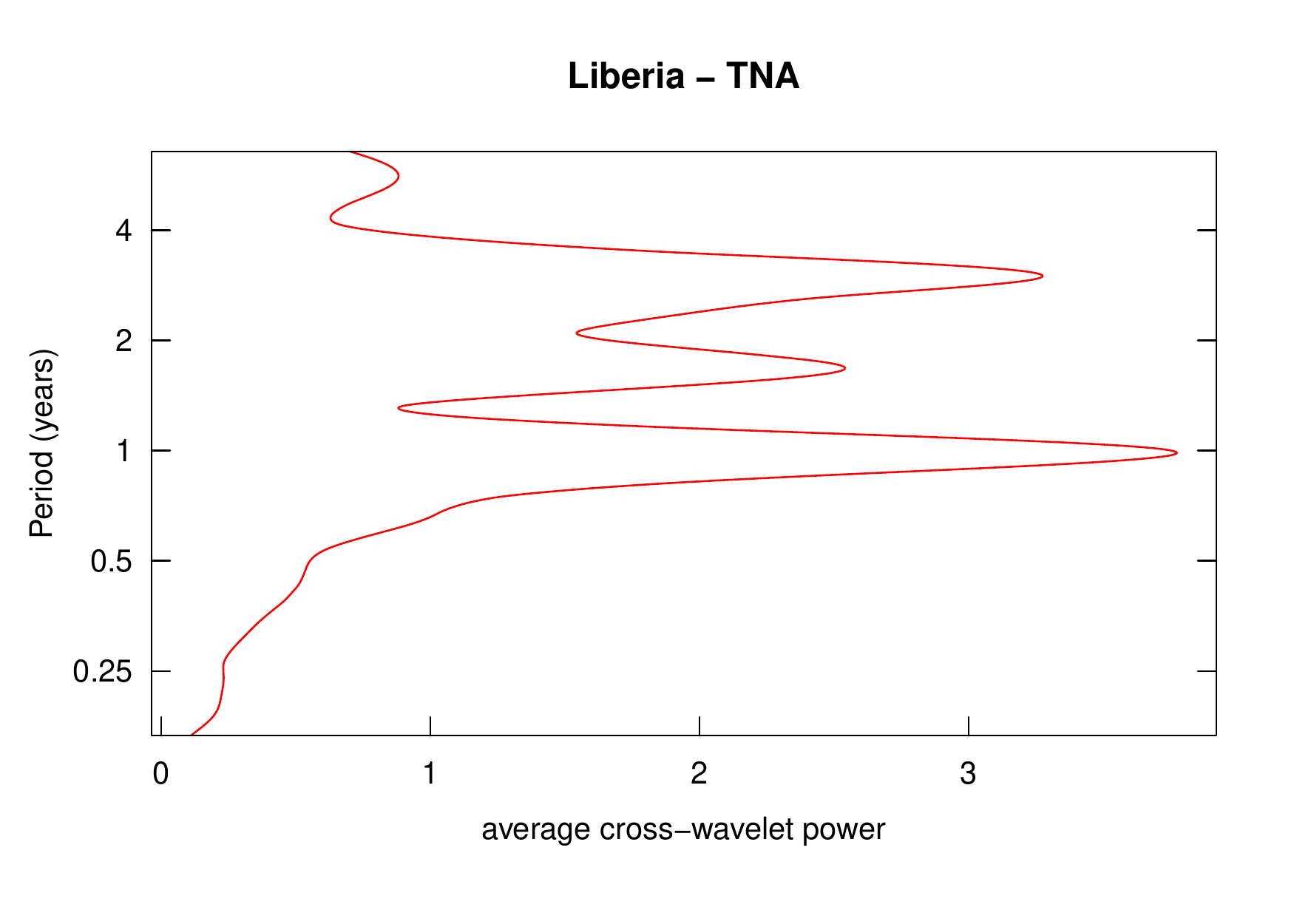}}\vspace{-0.15cm}\\
\subfloat[]{\includegraphics[scale=0.23]{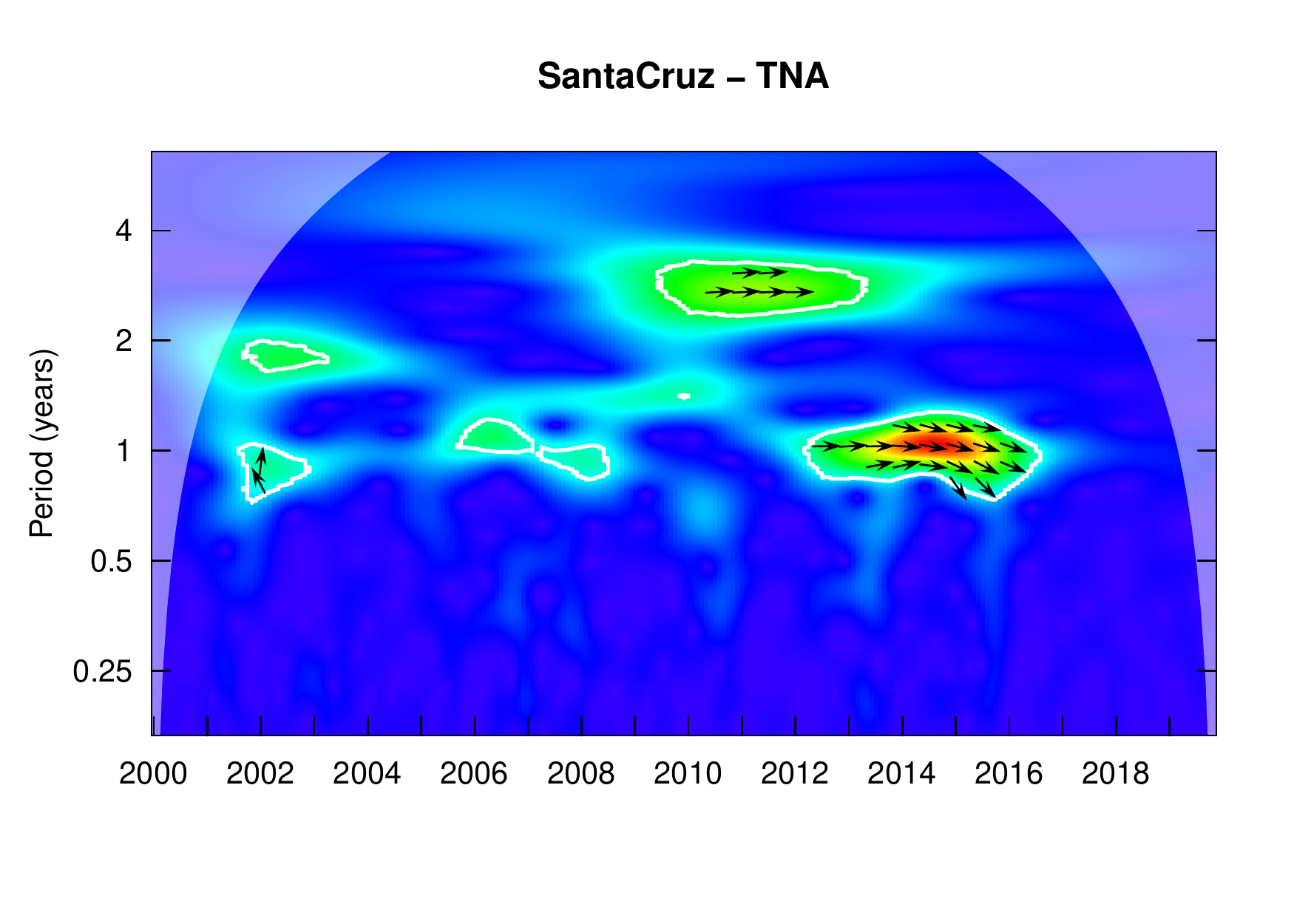}}\vspace{-0.15cm}%
\subfloat[]{\includegraphics[scale=0.23]{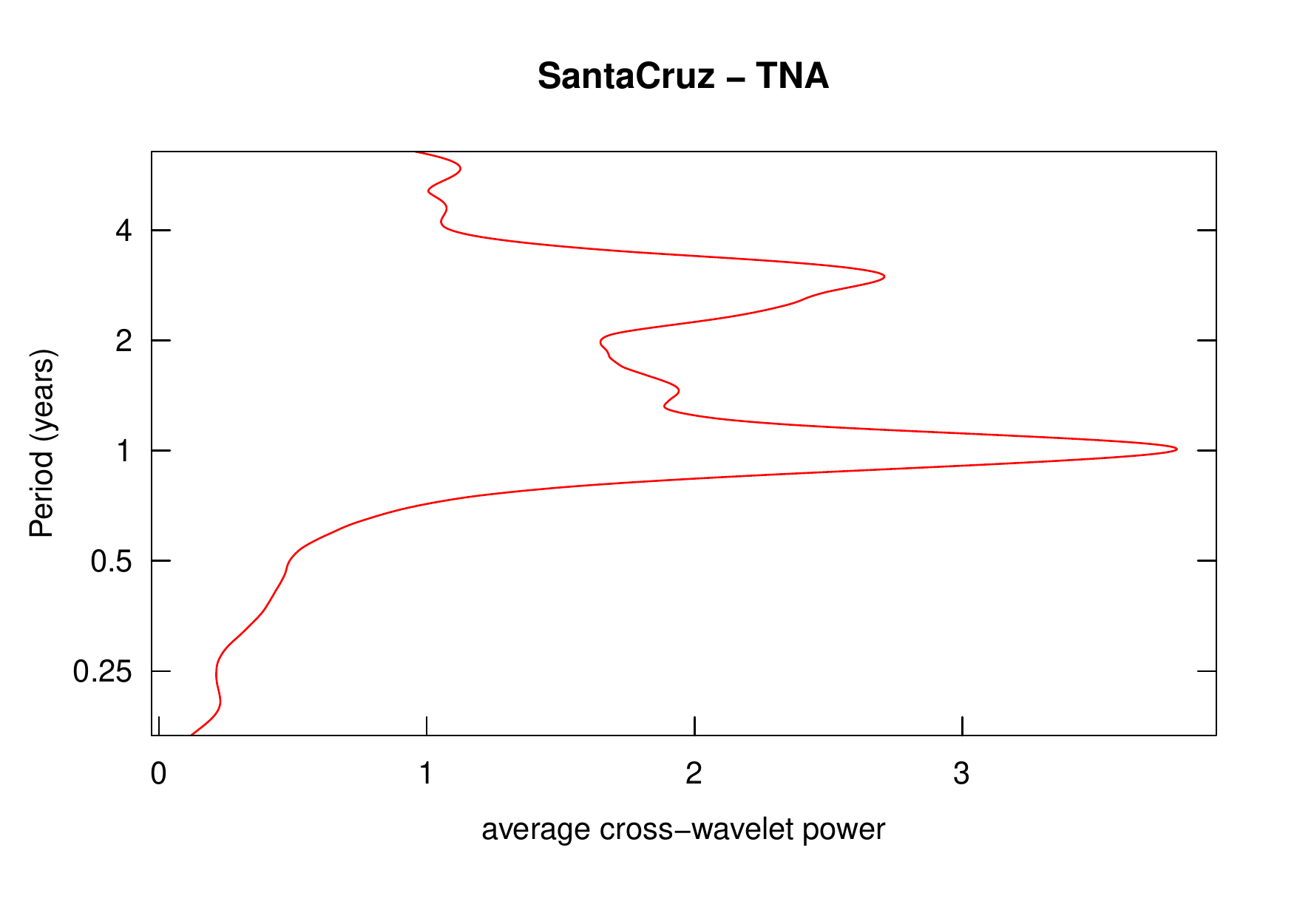}}\vspace{-0.15cm}%
\subfloat[]{\includegraphics[scale=0.23]{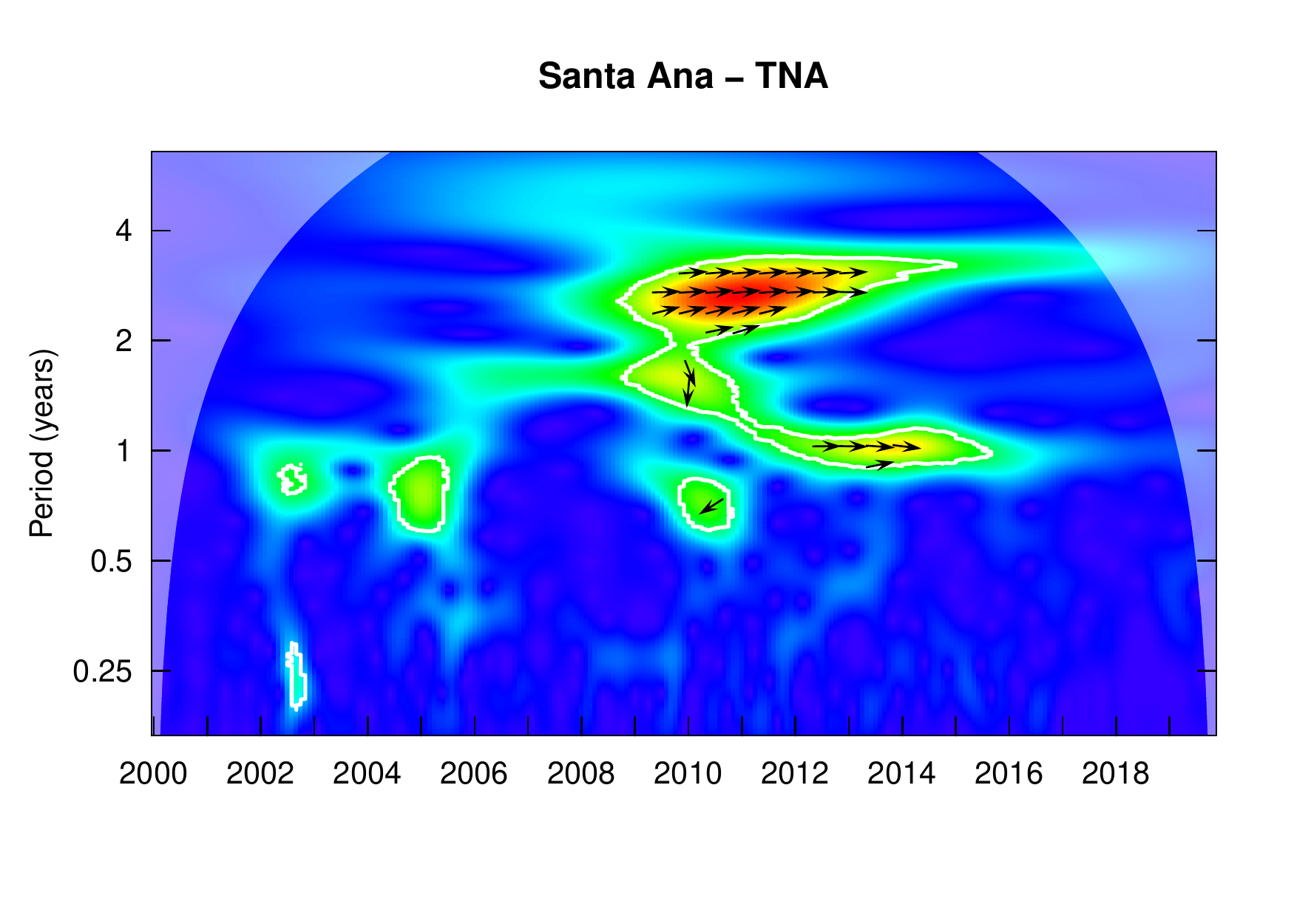}}\vspace{-0.15cm}%
\subfloat[]{\includegraphics[scale=0.23]{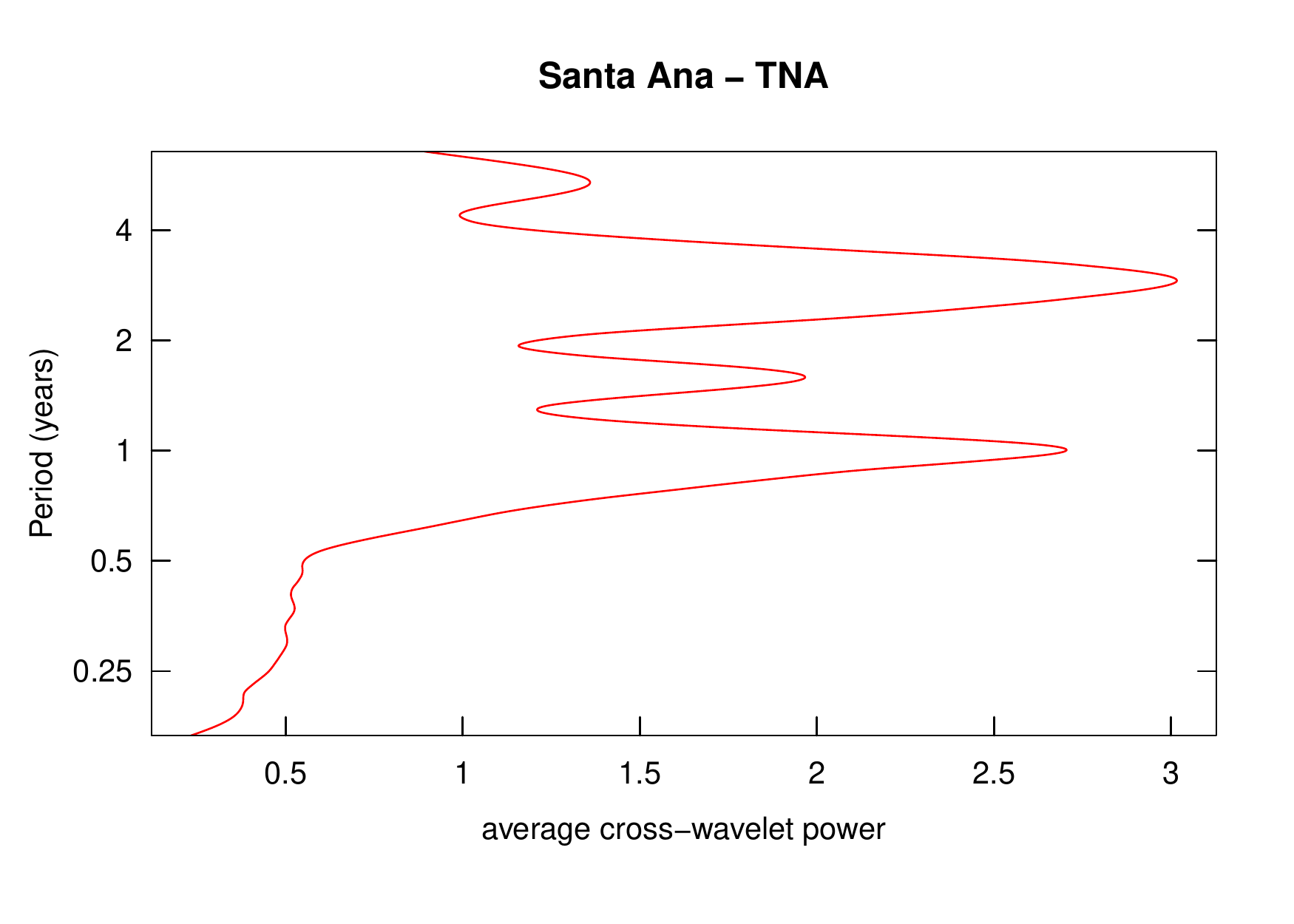}}\vspace{-0.15cm}\\
\subfloat[]{\includegraphics[scale=0.23]{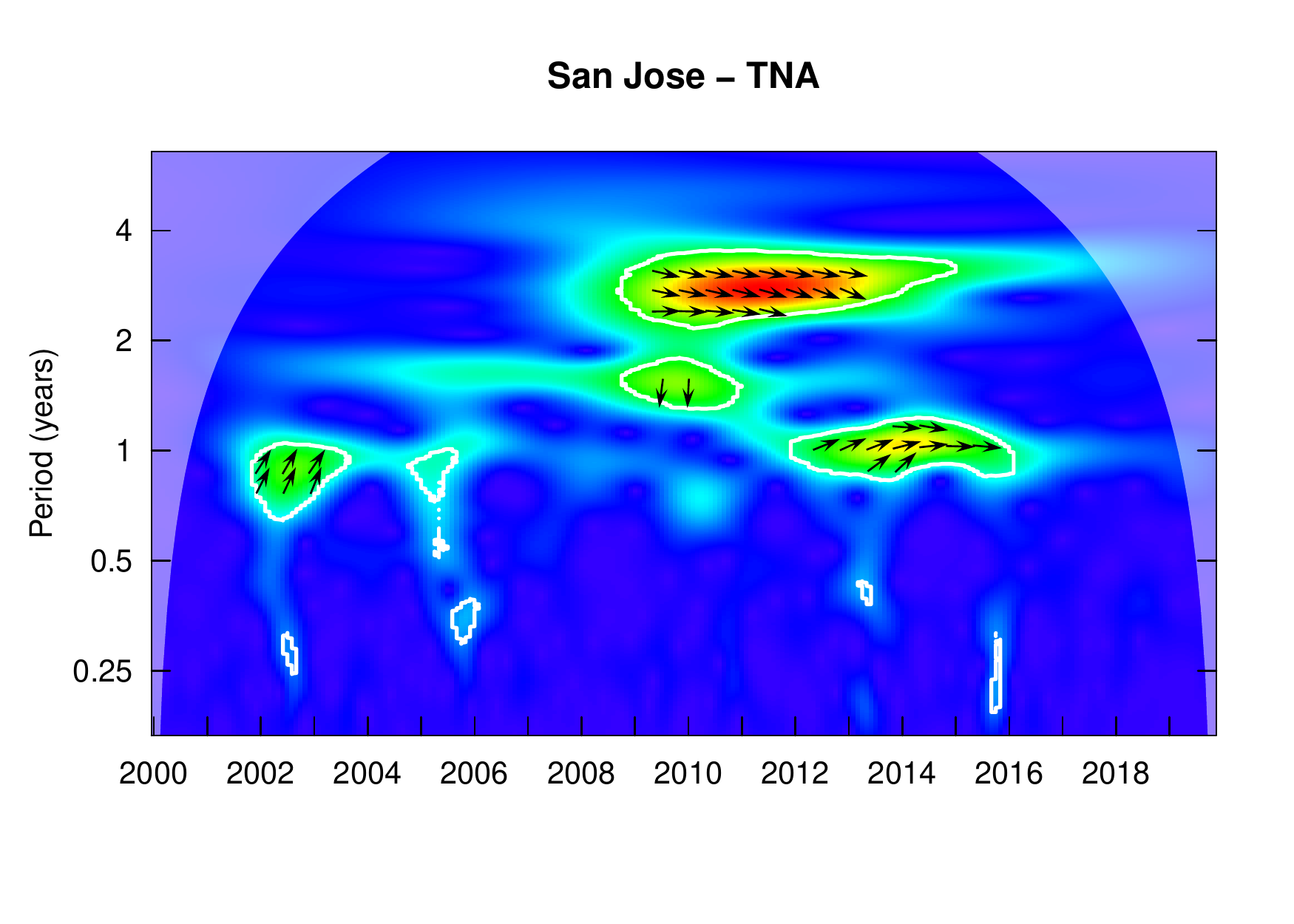}}\vspace{-0.15cm}%
\subfloat[]{\includegraphics[scale=0.23]{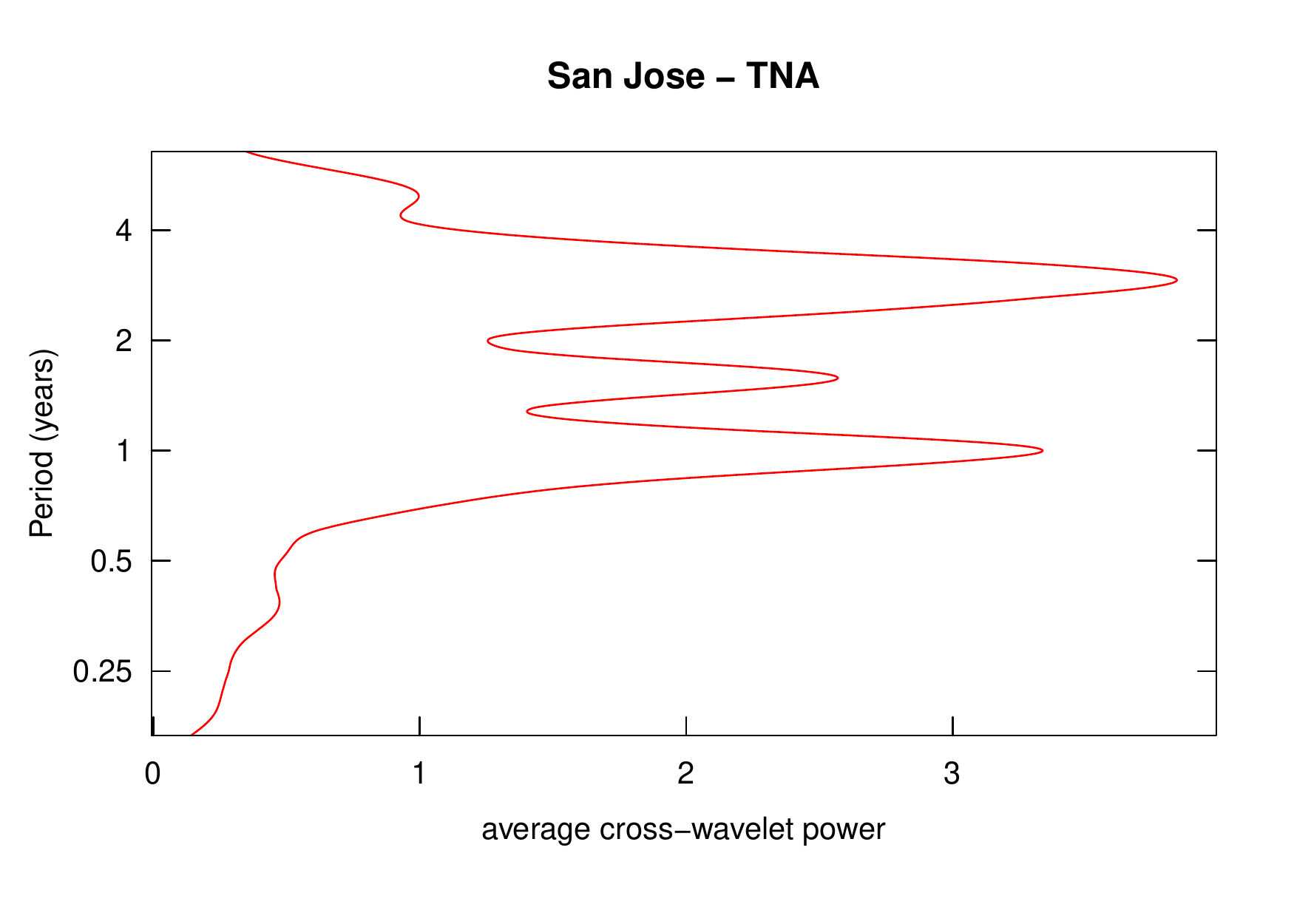}}\vspace{-0.15cm}%
\subfloat[]{\includegraphics[scale=0.23]{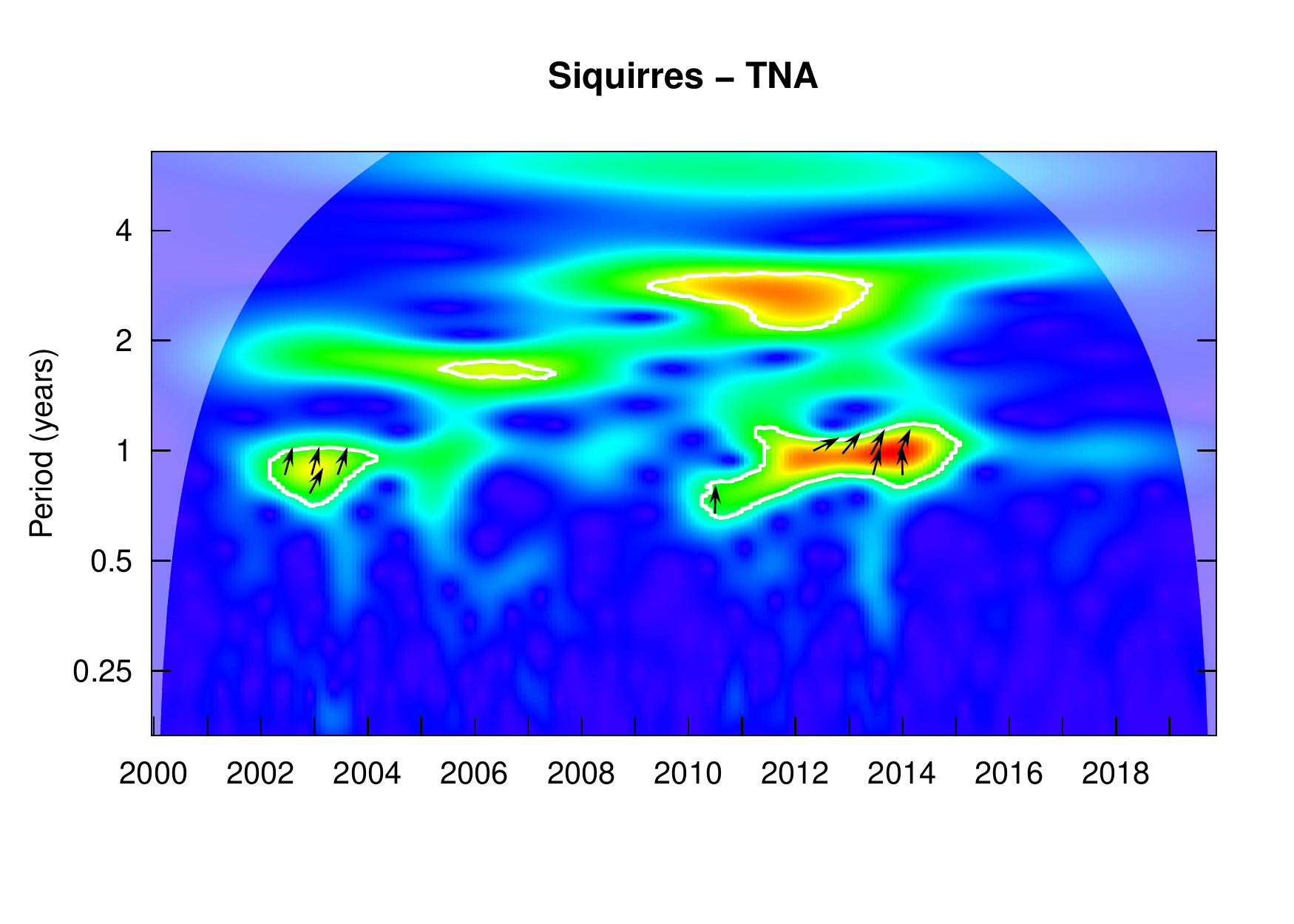}}\vspace{-0.15cm}%
\subfloat[]{\includegraphics[scale=0.23]{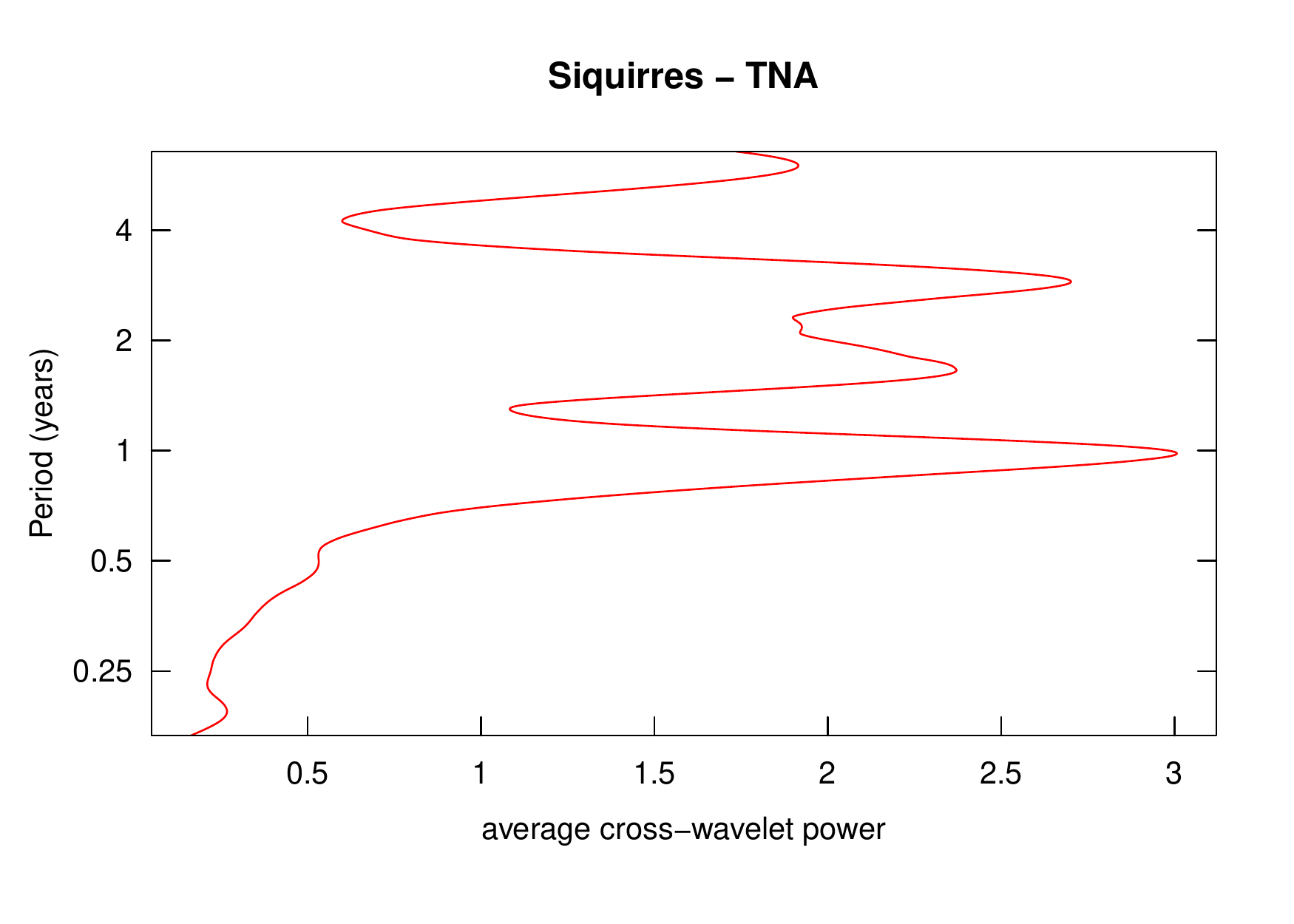}}\vspace{-0.15cm}\\
\subfloat[]{\includegraphics[scale=0.23]{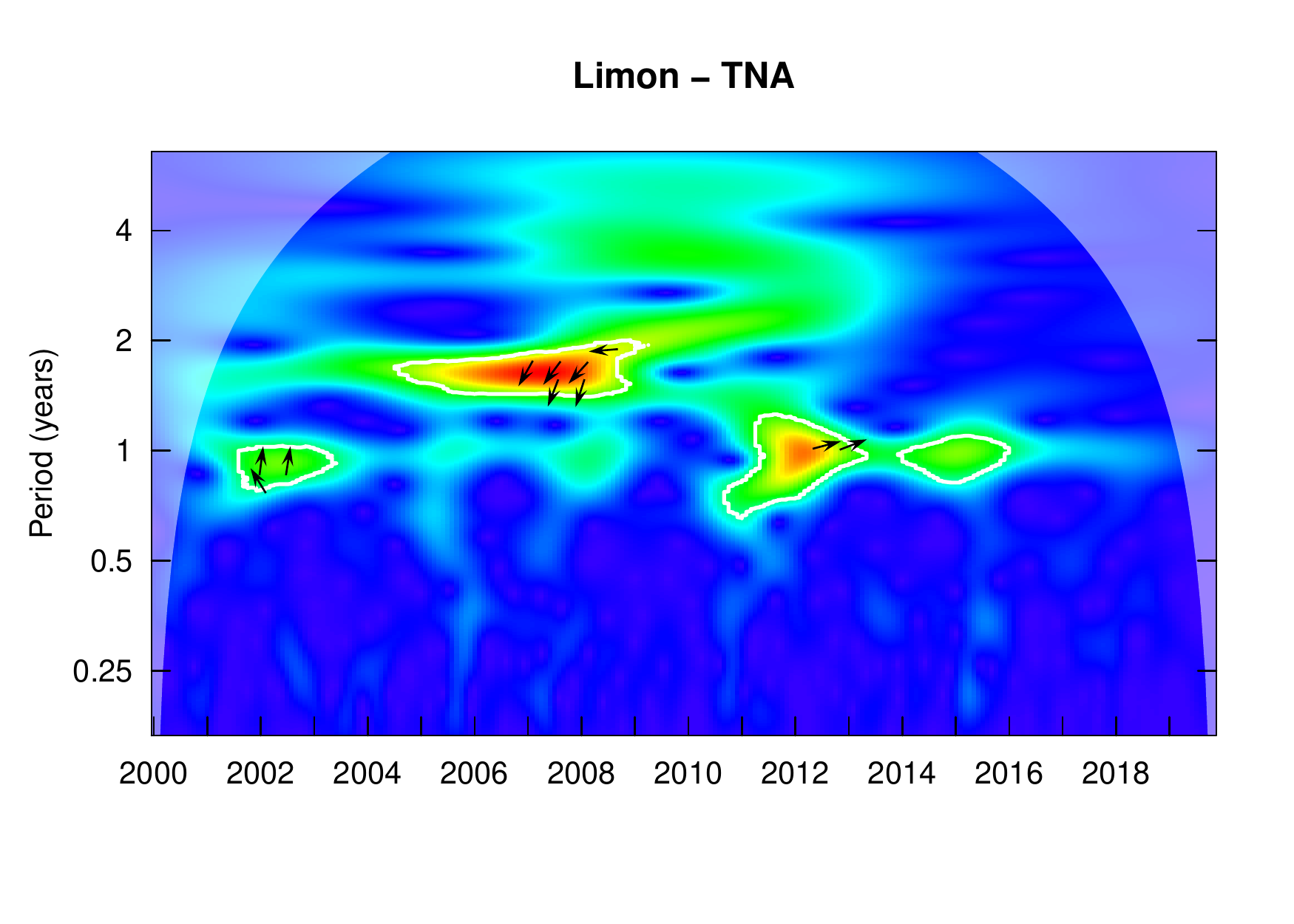}}\vspace{-0.15cm}%
\subfloat[]{\includegraphics[scale=0.23]{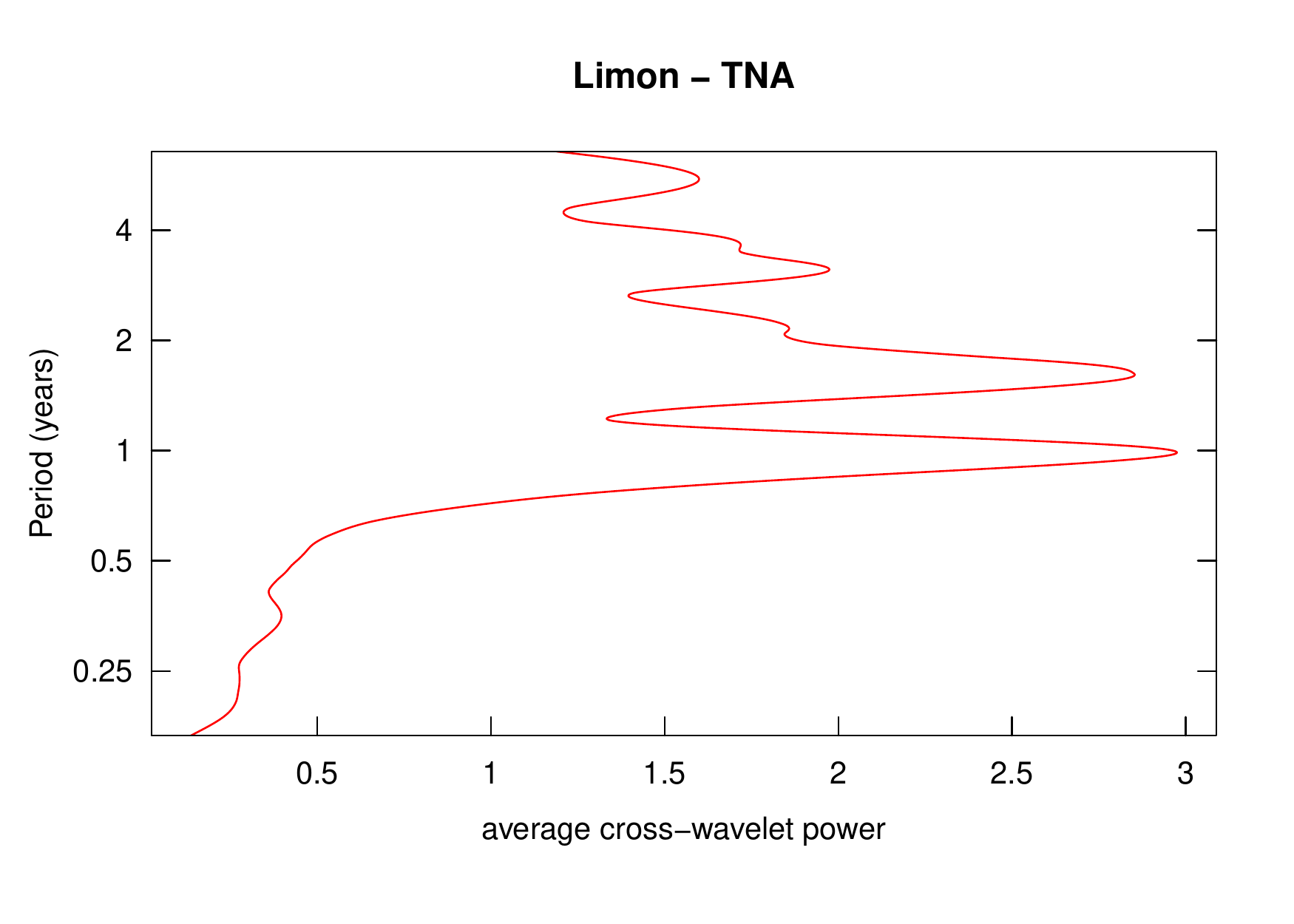}}\vspace{-0.15cm}%
\subfloat[]{\includegraphics[scale=0.23]{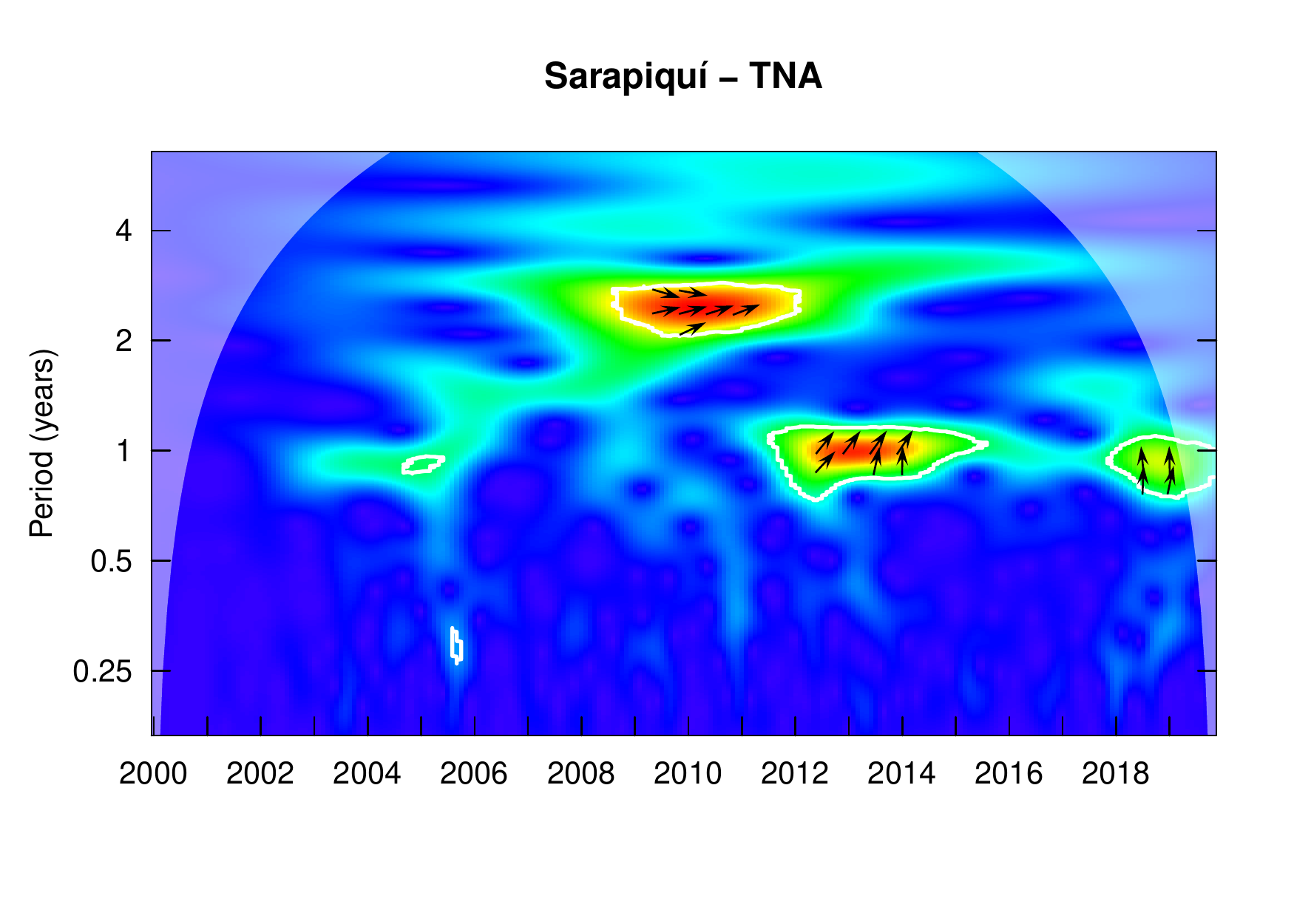}}\vspace{-0.15cm}%
\subfloat[]{\includegraphics[scale=0.23]{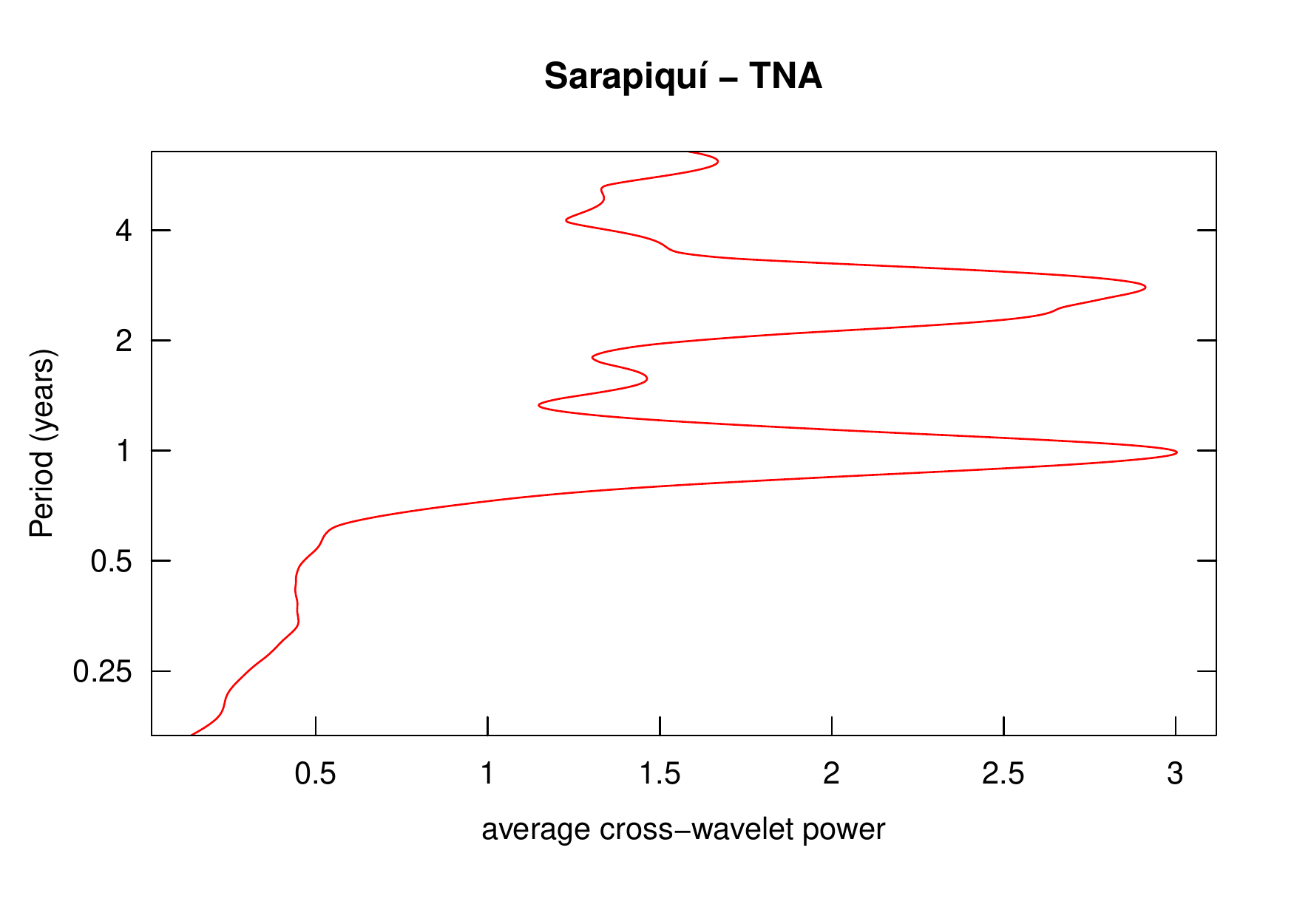}}\vspace{-0.15cm}\\
\caption*{}
\end{figure}

\begin{figure}[H]
\captionsetup[subfigure]{labelformat=empty}
\subfloat[]{\includegraphics[scale=0.23]{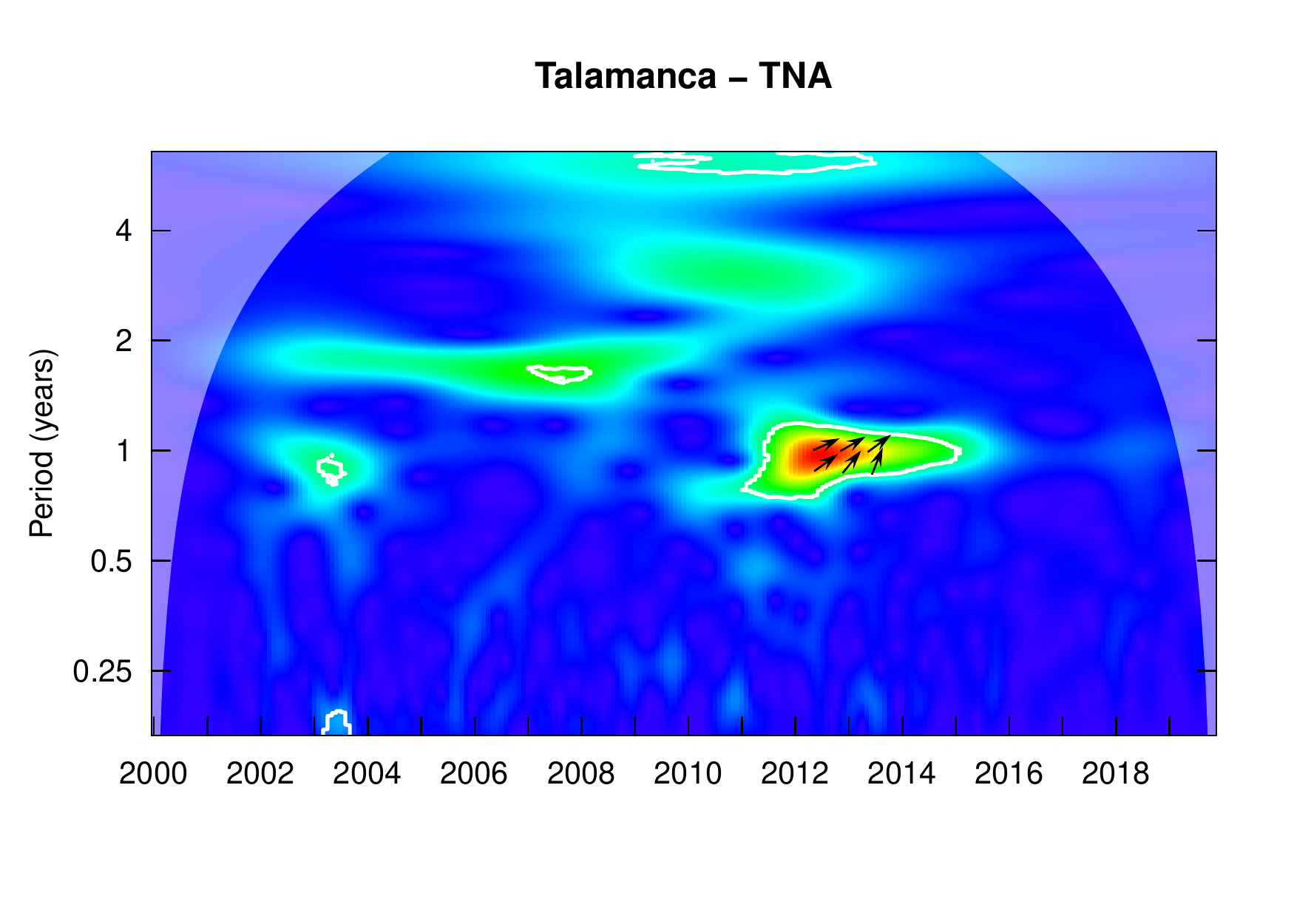}}\vspace{-0.15cm}%
\subfloat[]{\includegraphics[scale=0.23]{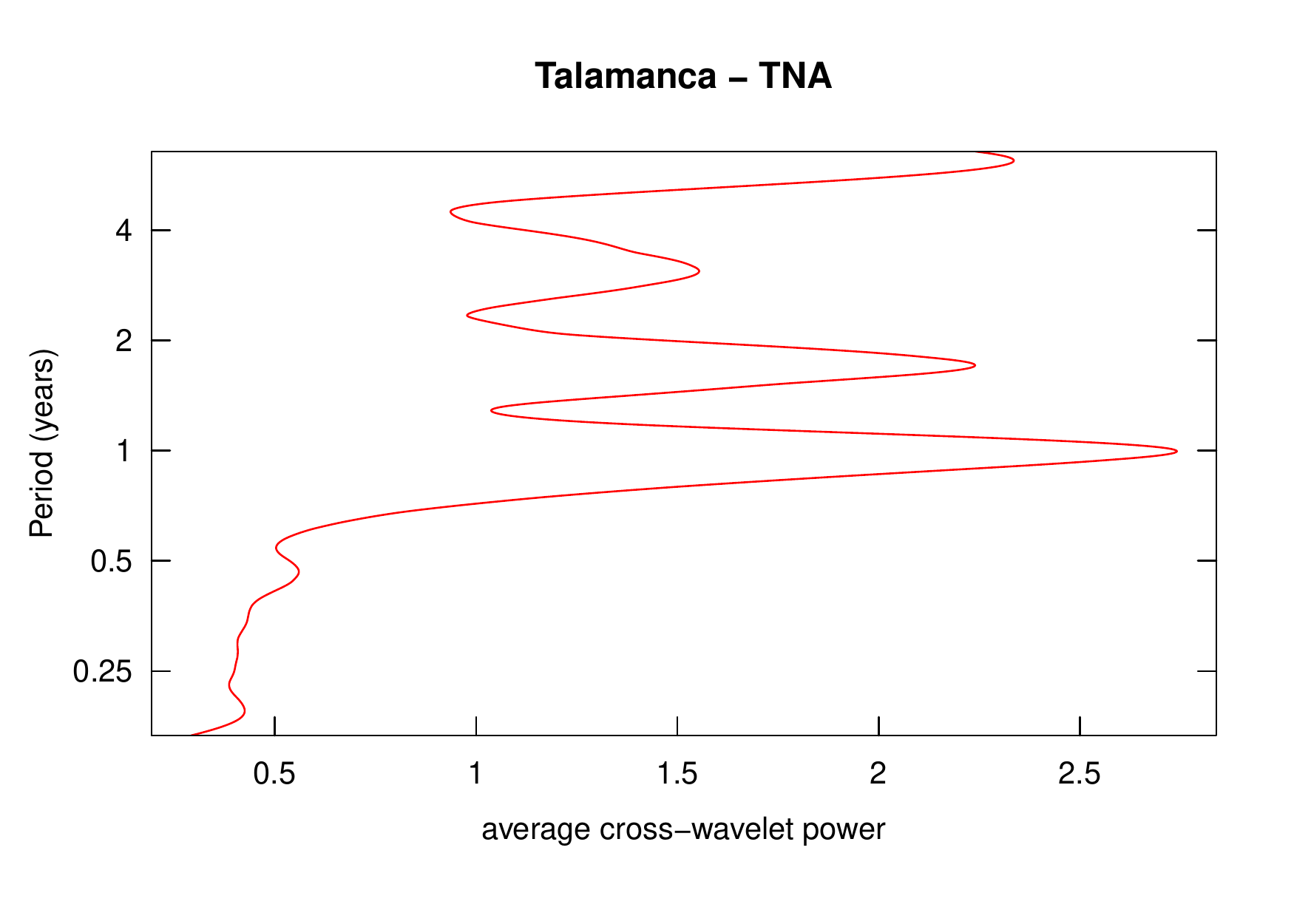}}\vspace{-0.15cm}%
\subfloat[]{\includegraphics[scale=0.23]{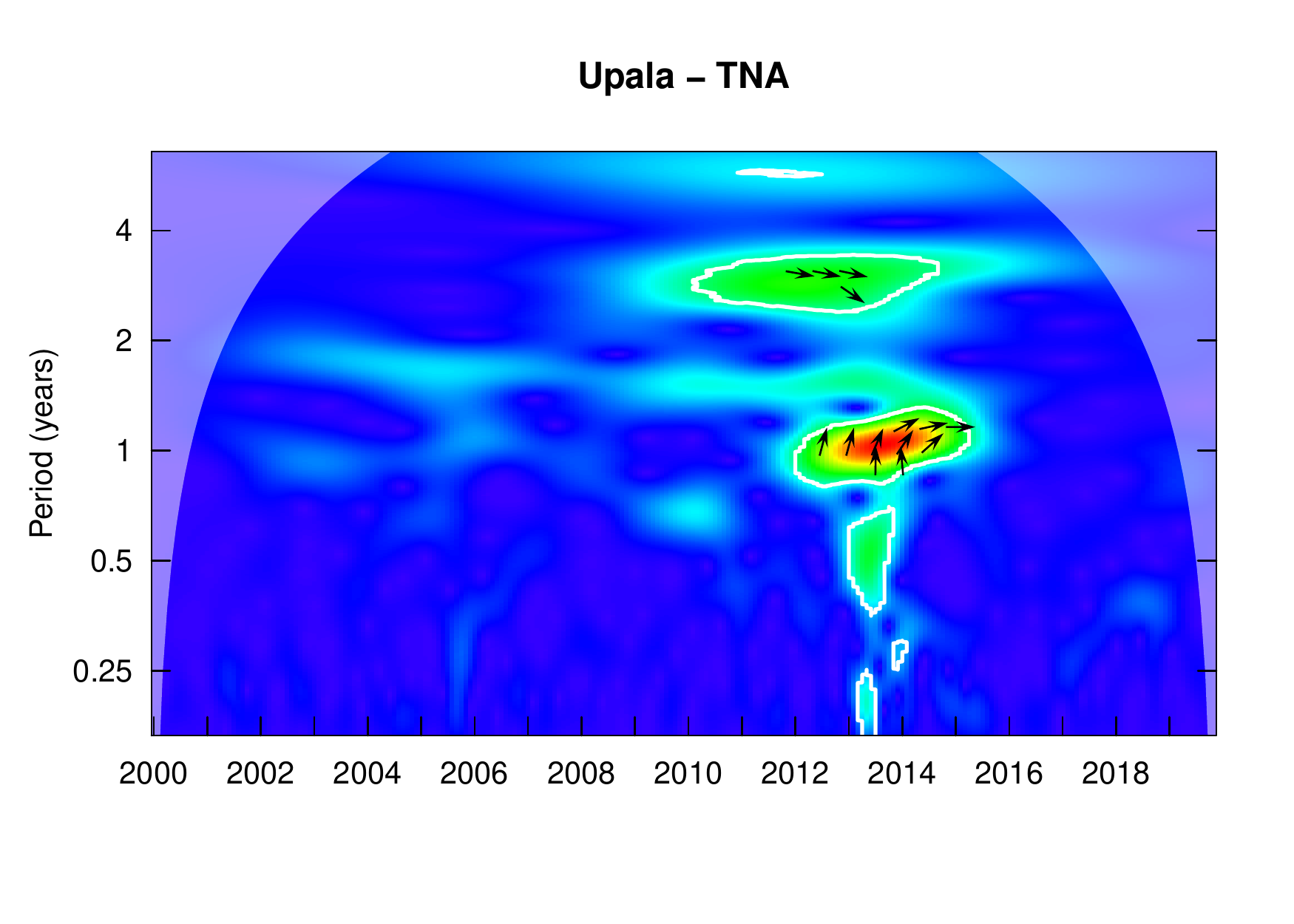}}\vspace{-0.15cm}%
\subfloat[]{\includegraphics[scale=0.23]{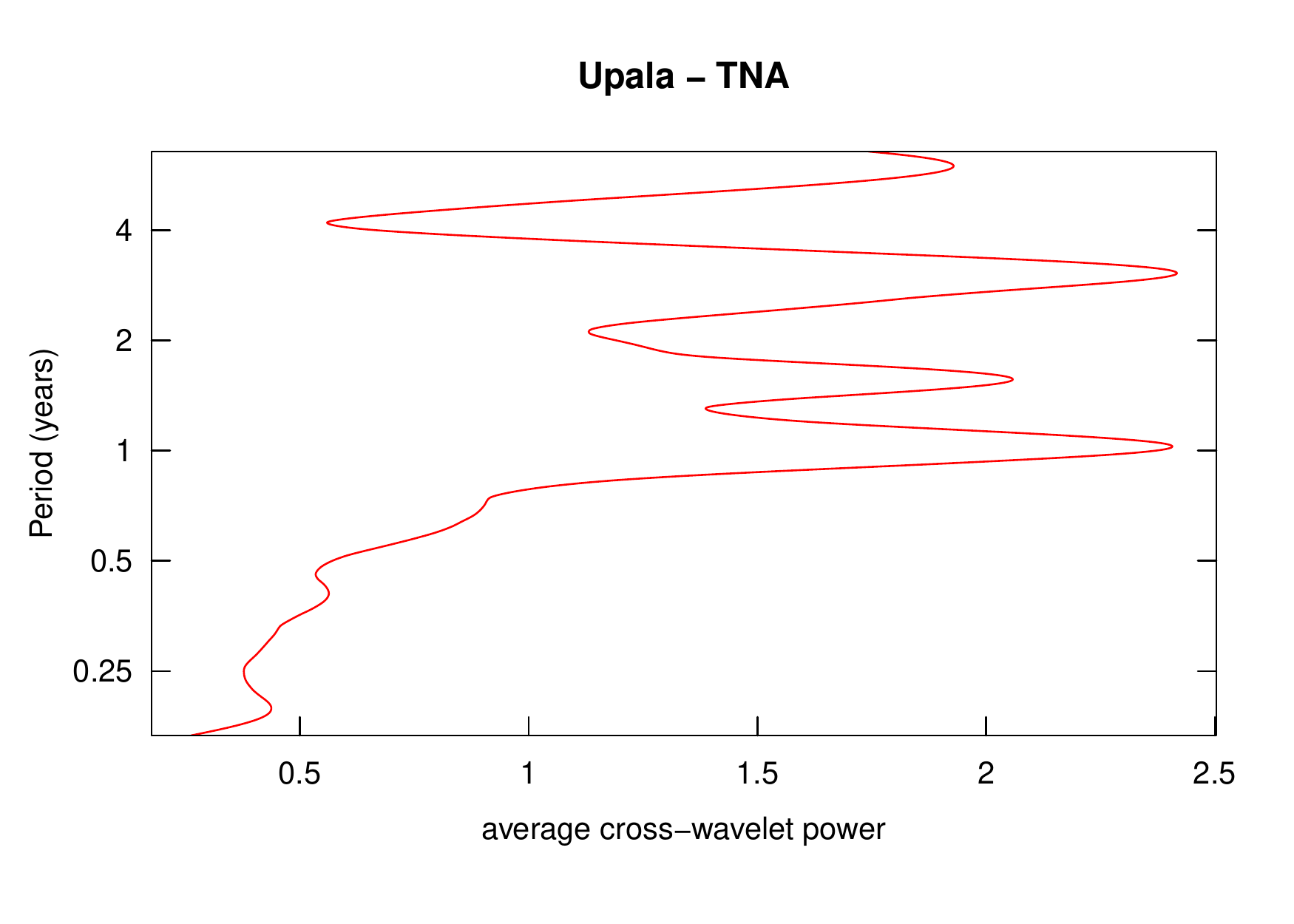}}\vspace{-0.15cm}\\
\subfloat[]{\includegraphics[scale=0.23]{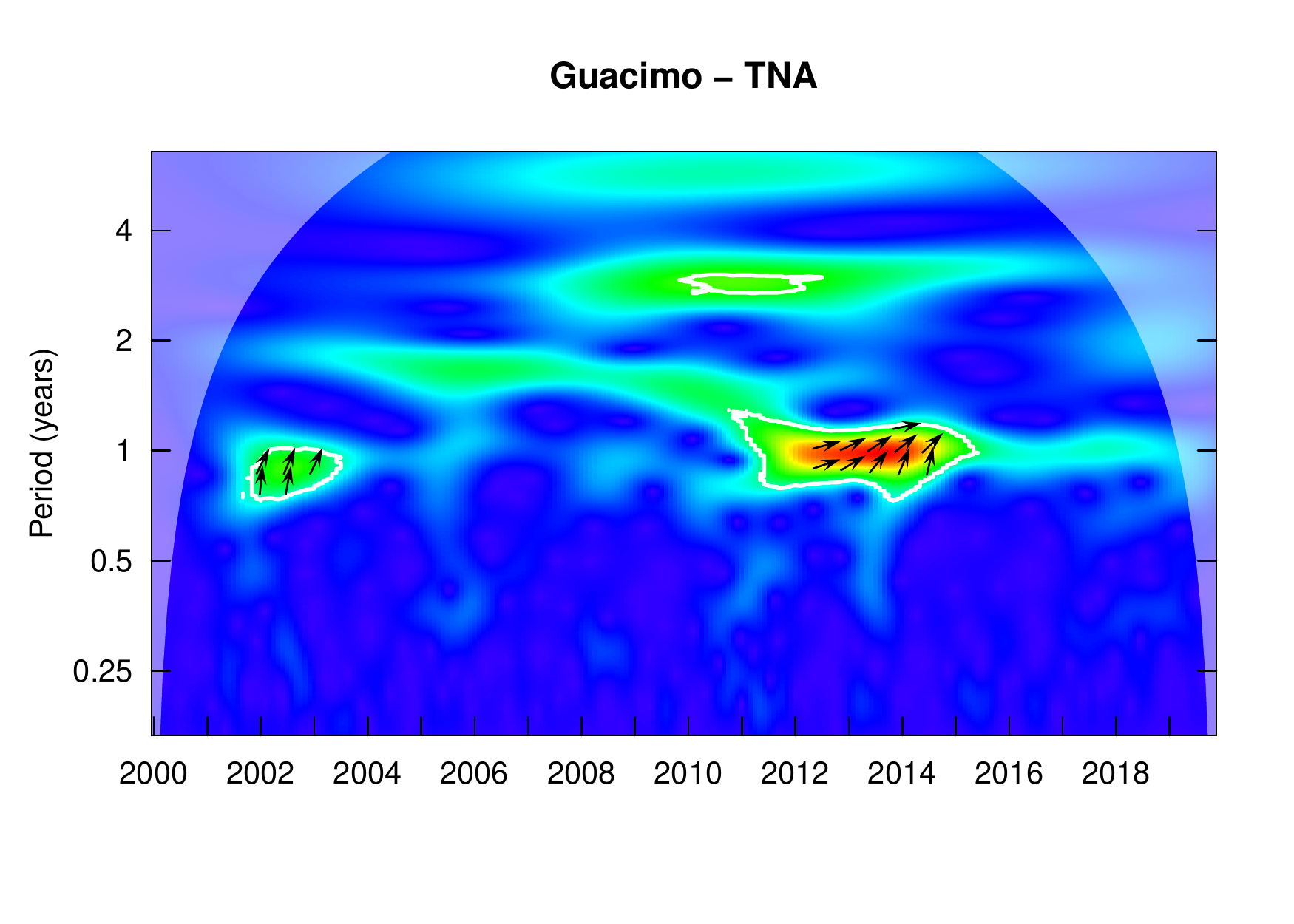}}\vspace{-0.15cm}%
\subfloat[]{\includegraphics[scale=0.23]{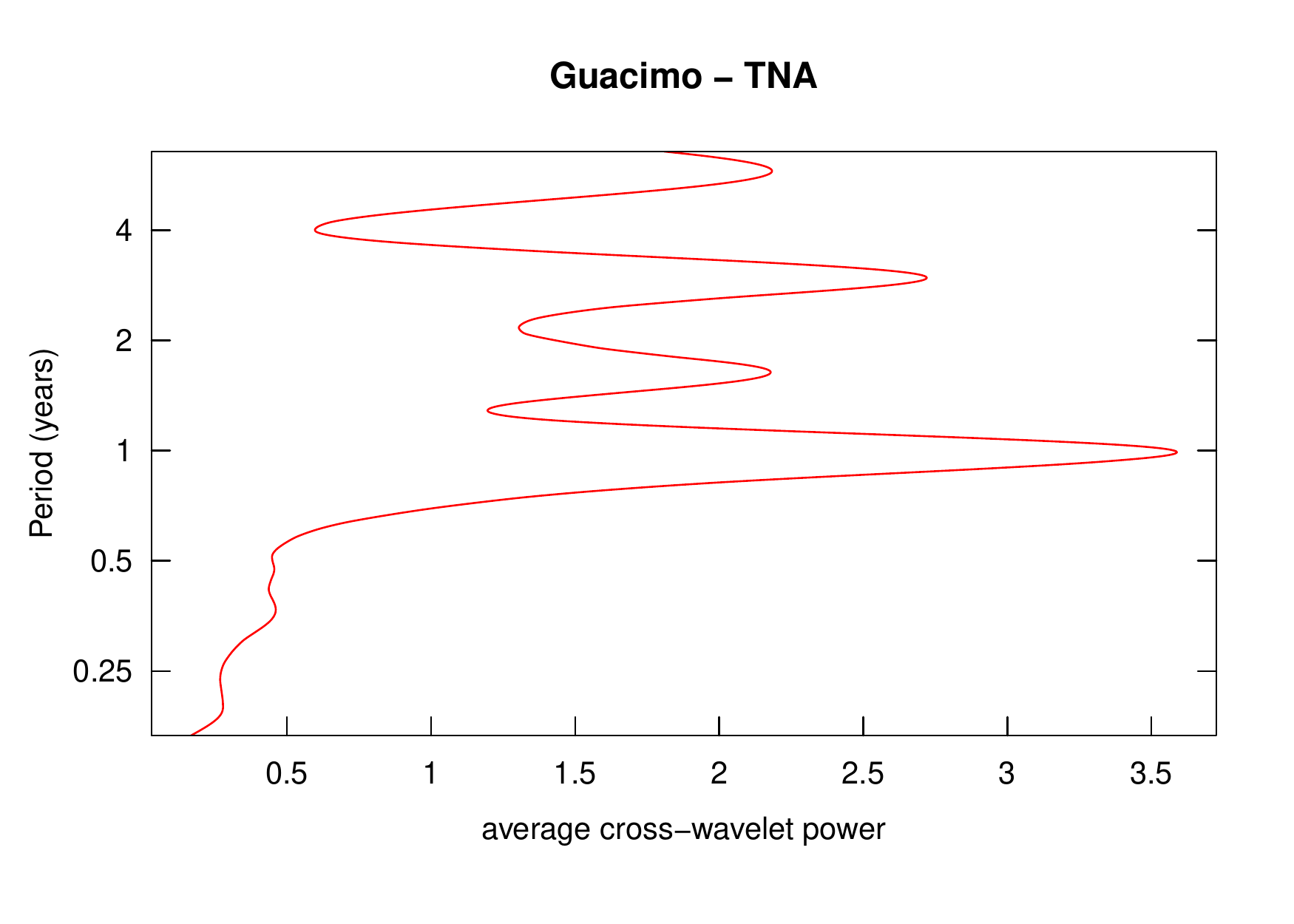}}\vspace{-0.15cm}%
\subfloat[]{\includegraphics[scale=0.23]{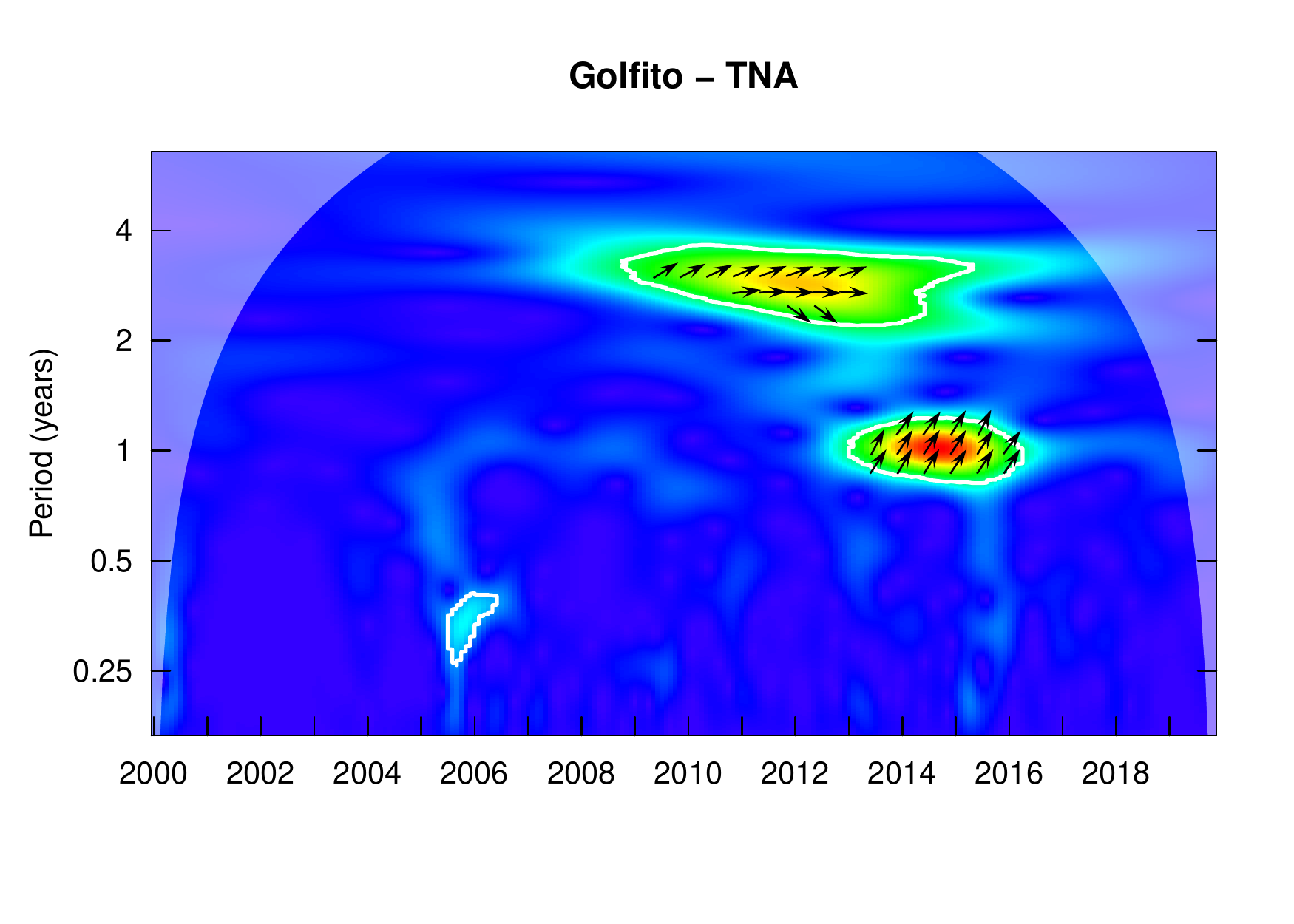}}\vspace{-0.15cm}%
\subfloat[]{\includegraphics[scale=0.23]{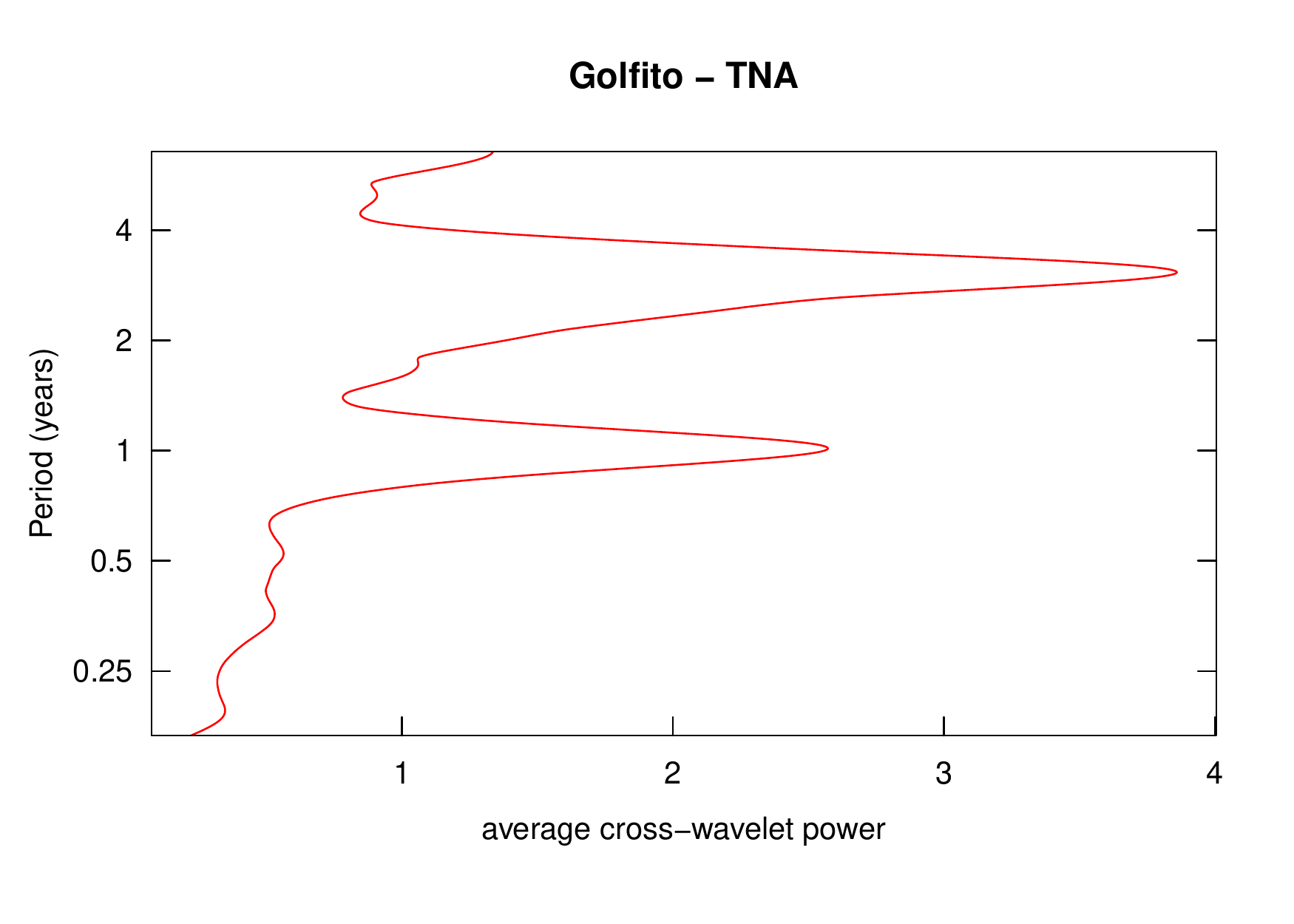}}\vspace{-0.15cm}\\
\subfloat[]{\includegraphics[scale=0.23]{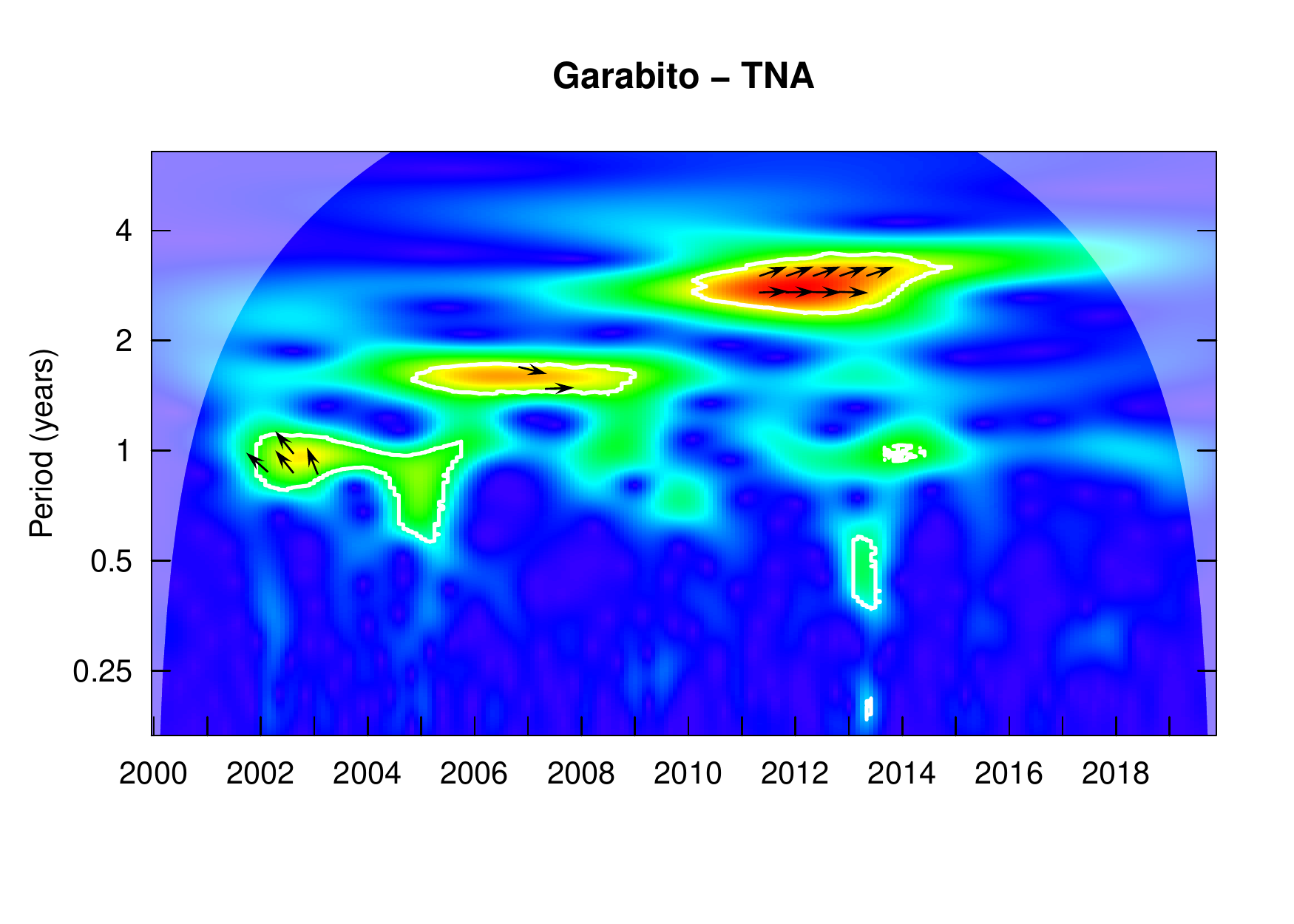}}\vspace{-0.15cm}%
\subfloat[]{\includegraphics[scale=0.23]{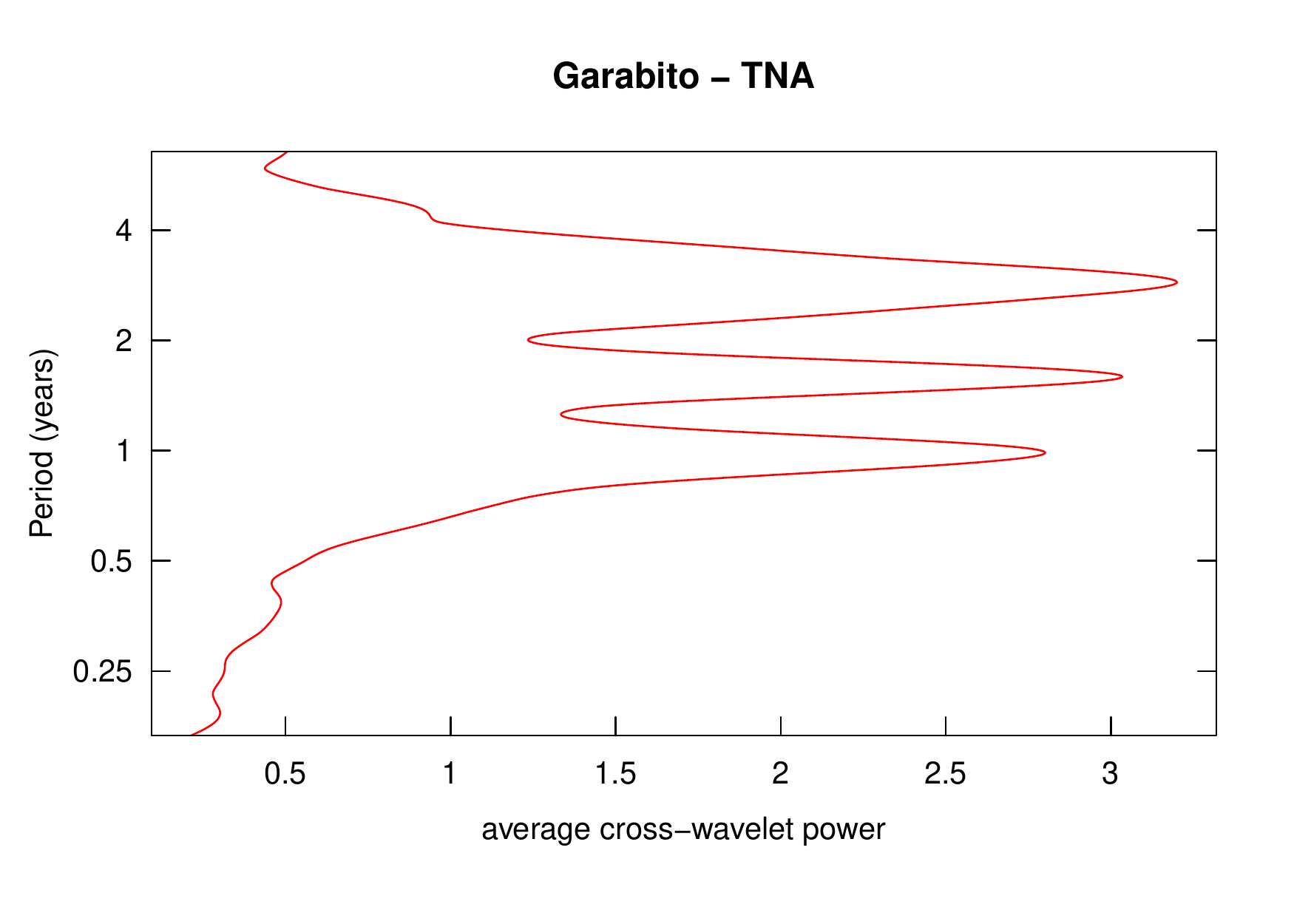}}\vspace{-0.15cm}%
\subfloat[]{\includegraphics[scale=0.23]{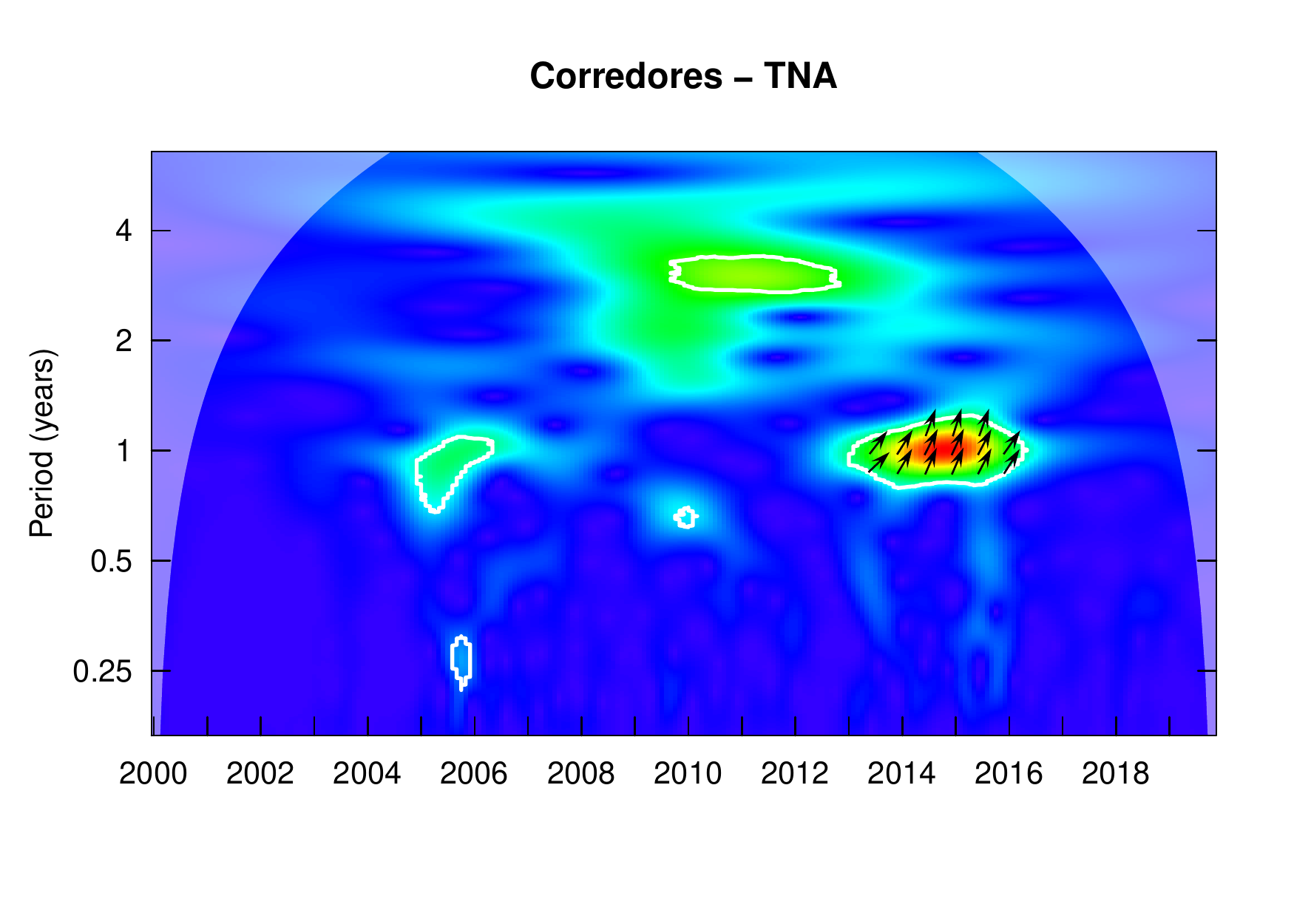}}\vspace{-0.15cm}%
\subfloat[]{\includegraphics[scale=0.23]{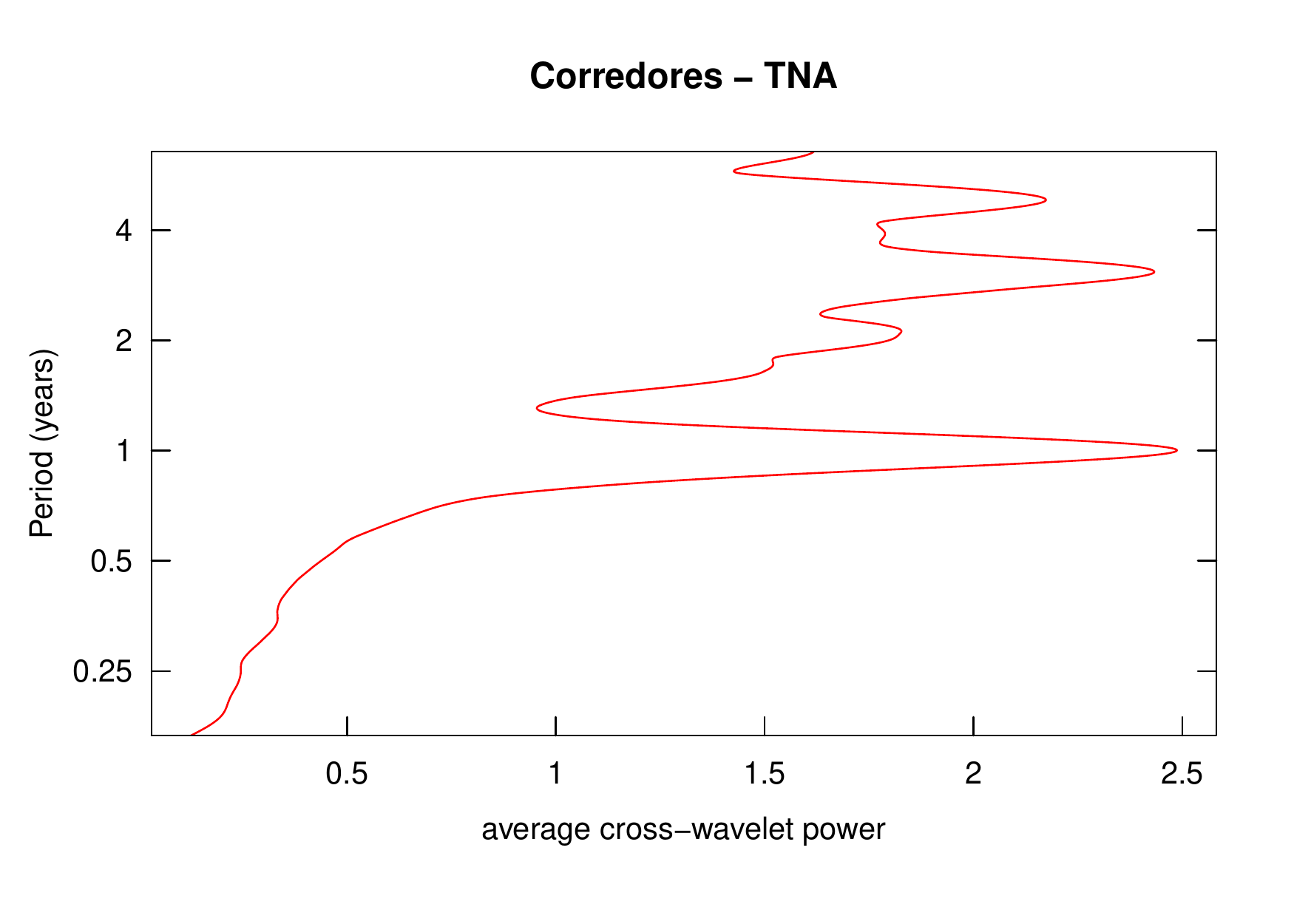}}\vspace{-0.15cm}\\
\subfloat[]{\includegraphics[scale=0.23]{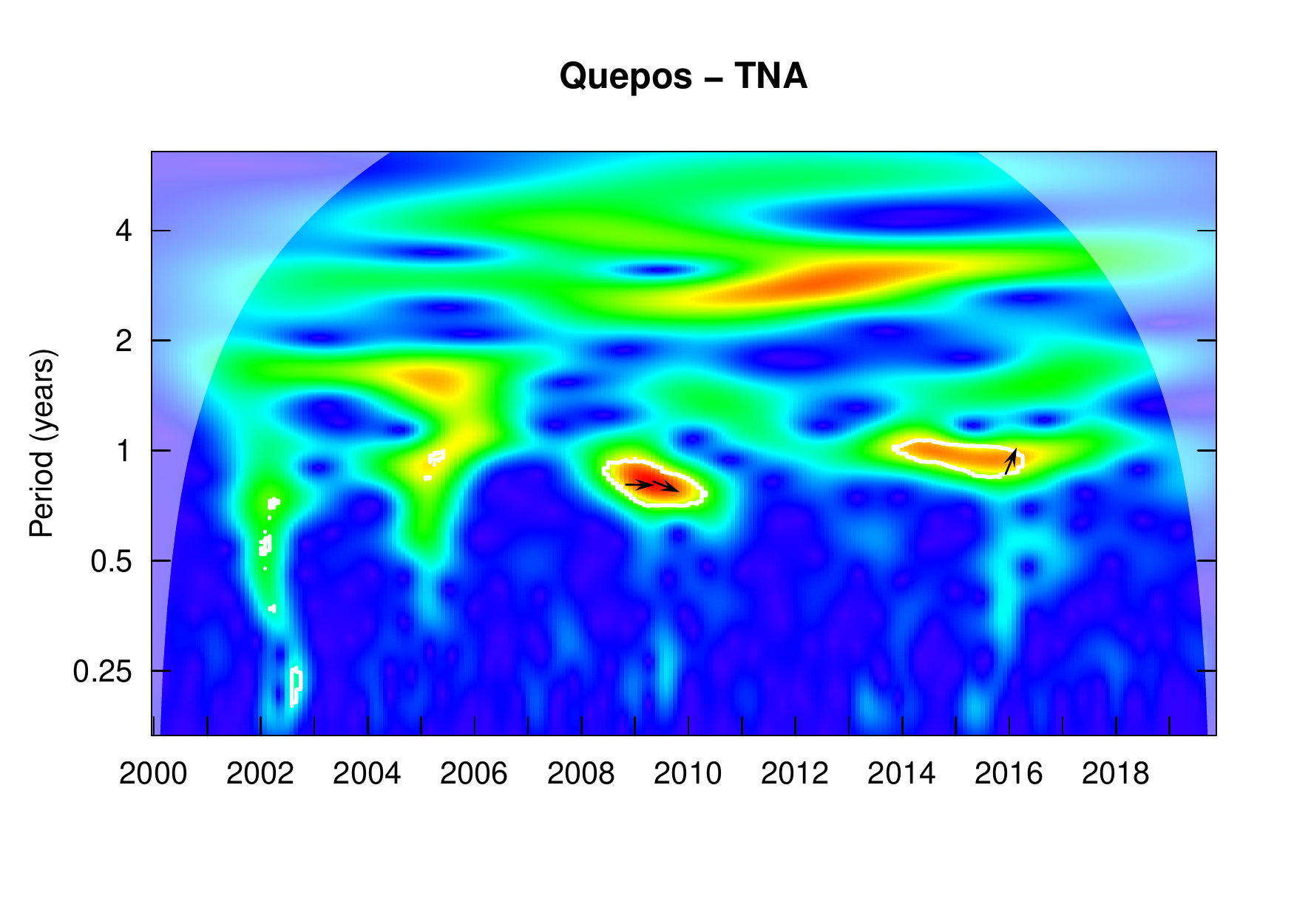}}\vspace{-0.15cm}%
\subfloat[]{\includegraphics[scale=0.23]{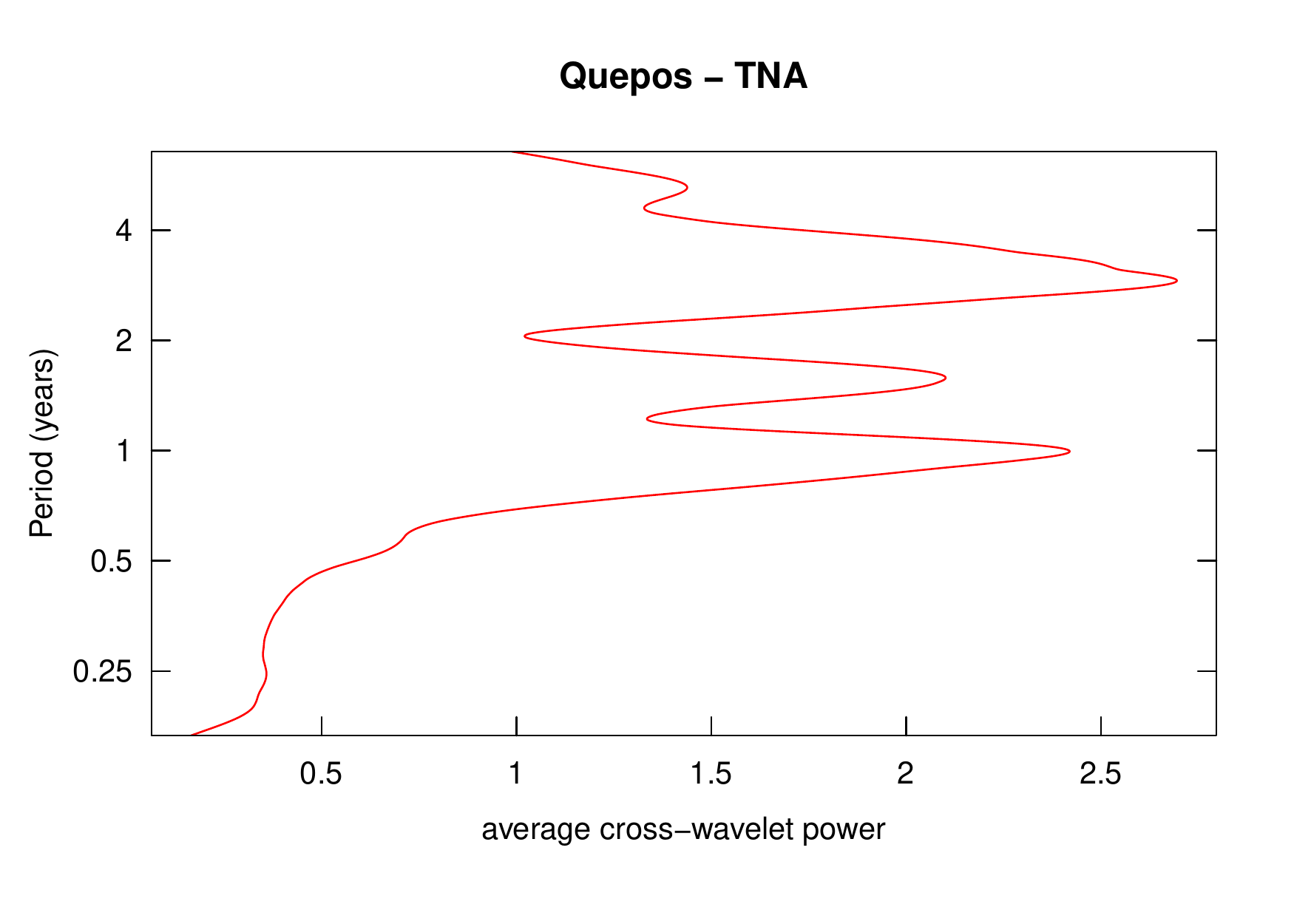}}\vspace{-0.15cm}%
\subfloat[]{\includegraphics[scale=0.23]{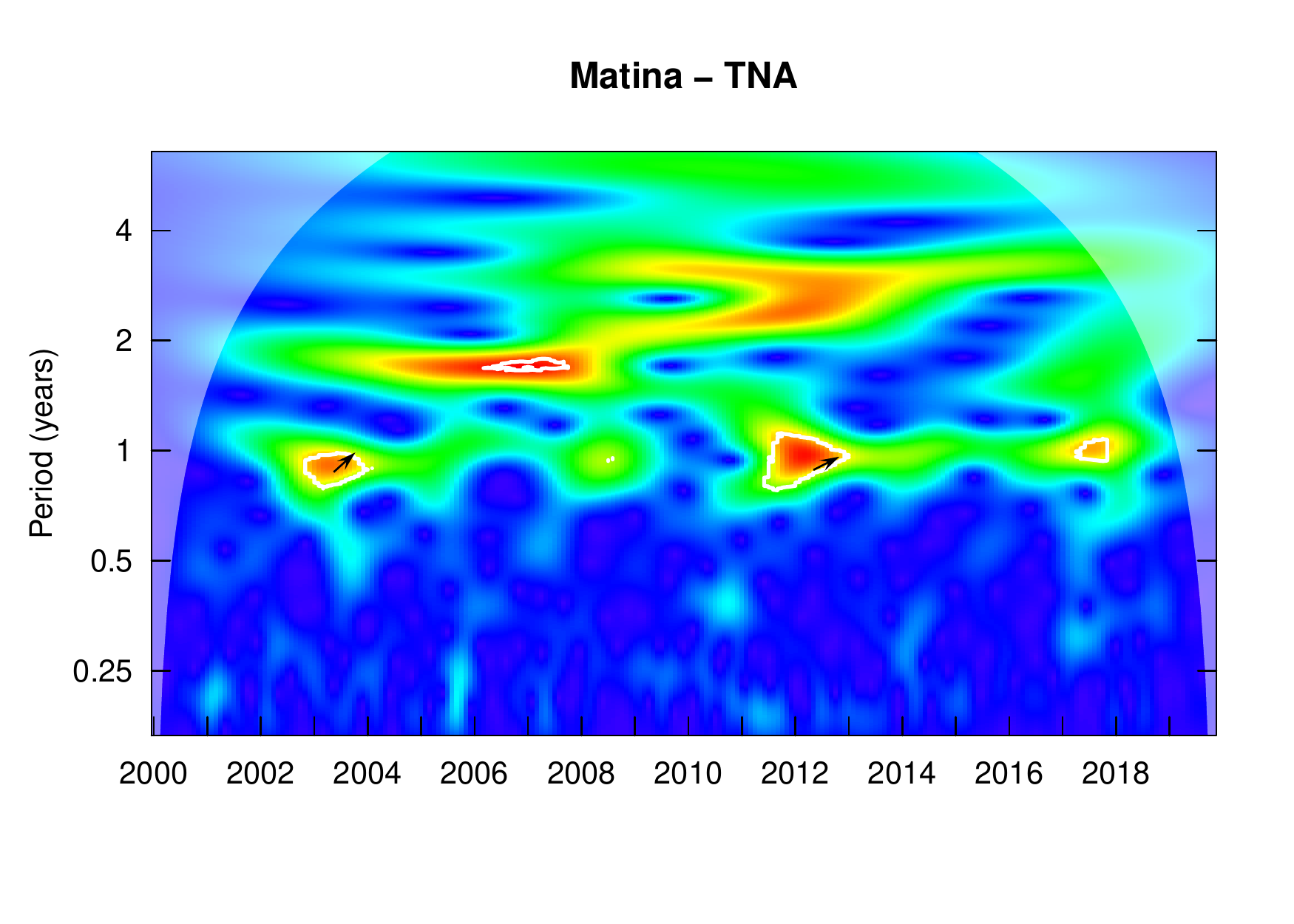}}\vspace{-0.15cm}%
\subfloat[]{\includegraphics[scale=0.23]{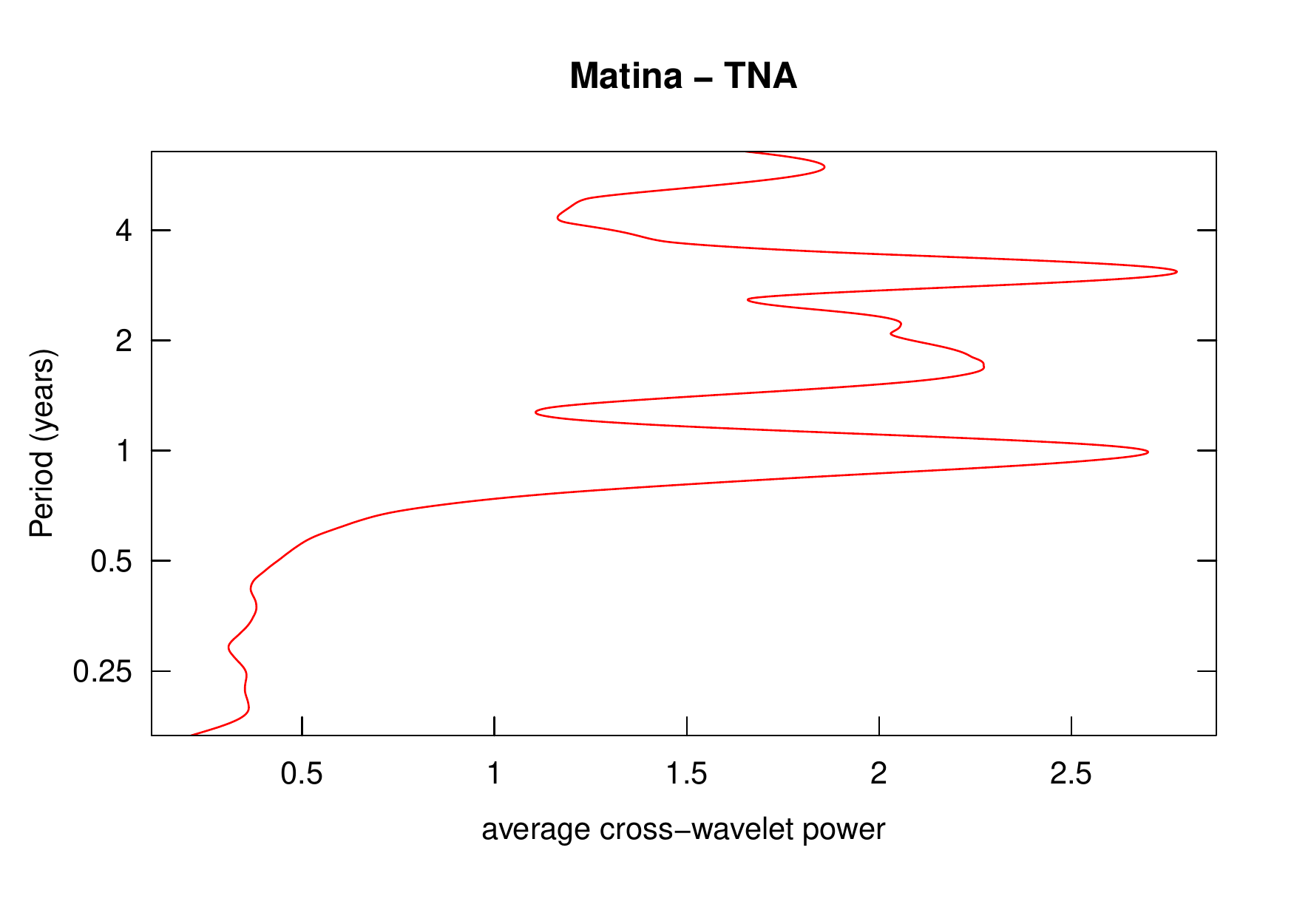}}\vspace{-0.15cm}\\
\subfloat[]{\includegraphics[scale=0.23]{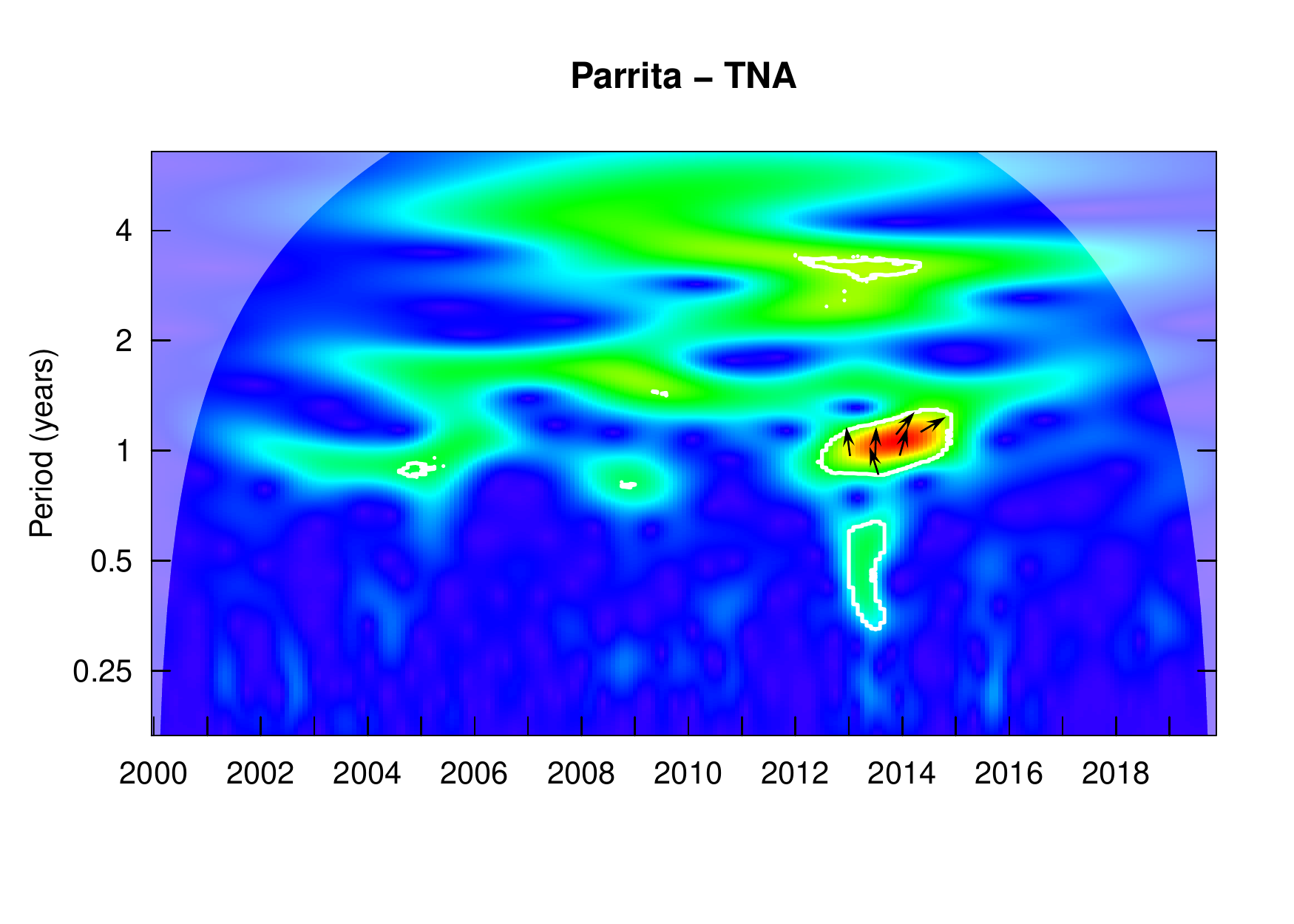}}\vspace{-0.15cm}%
\subfloat[]{\includegraphics[scale=0.23]{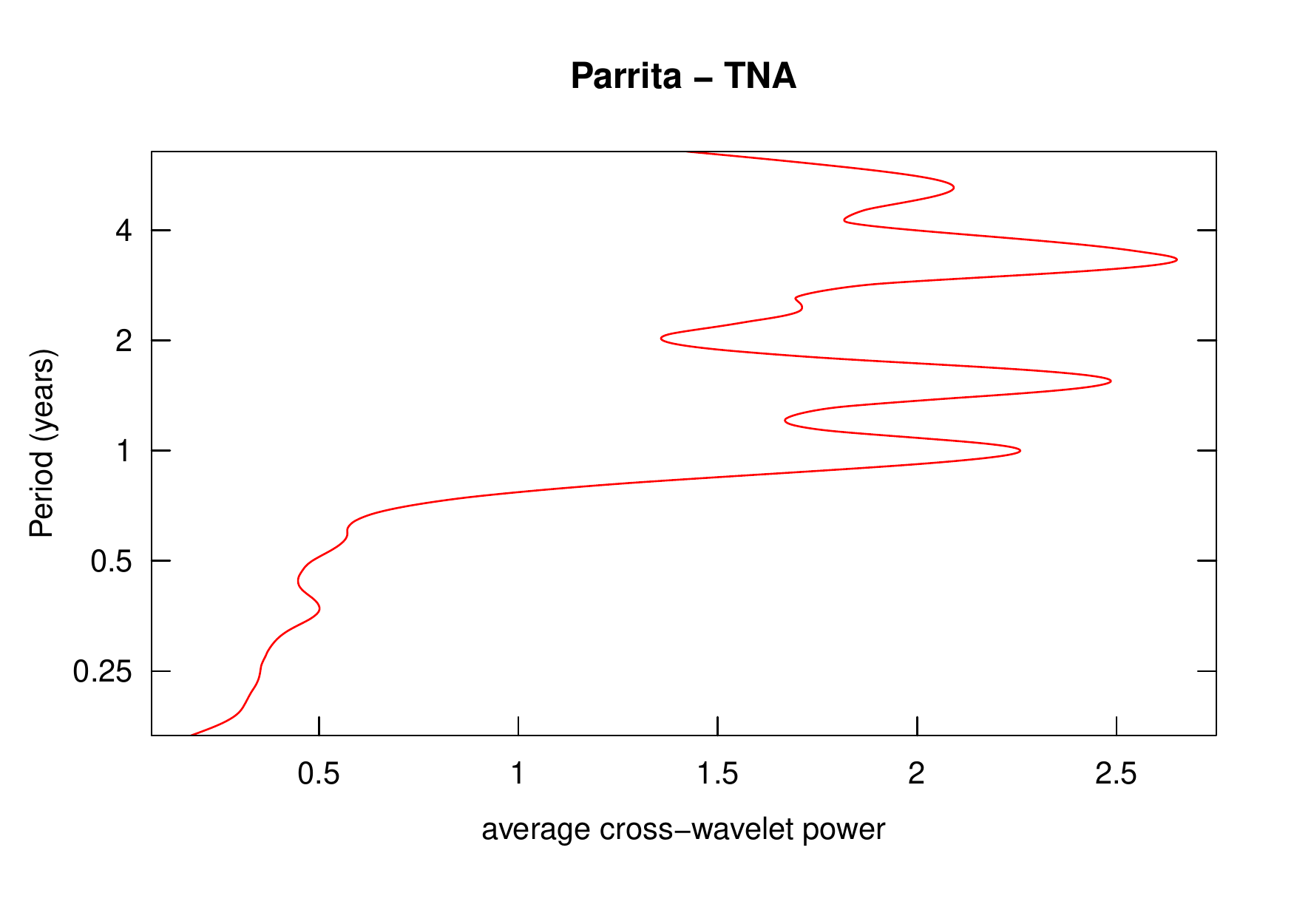}}\vspace{-0.15cm}%
\subfloat[]{\includegraphics[scale=0.23]{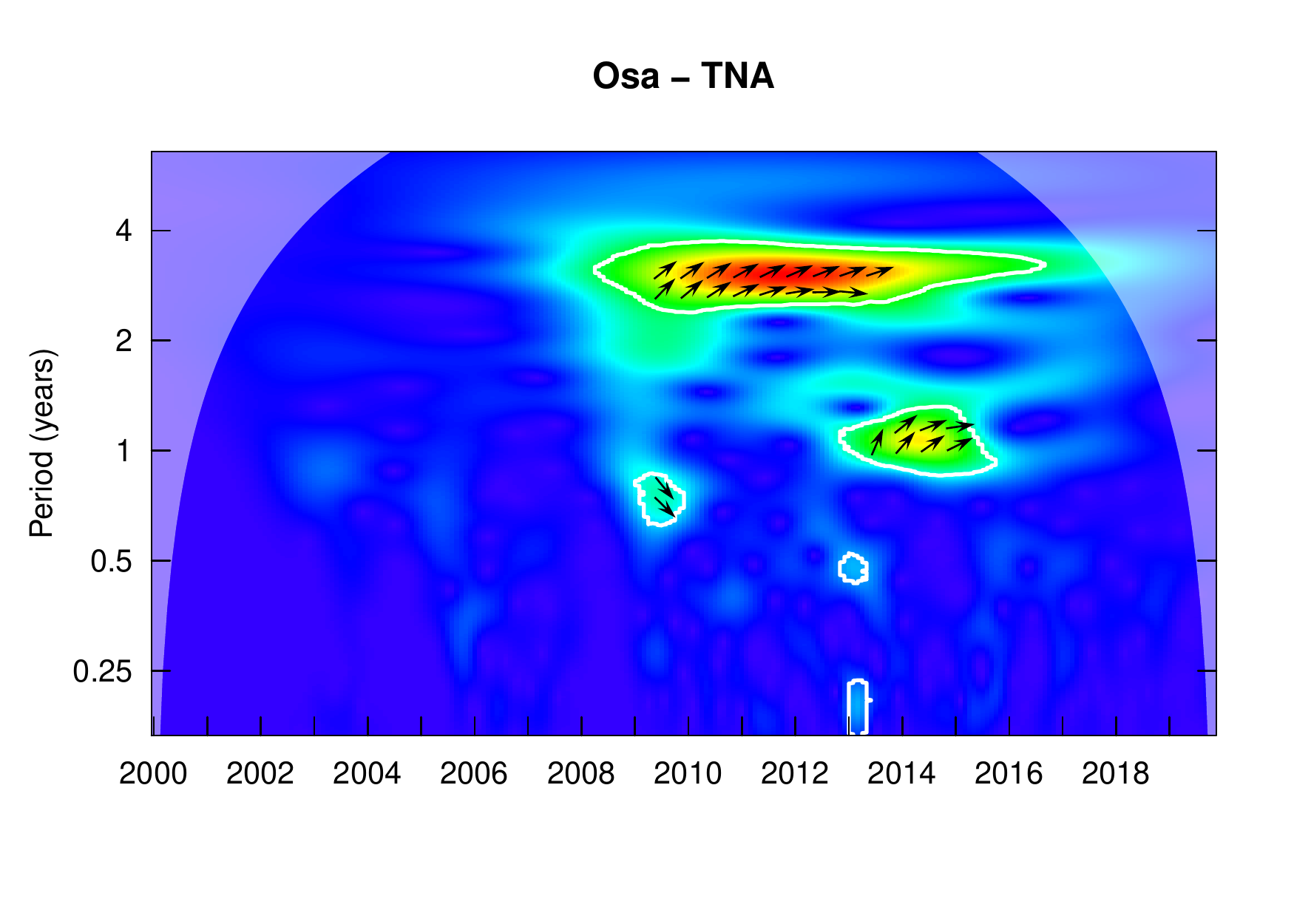}}\vspace{-0.15cm}%
\subfloat[]{\includegraphics[scale=0.23]{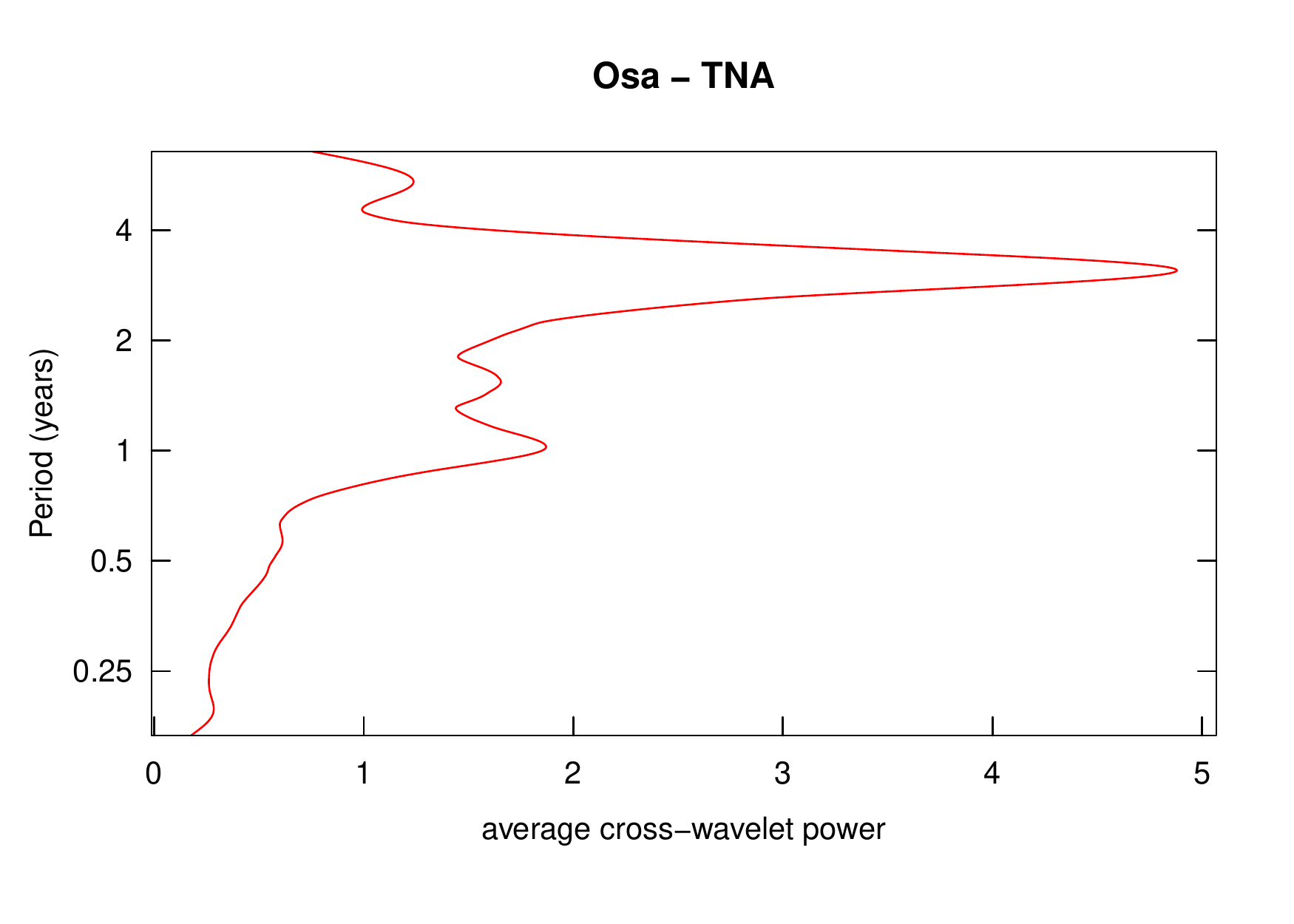}}\vspace{-0.15cm}
\caption*{}
\end{figure}


\section*{Wavelet coherence and average cross-wavelet power between dengue incidence and Ni\~no 1+2}

\begin{figure}[H]
\captionsetup[subfigure]{labelformat=empty}
\caption*{\textbf{Figure S7:} Wavelet coherence (color map) between dengue incidence from 2000 to 2019, and Ni\~no 1+2 in 32 municipalities of Costa Rica (periodicity on y-axis, time on x-axis). Colors code for increasing power intensity, from blue to red; $95\%$ confidence levels are encircled by white lines, and shaded areas indicate the presence of significant edge effects. On the right side of each wavelet coherence is the average cross-wavelet power (Red line). The arrows indicate whether the two series are in-phase or out-phase.}
\subfloat[]{\includegraphics[scale=0.23]{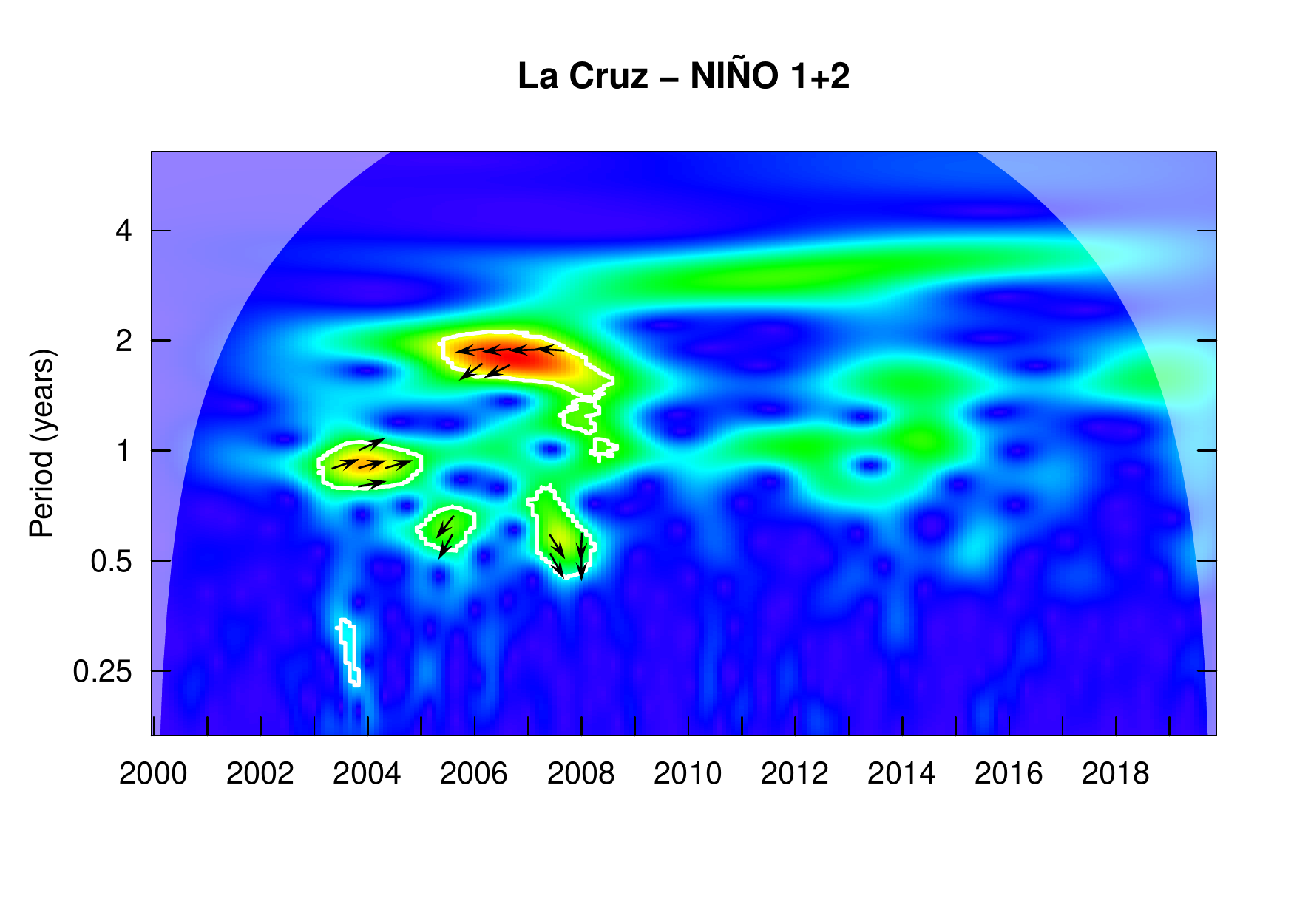}}\vspace{-0.15cm}%
\subfloat[]{\includegraphics[scale=0.23]{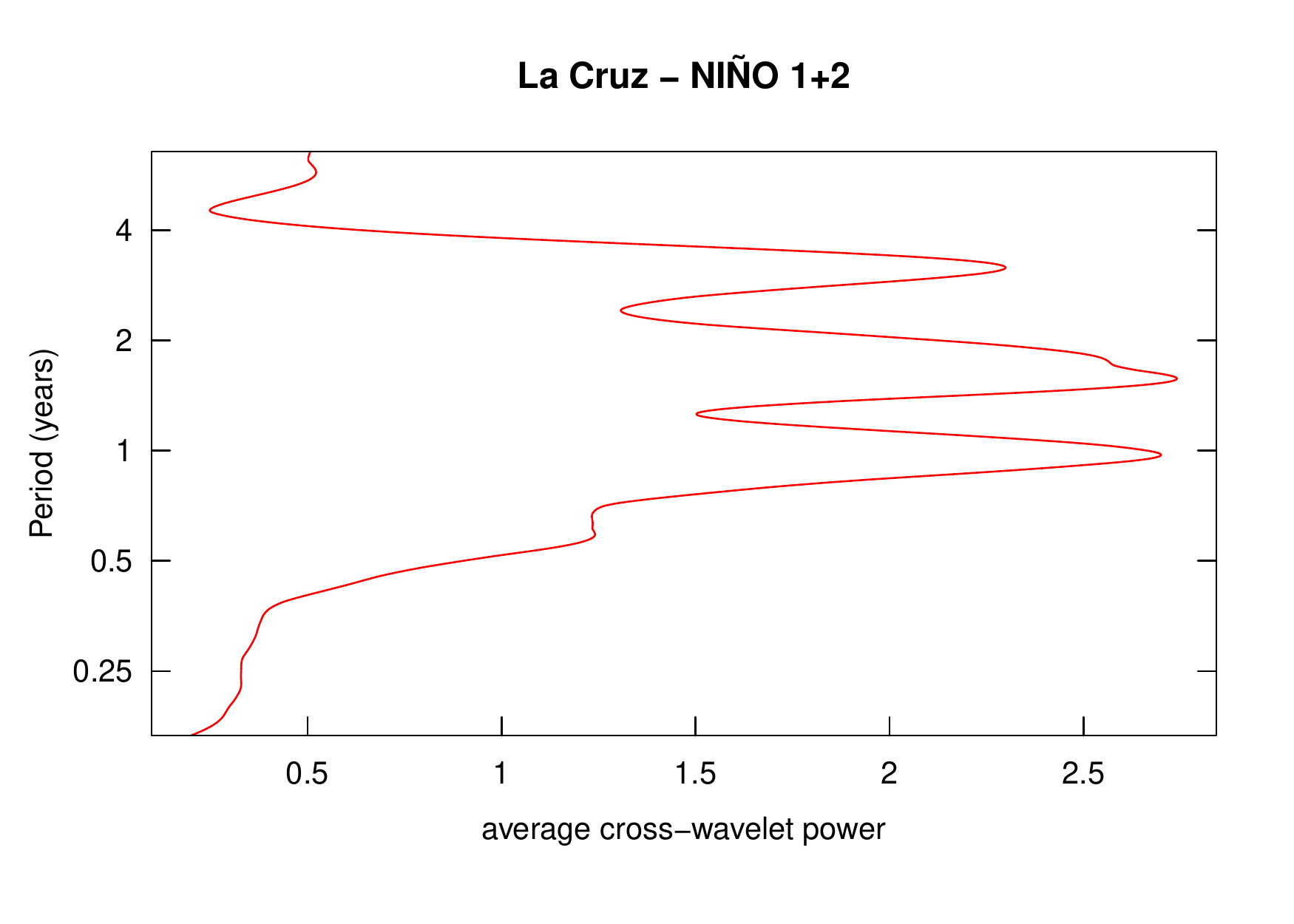}}\vspace{-0.15cm}%
\subfloat[]{\includegraphics[scale=0.23]{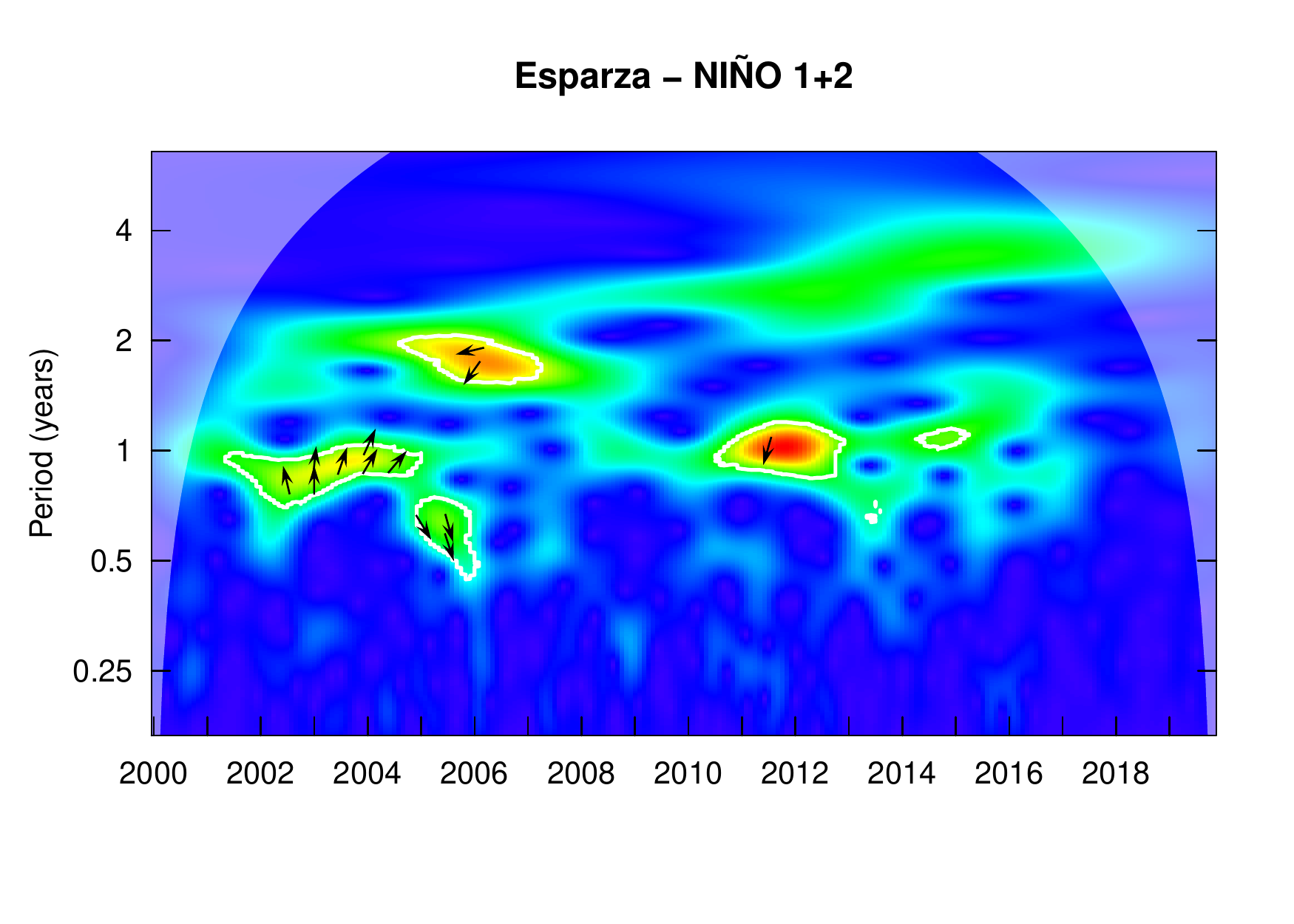}}\vspace{-0.15cm}%
\subfloat[]{\includegraphics[scale=0.23]{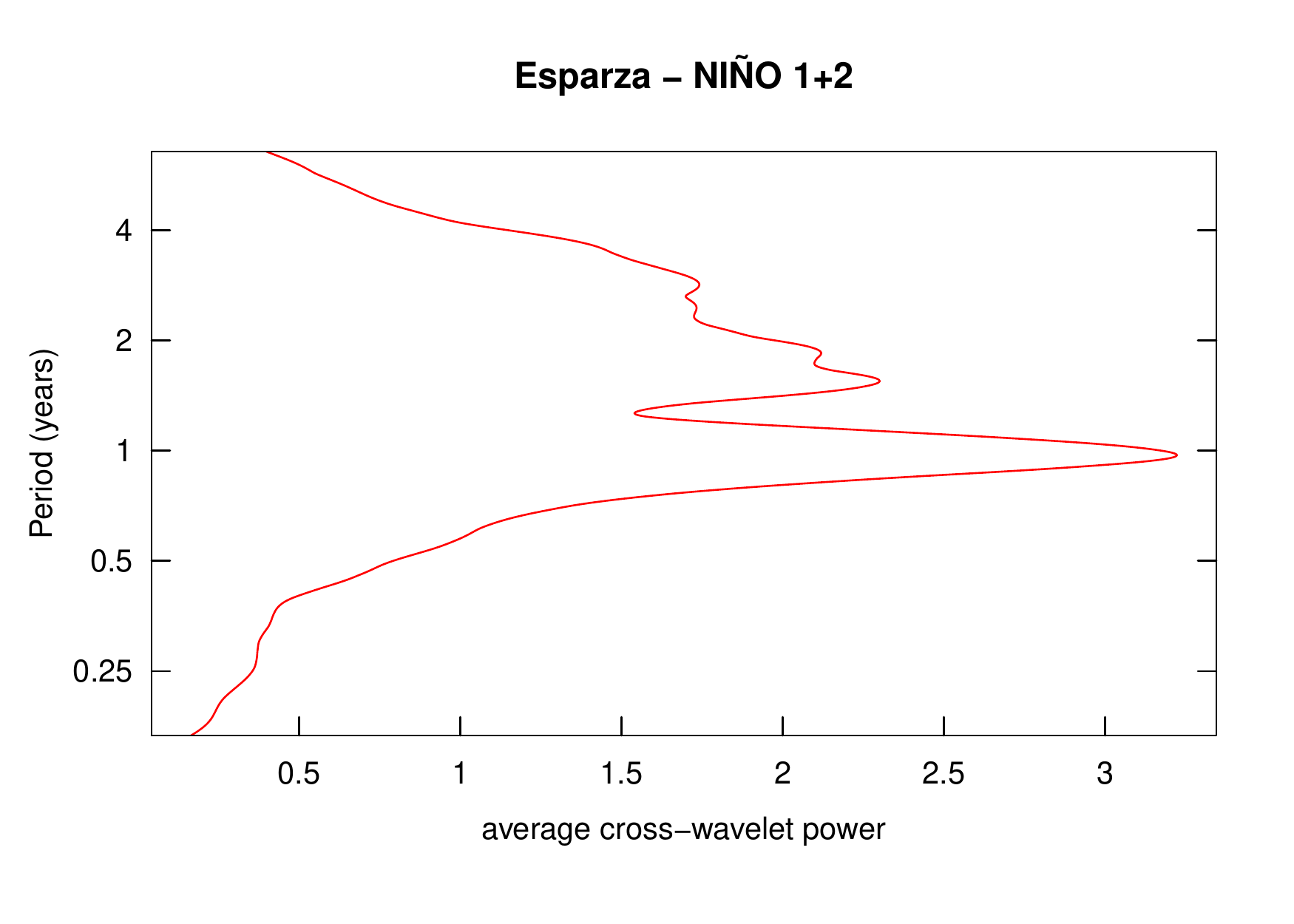}}\vspace{-0.15cm}\\
\subfloat[]{\includegraphics[scale=0.23]{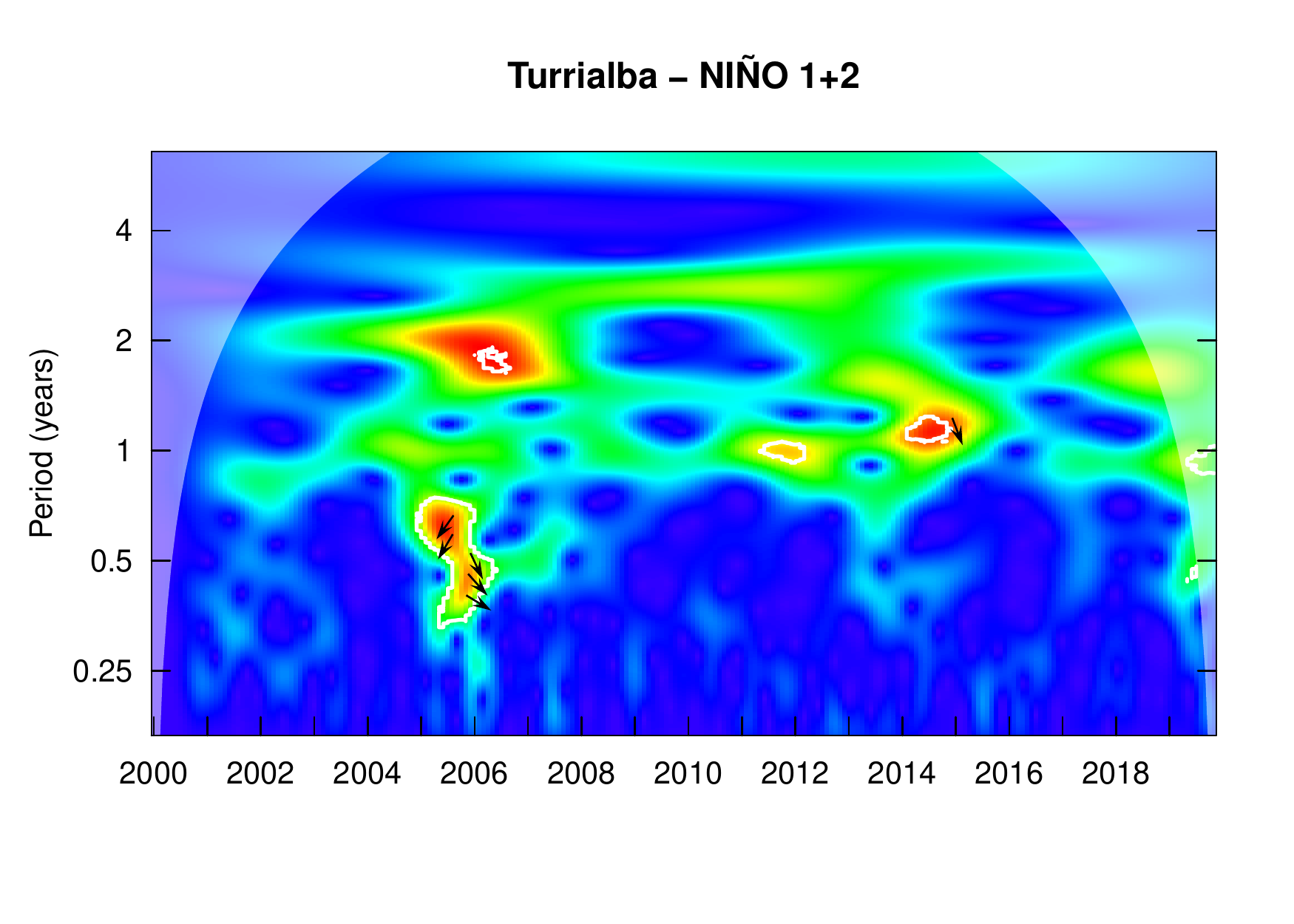}}\vspace{-0.15cm}%
\subfloat[]{\includegraphics[scale=0.23]{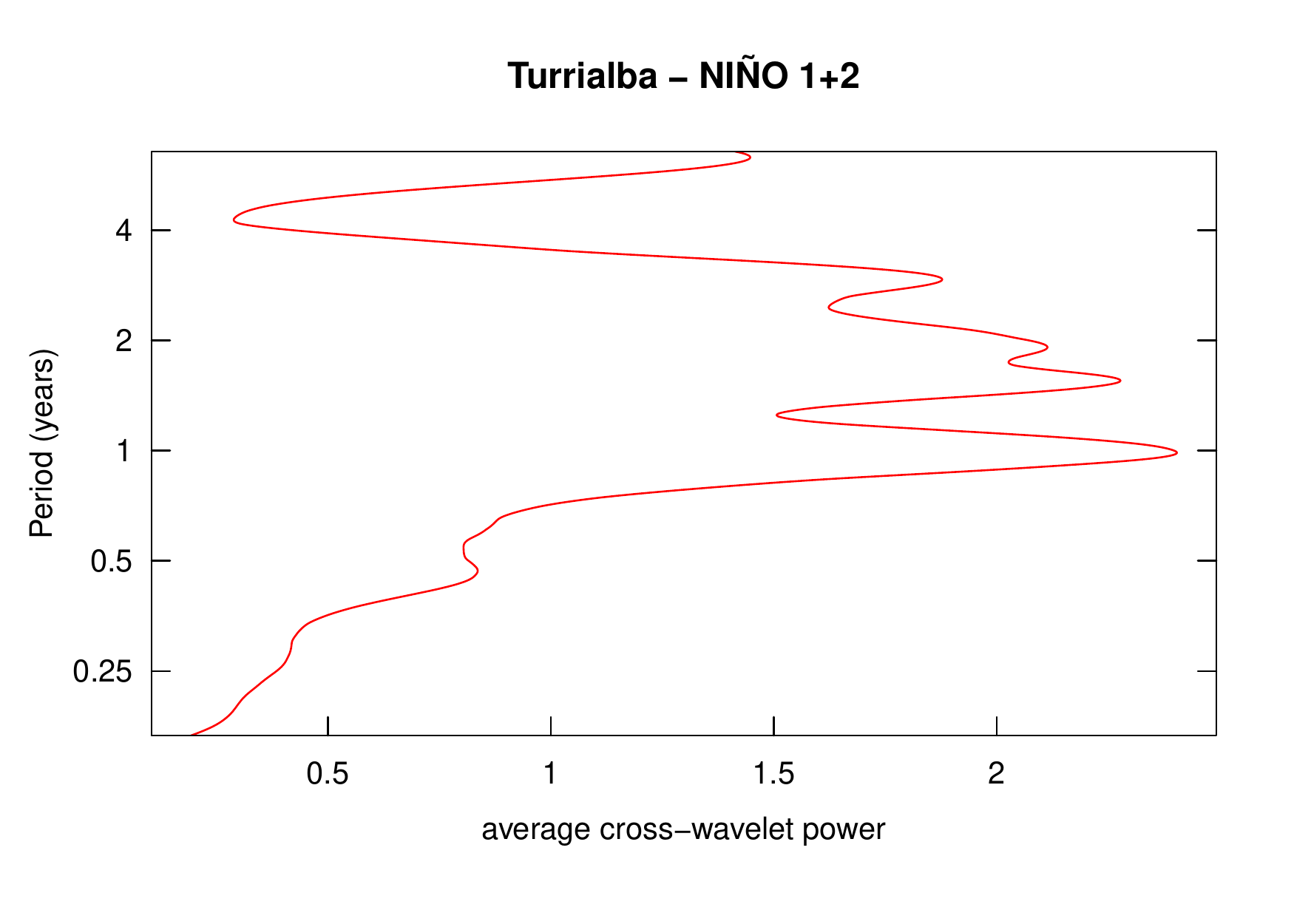}}\vspace{-0.15cm}%
\subfloat[]{\includegraphics[scale=0.23]{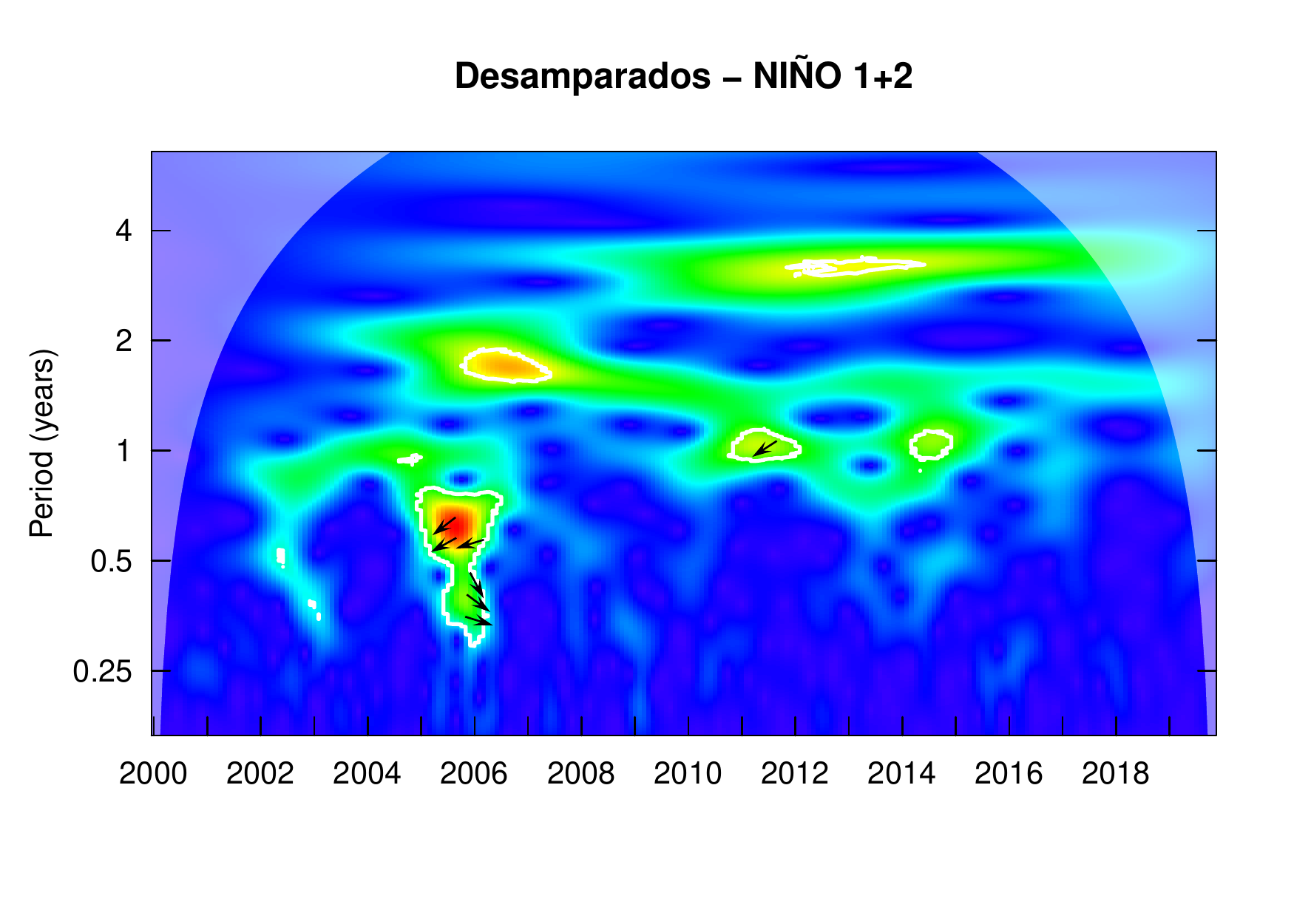}}\vspace{-0.15cm}%
\subfloat[]{\includegraphics[scale=0.23]{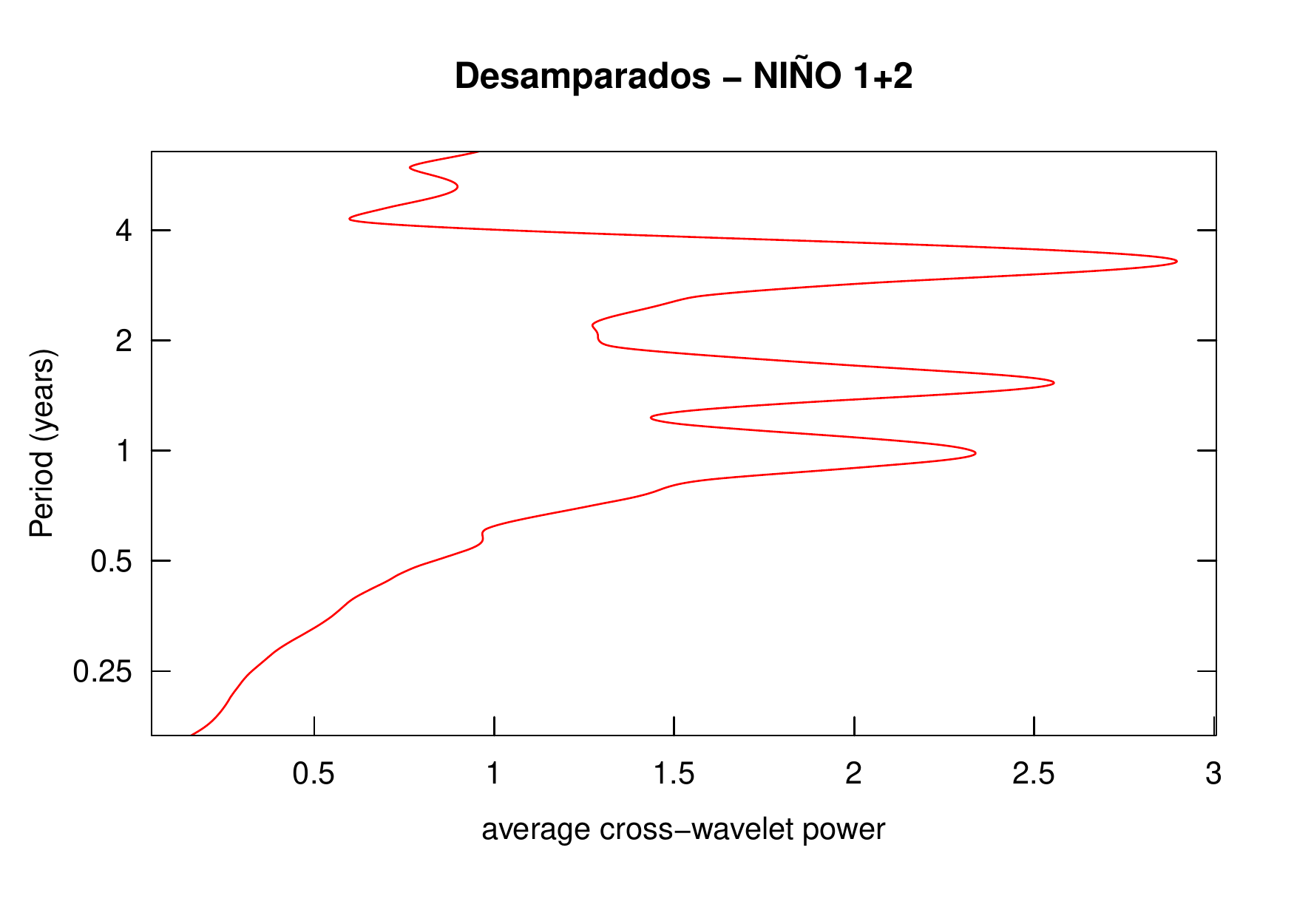}}\vspace{-0.15cm}\\
\subfloat[]{\includegraphics[scale=0.23]{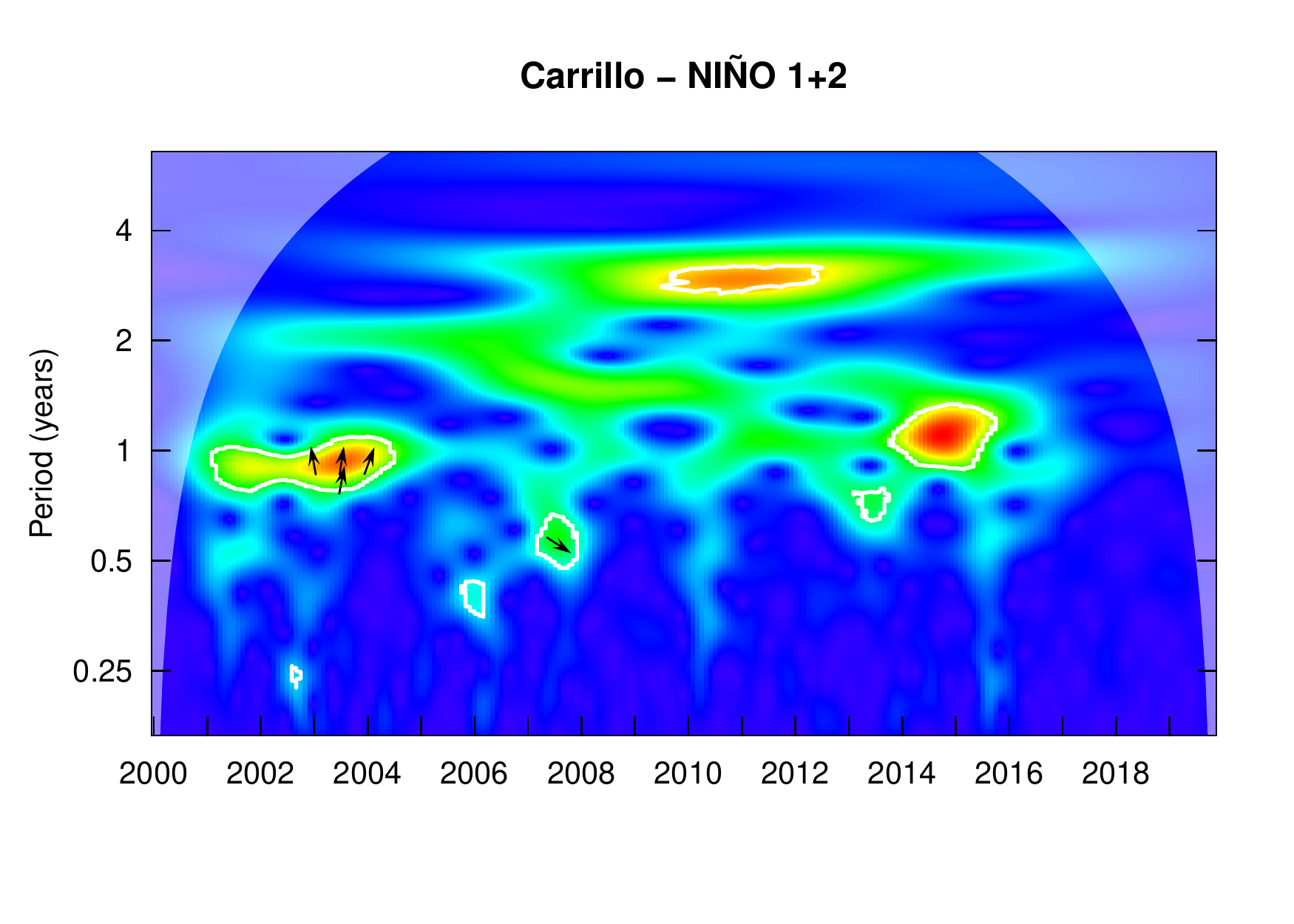}}\vspace{-0.15cm}%
\subfloat[]{\includegraphics[scale=0.23]{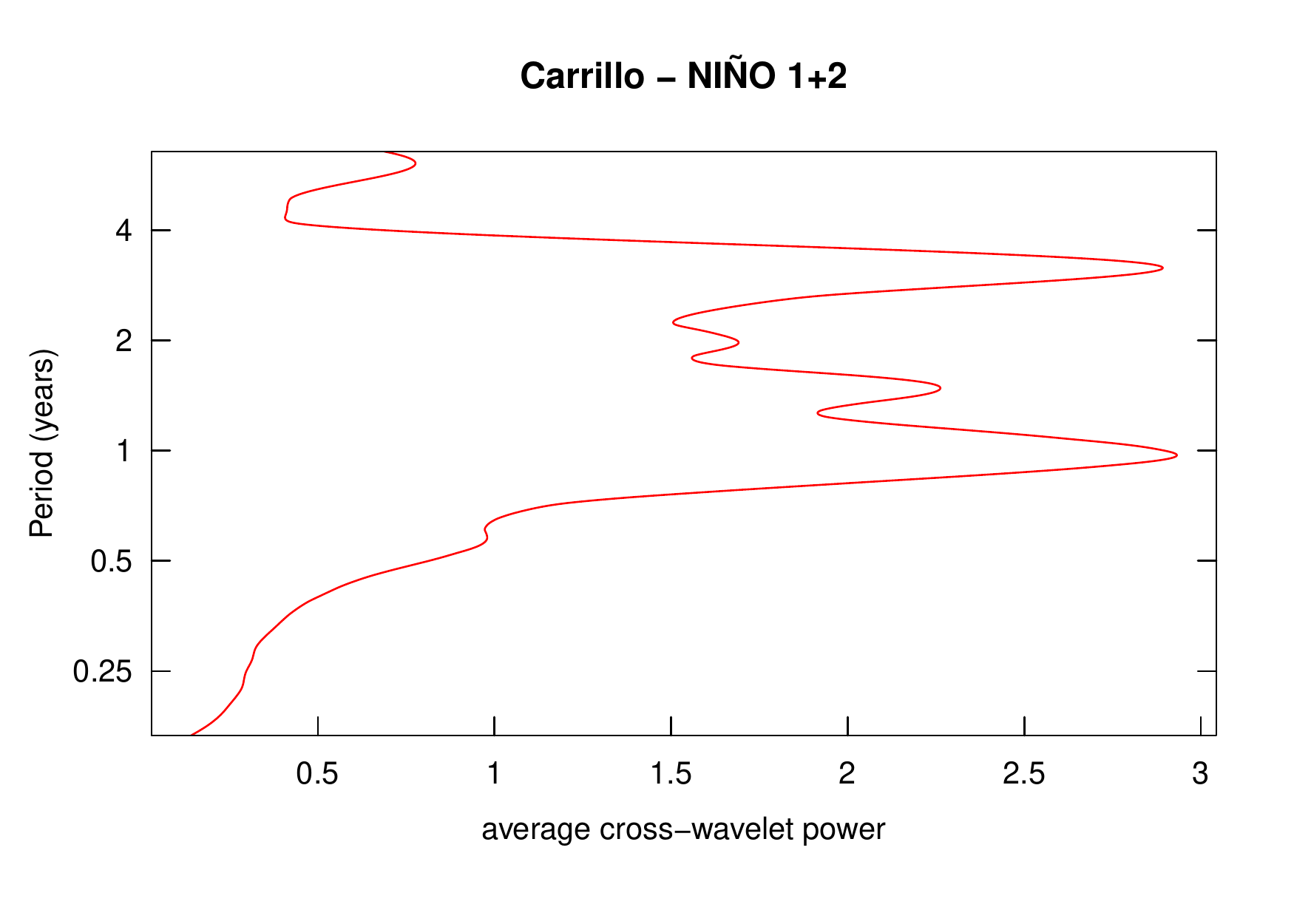}}\vspace{-0.15cm}%
\subfloat[]{\includegraphics[scale=0.23]{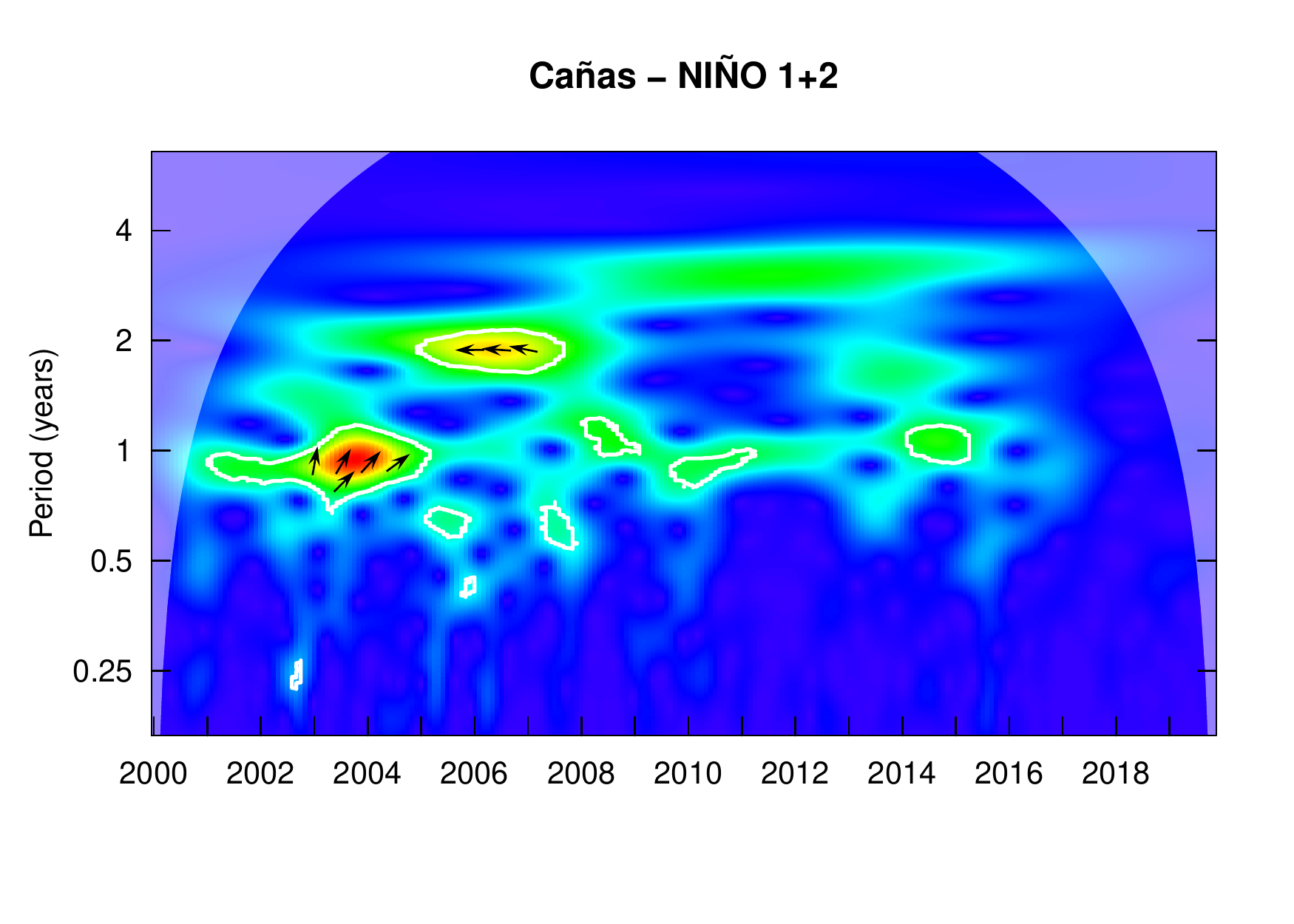}}\vspace{-0.15cm}%
\subfloat[]{\includegraphics[scale=0.23]{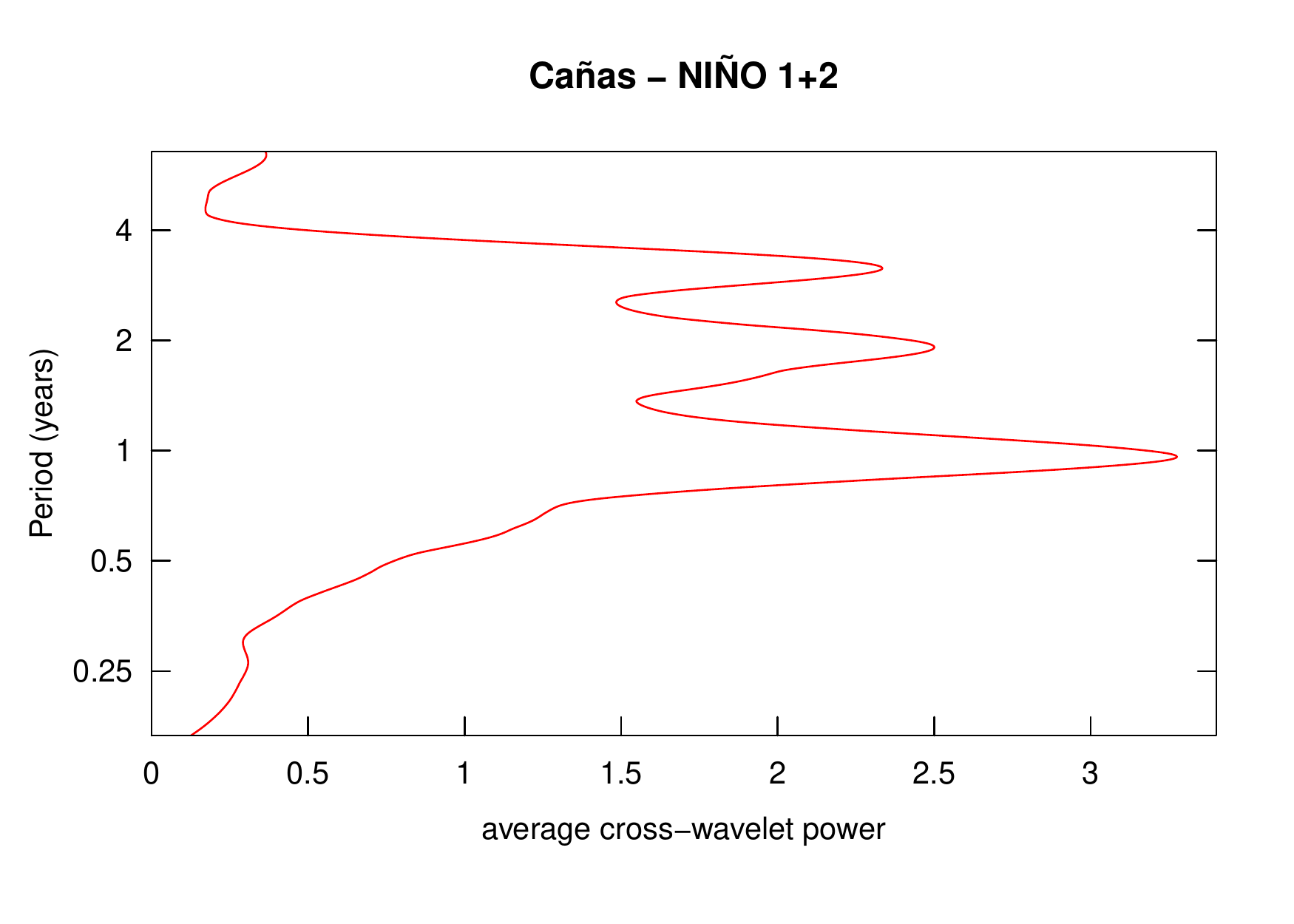}}\vspace{-0.15cm}\\
\subfloat[]{\includegraphics[scale=0.23]{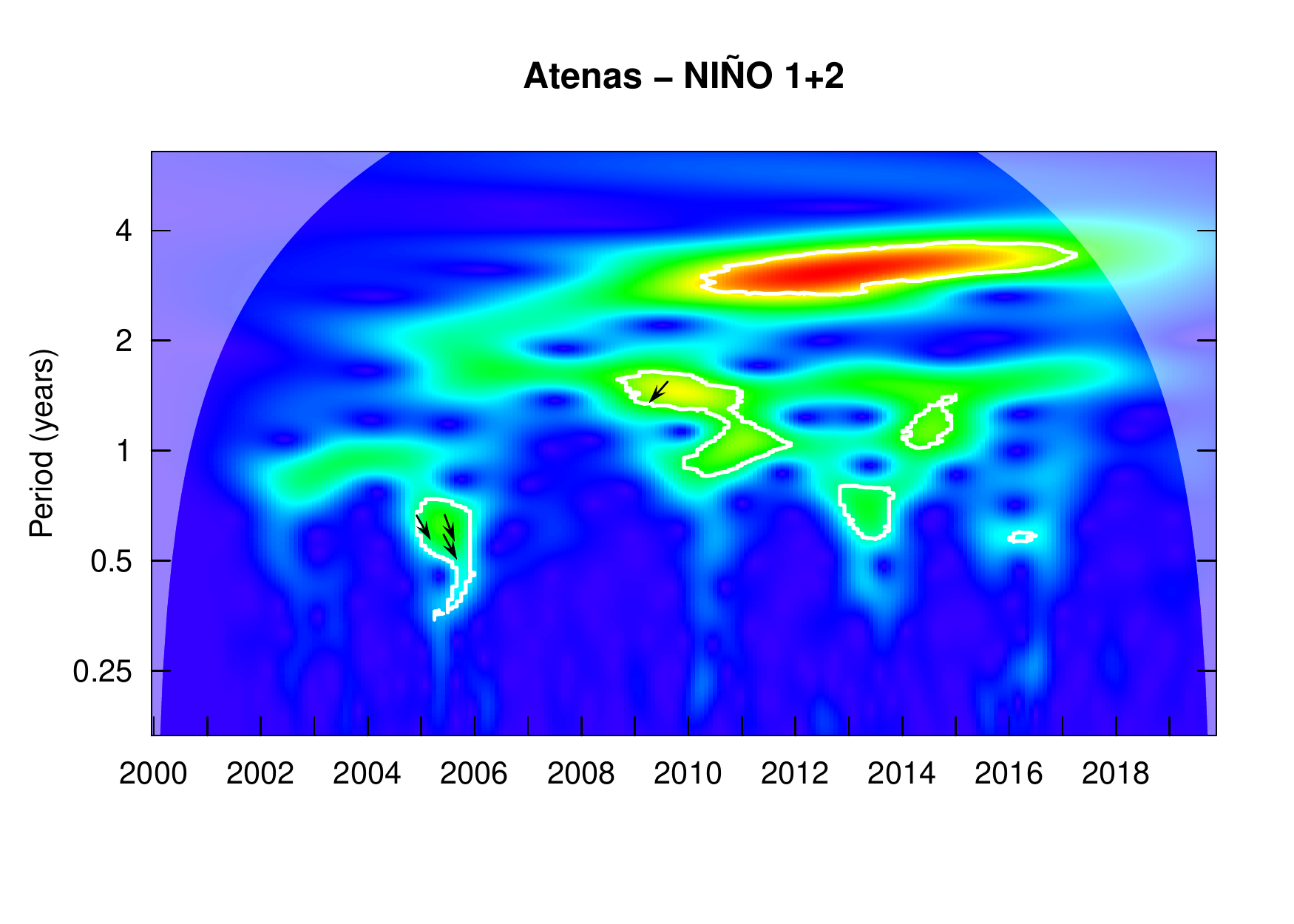}}\vspace{-0.15cm}%
\subfloat[]{\includegraphics[scale=0.23]{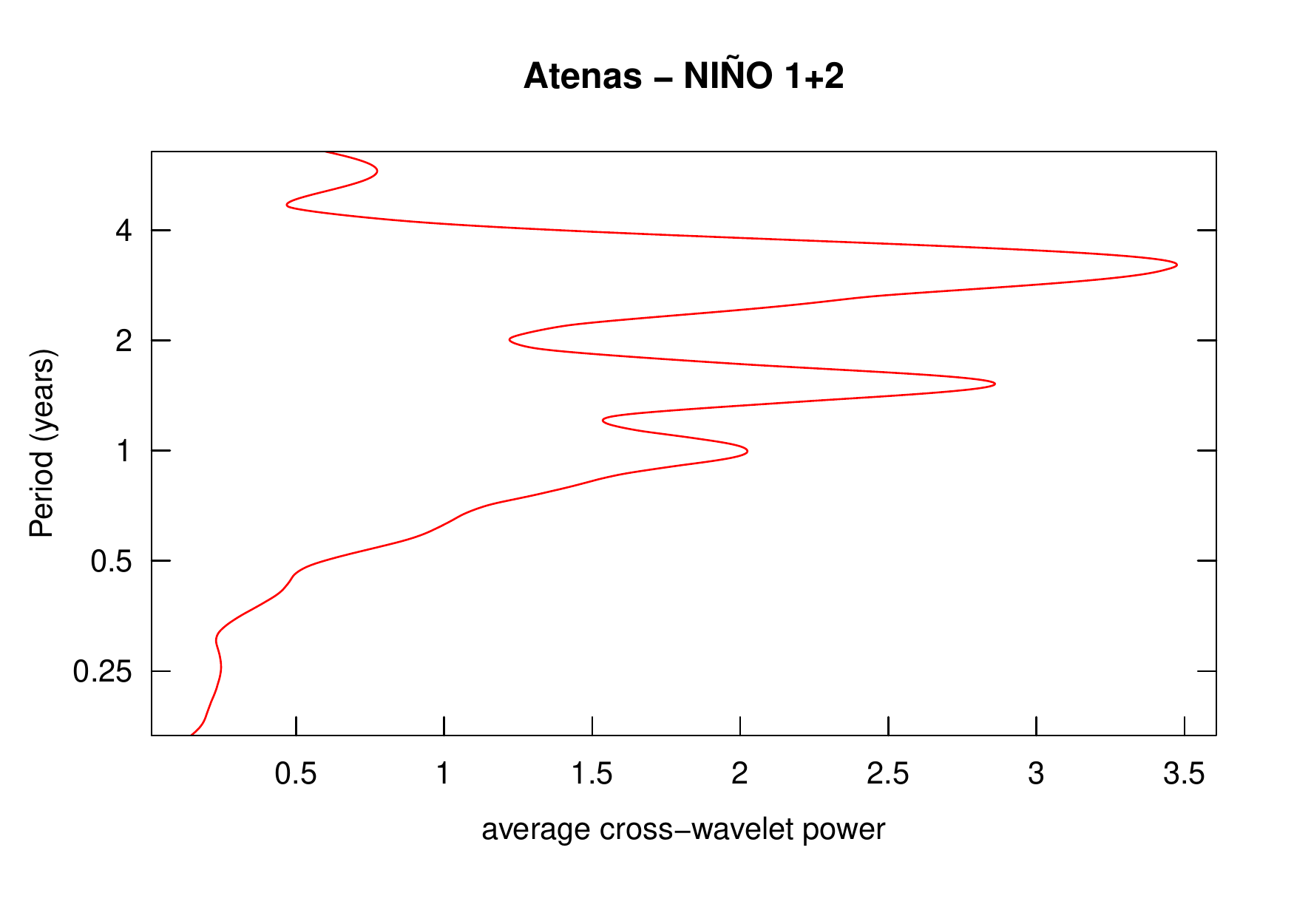}}\vspace{-0.15cm}%
\subfloat[]{\includegraphics[scale=0.23]{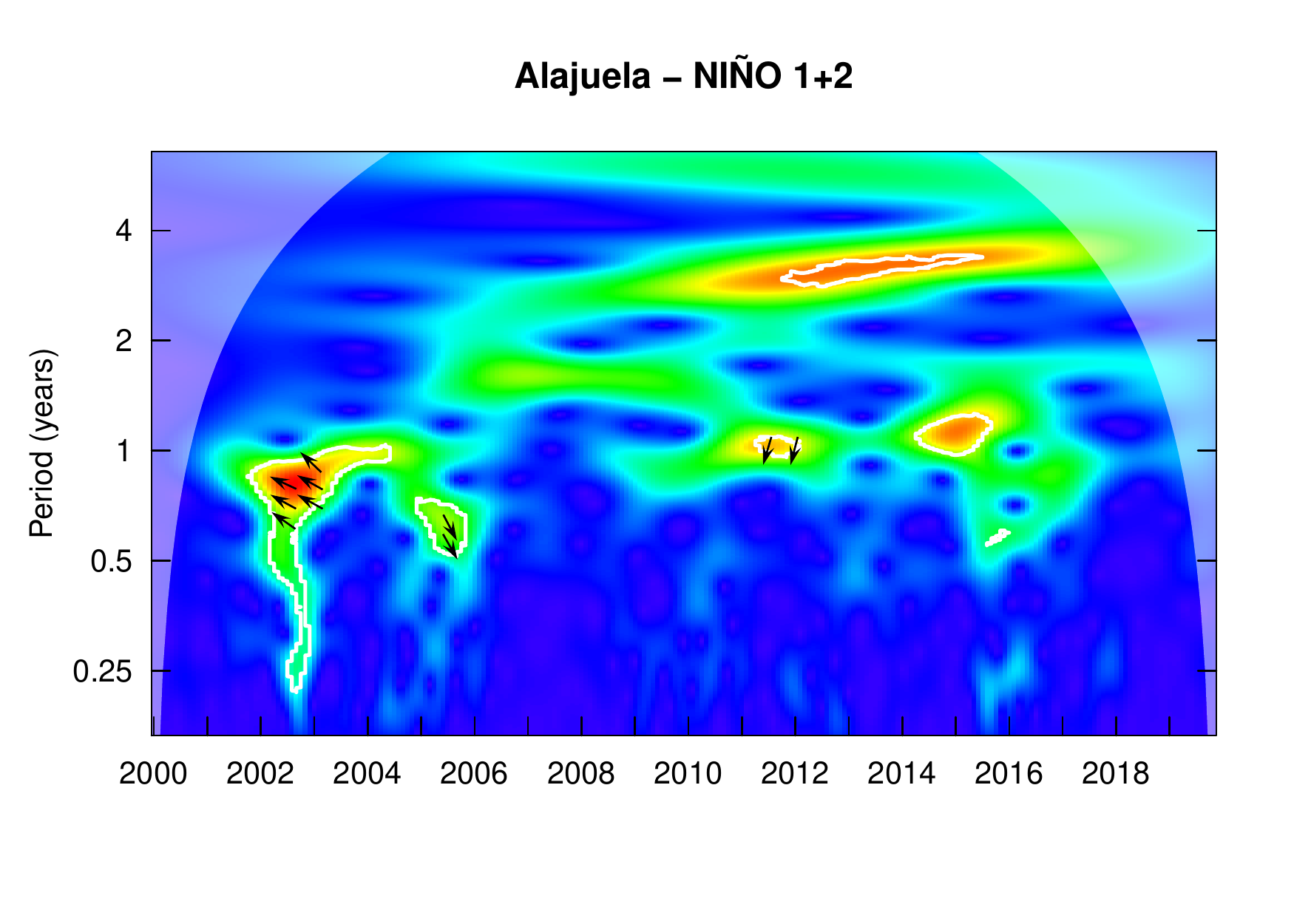}}\vspace{-0.15cm}%
\subfloat[]{\includegraphics[scale=0.23]{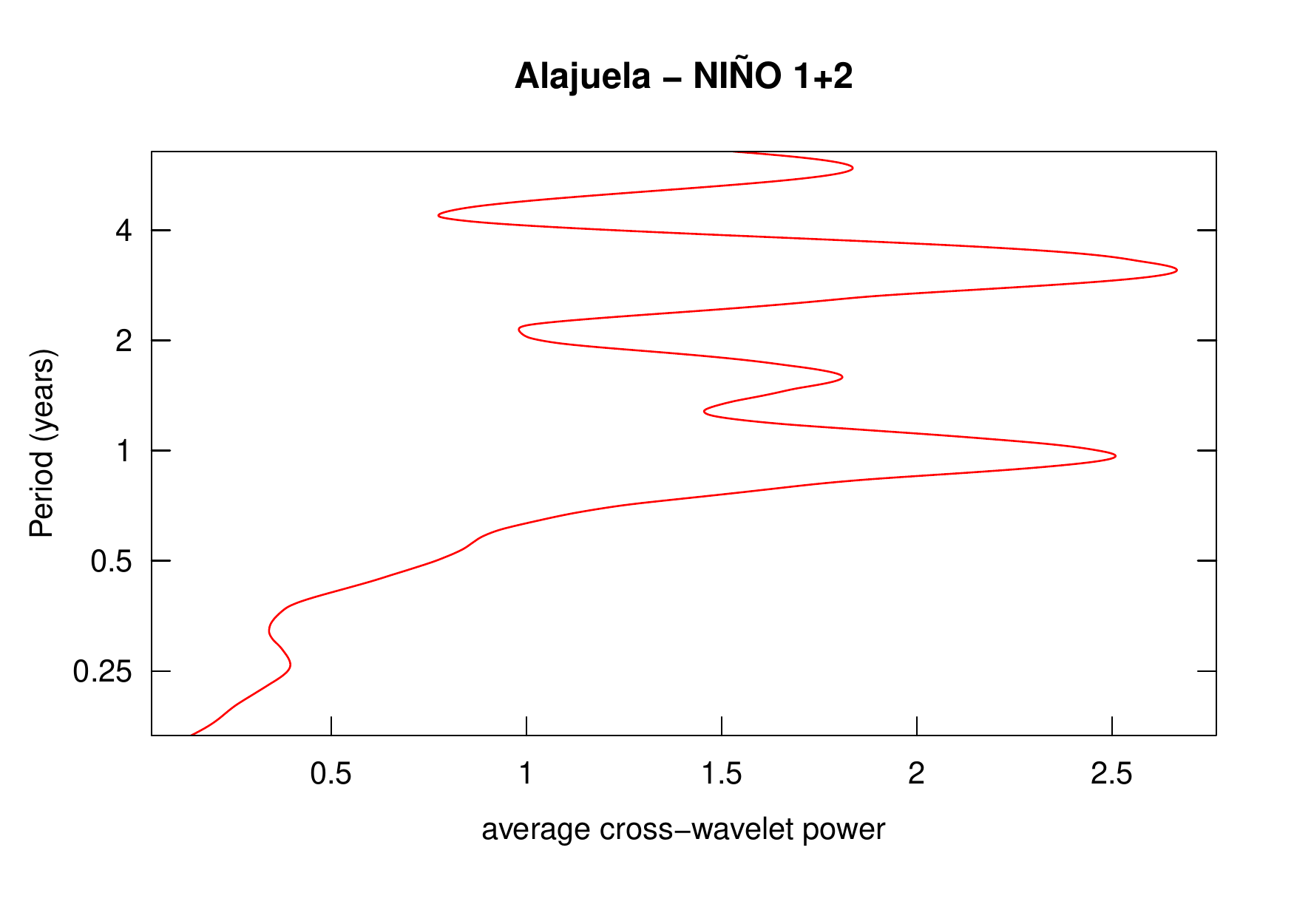}}\vspace{-0.15cm}\\
\subfloat[]{\includegraphics[scale=0.23]{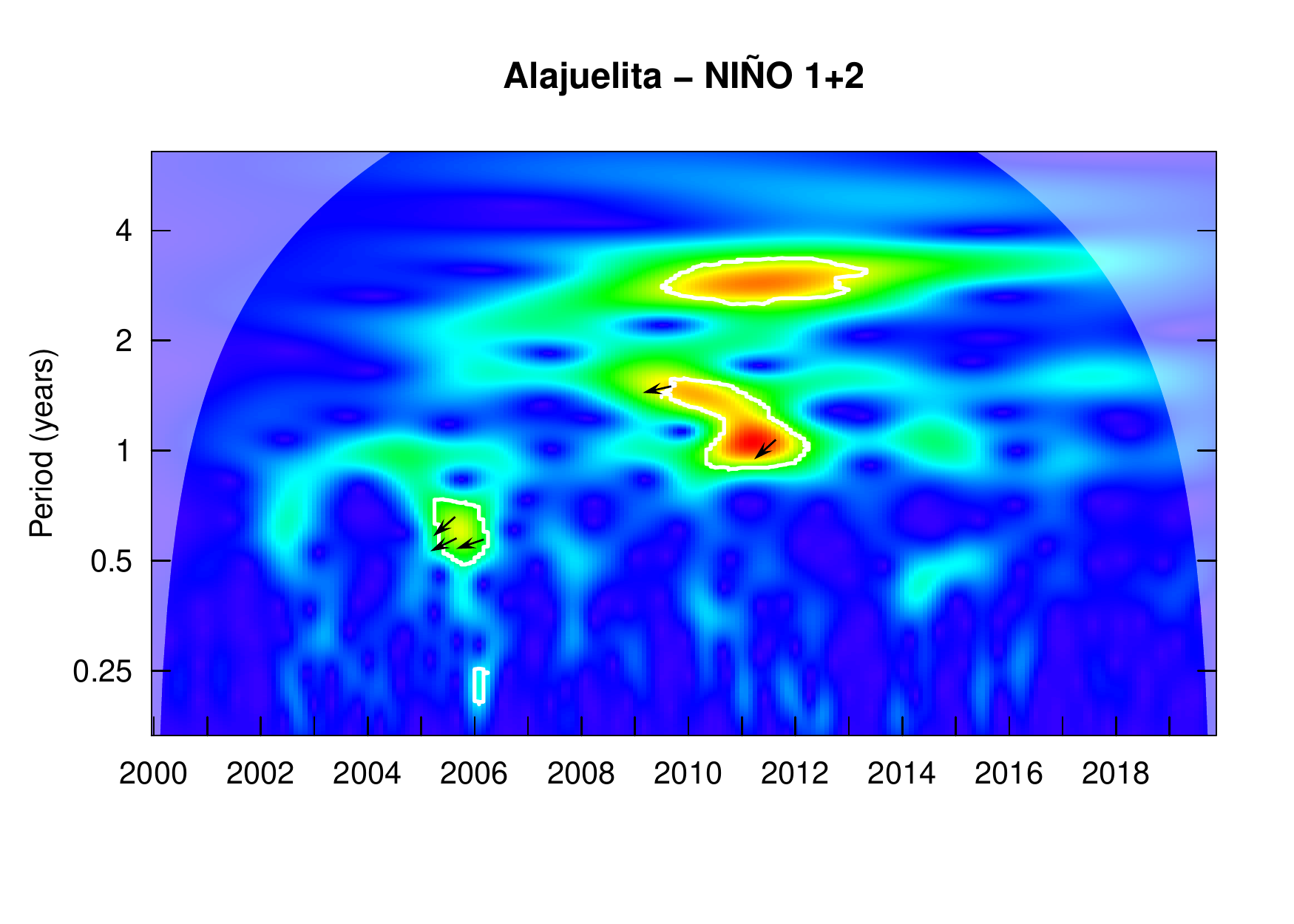}}\vspace{-0.15cm}%
\subfloat[]{\includegraphics[scale=0.23]{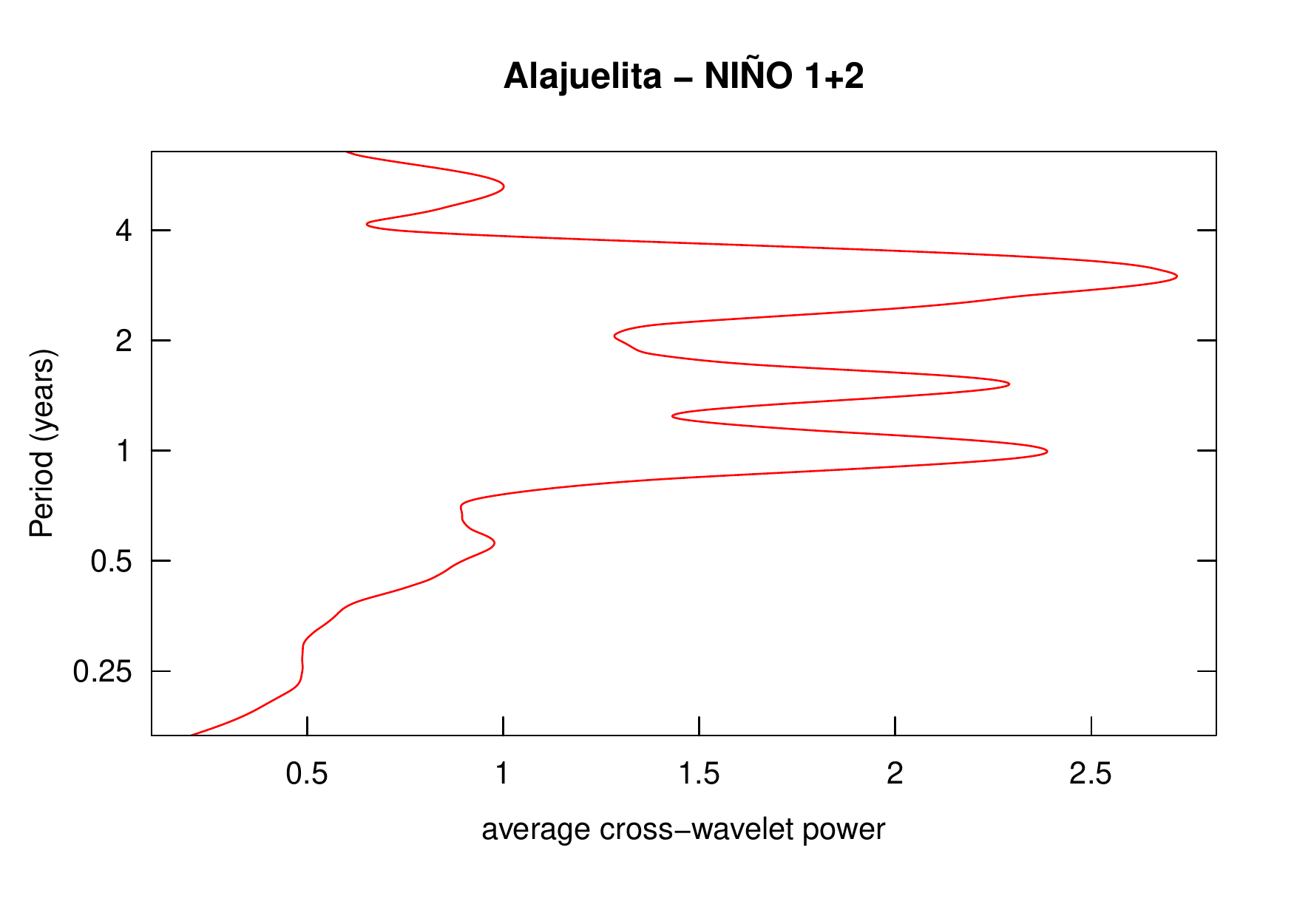}}\vspace{-0.15cm}%
\subfloat[]{\includegraphics[scale=0.23]{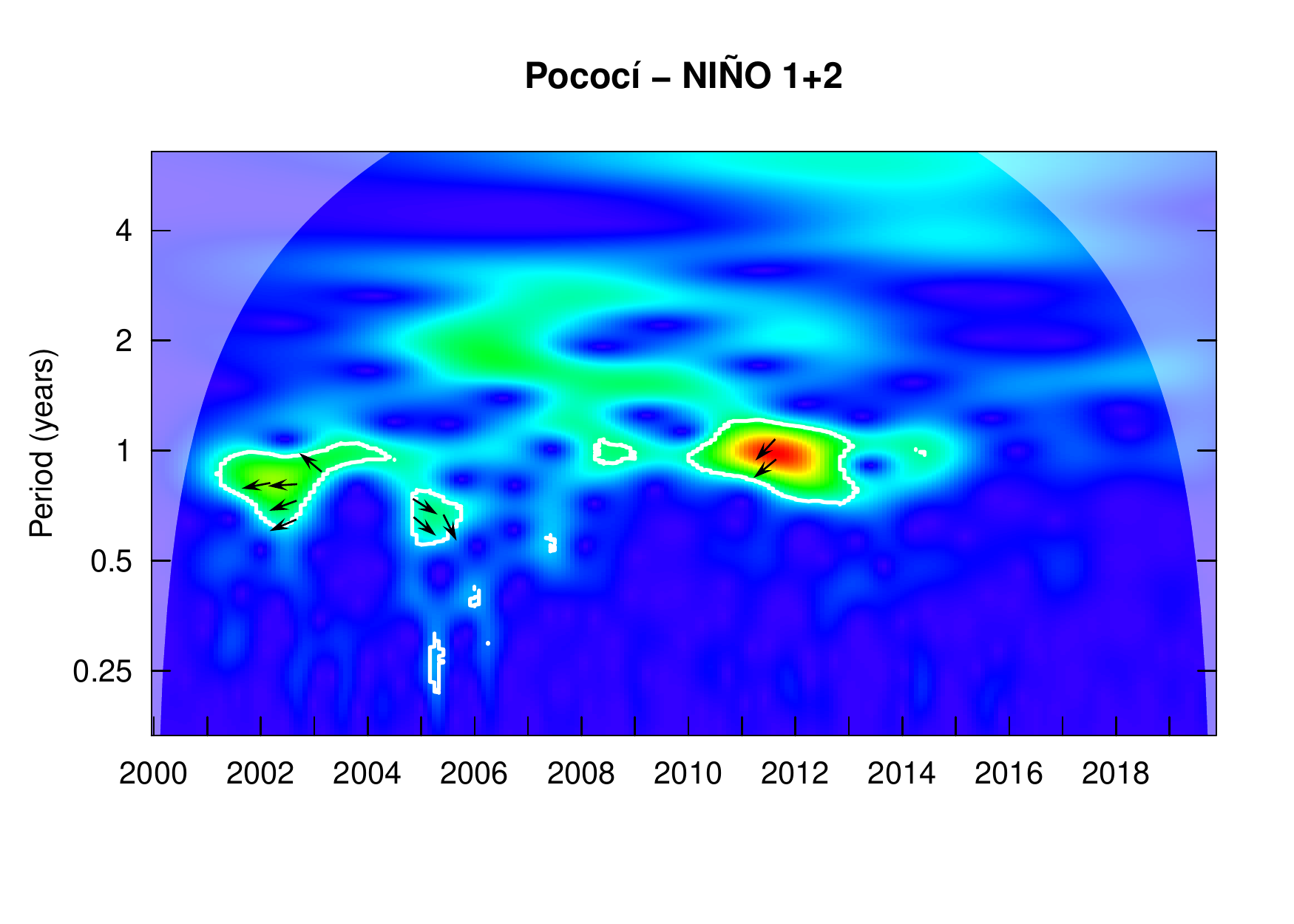}}\vspace{-0.15cm}%
\subfloat[]{\includegraphics[scale=0.23]{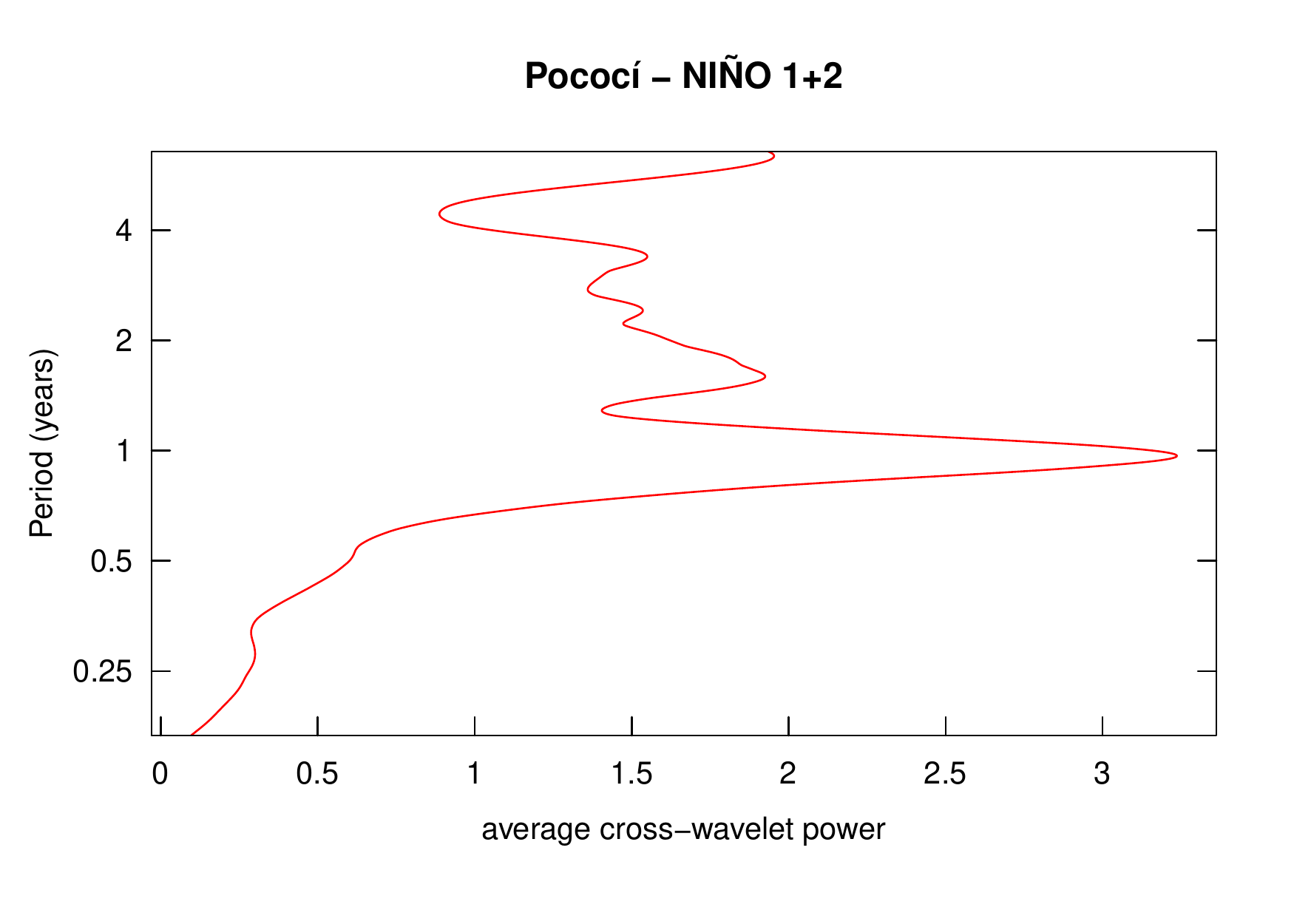}}\vspace{-0.15cm}\\
\subfloat[]{\includegraphics[scale=0.23]{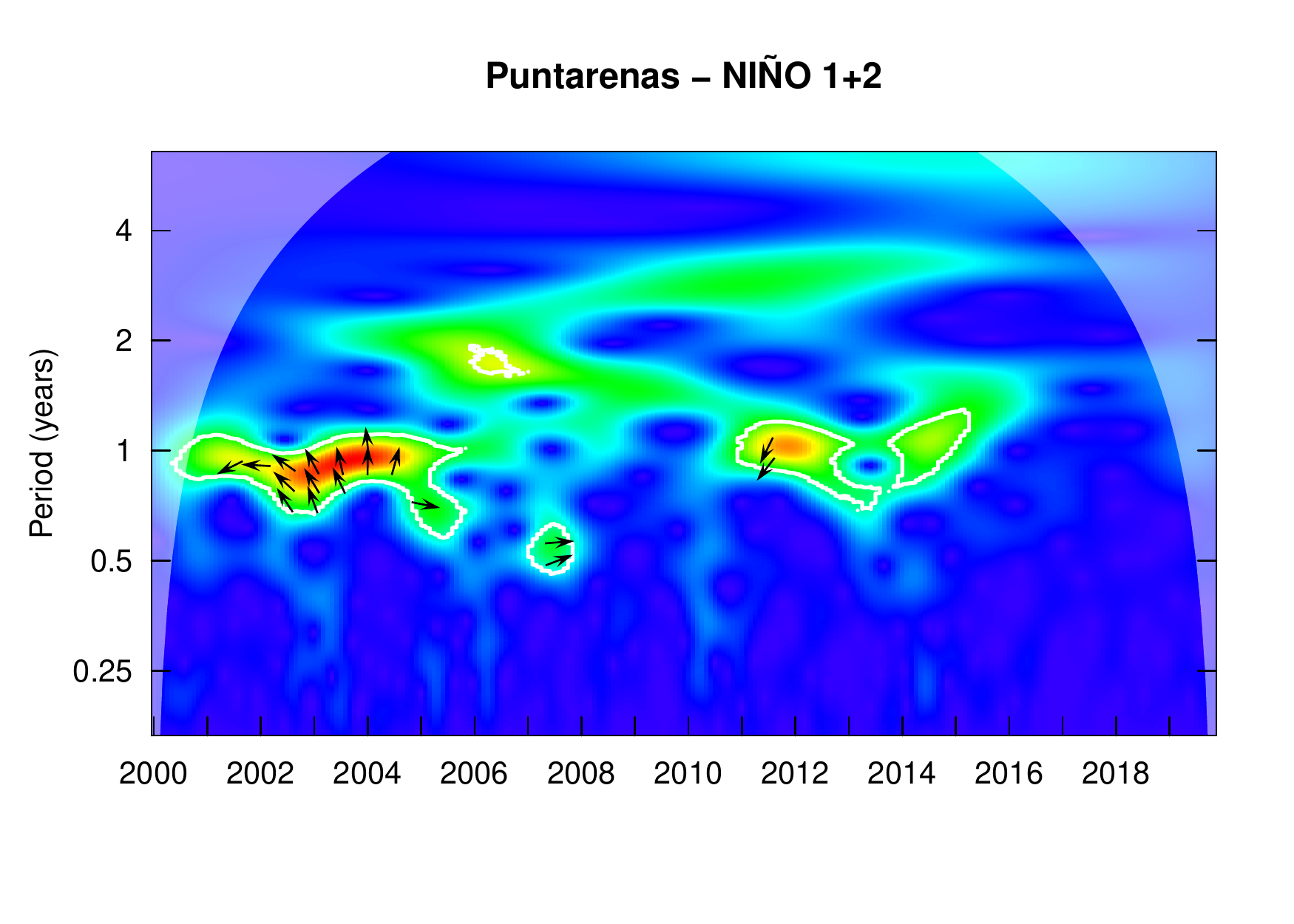}}\vspace{-0.15cm}%
\subfloat[]{\includegraphics[scale=0.23]{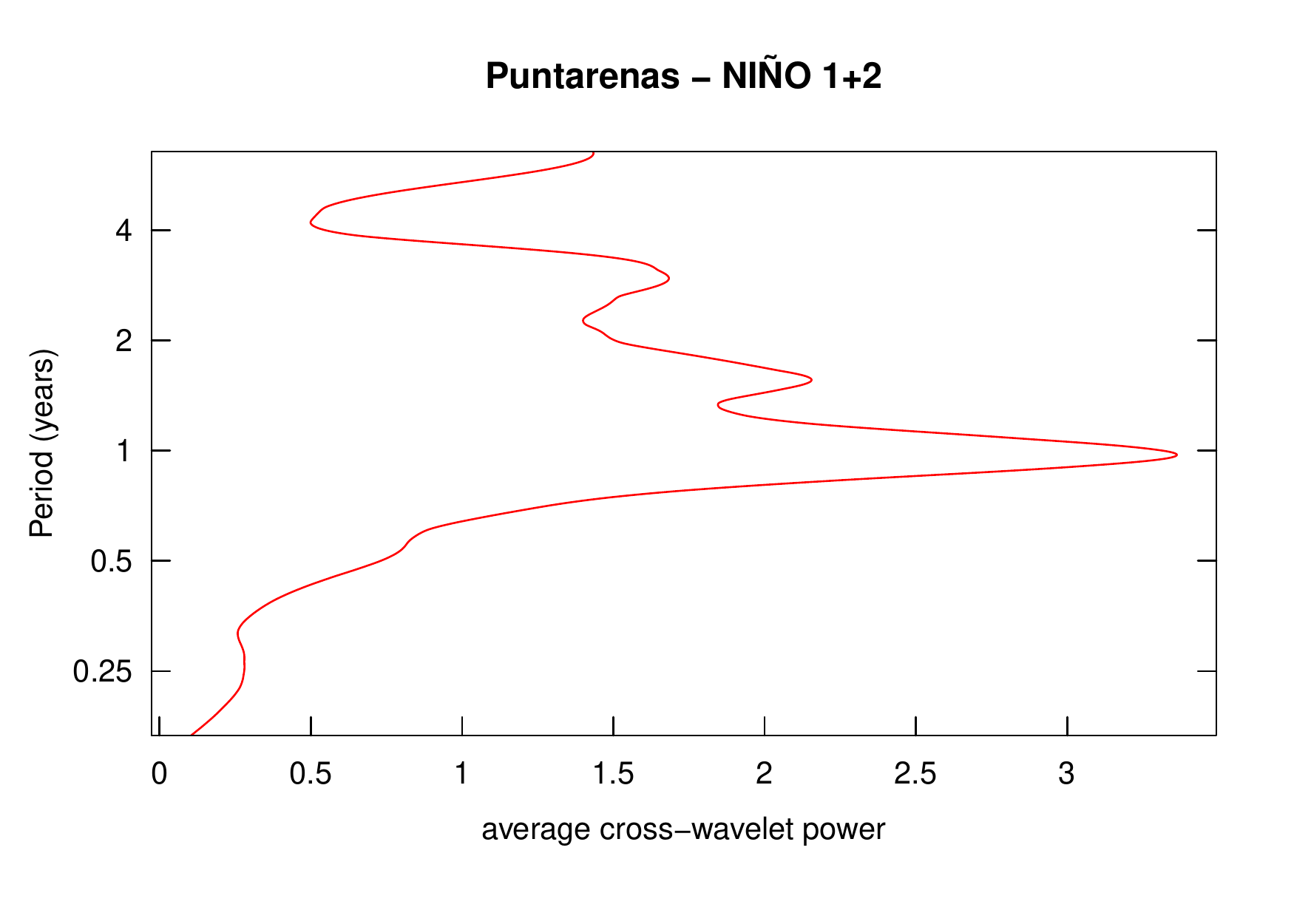}}\vspace{-0.15cm}%
\subfloat[]{\includegraphics[scale=0.23]{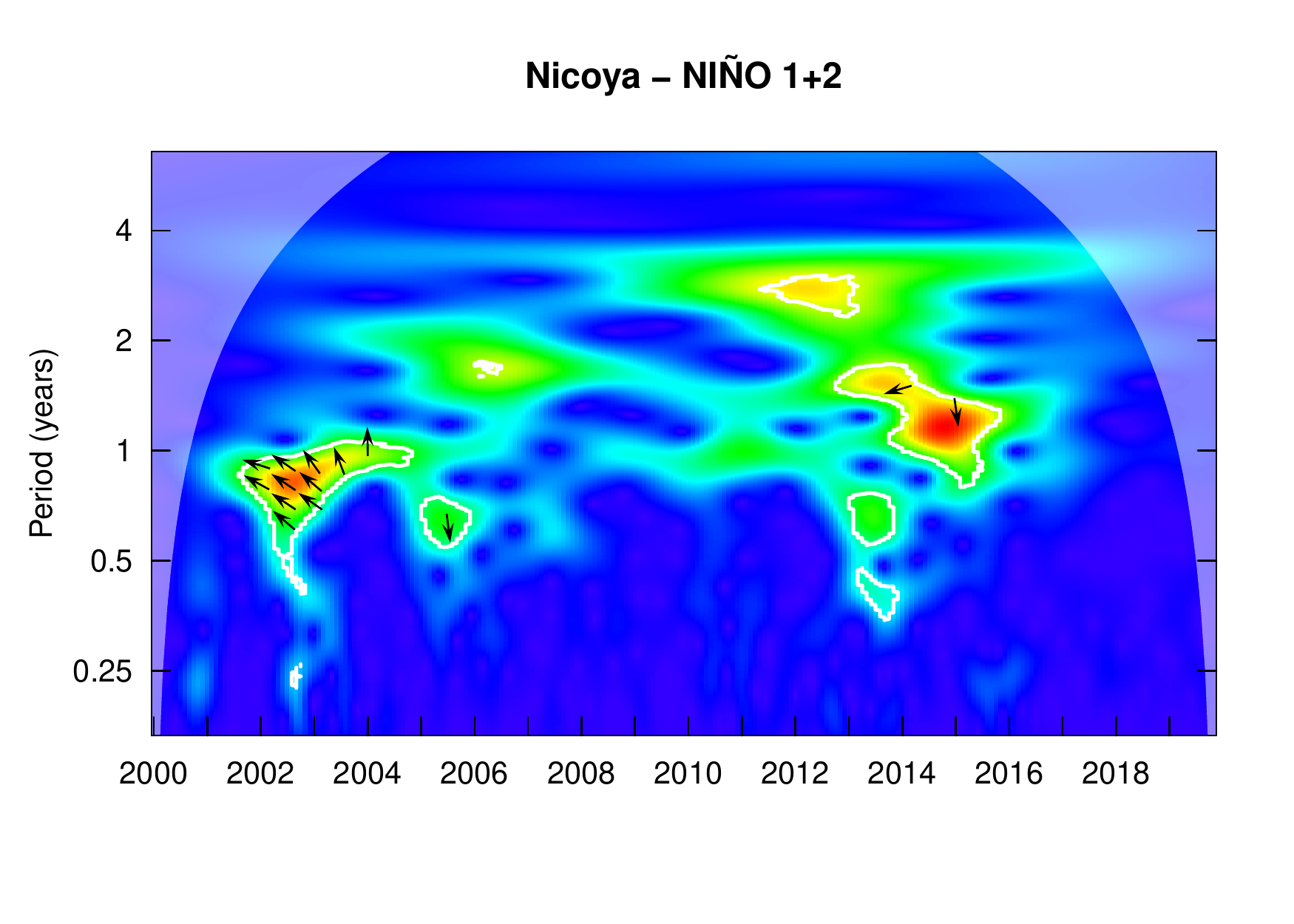}}\vspace{-0.15cm}%
\subfloat[]{\includegraphics[scale=0.23]{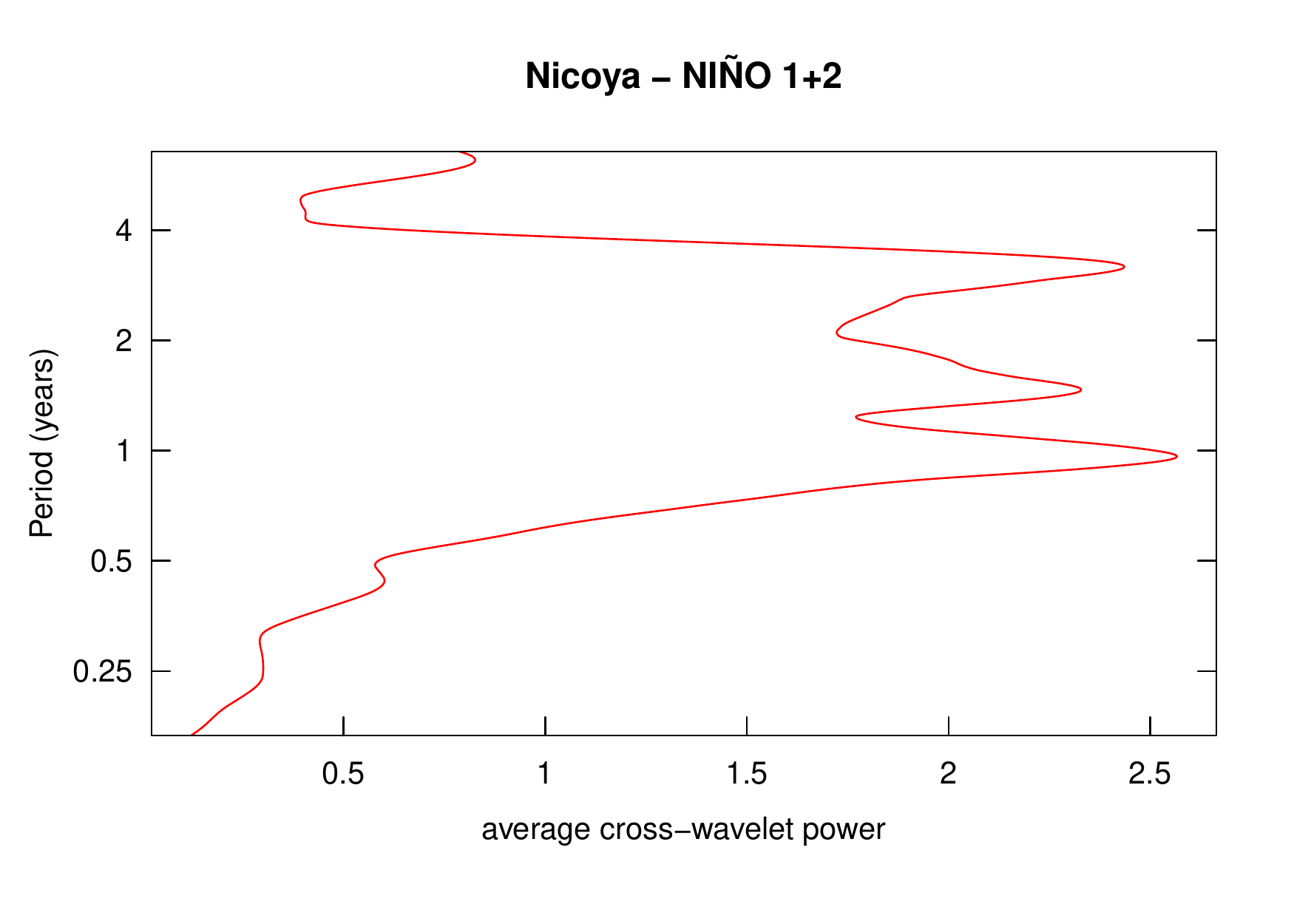}}\vspace{-0.15cm}
\caption*{}
\end{figure}

\begin{figure}[H]
\captionsetup[subfigure]{labelformat=empty}
\subfloat[]{\includegraphics[scale=0.23]{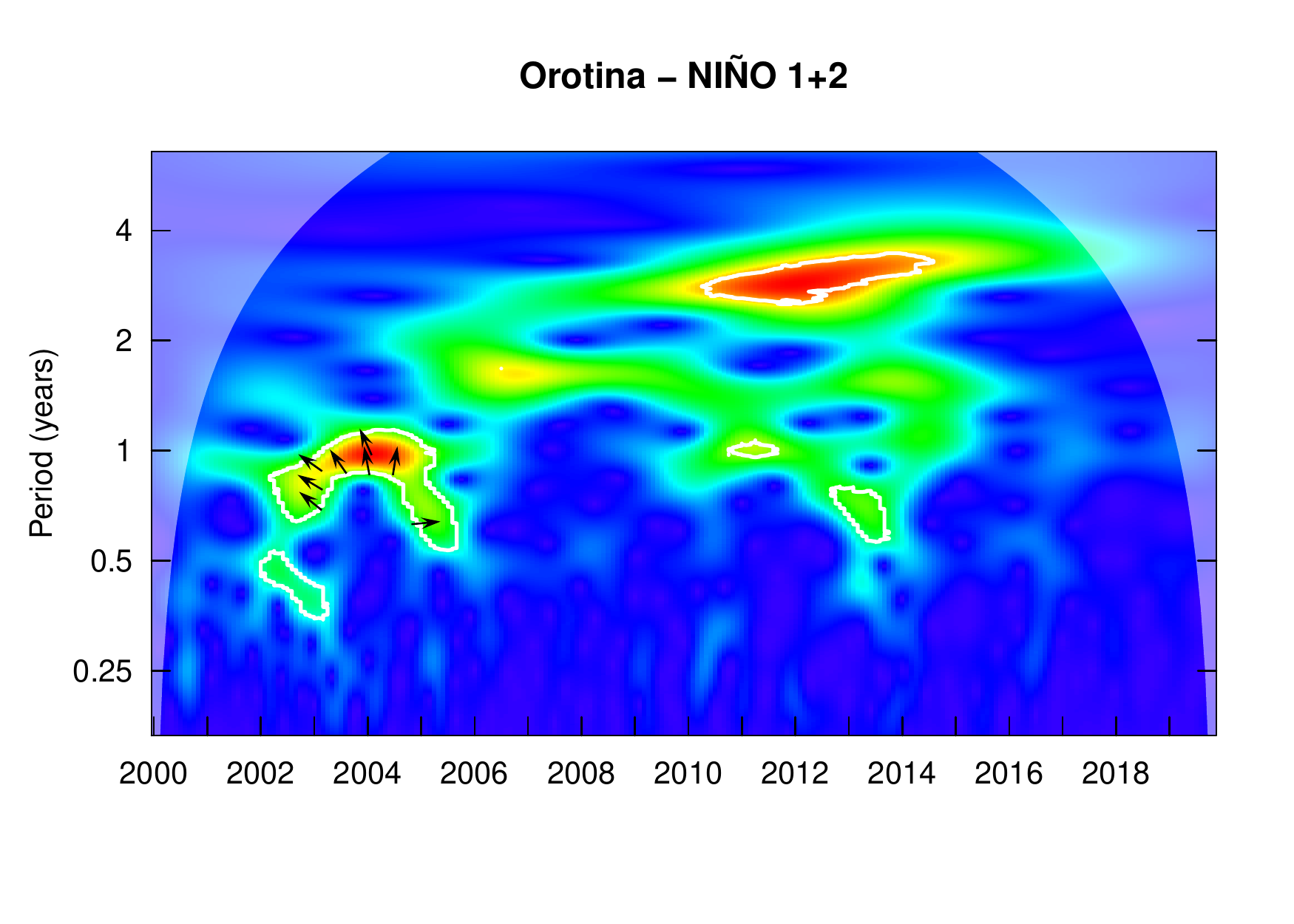}}\vspace{-0.15cm}%
\subfloat[]{\includegraphics[scale=0.23]{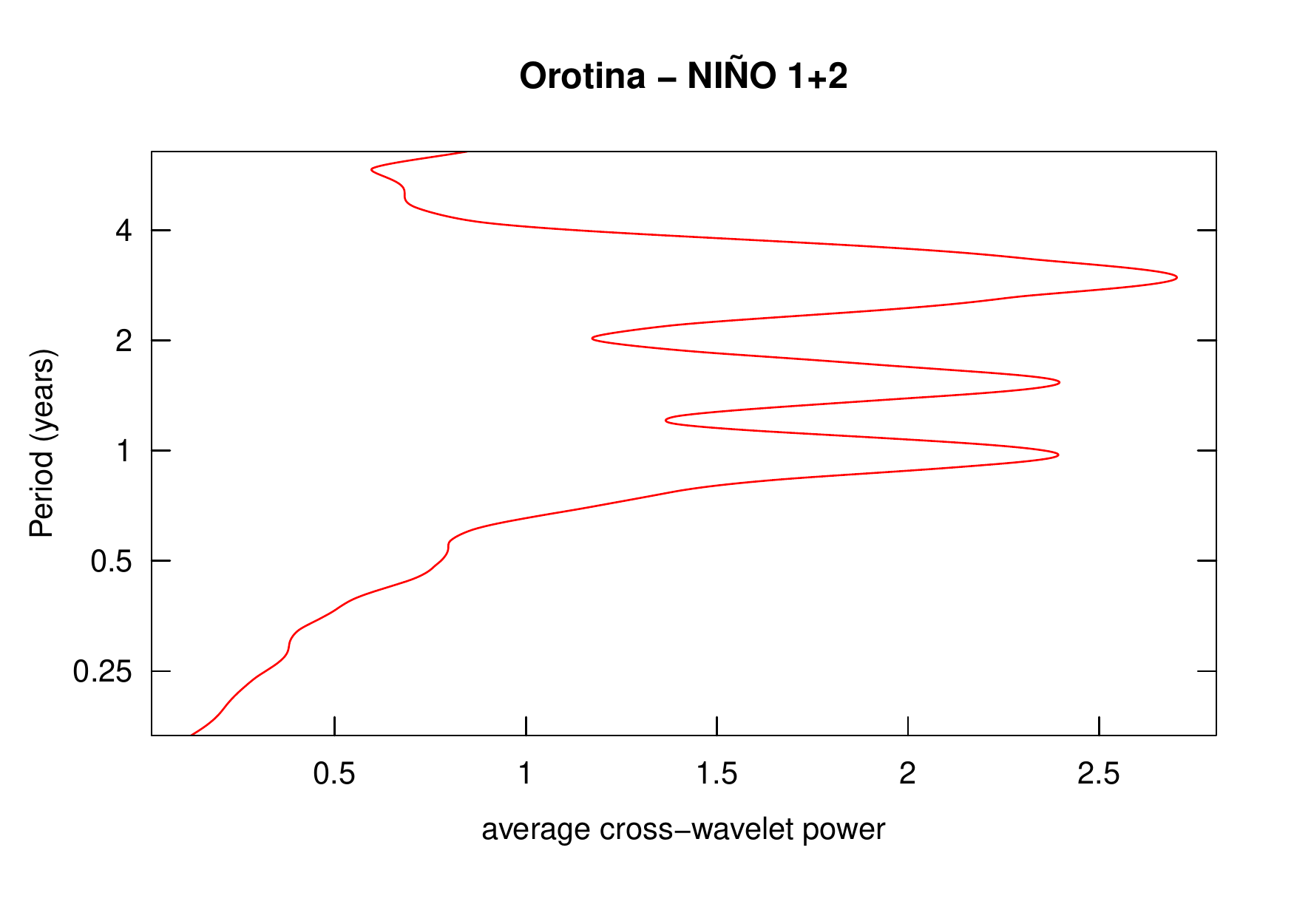}}\vspace{-0.15cm}%
\subfloat[]{\includegraphics[scale=0.23]{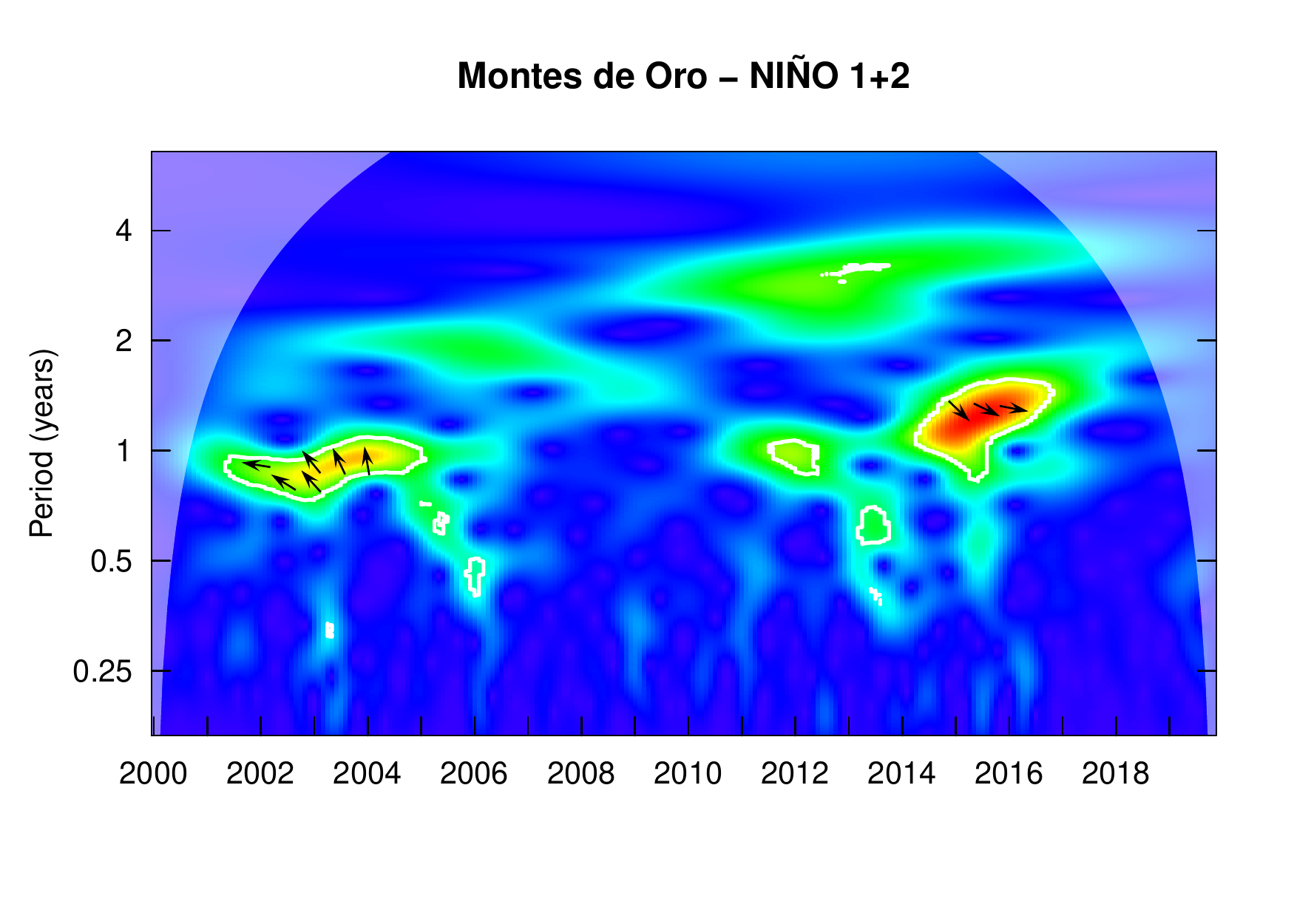}}\vspace{-0.15cm}%
\subfloat[]{\includegraphics[scale=0.23]{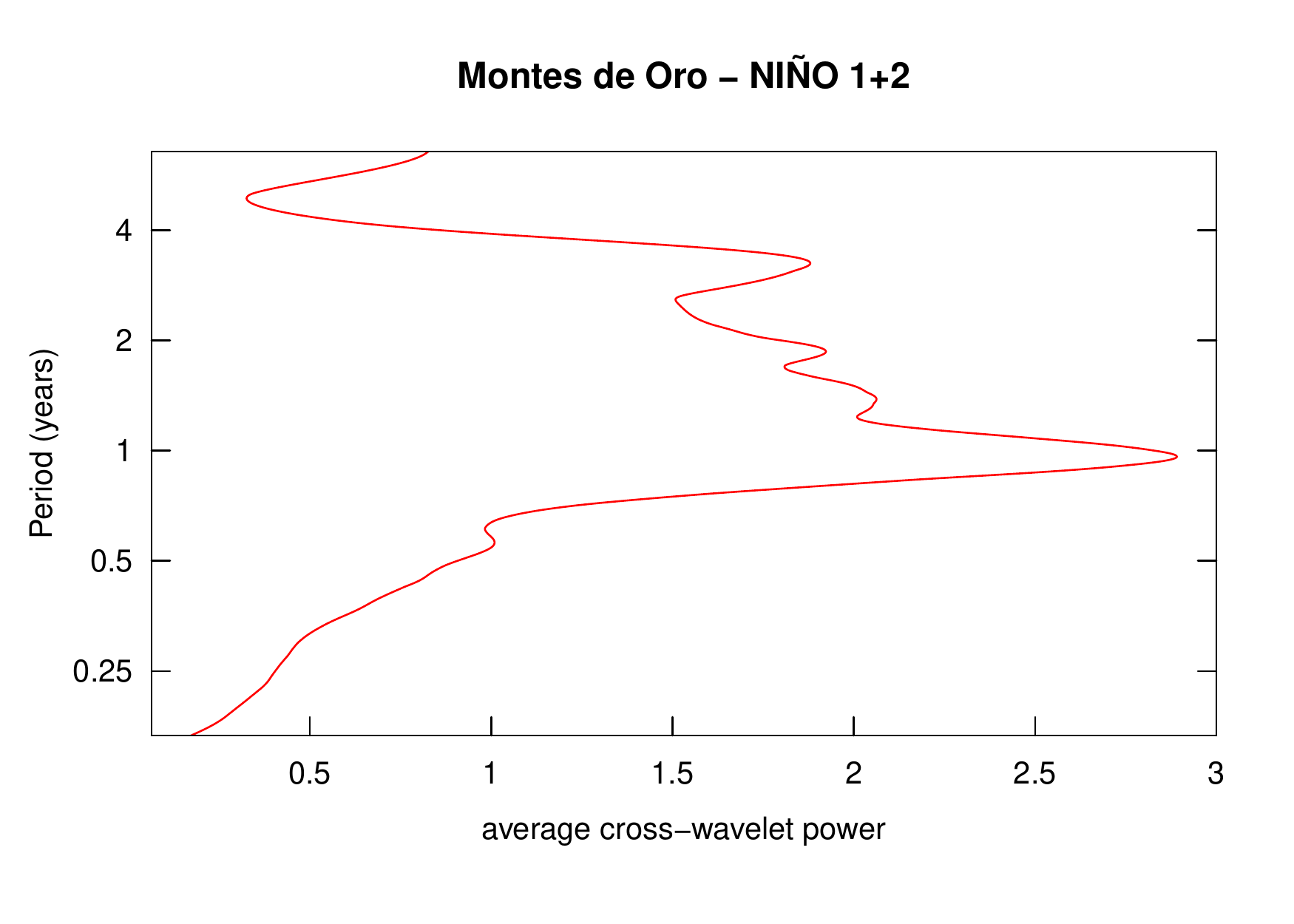}}\vspace{-0.15cm}\\
\subfloat[]{\includegraphics[scale=0.23]{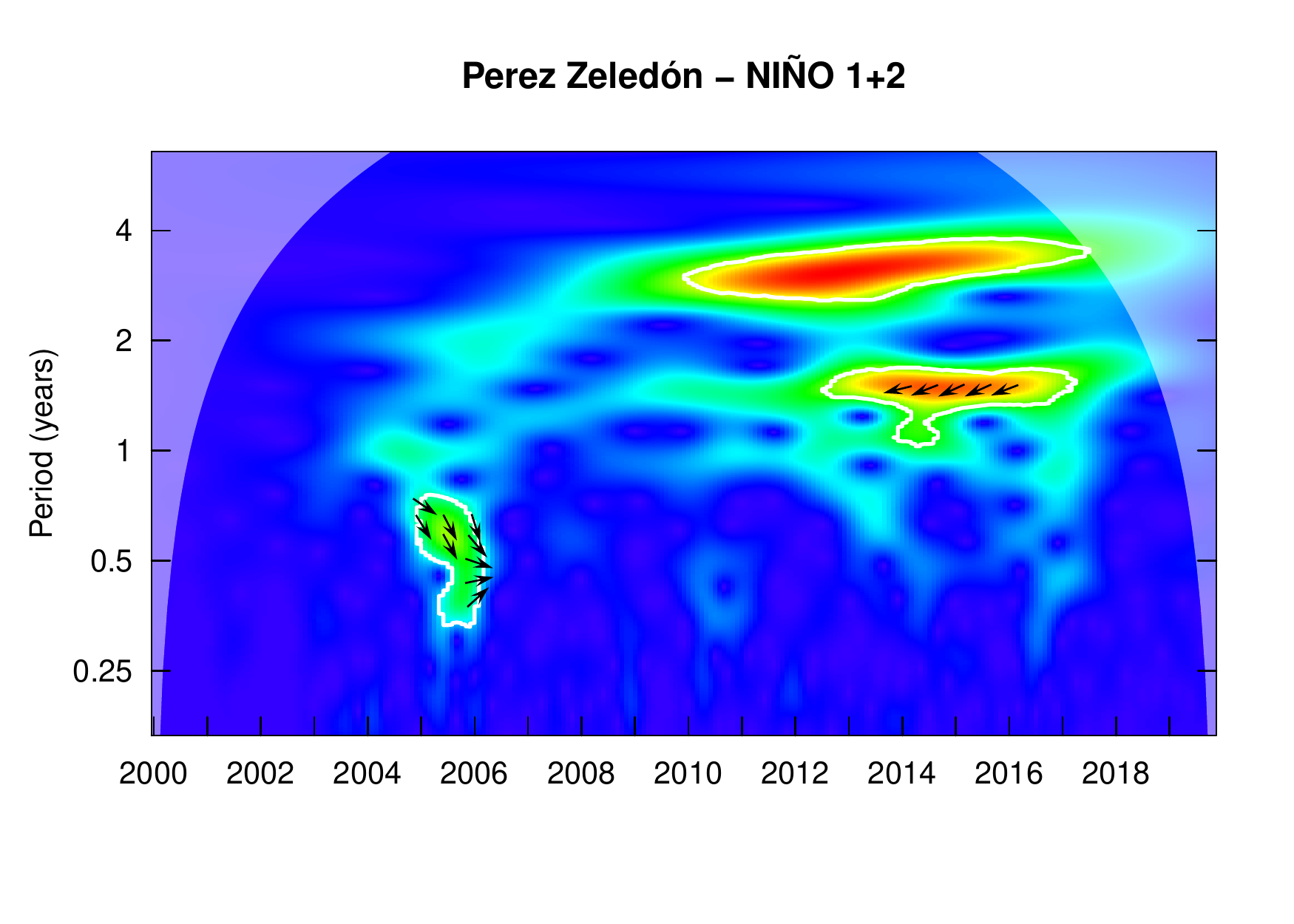}}\vspace{-0.15cm}%
\subfloat[]{\includegraphics[scale=0.23]{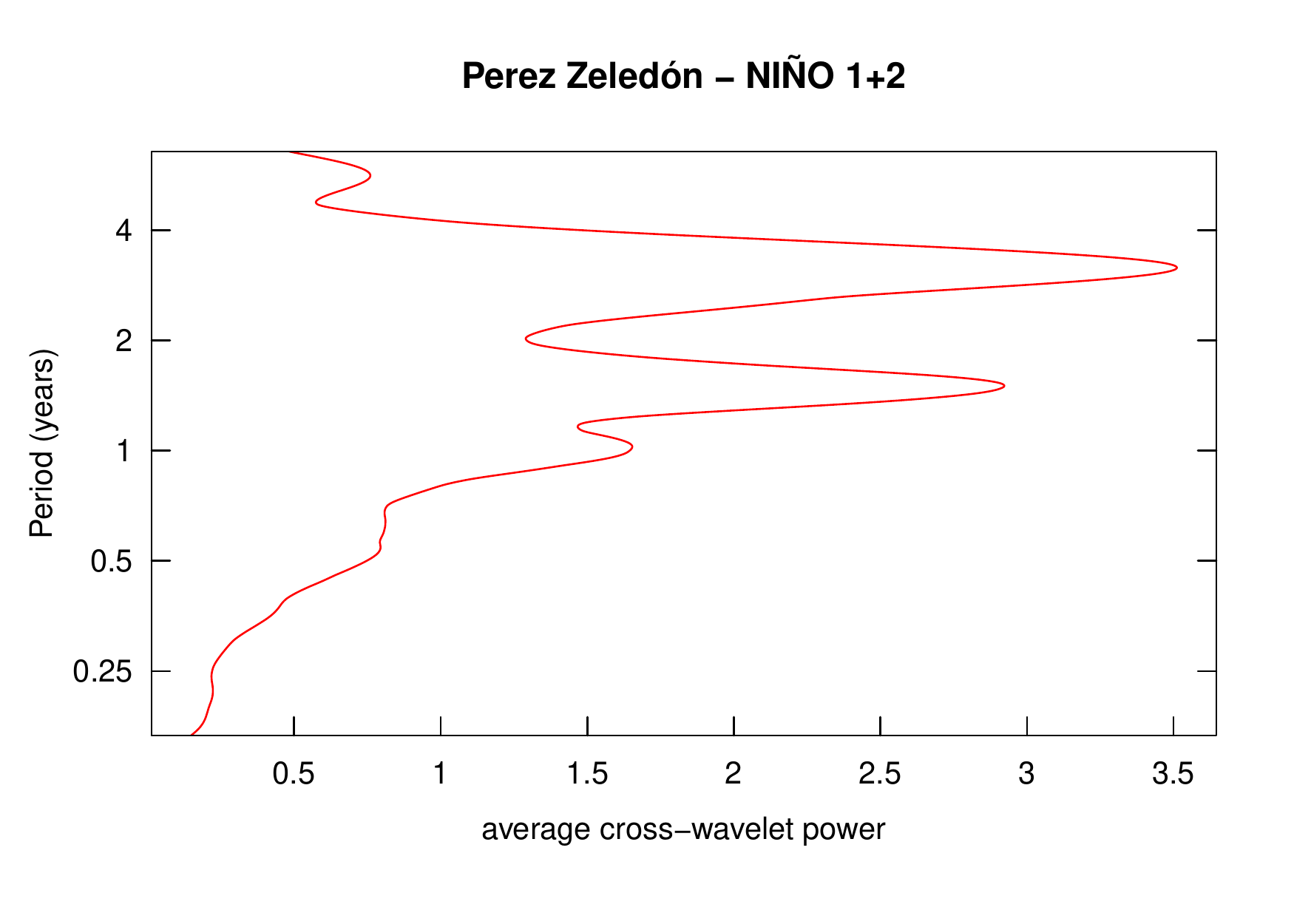}}\vspace{-0.15cm}%
\subfloat[]{\includegraphics[scale=0.23]{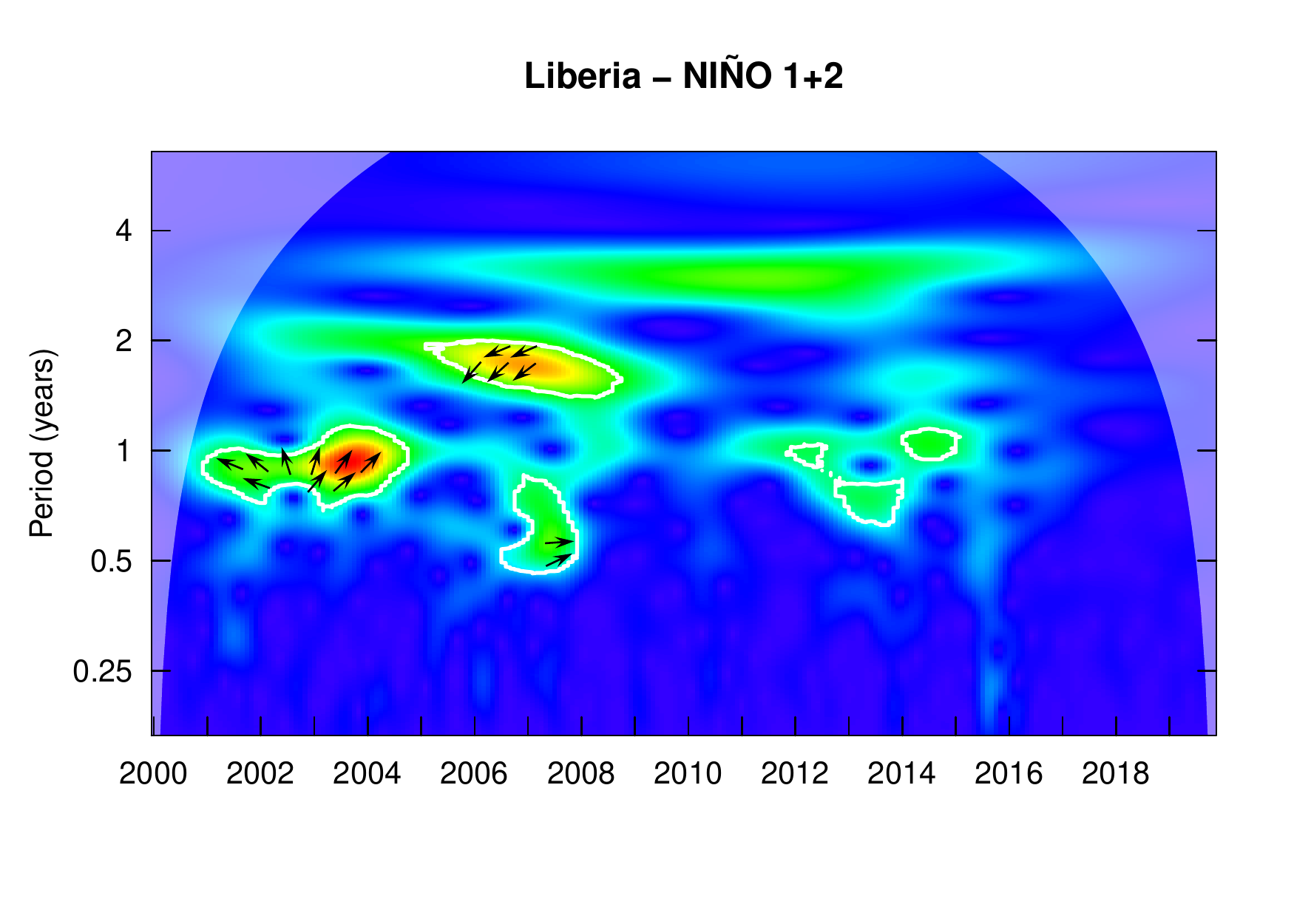}}\vspace{-0.15cm}%
\subfloat[]{\includegraphics[scale=0.23]{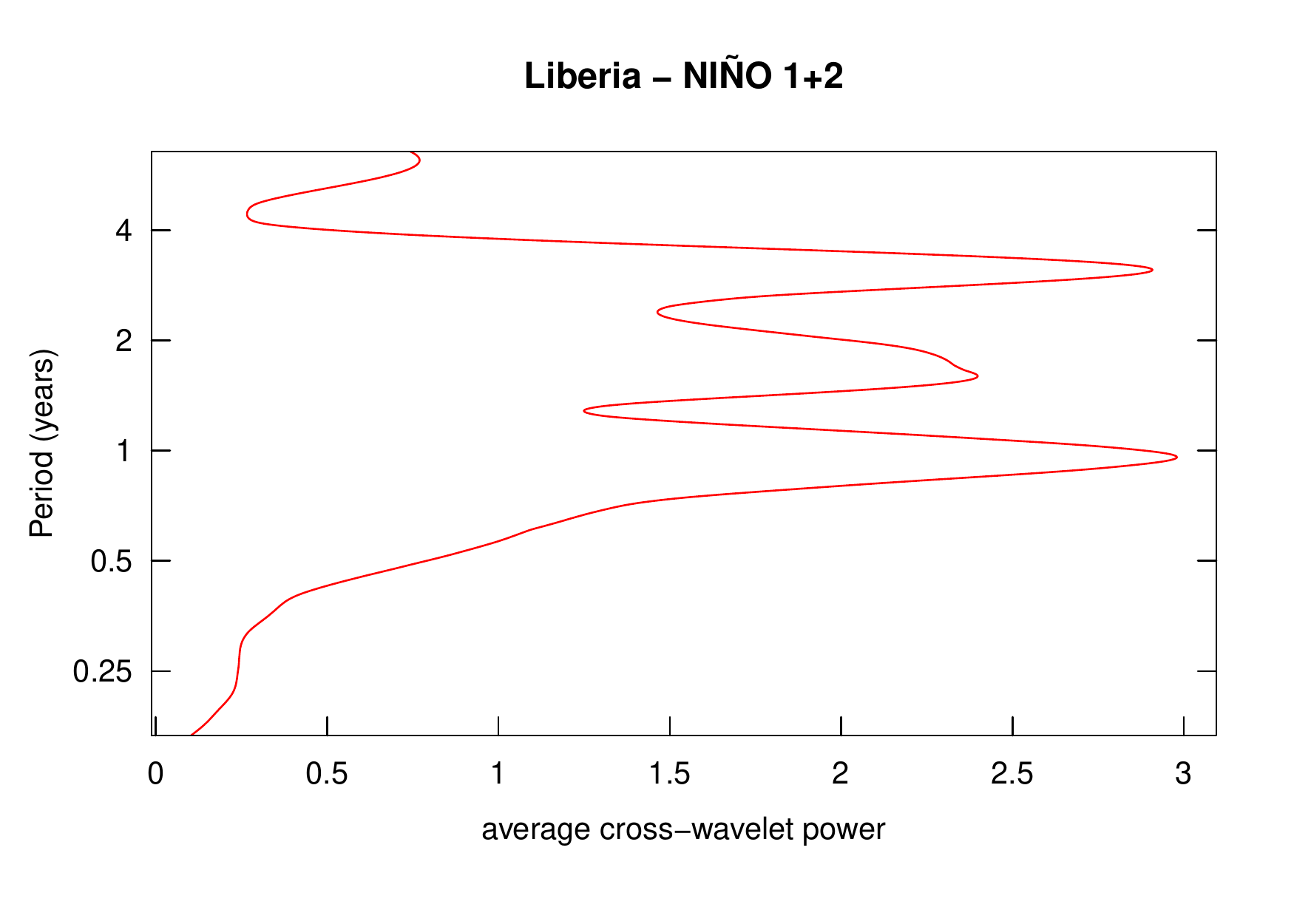}}\vspace{-0.15cm}\\
\subfloat[]{\includegraphics[scale=0.23]{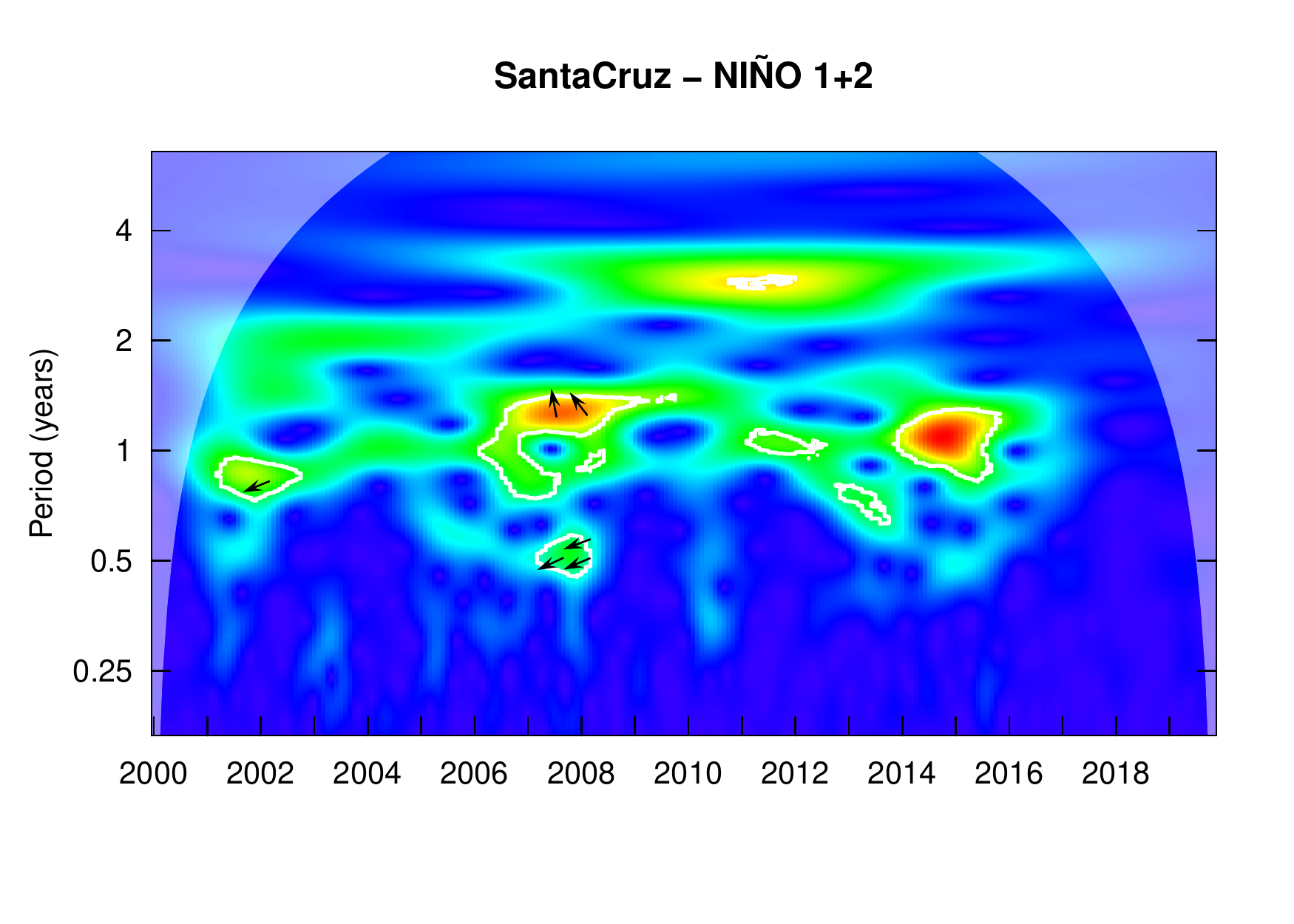}}\vspace{-0.15cm}%
\subfloat[]{\includegraphics[scale=0.23]{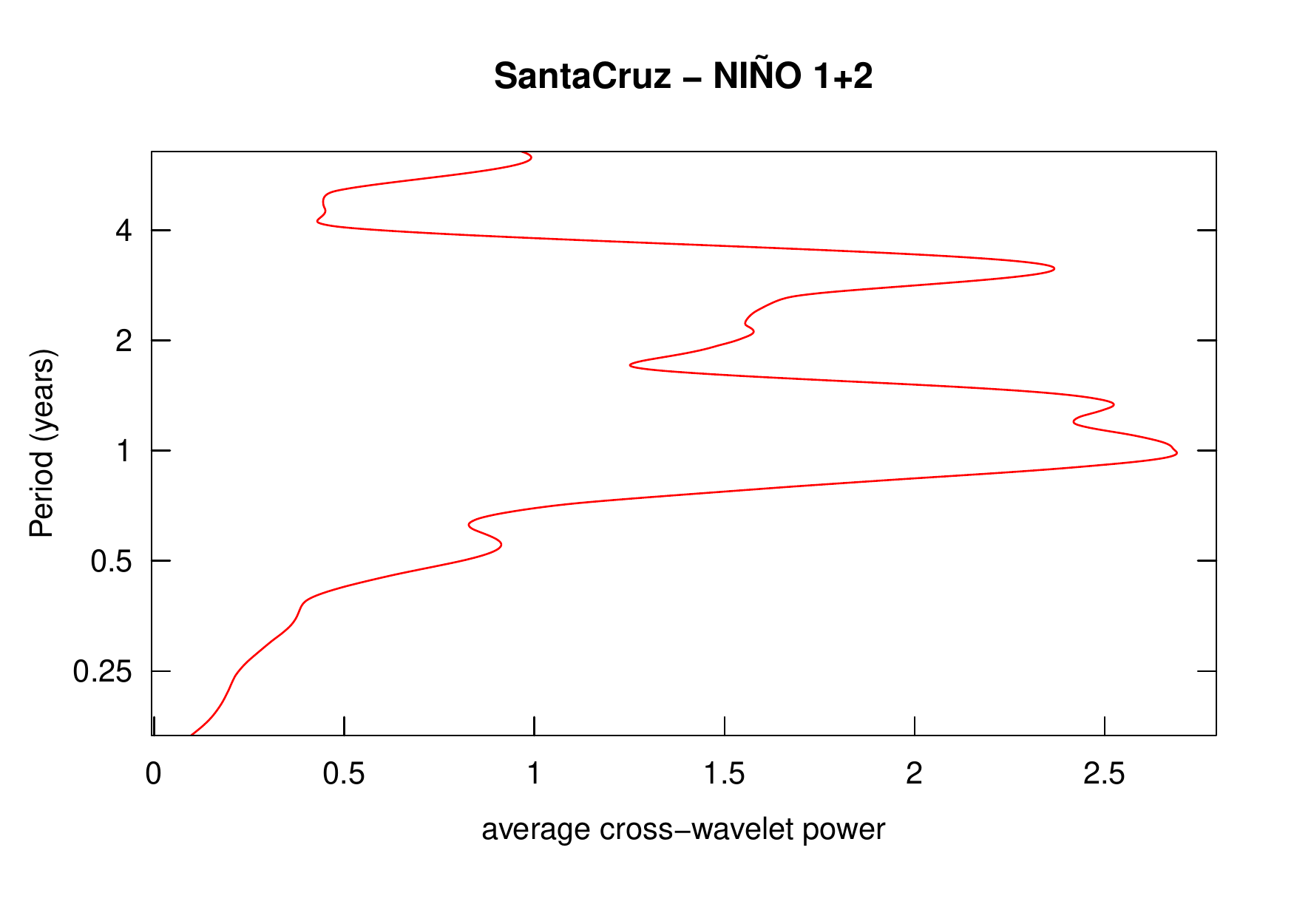}}\vspace{-0.15cm}%
\subfloat[]{\includegraphics[scale=0.23]{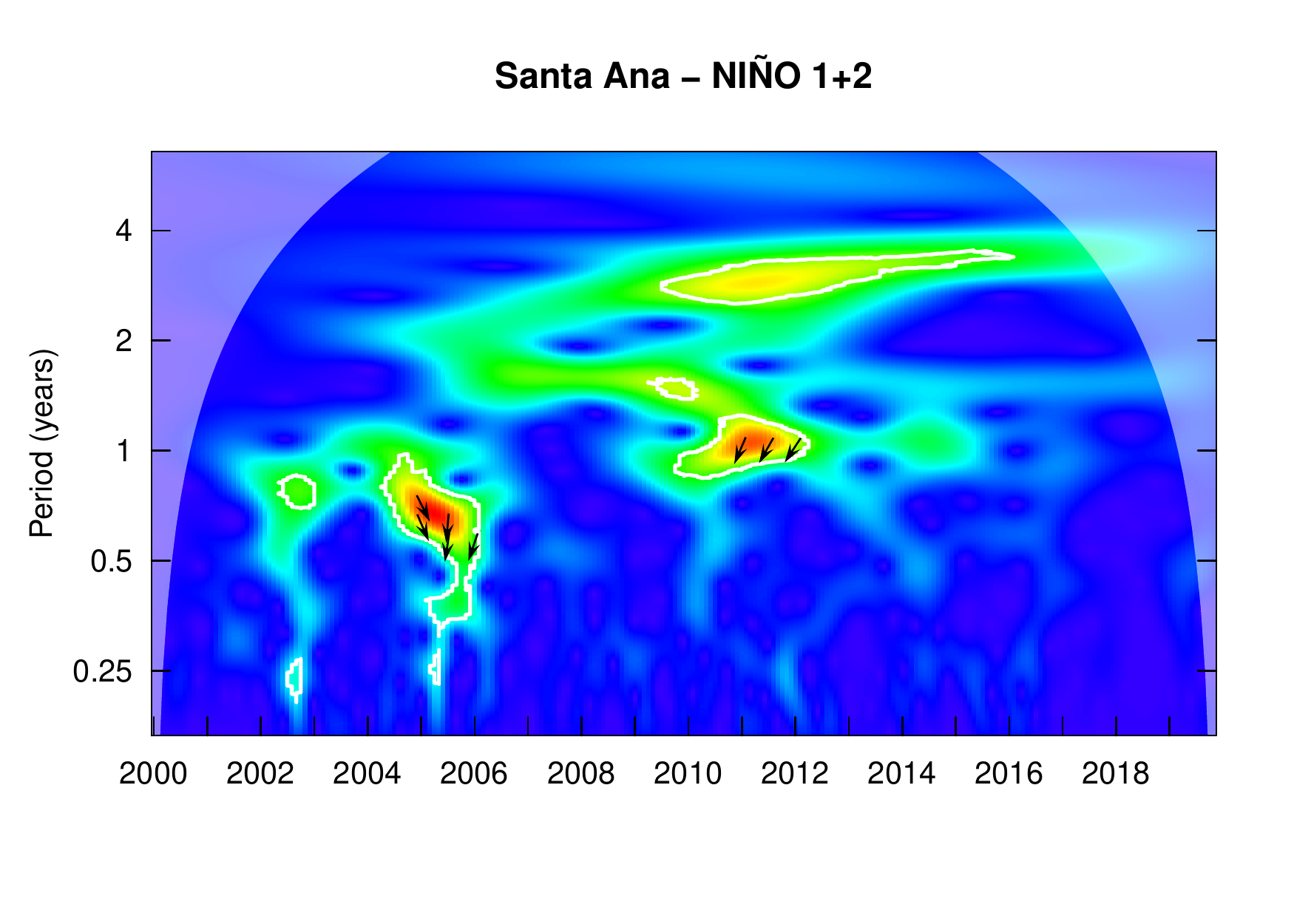}}\vspace{-0.15cm}%
\subfloat[]{\includegraphics[scale=0.23]{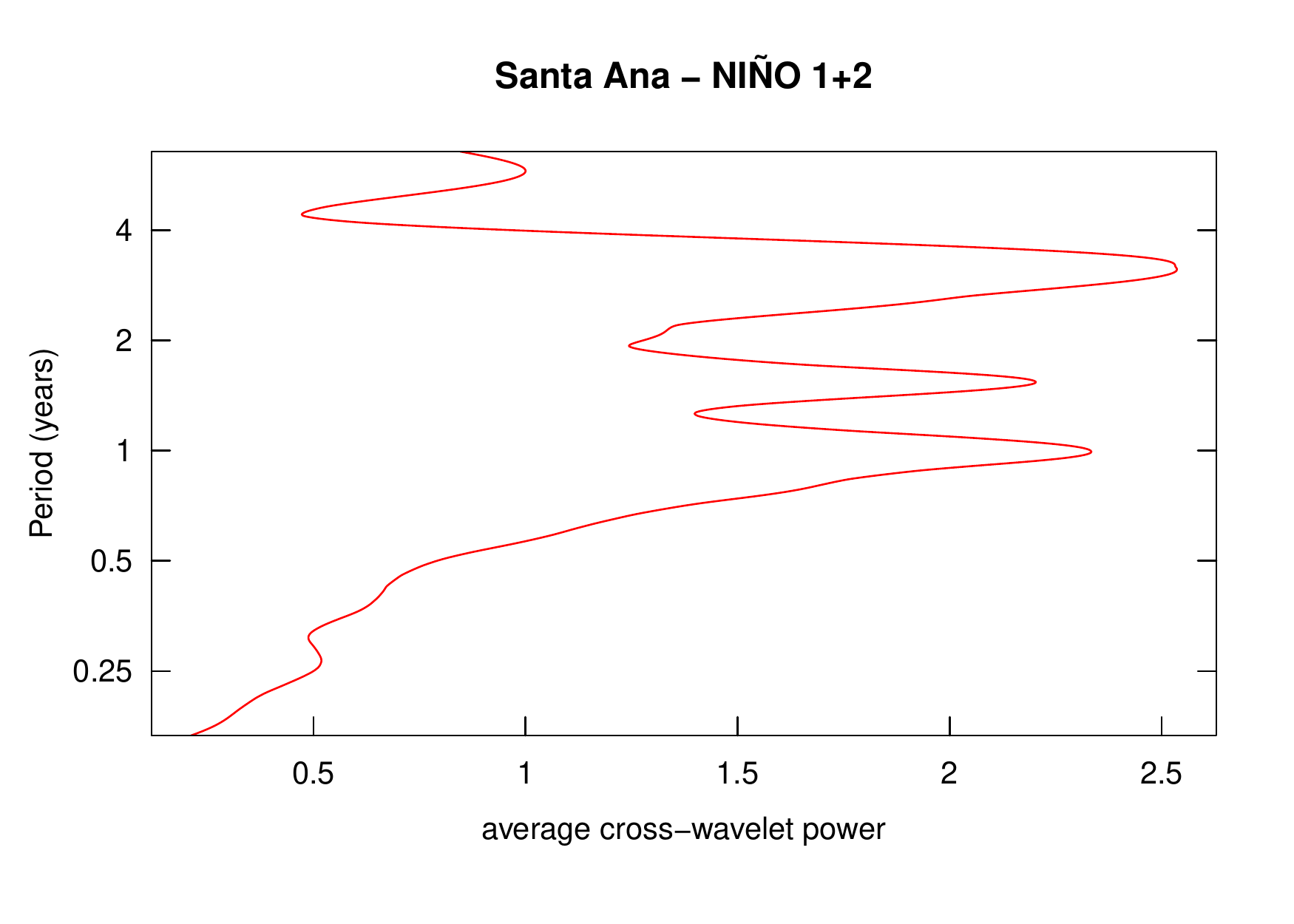}}\vspace{-0.15cm}\\
\subfloat[]{\includegraphics[scale=0.23]{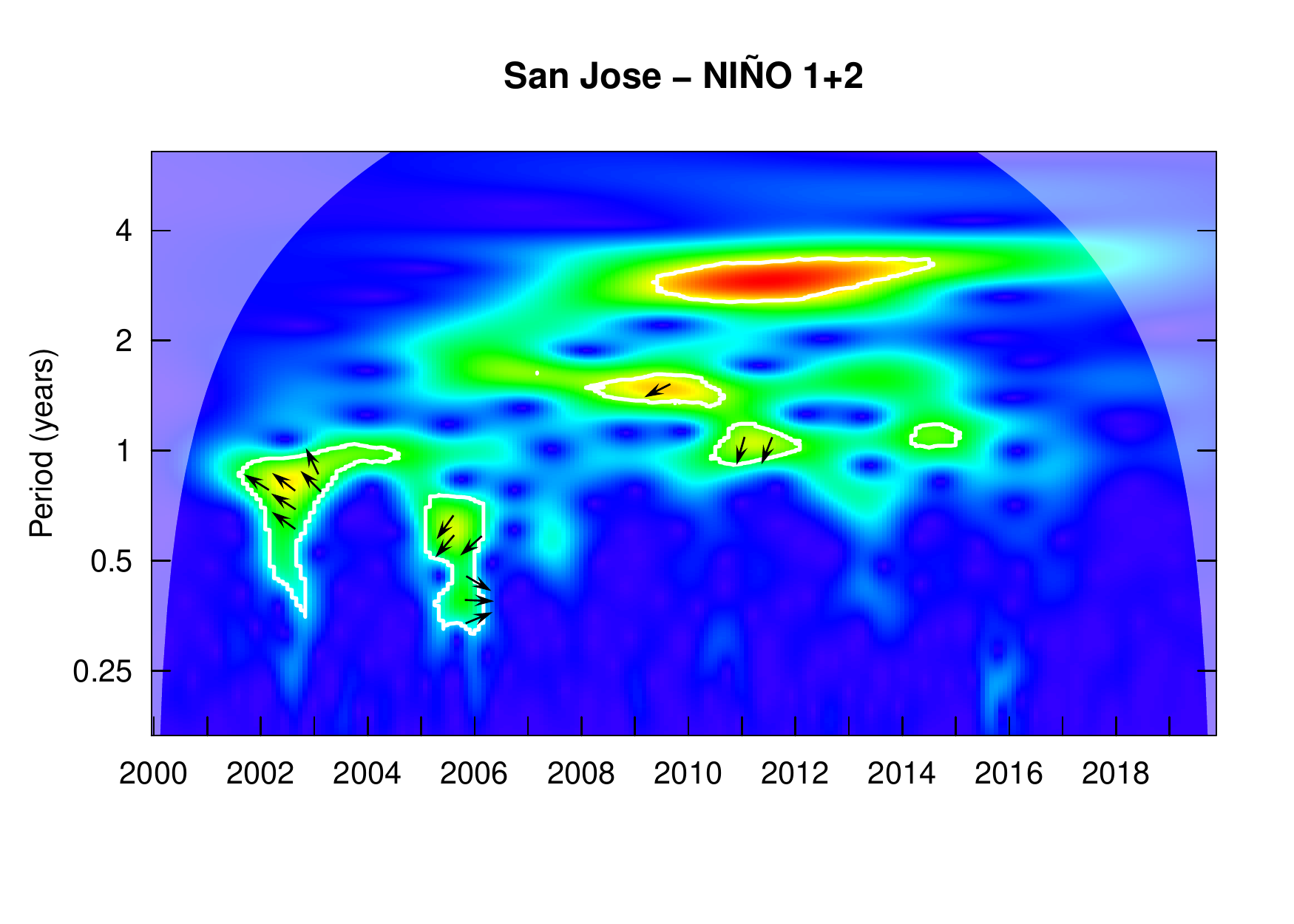}}\vspace{-0.15cm}%
\subfloat[]{\includegraphics[scale=0.23]{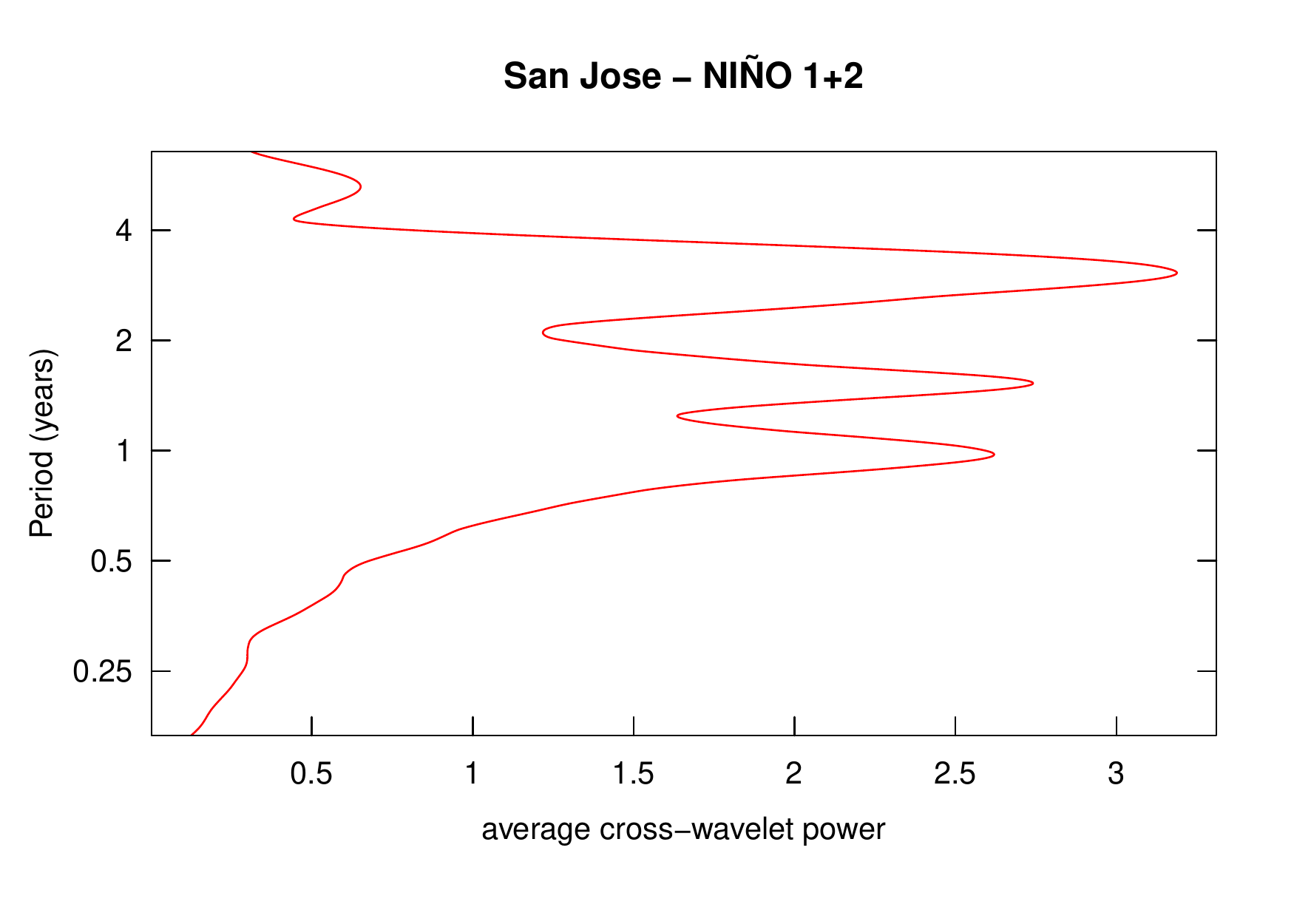}}\vspace{-0.15cm}%
\subfloat[]{\includegraphics[scale=0.23]{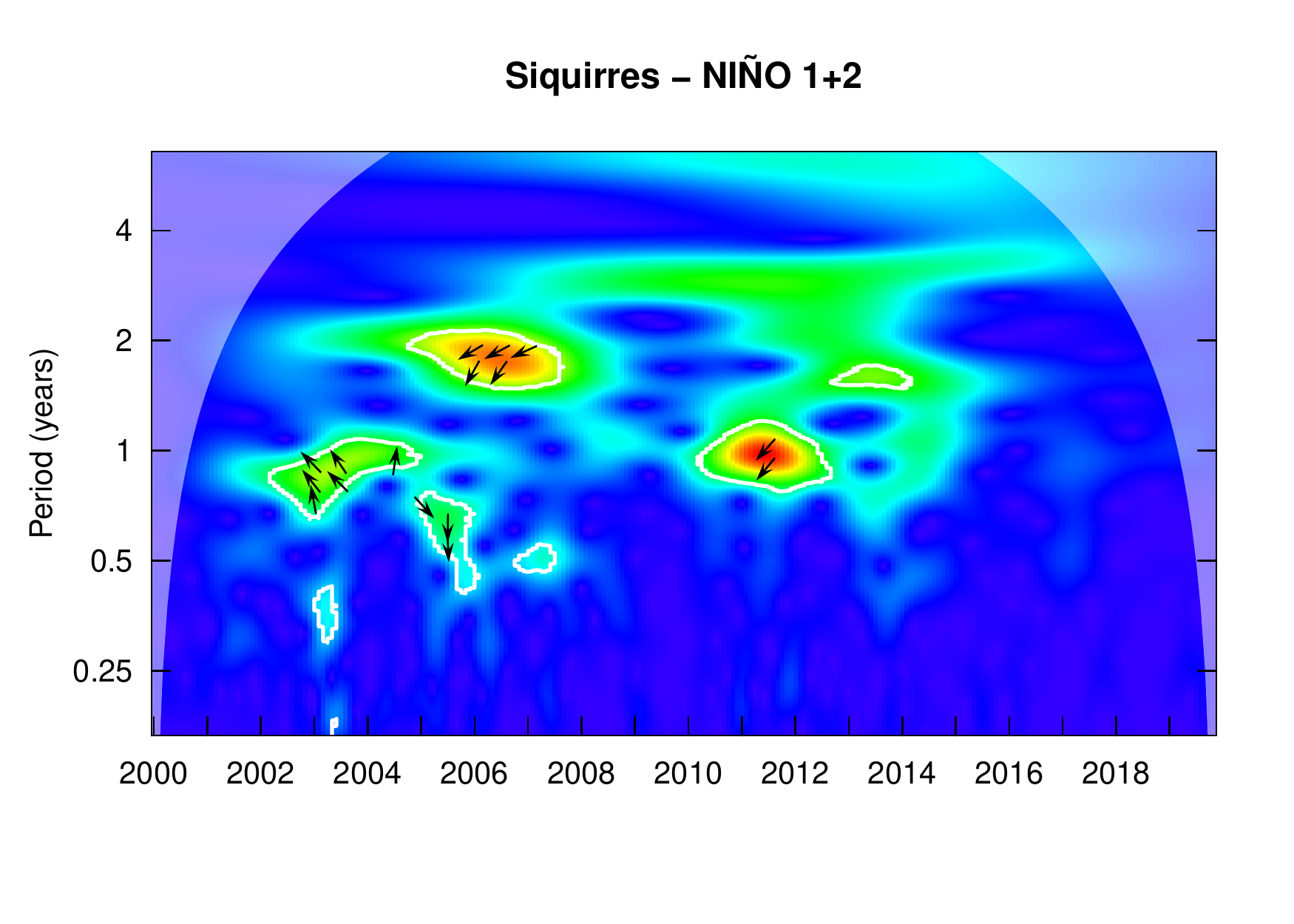}}\vspace{-0.15cm}%
\subfloat[]{\includegraphics[scale=0.23]{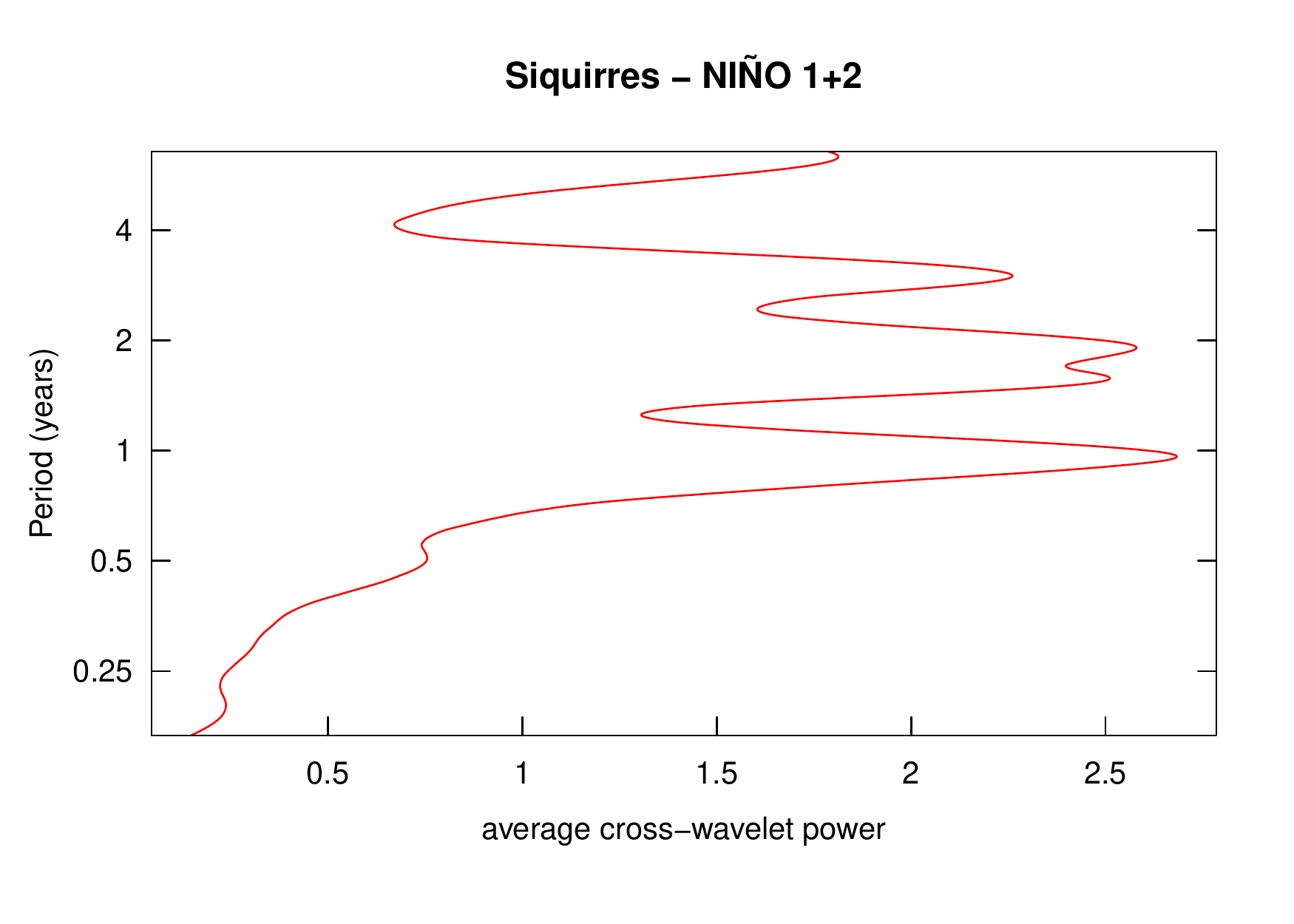}}\vspace{-0.15cm}\\
\subfloat[]{\includegraphics[scale=0.23]{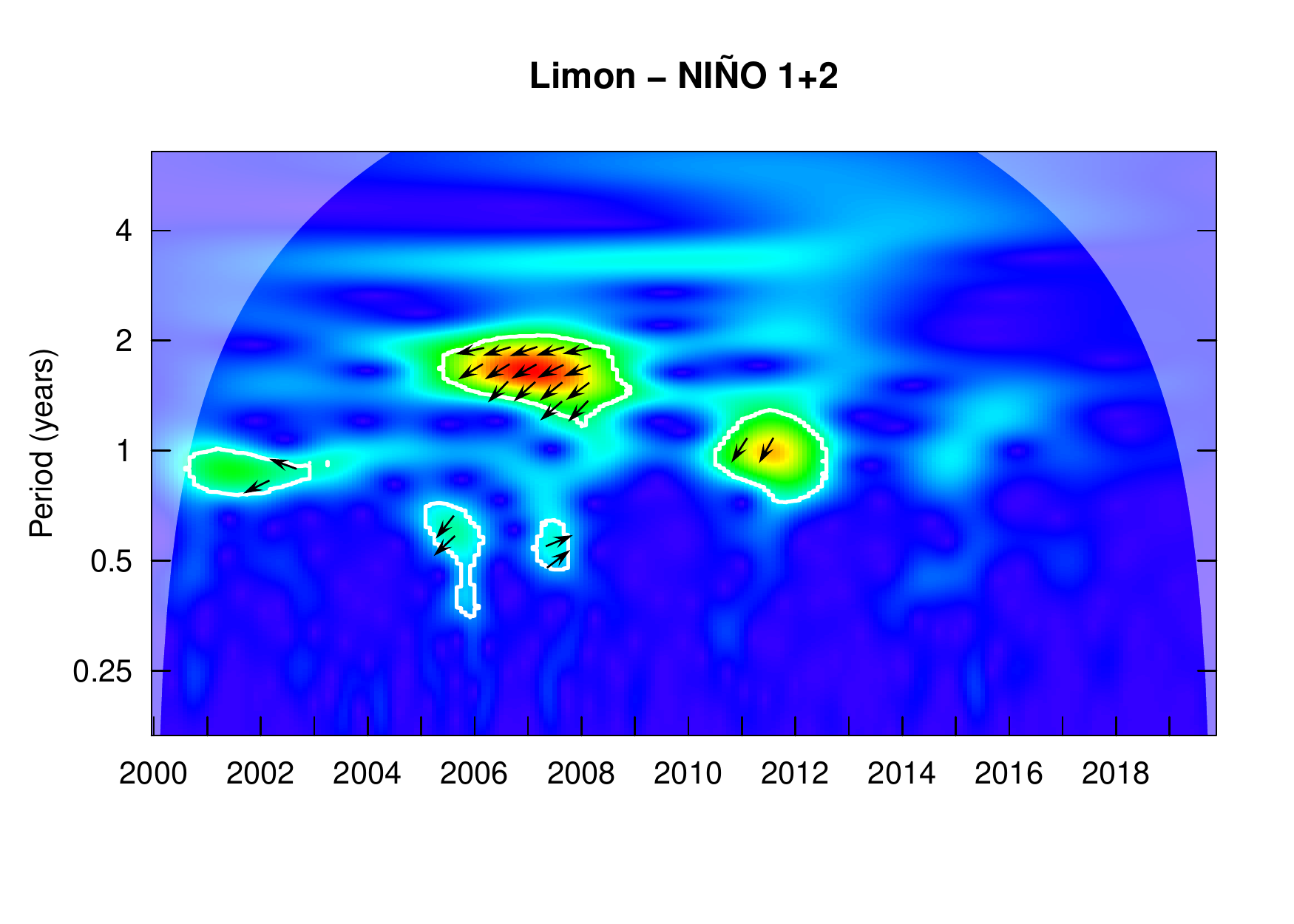}}\vspace{-0.15cm}%
\subfloat[]{\includegraphics[scale=0.23]{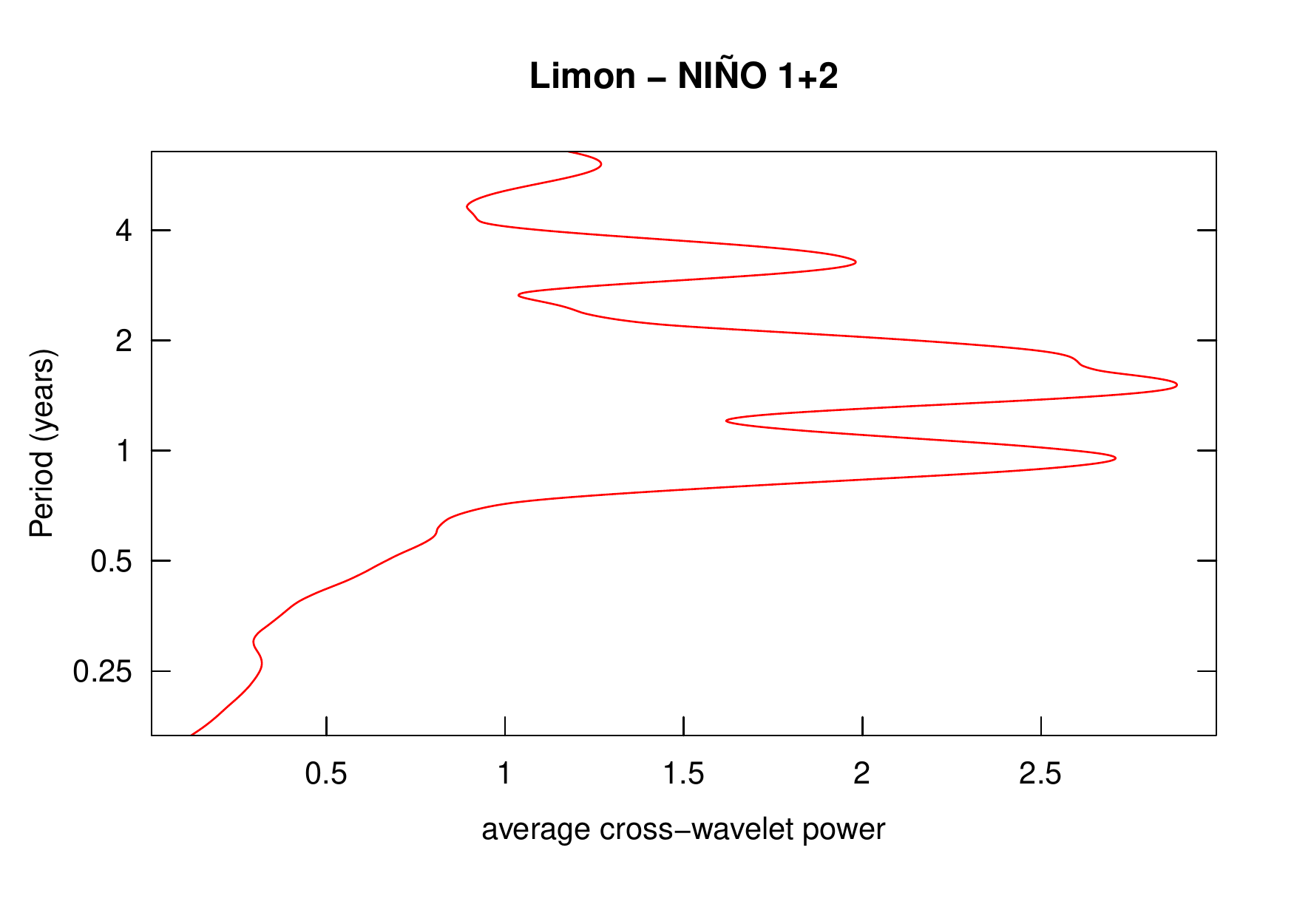}}\vspace{-0.15cm}%
\subfloat[]{\includegraphics[scale=0.23]{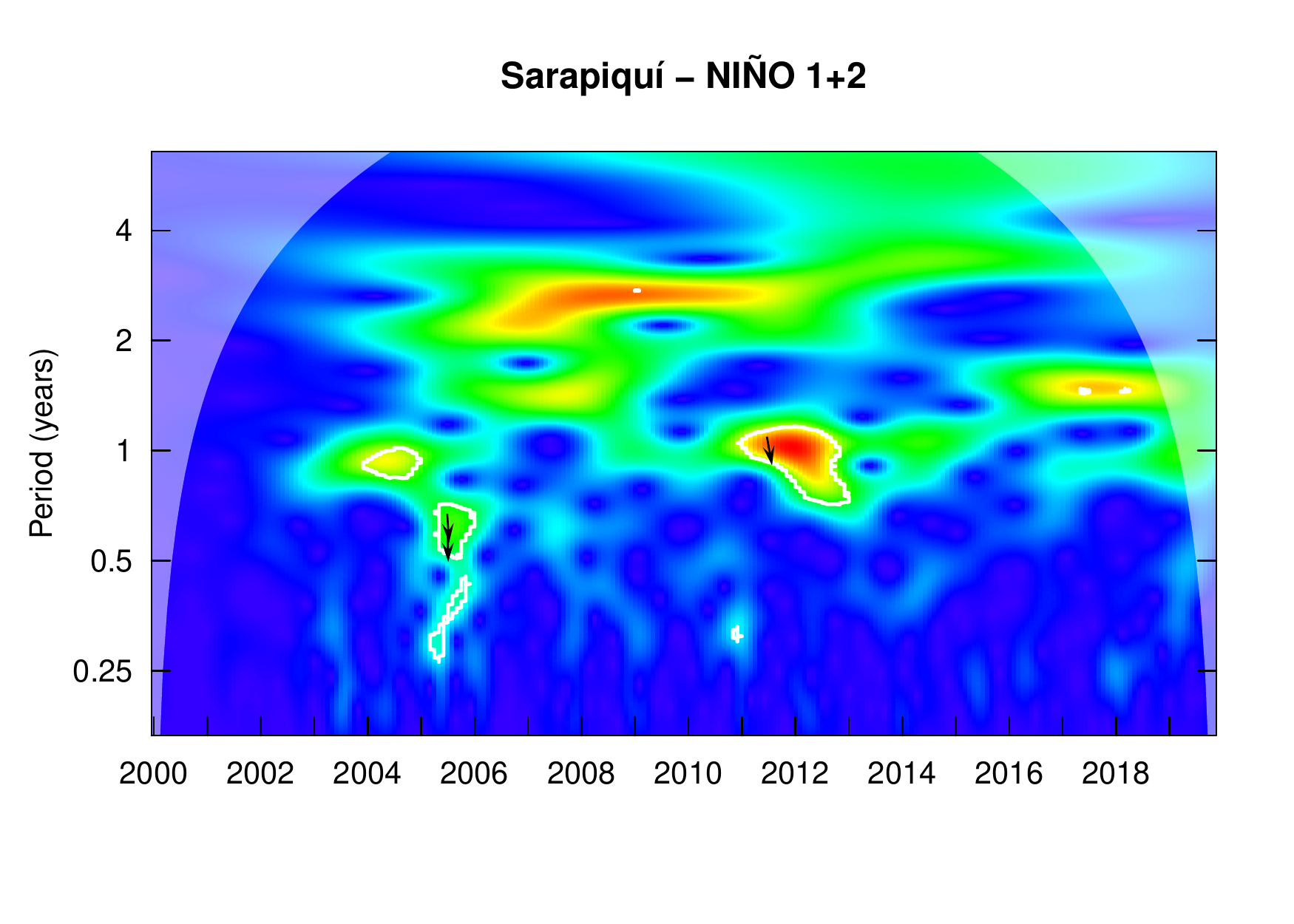}}\vspace{-0.15cm}%
\subfloat[]{\includegraphics[scale=0.23]{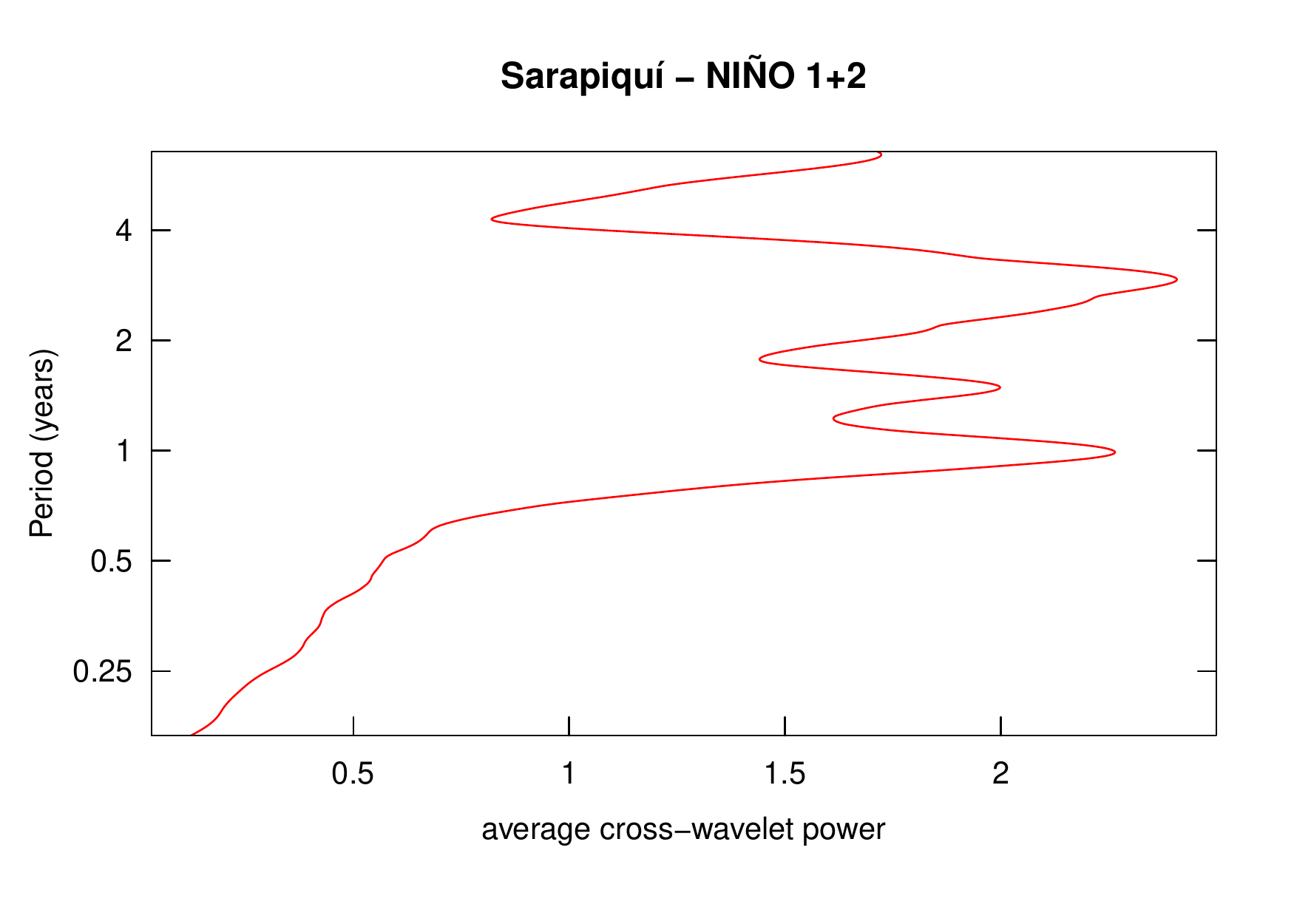}}\vspace{-0.15cm}\\
\subfloat[]{\includegraphics[scale=0.23]{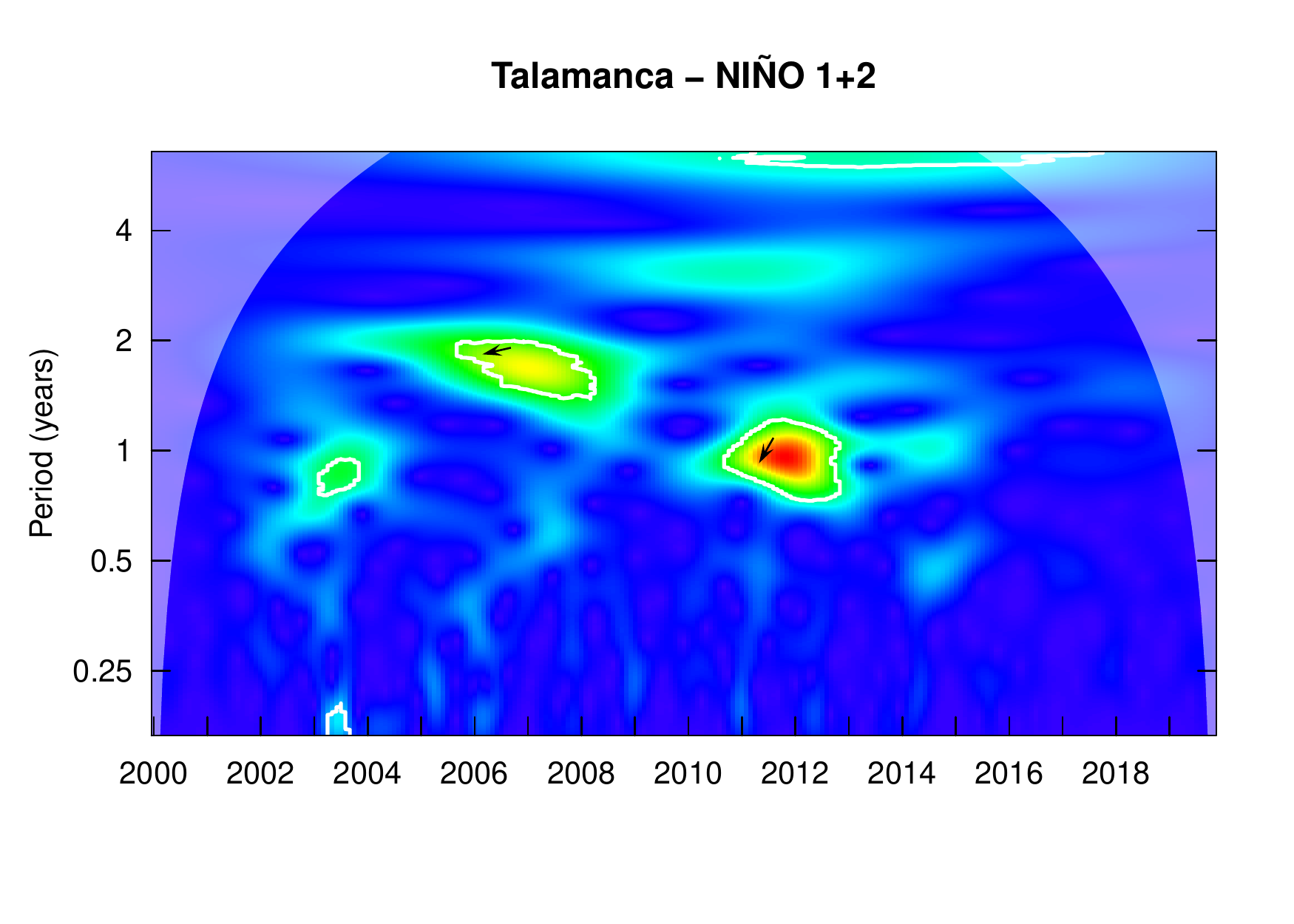}}\vspace{-0.15cm}%
\subfloat[]{\includegraphics[scale=0.23]{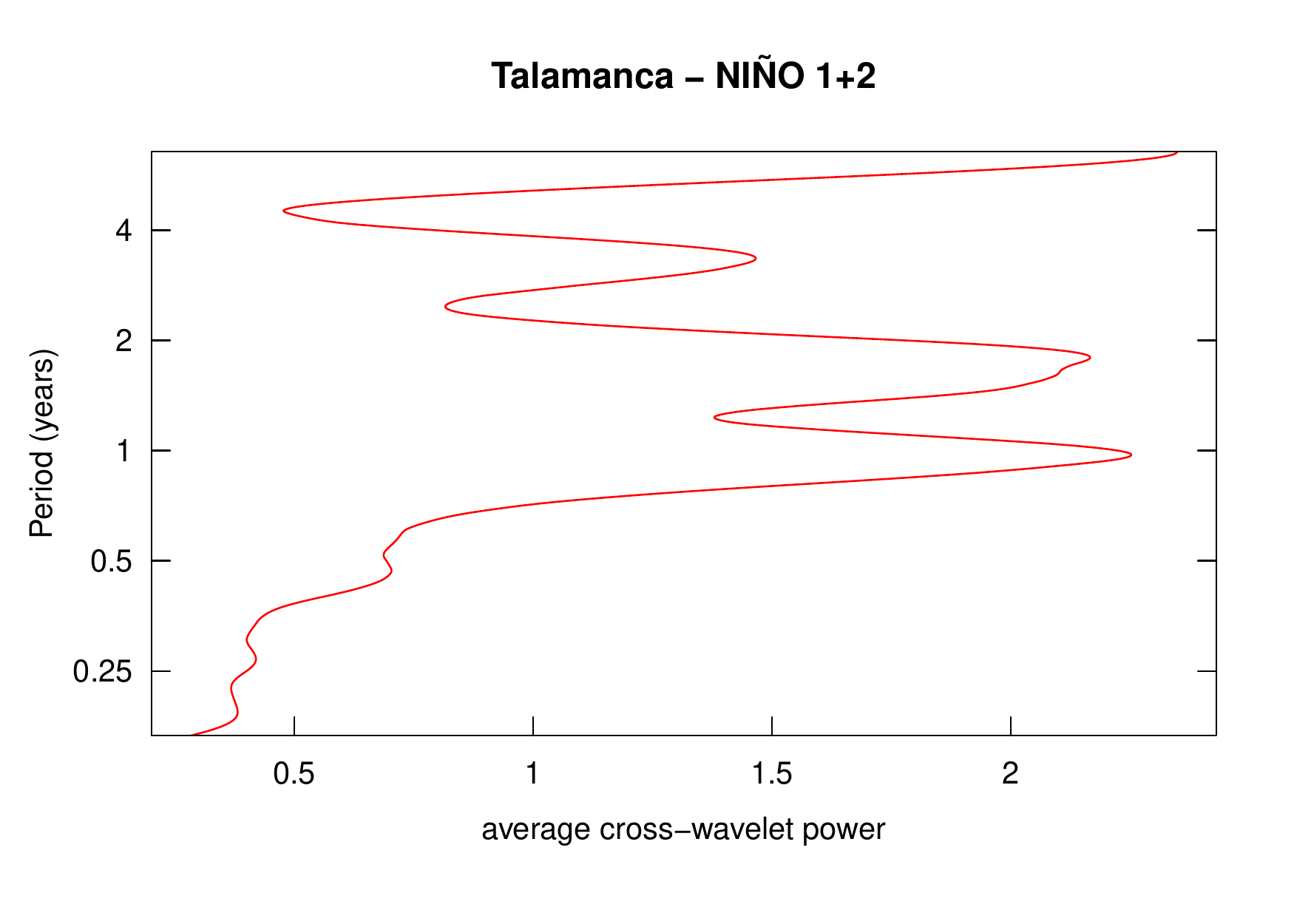}}\vspace{-0.15cm}%
\subfloat[]{\includegraphics[scale=0.23]{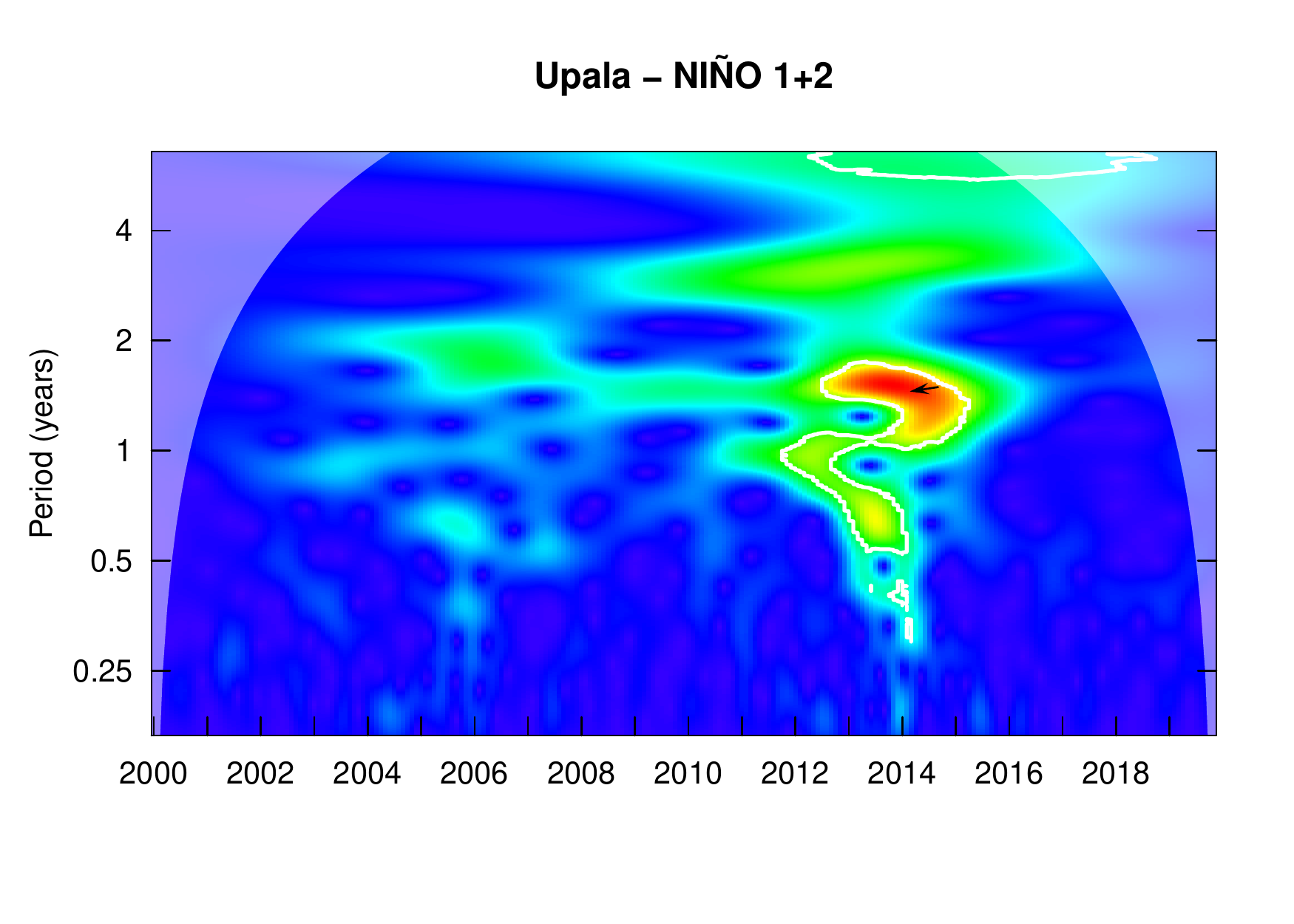}}\vspace{-0.15cm}%
\subfloat[]{\includegraphics[scale=0.23]{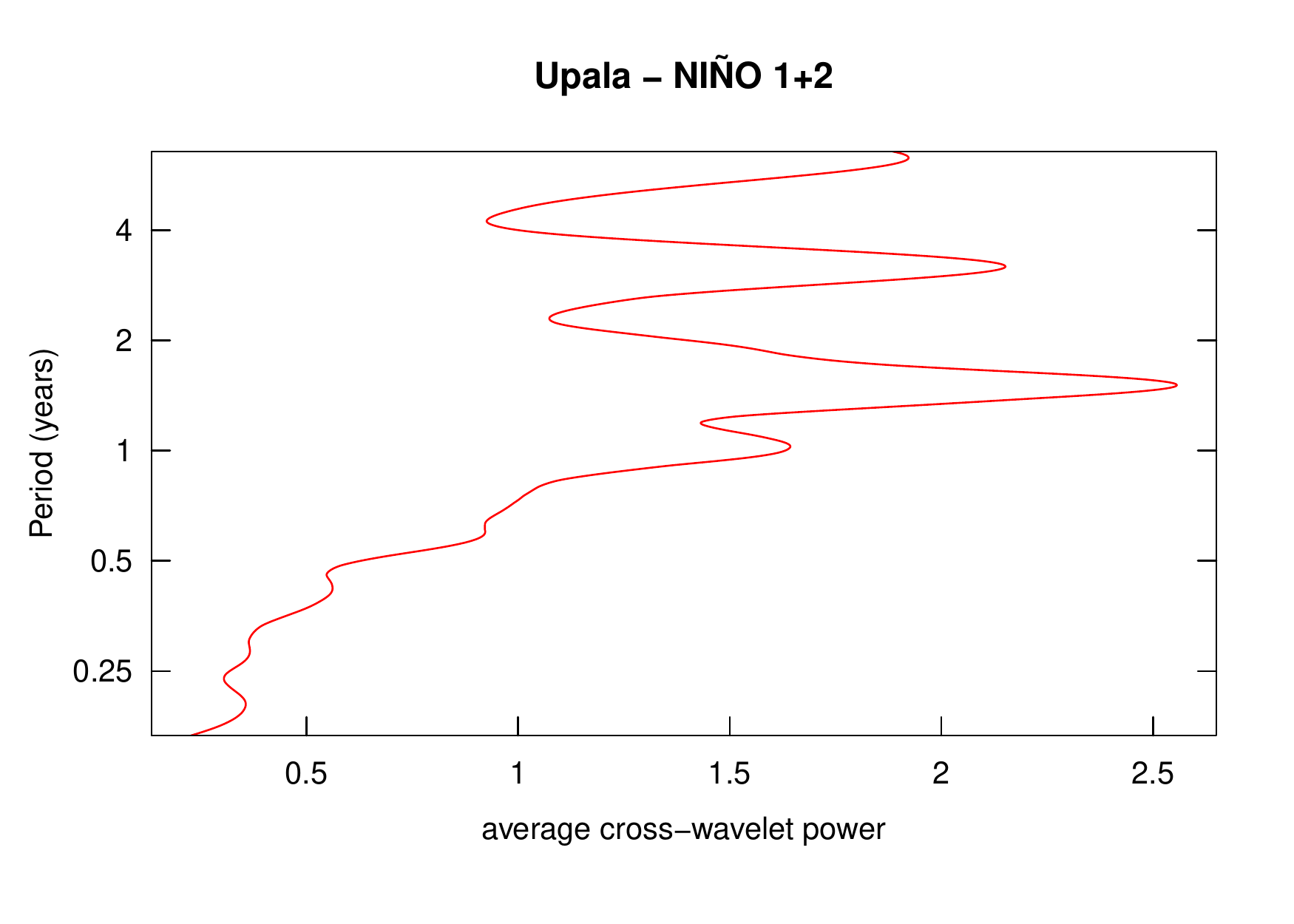}}\vspace{-0.15cm}\\
\subfloat[]{\includegraphics[scale=0.23]{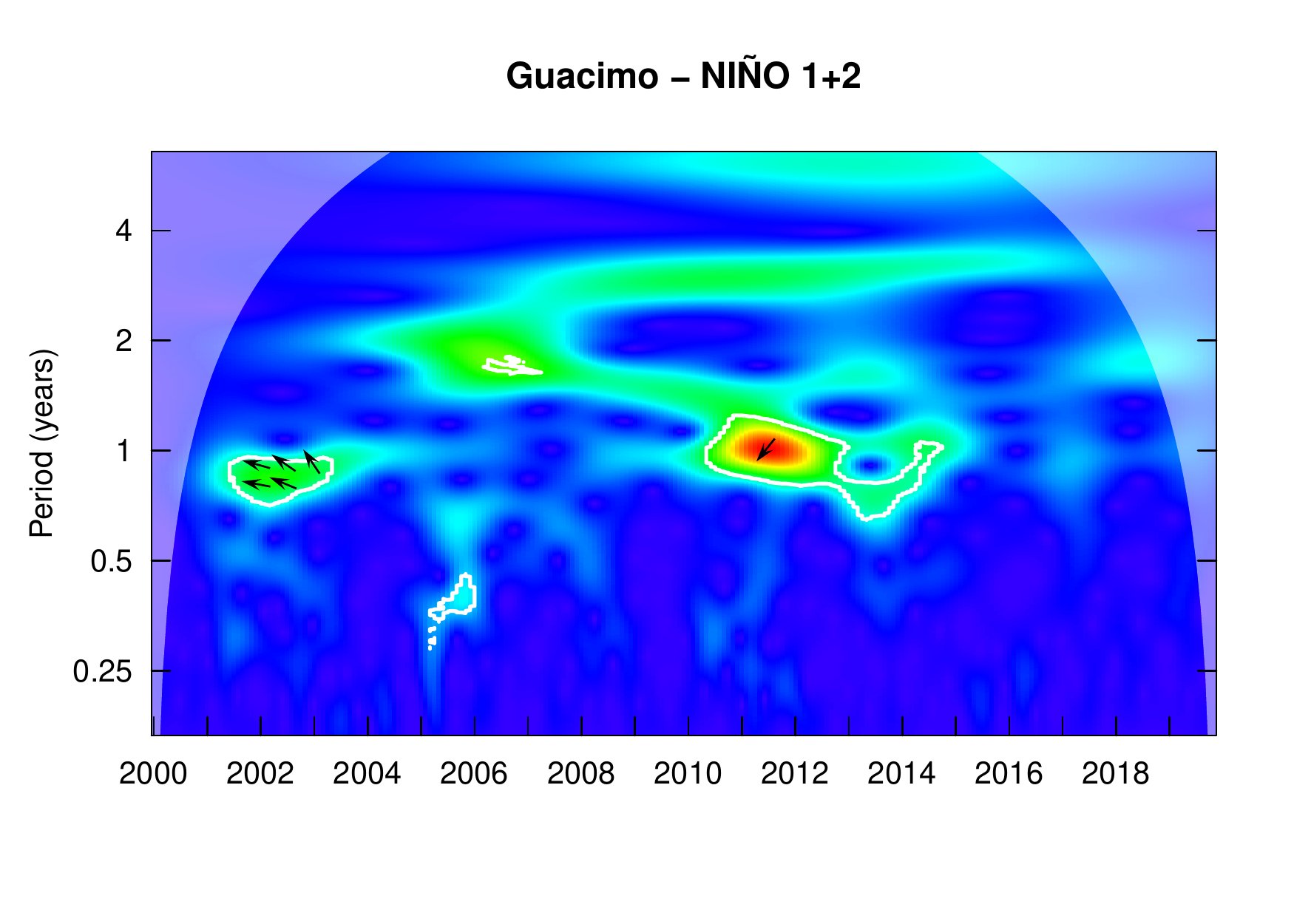}}\vspace{-0.15cm}%
\subfloat[]{\includegraphics[scale=0.23]{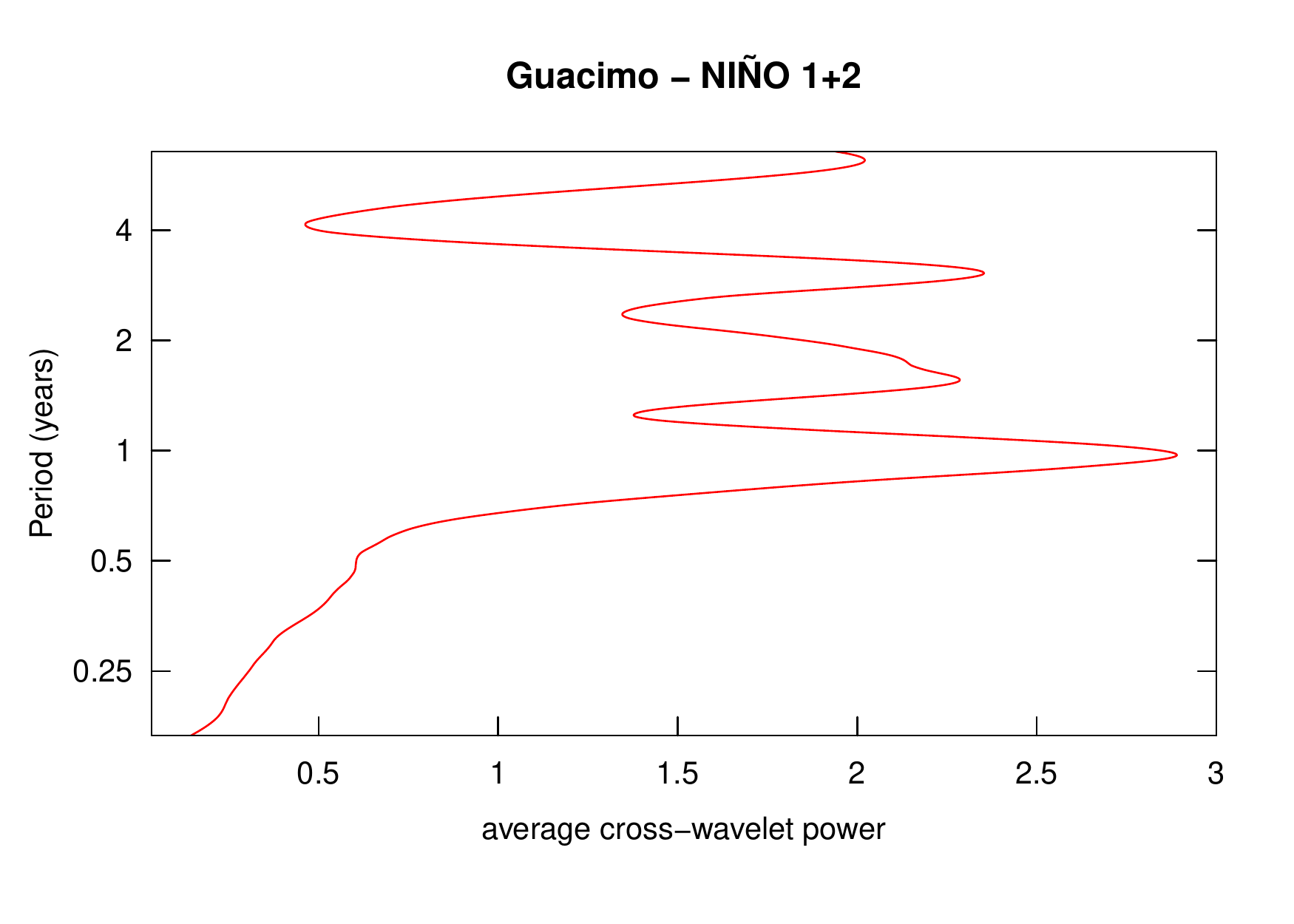}}\vspace{-0.15cm}%
\subfloat[]{\includegraphics[scale=0.23]{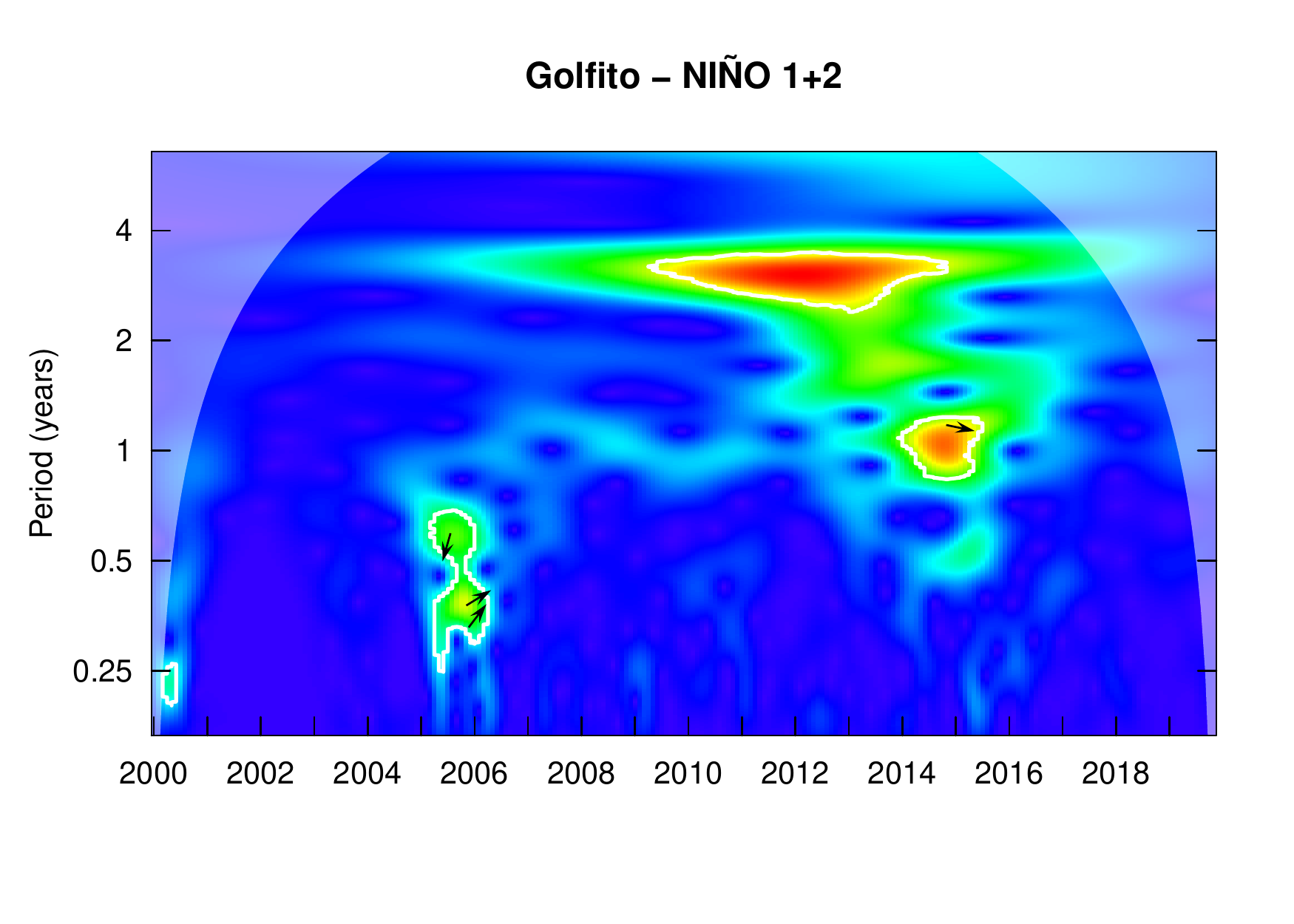}}\vspace{-0.15cm}%
\subfloat[]{\includegraphics[scale=0.23]{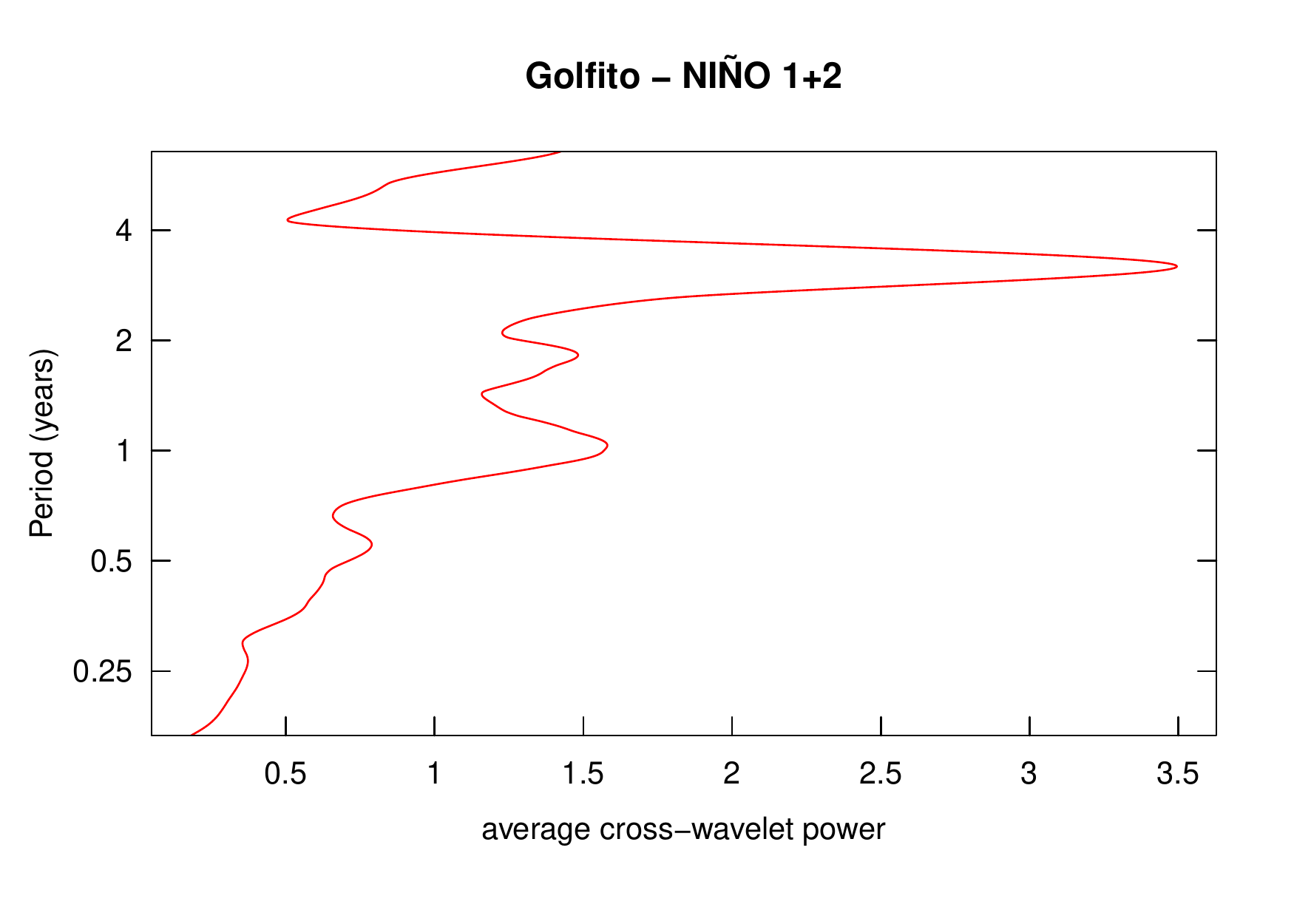}}\vspace{-0.15cm}\\
\subfloat[]{\includegraphics[scale=0.23]{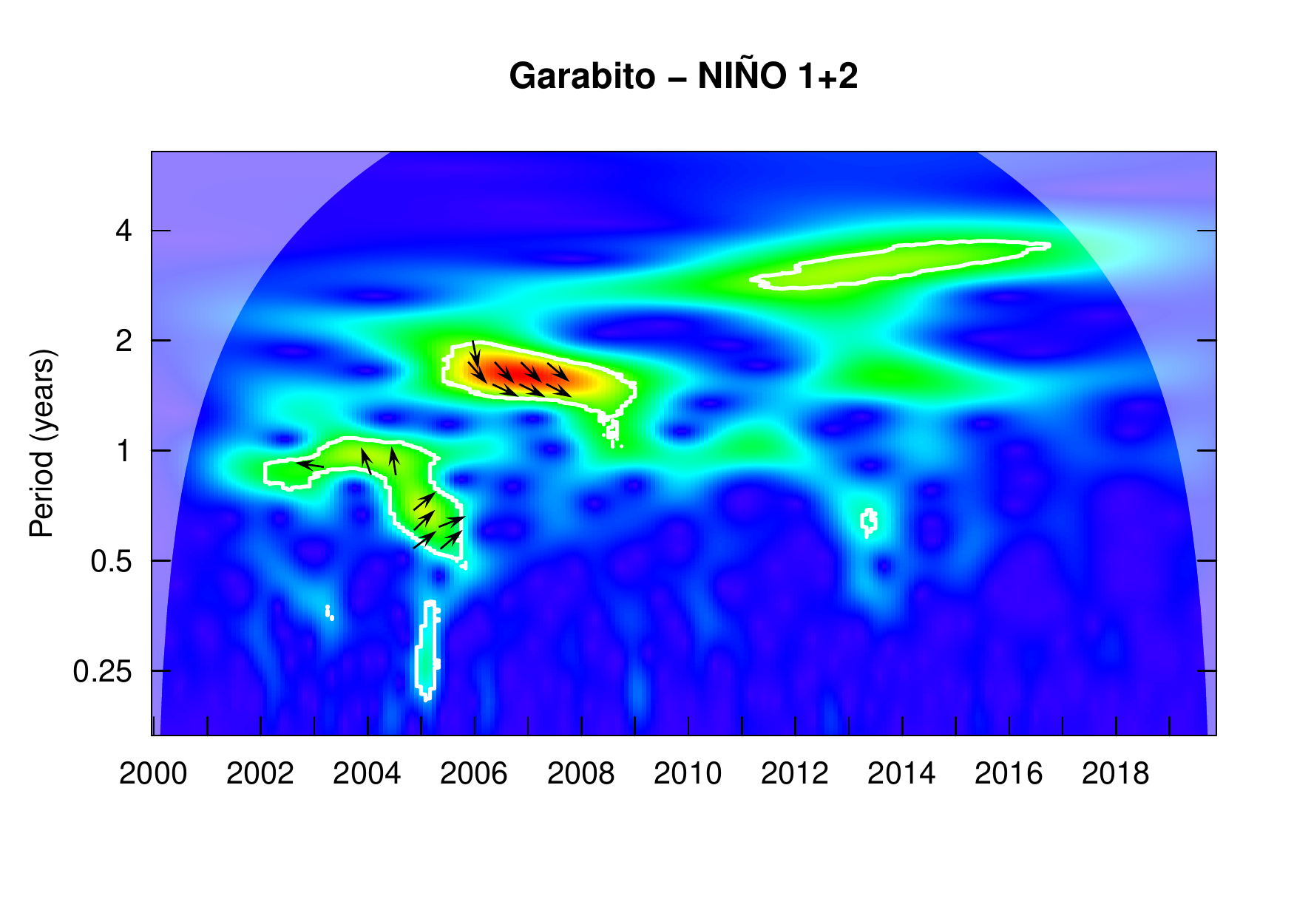}}\vspace{-0.15cm}%
\subfloat[]{\includegraphics[scale=0.23]{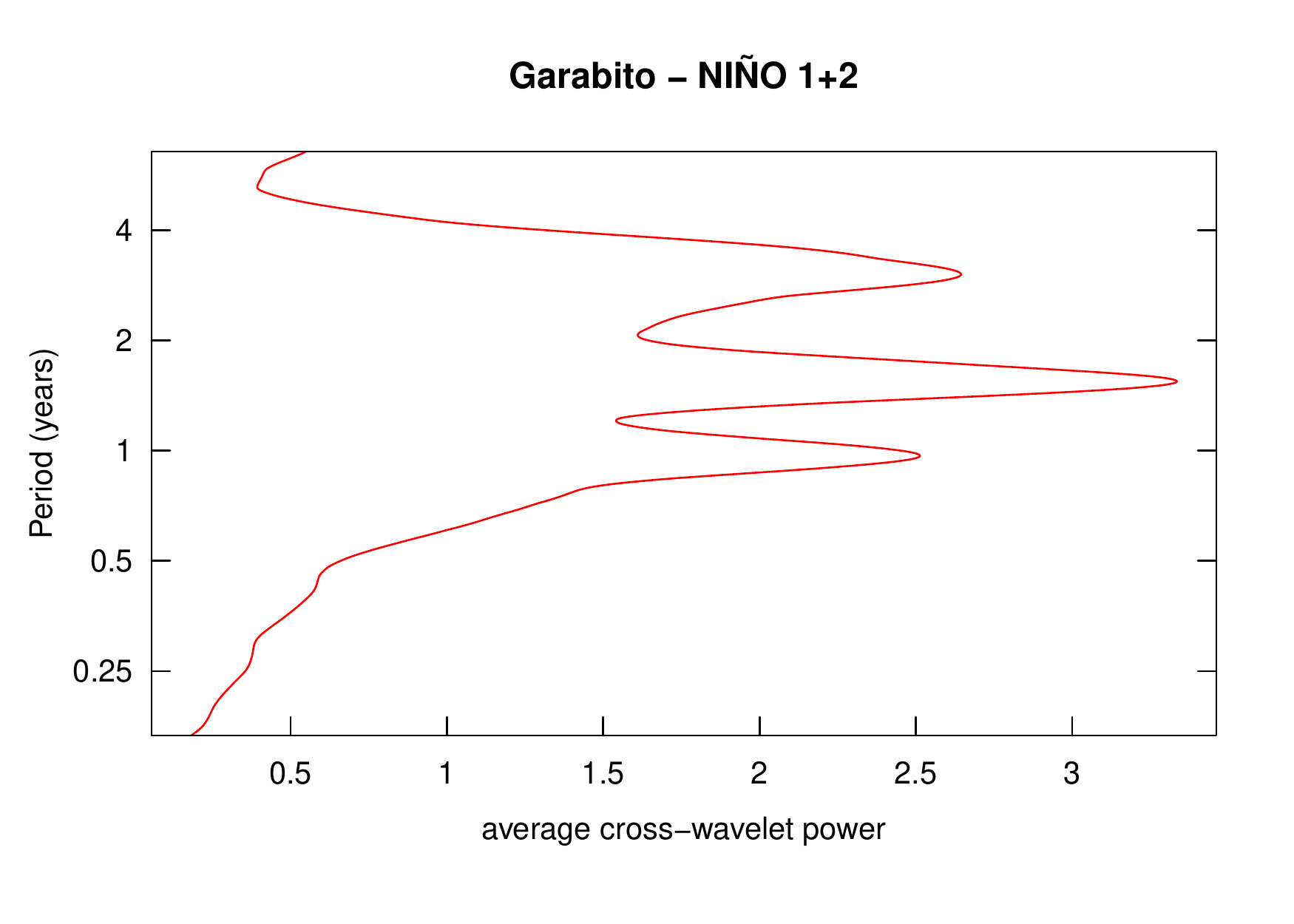}}\vspace{-0.15cm}%
\subfloat[]{\includegraphics[scale=0.23]{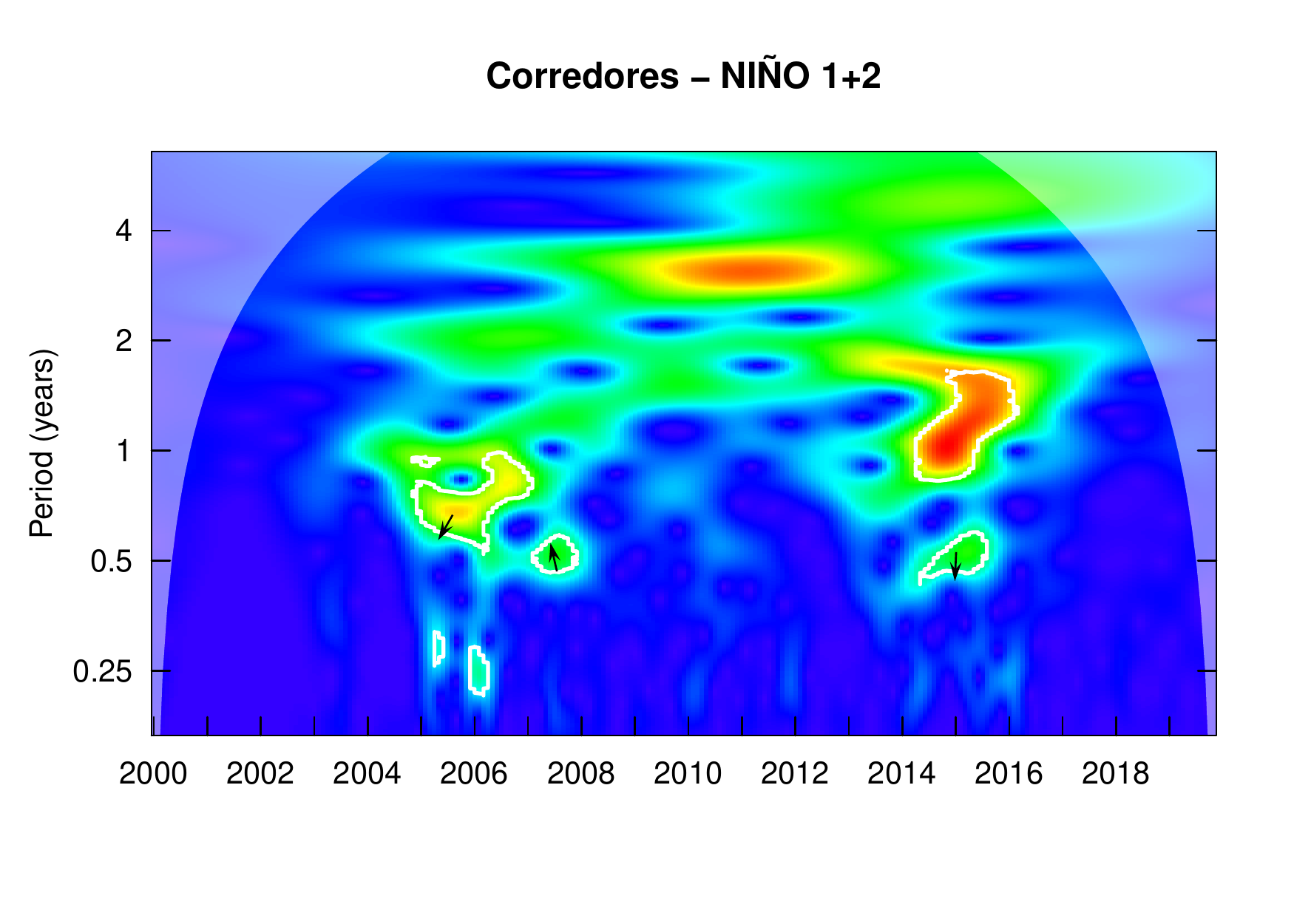}}\vspace{-0.15cm}%
\subfloat[]{\includegraphics[scale=0.23]{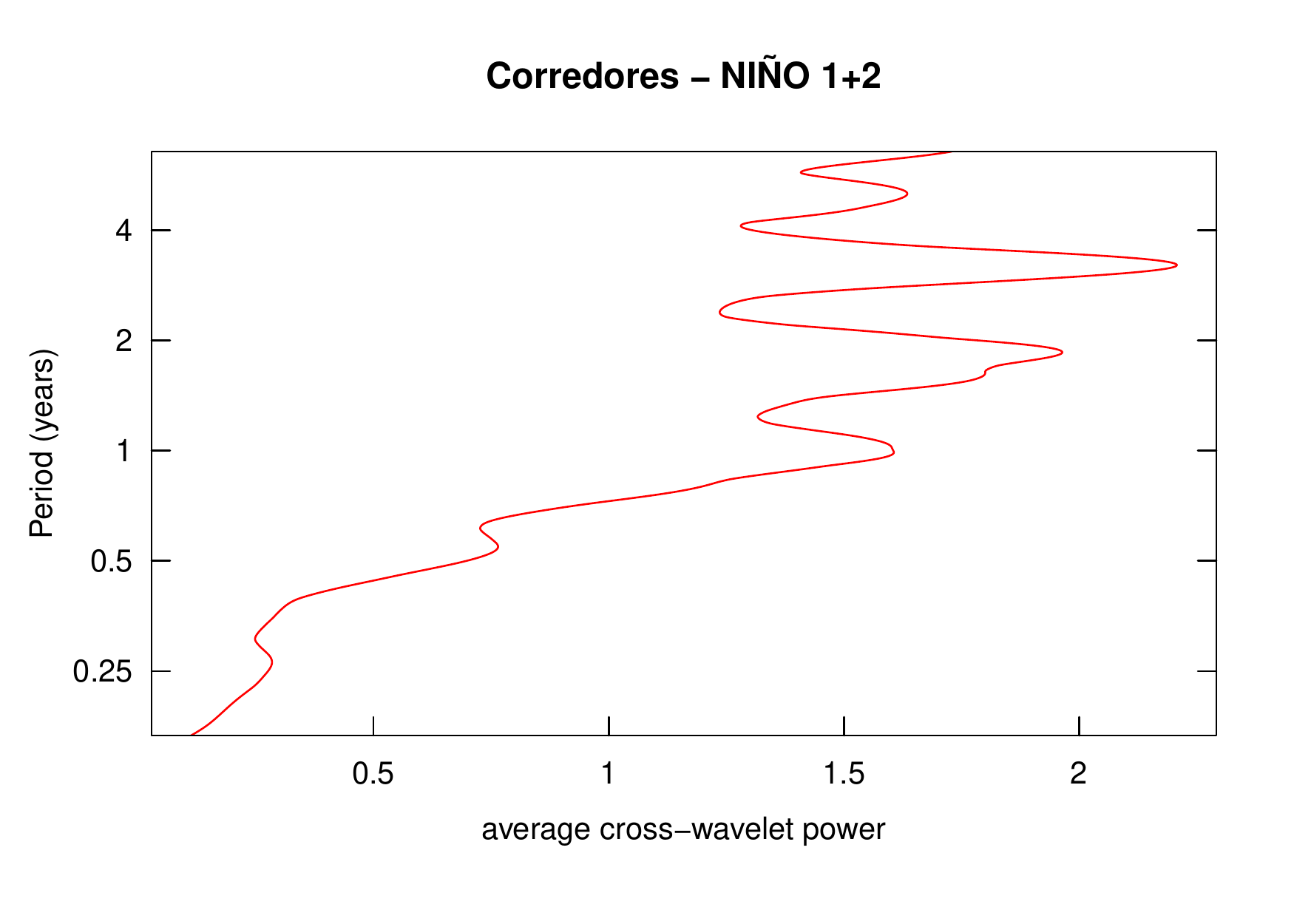}}\vspace{-0.15cm}
\caption*{}
\end{figure}

\begin{figure}[H]
\captionsetup[subfigure]{labelformat=empty}
\subfloat[]{\includegraphics[scale=0.23]{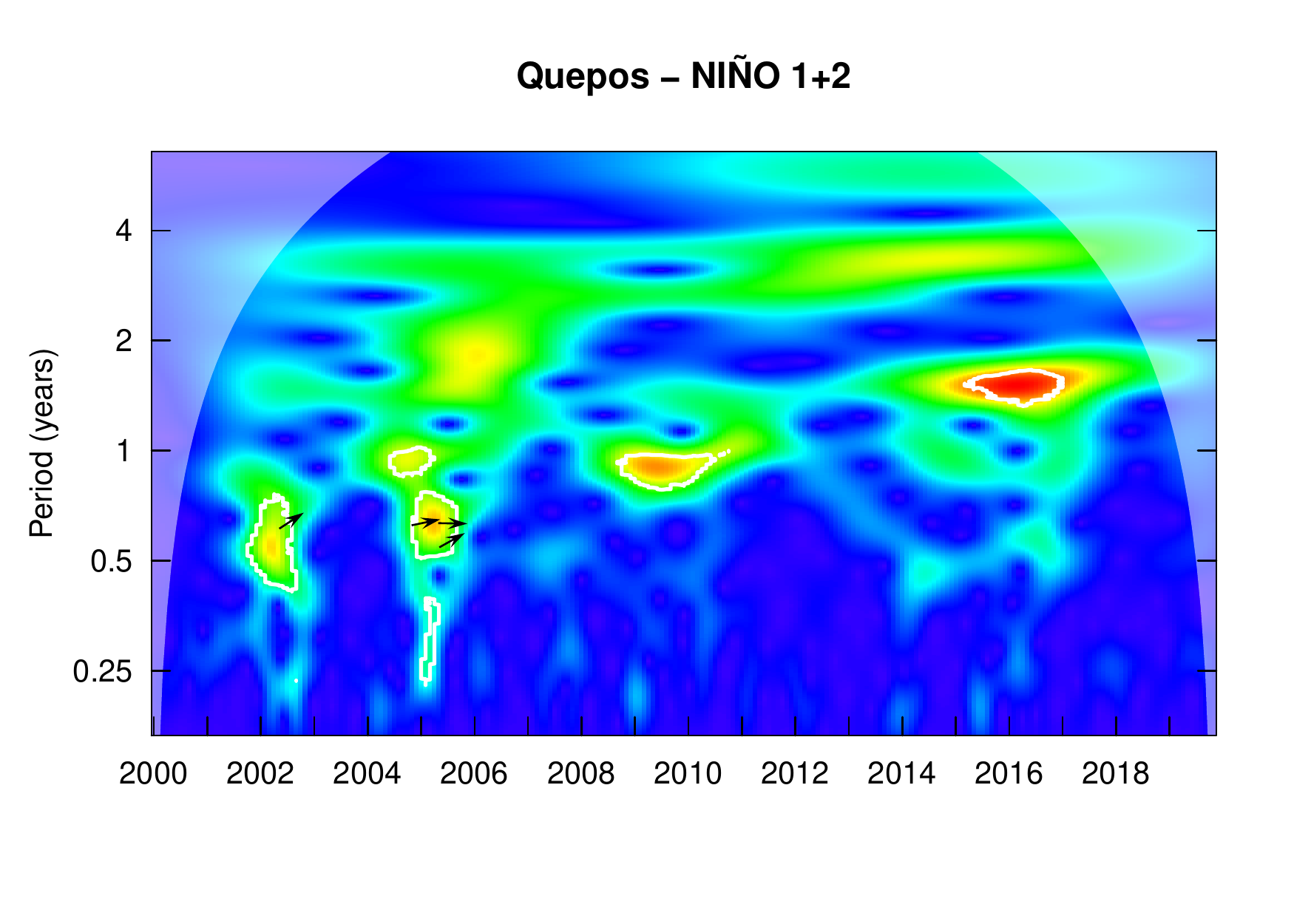}}\vspace{-0.15cm}%
\subfloat[]{\includegraphics[scale=0.23]{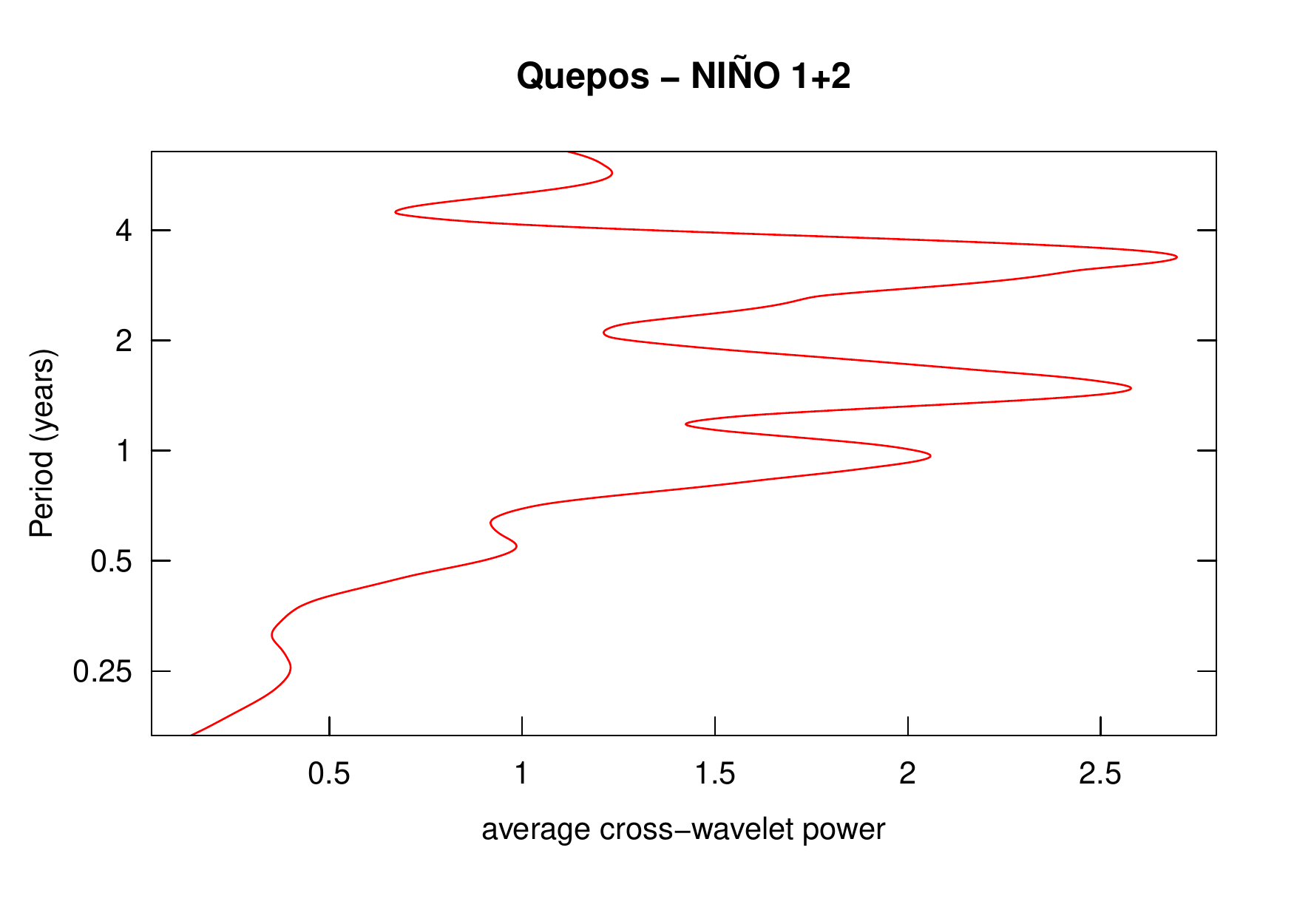}}\vspace{-0.15cm}%
\subfloat[]{\includegraphics[scale=0.23]{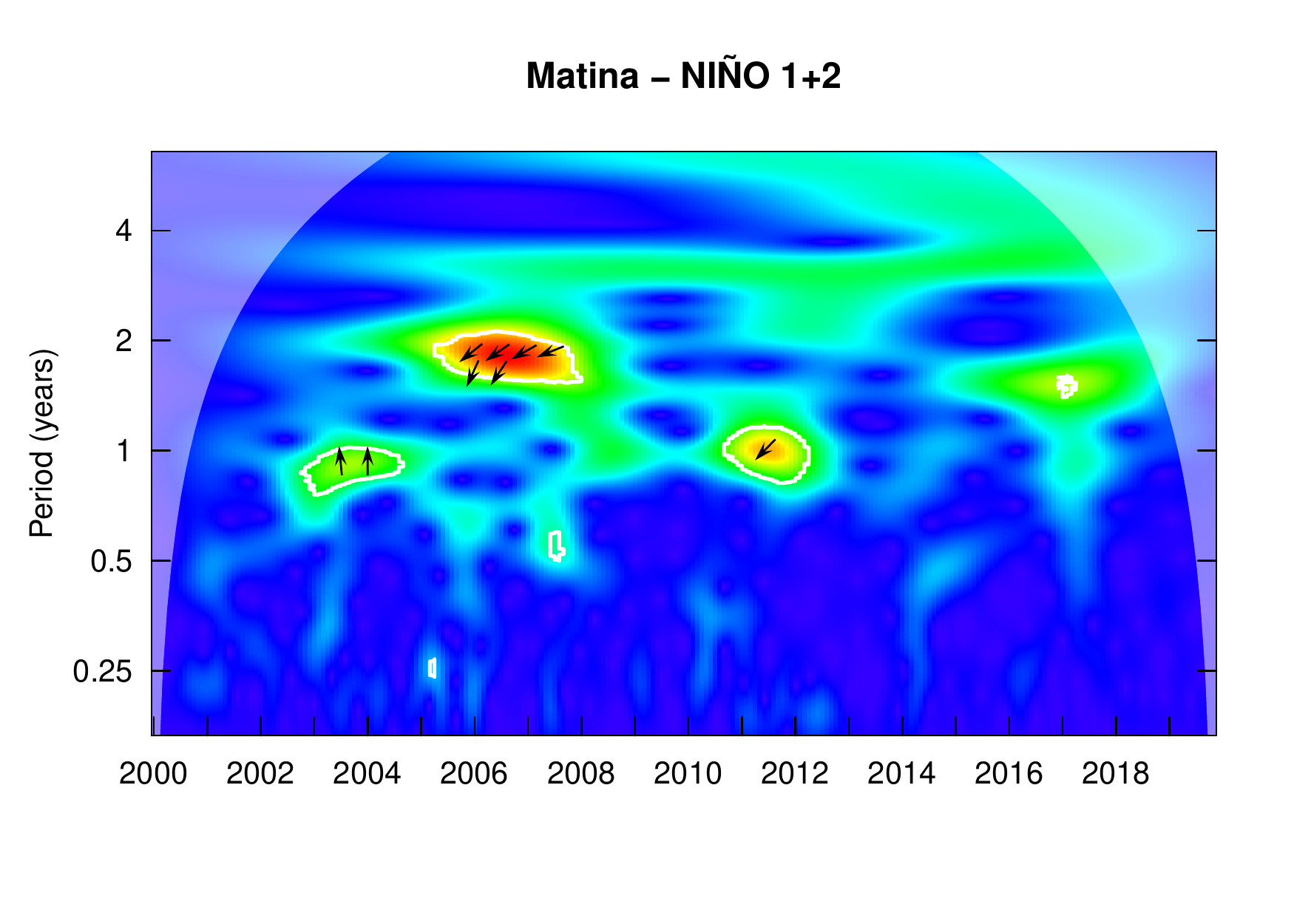}}\vspace{-0.15cm}%
\subfloat[]{\includegraphics[scale=0.23]{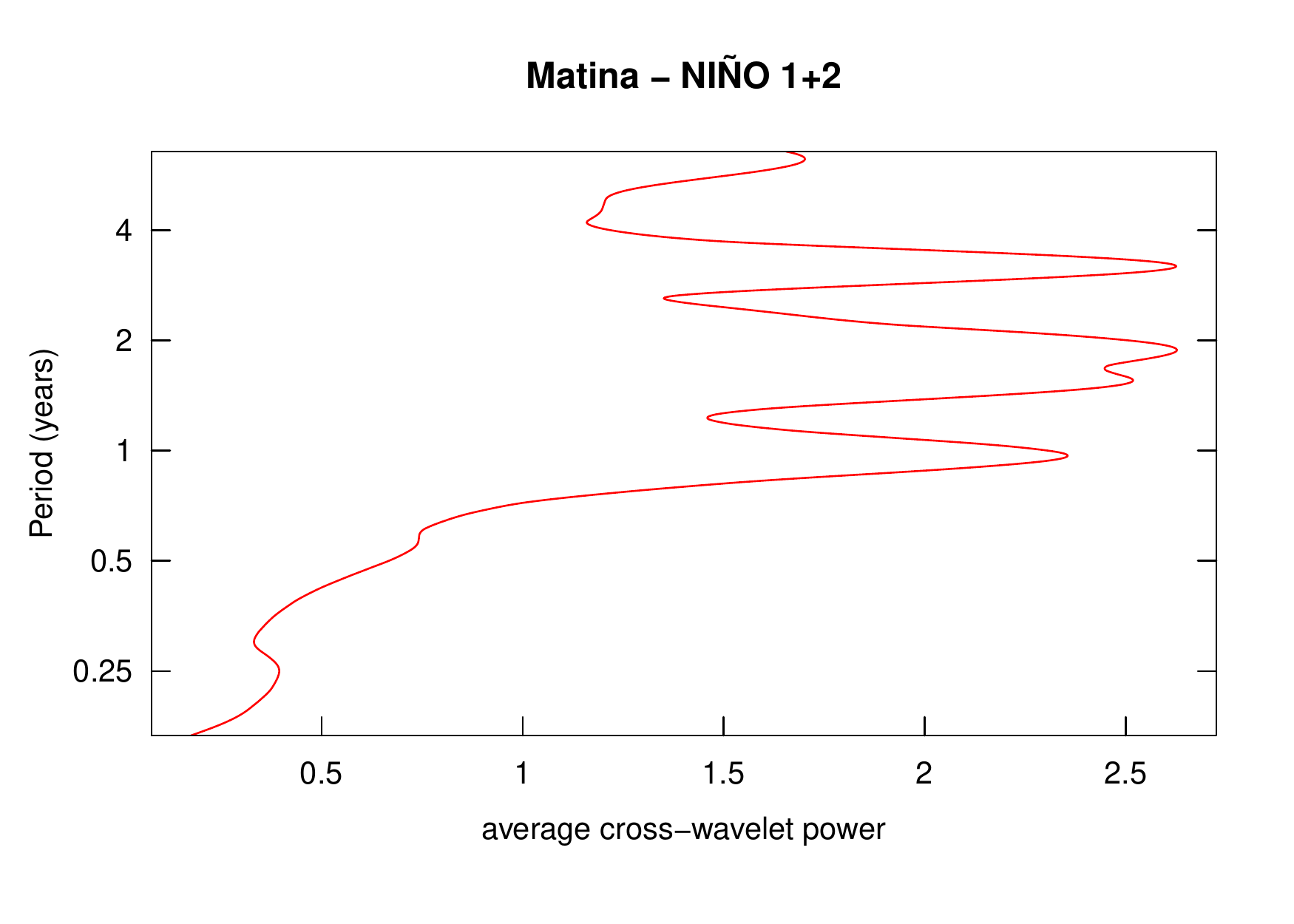}}\vspace{-0.15cm}\\
\subfloat[]{\includegraphics[scale=0.23]{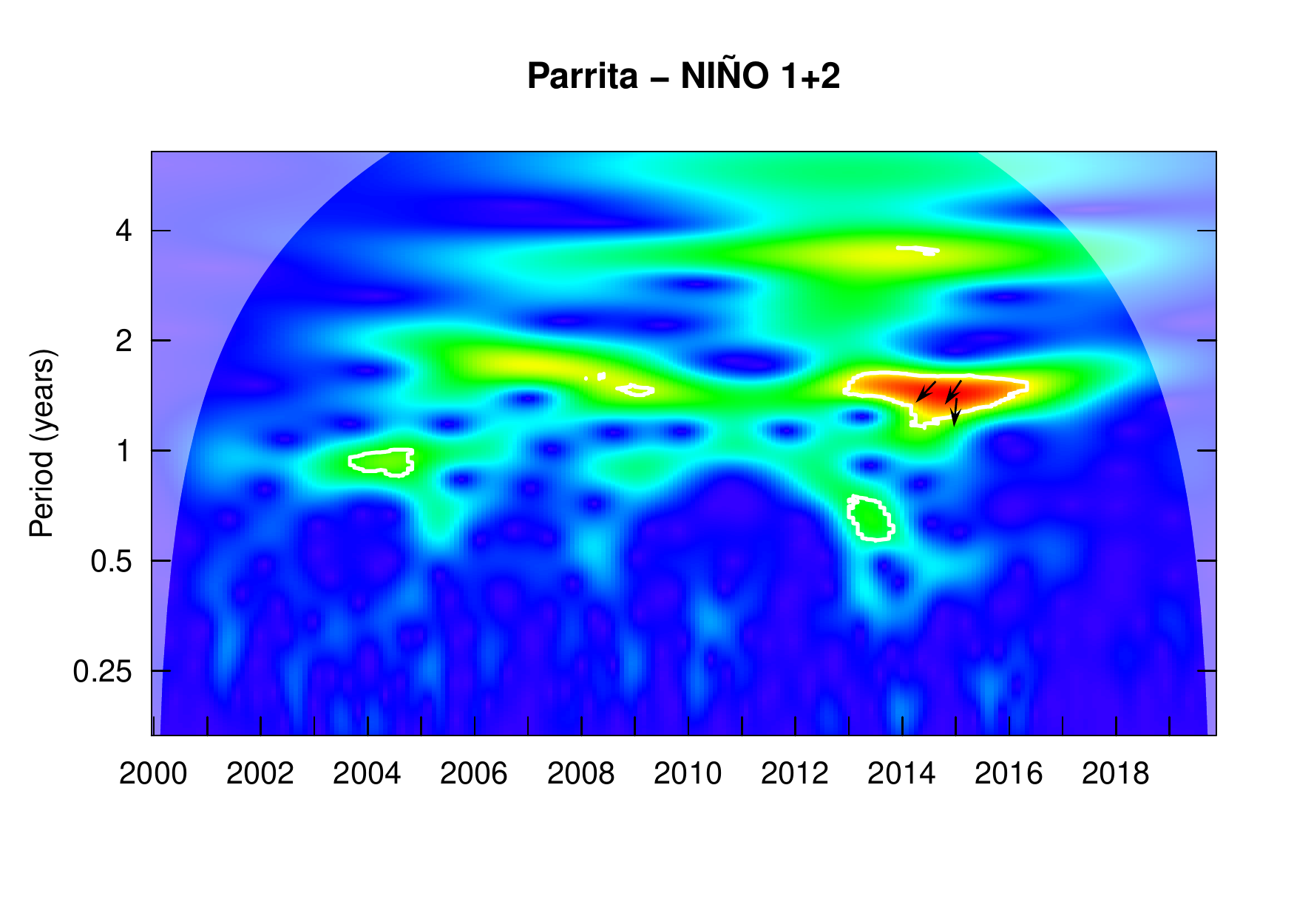}}\vspace{-0.15cm}%
\subfloat[]{\includegraphics[scale=0.23]{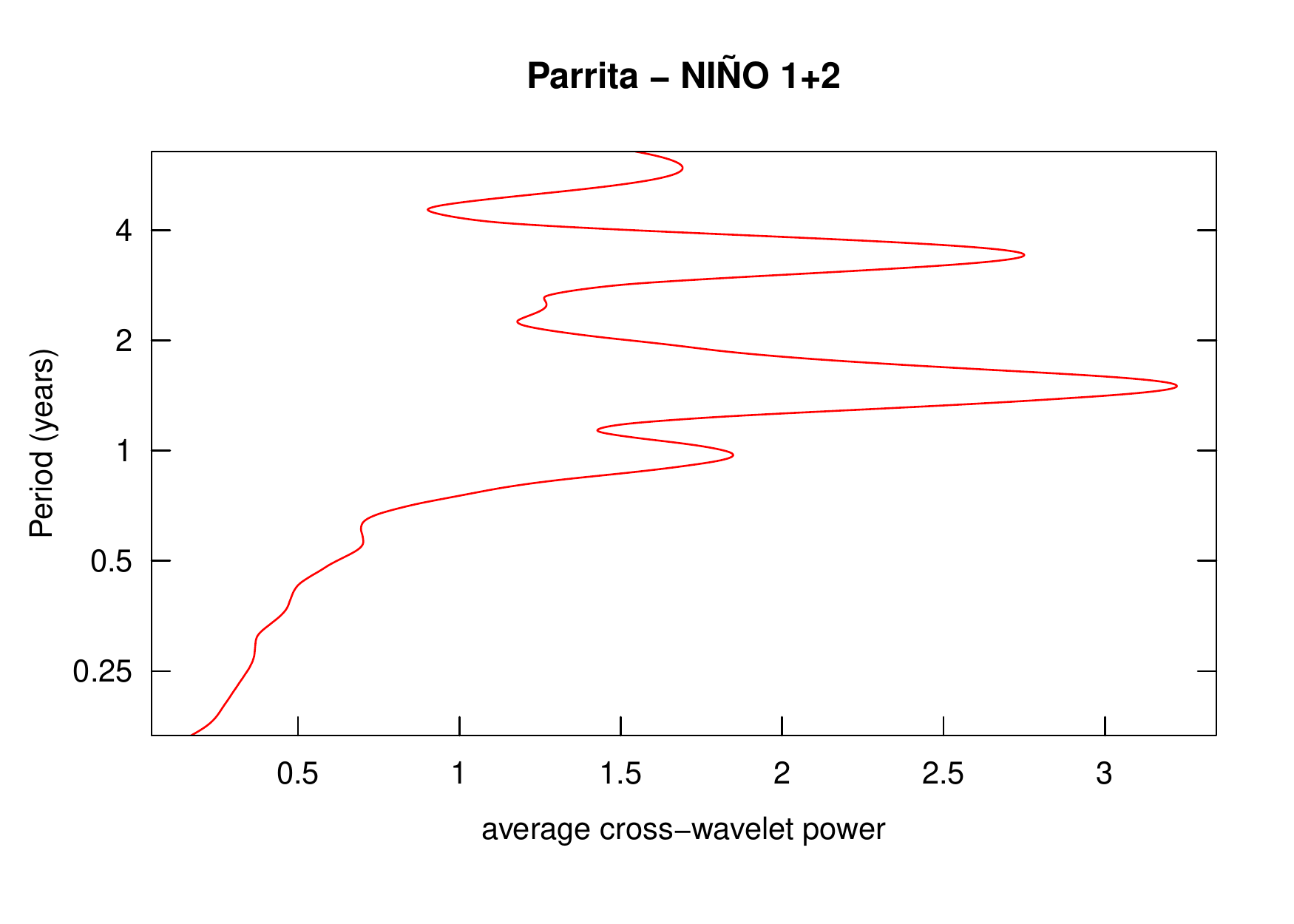}}\vspace{-0.15cm}%
\subfloat[]{\includegraphics[scale=0.23]{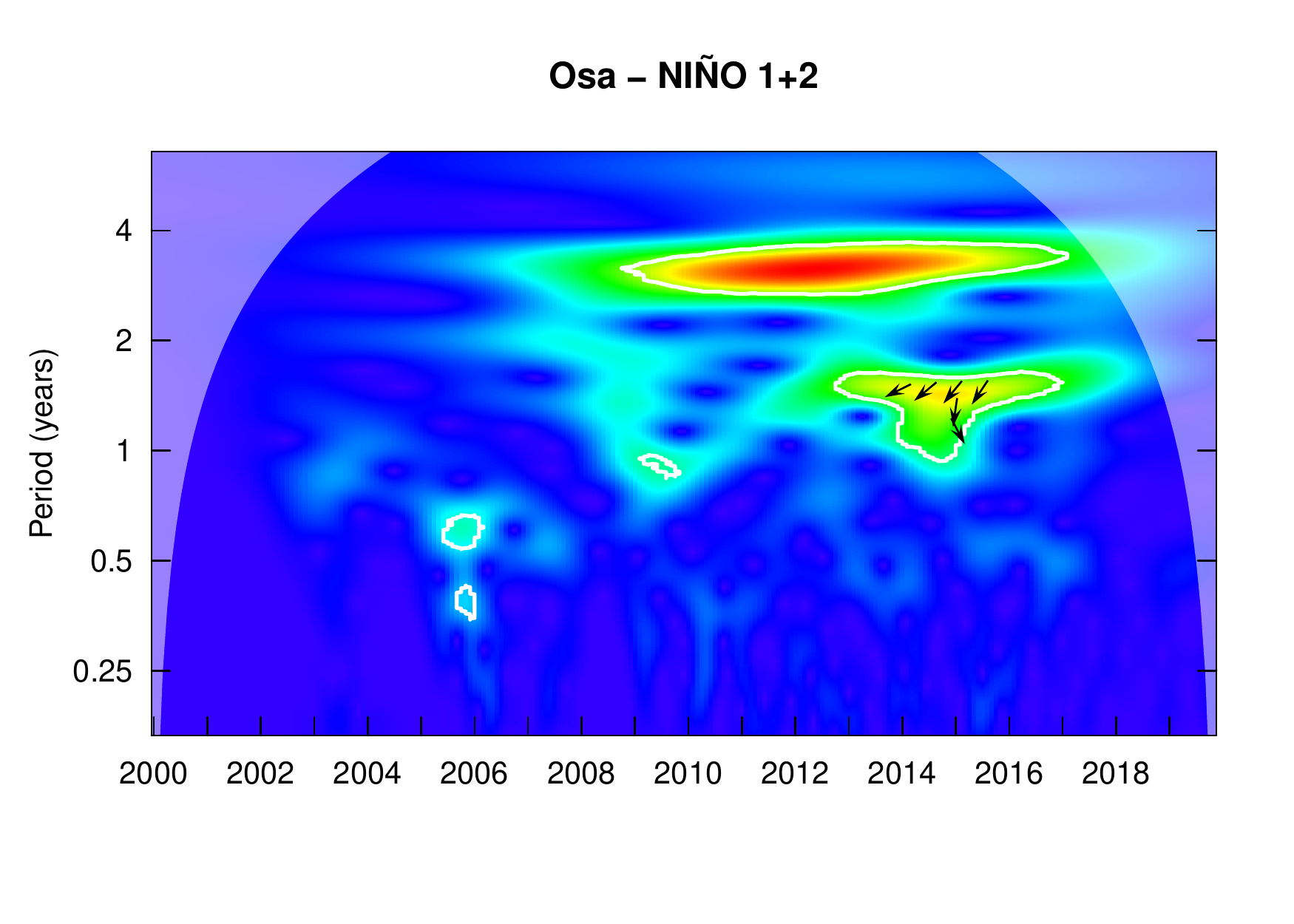}}\vspace{-0.15cm}%
\subfloat[]{\includegraphics[scale=0.23]{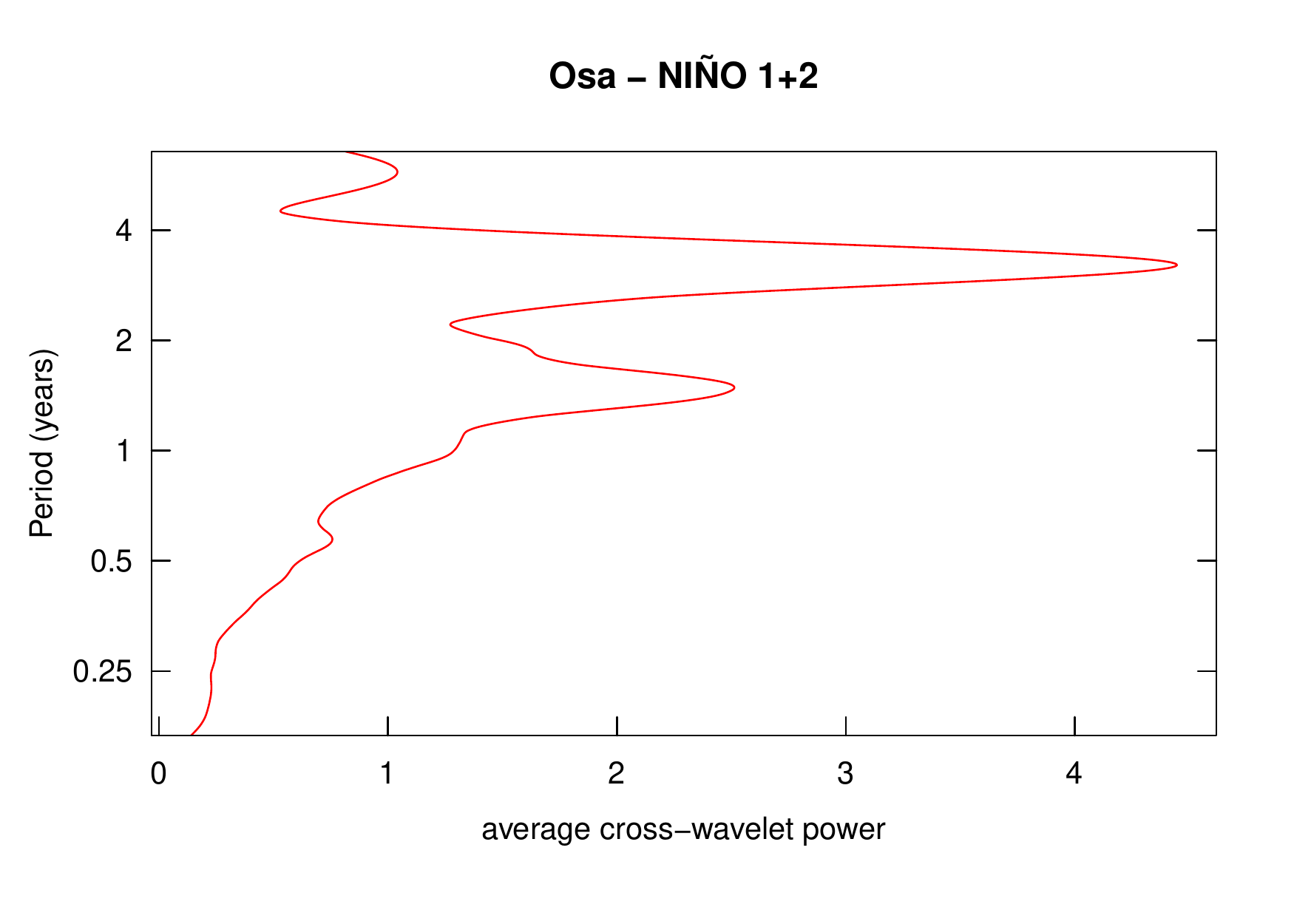}}\vspace{-0.15cm}
\caption*{}
\end{figure}

\section*{Wavelet coherence and average cross-wavelet power between dengue incidence and Ni\~no 3}

\begin{figure}[H]
\captionsetup[subfigure]{labelformat=empty}
\caption*{\textbf{Figure S8:} Wavelet coherence (color map) between dengue incidence from 2000 to 2019, and Ni\~no 3 in 32 municipalities of Costa Rica (periodicity on y-axis, time on x-axis). Colors code for increasing power intensity, from blue to red; $95\%$ confidence levels are encircled by white lines, and shaded areas indicate the presence of significant edge effects. On the right side of each wavelet coherence is the average cross-wavelet power (Red line). The arrows indicate whether the two series are in-phase or out-phase.}
\subfloat[]{\includegraphics[scale=0.23]{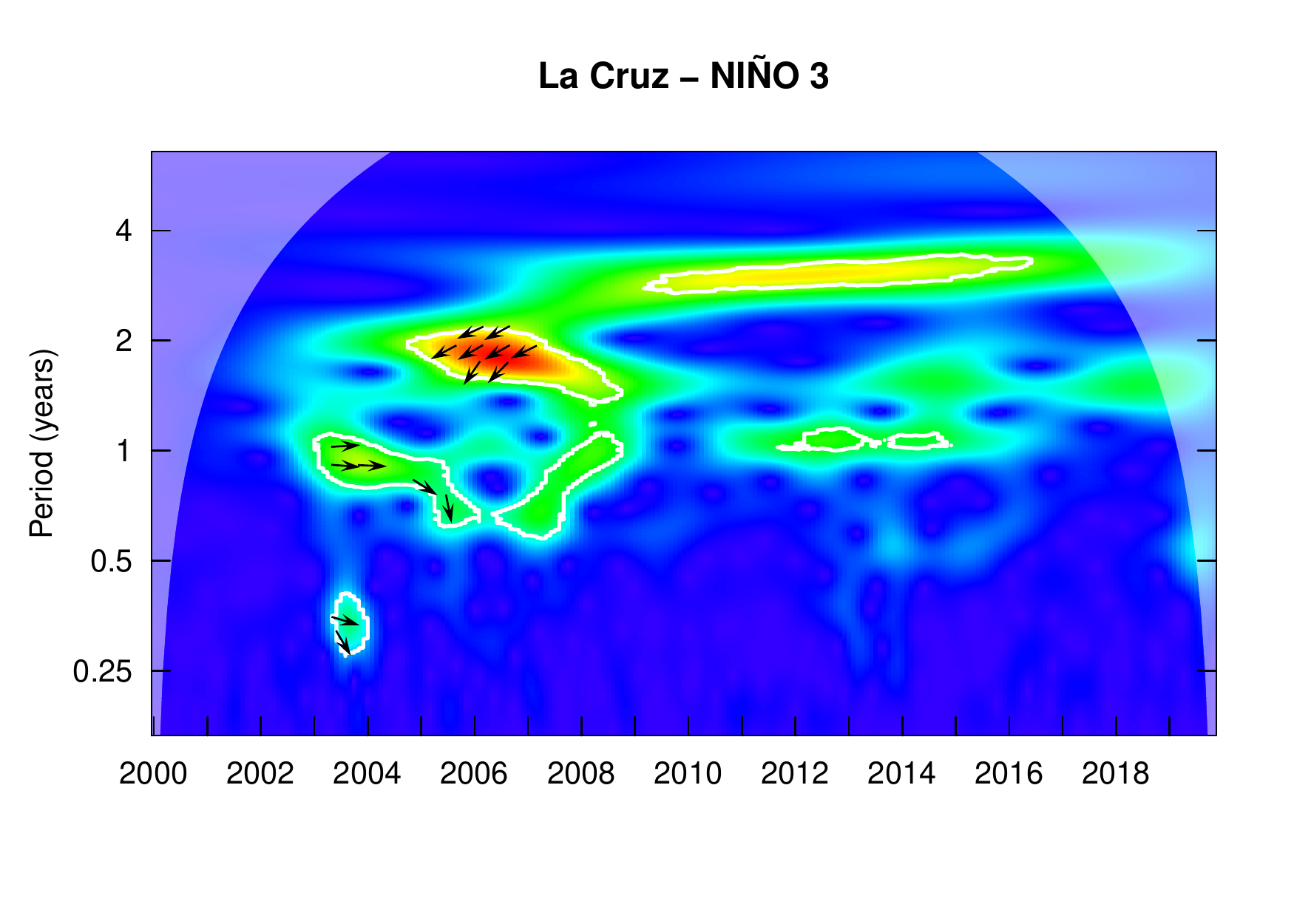}}\vspace{-0.15cm}%
\subfloat[]{\includegraphics[scale=0.23]{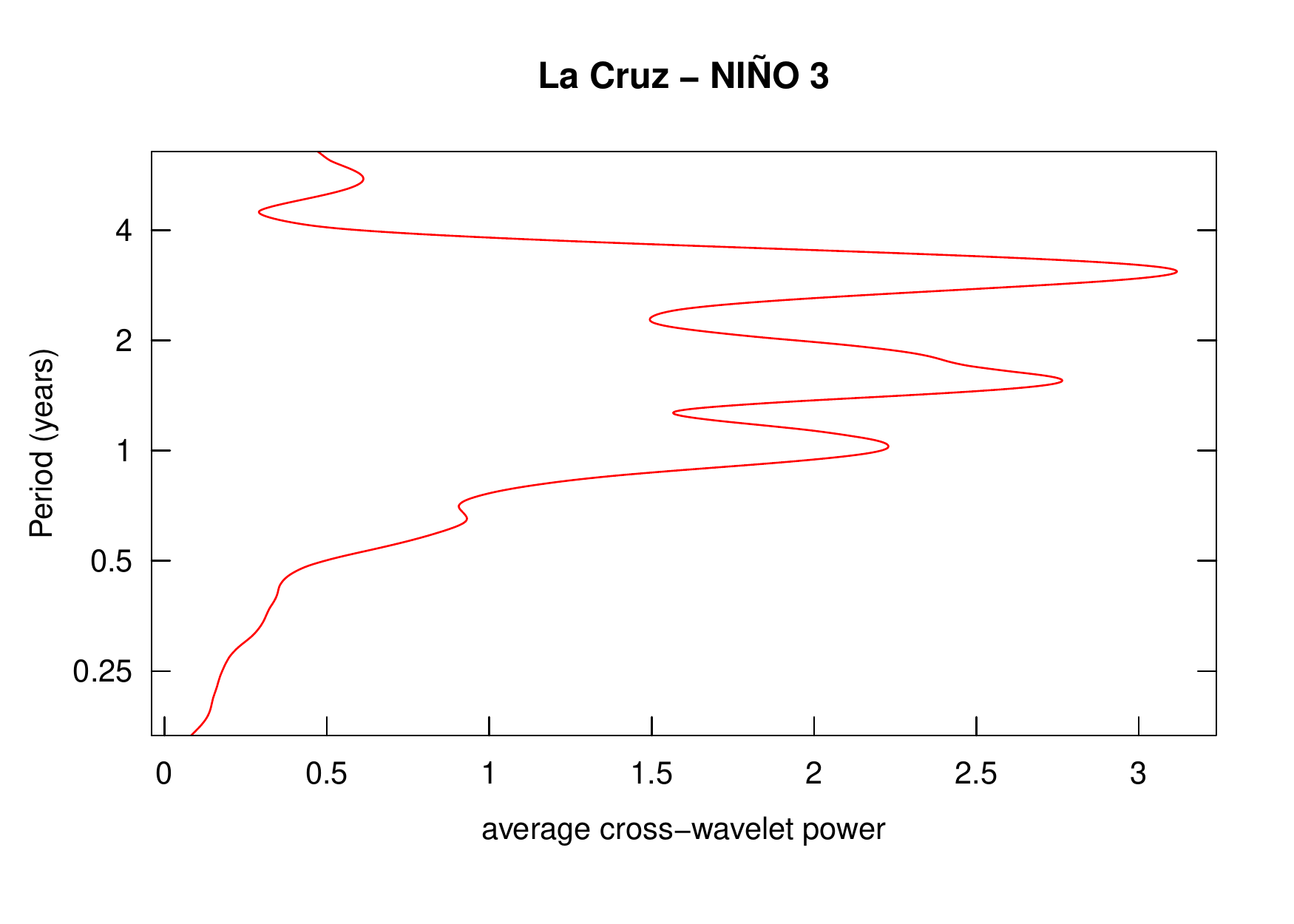}}\vspace{-0.15cm}%
\subfloat[]{\includegraphics[scale=0.23]{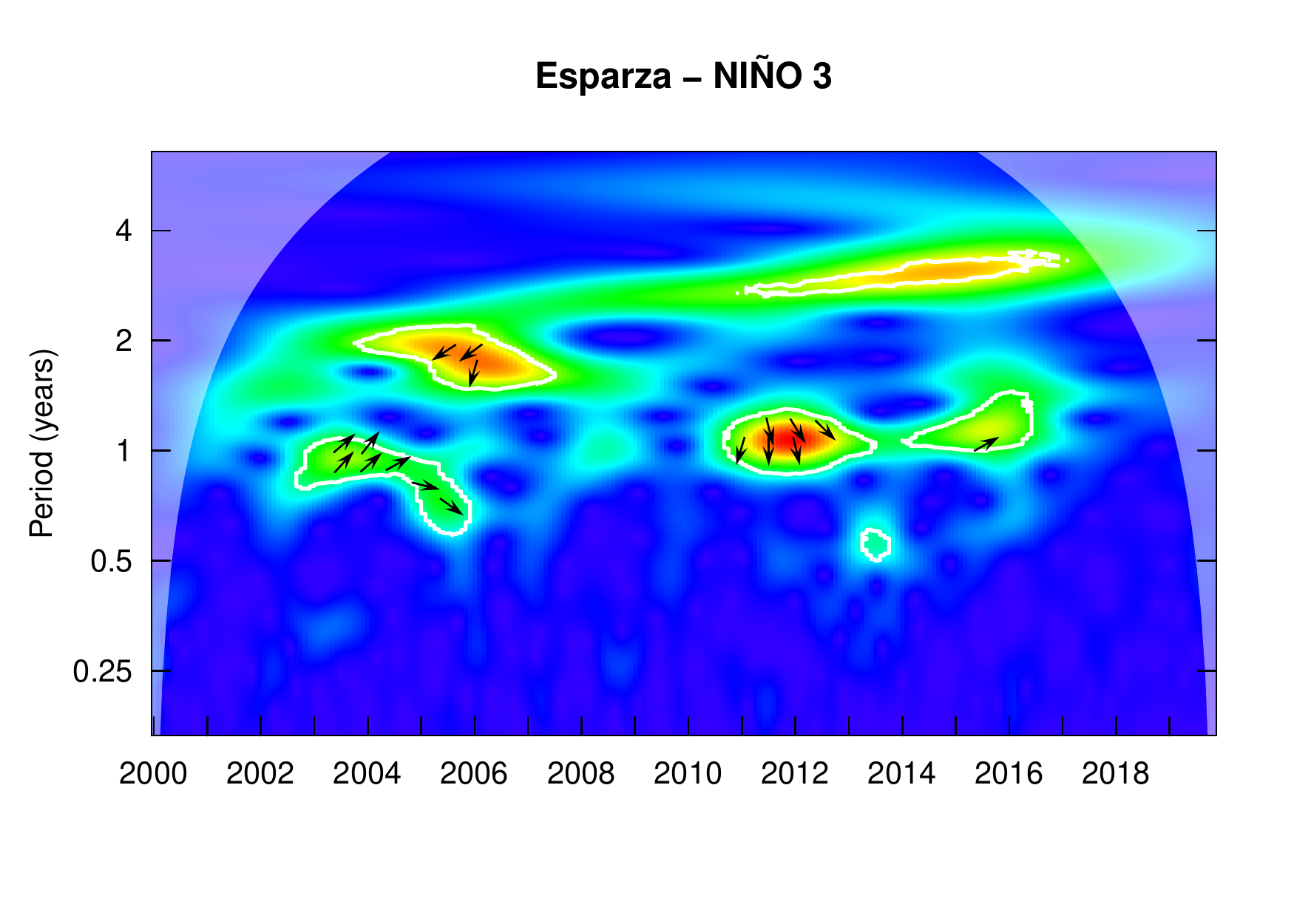}}\vspace{-0.15cm}%
\subfloat[]{\includegraphics[scale=0.23]{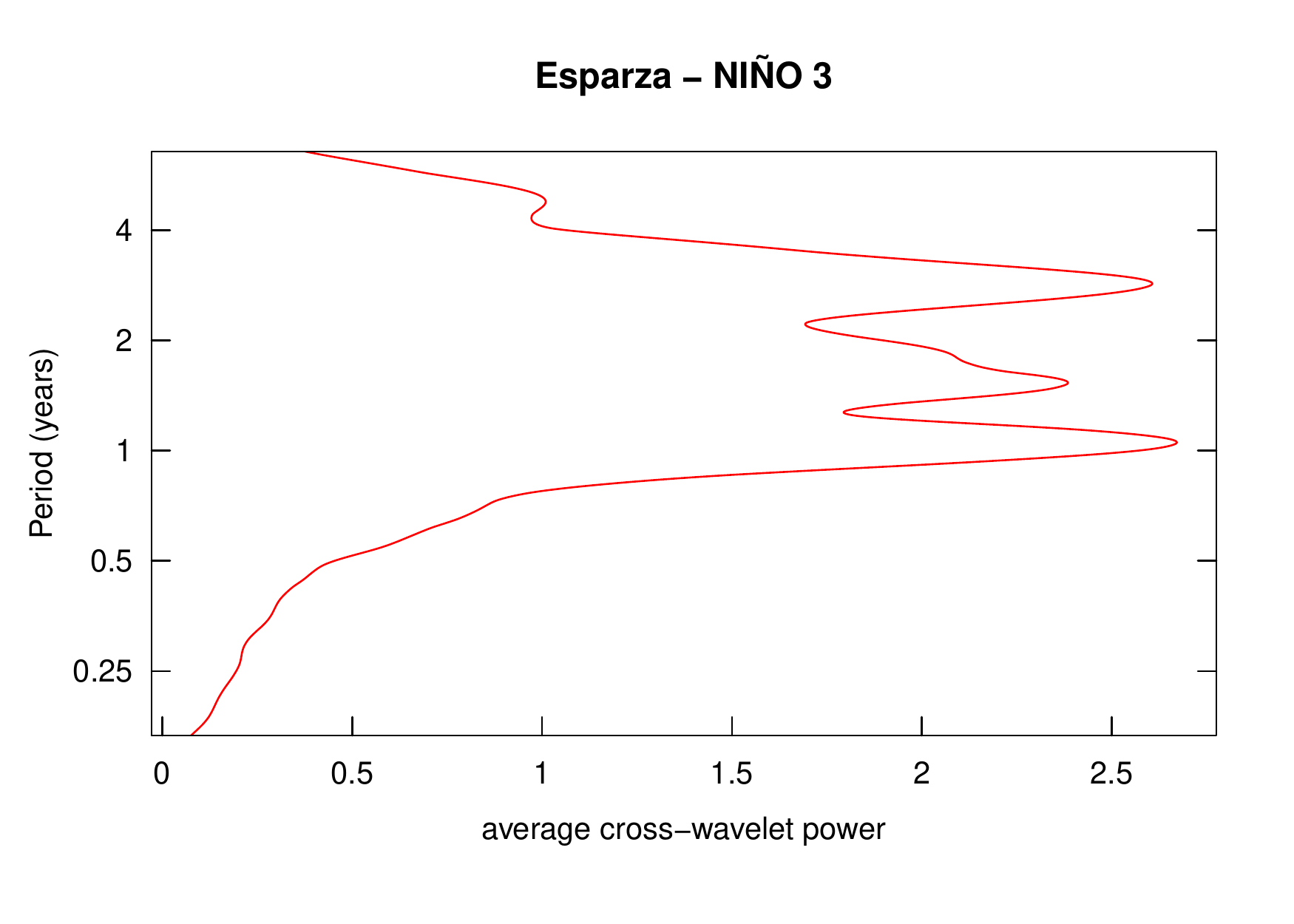}}\vspace{-0.15cm}\\
\subfloat[]{\includegraphics[scale=0.23]{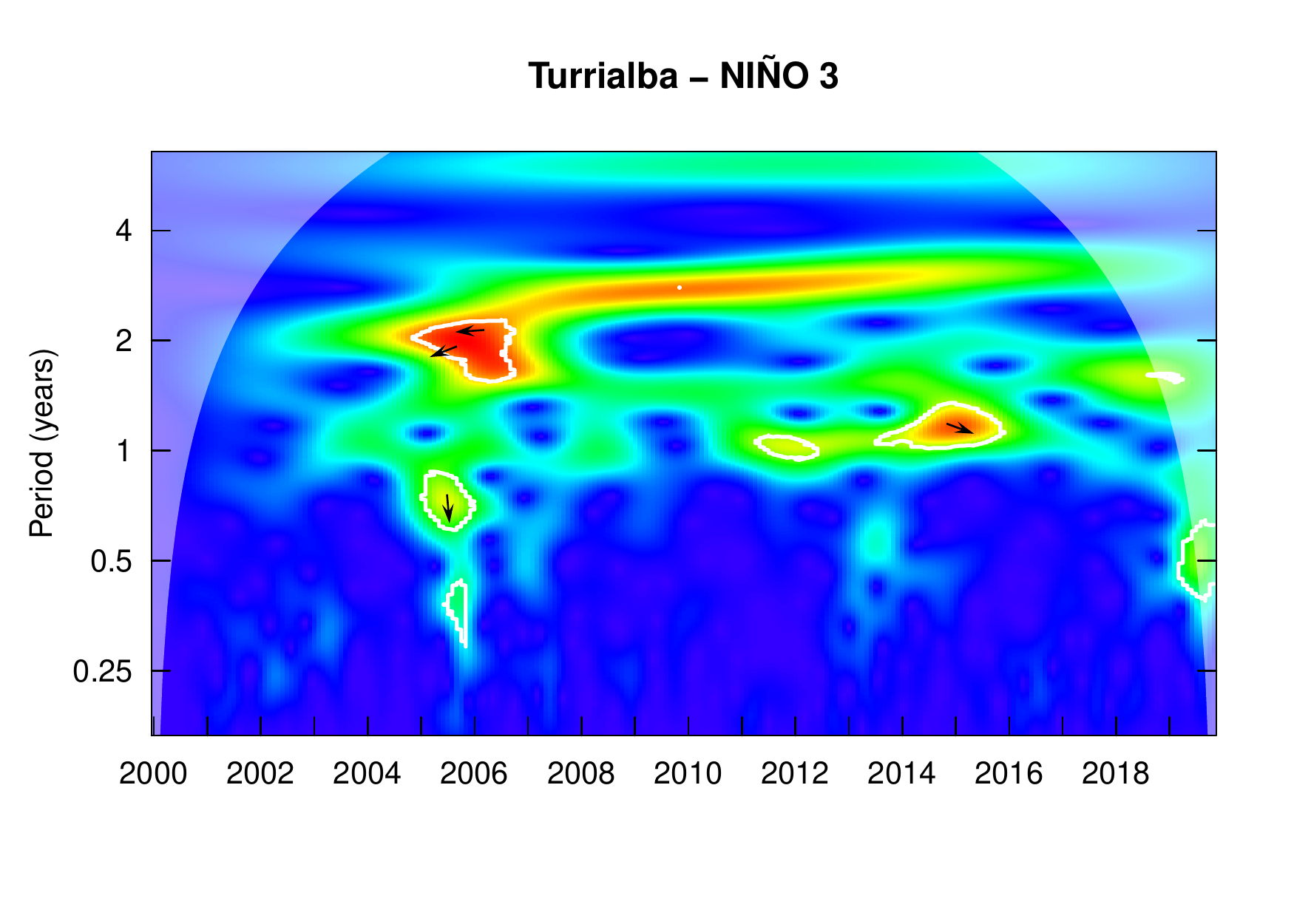}}\vspace{-0.15cm}%
\subfloat[]{\includegraphics[scale=0.23]{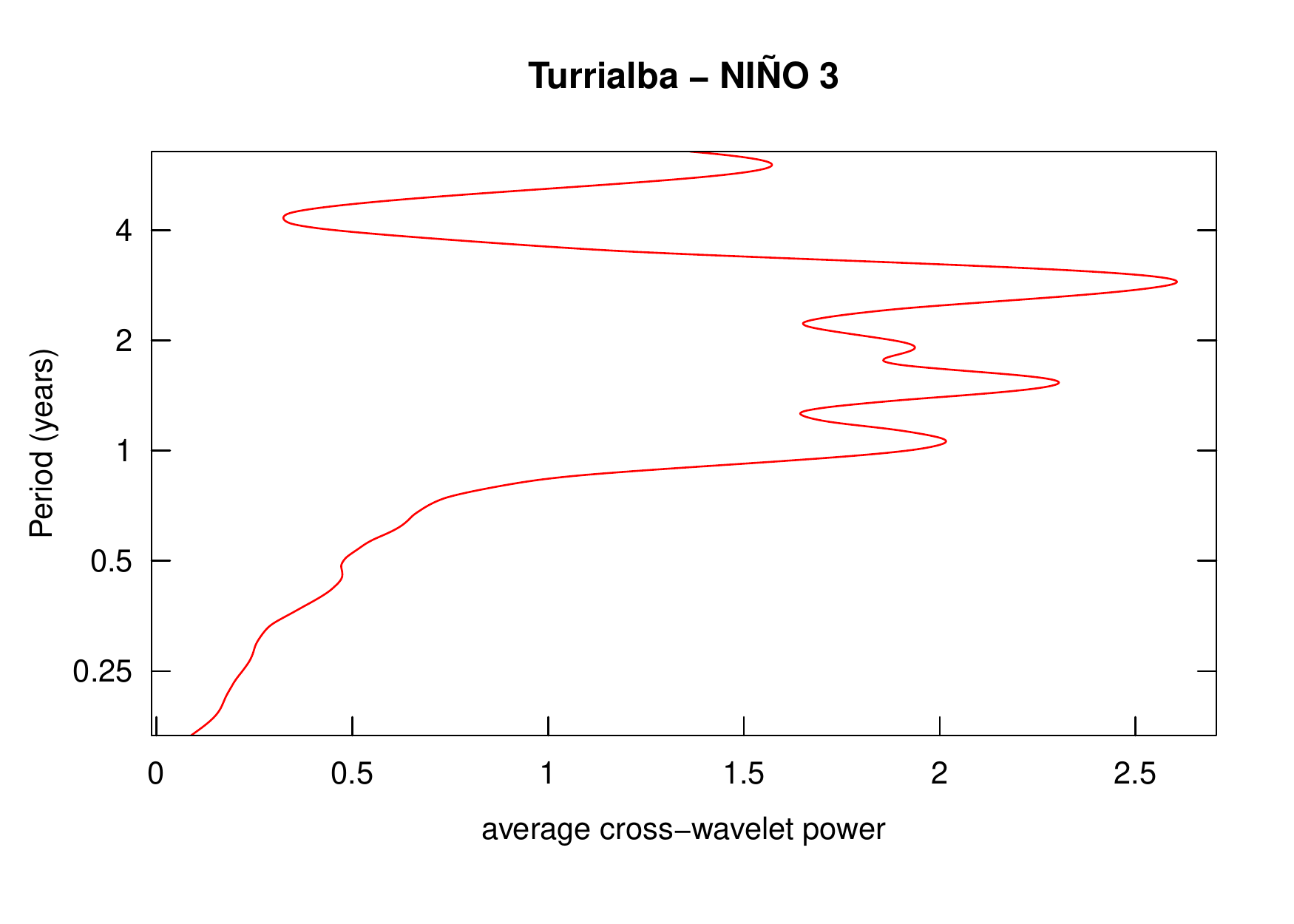}}\vspace{-0.15cm}%
\subfloat[]{\includegraphics[scale=0.23]{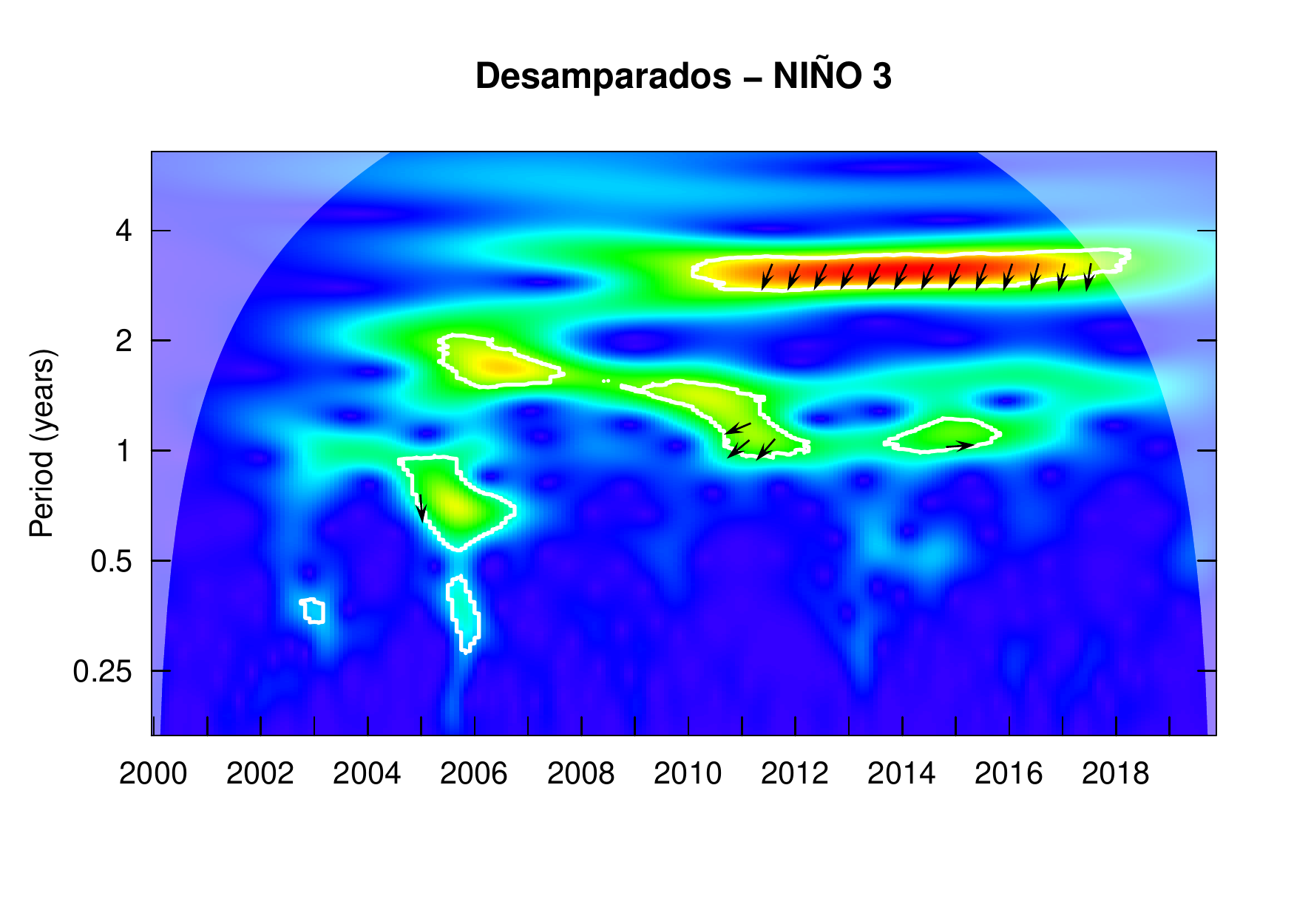}}\vspace{-0.15cm}%
\subfloat[]{\includegraphics[scale=0.23]{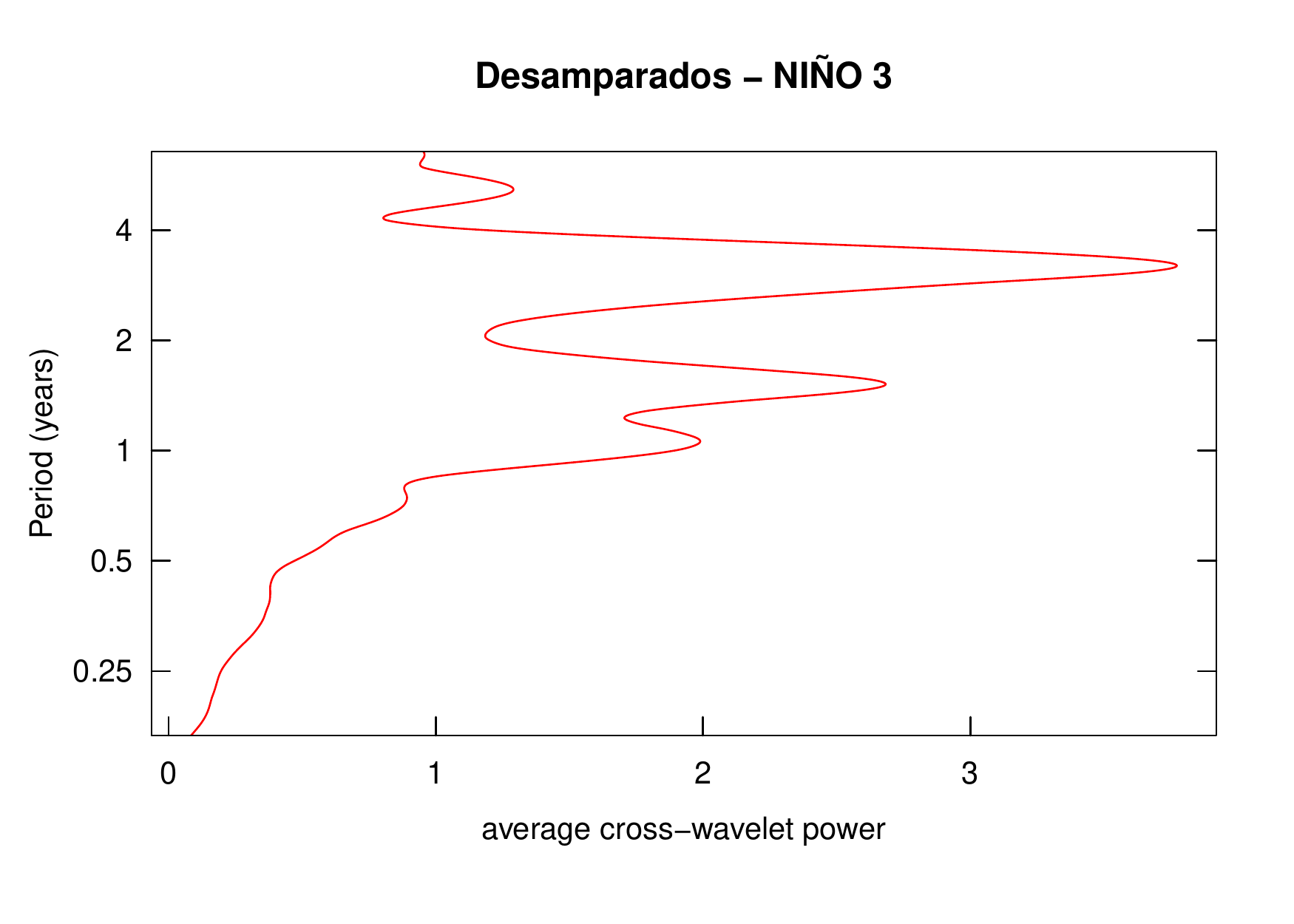}}\vspace{-0.15cm}\\
\subfloat[]{\includegraphics[scale=0.23]{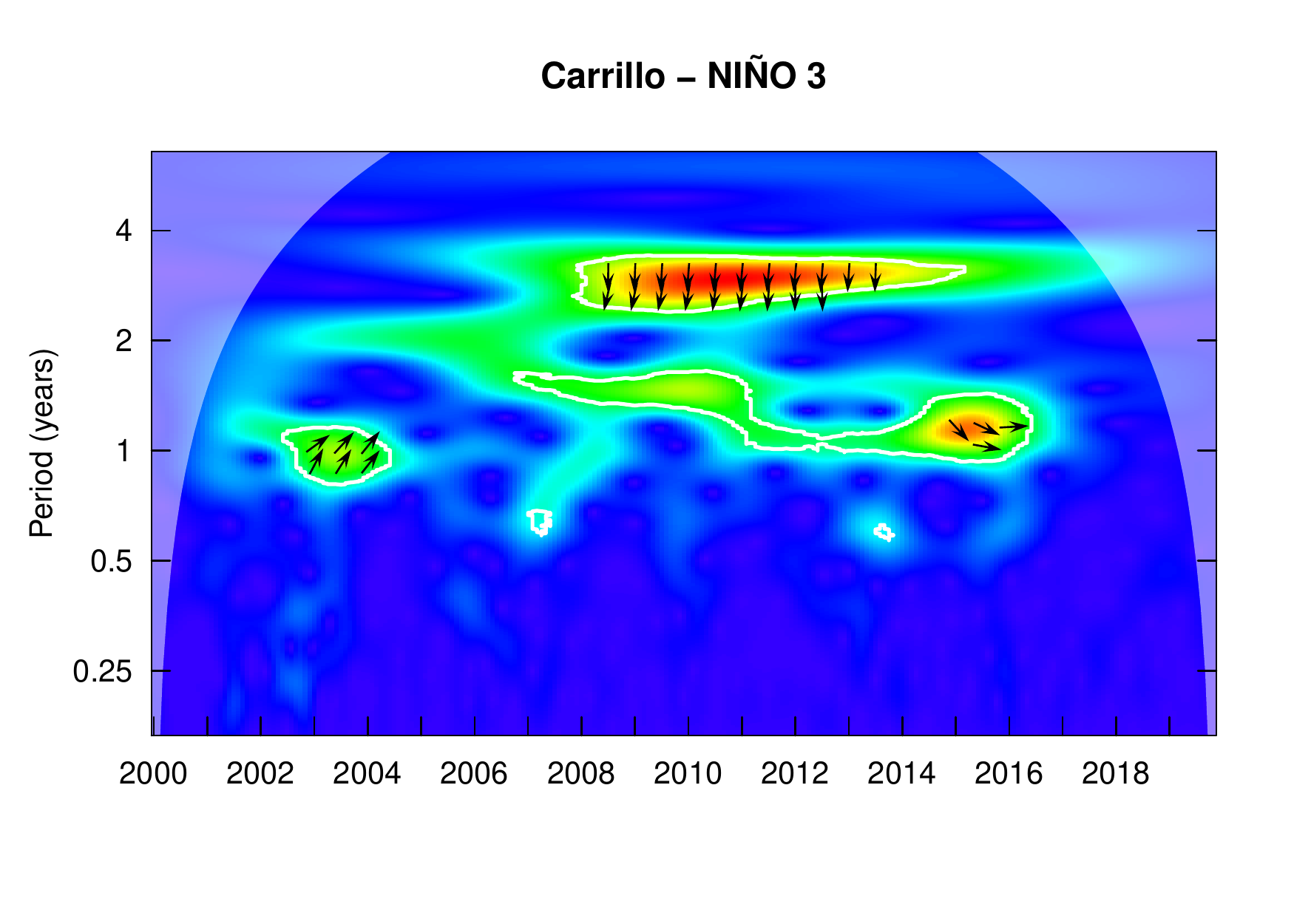}}\vspace{-0.15cm}%
\subfloat[]{\includegraphics[scale=0.23]{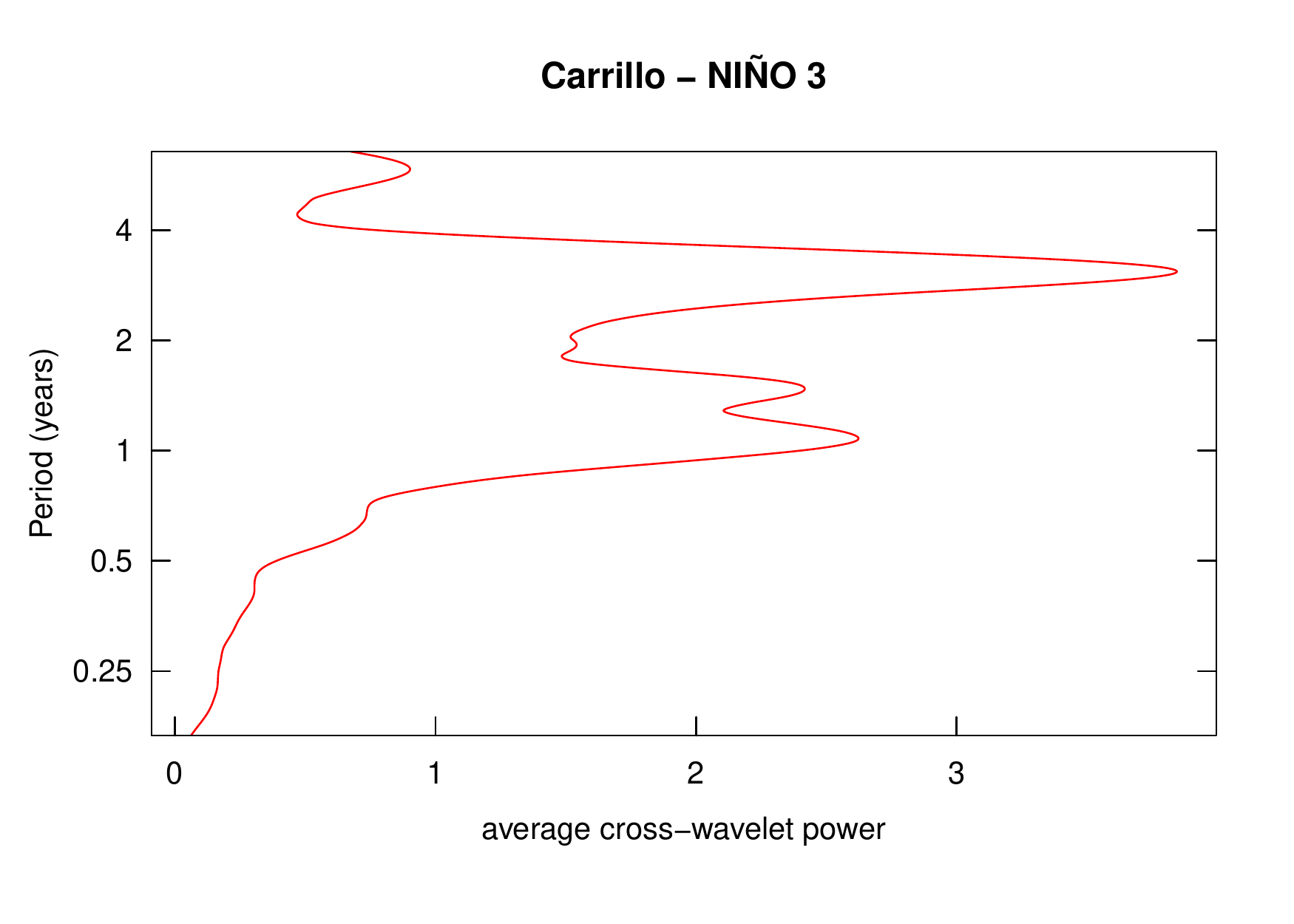}}\vspace{-0.15cm}%
\subfloat[]{\includegraphics[scale=0.23]{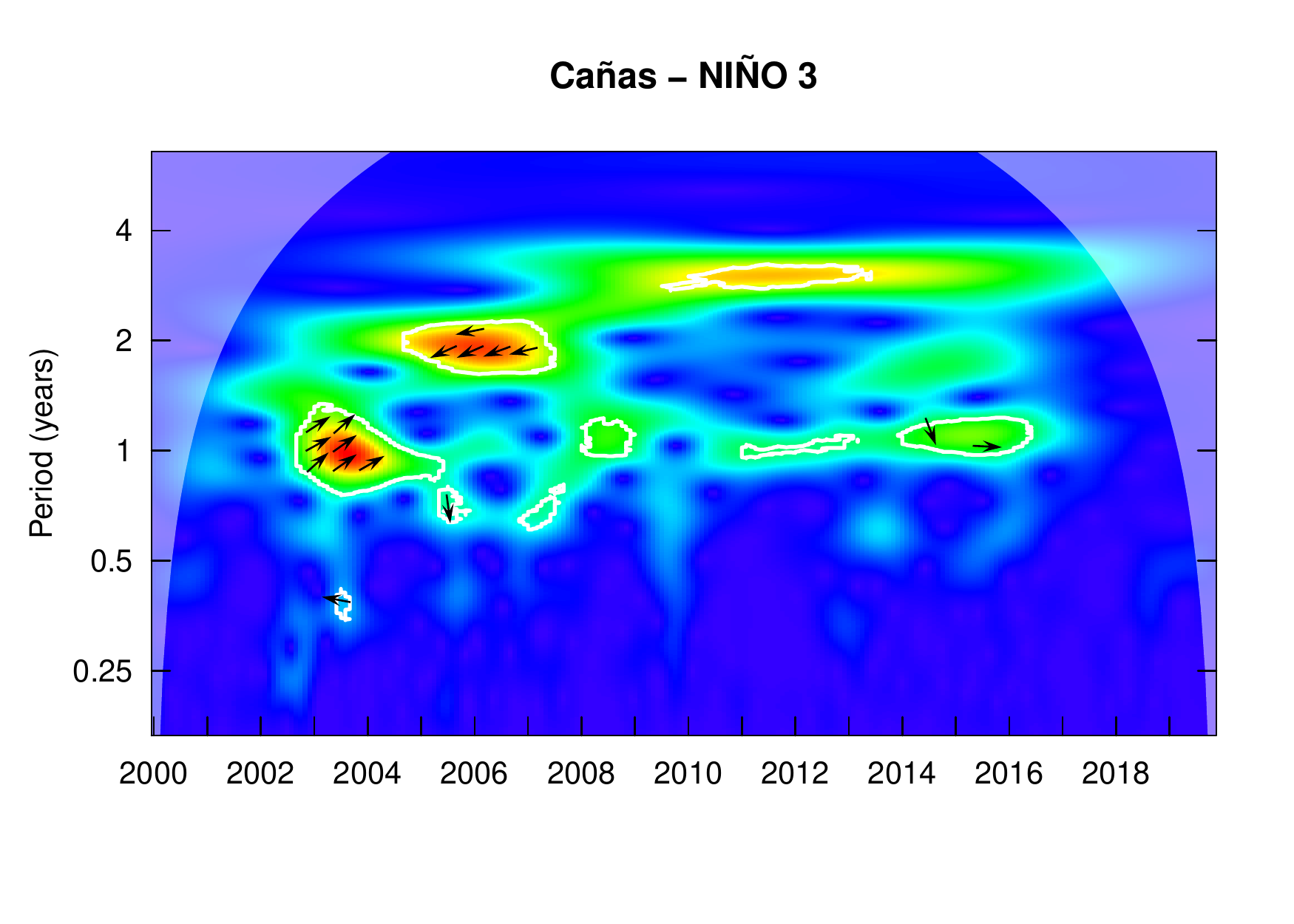}}\vspace{-0.15cm}%
\subfloat[]{\includegraphics[scale=0.23]{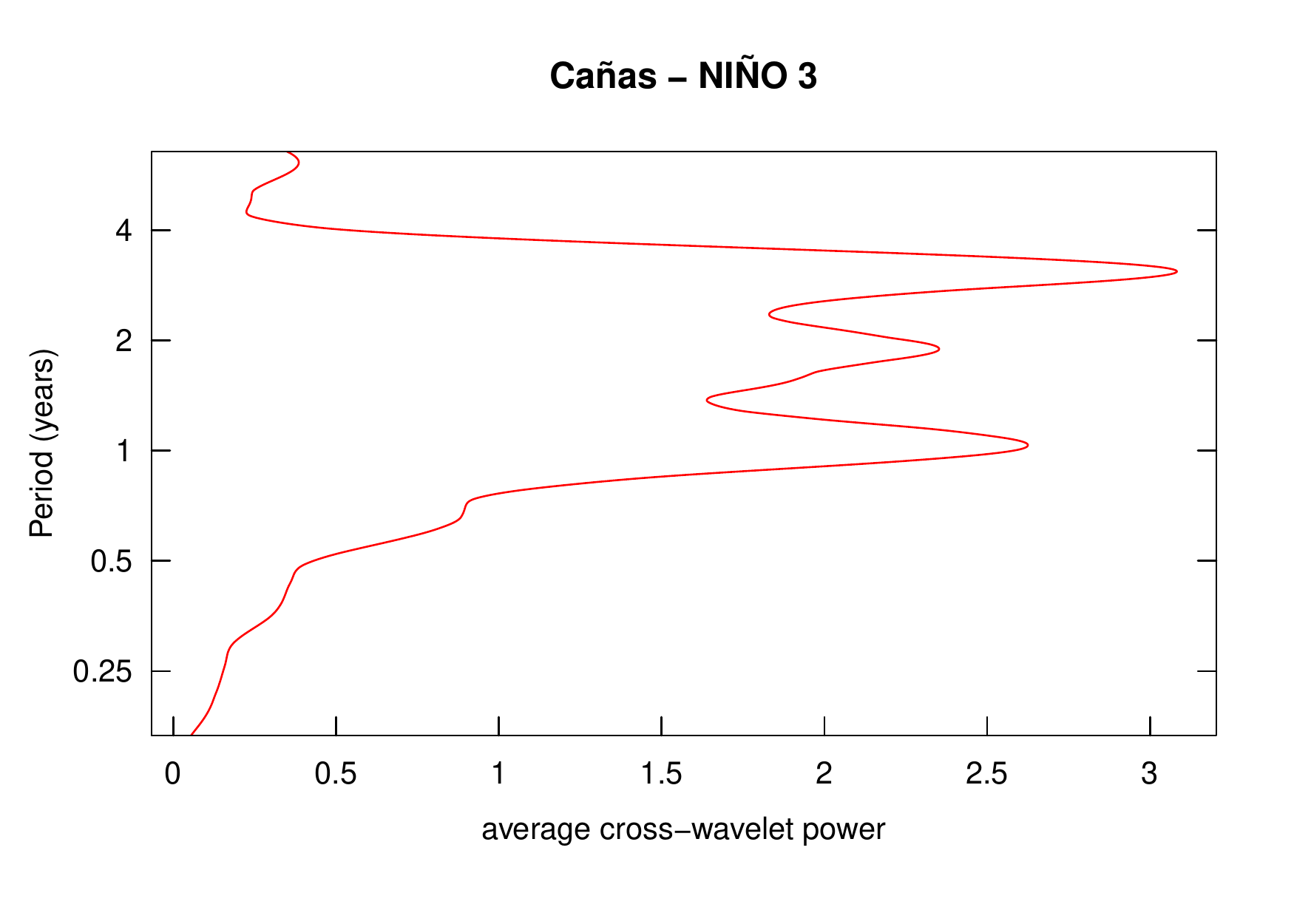}}\vspace{-0.15cm}\\
\caption*{}
\end{figure}

\begin{figure}[H]
\captionsetup[subfigure]{labelformat=empty}
\subfloat[]{\includegraphics[scale=0.23]{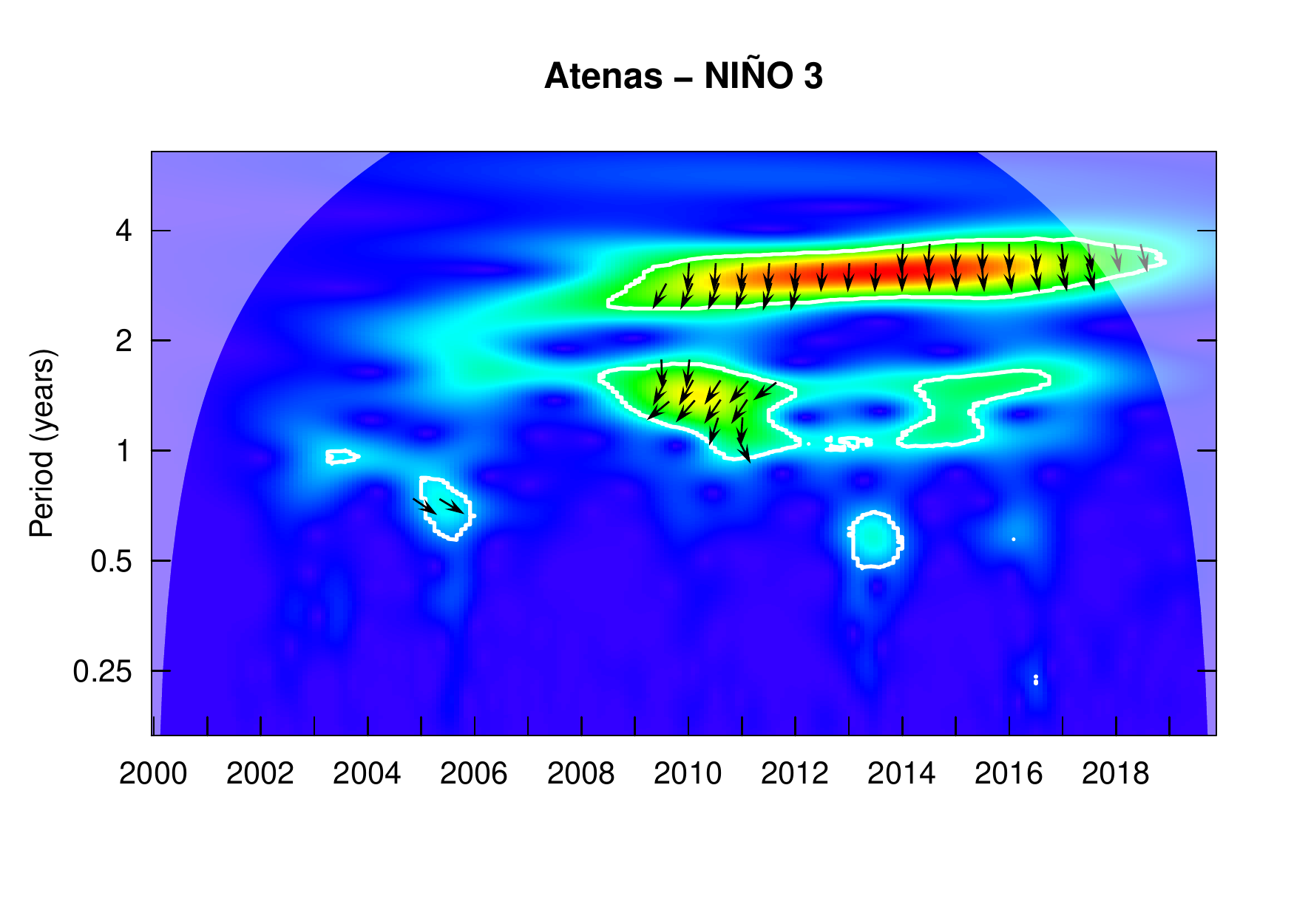}}\vspace{-0.15cm}%
\subfloat[]{\includegraphics[scale=0.23]{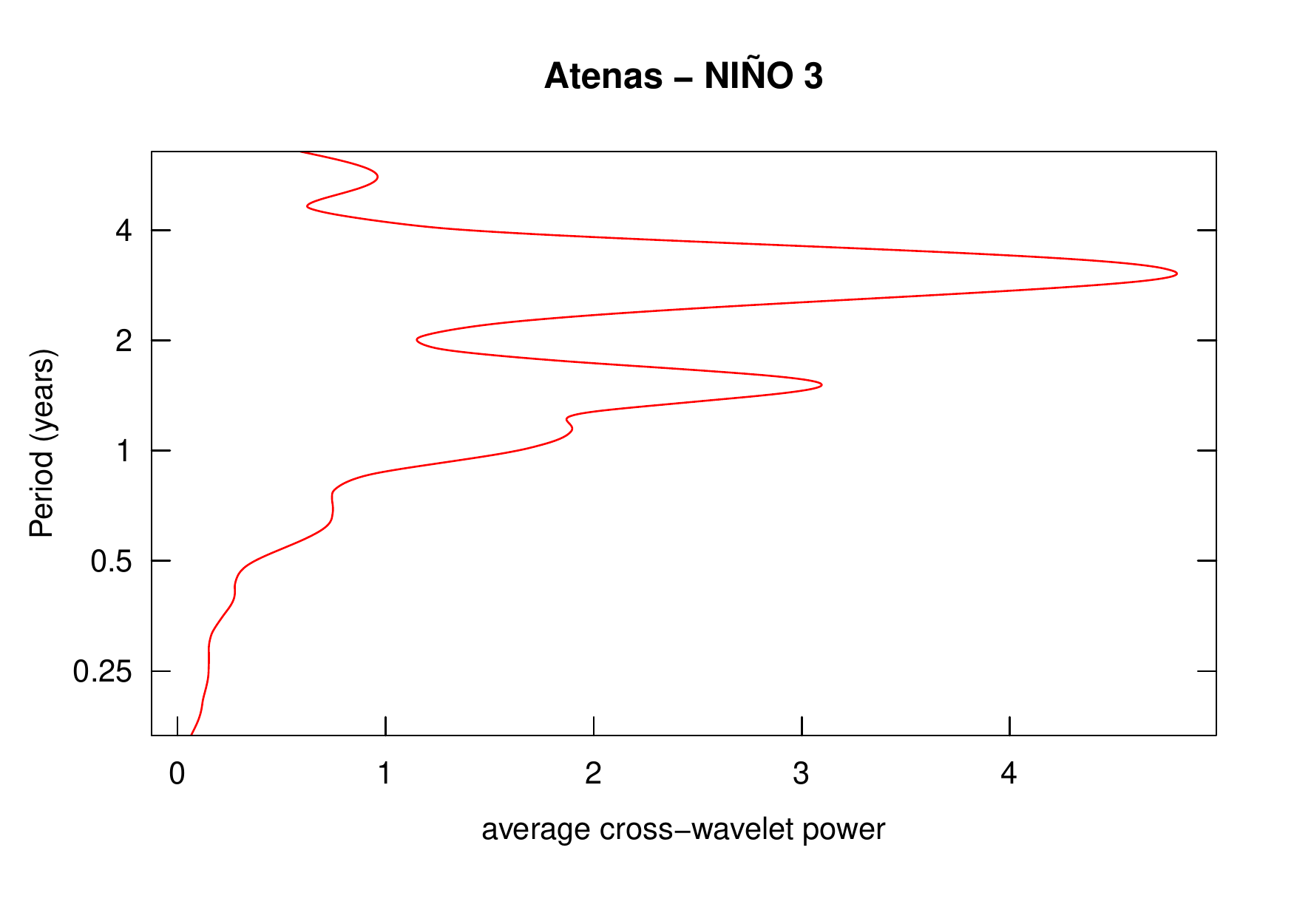}}\vspace{-0.15cm}%
\subfloat[]{\includegraphics[scale=0.23]{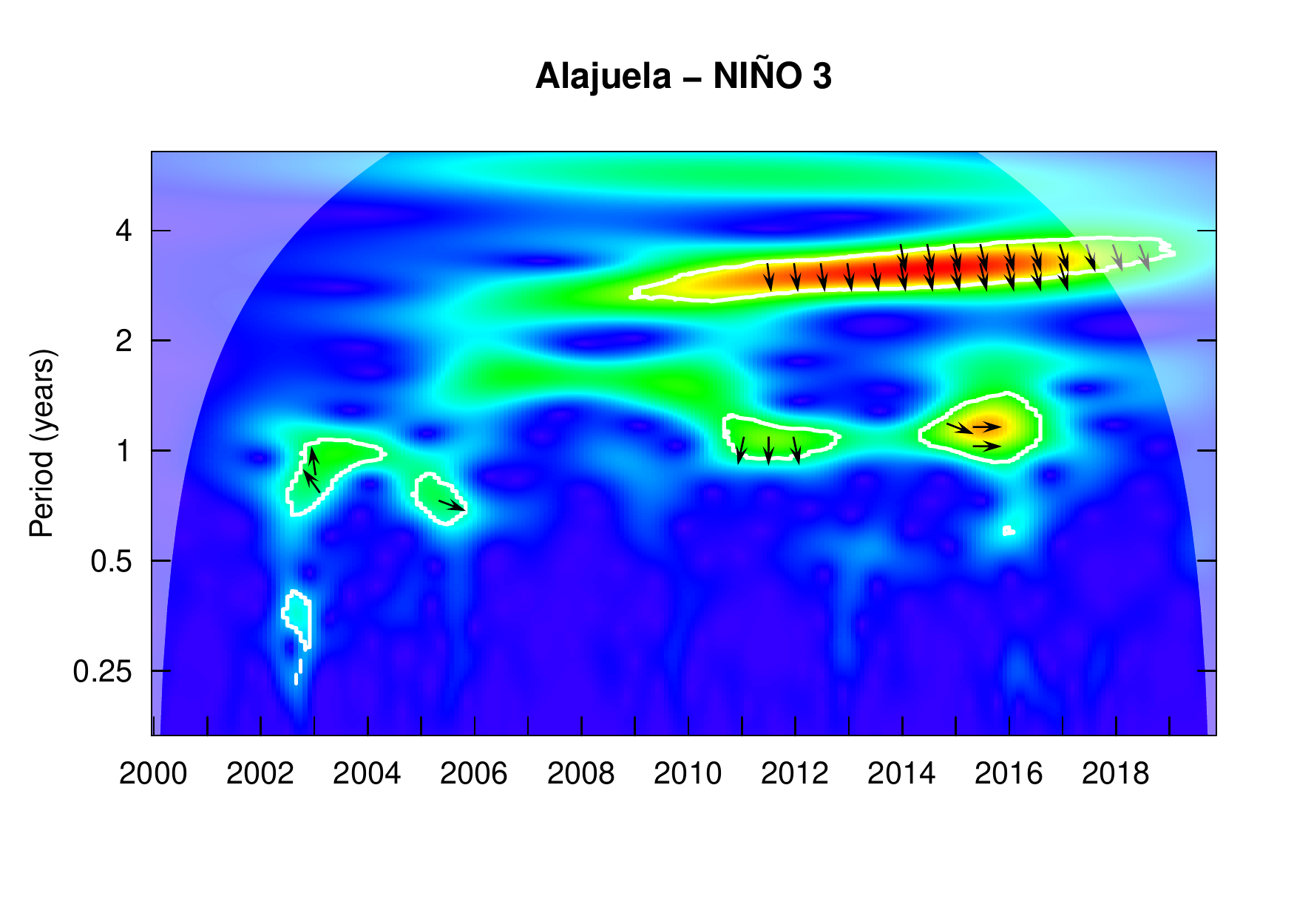}}\vspace{-0.15cm}%
\subfloat[]{\includegraphics[scale=0.23]{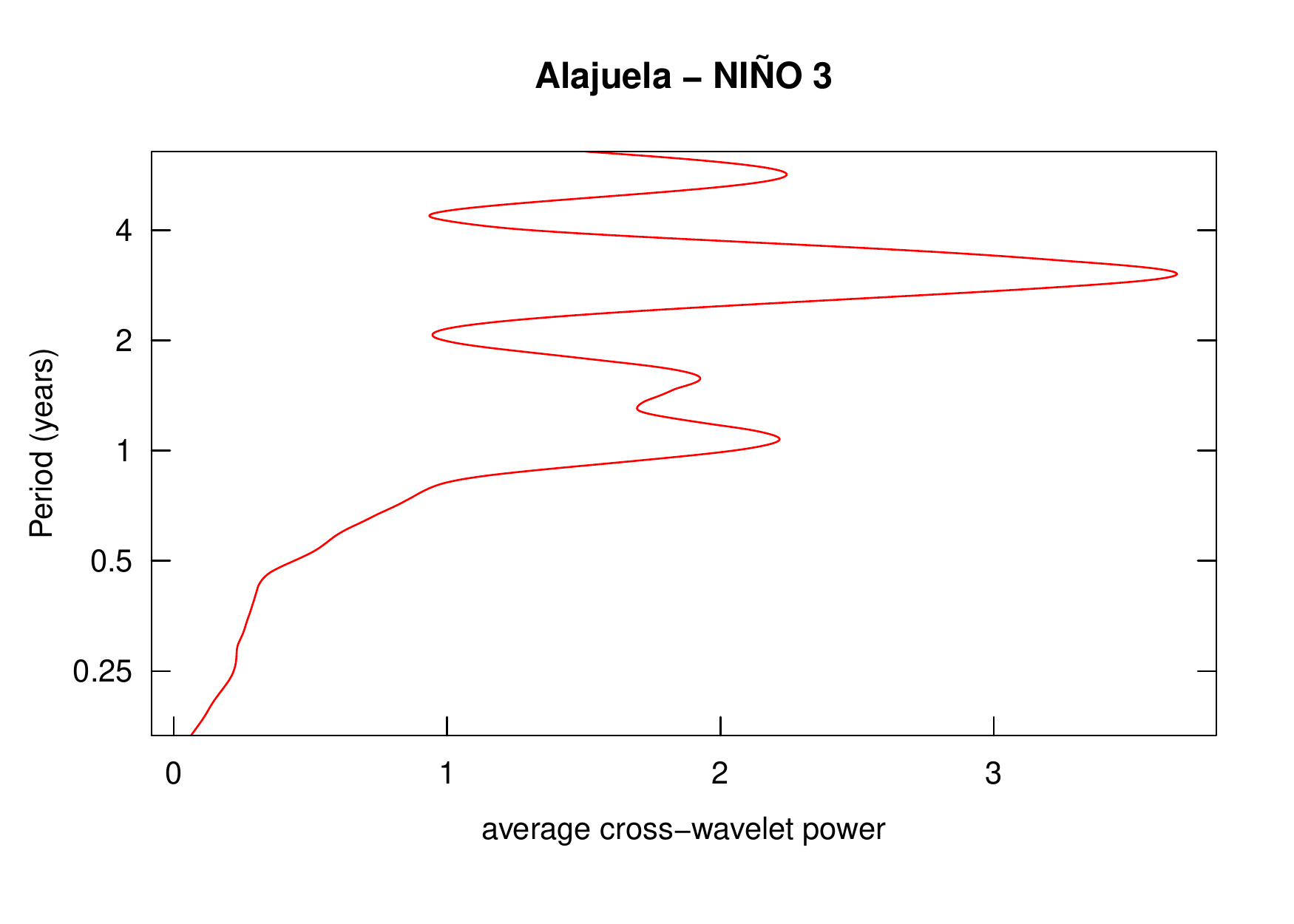}}\vspace{-0.15cm}\\
\subfloat[]{\includegraphics[scale=0.23]{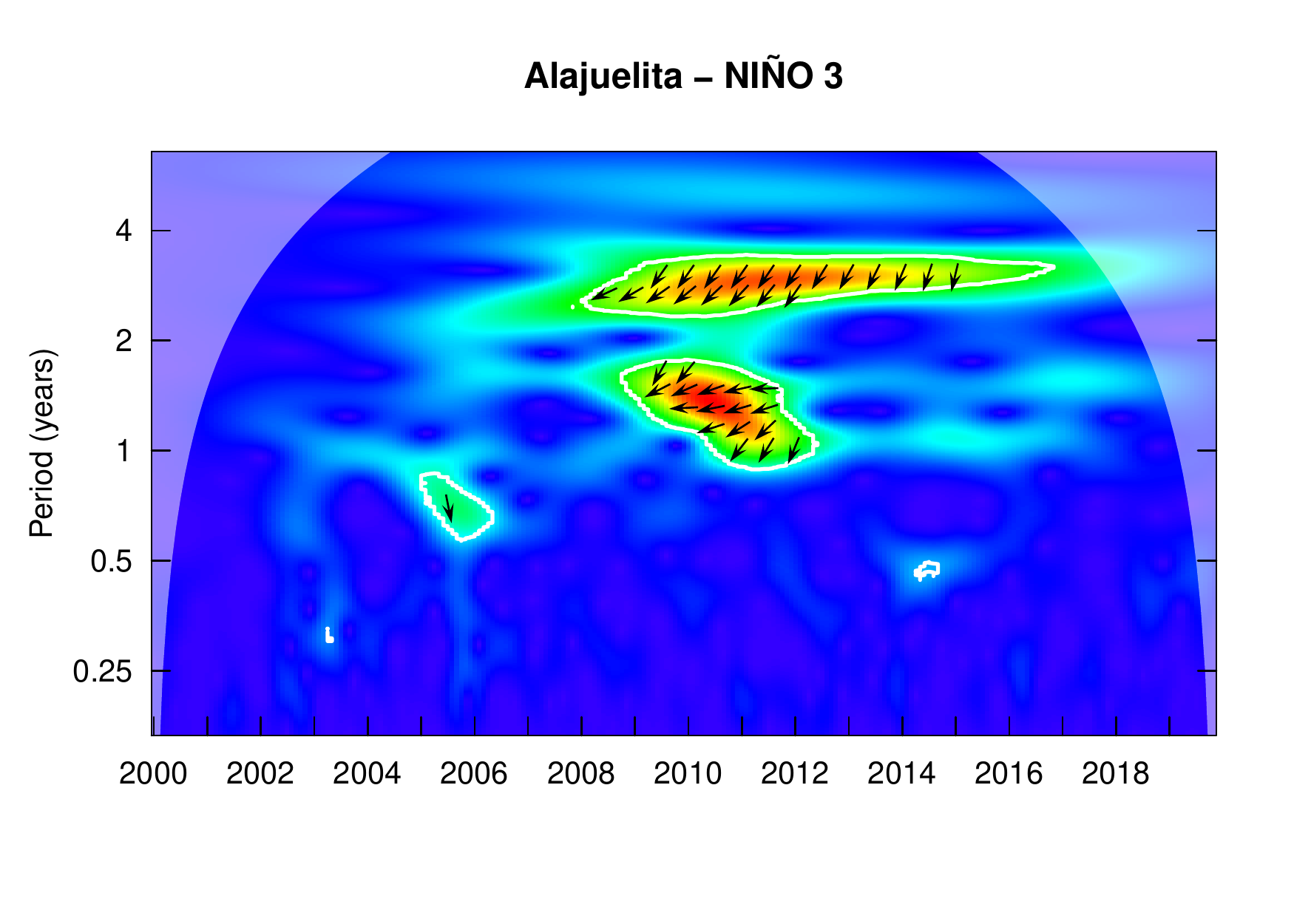}}\vspace{-0.15cm}%
\subfloat[]{\includegraphics[scale=0.23]{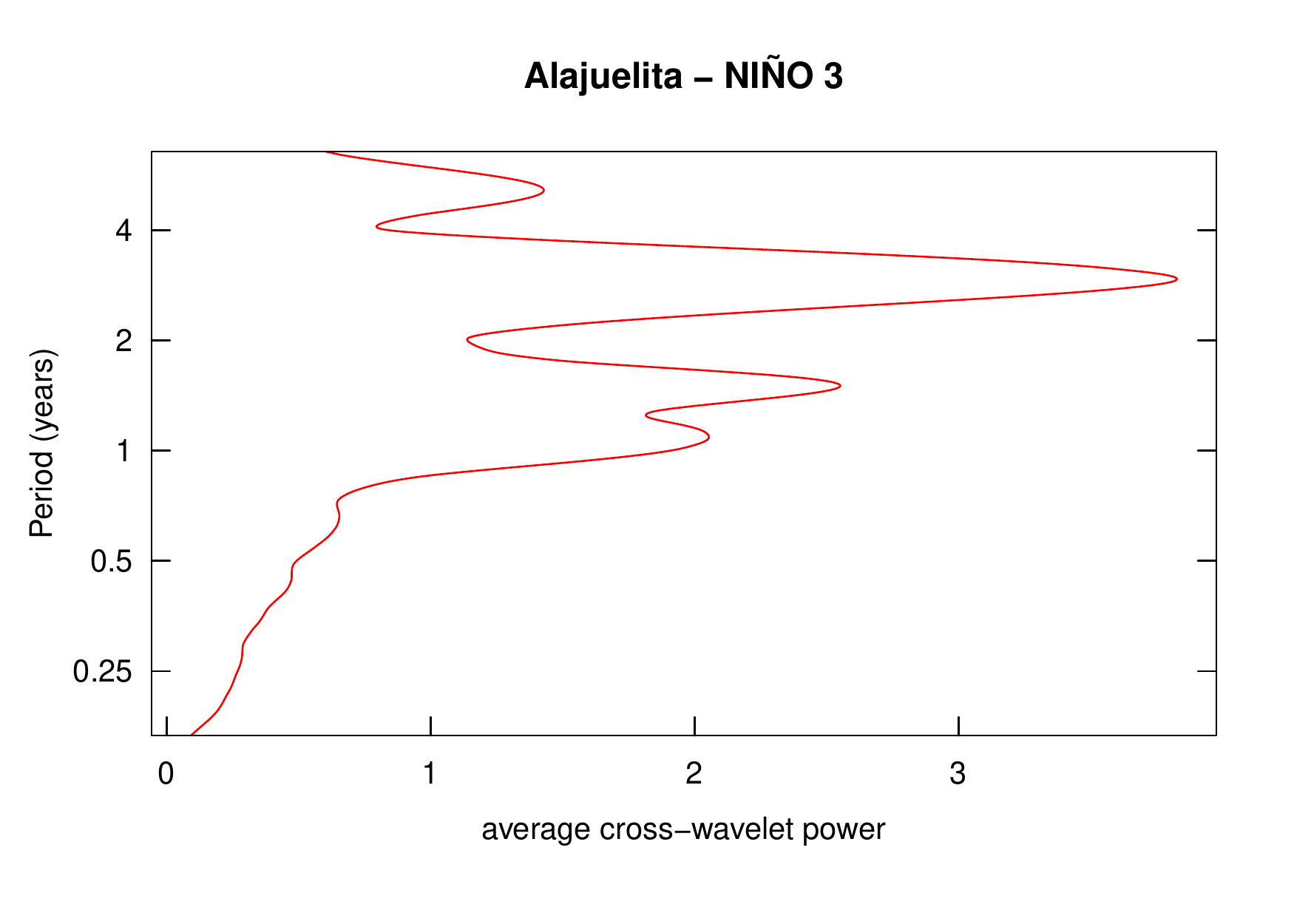}}\vspace{-0.15cm}%
\subfloat[]{\includegraphics[scale=0.23]{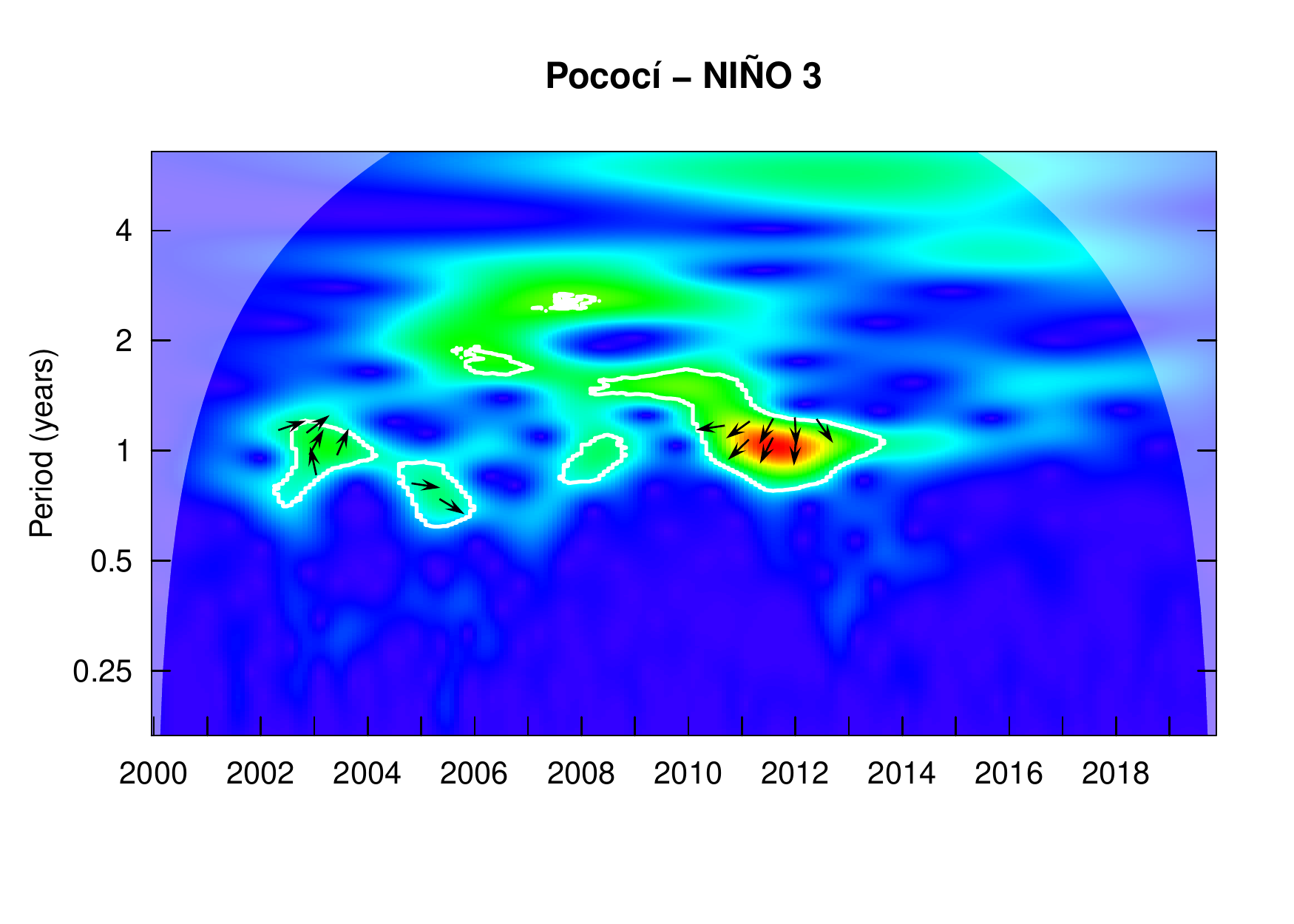}}\vspace{-0.15cm}%
\subfloat[]{\includegraphics[scale=0.23]{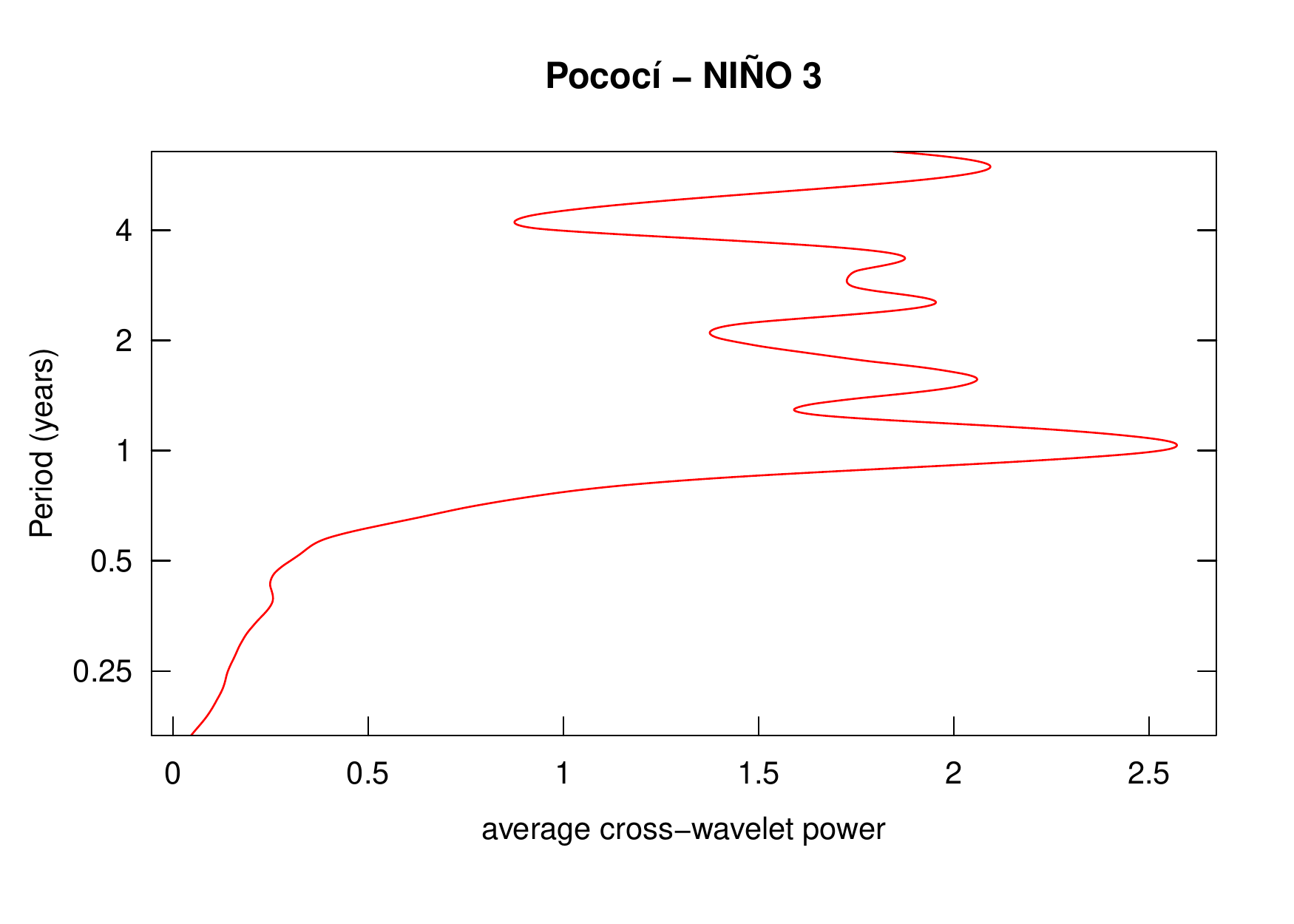}}\vspace{-0.15cm}\\
\subfloat[]{\includegraphics[scale=0.23]{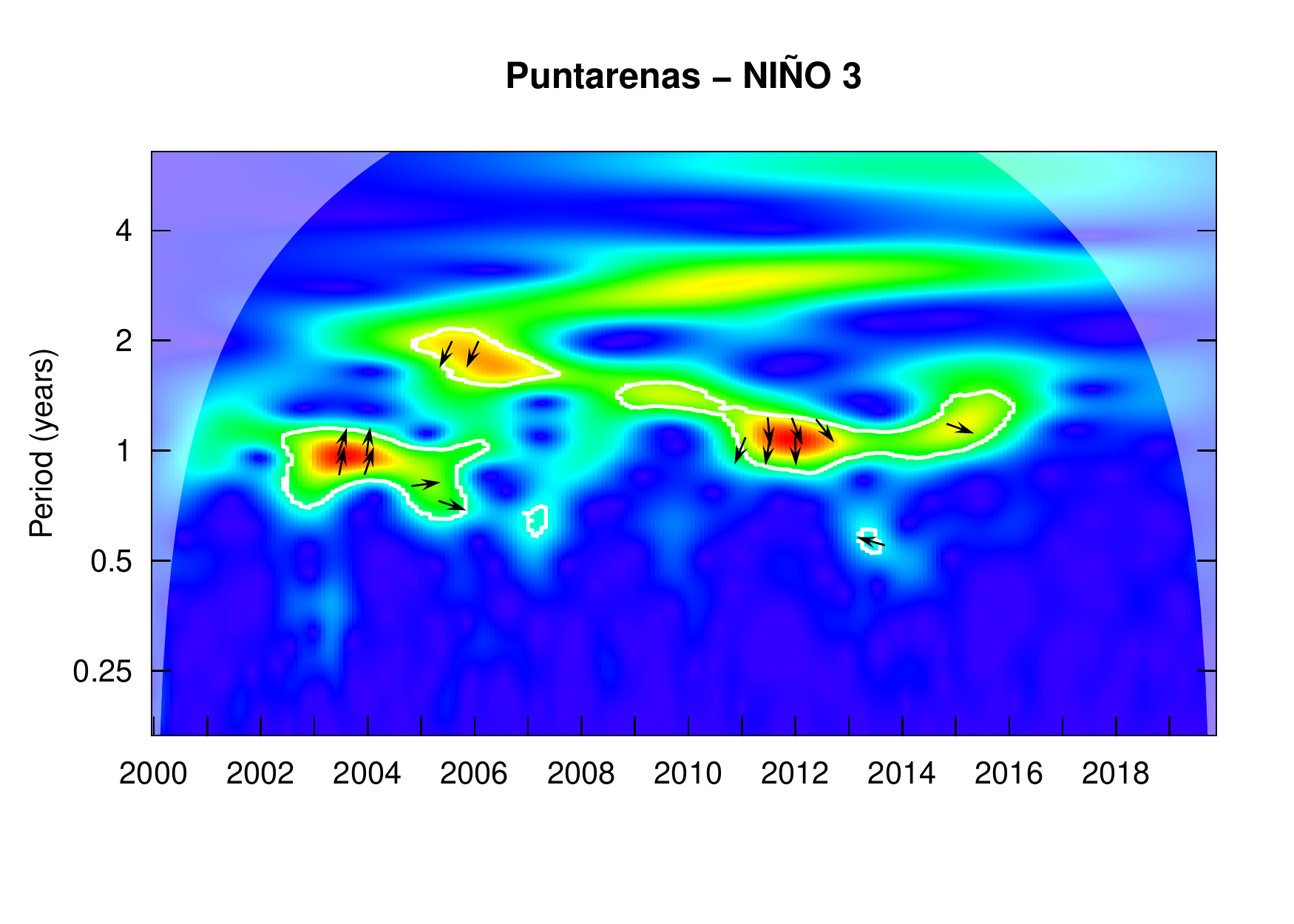}}\vspace{-0.15cm}%
\subfloat[]{\includegraphics[scale=0.23]{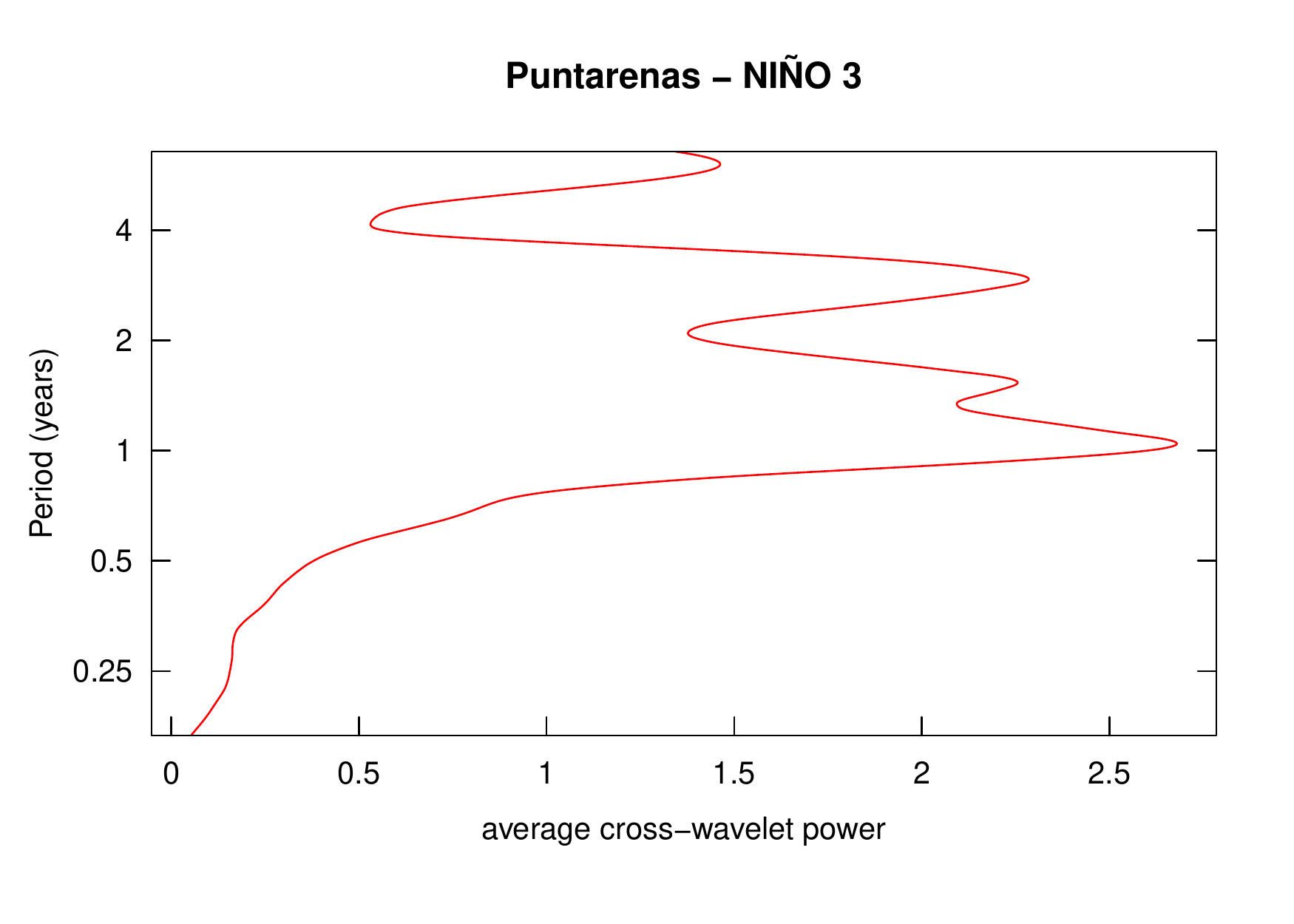}}\vspace{-0.15cm}%
\subfloat[]{\includegraphics[scale=0.23]{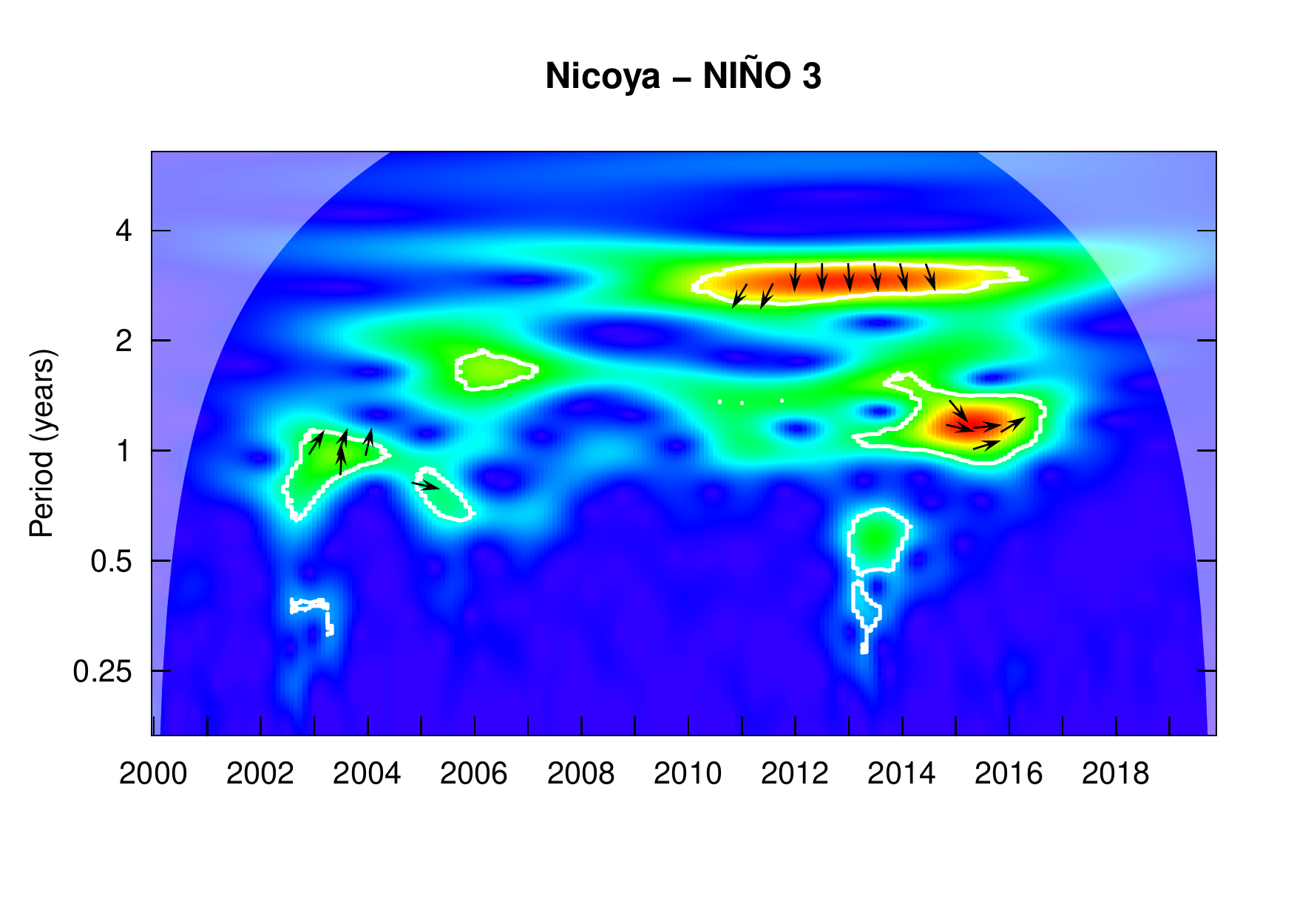}}\vspace{-0.15cm}%
\subfloat[]{\includegraphics[scale=0.23]{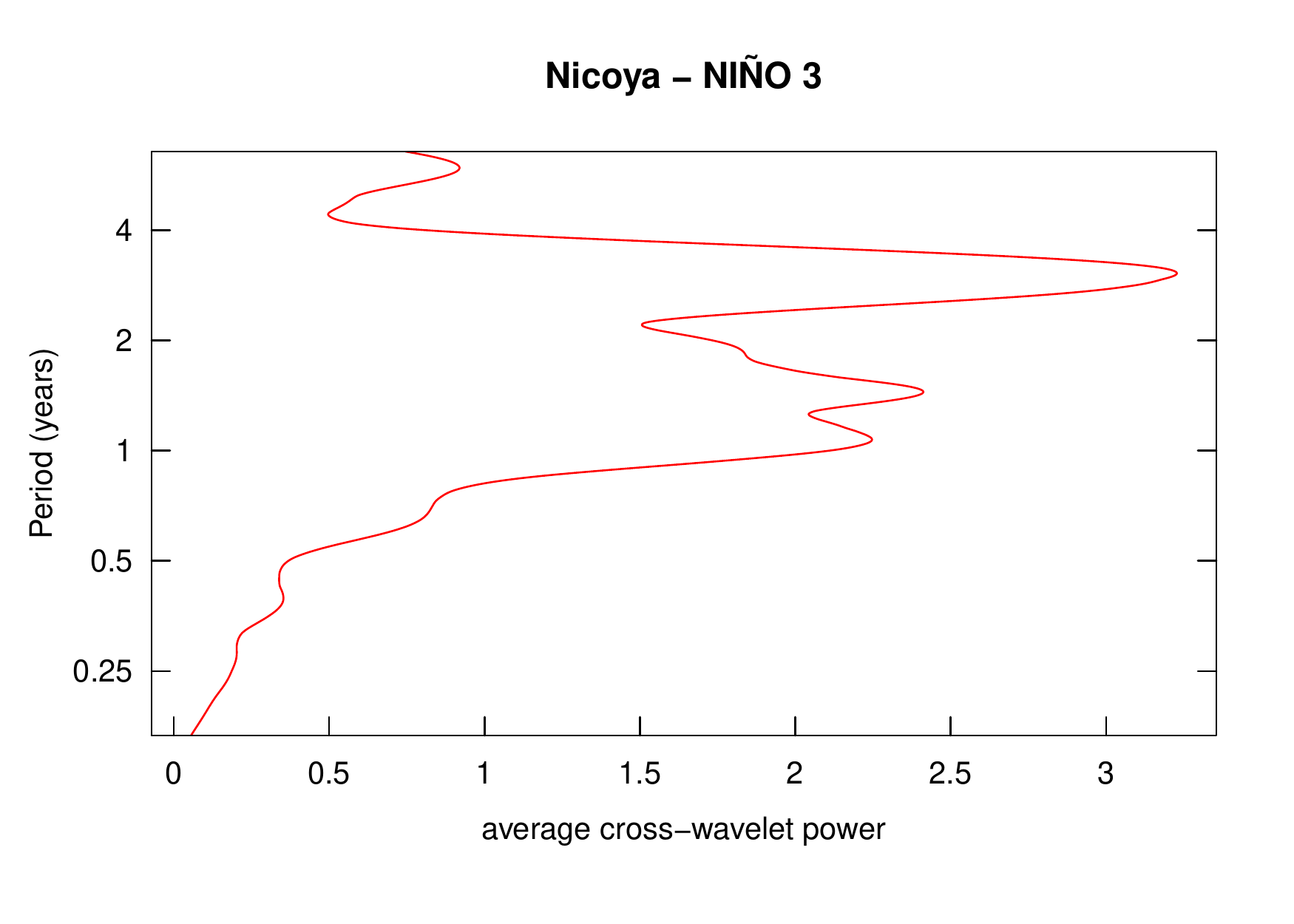}}\vspace{-0.15cm}\\
\subfloat[]{\includegraphics[scale=0.23]{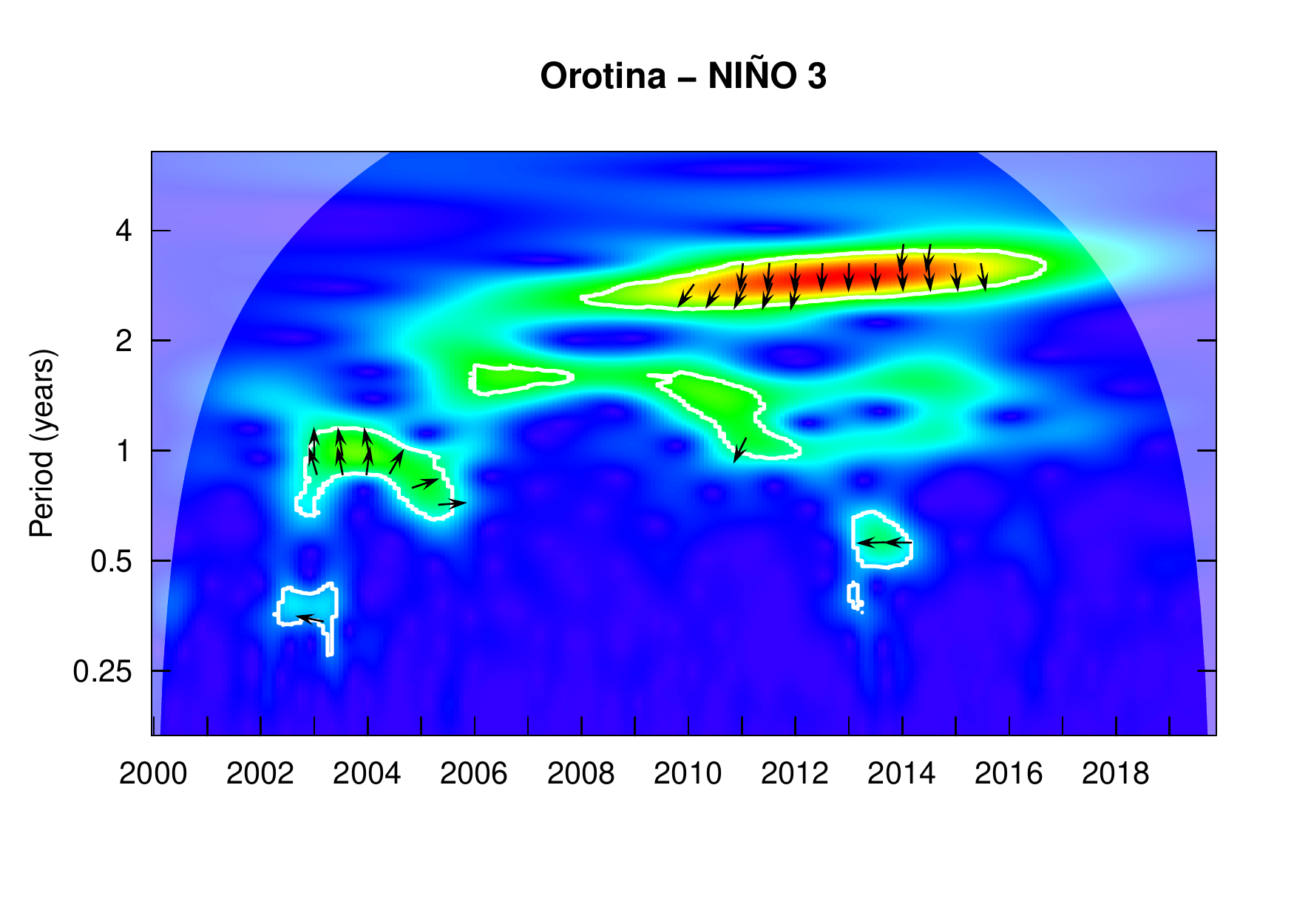}}\vspace{-0.15cm}%
\subfloat[]{\includegraphics[scale=0.23]{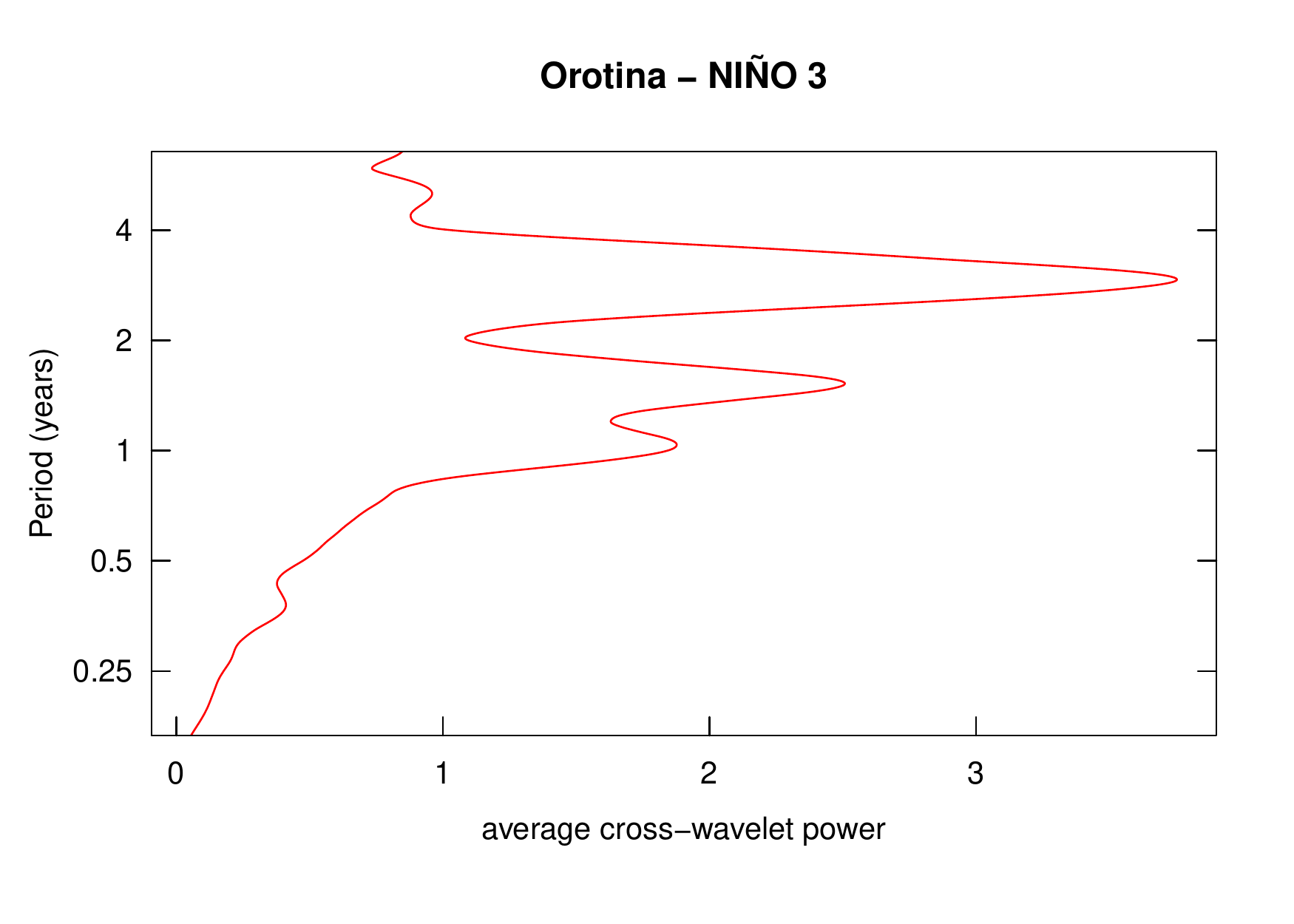}}\vspace{-0.15cm}%
\subfloat[]{\includegraphics[scale=0.23]{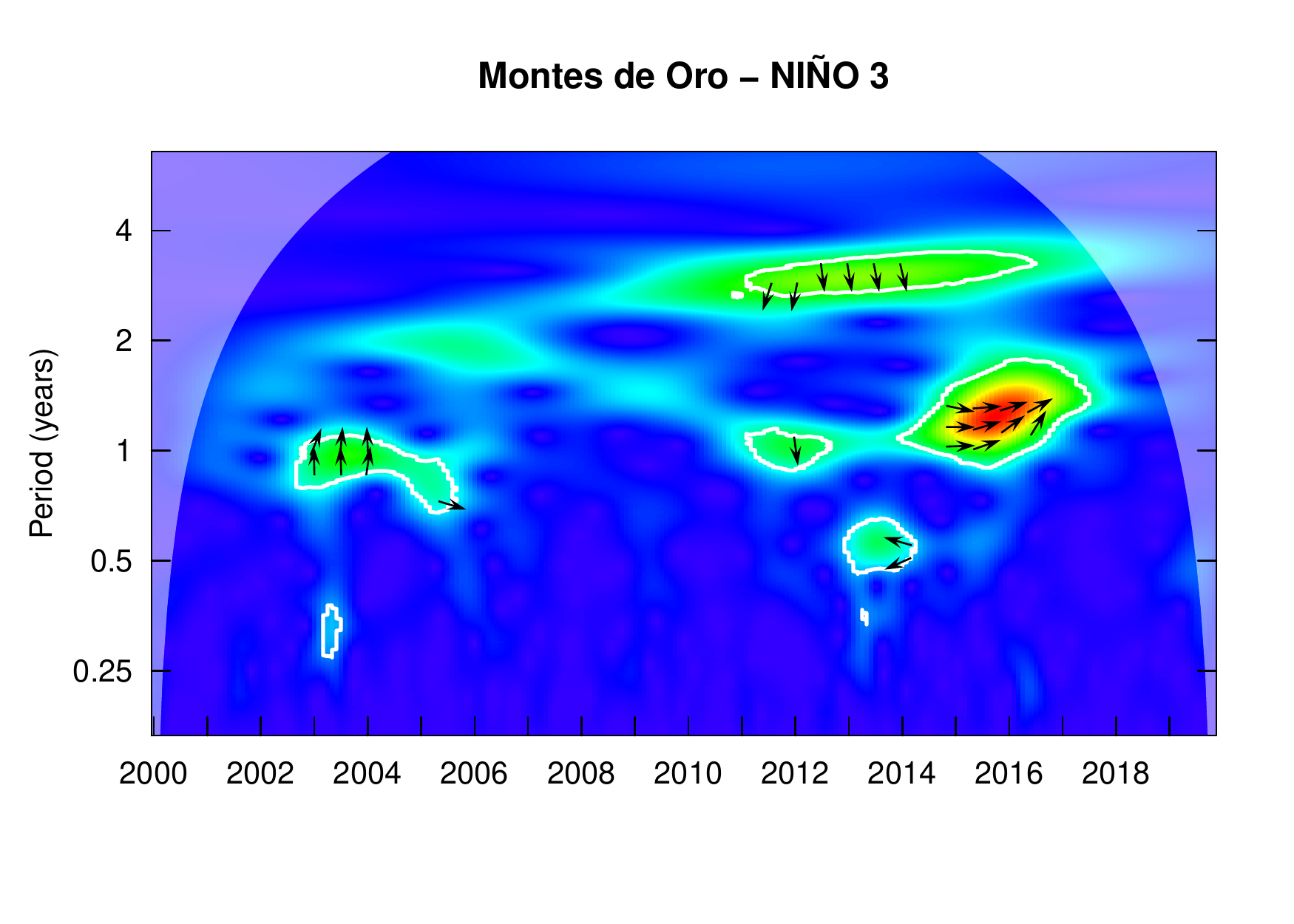}}\vspace{-0.15cm}%
\subfloat[]{\includegraphics[scale=0.23]{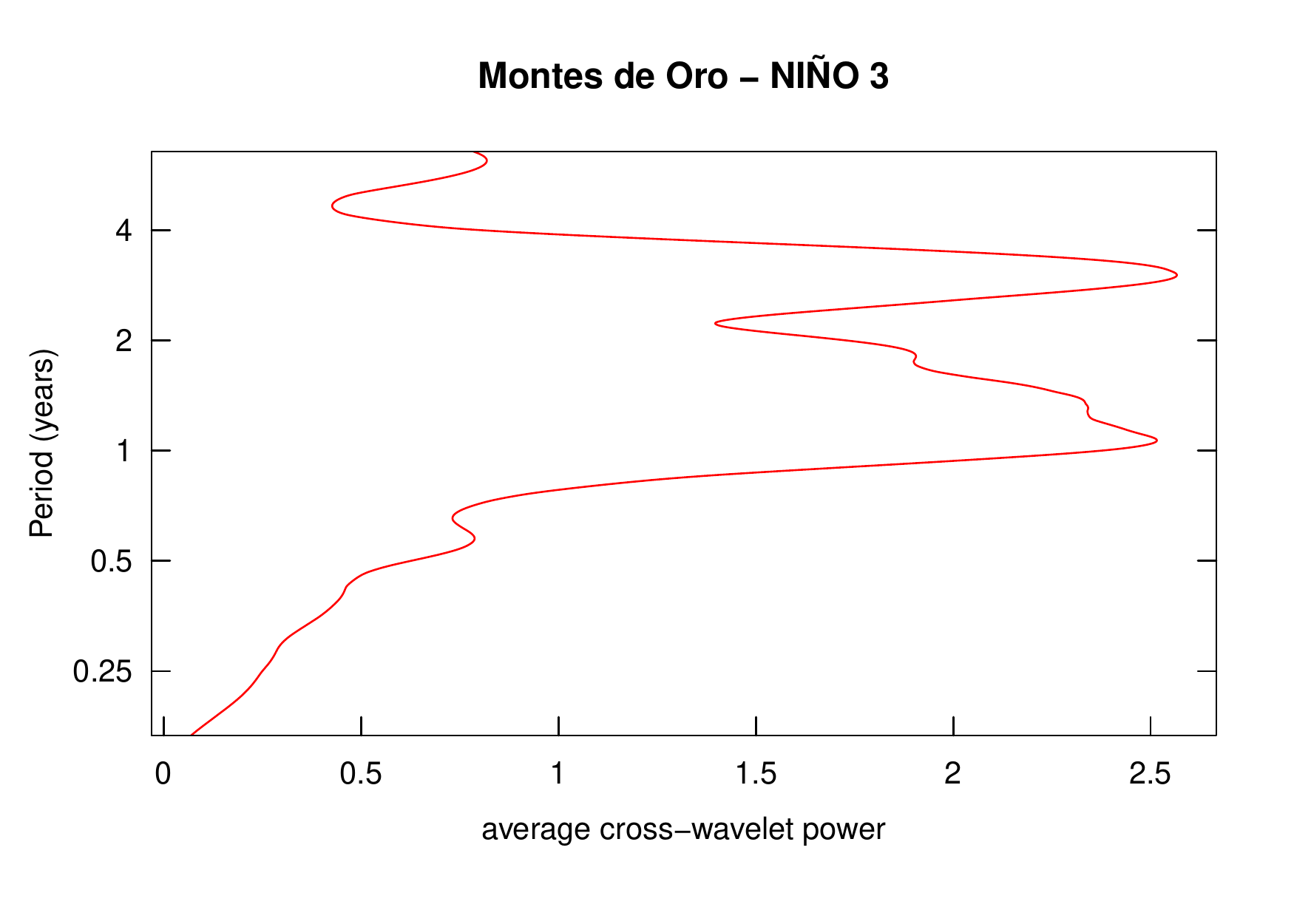}}\vspace{-0.15cm}\\
\subfloat[]{\includegraphics[scale=0.23]{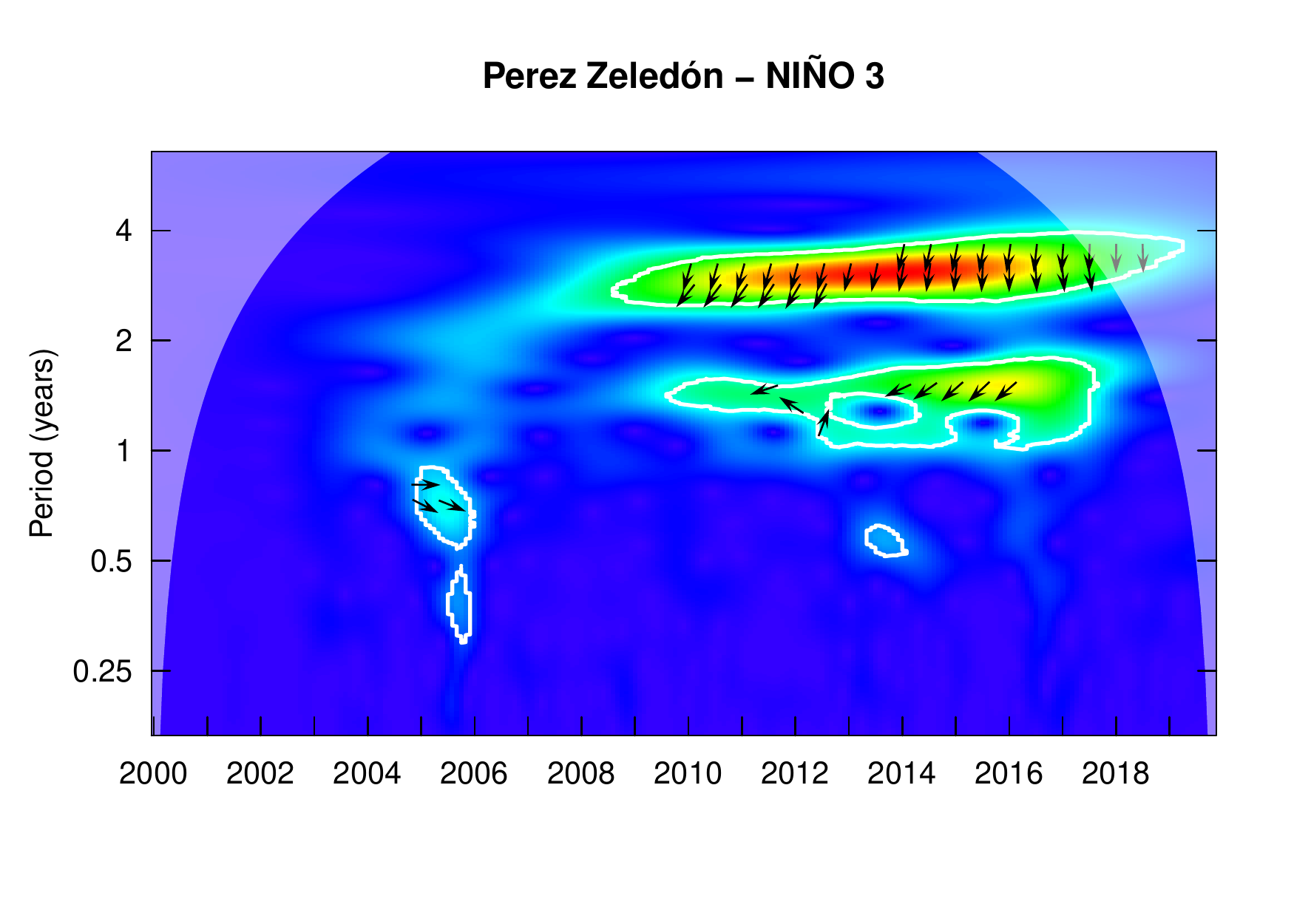}}\vspace{-0.15cm}%
\subfloat[]{\includegraphics[scale=0.23]{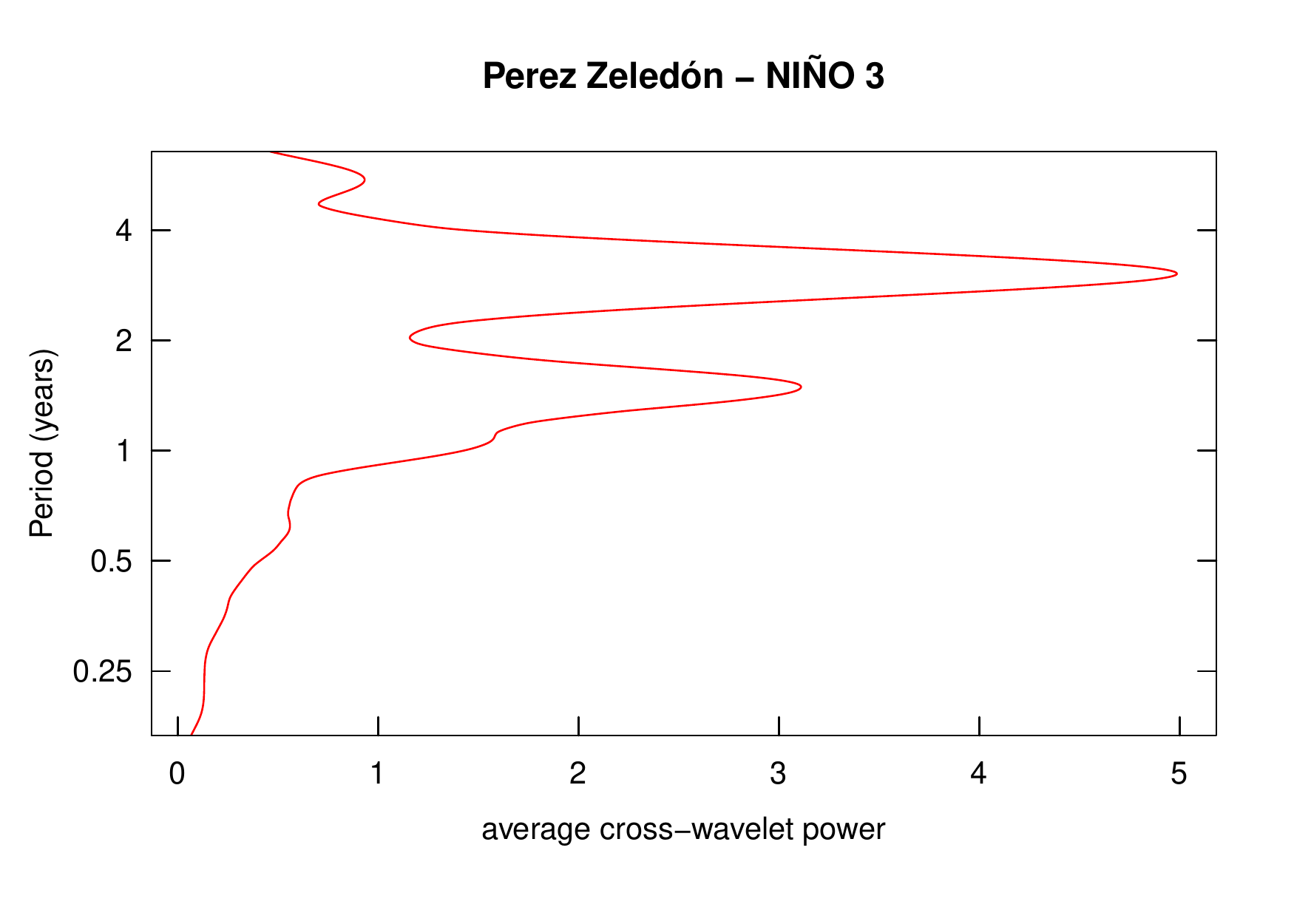}}\vspace{-0.15cm}%
\subfloat[]{\includegraphics[scale=0.23]{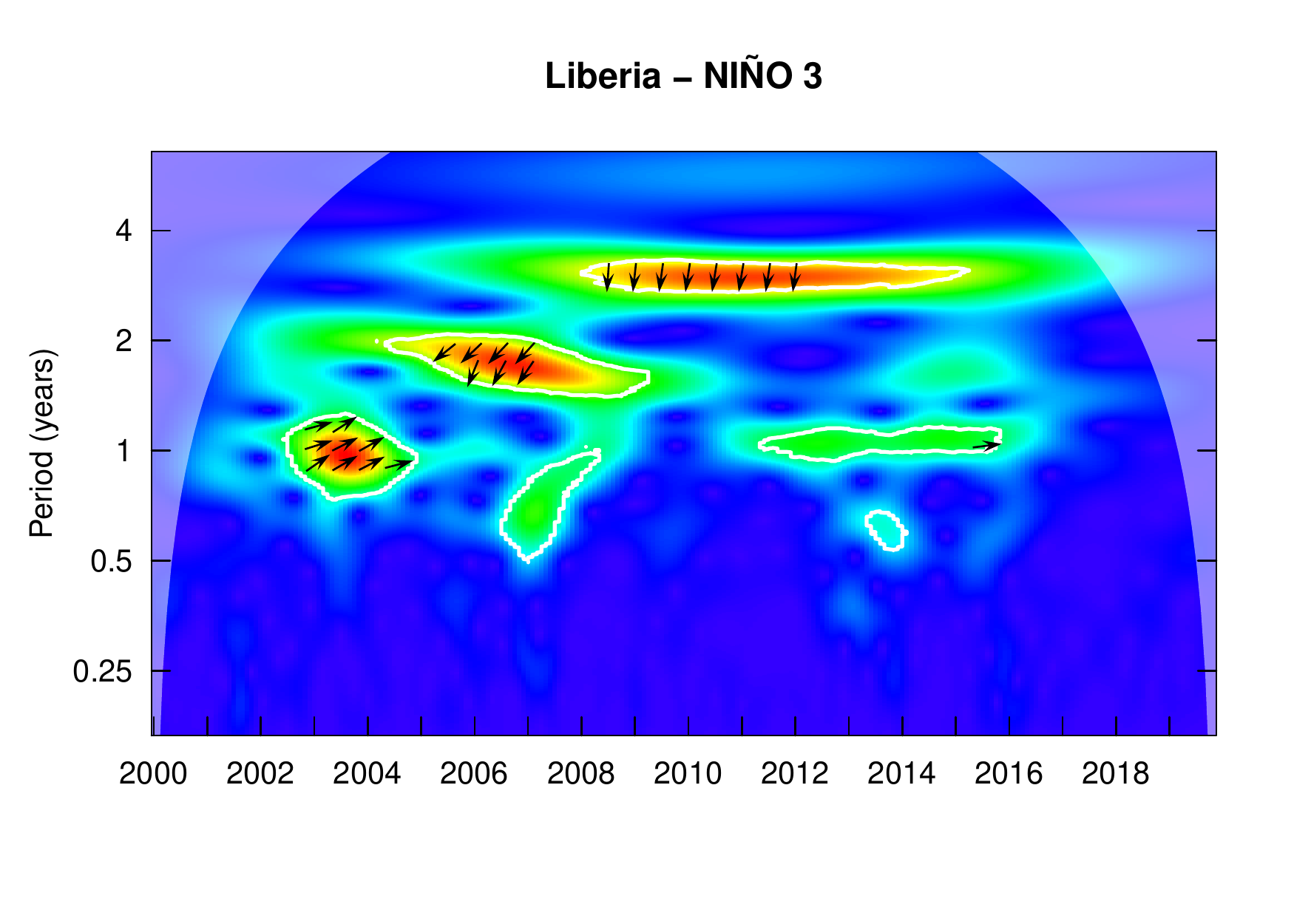}}\vspace{-0.15cm}%
\subfloat[]{\includegraphics[scale=0.23]{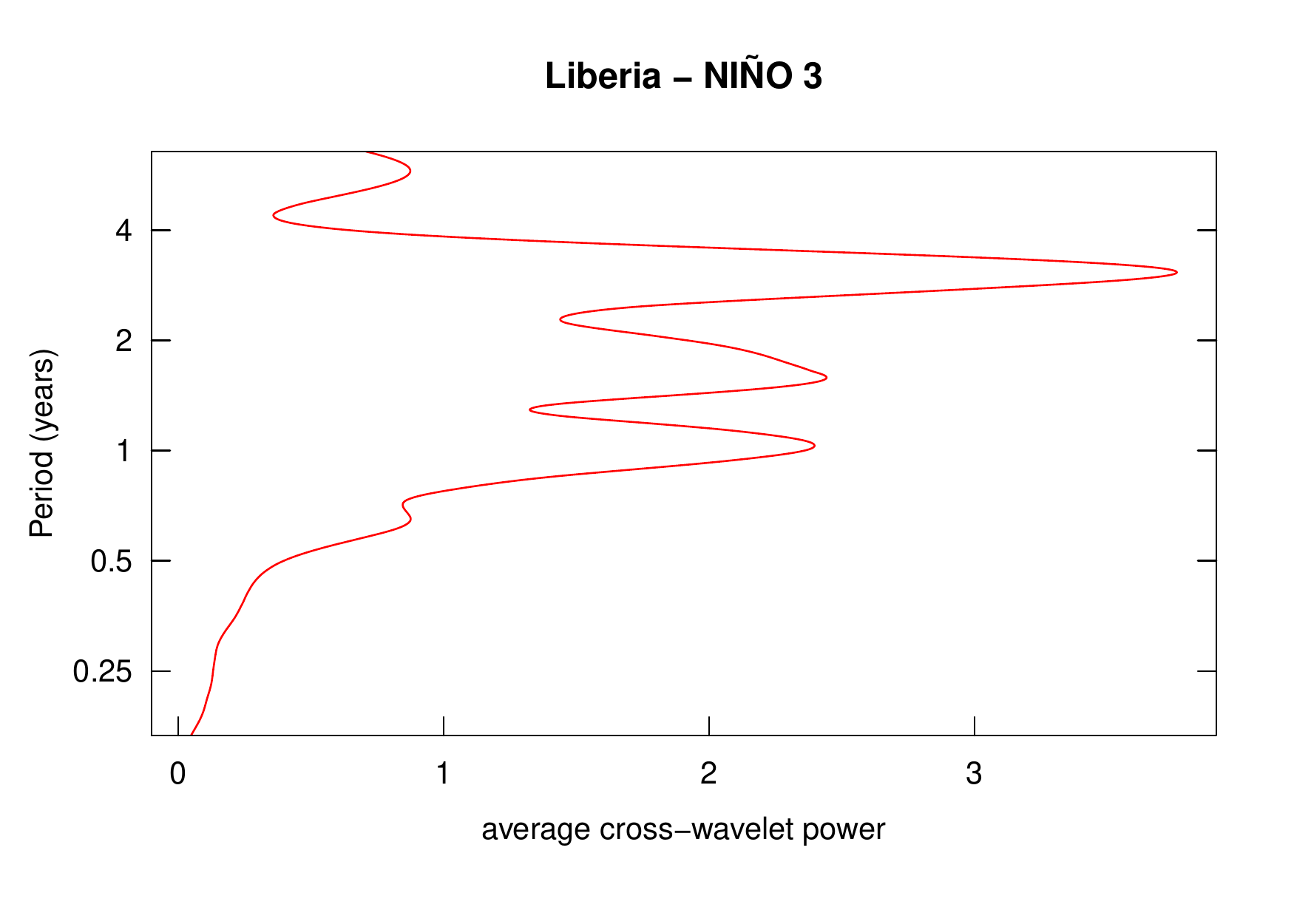}}\vspace{-0.15cm}\\
\subfloat[]{\includegraphics[scale=0.23]{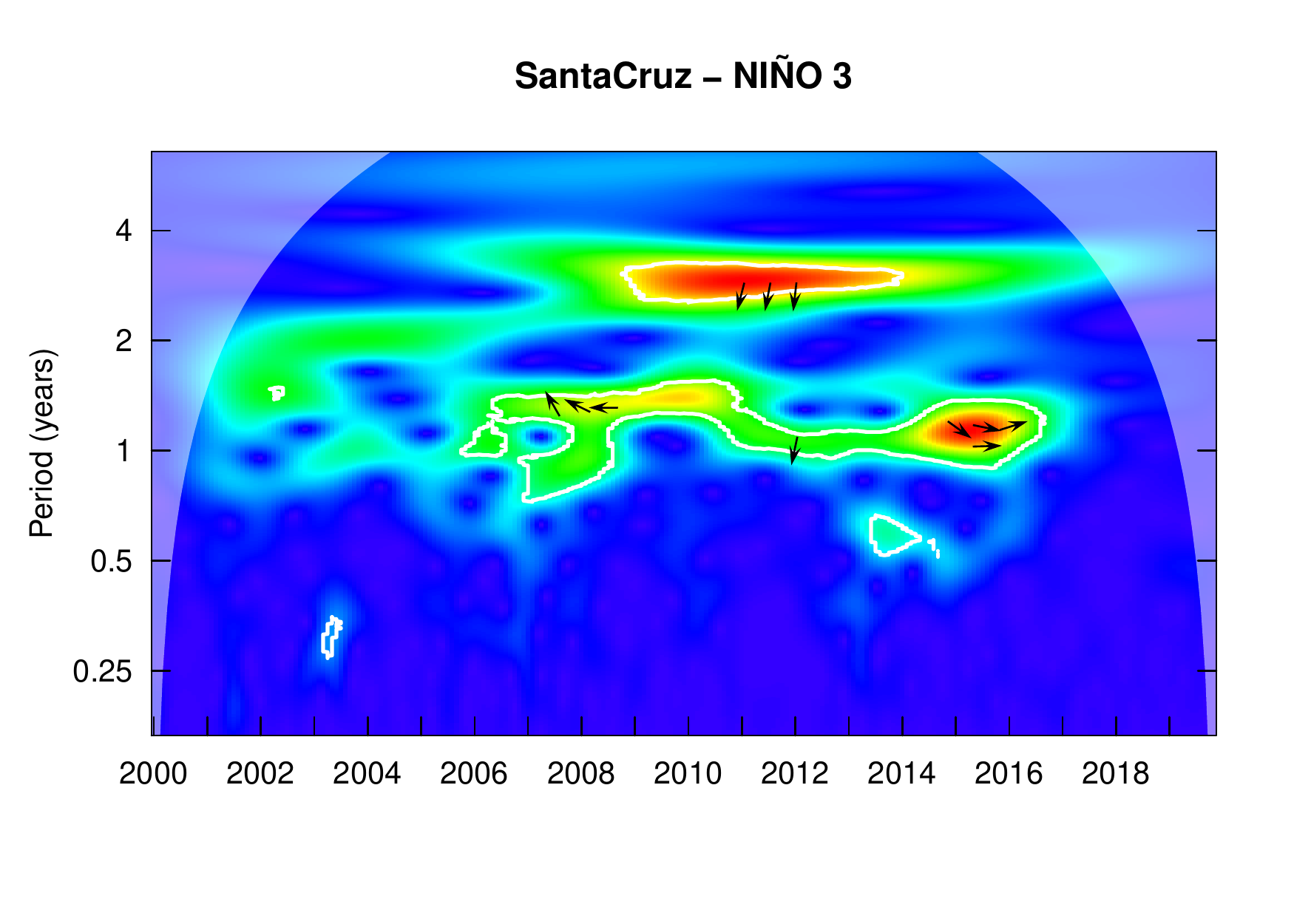}}\vspace{-0.15cm}%
\subfloat[]{\includegraphics[scale=0.23]{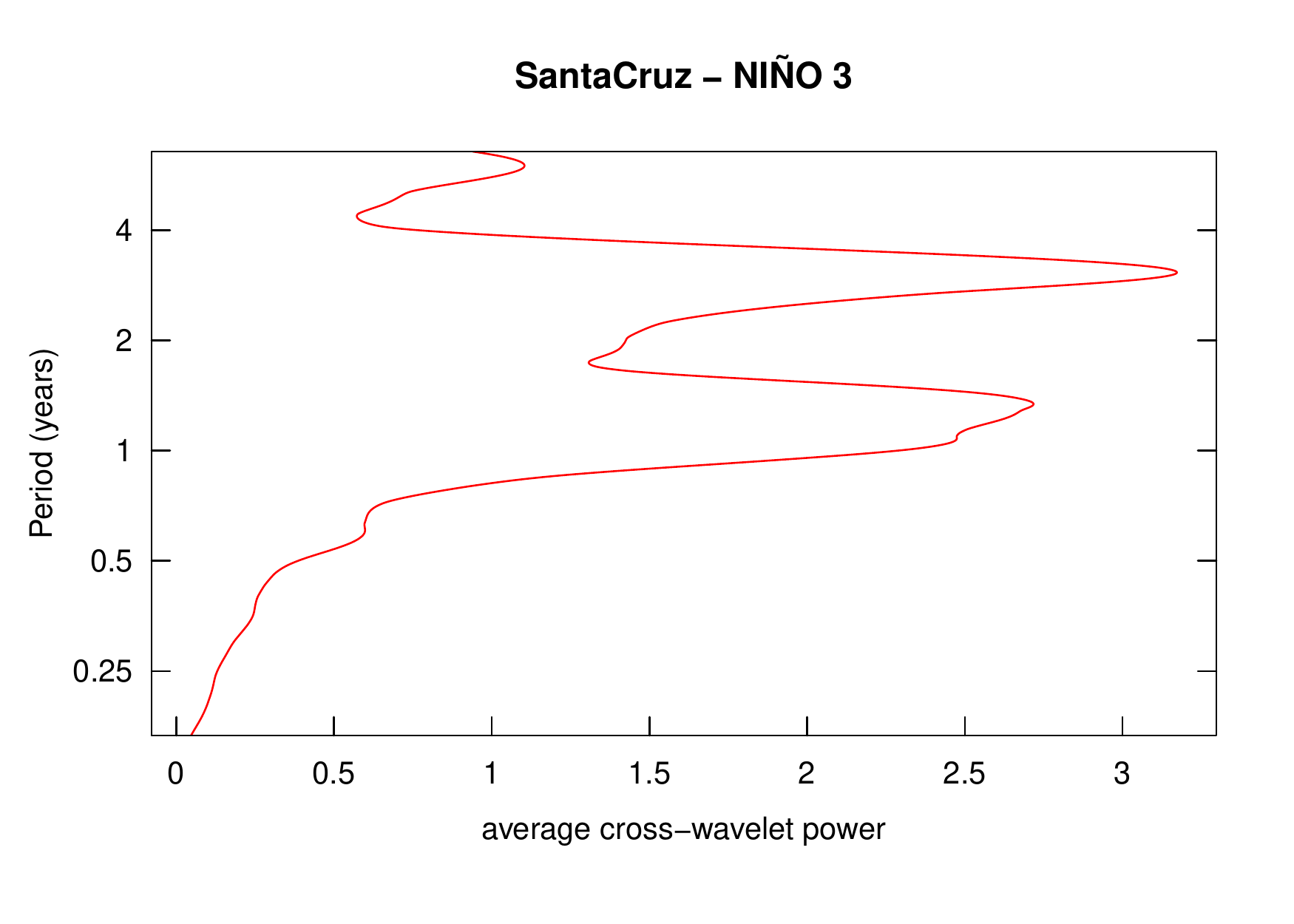}}\vspace{-0.15cm}%
\subfloat[]{\includegraphics[scale=0.23]{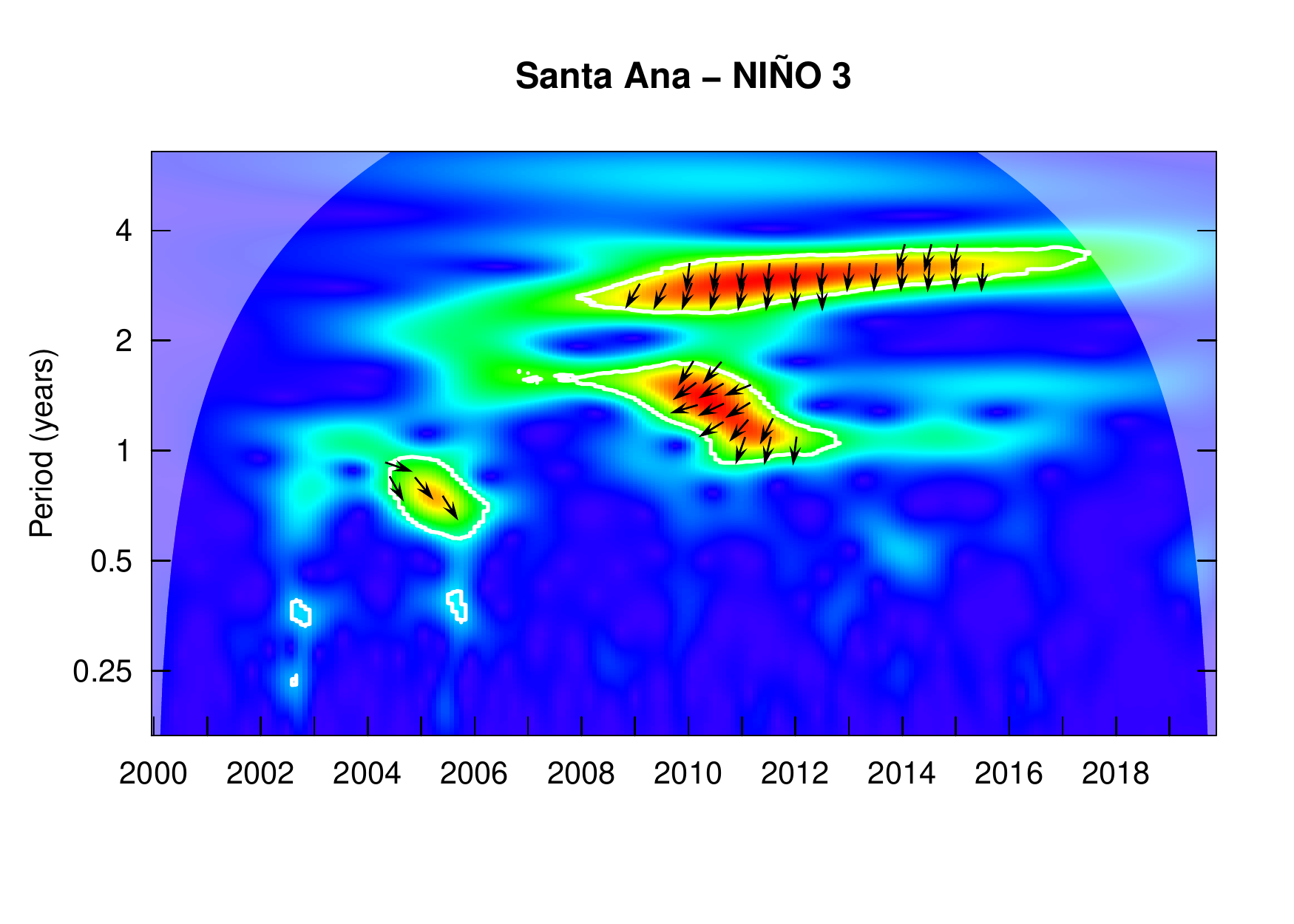}}\vspace{-0.15cm}%
\subfloat[]{\includegraphics[scale=0.23]{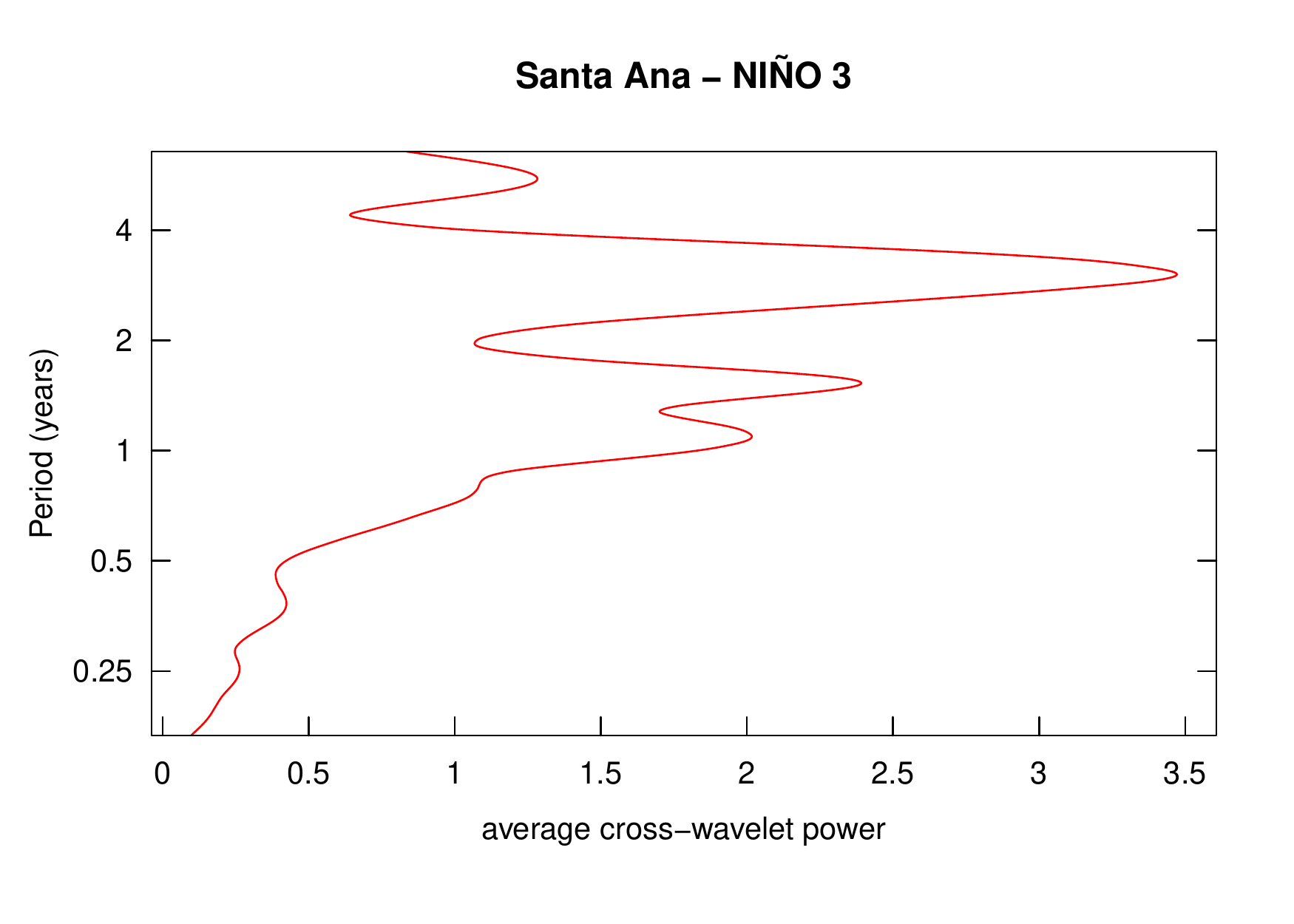}}\vspace{-0.15cm}\\
\subfloat[]{\includegraphics[scale=0.23]{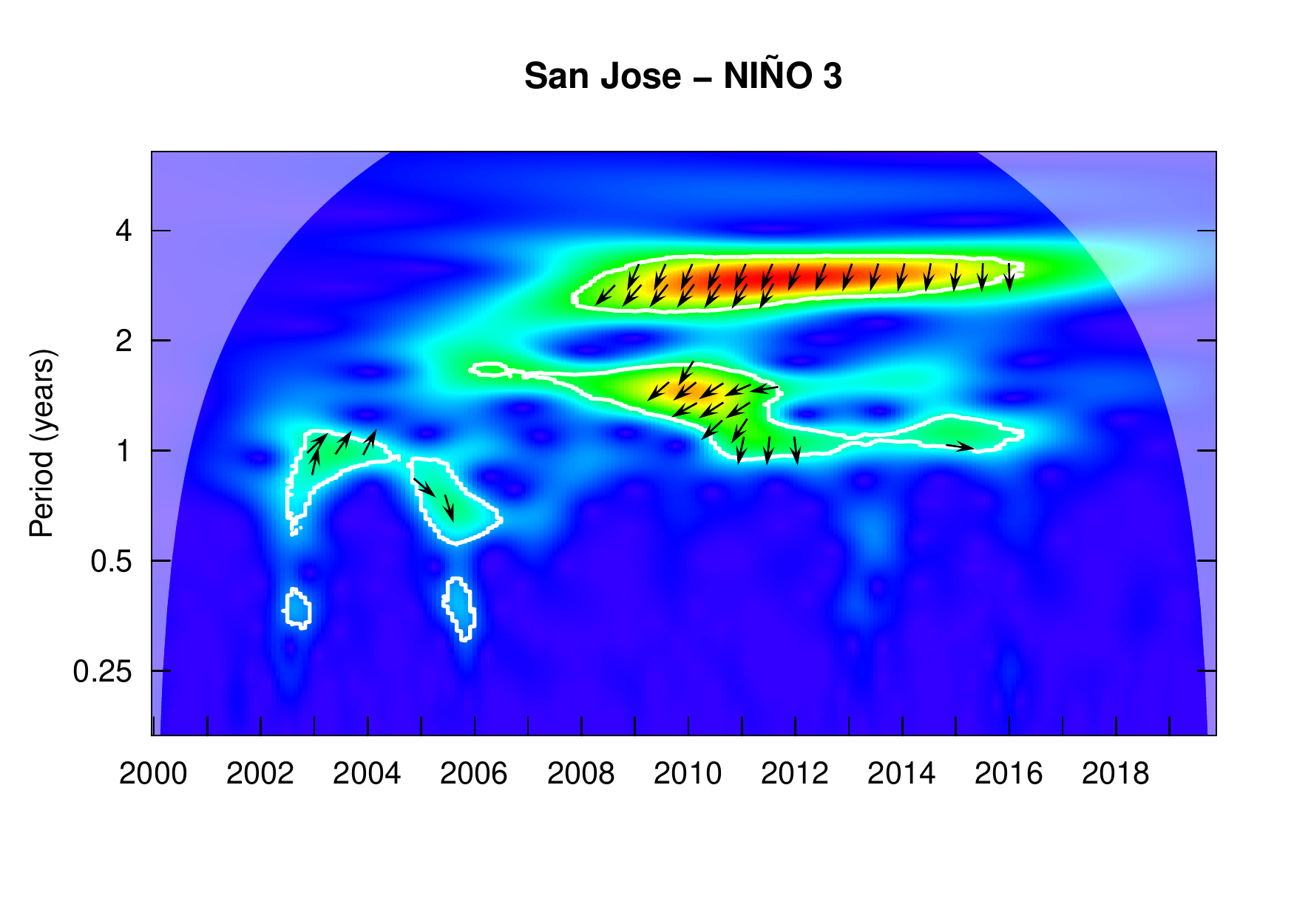}}\vspace{-0.15cm}%
\subfloat[]{\includegraphics[scale=0.23]{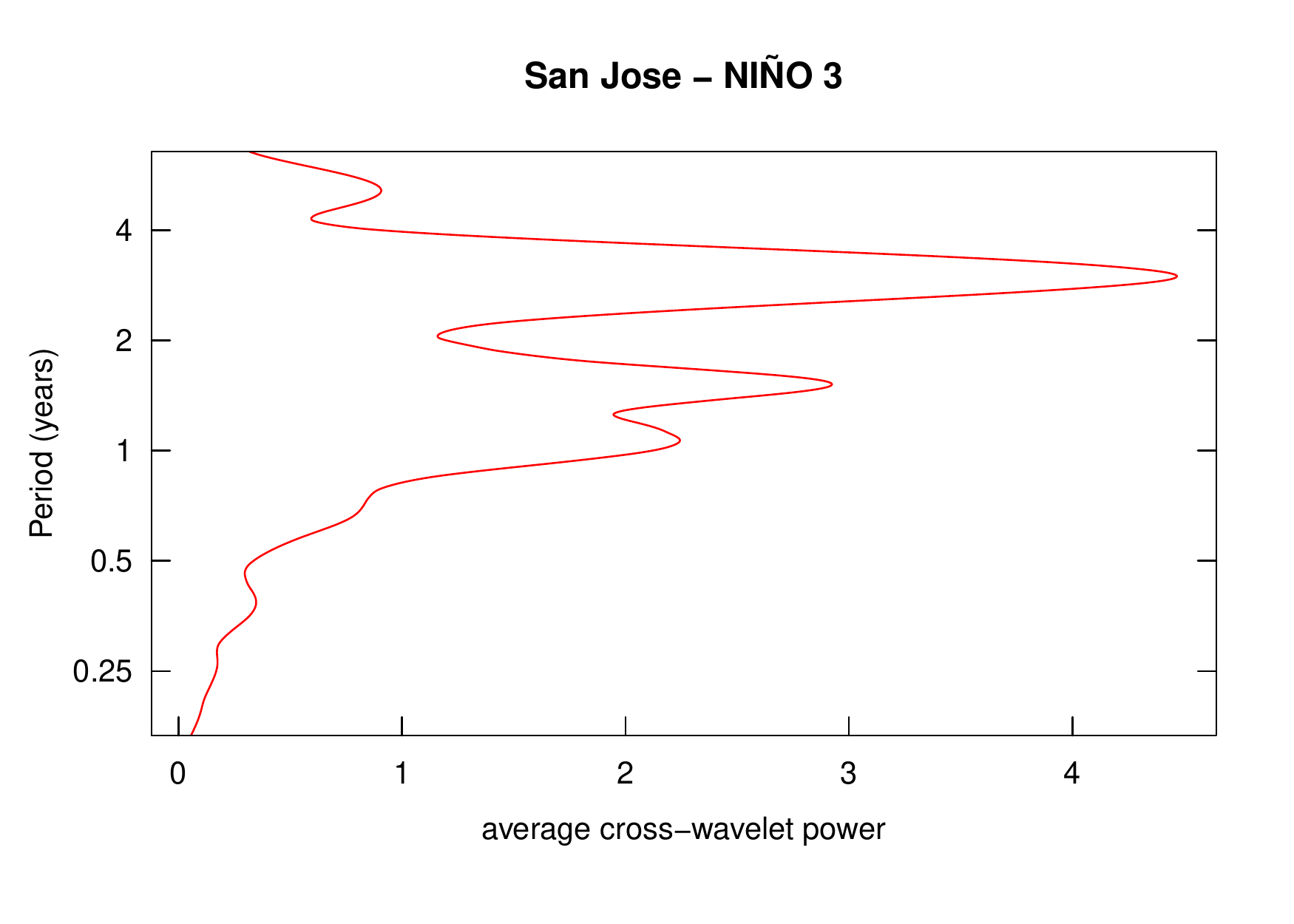}}\vspace{-0.15cm}%
\subfloat[]{\includegraphics[scale=0.23]{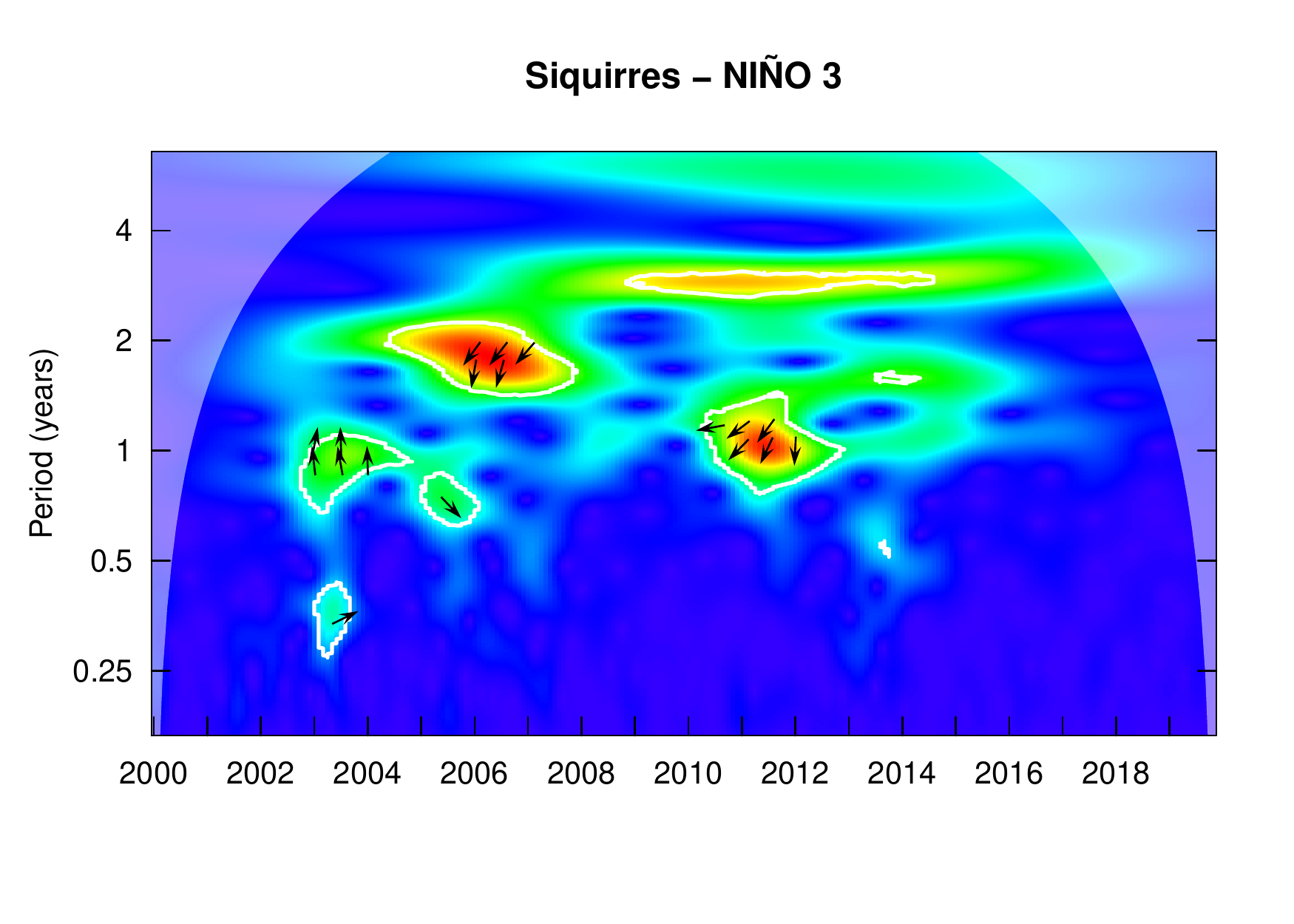}}\vspace{-0.15cm}%
\subfloat[]{\includegraphics[scale=0.23]{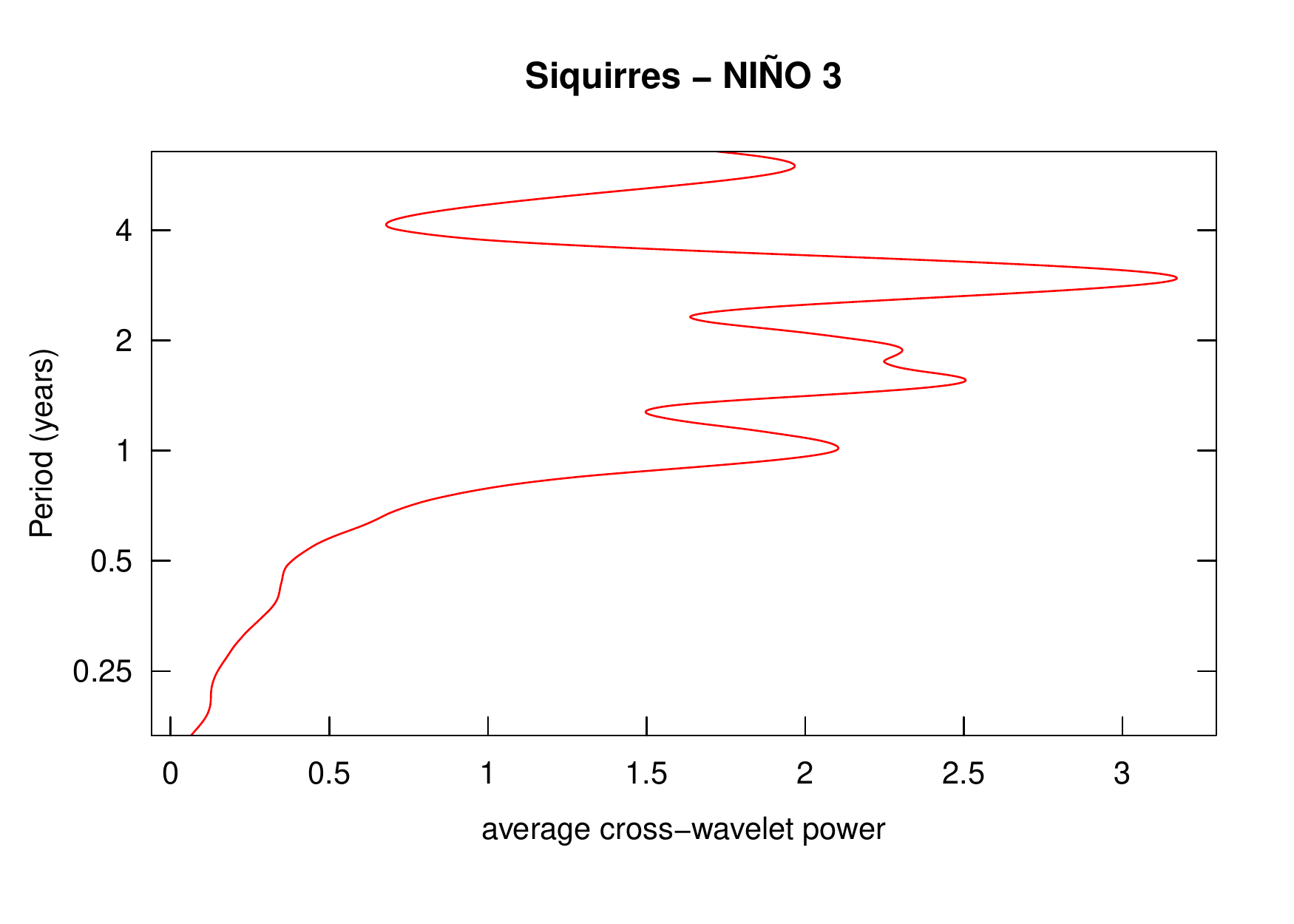}}\vspace{-0.15cm}\\
\subfloat[]{\includegraphics[scale=0.23]{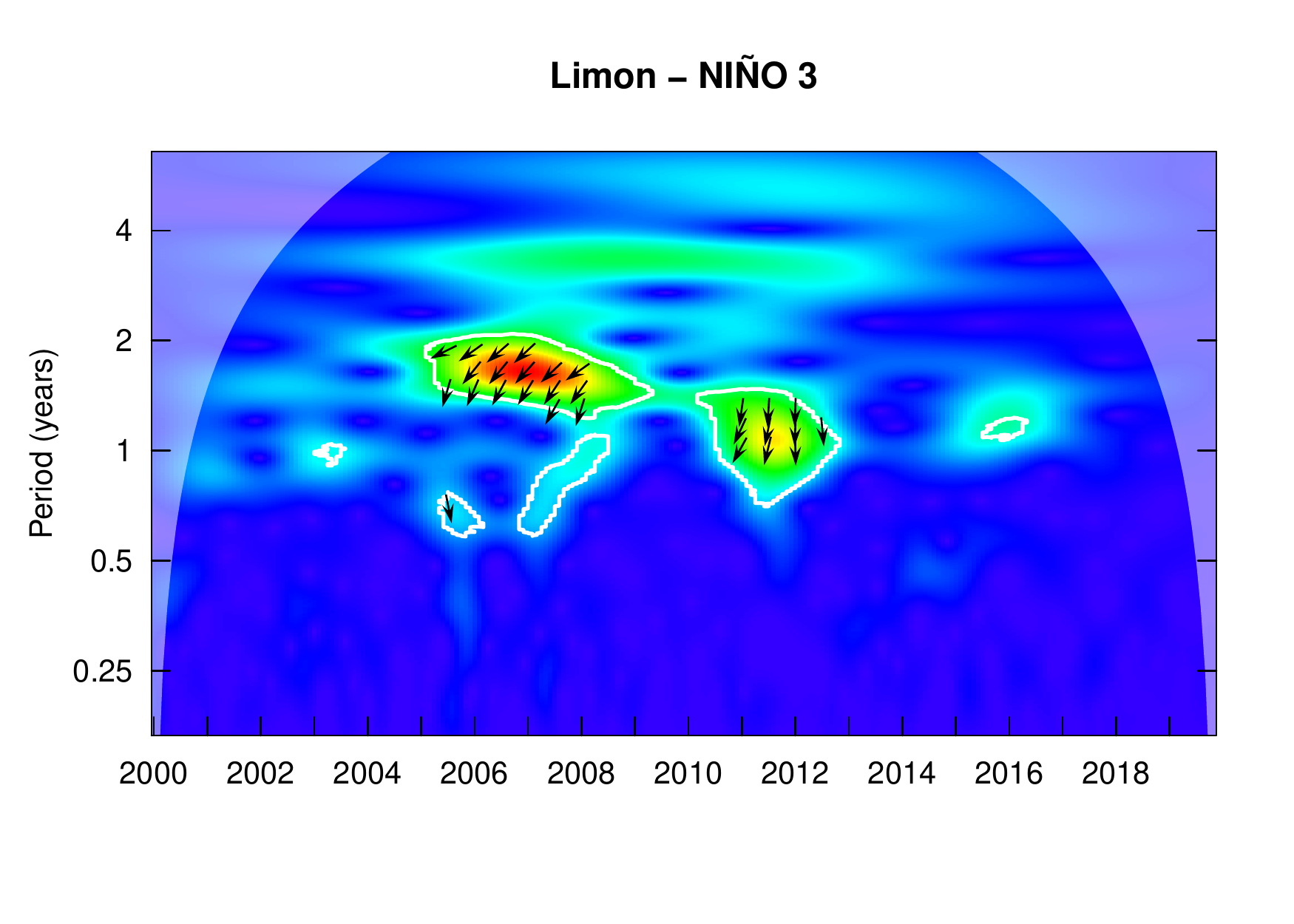}}\vspace{-0.15cm}%
\subfloat[]{\includegraphics[scale=0.23]{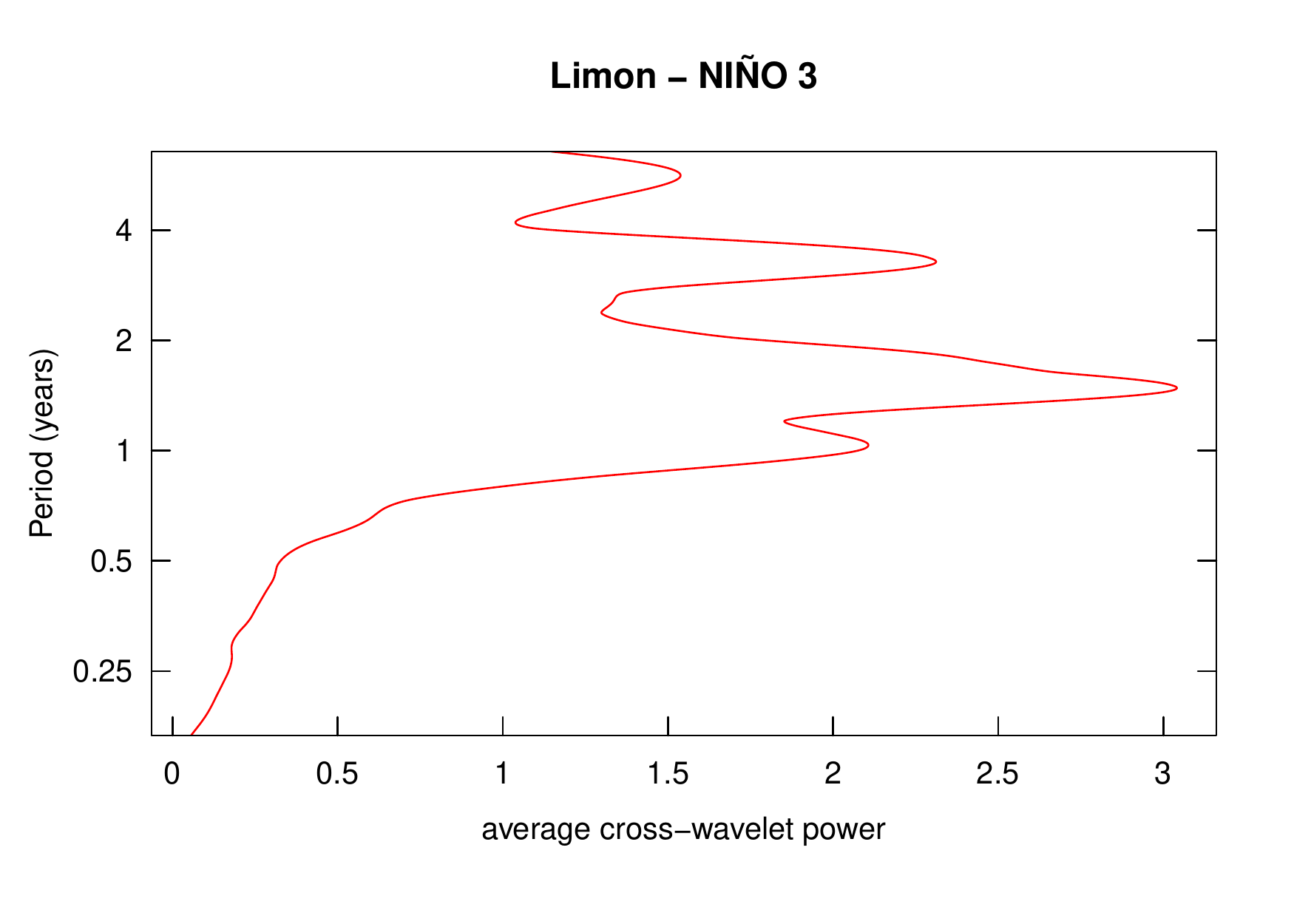}}\vspace{-0.15cm}%
\subfloat[]{\includegraphics[scale=0.23]{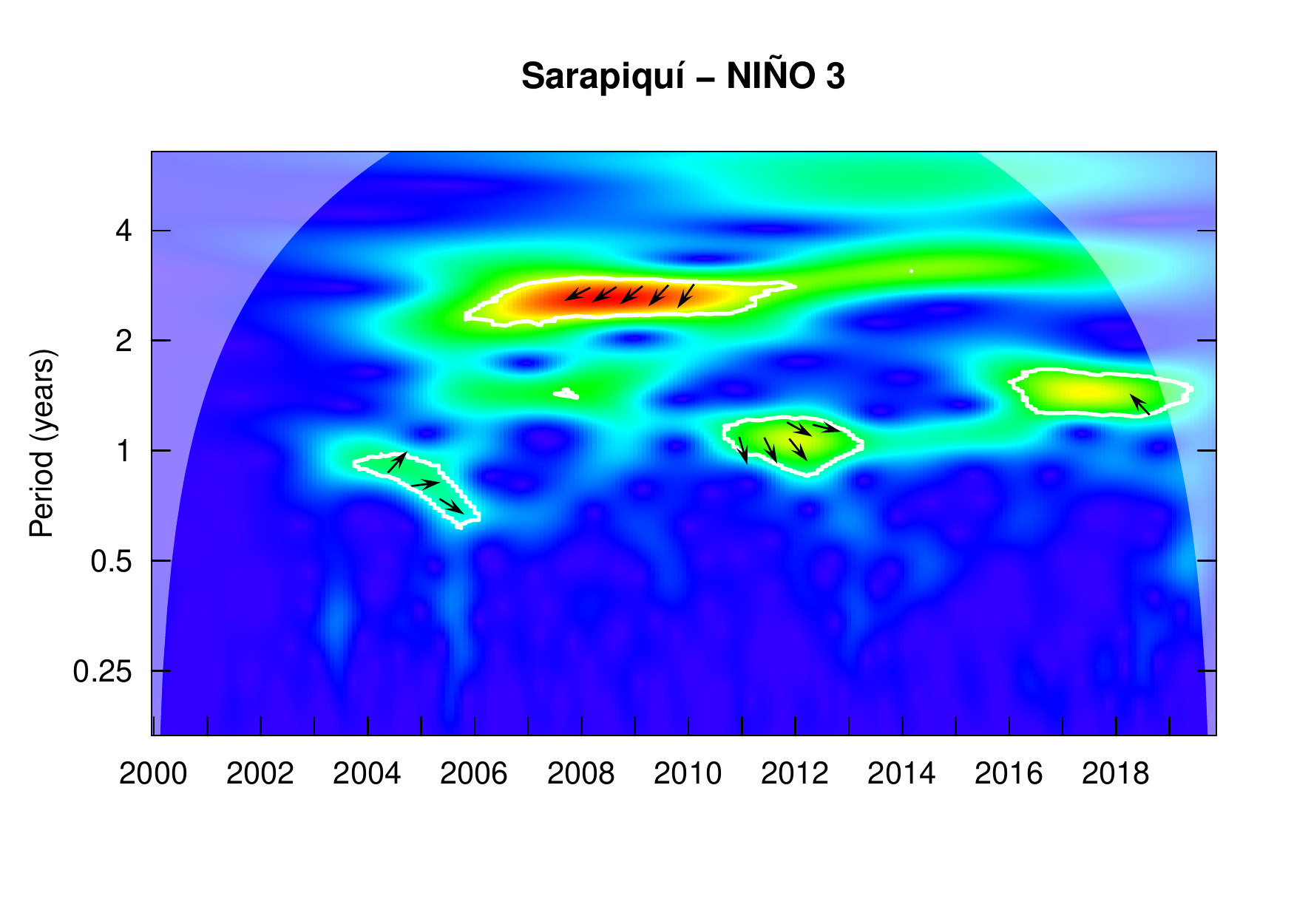}}\vspace{-0.15cm}%
\subfloat[]{\includegraphics[scale=0.23]{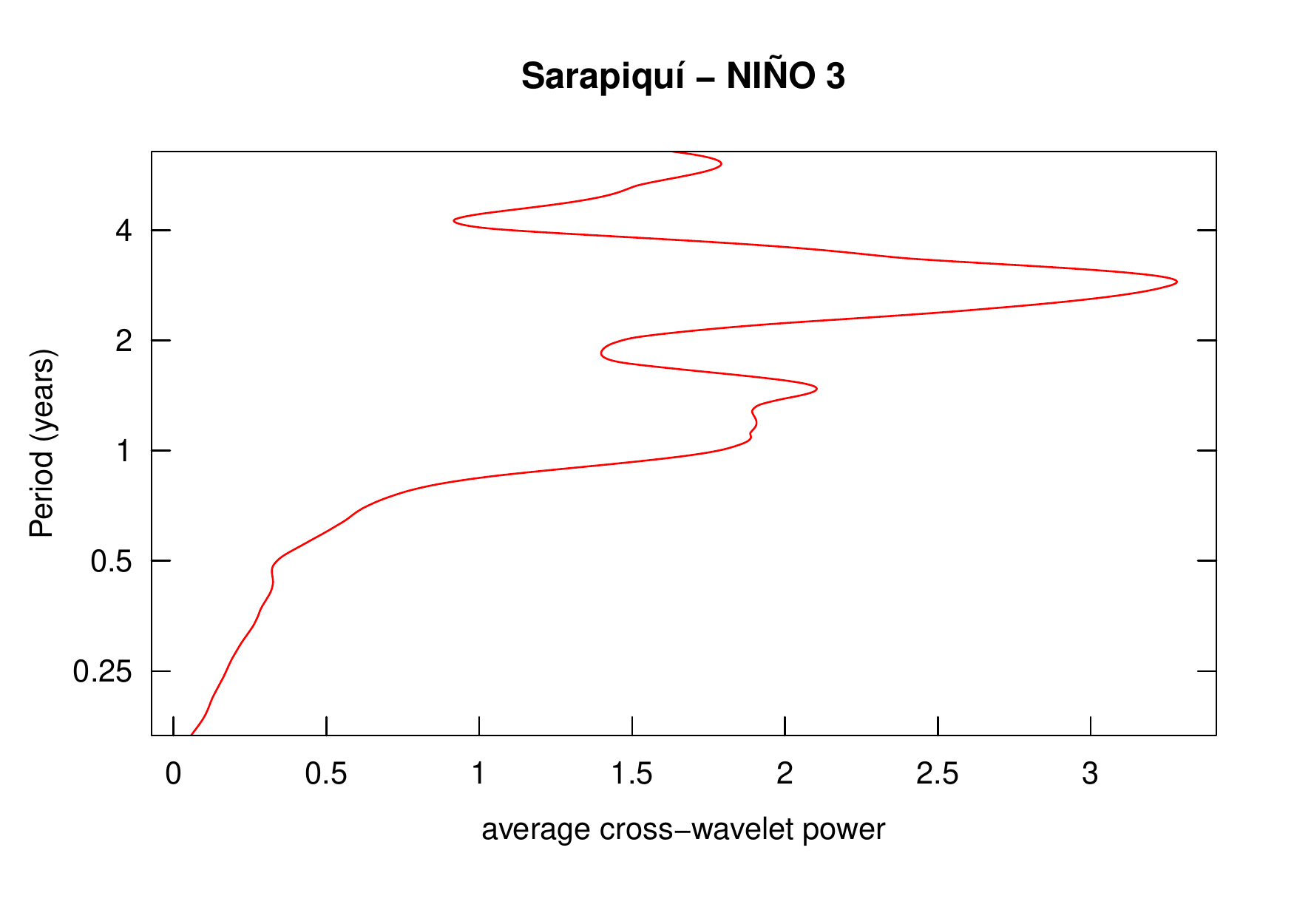}}\vspace{-0.15cm}
\caption*{}
\end{figure}

\begin{figure}[H]
\captionsetup[subfigure]{labelformat=empty}
\subfloat[]{\includegraphics[scale=0.23]{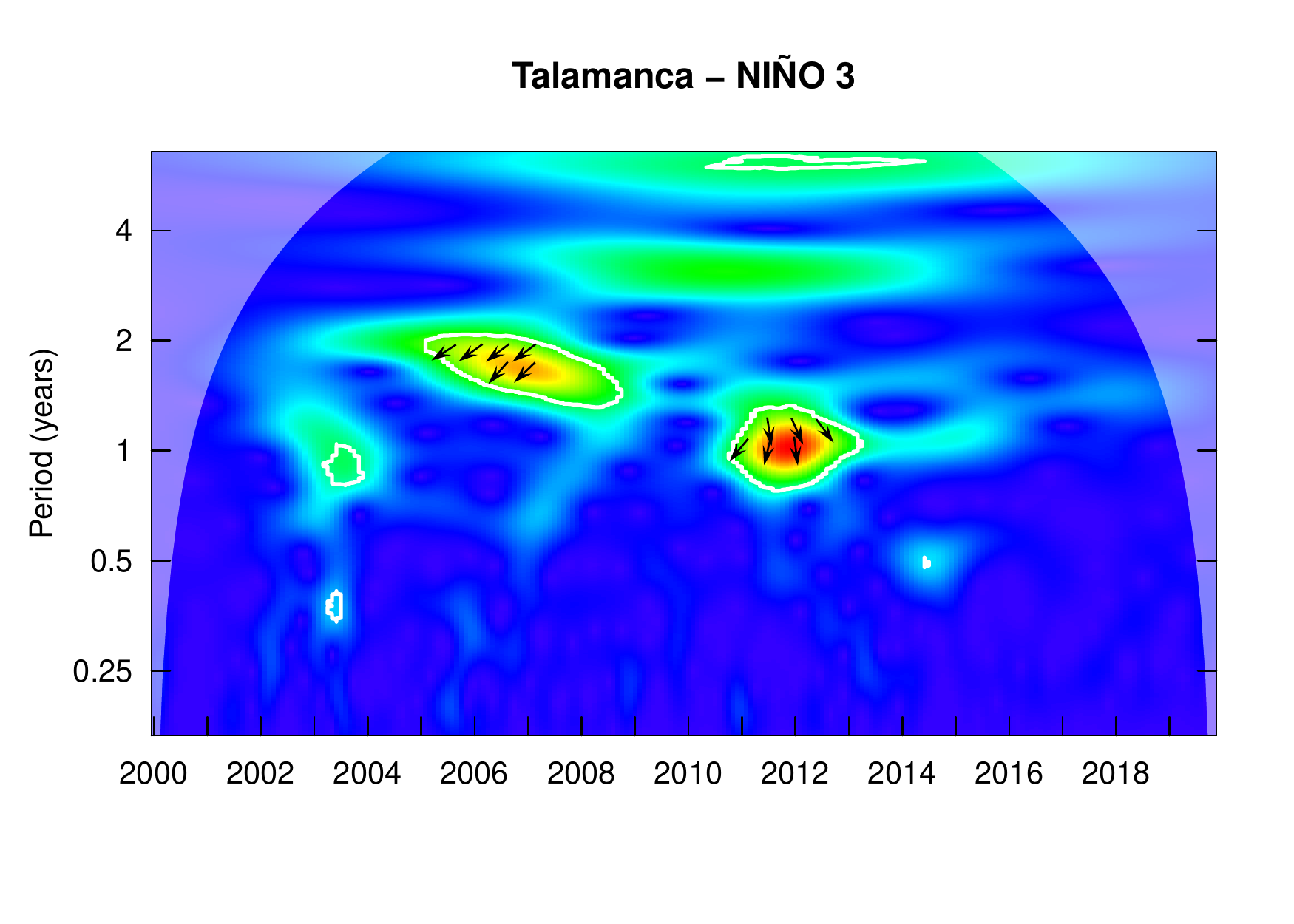}}\vspace{-0.15cm}%
\subfloat[]{\includegraphics[scale=0.23]{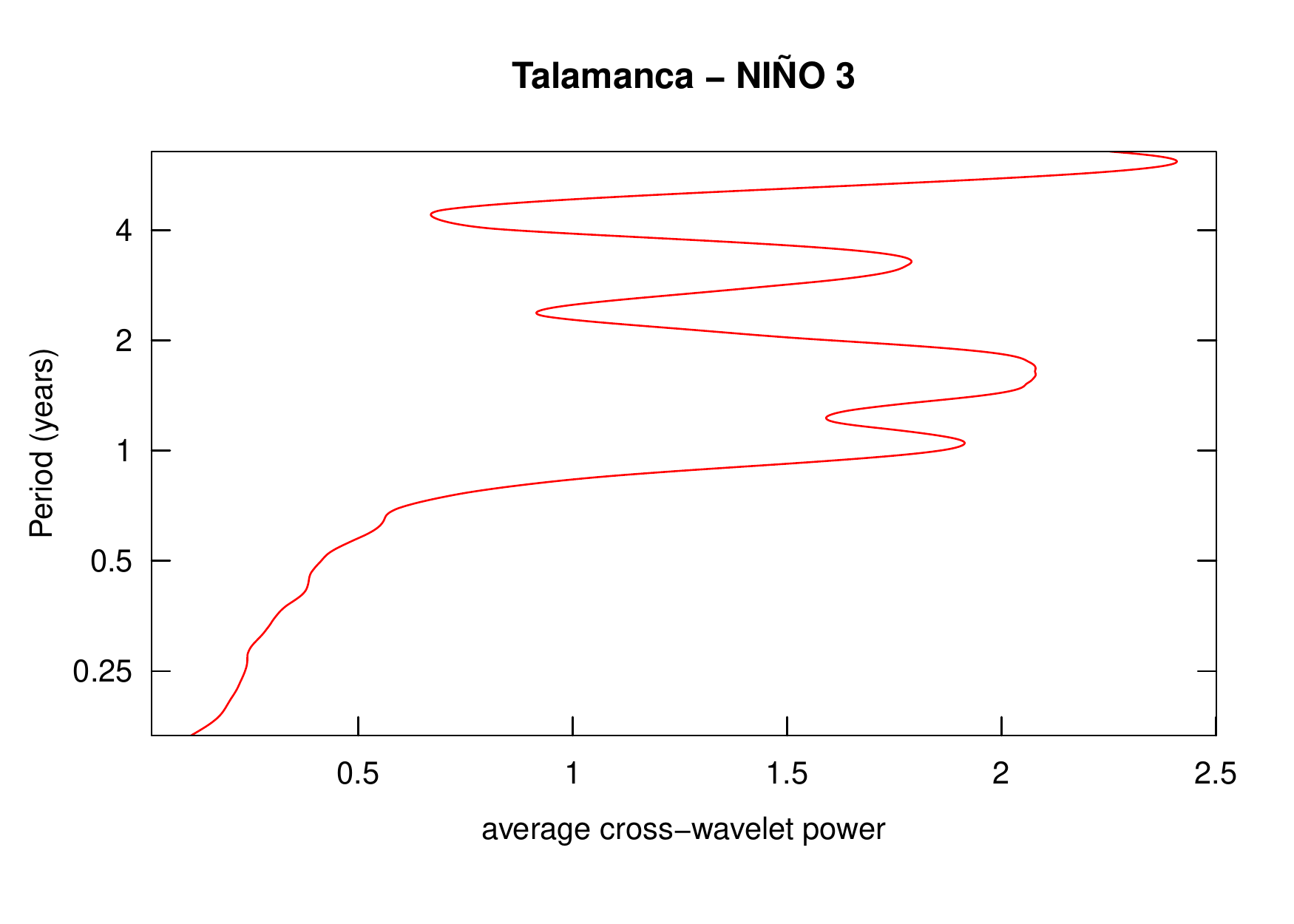}}\vspace{-0.15cm}%
\subfloat[]{\includegraphics[scale=0.23]{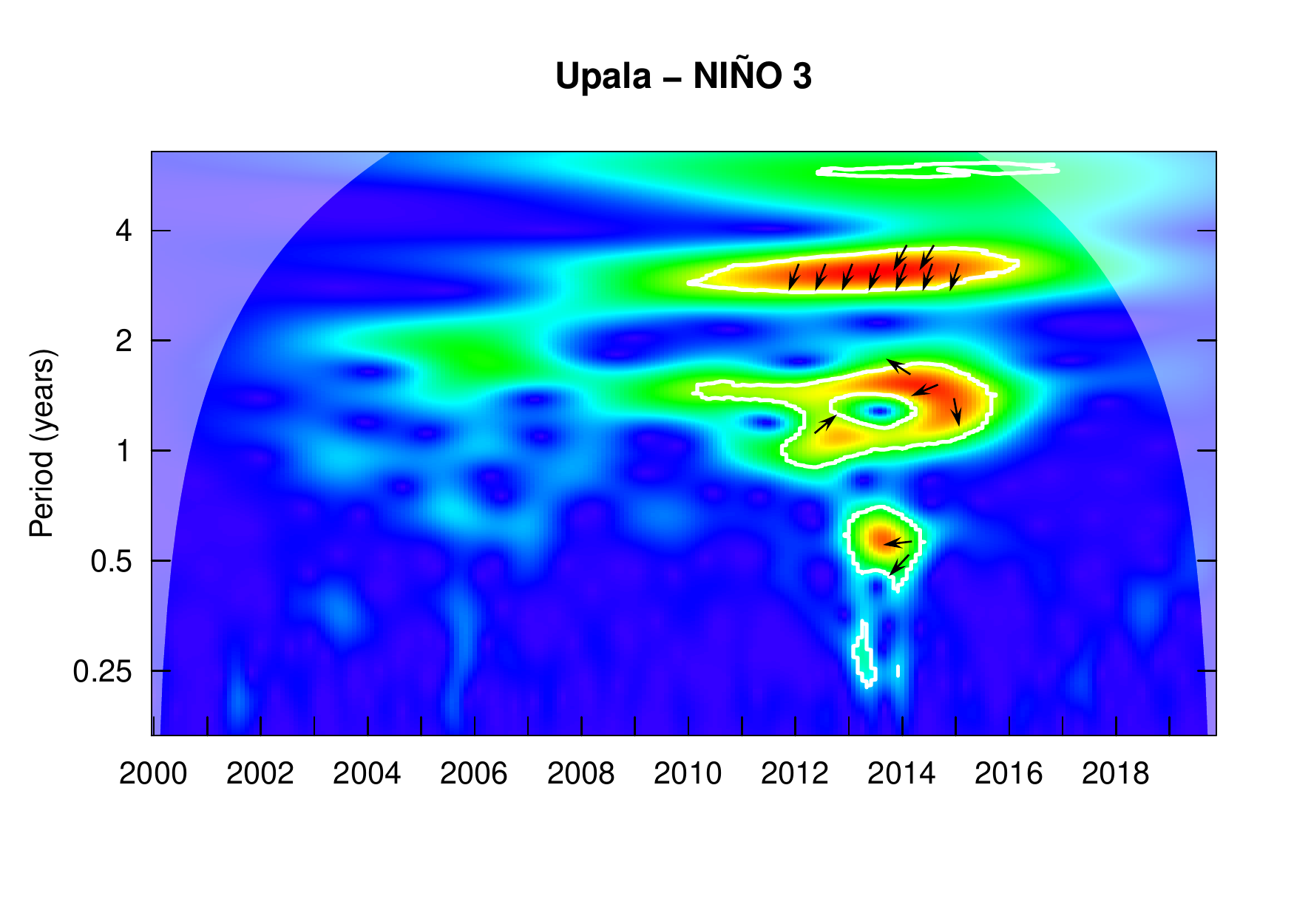}}\vspace{-0.15cm}%
\subfloat[]{\includegraphics[scale=0.23]{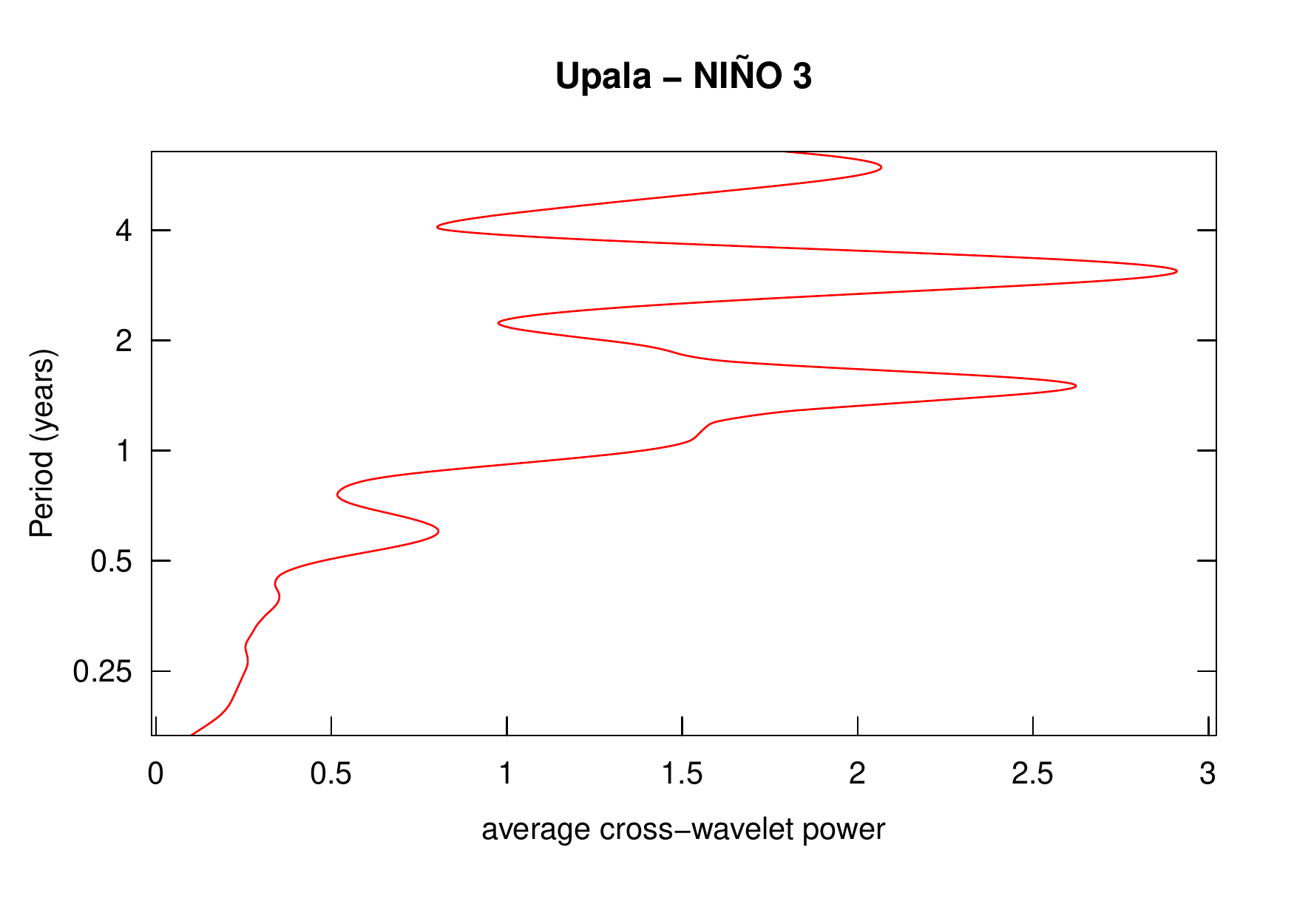}}\vspace{-0.15cm}\\
\subfloat[]{\includegraphics[scale=0.23]{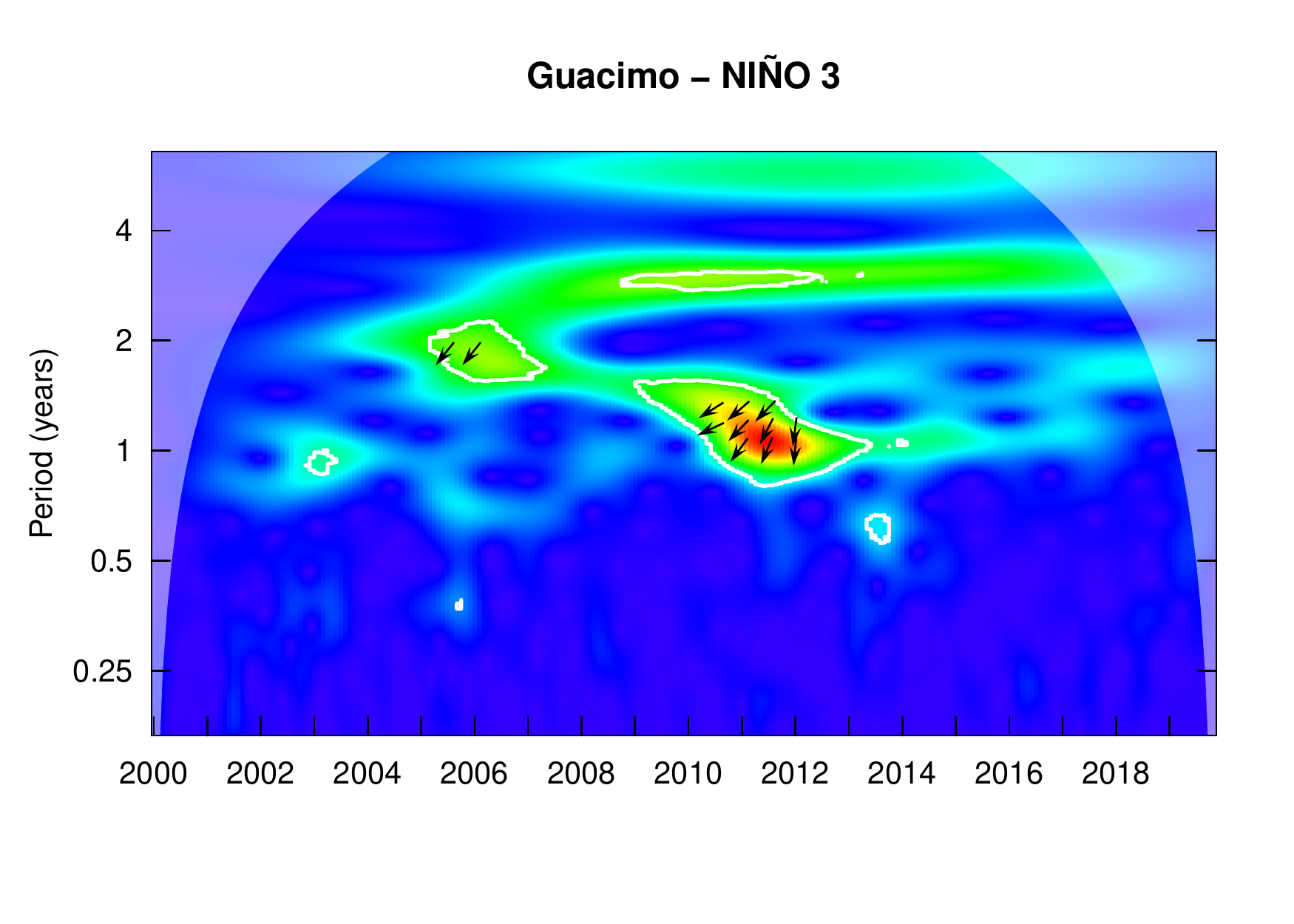}}\vspace{-0.15cm}%
\subfloat[]{\includegraphics[scale=0.23]{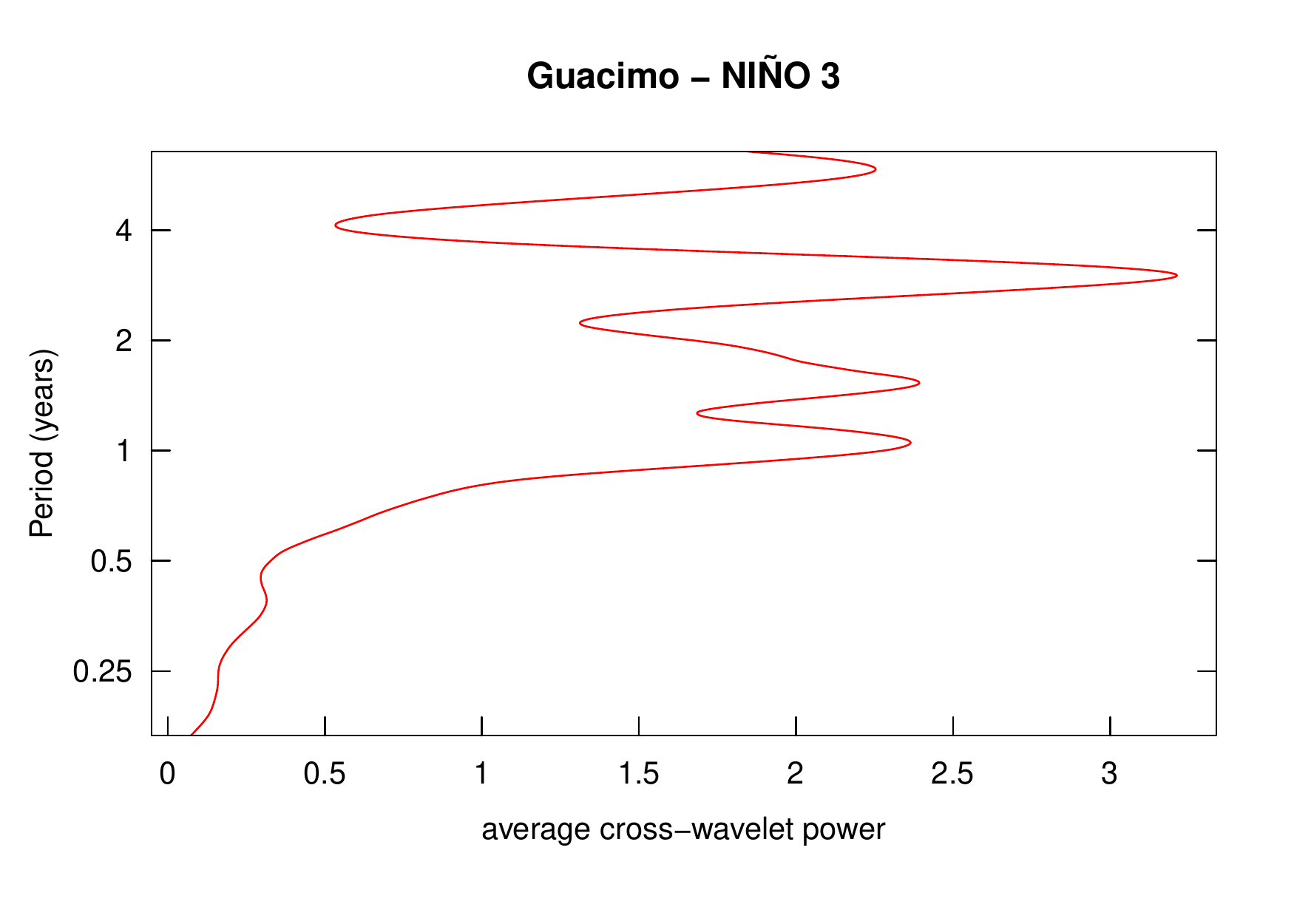}}\vspace{-0.15cm}%
\subfloat[]{\includegraphics[scale=0.23]{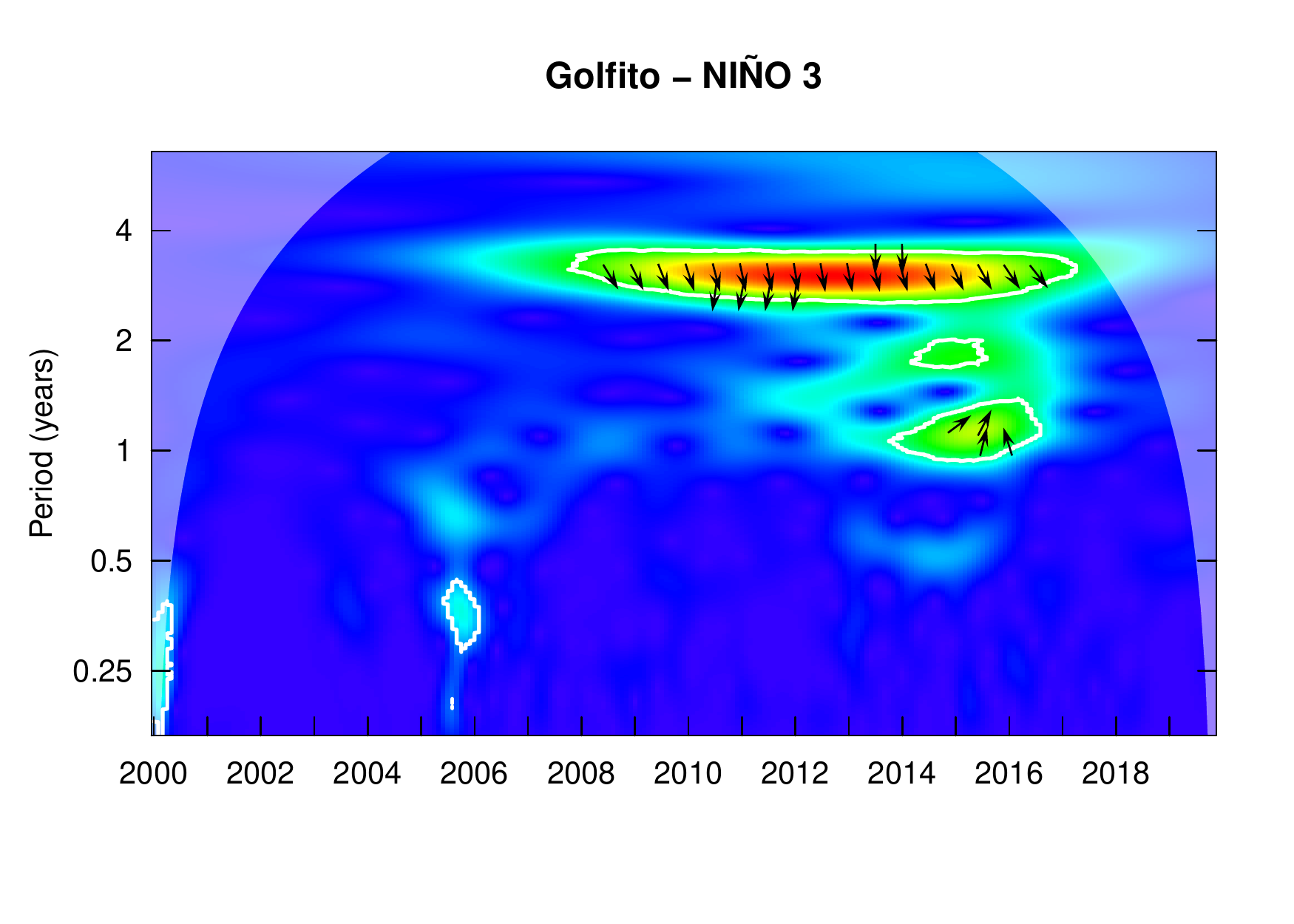}}\vspace{-0.15cm}%
\subfloat[]{\includegraphics[scale=0.23]{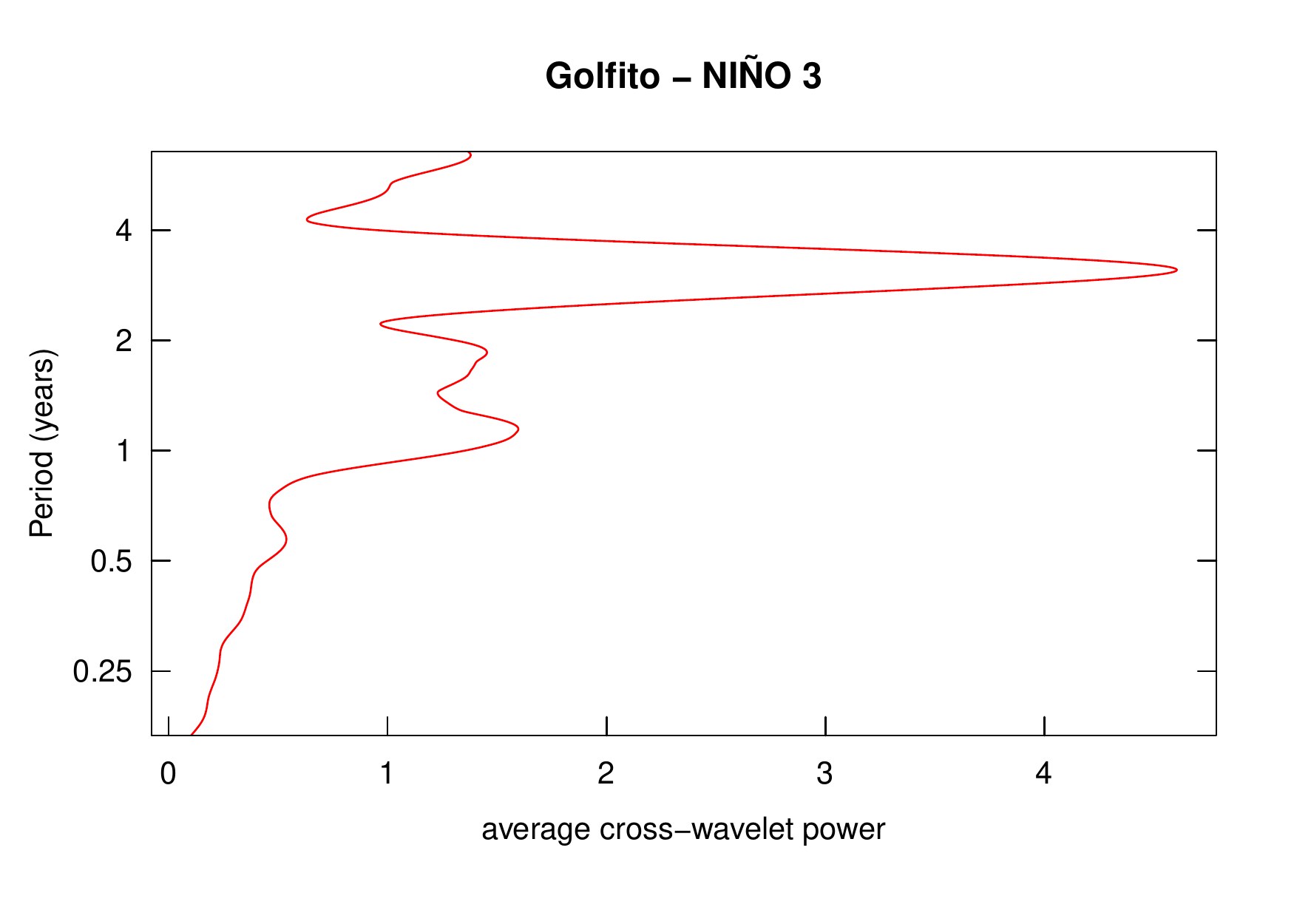}}\vspace{-0.15cm}\\
\subfloat[]{\includegraphics[scale=0.23]{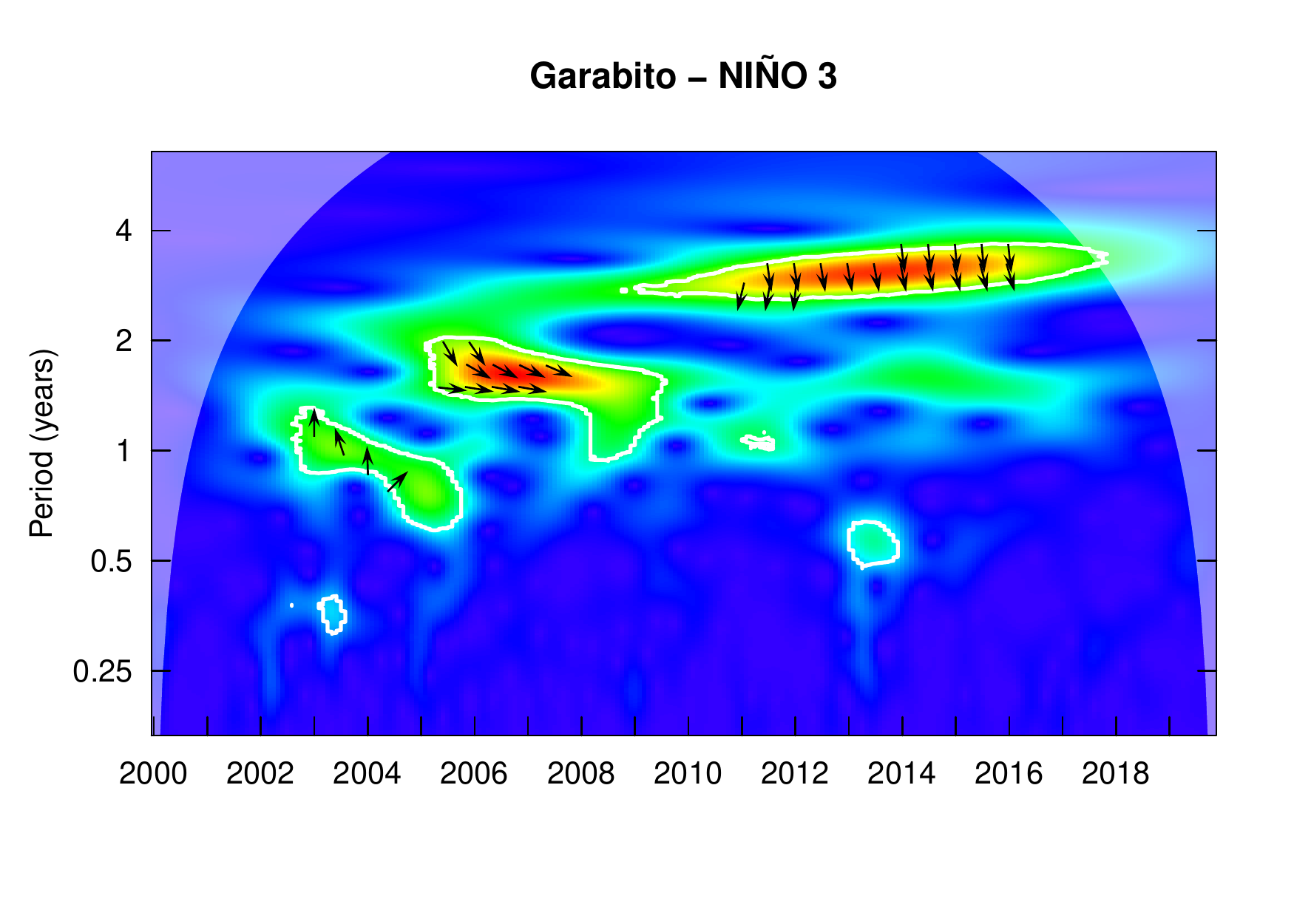}}\vspace{-0.15cm}%
\subfloat[]{\includegraphics[scale=0.23]{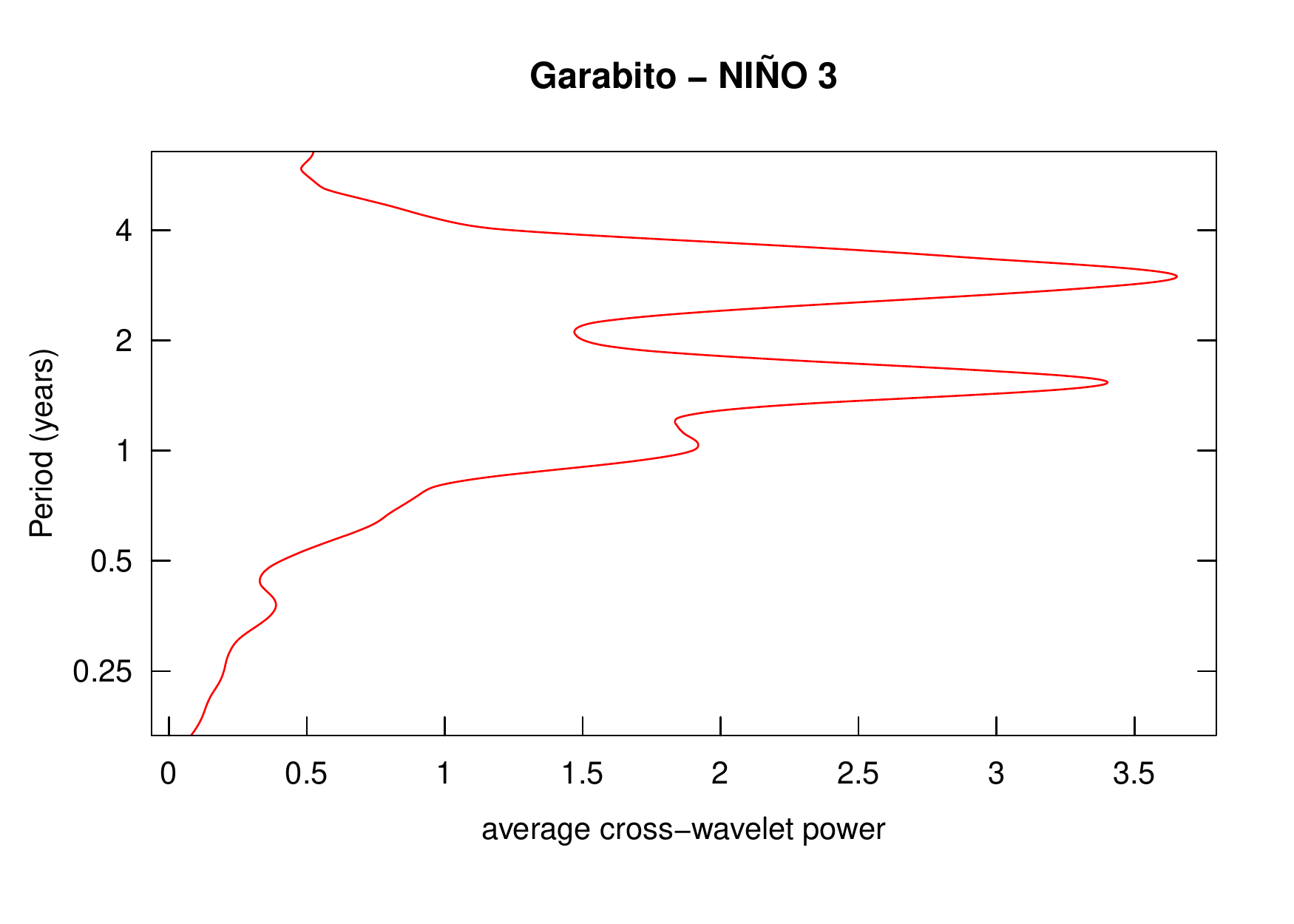}}\vspace{-0.15cm}%
\subfloat[]{\includegraphics[scale=0.23]{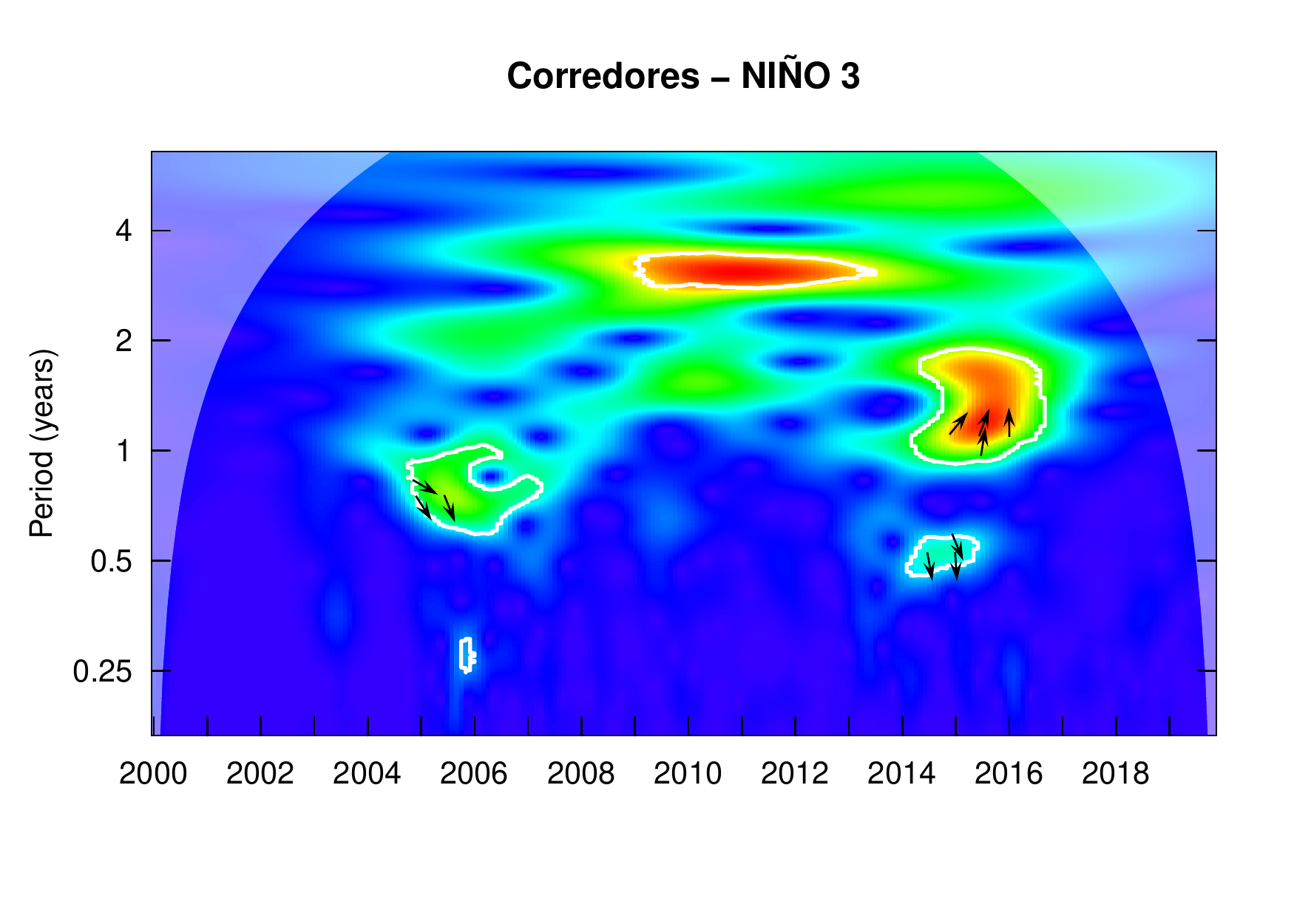}}\vspace{-0.15cm}%
\subfloat[]{\includegraphics[scale=0.23]{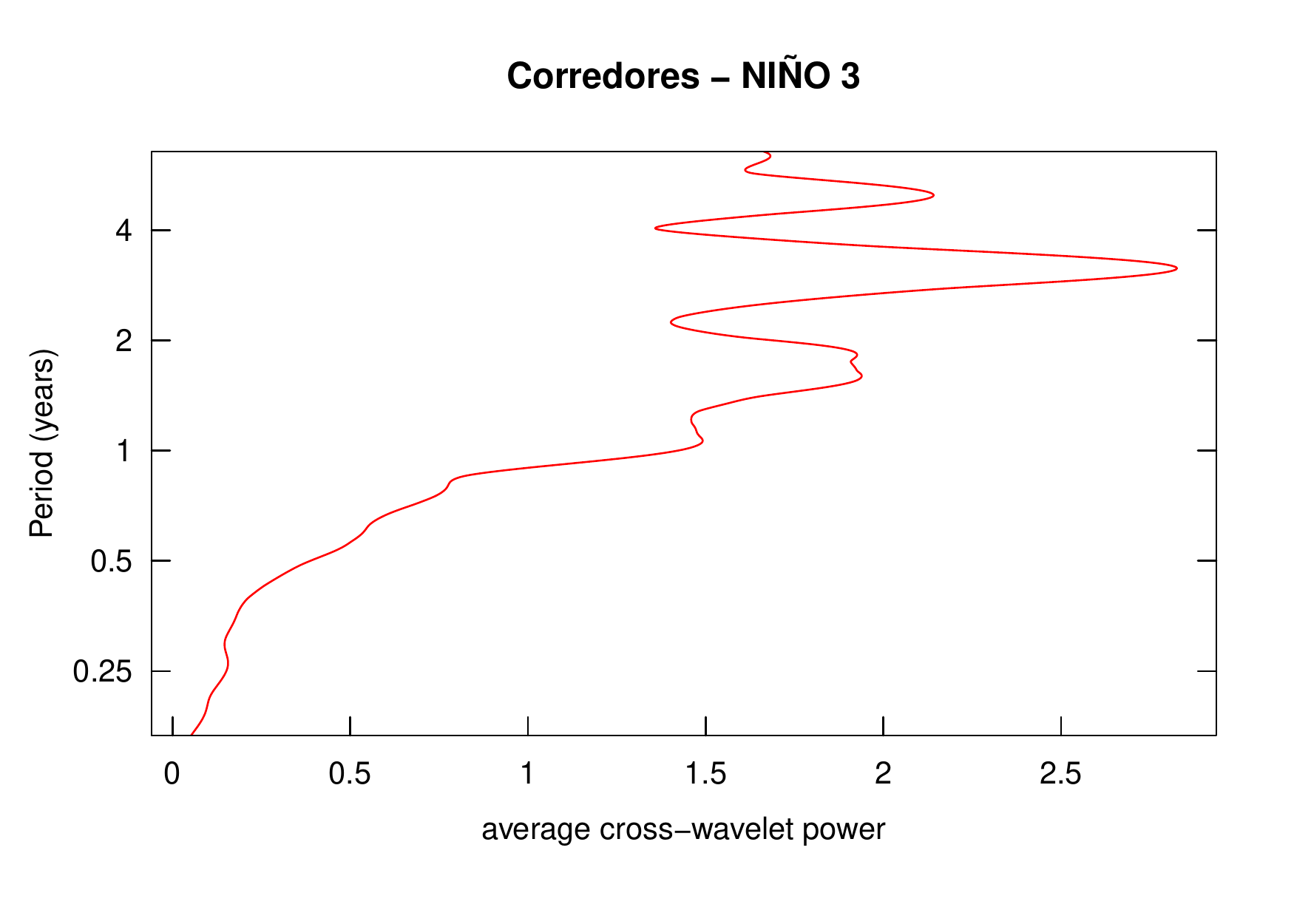}}\vspace{-0.15cm}\\
\subfloat[]{\includegraphics[scale=0.23]{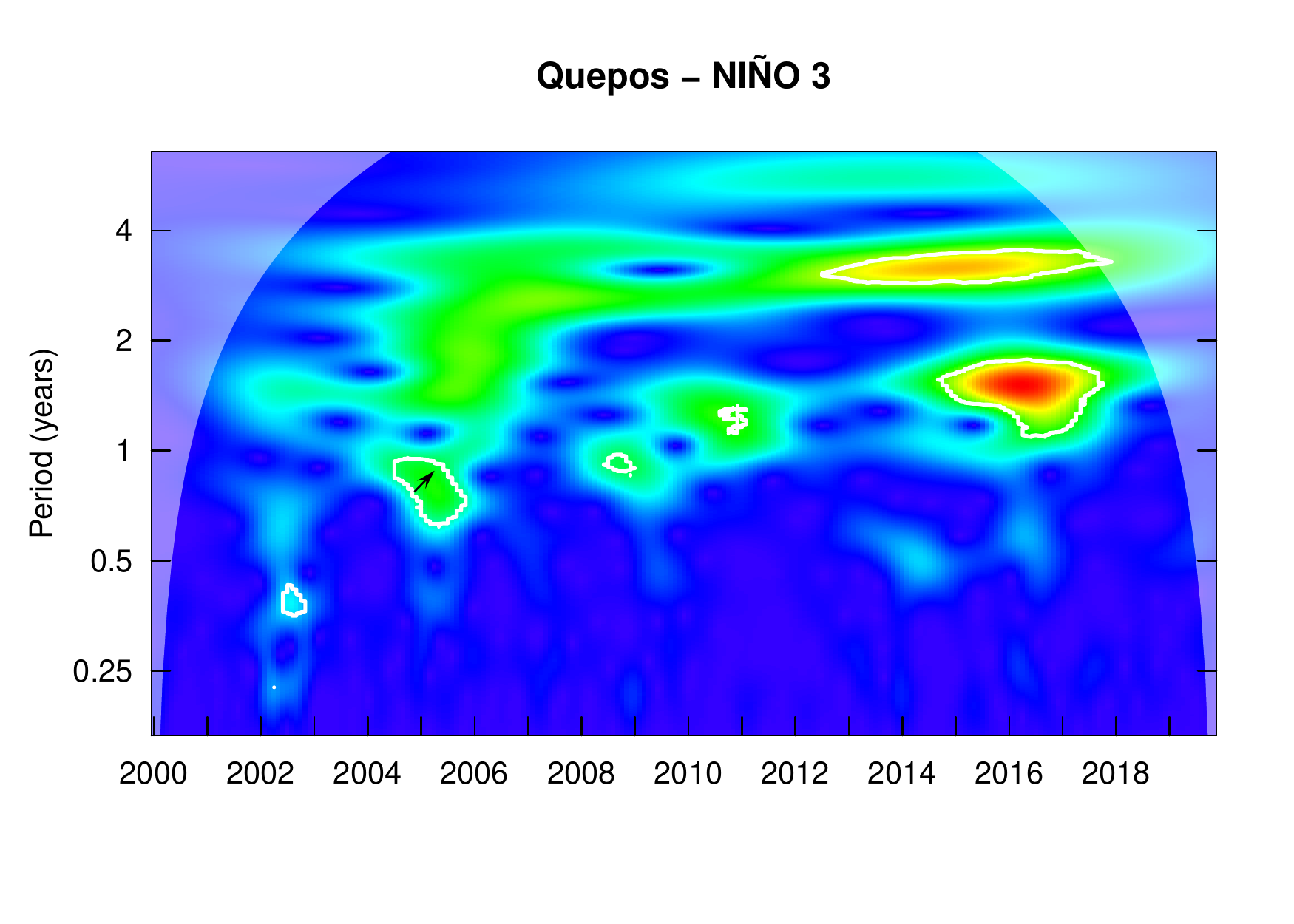}}\vspace{-0.15cm}%
\subfloat[]{\includegraphics[scale=0.23]{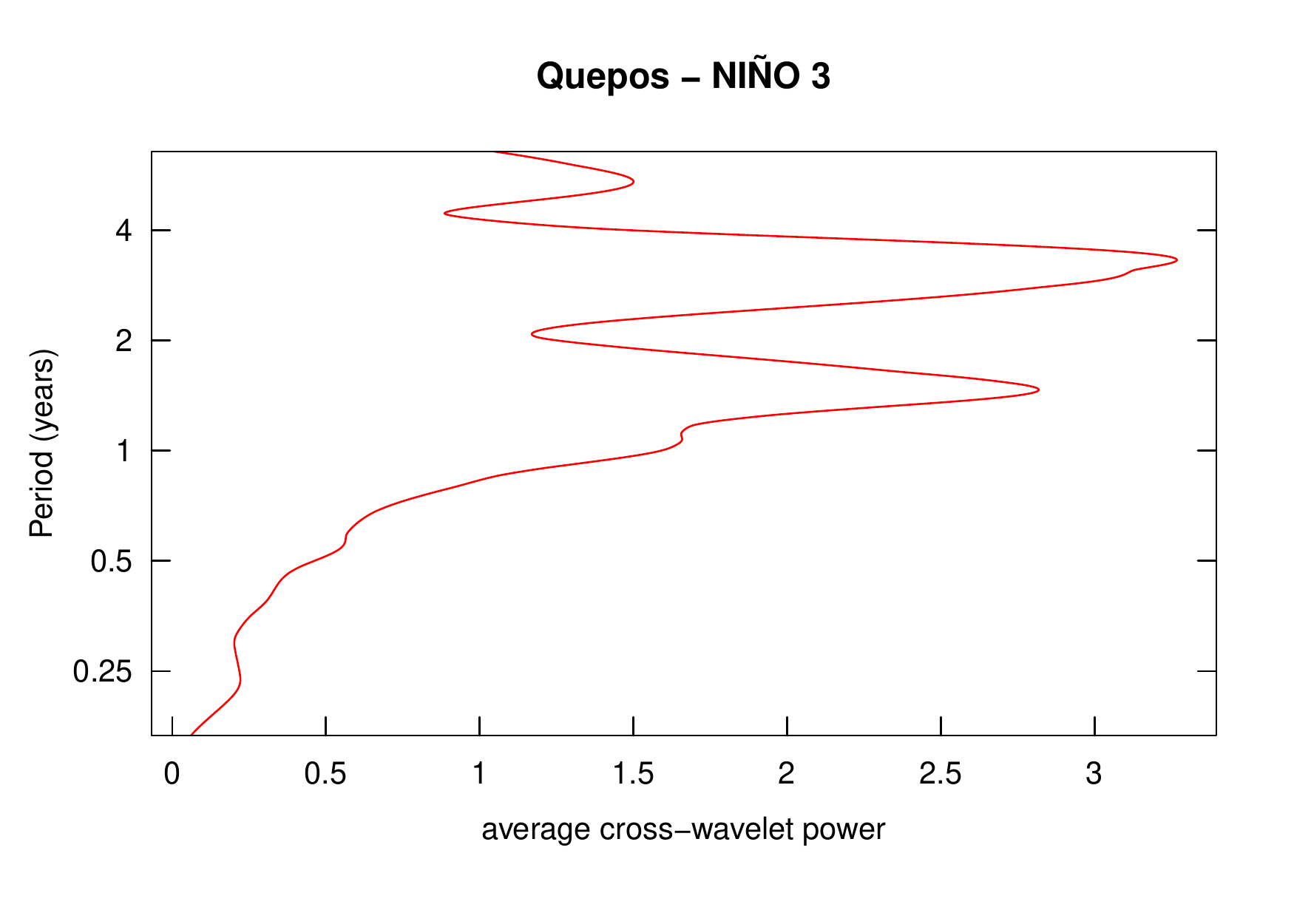}}\vspace{-0.15cm}%
\subfloat[]{\includegraphics[scale=0.23]{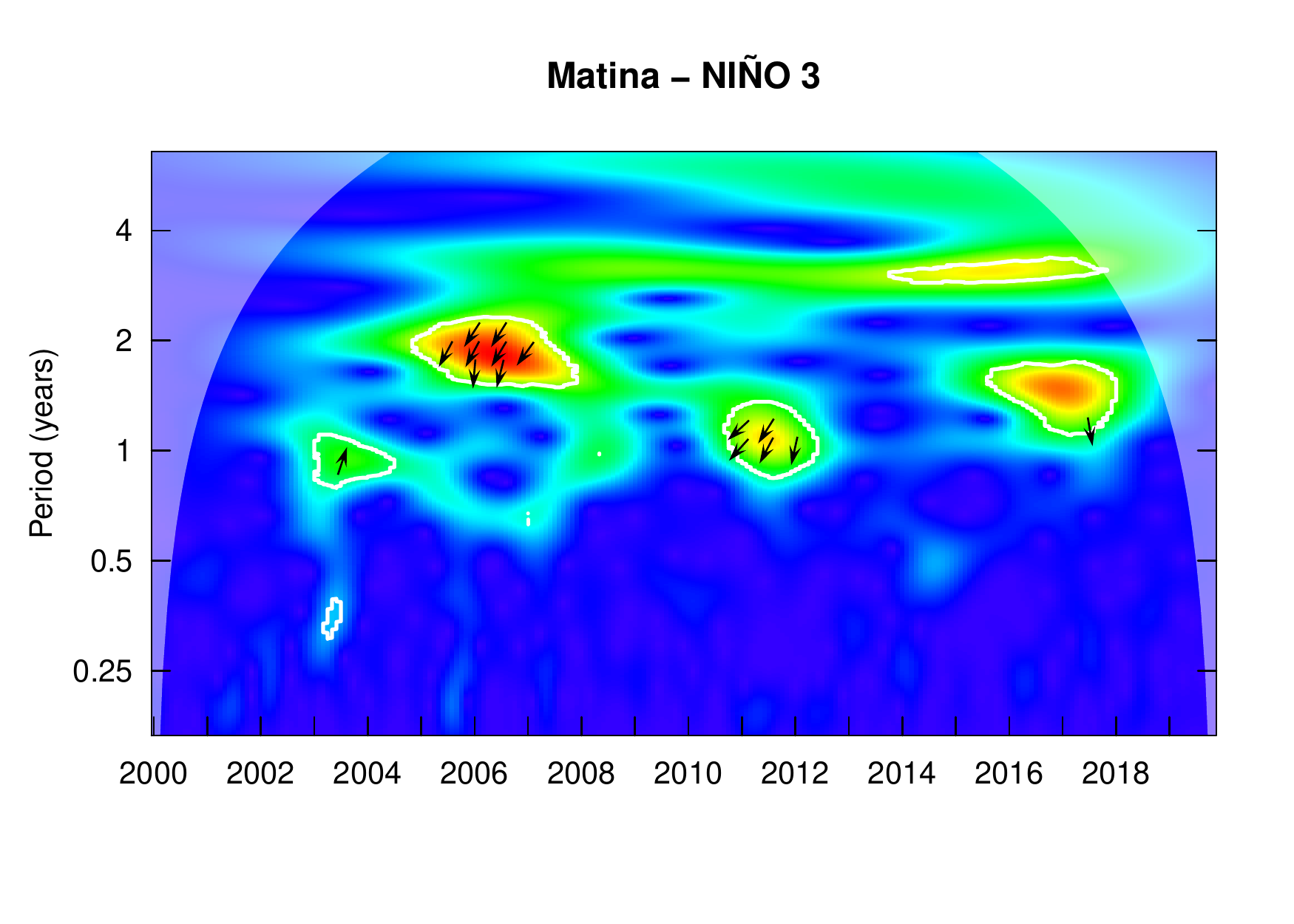}}\vspace{-0.15cm}%
\subfloat[]{\includegraphics[scale=0.23]{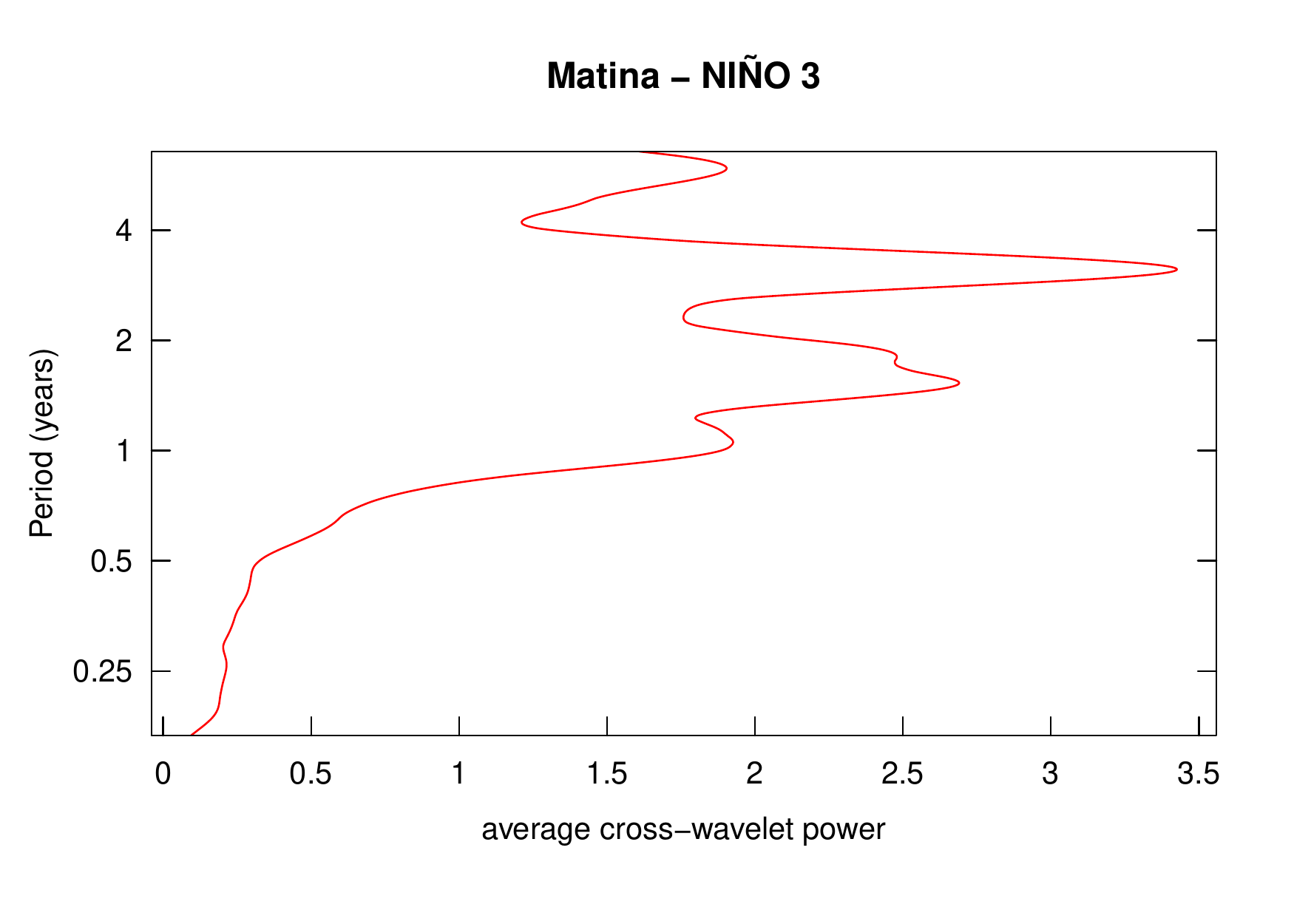}}\vspace{-0.15cm}\\
\subfloat[]{\includegraphics[scale=0.23]{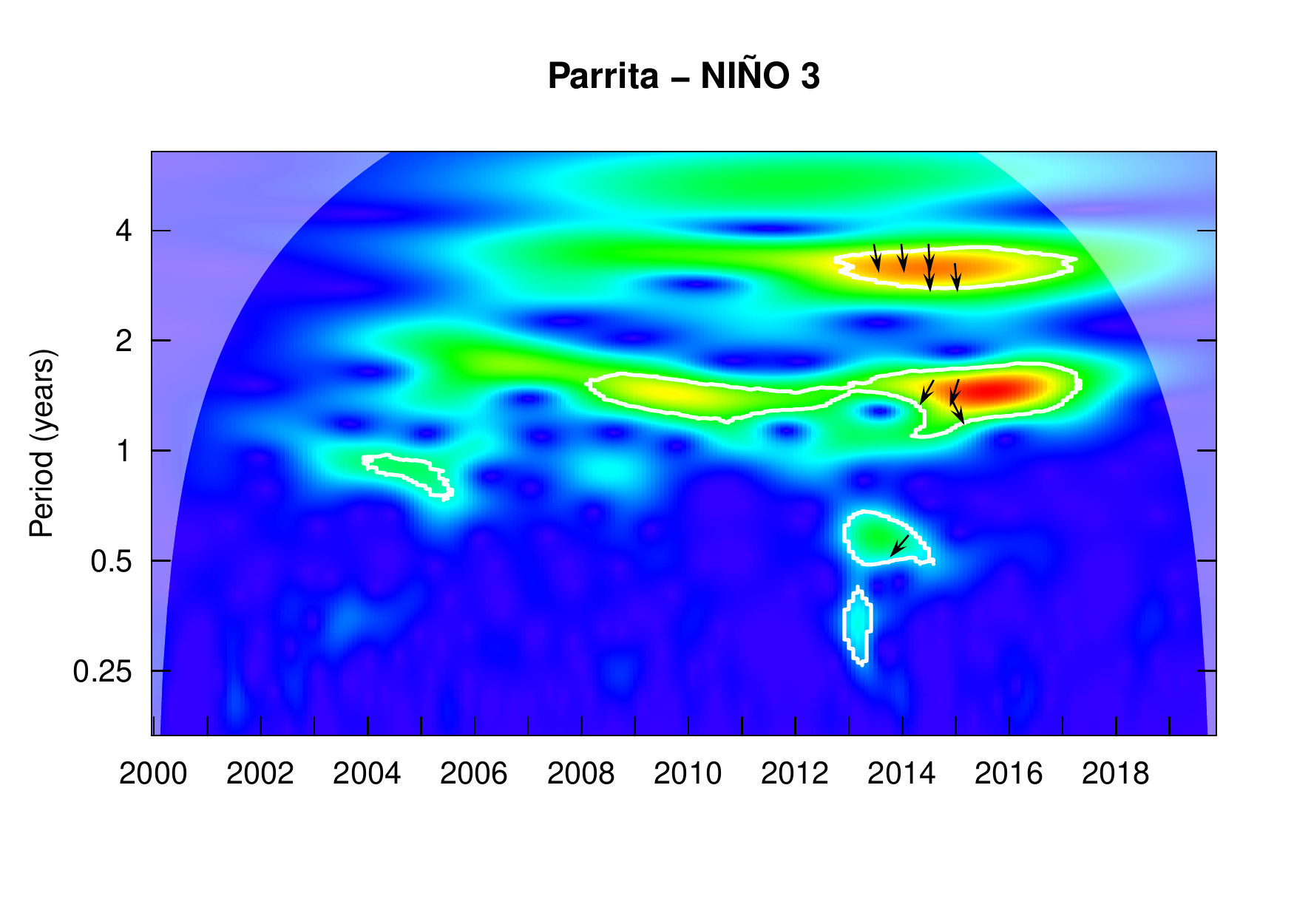}}\vspace{-0.15cm}%
\subfloat[]{\includegraphics[scale=0.23]{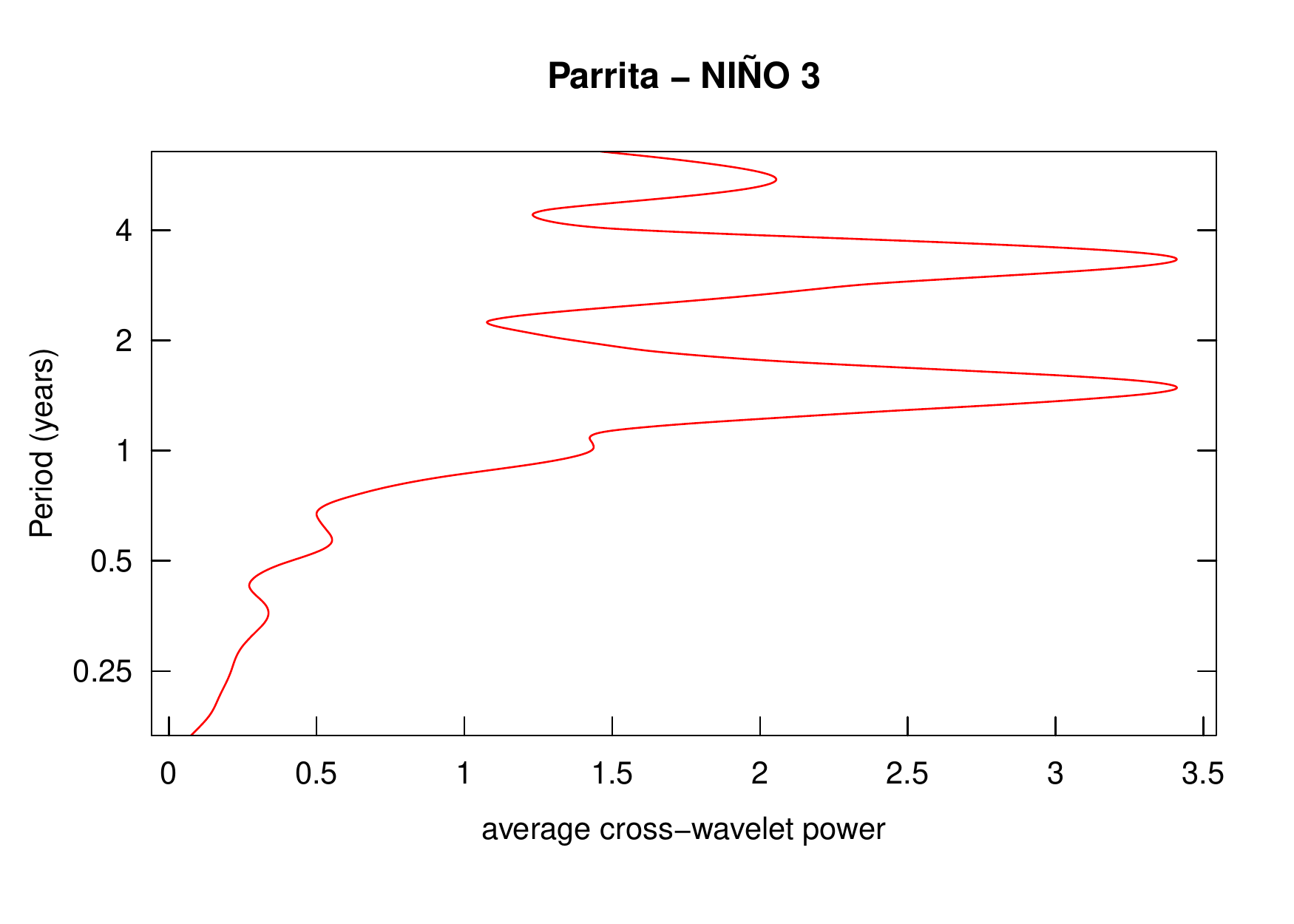}}\vspace{-0.15cm}%
\subfloat[]{\includegraphics[scale=0.23]{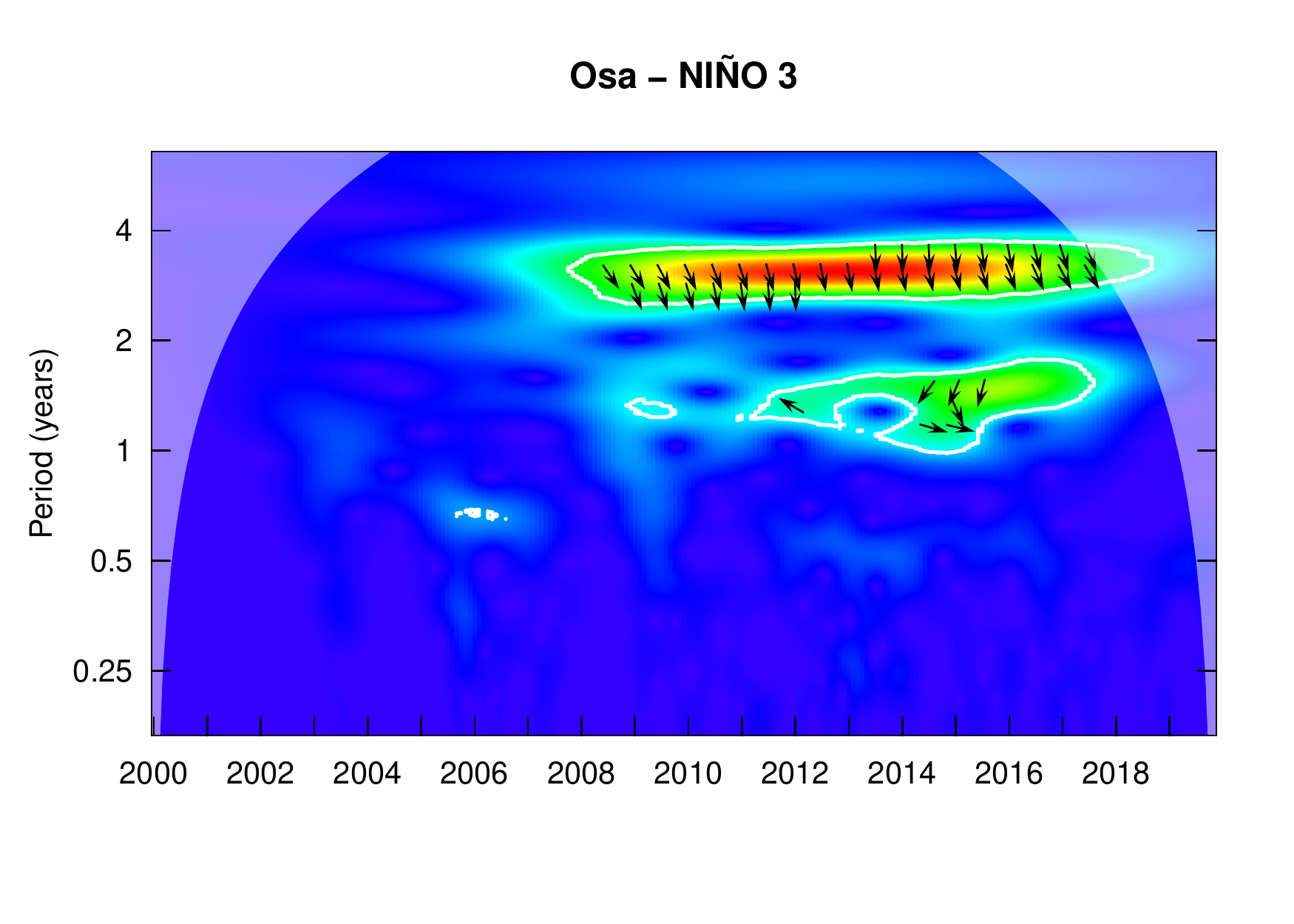}}\vspace{-0.15cm}%
\subfloat[]{\includegraphics[scale=0.23]{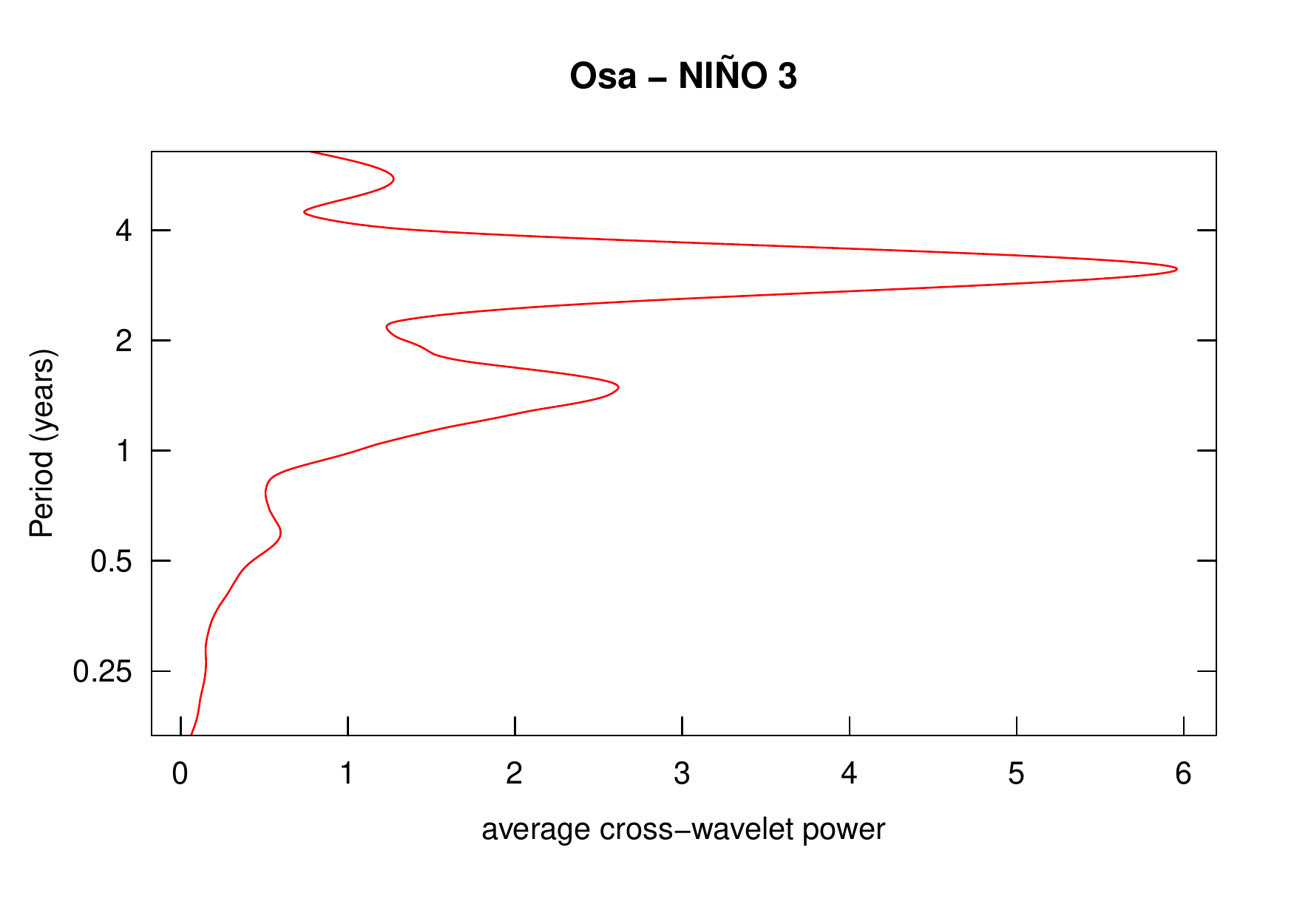}}\vspace{-0.15cm}
\caption*{}
\end{figure}

\section*{Wavelet coherence and average cross-wavelet power between dengue incidence and Ni\~no 3.4}

\begin{figure}[H]
\captionsetup[subfigure]{labelformat=empty}
\caption*{\textbf{Figure S9:} Wavelet coherence (color map) between dengue incidence from 2000 to 2019, and Ni\~no 3.4 in 32 municipalities of Costa Rica (periodicity on y-axis, time on x-axis). Colors code for increasing power intensity, from blue to red; $95\%$ confidence levels are encircled by white lines, and shaded areas indicate the presence of significant edge effects. On the right side of each wavelet coherence is the average cross-wavelet power (Red line). The arrows indicate whether the two series are in-phase or out-phase.}
\subfloat[]{\includegraphics[scale=0.23]{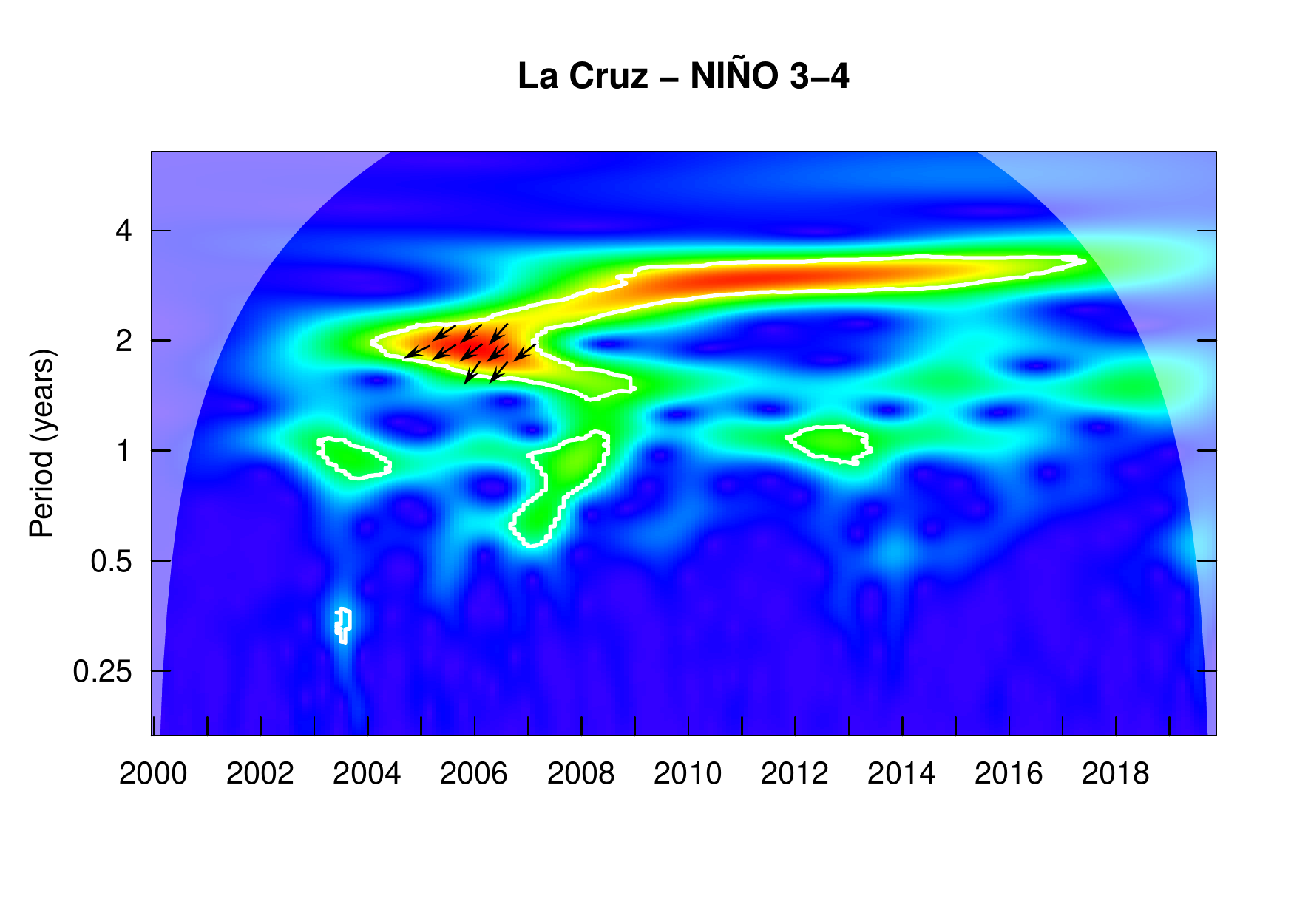}}\vspace{-0.15cm}%
\subfloat[]{\includegraphics[scale=0.23]{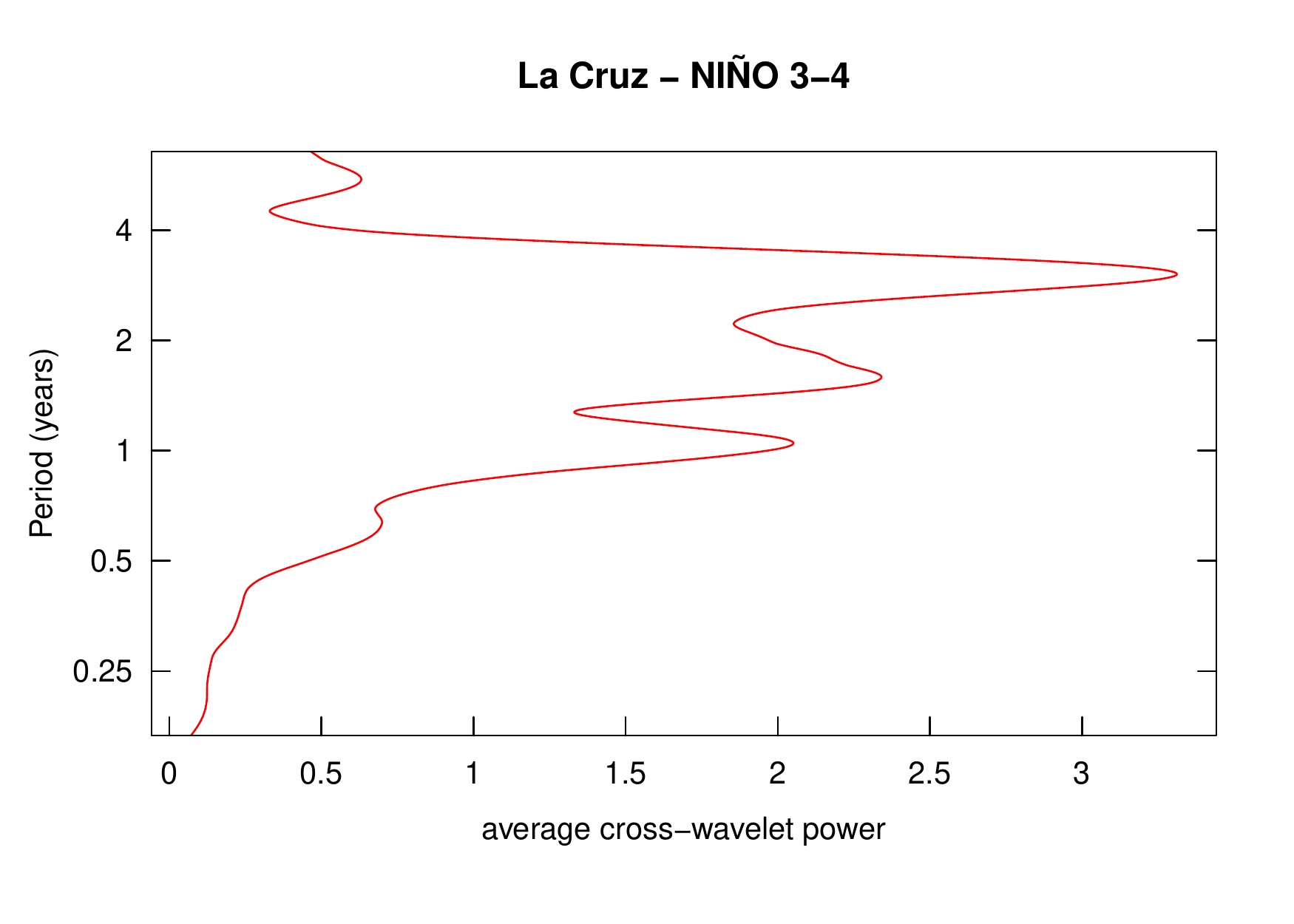}}\vspace{-0.15cm}%
\subfloat[]{\includegraphics[scale=0.23]{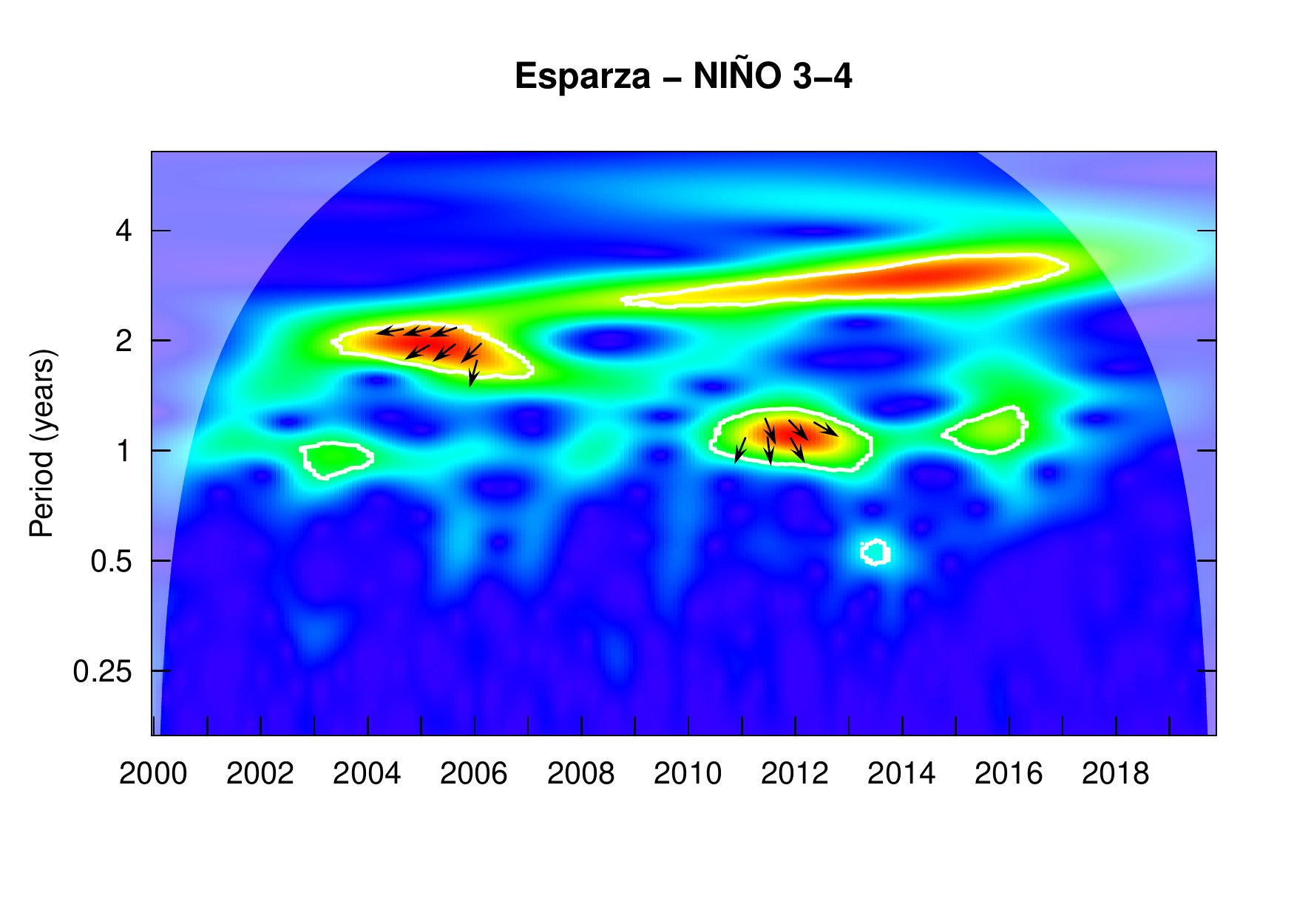}}\vspace{-0.15cm}%
\subfloat[]{\includegraphics[scale=0.23]{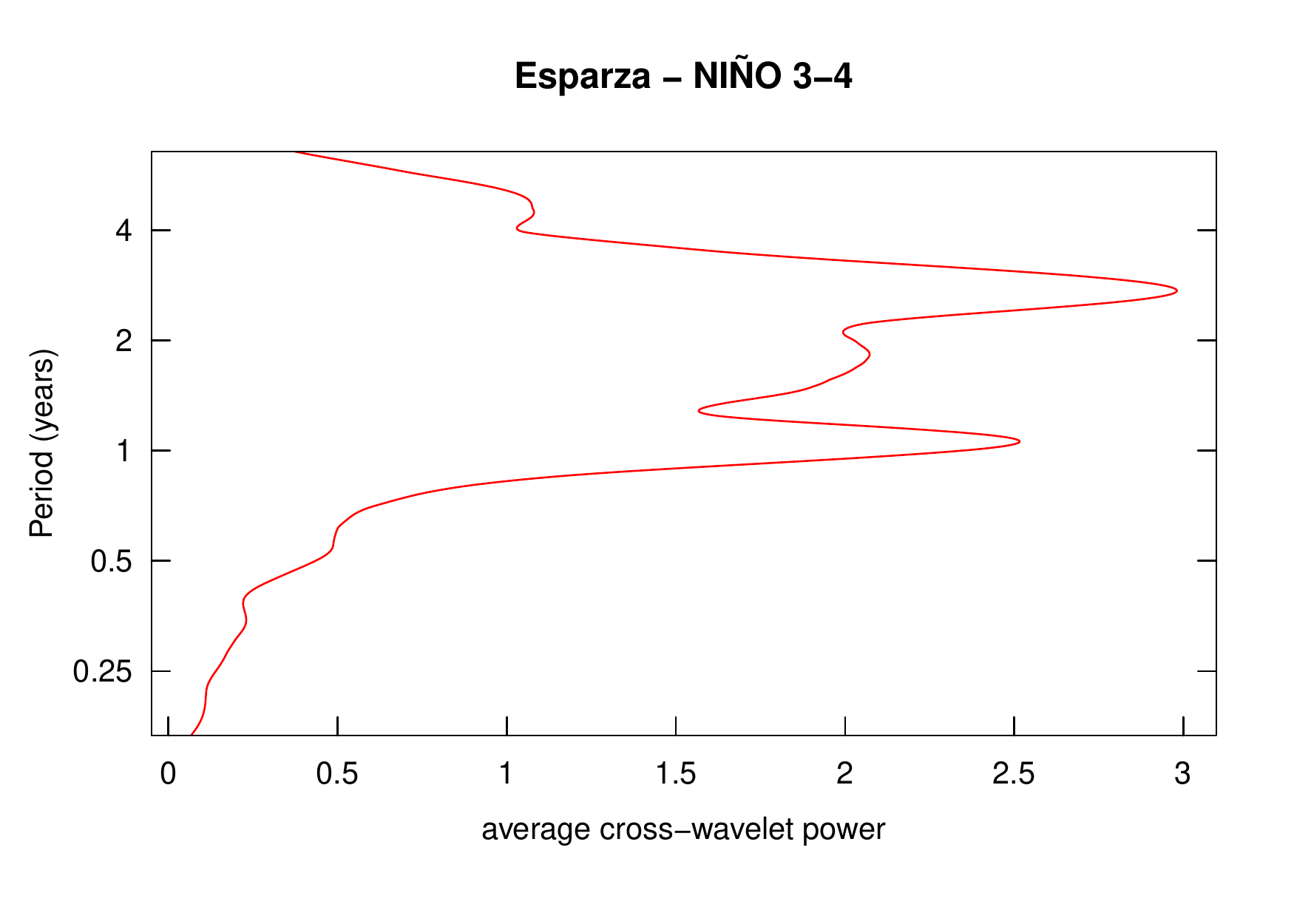}}\vspace{-0.15cm}\\
\subfloat[]{\includegraphics[scale=0.23]{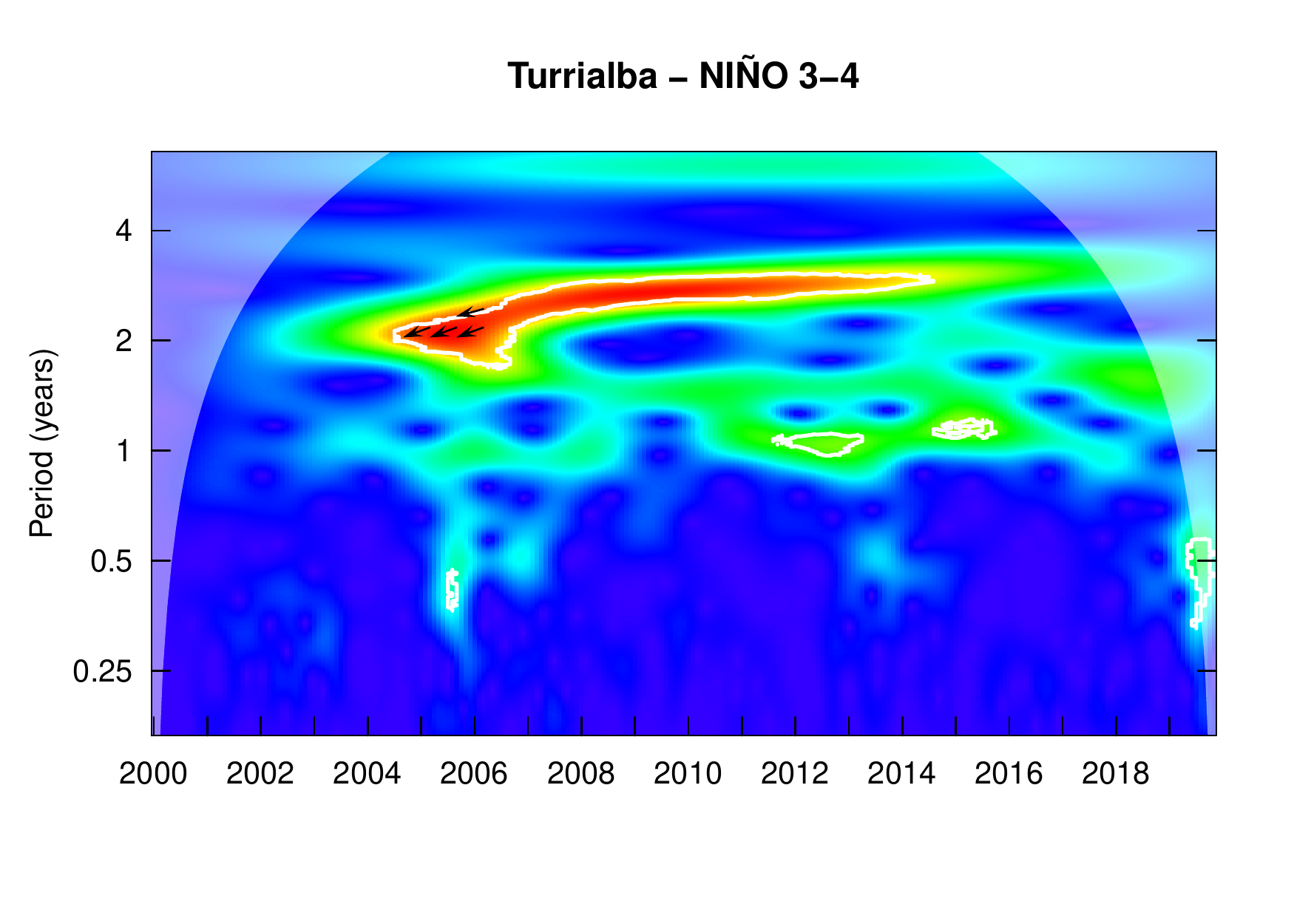}}\vspace{-0.15cm}%
\subfloat[]{\includegraphics[scale=0.23]{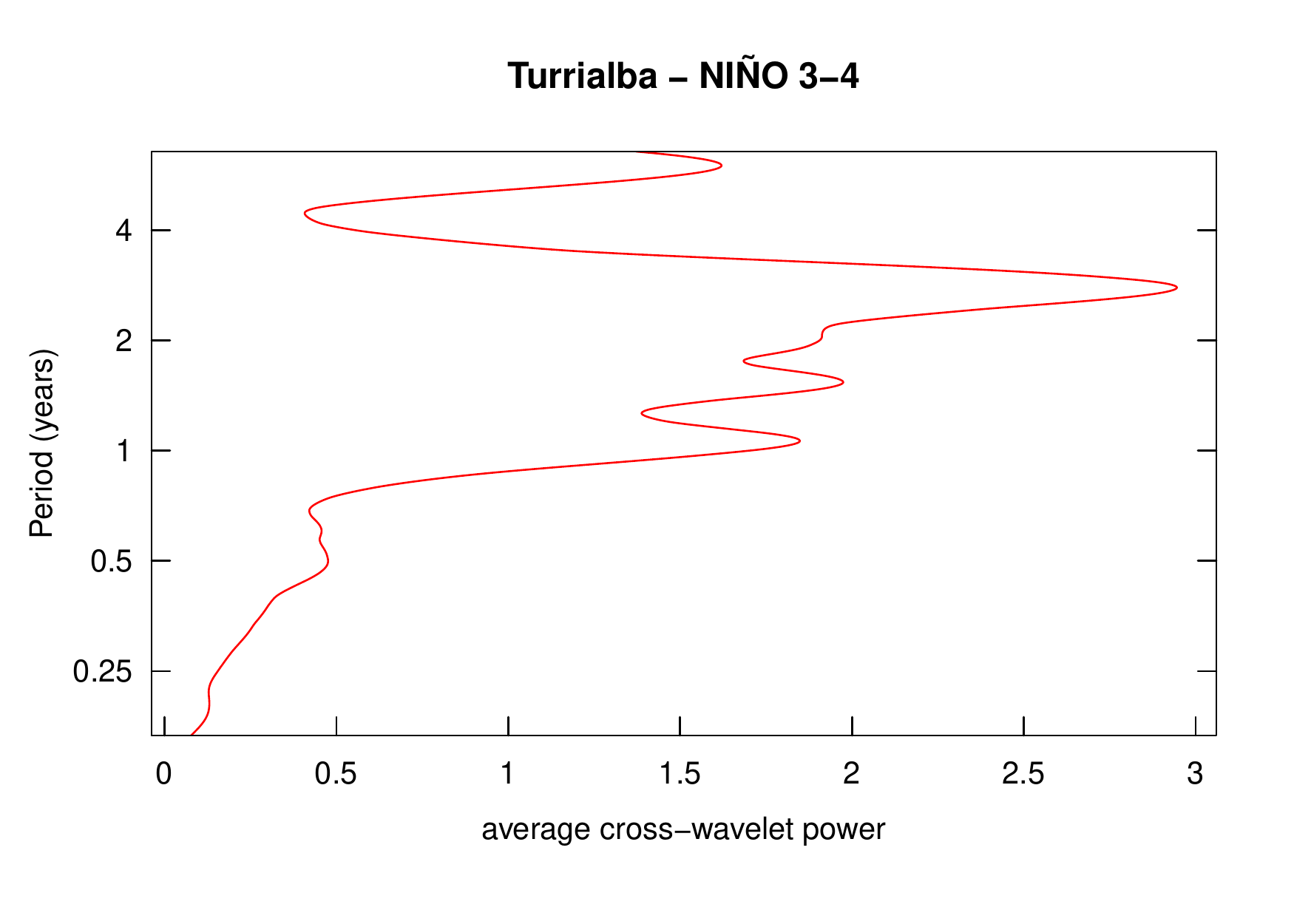}}\vspace{-0.15cm}%
\subfloat[]{\includegraphics[scale=0.23]{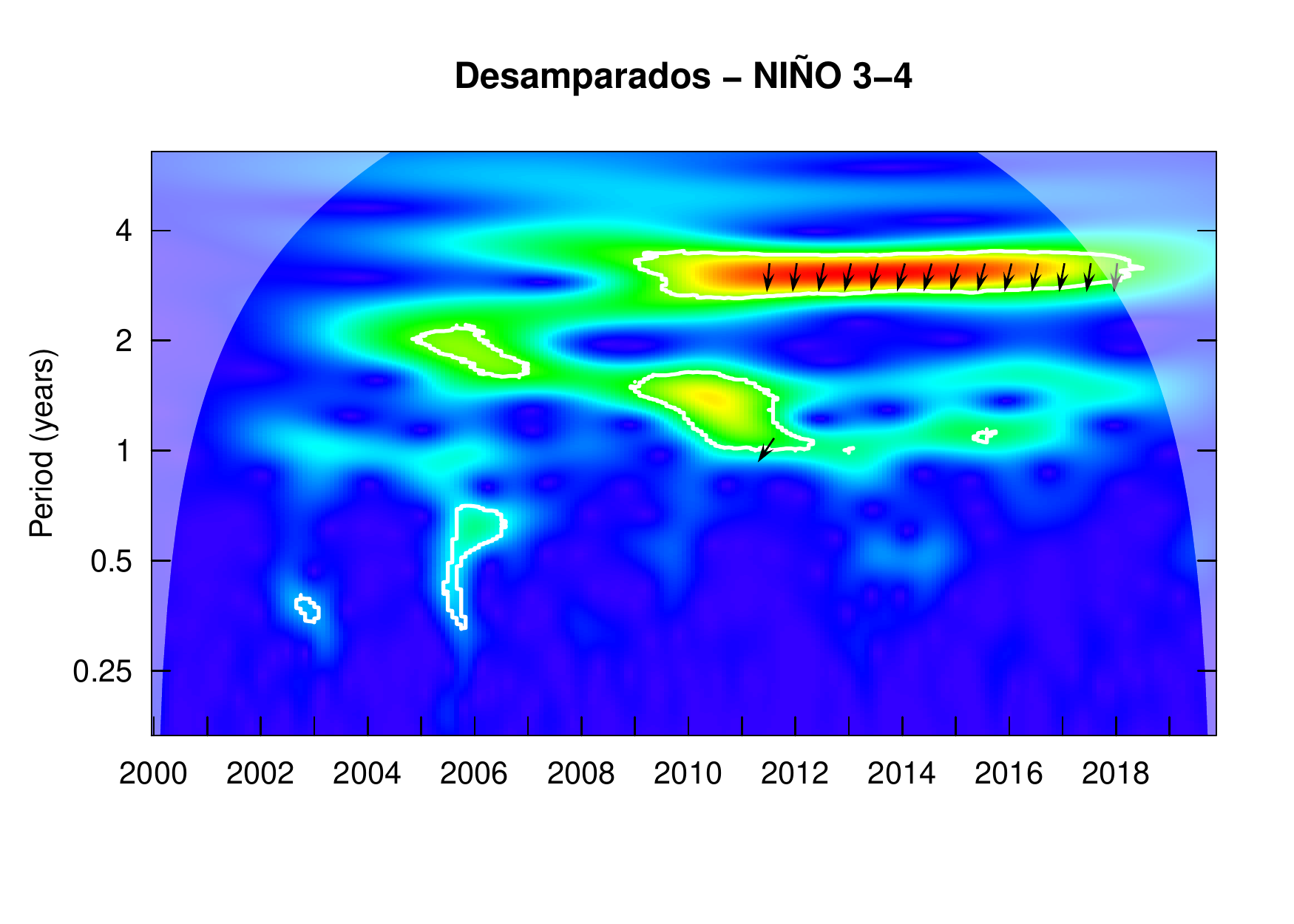}}\vspace{-0.15cm}%
\subfloat[]{\includegraphics[scale=0.23]{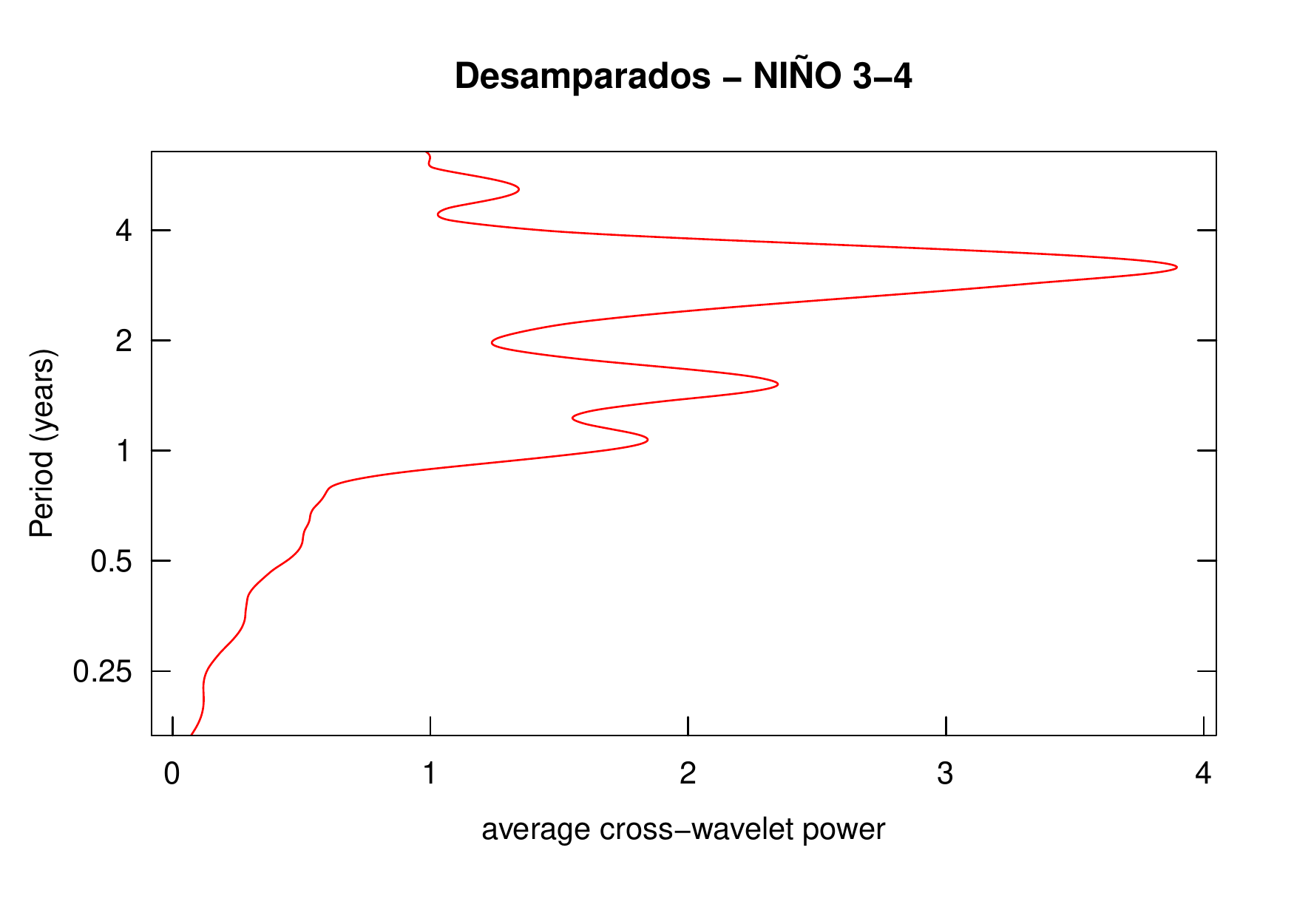}}\vspace{-0.15cm}\\
\subfloat[]{\includegraphics[scale=0.23]{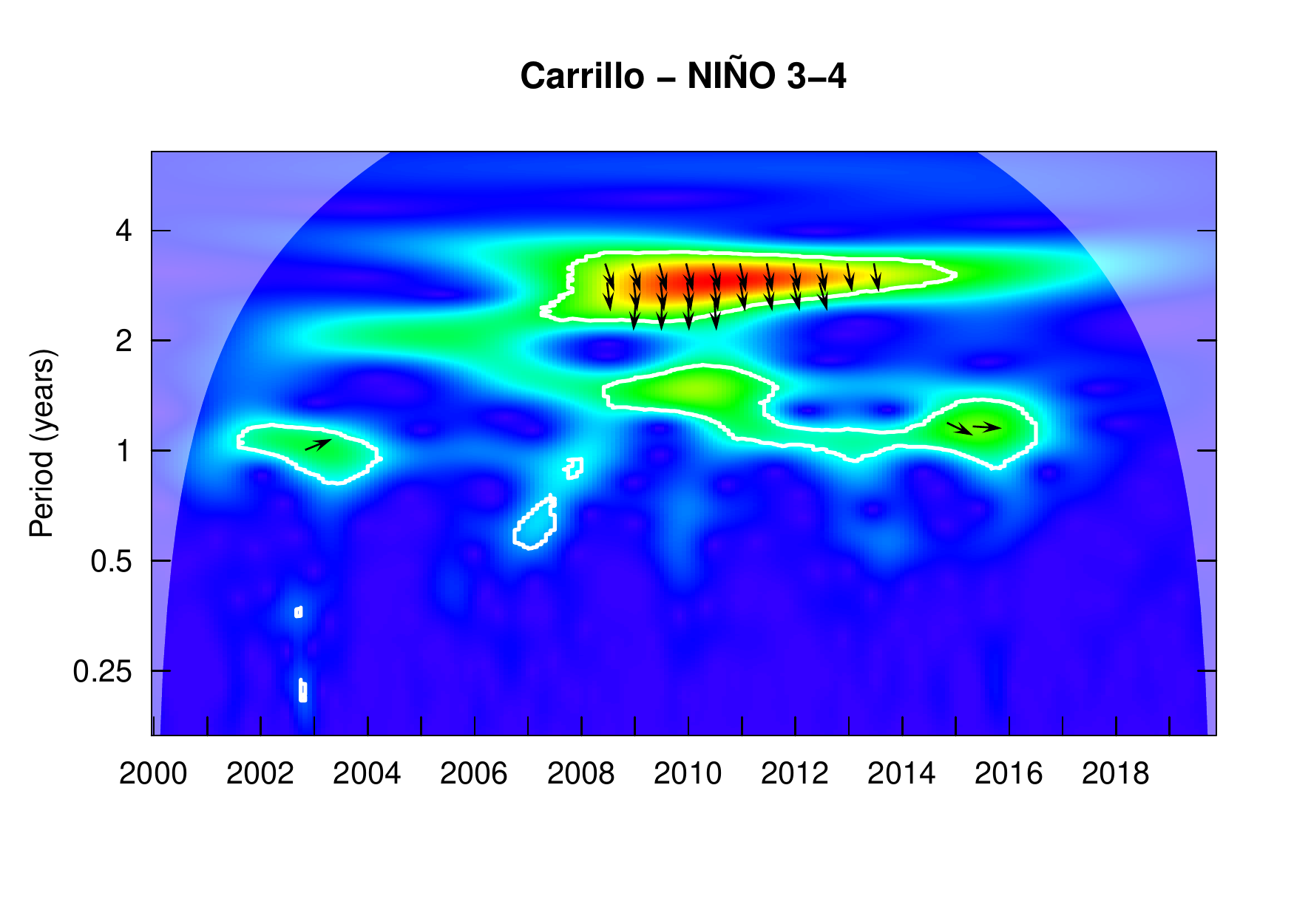}}\vspace{-0.15cm}%
\subfloat[]{\includegraphics[scale=0.23]{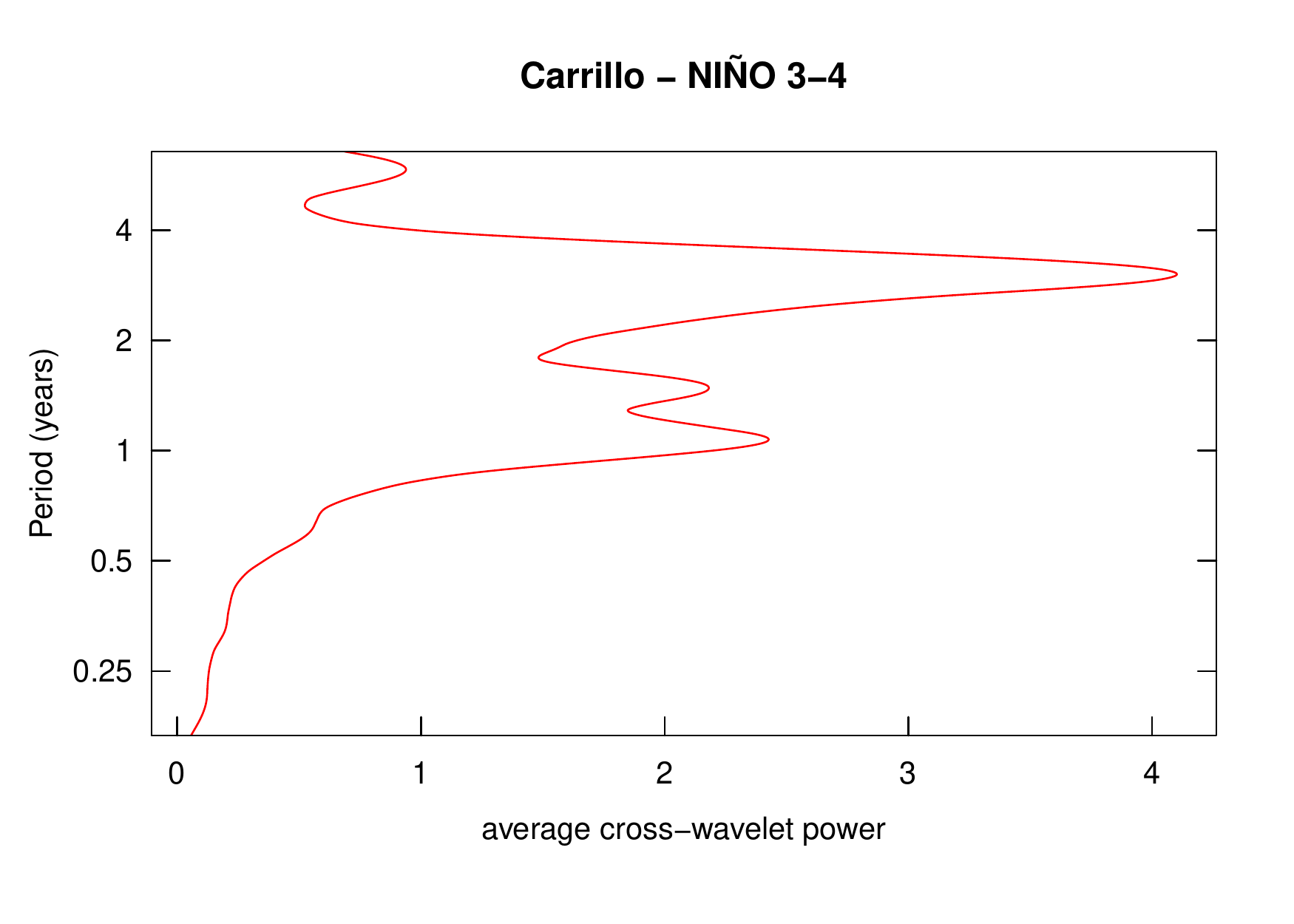}}\vspace{-0.15cm}%
\subfloat[]{\includegraphics[scale=0.23]{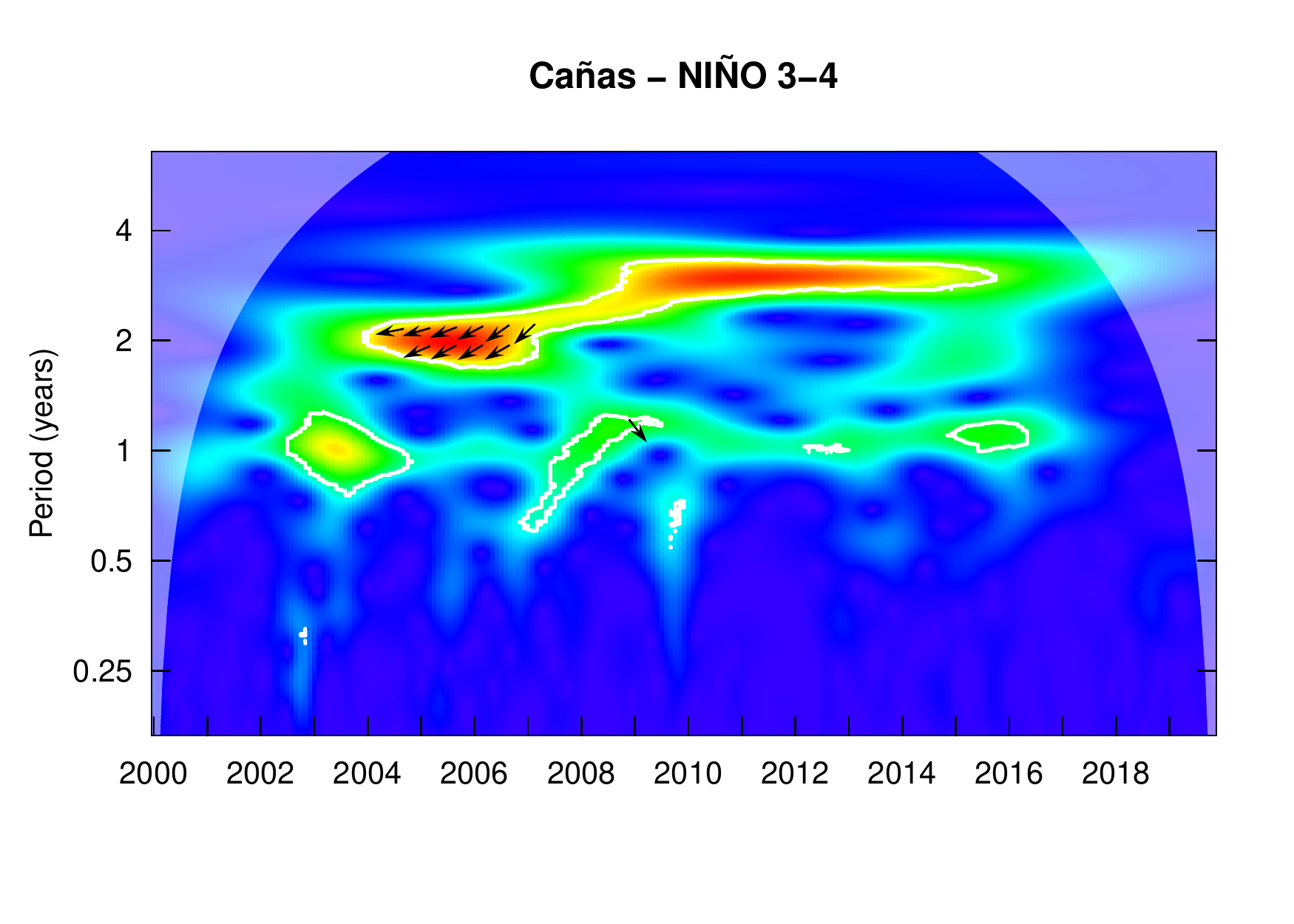}}\vspace{-0.15cm}%
\subfloat[]{\includegraphics[scale=0.23]{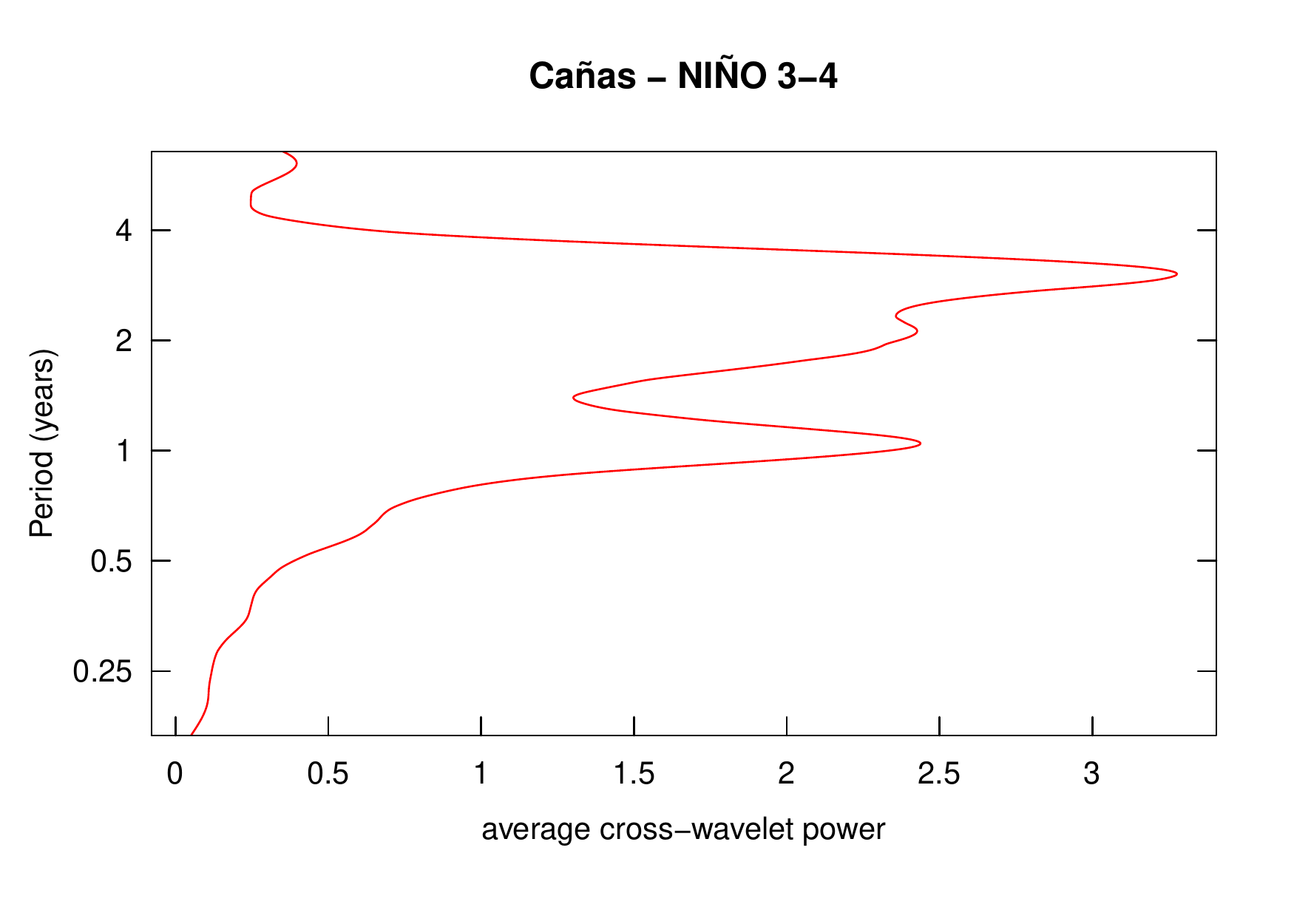}}\vspace{-0.15cm}\\
\subfloat[]{\includegraphics[scale=0.23]{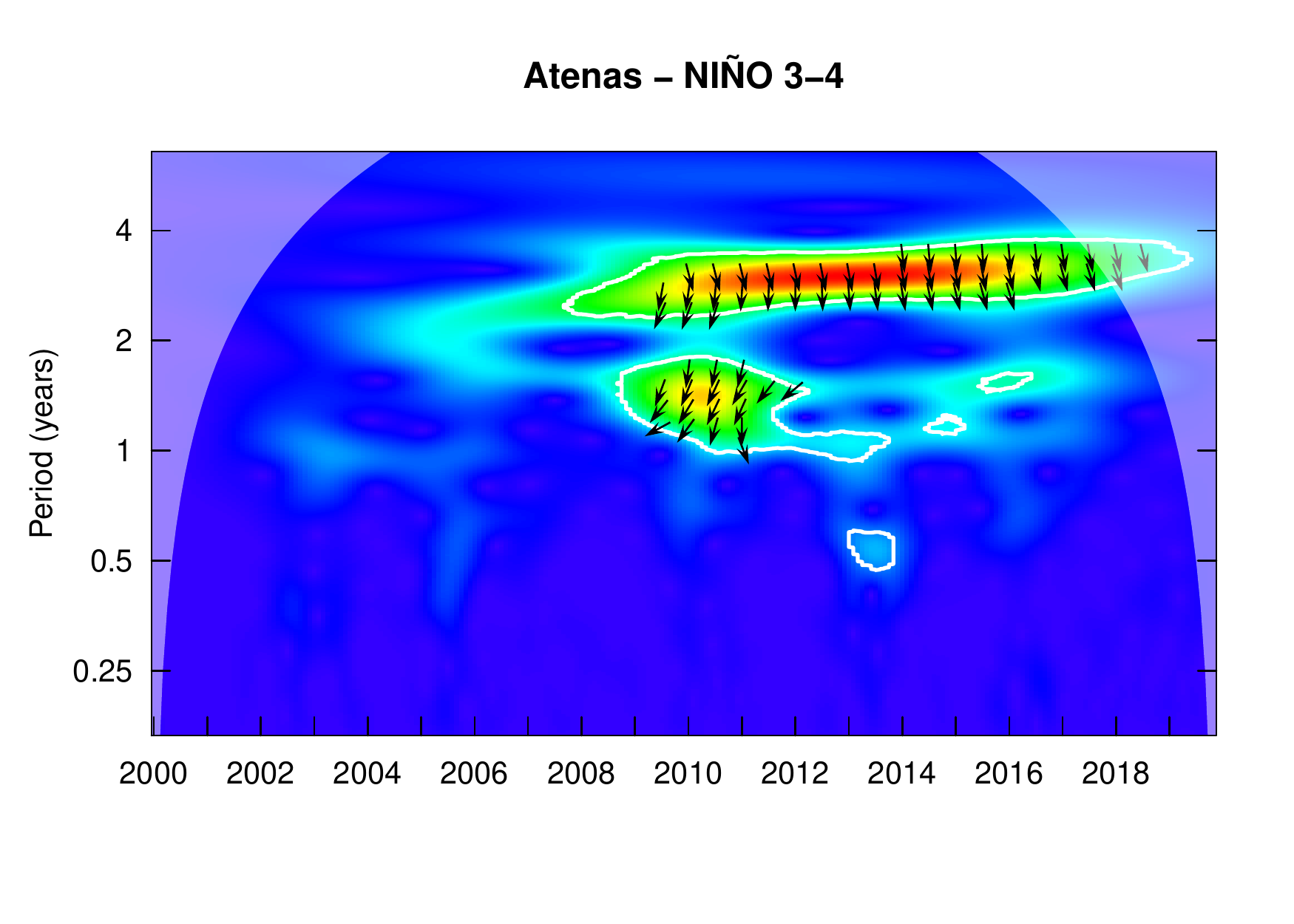}}\vspace{-0.15cm}%
\subfloat[]{\includegraphics[scale=0.23]{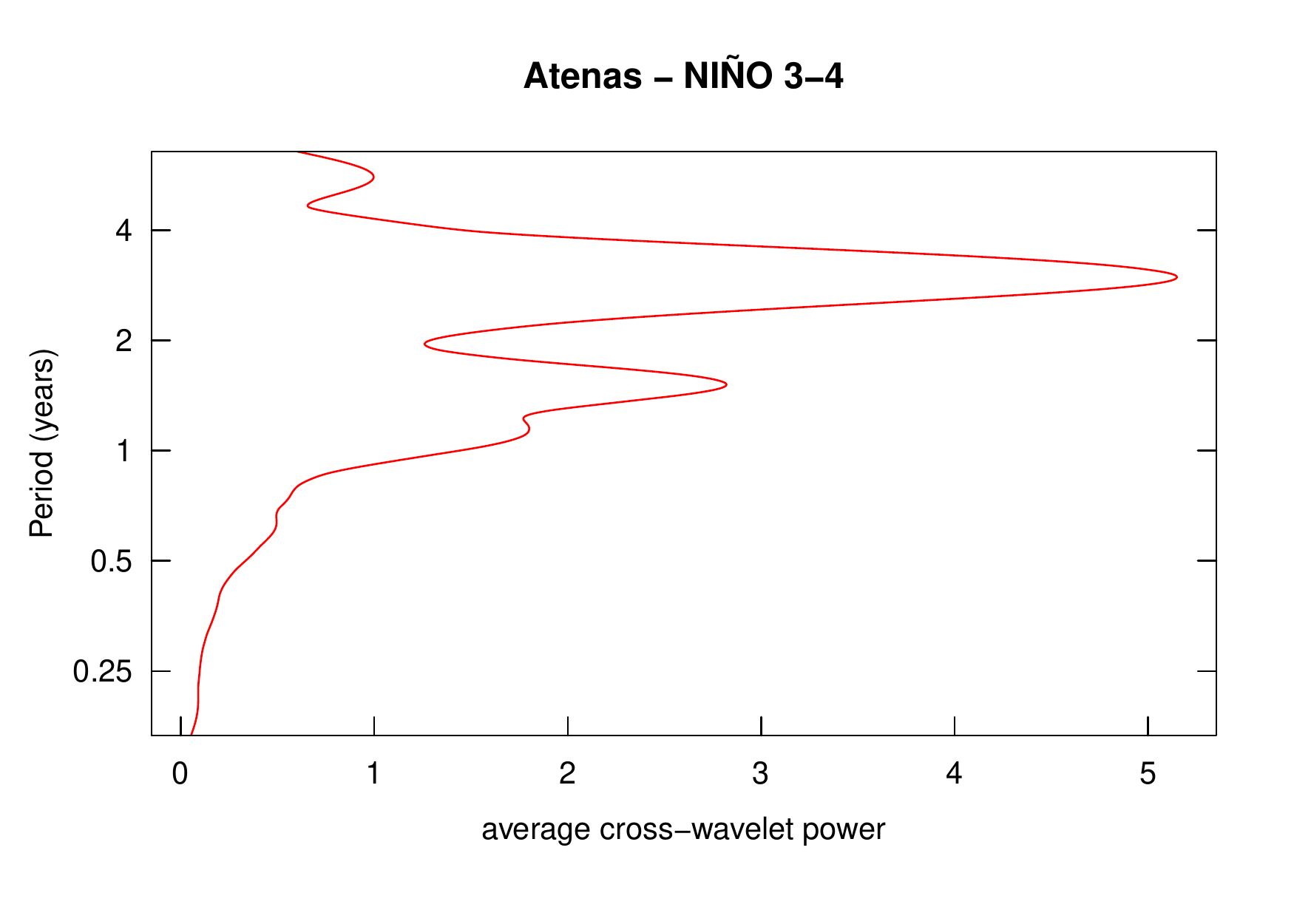}}\vspace{-0.15cm}%
\subfloat[]{\includegraphics[scale=0.23]{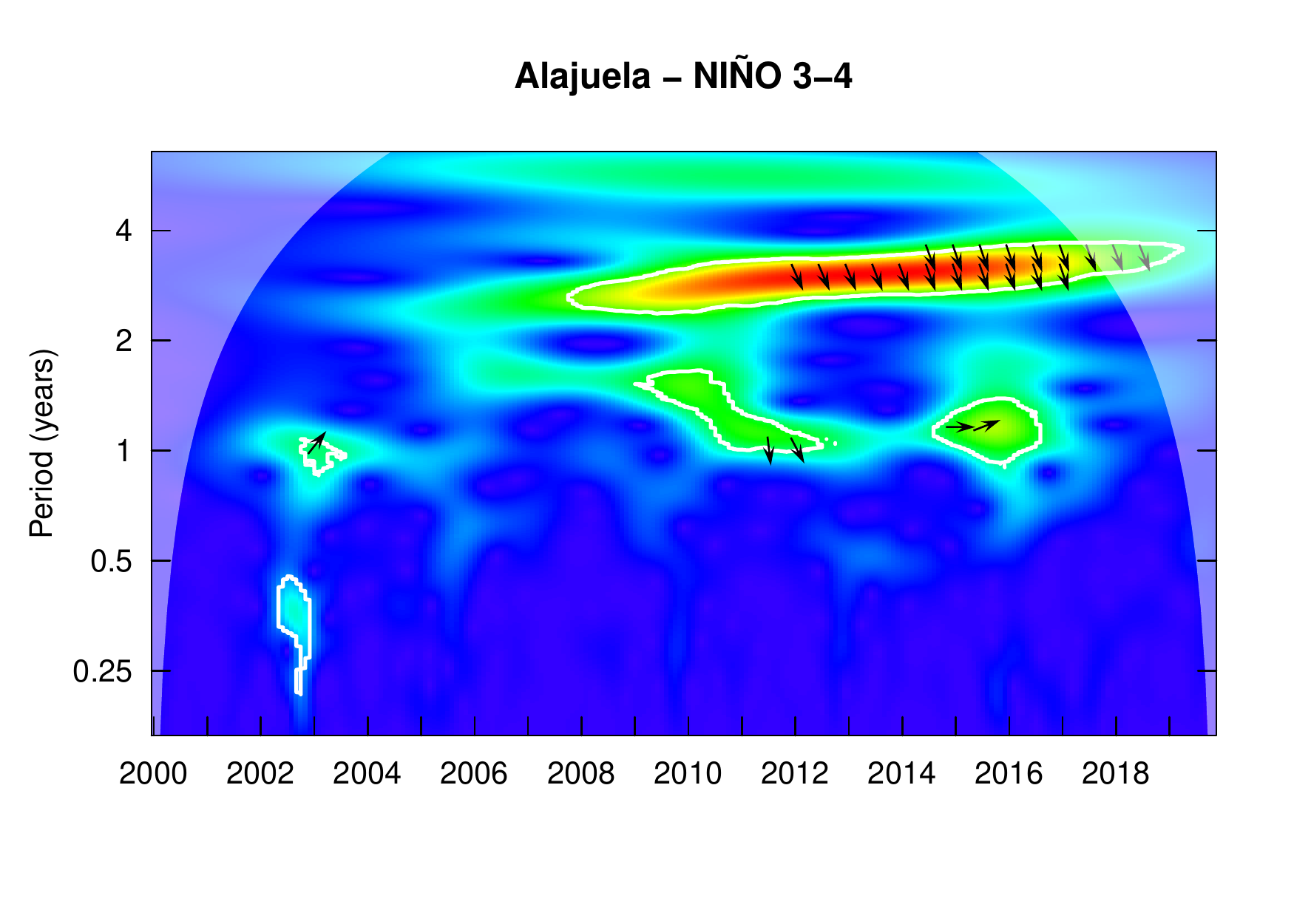}}\vspace{-0.15cm}%
\subfloat[]{\includegraphics[scale=0.23]{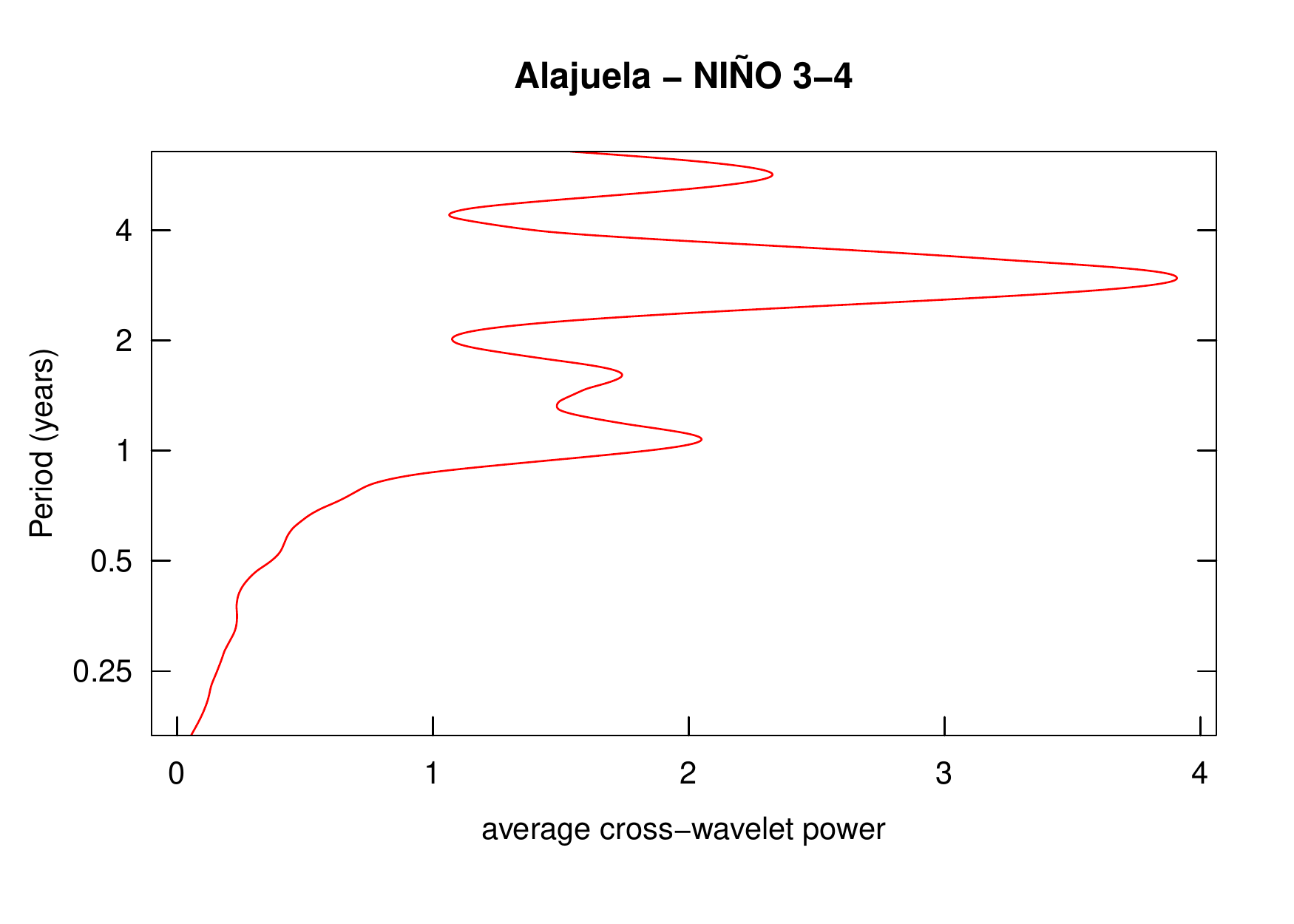}}\vspace{-0.15cm}\\
\subfloat[]{\includegraphics[scale=0.23]{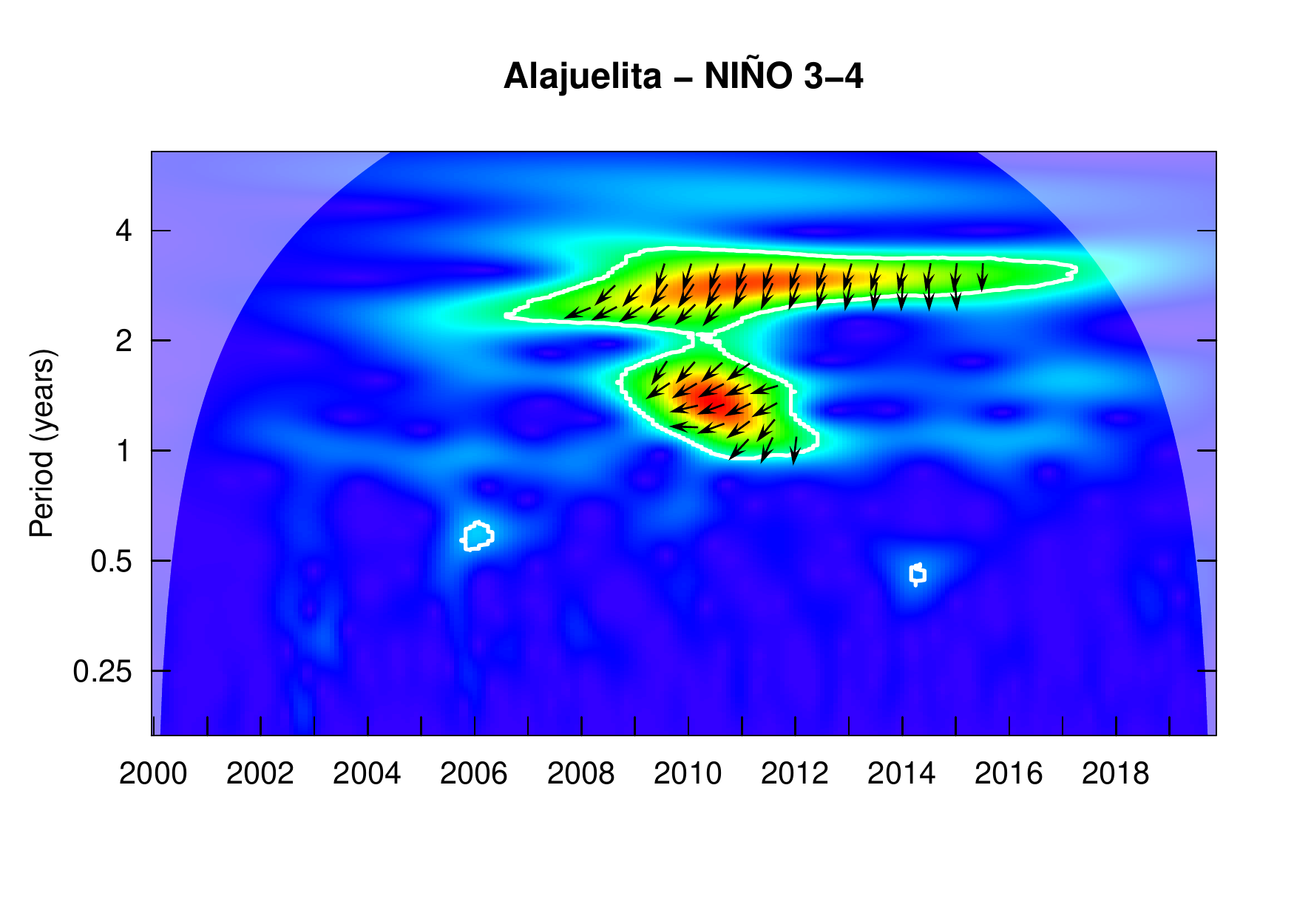}}\vspace{-0.15cm}%
\subfloat[]{\includegraphics[scale=0.23]{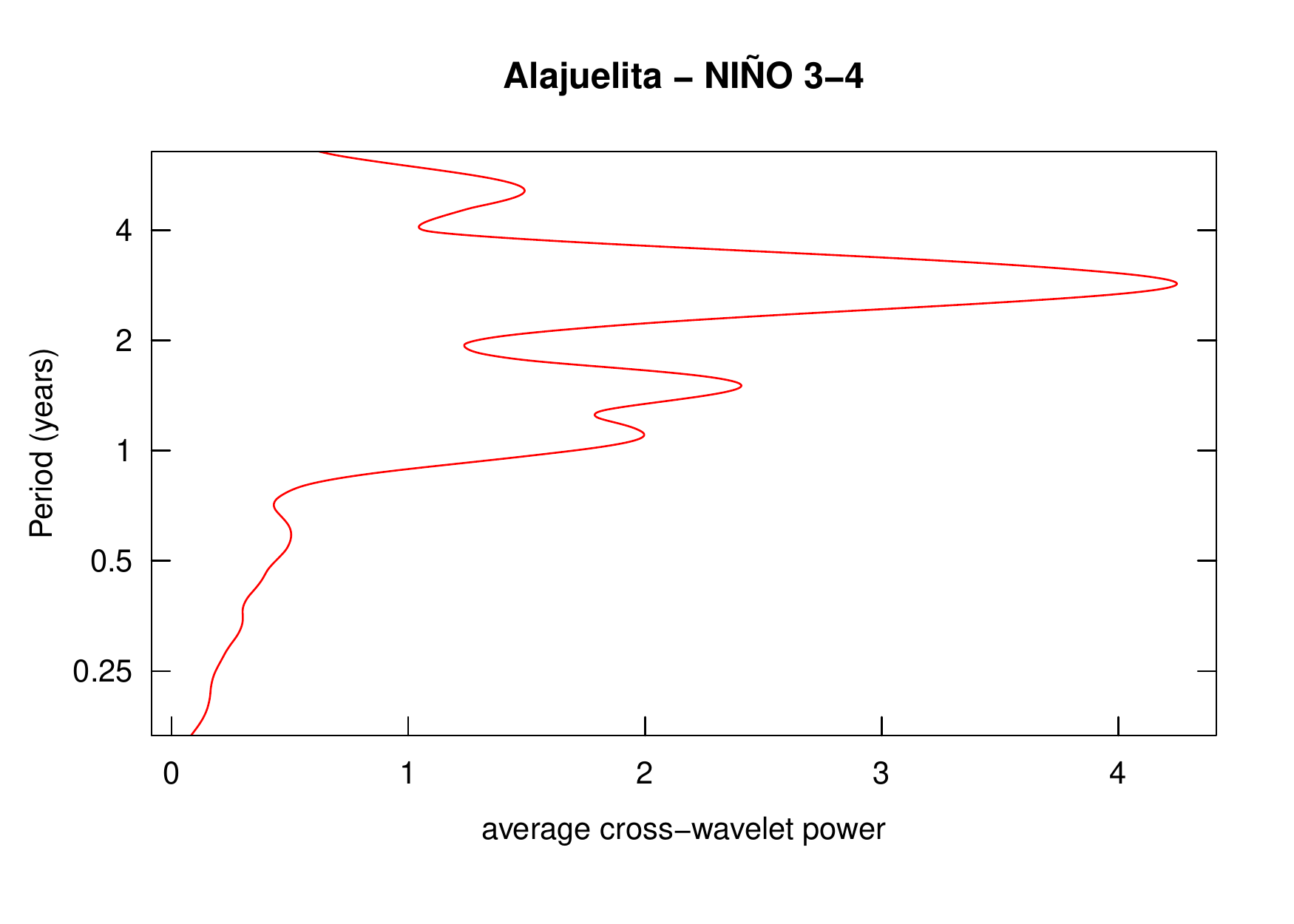}}\vspace{-0.15cm}%
\subfloat[]{\includegraphics[scale=0.23]{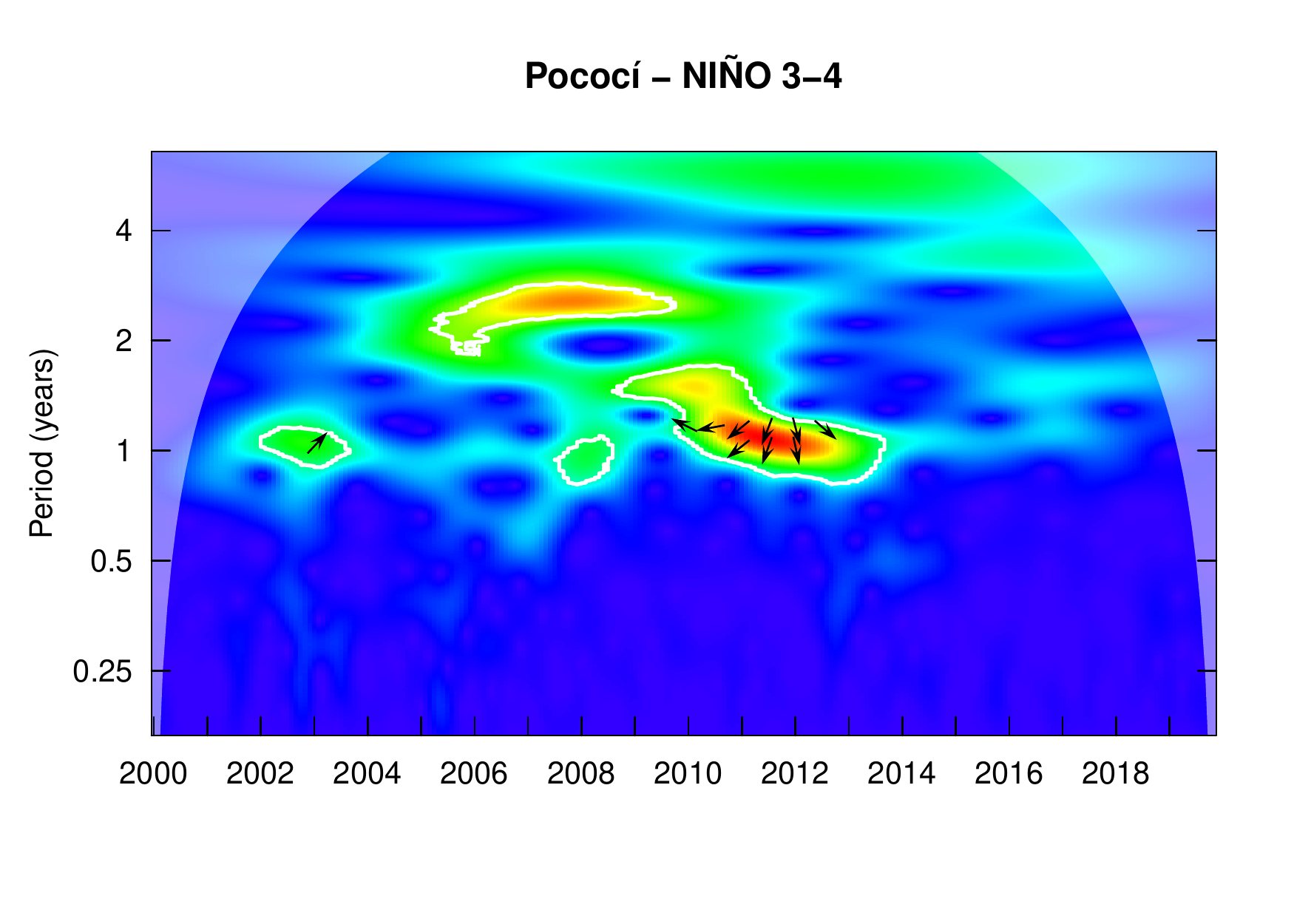}}\vspace{-0.15cm}%
\subfloat[]{\includegraphics[scale=0.23]{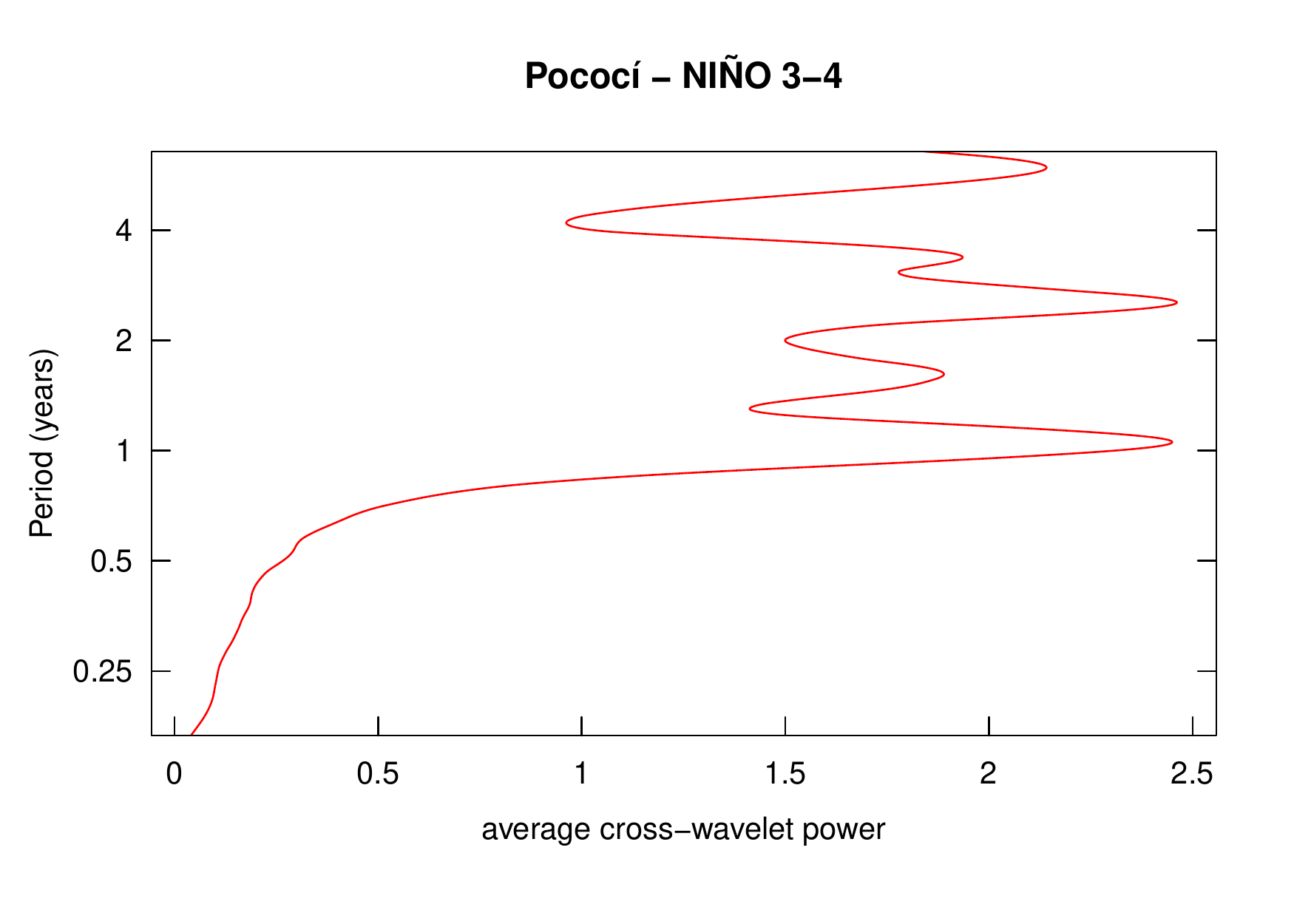}}\vspace{-0.15cm}\\
\subfloat[]{\includegraphics[scale=0.23]{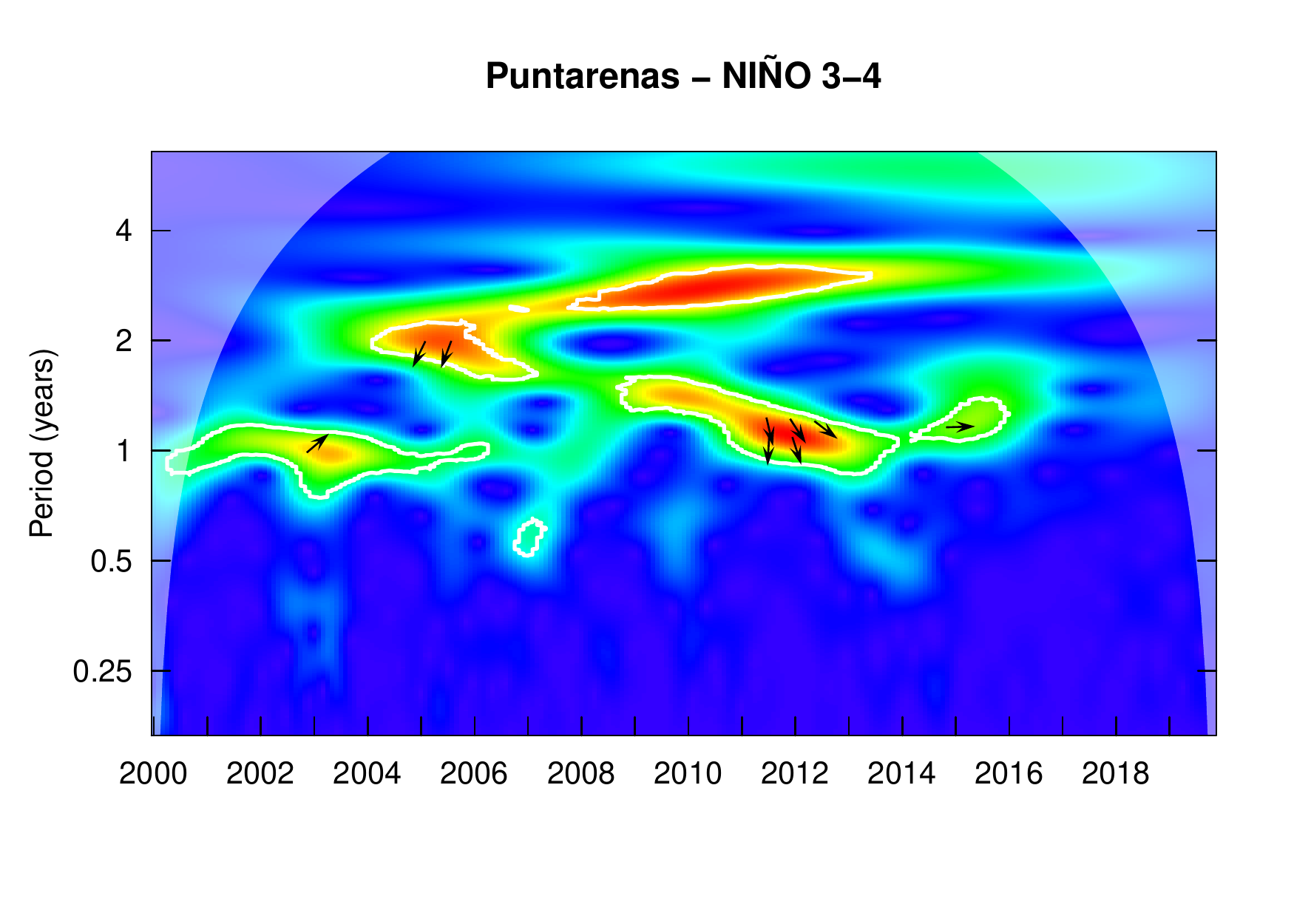}}\vspace{-0.15cm}%
\subfloat[]{\includegraphics[scale=0.23]{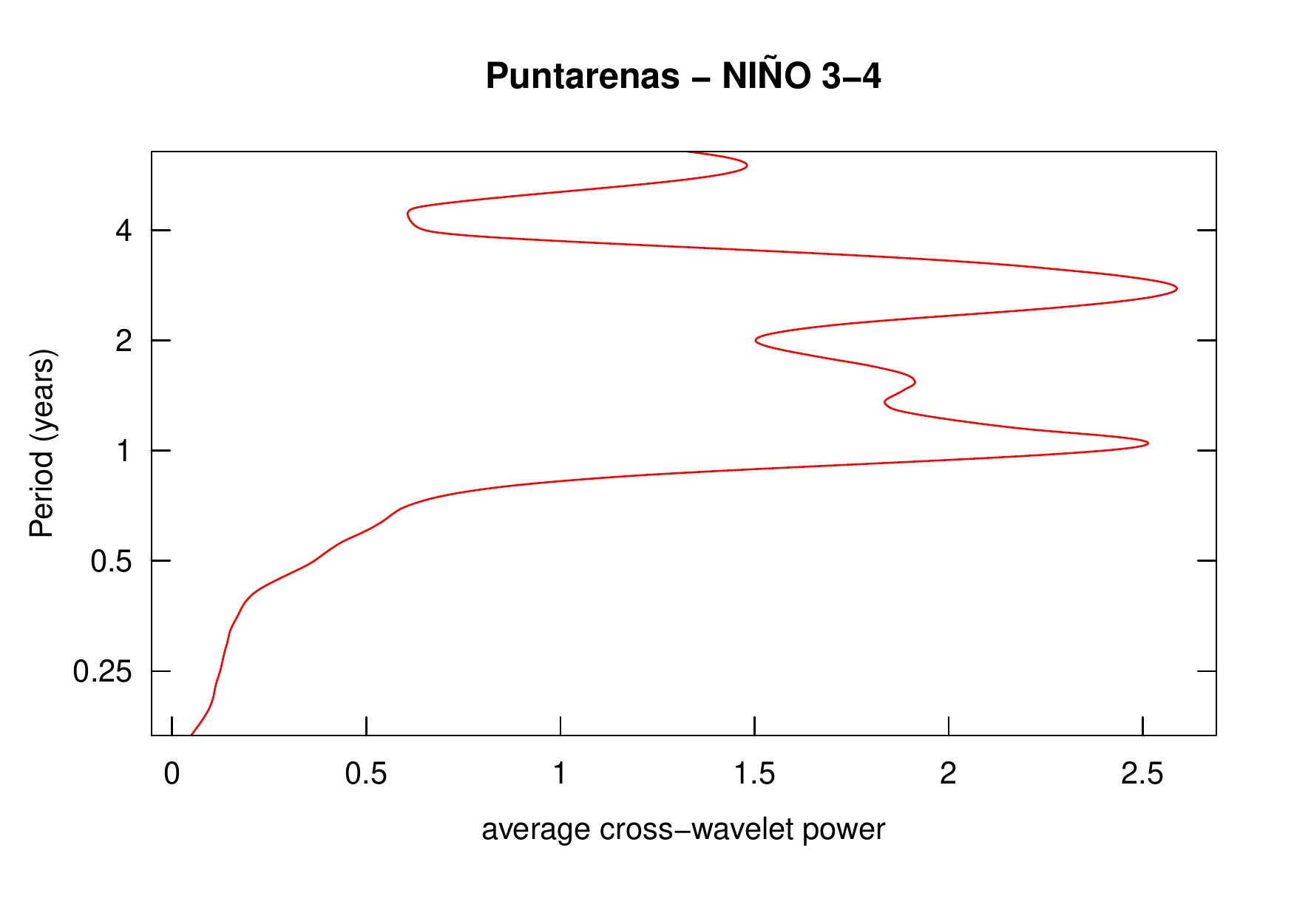}}\vspace{-0.15cm}%
\subfloat[]{\includegraphics[scale=0.23]{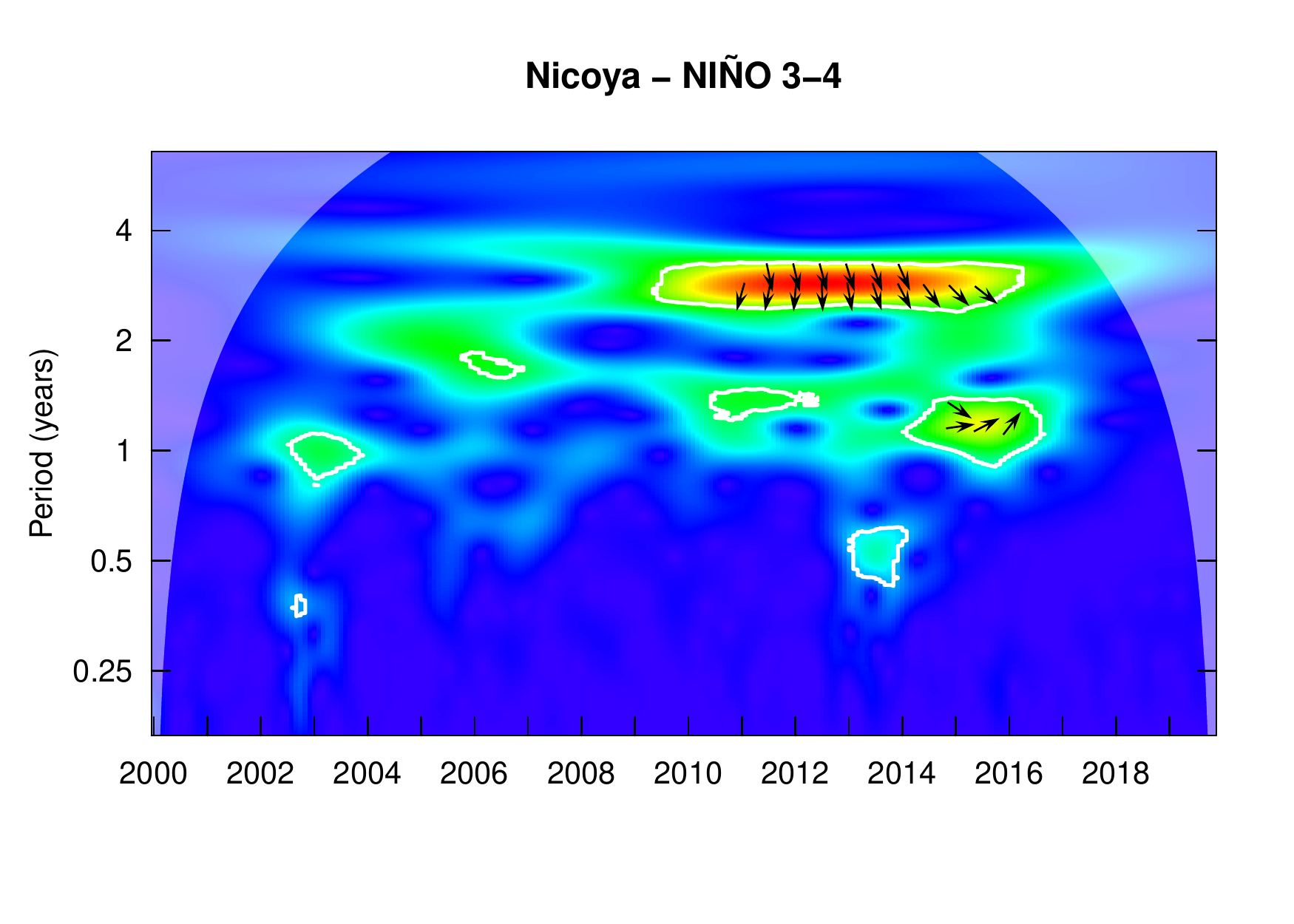}}\vspace{-0.15cm}%
\subfloat[]{\includegraphics[scale=0.23]{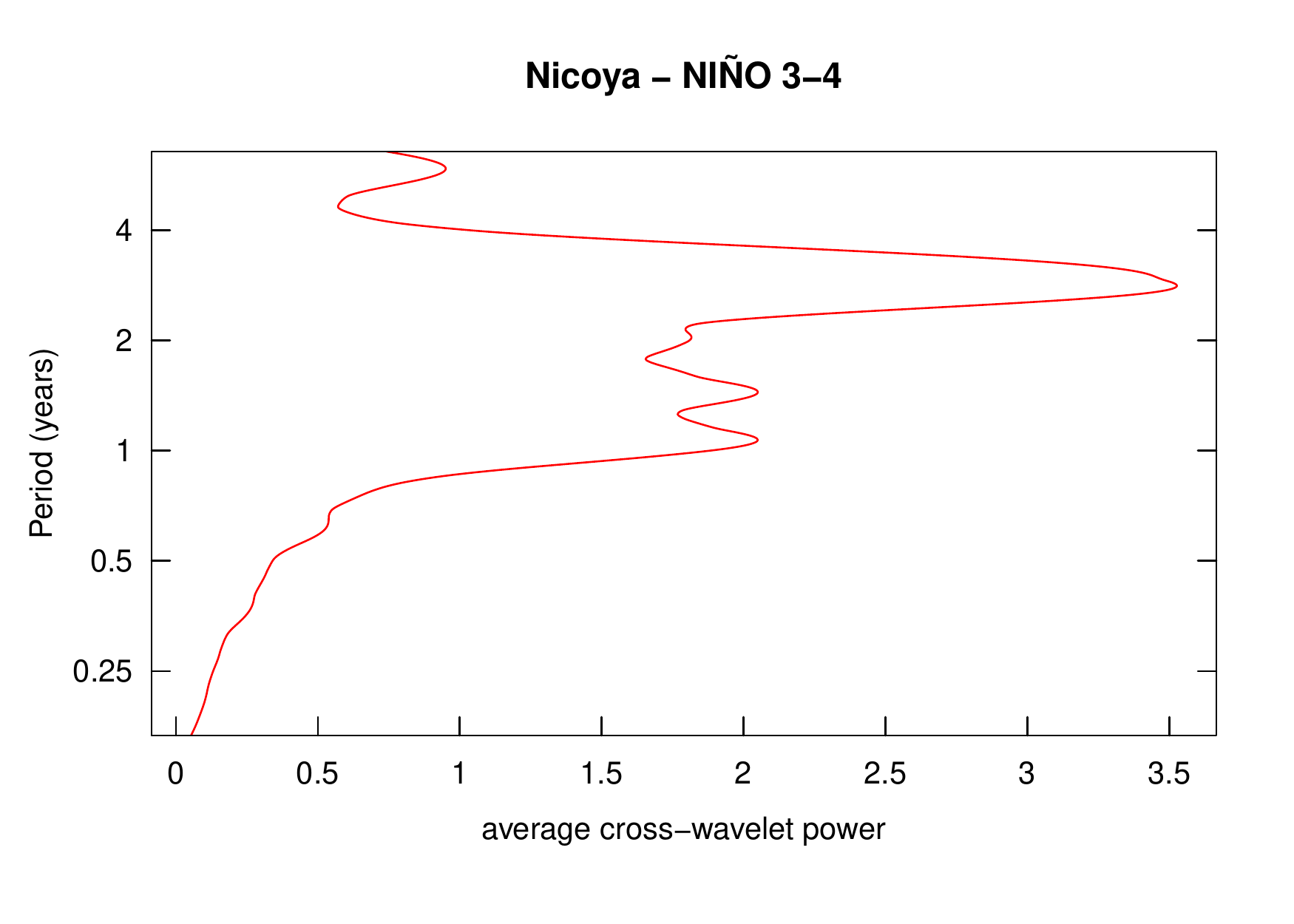}}\vspace{-0.15cm}\\
\caption*{}
\end{figure}

\begin{figure}[H]
\captionsetup[subfigure]{labelformat=empty}
\subfloat[]{\includegraphics[scale=0.23]{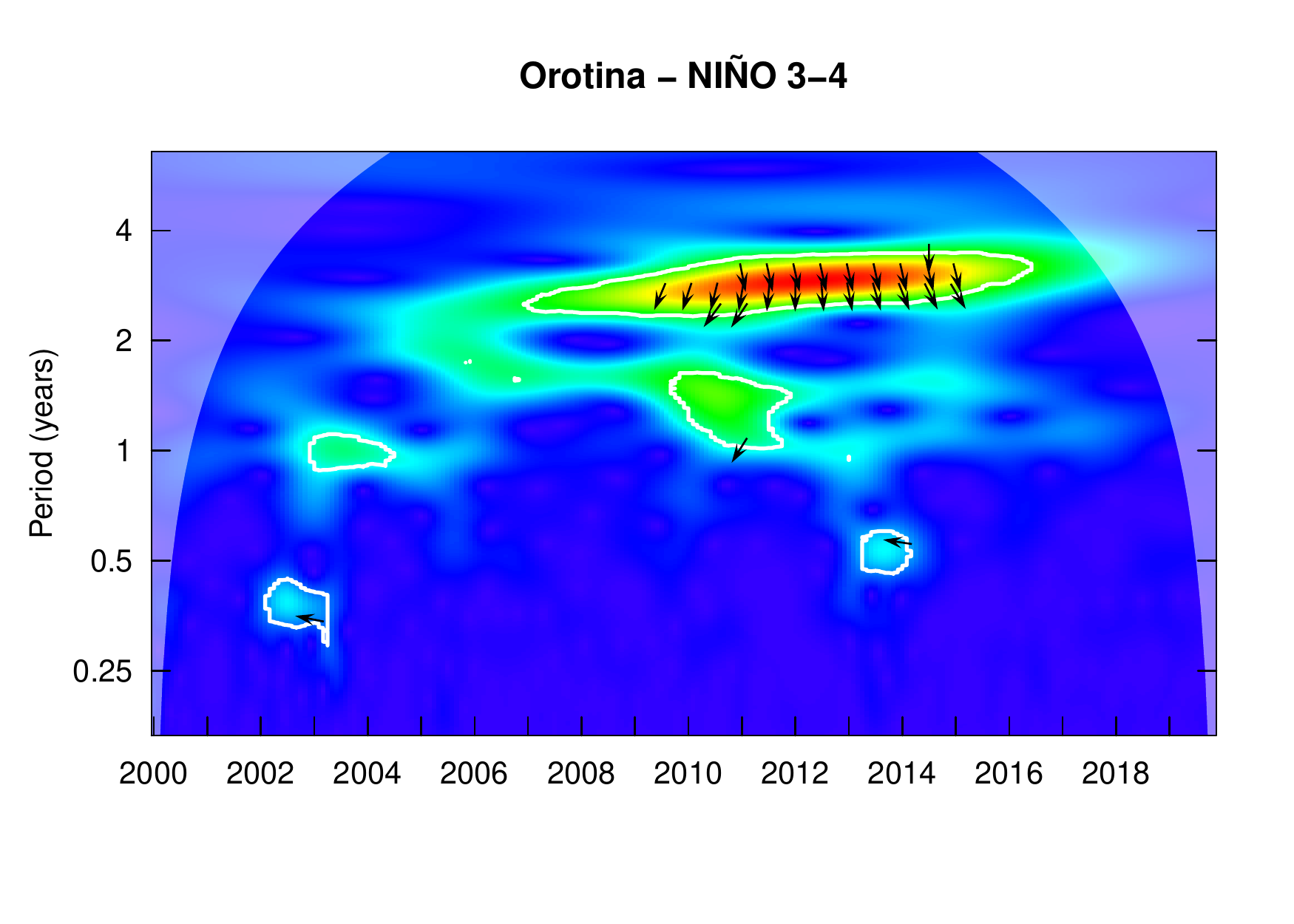}}\vspace{-0.15cm}%
\subfloat[]{\includegraphics[scale=0.23]{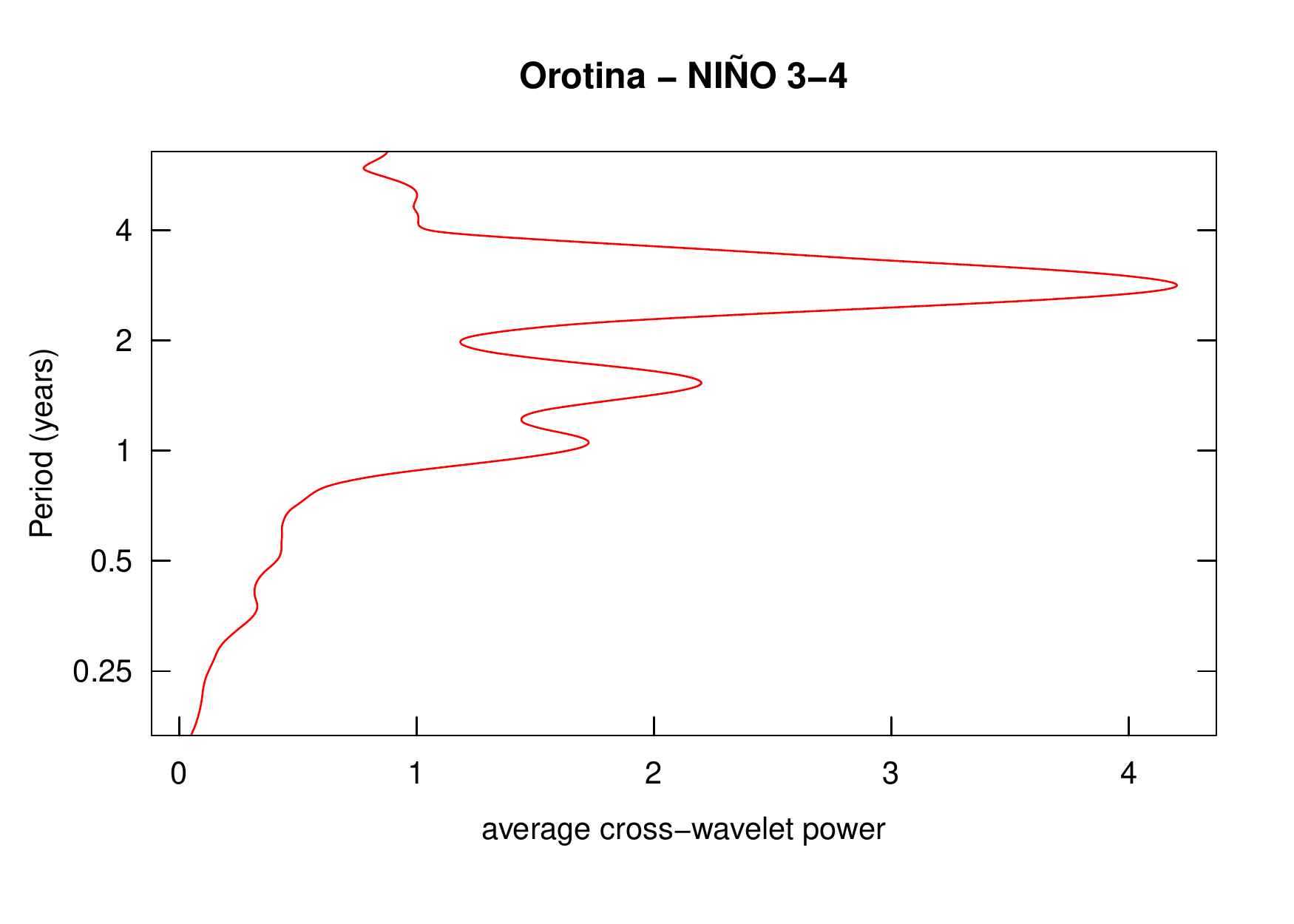}}\vspace{-0.15cm}%
\subfloat[]{\includegraphics[scale=0.23]{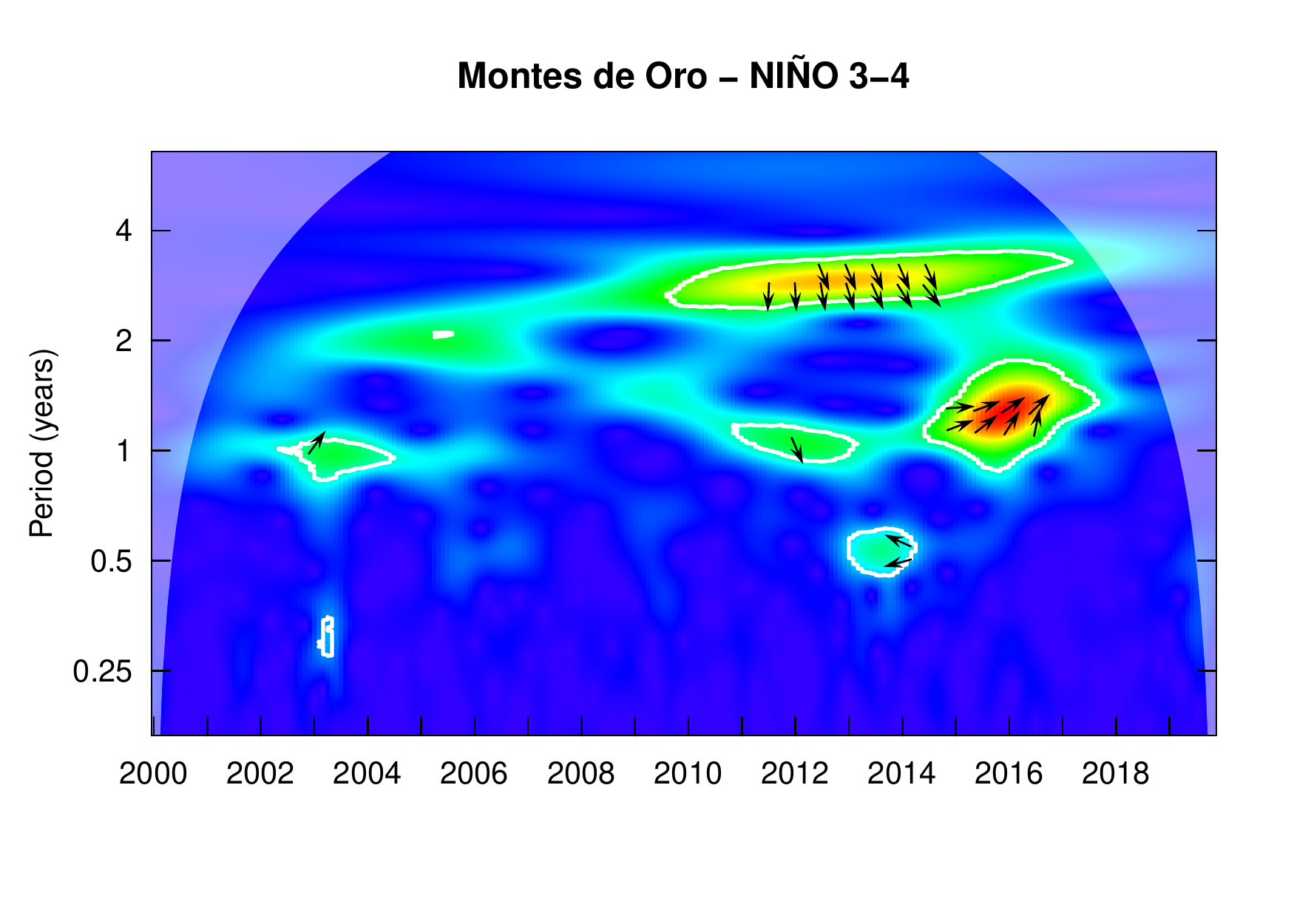}}\vspace{-0.15cm}%
\subfloat[]{\includegraphics[scale=0.23]{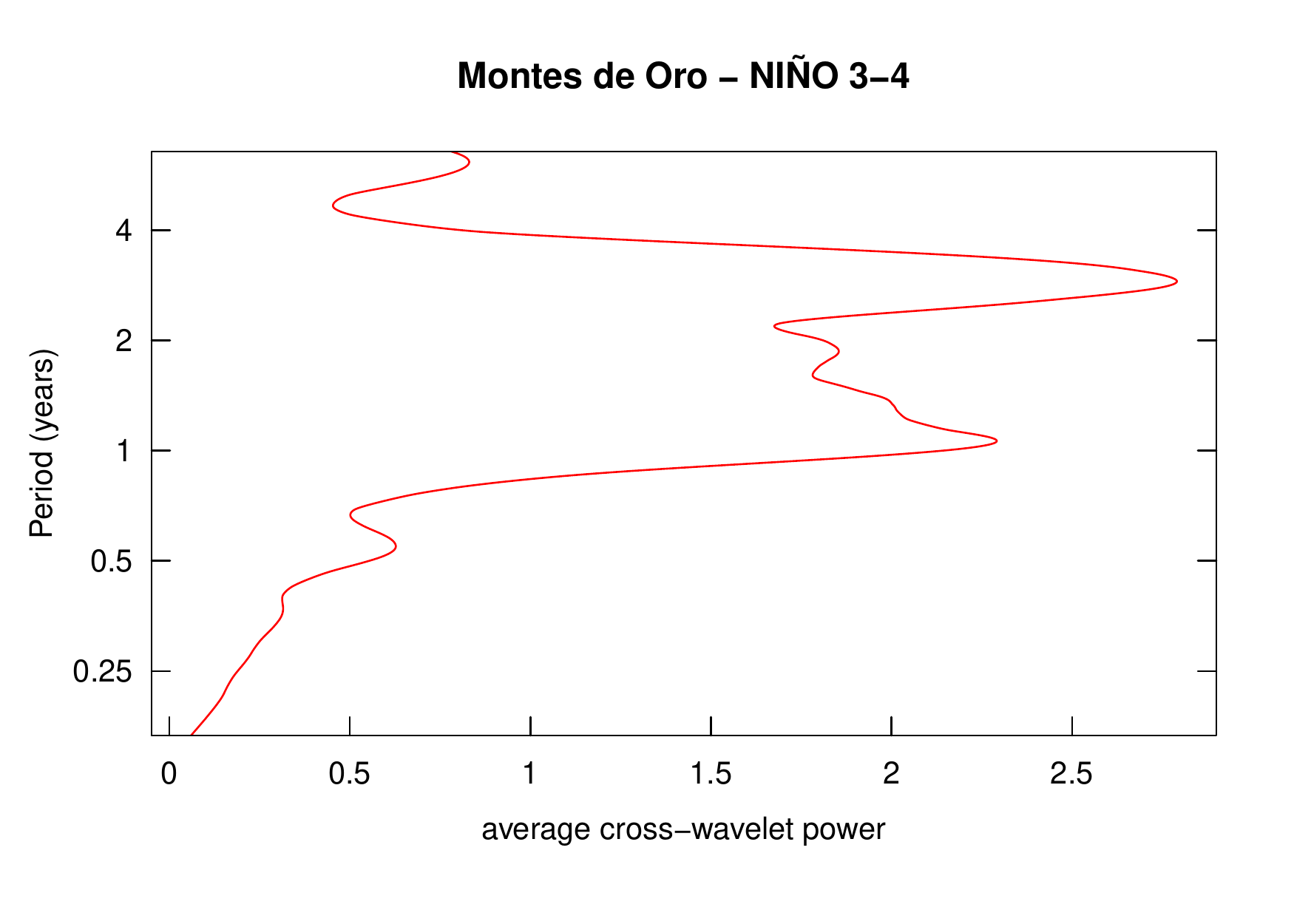}}\vspace{-0.15cm}//
\subfloat[]{\includegraphics[scale=0.23]{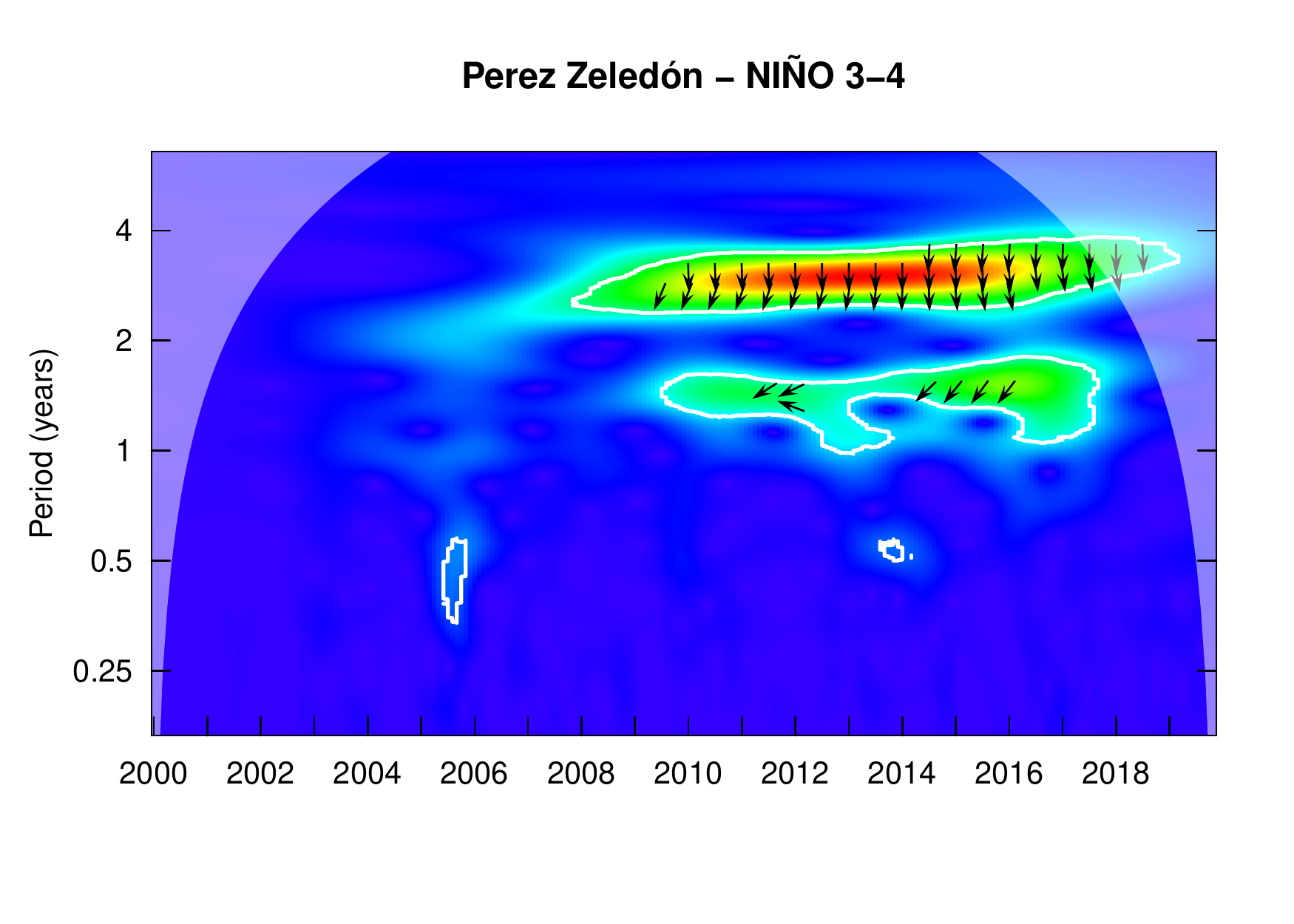}}\vspace{-0.15cm}%
\subfloat[]{\includegraphics[scale=0.23]{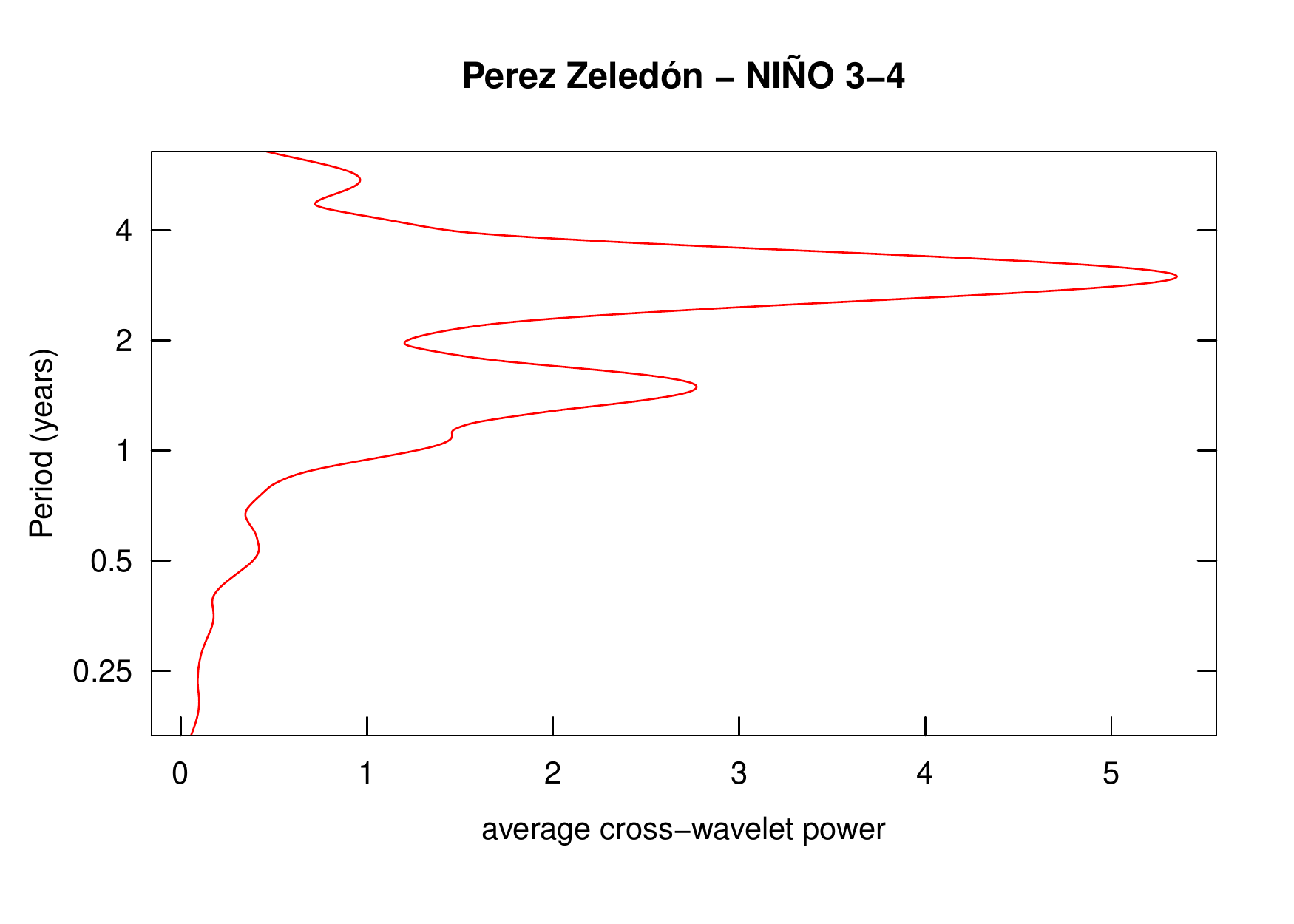}}\vspace{-0.15cm}%
\subfloat[]{\includegraphics[scale=0.23]{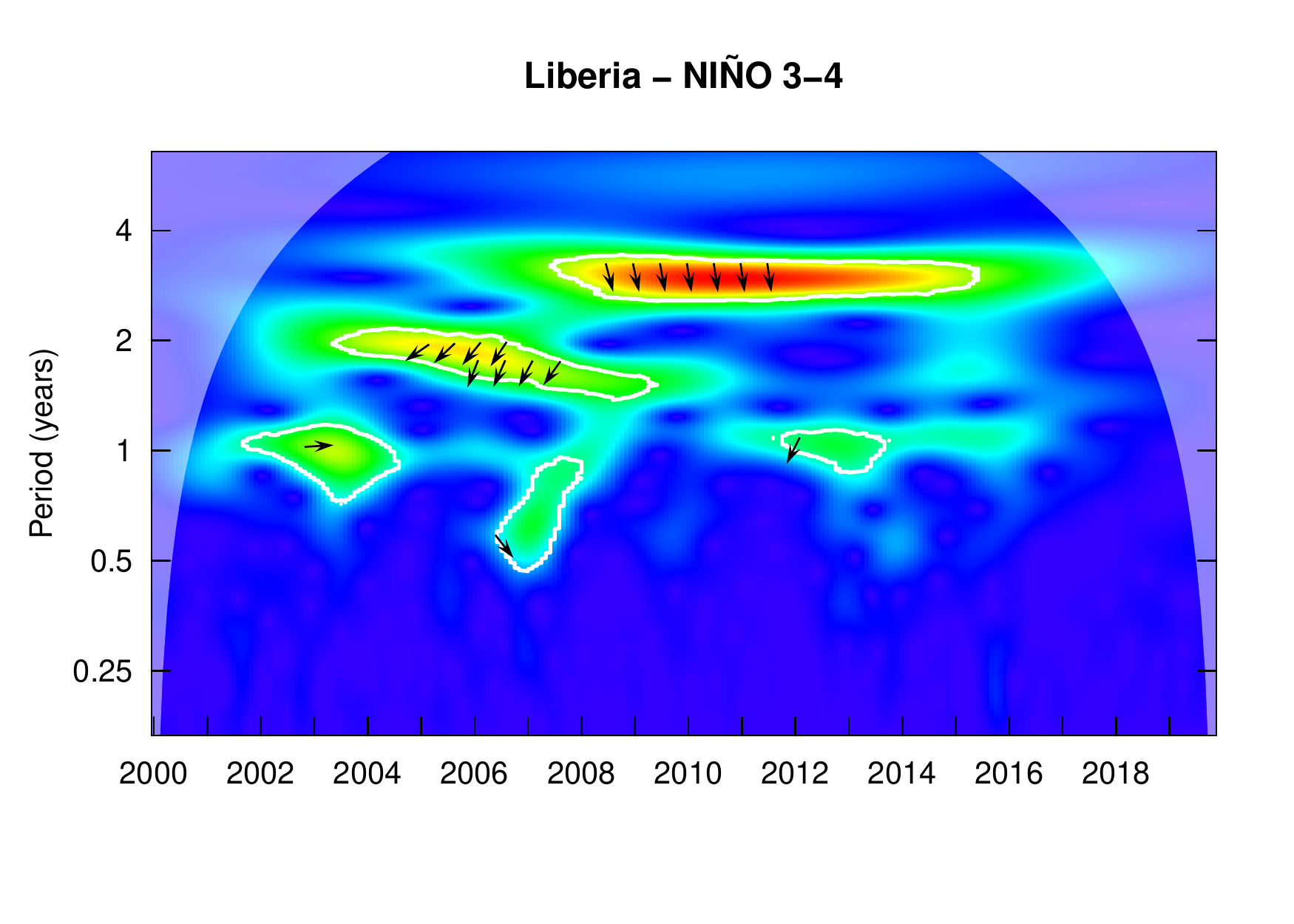}}\vspace{-0.15cm}%
\subfloat[]{\includegraphics[scale=0.23]{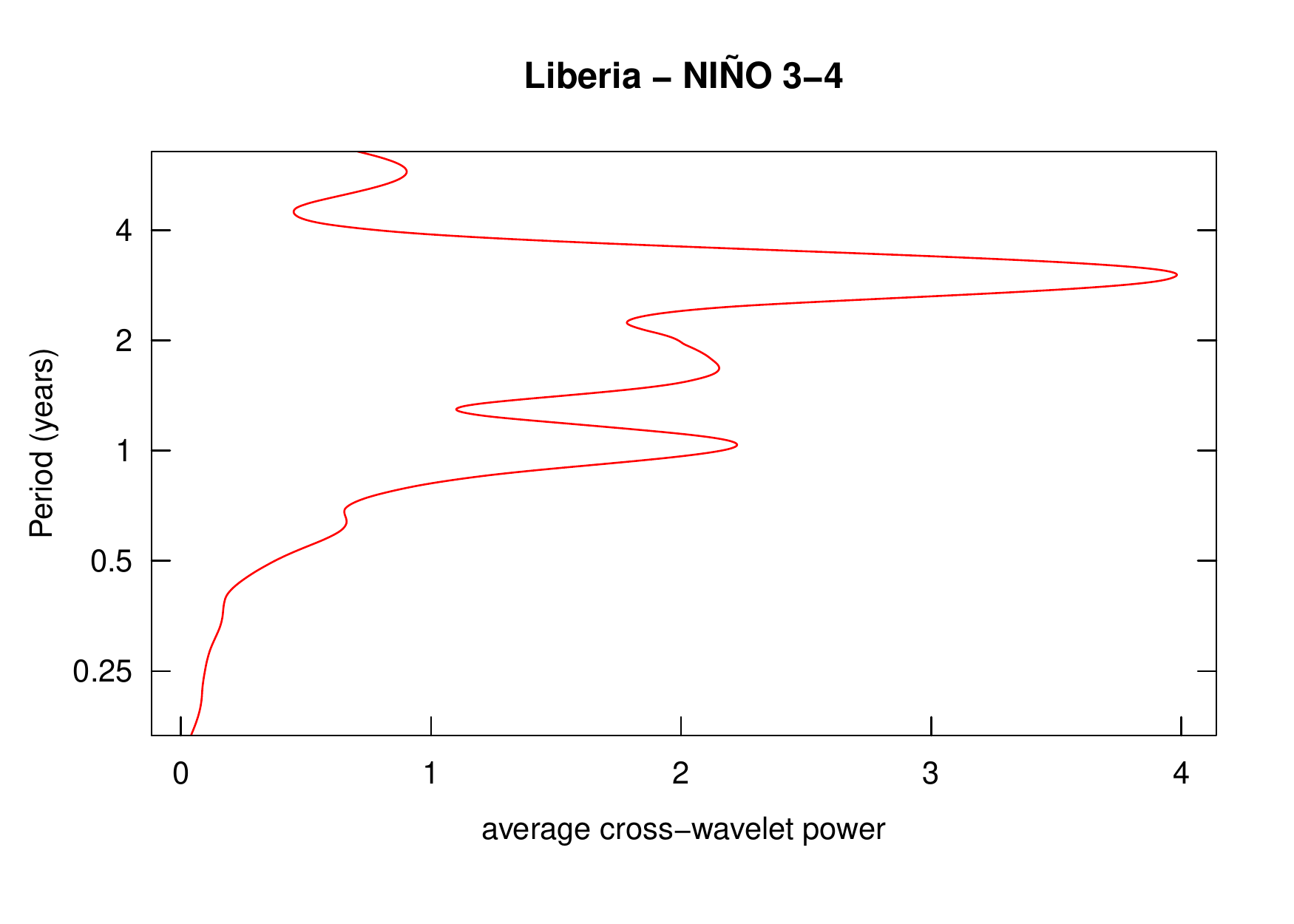}}\vspace{-0.15cm}
\subfloat[]{\includegraphics[scale=0.23]{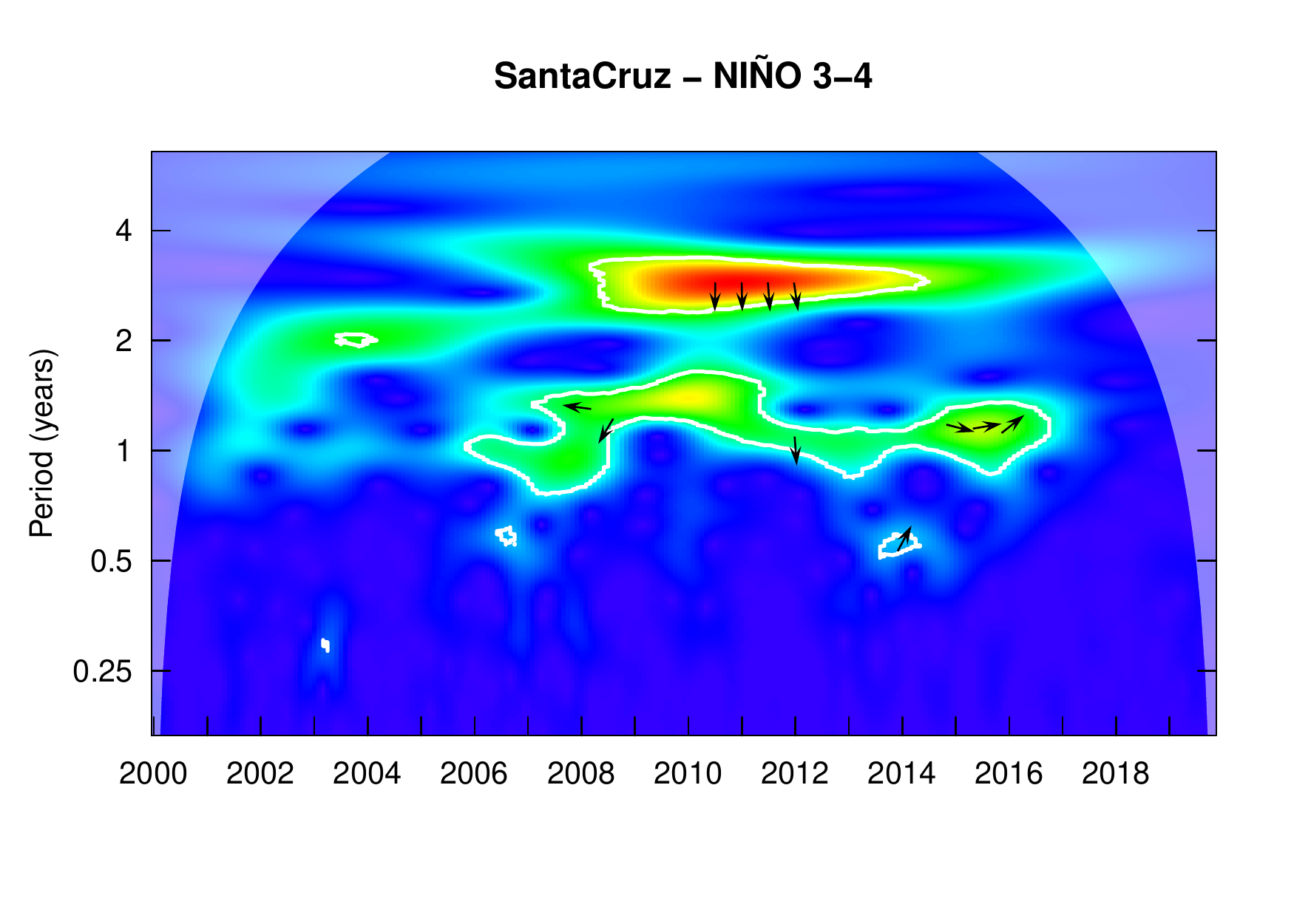}}\vspace{-0.15cm}%
\subfloat[]{\includegraphics[scale=0.23]{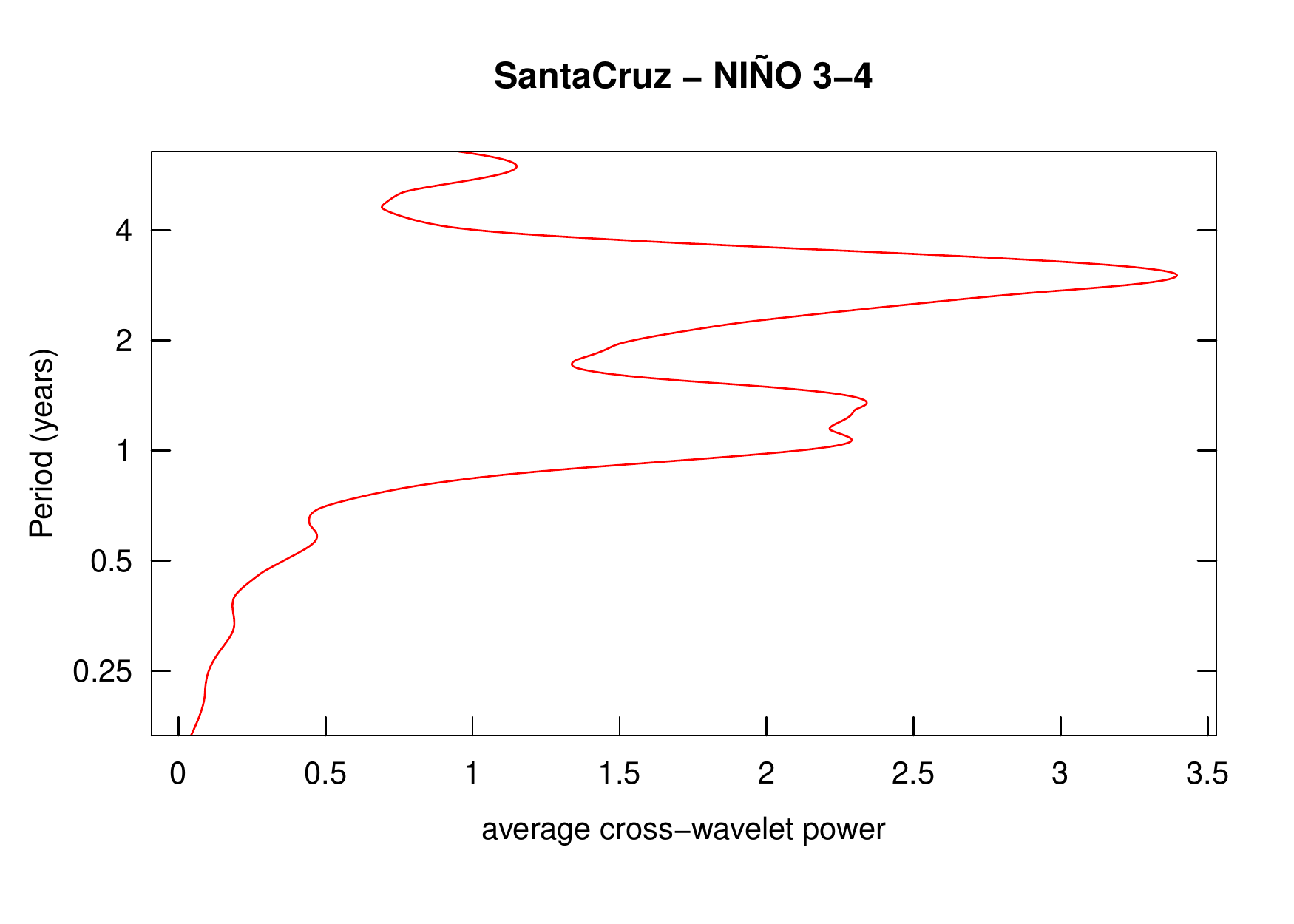}}\vspace{-0.15cm}%
\subfloat[]{\includegraphics[scale=0.23]{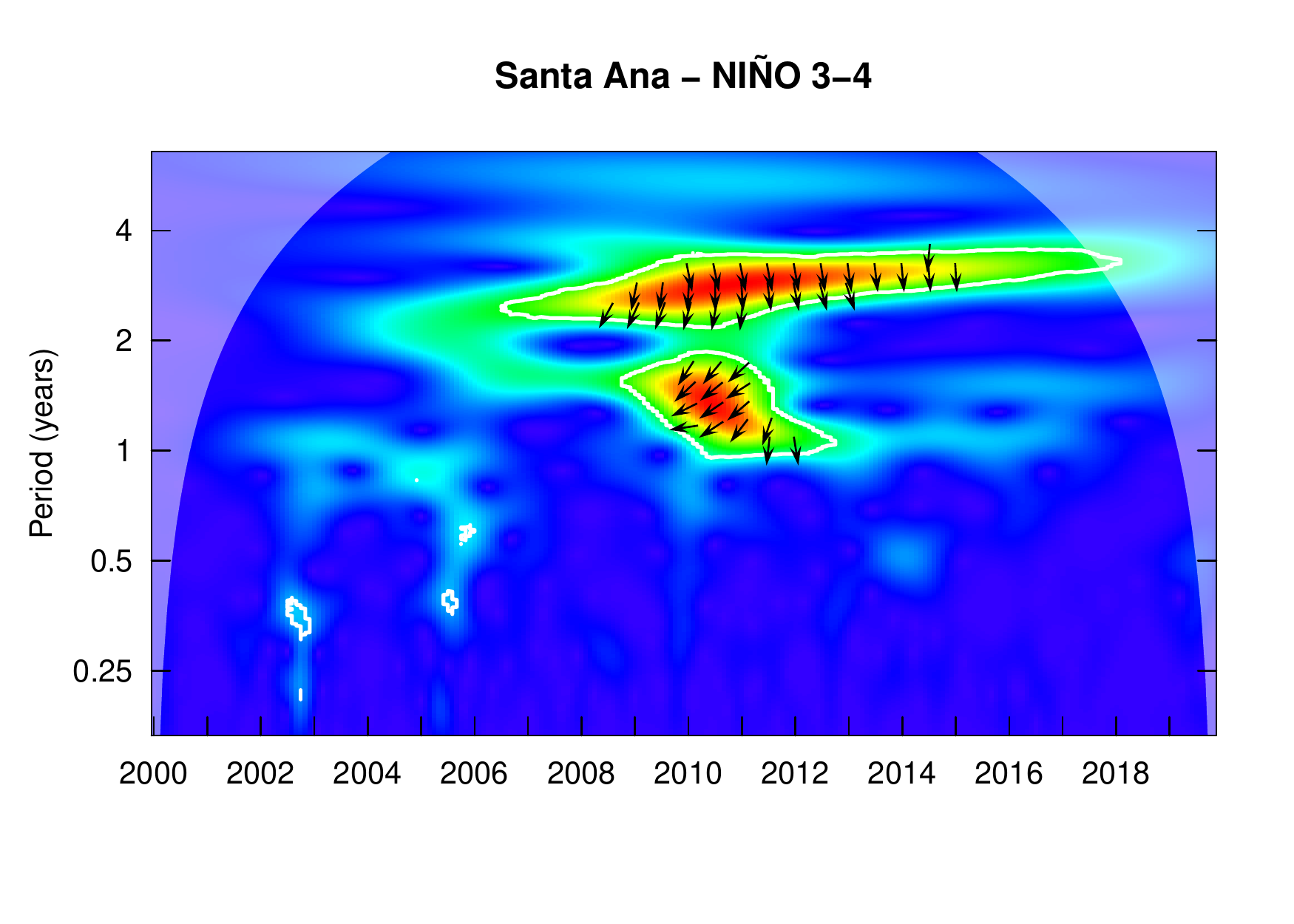}}\vspace{-0.15cm}%
\subfloat[]{\includegraphics[scale=0.23]{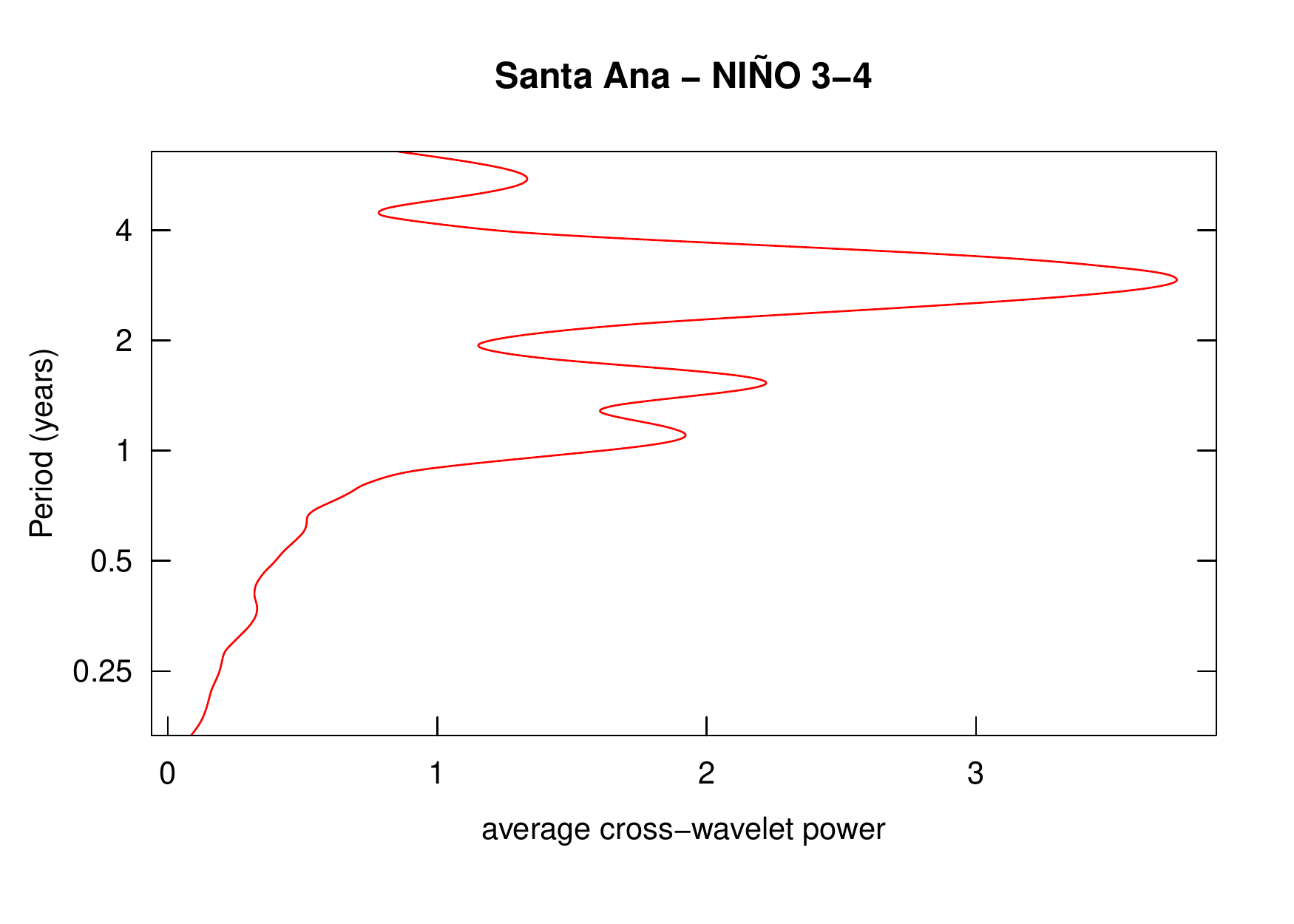}}\vspace{-0.15cm}\\
\subfloat[]{\includegraphics[scale=0.23]{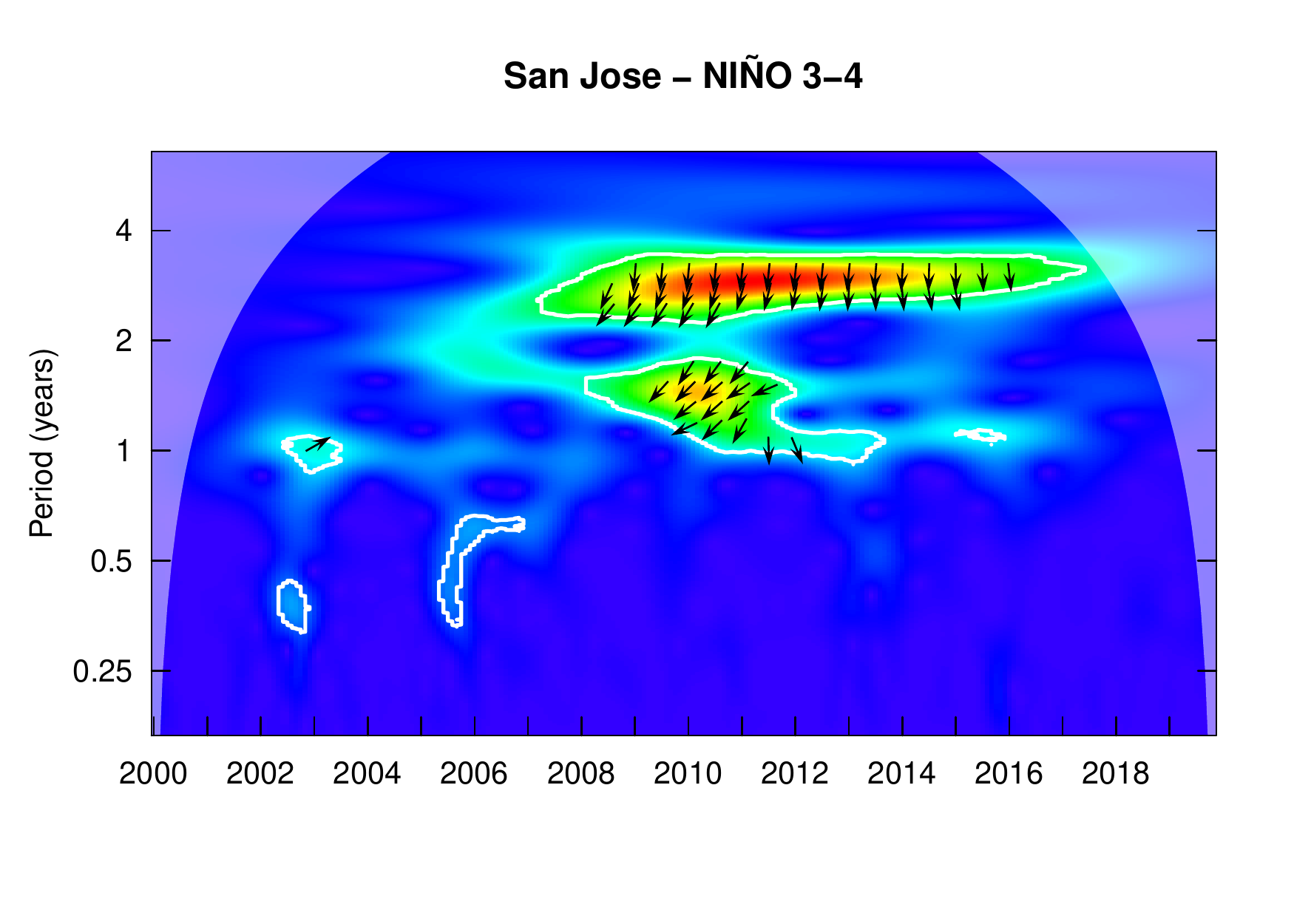}}\vspace{-0.15cm}%
\subfloat[]{\includegraphics[scale=0.23]{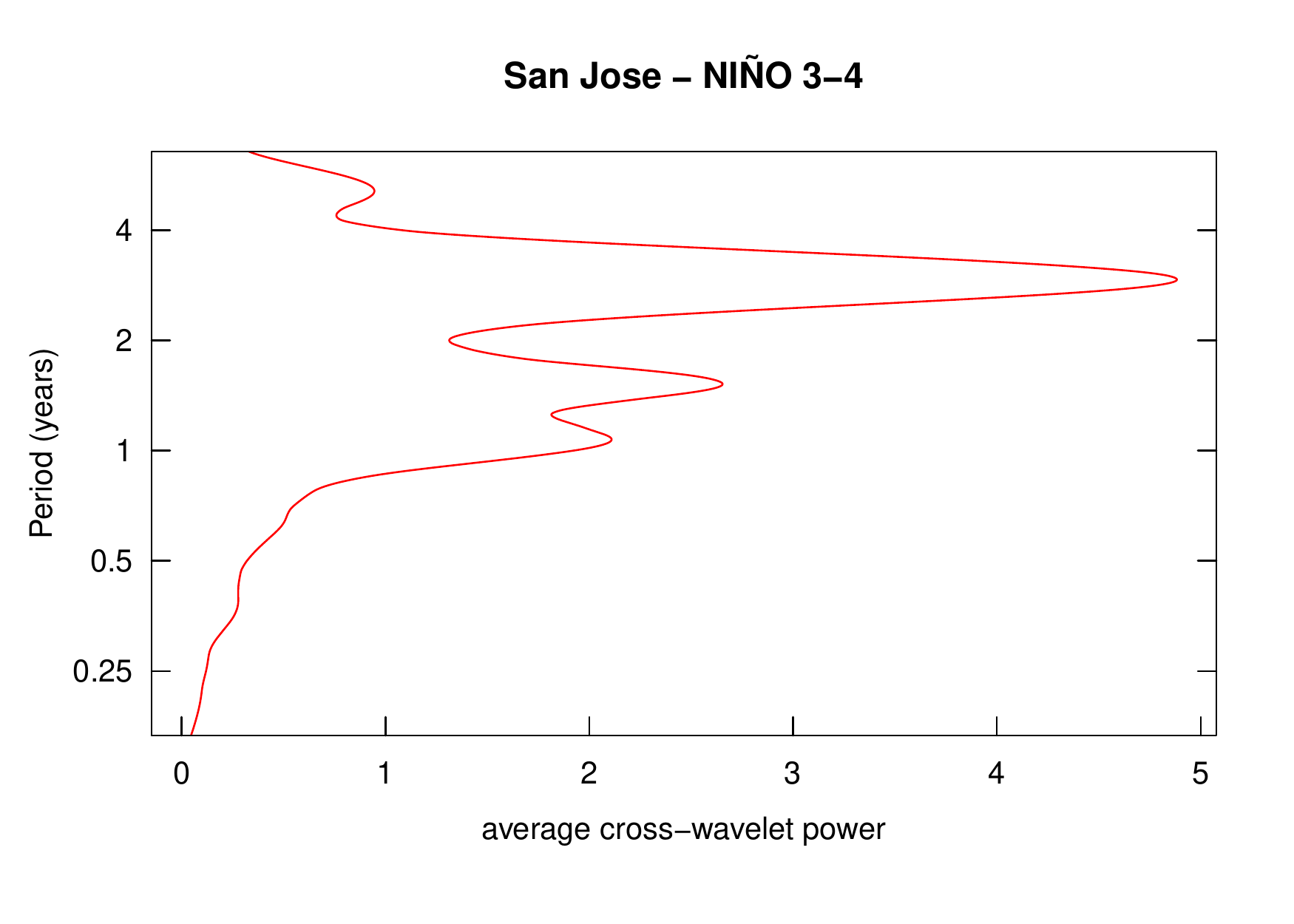}}\vspace{-0.15cm}%
\subfloat[]{\includegraphics[scale=0.23]{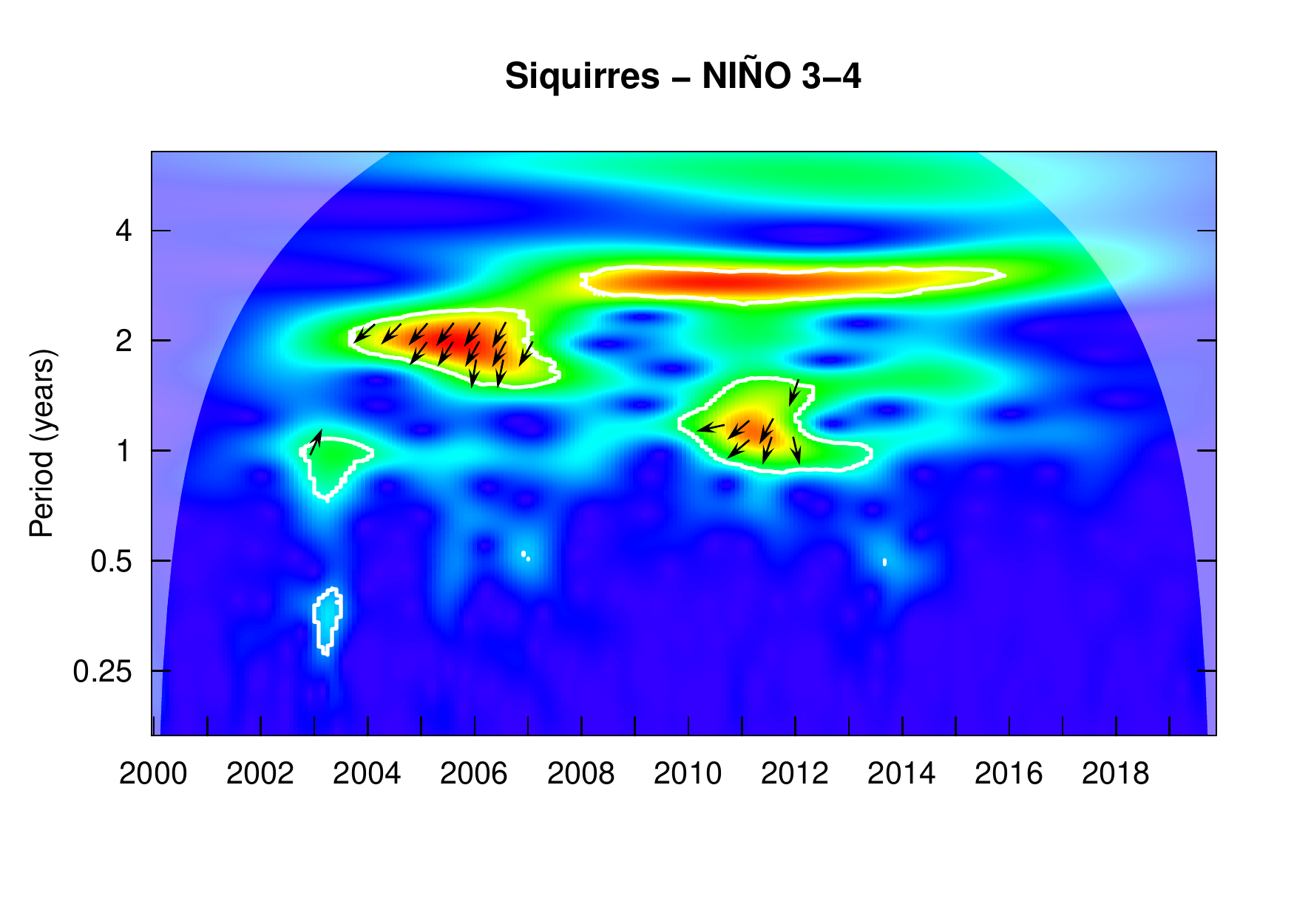}}\vspace{-0.15cm}%
\subfloat[]{\includegraphics[scale=0.23]{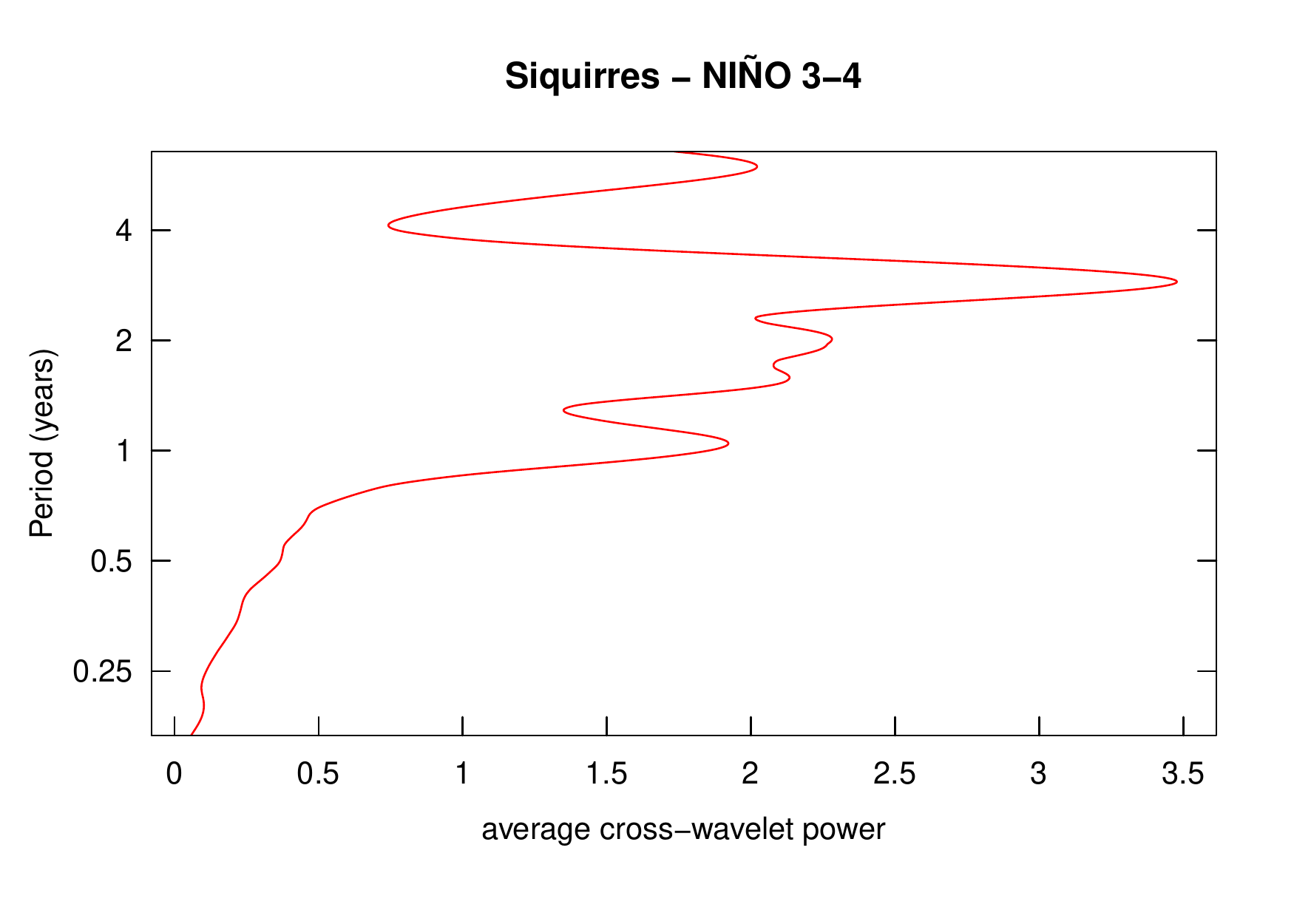}}\vspace{-0.15cm}\\
\subfloat[]{\includegraphics[scale=0.23]{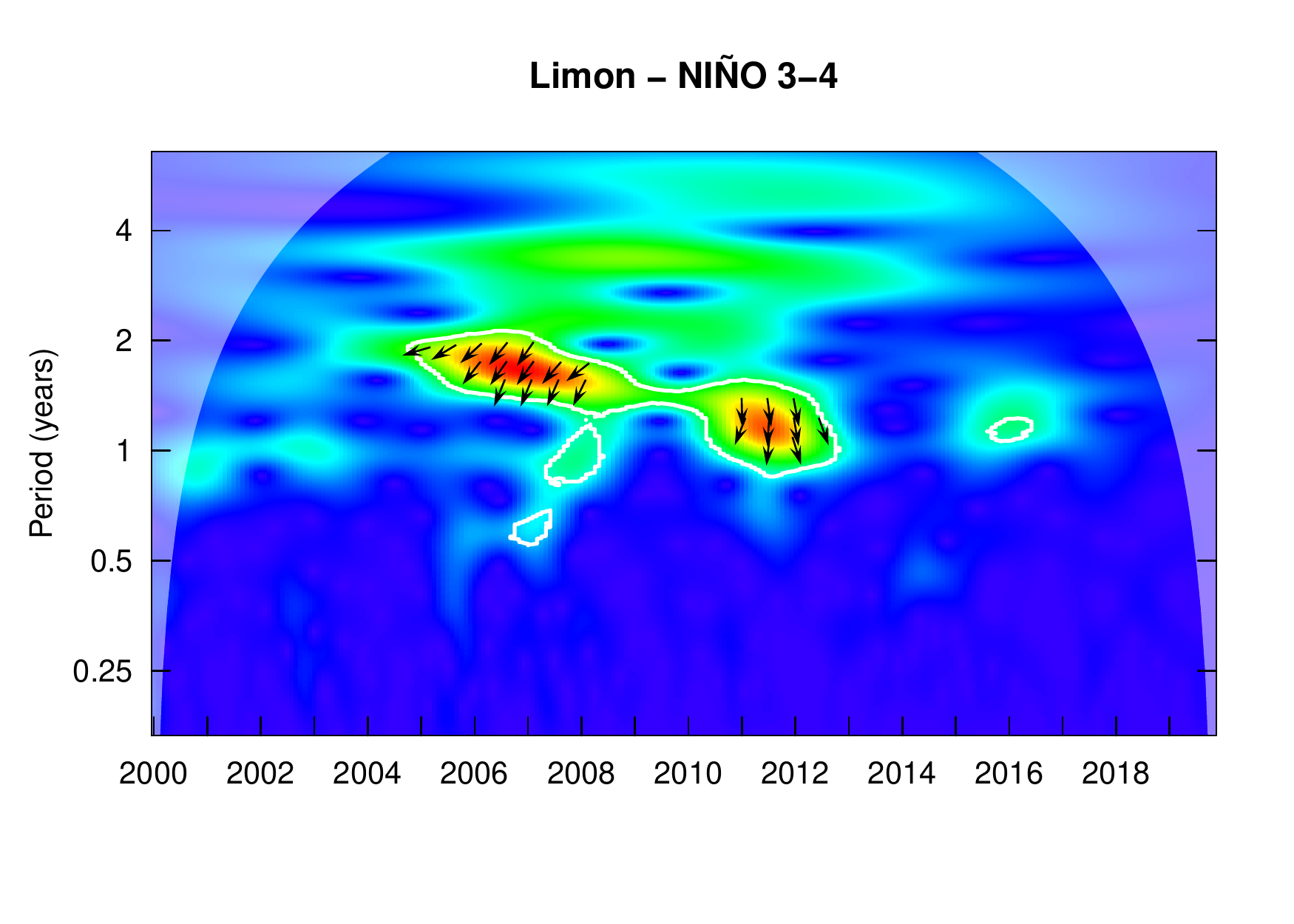}}\vspace{-0.15cm}%
\subfloat[]{\includegraphics[scale=0.23]{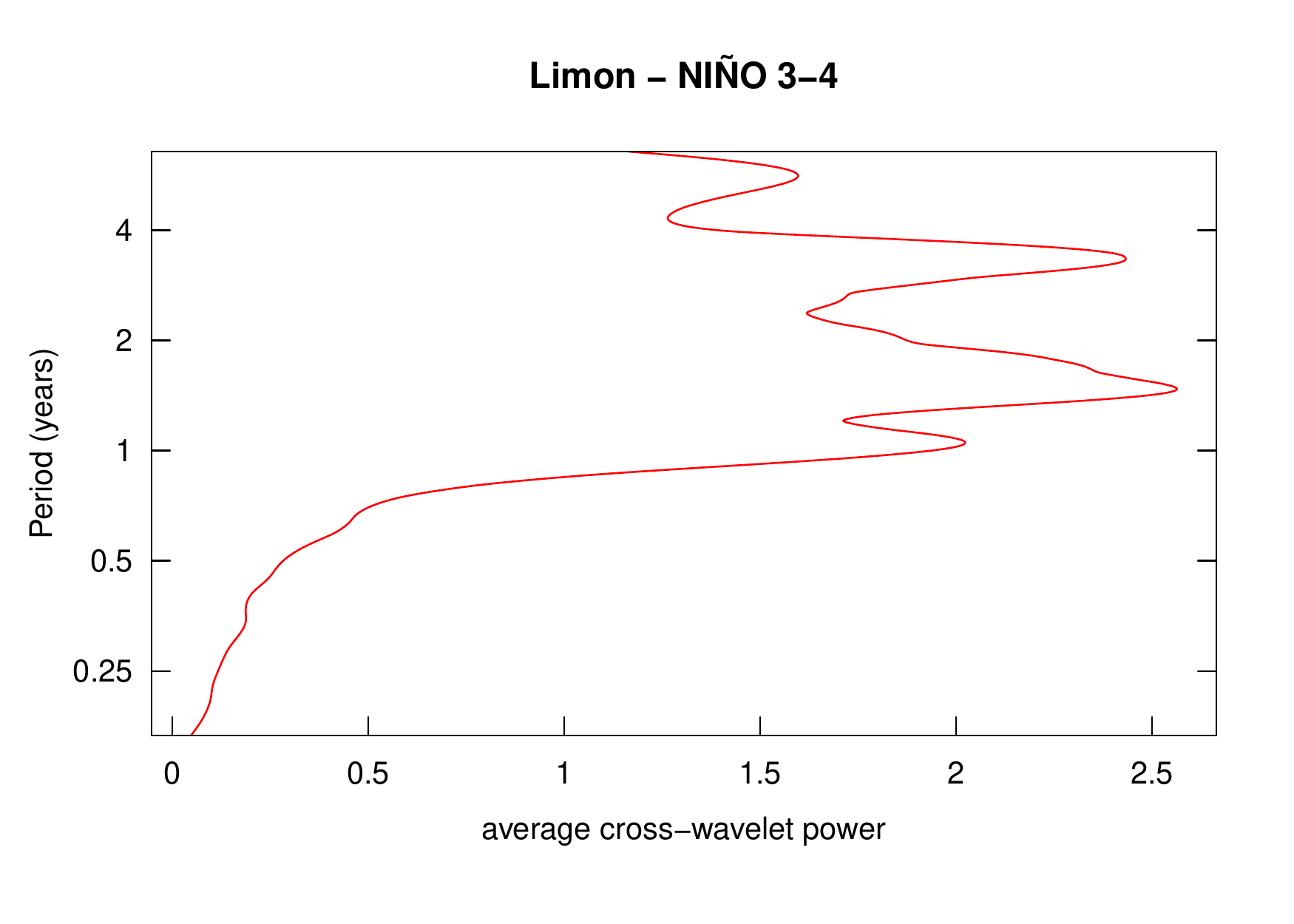}}\vspace{-0.15cm}%
\subfloat[]{\includegraphics[scale=0.23]{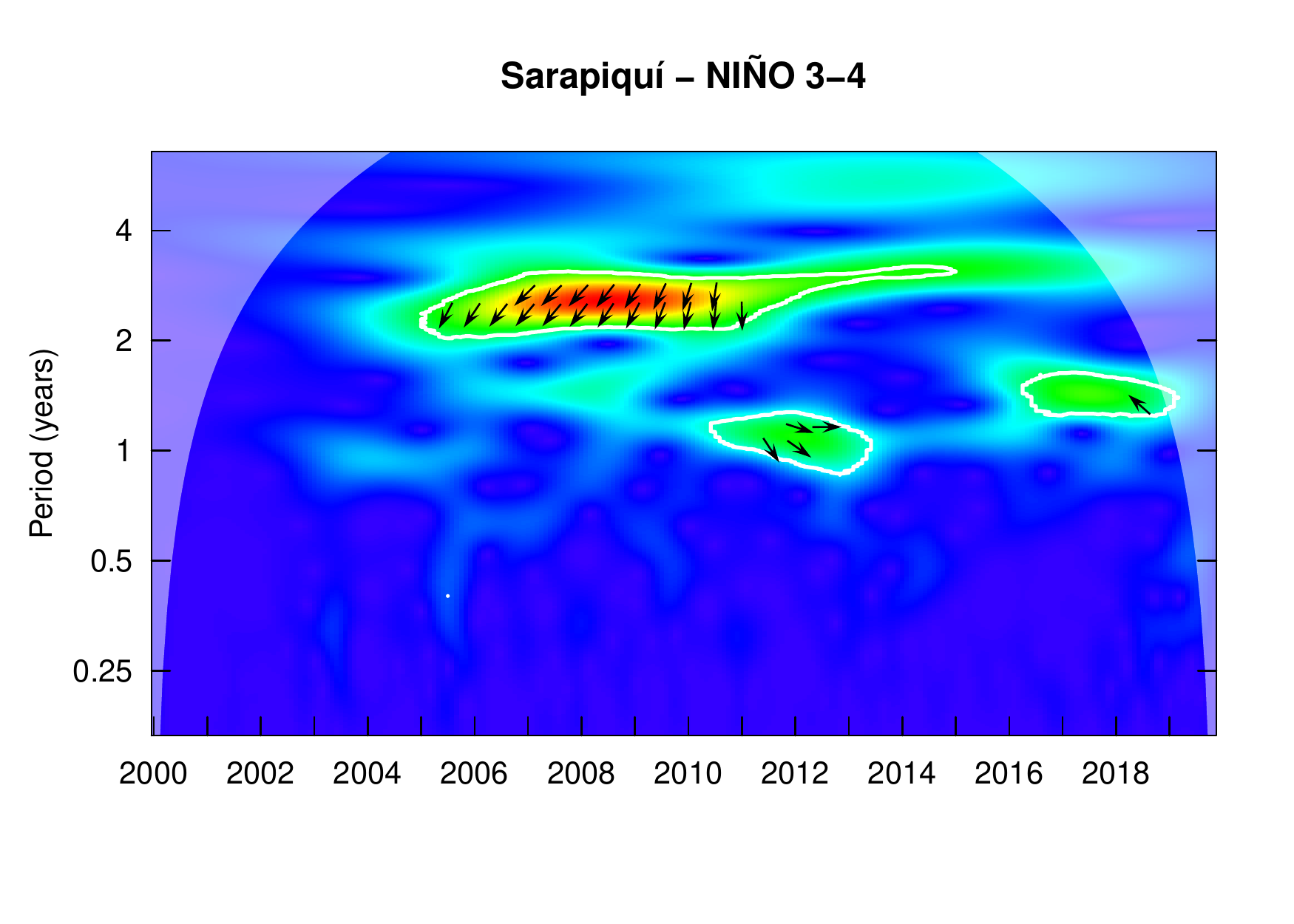}}\vspace{-0.15cm}%
\subfloat[]{\includegraphics[scale=0.23]{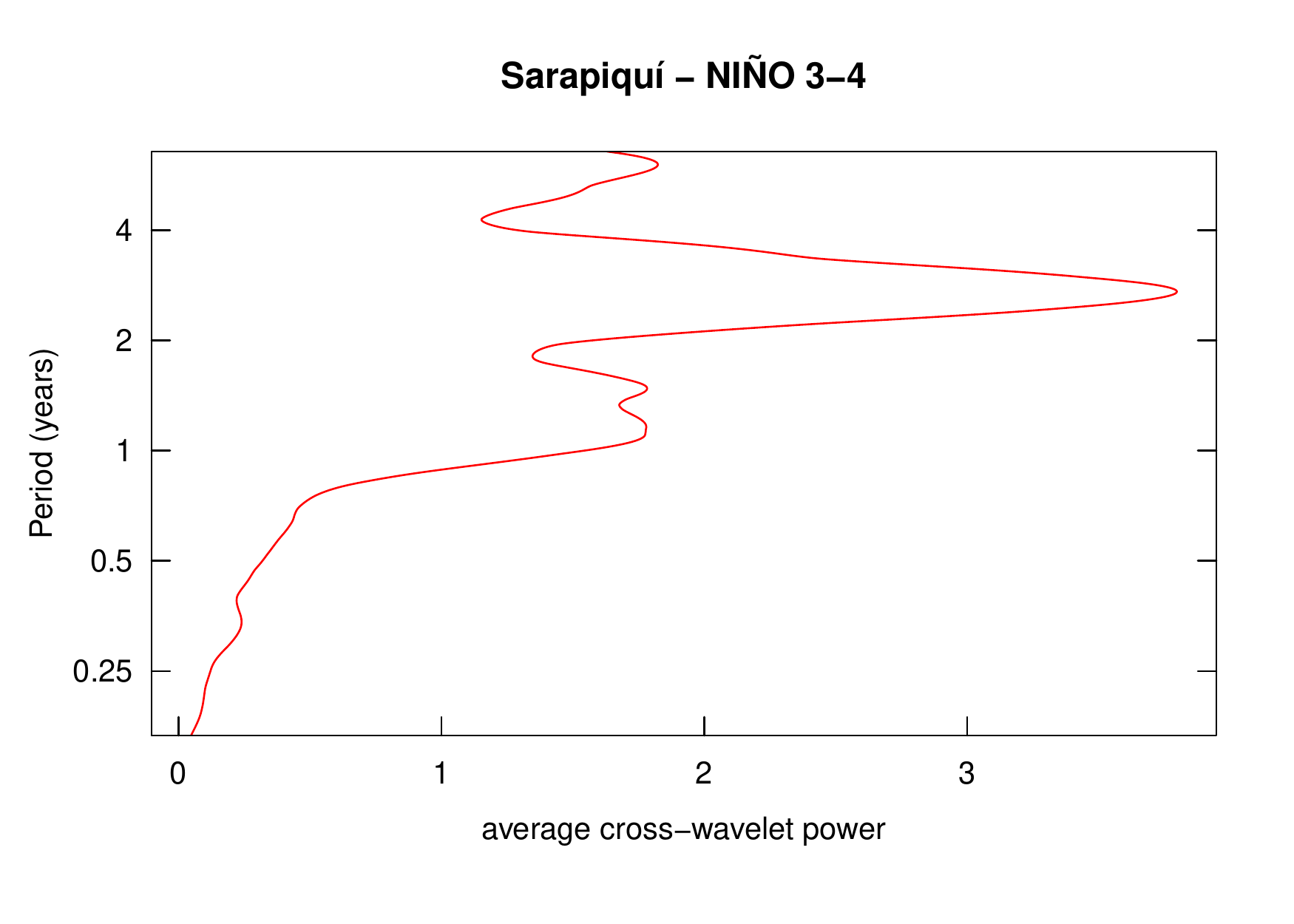}}\vspace{-0.15cm}\\
\subfloat[]{\includegraphics[scale=0.23]{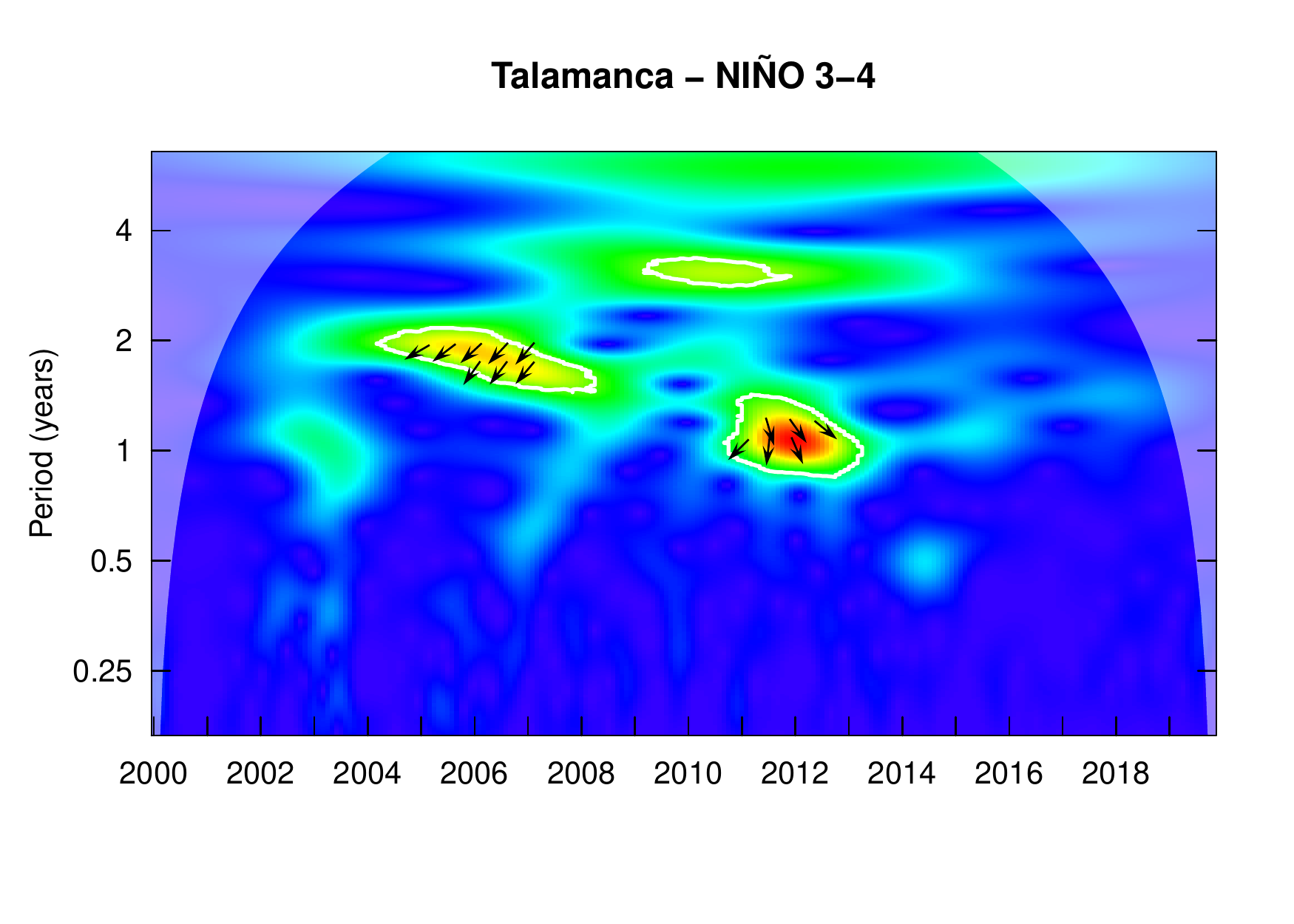}}\vspace{-0.15cm}%
\subfloat[]{\includegraphics[scale=0.23]{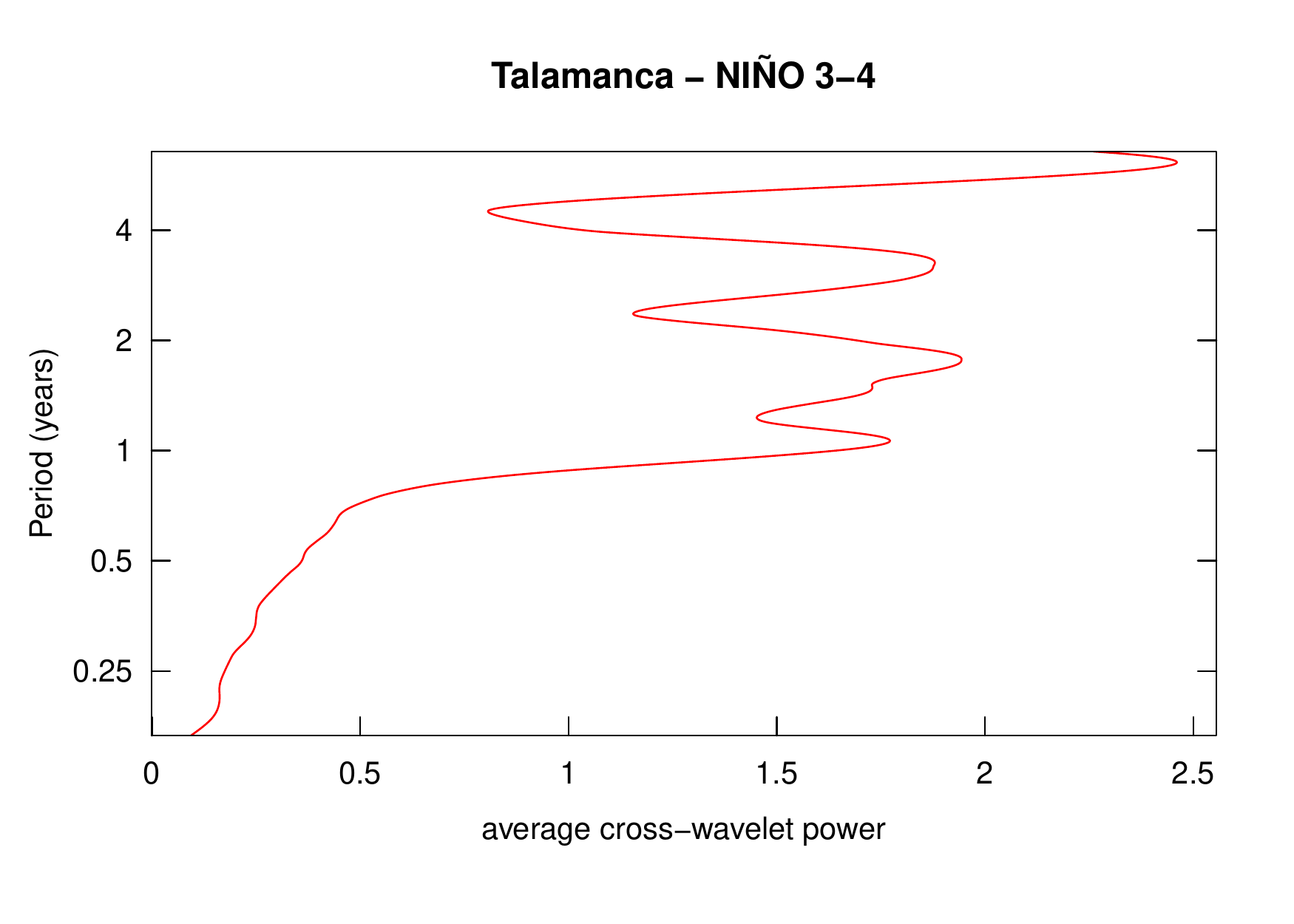}}\vspace{-0.15cm}%
\subfloat[]{\includegraphics[scale=0.23]{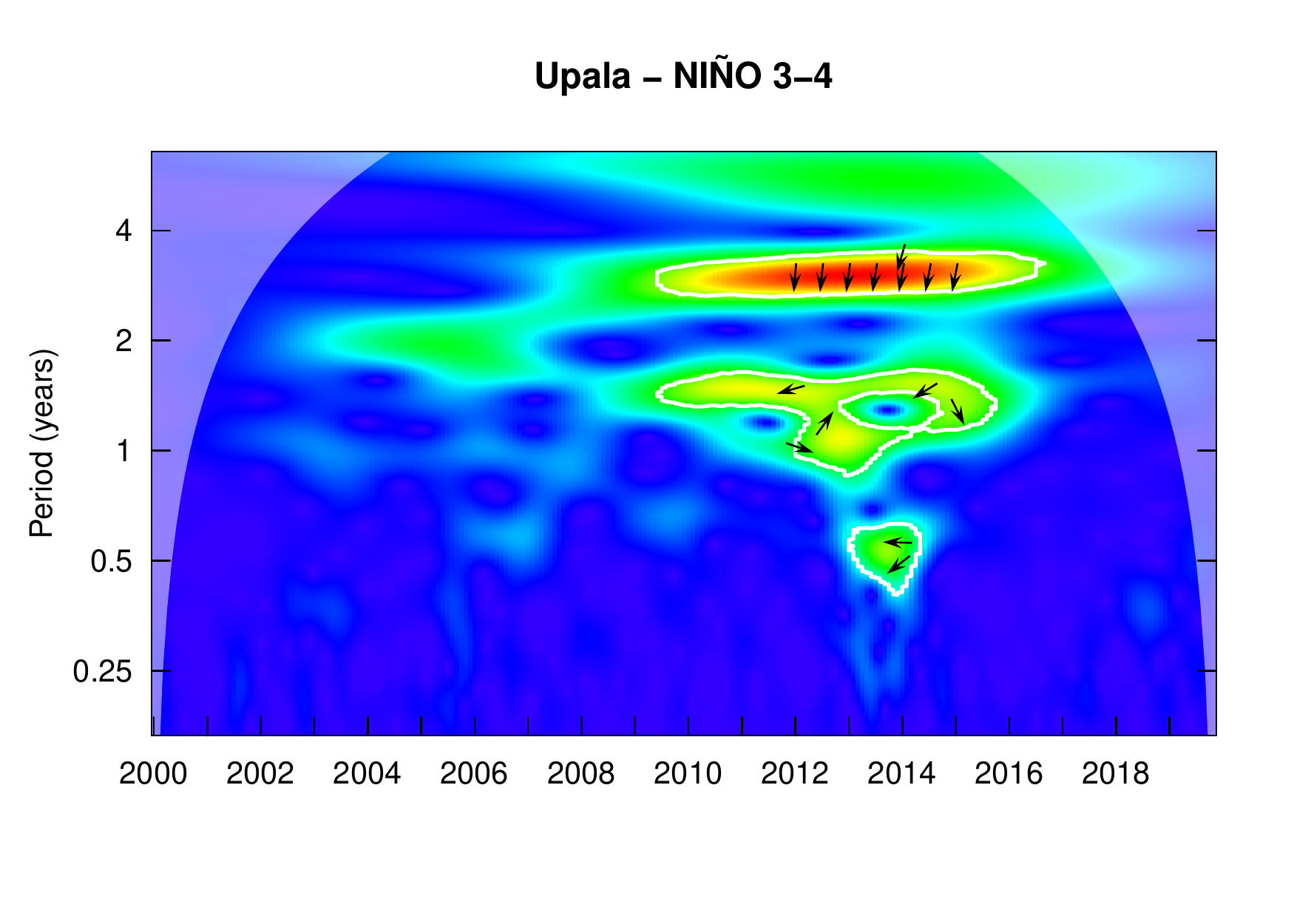}}\vspace{-0.15cm}%
\subfloat[]{\includegraphics[scale=0.23]{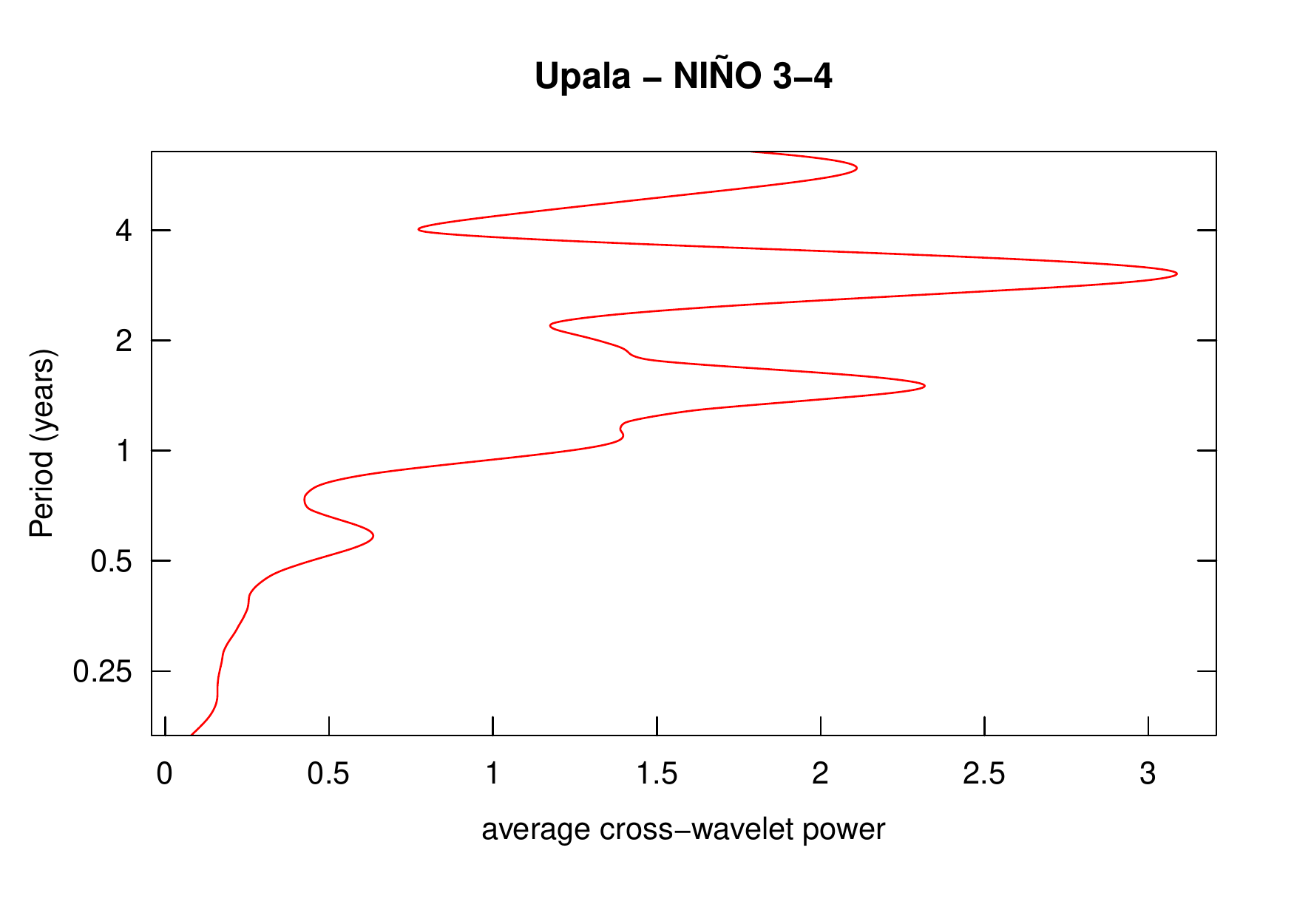}}\vspace{-0.15cm}\\
\subfloat[]{\includegraphics[scale=0.23]{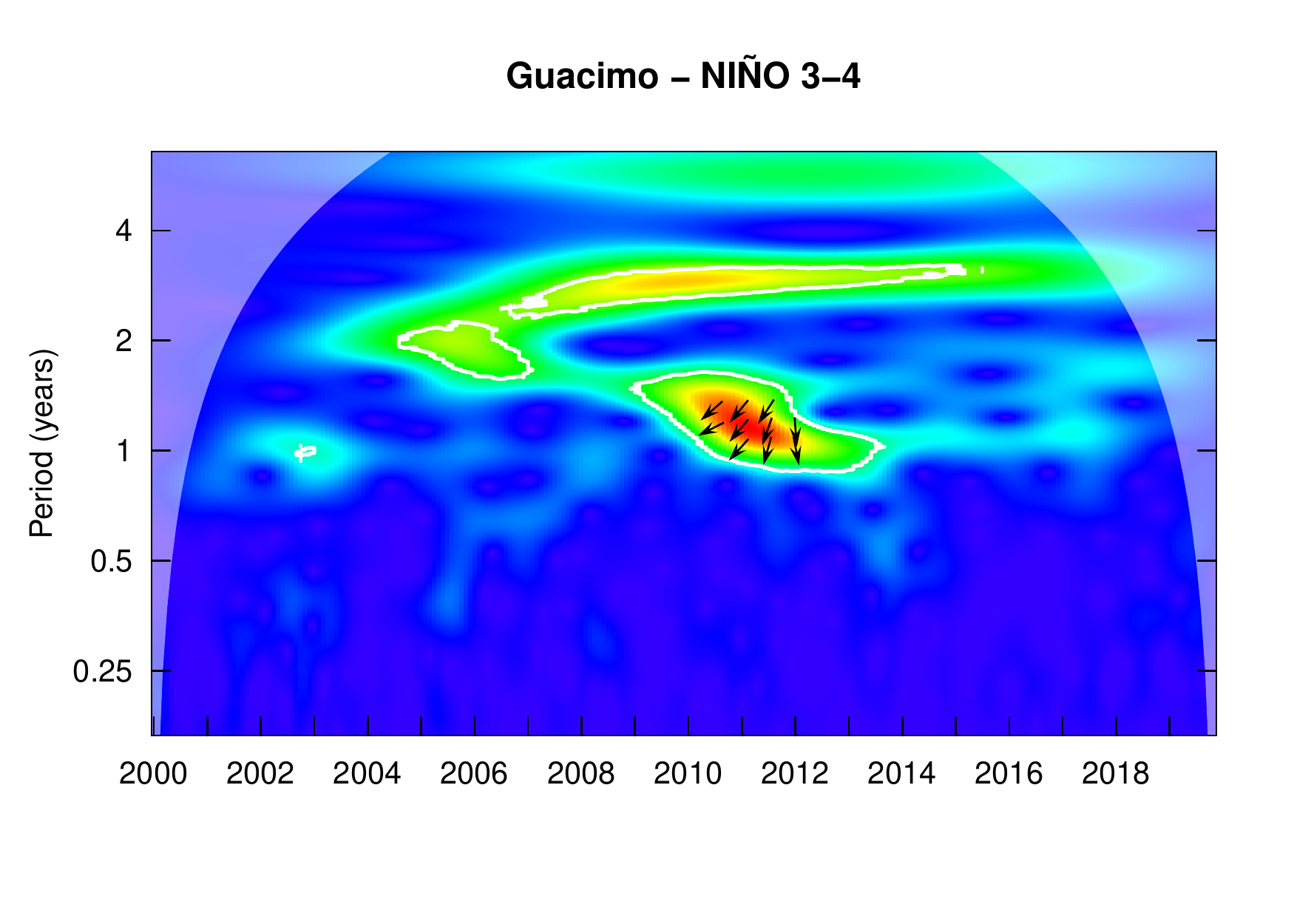}}\vspace{-0.15cm}%
\subfloat[]{\includegraphics[scale=0.23]{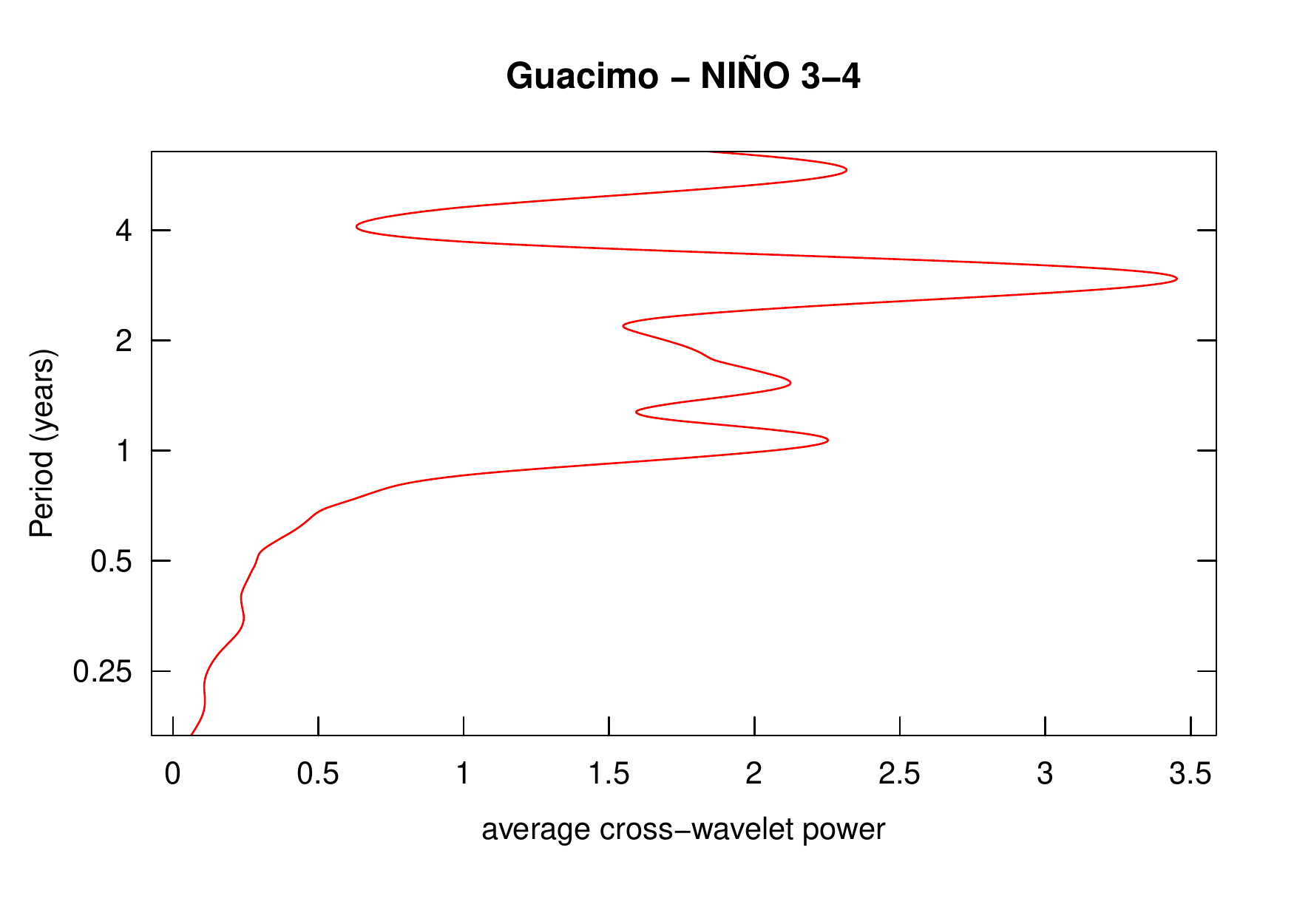}}\vspace{-0.15cm}%
\subfloat[]{\includegraphics[scale=0.23]{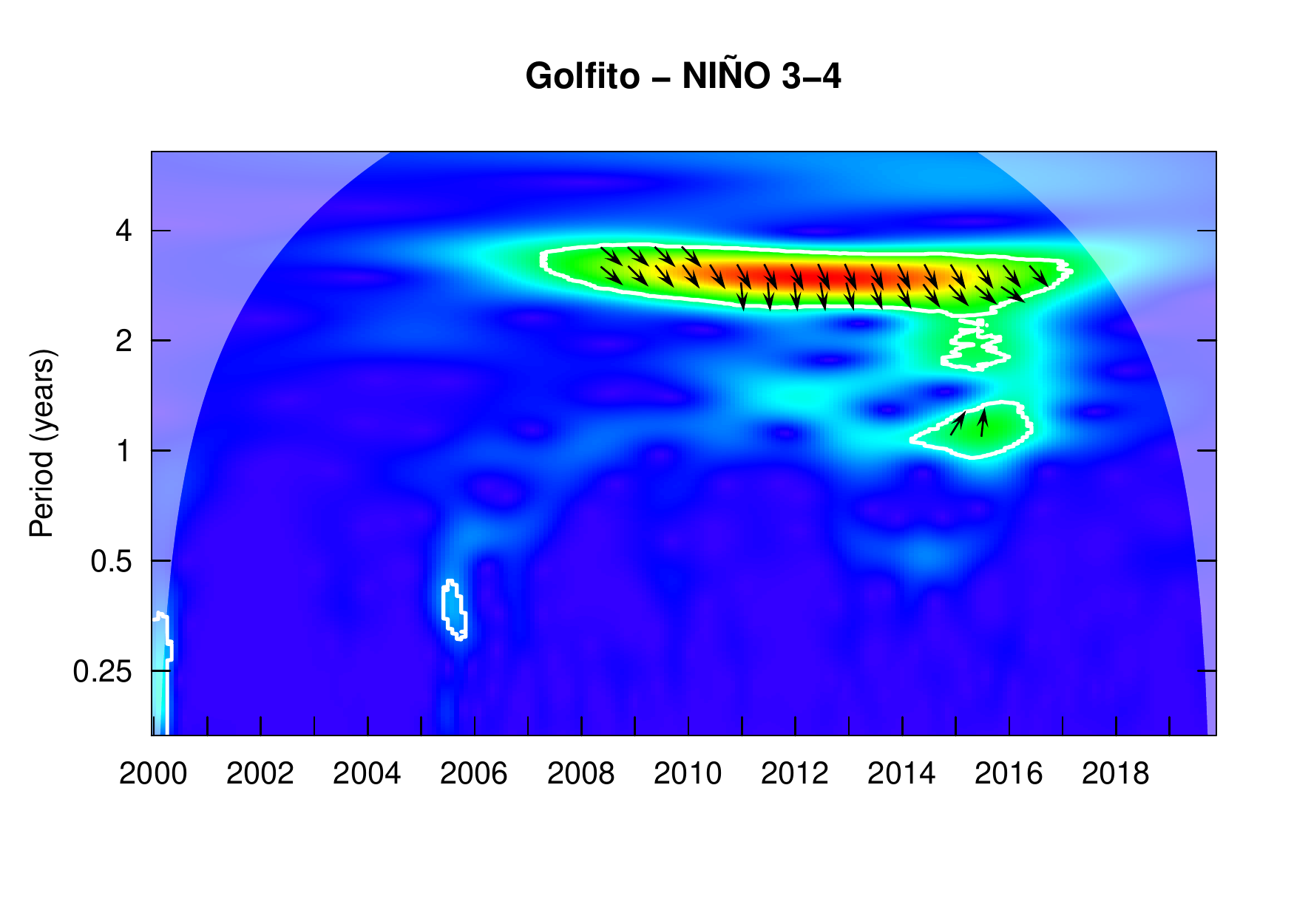}}\vspace{-0.15cm}%
\subfloat[]{\includegraphics[scale=0.23]{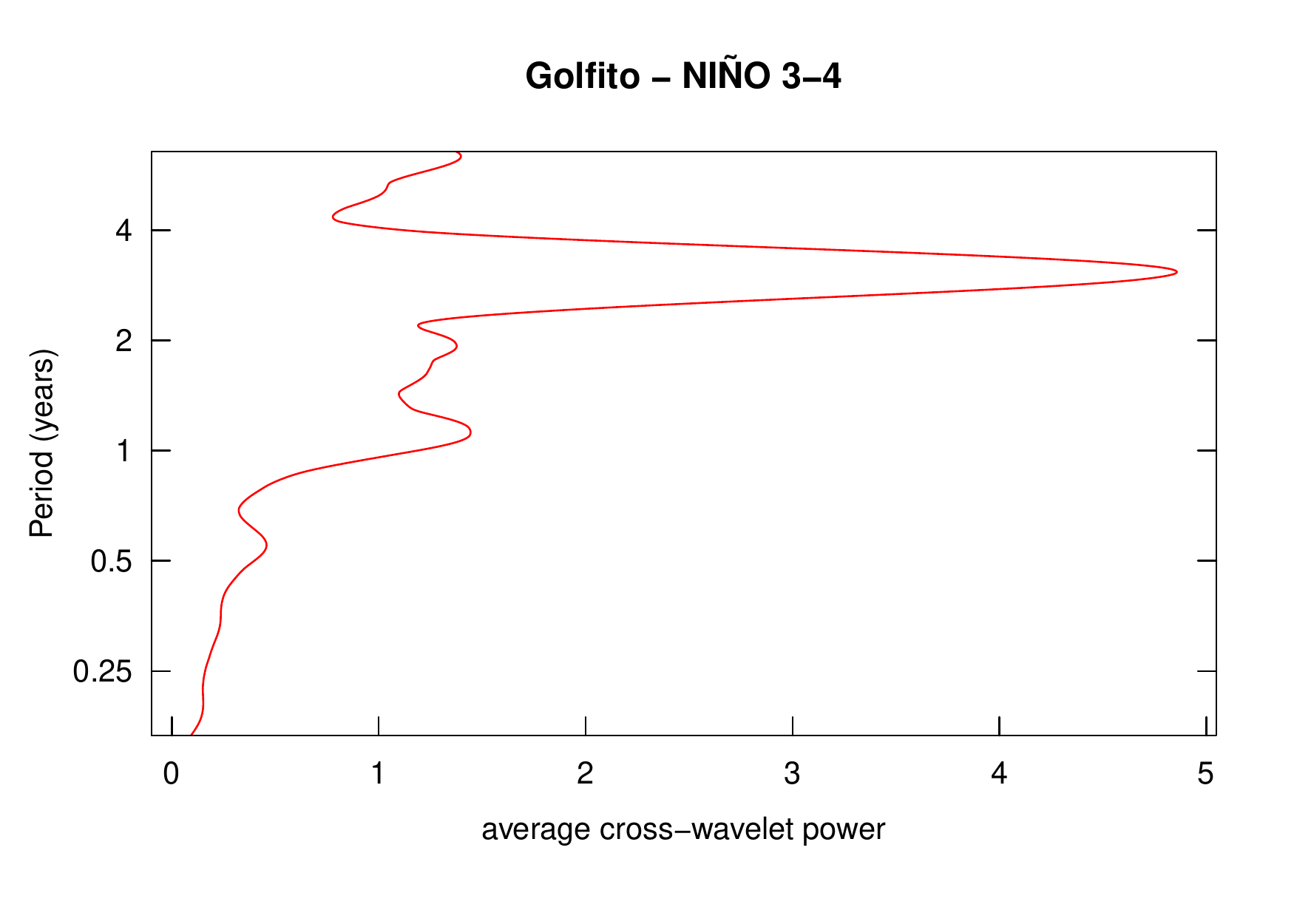}}\vspace{-0.15cm}\\
\subfloat[]{\includegraphics[scale=0.23]{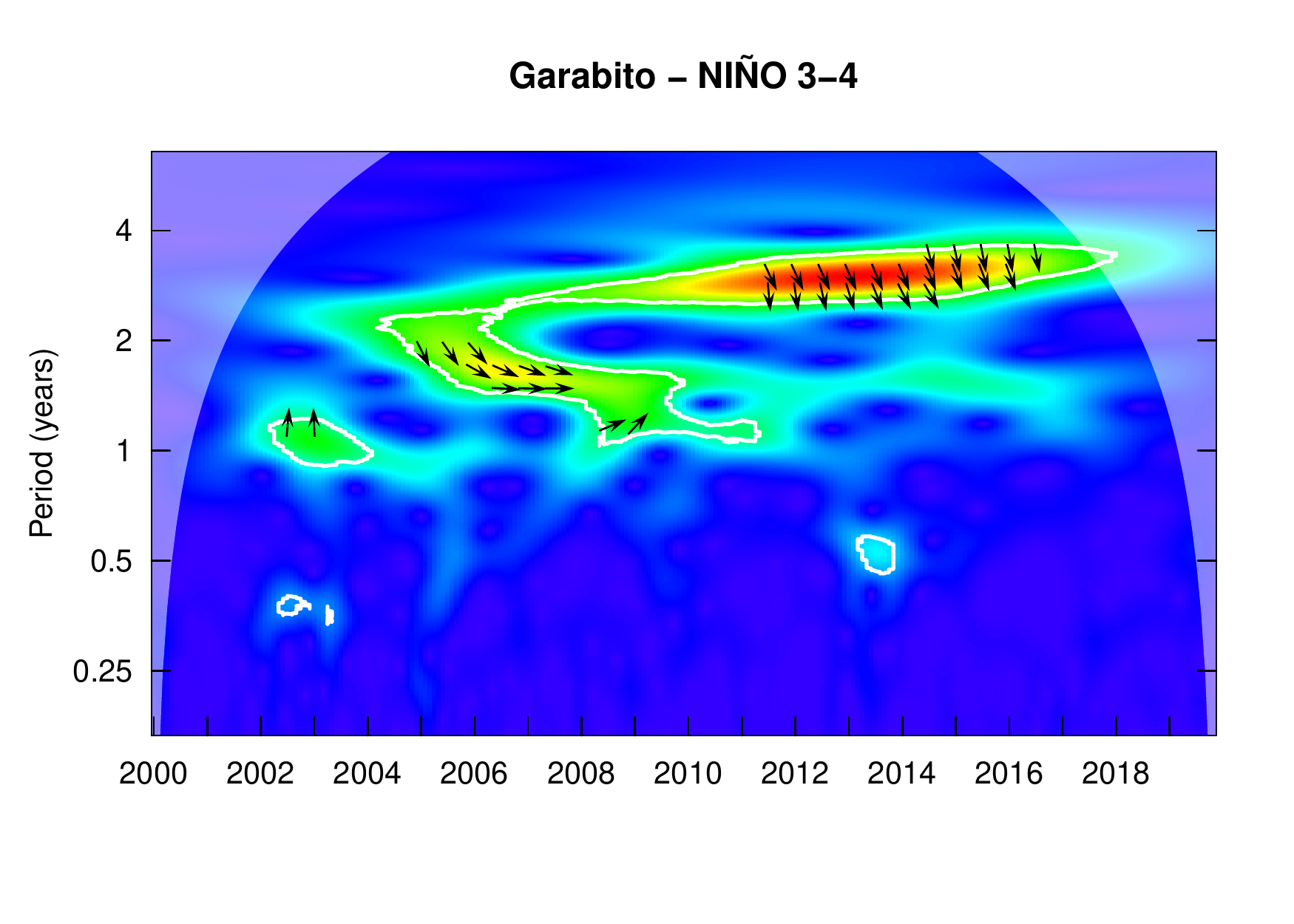}}\vspace{-0.15cm}%
\subfloat[]{\includegraphics[scale=0.23]{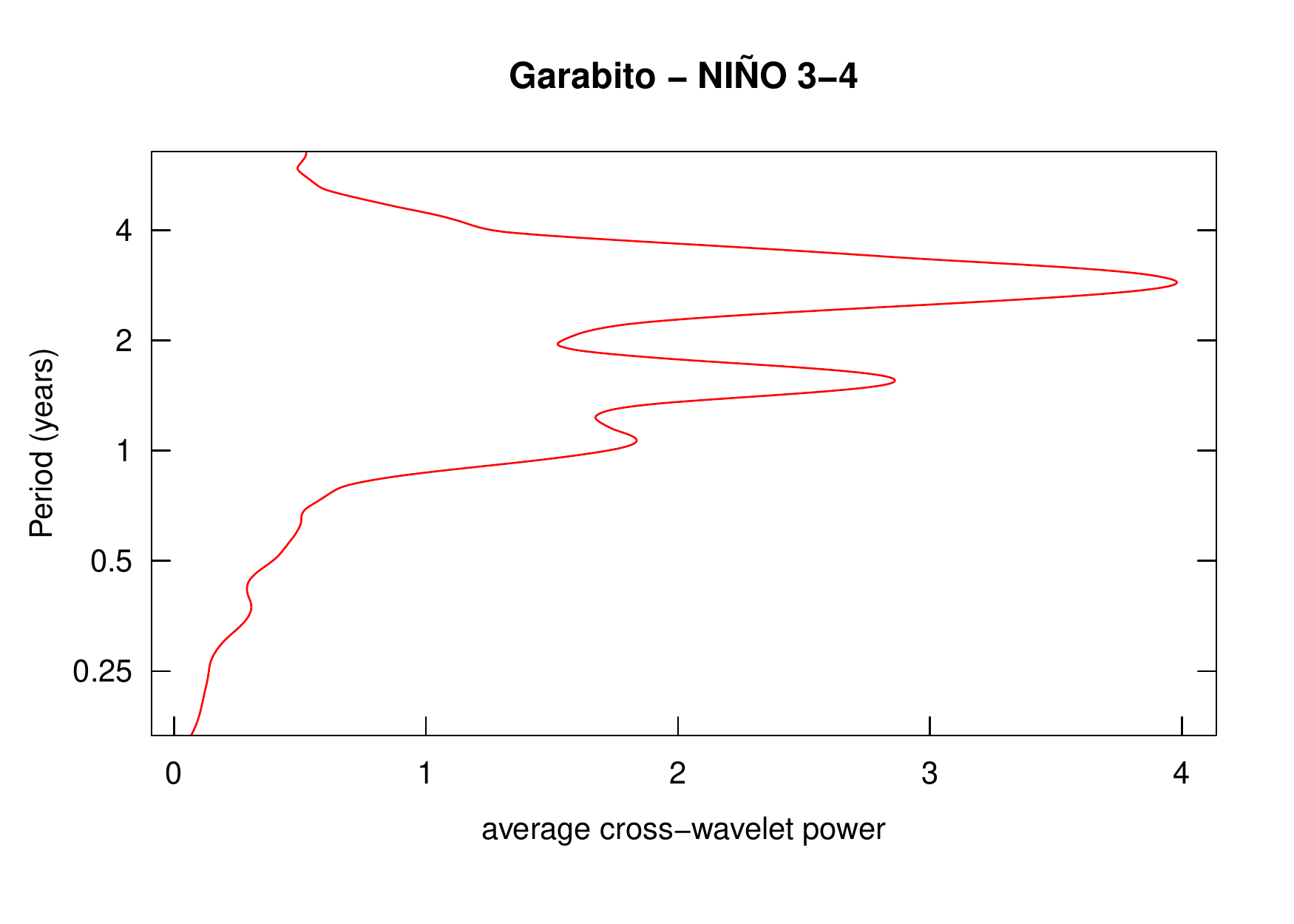}}\vspace{-0.15cm}%
\subfloat[]{\includegraphics[scale=0.23]{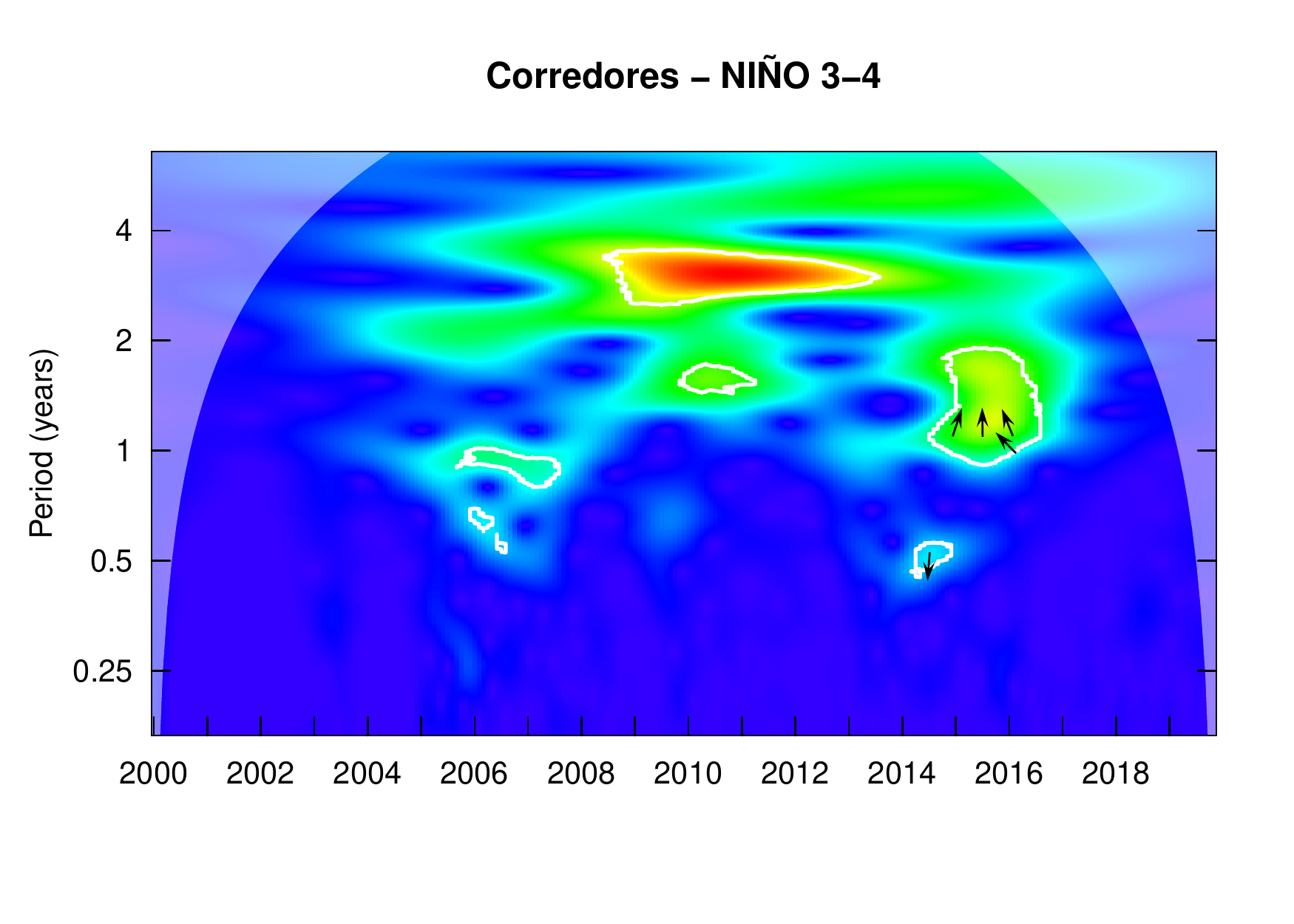}}\vspace{-0.15cm}%
\subfloat[]{\includegraphics[scale=0.23]{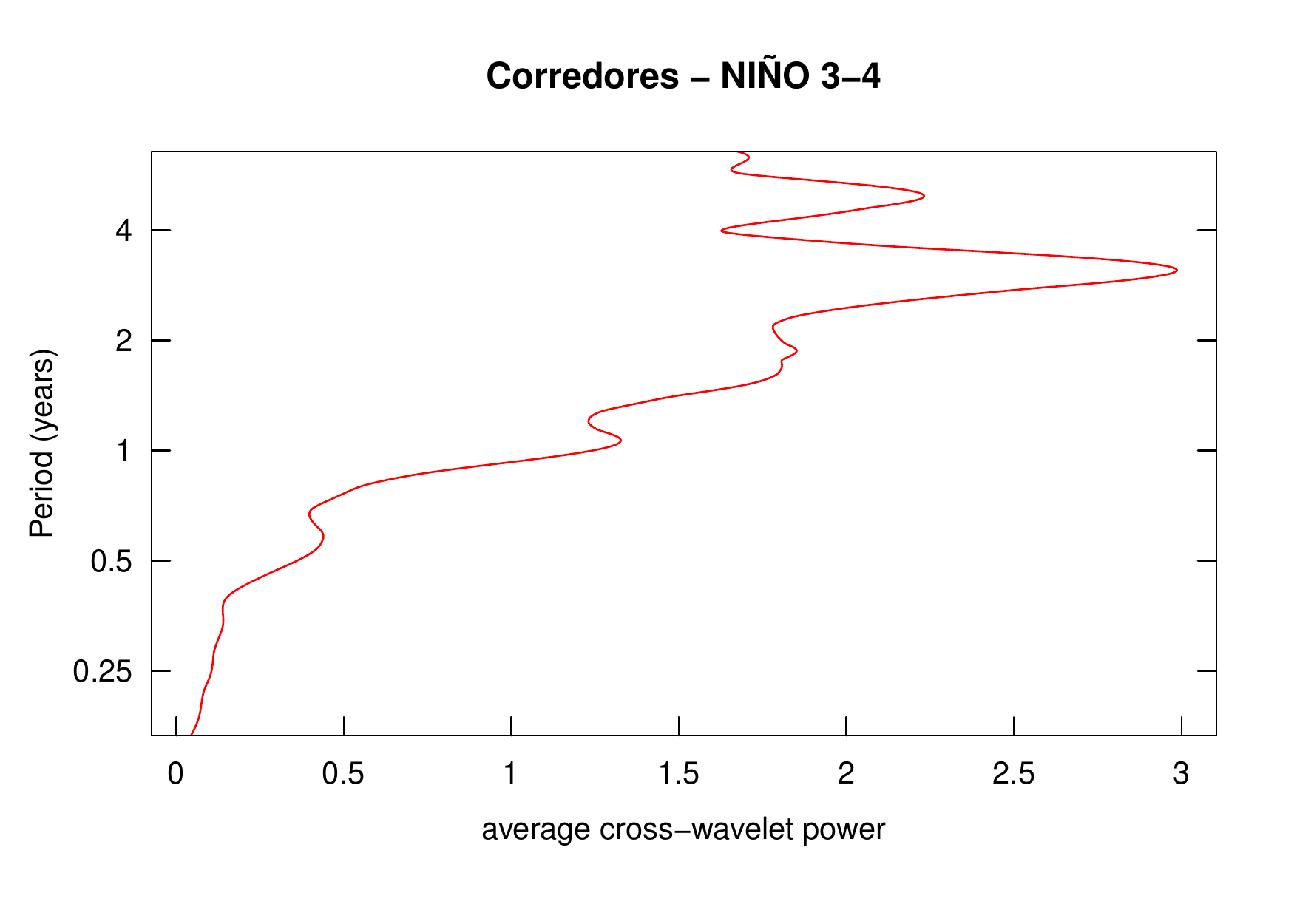}}\vspace{-0.15cm}
\caption*{}
\end{figure}

\begin{figure}[H]
\captionsetup[subfigure]{labelformat=empty}
\subfloat[]{\includegraphics[scale=0.23]{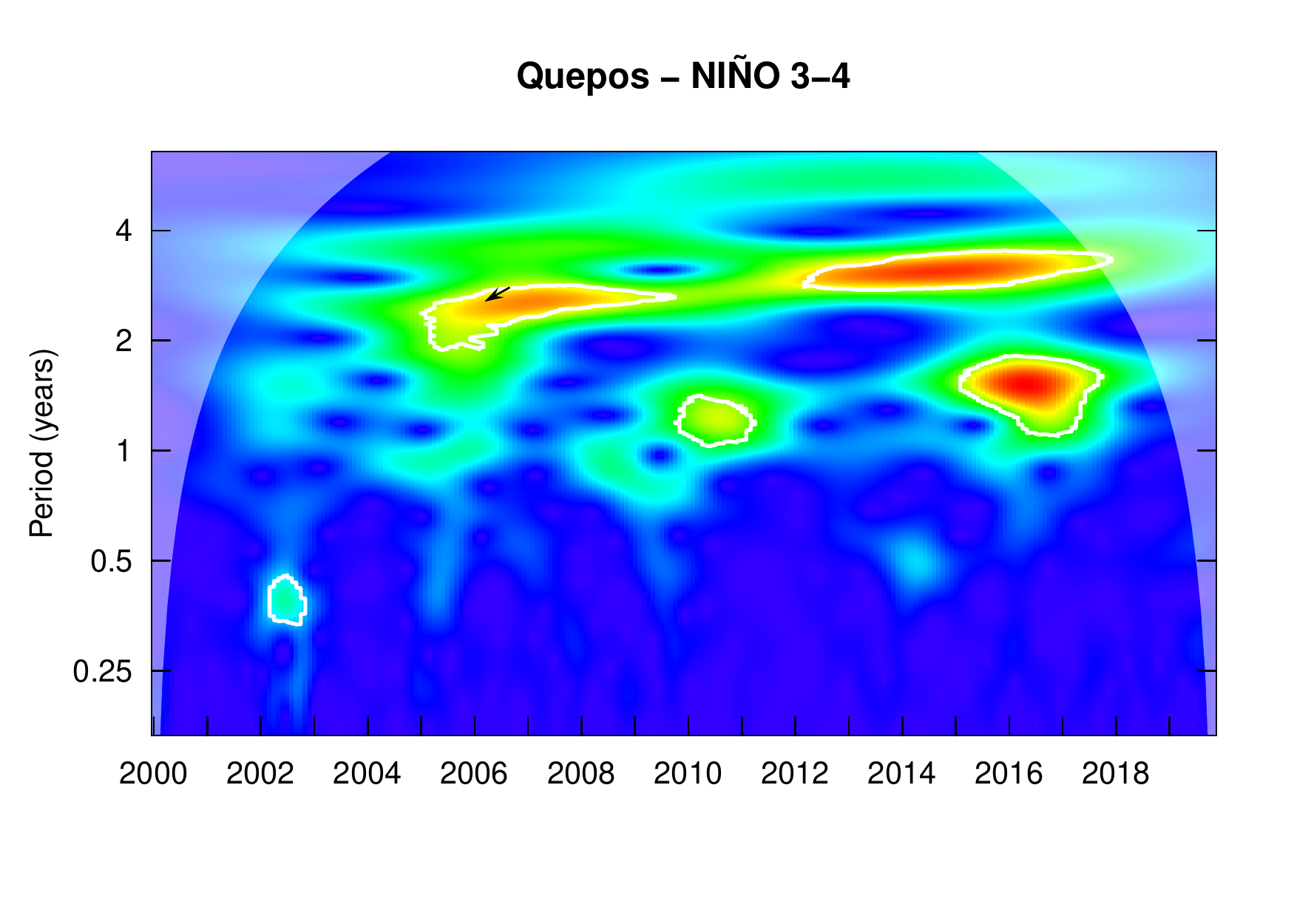}}\vspace{-0.15cm}%
\subfloat[]{\includegraphics[scale=0.23]{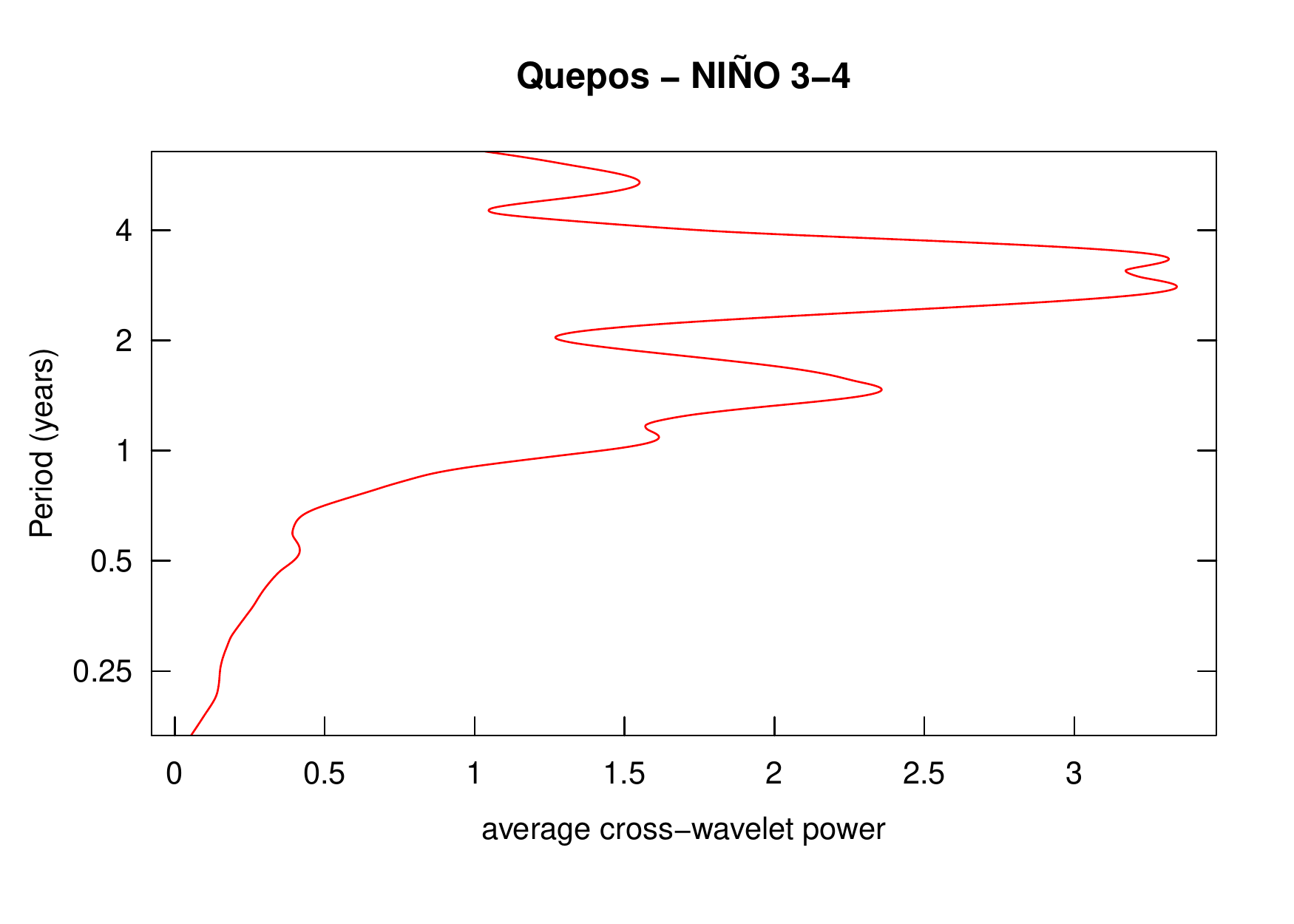}}\vspace{-0.15cm}%
\subfloat[]{\includegraphics[scale=0.23]{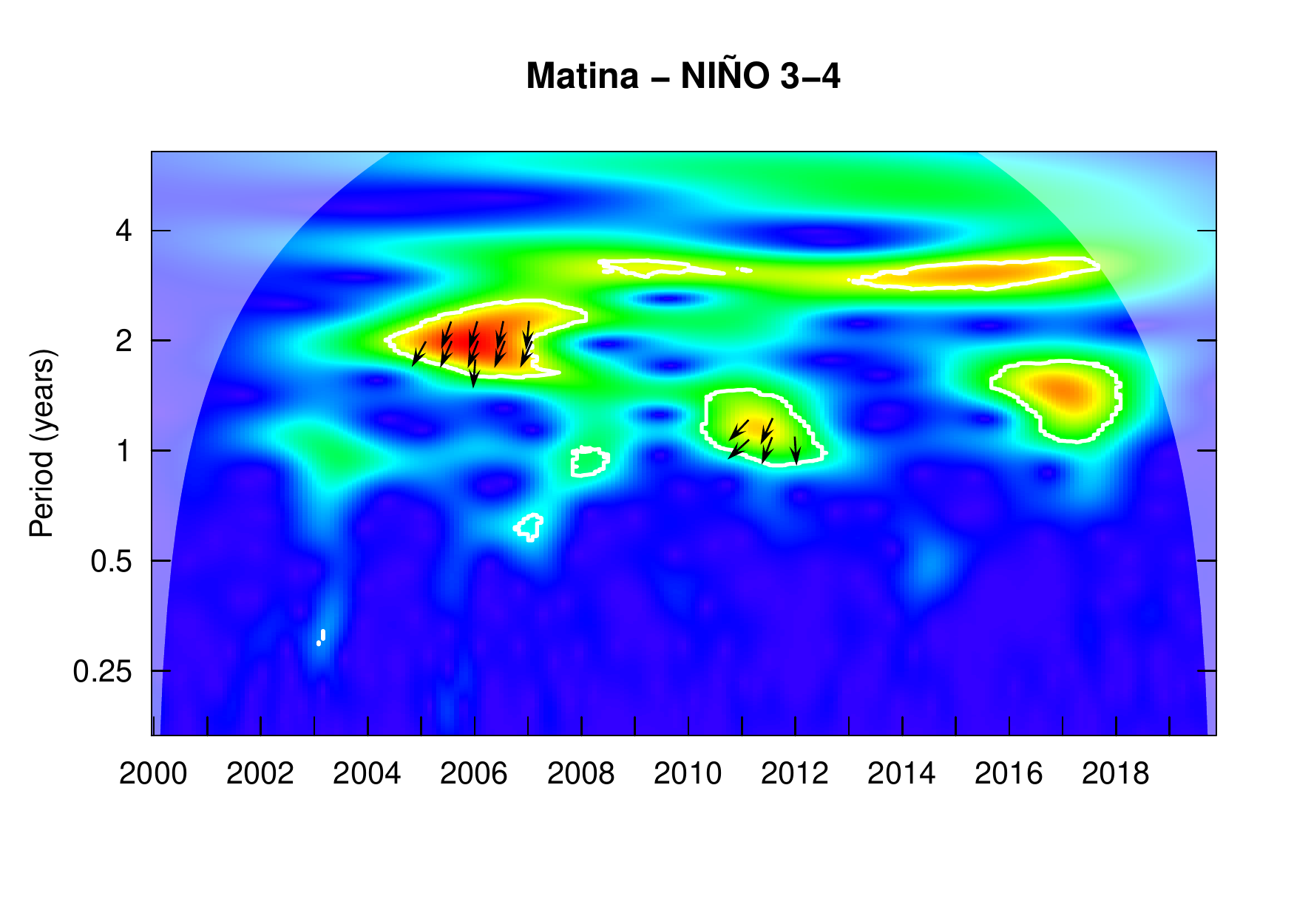}}\vspace{-0.15cm}%
\subfloat[]{\includegraphics[scale=0.23]{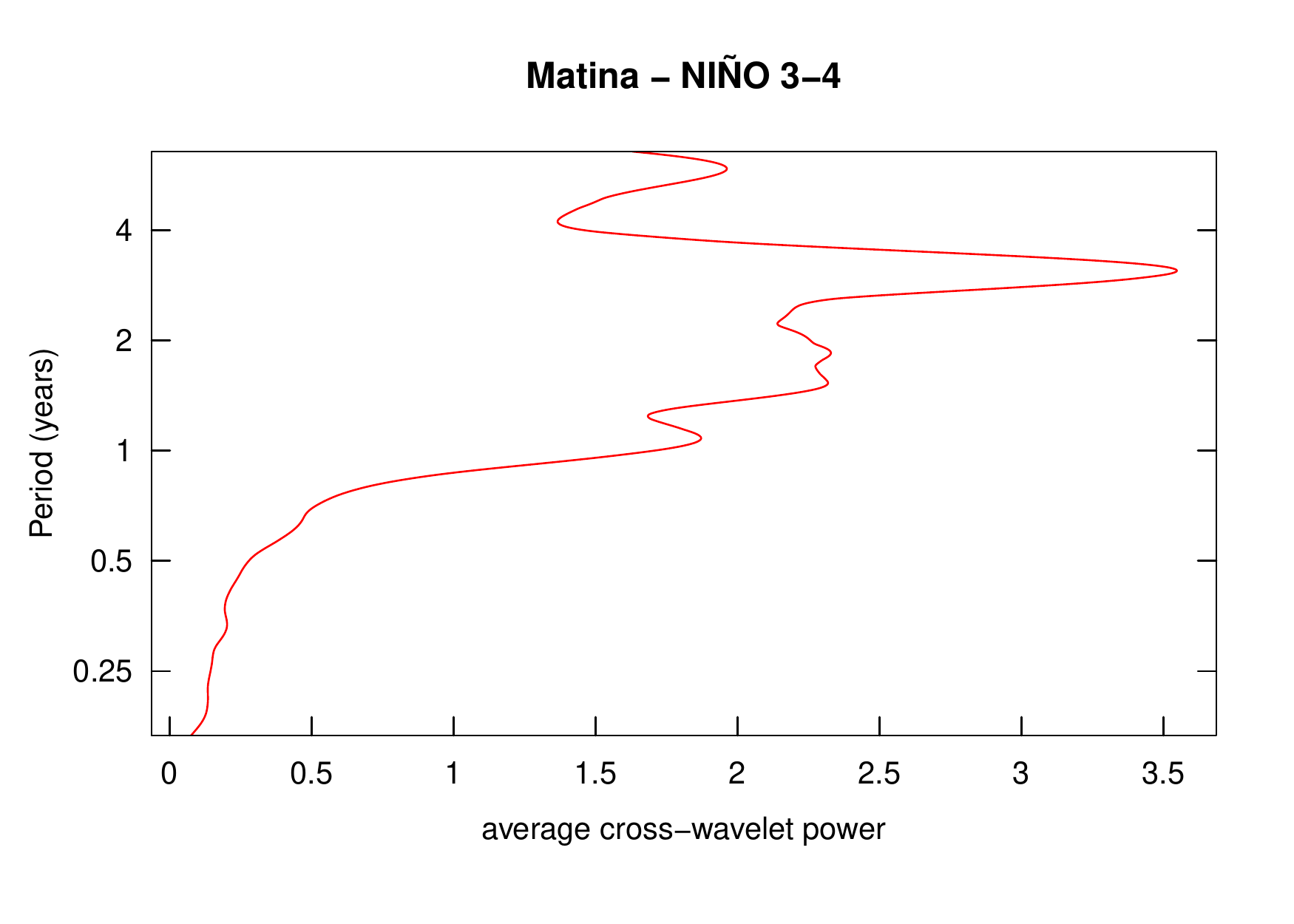}}\vspace{-0.15cm}\\
\subfloat[]{\includegraphics[scale=0.23]{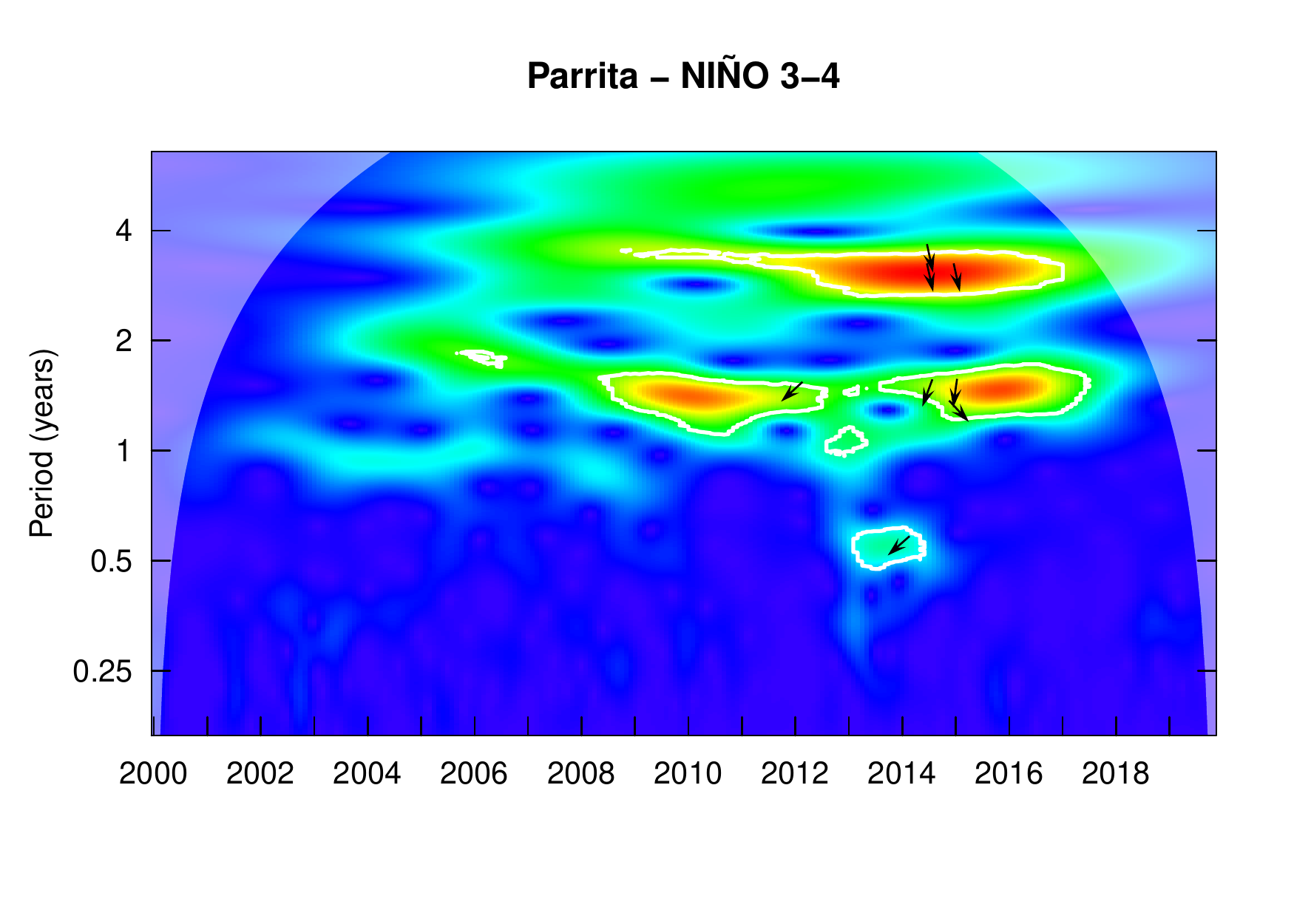}}\vspace{-0.15cm}%
\subfloat[]{\includegraphics[scale=0.23]{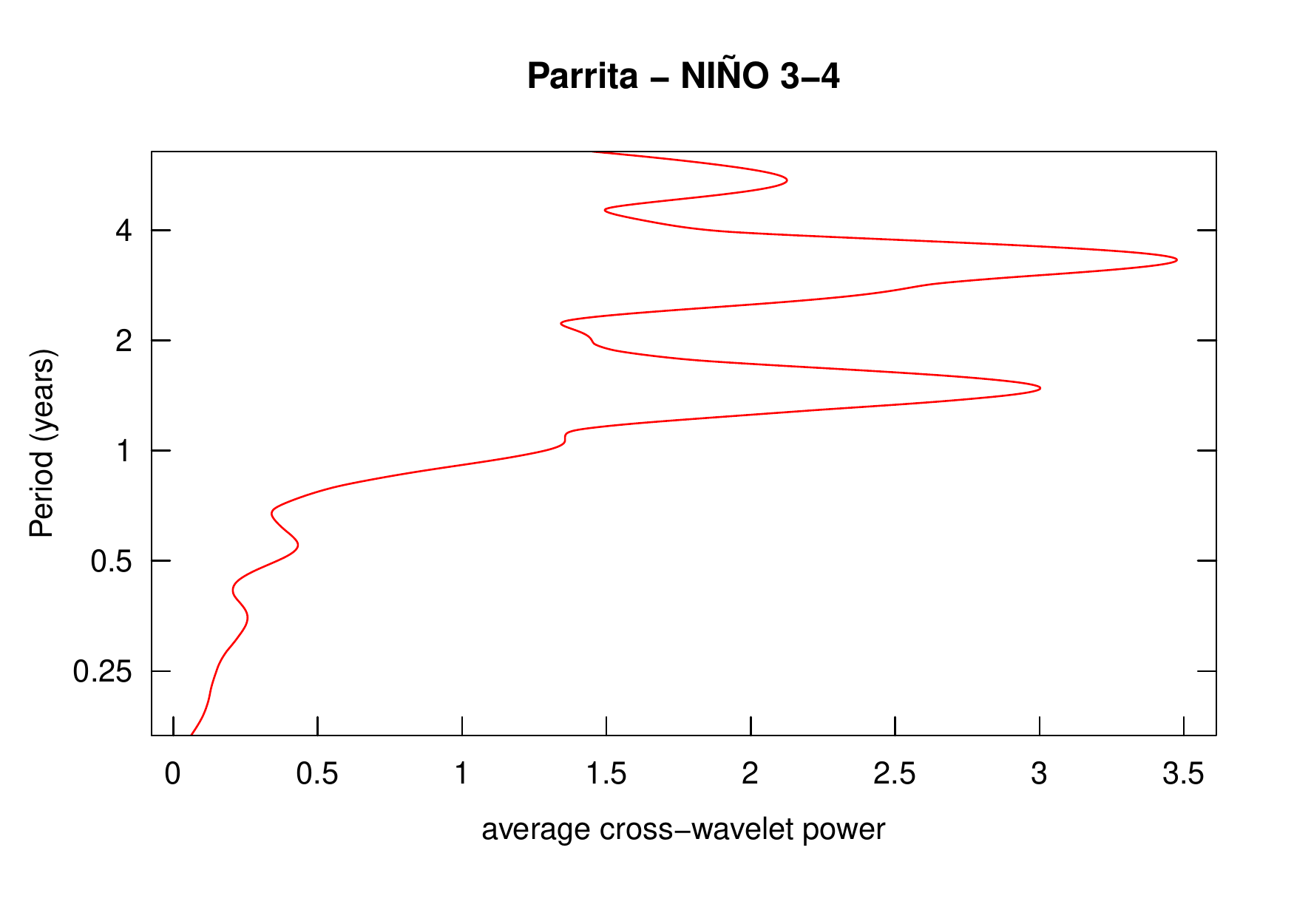}}\vspace{-0.15cm}%
\subfloat[]{\includegraphics[scale=0.23]{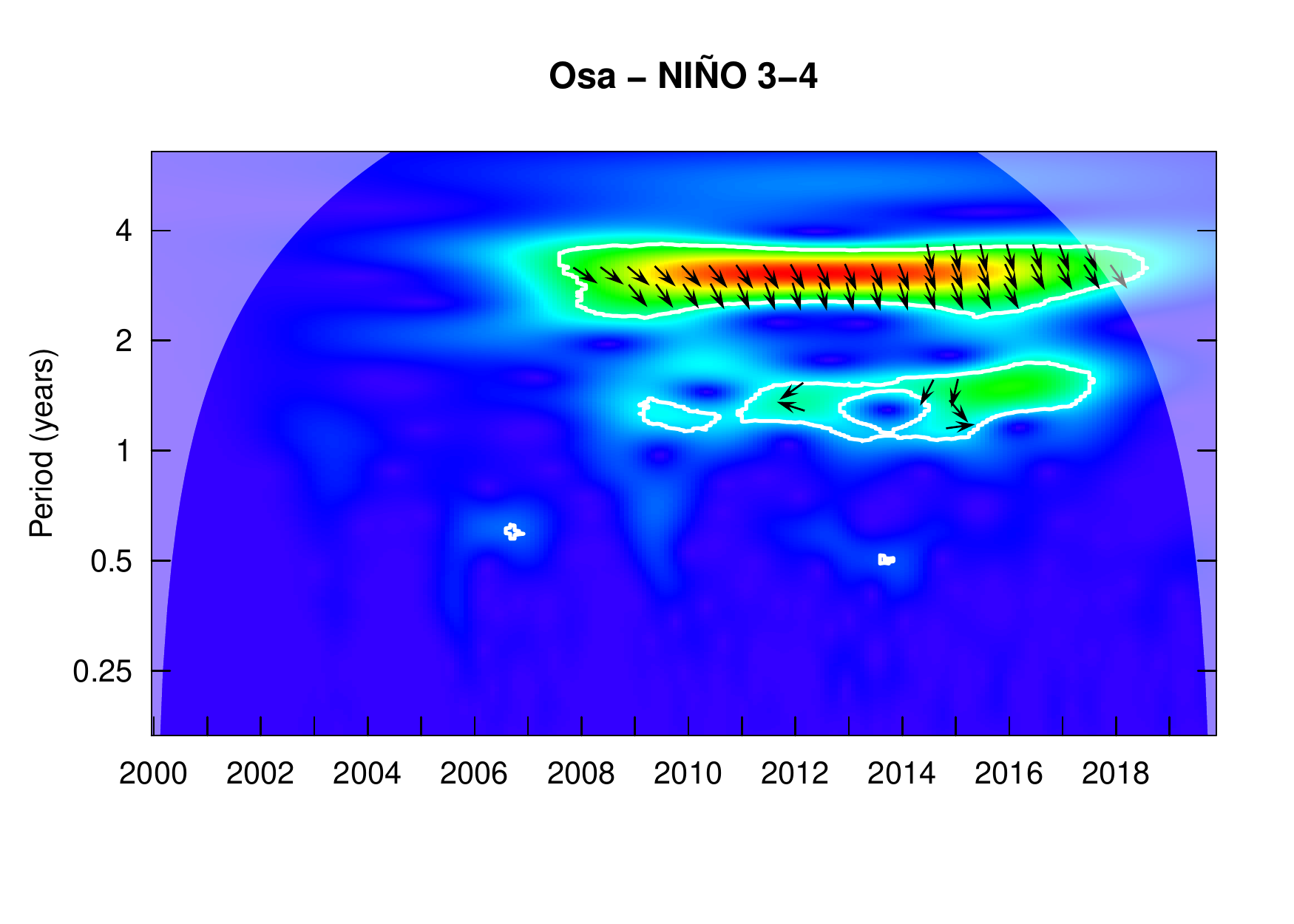}}\vspace{-0.15cm}%
\subfloat[]{\includegraphics[scale=0.23]{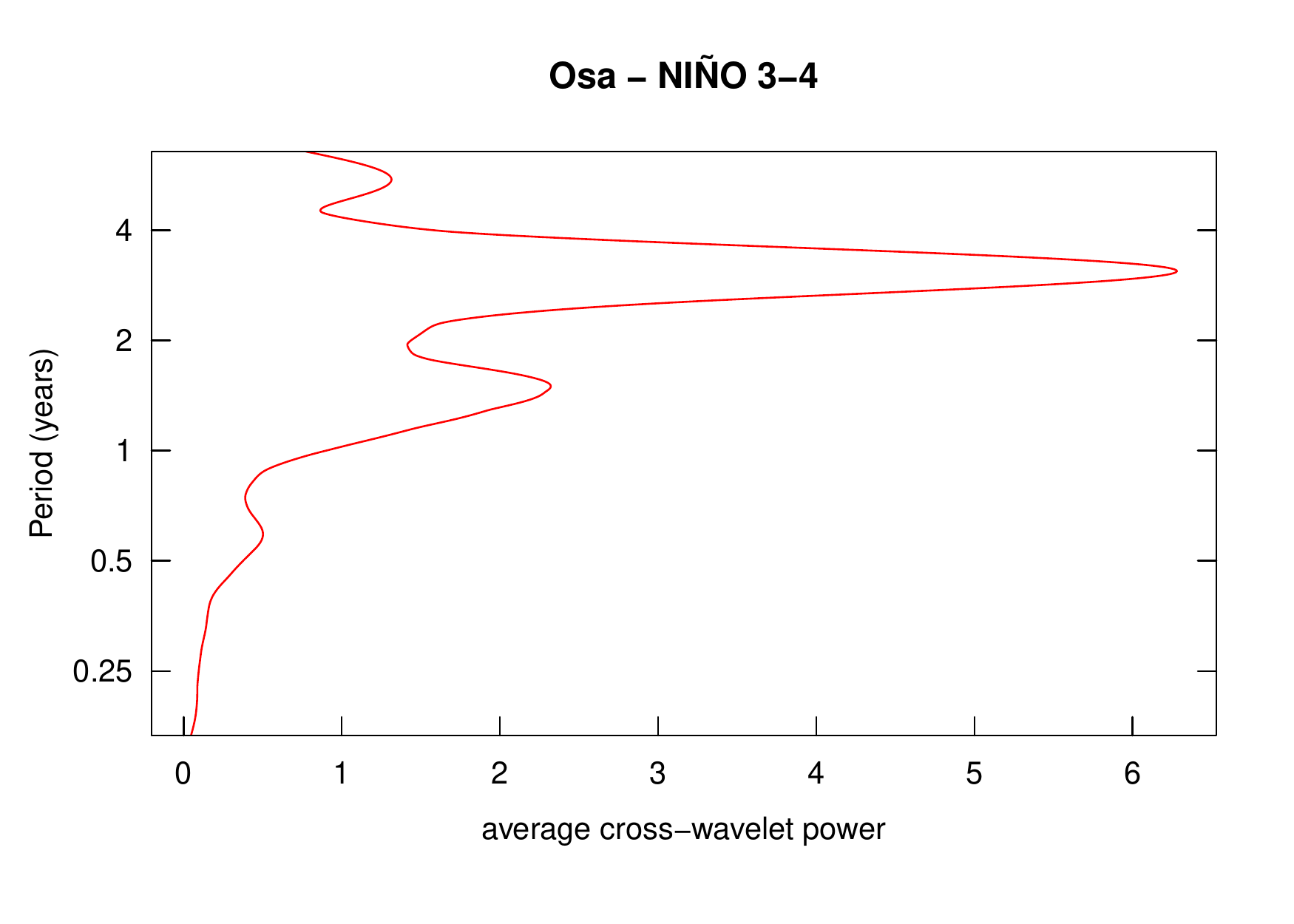}}\vspace{-0.15cm}%
\caption*{}
\end{figure}

\section*{Wavelet coherence and average cross-wavelet power between dengue incidence and Ni\~no 4}

\begin{figure}[H]
\captionsetup[subfigure]{labelformat=empty}
\caption*{\textbf{Figure S10:} Wavelet coherence (color map) between dengue incidence from 2000 to 2019, and Ni\~no 4 in 32 municipalities of Costa Rica (periodicity on y-axis, time on x-axis). Colors code for increasing power intensity, from blue to red; $95\%$ confidence levels are encircled by white lines, and shaded areas indicate the presence of significant edge effects. On the right side of each wavelet coherence is the average cross-wavelet power (Red line). The arrows indicate whether the two series are in-phase or out-phase.}
\subfloat[]{\includegraphics[scale=0.23]{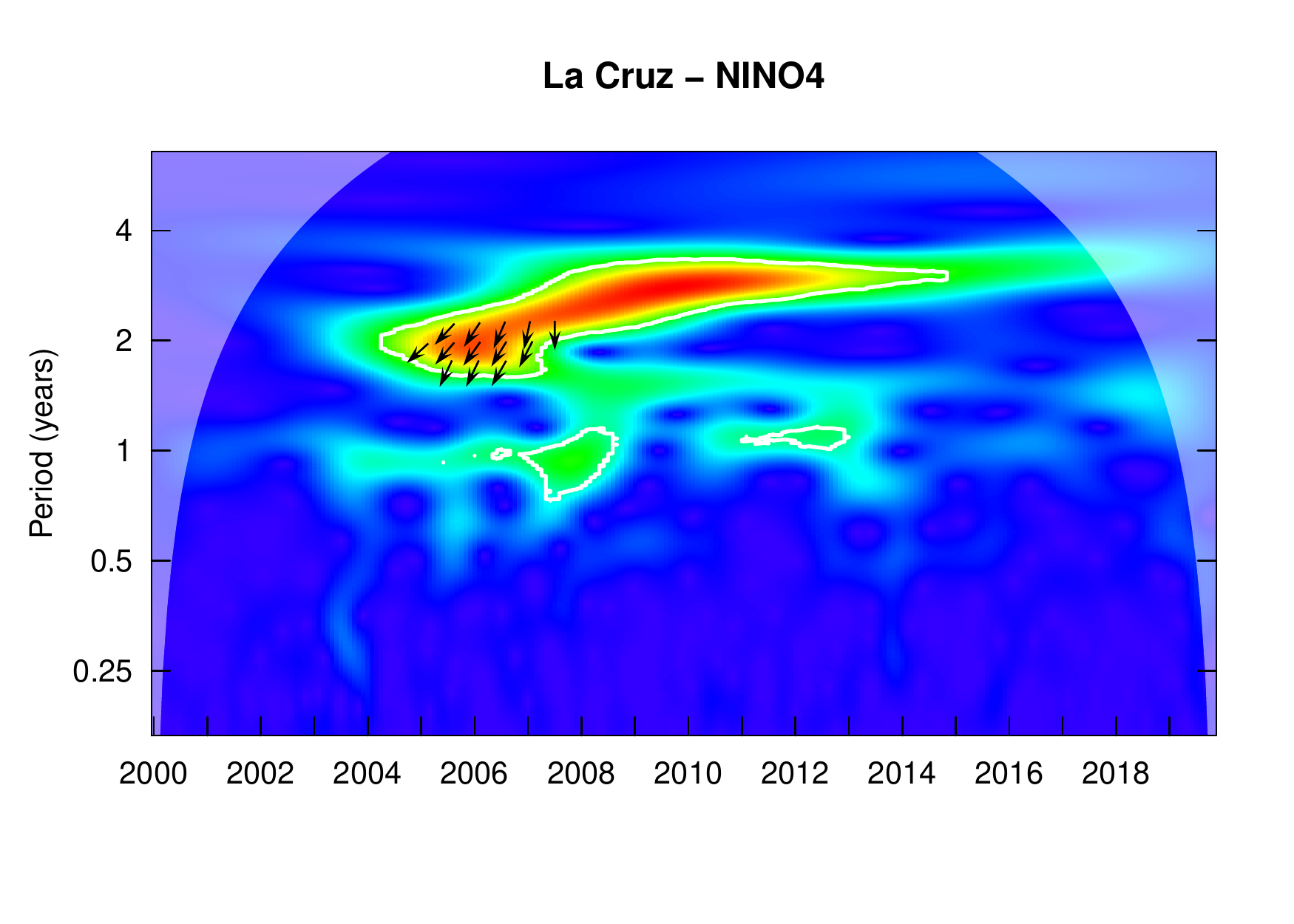}}\vspace{-0.15cm}%
\subfloat[]{\includegraphics[scale=0.23]{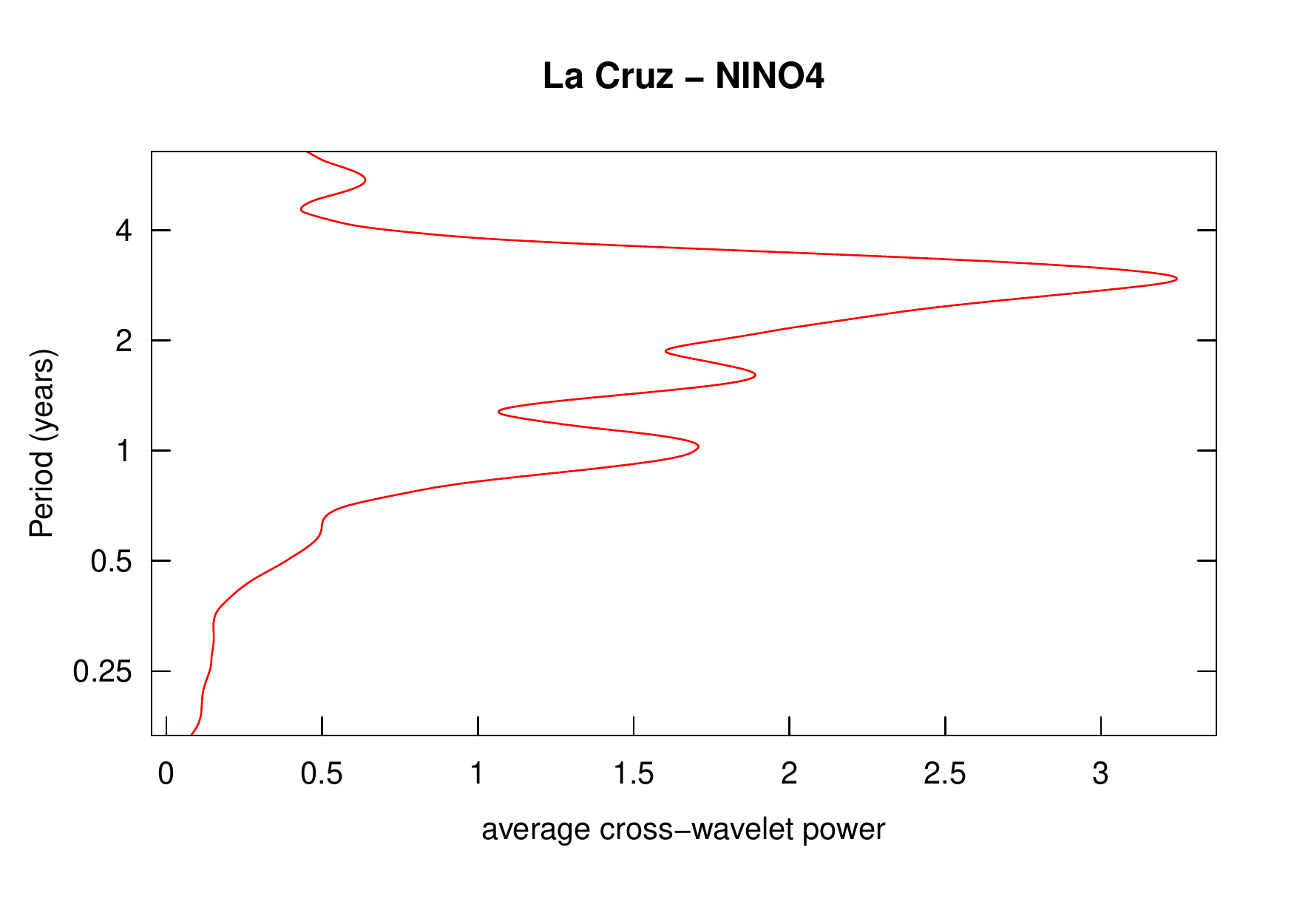}}\vspace{-0.15cm}%
\subfloat[]{\includegraphics[scale=0.23]{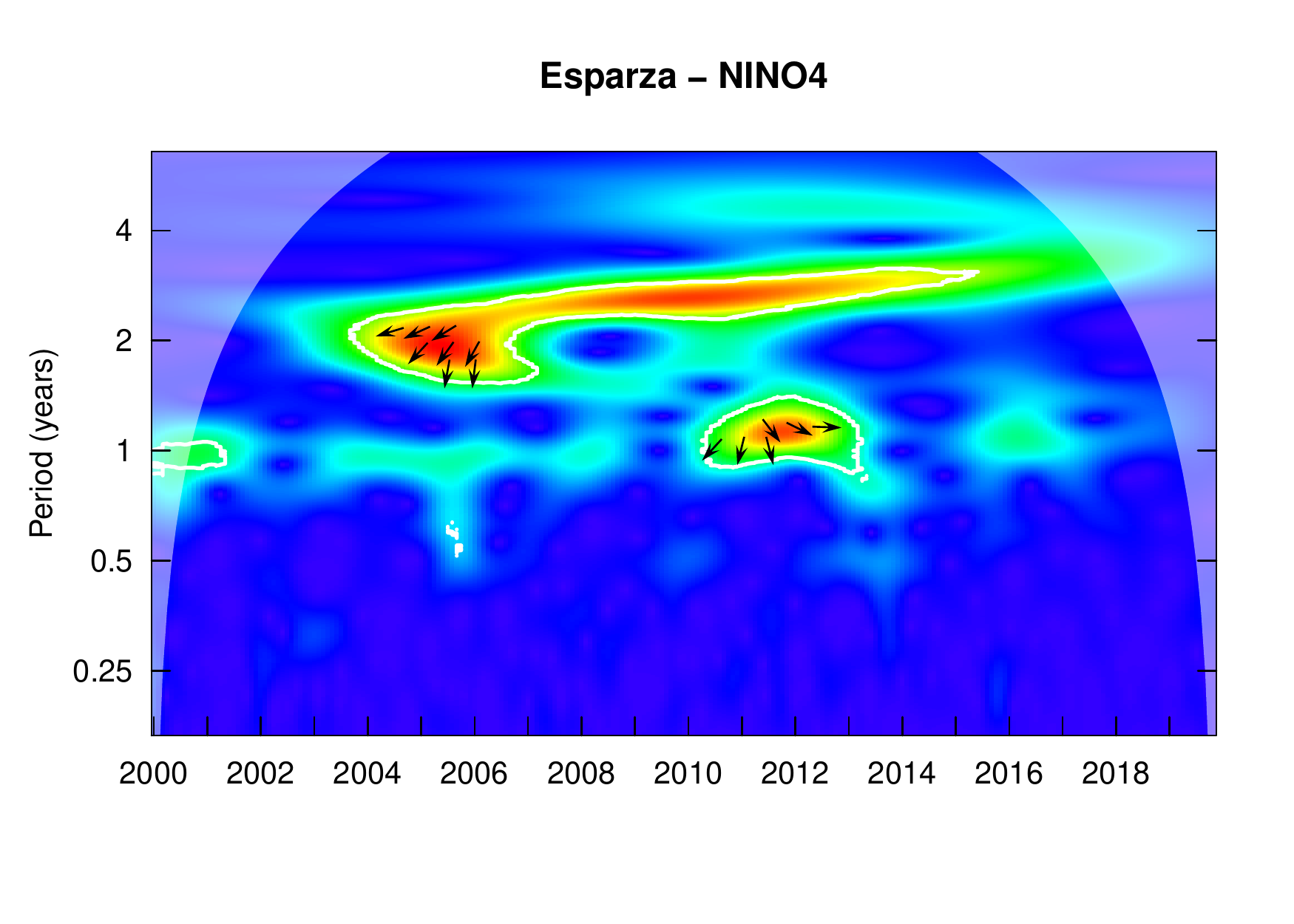}}\vspace{-0.15cm}%
\subfloat[]{\includegraphics[scale=0.23]{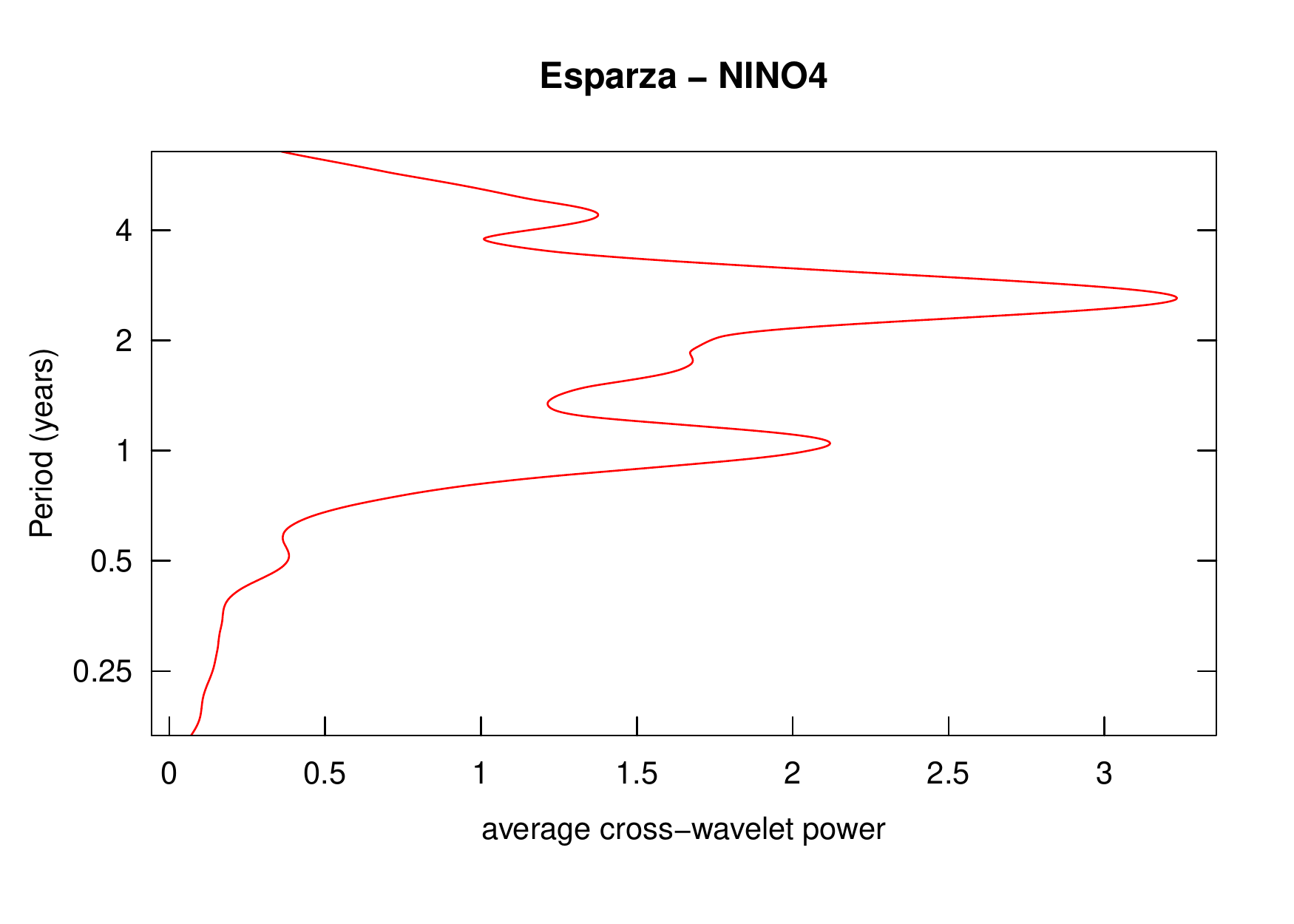}}\vspace{-0.15cm}\\
\subfloat[]{\includegraphics[scale=0.23]{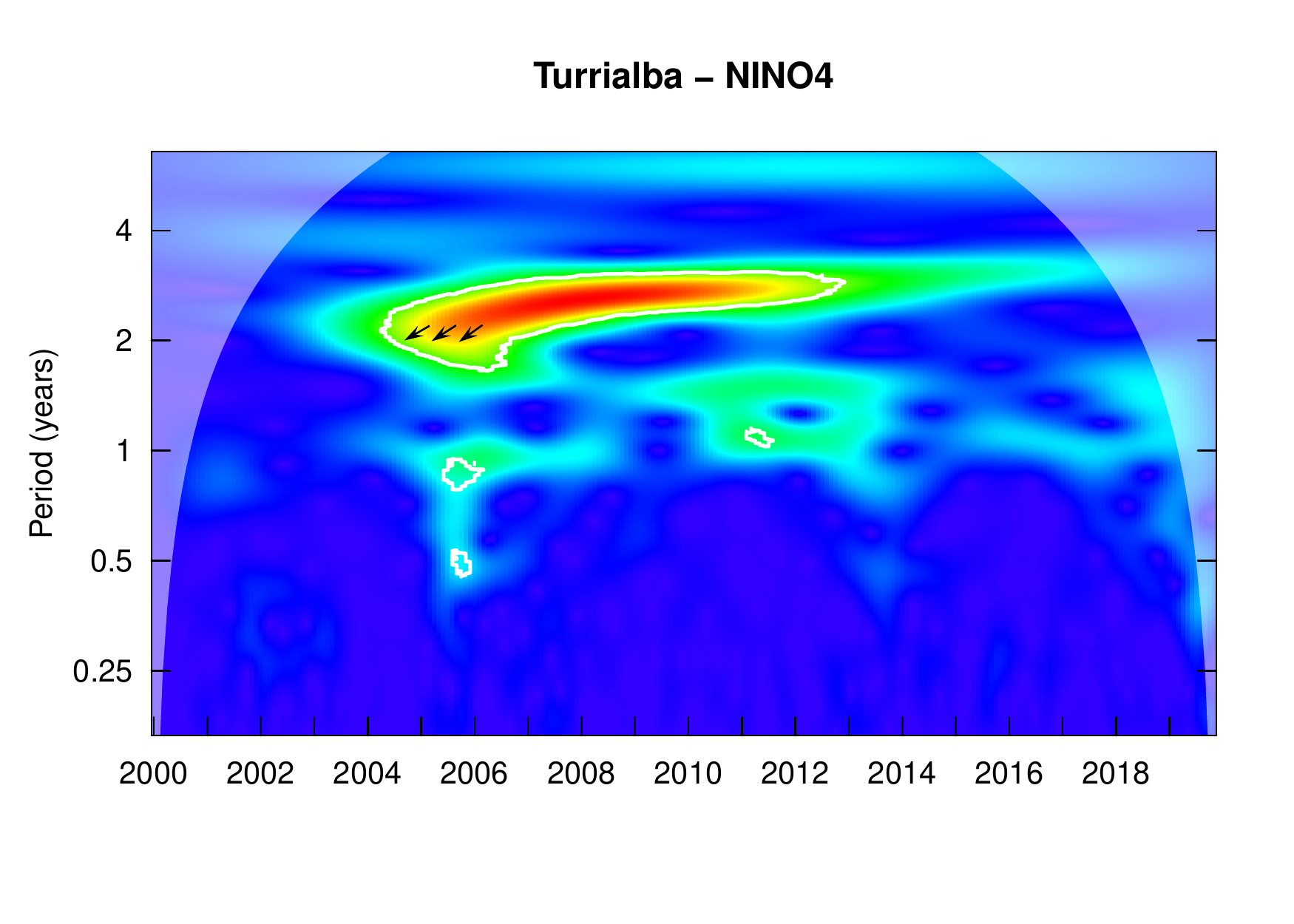}}\vspace{-0.15cm}%
\subfloat[]{\includegraphics[scale=0.23]{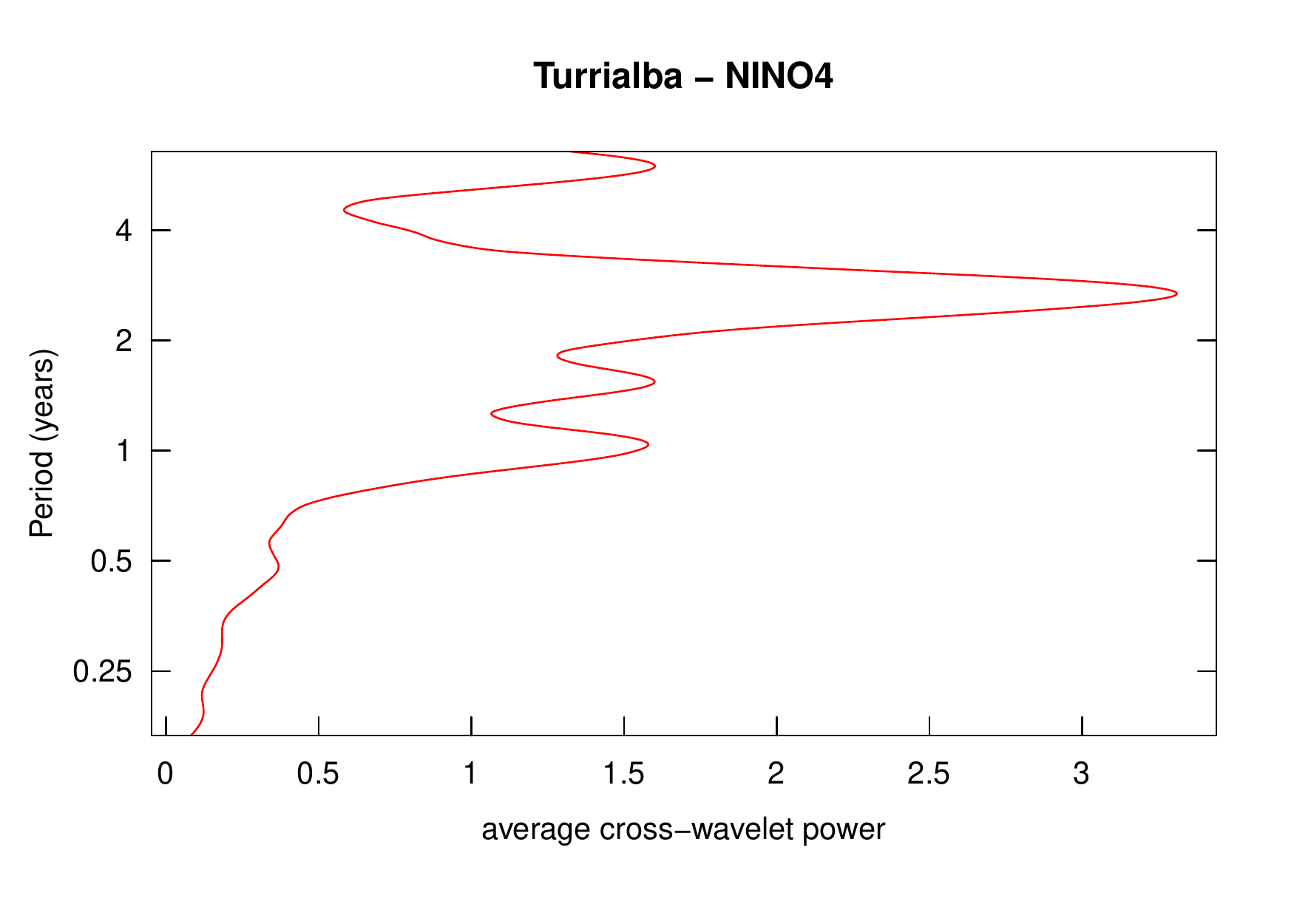}}\vspace{-0.15cm}%
\subfloat[]{\includegraphics[scale=0.23]{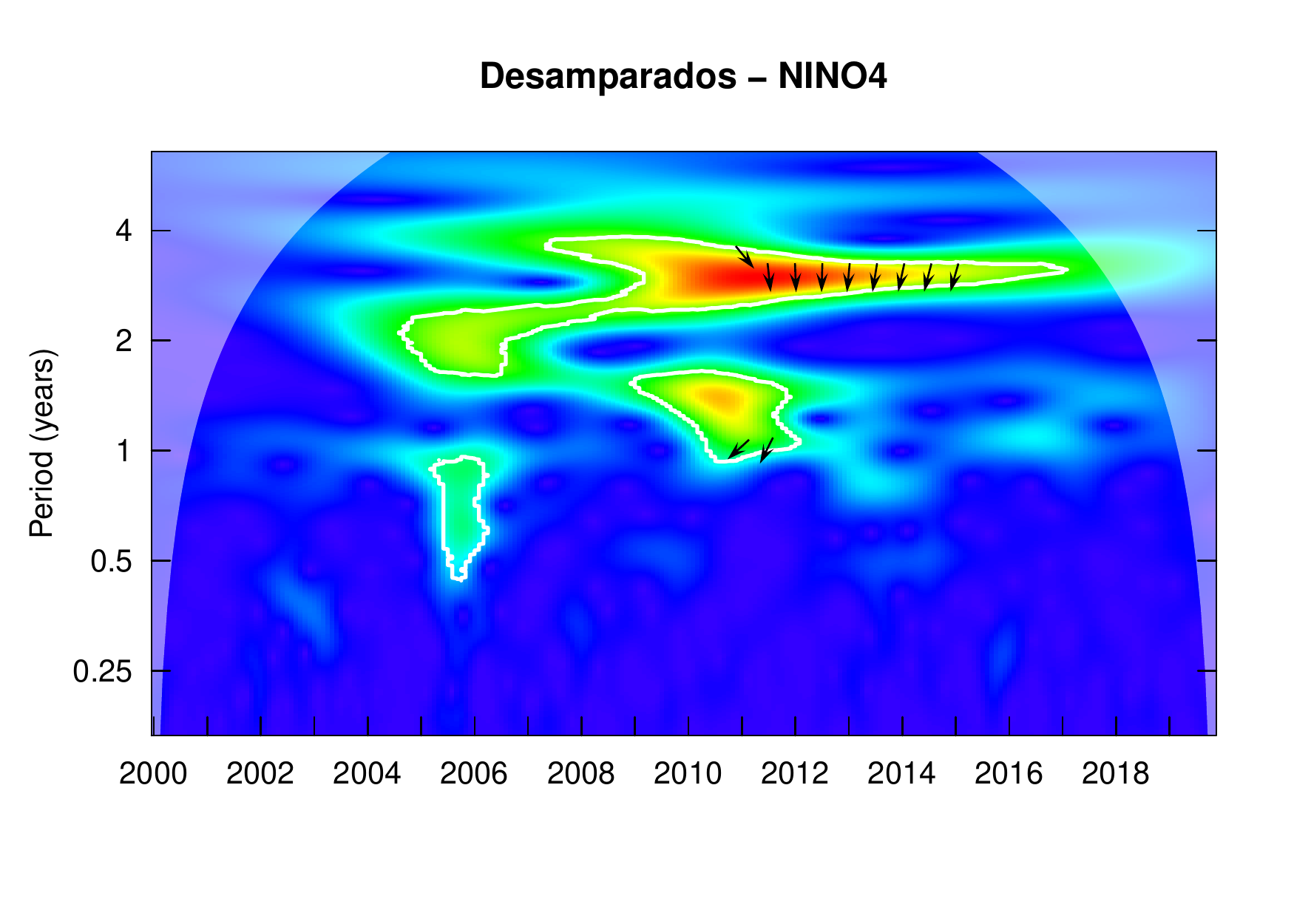}}\vspace{-0.15cm}%
\subfloat[]{\includegraphics[scale=0.23]{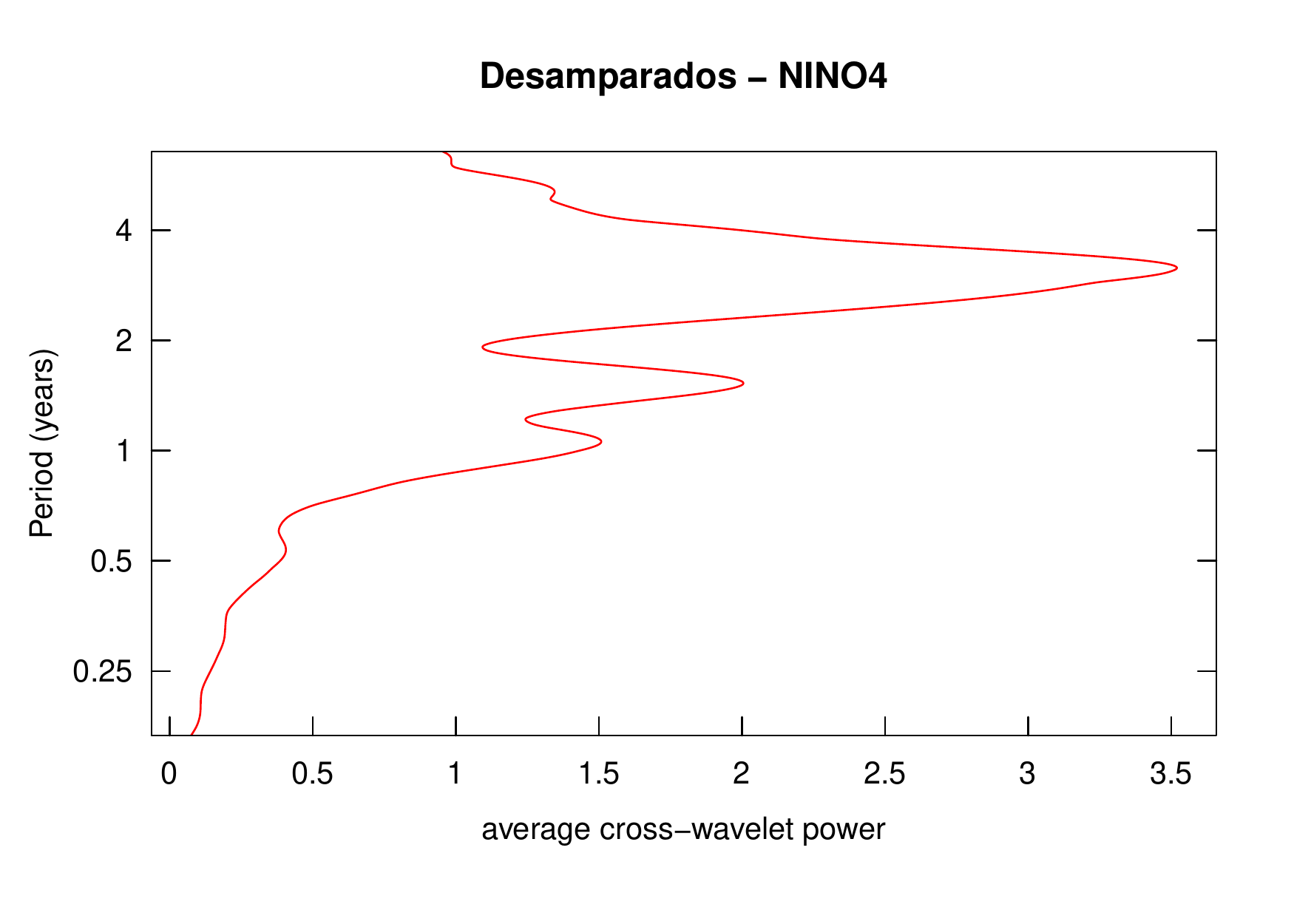}}\vspace{-0.15cm}\\
\subfloat[]{\includegraphics[scale=0.23]{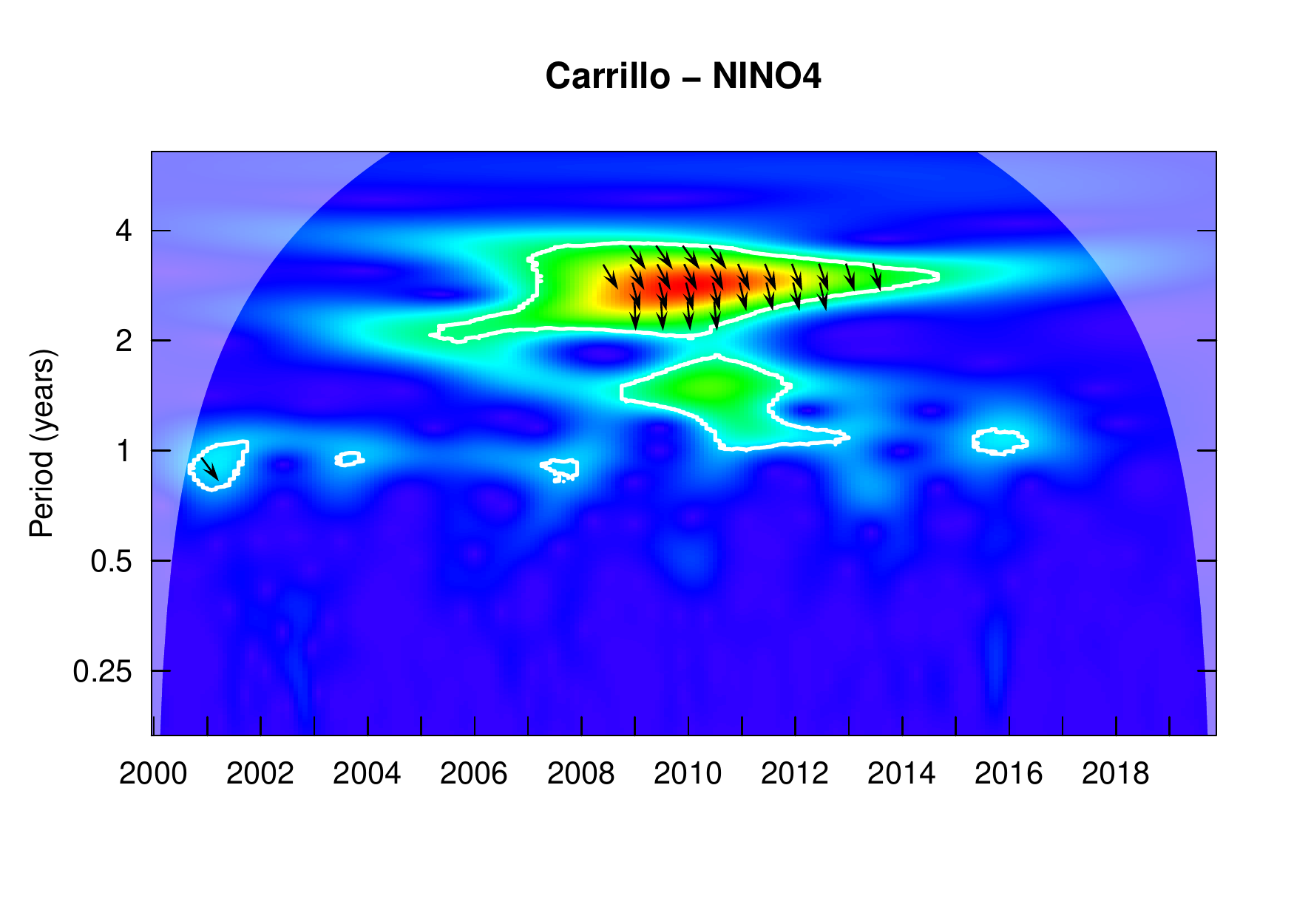}}\vspace{-0.15cm}%
\subfloat[]{\includegraphics[scale=0.23]{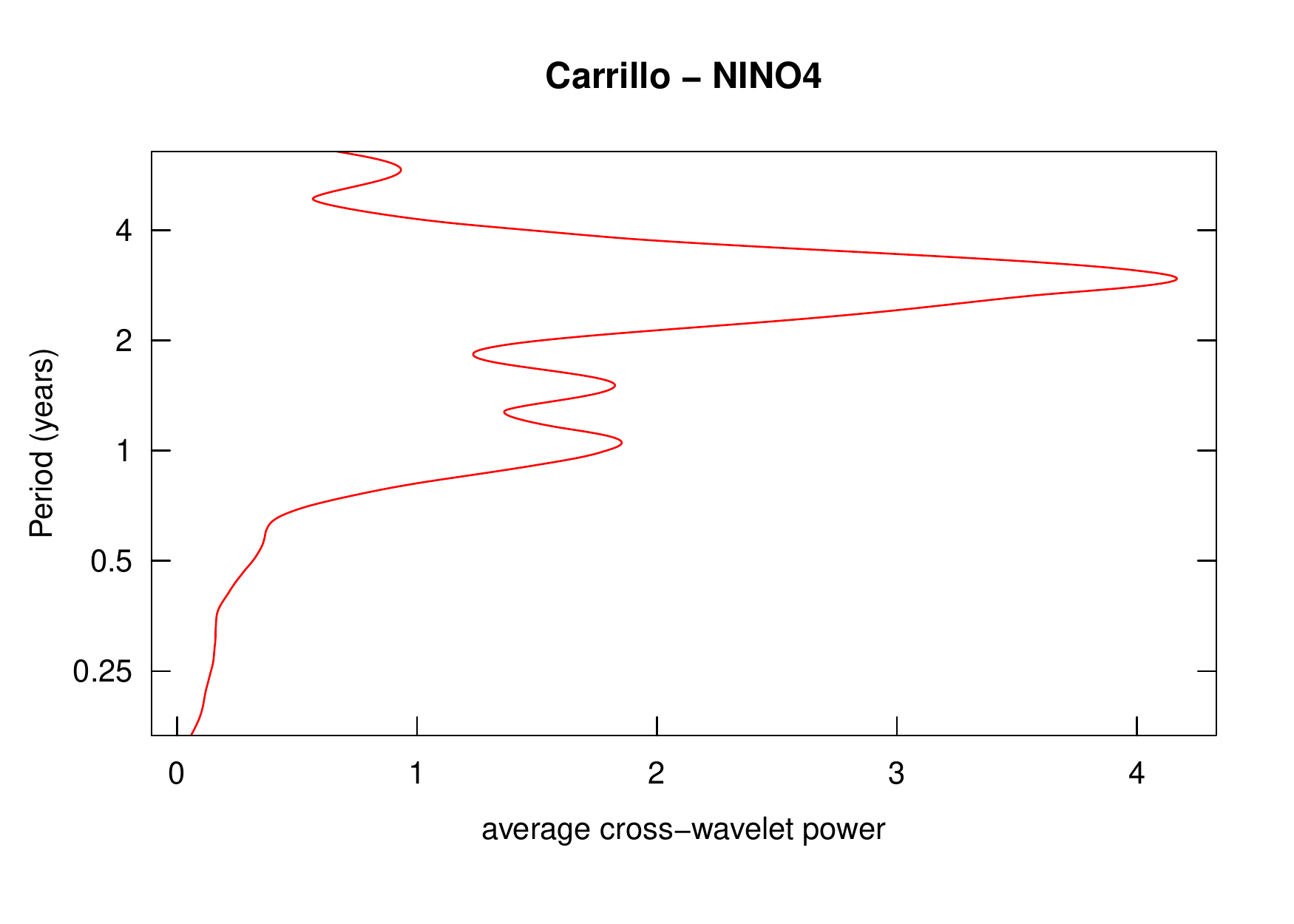}}\vspace{-0.15cm}%
\subfloat[]{\includegraphics[scale=0.23]{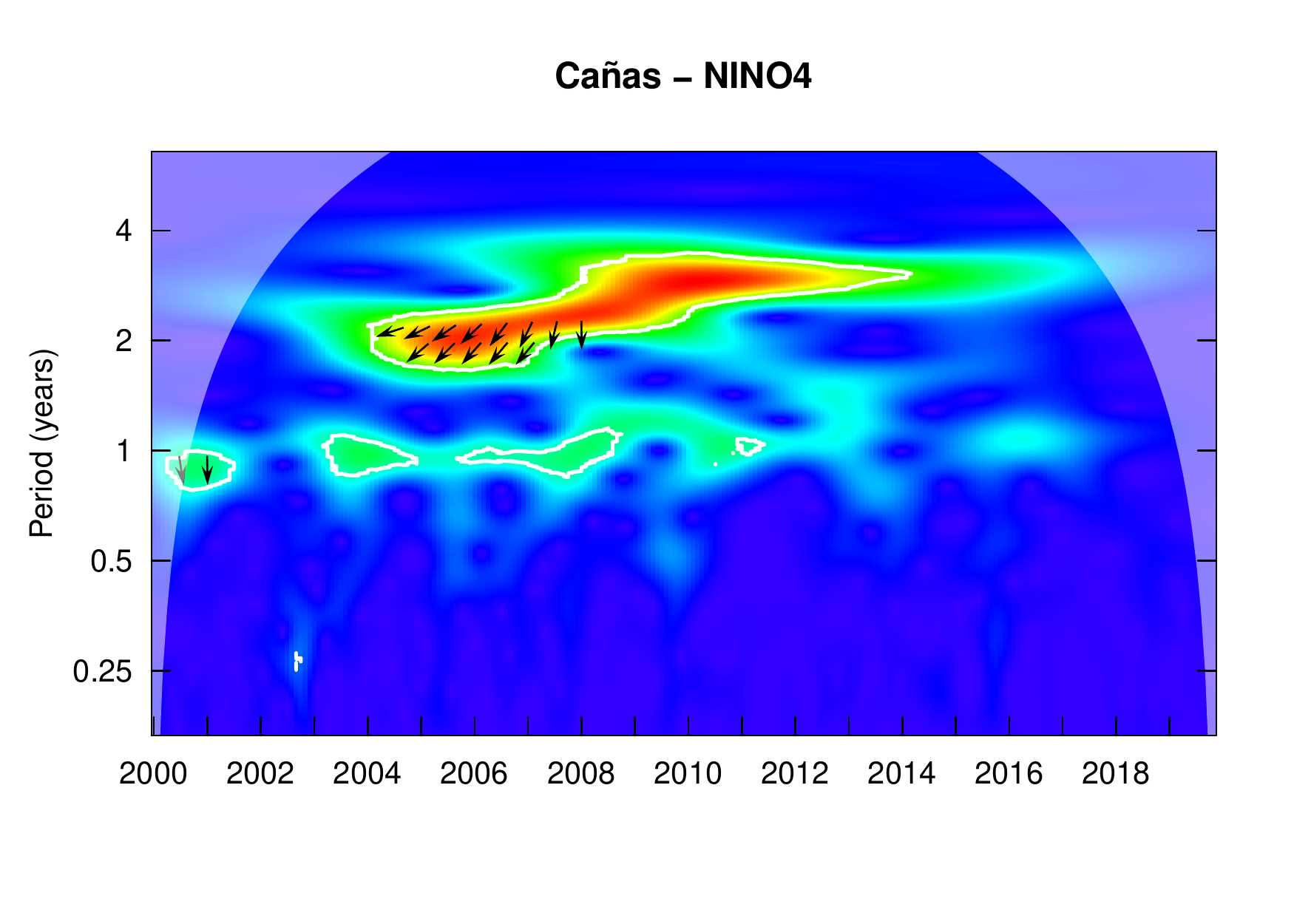}}\vspace{-0.15cm}%
\subfloat[]{\includegraphics[scale=0.23]{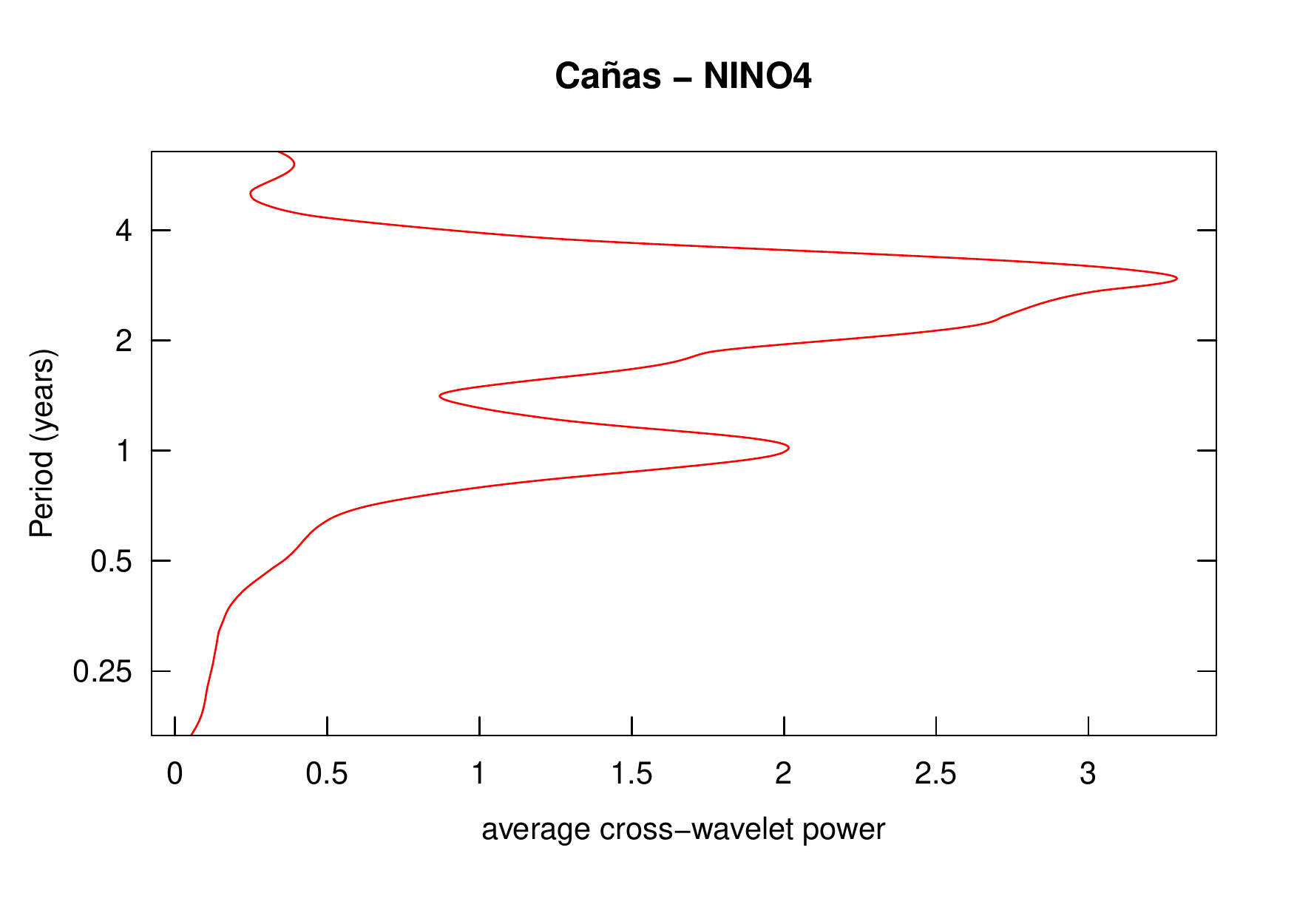}}\vspace{-0.15cm}\\
\caption*{}
\end{figure}

\begin{figure}[H]
\captionsetup[subfigure]{labelformat=empty}
\subfloat[]{\includegraphics[scale=0.23]{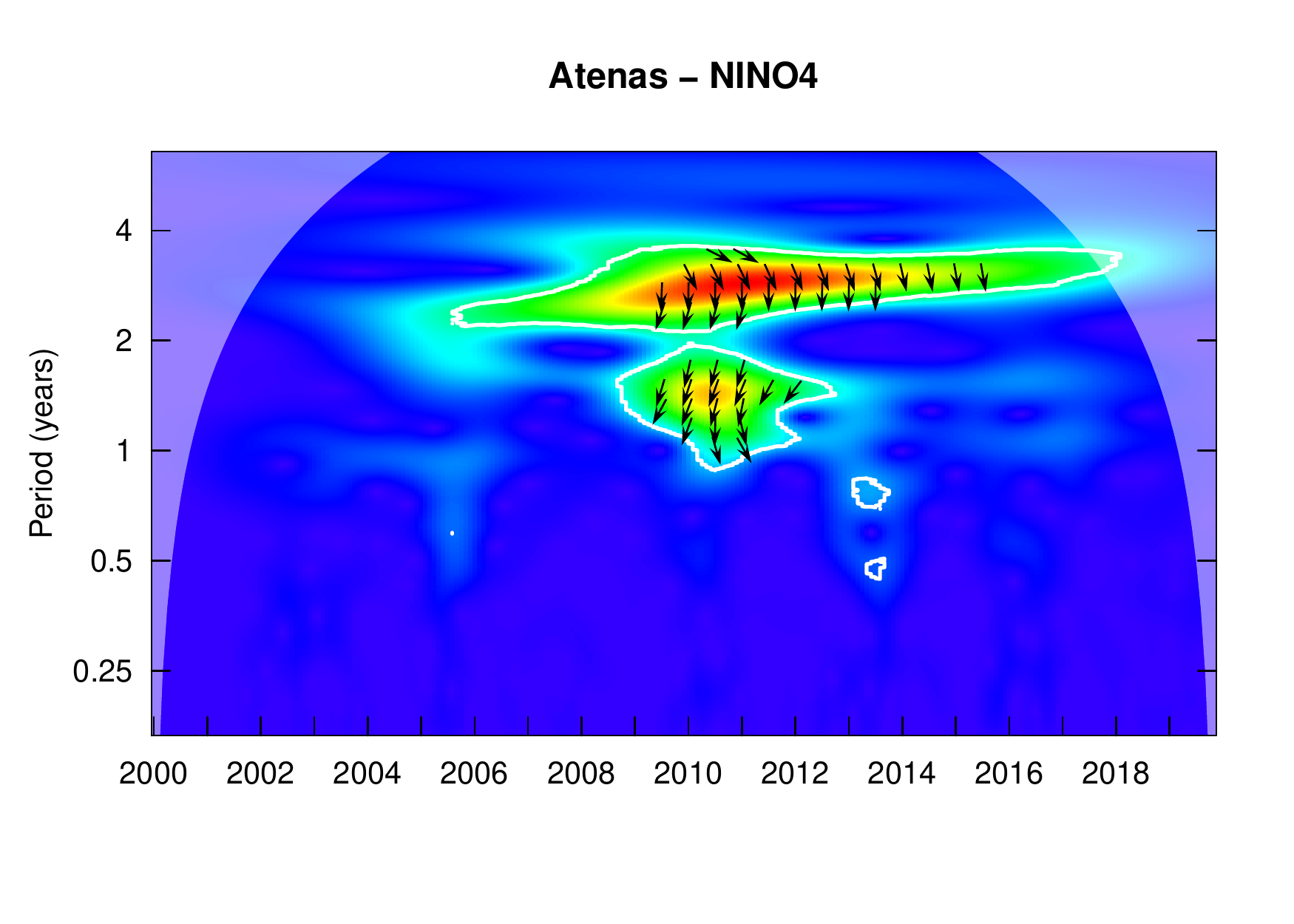}}\vspace{-0.15cm}%
\subfloat[]{\includegraphics[scale=0.23]{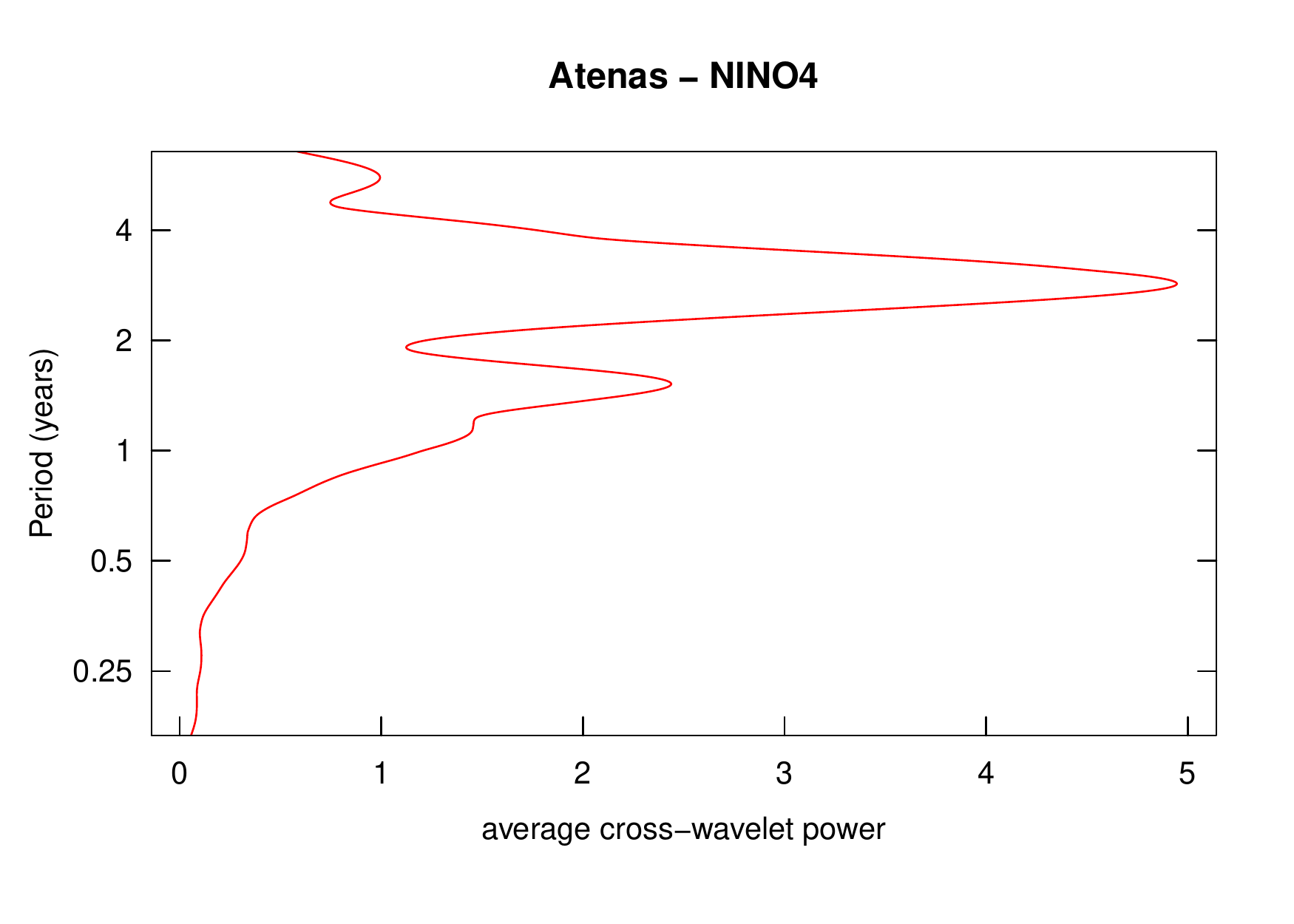}}\vspace{-0.15cm}%
\subfloat[]{\includegraphics[scale=0.23]{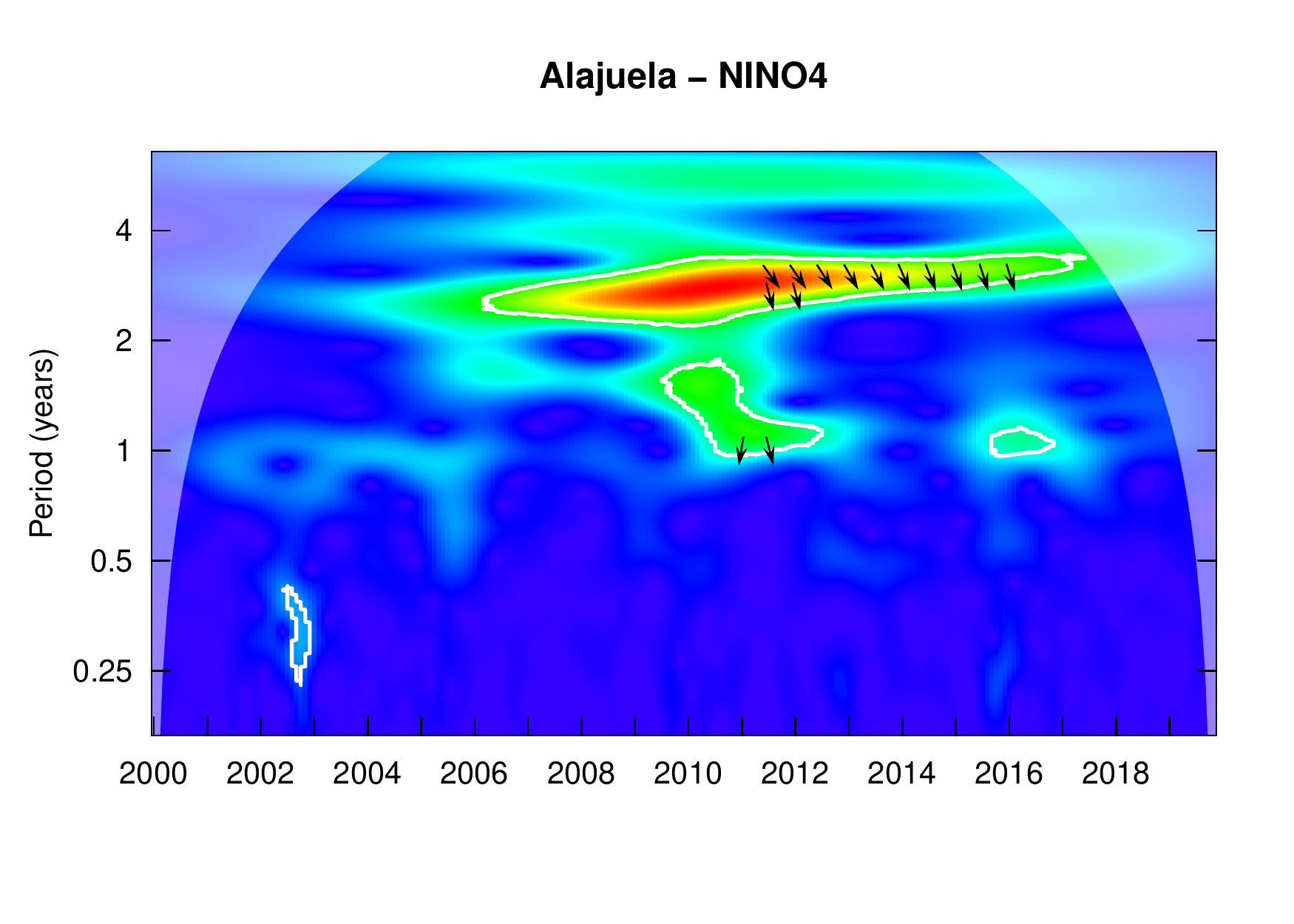}}\vspace{-0.15cm}%
\subfloat[]{\includegraphics[scale=0.23]{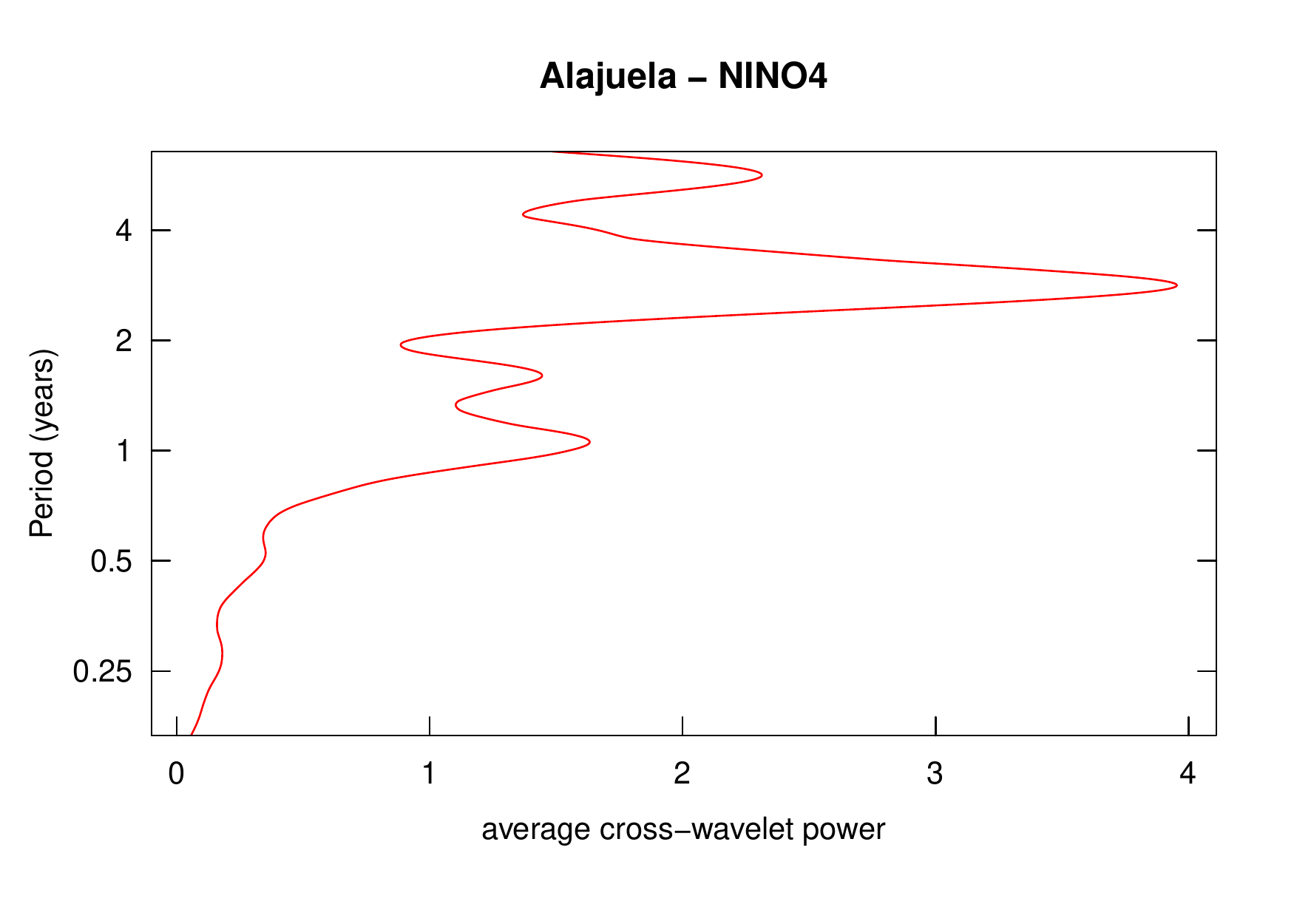}}\vspace{-0.15cm}\\
\subfloat[]{\includegraphics[scale=0.23]{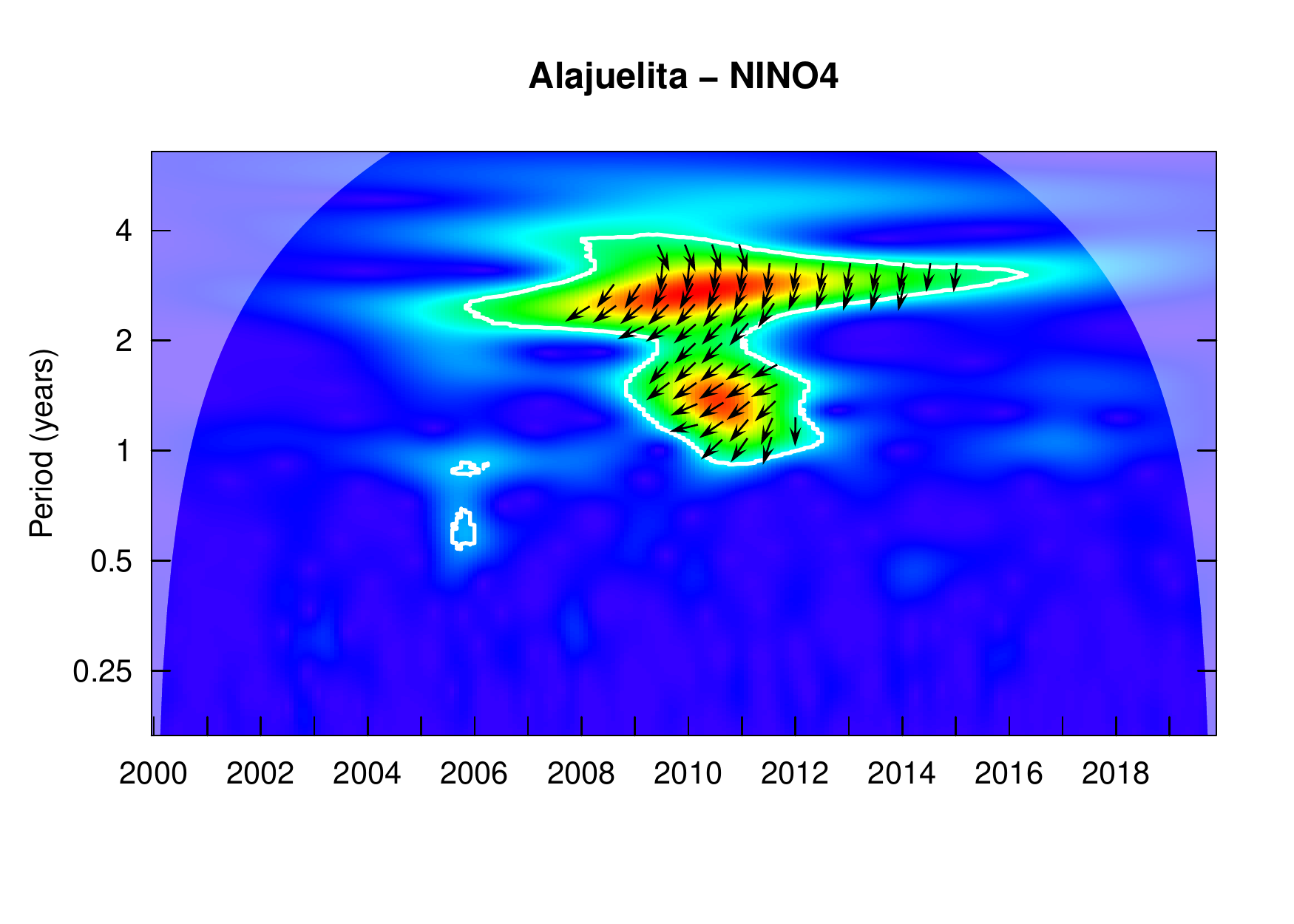}}\vspace{-0.15cm}%
\subfloat[]{\includegraphics[scale=0.23]{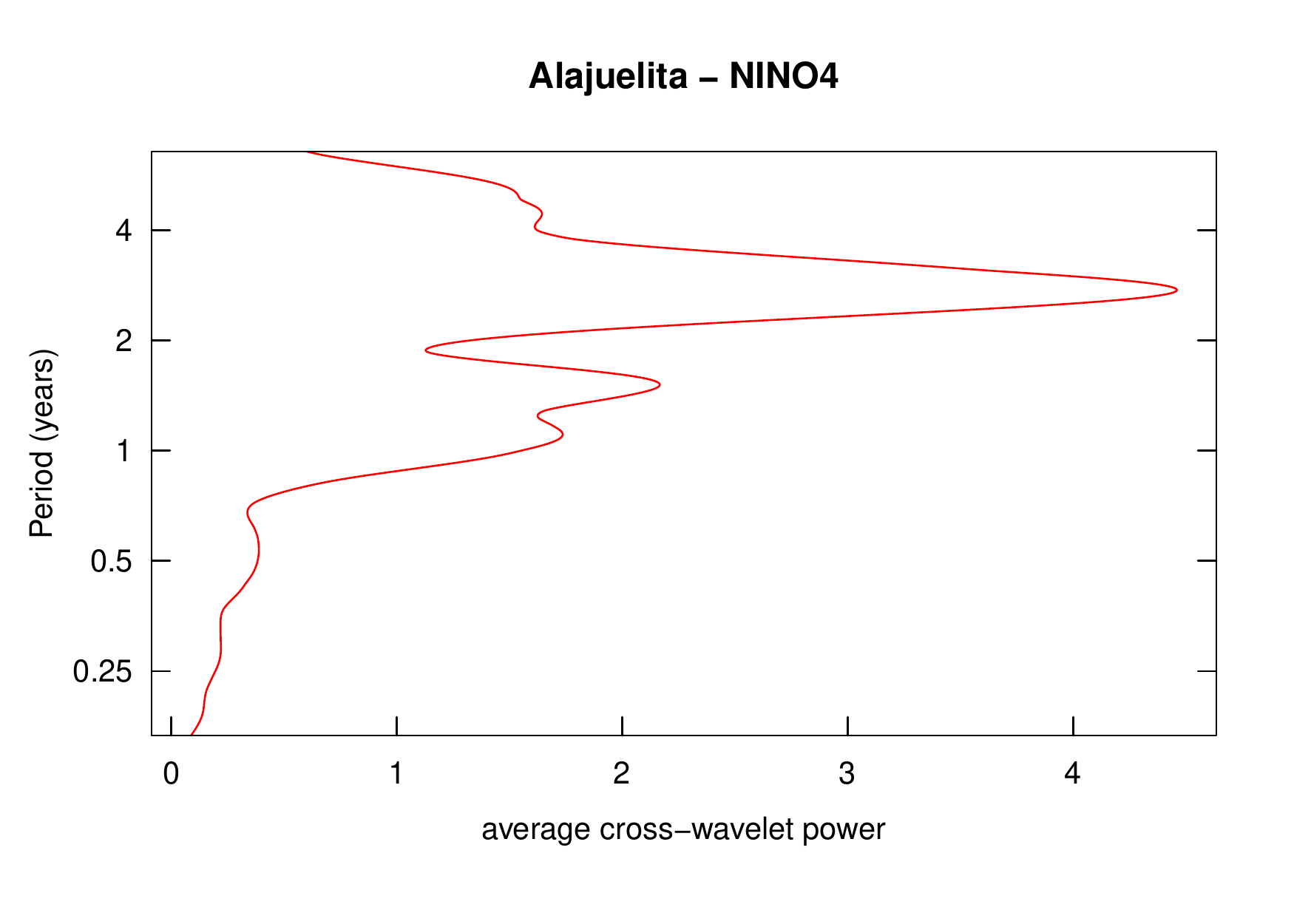}}\vspace{-0.15cm}%
\subfloat[]{\includegraphics[scale=0.23]{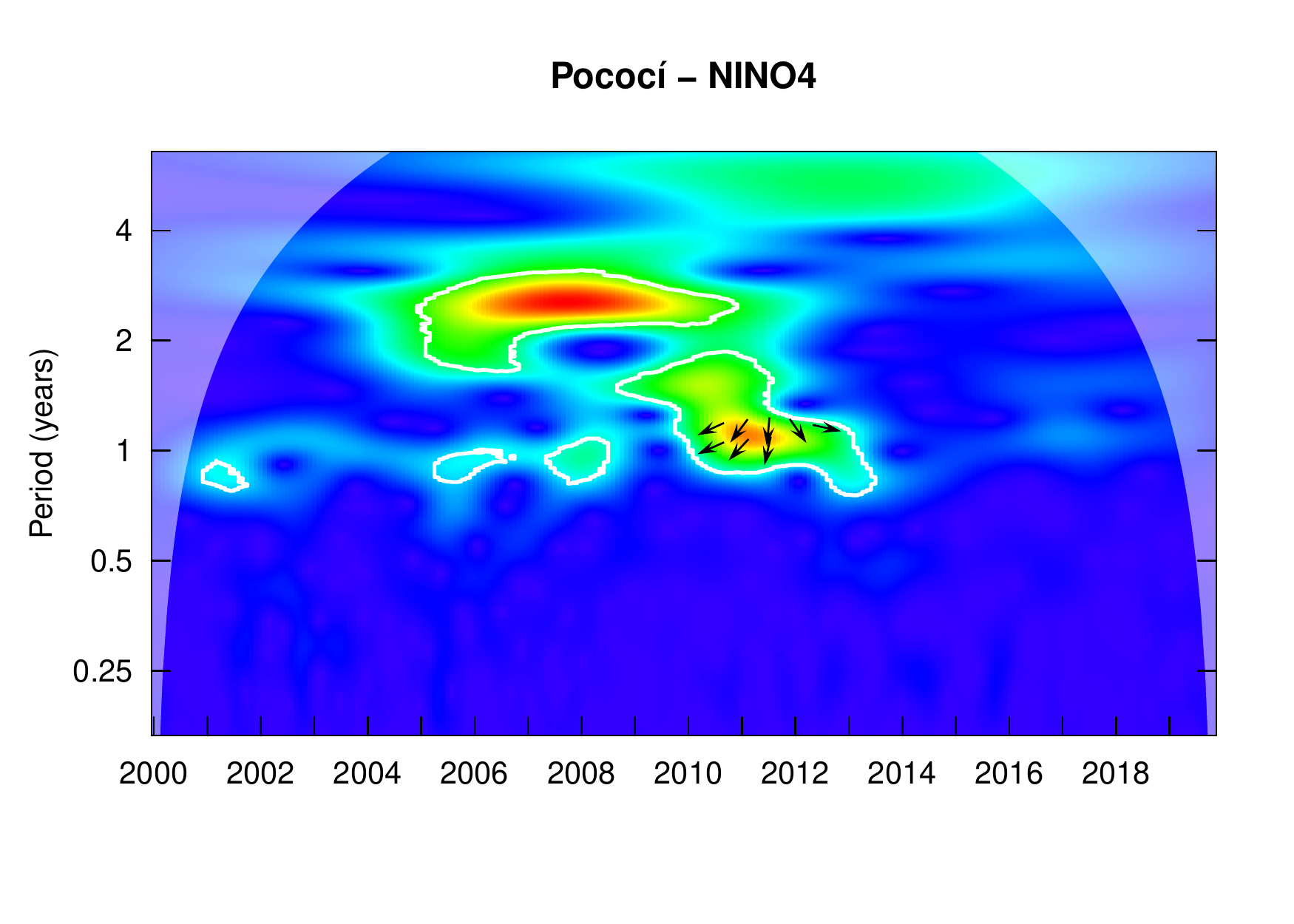}}\vspace{-0.15cm}%
\subfloat[]{\includegraphics[scale=0.23]{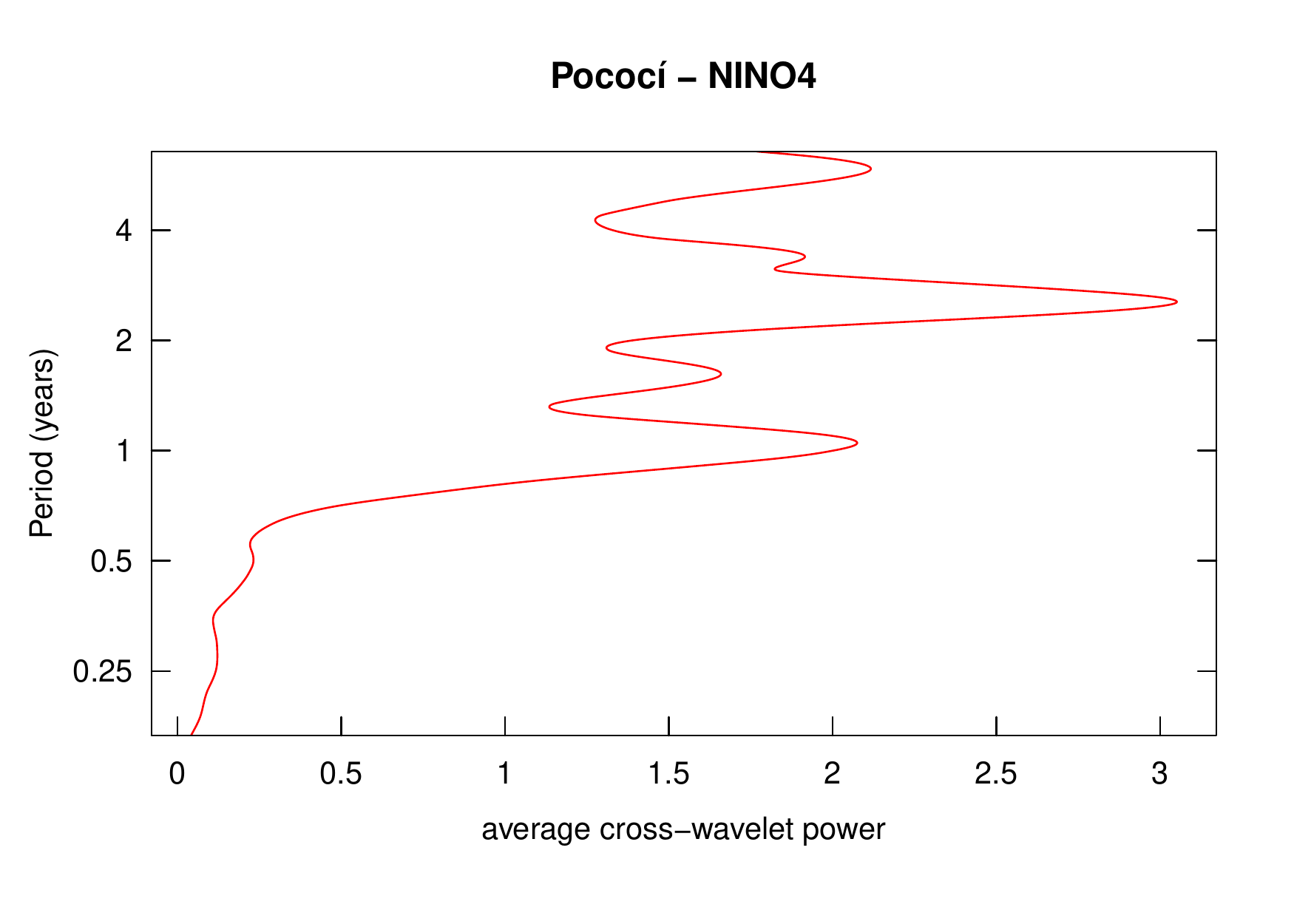}}\vspace{-0.15cm}\\
\subfloat[]{\includegraphics[scale=0.23]{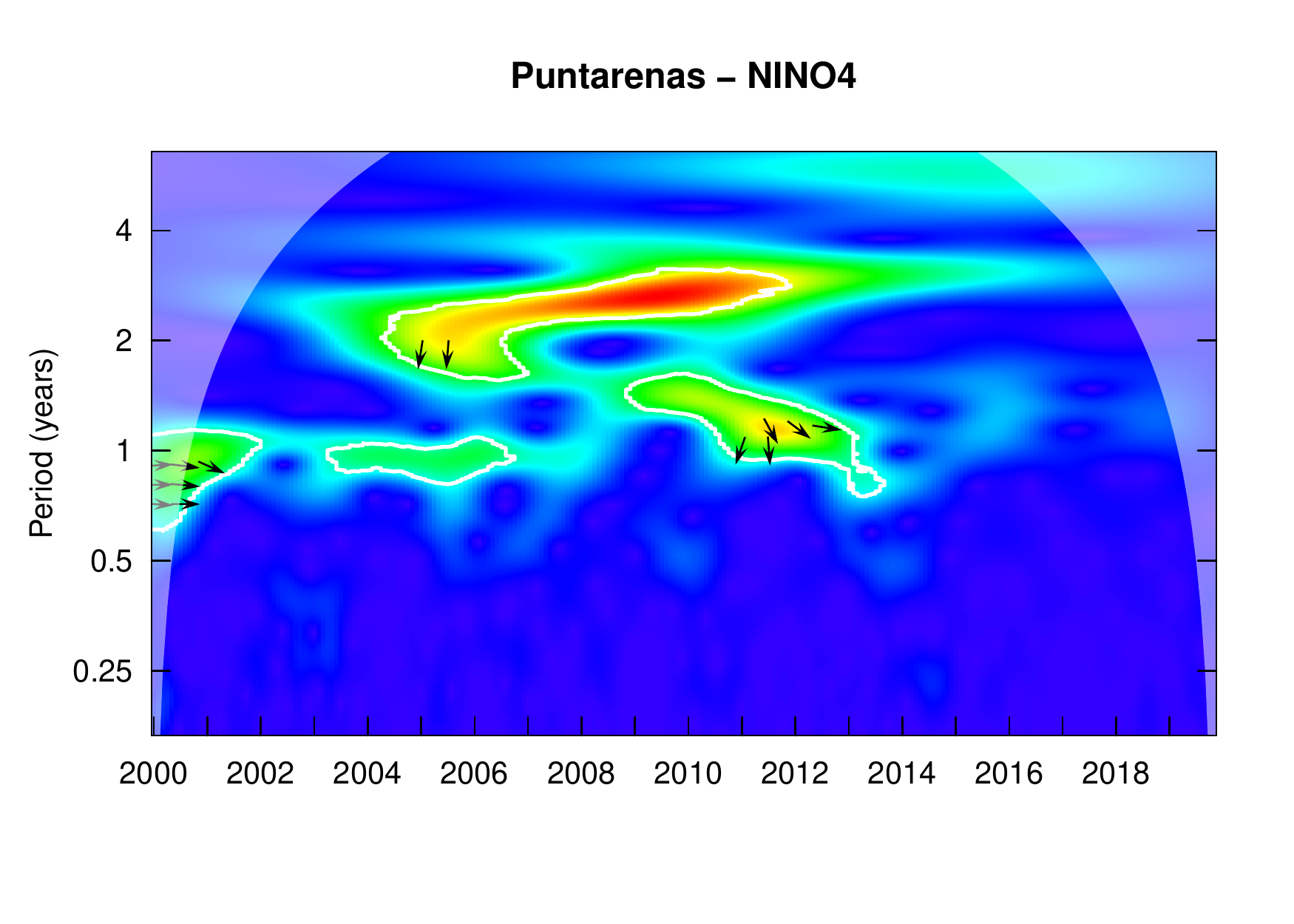}}\vspace{-0.15cm}%
\subfloat[]{\includegraphics[scale=0.23]{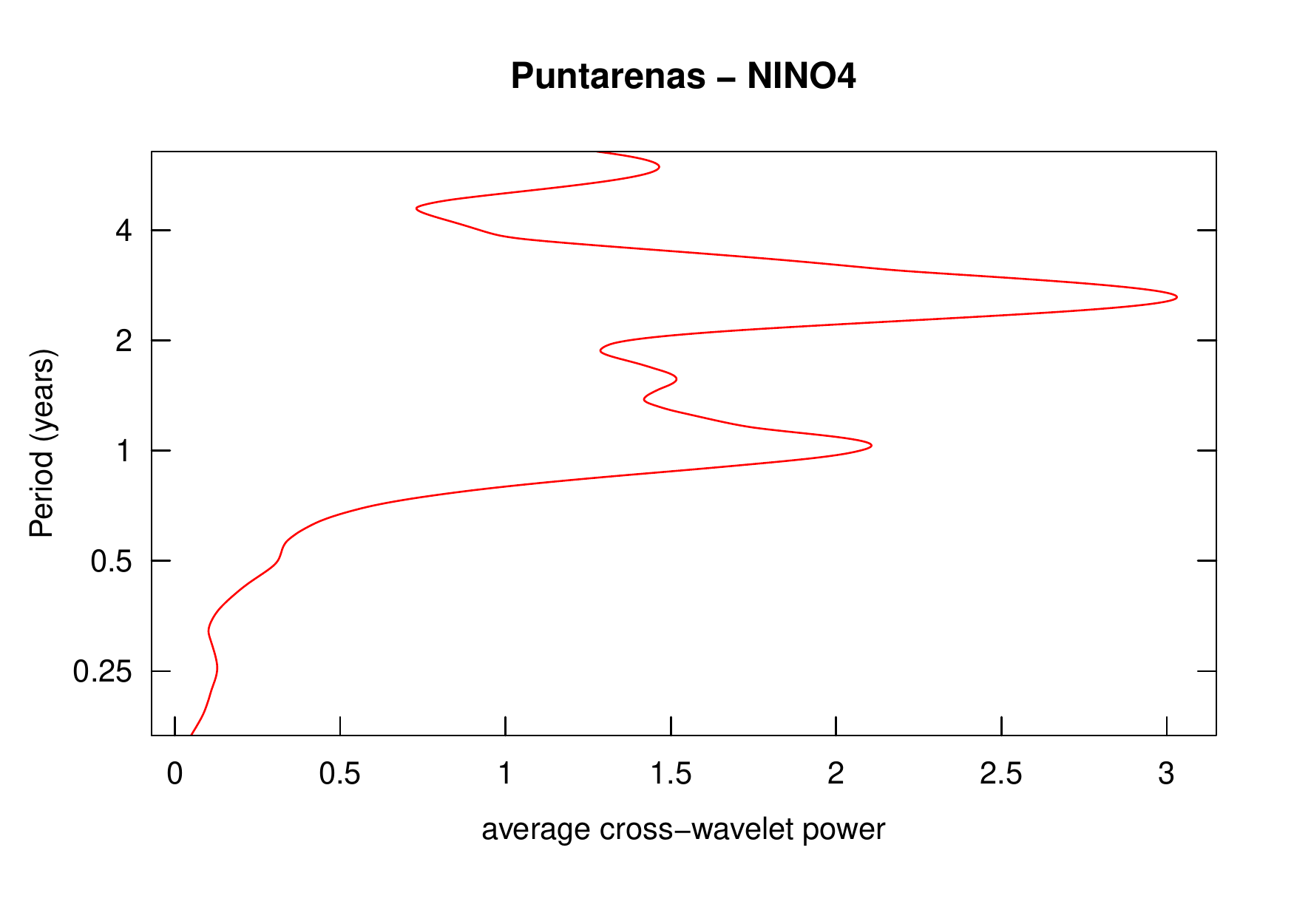}}\vspace{-0.15cm}%
\subfloat[]{\includegraphics[scale=0.23]{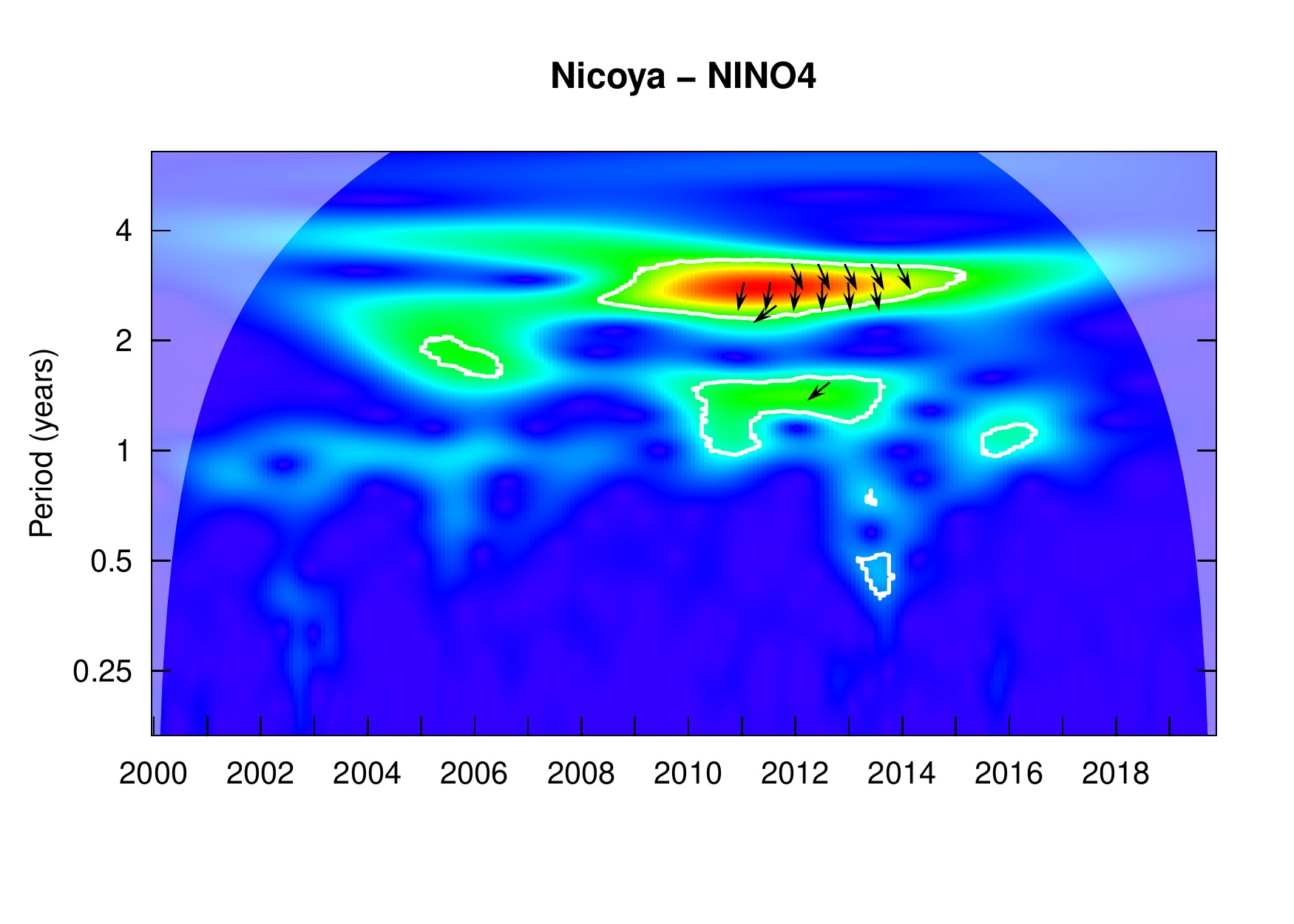}}\vspace{-0.15cm}%
\subfloat[]{\includegraphics[scale=0.23]{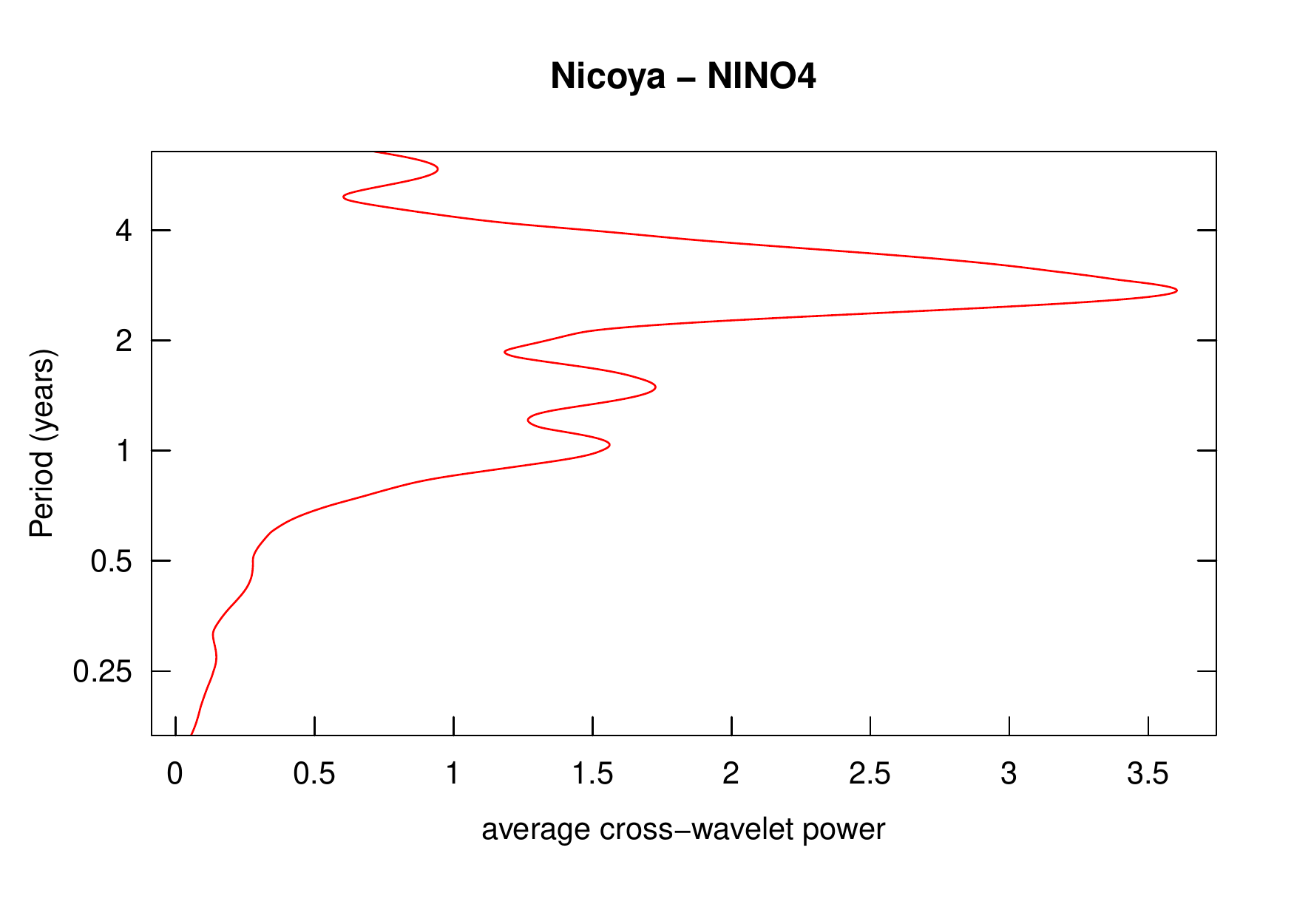}}\vspace{-0.15cm}\\
\subfloat[]{\includegraphics[scale=0.23]{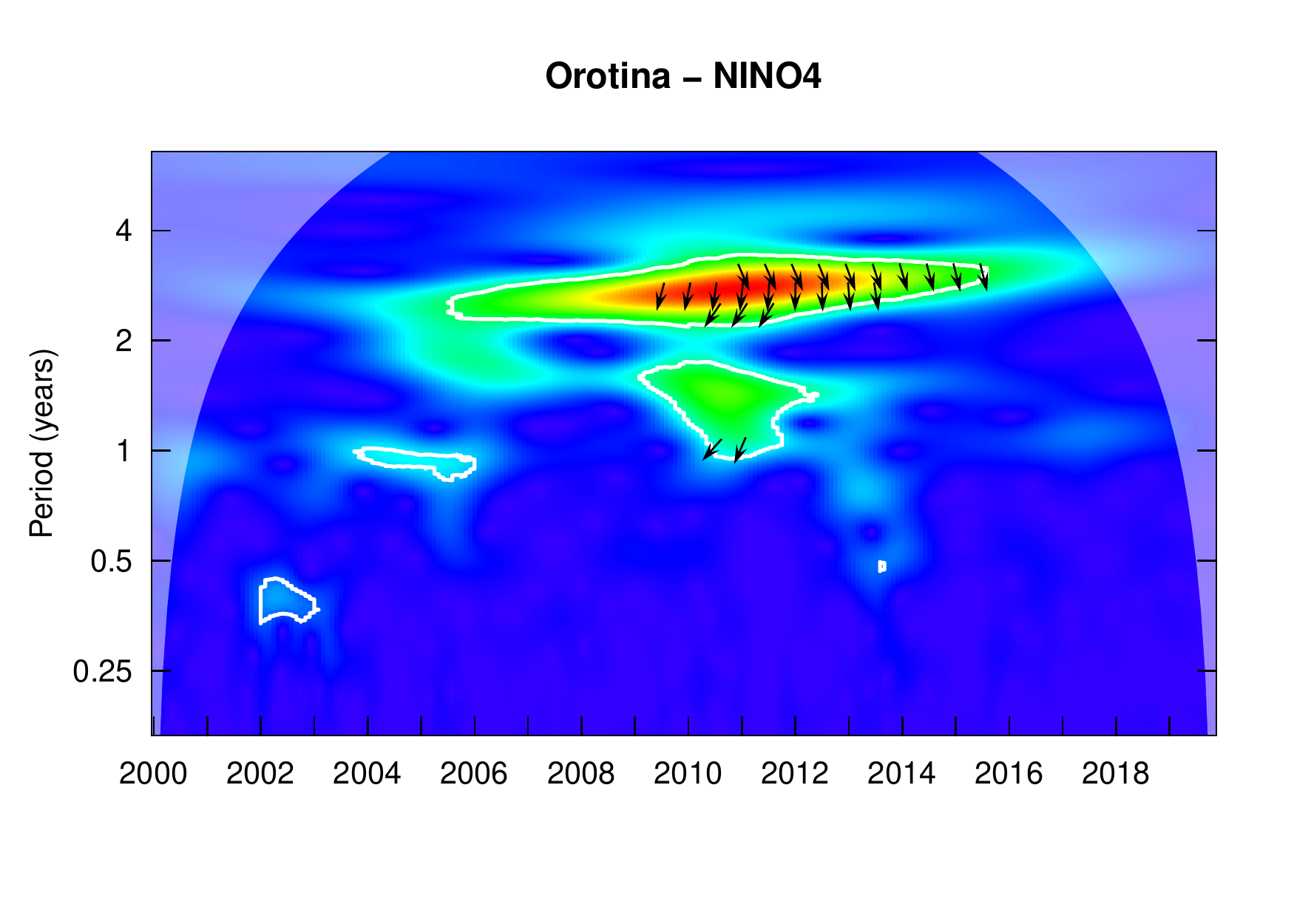}}\vspace{-0.15cm}%
\subfloat[]{\includegraphics[scale=0.23]{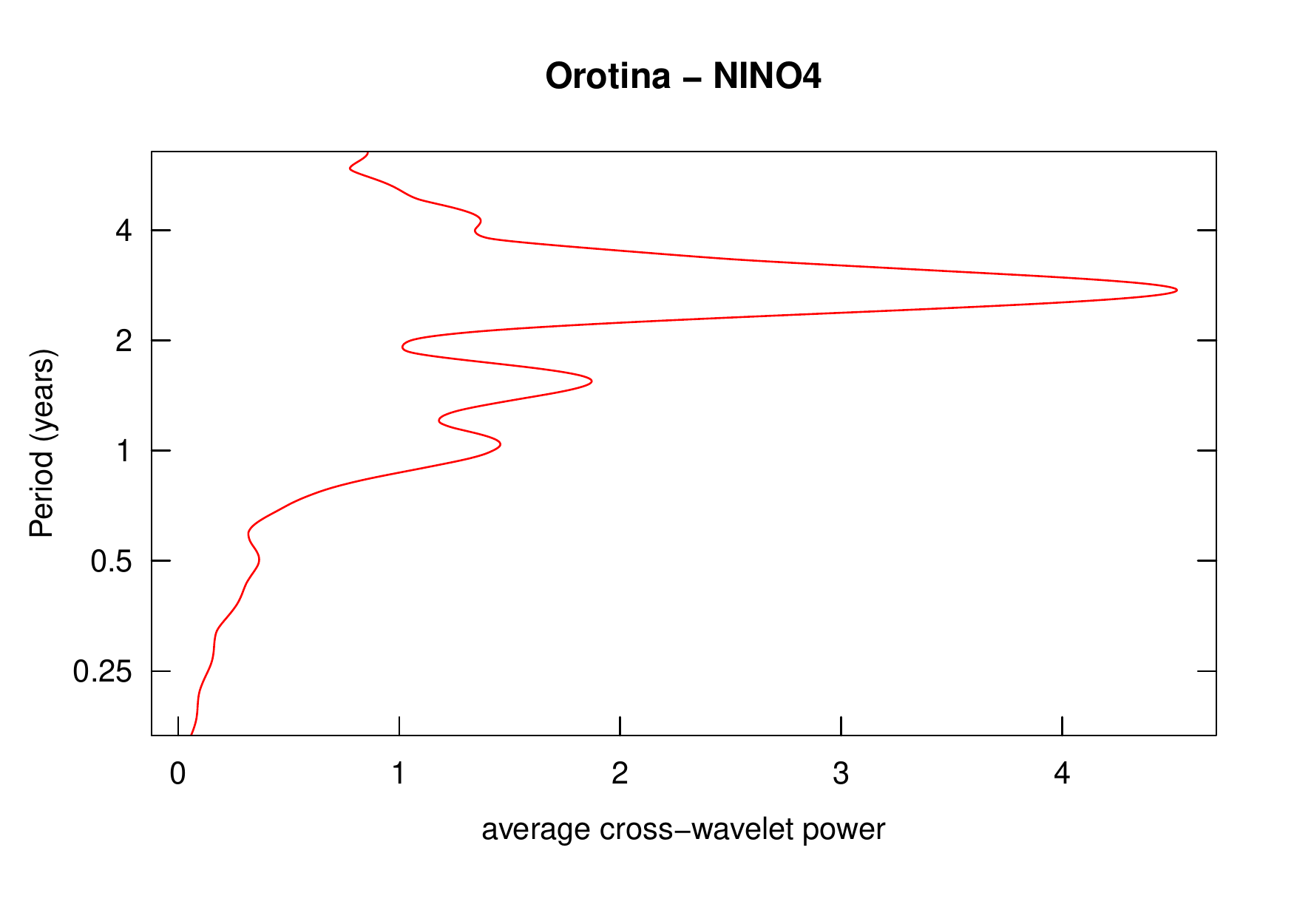}}\vspace{-0.15cm}%
\subfloat[]{\includegraphics[scale=0.23]{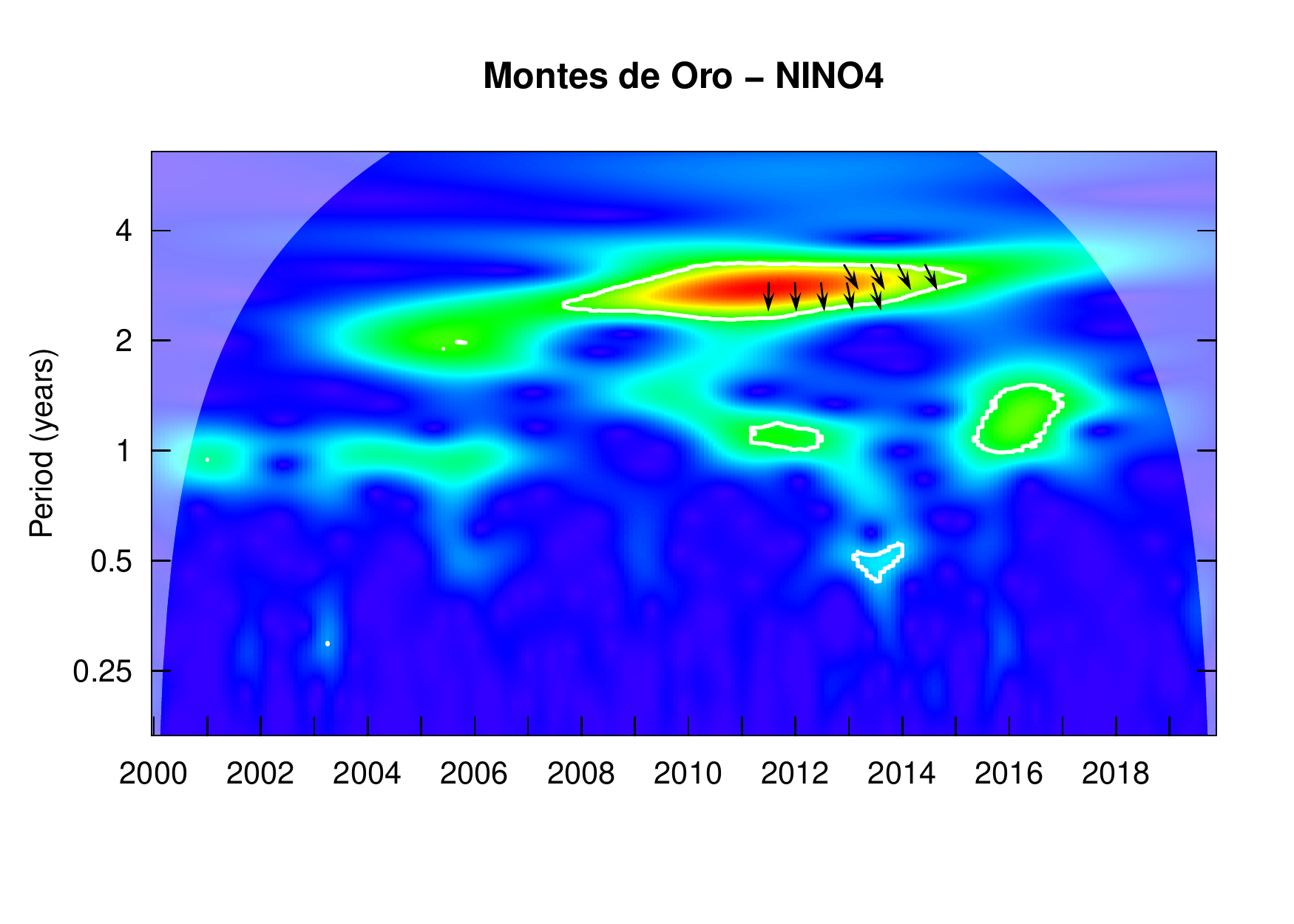}}\vspace{-0.15cm}%
\subfloat[]{\includegraphics[scale=0.23]{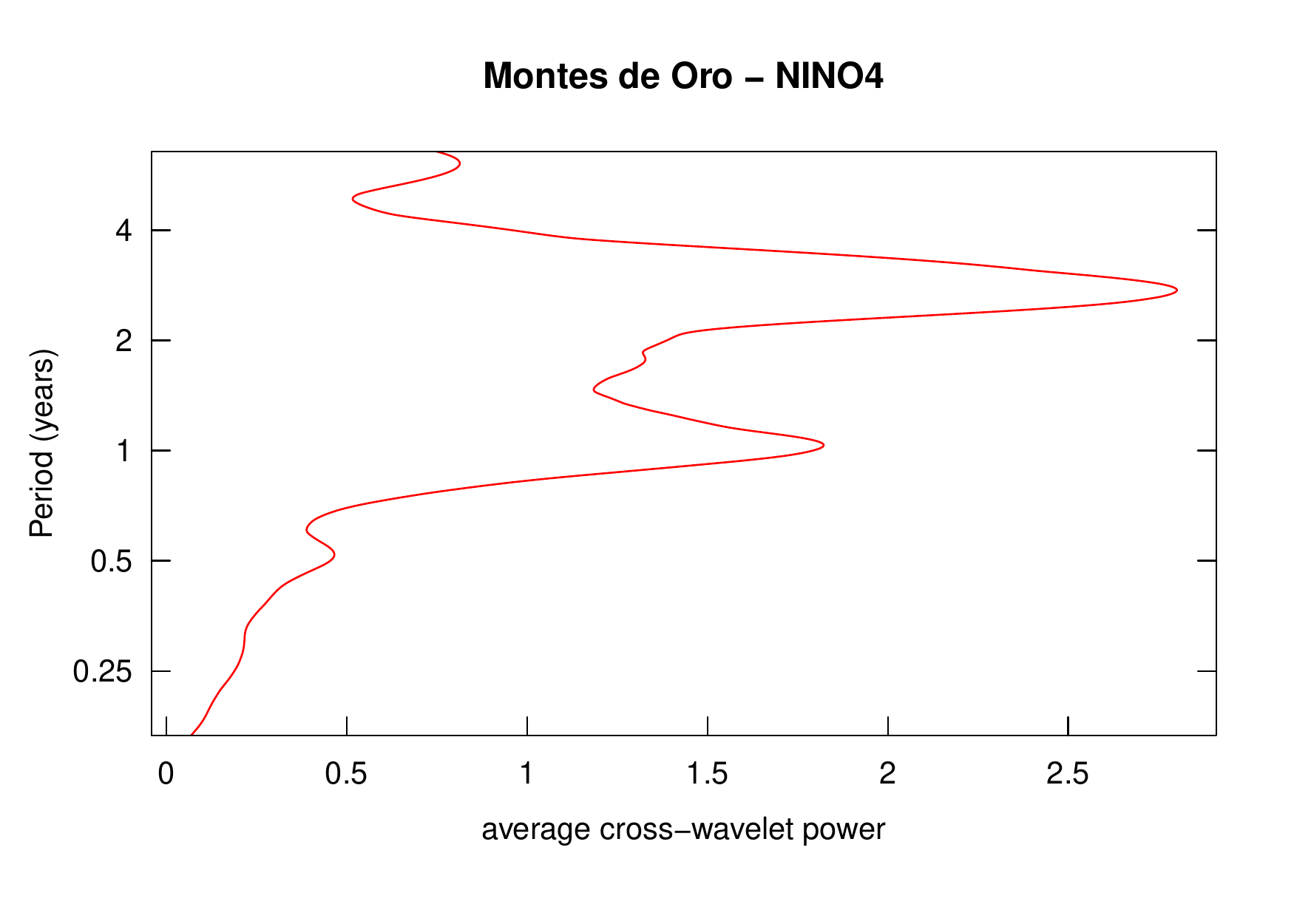}}\vspace{-0.15cm}\\
\subfloat[]{\includegraphics[scale=0.23]{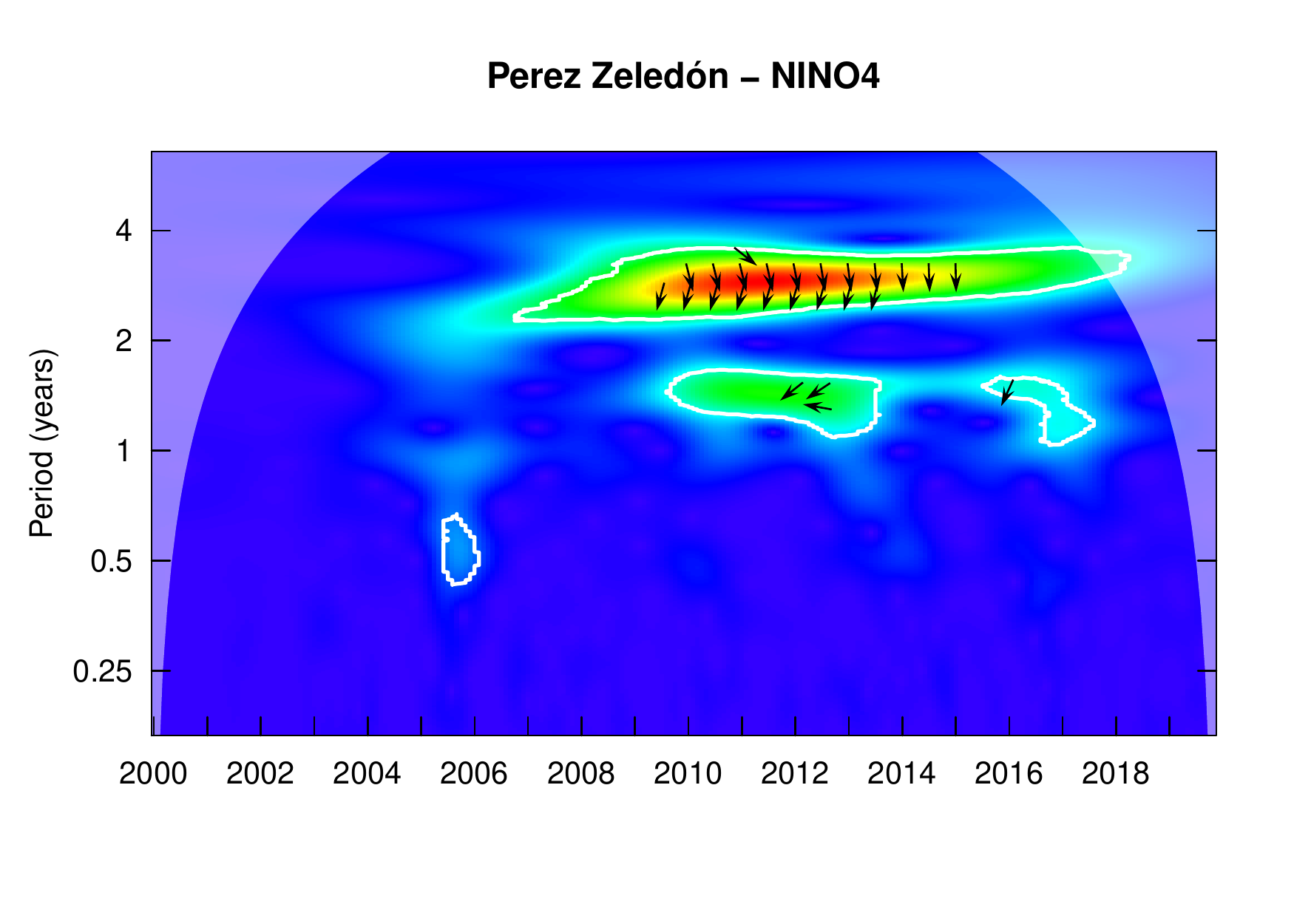}}\vspace{-0.15cm}%
\subfloat[]{\includegraphics[scale=0.23]{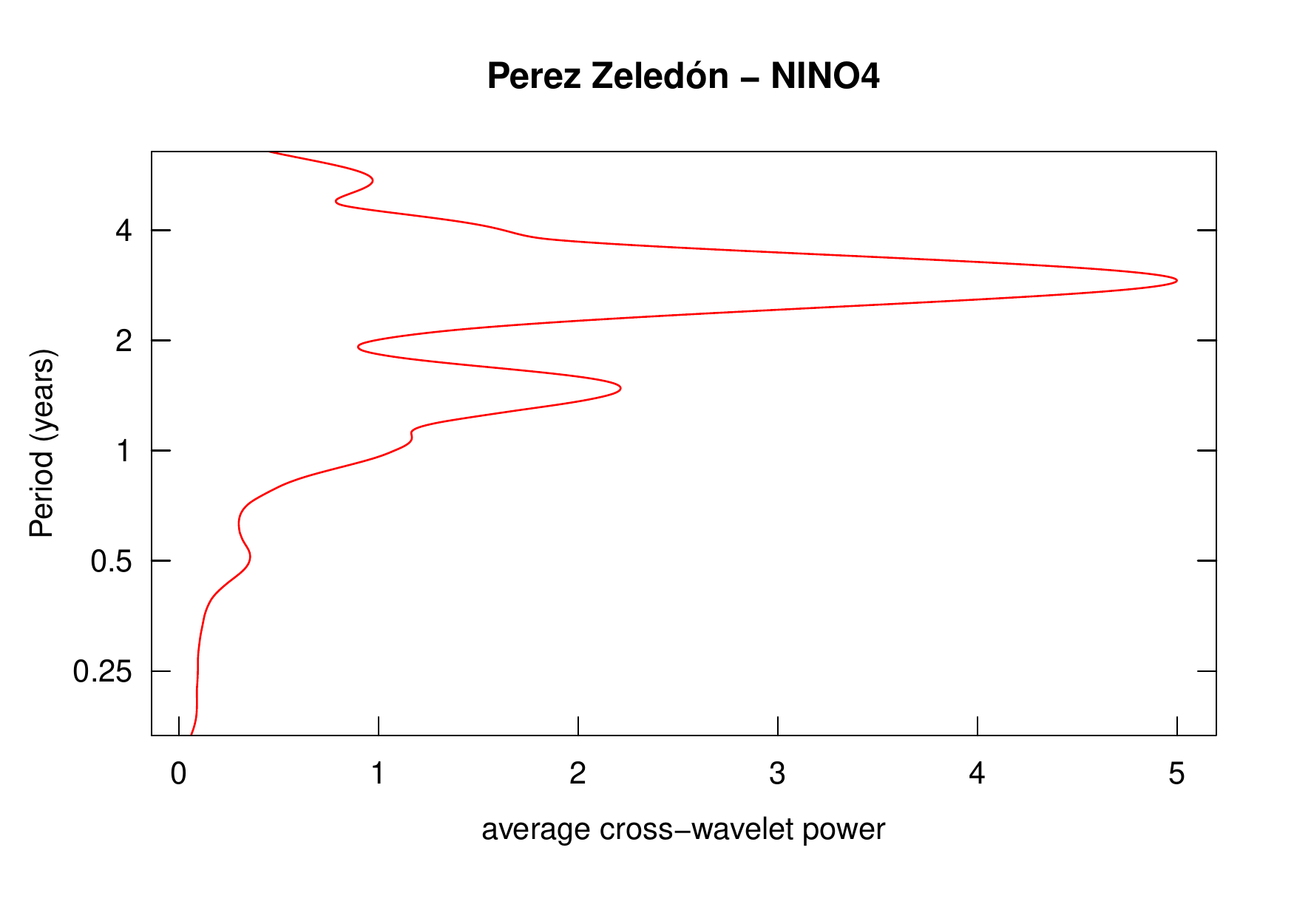}}\vspace{-0.15cm}%
\subfloat[]{\includegraphics[scale=0.23]{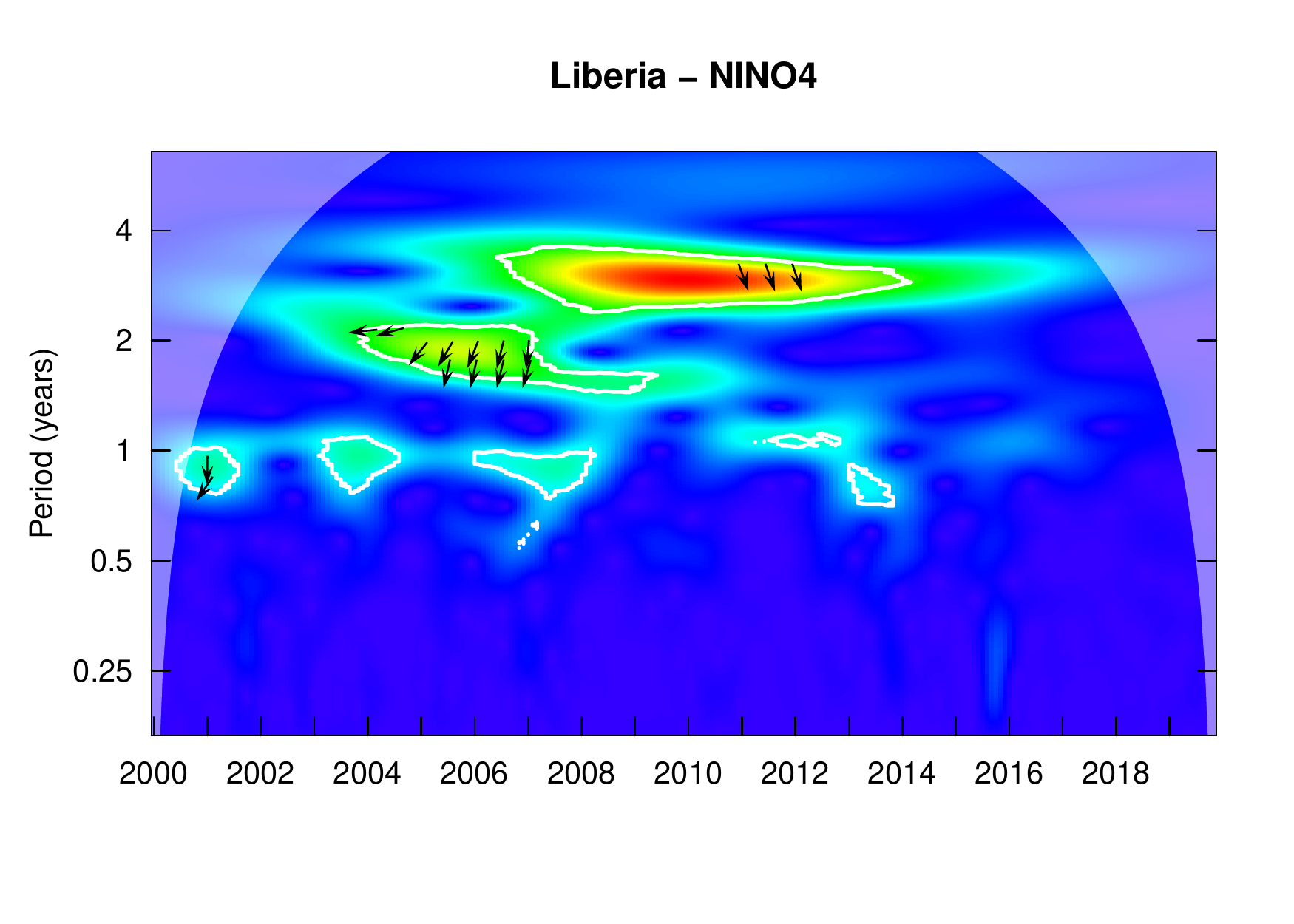}}\vspace{-0.15cm}%
\subfloat[]{\includegraphics[scale=0.23]{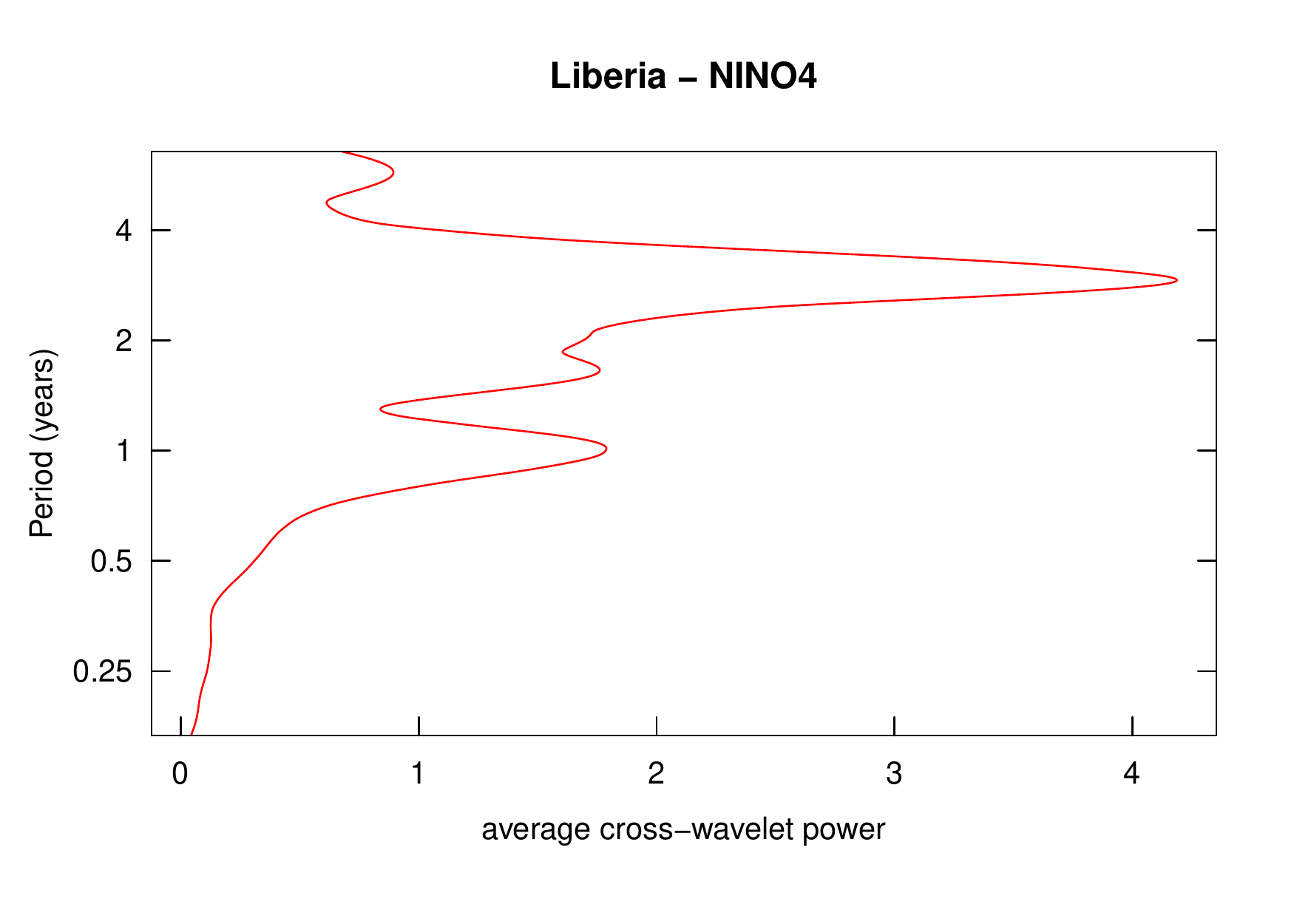}}\vspace{-0.15cm}
\subfloat[]{\includegraphics[scale=0.23]{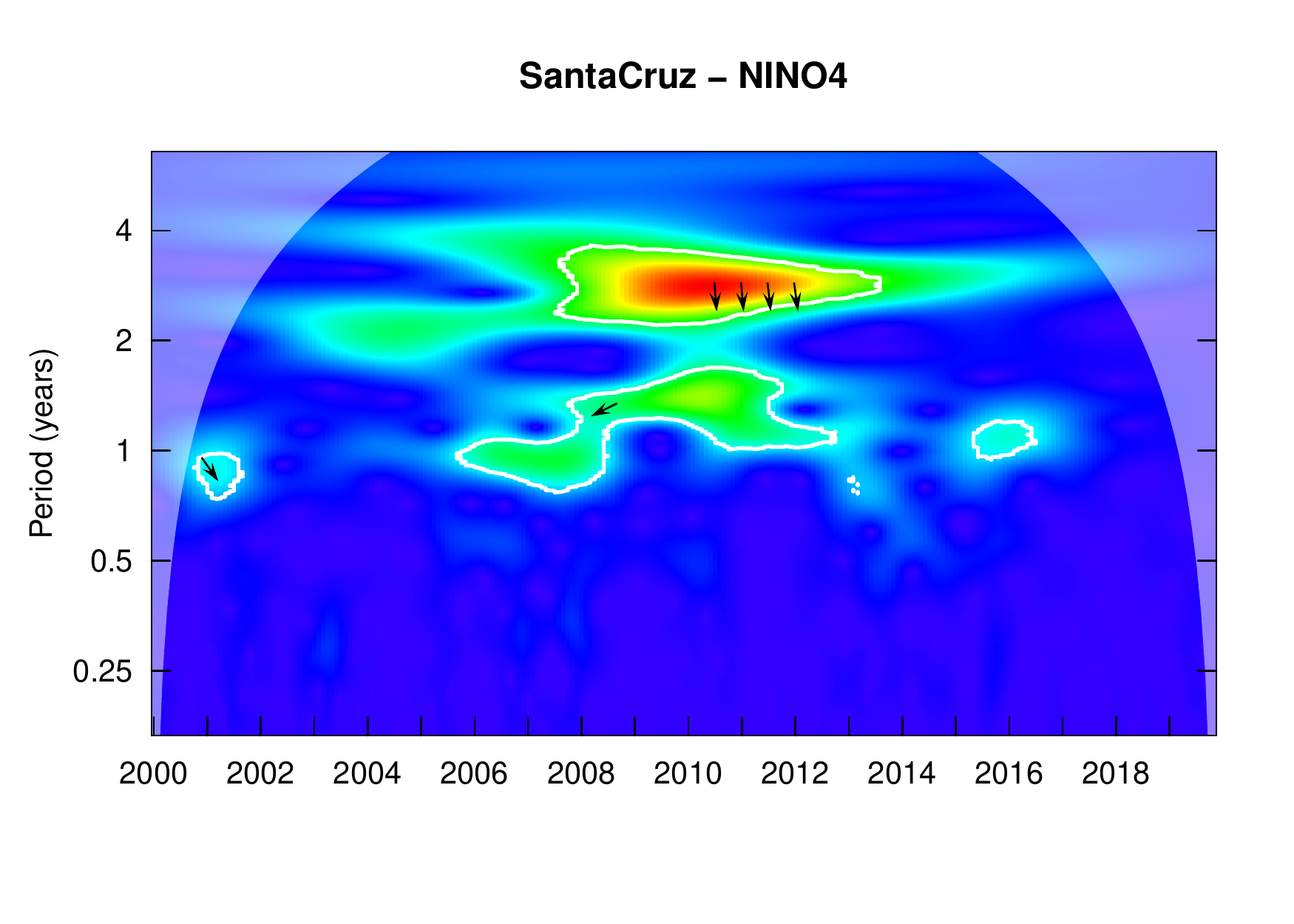}}\vspace{-0.15cm}%
\subfloat[]{\includegraphics[scale=0.23]{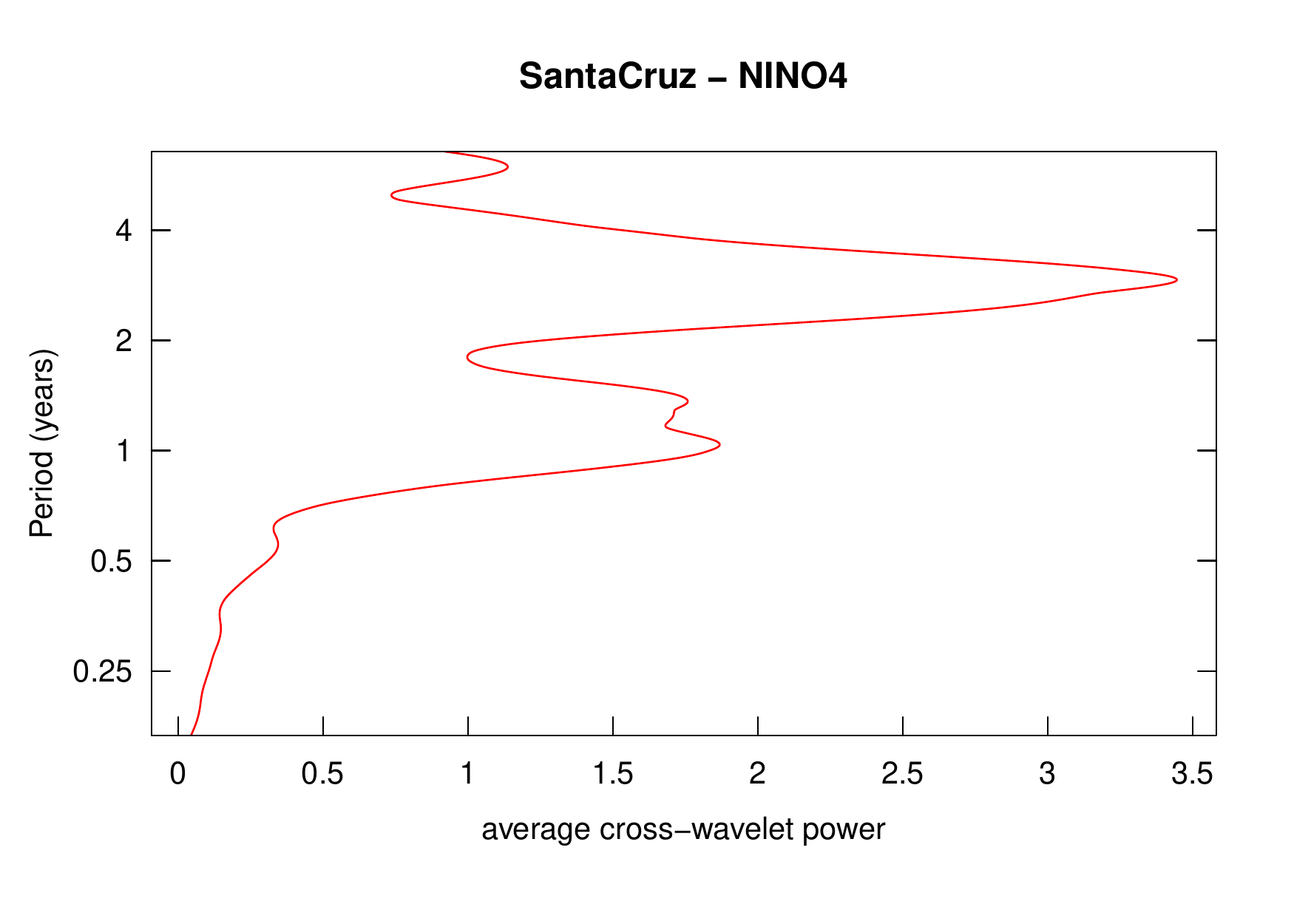}}\vspace{-0.15cm}%
\subfloat[]{\includegraphics[scale=0.23]{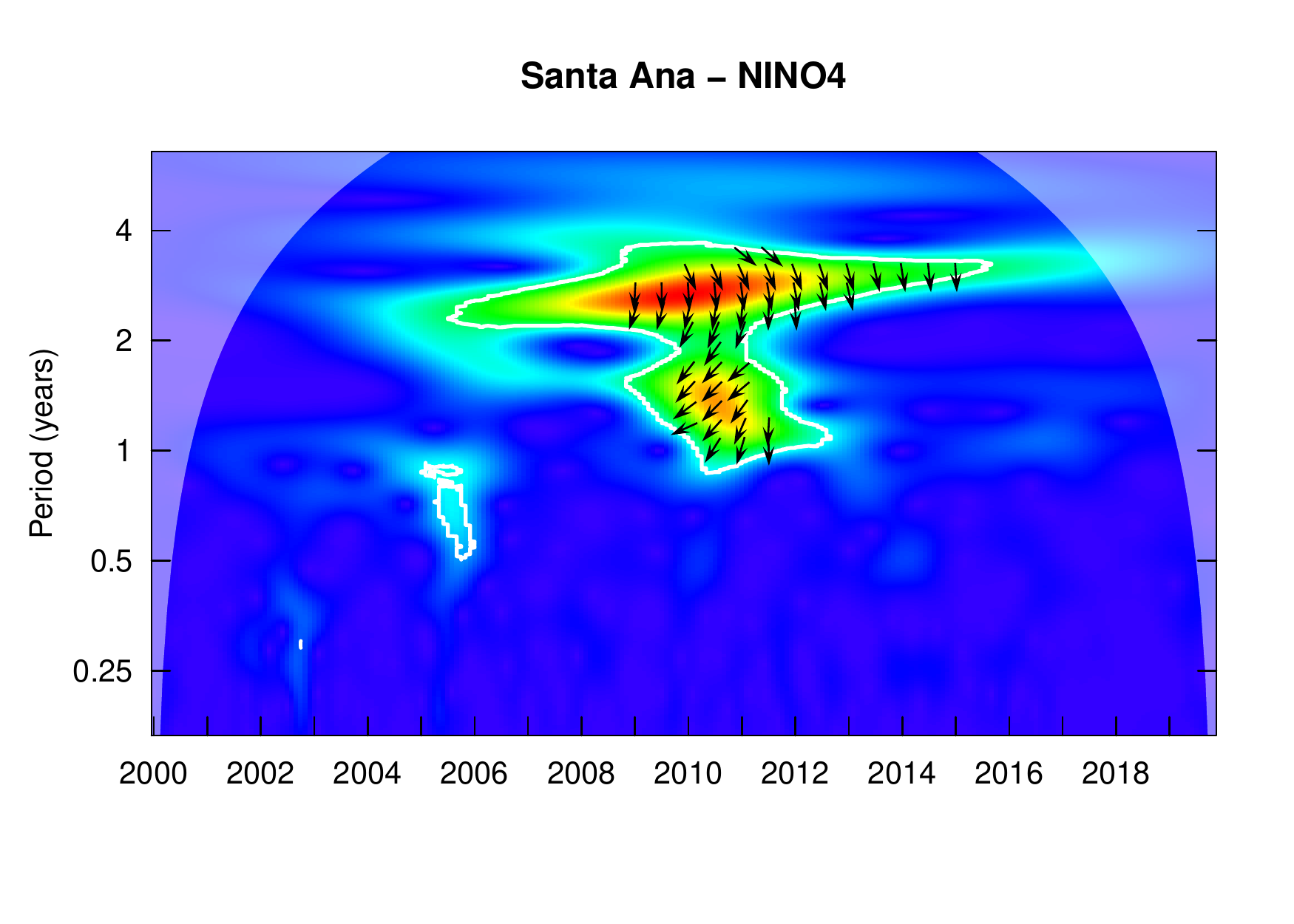}}\vspace{-0.15cm}%
\subfloat[]{\includegraphics[scale=0.23]{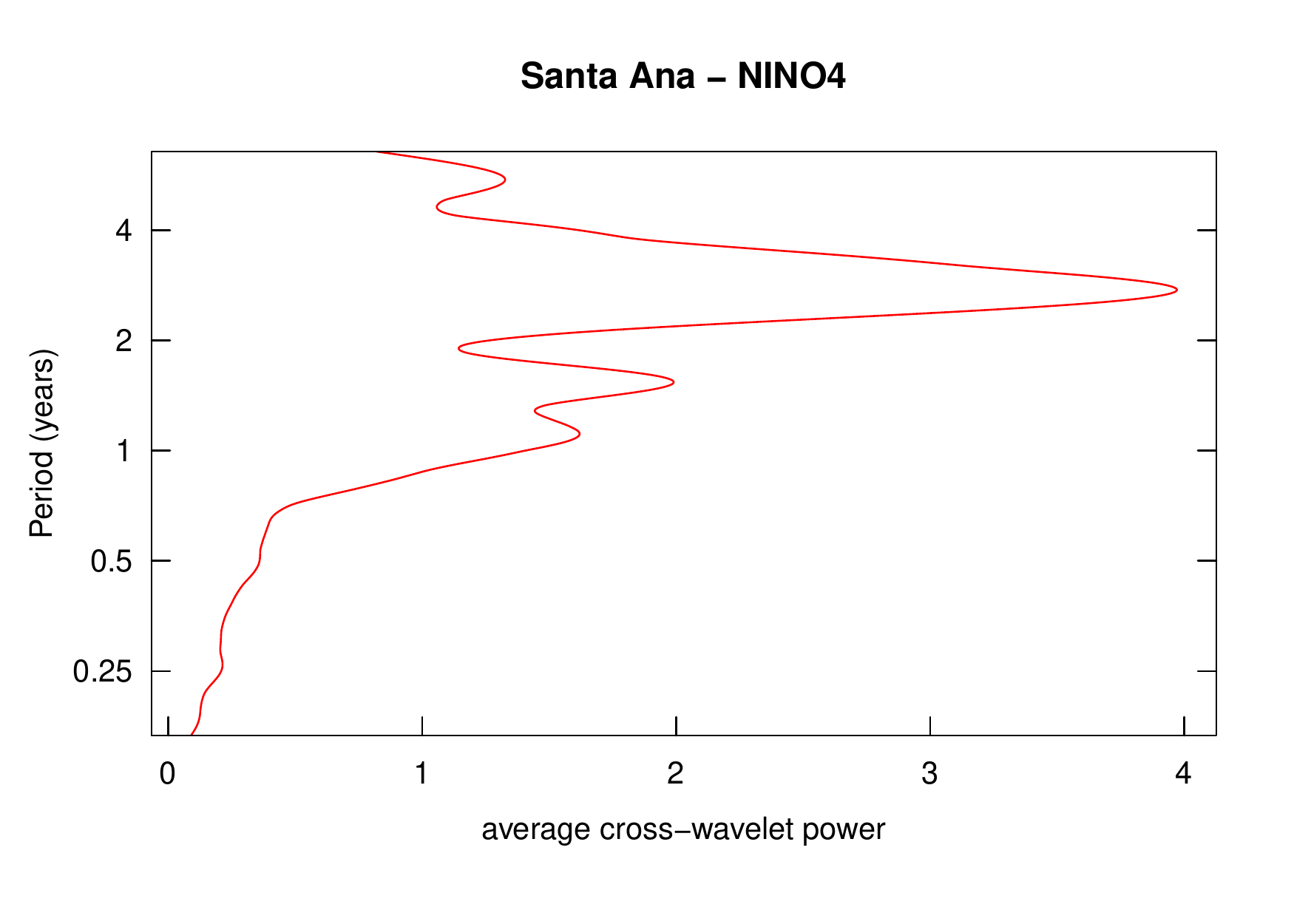}}\vspace{-0.15cm}\\
\subfloat[]{\includegraphics[scale=0.23]{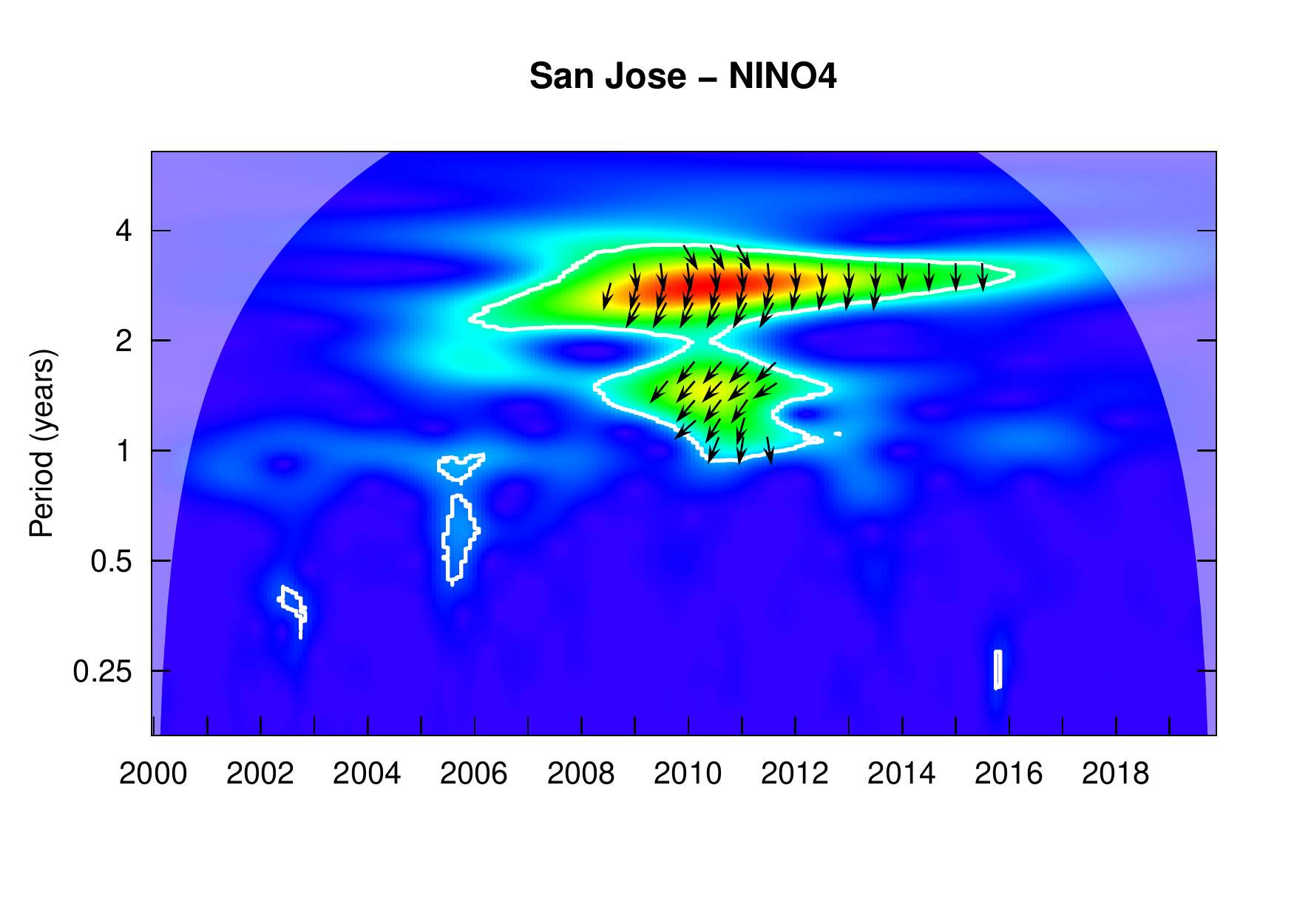}}\vspace{-0.15cm}%
\subfloat[]{\includegraphics[scale=0.23]{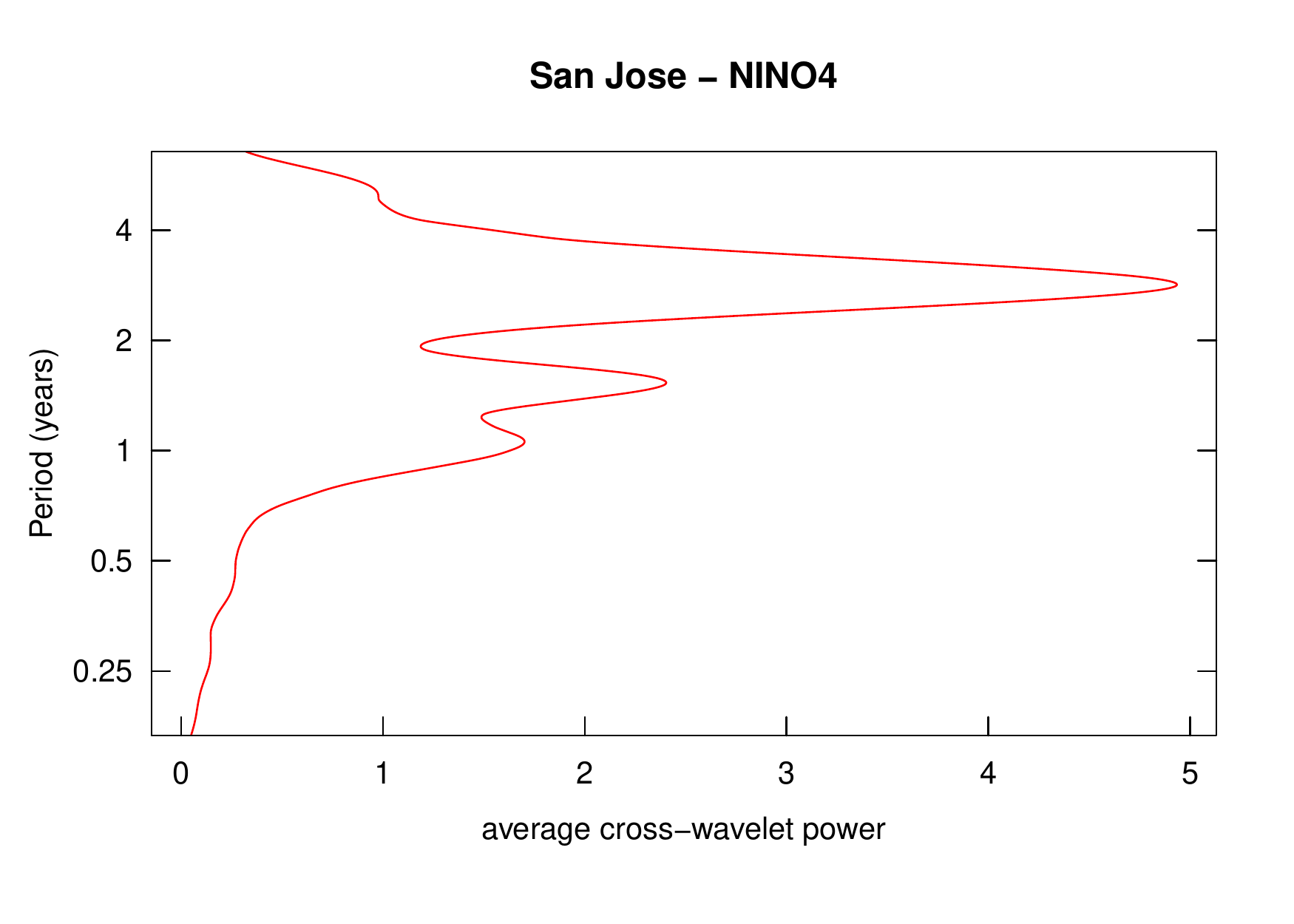}}\vspace{-0.15cm}%
\subfloat[]{\includegraphics[scale=0.23]{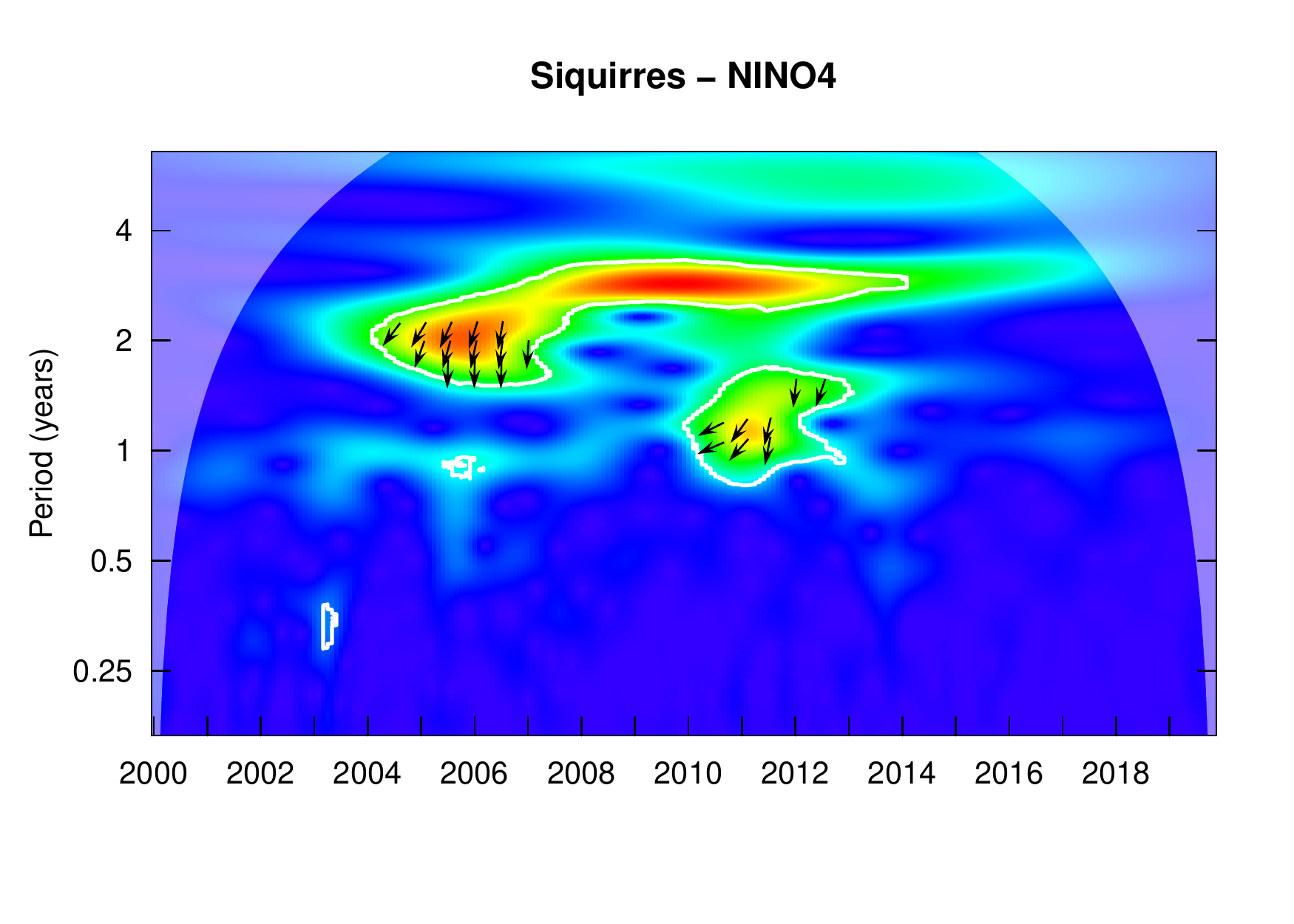}}\vspace{-0.15cm}%
\subfloat[]{\includegraphics[scale=0.23]{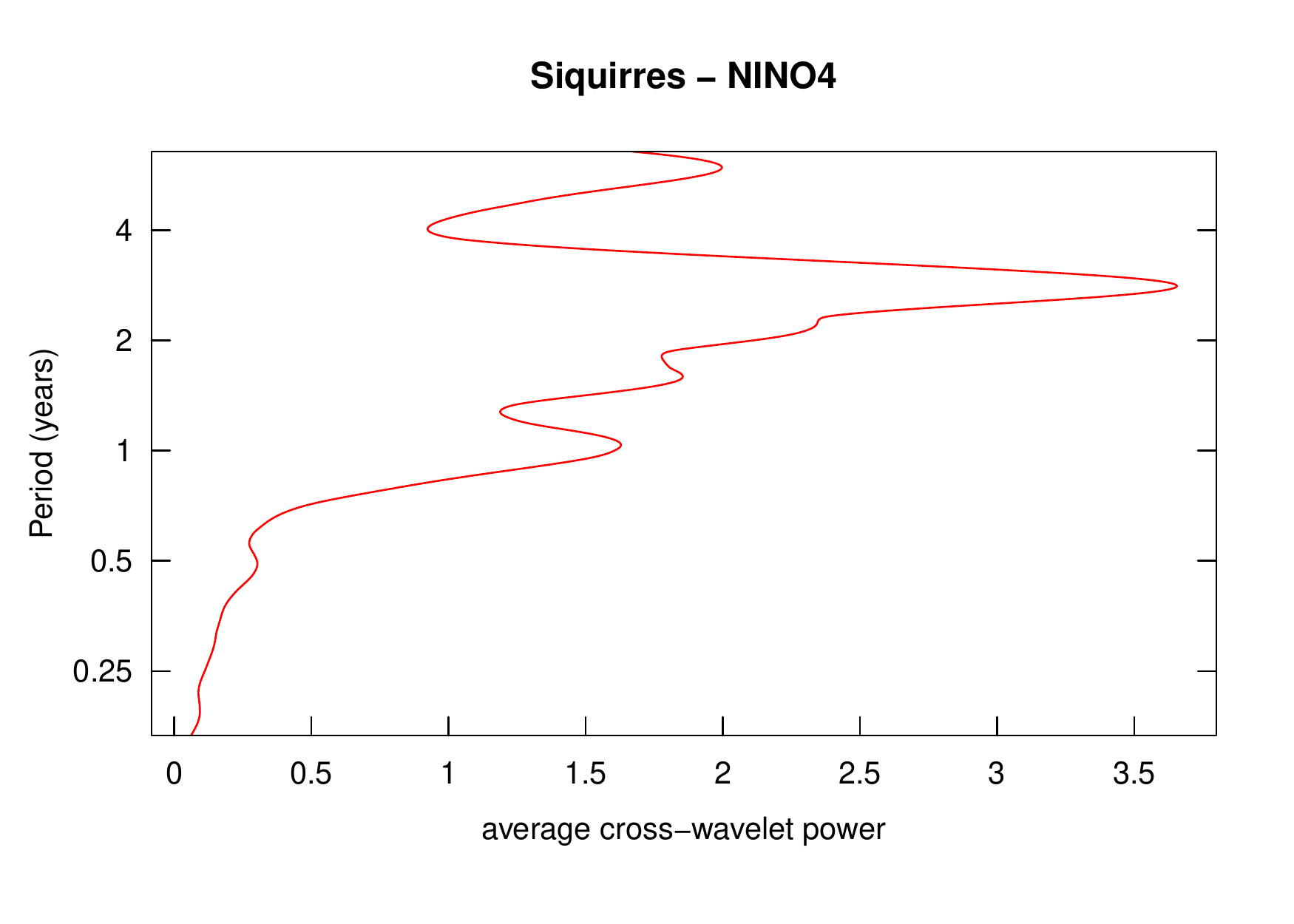}}\vspace{-0.15cm}\\
\subfloat[]{\includegraphics[scale=0.23]{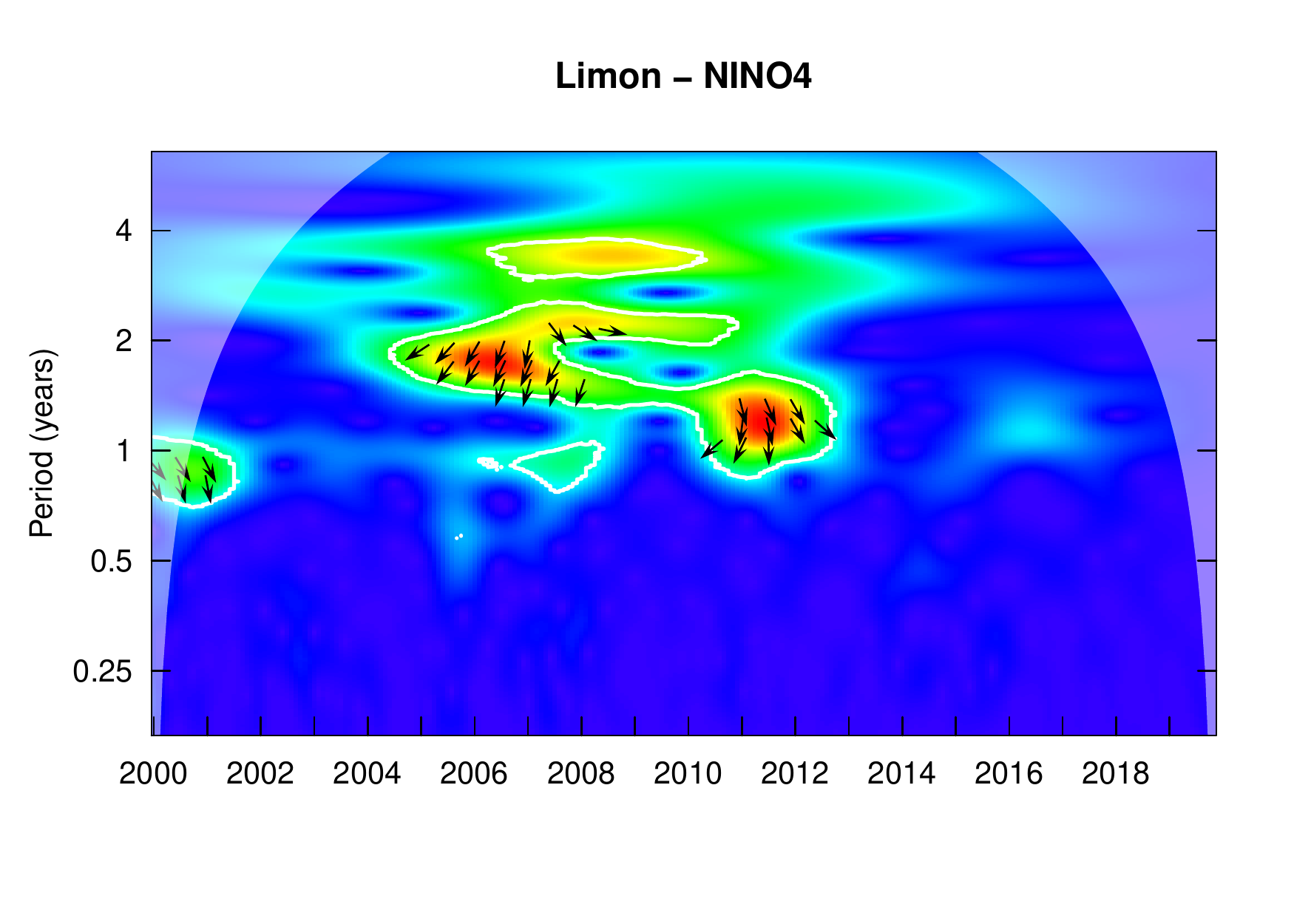}}\vspace{-0.15cm}%
\subfloat[]{\includegraphics[scale=0.23]{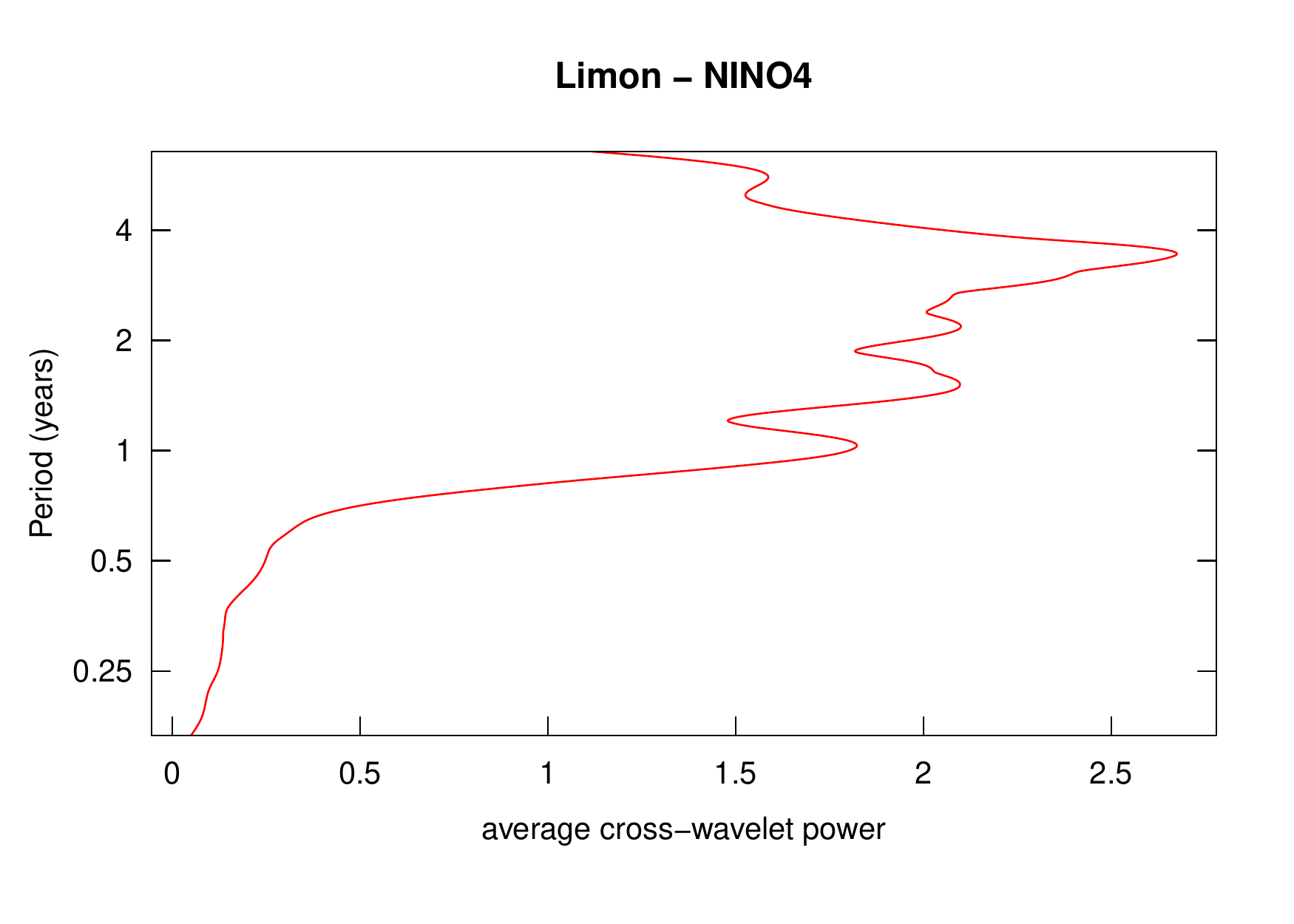}}\vspace{-0.15cm}%
\subfloat[]{\includegraphics[scale=0.23]{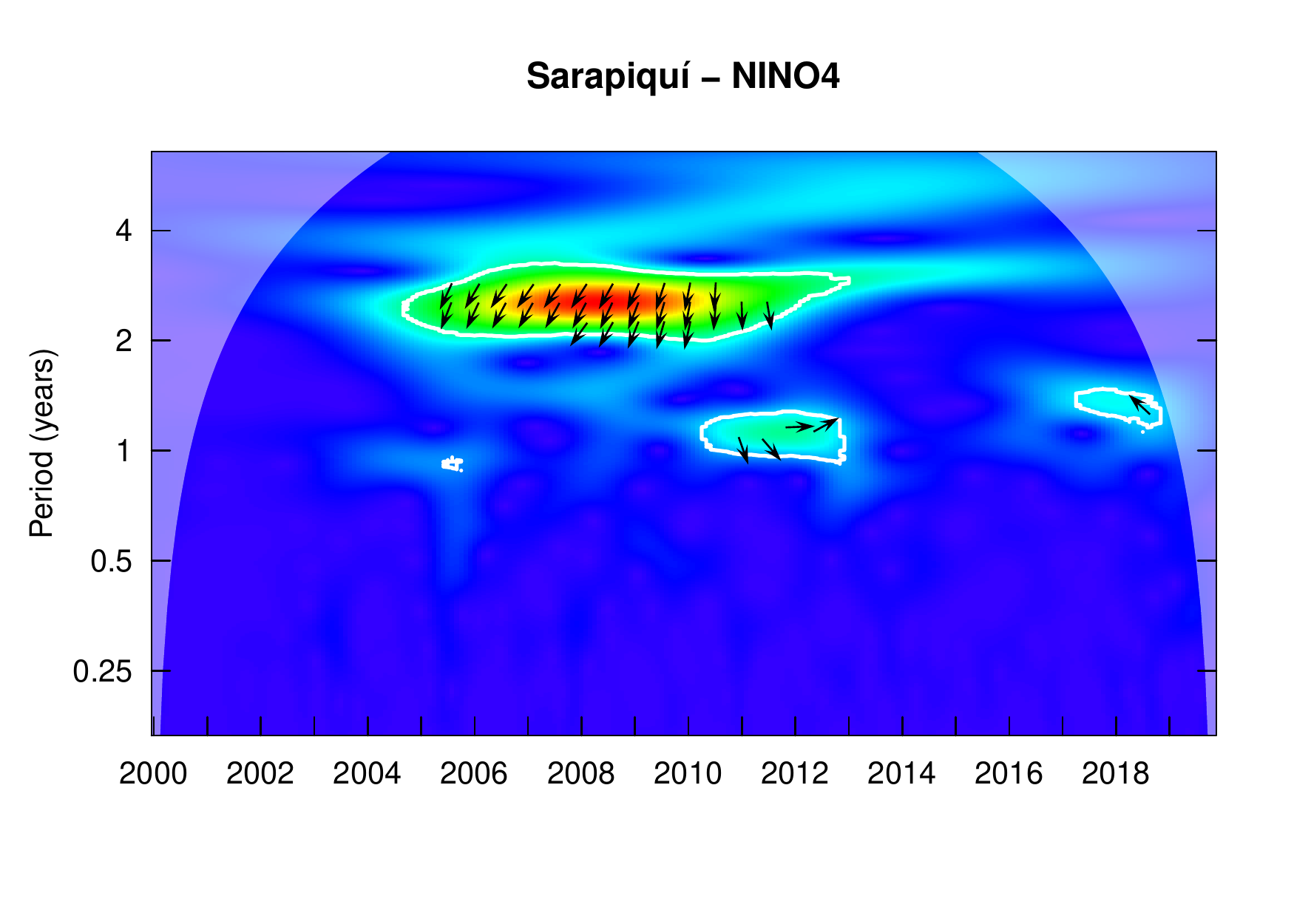}}\vspace{-0.15cm}%
\subfloat[]{\includegraphics[scale=0.23]{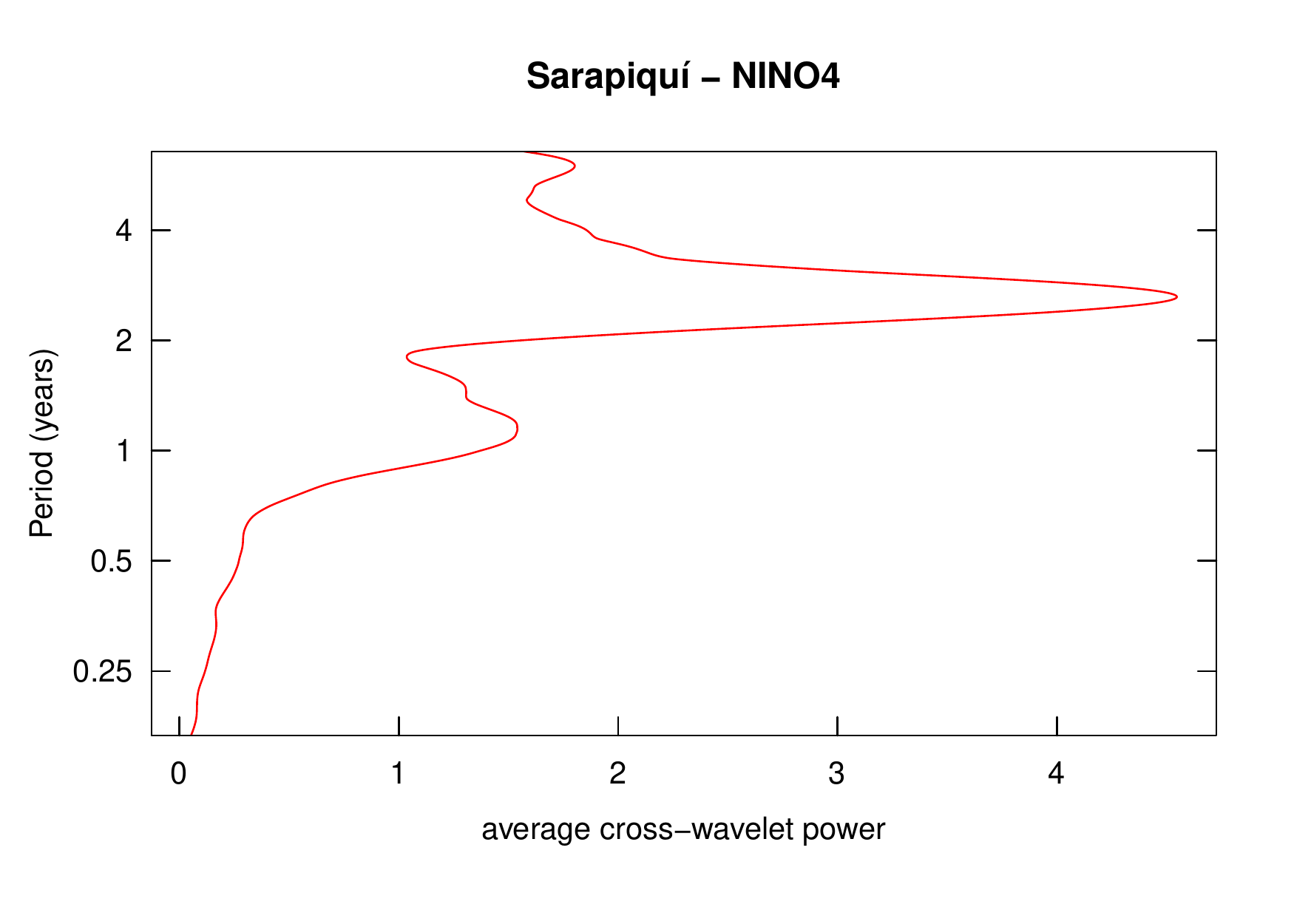}}\vspace{-0.15cm}\\
\caption*{}
\end{figure}

\begin{figure}[H]
\captionsetup[subfigure]{labelformat=empty}
\subfloat[]{\includegraphics[scale=0.23]{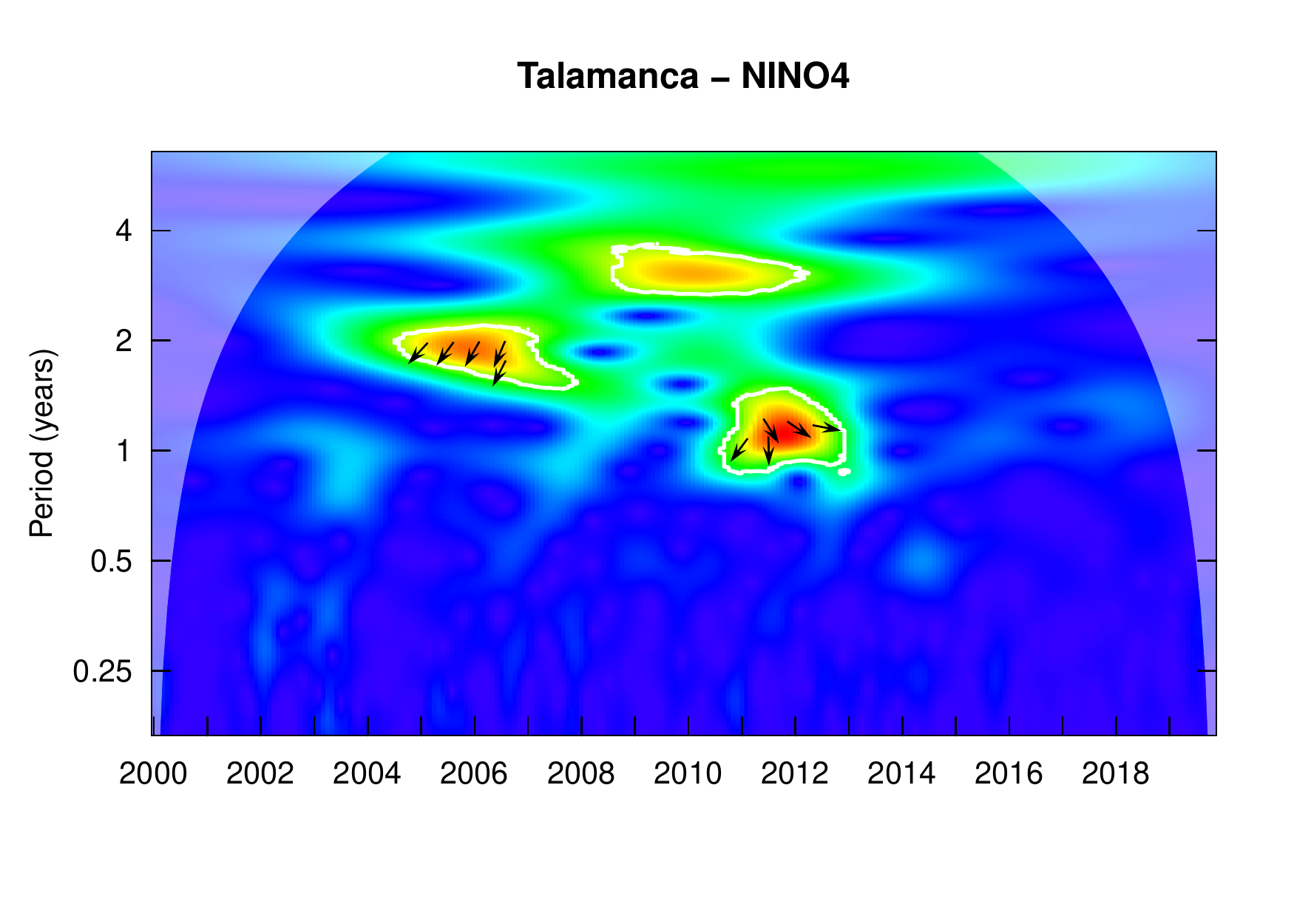}}\vspace{-0.15cm}%
\subfloat[]{\includegraphics[scale=0.23]{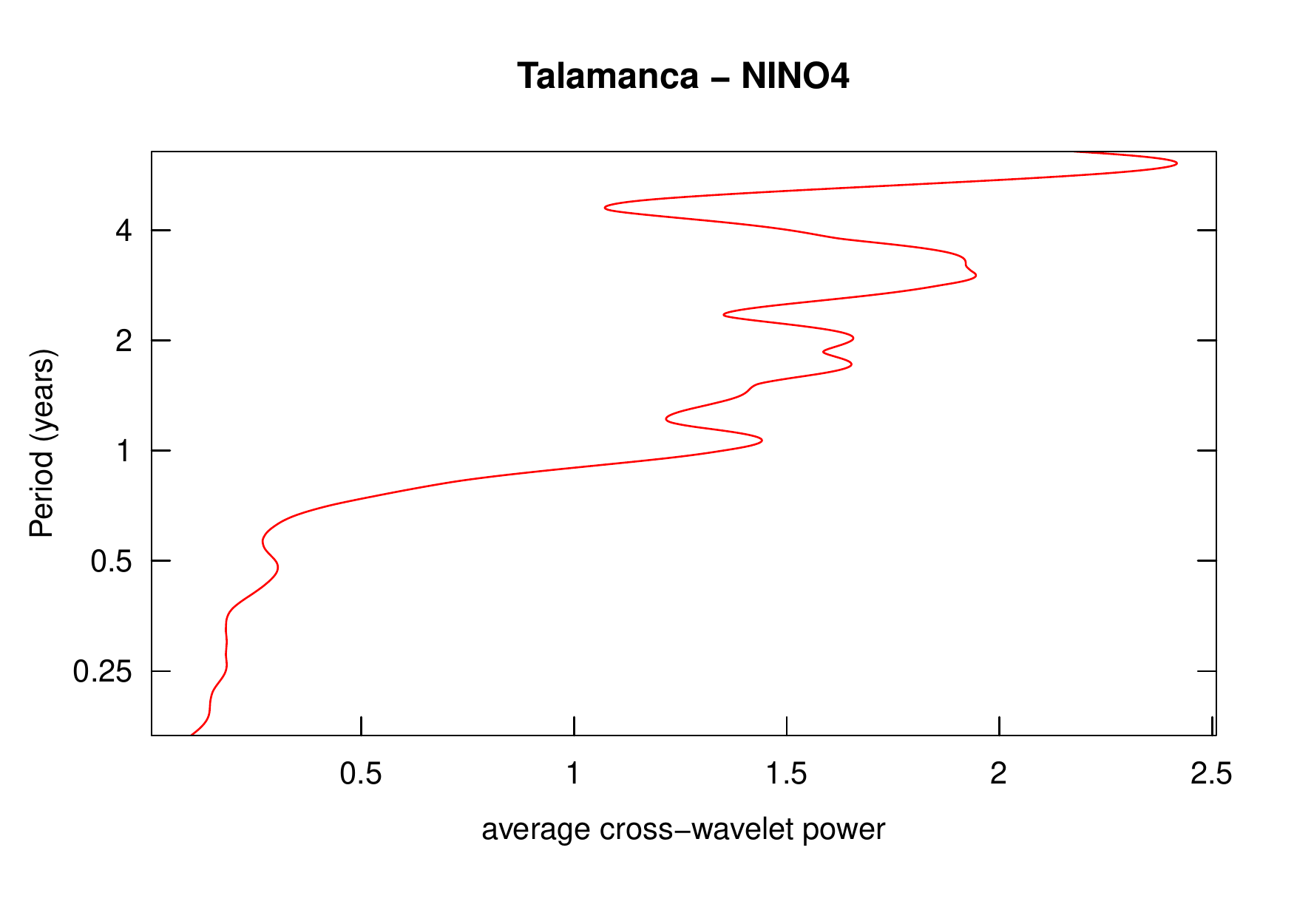}}\vspace{-0.15cm}%
\subfloat[]{\includegraphics[scale=0.23]{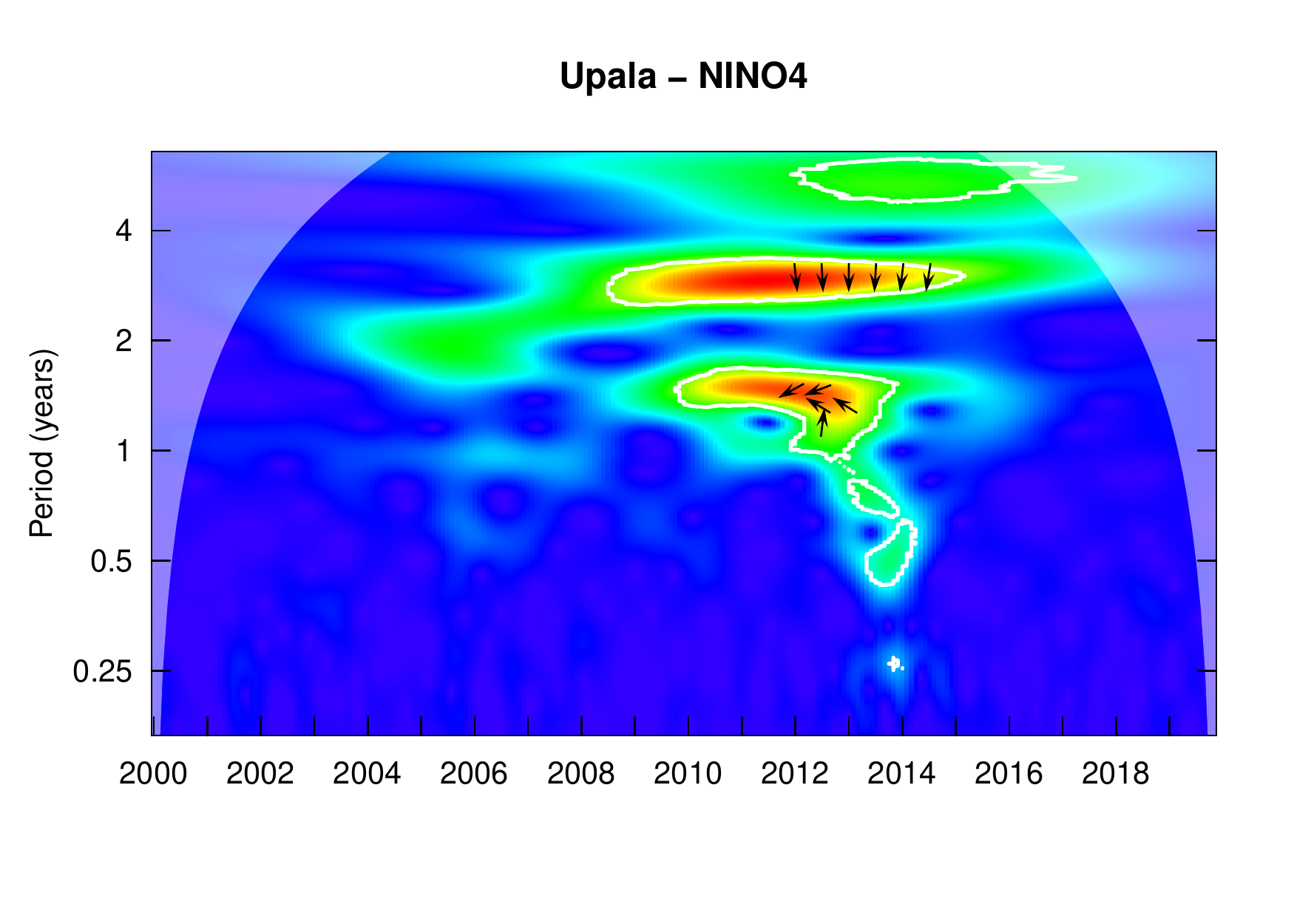}}\vspace{-0.15cm}%
\subfloat[]{\includegraphics[scale=0.23]{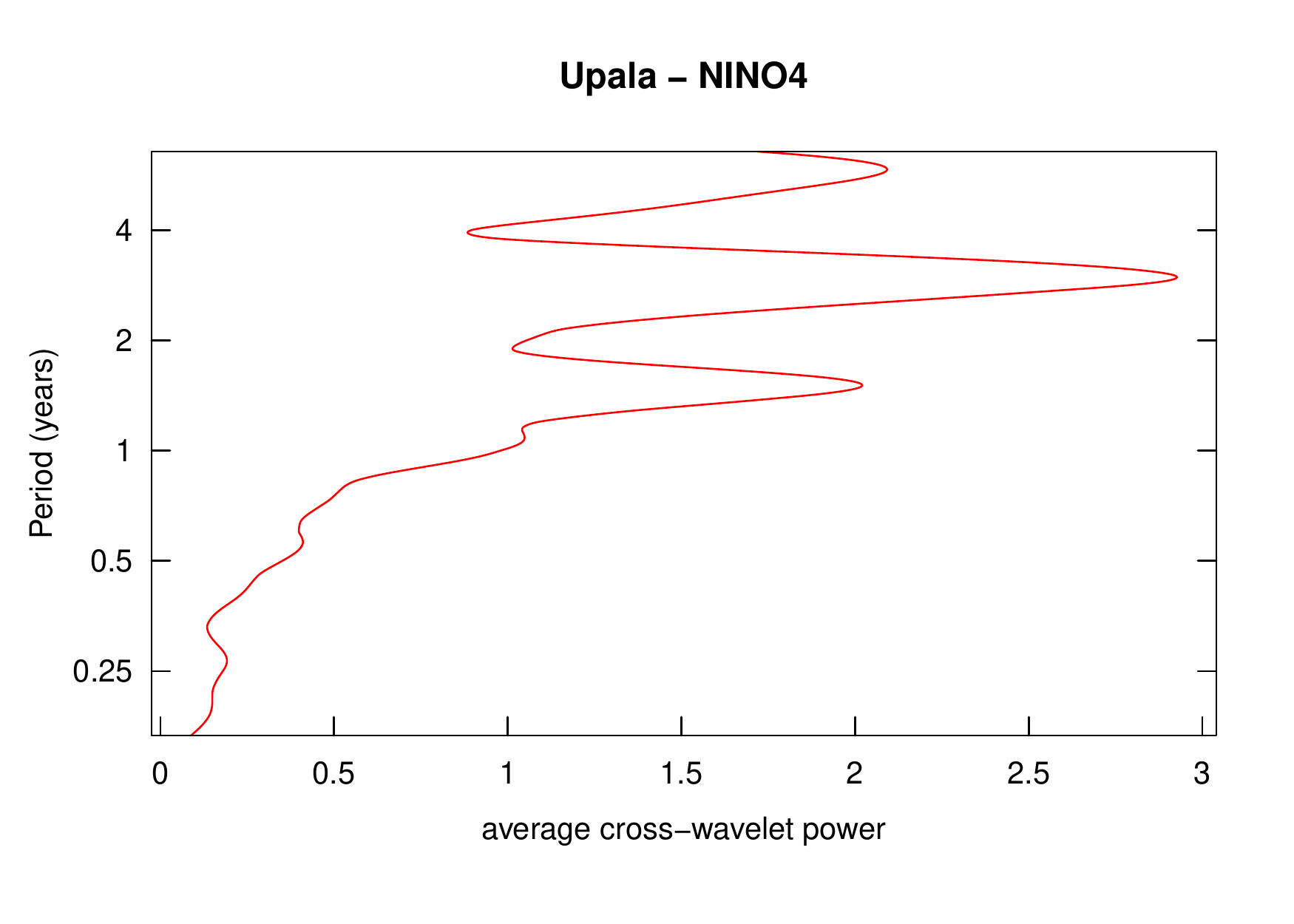}}\vspace{-0.15cm}\\
\subfloat[]{\includegraphics[scale=0.23]{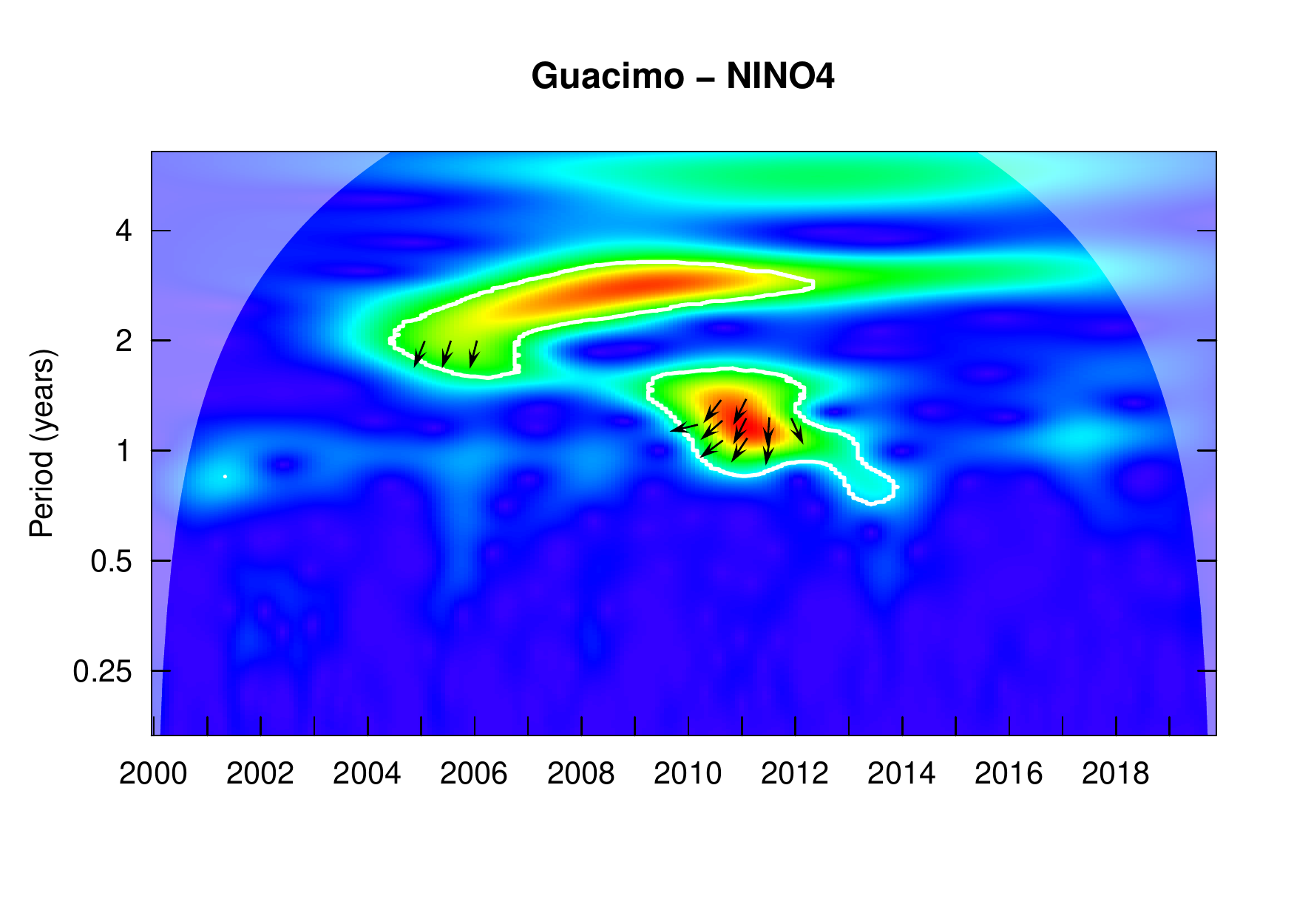}}\vspace{-0.15cm}%
\subfloat[]{\includegraphics[scale=0.23]{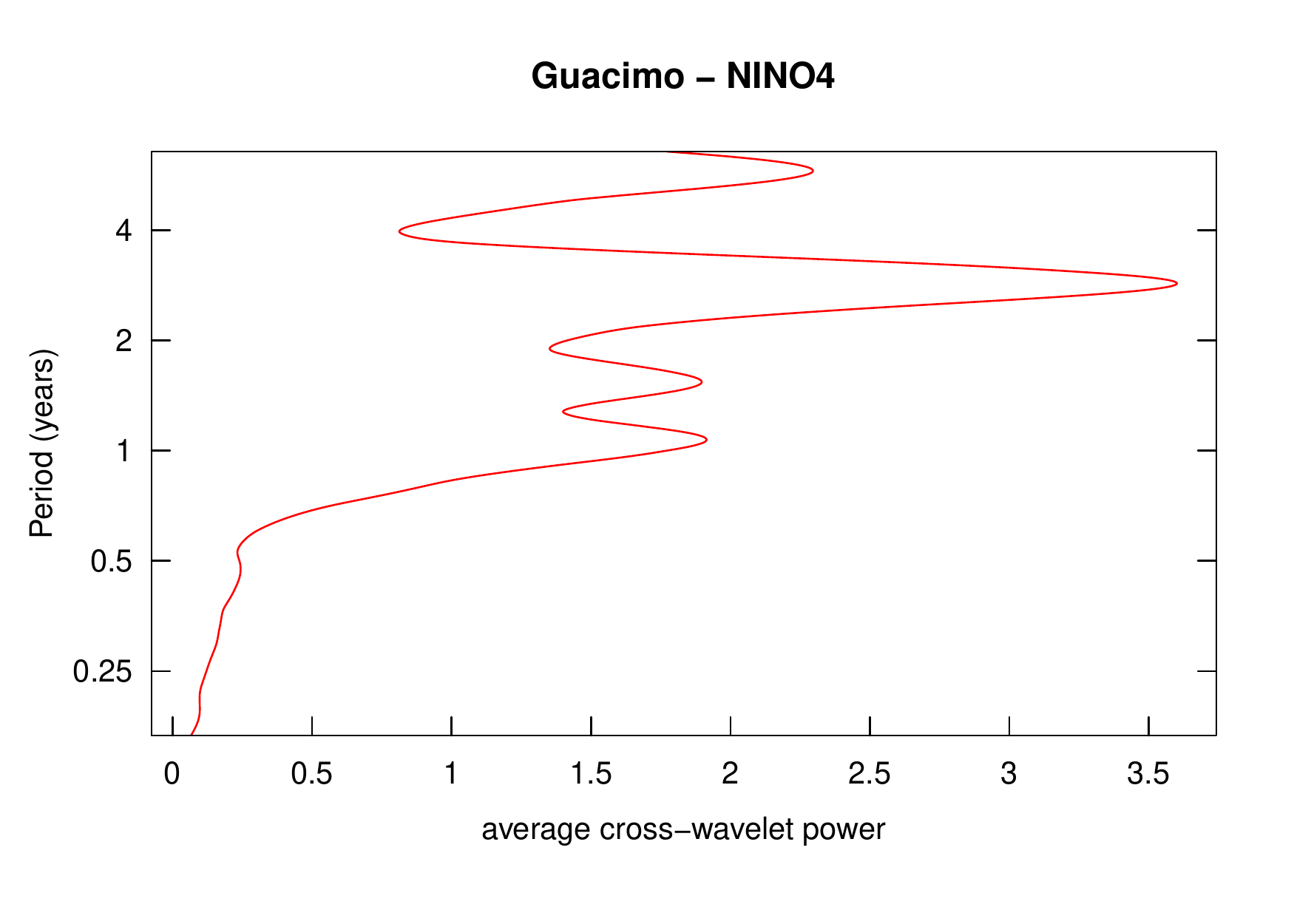}}\vspace{-0.15cm}%
\subfloat[]{\includegraphics[scale=0.23]{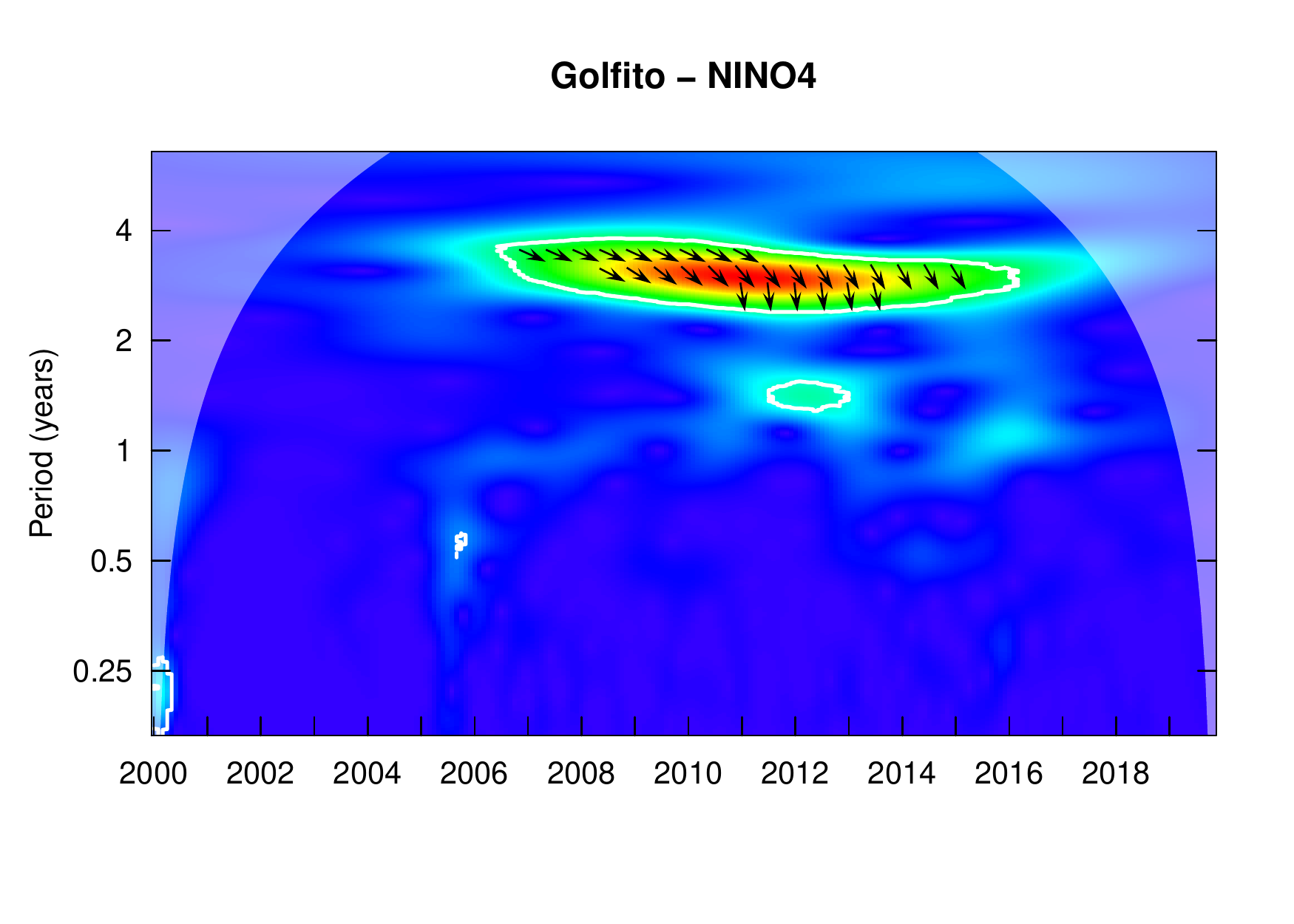}}\vspace{-0.15cm}%
\subfloat[]{\includegraphics[scale=0.23]{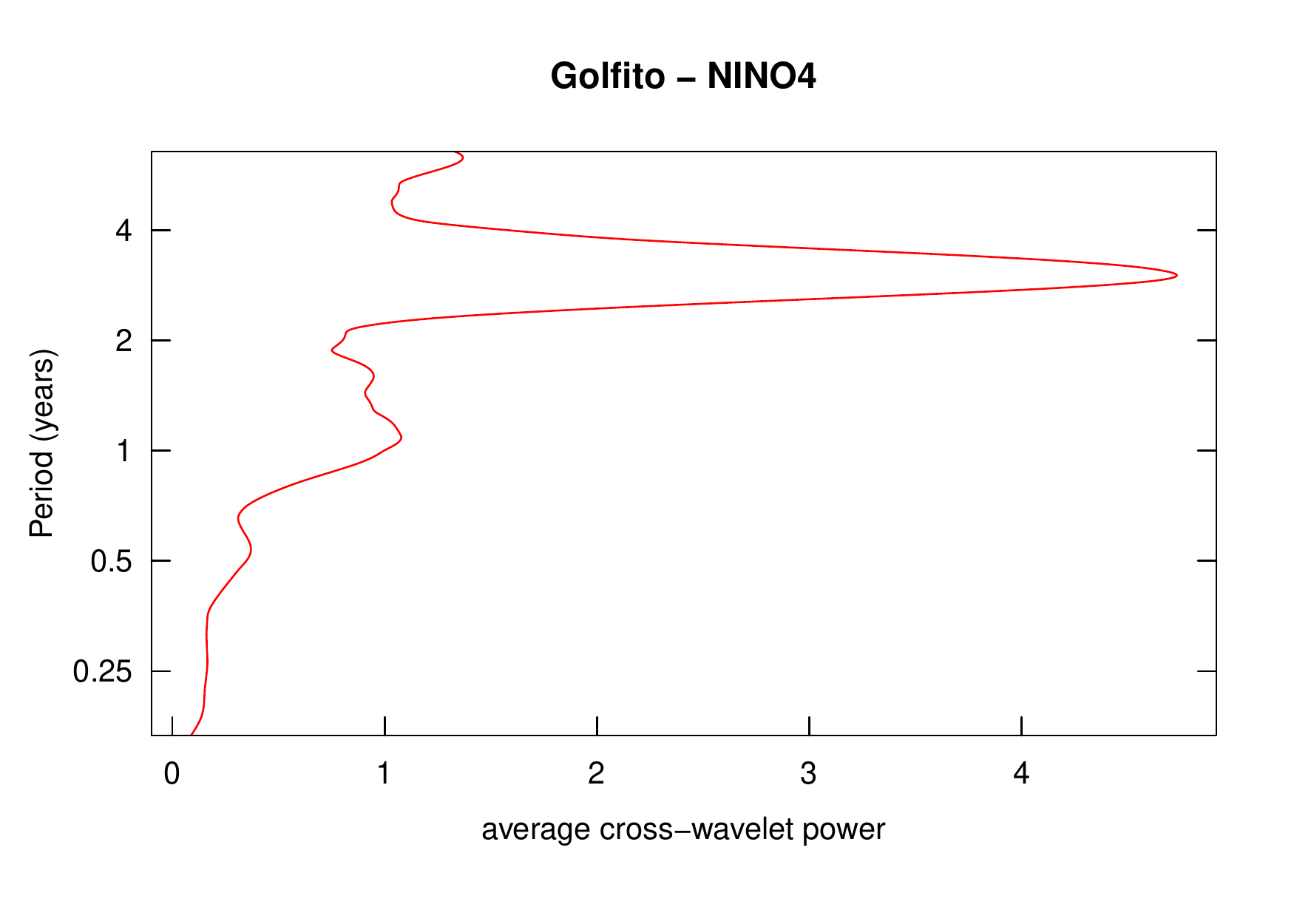}}\vspace{-0.15cm}\\
\subfloat[]{\includegraphics[scale=0.23]{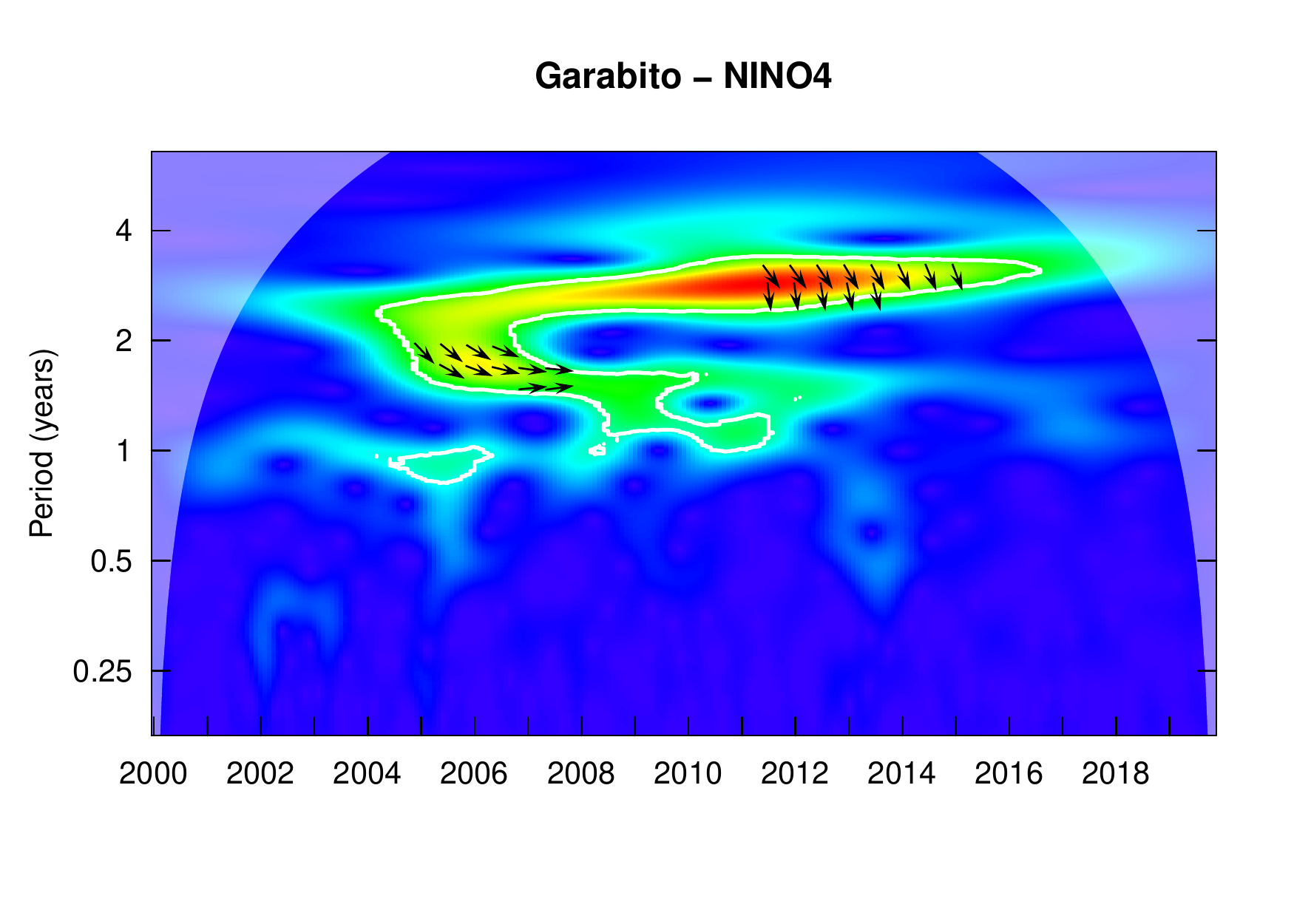}}\vspace{-0.15cm}%
\subfloat[]{\includegraphics[scale=0.23]{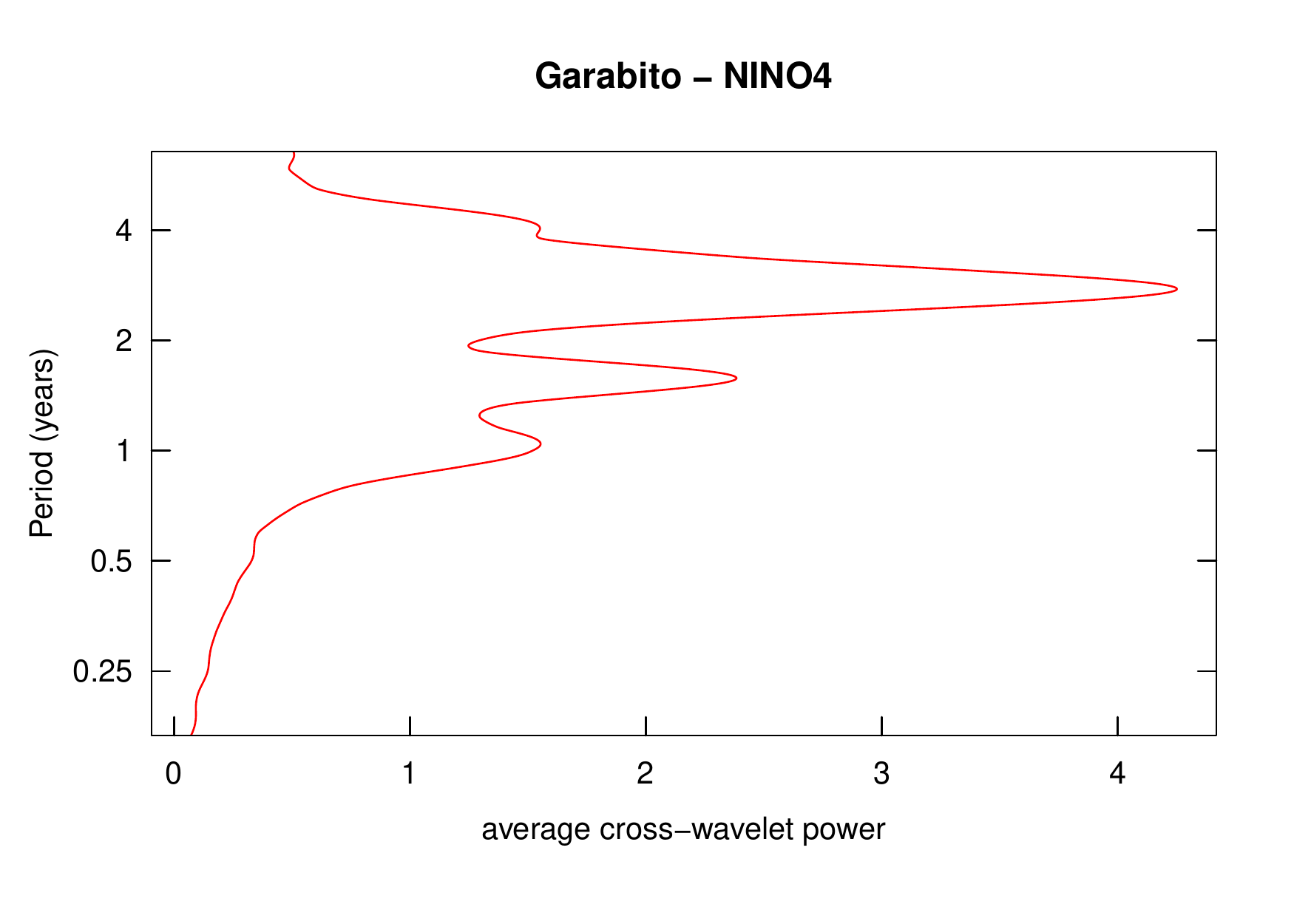}}\vspace{-0.15cm}%
\subfloat[]{\includegraphics[scale=0.23]{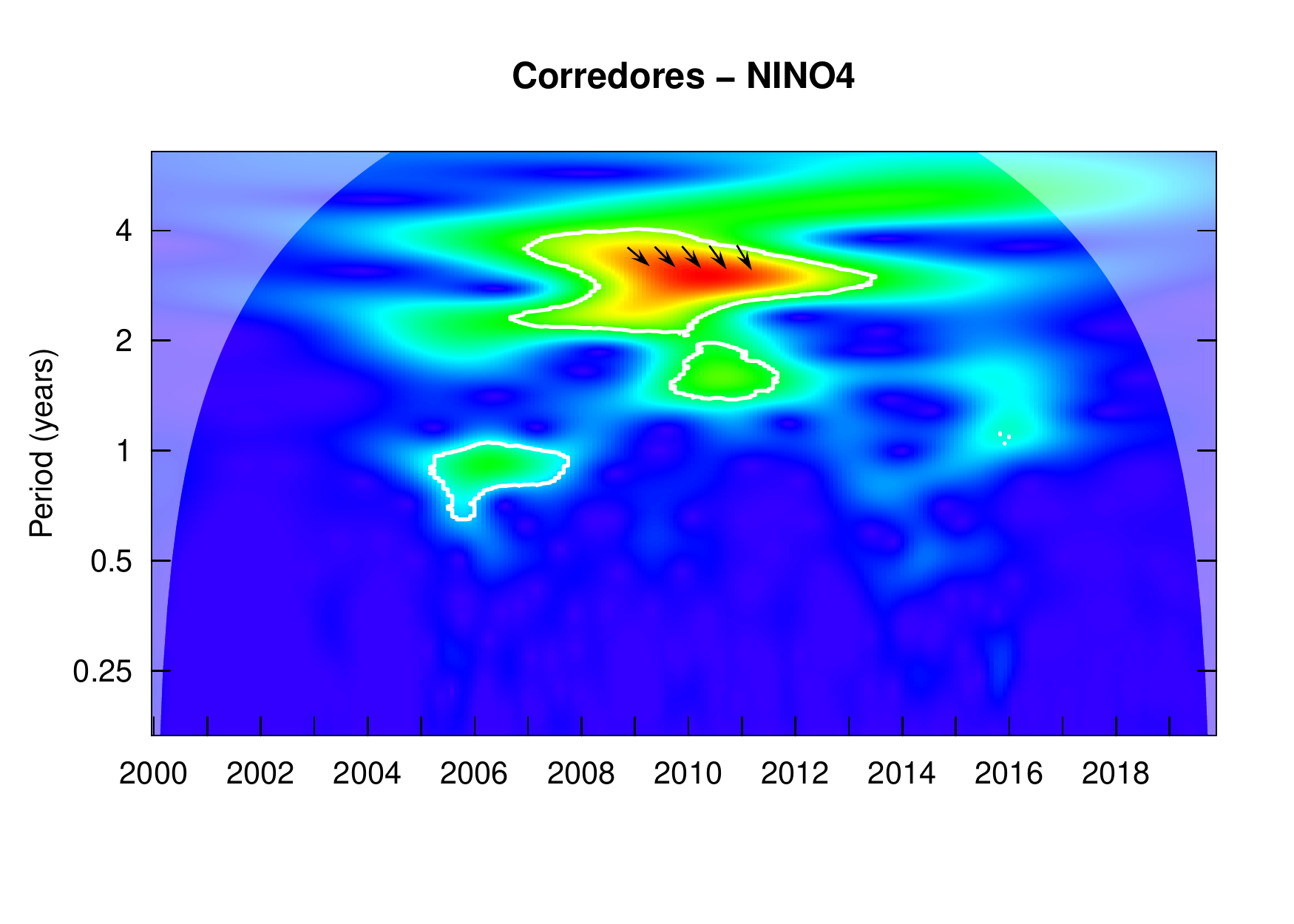}}\vspace{-0.15cm}%
\subfloat[]{\includegraphics[scale=0.23]{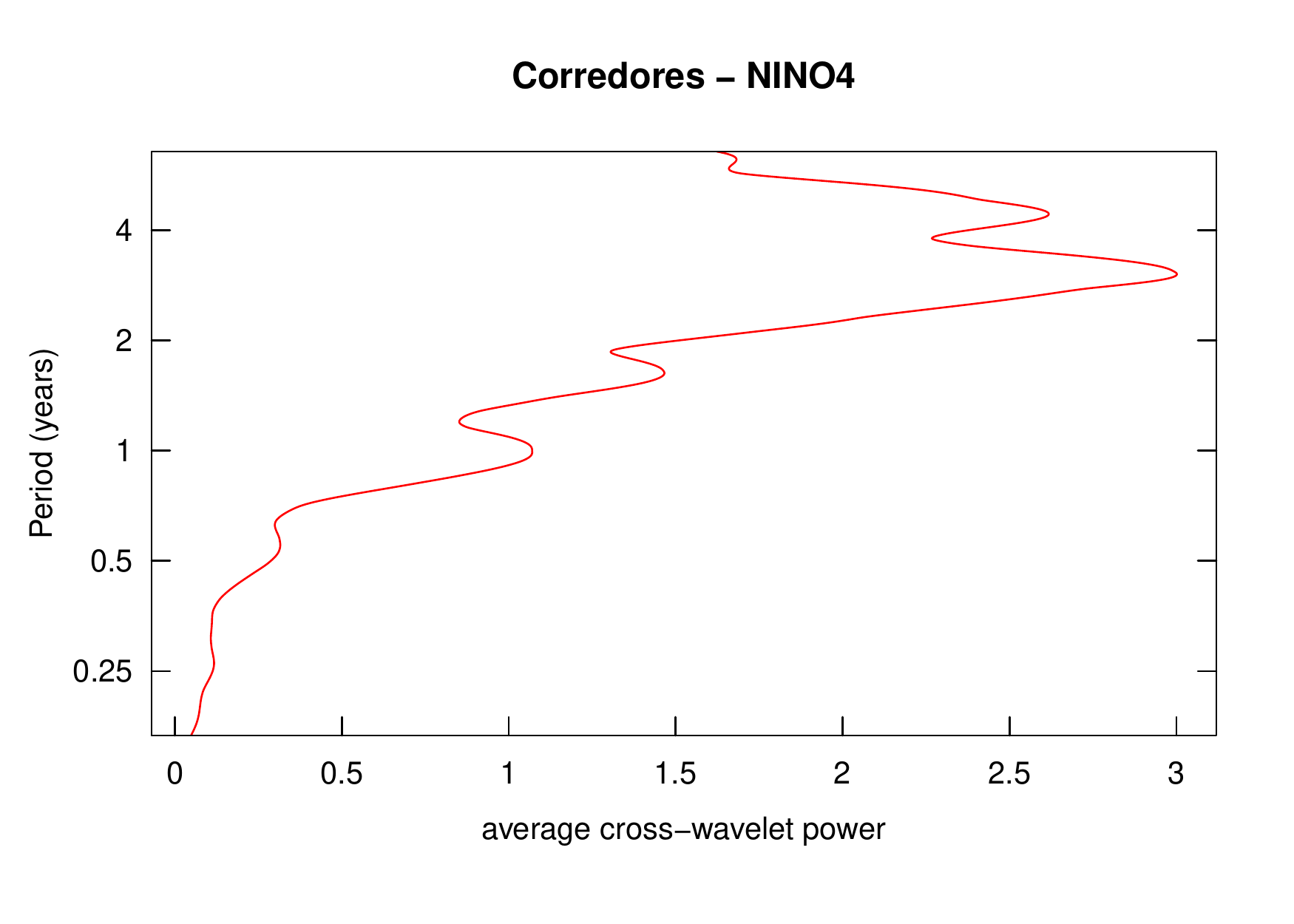}}\vspace{-0.15cm}\\
\subfloat[]{\includegraphics[scale=0.23]{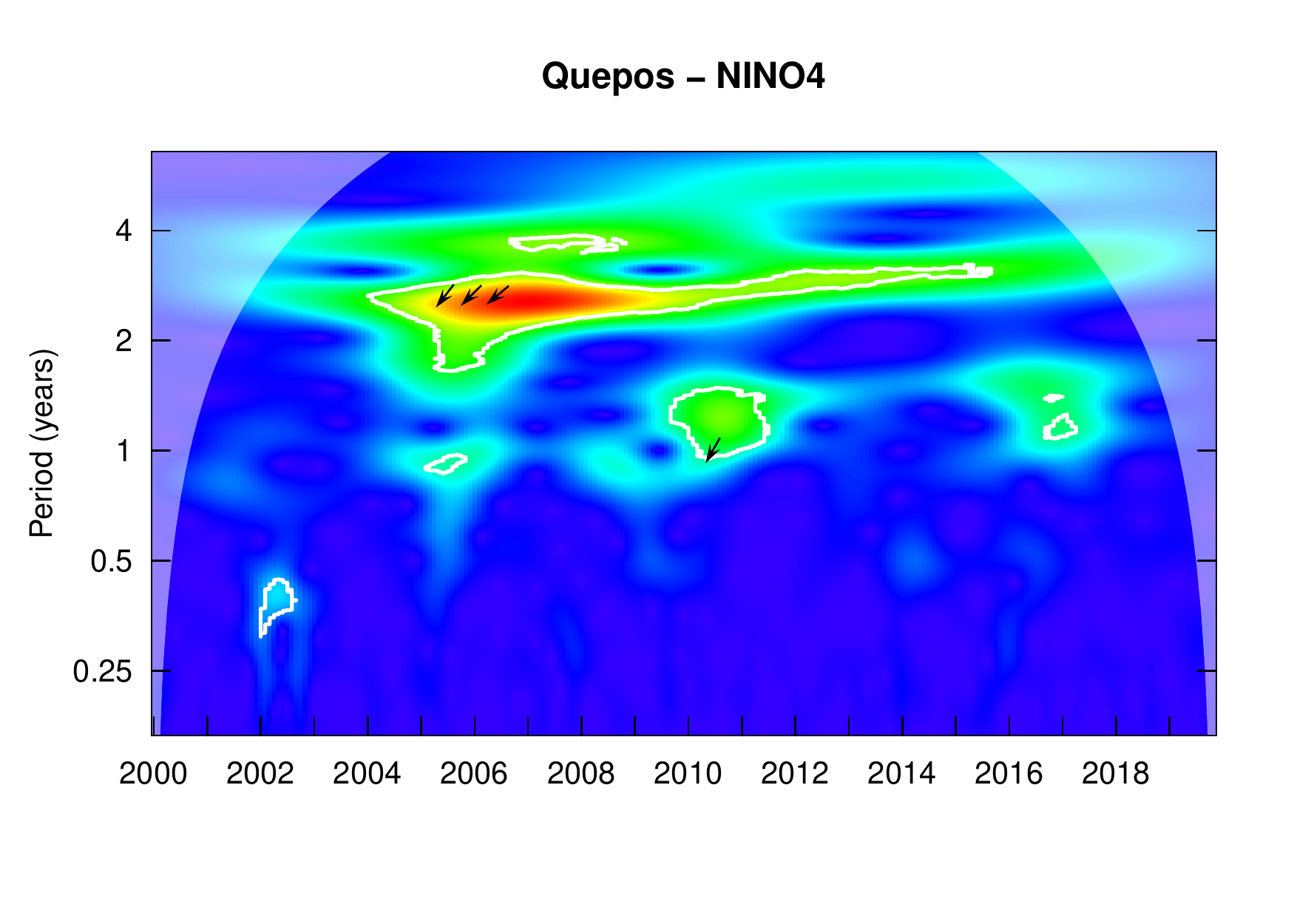}}\vspace{-0.15cm}%
\subfloat[]{\includegraphics[scale=0.23]{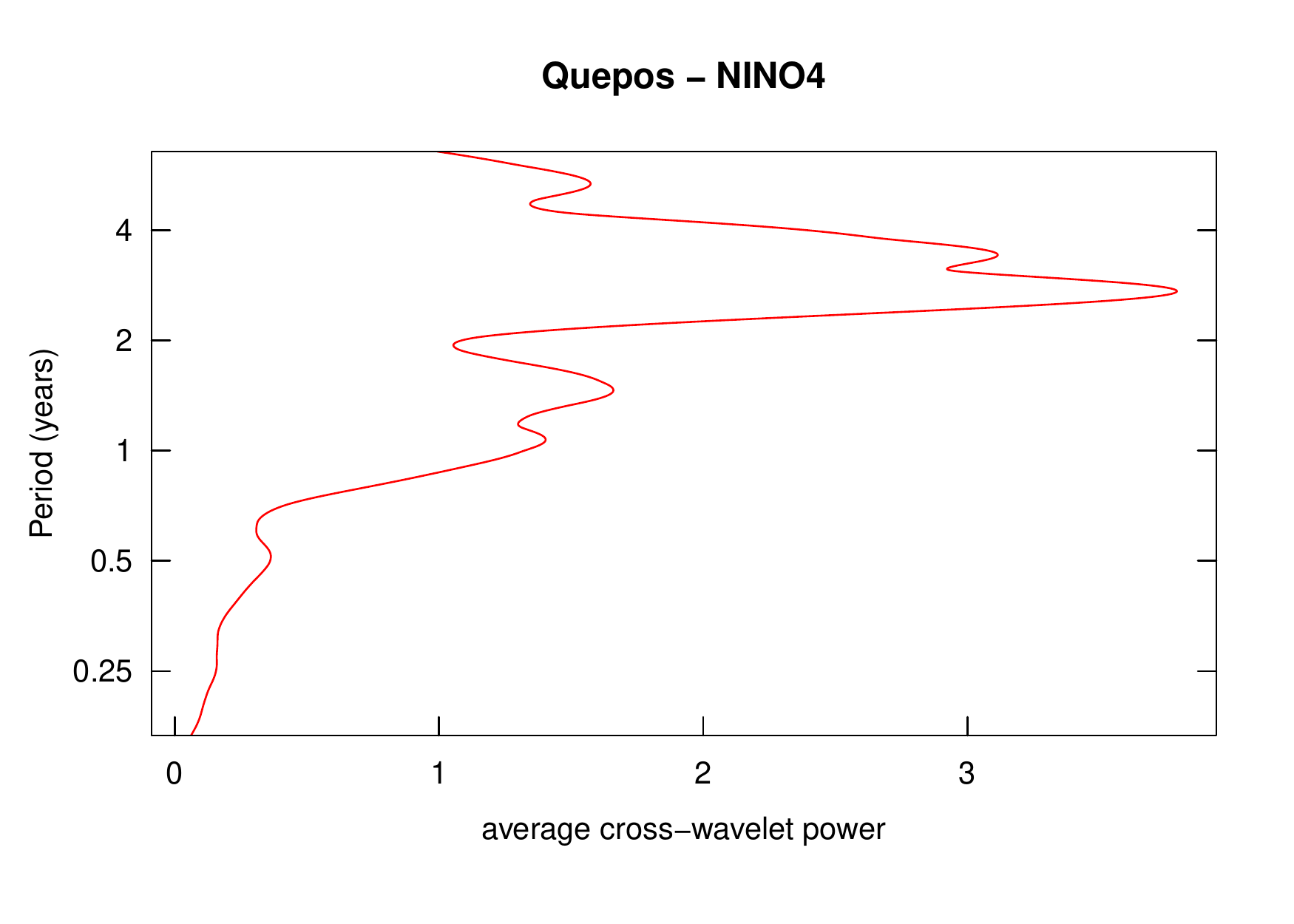}}\vspace{-0.15cm}%
\subfloat[]{\includegraphics[scale=0.23]{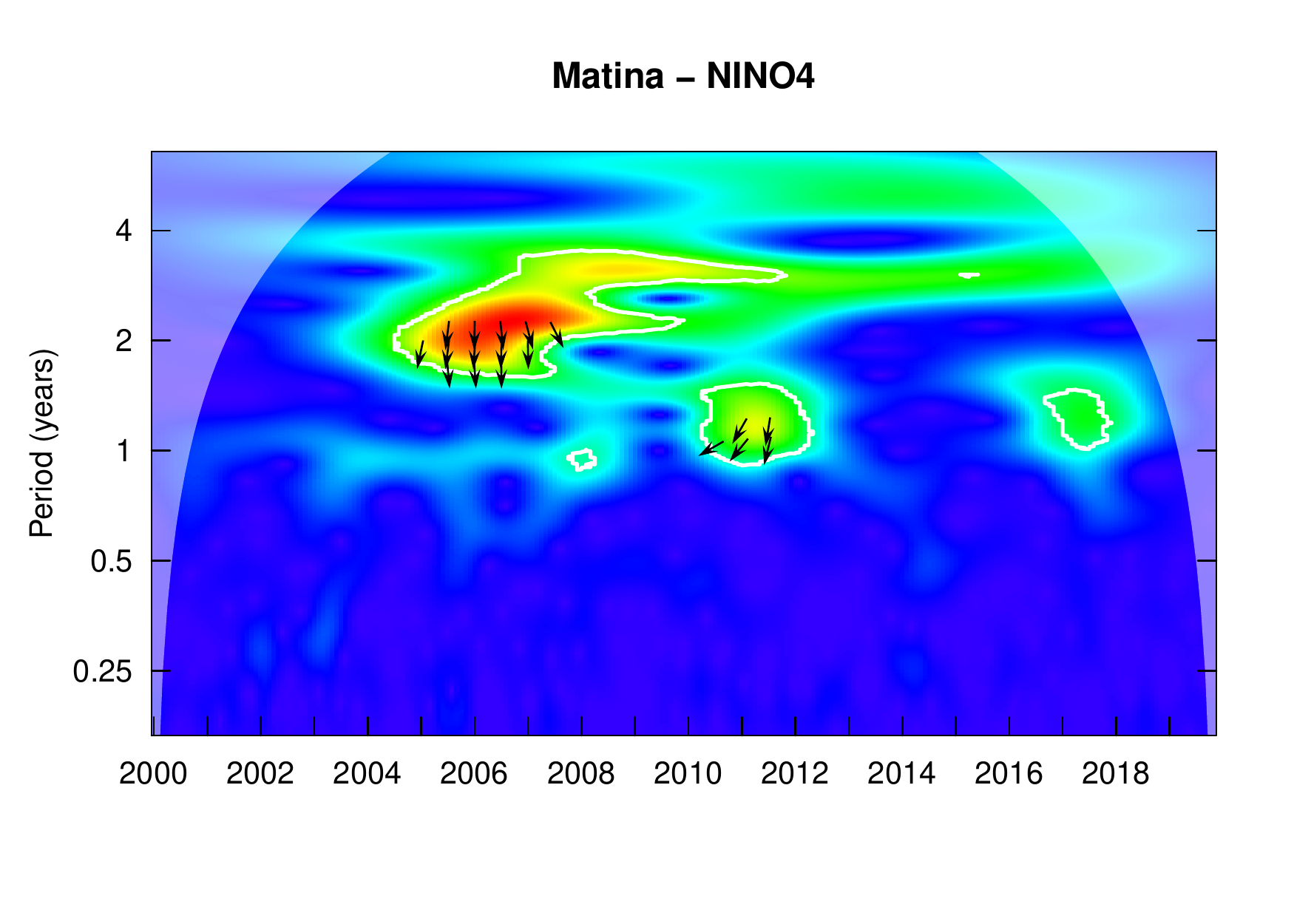}}\vspace{-0.15cm}%
\subfloat[]{\includegraphics[scale=0.23]{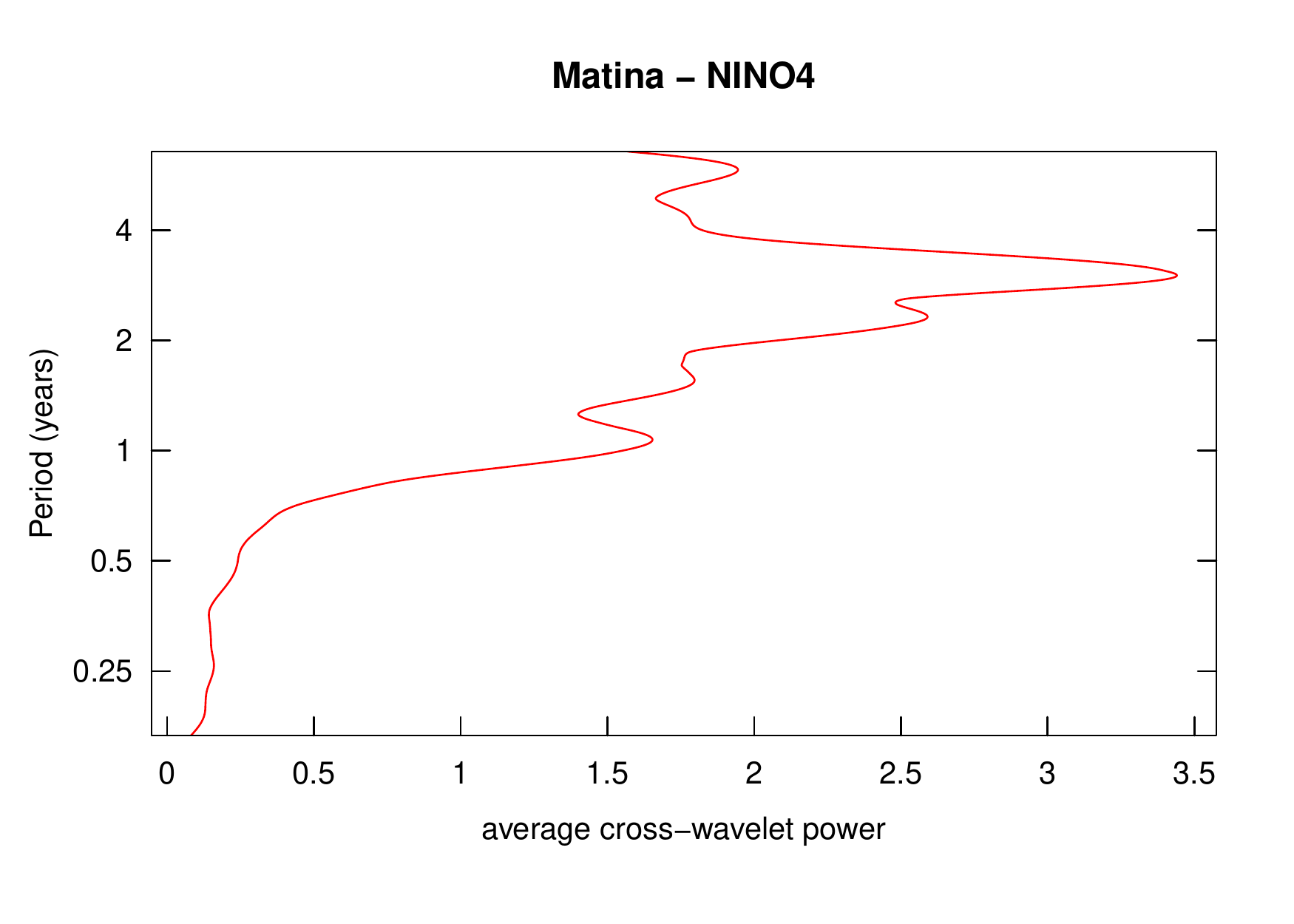}}\vspace{-0.15cm}\\
\subfloat[]{\includegraphics[scale=0.23]{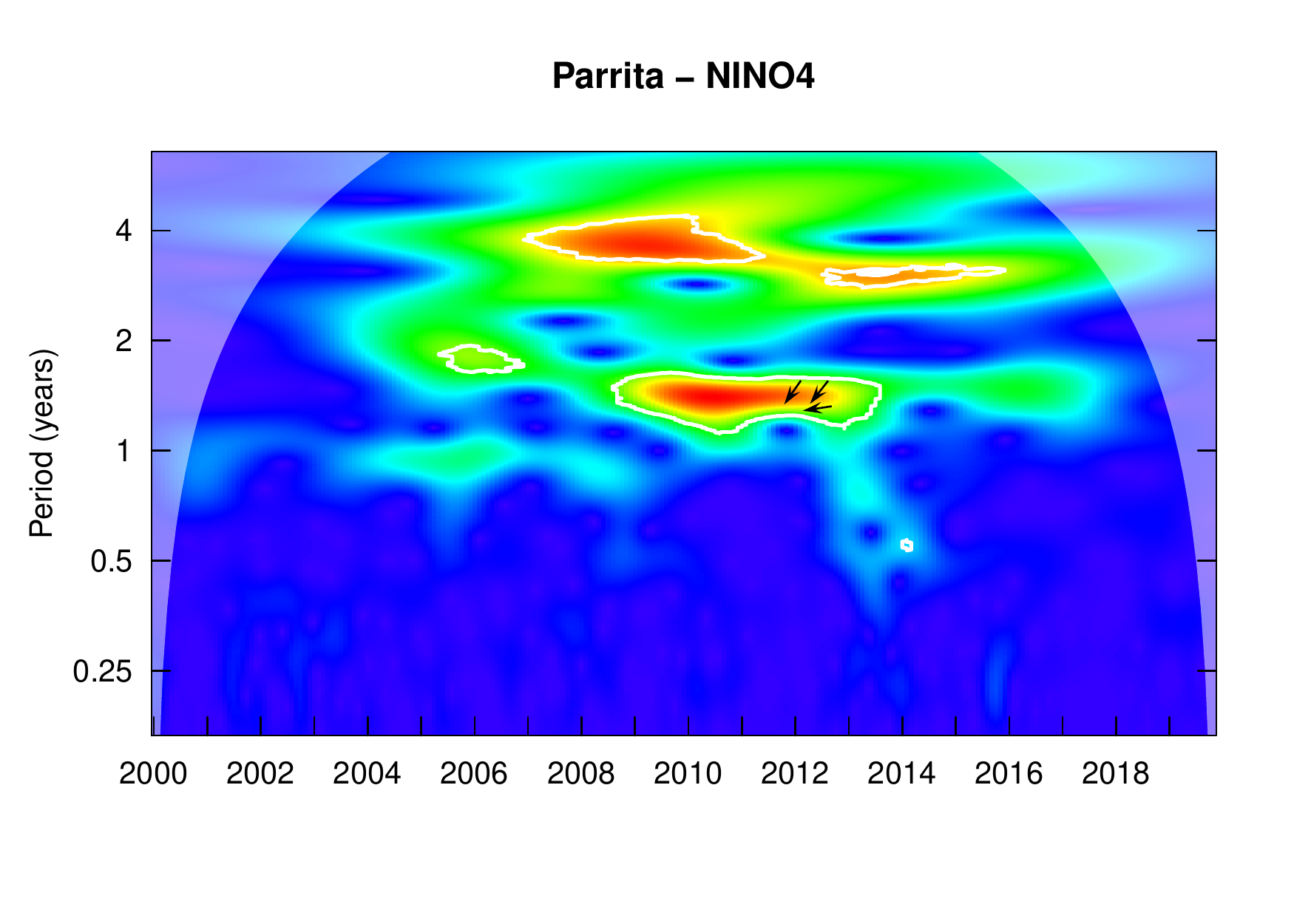}}\vspace{-0.15cm}%
\subfloat[]{\includegraphics[scale=0.23]{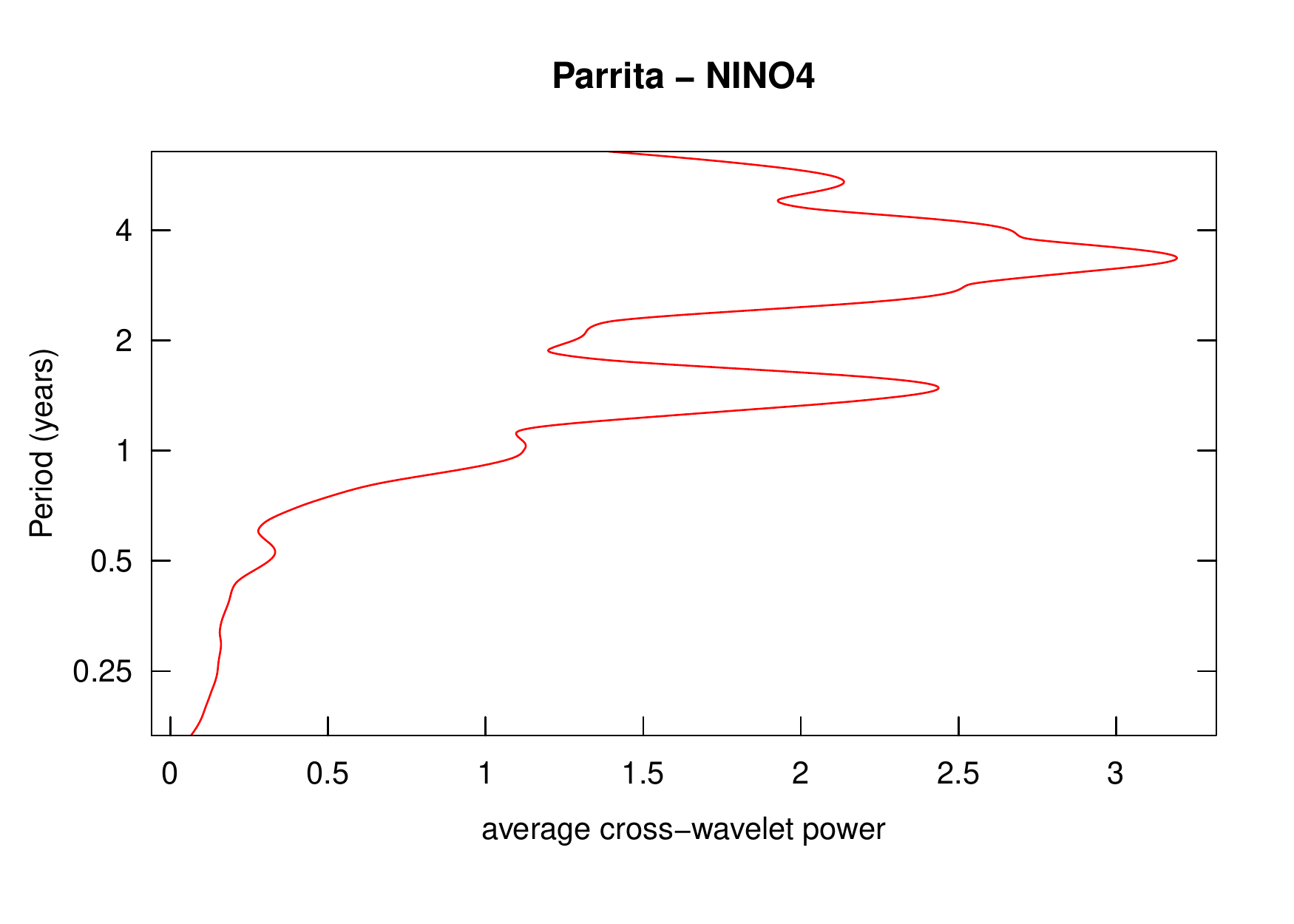}}\vspace{-0.15cm}%
\subfloat[]{\includegraphics[scale=0.23]{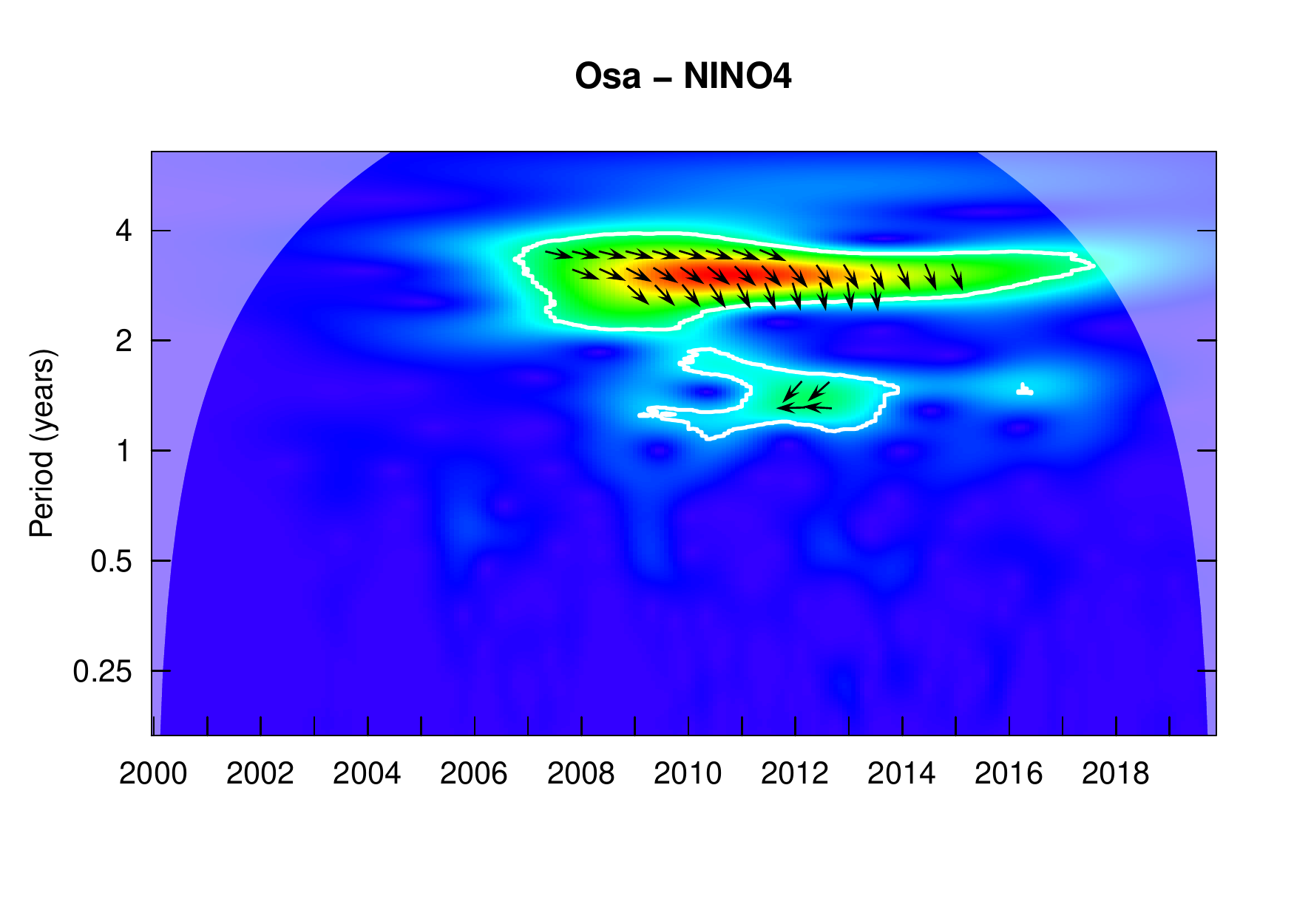}}\vspace{-0.15cm}%
\subfloat[]{\includegraphics[scale=0.23]{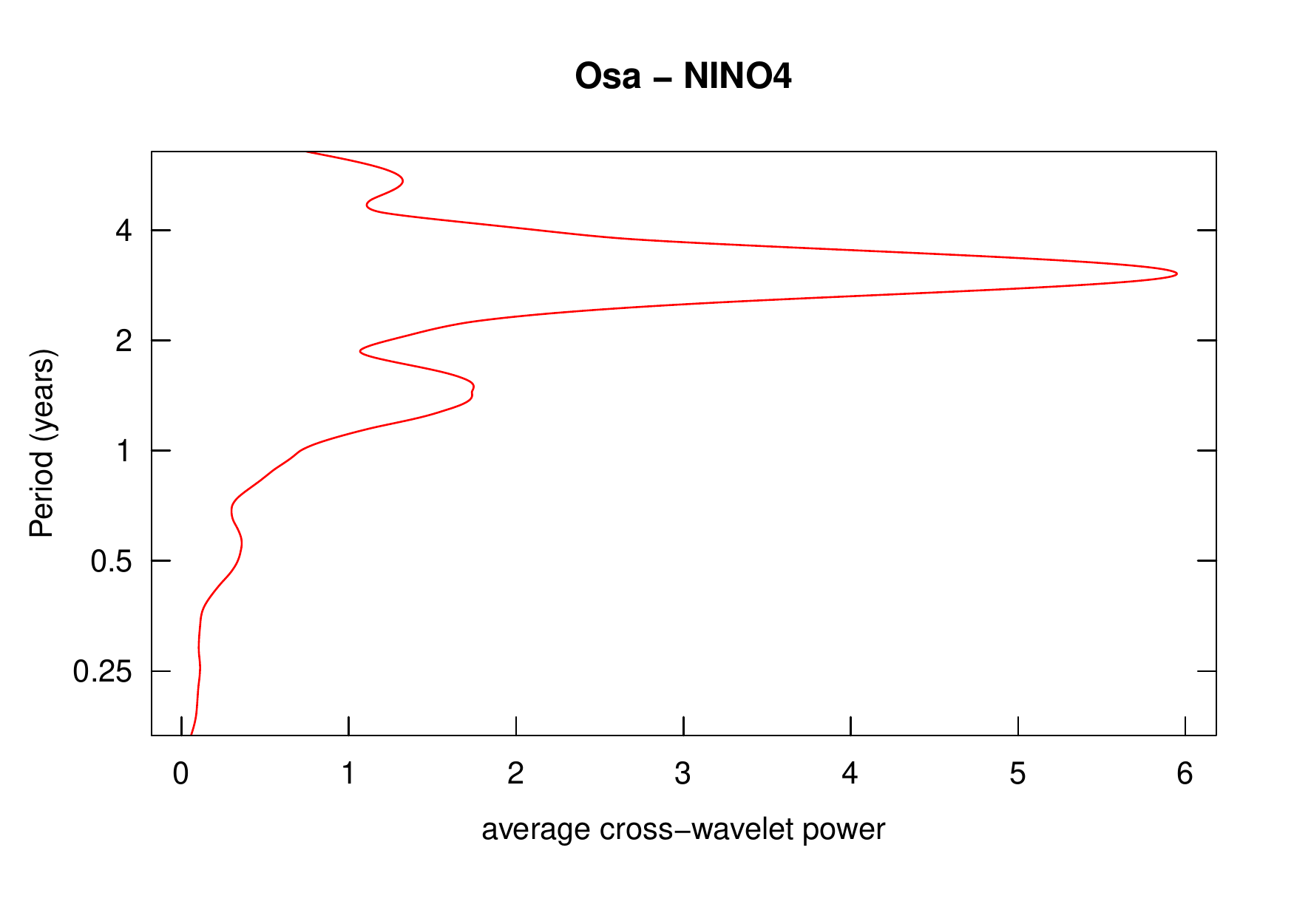}}\vspace{-0.15cm}%
\caption*{}
\end{figure}

\section*{Phase difference}

Phase difference provides information about time series synchronization \cite{rosch2016waveletcomp}. An absolute value less (larger) than $\pi/2$ or the equivalent period (days, weeks, months, or years) indicates that the two series move in phase (anti-phase, respectively). The sign of the phase difference shows which series is the leading one in this relationship. In the figures below, the horizontal lines indicate the points corresponding to $-\pi/2$ and $\pi/2$.

The left panels in Figures S11, S12, S13 show that the incidence of dengue and the time series for EVI, NDVI, and NDWI are in phase. However, the leading and lagging series change over time. When the difference is negative, the climate and vegetation index is the leading series, if it is positive, the incidence of dengue is the leading one. The lag range between 0 to 3 months. On the other hand, the panels on the right present the series without consistent behavior over time.

\begin{figure}[H]
\captionsetup[subfigure]{labelformat=empty}
\subfloat[]{\includegraphics[scale=0.5]{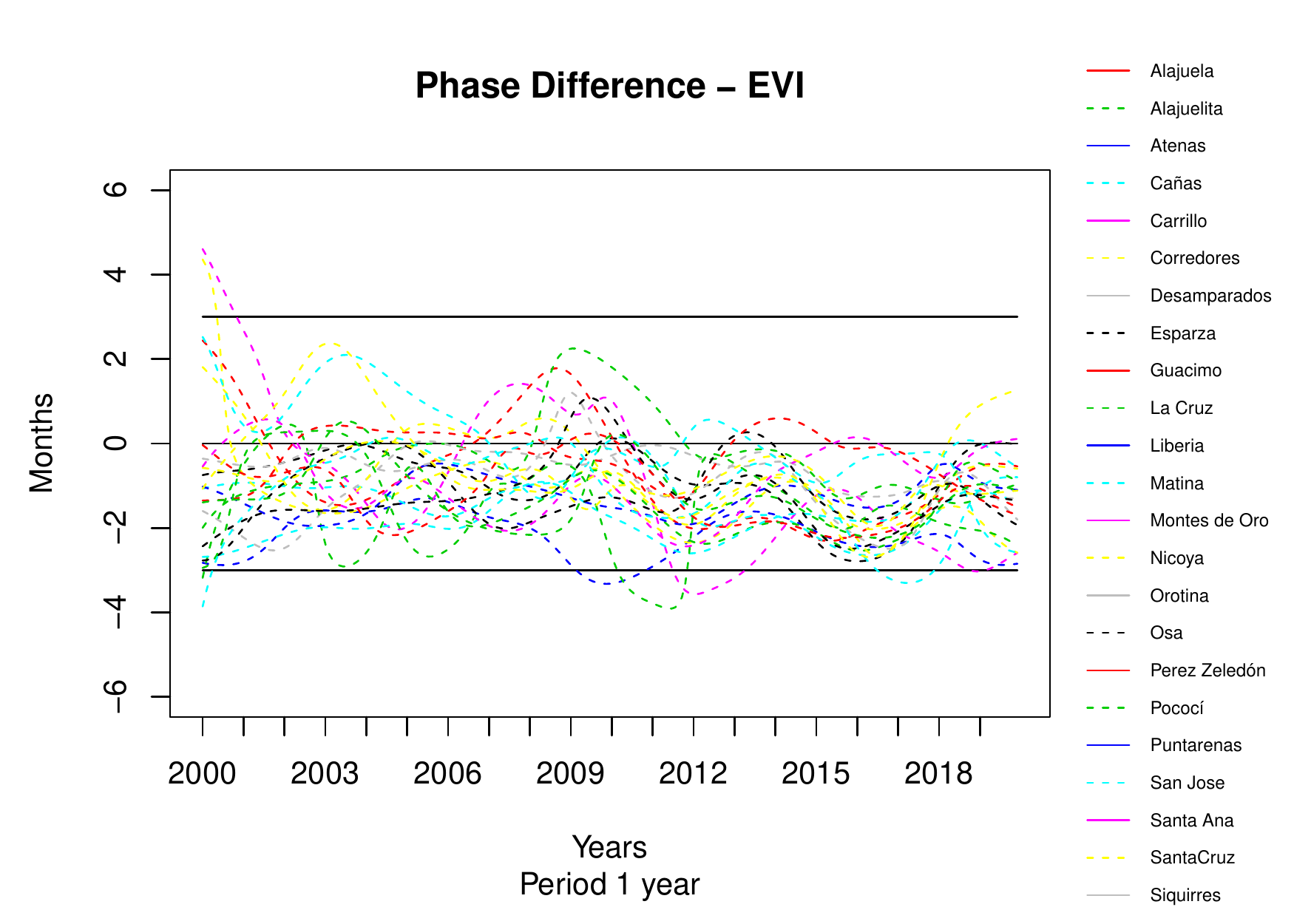}}\vspace{-0.15cm}%
\subfloat[]{\includegraphics[scale=0.5]{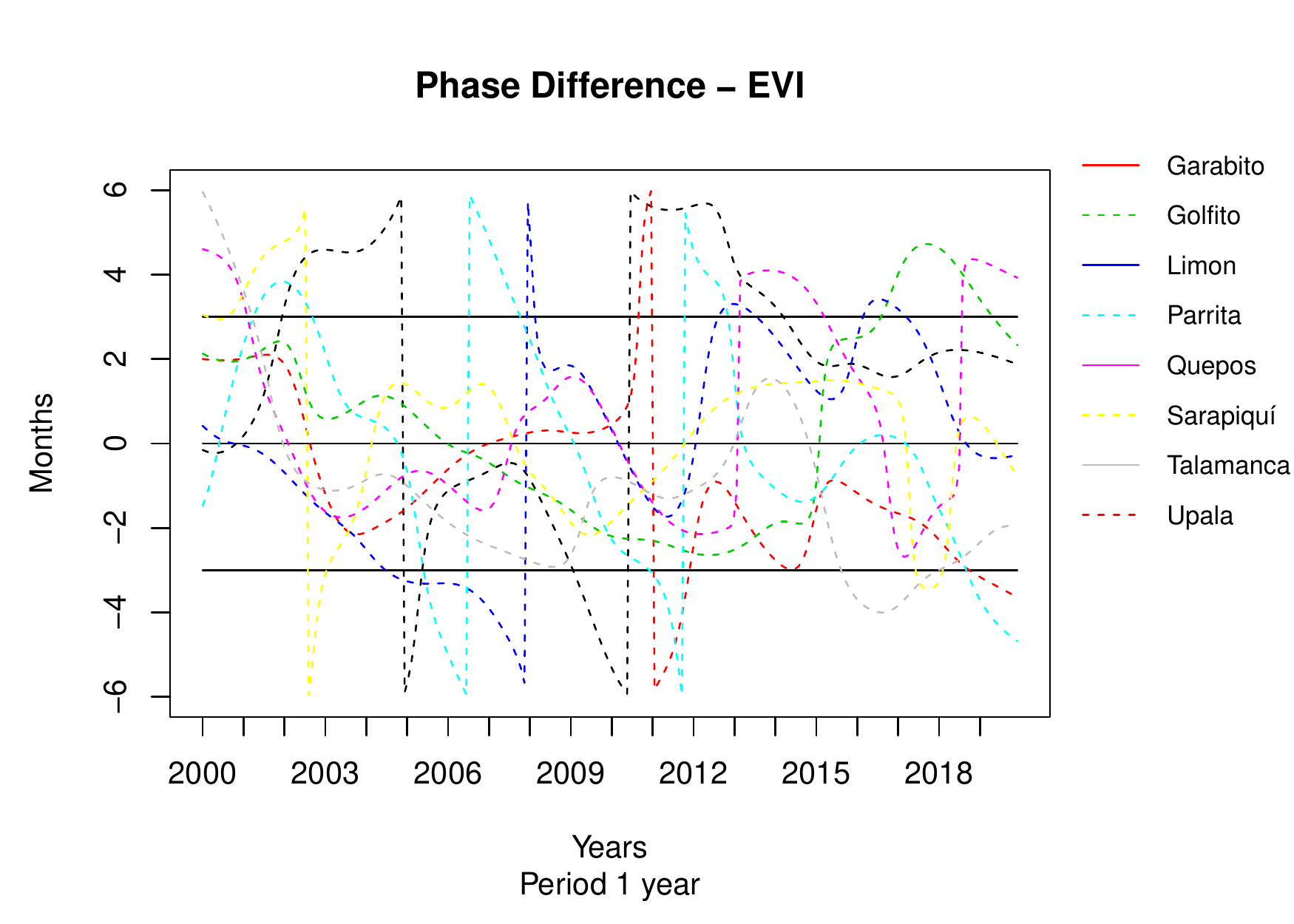}}\vspace{-0.15cm}%
\caption*{\textbf{Figure S11:} Phase difference at period 1-yr. Left panel: cantons in-phase. Right panel: cantos with an irregular behaviour over time.}
\end{figure}

\begin{figure}[H]
\captionsetup[subfigure]{labelformat=empty}
\subfloat[]{\includegraphics[scale=0.5]{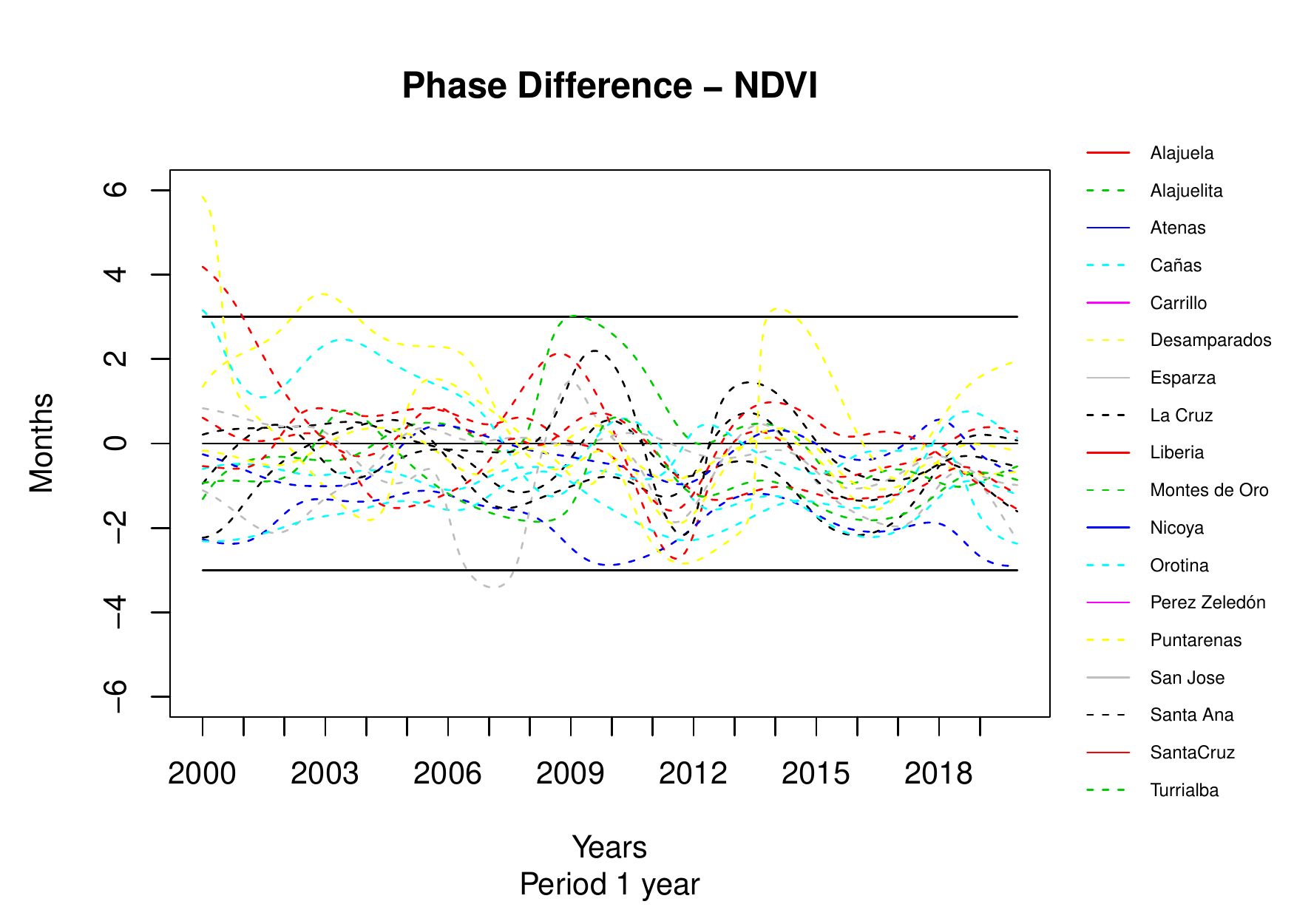}}\vspace{-0.15cm}%
\subfloat[]{\includegraphics[scale=0.5]{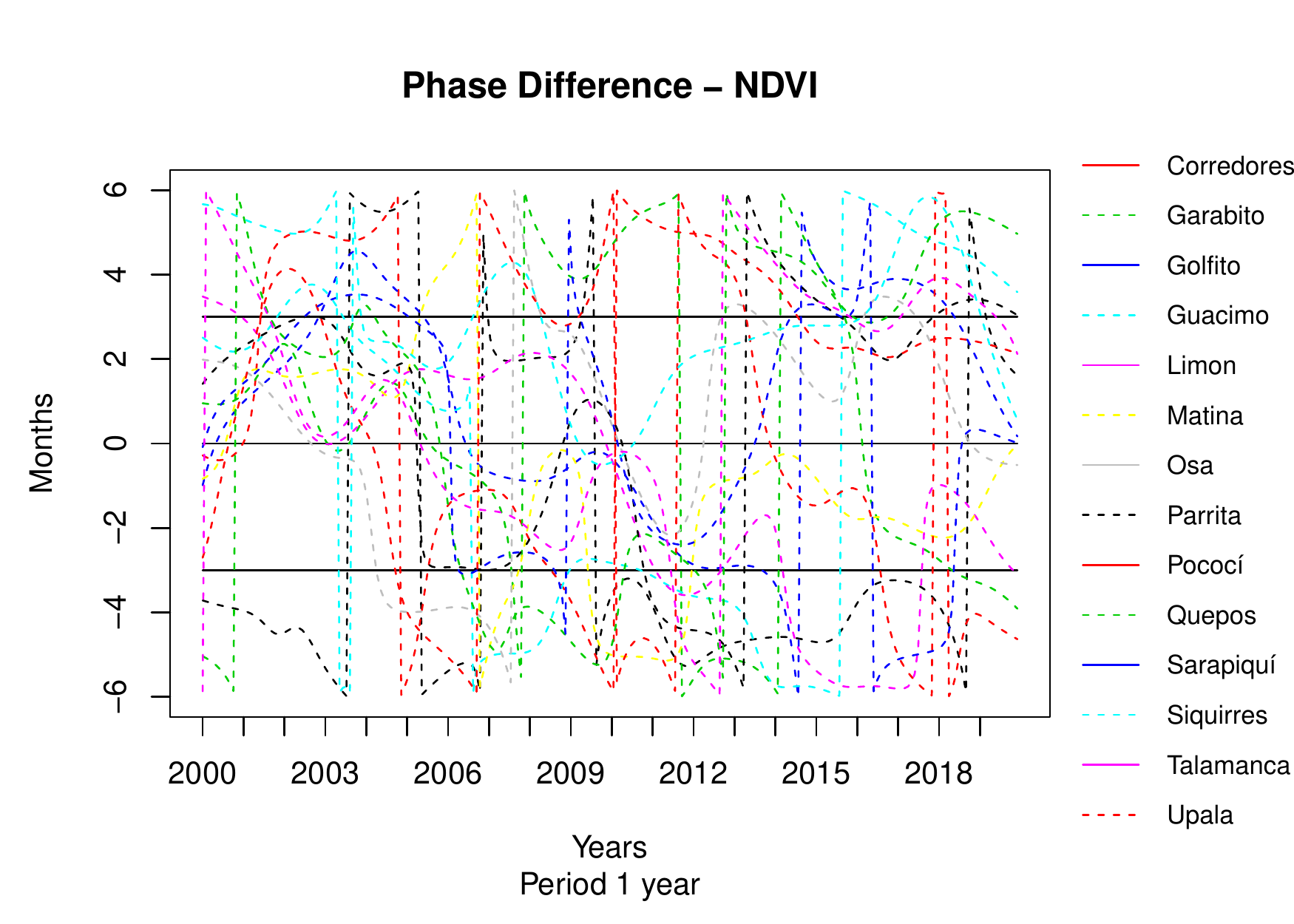}}\vspace{-0.15cm}%
\caption*{\textbf{Figure S12:} Phase difference at period 1-yr. Left panel: cantons in-phase. Right panel: cantos with an irregular behaviour over time.}
\end{figure}

\begin{figure}[H]
\captionsetup[subfigure]{labelformat=empty}
\subfloat[]{\includegraphics[scale=0.5]{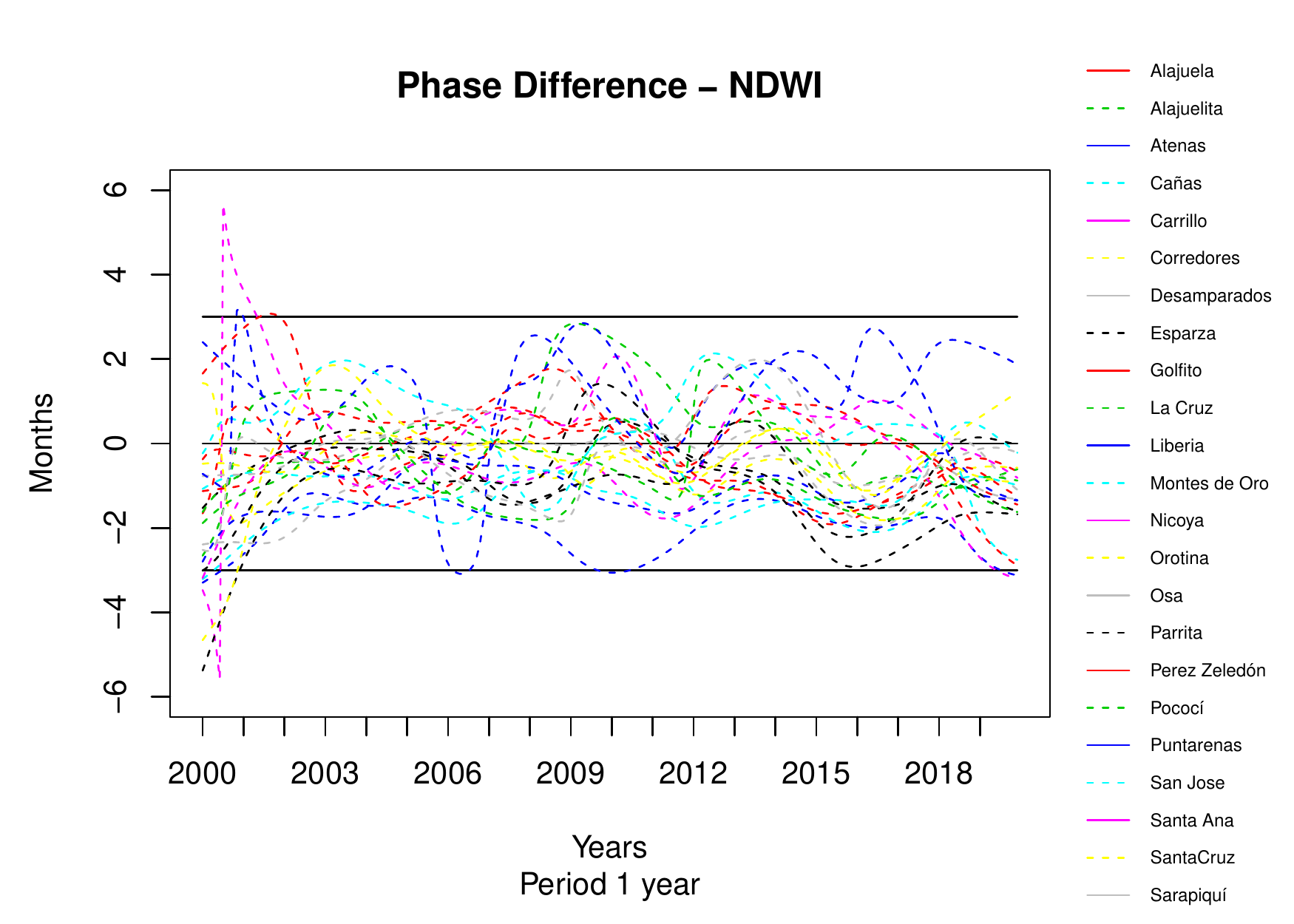}}\vspace{-0.15cm}%
\subfloat[]{\includegraphics[scale=0.5]{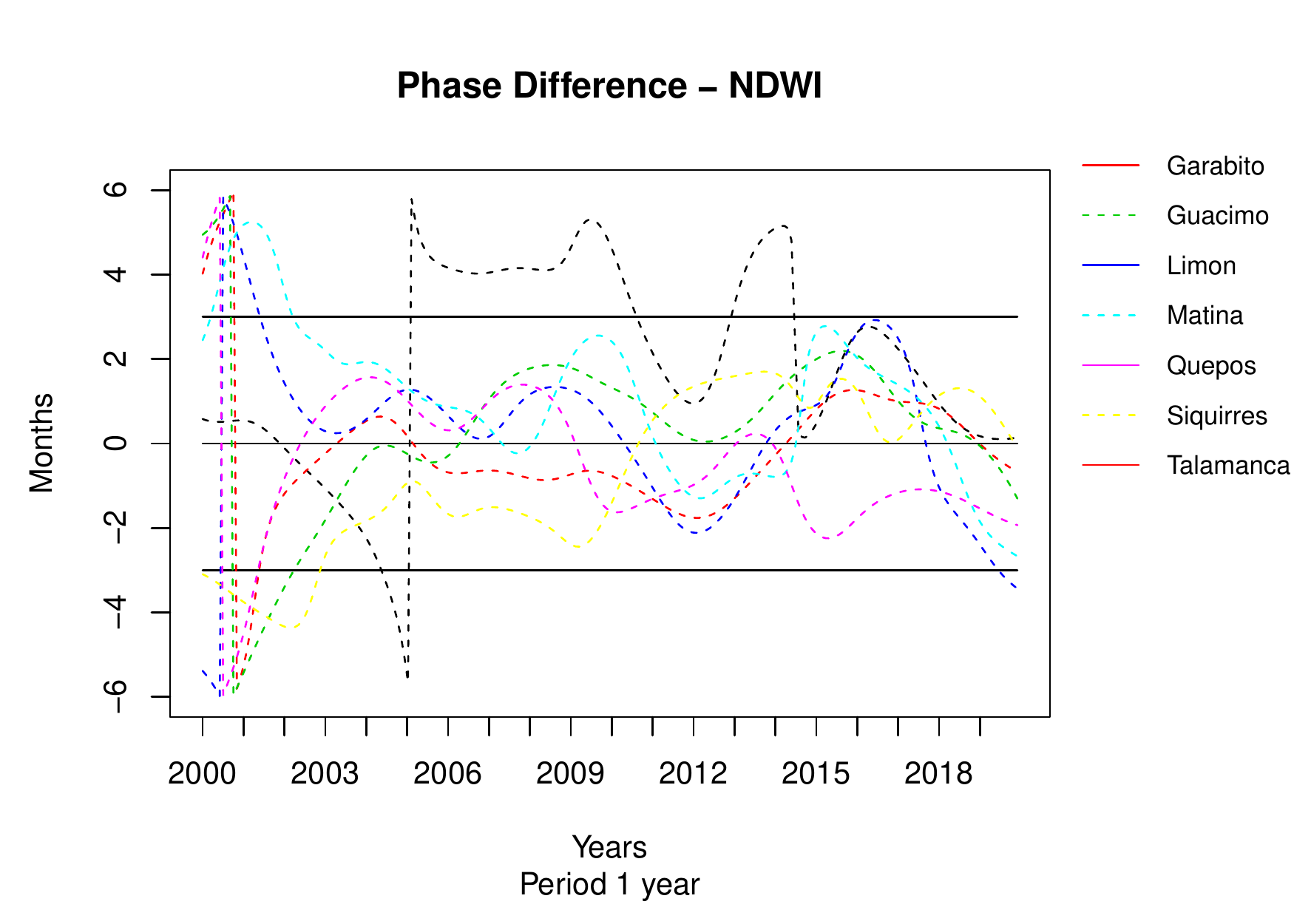}}\vspace{-0.15cm}%
\caption*{\textbf{Figure S13:} Phase difference at period 1-yr. Left panel: cantons in-phase. Right panel: cantos with an irregular behaviour over time.}
\end{figure}

Figure S14 show that the LSD and LSN are out-phase with dengue incidence. Positive values indicate that LSD and LSN are the lagging series with a lag period from 3 to 6 months.

\begin{figure}[H]
\captionsetup[subfigure]{labelformat=empty}
\subfloat[]{\includegraphics[scale=0.5]{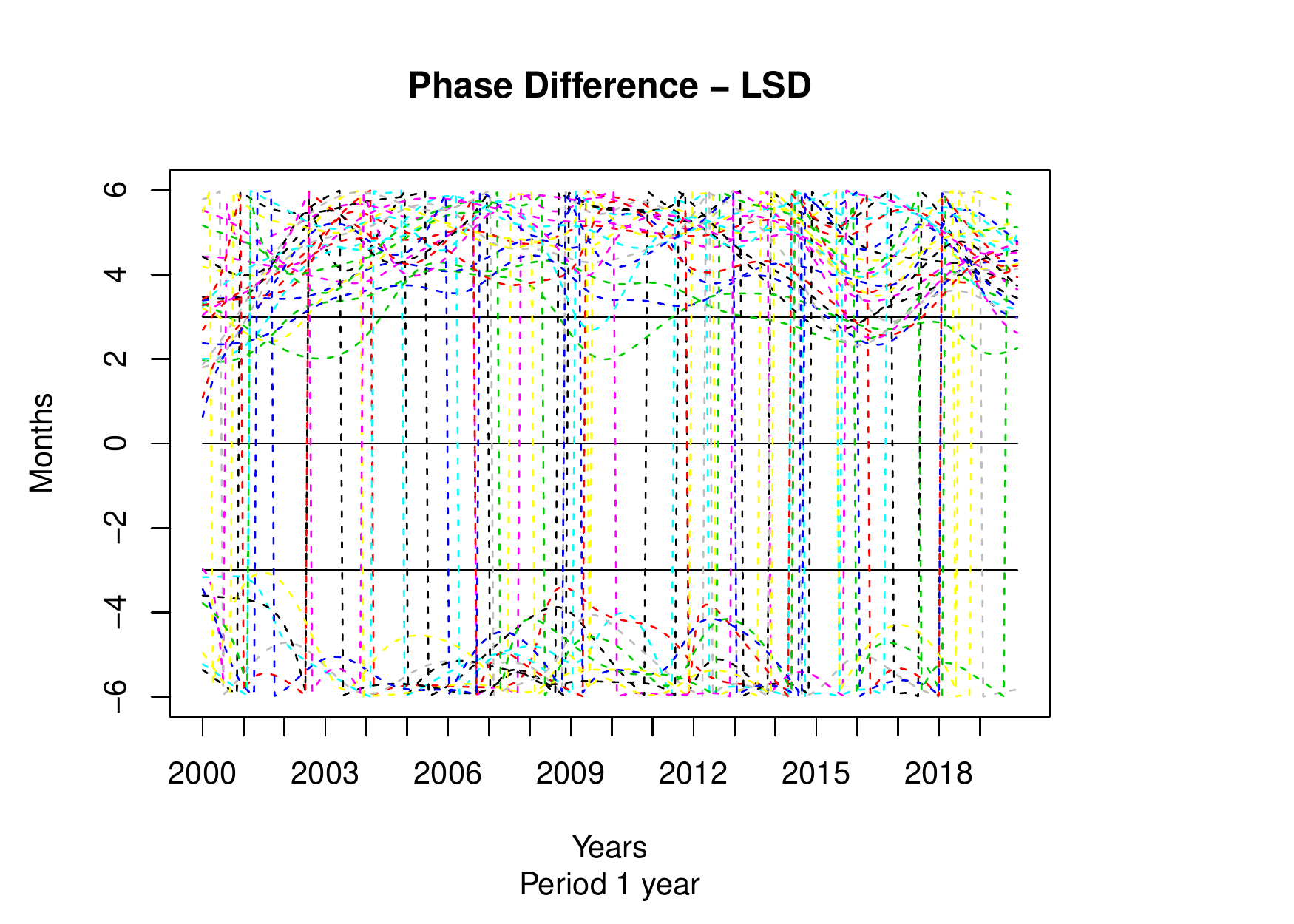}}\vspace{-0.15cm}%
\subfloat[]{\includegraphics[scale=0.5]{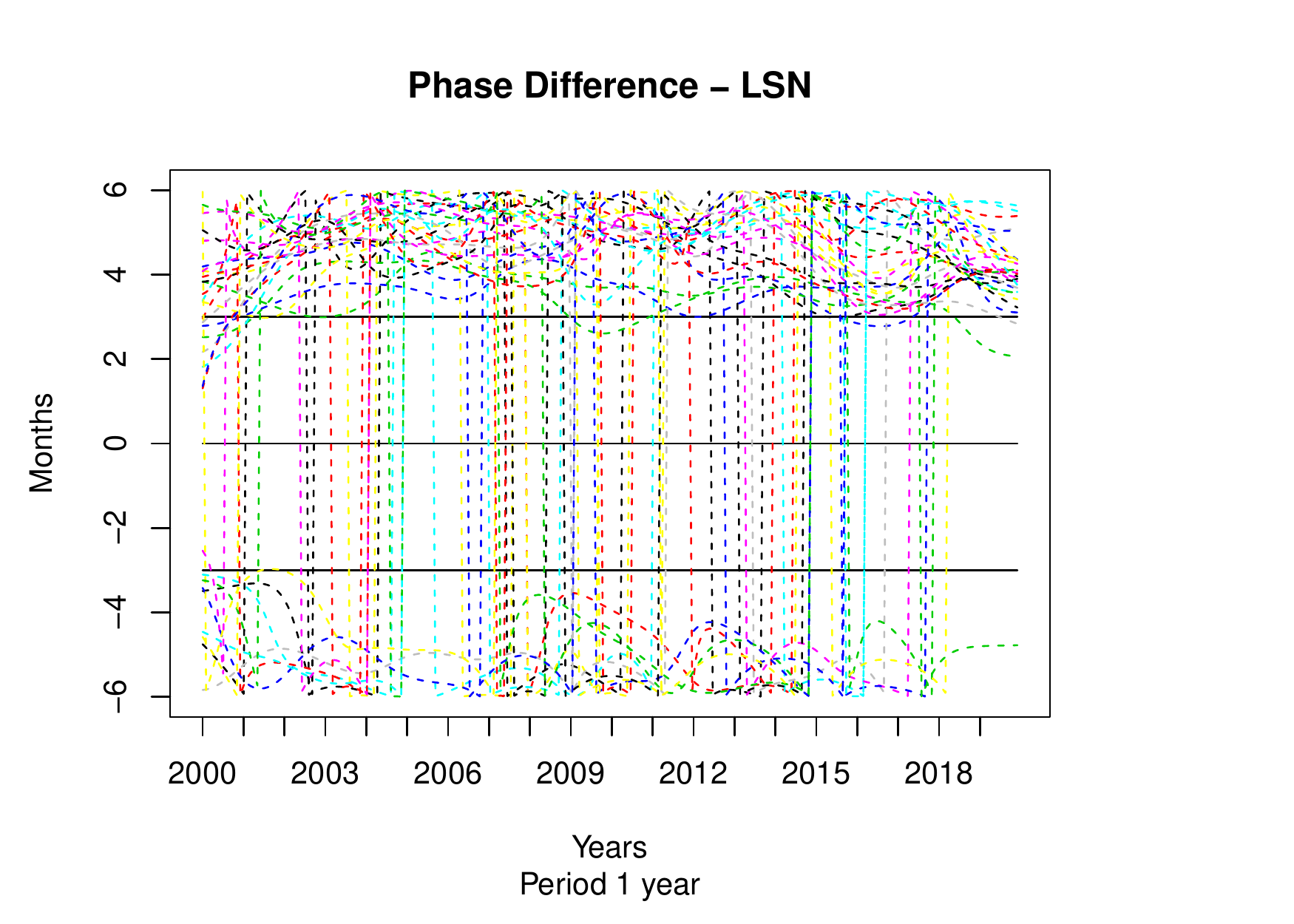}}\vspace{-0.15cm}%
\caption*{\textbf{Figure S14:} Phase difference at period 1-yr. Left panel: cantons in-phase. Right panel: cantos with an irregular behaviour over time.}
\end{figure}

Figures S15 and S16 shows that most of the cantons are in phase with the Ni\~no 3, Ni\~no 3.4, and Ni\~no 4 in the last years.

\begin{figure}[H]
\captionsetup[subfigure]{labelformat=empty}
\subfloat[]{\includegraphics[scale=0.5]{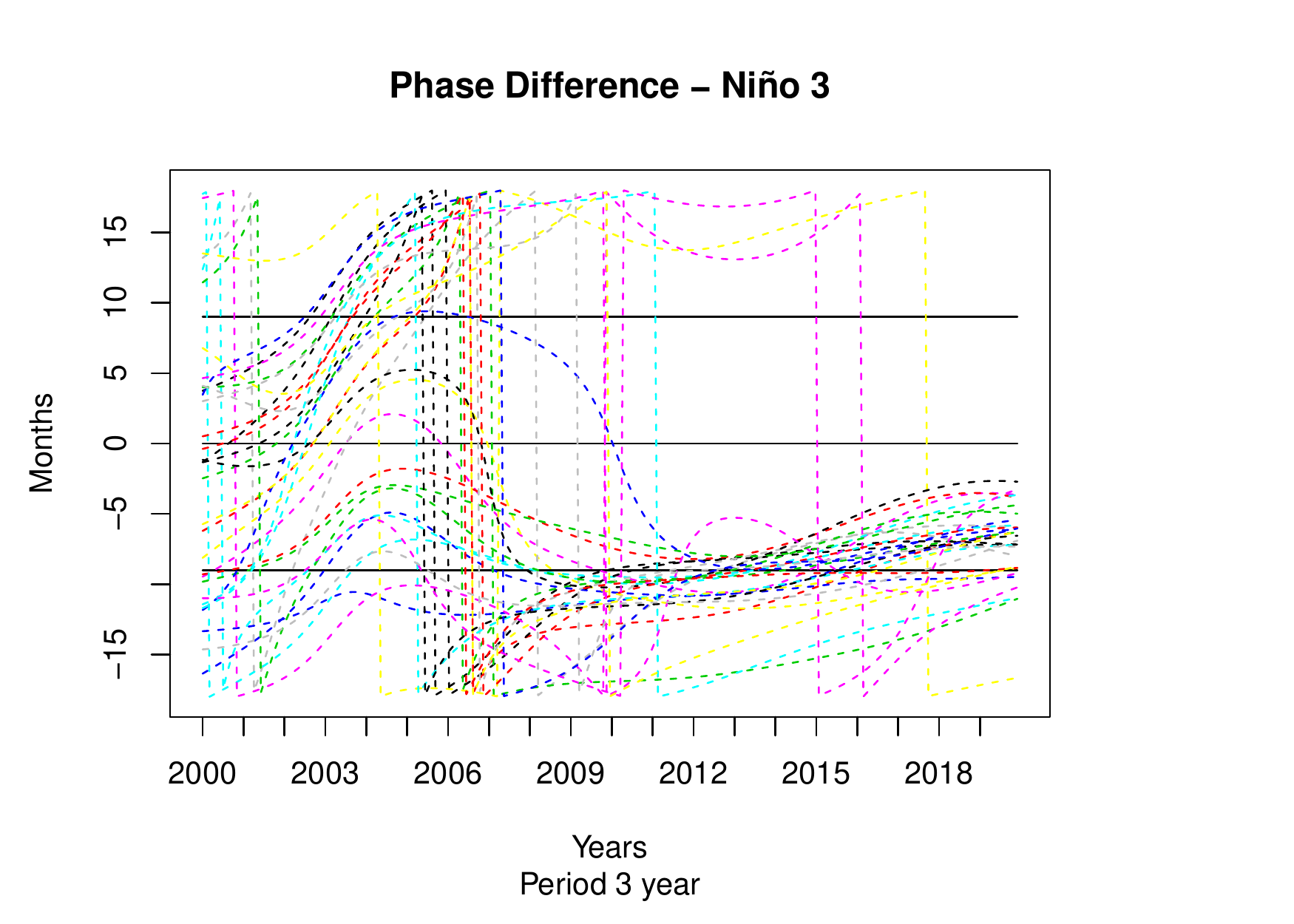}}\vspace{-0.15cm}%
\subfloat[]{\includegraphics[scale=0.5]{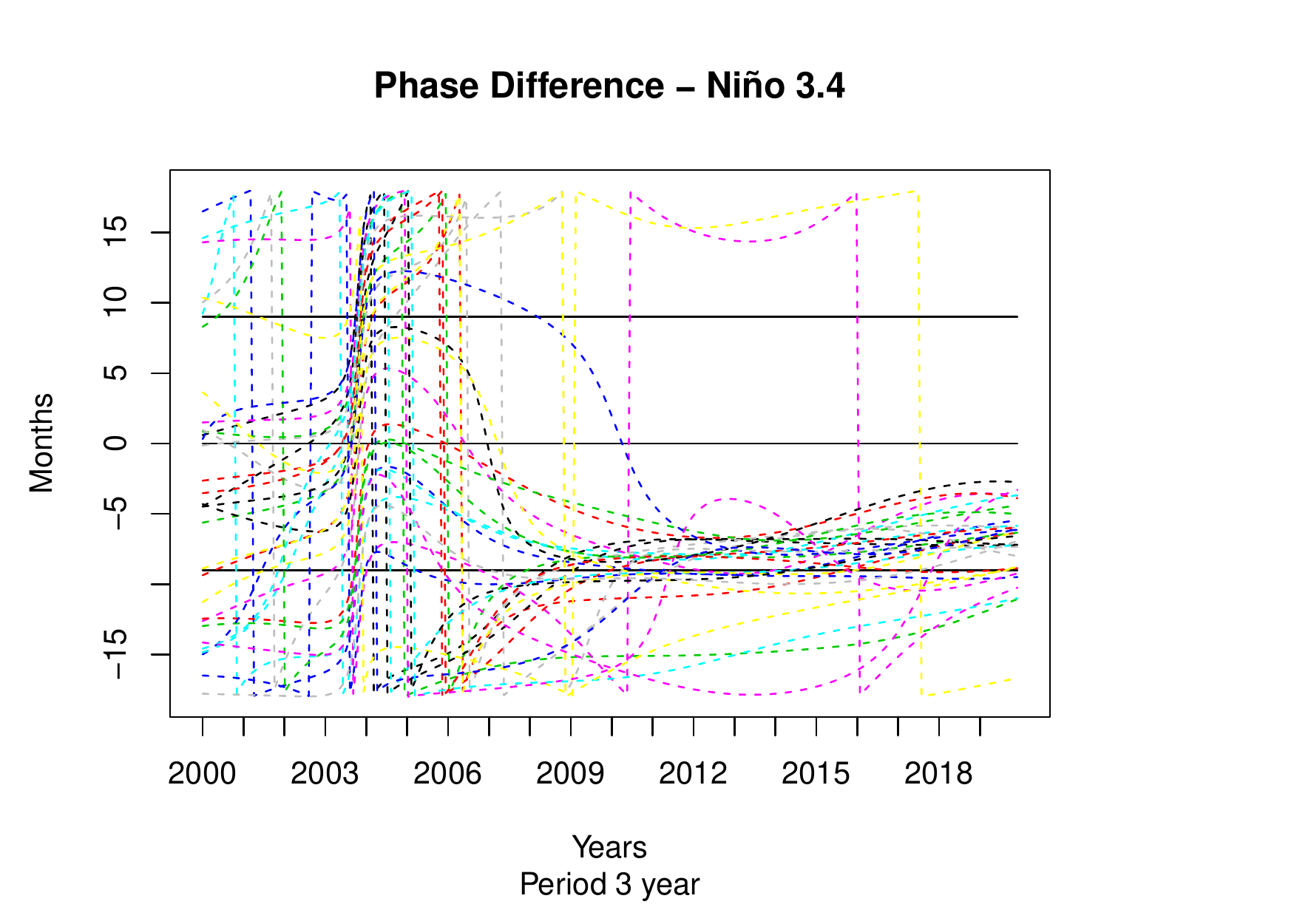}}\vspace{-0.15cm}%
\caption*{\textbf{Figure S15:} Phase difference at period 3-yr. Left panel: cantons in-phase. Right panel: cantos with an irregular behaviour over time.}
\end{figure}

\begin{figure}[H]
\centering
\includegraphics[scale=0.5]{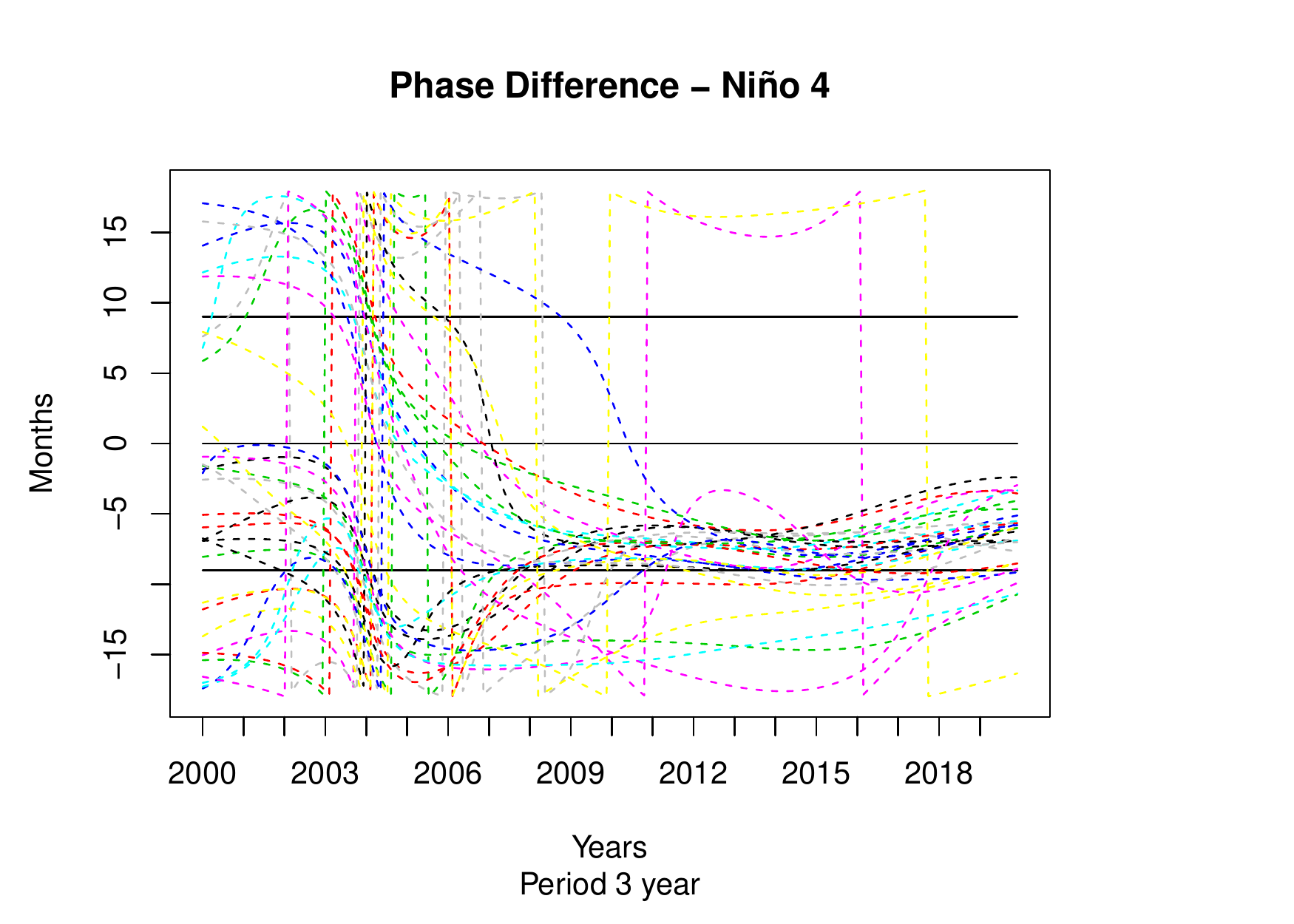}\vspace{-0.15cm}
\caption*{\textbf{Figure S16:} Phase difference at period 3-yr. Left panel: cantons in-phase. Right panel: cantos with an irregular behaviour over time.}
\end{figure}

\begin{table}[H]
\begin{adjustbox}{width=1\textwidth}
\begin{tabular}{llllllllll}
\rowcolor{gray}
\hline
\centering \textbf{Canton} & \centering\textbf{NDVI} & \centering \textbf{NDVI} & \centering \textbf{NDWI}& \textbf{LSD}& \textbf{LSN}&\textbf{TNA} & \textbf{NI\~NO 3}&\centering \textbf{NI\~NO 34}& \textbf{NI\~NO 4}\\
\hline
La Cruz& x&	x&	x& x& x&P1-P1.5-P3&	<P1-P1.5-P3&P1-P1.5-P3&	P3\\
\rowcolor{gray!8}
Esparza& x&	x&	x& x& x&P1-P1.5-P3&	P1-P1.5-P3 &P1-P3 &P1-P3\\
Puntarenas&	x&	x& x& x& x&	P1-P3&	P1&	P1-P3&	P1-P3\\
\rowcolor{gray!8}
Carrillo  &	x&	x& x& x& x&	P1-P3&	P1-P3&	P1-P1.5-P3&	P3\\
canas	  & x&	x& x& x& x&	P1&	P1-P2-P3&	P1-P3&	P1-P3\\
\rowcolor{gray!8}
Atenas	  & x&	x& x& x& x&	P1-P3&	P3&	P1.5-P3&	P1-P3\\
Alajuela  & x&	x& x& x& x&	P1-P3&	P3&	P1-P1.5-P3&	P3\\
\rowcolor{gray!8}
Nicoya	  & x&	x& x& x& x&	P1-P3&	P1-P3&	P3&	P3\\
Orotina   & x&	x& x& x& x&	P1-P3&	P3&	P3&	P3\\
\rowcolor{gray!8}
Liberia	  & x&	x& x& x& x&	P1-P2-P3&P1-P2-P3&P1-P2-P3 &P1-P3\\
Santa Cruz& x&	x& x& x& x&	P1-P3&	P1-P3&	P1-P3&	P1-P3\\
\rowcolor{gray!8}
Santa Ana &	x&	x& x& x& x&	P1-P3&	P1-P3&	P1-P3-P1.5&	P1.5-P3\\
San José& x& x& x&	x&	x& P1-P3& P1-P1.5-P3& P1-P1.5-P3&	P1.5-P3\\
\rowcolor{gray!8}
Alajuelita&x &  & x& x& x& P1-P1.5-P3&	P3&	P1.5-P3& P1.5-P3\\
Matina	  &x &  &  & x& x& &P1-P2&	P1-P2-P3&	P3\\
\rowcolor{gray!8}
Turrialba &x &  & x& x& x&	P1-P3&	P1.5&	P3&	P3\\
Siquirres & x& 2011/2013 &  2003/2008-2010/2014& x& x&	P1-P3&	P1-P1.5-P3&	P1-P2-P3&	P1-P3\\
\rowcolor{gray!8}
Desamparados&x	&2012/2017&	x&	x&	x&	P1-P3& P3& P3&	P1.5-P3\\
Perez Zeledón&x	&2004/2006-2013/2018& x& 2004/2006-2013/2018& x&	P1-P3&	P1-P3&	P1.5-P3& P1-P3\\
\rowcolor{gray!8}
Pococí&	x& 2001/2003-2011/2014&	x&	x&	x&P1& P1&P1-P2& P1-P3\\
Montes de Oro &x& 2000/2003-2013/2017& x& x& x&	P1-P3& P1-P3&	P1-P3&	P1-P3\\
\rowcolor{gray!8}
Corredores &2005/2007 - 2013/2017&	2013/2015&	2005/2008-2015/2017&	2004/2008-2013/2017&	2004/2008-2013/2017&	P1&	P1.5-P3&	P3&	P3\\
Limón &	2010/2013  &      & 2007/2013&	x &	x &P1-P1.5 & P1-P1.5 &	P1-P1.5	& P1-P1.5-P3\\
\rowcolor{gray!8}
Talamanca&	2011/2014&	2011/2014&  &2010/2015& 2010/2015& P1&	P1-P2 &	P1-P1.5-P3 & P1-P1.5-P3\\
Sarapiquí&	2011/2014&	2012/2015&	2011/2015&	x&	x&	P1-P3&	P1-P2&	P1-P3&	P1-P3\\
\rowcolor{gray!8}
Guacimo	&2010/2015&	2002-2011/2014&	2010/2014&	2010/2015&	x&	P1&	P1&	P1-P1.5-P3&	P1-P1.5-P3\\
Osa	&2012/2017 & &2013/2016 & 2008/2011 - 2013/2015&	2008/2011-2013/2018&P1-P3 &	P3	&P1.5-P3 &	P3\\
\rowcolor{gray!8}
Quepos& &  & &2004/2007 - 2009/2011 &x & &P1-P3 &P1.5-P3&	P3\\
Parrita	& & &2013/2014	&2004/2007 - 2013/2015&	x &P1 &	P1.5-P3 &P1.5-P3 &P1.5-P3\\
\rowcolor{gray!8}
Upala & &2013 &	2012/2015 &	2012/2016 & 2012/2016 &P1-P3 &P3&	P1.5-P3	&P1.4-P3\\
Golfito	& & & 2013/2017	&2012/2017 &2012/2017 &P1-P3& P3& P3&P3\\
\rowcolor{gray!8}
Garabito& &  & &2003/2007 &x	&P1.5-P2-P3	&P1-P2-P3&	P1.5-P3&	P1.5-P3\\
\hline
\end{tabular}
\end{adjustbox}
\caption{Summary results: (x) represent those places where there is a significant correlation between the incidence of dengue and the corresponding index. When the correlation time interval is short, the dates for the area of high significance are indicated. P1 represents places where there is a notable correlation in the 1-yr band for the incidence of dengue and TNA after 2011. P2 and P3 corresponds to those places where there is a strong correlation between the ENSO variables and the incidence of dengue in a band of 2-yr and 3-yr, respectively.}
\end{table}

\end{document}